# CEPC Reference Detector

## Technical Design Report

**Version:** v1.0.0 build: 2025-10-08 00:09:14Z

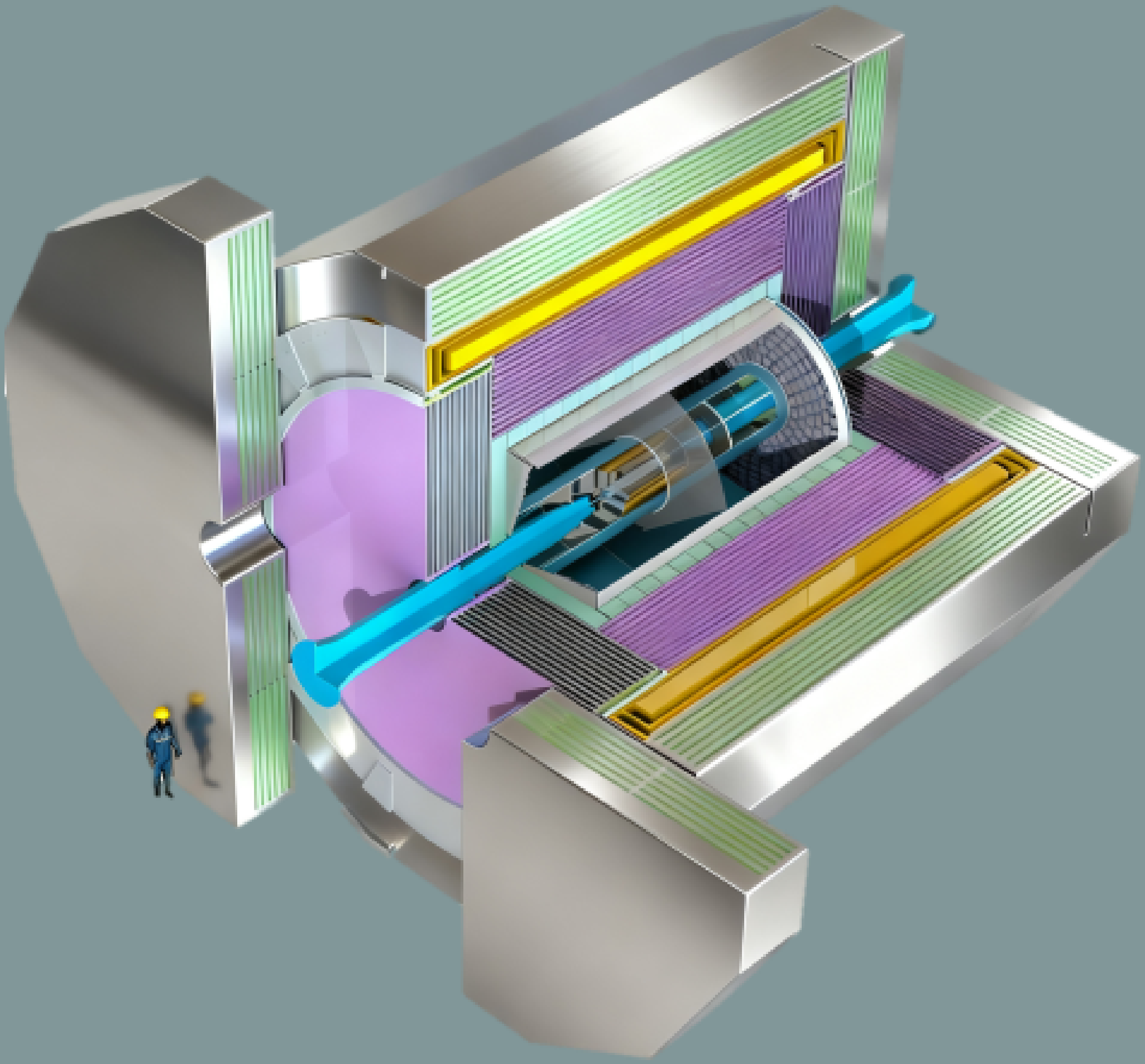

*The CEPC Study Group*

*October, 2025*

# CEPC Technical Design Report

## Reference Detector

Souvik Priyam Adhya[132], Konstantin Afanaciev[135], Shakeel Ahmad[200], Ijaz Ahmed[65], Xiaocong Ai[351], Mohammad Alhakami[43], Simone Alioli[109, 277], Azzah Alshehri[56, 62], Numa Althubiti[45], Atif Rahman Alvi[290], Haipeng An[268], Fenfen An[145], Liupan An[202], Attilio Andreazza[107, 273], Giuseppe Andronico[103], Yasuo Arai[87], Andrej Arbuzov[158], Nima Arkani-Hamed[119], Akram Artikov[158, 215], Burin Asavapibhop[244], Saiyad Ashanujjaman[124], Nikolay Atanov[158], Tagir Aushev[189], Murad Badshah[1], Naveen Kumar Baghel[306], Yang Bai[330, 9], Sha Bai[126], Yu Bai[245], Vipul Bairathi[138], Olga Bakina[158], Csaba Balazs[182], Philip Bambade[93], Yong Ban[202], Triparno Bandyopadhyay[44], Sudeshna Banerjee[255], Swagato Banerjee[306], Chentao Bao[350], Qinyu Bao[40], Vladimir Baranov[158], Joao Barreiro Guimaraes da Costa[126], Varvara Batozskaya[126], Bilguun Bayarsaikhan[128, 292], Vadim Bayev[135], Khaled Belkadhi[179], Khaled Belkadhi[243], Alain Bellerive[19], Alexander Belyaev[325, 251], Driss Benchekroun[80], Jorge Berenguer Antequera[294], Srimoy Bhattacharya[326], Nazima Bi[126, 292], Qi Bi[10], Shalva Bilanishvili[163], Jyoti Prakash Biswal[251], Mustapha Biyabi[126, 292], Simon Blyth[126], Denis Bodrov[239, 189], Anton Bogomyagkov[17], Abhishek Bohare[295], Aleksandr Boikov[158], Stewart Boogert[307, 37], Maarten Boonekamp[38], Eduard Boos[238], Marcello Borri[269], Vincent Boudry[166], Mohammed Boukidi[130], Igor Boyko[158], Ivanka Bozovic[335, 284], Giuseppe Bozzi[102, 288], Jean-Claude Brient[166], Dzmitry Budkouski[158], Anastasiia Budzinskaya[123], Masroor Bukhari[152, 333], Viacheslav Bunichev[238], Dario Buttazzo[111], Vladimir Bytev[158], Giacomo Cacciapaglia[173, 240], Mengke Cai[126], Xiao Cai[126], Yizhou Cai[185], Yuman Cai[126], Yanbing Cai[73], Yuchen Cai[101], Huacheng Cai[24], Lorenzo Calibbi[187], Junsong Cang[264], Qing-Hong Cao[202], Guofu Cao[126], Ning Cao[126], Tianyu Cao[126], Raimon Casanova[118, 42], Cesar Ceballos Sanchez[158], Serkant Ali Cetin[148], Vongani Chabalala[326], Marina Chadeeva[199], Yue Chang[20], Zheng-Ze Chang[126], Wei Chao[13], Auttakit Chatrabhuti[33], Yuzhi Che[126], Yimin Che[185], Guorong Che[187], Ye Chen[126], Boping Chen[126], Gang Chen[194], Hua-Xing Chen[245], Ying Chen[126], Gang Chen[126], Hai Chen[350], Ji-Yuan Chen[233], Xuan Chen[221], Long Chen[221], Xiang Chen[233], Ning Chen[187], Wei Chen[222], Jing Chen[321], Shiqiang Chen[185], Ennan Chen[187], Kai Chen[20], Mingshui Chen[126], Shanzhen Chen[126], Mali Chen[126], Jiaolong Chen[126], Shaomin Chen[268], Bin Chen[126], Tong Chen[126], Yebo Chen[126], Jiahua Chen[202], Kailu Chen[126], Xin Chen[268], Jinhui Chen[66], Huacong Chen[71], Shan Cheng[92], Huajie Cheng[192], Tongguang Cheng[10], Kun Cheng[317], Huitong Cheng[85], Sibo Cheng[202], Hridey Chetri[97], Pietro Chimenti[272], Mokhtar Chmeissani[118], Min-Huan Chu[4], Ming-Chung Chu[258], Xiaotong Chu[126], Gianluigi Cibinetto[104], Rafael Coelho Lopes De Sa[309], Guglielmo Coloretti[332], Patricia Conde Muíño[142, 167], Fernando Cornet[294], Andreas Crivellin[332], Yuxin Cui[126], Tao Cui[126], Han Cui[71], Hanhua Cui[126], Xiangyi Cui[233], Manuel Da Rocha Rolo[115],

Alina Dadashova[158], Hongliang Dai[126], Lingyun Dai[92], Zilin Dai[202], Sridhara Dasu[330], Yuri Davydov[158], Nicola De Filippis[100, 204], Francesca De Mori[115, 278], Dmitrii Dementev[158], Mehmet Demirci[49], Binwei Deng[90], Yunqi Deng[126], Siqi Deng[126], Jianpeng Deng[350], Zhi Deng[268], Ziyan Deng[126], Ansh Desai[314, 121], Biagio Di Micco[114, 275], Ran Ding[7], Houqian Ding[185], Xuefeng Ding[126], Marisilvia Donadelli[271], Sheng Dong[126], Jianing Dong[210], Xu Dong[57], Zhenyu Dong[202], Mingyi Dong[126], Chao Dong[126], Yabo Dong[85], Milos Dordevic[335, 284], Eugene Doroshkevich[123], Marcos Dracos[146], Marco Drewes[274, 257], Alexey Drutskoy[199], Xiaokang Du[134, 84], Yong Du[128], Dongshuo Du[128], Dejing Du[126], Baoning Du[191], Mengchuan Du[126], Yanyan Du[344], Zhe Duan[126], Chenfeng Duan[126], Zonghuan Duan[185], Lev Dudko[238], Sajan Easo[251], Ulrik Egede[182], Malika El Aslani[80], Jaouad El Falaki[174], Faizan Elahi[126, 52], Walaa Elmetenawee[100], Iouri Ershov[158], Ghizlane Ez-Zobayr[181], Zhouyue Fan[126], Yunyun Fan[126], Lei Fan[126], Jun Fan[85], Shuangshi Fang[126], Wenxing Fang[126], Yaquan Fang[† 126], George Fanourakis[129], Muhammad Farooq[312], Almaz Fazliakhmetov[123], Marco Fedele[303, 176], Jiale Fei[339], Mingjie Feng[126, 292], Xu Feng[202], Tai-Fu Feng[47], Feng Feng[31], Chong Feng[40], Yanan Feng[241], Siya Feng[126], Yue Feng[145], Bo Feng[136], Laurent Forthomme[6], Andrew Fowlie[341], Harald Fox[168], Jinyu Fu[126], Hai-Bing Fu[74], Chengdong Fu[126], Yisheng Fu[126], Dawei Fu[202], Benjamin Fuks[173, 240], Emidio Gabrielli[54], Frank Gaede[57], Nan Gan[126], Li Gang[7], Jie Gao[126], Yu Gao[126], Yanyan Gao[295], Yuanning Gao[202], Lin-Qing Gao[25], Zhanxiang Gao[185], Hong Gao[10], Jun Gao[233], Christina Gao[246], Ze Gao[126], Zhixing Gao[233], Yunfei Gao[126], Zihan Gao[202], Tianqi Gao[260], Dmitrii Gavrilov[189], Shao-Feng Ge[233], Kun Ge[11], Jieru Geng[126], Diwash Ghimire[233], Swagata Ghosh[98], Eduard Ginya[158], Francesco Giuli[112, 319], Leonid Gladilin[238], Mpho Gololo[303], Aleksandr Golunov[158], Ricardo Goncalo[293, 167], Jiading Gong[154], Wenxuan Gong[126], Ti Gong[85], Shaokun Gong[202], Xue Gong[266], Quanbu Gou[126], Boxing Gou[128], Francesco Grancagnolo[106], Gérald Grenier[280], Ambra Gresele[347], Sebastian Grinstein[118, 117], Zhihang Gu[185], Jiayin Gu[66], Sanele Gumede[326, 149], Fangyi Guo[126, 27], Jie Guo[35, 175], Jun Guo[233], Jialin Guo[126], Yu-Chen Guo[170], Hong Guo[292], Lei Guo[225], Qianying Guo[202], Yuping Guo[66], Alexey Guskov[158], Alejandro Gutierrez-Rodriguez[270], Noman Habib[126], Fazlollah Hajkarim[313], Rabia Hameed[126, 77], Jifeng Han[237], Chengcheng Han[222], Shuo Han[126], Huayong Han[73], Qundong Han[316], Tao Han[317], Tingting Han[228], Zhi-Long Han[302], Xue-Ying Han[126], Mahmoud Hanafy[15, 62], Xiqing Hao[84], Jiankui Hao[202], Changhua Hao[185], Kirill Haritonov[238], Miao He[126], Hong-Jian He[233], Zhenqiang He[126], Jibo He[292], Bingkun He[126], Sven Heinemeyer[140], Yuekun Heng[126, 292], Maria A. Hernandez-Ruiz[270], Xiantao Hou[126], Suen Hou[3], George Wei-Shu Hou[190], Zhilong Hou[126], Shaojing Hou[126], Baorui Hou[126], Xiaonan Hou[352], Yu Hu[126], Kun Hu[210], Tao Hu[126], Chen Hu[231], Dongdong Hu[321], Zhen Hu[268], Peng Hu[29], Jun Hu[126], Fan Hu[126], Xiao-Hui Hu[31], Fei Huang[171, 126], Xu-Guang Huang[66], Xingtao Huang[210], Fei Huang[302], Fa Peng Huang[180], Yongsheng Huang[222], Guangshun Huang[321], Suyun Huang[126], Tian Huang[197],

Xuelei Huang[197], Zixun Huang[66], Kairui Huang[66], Zeyu Huang[66], Yanping Huang[126], He Huang[40], Xinhui Huang[126], Wu Huimin[197], Nazim Huseynov[158], Fabio Iemmi[126], Pelevanyuk Igor[158], Ara Ioannisian[125], Munawar Iqbal[126, 22], Dubovskaya Irina[14], Ammad Ul Islam[316, 114], Wasikul Islam[330], Paul Jackson[5], Khawla Jaffel[188, 24], Tahir Javaid[10, 126], Daniel Jeans[87], Xiaolu Ji[126], Wang Ji[126], Quan Ji[126], Qingping Ji[84], Xiaobin Ji[126], Zekun Jia[172, 240], Huayu Jia[254], Yu Jia[126], Wentao Jia[126], Xuewei Jia[178], Junji Jia[339], Sen Jia[245], Gong Jiabao[126], Jiang Jianfeng[126, 292], Jun Jiang[221], Cheng Jiang[295], Houbing Jiang[339], Xiaowei Jiang[126], Ruobing Jiang[202], Xiaojie Jiang[126], Junjie Jiang[233], Xuhui Jiang[126], Yuanhong Jiao[197], Jianbin Jiao[210], Shen Jibei[185], Wenyi Jin[126], Dapeng Jin[126], Mingjie Jin[156], Yi Jin[302], Shan Jin[185], Liangchenglong Jin[126], Liyan Jin[210], Shengjie Jin[126], Shuzhu Jin[126], Yujing Jin[210], Xiaoping Jing[126], Qin Junkai[268], Vuyolwethu Happyboy Kakancu[326], Giorgi Kakhniashvili[88], Lidia Kalinovskaya[158], Jochen Kaminski[286], Aleksei Kampf[158], Shinya Kanemura[262], Wen Kang[126], Xian-Wei Kang[13], Zizi Kang[187], Daekyoung Kang[66], Vladimir Karjavine[158], Biswajit Karmakar[324], Vakhtang Kartvelishvili[88], Yukihiro Kato[160], Chaoyi Ke[77, 126], Sameen Ahmed Khan[58], Hamzeh Khanpour[6, 53, 219], Alexey Kharlamov[17], Tatyana Kharlamova[17], Wolfgang Kilian[323], Viktoriia Kiseeva[158], Michael Klasen[311], Byeong Rok Ko[162], Sanghyun Ko[24], Olga Kodolova[158], Onur Bugra Kolcu[148], Kyoungchul Kong[304], Ujwal Konissery Santhosh[69], Ivan Koop[17], Sergey Korpachev[199], Iurii Korsakov[158], Zixun Kou[202], Dmitrii Kozlov[158], Mohamed Krab[190], Alexander Krasnov[17], Yuri Kulchitsky[158, 133], Mukesh Kumar[326], Andrey Kupich[17], Youngjoon Kwon[346], François Sylvain René Lagarde[321], Pei-Zhu Lai[126, 292], Haojing Lai[233], Imad Laktineh[280], Alexander Lanyov[158], Lia Lavezzi[115], Catarina Leal[205, 177], Kyeongpil Lee[34], Sehwook Lee[164], Sheldon Lee Glashow[16], Ge Lei[126], Marco Leite[320], Nuno Leonardo[142], Boris Levchenko[238], Eugene Levichev[17], Qiang Li[126, 30], Jiarong Li[268, 126], Tianyou Li[187, 126], Yuji Li[66, 126], Hai-Jun Li[137, 143], Hengne Li[241], Dikai Li[236], Fei Li[126], Tong Li[187], Honglei Li[302], Yiming Li[126], Teng Li[210], Leyan Li[202], Minqia Li[126], Ya Li[184], Haifeng Li[210], Yanfeng Li[126], Yufeng Li[126], Zhan Li[126], Haopeng Li[187], Gang Li[207], Hai Tao Li[221], Chun-Yuan Li[229], Gang Li[126], Zhihao Li[126], Leyi Li[210], Cheng Li[321], Huijing Li[84], Pengxu Li[350], Xiaoting Li[126], Fei Li[126], Yujie Li[126], Minxian Li[126], Yuan-Zhen Li[274], Mengzhao Li[126], Bingzhi Li[349], Long Li[285], Shi-Yuan Li[221], Hai-Bo Li[126], Mei Li[126], Xin-Qiang Li[20], Shu Li[233], Chunhua Li[184], Shuo Li[185], Ke Li[126], Qiming Li[126], Huaishen Li[126], Mujin Li[126], Lei Li[209], Chunkai Li[187], Ying-Ying Li[126], Yulong Li[126], Mengran Li[126], Changqiao Li[178], Weidong Li[126], Haibo Li[126], Pei-Rong Li[169], Qinze Li[126], Zhao Li[126], Congqiao Li[202], Canming Li[126], Peiran Li[310], Bing Li[210], Yaxuan Li[187], Liang Li[233], Lankun Li[126], Tiange Li[92], Wenlong Li[66], Jialin Li[233], Tianqi Li[242], Xiaoyu Li[233], Xiang Li[126], Hui Li[187], Qiang Li[197], Yuhong Li[187], Yubo Li[342], Xinbai Li[321], Yubo Li[342], Ruhui Li[126], Shuqi Li[126], Bo Li[344], Meijian Li[141], Xiaomei Li[28], Zhijun Liang[126], Jinhan Liang[241], Xiao Liang[229], Jian Liang[241], Zhengchen Liang[268], Chengxin Liao[126], Jiajun Liao[222], Guangrui Liao[71], Yi Liao[241],

Yipu Liao[126], Libo Liao[225], Hongbo Liao[126], Ayut Limphirat[253], Tao Lin[126], Yuming Lin[202], Yugen Lin[137], Weiping Lin[237], Jacob Linacre[251], Jenny List[57], Chen Litong[233], Yong Liu[217, 144], Chao Liu[197, 268], Guoming Liu[241], Wei Liu[186], Yang Liu[225], Fu-Hu Liu[235], Tao Liu[126], Xiang Liu[169], Zhiqing Liu[210], Xin Liu[153], Yao-Bei Liu[83], Jia Liu[202], Xiao-Hai Liu[265], Hang Liu[234], Zhenchao Liu[342], Yi Liu[351], Yang Liu[343], Zhaofeng Liu[126], Bo Liu[187], Tao Liu[259], Manwen Liu[127], Kun Liu[233], Liqiang Liu[256], Dong Liu[126], Yanlin Liu[210], Mingyi Liu[321], Jianbei Liu[321], Yan-Rui Liu[221], Beijiang Liu[126], Xingquan Liu[237], Geliang Liu[126], Rong Liu[13], Zhen Liu[113], Xinglin Liu[66], Zhen Liu[310], Tong Liu[126], Ziying Liu[187], Dong Liu[210], Xinyan Liu[210], Yiming Liu[295], Shulin Liu[126], Yonglu Liu[191], Yong Liu[126], Difei Liu[350], Shuxian Liu[185], Canwen Liu[268], Qibin Liu[233], Zhongxiu Liu[126], Hongbang Liu[72], Baiqi Liu[126], Shidong Liu[207], Kang Liu[233], Tianfeng Liu[126], Yichen Liu[232], Danning Liu[233], Jian Liu[305], Zhi-Feng Liu[350], Alexander Lobko[122], Bingzhi Long[241], Xinchou Lou[126], Miaoran Lu[281, 93], Hancen Lu[126], Weiguo Lu[126], Peng-Cheng Lu[221], Yunpeng Lu[126], Zhijun Lu[126], Yu Lu[21], Xianguo Lu[263], Cai-Dian Lu[126], Meng Lu[195], Bayarto Lubsandorzhiev[123], Sultim Lubsandorzhiev[123], Marco Toliman Lucchini[109, 277], Fabio Lucio Alves[185], José Luis Ávila[294], Arslan Lukanov[123], Shoudong Luo[350], Guang Luo[225], Xiaolan Luo[126], Jiashun Luo[126], Wuming Luo[126], Min Luo[78], Feng Lyu[126], Xiao-Rui Lyu[292], Jiaxin Lyu[126], Kunfeng Lyu[313], Vladimir Lyubushkin[158], Tatiana Lyubushkina[158], Bo-Qiang Ma[351, 202], Yang Ma[274], Junli Ma[126], Hong-Hao Ma[71], Xiaoyan Ma[126], Si Ma[126], Yanhui Ma[221], Zhongjian Ma[126], Chenglong Ma[126], Hailong Ma[126], Lianliang Ma[210], Xiaotian Ma[126], Xiao-Dong Ma[241], Yongsheng Ma[126], Wenbo Ma[315], Marco Maggiora[115, 278], Siddharth Prasad Maharathy[326, 95], Farvah Mahmoudi[280], Vladimir Makarenko[122], Kutlwano Makgetha[326], Vladimir Malyshev[158], Lining Mao[233], Carlos Marinas[139], Samuele Mariotto[108, 273], Phodiso Maroeshe[326], Verena Ingrid Martinez Outschoorn[309], Diego Martinez Santos[23], Haidar Mas'ud Alfanda[233], Shigeki Matsumoto[159], Kentarou Mawatari[150], Rachid Mazini[326], Njokweni Mbuyiswa[326], David Mchedlishvili[88], Krzysztof Mekala[328, 57], Mohamed Reda Mekouar[126, 292], Dmitri Melikhov[238], Bruce Mellado[126, 326], Lingxin Meng[168], Weiqi Meng[187], Cai Meng[126], Yu Meng[351], Ji Mengyang[126], Mikhail Merkin[238], Dexing Miao[126, 292], Mauro Migliorati[318], Catia Milardi[99], Jovan Milosevic[335], Dayanand Mishra[280], Irma Mklavashvili[88], Cen Mo[233], Gagan Mohanty[255], Arthur Moraes[337], Karabo Mosala[326], Viktoriia Moskalenko[158], Chuene Mosomane[149], Zhihui Mu[126], Zeqing Mu[202], Pankaj Munbodh[289, 216], Brenton Munhungewarwa[303], Yuri Murin[158], William Murray[263, 251], Dmitrii Nanzanov[123], Shinya Narita[150], Baballo-Victor Ndhlovu[326], Siavash Neshatpour[280], Matthias Neubert[176, 157], Phumlani Zipho Ngcobo[331], Donald Ngobeni[326], Alexander Nikitenko[158], Feipeng Ning[126], Yujie Niu[126, 292], Juan-Juan Niu[71], Yan Niu[210], Edward Nkadimeng[149, 250], Takaaki Nomura[237], Kgothatso Ntumbe[149, 326], Hideki Okawa[126], Emmanuel Olaiya[251], Danila Oleynik[158], Sebastian Olivares[63], Mohamed Ouchemhou[211], Yassine Oulad Aiad[80], Qun Ouyang[126], Carlo Pagani[107], Stathes Paganis[190], Pavel Pakhlov[189], Wenyu Pan[126],

Song Pan[233], Yiyang Pan[126], Papia Panda[300], Saraswati Pandey[55], Hyangkyu Park[162], Roman Pasechnik[50], Tanguy Pasquier[280], Hua Pei[20], Junle Pei[134], Yatian Pei[126], Zhaoyuan Peng[126], Michael Peskin[248], Nikolai Peters[189], Artem Petrosyan[158], Lorenzo Pezzotti[101], Lungisani Phakathi[326], Mohan Pi[187], Thabo Pilusa[326], Ronggang Ping[126], Jialun Ping[184], Krzysztof Piotrzkowski[6], Marco Poli Lener[99], Karolos Potamianos[263], Vindhyawasini Prasad[154], Alan Price[151], Francesco Massimiliano Procacci[100, 204], Andrei Prokhorov[158], Baohua Qi[126], Huirong Qi[126], Binbin Qi[321], Ming Qi[185], Fazhi Qi[126], Xiaohui Qian[126], Sen Qian[126], Qin Qin[89], Zhonghua Qin[126], Xiaoping Qin[170], Tong Qiu[295], Guofeng Qu[237], Muhammad Ramzan[22], Phuti Rapheeha[326], Shabbar Raza[65], Matthew Reece[79], Vladimir Rekovic[335, 284], Jing Ren[77], Jiayi Ren[126], Zan Ren[292], Juergen Reuter[57], Barbara Ricci[297, 104], Jack Rolph[126], Nikolaos Rompotis[305], Subhojit Roy[9], Leonid Rumyantsev[158], Rahal Saaidi[282], Renat Sadykov[158], Romualdo Santoro[107, 276], Juan José Sanz-Cillero[39, 131], Miroslav Saur[169], Mariia Savina[158], Nadezhda Savkova[238], Ricardo Schiappa[142], Michael A. Schmidt[261], Matthias Schott[286], Lakhdar Sek[296], Vladimir Sergeychik[268], Ronald Settles[178], Vladislav Shalaev[158], Andrey Shalyugin[158], Lianyou Shan[126], Wei Shan[91], Bofeng Shang[351], Jianfeng Shangguan[76], Revaz Shanidze[88], Hua-Sheng Shao[173], Dingyu Shao[66], Ligang Shao[126], Xin She[126], Qiuping Shen[126], Chengping Shen[66], Peixun Shen[126], Hong-Fei Shen[126], Yaxin Shen[126], Jie Sheng[233], Shuqi Sheng[61], Jun Shi[241, 70], Zhan Shi[40], Xin Shi[126], Shusu Shi[20], Hua Shi[126], Haoyu Shi[126], Jialei Shi[26], Tianyu Shi[126], Jingyan Shi[126], Shuo Shi[126], Qi Shi[292], Gary Shiu[330], Sergei Shmatov[158], Jing Shu[202, 12], Zong-Guo Si[221], Meiyu Si[225], Elias Sideras-Haddad[326], Andrei Sidorenkov[123, 126], Aleksandr Simonenko[158], Tatiana Skorodko[123], Kirill Slizhevskii[158], Ivan Smiljanić[335, 284], Firdaus Soberi[295], Elena Solovieva[199], Weizheng Song[126, 292], Weimin Song[154], Siyuan Song[213], Yanchang Song[256], Shaowei Song[126], Huayang Song[120], Stefano Sorti[108, 273], Anna Staritsyna[199], Yongquan Su[232], Shufang Su[283], Amir Subba[96, 94], Junfeng Sun[84], Xiaohu Sun[202], Shengsen Sun[126], Yanjun Sun[196], Hao-Kai Sun[126], Xilei Sun[126], Xin-Yuan Sun[48], Xingyang Sun[185], Xiangming Sun[20], Weiyi Sun[126], Ziyang Sun[233], Jin Sun[120], Fei Sun[126], Zheng Sun[237], Mingkai Sun[202], Igor Suslov[158], Narumon Suwonjandee[33], Hirmans Tabaharizato[126], Mirian Tabidze[88], Michele Tammaro[67], Longsheng Tan[10], Xiao-Ze Tan[66], Yanbang Tang[126], Guangyi Tang[126], Ying-Ao Tang[339], Yulong Tang[126], Jian Tang[222], Junquan Tao[126], Ze Tao[210], Khuram Tariq[126], Giovanni Francesco Tassielli[106, 279], Karabo Tau[326], Abdel Nasser Tawfik[147, 64, 62], Yahya Tayalati[282, 181], Geoffrey Taylor[223], Valery Telnov[17, 198], Jia Jian Teoh[126], Stefano Terzo[118], Pilong Tian[126], Victor Tikhomirov[122], Andrey Timoshchenko[14], Rahul Tiwary[267], Riccardo Torre[105], Adel Trabelsi[327, 60], Anastasiia Tropina[158], Edisher Tskhadadze[88], Toru Tsuboyama[87], Dmitri Tsybychev[252], Yanjun Tu[299], Shengquan Tuo[334], Michael Tytgat[336], Ghalib Ul Islam[46], Nikita Ushakov[123], Fuat Ustuner[295], German Valencia[182], Egor Vasenin[199], Ilya Vasilyev[158], Nikolaos Vassilopoulos[126, 30], Jaap Velthuis[287], Ivana Vidakovic[335], Natascia Vignaroli[214, 106],

Raymond Volkas[223], Dmitry Voronin[123], Nikolay Voytishin[158], Natasa Vukasinovic[335, 284], Carlos Wagner[291, 203], Jiawei Wan[185], Xuying Wan[206], Xia Wan[226], Qun Wang[321, 8], Yuexin Wang[126, 30], Chuanye Wang[185, 126], Hengyu Wang[126, 292], Han Wang[126, 292], Yifang Wang[126], Fei Wang[351], Jia Wang[197], Jianchun Wang[126], En Wang[351], Dou Wang[126], Shudong Wang[126], Wei Wang[126], Zeren Simon Wang[81], Guangyu Wang[233], Qian Wang[241], Xiongfei Wang[169], Yu-Ming Wang[187], Wei Wang[233], Hui Wang[184], Yaqian Wang[47], Zihui Wang[193], Zeqiang Wang[274], Sheng-Quan Wang[74], Hanwen Wang[126], Biao Wang[301], Congcong Wang[126], Xiaolong Wang[126], Jian Wang[221], Dayong Wang[202], Guo-Li Wang[47], Tianyang Wang[348], Anqi Wang[292], Zirui Wang[66], Hongbo Wang[227], Xujian Wang[126], Xiao-Ping Wang[10], Feng Wang[126], Meng Wang[210], Xiayu Wang[197], Kechen Wang[51], Fasheng Wang[40], Kun Wang[322], Ke Wang[126], Liguo Wang[41], Jin-Wei Wang[224], Yiwei Wang[126], Jin Wang[126], Zhi Wang[187], Yi Wang[126], Mengliang Wang[73], Haijun Wang[154], Xinzhu Wang[233], Zebing Wang[126], Xiaolian Wang[321], Hao-Lin Wang[241], Jian Wang[126], Xingquan Wang[220], Xiyang Wang[66], Bo Wang[210], Boxin Wang[126], Tianhong Wang[78], Yufeng Wang[233], Chengtao Wang[126], Shengpei Wang[126], Lian-Tao Wang[291], Zhen Wang[321], Menglin Wang[126], Xiaolong Wang[66], Enke Wang[241], Ting Wang[66], Yaping Wang[20], Huayang Wang[233], Xinyue Wang[241], Youkai Wang[226], Zheng Wang[126], Yaoguang Wang[221], Chenxu Wang[78], Anqing Wang[210], Weiping Wang[157], Yi Wang[126], Di Wang[57], Chengwei Wang[126], Yi Wang[268], Xining Wang[268], Yuhao Wang[202], Takashi Watanabe[161], Yingjie Wei[298], Wei Wei[126], Xiaomin Wei[197], He Wei[126], Yajing Wei[202], Yang Weiming[197], Liangjian Wen[126], Martin White[5], Ulrich Wiedner[212], Alasdair Winter[285], John Womersley[295], Sam S. C. Wong[36], Zuofei Wu[185, 126], Xuehong Wu[217, 144], Jinfei Wu[126], Yaru Wu[126], Lei Wu[184], Qun Wu[221], Yongcheng Wu[184], Zhi Wu[126], Lianjin Wu[84], Tianya Wu[183], Linghui Wu[126], Weihao Wu[233], Peiwen Wu[245], Minlin Wu[225], Jiahui Wu[233], Zhenyu Wu[232], Libo Wu[126], Xin Wu[321], Shaun Wyngaardt[250], Ligang Xia[185], Lei Xia[321], Xin Xia[126], Shang Xia[126], Shu Xian[241], Zhiyu Xiang[21], Yunlong Xiao[66], Zhen-Jun Xiao[184], Suyu Xiao[227], Rui-Qing Xiao[202], Ying Xiao[40], Chu-Wen Xiao[71], Xiang Xiao[222], Meng Xiao[350], Yuehong Xie[20], Yuguang Xie[126], Shiqing Xie[66], Wan Xie[126], Jiale Xie[241], Xinhao Xie[197], Du Xingrui[66], Yujie Xiong[233], Wang Xiuku[126], Xiajian Xu[35, 175], Weiwei Xu[227, 221], Zijun Xu[126], Fanrong Xu[155], Wei Xu[126], Qingjin Xu[126], Zhongyukun Xu[295], Meihang Xu[126], Haisheng Xu[126], Ling-Xiao Xu[2], Jingkang Xu[183], Chang Xu[126], Yuyao Xu[187], Da Xu[126], Ji Xu[169], Fang Xu[66], Yue Xu[329], Qing-Jun Xu[76], Zhen Xu[350], Guojun Xu[233], Xinping Xu[239], Feifei Xue[197], Youwen Xue[187], Tian Yan[126], Qi Yan[126], Xiongbo Yan[126], Wenbiao Yan[321], Liang Yan[66], Baojun Yan[126], Bingchen Yan[126], Zelin Yan[233], Yupeng Yan[253], Haozhong Yang[226, 126], Chao Yang[126, 292], Yueling Yang[84], Haijun Yang[233], Xing-Hua Yang[229], Heng Yang[197], Shuo Yang[170], Zhongjuan Yang[302], Qiaoli Yang[155], Lan Yang[185], Lin Yang[233], Ping Yang[20], Junling Yang[256], Tong-Zhi Yang[241], Hao Yang[341], Chi Yang[210], Li Lin Yang[350], Xuan Yang[126], Ziqiu Yang[126], Changgen Yang[126], Xinsheng Yang[247], Ji-Chong Yang[170], Fengfan Yang[126], Zhenwei Yang[202], Xueting Yang[249],

Lekai Yao[268], Mei Ye[126], Jingbo Ye[126], Lei Ye[206], Vitaly Yermolchyk[158], Li Yi[210], Kai Yi[184], Hang Yin[20], Junhao Yin[187], Zhongbao Yin[20], Peng-Fei Yin[126], Haoliang Ying[202], Hou Yingqi[126], Dong Yipei[206], Jiang Yitonh[187], Hwi Dong Yoo[345], Zhengyun You[222], Issam Youbi[118], Felix Yu[157], Zhao-Huan Yu[222], Boxiang Yu[126], Dian Yu[233], Jiesheng Yu[92], Chunxu Yu[187], Jinqing Yu[92], Kang Yu[233], Liwen Yu[66], Taozhe Yu[126], Yinghao Yu[126], Mingkuan Yuan[66, 128], Jiarong Yuan[126, 292], Changzheng Yuan[126], Xuhao Yuan[126], Xing-Bo Yuan[20], Rui Yuan[233], Li Yuan[10], Yuan Yuan[126], Ye Yuan[126], Hao Yuan[126], Dongyu Yuan[171], Chong Xing Yue[170], Yunkai Yue[256], Shi Yukun[166, 321], Gao Yun-Fei[126], Anwar Zada[126, 292], Riccardo Zanottera[107, 273], Aleksander Filip Zarnecki[328], Abdelkrim Zeghari[282], Hao Zeng[126, 292], Jun Zeng[75], Cheng Zeng[126], Wangmei Zha[321], Viktor Zhabin[17], Tianzheng Zhan[126], Licheng Zhang[308, 24], Kaili Zhang[126, 30], Zhong Zhang[245, 59], Likai Zhang[126, 221], Hao Zhang[126, 292], Yang Zhang[126, 292], Yizhou Zhang[126], Chenguang Zhang[126], Jialiang Zhang[185], Di Zhang[86], Rao Zhang[237], Yu Zhang[81], Mu-Hua Zhang[233], Wen-Chao Zhang[226], Zhandong Zhang[126], Jielei Zhang[85], Hongyu Zhang[126], Zhi-Qing Zhang[86], Jiehao Zhang[126], Zhuo Zhang[126], Liang Zhang[210], Jian-Rong Zhang[191], Hui Zhang[126], Tianyuan Zhang[126], Wei Zhang[338], Yinhong Zhang[126], Ying Zhang[126], Zhaoru Zhang[126], Xin Zhang[170], Yongchao Zhang[245], Qiqi Zhang[197], Huaqiao Zhang[126], Zhenyu Zhang[339], Jie Zhang[126], Yu Zhang[208], Xiangzhen Zhang[126], Xiyuan Zhang[126], Hongjun Zhang[321], Zhicai Zhang[268], Lei Zhang[185], Yumei Zhang[222], Rui Zhang[185], Guangyi Zhang[84], Quanzheng Zhang[292], Xuantong Zhang[126], Hangchang Zhang[126], Zhen-Hua Zhang[218], Hanlin Zhang[187], Liming Zhang[268], Bolun Zhang[126], Jianyu Zhang[292], Wang Zhang[66], Hongyu Zhang[66], Jie Zhang[126], Shulei Zhang[92], Yao Zhang[126], Xiaomei Zhang[126], Hong-Hao Zhang[222], Qiuyan Zhang[154], Zhiqing Zhang[93], Kairui Zhang[313], Jinxian Zhang[126], Jingbo Zhang[78], Dongliang Zhang[20], Yihan Zhang[126], Mengchao Zhang[155], Sijing Zhang[165], Yixiang Zhang[185], Danyang Zhang[241], Peng Zhang[126], Yongpeng Zhang[126], Jingxu Zhang[169], Jingqing Zhang[184], Lei Zhang[126], Wenxing Zhang[47], Yaosheng Zhang[78], Mingxuan Zhang[126], Jin Zhang[225], Zhaoling Zhang[154], Xiaoming Zhang[20], Mingxuan Zhang[202], Yanxi Zhang[202], Xiao Zhao[126, 201], Ling Zhao[126], Jingyi Zhao[126], Luyang Zhao[126], Zexuan Zhao[197], Yu Zhao[197], Guang Zhao[126], Zhiyu Zhao[233], Jingzhou Zhao[126], Mei Zhao[126], Xinhao Zhao[187], Zhengyang Zhao[225], Qiang Zhao[126], Sen Zhao[126], Jie Zhao[126], Zhenyu Zhao[321], Xiang Zhao[187], Minggang Zhao[187], Yubin Zhao[126], Ruiguang Zhao[197], Jiayao Zhao[197], Yan Zhao[40], Yu Zhao[340], Alexey Zhemchugov[158], Hongjuan Zheng[126], Xu-Chang Zheng[32], Wei Zheng[197], Ran Zheng[197], Wei Zheng[126], Lu Zheng[210], Linyue Zheng[241], Ya-Juan Zheng[262], Yingsen Zheng[126], Ilia Zhizhin[158], Yuzhang Zhong[126, 221], Yi-Ming Zhong[36], Bin Zhong[184], Hang Zhou[82, 221], Yang Zhou[126], Si-Hong Zhou[116], Jianxin Zhou[126], Jia Zhou[126], Shiyu Zhou[126], Xiang Zhou[339], Zusheng Zhou[126], Yi Zhou[321], Xiaokang Zhou[20], Chen Zhou[202], Jian Zhou[210], Baihong Zhou[233], Daicui Zhou[20], Qi-Dong Zhou[210], Shun Zhou[126], Yilin Zhou[66], Junjie Zhou[340], Yu-Feng Zhou[137], Qiusheng Zhu[220, 126], Rui Zhu[137, 292], Ruilin Zhu[184],

Kai Zhu[126], Yingshun Zhu[126], Yongfeng Zhu[126], Pengxuan Zhu[5], Huaxing Zhu[202], Jingya Zhu[85], Zian Zhu[321], Tongru Zhu[187], Chunxiang Zhu[233], Kejun Zhu[126], Hao Zhu[126], Hongbo Zhu[350], Yifan Zhu[233], Keyu Zhu[126], Zenghao Zhu[40], Xingwang Zhu[230], Ren-Yuan Zhu[18], Ren-Yuan Zhu[18], Xuai Zhuang[126], Wanyi Zhuang[66], Ilia Zimin[158], Mikhail Zobov[99], Shiming Zou[66], Jiaheng Zou[126], Jiahui Zou[233], Davide Zuliani[316, 110], Vladimir Zykunov[158, 68],

[1] Abdul Wali Khan University Mardan, Mardan.
[2] Abdus Salam International Centre for Theoretical Physics (ICTP), Trieste.
[3] Academia Sinica, Taipei.
[4] Adam Mickiewicz University, Poznań.
[5] Adelaide University, Adelaide.
[6] AGH University, Krakow.
[7] Anhui university, HeFei, Anhui.
[8] Anhui University of Science and Technology, Hefei, Anhui.
[9] Argonne National Laboratory, Lemont, IL.
[10] Beihang University, Beijing.
[11] Beijing Glass Research Institute, Beijing.
[12] Beijing Laser Acceleration Innovation Center, Huairou, Beijing.
[13] Beijing Normal University, Beijing.
[14] Belarusian State University, Minsk.
[15] Benha University, Faculty of Science, Benha.
[16] Boston University, Boston, MA.
[17] Budker Institute of Nuclear Physics (BINP), Novosibirsk.
[18] California Institute of Technology, Pasadena, CA.
[19] Carleton University, Ottawa.
[20] Central China Normal University, Wuhan, Hubei.
[21] Central South University, Changsha, Hunan.
[22] Centre for High Energy Physics, University of the Punjab, Lahore.
[23] Centro de Investigacion en Tecnologias Navales e Industriales, Univeraidade da Coruna, A Coruna.
[24] CERN, Geneva.
[25] Changsha University of Science and Technology, Changsha, Hunan.
[26] Chengdu University of Technology, Chengdu, Sichuan.
[27] China Center of Advanced Science and Technology, Beijing.
[28] China Institute of Atomic Energy, Beijing.
[29] China Nuclear (Beijing) Nuclear Instrument CO.,LTD, Beijing.

[30] China Spallation Neutron Source Science Center, Dongguan, Guangdong.
[31] China University of Mining and Technology, Beijing.
[32] Chongqing University, Chongqing.
[33] Chulalongkorn University, Bangkok.
[34] Chung-ang University, Seoul.
[35] Citic Heavy Industries Co., Ltd., luoyang, Henan.
[36] City University of Hong Kong, Hong Kong.
[37] Cockcroft Insitute, Daresbury.
[38] Commissariat à l'énergie atomique (CEA), Saclay.
[39] Complutense University of Madrid, Madrid.
[40] Dalian Minzu University, Dalian, Liaoning.
[41] Dalian University, Dalian, Liaoning.
[42] Department of Microelectronics and Electronics Systems, Autonomous University of Barcelona (UAB), Barcelona.
[43] Department of Physics and Astronomy, College of Science, King Saud University, Riyadh.
[44] Department of Physics and Nanotechnology, College of Engineering and Technology, SRM Institute of Science and Technology, Kattankulathur.
[45] Department of Physics, College of Science and Humanities, Prince Sattam bin Abdulaziz University, Al-Kharj.
[46] Department of Physics, Division of Science and Technology, University of Education, Lahore.
[47] Department of Physics, Hebei University, Baoding, Hebei.
[48] Department of Physics, Jinggangshan University, Ji'an, Jiangxi.
[49] Department of Physics, Karadeniz Technical University, Trabzon.
[50] Department of Physics, Lund university, Lund.
[51] Department of Physics, School of Physics and Mechanics, Wuhan University of Technology, Wuhan, Hubei.
[52] Department of Physics, The University of Lahore, Punjab.
[53] Department of Physics, University of Science and Technology of Mazandaran, Behshahr.
[54] Department of Physics, University of Trieste, Trieste.
[55] Department of Physics, University of Virginia, Charlottesville, VA.
[56] Department of Science and Technology, University College at Nairiyah, University of Hafr Al Batin (UHB), Nairiyah.
[57] Deutsches Elektronen-Synchrotron DESY, Hamburg.
[58] Dhofar University, Salalah.
[59] Durham University, Durham.
[60] Ecole Polytechnique de Tunisie, La Marsa.
[61] École Polytechnique Fédérale de Lausanne, Lausanne.
[62] Egyptian Center for Theoretical Physics (ECTP), Giza.

[63] Facultad de Ingenieria, Universidad San Sebastian, Santiago.
[64] Faculty of Engineering, Basic Science Department, Ahram Canadian University (ACU), Giza.
[65] Federal Urdu University of Arts, Science and Technology, Karachi.
[66] Fudan University, Shanghai.
[67] Galileo Galilei Institute for Theoretical Physics, Firenze.
[68] Gomel State University, Gomel.
[69] GSI Helmholtz Centre for Heavy Ion Research, Darmstadt.
[70] Guangdong Basic Research Center of Excellence for Structure and Fundamental Interactions of Matter, Guangdong Provincial Key Laboratory of Nuclear Science, Guangzhou, Guangdong.
[71] Guangxi Normal University, Guilin, Guangxi.
[72] Guangxi University, Nanning, Guangxi.
[73] Guizhou Key Laboratory in Physics and Related Areas, Guizhou University of Finance and Economics, Guiyang, Guizhou.
[74] Guizhou Minzu University, Guiyang, Guizhou.
[75] Hainan Normal University, Haikou, Hainan.
[76] Hangzhou Normal University, Hangzhou, Zhejiang.
[77] Harbin Engineering University, Harbin, Heilongjiang.
[78] Harbin Institute of Technology, Harbin, Heilongjiang.
[79] Harvard University, Cambridge, MA.
[80] Hassan II University of Casablanca, Casablanca.
[81] Hefei University of Technology, Hefei, Anhui.
[82] Helmholtz Institute Mainz, Mainz.
[83] Henan Institute of Science and Technology, Xinxiang, Henan.
[84] Henan Normal University, Xinxiang, Henan.
[85] Henan University, Kaifeng, Henan.
[86] Henan University of Technology, Zhengzhou, Henan.
[87] High Energy Accelerator Research Organization (KEK), Tsukuba.
[88] High Energy Physics Institute, Tbilisi State University, Tbilisi.
[89] Huazhong University of Science and Technology, Wuhan, Hubei.
[90] Hubei Polytechnic University, Huangshi, Hubei.
[91] Hunan Normal University, Changsha, Hunan.
[92] Hunan University, Changsha, Hunan.
[93] IJCLab, Laboratoire de Physique des 2 Infinis Irène Joliot-Curie, Orsay.
[94] Indian Institute of Science Education and Research (Kolkata), Kolkata.
[95] Indian Institute of Science Education and Research (Pune), Pune.
[96] Indian Institute of Technology Guwahati, Guwahati.
[97] Indian Institute of Technology Hyderabad, Hyderabad.
[98] Indian Institute of Technology Kharagpur, Kharagpur.

[99] INFN Laboratori Nazionali di Frascati, Frascati.
[100] INFN Sezione di Bari, Bari.
[101] INFN Sezione di Bologna, Bologna.
[102] INFN Sezione di Cagliari, Cagliari.
[103] INFN Sezione di Catania, Catania.
[104] INFN Sezione di Ferrara, Ferrara.
[105] INFN Sezione di Genova, Genova.
[106] INFN Sezione di Lecce, Lecce.
[107] INFN Sezione di Milano, Milan.
[108] INFN Sezione di Milano - LASA, Milan.
[109] INFN Sezione di Milano-Bicocca, Milan.
[110] INFN Sezione di Padova, Padova.
[111] INFN Sezione di Pisa, Pisa.
[112] INFN Sezione di Roma 2, Rome.
[113] INFN Sezione di Roma Tor Vergata, Rome.
[114] INFN Sezione di Roma Tre, Rome.
[115] INFN Sezione di Torino, Torino.
[116] Inner Mongolia University, Hohhot, Inner Mongolia.
[117] Institucio Catalana de Recerca i Estudis Avancats, (ICREA), Barcelona.
[118] Institut de Física d'Altes Energies (IFAE), Barcelona.
[119] Institute for Advanced Study, Princeton, NJ.
[120] Institute for Basic Science (IBS), Daejeon.
[121] Institute for Fundamental Science, Eugene, OR.
[122] Institute for Nuclear Problems, Belarusian State University, Minsk.
[123] Institute for Nuclear Research, Russian Academy of Sciences, Moscow.
[124] Institute for Theoretical Particle Physics, Karlsruhe Institute of Technology, Karlsruhe.
[125] Institute for Theoretical Physics and Modeling, Yerevan.
[126] Institute of High Energy Physics, Chinese Academy of Sciences, Beijing.
[127] Institute of Microelectronics, Chinese Academy of Sciences, Beijing.
[128] Institute of Modern Physics, Chinese Academy of Sciences, Lanzhou, Gansu.
[129] Institute of Nuclear and Particle Physics, NCSR Demokritos, Agia Paraskevi, Attiki.
[130] Institute of Nuclear Physics, Polish Academy of Sciences, Krakow.
[131] Institute of Particle and Cosmos Physics, Madrid.
[132] Institute of Physics of the Czech Academy of Sciences, Praha.
[133] Institute of Physics of the National Academy of Sciences of Belarus, Minsk.
[134] Institute of Physics, Henan Academy of Sciences, Zhengzhou, Henan.
[135] Institute of Power Engineering of National Academy of Sciences of Belarus, Minsk.
[136] Institute of Quantum Matter, South China Normal University, Guangzhou.
[137] Institute of Theoretical Physics, Chinese Academy of Sciences, Beijing.

[138] Instituto de Alta Investigación, Universidad de Tarapacá, Arica.
[139] Instituto de Física Corpuscular, IFIC (CSIC/UV), Valencia.
[140] Instituto de Física Teórica (IFT), Consejo Superior de Investigaciones Científicas (CSIC), Madrid.
[141] Instituto Galego de Física de Altas Enerxías, Santiago de Compostela.
[142] Instituto Superior Técnico, University of Lisbon, Lisbon.
[143] International Centre for Theoretical Physics Asia-Pacific, Beijing.
[144] International Joint Laboratory of Energy Efficiency Conversion and Utilization of Henan Province, Zhengzhou, Henan.
[145] Iowa State University, Ames, IA.
[146] IPHC, Université de Strasbourg, CNRS/IN2P3, Strasbourg.
[147] Islamic University in Madinah, Madinah.
[148] Istinye University, Istanbul.
[149] iThemba Laboratory for Accelerator Based Sciences, Cape Town.
[150] Iwate University, Morioka.
[151] Jagiellonian University, Krakow.
[152] Jazan University, Jazan.
[153] Jiangsu Normal University, Xuzhou, Jiangsu.
[154] Jilin University, Changchun, Jilin.
[155] Jinan University, Guangzhou, Guangdong.
[156] Jingchu University of Technology, Jingmen, Hubei.
[157] Johannes Gutenberg University, Mainz.
[158] Joint Institute for Nuclear Research, Dubna.
[159] Kavli Institute for the Physics and Mathematics of the Universe, The University of Tokyo, Kashiwa.
[160] Kindai University, Osaka.
[161] Kogakuin University of Technology & Engineering, Shinjuku-ku, Tokyo.
[162] Korea University Sejong Campus, Sejong.
[163] Kutaisi International University, Kutaisi.
[164] Kyungpook National University, Daegu.
[165] L2IT/Université de Toulouse, CNRS/IN2P3, Toulouse.
[166] Laboratoire Leprince-Ringuet (LLR), CNRS, École polytechnique, Palaiseau.
[167] Laboratory for Instrumentation and Experimental Particle Physics (LIP), Lisbon.
[168] Lancaster University, Lancaster.
[169] Lanzhou University, Lanzhou, Gansu.
[170] Liaoning Normal University, Dalian, Liaoning.
[171] Liaoning University, Shenyang, Liaoning.
[172] LPNHE, Sorbonne Université, Paris Diderot Sorbonne Paris Cité, CNRS/IN2P3, Paris.
[173] LPTHE, Laboratoire de Physique Théorique et Hautes Energies, Paris.

[174] LPTHE, Physics Department, Faculty of Sciences, Ibnou Zohr University, Agadir.
[175] Luoyang Mining Machinery Engineering Design Institute CO.,LTD, Luoyang, Henan.
[176] Mainz Institute of Theoretical Physics (MITP), Mainz.
[177] Materials Research Center - Institute for Nanostructures, Nanomodelling and Nanofabrication / Faculdade de Ciências e Tecnologia, Universidade Nova de Lisboa, Monte da Caparica.
[178] Max Planck Institute for Physics, Munich.
[179] Mediterranean School of Business at South Mediterranean University, Tunis.
[180] MOE Key Laboratory of TianQin Mission, TianQin Research Center for Gravitational Physics & School of Physics and Astronomy, Frontiers Science Center for TianQin, Gravitational Wave Research Center of CNSA, Sun Yat-sen University, Zhuhai, Guangdong.
[181] Mohammed VI Polytechnic University, Benguerir.
[182] Monash University, Melbourne.
[183] NanChang University, Nanchang, Jiangxi.
[184] Nanjing Normal University, Nanjing, Jiangsu.
[185] Nanjing University, Nanjing, Jiangsu.
[186] Nanjing University of Science and Technology, Nanjing, Jiangsu.
[187] Nankai University, Tianjin.
[188] National Institute of Chemical Physics and Biophysics (NICPB), Tallinn.
[189] National Research University Higher School of Economics (HSE), Moscow.
[190] National Taiwan University, Taipei.
[191] National University of Defense Technology, Changsha, Hunan.
[192] Naval University of Engineering, Wuhan, Hubei.
[193] New York University, New York, NY.
[194] Niels Bohr Institute, Copenhagen.
[195] Northeastern University, Boston, MA.
[196] Northwest Normal University, Lanzhou, Gansu.
[197] Northwestern Polytechnical University, Xi'an, Shaanxi.
[198] Novosibirsk State University, Novosibirsk.
[199] P.N. Lebedev Physical Institute of the Russian Academy of Sciences, Moscow.
[200] Pakistan Institute of Nuclear Science and Technology, Islamabad.
[201] Paul Scherrer Institute, Villigen.
[202] Peking University, Beijing.
[203] Perimeter Institute for Theoretical Physics, Waterloo.
[204] Politecnico di Bari, Bari.
[205] Polytechnic University of Lisbon, Lisbon.
[206] PowerChina HuaDong Engineering Corporation Limited, Hangzhou, Zhejiang.
[207] Qufu Normal University, Qufu, Shandong.
[208] Qujing Normal University, Qujing, Yunnan.

[209] Renmin University of China, Beijing.
[210] Research Center for Particle Science and Technology, Shandong University, Qingdao, Shandong.
[211] Rudjer Boskovic Institute (RBT), Zagreb.
[212] Ruhr-University Bochum, Bochum.
[213] Rutgers University, New Brunswick, NJ.
[214] Salento University, Lecce.
[215] Samarkand State University, Samarkand.
[216] Santa Cruz Institute for Particle Physics, Santa Cruz, CA.
[217] School of Energy and Power Engineering, Zhengzhou University of Light Industry, Zhengzhou, Henan.
[218] School of Nuclear Science and Technology, University of South China, Hengyang, Hunan.
[219] School of Particles and Accelerators, Institute for Research in Fundamental Sciences (IPM), Tehran.
[220] School of Physics and Electronic Information, Gannan Normal University, Ganzhou, Jiangxi.
[221] School of Physics, Shandong University, Jinan, Shandong.
[222] School of Physics, Sun Yat-Sen University, Guangzhou, Guangdong.
[223] School of Physics, The University of Melbourne, Melbourne.
[224] School of Physics, University of Electronic Science and Technology of China, Chengdu, Sichuan.
[225] School of Science, Shenzhen Campus of Sun Yat-sen University, Shenzhen, Guangdong.
[226] Shaanxi Normal University, Xi'an, Shaanxi.
[227] Shandong Institute of Advanced Technology, Jinan, Shandong.
[228] Shandong Management University, Jinan, Shandong.
[229] Shandong University of Technology, Zibo, Shandong.
[230] Shanghai Industrial $\mu$Technology Research Institute, Shanghai.
[231] Shanghai Institute of Ceramics, Chinese Academy of Sciences, Shanghai.
[232] Shanghai Institute of Microsystem and Information Technology Chinese Academy of Sciences, Shanghai.
[233] Shanghai Jiao Tong University, Shanghai.
[234] Shanghai Normal University, Shanghai.
[235] Shanxi University, Taiyuan, Shanxi.
[236] Shenzhen Technology University, Shenzhen, Guangdong.
[237] Sichuan University, Chengdu, Sichuan.
[238] Skobeltsyn Institute of Nuclear Physics, Lomonosov Moscow State University, Moscow.
[239] Soochow University, Suzhou, Jiangsu.
[240] Sorbonne University, Paris.
[241] South China Normal University, Guangzhou, Guangdong.

[242] South China University of Technology, Guangzhou, Guangdong.
[243] South Mediterranean University, Tunis.
[244] Southeast Asian Ministers of Education Organization Regional Centre for STEM Education, Bangkok.
[245] Southeast University, Nanjing, Jiangsu.
[246] Southern University of Science and Technology, Shenzhen, Guangdong.
[247] Southwest Jiaotong University, Chengdu, Sichuan.
[248] Stanford University, Stanford, CA.
[249] State Key Laboratory of Nuclear Physics and Technology, Beijing.
[250] Stellenbosch University, Stellenbosch.
[251] STFC Rutherford Appleton Laboratory, Didcot.
[252] Stony Brook University, Stony Brook.
[253] Suranaree University of Technology, NakhonRatchasima.
[254] TaiYuan University of Technology, Taiyuan, Shanxi.
[255] Tata Institute of Fundamental Research, Mumbai.
[256] Technical Institute of Physics and Chemistry, Chinese Academy of Sciences, Beijing.
[257] Technical University of Munich, Munich.
[258] The Chinese University of Hong Kong, Hong Kong.
[259] The Hong Kong University of Science and Technology, Hong Kong.
[260] The University of Cambridge, Cambridge.
[261] The University of New South Wales, Sydney.
[262] The University of Osaka, Osaka.
[263] The University of Warwick, Coventry.
[264] Theoretical and Scientific Data Science, Scuola Internazionale Superiore di Studi Avanzati (SISSA), Trieste.
[265] Tianjin University, Tianjin.
[266] Tianjin University of Technology, Tianjin.
[267] Toshiko Yuasa Laboratory (TYL), Tsukuba.
[268] Tsinghua University, Beijing.
[269] UKRI STFC Daresbury, Warrington.
[270] Universidad Autónoma de Zacatecas, Zacatecas.
[271] Universidade do Estado do Rio de Janeiro, Rio de Janeiro.
[272] Universidade Estadual de Londrina (UEL), Londrina.
[273] Università degli Studi di Milano, Milan.
[274] Université catholique de Louvain, Louvain La Neuve.
[275] Università degli Studi di Roma Tre, Rome.
[276] Università dell'Insubria, Como.
[277] Università di Milano-Bicocca, Milan.
[278] Università di Torino, Turin.

[279] Università Mercatorum, Rome.
[280] Université Claude Bernard Lyon 1, Lyon.
[281] Université Paris-Saclay, Antony.
[282] University Mohammed V in Rabat, Rabat.
[283] University of Arizona, Tucson, AZ.
[284] University of Belgrade, Belgrade.
[285] University of Birmingham, Birmingham.
[286] University of Bonn, Bonn.
[287] University of Bristol, Bristol.
[288] University of Cagliari, Cagliari.
[289] University of California Santa Cruz, Santa Cruz, CA.
[290] University Of Central Punjab, Lahore.
[291] University of Chicago, Chicago, IL.
[292] University of Chinese Academy of Sciences, Beijing.
[293] University of Coimbra, Coimbra.
[294] University of Cordoba, Cordoba.
[295] University of Edinburgh, Edinburgh.
[296] University of El Oued, El Oued.
[297] University of Ferrara, Ferrara.
[298] University of Freiburg, Freiburg.
[299] University of Hong Kong, Hong Kong.
[300] University of Hyderabad, Hyderabad.
[301] University of Iowa, Iowa City, IA.
[302] University of Jinan, Jinan, Shandong.
[303] University of Johannesburg, Johannesburg.
[304] University of Kansas, Lawrence, KS.
[305] University of Liverpool, Liverpool.
[306] University of Louisville, Louisville, KY.
[307] University of Manchester, Manchester.
[308] University of Maryland, College Park, MD.
[309] University of Massachusetts Amherst, Amherst, MA.
[310] University of Minnesota, Minneapolis, MN.
[311] University of MÜnster, Münster.
[312] University of New Hampshire, Durham, NH.
[313] University of Oklahoma, Norman, OK.
[314] University of Oregon, Eugene, OR.
[315] University of Oxford, Oxford.
[316] University Of Padua, Padua .
[317] University of Pittsburgh, Pittsburgh, PA.

[318] University of Rome ”La Sapienza”, Rome.
[319] University of Rome Tor Vergata, Rome.
[320] University of Sao Paulo, Sao Paulo.
[321] University of Science and Technology of China, Hefei, Anhui.
[322] University of Shanghai for Science and Technology, Shanghai.
[323] University of Siegen, Siegen.
[324] University of Silesia, Katowice.
[325] University of Southampton, Southampton.
[326] University of the Witwatersrand, Johannesburg.
[327] University of Tunis El Manar, Tunis.
[328] University of Warsaw, Warsaw.
[329] University of Washington, Seattle, WA.
[330] University of Wisconsin-Madison, Madison, WI.
[331] University of Zululand, Richards Bay, Dlangezwa.
[332] University of Zurich, Zurich.
[333] US Jefferson National Laboratory, Newport News, VA.
[334] Vanderbilt University, Nashville, TN.
[335] Vinca Institute of Nuclear Sciences, Belgrade.
[336] Vrije Universiteit Brussel, Brussels.
[337] Westlake University, Hangzhou, Zhejiang.
[338] Wuhan Textile University, Wuhan, Hubei.
[339] Wuhan University, Wuhan, Hubei.
[340] Wuxi Toly Electric Works Co.,Ltd, Wuxi, Jiangsu.
[341] X-HEP Laboratory, Department of Physics, School of Mathematics and Physics, Xi’an Jiaotong-Liverpool University, Suzhou, Jiangsu.
[342] Xi’an Jiaotong University, Xi’an, Shaanxi.
[343] Xidian University, Xi’an, Shaanxi.
[344] Yantai University, Yantai, Shandong.
[345] Yonder University, Seoul.
[346] Yonsei University, Seoul.
[347] Zanon Research and Innovation, Vicenza.
[348] Zhangjiang Laboratory, Shanghai.
[349] Zhejiang Lab, Hangzhou, Zhejiang.
[350] Zhejiang University, Hangzhou, Zhejiang.
[351] Zhengzhou University, Zhengzhou, Henan.
[352] Zhongyuan University of Technology, Zhengzhou, Henan.
[†] Deceased.

corresponding author: Joao Barreiro Guimaraes da Costa, guimaraes@ihep.ac.cn

# Acknowledgements

The completion of the CEPC Reference Detector Technical Design Report (TDR) owes its success to the diligent efforts of the CEPC detector research team, spearheaded by the Institute of High Energy Physics (IHEP) of the Chinese Academy of Sciences (CAS). Collaborations with both domestic and international institutes played pivotal roles by offering invaluable advice and support, contributing significantly to the TDR's creation.

Over the past years of prototype's R&D and physics design optimization, the International Advisory Committee (IAC), and more recently the CEPC International Detector Review Committee (IDRC), have been indispensable. Their generous support and insightful suggestions have significantly shaped the CEPC Reference Detector TDR. This document stands as a testament to the collaborative efforts of hundreds of scientists and engineers from the host institute, IHEP, along with numerous domestic and international institutes.

**The International Detector Review Committee:**

- Daniela Bortoletto, University of Oxford, Oxford (Chair)
- Jim Brau, University of Oregon, Eugene, OR
- Anna Colaleo, INFN, Bari
- Paul Colas, CEA, Paris
- Christophe De La Taille, OMEGA Laboratory, CNRS, Orsay
- Cristinel Diaconu, CPPM, Marseille
- Frank Gaede, DESY, Hamburg
- Colin Gay, University of British Columbia, Vancouver
- Liang Han, University of Science and Technology of China, Hefei, Anhui
- Bob Kowalewski, University of Victoria, Victoria
- Gregor Kramberger, Jožef Stefan Institute, Ljubljana
- Roman Poeschl, IJCLab, Orsay
- Burkhard Schmidt, CERN, Geneva
- Tommaso Tabarelli de Fatis, INFN, Milan
- Roberto Tenchini, INFN, Pisa
- Maxim Titov, CEA, Paris
- Ivan Vila Alvarez, University of Cantabria, Santander
- Akira Yamamoto, KEK, Tsukuba
- Hitoshi Yamamoto, Tohoku University, Sendai

**The International Advisory Committee:**

- Eugene Levichev, BINP, Novosibirsk
- Barry Barish, Caltech, CA

- Lucie Linssen, CERN, Geneva
- Michelangelo Mangano, CERN, Geneva
- Steinar Stapnes, CERN, Geneva
- Rohini Godbole[†], CHEP, Bangalore
- Beate Heinemann, DESY, Hamburg
- Tatsuya Nakada, EPFL, Lausanne
- Joe Lykken, Fermilab, IL
- Andrew Cohen, Hong Kong University of Science and Technology, Hong Kong
- Michael Davier, IJCLab, Orsay
- Maria Enrica Biagini, INFN Frascati, Rome
- Hongwei Zhao, Institute of Modern Physics, CAS, Lanzhou, Gansu
- Yuan-Hann Chang, Institute of Physics, Academia Sinica, Taipei
- Hitoshi Murayama, IPMU/UC Berkeley, CA
- Akira Yamamoto, KEK, Tsukuba
- Marcel Demarteau, ORNL, Oak Ridge, TN
- Brian Foster, University of Oxford, Oxford (Chair)
- Ian Shipsey[†], University of Oxford/DESY, Oxford
- Geoffrey Taylor, U. Melbourne, Melbourne
- Luciano Maiani, U. Rome, Rome
- David Gross, UC Santa Barbara, CA
- Karl Jakobs, University of Freiburg/CERN, Freiburg

**Funding**

This study received unwavering support from diverse funding sources, including the National Key Program for S&T Research and Development of the Ministry of Science and Technology (MOST), the CAS Key Foreign Cooperation Grant, the National Natural Science Foundation of China (NSFC), Beijing Municipal Science & Technology Commission, the CAS Focused Science Grant, the IHEP Innovation Grant, the CAS Lead Special Training Program, the CAS Center for Excellence in Particle Physics, the CAS International Partnership Program, and the CAS/SAFEA International Partnership Program for Creative Research Teams.

**Conflict of interest**

The authors declare that they have no conflict of interest.

[†]Deceased.

# Executive summary

The discovery of the Higgs boson in 2012 at the Large Hadron Collider (LHC) has ushered a new era in particle physics. Soon after, the Circular Electron Positron Collider (CEPC) was proposed, by Chinese scientists, as a Higgs factory to test with high precision the fundamental physics principles of the Standard Model (SM) and to search for new physics beyond the SM (BSM).

The CEPC is designed to operate at between 91 to 240 GeV as a powerful factory of clean Higgs, $Z$ and $W$ bosons, with an upgrade plan to run at the top-quark pair threshold. In a tentative baseline ”15-4-1-5” operational plan, the CEPC will operate with a synchrotron radiation power of up to 30 MW. Firstly as a Higgs boson factory for 15 years, followed by four years of operation as a Super $Z$ factory, and then one year as a $W$ factory. One experiment will collect 2 million Higgs particles, over 0.6 trillion $Z$ bosons, and approximately 150 million $W$-boson pairs. An upgraded scenario enhances the synchrotron radiation power to 50 MW for higher luminosities that will deliver 4.3 million Higgs events. The CEPC accelerator Technical Design Report was completed in December 2023, demonstrating the readiness for the CEPC construction phase that is expected to begin around 2027-2028 during China's 15th Five-Year Plan, pending government approval. Upon the approval of the CEPC, two international collaborations will be established to design, build, and operate two advanced detectors to fully explore the CEPC physics potential.

A CEPC detector is a critical instrument for scientists to perform experimental measurements with the colliding beam data delivered by the electron positron collider. Given the operating environment and the physics objectives, the capability of the CEPC detector represents a monumental leap in particle physics instrumentation in order to carry out extremely challenging experimental tasks. The current design of the CEPC Reference Detector was primarily initiated with the intention of submission to the government for the CEPC project approval. It has been strengthened by dedicated R&D efforts within the CEPC community, expertise gained from LHC detector upgrades, and advances in semiconductor technology, photonics, and communications. Incorporating these developments, the CEPC Reference Detector integrates state-of-the-art technologies and is designed to deliver unprecedented precision in the study of Higgs bosons, electroweak interactions, and potential physics beyond the Standard Model.

The CEPC Reference Detector comprises a silicon pixel vertex detector, a silicon inner tracker, and a Time Projection Chamber surrounded by a silicon external tracker equipped with AC-coupled Low Gain Avalanche Detector technology that provides Time-of-Flight information. The calorimetry system consists of a crystal electromagnetic calorimeter and a steel–glass-scintillator sampling hadronic calorimeter, all enclosed within a 3-Tesla superconducting solenoid. A muon detector is embedded in the magnetic flux-return yoke. To extend tracking acceptance, five pairs of silicon tracking disks are installed in the forward regions on both sides of the interaction point. Innovative designs of the tracking and calorimeter systems achieve outstanding vertex and spatial resolutions, $\gamma$ and $\pi^0$ energy resolutions, and excellent Particle Flow Algorithm capability and hadronic jet reconstruction. The associated electronics, the trigger and data acquisition system, and the installation plan have been designed. While most of the detector technologies have already been established, some technical challenges remain under study and are expected to be successfully addressed in the near future.

The CEPC Reference Detector has been designed and optimized primarily for the Higgs program, and for the low luminosity $Z$ runs aimed at initial physics and calibrations, within the first 20 years of the CEPC operation. The current Reference Detector design is not adequate for the high luminosity CEPC tera-$Z$ program beyond the baseline operation. Some of the detector components will likely be upgraded or replaced with future technologies to fully explore the physics potential of the tera-$Z$ program offered by the CEPC accelerator.

This report presents a detailed technical design study of the CEPC Reference Detector including subsystem performance, overall detector capabilities, and sensitivity to selected key physics benchmark processes. The results demonstrate that the detector meets or exceeds its design goals. A cost estimate for the Reference Detector and the future development plan are provided. Also included are the International Large Detector, ILD, which is reconfigured for the circular $e^+e^-$ collider, and the Innovative Detector for Electron-positron Accelerator, IDEA, originally proposed for the FCC-ee, both of which are well suited for the CEPC.

The CEPC Reference Detector will be an excellent instrument that can be built and put into operation in less than a decade and can provide an excellent foundation for one of the two international collaborations to be formed in the future.

# Preamble

This document is the second volume of the Circular Electron Positron Collider (CEPC) Technical Design Report. It presents a comprehensive overview of the CEPC Reference Detector technical design. The first volume [1], published in December 2023, describes the technical design of the CEPC accelerator complex and its associated civil engineering. This second volume is split into two parts. The first part, presents a brief summary of the physics case for the CEPC project, describes the technical details of the Reference Detector, and its technological options, highlights the expected detector and physics performance, and discusses future plans for detector development. The second part of this document, presents two other detector concepts, International Large Detector (ILD) and Innovative Detector for Electron-positron Accelerator (IDEA), that have been developed for future electron-positron colliders and put forward by the international community. They cover a similar physics program as the CEPC Reference Detector and could equip the CEPC second interaction point.

# Contents

# Part I

# CEPC Reference Detector

# Chapter 1 Introduction

The discovery of the Higgs boson in 2012 by the ATLAS and CMS Collaborations at the Large Hadron Collider (LHC) of the European Organization for Nuclear Research (CERN) [1, 2] marked the beginning of a new era in particle physics. The Higgs boson represents a cornerstone of the Standard Model (SM) of particle physics. It lies at the heart of some of the most profound mysteries in modern particle physics, including the significant hierarchy between the electroweak and Planck scales, and the nature of the electroweak phase transition. Precise determination of the Higgs boson properties, in conjunction with those of the $W$ and $Z$ bosons, offers essential tests of the fundamental physical principles underpinning the Standard Model. Moreover, such measurements are of paramount importance to explore physics Beyond the Standard Model (BSM).

Owing to its relatively low mass, Higgs bosons can be generated within the clean environment of a circular electron-positron collider with realistic luminosities and at a reasonable cost. The Circular Electron Positron Collider (CEPC) [3], proposed by China's particle physics community, is a key international facility centered on Higgs research around 240 GeV, where it will operate as a high-luminosity Higgs boson factory. Producing a few million Higgs bosons, it will enable precision measurements of decay modes and interactions, while searching for deviations from the Standard Model that could reveal BSM physics, such as dark matter or new interactions.

While the Higgs boson program with $e^+e^-$ collisions at center-of-mass energy ($\sqrt{s}$) of 240 GeV is central, the CEPC is designed for versatile operation across different energies. This includes running as a $Z$ factory at $\sqrt{s}$ = 91 GeV to refine SM parameters, and at the $WW$ production threshold ($\sim$ 160 GeV) to study electroweak dynamics. A potential upgrade to $\sqrt{s}$ = 360 GeV can further explore top quark physics. The CEPC's multi-energy capabilities, including $B/\tau$-charm physics from $Z$ decays, enhance its role as both a SM precision tool and a gateway to uncovering new physics, with Higgs boson research at its core. Such a comprehensive physics program will constitute a crucial element in any strategic road-map for particle physics research over the next several decades.

In the baseline operational scenario the CEPC will operate with a synchroton radiation power of up to 30 MW and one experiment. If additional resources become available, an upgraded scenario increases the synchroton radiation power to 50 MW, for higher luminosities, and enables two active experiments.

This document is the second volume of the CEPC Technical Design Report (TDR). The first volume [3], published in December 2023, describes the technical design of the CEPC accelerator complex and its associated civil engineering. This second volume presents a comprehensive overview of the CEPC Reference Detector technical design. The detector proposed here was conceived to fully explore the CEPC physics potential and it represents a significant evolution of the concepts introduced in the CEPC Detector Conceptual Design Report (CDR) [4].

This chapter starts by introducing the CEPC physics motivation, highlighting the precision measurements of the Higgs boson and physics potential around the $Z$ pole and the $WW$ threshold. It then briefly describes the detector challenges and performance requirements, followed by a definition of the benchmark processes that drove the design and demonstrate the detector performance. Chapter 2 presents the overall detector concept, deriving the design framework from the physics requirements. The global tracking system (from vertex to timing layer) and the PFA-optimized calorimeter system, among the other systems, are introduced. Chapter 3 addresses the Machine-Detector Interface (MDI) and luminosity measurement, including beam-induced background estimations essential for the design of several detector components.

Most of the sub-detector chapters follow a common structure: they begin with a set of physics and detector requirements, followed by a design overview and then a discussion of technical details. Past R&D is provided to justify the technical choices, as well as performance studies. Alternative detector designs are discussed when applicable. Chapters 4-6 cover the tracking system: Chapter 4 details the vertex detector design; Chapter 5 discusses the silicon tracker (including a timing layer); and Chapter 6 describes the Time Projection Chamber (TPC), and a drift chamber as a possible alternative. The calorimetry system is covered in two subsequent chapters: Chapter 7 presents the crystal-based Electromagnetic Calorimeter (ECAL), while Chapter 8 details the glass-based Hadronic Calorimeter (HCAL). Finally, Chapter 9 discusses the plastic-scintillator muon detector.

Chapter 10 introduces the detector superconducting solenoid magnet and the supporting infrastructure. Chapter 11 covers common electronics, and Chapter 12 addresses the Trigger and Data Acquisition system. Chapter 13 presents the common offline software framework and computing resources, including simulation, reconstruction, and analysis tools. Chapter 14 details mechanical integration, services, installation, and underground hall integration. The Reference Detector design is validated through physics performance studies in Chapter 15. Chapter 16 presents an estimate of the detector overall cost. Chapter 17 provides the project timeline, future R&D priorities, and the challenges ahead. CEPC project approval is expected shortly after the publication of this TDR. Two international collaborations will form concurrently with the start of construction, targeting to begin data-taking in 2037.

The CEPC embodies a vision of scientific excellence and technological innovation. By combining cutting-edge detector design, global collaboration, and a robust physics program, the project is poised to redefine our understanding of fundamental interactions and matter. Beyond its immediate scientific goals, CEPC will catalyze technological advancements, inspire educational initiatives, and foster international cooperation, leaving a lasting legacy as a symbol of humanity's pursuit of knowledge. As the detector takes shape in the coming decade, CEPC will open a new frontier in particle physics, where precision meets discovery, and the mysteries of the universe are unraveled one collision at a time.

## 1.1 Physics case

The discovery of the Higgs boson at the LHC in 2012 marked a monumental achievement in particle physics, completing the SM while opening new avenues to probe fundamental physics. Nevertheless, critical questions regarding the Higgs boson's nature, its role in electroweak symmetry breaking, and potential connections to BSM physics remain unresolved. A circular electron-positron collider operating at a center-of-mass energy of 240GeV within an approximately 100-km circumference tunnel presents a compelling opportunity to address these questions with unprecedented precision. Functioning as a "Higgs factory," such machine can achieve extremely high instantaneous luminosities, enabling high-precision measurements of Higgs boson properties including couplings to SM particles, mass, width, and potential deviations from SM predictions.

With clean experimental conditions, low backgrounds, and well-defined initial states, an electron-positron collider provides a superior environment compared to hadron colliders for studying the Higgs boson. For instance, it would allow measurements of the Higgs boson couplings to better than 1% precision, shedding light on whether the Higgs boson is a fundamental scalar or part of a broader BSM framework, such as supersymmetry, composite Higgs models, or dark matter interactions. A 240-GeV $e^+e^-$ collider would serve as a powerful probe of new physics. It could search for rare processes, such as invisible Higgs boson decays or new particles like dark photons, $Z'$ bosons, or light scalars, with sensitivity to energy scales far beyond its center-of-mass energy through indirect effects.

Beyond Higgs boson characterization, a 240-GeV $e^+e^-$ collider serves as a powerful instrument for new physics searches. It enables investigations of rare processes including invisible Higgs boson decays and production of exotic particles like dark photons, $Z'$ bosons, or light scalars, with sensitivity to energy scales far exceeding its center-of-mass collision energy through precision indirect measurements.

Hence, the scientific case for constructing this machine rests not only on its capacity to deepen understanding of the Higgs boson mechanism but also on its potential to resolve outstanding mysteries including dark matter, matter-antimatter asymmetry, and the hierarchy problem. By combining precision Higgs boson studies with broad BSM discovery potential, a 240-GeV electron-positron collider represents a critical next step in particle physics, complementing and extending the LHC's reach. Furthermore, the large-scale colliding ring enables a comprehensive research program encompassing $W$, and $Z$ boson studies, Quantum ChromoDynamics (QCD) and flavor physics. Potential future upgrades for $t\bar{t}$ production at 360 GeV, and to a 100 TeV proton-proton collider further expand the physics potential.

### 1.1.1 Scope of the physics program

As an electron-positron Higgs boson factory, the primary physics objective is the systematic and precise measurement of the Higgs boson's properties. However, from the broader perspective of performing precision tests of the Standard Model and probing for physics beyond the Standard Model (BSM), the optimal strategy is to operate over a wide center-of-mass energy range, spanning from the $Z$-pole at $\sqrt{s} = 91.2$ GeV to $\sqrt{s} = 360$ GeV, above the $t\bar{t}$ production threshold.

**Primary objective: maximizing Higgs boson production** An electron-positron Higgs boson factory must prioritize producing the largest possible Higgs boson dataset. This requires operating at energies where the Higgs boson production cross-section peaks — specifically, the 240–260 GeV range for the dominant $e^+e^- \to ZH$ process. With a high Higgs boson yield, precision measurements of its properties become feasible. For instance, the decay branching ratios to various final states can be determined with unprecedented accuracy, shedding light on the Higgs boson's interactions and fundamental nature.

**Essential role of $Z$-pole experiments** Experiments at the $Z$ pole are indispensable for both physics research and experimental calibration. Their importance stems from two key advantages: (1) The $Z$-pole cross-section is exceptionally large, enabling rapid accumulation of lepton and jet data for detector and software calibration; (2) The dataset collected even in brief runs will surpass the total integrated luminosity of Large Electron-Positron Collider (LEP), facilitating comprehensive $Z$ boson studies, from precision electroweak tests to searches for rare decays, and an extensive flavor physics program.

**Strategies for $W$ boson studies** The 91 – 240 GeV energy range is ideal for $W$ boson research. Near the threshold ($\sim$ 160 GeV), dedicated energy scans can yield high-precision measurements of the $W$ mass and width. Simultaneously, the copious $W$-pair production at 240 GeV enables systematic studies of $W$ decays and production dynamics.

**Future-proofing with top quark capabilities** With an energy and luminosity upgrade, such a collider could extend operations to $\sqrt{s} = 360$ GeV, unlocking two critical opportunities: (1) Threshold scans of $t\bar{t}$ production to measure the top quark mass and width with minimal systematic uncertainties; (2) Clean, high-statistics $t\bar{t}$ samples for detailed studies of top quark decays and couplings—complementing LHC results in a low-background environment.

### 1.1.2 Key physics opportunities

The CEPC study group has published several white papers on the Higgs boson, flavor physics, and new physics [5, 6, 7], describing the physics goals of the CEPC in detail. Those physics studies, among others, are briefly reviewed in this section.

#### 1.1.2.1 Higgs boson physics and electroweak physics

The CEPC aims to perform the most precise measurements of Higgs boson properties ever achieved. As an electron-positron Higgs boson factory, the CEPC will identify Higgs boson signals via the recoil mass method, enabling model-independent measurements. The clean collision environment allows nearly complete reconstruction of Higgs boson events, with up to four million Higgs bosons expected during nominal operation. This large dataset will enable comprehensive characterization of Higgs boson properties through innovative detector design and advanced reconstruction techniques.

The CEPC will determine Higgs boson couplings with unprecedented precision, reaching relative accuracies from percentage to permille levels - one order of magnitude better than the ultimate precision of the HL-LHC [5]. Of particular importance is the measurement of the $HZZ$ coupling ($g_{HZZ}$) at permille-level accuracy, which provides critical insight into the origin of matter. This precision complements gravitational wave searches for strong first-order phase transitions in early universe models, creating a powerful synergy between collider and cosmic frontier experiments in probing baryogenesis scenarios. In addition, the CEPC's Higgs boson program includes several key measurements: Higgs boson mass determination with uncertainty better than 10 MeV [8]; Higgs boson width measurement at percentage-level accuracy, sensitive to potential beyond-SM decay channels; and precision determination of CP properties surpassing HL-LHC capabilities [6]. These measurements will significantly advance our understanding of electroweak symmetry breaking and vacuum stability. The coupling measurements will test the SM with particular sensitivity to the top-Higgs system, where deviations could indicate new physics. The CEPC also excels in detecting dark matter candidates ($0.1 - 100$ GeV mass range) and probing new physics through Higgs boson portal interactions, especially via exotic decay searches.

The comprehensive Higgs boson coupling measurements will test fermionic (quark and lepton) and bosonic ($W$, $Z$) interactions at percent-level precision. These measurements provide powerful constraints on potential SM deviations, with particular sensitivity to anomalies in the top-Higgs sector that could reveal new particles or interactions.

#### 1.1.2.2 New physics searches

The CEPC will conduct new physics searches through diverse channels, including hunting for new particles produced in association with the Higgs boson, signals emerging from heavy particle decays, or indirect effects revealed through high-precision measurements of known processes. Even if new particles exceed CEPC's direct production kinematic threshold, their effects might appear as tiny deviations in properties of the Higgs boson, $W/Z$ bosons, or top quarks. An overview of the CEPC reach to BSM physics has recently been published [6].

The Higgs boson could play a crucial role in addressing open questions in the SM, such as the origin of dark matter and neutrino masses, the stability of the Universe, and the self-consistency of the particle physics theory at quantum level. Studying the Higgs boson with the highest possible precision is therefore a promising path toward deeper insights. The CEPC can provide unprecedented opportunities for making fundamental discoveries of BSM physics: For instance, it could help identify the origin of matter, especially the mechanism related to the first-order phase transition in the early Universe. CEPC has the potential to cover most of the theoretical phase space predicted by relevant new physics models, such as electroweak baryogenesis [9]. A first-order electroweak phase transition could also generate detectable gravitational wave signals. CEPC measurements will coincide with those expected from the next generation of gravitational wave detectors such as LISA [10], Taiji [11, 12], and TianQin [13, 14], offering a powerful synergy between terrestrial and astrophysical probes [15]. The CEPC could discover dark matter, particularly dark matter particles with a mass between one tenth and 100 times the proton mass [16, 17, 18, 19]. If a new "dark force" exists between dark matter particles, studies have also demonstrated that a CEPC-like collider is particularly powerful in testing scenarios where the dark force carrier is relatively light. In this regard, the CEPC is highly complementary to many low-background, deep underground dark matter direct detection experiments. The CEPC could observe an array of new physics smoking guns, with sensitivities orders of magnitude better than those of existing facilities, including: long-lived particles [20]; exotic decays of the Higgs and $Z$ bosons [21, 22]; rare or SM–forbidden decays of heavy quarks and leptons [7]; exotic mono-photon events; and many others. In addition, the CEPC provides excellent discovery potential for heavy neutrino partners and new light particles such as axion-like particles [23] and dark photons. Such discoveries could point toward a more complete fundamental theory, such as supersymmetry [24]. Together, these endeavors aim to surpass the Standard Model, addressing fundamental puzzles: the genesis of matter, the nature of dark matter, and the revelation of uncharted physical phenomena that could redefine our understanding of the cosmos.

#### 1.1.2.3 Flavor physics

The large $Z$ boson sample produced at the CEPC will enable precision heavy-flavor measurements that surpass current statistical limitations. Its clean experimental environment and advanced detectors facilitate unprecedented precision in flavor physics studies [7].

The CEPC's flavor program will conduct rigorous Standard Model tests, particularly in resolving the persistent discrepancies between inclusive and exclusive determinations of CKM matrix elements $|V_{cb}|$ and $|V_{ub}|$ through complementary measurements of semileptonic $B$ decays and direct $W$-boson decays at the $WW$ threshold [25, 26, 27]. The program will precisely measure $CP$-violating phases in mixing-induced and direct decays, test CKM unitarity [28, 29, 30], and examine lepton flavor universality via ratios like $R(D^*)$ in $B$-meson decays [31], providing sensitive probes for new physics.

A key strength of the CEPC lies in its sensitivity to rare processes, including searches for flavor-changing neutral currents in $B_s \to \phi\nu\nu$ and $B \to K^{(*)}\tau^+\tau^-$ decays [32, 33], lepton flavor violation in $\tau \to \mu\gamma$ and $Z \to \ell\ell'$ transitions [34], and tests of baryon number conservation through exotic decays such as $\Lambda_b^0 \to \pi^-\ell^+$.

The CEPC will significantly impact heavy-flavor spectroscopy by producing over $10^{11}$ bottom and charm hadrons, enabling systematic studies of exotic states including tetraquarks ($T_{bb}$, $T_{cc}$) [35, 36, 37, 38] and doubly-heavy baryons ($\Xi_{cc}$, $\Xi_{bb}$) [39, 40]. These investigations will clarify the nature of exotic hadrons and their production mechanisms.

Extended physics reach comes from the CEPC's higher-energy operations, including searches for flavor-violating Higgs boson decays ($H \to bs$, $H \to \tau\mu$) [41, 42] and studies of rare top decays ($t \to cH$) sensitive to new physics in top-Higgs couplings [43, 44, 45].

The CEPC's unique advantages — such as negligible pileup, complete collision energy knowledge, and superior vertexing — provide unmatched capabilities for final states containing neutrals, soft particles, or multiple bodies. By combining $Z$-pole precision with Higgs boson factory and top-quark capabilities, the CEPC offers a comprehensive flavor physics program with discovery potential extending to mass scales beyond 10 TeV [7].

#### 1.1.2.4 QCD Physics

The CEPC offers unique capabilities for exploring both perturbative and non-perturbative aspects of QCD through its high luminosity and precision detector systems. The facility enables comprehensive studies ranging from fundamental strong interaction dynamics to precision tests of Standard Model predictions, addressing critical challenges in hadronic physics. Key measurements include determinations of the strong coupling constant $\alpha_S$ with unprecedented accuracy [46], studies of fragmentation processes, detailed jet substructure analyses [47], and searches for novel hadronic states.

The CEPC's QCD research program focuses on four interconnected areas of investigation. First, it will probe hadronization mechanisms through jet substructure measurements and searches for exotic states such as charmonium-like hybrids, providing crucial insights into non-perturbative QCD phenomena. Second, the collider will test high-energy QCD dynamics, including gluon density evolution, against theoretical predictions. Third, the program will examine QCD-electroweak interference effects through precision measurements of $Z/W$ boson decays and flavor observables, enabling improved determinations of electroweak parameters such as the top quark mass. Finally, the CEPC will pioneer innovative approaches including AdS/QCD modeling [48] and machine learning-enhanced jet reconstruction techniques to connect theoretical predictions with experimental observations.

These studies will be further enhanced through synergies with complementary facilities like the Electron-Ion Collider (EIC), particularly in constraining parton distribution functions and understanding QCD phenomena across different energy scales. By resolving outstanding questions in strong interaction dynamics, exotic matter formation, and high-temperature QCD thermodynamics, the CEPC will significantly advance our understanding of fundamental forces while driving technological progress in computational QCD, detector development, and data analysis methods. The full realization of this program's potential will require close collaboration between experimentalists and theorists across the international particle physics community.

#### 1.1.2.5 Two photon collision physics

The CEPC's high-luminosity electron-positron collisions enable comprehensive studies of two-photon interactions, providing a unique probe of both Quantum ElectroDynamics (QED) processes and BSM phenomena [49]. In this process, virtual photons emitted by the colliding beams interact to produce diverse final states without requiring direct beam-particle annihilation. The CEPC's tunable collision energy and ultra-high integrated luminosity, exceeding $10^{35}$ cm$^{-2}$s$^{-1}$ at optimized configurations, make it particularly well-suited for this physics program.

The two-photon research program encompasses several key areas of investigation. First, it enables precision lepton and meson spectroscopy through measurements of lepton pair production processes such as $\gamma\gamma \to \mu^+\mu^-$ and $\tau^+\tau^-$, which test QED predictions at high invariant masses while searching for lepton flavor violation and anomalous magnetic moments. Second, the program includes studies of exotic mesons including glueball candidates and hybrid states via $\gamma\gamma \to \eta, \eta', \chi_c$ channels, resolving their quantum numbers and decay dynamics to constrain QCD confinement models.

Third, the CEPC will investigate Higgs boson production through $\gamma\gamma$ fusion. While Higgs boson production at 240 GeV primarily occurs via $e^+e^- \to ZH$, the rare $\gamma\gamma \to H$ channel mediated by loop processes offers an independent verification of Higgs boson couplings to photons and fermions. This complementary approach enhances sensitivity to BSM contributions in the Higgs boson sector. Fourth, the program includes searches for BSM physics such as axion-like particles and dark sector mediators through $\gamma\gamma \to$ ALP $\to \gamma\gamma/l^+l^-$ resonances, leveraging the CEPC's low-background environment to access sub-GeVmass ranges with high coupling precision.

The successful execution of this research program puts strict demands on the CEPC's technical capabilities, including near-$4\pi$ detector coverage, finely segmented calorimeters, and vertex detectors that enable precise reconstruction of $\gamma\gamma$-induced events with minimal pileup contamination. Addressing challenges such as beam-related backgrounds from synchrotron radiation photons requires advanced beam optics tuning and real-time veto algorithms. Integrating $\gamma\gamma$ physics into a staged run plan through dedicated low-energy runs could further enhance the discovery potential, positioning the CEPC as a multi-purpose facility that bridges QED, QCD, and BSM physics frontiers.

#### 1.1.2.6 Top physics

Based on the planned physics program following the CEPC energy upgrade, comprehensive top-quark studies will constitute a key component, conducted through two primary phases.

The initial phase will feature a precision threshold scan around 345 GeV to perform a model-independent measurement of the top-quark mass and total width with $\mathcal{O}(10)$ MeV accuracy [50]. This dramatically improved mass precision, when combined with excellent measurements of the Higgs boson mass and electroweak observables (particularly $m_W$), will powerfully constrain the oblique parameters $S$ and $T$ in the global electroweak fit [4], providing sensitive probes for beyond Standard Model physics. Furthermore, the same high-precision threshold dataset enables investigations of toponium, the $t\bar{t}$ quarkonium bound state, properties and stringent tests of CP violation in the top-quark sector [51].

Subsequent data collection at 360 GeV, where the top-quark pair production cross-section is substantially larger, will accumulate a significant sample of clean $t\bar{t}$ events. This dataset facilitates detailed studies of the top quark's production mechanism and decay properties, particularly its couplings to $\gamma$ and $Z$ bosons, which have high sensitivity to potential deviations from Standard Model predictions [52, 53].

Collectively, these measurements will determine the top-quark mass with $\mathcal{O}(10)$ MeV precision while providing multiple stringent tests of the Standard Model framework. Concurrently, operation at 360 GeV offers significant benefits for Higgs boson physics. The larger center-of-mass energy substantially enhances the Higgs boson production cross-section via vector boson fusion, particularly the $W$-fusion process. The large integrated luminosity accumulated during top-quark data collection will therefore enable precise determination of the Higgs boson width and other key properties, advancing our understanding of the Higgs boson sector.

## 1.2 CEPC operation scenarios

The CEPC is designed to operate at center of mass energies that scale from the $Z$-pole at 91 GeV to $WW$ at 160 GeV, $ZH$ at 240 GeV, and up to $t\bar{t}$ at 360 GeV. The project employs a two-phase operational strategy – baseline and upgraded – to accelerate its readiness and maximize its scientific output. The baseline scenario operates at center-of-mass energies from 91 GeV to 240 GeV with a Synchrotron Radiation (SR) power up to 30 MW, while the upgraded scenario increases this power to 50 MW for higher luminosities and potentially extends the energy reach to 360 GeV.

The baseline operational plan is described in Table 1.1. It allows to achieve good physics results with minimum resources by utilizing one single detector operating during an extensive period of time. In this baseline scenario, the CEPC will first operate with synchrotron radiation power of 30 MW at $\sqrt{s}$ = 240 GeV for 15 years to produce 2.0 million Higgs boson particles, followed by four years of operation with synchrotron radiation power of 12.1 MW as a Super $Z$ factory to create $5.6\times10^{11}$ $Z$ bosons, and then one year dedicated to a $W^+W^-$ production threshold scan between 155–170 GeV to refine the $W$-mass measurement to a precision of $\sim$ 1 MeV using energy-dependent cross-section profiles. Short operation periods at $\sqrt{s}$ = 91 GeV interleaved within the initial $ZH$ operation will provide detector calibration and aid on background modeling. The key parameters of the baseline accelerator design [3], such as synchrotron radiation power and peak luminosity, are summarized in Table 1.2. The peak instantaneous luminosity achievable at the $Z$ pole operation, $26 \times 10^{34}$ cm$^{-2}$s$^{-1}$ is dependent on the detector solenoid magnetic field. The value reported here assumes a 3 Tesla solenoid.

**Table 1.1:** CEPC baseline operation plan at different center-of-mass energies ($\sqrt{s}$), and corresponding anticipated peak instantaneous luminosity ($\mathcal{L}$), total integrated luminosity ($\int \mathcal{L}$) and event yields. The integrated luminosity assumes only one operational Interaction Point (IP). (*) The maximum instantaneous luminosity achievable at the $Z$ pole operation is dependent on the detector solenoid magnetic field. The value reported here assumes a 3 Tesla solenoid. For a 2 Tesla magnet, the luminosity will be 78% higher. (†) The cross section of $W^+W^-$ at threshold is about 8 pb. An additional $1.5 \times 10^8$ $W^+W^-$ events will be produced during the Higgs boson factory operation.

| Operation mode | $\sqrt{s}$ (GeV) | SR power (MW) | $\mathcal{L}$ ($10^{34}$ cm$^{-2}$s$^{-1}$) | $\int \mathcal{L}$/year (ab$^{-1}$) | Years | Total $\int \mathcal{L}$ (ab$^{-1}$) | Event yields |
|---|---|---|---|---|---|---|---|
| $H$ | 240 | 30 | 5 | 0.65 | 15 | 10 | $2.0 \times10^6$ |
| $Z$ | 91 | 12.1 | 26(*) | 3.2 | 4 | 13 | $5.6 \times 10^{11}$ |
| $W^+W^-$ | 155-170 | 30 | 16 | 1.2 | 1 | 1.2 | $1.0 \times 10^7$(†) |

The upgraded scenario (Table 1.3) can be deployed as soon as resources are identified. Two detectors installed at the interaction points will allow for a more compact operational plan, as detailed in Table 1.4, while accumulating a larger total integrated luminosity. In such case, the CEPC will operate with synchrotron radiation power of 50 MW at $\sqrt{s}$ = 240 GeV to deliver 4.3 million Higgs boson events (21.6 ab$^{-1}$) in 10 years, enabling stronger studies of rare decays ($H \to \mu^+\mu^-$, invisible modes) and constraints on the Higgs boson self-coupling. This energy also provides high statistics ($3.2\times10^8$ pairs) of $W^+W^-$ events for electroweak studies. Operation at the $Z$-pole (91 GeV) is designed to accumulate up to $4.1 \times 10^{12}$ $Z$-boson events in two years, providing unparalleled sensitivity to flavor physics (e.g., $Z \to b\bar{b}$, $c\bar{c}$ asymmetries), invisible decays, and searches for dark photons. The 50-MW power mode will also allow to increase the integrated luminosity for the $WW$ threshold scan to 6.9 ab$^{-1}$.

Following the successful run of this program, an additional optional energy upgrade to and above the $t\bar{t}$ threshold (340–360 GeV) would provide $0.6 \times 10^6$ $t\bar{t}$ events during a 5-year period. These can enable a top-quark mass measurement with a precision of $O(10)$ MeV with the scan method and probe the deviations of top-quark couplings from the SM.

**Table 1.2:** Main beam parameters for the CEPC baseline operation at three center-of-mass energies, based on the updated designs in [54] and [55]. (*) Maximum instantaneous luminosity achievable at the $Z$ factory operations depends on the detector solenoid magnetic field. The value reported here assumes 3 Tesla solenoid.

| | **Operational mode** | | |
|---|---|---|---|
| | **Higgs** | $\boldsymbol{Z}$ | $\boldsymbol{W^+W^-}$ |
| Number of IPs | | 1 | |
| Circumference (km) | | 99.955 | |
| SR power per beam (MW) | 30 | 12.1 | 30 |
| Beam energy (GeV) | 120 | 45.5 | 80 |
| Number of bunches | 268 | 3978 | 1297 |
| Bunch spacing (ns) | 553.9 | 69.2 | 184.6 |
| Train gap (%) | 55 | 17 | 9 |
| Beam current (mA) | 16.7 | 325.0 | 84.1 |
| Beam size at IP $\sigma_x/\sigma_y$ ($\mu$m/nm) | 14/36 | 6/72 | 13/42 |
| Energy spread (natural/total) (%) | 0.10/0.17 | 0.04/0.13 | 0.07/0.14 |
| Half crossing angle at IP (mrad) | | 16.5 | |
| RF frequency (MHz) | | 650 | |
| Peak luminosity ($10^{34}$cm$^{-2}$s$^{-1}$) | 5.0 | 26(*) | 16 |

**Table 1.3:** Main beam parameters for the CEPC upgraded scenario with maximize luminosity and extended center-of-mass energy. Values for four center-of-mass energies are provided. (*) Maximum instantaneous luminosity achievable at the $Z$ factory operations depends on the detector solenoid magnetic field. The value reported here assumes a 2 Tesla solenoid.

| | **Operational mode** | | | |
|---|---|---|---|---|
| | **Higgs** | $\boldsymbol{Z}$ | $\boldsymbol{W^+W^-}$ | $\boldsymbol{t\bar{t}}$ |
| Number of IPs | | 2 | | |
| Circumference (km) | | 99.955 | | |
| SR power per beam (MW) | | 50 | | |
| Beam energy (GeV) | 120 | 45.5 | 80 | 180 |
| Number of bunches | 446 | 13104 | 2162 | 58 |
| Bunch spacing (ns) | 277.0 | 23.1 | 138.5 | 2585.0 |
| Train gap (%) | 63 | 9 | 10 | 55 |
| Beam current (mA) | 27.8 | 1340.9 | 140.2 | 5.5 |
| Beam size at IP $\sigma_x/\sigma_y$ ($\mu$m/nm) | 14/36 | 6/35 | 13/42 | 39/113 |
| Energy spread (natural/total) (%) | 0.10/0.17 | 0.04/0.15 | 0.07/0.14 | 0.15/0.20 |
| Half crossing angle at IP (mrad) | | 16.5 | | |
| RF frequency (MHz) | | 650 | | |
| Peak luminosity ($10^{34}$cm$^{-2}$s$^{-1}$) | 8.3 | 192(*) | 26.7 | 0.8 |

**Table 1.4:** CEPC upgraded operation plan at different center-of-mass energies ($\sqrt{s}$), and corresponding anticipated peak instantaneous luminosity ($\mathcal{L}$), total integrated luminosity ($\int \mathcal{L}$) and event yields. (*) The maximum instantaneous luminosity achievable at the Z pole operation is dependent on the detector solenoid magnetic field. The value reported here assumes a 2 Tesla solenoid. (†) The cross section of $W^+W^-$ at threshold is about 8 pb. An additional $3.2 \times 10^8$ $W^+W^-$ events will be produced during the Higgs boson factory operation.

| Operation mode | $\sqrt{s}$ (GeV) | SR power (MW) | $\mathcal{L}$ ($10^{34}$ cm$^{-2}$s$^{-1}$) | $\int \mathcal{L}$/year (ab$^{-1}$, 2 IPs) | Years | Total $\int \mathcal{L}$ (ab$^{-1}$, 2 IPs) | Event yields |
|---|---|---|---|---|---|---|---|
| $H$ | 240 | 50 | 8.3 | 2.2 | 10 | 21.6 | $4.3 \times 10^6$ |
| $Z$ | 91 | 50 | 192(*) | 50 | 2 | 100 | $4.1 \times 10^{12}$ |
| $W^+W^-$ | 155-170 | 50 | 26.7 | 6.9 | 1 | 6.9 | $5.5 \times 10^7$ |
| $t\bar{t}$ | 360 | 50 | 0.8 | 0.2 | 5 | 1.0 | $0.6 \times 10^6$ |

## 1.3 CEPC Reference Detector

The design of the CEPC detector faces significant challenges imposed by the collider's operating environment while meeting stringent performance requirements needed to deliver a precision physics program that tests the SM and searches for new physics. These specifications include large and precisely defined solid angle coverage, excellent particle identification, precise energy and momentum measurements, efficient vertex reconstruction, and excellent jet reconstruction with flavor tagging capabilities.

The design of the CEPC detectors has evolved since the Preliminary Conceptual Design Report [56] in 2015, culminating in the Reference Detector described here. The work benefitted extensively from international collaboration with a broad community of physicists that have for long been dedicated to study future electron-positron collider detectors, such as for the International Linear Collider (ILC) [57].

Two primary detector concepts were studied at the time of the CDR [4]. A baseline detector concept, with two approaches to the tracking system, incorporated the particle flow principle with a precision vertex detector, a TPC, a silicon tracker, a 3-Tesla solenoid, and a high granularity calorimeter followed by a muon detector. A variant of the baseline detector concept incorporated a full silicon tracker. These were derived from the ILD and Silicon Detector (SiD) detector concepts for the ILC. An evolution of ILD [58] now being proposed for the CEPC is discussed in Chapter 18.7.

At the time of the CDR, the alternative detector concept, IDEA, relied on a different strategy for meeting the jet resolution requirements. It was based on dual-readout calorimetry with a precision vertex detector, a drift chamber tracker, a 2-Tesla solenoid, and a muon detector. An evolution of the IDEA [59] for the CEPC is discussed in Chapter 19, which is also being proposed for the Future Circular Collider (FCC).

Since then, different technologies for each detector subsystem have been actively pursued with R&D programs, while new ideas have emerged. The CEPC Reference Detector benefits from this extensive work, leveraging leading advances in detector development in the past few years, to introduce innovative solutions.

### 1.3.1 Requirements on detector design

The physics program demands that all possible final states from the decays of the $W$, $Z$, and Higgs bosons be separately identified and reconstructed with high resolution. To clearly discriminate between $H \to ZZ^* \to 4j$ and $H \to WW^* \to 4j$ final states, the calorimetry system must achieve energy resolutions for hadronic jets beyond current limits. The $H \to \gamma\gamma$ decay channel and searches for $H \to$ invisible decays impose additional requirements on energy and missing energy measurement resolutions. Measuring the Higgs boson coupling to the charm quark requires efficient discrimination between $b$-jets, $c$-jets, and light jets, while excellent sensitivity for the $H \to \mu\mu$ decay demands momentum resolution at the per mille level, driving the performance requirements for the vertex detector and tracking systems.

The experimental environment is characterized by significant beam induced backgrounds that present substantial design challenges. These include synchrotron radiation photons from bending magnets, incoherent $e^+e^-$ pair production from beamstrahlung, and off-energy particles from processes like radiative Bhabha scattering causing high hit densities and radiation doses. These conditions necessitate a detector capable of high-efficiency reconstruction and particle identification for electrons, muons, and jets, with excellent track momentum resolution. Achieving near-$4\pi$ geometric coverage is critical for accurate momentum and energy measurements and complete event topology reconstruction. The detector must resolve closely spaced particles in high-multiplicity events containing up to approximately 100 charged particles at 240 GeV, provide hadron identification for flavor physics, and measure integrated luminosity to precisions of 0.1‰ for Higgs boson studies and 0.01‰ for $Z$ boson measurements, with beam energy determination to 1 MeV for Higgs boson and 100 keV for $Z/W$ bosons.

Beam induced background mitigation strategies will be essential for maintaining detector performance. The detector requires real-time beam diagnostics including high-precision beam position monitors to correct orbit instabilities and beam loss monitors to detect and mitigate beam halo interactions with collimators. These systems need to work in tandem with physical masks to stop off-beam and synchrotron radiation particles, reducing background rates and improving signal-to-noise ratios.

A compact barrel-endcap configuration is required to minimize dead zones, while low-scattering materials before calorimeters are needed to reduce particle interactions before detection. An outermost muon system is required to ensure no high-momentum particles escape undetected, which is particularly vital for reconstructing multi-particle final states and suppressing background-induced fake signals.

The CEPC detector must therefore balance background suppression, geometric coverage,

and subsystem performance to unlock the collider's physics potential. Integrating advanced tracking, calorimetry, and particle identification technologies within a hermetic design ensures robust operation across the diverse CEPC physics program, from precision Higgs boson measurements to QCD phenomena and flavor physics studies.

### 1.3.2 Reference Detector design

The CEPC detector is a comprehensive apparatus designed for ultra-precise particle reconstruction, central to its Higgs boson and electroweak physics program. The tracking system integrates a silicon pixel vertex detector for precise vertexing, an inner silicon tracker providing good timing for pile-up rejection, and a large gaseous TPC for continuous tracking and excellent particle identification. The outer tracker employs innovative AC-coupled Low Gain Avalanche Detector (AC-LGAD) technology, providing both good spatial resolution and precise timing, playing double roles of aiding tracking and complementing particle identification. The system is engineered with ultra-low material budget to minimize multiple scattering. The calorimeter system is explicitly designed for PFA, featuring a high-resolution Bismuth Germanate (BGO) crystal ECAL and a high-granularity glass scintillator HCAL to achieve superior jet energy resolution. The detector incorporates a 3 T superconducting solenoid, which, along with the integrated iron return yoke, provides the magnetic field for momentum measurement and serves as the infrastructure for a muon identification system with an efficiency exceeding 95%. The detector solenoid magnetic field can be reduced to 2 T during the high-luminosity $Z$ factory operation to maximize instantaneous luminosity.

## 1.4 Performance and physics benchmarks

Physics benchmarks are essential for evaluating the CEPC detector's performance and the feasibility of its physics program. These benchmarks are defined based on key processes central to the CEPC's scientific goals, including Higgs, electroweak, flavor, and new physics studies. They establish clear performance targets that drive the optimization of the detector design.

For Higgs boson studies, benchmarks include achieving a mass resolution better than 10 MeVusing the model-independent process $e^+e^- \to ZH \to \ell^+\ell^- H$ ($\ell = e, \mu$), which requires excellent calorimeter energy resolution and precise tracking. The detector must also precisely measure the Higgs boson width and production rates via $W$-fusion, as well as its branching ratios to various final states. Key channels include $H \to \gamma\gamma$, demanding high photon energy and angular resolution, and hadronic decays such as $H \to b\bar{b}/c\bar{c}/s\bar{s}/gg/(WW^*)_h/(ZZ^*)_h$, which impose stringent requirements on jet energy resolution and flavor identification capabilities.

Electroweak precision tests require measuring the $W$ boson mass with an uncertainty of 1 MeVor better through dedicated threshold scans, and the $Z$ boson mass with an uncertainty below 0.1 MeV. The detector must also provide high-precision measurements of $W$ and $Z$ decay properties, including branching ratios to leptons and quarks. The ratio of $Z \to b\bar{b}$ to all hadronic decays serves as a key observable sensitive to the Higgs-top Yukawa coupling and potential beyond the Standard Model (BSM) contributions.

Flavor physics benchmarks are critical for testing the Standard Model and probing new physics through heavy-flavor hadron decays. Benchmark channels such as $D^0 \to \pi^+\pi^-\pi^0$ and $D^0 \to K^+\pi^-\pi^0$ are used to validate detector performance in secondary vertex reconstruction, charged/neutral pion discrimination, $K/\pi$ separation via $\mathrm{d}E/\mathrm{d}x$ or time-of-flight, and $\pi^0$ reconstruction with high photon energy resolution.

New physics benchmarks drive the detector design toward maximum sensitivity to BSM phenomena. These include searches for long-lived particles (LLPs), requiring precise vertex reconstruction and timing; measurements of Higgs boson invisible decays ($H \to$ invisible), demanding exceptional hermeticity and energy resolution to identify missing energy; and supersymmetric slepton production ($e^+e^- \to \tilde{\mu}^+\tilde{\mu}^-$), which requires excellent lepton identification and particle flow performance. These benchmarks collectively optimize detector parameters such as calorimeter granularity, coverage, and track momentum resolution to ensure broad sensitivity to new physics across multiple scales.

# Chapter 2 CEPC Reference Detector concept

The CEPC detector is designed to study Higgs bosons, electroweak interactions, and search for physics beyond the Standard Model (BSM). It is optimized primarily for the Higgs program, and for the low luminosity $Z$ runs dedicated to initial physics studies and calibrations, during the first 20 years of the CEPC operation. The design adopts many recently developed advanced technologies, yet some of detector components will likely be upgraded or replaced with future technologies to fully explore the physics potential of the tera-$Z$ program offered by the CEPC accelerator.

This chapter starts with the discussion in Section 2.1 of the physics requirements that drive the design of the Reference Detector. These are divided into requirements for detector subsystems related to tracking, leptons, photons, hadronic jets and missing energy.

The Reference Detector design is described in Section 2.2, including the tracking system, PFA-oriented calorimeter system, muon detector, magnet system, machine detector interface, and luminosity monitor. Section 2.3 briefly summarizes the major innovations and impact of the detector system.

## 2.1 Physics requirements

Operating as $H$, $W$ and $Z$ boson factories in an exceptionally clean environment, the CEPC places unprecedented demands on the detector: every decay product must be reconstructed, identified and measured with simultaneously high efficiency, purity and precision.

The CEPC physics program also requires the precise determination of both luminosities and beam energies. The integrated luminosity should be measured with a relative accuracy of 0.1% for the Higgs factory operation, and $\mathcal{O}(0.01\%)$ for the $Z$ factory operation. The beam energy must be controlled and known with an accuracy of the order of 1 MeV for the Higgs factory operation and 100 keV for the $Z$ factory operation and the $WW$ threshold scan. The requirements for reconstructed physics objects are summarized in Table 2.1.

### 2.1.1 Tracking

The detector should have excellent track finding efficiency and track momentum resolution. For tracks within the detector acceptance and transverse momenta above 1 GeV/$c$, a track-finding efficiency of better than 99% is required. To measure the $H \to \mu^+\mu^-$ signal and precisely reconstruct the Higgs boson mass from the recoil mass distribution with $ZH \to \ell^+\ell^- H$ events, a relative momentum resolution at the per mille level for momenta below 100 GeV/$c$is required. Silicon microstrip and pixel detectors provide sub-10 μm impact parameter resolution, enabling efficient $b/c-$ jet tagging through secondary vertex reconstruction. The $ZH$ process has a flat $\cos\theta$ distribution, whereas the background processes are concentrated in the forward region. A large solid angle coverage is essential for extensive acceptance and for separating different processes; hence, a coverage of up to $\lvert\cos(\theta)\rvert = 0.99$ is required.

**Table 2.1:** Physics objects and key observables used as benchmarks for setting the requirements and the optimization of the CEPC detector.

| **Physics objects** | **Measurands** | **Detector subsystem** | **Performance requirement** |
|---|---|---|---|
| Tracking | Coverage<br>Recon. efficiency<br>Resolution in barrel<br>Resolution in endcap | Tracker | $\lvert\cos\theta\rvert \le 0.99$<br>$\ge 99\%$ ($p_T > 1$ GeV/$c$)<br>$\sigma_{p_T}/p_T < 0.3\%$ ($\lvert\cos\theta\rvert \le 0.85$ )<br>$\sigma_{p_T}/p_T < 3\%$ ($\lvert\cos\theta\rvert > 0.85$ ) |
| Leptons ($e$, $\mu$) | PID efficiency<br>Mis-ID rate | Tracker, ECAL<br>HCAL, Muon | $\ge 99\%$ ($p > 5$ GeV/$c$, isolated)<br>$\le 2\%$ ($p > 5$ GeV/$c$, isolated) |
| Photons | PID efficiency<br>Mis-ID rate<br>Energy resolution | ECAL, HCAL | $\ge 95\%$ ($E > 3$ GeV, isolated)<br>$\le 5\%$ ($E > 3$ GeV, isolated)<br>$\sigma_E/E \le 3\%/\sqrt{E(\text{GeV})} \oplus 1\%$ |
| Vertex | Position resolution | Vertex | $\sigma_{r\phi} = 5 \oplus \frac{10}{p(\text{ GeV})\times\sin^{3/2}\theta}(\mu\text{m})$ |
| Hadronic jets | Energy resolution<br>Mass resolution | Tracker<br>ECAL, HCAL | $\sigma_E/E \sim 30\%/\sqrt{E(\text{GeV})} \oplus 4\%$<br>BMR $\le 4\%$ |
| Jet flavor tagging | b-tagging efficiency<br>c-tagging efficiency | Full detector | $\sim 80\%$, mis-ID of uds $< 0.3\%$<br>$\sim 50\%$, mis-ID of uds $< 1\%$ |
| Charged kaon | PID efficiency, purity | Tracker, TOF | $\ge 90\%$ (inclusive $Z$ sample) |

### 2.1.2 Leptons

Leptons (electrons and muons, denoted as $\ell$) are among the most important physics signatures and play a crucial role in the classification of different physics events. Efficient lepton identification with high purity is fundamental to the physics program. In particular $e^+e^- \to ZH \to \ell^+\ell^- H$ events are ideal for the recoil mass analysis.

Specifically, the detector must achieve lepton-identification efficiency above 99% with a misidentification rate below 2% for isolated leptons with momenta above 5 GeV/$c$.

These requirements are essential for identifying $H \to \tau^+\tau^-$ events and the semileptonic or leptonic decay modes of $H \to ZZ^*, WW^*$ events. Furthermore, leptons from heavy quark decays are important for jet flavor tagging and charge measurement, making accurate identification of leptons inside jets indispensable. Finally, achieving precise measurements requires a transverse momentum resolution at the 0.1% level for leptons with momenta below 100 GeV/$c$.

### 2.1.3 Photons

Photons are crucial for $H \rightarrow \gamma\gamma$ and $\pi^0 \rightarrow \gamma\gamma$ measurements, jet energy resolution, studies of radiative processes and $\tau$-leptons final states. The electromagnetic calorimeter is required to achieve photon energy resolution of $\sigma_E/E \leq 3\%/\sqrt{E(\mathrm{GeV})} \oplus 1\%$, essential for both $H \rightarrow \gamma\gamma$ and $\pi^0 \rightarrow \gamma\gamma$ reconstruction. For the successful reconstruction of unconverted and isolated photons with energies exceeding 3 GeV, an identification efficiency greater than 95% and a misidentification rate smaller than 5% are required.

### 2.1.4 Particle identification

Particle identification, particularly the identification of charged kaons, is important for the flavor physics program. Similar to the leptons in jets, identifying charged kaons is essential for jet flavor tagging and charge measurement. For an inclusive $Z$ boson sample, both the efficiency and purity of kaon identification must exceed 90%. Charged particle identification relies on a combination of Time-of-Flight (ToF) detectors ($\Delta t < 50$ ps) and gaseous detector TPC with $\mathrm{d}N_p/\mathrm{d}x$ measurement, using machine-learning techniques to count primary ionization clusters. This enables $\pi/K/p$ separation at $3\sigma$ level up to 20 GeV/$c$, thereby enhancing jet flavor tagging and enabling precise studies of QCD jet composition.

### 2.1.5 Hadronic jets and missing energy

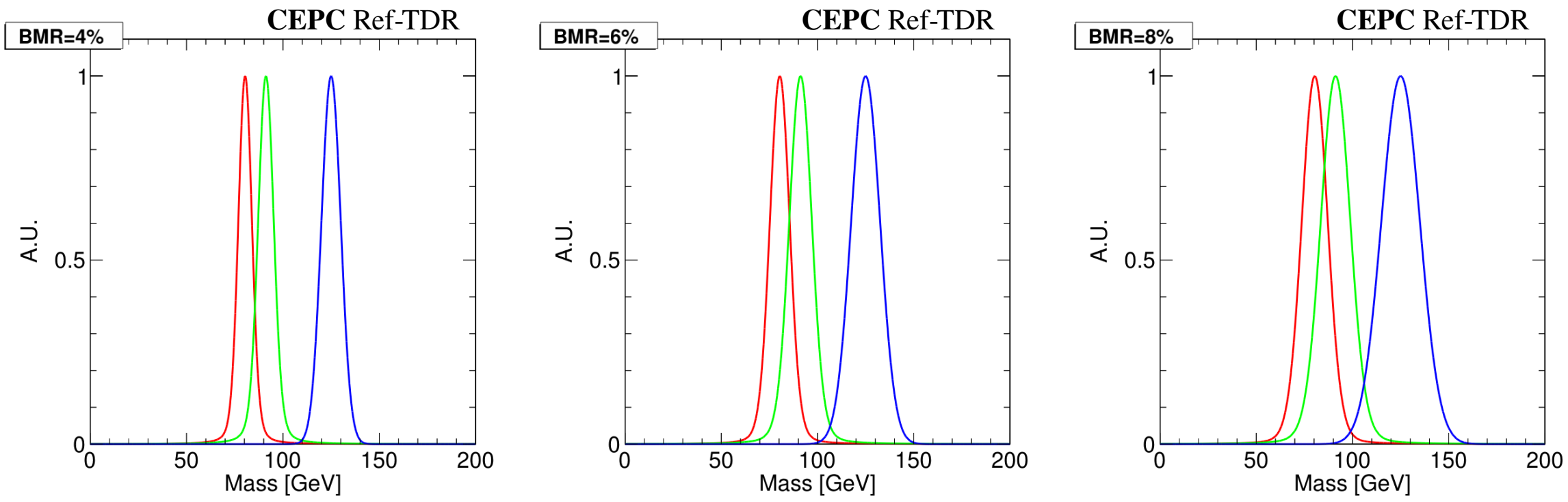

**(a)** **(b)** **(c)**

**Figure 2.1:** Separation of $W$ (red), $Z$ (green) and $H$ (blue) bosons in their hadronic decays with different BMR: (a) 4%, (b) 6% and (c) 8%, respectively.

The majority of the $H$, $W$ and $Z$ bosons decay into quarks or gluons, which then fragment into hadronic final states. These final states are traditionally reconstructed using jet algorithms. In events such as $ZZ \rightarrow \nu\nu qq$ and $ZH \rightarrow \nu\nu(bb, cc, gg)$, the invariant mass of the hadronic system serves as the primary discriminant variable. A physics observable called Boson Mass Resolution (BMR) [1], defined as the boson mass resolution of the hadronic system, is introduced to quantify detector performance. The hadronic system can also be reconstructed in events with leptons from $W$ and $Z$ boson decays and with high energy photons from radiation, as these leptons and photons can be cleanly identified and removed.

BMR is also a key metric to evaluate missing energy performance in electron-positron collisions, as missing energy resolution largely depends on hadronic energy measurements. For example, in $ZH$ production, the hadronic final states, $Z \to qq$, provides the most sensitive channel for searching for invisible Higgs decays. The dominant background arises from $ZZ \to \nu\nu qq$ production. Although the signal and background events are expected to share the same hadronic mass, they differ in missing mass (or the dijet recoil mass).

Separation of $W$, $Z$ and $H$ bosons in their hadronic decays with different BMR of 4%, 6% and 8% are shown in Figure 2.1a, Figure 2.1b and Figure 2.1c, respectively. As illustrated in the Figure, if the BMR exceeds 4%, the $ZZ$ background starts to overlap significantly with the $ZH$ signal. Therefore, a BMR of 4% or better is an essential benchmark, it is equivalent to a jet energy resolution of $30\%/\sqrt{E(\text{GeV})} \oplus 4\%$. This represents an energy resolution about a factor of two better than that achieved at LEP, the Stanford Linear Collider (SLC) and current LHC detectors – an exceptionally challenging performance target.

## 2.2 Reference Detector design

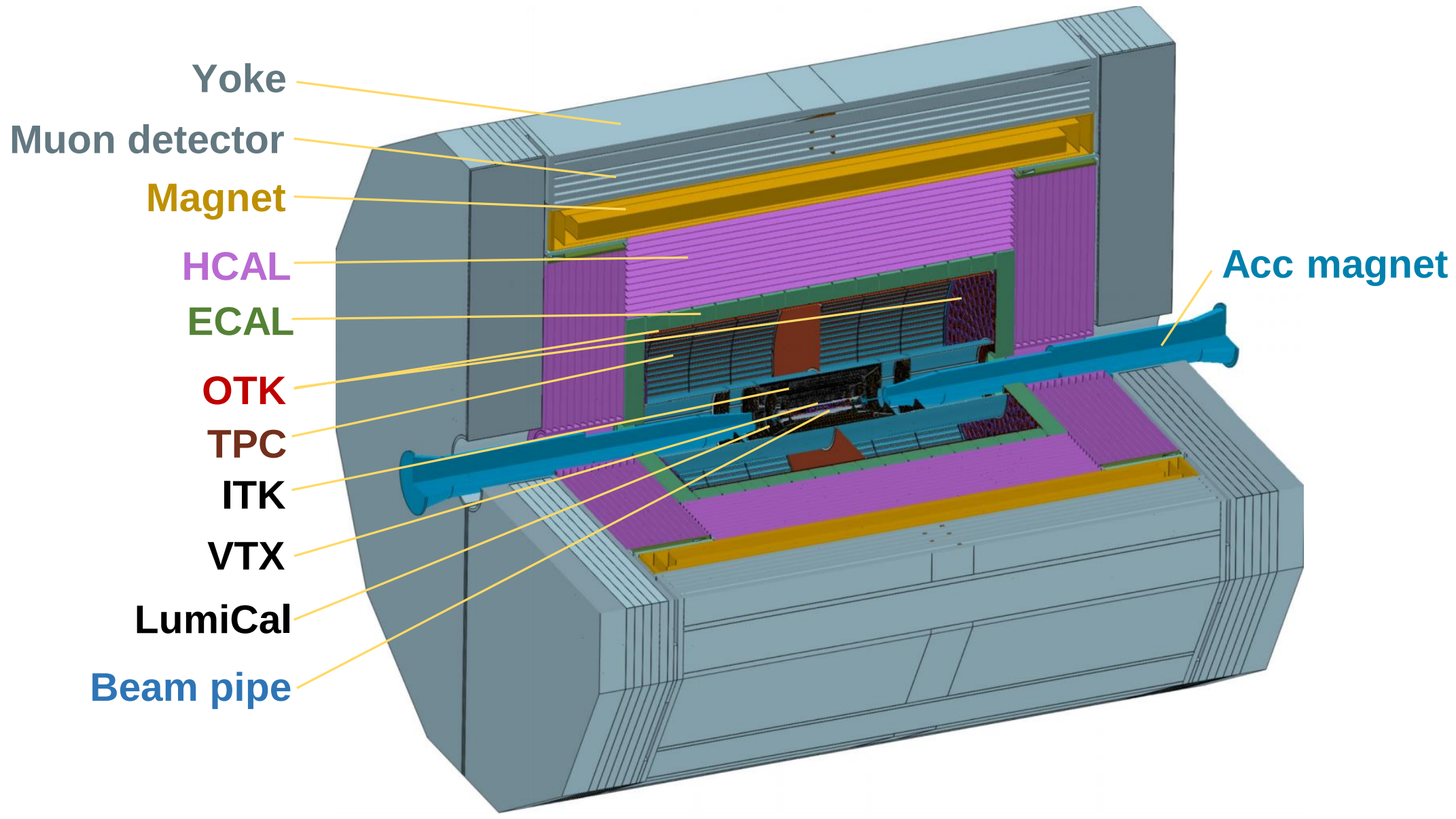

**Figure 2.2:** Schematic diagram of the Reference Detector design, illustrating its core subsystems. From the innermost to outermost, it includes a silicon vertex, an inner silicon tracker, a TPC tracker, an outer silicon tracker, ECAL, HCAL, superconducting solenoid, return yoke with a muon detector, and a dedicated calorimeter for luminosity monitoring which is mounted next to the beampipe.

Taking CEPC detector conceptual designs as a foundation [2], the Reference Detector has been further optimized and integrated with state-of-the-art, newly developed technologies, marking a significant advancement in instrumentation of particle physics. It is designed to achieve unprecedented precision in the study of Higgs bosons, electroweak interactions, and in searches for BSM phenomena as described in Chapter 1. At its core is PFA[3, 4, 5], a revolutionary approach that combines tracking and calorimetry data to reconstruct particle showers with unparalleled precision.

**Table 2.2:** Key technologies and parameters of the CEPC reference detector, the dimensions listed in the table are for the sensitive parts.

| Detector | Technology | In (mm) | Out (mm) | Comment |
|---|---|---|---|---|
| Vertex | Silicon pixel | $r_{in}$ = 11.1 | $r_{out}$ = 47.9 | 4 single, 2 double layers |
| ITK (barrel) | Silicon pixel | $r_{in}$ = 235.0 | $r_{out}$ = 555.6 | 3 layers |
| ITK (endcap) | Silicon pixel | $\|z_{in}\|$ = 505.0 | $\|z_{out}\|$ = 1489 | 4 disks |
| LumiCAL (front) | Silicon pixel | $r_{in}$ = 12<br>$\|z\|$ = 560 | $r_{out}$ = 42 | 1 double layer disk |
| LumiCAL (back) | Silicon pixel | $r_{in}$ = 12<br>$\|z\|$ = 640 | $r_{out}$ = 51 | 1 double layer disk |
| OTK (barrel) | Silicon microstrip | $r$ = 1800<br>$\|z\| \leq 2840$ | | 1 layer |
| OTK (endcap) | Silicon microstrip | $r_{in}$ = 406<br>$\|z\|$ = 2910 | $r_{out}$ = 1816 | 1 disk |
| TPC | Gaseous tracking | $r_{in}$ = 600<br>$\|z\| \leq 2900$ | $r_{out}$ = 1800 | MPGD high-granularity readout |
| ECAL (barrel) | Crystal + SiPM | $r_{in}$ = 1830<br>$\|z\| \leq 2900$ | $r_{out}$ = 2130 | 18 layers, 32 sectors |
| ECAL (endcap) | Crystal + SiPM | $r_{in}$ = 350<br>$\|z_{in}\|$ = 2930 | $r_{out}$ = 2130<br>$\|z_{out}\|$ = 3230 | $1.5 \times 1.5 \times 3$ cm$^3$ 3D granularity<br>18 layers, 24 $X_0$, 1.2 $\lambda_I$ |
| LumiCAL (front) | LYSO + SiPM | $r_{in}$ = 12<br>$\|z_{in}\|$ = 647 | $r_{out}$ = 56<br>$\|z_{out}\|$ = 670 | 14 layers<br>2 $X_0$ |
| LumiCAL (back) | LYSO + SiPM | $r_{in}$ = 12<br>$\|z_{in}\|$ = 800 | $r_{out}$ = 110<br>$\|z_{out}\|$ = 950 | 10 layers<br>13.4 $X_0$ |
| HCAL (barrel) | Glass scintillator + SiPM | $r_{in}$ = 2140<br>$\|z\| \leq 3230$ | $r_{out}$ = 3455 | 48 layers, 16 sectors |
| HCAL (endcap) | Glass scintillator + SiPM | $r_{in}$ = 400<br>$\|z_{in}\|$ = 3260 | $r_{out}$ = 3385<br>$\|z_{out}\|$ = 4575 | $4 \times 4 \times 1$ cm$^3$ pixel<br>48 layers, 6 $\lambda_I$ |
| Muon (barrel) | Plastic scintillator + SiPM | $r_{in}$ = 4245<br>$\|z\| \leq 4475$ | $r_{out}$ = 5185 | 6 layers, 12 sectors |
| Muon (endcap) | Plastic scintillator + SiPM | $r_{in}$ = 550<br>$\|z_{in}\|$ = 4635 | $r_{out}$ = 5085<br>$\|z_{out}\|$ = 5875 | |

The Reference Detector as shown in Figure 2.2, is composed of a silicon pixel Vertex Detector (VTX), an inner silicon tracker, a TPC tracker surrounded by an outer silicon tracker, a crystal ECAL, a steel-glass scintillator sampling HCAL, a 3 Tesla superconducting solenoid, and a flux return yoke embedded with a muon detector. In addition, five pairs of silicon tracking disks are placed in the forwarded regions on both sides of the Interaction Point (IP) to extend the tracking acceptance to $|\cos\theta| < 0.99$. The key technologies and parameters of the Reference

Detector are summarized in Table 2.2. The following sections provide brief overview of the detector's subsystems, emphasizing its innovative design choices, materials, geometries, and performance metrics.

### 2.2.1 Machine detector interface and luminosity measurement

The MDI bridges the collider and detector, minimizing interference between beam dynamics and experimental measurements. Its centerpiece is an ultra-thin beryllium beampipe, with 10 mm inner radius, engineered to withstand vacuum pressures while minimizing multiple scattering. LumiCal detectors are positioned at ±560 mm from the IP, combining AC-LGAD silicon layers and Lutetium Yttrium Oxyorthosilicate (LYSO) crystals for luminosity measurement.

The LumiCal's dual functionality, tracking beam profiles via AC-LGAD strips and measuring integrated luminosity via LYSO energy deposition, exemplifies integration of multiple technologies into cohesive systems. Precision alignment mechanisms and radiation-hardened electronics ensure that MDI operates seamlessly under the demanding conditions of the collider. A detailed description of the MDI is shown in Chapter 3.

### 2.2.2 Tracking system

The tracking system is a crucial part of the detector. It is composed of a silicon pixel detector, the VTX, a silicon tracker, the Inner Silicon Tracker (ITK), a TPC tracker and an outer silicon layer, the Outer Silicon Tracker (OTK). The main parameters and layout of the tracking system are listed in Table 2.3 and shown in Figure 2.3. The system requires an overall track momentum resolution at the per mille level or better for momenta below 100 GeV/$c$, enabling precise measurements of Higgs properties and other key physics objectives.

Accurate position measurements from the OTK allow charged-particle tracks to be precisely extrapolated to the ECAL, improving the performance of PFA. Additionally, effective Particle Identification (PID) is crucial for distinguishing various particle types, supporting flavor physics studies and measurement of jet substructures. This can be achieved with ToF measurements in the OTK, providing a time precision of about 50 ps and enabling robust $\pi$/K/p separation below a few GeV/$c$.

#### 2.2.2.1 Silicon pixel vertex detector

The silicon pixel VTX is a crucial component of the CEPC detector system. It consists of multiple layers of silicon pixel sensors, which are capable of precisely measuring the position and trajectory of charged particles produced in the collisions. Positioned closest to IP, VTX is essential to resolve short-lived particles such as B mesons. Its compact geometry comprises six concentric layers of silicon pixel sensors, arranged to provide three-dimensional tracking. Each layer is required to measure charged particles with a spatial resolution of about 5 µm.

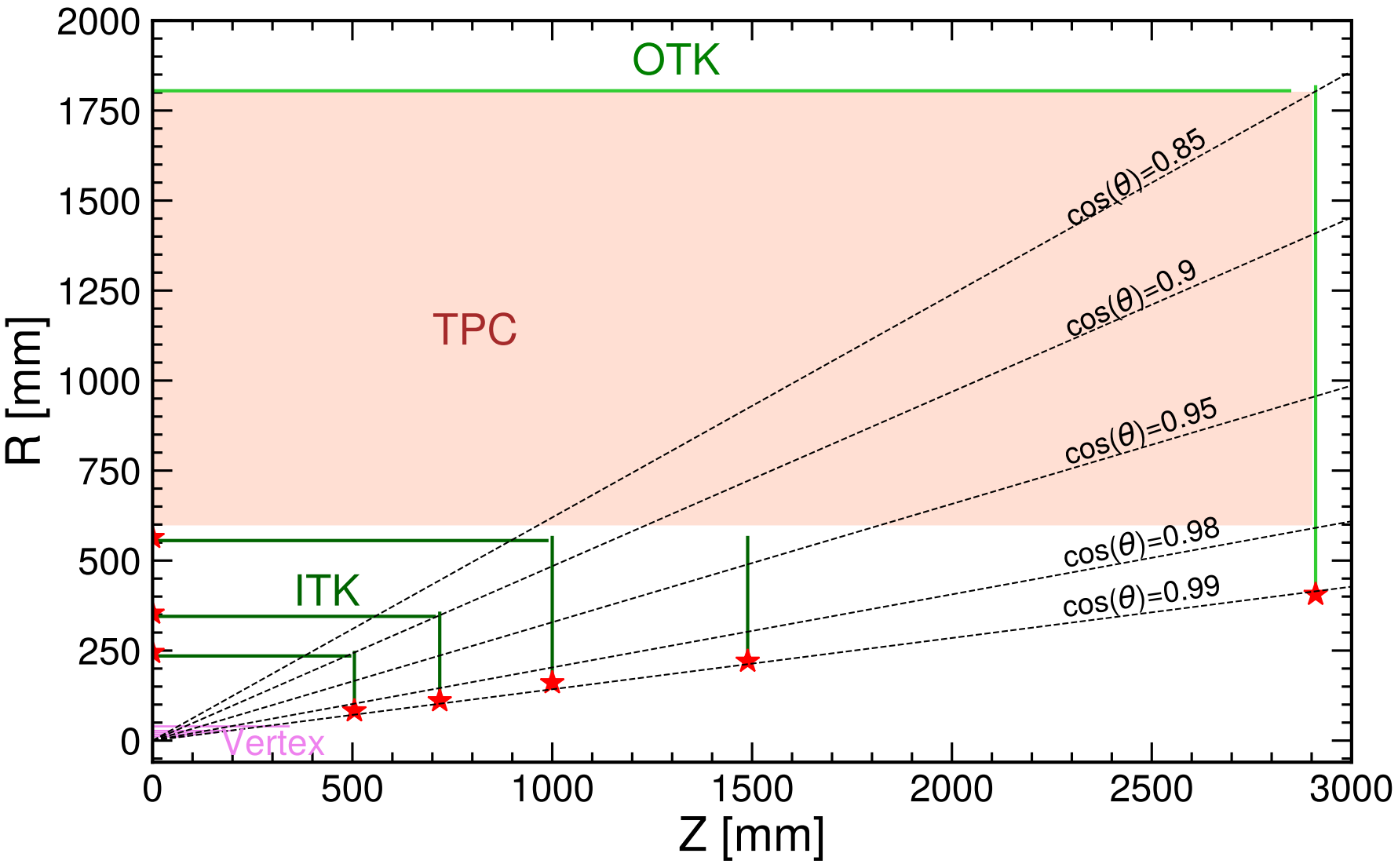

**Figure 2.3:** Layout of the tracking system of the Reference Detector. The innermost silicon vertex detector, shown in magenta lines, consists of six barrel layers. The green lines represent the ITK and OTK, which together comprise four barrel layers and five endcap disks on each side. The OTK provides both high-precision spatial (~ 10 µm) and timing (~ 50 ps) measurements, enabling ToF capability. The light orange area represents the TPC, which is embedded within the silicon tracker. The $z$-axis is defined to align with the beam direction, the $x - y$ plane is perpendicular to the beam direction, this plane spans horizontally and vertically across the detector's center (0,0,0).

This precision is critical for identifying secondary vertices from heavy-flavor decays such as B-hadrons.

The total material budget of the VTX (6 layers) is minimized to 0.8% of a radiation length $X_0$ with ultra-thin carbon fiber support structures and radiation-tolerant readout electronics. The design also incorporates novel stitching technology in the first 4 layers to further minimize the material budget. This mitigation of multiple scattering improves the measurements of low-momentum particle trajectories. Power consumption is optimized via Monolithic Active Pixel CMOS (MAPS) technology, which integrates sensor and readout circuitry onto a single chip, reducing heat generation and enabling dense pixel arrangements.

Further innovations in the VTX include an advanced air-cooling system that ensures thermal stability during operation. The vertex detector's configuration maximizes hermeticity, with overlapping sensor coverage to eliminate dead zones. These features collectively enable the VTX to achieve tracking efficiencies above >99% for particles originating near the IP, setting a new benchmark for vertex detection in collider experiments. A detailed description of the silicon pixel vertex detector is provided in Chapter 4.

**Table 2.3:** Parameters and layout of the CEPC tracker including VTX, ITK, TPC, and OTK. The column labeled $\pm z$ shows the half-length of the barrel layers, and the $z$ position of the endcap disks. The column labeled $\sigma_\phi$ and $\sigma_t$ represent the spatial resolution in the bending direction and time resolution, respectively.

| **Components** | | **Radius $R$** [mm] | | $\pm z$ [mm] | **Material** [$X_0$] | $\sigma_\phi$ [µm] | $\sigma_t$ [ns] |
|---|---|---|---|---|---|---|---|
| **Beam pipe** | | 10.55 | | 85.0 | 0.454% | — | — |
| **VTX** | Layer 1 | 11.1 | | 80.7 | 0.067% | 5 | 100 |
| | Layer 2 | 16.6 | | 121.1 | 0.059% | 5 | 100 |
| | Layer 3 | 22.1 | | 161.5 | 0.058% | 5 | 100 |
| | Layer 4 | 27.6 | | 201.9 | 0.061% | 5 | 100 |
| | Layer 5 | 39.5 | | 341.0 | 0.280% | 5 | 100 |
| | Layer 6 | 47.9 | | 341.0 | 0.280% | 5 | 100 |
| **Beam pipe protective cylinder** | | 66.3 | | 462.0 | 0.19% | — | — |
| **ITK Barrel** | Layer 1 (ITKB1) | 235.0 | | 493.3 | 0.68% | 8 | 3–5 |
| | Layer 2 (ITKB2) | 345.0 | | 704.8 | 0.68% | 8 | 3–5 |
| | Layer 3 (ITKB3) | 555.6 | | 986.6 | 0.68% | 8 | 3–5 |
| **OTK Barrel** | Layer 4 (OTKB) | 1,800.0 | | 2,840.0 | 1.58% | 10 | 0.05 |
| **TPC** | Inner wall | 600.0 | | 2,900.0 | 0.45% | — | — |
| | Gas | 605.4–1,779.6 | | 2,750.0 | 1.00% | 110–144 | — |
| | Outer wall | 1,800.0 | | 2,900.0 | 0.69% | — | — |
| | | $R_{\text{in}}$ | $R_{\text{out}}$ | | | | |
| **ITK Endcap** | Disk 1 (ITKE1) | 82.5 | 244.7 | 505.0 | 0.76% | 8 | 3–5 |
| | Disk 2 (ITKE2) | 110.5 | 353.7 | 718.5 | 0.76% | 8 | 3–5 |
| | Disk 3 (ITKE3) | 160.5 | 564.0 | 1,000.0 | 0.76% | 8 | 3–5 |
| | Disk 4 (ITKE4) | 220.3 | 564.0 | 1,489.0 | 0.76% | 8 | 3–5 |
| **OTK Endcap** | Disk 5 (OTKE) | 406.0 | 1,816.0 | 2,910.0 | 1.37% | 10 | 0.05 |

#### 2.2.2.2 Inner silicon tracker

Surrounding the VTX, the ITK comprises three barrel layers of single-sided silicon sensors at radii of 235 mm, 345 mm and 555.6 mm, complemented by four pairs of endcaps positioned along the beam axis, with a spatial resolution of $8\mu\text{m} \times 40\mu\text{m}$ per layer.

The ITK's design prioritizes precise time resolution to address the challenges of pile-up events at high collision rates. Advanced HV-CMOS sensors achieve a timing precision of 3–5 ns, enabling accurate track-to-bunch crossing association. The sensor performance is supported by low-noise DC-DC converters and other dedicated electronic components integrated onto Flexible Printed Circuit (FPC).

Material optimization is central to the ITK's success. Each stave consists of 7 to 14 modules (each containing 14 monolithic pixel sensors) and carbon fiber trusses with water-cooled embedded polyimide pipes, achieving a total material budget of 0.7% $X_0$. The staves are arranged in staggered configurations to ensure hermetic coverage, with 44, 64, and 102 staves populating the three barrel layers. Endcaps use honeycomb carbon fiber disks to support pixel modules, with front and back sensors offset to minimize dead areas. This design minimizes particle scattering while maintaining high tracking efficiency over the ITK's active area of approximately 20 $m^2$. A detailed description of the silicon inner tracker is shown in Chapter 5.

#### 2.2.2.3 Outer silicon tracker with precision timing

The outermost tracking layer, the OTK, integrates momentum measurement and ToF detection into a single subsystem. Its geometry comprises a single barrel layer (3.6 m diameter) and two endcap disks, covering a total area of $\sim$ 85 $m^2$. The OTK design maximizes the lever arm to enhance momentum resolution while minimizing extrapolation errors between the tracking system and the calorimeter.

The OTK's innovation lies in the use of AC-LGAD microstrip sensors. These devices combine 100 μm pitch strips with intrinsic gain mechanisms, enabling simultaneous position and time measurements. The spatial resolution reaches 10 μm, while the ToF resolution of 50 ps allows for the precise determination of the velocity of the particles that traverse the detector. This capability is crucial for particle identification and for searches of long-lived particles.

Mechanically, the OTK employs carbon fiber staves with embedded cooling channels, maintaining a material budget of 1.6% $X_0$. To match the length (5,680 mm) of the OTK barrel, 8 ladders are mounted on the TPC outer barrel in a row: 4 short ladders (700 mm) and 4 long ladders (720 mm). Each ladder comprises 8 modules, each module consists of four sensors arranged in a 2 × 2 configuration, together with a low-mass PCB that integrates DC-DC converters, readout ASICs, and data aggregation chips. Each endcap is segmented into 16 sectors. This design ensures uniform performance across the detector while accommodating the stringent spatial constraints for the assembly. A detailed description of the silicon outer tracker is shown in Chapter 5.

#### 2.2.2.4 Time projection chamber

The TPC serves as the gaseous tracker, offering continuous three-dimensional reconstruction of charged-particle trajectories. Housed within a cylindrical volume of an outer diameter of 3.6 m and length of 5.9 m, the TPC is divided into two drift regions by a central cathode held at -63 kV. Charged particles ionize the gas mixture (T2K gas), with liberated electrons drifting toward the readout planes under the influence of a uniform electric field.

One of the key parameters of the TPC is its inner and outer walls, which are constructed from lightweight composite materials, with a total material budget of merely 1.14% $X_0$. This design minimizes energy loss for traversing particles while maintaining mechanical stability. The readout planes use double-mesh Micromegas detectors and high-granularity readout with a pad size of 500 μm $\times$ 500 μm to achieve high spatial resolution, high rate capability, and excellent PID performance.

Particle identification is further enhanced by the TPC's ability to measure $dN_p/dx$ (the number of activated pads along the track segments) with a resolution better than 3%. This capability is bolstered by a laser calibration system that periodically aligns the drift field, ensuring consistent performance over time. The TPC's modular endplates, each hosting 244 readout modules, facilitate easy maintenance and upgrades, reflecting a design philosophy that balances cutting-edge performance with operational practicality. A detailed description of the TPC is shown in Chapter 6.

### 2.2.3 PFA-oriented calorimeter system

The calorimeter system consists of an electromagnetic calorimeter (ECAL) and a hadronic calorimeter (HCAL), designed to provide precise energy measurements of final-state particles such as electrons, photons, hadrons, and jets. Conventional calorimeters face performance limitations due to hadronic shower fluctuations and jet fragmentation effects. To fully exploit the physics potential—particularly in studying hadronic decays of the $H$, $W$ and $Z$ bosons—the BMR for jet final states should reach 3–4%. Achieving this represents a major technical challenge but is essential for precision Higgs and electroweak physics.

The PFA method [3, 4, 5] addresses this challenge by improving energy resolution through identification and separation of the responses of different particle components within a jet. This approach combines precise position and momentum measurements from the tracking system with the separation of charged and neutral deposits in the ECAL and HCAL. High granularity of the calorimeters is a key ingredient for PFA.

#### 2.2.3.1 Electromagnetic calorimeter

Several sampling ECAL prototypes have been built, typically achieving an energy resolution of about $15\%/\sqrt{E(\text{GeV})}$ using different technologies options. The baseline ECAL design employs BGO crystals as a full-absorption calorimeter, which significantly improves energy resolution, essential for optimal detection of processes such as $H \to \gamma\gamma$.

Its geometry features 18 longitudinal layers of BGO crystal bars, arranged orthogonally in successive layers with a transverse segmentation of $1.5 \times 1.5\ \text{cm}^2$. Each crystal bar spans 40 cm in length, providing at least 24 $X_0$ to maximize containment of electromagnetic showers.

BGO crystals were selected for their high density (7.13 g/cm$^3$) and superior light yield, enabling an energy resolution of $< 3\%/\sqrt{E(\text{GeV})}$ for electrons and photons. Silicon Photomultipliers (SiPMs) coupled to the crystals provide fast, low-noise readout, ensuring precise spatial reconstruction of shower profiles. Mechanical integration is achieved via carbon fiber modules housing the crystal bars, cooled by a dedicated liquid-based system to mitigate thermal noise. A detailed description of the ECAL is given in Chapter 7.

While BGO is the baseline option, backup options include silicon-tungsten [6, 7, 8, 9] and scintillator-tungsten designs [10, 11]. These alternatives, though less performant ($\sim$ $15\%/\sqrt{E(\mathrm{GeV})}$), offer proven reliability and faster response time, ensuring flexibility for detector construction. The ECAL's orthogonal crystal alignment and ultra-fine granularity represent a paradigm shift in calorimetry, enabling the separation of overlapping showers in dense jet environments — a critical requirement for Higgs boson studies.

#### 2.2.3.2 Hadronic calorimeter

The HCAL targets the measurement of hadronic jets with resolutions previously unattainable in collider experiments. Its geometry comprises a 48 layers barrel section (1.3 m thick) and two endcap disks, segmented into $4 \times 4$ cm$^2$ cells to satisfy the PFA's granularity requirements. The HCAL employs high-density glass scintillators (6 g/cm$^3$) as active material, chosen for their exceptional light yield (1000–2000 photons/ MeV) and short decay time ($O(500)$ ns). SiPMs read out the scintillation light, with approximately 5.3 million channels providing hermetic coverage across the barrel and endcaps.

HCAL prototypes built using plastic scintillator or RPC as active material [12, 13, 14, 15, 16, 17, 18, 19, 20], typically achieving a sampling fraction of about 1.5%, and single hadron energy resolution of $\sim 60\%/\sqrt{E(\mathrm{GeV})}$. In contrast, high density glass scintillator (GS) as thicker active layers dramatically improve the sampling ratio by a factor of about 20, significantly improving the single hadron energy resolution by a factor of 2, achieving $\sim 30\%/\sqrt{E(\mathrm{GeV})}$.

Mechanical design challenges are addressed via steel-reinforced support structures and a distributed cooling system to prevent thermal expansion-induced misalignment. The HCAL's barrel weighs 955 tons, with each endcaps weighing 367 tons, representing a balance between robustness and precision. This ensures containment of high-energy hadronic showers while maintaining the spatial resolution needed for jet substructure studies. A detailed description of the HCAL is presented in Chapter 8.

### 2.2.4 Solenoid magnet and return yoke

The detector employs a superconducting solenoid magnet, which can operate at either 3 T or 2 T (for $Z$-pole run). The coil is wound from NbTi/Cu Rutherford cables, spanning 9.05 m in length with inner and outer diameters of 7.07 m and 8.47 m, respectively. A thermal siphon cooling system maintains the magnet at 4.5 K. The solenoid provides excellent field uniformity, enabling precise momentum measurements.

Surrounding the solenoid, the 3,010-ton return yoke not only channels magnetic flux but also hosts the muon detection layers. This design optimizes muon track sampling while accommodating mechanical and thermal stresses. The modular construction allows phased installation and future upgrades.

An anti-solenoid magnet cancels the main solenoid's magnetic field to prevent beam distortion, ensuring a near-zero net field and less than 170 Gauss in quadrupoles. It is segmented into over 100 sections, reaching fields of 8.7 T (on-axis) and 10.7 T (in-coil). While critical for accelerator performance, it introduces local field inhomogeneities in the TPC (up to 0.03 T) and ITK (up to 0.6 T), affecting detector accuracy. The proximity of the anti-solenoid to the ITK exacerbates these effects, requiring careful simulation. Future studies will integrate quadrupole magnet effects to fully assess field homogeneity. Balancing field cancellation with detector accuracy remains the primary design challenge. A detailed description of the magnet system is shown in Chapter 10.

### 2.2.5 Muon detector

The Muon Detector is the final element in the particle identification chain, dedicated to reconstruct muons and distinguish them from hadronic punch-through particles. It consists of six concentric barrel layers and four endcap sectors, covering a total area of 4,800 $m^2$. Each layer employs extruded plastic scintillators read out via wavelength-shifting fibers and SiPMs, achieving a per-layer efficiency >95% and a time resolution of 1 ns.

The solenoid magnet return yoke is constructed from iron laminations, minimizing dead zones while providing the magnetic flux return path. Muon tracks traverse multiple scintillator layers embedded within the yoke, raising the overall muon identification efficiency to more than 99%. The detector's 43,000 channels are calibrated using cosmic ray muons, ensuring consistent performance across the full system. A detailed description of the muon detector is described in Chapter 9.

## 2.3 Summary

The CEPC detector design integrates a wide range of technological advancements. Stitching silicon technology in the vertex detector, together with the TPC and silicon tracker for tracking and PID, defines state-of-the-art tracking performance. AC-LGAD ToF devices redefine timing resolution in tracking detectors, while BGO crystal ECAL set new standards in electromagnetic calorimetry. Advanced high-density glass scintillators in the HCAL, and HV-CMOS integration in the ITK exemplifies miniaturization without sacrificing performance.

This detector concept is a sophisticated and comprehensive system that leverages advanced technologies to enable high-precision measurements of particles produced in electron-positron collisions. The silicon pixel vertex detector, silicon and TPC tracker, PFA electromagnetic and hadronic calorimeters, muon detector, and magnets are all key components of the system, each playing a crucial role in achieving the scientific goals of the project.

In summary, these innovations collectively enable CEPC to achieve its physics goals: measuring Higgs couplings with sub-1% precision, resolving $W/Z$ boson masses to MeV levels, and probing BSM phenomena at energy scales beyond current colliders. By harmonizing cutting-edge materials, geometries, and algorithms, this detector design stands as a testament to the ingenuity of modern particle physics — a tool not just for discovery, but for redefining the boundaries of human knowledge and the frontiers of technology.

# Chapter 3 Machine Detector Interface and luminosity measurement

The design of the Machine-Detector Interface (MDI) represents one of the most critical challenges for the CEPC project. The MDI design must be compact and complex to meet the constraints imposed by both the machine and the detector, enhancing the performance of all components. This chapter covers several critical subjects. These include the layout of the interaction region together in Section 3.1, the designs of the key components like the beam pipes and the final focusing magnets in Section 3.2, the estimate of the Beam Induced Backgrounds (BIBs) in Section 3.3 and the luminosity measurements in Section 3.4. The ongoing optimization efforts and future plans are also discussed in Section 3.5. The study on the design of the MDI, particularly the estimate and mitigation of BIBs, is still in a very preliminary phase. Current results serve as a reference for designers of sub-detectors and other related systems. The optimization of these background estimate and mitigation techniques is an ongoing process, in parallel with the optimization of the detector and accelerator designs, throughout the construction, commissioning, and even data-taking phases. This is a common and necessary practice for all lepton collider experiments.

## 3.1 Interaction Region

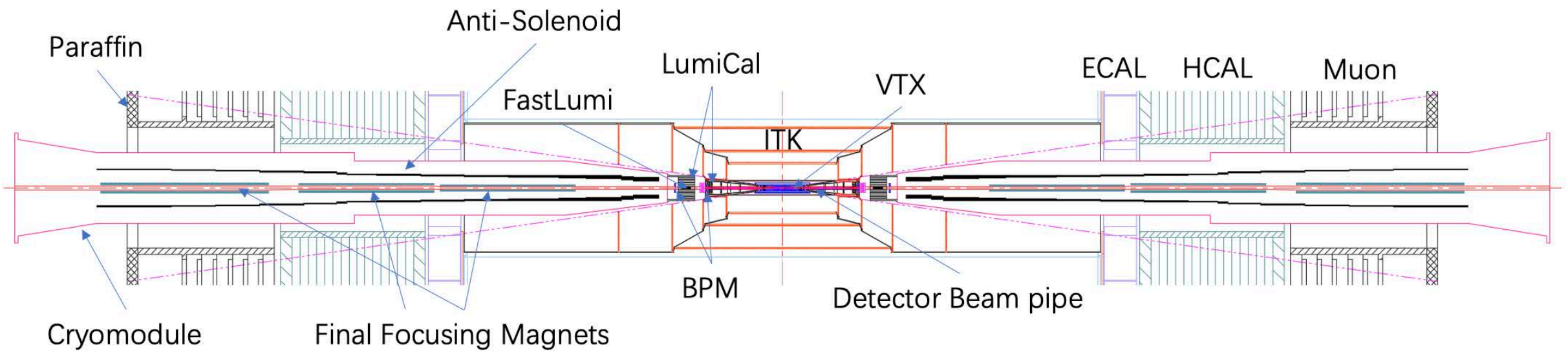

**Figure 3.1:** Layout of the CEPC Interaction Region (IR). The dash line shows the detector acceptance of $|\cos\theta| = 0.99$. All sub-detectors are located within the detection angle of the experiment. The key components shown include the Inner Silicon Tracker (ITK), the Electromagnetic Calorimeter (ECAL), the Hadron Calorimeter (HCAL), and the Muon Detector (Moun). The MDI components, including the detector beam pipe, the Vertex Detector (VTX), the anti-solenoids, the final focusing magnets, the accelerator cryomodule, the shielding of detectors, and the luminosity measurement detectors are also shown, while most of them are beyond the detector acceptance.

The Interaction Region (IR) layout is depicted in Figure 3.1, with its design optimized to accommodate different CEPC operation modes. This region spans 14 m in length, extending ± 7 m from the IP. It includes the whole main detector, which is within the detector's acceptance of $|\cos\theta| = 0.99$, and several key components beyond the acceptance, such as the detector beam pipe, the final focusing magnets, the luminosity measurement detectors and supporting structures. The crossing angle between the electron and positron beams is designed as 33 mrad in the horizontal plane.

The detector beam pipe, with a total length of 1400 mm, interfaces with bellows and the accelerator vacuum chamber at both ends. Its central beryllium part features a 20 mm inner diameter and a double-layer design, contributing a total thickness of 0.454% radiation length. This part transitions into an extending aluminum pipe, which adopts a racetrack cross-section.

Beyond the flange and bellow interfaces, the luminosity measurement system incorporates the Luminosity Calorimeter (LumiCal), the Fast Luminosity Monitor (FastLumi) and the Beam Position Monitor (BPM). The LumiCal comprises two silicon detectors for position measurement and both short and long Lutetium Yttrium Oxyorthosilicate (LYSO) crystals functioning as calorimeters. The Si-wafers and the short LYSO crystal are situated within the flange, while the long LYSO crystal is positioned externally. These components occupy vertical space relative to the beam pipe. The FastLumi is co-located with the long LYSO crystal, utilizing the space between the two LumiCal modules, and is mounted directly onto the beam pipe. One BPM is integrated within the flange, while two 4-button BPMs is incorporated into the accelerator vacuum chamber.

The accelerator vacuum chamber primarily consists of double-walled pipes with varying internal diameters along its length. These vacuum chambers are enveloped by superconducting final focusing quadrupoles and anti-solenoid magnets, all of which necessitate housing within cryomodules. The final focusing magnets are situated starting from $|z| = 1.9$ m away from the IP. The cryomodule shells, constructed from 15 mm thick stainless steel, provide shielding against BIBs, augmented by a 10 mm tungsten shell preceding the cryomodule and 10 cm of paraffin on the yoke surface.

## 3.2 Detector beam pipe and final focusing system

The following sections will focus on two primary components, the detector beam pipe and the final focusing magnets. The detector beam pipe is discussed in detail, while the final focusing magnets are discussed briefly. The detailed design of these magnets and other accelerator components within the IR can be found in Ref. [1].

### 3.2.1 Detector beam pipe

The detector beam pipe must be designed with the smallest possible radius while still meeting the Beam Stay Clear (BSC) tolerance requirements. Its small radius and minimal material budget are crucial for efficiently reconstructing charged-particle trajectories, which are essential for physics analysis. The detector beam pipe has a total length of 1400 mm. Its design is illustrated in Figure 3.2. It connects to the accelerator vacuum tube on both ends and comprises several segments, including the central beryllium pipe, the transition pipe between beryllium and aluminum, the extending aluminum pipe, and the connection pipe to the flange.

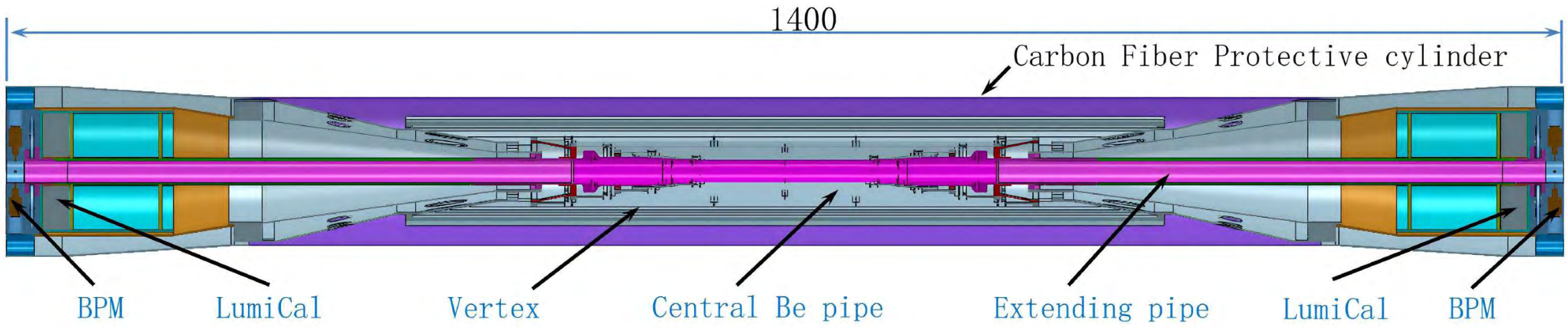

**Figure 3.2:** The side view of the detector beam pipe. The detector beam pipe consists of the central beryllium pipe, the extending pipe on both ends and the carbon fiber protective cylinder. It serves as the installation and support structure for several critical components, including the BPM, the LumiCal and the VTX.

The central beam pipe is designed with an inner diameter of 20 mm and is fabricated with beryllium to lower the material budget. This central beam pipe has a total length of 380 mm and features a double-layer design to maintain mechanical stability, as shown in Figure 3.3. The inner layer is 220 mm long and 0.2 mm thick, while the outer layer is 170 mm long and 0.15 mm thick. Two coolant options are being considered for the gap between the layers, water and paraffin. Water is chosen as the baseline coolant due to its superior cooling efficiency, which allows for a thinner coolant layer compared to paraffin. Paraffin is considered as an alternative option. The coolant flows into the left side of the pipe, passes through the gap, and exits through the right side. The thickness of the gap is designed as 0.2 mm when using water, and 0.3 mm when using paraffin. Although water can be corrosive to beryllium, preliminary studies indicate that a thin-film coating of oxide material, such as alumina, can protect the beryllium from water-induced corrosion. Study on corrosion prevention for the beryllium layer is still ongoing. To shield the SR photons, a layer of 10 µm gold coating is implemented on the inner layer. The details are discussed in Section 3.3.4.

The beryllium beam pipe for the Beijing Spectrometer (BESIII) experiment was successfully manufactured and has operated stably for nearly two decades [2]. Critical technologies including beryllium powder production, cold or hot isostatic pressing and heat treatment were fully developed. In particular, the refinement process of beryllium powder granulometry was optimized to meet the low leakage rate requirement. This accumulated experience serves as a solid basis for designing the more challenging beryllium beam pipe for the CEPC project. With available technologies, it is already feasible to manufacture beryllium pipes with a wall thickness of 0.20 mm and a length of 100 mm. Although extending the fabrication to 220 mm-long pipes as required by CEPC does not pose fundamental technical challenges, additional R&D efforts are still needed to fully meet the project specifications.

Two extending aluminum pipes are arranged on either side of the central beryllium pipe using a gradual racetrack shape and cooled by water. To minimize the material budget in front of LumiCal, beryllium plates are embedded on the upper and lower surfaces of the extending aluminum pipes, as shown in the green part in Figure 3.4. Welding beryllium to aluminum requires high-energy beam welding techniques with low thermal input and minimal deformation, such as laser welding or vacuum electron beam welding. Experience from the BESIII experiment has shown that electron beam welding effectively prevents crack formation during the process and ensures superior welding quality [2]. This technology is applicable to the beryllium-aluminum welding required for the CEPC detector beam pipe, although specific process optimizations are still needed. The detector beam pipe needs to be mounted at both ends of the vacuum flanges in an ultra-high vacuum environment. These flanges not only connect the front and rear accelerator vacuum chambers but also provide space for the outlet of water cooling channels, air cooling channels, detector cables, and several other components of the detector beam pipe.

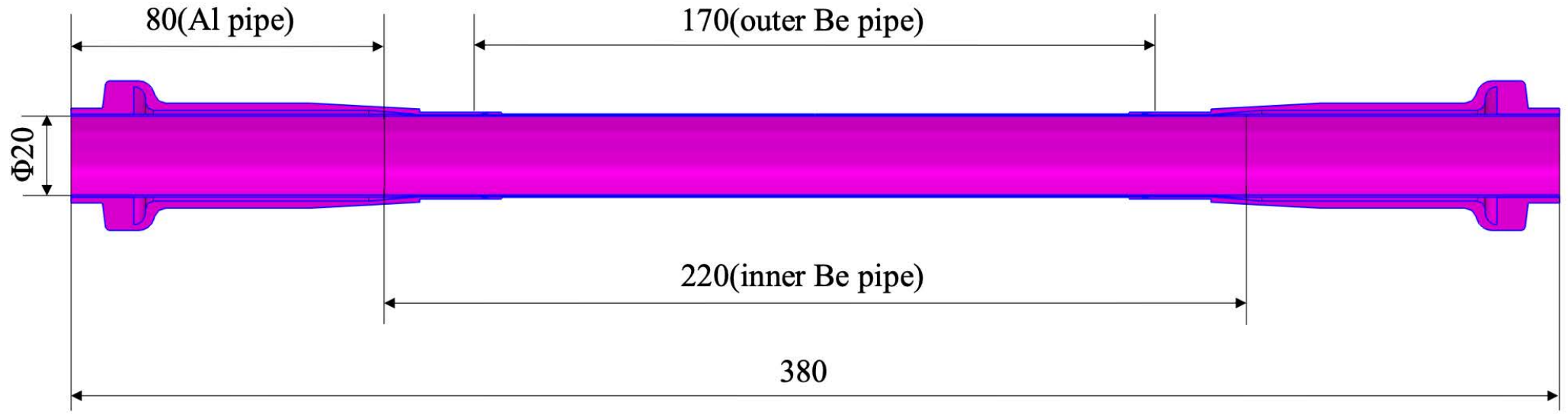

**Figure 3.3:** The central beryllium beam pipe features a double-layer structure with an inner diameter of 20 mm. The outer layer is 170 mm in length, while the inner layer extends to 220 mm. The entire central beam pipe is 380 mm long, with an 80 mm long aluminum pipe connected to each end.

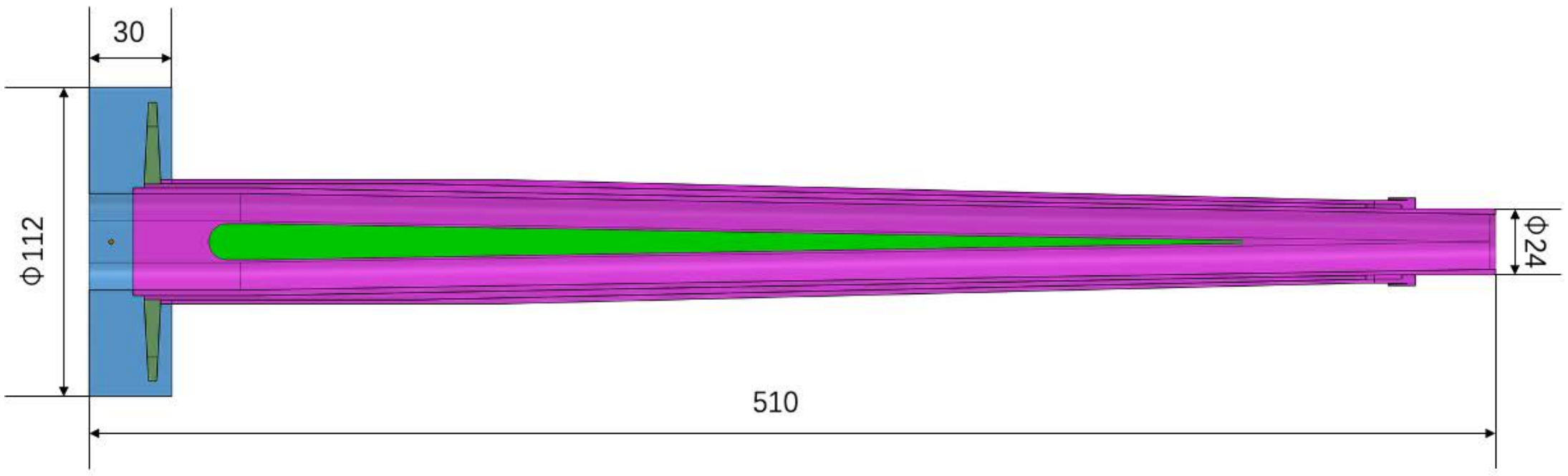

**Figure 3.4:** The top view of the structure of the extending aluminum pipe on one side. It is 510 mm in length and features a gradual racetrack shape. The upper and lower track-shaped planes are embedded with 2 mm thick beryllium plates shown in green.

Based on the overall strength analysis of the detector beam pipe, when one end of the beam pipe is fixed and the other end is cantilevered, the maximum deformation of the beam pipe is

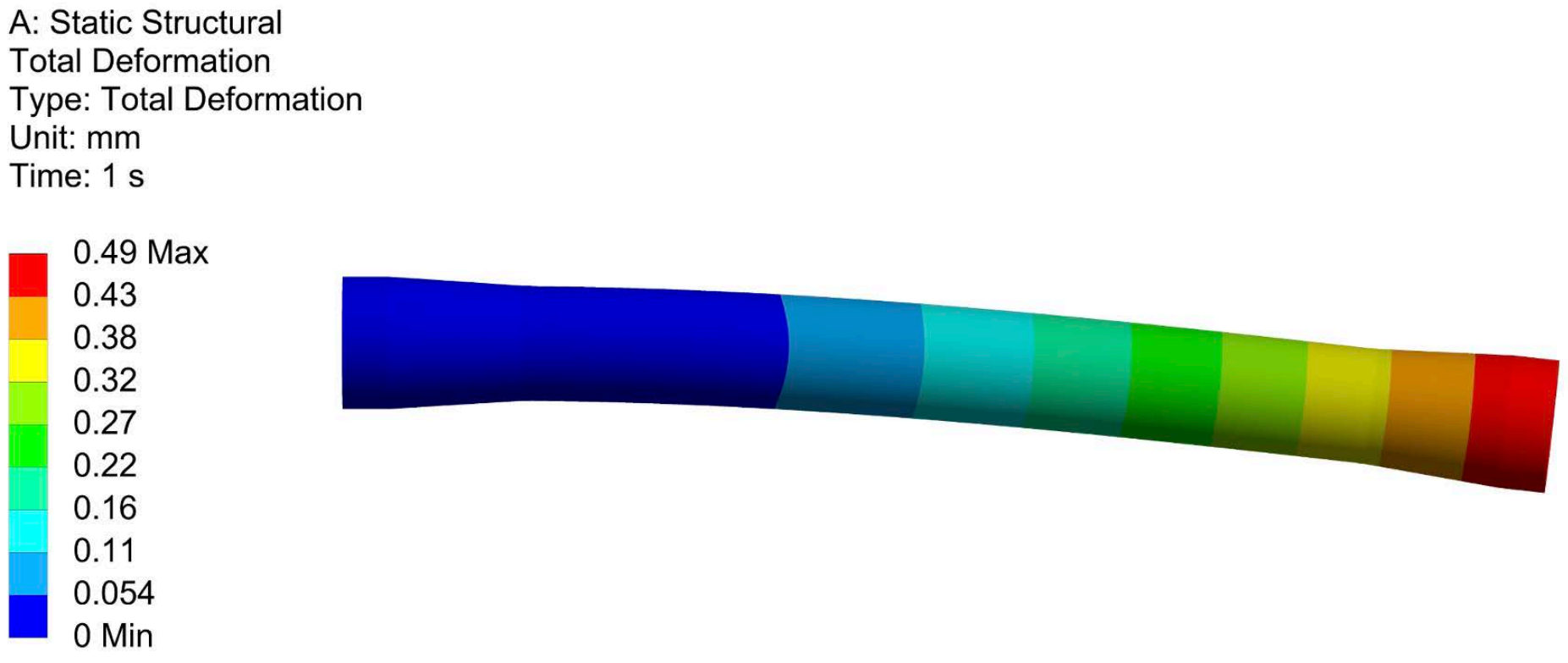

**(a)**

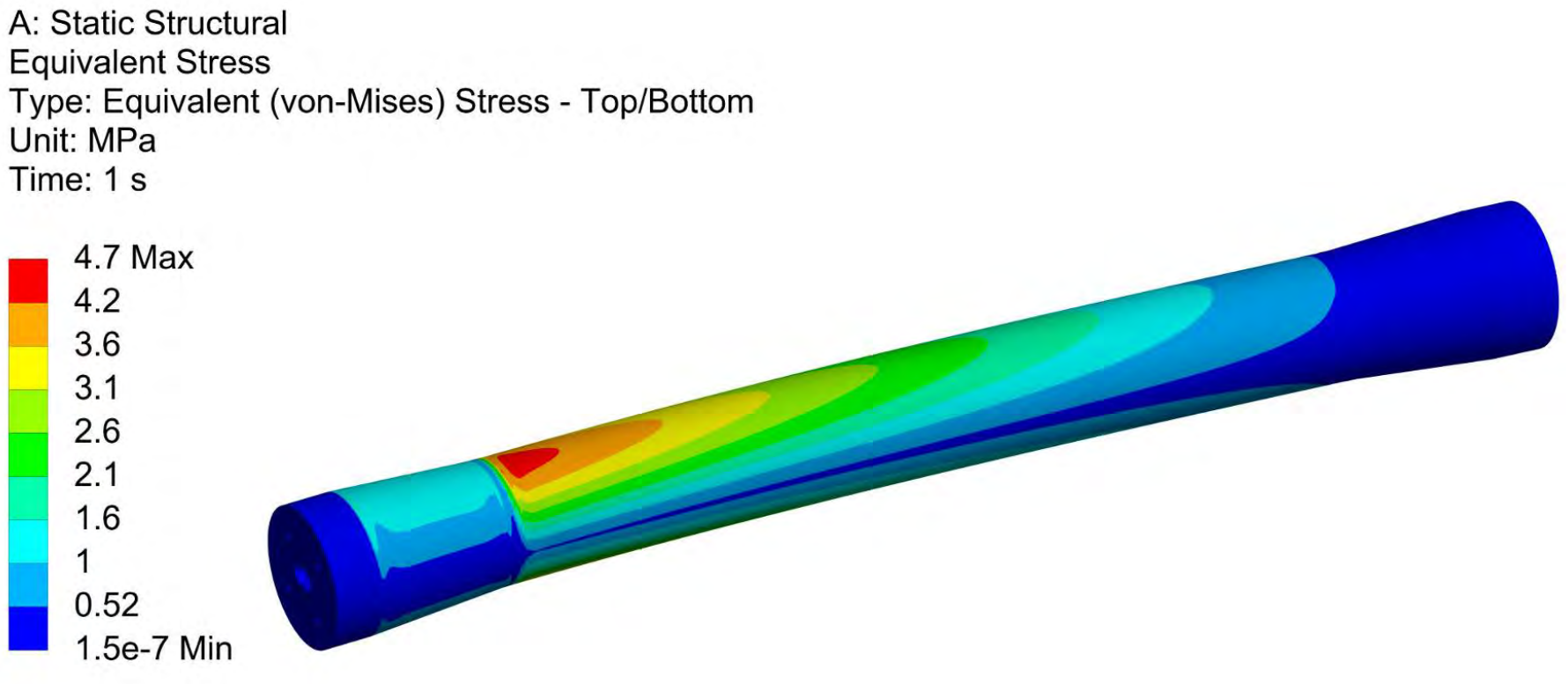

**(b)**

**Figure 3.5:** Stress structure analysis of the beam pipe when one end is fixed and the other end is cantilevered. a) Total deformation of beam pipe, with maximum deformation of 0.49 mm at the cantilevered end flange; b) Equivalent stress (von-mises), showing the maximum stress at the outer beryllium pipe to be 4.7 MPa.

0.49 mm. And the maximum stress at the joint between the beryllium pipe and the extending aluminum pipe is 4.7 MPa, as shown in Figure 3.5. When both ends of the beam pipe are fixed, the maximum deformation and equivalent stress is smaller than the previous case. These simulation results demonstrate that under both conditions, the beam pipe's deformation is well within the required limits, and the maximum stress is significantly lower than the yield strength of aluminum and carbon fiber. Therefore, the design satisfies the structural safety requirements.

**Table 3.1:** High Order Mode (HOM) heat load on the detector beam pipe as input to the thermal analysis in high luminosity $Z$ mode, which has the maximum value of heat deposition.

| Part | Start-End[mm] | Material | Length[mm] | Heat[W] | Heat Flux(Density)[W/cm$^2$] |
|---|---|---|---|---|---|
| Central Be Pipe | 0-110 | Be | 110 | 55 | 1.35 |
| Be Pipe Transition | 110-190 | Al | 80 | 29.28 | 0.49 |
| Extending Pipe | 190-655 | Al | 465 | 341.56 | 0.83 |
| Flange Connection | 655-700 | Al | 45 | 29.28 | 0.70 |

Thermal analysis was conducted to validate the design. Table 3.1 shows the expected heat load generated by High Order Mode (HOM) on the detector beam pipe. The outer beryllium pipe is in direct thermal contact with the vertex detector, which imposes a strict operational constraint that its surface temperature must remain below 20 °C. Concurrently, the upper and lower planes of the extending aluminum pipe interface with the LumiCal detectors, which also require temperature regulation below 20 °C.

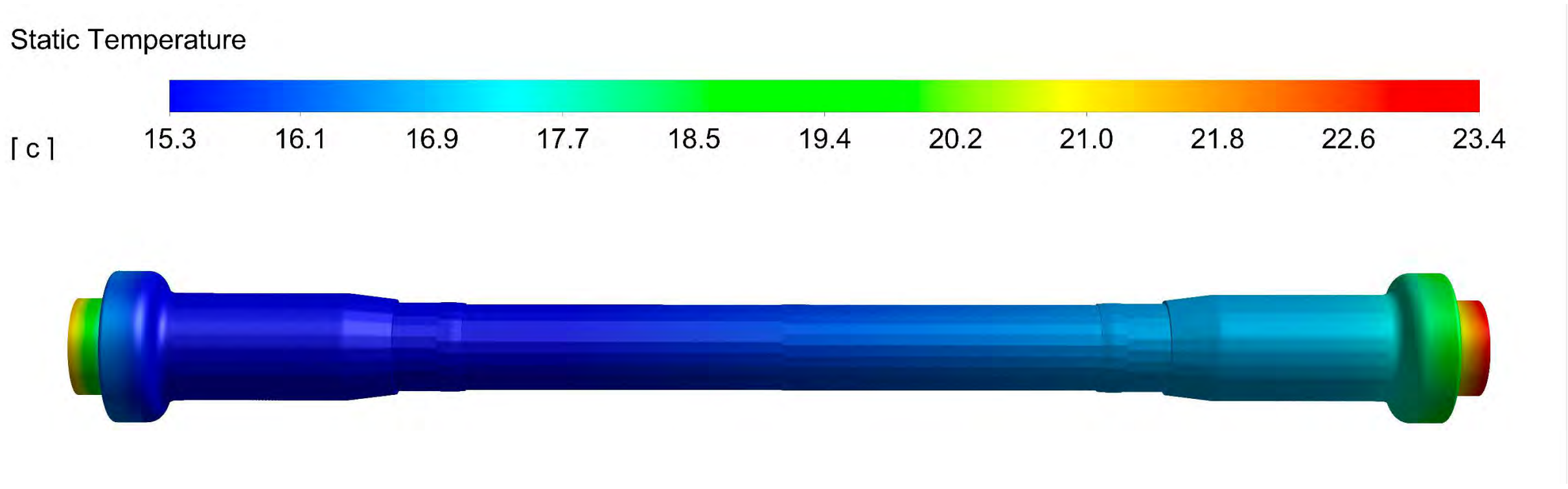

**(a)**

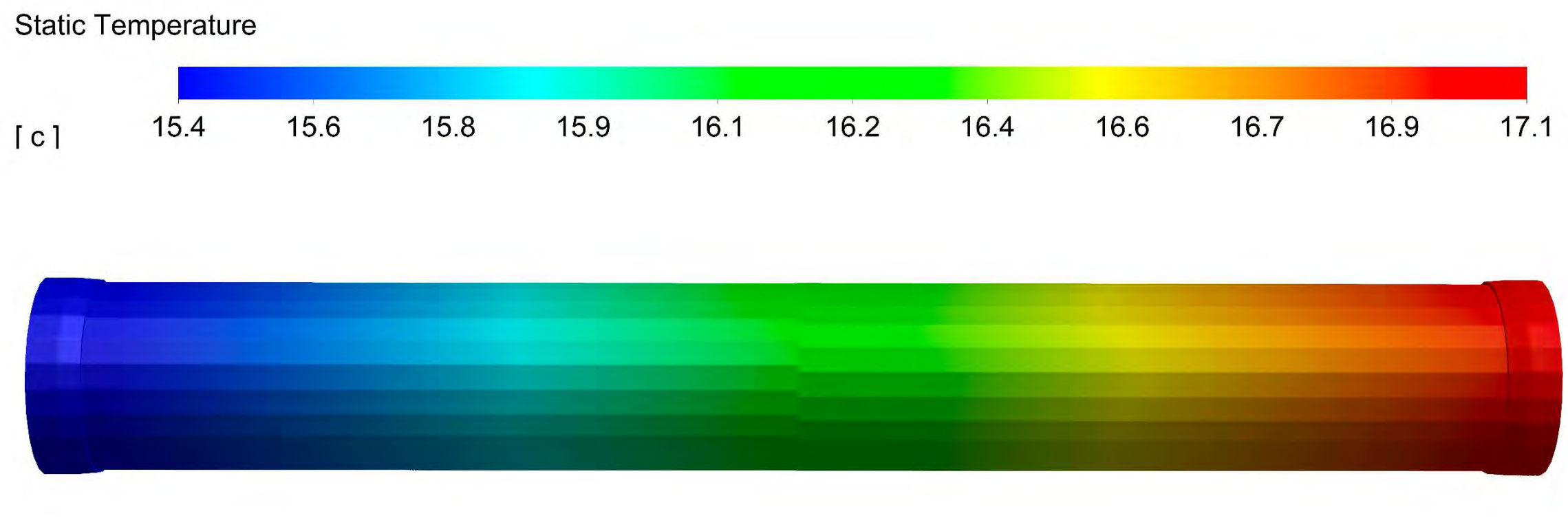

**(b)**

**Figure 3.6:** Temperature distribution of the central beryllium pipe when water cooled. a) Shows the inner beryllium pipe with a maximum temperature of 23.4 °C, while b) shows the outer beryllium pipe with a maximum temperature of 17.1 °C.

To manage the thermal load, water cooling is implemented for both the central beryllium and the extending aluminum beam pipe. The central beryllium pipe operates with a coolant inflow rate of 1.36 L/min, while the extending aluminum pipe utilizes a higher flow rate of 6.1 L/min. Both cooling circuits maintain an identical inlet coolant temperature of 15 °C. As shown in Figure 3.6a, the maximum temperature of the beam pipe is 23.4 °C at the short aluminum section of the cooling water outlet. The inlet-outlet pressure drop is simulated to be 56.9 kPa. The maximum temperature of the outer beryllium pipe is 17.1 °C, with a temperature differential of 1.7 °C, as shown in Figure 3.6b. These temperatures meet the operational thermal requirements. The joint part of the detector pipe reaches a maximum temperature of 25.7 °C, as shown in Figure 3.7. This temperature will not affect the VTX, as discussed in Chapter 4.

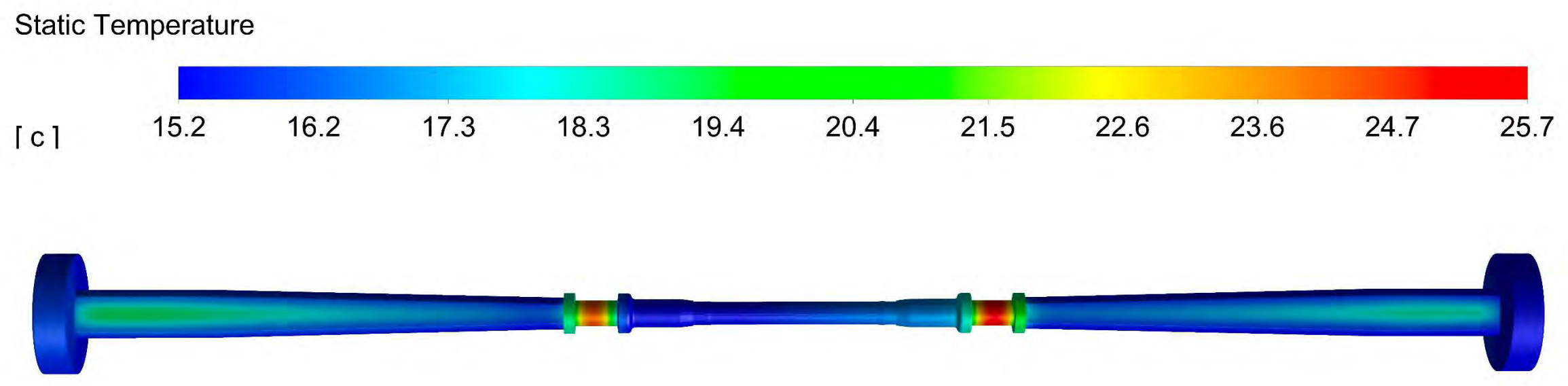

**Figure 3.7:** Temperature distribution of the beam pipe. The maximum temperature of the beam pipe reaches 25.7 °C at the junction between the cooling outlet end of the central beryllium pipe and the extending aluminum pipe.

### 3.2.2 Final focusing magnets

To achieve high luminosity, high-gradient final focusing magnets are required on both sides of the IP in the IR. These magnets are superconducting dual-aperture quadrupole magnets, namely Q1a, Q1b and Q2. They operate entirely within the detector solenoid, which has a central field strength of either 3.0 T or 2.0 T depending on the machine operation mode. To minimize the impact of the solenoid field on the beam particles, anti-solenoids must be positioned before and outside the Q1a, Q1b and Q2 magnets. The field direction of these anti-solenoids is opposite to that of the detector solenoid, and the total longitudinal integral field generated by the detector solenoid and anti-solenoid coils must be zero. Additionally, it is necessary to keep the total solenoid field inside the apertures of Q1a, Q1b, and Q2 as close to zero as possible. Given the stringent space constraints, superconducting technology is employed for both the final focusing and anti-solenoid magnets. The basic design parameters of the quadrupole magnets are presented in Table 3.2.

**Table 3.2:** Basic design parameters of the CEPC superconducting quadrupole magnets

| | **Q1a** | **Q1b** | **Q2** |
|---|---|---|---|
| Maximum field gradient [T/m] | 142.3 | 85.4 | 96.7 |
| Magnet length[m] | 1.21 | 1.21 | 1.5 |
| Reference radius [mm] | 7.46 | 9.09 | 12.24 |
| Minimum distance between two aperture centerlines [mm] | 62.71 | 105.28 | 155.11 |
| High order harmonics | $< 3\times10^{-4}$ | | |

The final focusing and anti-solenoid magnets are integrated into cryomodules. Each cryomodule consists of several components arranged from the inside out: the beam chamber, the inner thermal shield, the inner wall of the liquid helium tank, the magnet coils, the outer wall of the liquid helium tank, the outer thermal shield, and the ambient vacuum cylinder. The thermal shields are made of aluminum or copper, while the rest of the cryomodule is constructed from stainless steel. Given the limited radial space, the magnets are cooled using supercritical helium. This coolant flows through both the outer and inner channels, absorbing heat through conduction and thermal radiation.

One cryomodule is installed on each side of the IP. Each cryomodule is supported radially and axially by stainless-steel supports that are welded to the outer vacuum shell layer to minimize heat transfer. The contact points for these supports are made of G-10 material, which also serves to support the helium thermal shielding. Both heat transfer optimization and structural integrity are carefully examined in the design of these support points. In the current design, each magnet weighs approximately 0.4 tons. The total weight of a cryogenic module is about 2.5 tons, with a total length of 5.6 m. The stress structure analysis was performed. The maximum stress of the helium tank and the vacuum vessel is 96 MPa and 126 MPa, respectively, which are both within the allowable stress of 205 MPa.

## 3.3 Beam-induced backgrounds

A comprehensive understanding and effective mitigation of BIBs are essential for achieving optimal detector and machine performance, and ultimately for realizing the precision physics program[3, 4]. Various BIB sources can produce primary particles that either directly enter the detector or generate secondary particles that eventually reach the detector. These particles can cause radiation damage to the detector and its electronic components, leading to degraded detector performance. Moreover, BIB particles can interfere with the collision data of physics events of interest. Such impacts must be carefully evaluated with BIB simulation, and the results can be used to optimize the designs of the detector and the accelerator.

BIB simulations typically begin either by generating BIB particles directly in the IR or by propagating them to a region close enough to the IR. The interactions of these particle with detector components are also simulated, and the energy deposition within the detector components are recorded for further evaluation. Although BIBs at the CEPC are expected to be less severe than those at the (HL-)LHC[5], they still impose significant constraints on the design of both the accelerator and the detector. The design parameters of CEPC accelerator used in this study are shown in Table 3.3. The parameters of 50 MW Higgs mode and high luminosity $Z$ mode are taken from Ref.[1] and Ref. [6], while the parameters of 12.1 MW low luminosity $Z$ mode are taken from Ref.[7]. The BIB estimates discussed in this chapter mainly concentrate on the Higgs and low luminosity $Z$ mode. These estimates are subject to change as the accelerator design continues to be refined in the future.

### 3.3.1 Sources of beam-induced backgrounds

BIBs can be simply classified into two categories: single-beam backgrounds and luminosity-related backgrounds. Single-beam backgrounds arise from processes involving one single beam and include SR, Beam-Gas Scattering (BGS), Beam Thermal photon scattering (BTH) and Touscheck Scattering (TSK) processes. In contrast, luminosity-related backgrounds result from interactions between the two beams at IP and mainly include Radiative Bhabha Scattering (RBB) and Beamstrahlung processes [8].

**Table 3.3:** Design parameters of the CEPC accelerator used in beam-induced background study.

| **Parameters** | **Higgs** | **Low-Lumi-$Z$** | **High-Lumi-$Z$** |
|---|---|---|---|
| Number of IPs | 2 | | |
| Half crossing angle at IP (mrad) | 16.5 | | |
| Solenoid Magnet[T] | 3 | | 2 |
| SR power per beam [MW] | 50 | 12.1 | 50 |
| Energy [GeV] | 120 | 45.5 | 45.5 |
| Bunch number | 446 | 3978 | 13104 |
| Bunch spacing [ns] | 277.0 | 69.2 | 23.1 |
| Train gap [%] | 63 | 17 | 9 |
| Bunch population [$10^{11}$] | 1.3 | 1.7 | 2.1 |
| Beta functions at IP $\beta_x^*/\beta_y^*$ (m/mm) | 0.3/1 | 0.2/1.0 | 0.13/0.9 |
| Emittance $\epsilon_x/\epsilon_y$ (nm/pm) | 0.64/1.3 | 0.27/5.1 | 0.27/1.4 |
| Beam size at IP $\sigma_x/\sigma_y$ [um/nm] | 14/36 | 6/72 | 6/35 |
| Bunch length (total) [mm] | 4.1 | 8.8 | 10.6 |
| Energy spread (total) [%] | 0.17 | 0.13 | 0.13 |
| Energy acceptance (DA) [%] | 1.6 | 1.0 | 1.0 |
| Luminosity per IP [$\times10^{34}$cm$^{-2}$s$^{-1}$ ] | 8.3 | 26 | 192 |

SR is a significant contributor to BIBs. As the beam traverses the accelerator's magnetic elements, charged particles are deflected by the magnetic force and emit SR photons tangentially to their trajectory. When SR occurs near IR, especially upstream of it, the emitted SR photons may enter IR, penetrate the beam pipe and reach the detector. The design of the IR must be optimized to prevent SR photons from directly hitting the detector beam pipe. The dipole components surrounding the final focusing magnets are maintained at less than 300 Gs. Moreover, the last bending magnet upstream of the IR should be ensured to have a critical energy of the SR photons less than 60 keV.

BTH arises from the scattering between beam particles and thermal photons emitted by the beam pipe. Components like beam pipes in the accelerator emit a large number of thermal photons of varying energies uniformly in all directions due to thermal radiation. These photons can undergo Compton scattering with charged particles in the beam, causing the beam particles to lose part of their energy.

BGS occurs between beam particles and residual gas molecules in the accelerator ring. Even when an ultra-high vacuum is achieved in the particle accelerator, a small amount of air molecules (such as $H_2O$, CO, etc.) may still remain in the vacuum pipe. As beam particles travel through, they can be scattered by these residual gas molecules. Both elastic scattering (Beam-Gas Coulomb scattering (BGC)) and inelastic scattering (Beam-Gas Bremsstrahlung (BGB)) should be considered. Elastic scattering does not change the energy of scattered charged particles but imparts a scattering angle relative to their original direction of motion. Inelastic scattering with the nuclei or electrons of atoms does not change the direction of the scattered charged particles but results in the emission of a photon. Consequently, the final-state charged particles exhibit a certain energy spread.

TSK is an intra-bunch scattering process. As a bunch of particles travels in an accelerator, particles within it may undergo elastic scattering due to transverse oscillations. This process converts the transverse energy of the particles into longitudinal energy, resulting in a certain energy spread of the final-state particles. All these scattering processes can contribute to BIBs and require thorough investigation. In the following estimation, the pressure of the beam pipe is assumed to be $10^{-7}$ Pa, and the temperature is assumed to be 300 K.

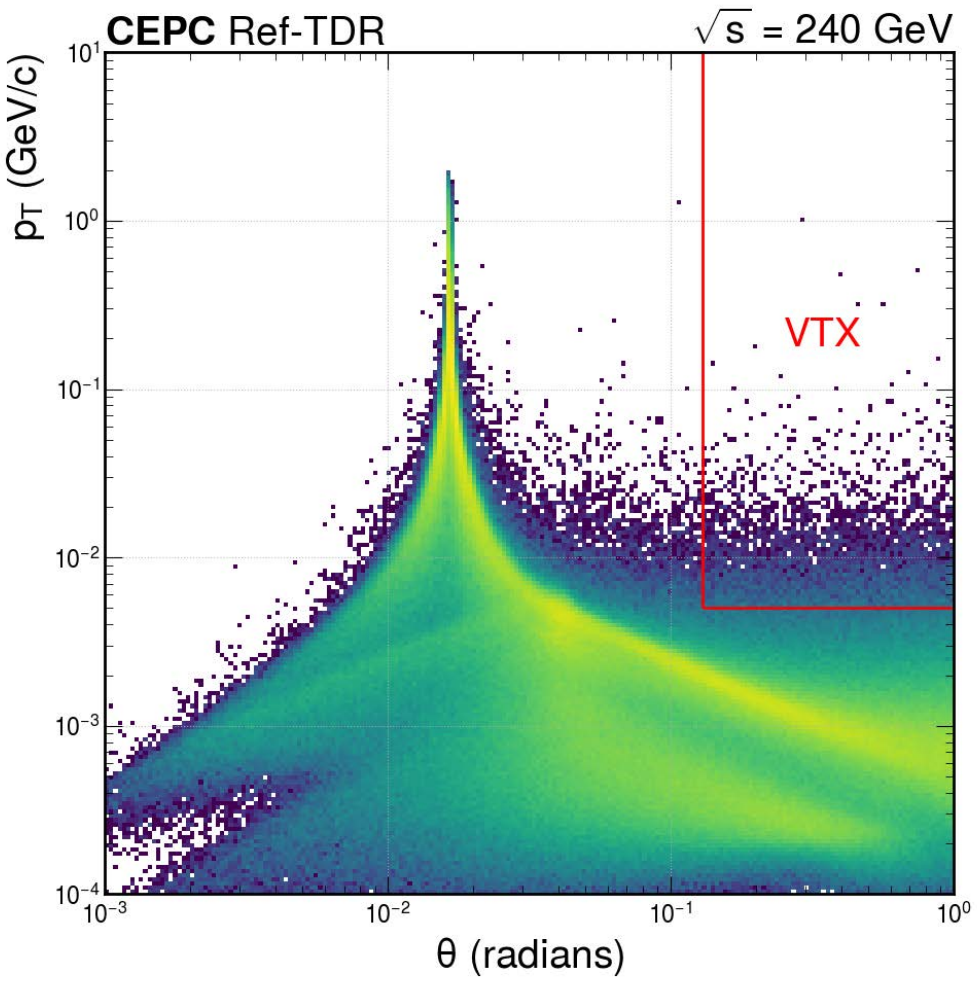

**Figure 3.8:** Energy and polar angle distribution of particles out of the pair production beam-induced background at Higgs mode. VTX stands for the vertex detector.

Among the luminosity-related backgrounds, RBB interactions can impact the superconducting quadrupole magnets, potentially generating showers of secondary particles that may degrade detector performance. Current study is concentrating on developing effective shielding designs for these magnets. Additionally, beam-beam interactions can lead to the emission of Beamstrahlung photons due to the pinch effect [9] and the large crossing angle. Consequently, pairs of electrons and positrons can be produced either coherently or incoherently from interactions involving real and/or virtual photons. Previous studies [10] suggested that coherent pair production is negligible, the focus must therefore be on the incoherent pair production. It should be pointed out that most Beamstrahlung photons, as well as electrons and positrons, are produced in the forward direction, leaving the interaction region without striking the beam pipe. However, there would be a small fraction of these particles that, with sufficient energy and large polar angle, can enter the detector, as illustrated in Figure 3.8. In addition, to prevent these particles from hitting beam-line elements and to avoid back-scattering photons towards the central detector, Beamstrahlung photons must be effectively absorbed. The power of the Beamstrahlung radiation at the IP is substantial, approximately 100 kW per beam, and photons are emitted in a narrow cone. This characteristic allows for the use of an exit window and a dedicated Beamstrahlung dump system, which is currently being designed. Moreover, Beamstrahlung photons can be utilized for beam diagnostics and as a source for a very high $\gamma$-fluence irradiation facility.

### 3.3.2 Simulation methods

The simulation of BIB studies involves three main steps: generation, (accelerator-)tracking, and detector simulation [11, 12].

**Table 3.4:** Tools used in the background simulation.

| Background | Generation | Tracking | Detector Simulation |
|---|---|---|---|
| Synchrotron Radiation | Geant4 | Geant4 | |
| Beamstrahlung/Pair Production | Guinea-Pig++ | | |
| Beam-Thermal Photon | PyBTH | | |
| Beam-Gas Bremsstrahlung | PyBGB | | CEPCSW/FLUKA |
| Beam-Gas Coulomb | BGC in SAD | SAD | |
| Radiative Bhabha | BBBREM | | |
| Touschek | PyTSK with SAD | | |

In the step of generation, each source of BIBs is generated separately using different tools, as summarized in Table 3.4. Inputs are based on the beam parameters of the accelerator, which are listed in Table 3.3. The processes of BGS, BTH and TSK are simulated using formula-based Monte Carlo (MC) codes [13], which using the same models as SuperKEKB[14]. SR photons are generated and tracked using Geant4 [15, 16] directly. The Beamstrahlung process is simulated with GUINEA-PIG++ [17, 18], with modifications to incorporate the external magnetic field. This adaptation allows this linear collider based code also applicable to the circular collider.

Following the generation step, files containing particle's ID, energy, initial position, direction, and other relevant information are saved. Luminosity related backgrounds are generated directly at the interaction point, while single-beam backgrounds are generated along the entire ring. Subsequently, these beam particles are tracked as they travel along the ring for up to 1000 turns.

In the tracking step, SAD [19, 20] is employed in the accelerator. Emission of SR photons and acceleration by radio frequency cavities are enabled in the simulation. Some built-in functions of SAD are used to generate a loss map for beam particles. Instead of taking the aperture as a parameter of accelerator elements, SAD considers the aperture as an element-like marker that could only be inserted at the ends of the elements. SAD compares the transverse position of the particle in tracking with the boundary of the aperture to determine whether the particle could survive or not. Currently, the loss position is simply moved to the inner surface of the beam pipe, and all particle information is saved. The field of detector solenoid and anti-solenoid shown in Figure 10.14 is not yet included, and the impact will be studied in future.

BIB files are then exported to the CEPC Software (CEPCSW) for the full detector simulation, with the uniform distribution of the detector solenoid currently, instead of the real distribution shown in Figure 10.14. The BIB estimate involves two key aspects: the detector noise level and the radiation dose level. The detector noise level is quantified using hit rate and detector occupancy, while the radiation dose level is assessed through the distributions of Total Ionizing Dose (TID) and Non-Ionizing Energy Loss (NIEL) (typically expressed as 1 MeV neutron equivalent fluence), as well as the ambient dose equivalent. For the noise level estimation, the CEPCSW is utilized, while FLUKA [21, 22] is employed for the radiation level estimation.

### 3.3.3 Beam-induced background studies with BEPCII/BESIII

The operational Beijing Electron-Positron Collider (BEPCII)/Beijing Spectrometer (BESIII) facility provides invaluable experimental data that can greatly improve the precision of simulations. A concise summary of dedicated BIB studies is provided here, with further details available in Ref. [23].

The two main sources of BIBs at the BEPCII/BESIII are TSK and BGS. Experiments were conducted in the accelerator's coasting mode, where the beams were allowed to decay naturally. The beam background rates under different conditions were measured using the Main Drift Chamber (MDC) of the BESIII detector, which is the innermost sub-detector surrounding the beam pipe. The collected BIBs were analyzed to extract contributions from different background sources and were compared with results from the MC simulations. The beam background simulation utilized a framework based on the SAD and Geant4 programs for modeling particle interactions with the BESIII detector following a similar procedure of the CEPC project. The residual gas was modeled as a mixture of 80% hydrogen and 20% carbon monoxide, maintaining a uniform pressure throughout the entire ring. Parameters used in the generators were with those from the experimental setup.

The data/MC ratios for different MDC layers are shown in Figure 3.9, after applying the scale factors derived from the beam lifetime to the simulation results. These ratios revealed that the simulated TSK rates were one to two orders of magnitude higher than those from the experiment. Such discrepancy was mainly due to the incomplete particle tracking and insufficiently modeled interactions involving complex components in IR in the simulation. On the other hand, significant fluctuations in the data/MC ratio for the BGS were observed, primarily due to the low count rate and the indirect measurements in the experiment.

In addition to the bunch current, which is the primary means of controlling beam parameters in the aforementioned experiment, the beam size can also affect the TSK background rate. The beam may expand due to the blowup effect, and the TSK rate may decrease with increasing bunch current. The beam-beam effect can significantly impact the beam distribution and the TSK cross section. More comprehensive study on the beam background with the beam-beam effects was conducted [24]. A new beam dynamics code, APES-T [25], which supports strong-strong beam-beam interactions in realistic lattice tracking, was used to simulate the beam background. The simulation results indicate that incorporating the beam-beam effect leads to a reduction of the BIB, which aligns with the experimental observations. To better understand and mitigate the remaining discrepancies between the data and MC results, additional dedicated experiments are planned at the BEPCII/BESIII facility.

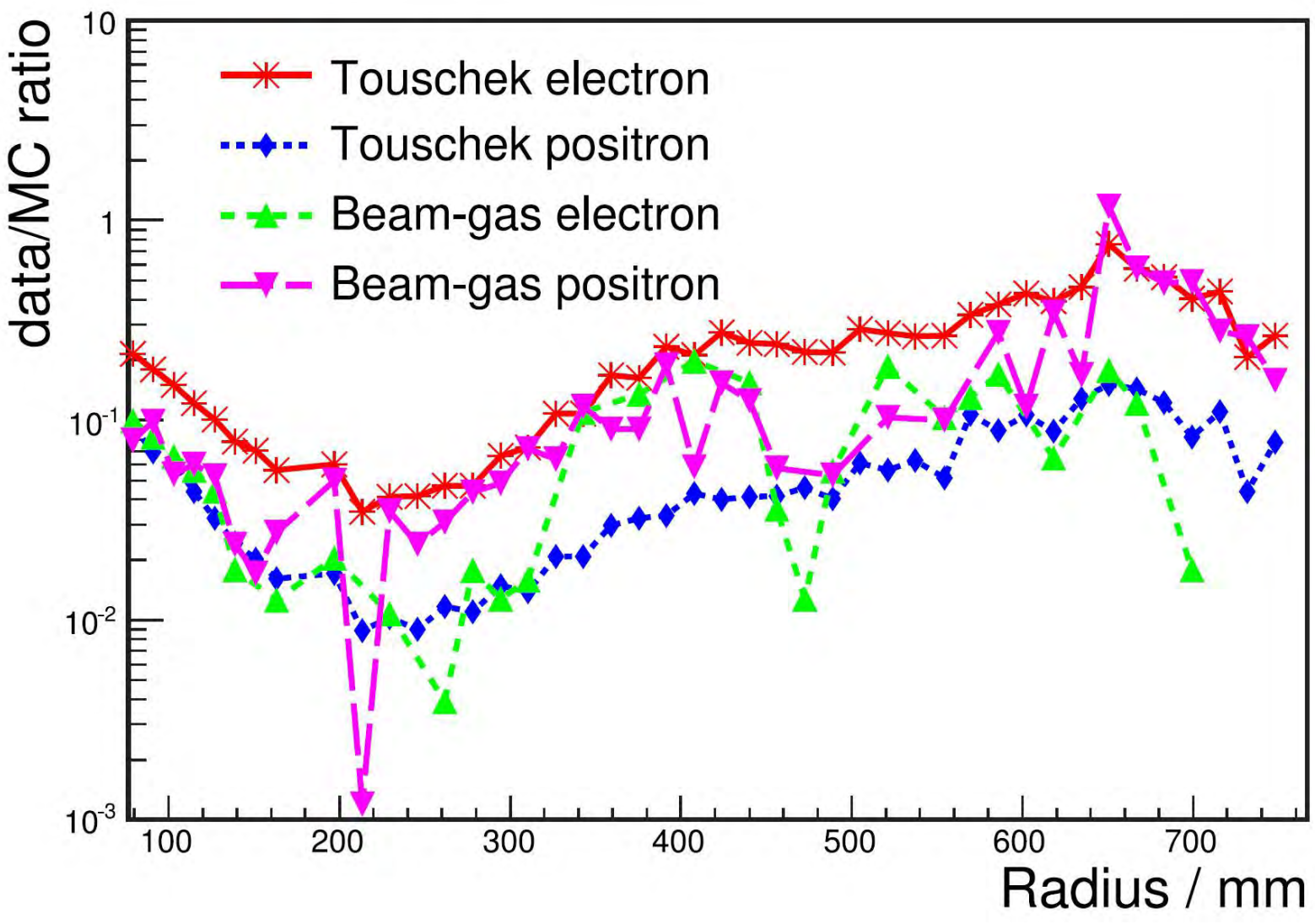

**Figure 3.9:** Data to Monte Carlo (MC) ratios for both electron and positron beams are shown for Touscheck Scattering (TSK) and Beam-Gas Scattering (BGS) backgrounds, at different layers of the Main Drift Chamber (MDC) in BESIII (taken from Ref. [23]).

### 3.3.4 Mitigation methods

BIBs can be reduced by using collimators or masks to block primary and secondary particles, as well as through shielding in the interaction region. Results presented here are just preliminary and should be considered as up limits to give a guidance to the following detector design. Studies are still on going to optimize the location, shape and number of masks, collimators and shielding. Further improvement is expected.

Current design to mitigate the SR background uses masks inside the beam pipe and gold coating to the Be beam pipe. One mask, made of tungsten with a height of four mm and a length of 50 mm, is positioned upstream of the IR at a distance of 1.9 m for each ring. The gold coating on the inner surface of the central beryllium pipe has a thickness of 10 µm. Experience at BESIII shows that both electroless plating and magnetron sputtering can achieve needed uniformity of gold deposition, with the later ultimately chosen. However, the much smaller inner diameter of the beam pipe for CEPC imposes technical challenges for magnetron sputtering, and further studies are needed to select and optimize the appropriate coating method. A total of $10^9$ primary electrons were simulated to generate primary SR photons. On average, one electron or positron generats three SR photons when traveling through the last bending magnet at Higgs mode. With the implementation of the mask and gold coating, the number of SR photons were substantially reduced, allowing only 112 to exit the beam pipe. Most of these photons have energies below 400 keV, which corresponds to an exit rate of 40.8 GHz. These photons are then sampled to perform a full detector simulation. With a statistics of $10^7$ photons, the total hit rate on the VTX is 5.6 kHz for both beams for the Higgs mode. Due to the limited MC statistics for low luminosity $Z$ mode, no photons can exit the beam pipe, resulting in an up limit of rates well below 12.2 GHz. Further study will be performed to incorporate the real magnetic field distribution, especially the strong anti-solenoid, into the simulation. Considering this, a safety factor of 10 dedicated for SR simulation was added, thus the hit rate on the VTX is 56 kHz for both beams. Further optimization on mitigation methods will be conducted, such as determining the optimal number and position of masks, adjusting collimators to absorb SR photons (or designing dedicated SR collimators), and implementing reweighting methods to increase statistics. These methods are essential for obtaining more reliable and accurate results.

Collimators need to be installed in the beam pipe to minimize single-beam-loss backgrounds. The aperture of these collimators must be sufficiently small to intercept as many beam lost particles as possible, while still permitting the beam to pass through. Two distinct types of collimators are implemented. The first one is specifically designed to reduce the particle loss rate in the IR, while the second one serves as part of the passive machine protection system, which also contributes to lowering the particle loss rate in IR.

**Table 3.5:** Key design parameters of the MDI collimators at Higgs mode per IP per ring.

| Collimator Name | Distance to IP[m] | Range of half width allowed [mm] |
|---|---|---|
| APTX.1 | 4969.5 | 1.4-7.3 |
| APTX.2 | 4891.4 | 1.4-7.3 |
| APTX.3 | 4121.2 | 1.4-7.3 |
| APTX.4 | 3961.1 | 1.4-7.3 |
| APTX.5 | 216.4 | 2.1-7.8 |
| APTX.6 | 19.07 | 2.6-11.0 |
| APTY.1 | 4723.69 | 0.14-3.5 |
| APTY.2 | 4669.01 | 0.14-3.5 |
| APTY.3 | 285.85 | 0.14-3.5 |
| APTY.4 | 141.90 | 0.14-3.5 |

The collimators near the IR are designed in accordance with beam clearance and impedance rules. A total of 40 sets of collimators are implemented to mitigate the BIBs, 10 sets of collimators per IP per ring. The settings at the Higgs mode is shown in Table 3.5. Currently in simulation, the collimator labeled as ”APTX” has an aperture of 8 mm, while the collimator labeled as ”APTY” has an aperture of 6 mm. The preliminary mechanical design of the collimators can be found in Ref.[1]. Figure 3.10 illustrates the effectiveness of the these collimators when operating at Higgs mode. The implementation of MDI collimators has resulted in a significant reduction in the loss rate in the IR. The SR mask is also subjected to single-beam loss particles, the effects of which are incorporated into the simulation. However, the potential risk to the detector still remains. To mitigate SR effects effectively, dedicated SR collimators may need to be installed further upstream rather than relying solely on the mask at current position. Further mitigation may require optimization on the collimator settings or fundamental modifications on this region. Collaborative efforts with our accelerator colleagues are currently underway.

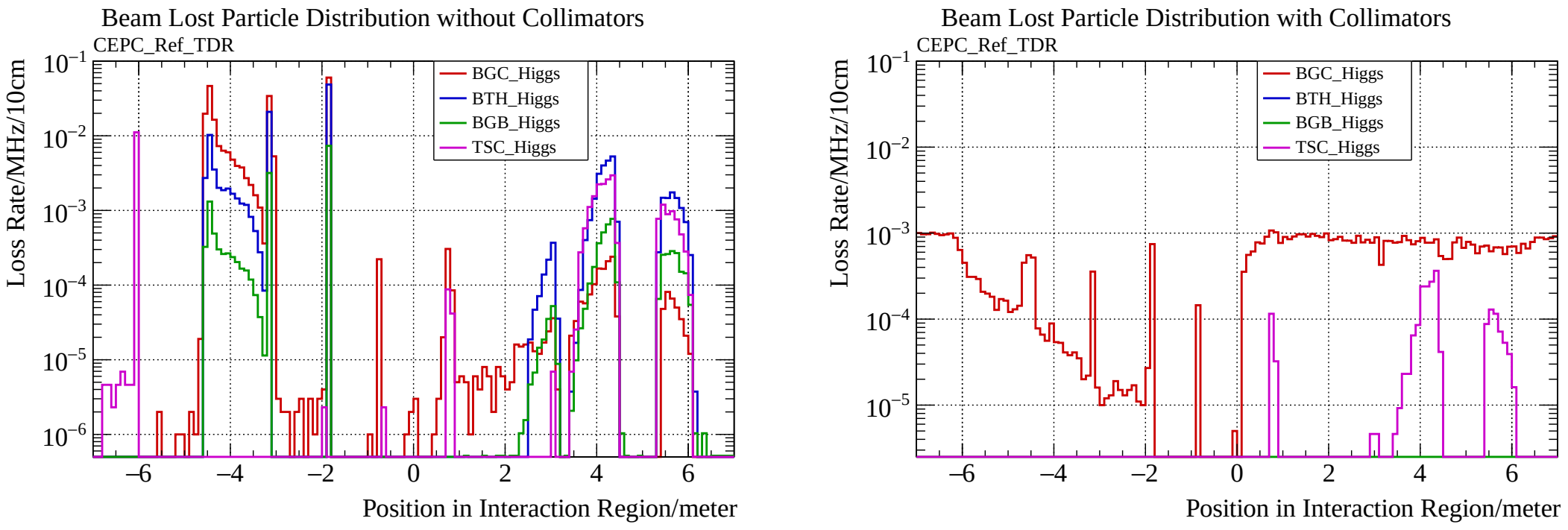

**(a)** **(b)**

**Figure 3.10:** The effectiveness of the MDI collimators when operating at the Higgs mode. a) the beam loss rates considering the BGC, the BGB, the BTH and the TSK with no MDI collimators; b) the beam loss rates considering the same beam induced background sources with the MDI collimators.

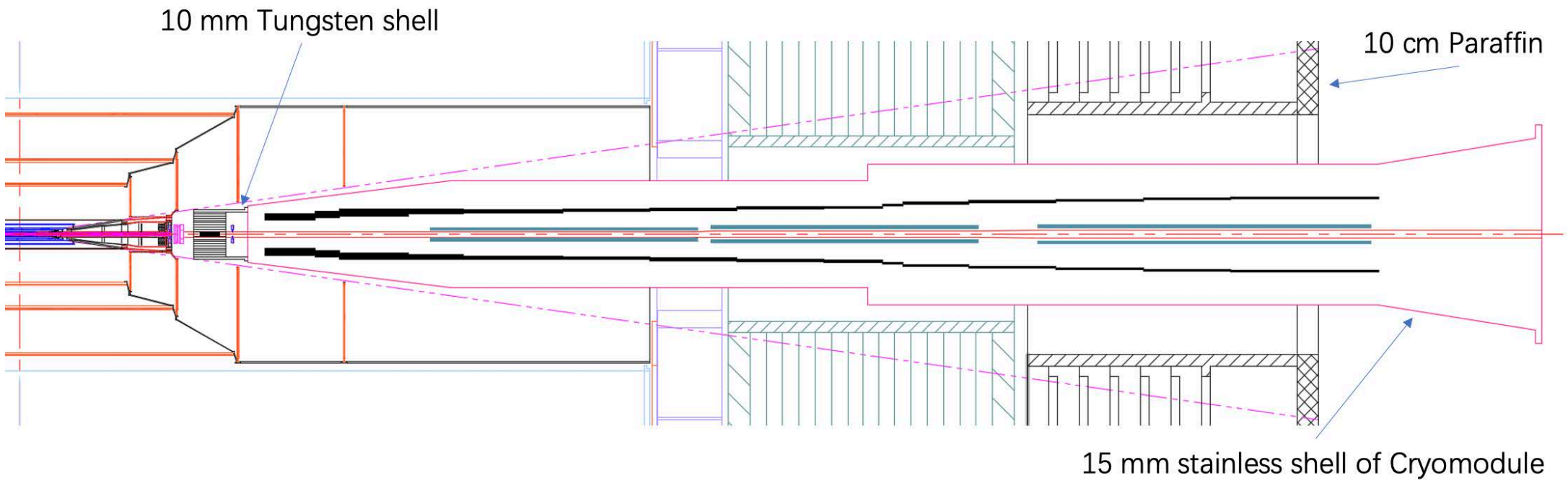

**Figure 3.11:** Illustration of the shielding surrounding the IR cryomodule. A 10 mm-thick tungsten shell serves as the outer layer for both the LumiCal and the IP BPM, while a 15 mm-thick stainless steel shell is used for shielding.

The luminosity-related backgrounds are hard to mitigate, especially for the primary background particles directly generated at the IP together with the signal. The masks and collimators are ineffective for them, leaving shielding as the only option. However, for the primary background particles, even shielding is almost impossible, since it will also shield the signal. After the implementation of the mask, gold coating and the collimaters, the contributions are dominated by pair production. Though it would be almost impossible to shield the primaries, there are still some methods to mitigate the secondaries like additional shielding beyond the detector acceptance. A 10 mm-thick tungsten shell serves as the outer layer for both the LumiCal and the IP BPM, while a 15 mm-thick stainless steel shell is used for shielding, as shown in Figure 3.11. The shielding is designed to protect against the secondary particles that are generated when the low angle pair production particles hit the heavy mental materials like the flange or the transition part of the beam pipe. Compared to the case without shielding, there is a 20% to 50% reduction in the hit rate across all operation modes, including the ITK, the TPC and the OTK. The average hit rate of the TPC running at Higgs mode was reduced by 33% from 1.8 kHz/cm$^2$ to 1.2 kHz/cm$^2$, while the barrel OTK was reduced by 50%, from 0.73 kHz/cm$^2$ to 0.37 kHz/cm$^2$. We are still working on the optimization of the shielding in the IR and with the goal of further reduction. Additionally, A 10 cm layer of paraffin has also been placed at both ends of the yoke to mitigate the neutrons generated by beam particles from the tunnel.

### 3.3.5 Beam-induced background results

The noise level estimation for sub-detectors was performed using the CEPCSW framework. For silicon detectors such as VTX, ITK, and OTK, a ”hit” is defined as the collection of signals produced by a single particle on the same chip. For calorimeters, a crystal records a hit when the sum of energy deposition within a time window exceeds a certain threshold, depending on the specific detector. In the case of TPC, the accumulation of slowly drifting ions in the gas volume can affect the homogeneity of the electric field and, consequently, the spatial resolution. In addition to hit rate, field distortion caused by space charge effects is also calculated using a built-in tool in CEPCSW. For detectors outside the VTX, most BIBs originate from secondary particles produced by primary particles interacting with the components like the beam pipe or flange.

Table 3.6 summarizes the BIB levels in the Higgs and low-luminosity $Z$ modes. The barrel part of the moun detector is omitted because the hit rate there is negligible. See Chapter 9 for more details. Currently, the simulation results indicate that pair production is the dominant background source, contributing over 90% of the total background, while the contribution from SR simulated with ideal beam orbit is negligible. The contribution from SR might be increased when take the actual magnetic field or the real beam orbit into account, which is on going incorporated with accelerator group, but also can be mitigated through proper methods. Some dedicated studies on the uncertainties and safety factor were performed. For single-beam-loss backgrounds, the simulation results exhibit an overestimate compared to the experimental data from the BESIII experiment (Sec. 3.3.3). Given that the same event generator and accelerator

**Table 3.6:** Beam-induced background levels in sub-detectors at Higgs and Low-Lumi-$Z$operation modes, including a safety factor of two.

| **Sub-Detectors** | **Ave. Hit Rate** | | **Max. Hit Rate** | | **Max. Occupancy [%]** | |
|---|---|---|---|---|---|---|
| | Higgs | Low-Lumi-Z | Higgs | Low-Lumi-Z | Higgs | Low-Lumi-Z |
| VTX [MHz/ cm$^2$] | 0.22 | 0.52 | 12 | 39 | 2.1×10$^{-2}$ | 1.3×10$^{-2}$ |
| ITK-Barrel [kHz/ cm$^2$] | 0.92 | 1.7 | 2.6 | 6.6 | 6.4×10$^{-3}$ | 1.3×10$^{-2}$ |
| TPC [kHz/ cm$^2$] | 2.4 | 5.2 | 26 | 24 | 0.15 | 0.14 |
| OTK-Barrel [kHz/ cm$^2$] | 0.74 | 1.3 | 1.2 | 2.2 | 4.2×10$^{-3}$ | 9.2×10$^{-4}$ |
| ECAL-Barrel [MHz/bar] | 1.4×10$^{-2}$ | 2.2×10$^{-2}$ | 1.7 | 0.66 | 1.6 | 0.4 |
| HCAL-Barrel [kHz/gs cell] | 4.6×10$^{-3}$ | 8.4×10$^{-3}$ | 14 | 24 | 8.0×10$^{-4}$ | 8.0×10$^{-4}$ |
| ITK-Endcap [kHz/ cm$^2$] | 3.0 | 5.4 | 24 | 50 | 2.4×10$^{-3}$ | 5.0×10$^{-3}$ |
| OTK-Endcap [kHz/ cm$^2$] | 1.9 | 3.1 | 8.2 | 13 | 7.4 ×10$^{-2}$ | 12×10$^{-2}$ |
| ECAL-Endcap [MHz/bar] | 0.062 | 0.10 | 7.2 | 13 | 7.0 | 1.8 |
| HCAL-Endcap [kHz/gs cell] | 0.24 | 0.24 | 640 | 340 | 8.0×10$^{-2}$ | 6.0×10$^{-3}$ |
| MD-Endcap [Hz/ cm$^2$] | 1.4 | 0.92 | 2.5 | 14 | 0.18 | 0.05 |

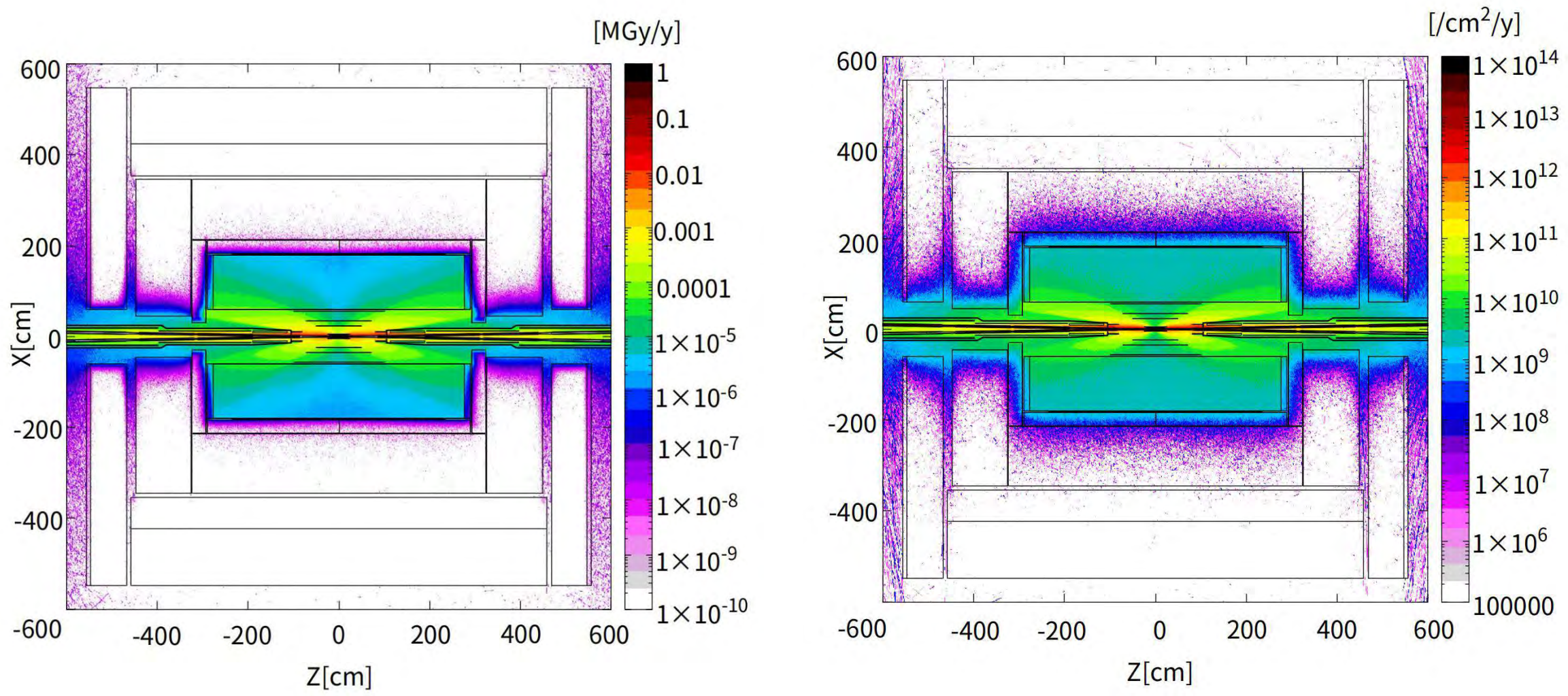

**(a)** **(b)**

**Figure 3.12:** The TID and NIEL distributions at Higgs mode on the CEPC detector. The highest TID is lower than 1 MGy per year as shown in a.), while the highest level of NIEL is in the order of $10^{13}(1\ MeVn_{eq})cm^{-2}$ per year as shown in b.).

tracking tool as used in the experiment was employed, our simulation can be regarded as conservative. For SR, a safety factor of 10 was applied to account for potential uncertainties. For the pair production, as discussed in the original paper of the event generator we employed (see Ref.[18]), the estimated uncertainty on the cross-section is lower than 43%. By varying certain beam conditions—such as introducing a transverse position offset within 100 µm or increasing the beam size—we observe that most variations reduce the luminosity and consequently the BIB rates. A few configurations lead to a marginal increase of less than 1%. Considering both the cross-section uncertainty and material effects, a safety factor of two for the rate estimation was adopted. This factor has already been incorporated into the background calculations presented in Table 3.6. Detailed distributions of hits across the detector are calculated. More in-depth discussions on the impact of BIBs on detectors are presented in the respective sub-detector chapters.

The distributions of TID and NIEL, caused by the considered BIB sources in Higgs mode are shown in Figure 3.12. These were obtained by simulating the absorbed dose and the 1-MeVsilicon equivalent fluence using FLUKA. The operational time is assumed to be 7000 hours per year. The highest TID in the whole detector region is lower than 1 MGy per year, while the highest level of NIEL is in the order of $10^{13}(1\ MeVn_{eq})cm^{-2}$ per year. These average dose value and the distribution of TID and NIEL based on a simplified FLUKA model have also been given to sub-detector designers for reference.

With the current IR design, background levels are acceptable for most sub-detectors. However, in high luminosity $Z$ mode, which has a luminosity almost two orders of magnitude higher than the Higgs mode, further optimization must be conducted. It is aimed to achieve a lightweight and efficient design through more optimized mitigation methods. It should be noted that the results and work presented have considered only the ideal beam conditions. In the future, more realistic scenarios must be carefully studied. These include non-ideal beam conditions (e.g., changes in beam orbit, tuning and commissioning phases, and magnet misalignment), as well as currently unaddressed items including injection backgrounds and failure cases like sudden beam loss.

## 3.4 Luminosity measurement

### 3.4.1 Introduction

The high event statistics foreseen at the CEPC require that the luminosity be measured with a precision better than $10^{-4}$ at the $Z$-pole for precision measurements of multiple SM processes. The $e^+e^-$collision luminosity is determined by measuring the rate of the Bhabha elastic scattering. The QED prediction for Bhabha scattering is with high precision, including complete Next-To-Leading Order (NLO) corrections, which encompass processes such as $e^+e^- \to e^+e^-\gamma$. The BHLUMI event generator, which was developed at the LEP for calculating low-angle Bhabha cross sections, has achieved a precision of 0.037% [26, 27]. Ongoing efforts aim to improve higher-order corrections, driven by the precision requirement of $10^{-4}$ at future $e^+e^-$ colliders. The leading-order Bhabha cross section, integrated over the angular coverage ($\theta_{min} < \theta < \theta_{max}$), is given by

$$\sigma = \frac{16\pi\alpha^2}{s}\left(\frac{1}{\theta_{min}^2} - \frac{1}{\theta_{max}^2}\right) \tag{3.1}$$

The integrated luminosity ($\mathcal{L}$) can be determined from the number of detected Bhabha events

$N_{acc}$, using the following equation:

$$\mathcal{L} = \frac{1}{\epsilon}\frac{N_{acc}}{\sigma}, \quad \frac{\Delta\mathcal{L}}{\mathcal{L}} \approx \frac{2\Delta\theta}{\theta_{min}} \tag{3.2}$$

where $\epsilon$ is the detection efficiency to be evaluated. The primary sources of systematic uncertainties are errors in $\theta_{min}$, which arise from mechanical alignment and detector resolution. These errors propagate to the luminosity measurement with a factor of two. To achieve the targeted $10^{-4}$ uncertainty in luminosity, assuming a $\theta_{min}$ threshold of 20 mrad for Bhabha events, the corresponding angular deviation $|\Delta\theta|$ is 1 μrad. At a distance of $|z| = 1$ m from the IP, this translates to a tolerance of 1 μm.

Bhabha events are identified by pairs of electrons and positrons that are back-to-back, each carrying a momentum equal to the beam energy. At the CEPC, these events are measured using the LumiCal, the BPM located near the interaction region (IP BPM), and the FastLumi. These systems are symmetrically installed on both sides of the IP (±z directions) and upstream of the cryomodule. A schematic layout on one side shown in Figure 3.13 illustrates their arrangement. The LumiCal, composed of two silicon detectors and two LYSO crystal parts, is installed inside the detector beam pipe. It covers the low-angle region, defined by polar angles $|\theta| < 100$ mrad in the laboratory frame. The FastLumi is a diamond-based detector positioned to cover the very forward region around $|\theta|$ 10 mrad. Two BPMs are deployed on each side, with one located inside the flange and the other on the dual beam pipes. In the following, the design of the three detectors is discussed.

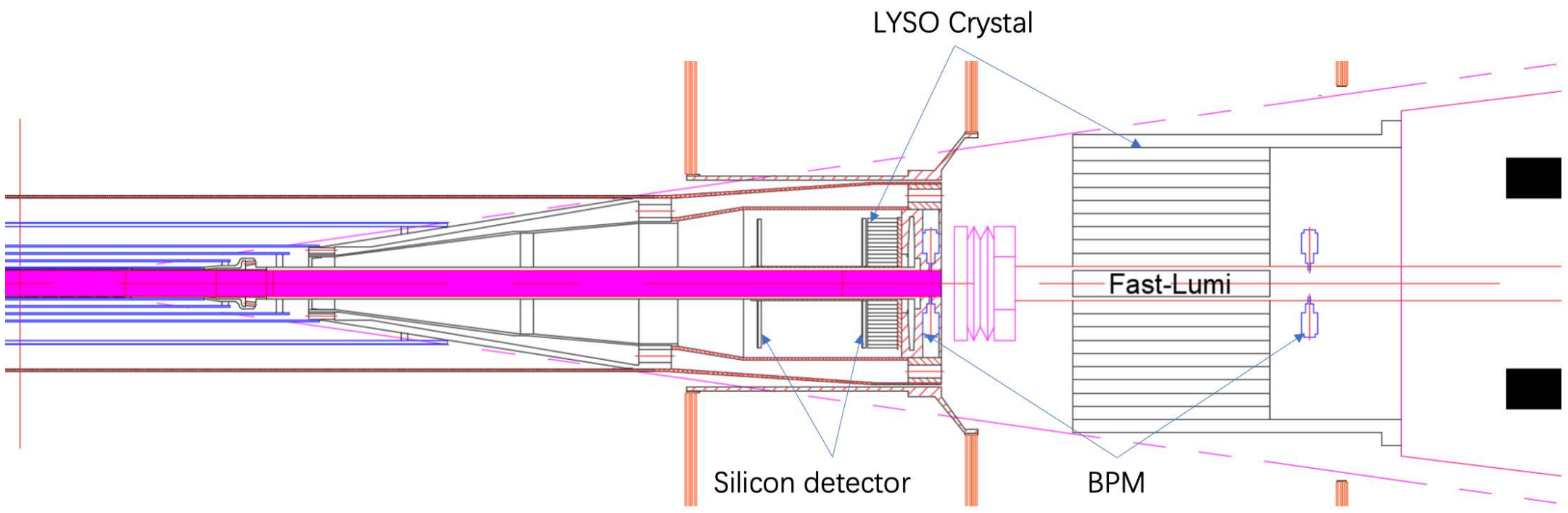

**Figure 3.13:** The luminosity measurement detectors are shown. Those detectors include the LumiCal, the FastLumi and the BPM. The LumiCal consists of two silicon detectors and two LYSO crystal parts. Before the flange, the Si-wafers combined with short LYSO crystals, provide the electron $\theta$ position with the $e/\gamma$ veto functionality. The long LYSO located behind the bellow is designed to identify beam electrons through energy deposits.

The location of LumiCal is favorable for a lower $\theta_{min}$ to increase the Bhabha cross section due to its $1/\theta^2$ dependence. The front LumiCal silicon sensor is located at $|z| = 560$ mm. The race-track beam pipe has an inner radius of 10 mm. Assuming the detector acceptance edge is 12 mm from the beam pipe center, the corresponding $\theta_{min}$ is 21.4 mrad. The outer radii of the first and second Si-wafers are 42 mm and 51 mm, respectively. The outer radii of the short and long LYSO modules are 56 mm and 110 mm, respectively. Figure 3.13 illustrates LumiCal, including the silicon wafers, and the LYSO crystals.

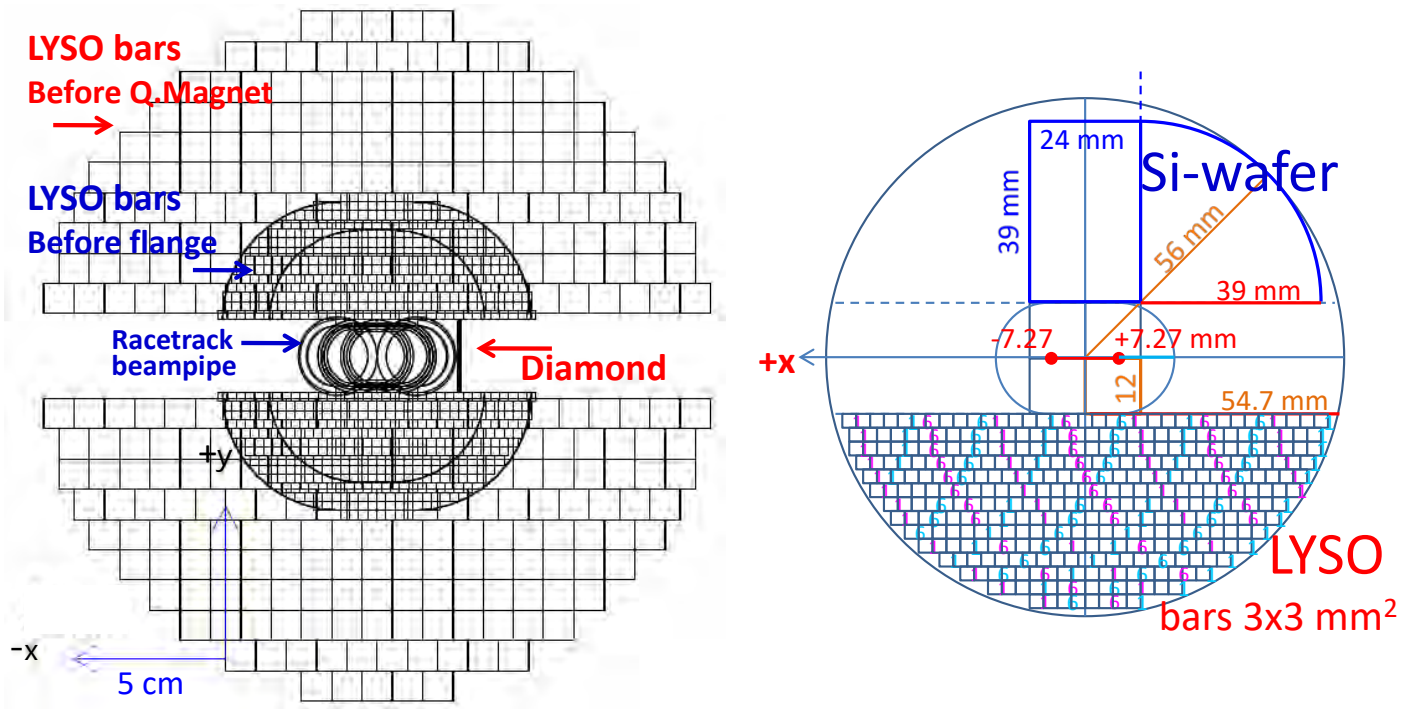

**(a)**

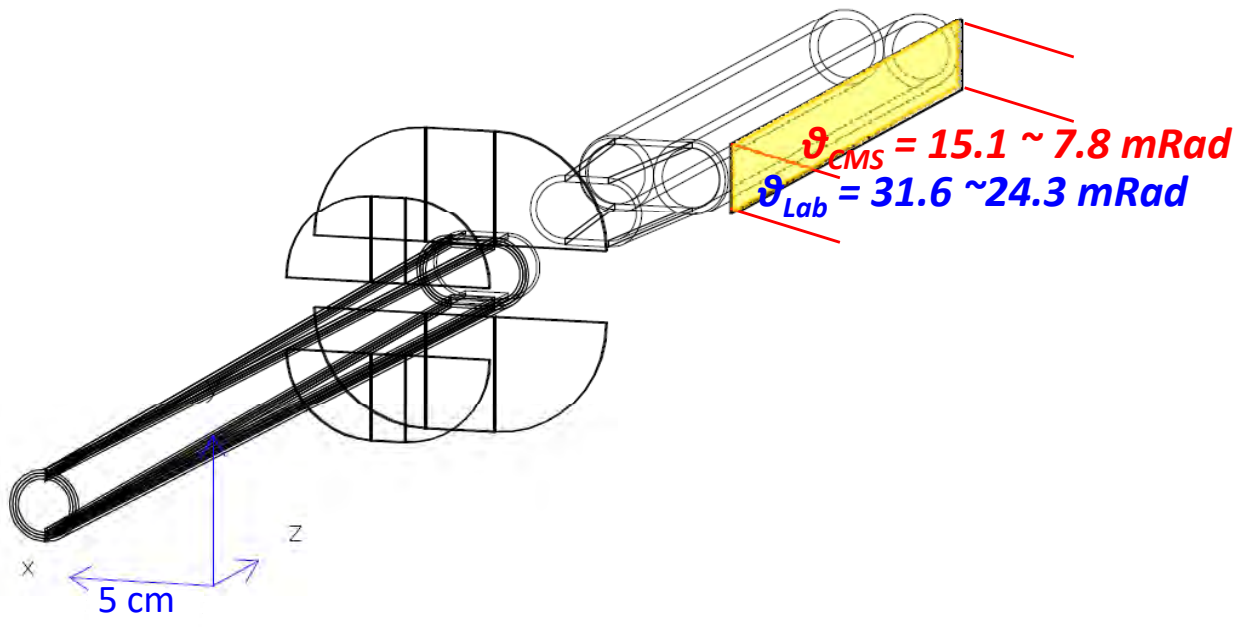

**(b)**

**Figure 3.14:** The projection of the LumiCal design is illustrated in (a), with the racetrack beam pipe extending from the IP to the quadrupole. The design of the second Si-wafer positioned at $|z|$ = 640 mm, and the 2 $X_0$ LYSO bars mounted on the flanges are also shown. The inner radius of the second Si-wafer is 12 mm, while the outer radius is 51 mm. The dimensions of the front LYSO bars are $3 \times 3 \times 23$ mm$^3$, while the outer radius of the overall module is 56 mm. The long LYSO modules located behind the bellow are segmented into $10 \times 10 \times 150$ mm$^3$ blocks. The outer radius of the long LYSO modules is about 110 mm. The fast monitoring diamond detectors, as shown in (b), are positioned between the long LYSO modules on the sides of electron boosted direction, where the Bhabha electron rate is the highest. The ”5 cm” in figure is used as the measurement scaling.

The latest design positions of the front surface of the at $|z|$ = 1050 mm, while the LYSO bars are located in the region of between $|z|$ = 800 mm and 950 mm. The $x$-$y$ projection of the LumiCal is shown in Figure 3.14a, depicting the race-track beam pipe extending from of a 20 mm diameter beryllium pipe at the IP into dual copper beam pipes with a thickness of 3 mm. The LumiCal modules are positioned above and below the beam pipe. The dimensions of the components upstream of the flange are illustrated in Figure 3.14b, which also shows the second layer of Si-wafers and the LYSO bars with a radiation length of 2 $X_0$.

The FastLumi detector is designed to detect single Bhabha electrons, which has the highest rate around the outgoing beam pipe. It is positioned on the $+x$ side of the beam pipe near the cryomodule, accepting particles at a polar angle of approximately 10 mrad. Utilizing a diamond detector, the FastLumi can withstand the high radiation levels in this region. By detecting the rate of Bhabha events, it provides online luminosity measurements. The location of the diamond FastLumi detector is shown in Figure 3.14b.

### 3.4.2 Fast luminosity monitor

The FastLumi is designed to provide real-time feedback of beam collisions, which is crucial for steering the beam position. It is located in the region with the highest $e^+e^-$ collision event rate, partly due to the beam crossing angle of 33 mrad. The FastLumi can assist in beam steering by observing the scattering of electrons on both $z$-sides. The integrated event rates along the $z$-positions can help monitor the $z$-position of the IP.

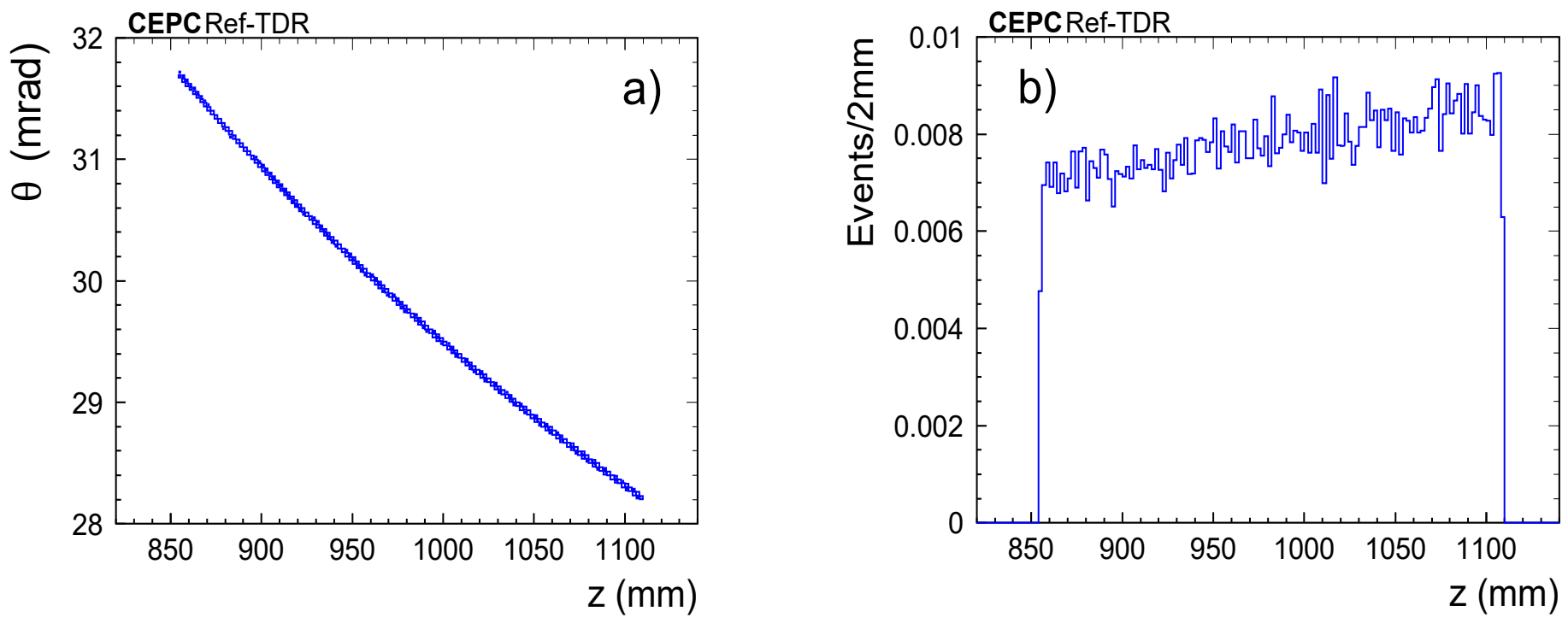

**(a)** **(b)**

**Figure 3.15:** Simulation of a 50 GeVelectron shower in a 3-mm-thick copper beam pipe. Results are shown for an electron incident at 10 mrad relative to the 16.5 mrad outgoing beam direction, interacting with a diamond slab ($24 \times 320$ mm$^2$) positioned 13 mm from the beam center. The distribution of hits on the slab is displayed as a function of the laboratory-frame coordinates $\theta$ and z in (a), while the normalized event rate along the $z$-axis is presented in (b).

Figure 3.15a illustrates a shower event of a 50 GeV electron traversing the 3-mm-thick copper beam pipe at 10 mrad. Figure 3.15b shows the Bhabha electrons from BHLUMI plotted around the beam pipe between $|z|$ = 800 to 950 mm. The FastLumi is positioned at the region of high radiation field. Given the high radiation levels, the most suitable detector is a diamond slab segmented into strips, which is able to detect electron showers generated in the copper beam pipe.

Diamond detectors are extensively used for beam monitoring in collider experiments [28, 29, 30]. These detectors are able to maintain low-noise operation at TID levels exceeding 10 MGy, with signal attenuation corrected through proper calibration. The newly developed synthetic diamond technology has significantly reduced defect densities, making low-cost diamonds more accessible for large-scale applications. Electronic-grade diamonds grown by Chemical Vapor

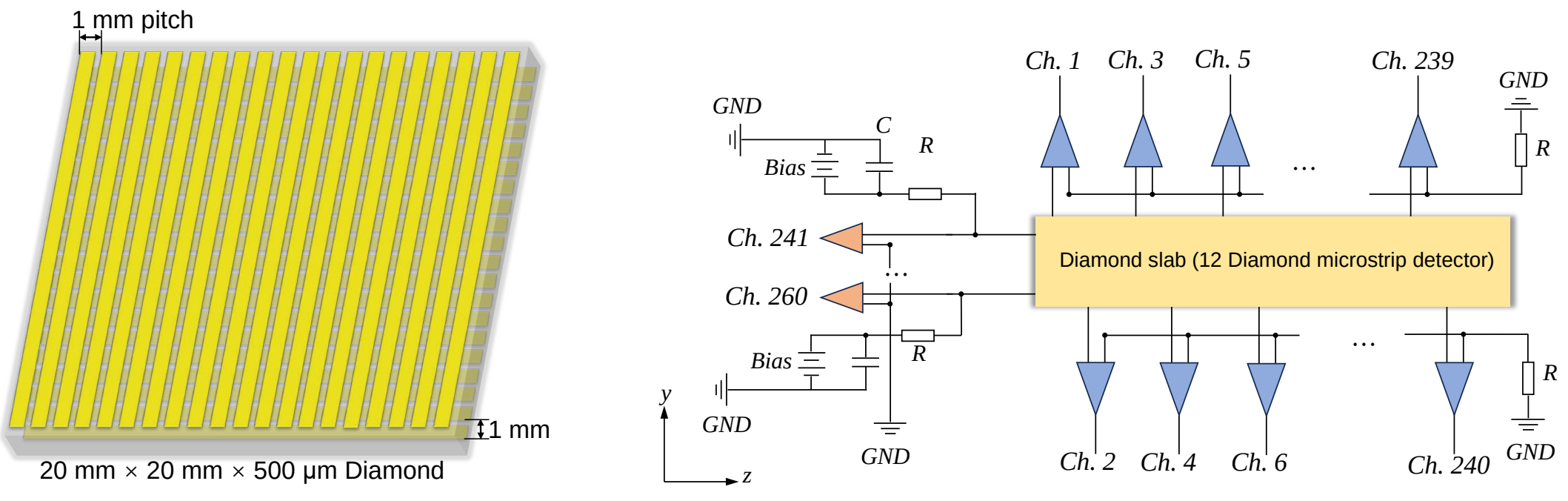

**(a)** **(b)**

**Figure 3.16:** (a) Schematic of a microstrip diamond sensor with orthogonal front-side and back-side microstrips, featuring a 1 mm pitch; (b) Illustration of the microstrip readout for a diamond slab. The front-side microstrips (channels 1–240) are connected to preamplifiers to measure $z$-axis IP offsets, whereas the back-side microstrips (channels 241–260) are connected to high voltage to measure $y$-axis IP offsets.

Deposition (CVD) are instrumented into strip diamond detectors. The prototype CVD-grown diamond wafer is fabricated with microstrip electrodes, on a $20 \times 20$ mm$^2$, 500 μm thick crystal. As shown in Figure 3.16a, the front face features twenty strips of $0.8 \times 19$ mm$^2$ with a 1 mm pitch and 0.2 mm gaps. The electrodes on the back-side have the same pattern but are oriented perpendicular to the front.

To achieve reliable Ohmic contact between the metal electrodes and the diamond crystal, the diamond surface must be oxygen-terminated [31]. The electrodes are created using a multilayer metallic stack through a series of photolithography, magnetron sputtering, and electron beam evaporation processes. The backside electrodes of the fabricated diamond detector are connected to a high voltage supply. The electric field can be raised to 2 V/μm, ensuring 100% ionization charge collection efficiency.

A luminosity monitor slab is proposed, consisting of 12 diamond detectors with a total of 240 strips. The readout circuits are illustrated in Figure 3.16b. The front-side electrodes (channels 1-240) are oriented perpendicular to the beam-pipe direction. The backside electrodes, with strips parallel to the beam-pipe, are interconnected to form 20 readout channels (channels 241–260). The center of the electron showers can be identified. By balancing the spectra of diamond slabs on both sides of the IP, the $x$- and $y$-positions of the electron beams and the IP position in the $z$-axis can be tuned and monitored with an angular precision of 1 μrad relative to the LumiCal.

In addition, a set of 1 cm × 3 cm SiC detectors is introduced to measure the electrons or positrons originated from the RBB events at dedicated positions for accelerator tuning, as discussed in Ref. [32, 33]. The design of both the sensor and the entire system is on-going.

### 3.4.3 Beam position monitor

The beam steering devices around the IP include the BPMs positioned inside flanges, and on dual pipes before the cryomodule, on both $z$-sides of the IP. The system operates by correlating the measured pulse heights from the BPM electrodes with the beam positions, allowing for the reconstruction of the true beam position in the $x$-$y$ plane through calibration. The system is designed to achieve a resolution of 1 μm, while the initial position accuracy is estimated to be $\sim$ 50 μm, which is primarily limited by the knowledge of the BPMs' true positions. The accuracy can be improved through precise surveys of the BPM positions.

The detection of Bhabha events will also rely on the BPM to determine the IP position. By combining a BPM position survey precision of 1 μm and a beam steering timing precision of 150 ps, the IP position can be determined with an accuracy of 1 μm, which is necessary for detecting Bhabha events.

The LumiCal's Si-wafers are mounted on the beam pipe, supported by same flange that houses the BPMs, while the BPMs are inserted inside the flanges. The center of the outgoing beam pipe at these flanges serves as the mechanical reference for the LumiCal. The online position deviation is consistent with the BPM output changes. During detector assembly, the focus will be on achieving high survey precision for the BPM and ensuring accurate relative positioning of the Si-wafers. For more detail design of the BPM, please refer to [1].

### 3.4.4 LumiCal

The LumiCal detector acceptance is optimized for elastic scattering of Bhabha events, where electrons are boosted towards the outgoing beam pipe direction. A lower $\theta$ fiducial angle relative to the beam center yields larger acceptance. The distribution of scattered electrons is simulated with BHLUMI. Figure 3.17a shows the $\theta$ distributions of scattered electrons in the center-of-mass frame and after being boosted into the laboratory frame, which accounts for the 33 mrad beam-crossing angle. The back-to-back angle distributions are plotted in Figure 3.17b. The boosted distributions in Figure 3.17 are selected based on the detector acceptance criteria: $r > 25$ mm from the beam centers at $|z| = 1000$ mm, corresponding to $\theta > 25$ mrad, and $|y| > 25$ mm with flat edges parallel to the race-track beam-pipe surfaces.

The cross-sections are calculated using BHLUMI within the fiducial region and are listed in Table 3.7. The cross-sections for detecting one or both electrons are provided for the lower $\theta$ cuts of 20 and 25 mrad. A slightly lower angle of 20 mrad results in a cross-section nearly twice as large as that at 25 mrad. The boost effect pushes events along the $x$-axis into the beam pipe, resulting in only about a 10% gain if requiring both electrons to be detected. For easier instrumentation, the cutoff band corresponds to $|y| < 25$ mm at $|z| = 1000$ mm.

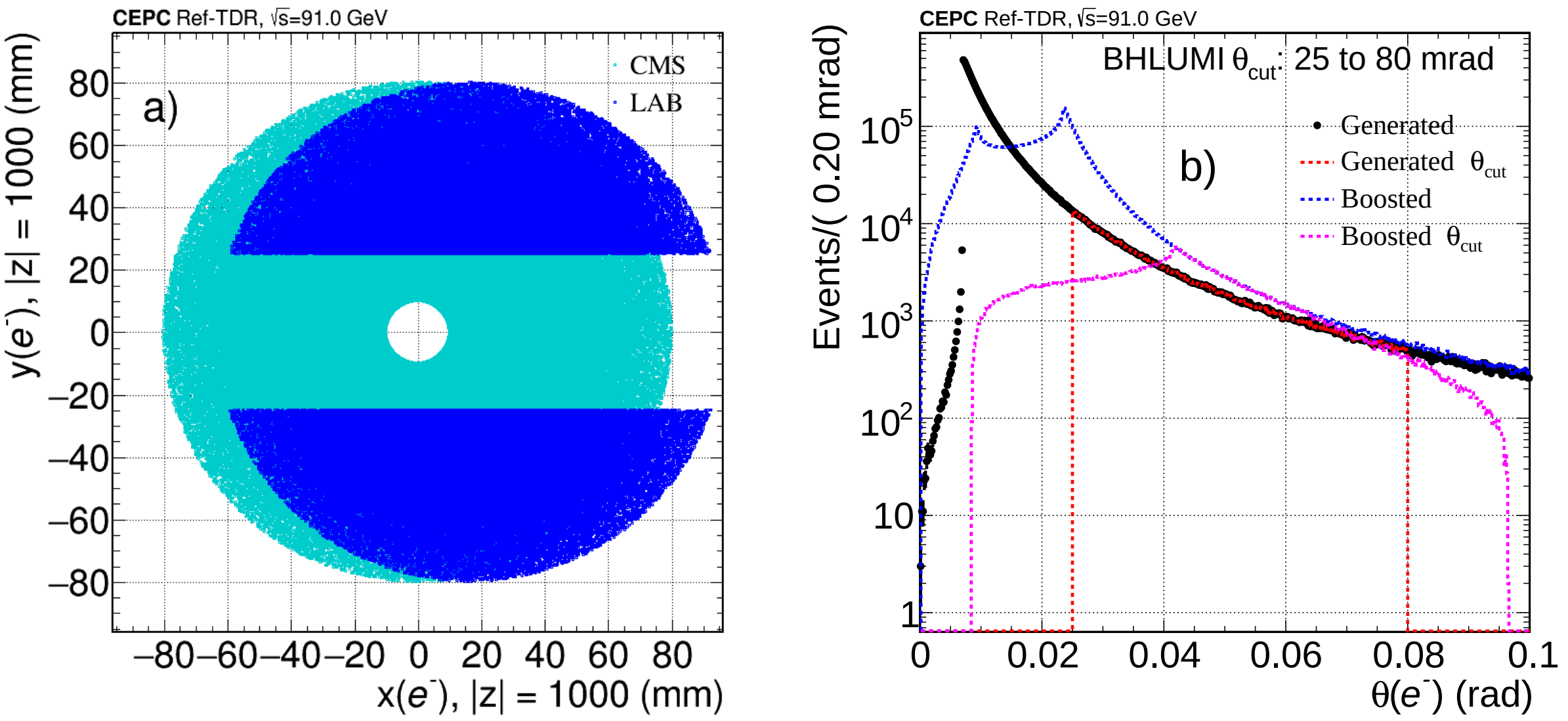

**(a)** **(b)**

**Figure 3.17:** (a) The distributions of scattered electron from BHLUMI simulations in the center-of-mass frame, after boosting for a 33-mrad beam crossing. Events are generated for $10 < \theta < 80$ mrad. The selections of the parameters are based on BHLUMI criteria for LEP applications. (b) The back-to-back angles between $e^+$ and $e^-$ are plotted for both the generated and boosted events, each with and without radiative photons, for events that pass the selection criteria.

**Table 3.7:** The acceptances of the LumiCal for Bhabha scattered electrons are simulated using BHLUMI. The cross-sections are listed for both $e^-$ and $e^+$ within the azimuthal angle range (Th1, Th2) in the center-of-mass frame, with the center-of-mass energy ratio $s'/s > 0.5$. The boosted $e^-$ or $e^+$ in the corresponding LumiCal acceptance of Th1< $r, |y|$ <Th2 at $|z| = 1$ m ($r$ radius to near beam-pipe center) are also listed.

| | CMS generated Th1, Th2 [mrad] | $e^-$ or $e^+$ to pipe centers [mm] | Single $e^-$ in [mm] | [mm] | Both $e^-$.and. $e^+$ in [mm] | [mm] |
|---|---|---|---|---|---|---|
| | 10 to 80 | $10 < r, \lvert y \rvert < 80$ | $r > 20$ | and $\lvert y \rvert > 20$ | $r > 20$ | .and. $\lvert y \rvert > 20$ |
| $\sigma$ [nb] | 1172.45 | 1217.45 | 217.20 | 137.46 | 140.74 | 131.21 |
| | 10 to 80 | $10 < r, \lvert y \rvert < 80$ | $r > 25$ | and $\lvert y \rvert > 25$ | $r > 25$ | and $\lvert y \rvert > 25$ |
| $\sigma$ [nb] | 1172.45 | 1217.45 | 133.42 | 82.20 | 85.89 | 78.08 |
| | 10 to 120 | $10 < r, \lvert y \rvert < 120$ | $r > 25$ | and $\lvert y \rvert > 25$ | $r > 25$ | and $\lvert y \rvert > 25$ |
| $\sigma$ [nb] | 1181.94 | 1226.98 | 144.59 | 91.34 | 96.94 | 87.18 |

#### 3.4.4.1 Detecting Radiative Bhabha events

The BHLUMI event generator provides a precise calculation of the Bhabha cross section, achieving a precision of 0.037%. The dominant uncertainties originate from missing photonic contributions at $O(\alpha^2 L)$ and hadronic corrections $\delta\alpha_{had}$. Future calculations incorporating NNLO corrections are expected to achieve a precision at the $10^{-4}$ level.

Radiative Bhabha events can alter the trajectories of the electrons that may smear the event rate crossing the $\theta_{min}$ fiducial edge. Therefore, measurement of $e^+e^-\gamma$ with Initial State Radiation (ISR) and Final State Radiation (FSR) photons is critical for assessing systematic uncertainties due to radiative Bhabha effects. The photons are simulated with the Yennie-Frautschi-Suura (YFS) exponentiation method [34] in BHLUMI, for the interaction of $e^+e^- \to e^+e^-(n\gamma)$.

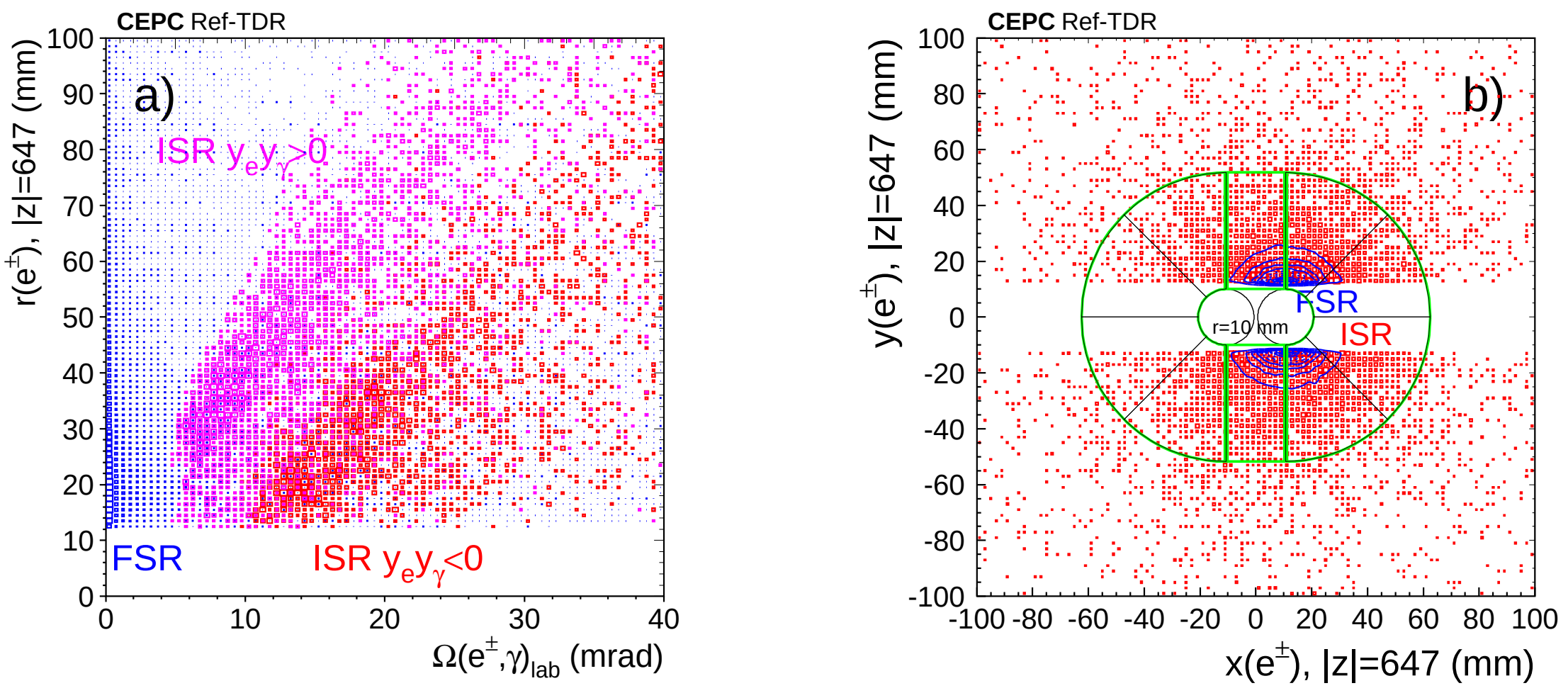

**(a)** **(b)**

**Figure 3.18:** Events with radiative photon ($E_\gamma > 0.1$ GeV) are plotted for $e, \gamma$ separation. (a) The positions of scattered electron in front of the LYSO detector(z=647 mm) are plotted for the laboratory frame, showing the $x$-$y$ radius versus the $e, \gamma$ opening angle within the same $z$-hemisphere. (b) The hit positions of electrons in the laboratory frame are plotted. For FSR, both $e$ and $\gamma$ are close to the beam pipe, whereas for ISR, $e$ and $\gamma$ are more widely separated.

The distinction between electrons and photons is achieved by combining information from the silicon wafer hits and the energy loss ($dE/dx$ ) in the segmented LYSO bars. The segmentation of the LYSO bars into $3 \times 3$ mm$^2$ corresponds to a pitch of 5 mrad. Figure 3.18a shows the radius of scattered electrons in the laboratory frame plotted against the opening angle to the photon within the same $z$-hemisphere. A key distinction between FSR and ISR photons is observed: FSR photons typically appear close to their parent electron, while ISR photons form distinct bands. The electron hit positions for events with ISR or FSR are illustrated in Figure 3.18b. For ISR events, the electron hits are more widely scattered towards larger radii.

#### 3.4.4.2 LumiCal detector simulation

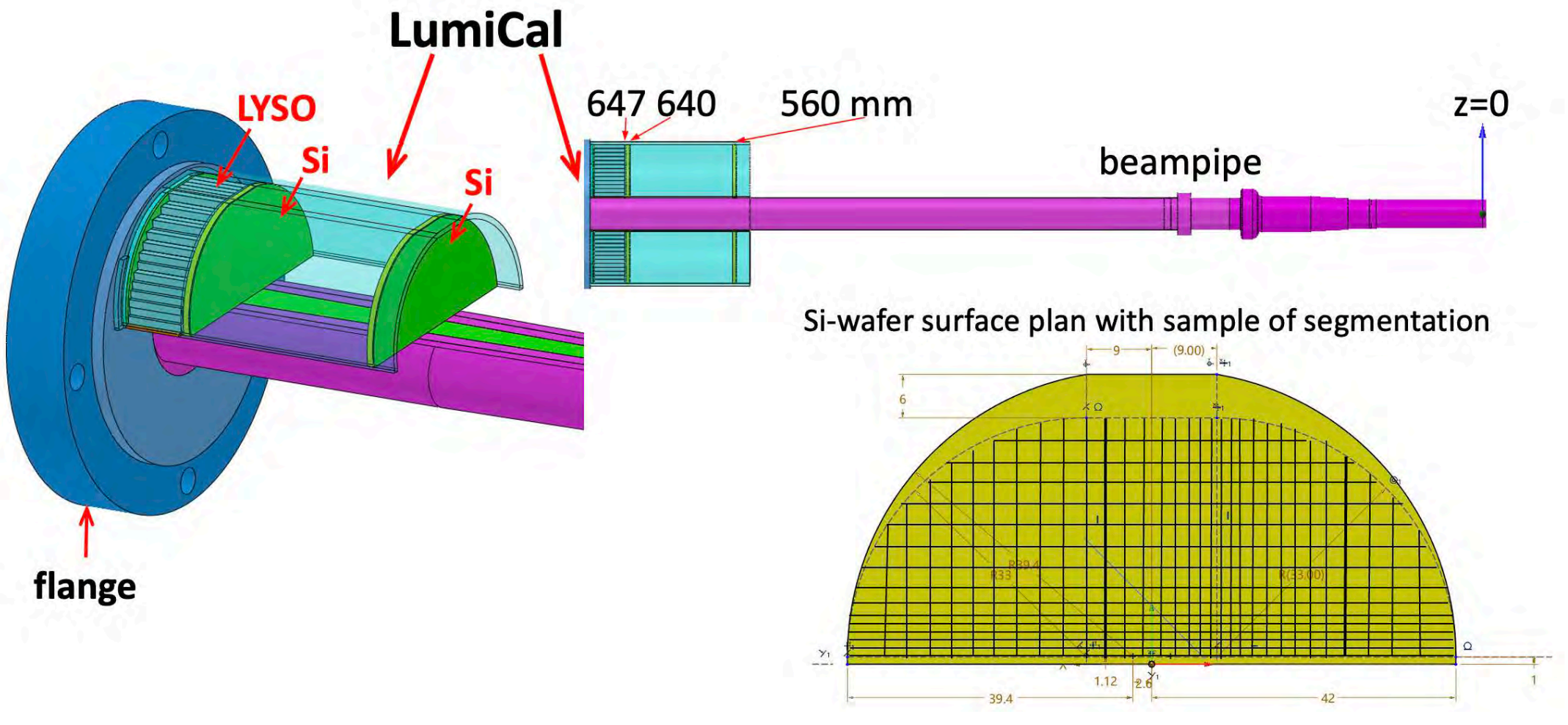

**(a)**

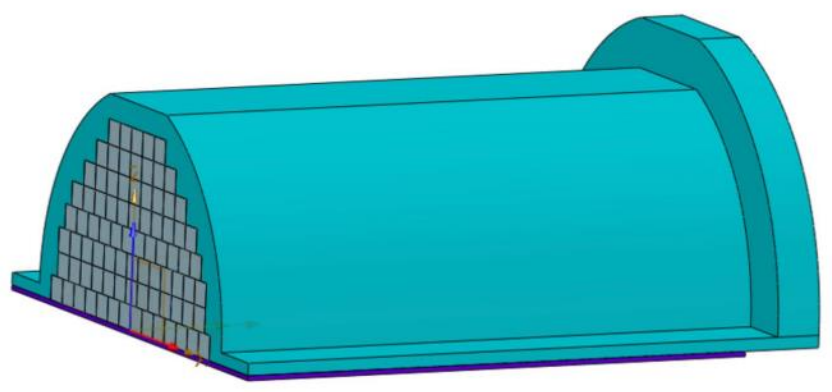

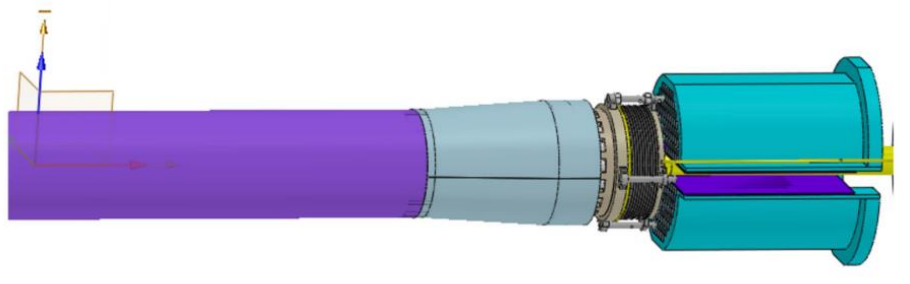

**(b)** **(c)**

**Figure 3.19:** (a) Mechanical drawings of the LumiCal modules before the flange of the race-track beam pipe are shown. The label is shown in mm. The two Si-wafers and 2 $X_0$ LYSO crystal bars are contained in half-circular tubes above and below the beam pipe. The cooling system for the beam pipe involves water injection from the flange towards the IP, passing through the double aluminum layers on the sides. The sample illustration of the segmentation in Si-wafer is presented. Currently, the design is planned to contain 128 segments in vertical and 320 segments in horizontal. The non-uniform segmentation symmetrized around the outgoing beam, with a higher density in regions of high signal activity (hotspots) and a sparser distribution in the periphery. The dimensions of each individual tile range from 50 to 300 μm in both length and height. (b) Mechanical drawings of the LumiCal modules beyond the flange, showing the 13.4 $X_0$ LYSO crystal bars enclosed within a tungsten shell. (c) The LumiCal LYSO modules are contained in half-circular tubes positioned above and below the beam pipe.

The LumiCal design must accommodate the limited space in the IR region while maximizing the $\theta$ coverage. Figure 3.19 illustrates the mechanical design of the LumiCal modules. The Si-wafers measuring the electron impact positions are located closer to the beam pipe to achieve the achieve coverage down to the smallest possible scattering angles (a low $\theta_{min}$ value), with minimum upstream beam-pipe materials. Si-wafers are segmented symmetrized around the outgoing beam with different size of strip pitches. The smallest pitch has a size of 50 μm that can achieve a resolution better than 5 μm. The 2 $X_0$ LYSO crystal covers a radius of 42 mm and has a length of 23 mm. The long LYSO modules consist of 10 layers of $10 \times 10 \times 150$ mm$^3$ bars. To minimize multiple scattering effects, a beryllium window is incorporated into the extending beam pipe, as described in Section 3.2.1.

The detector simulation using Geant4, based on the generated Bhabha electron distribution from BHLUMI, provides detailed information on multiple scattering and electromagnetic showers within the LumiCal. Bhabha electrons are predominantly distributed near the outgoing beam pipe. The smearing effects caused by multiple scattering are symmetric in the $x$-direction, both inside and outside the pipe region. The uncertainties in the mean scattering angles are less than 1 μrad. At the position of the first silicon wafer ($|z|$ = 560 mm), most electrons entering with $|y| > 12$ mm have scattering angles $\theta > 21$ mrad, corresponding to an angular resolution of $\Delta\theta = 1$ μrad for $\Delta\mathcal{L}/\mathcal{L} = 10^{-4}$ (Eq. 3.2).

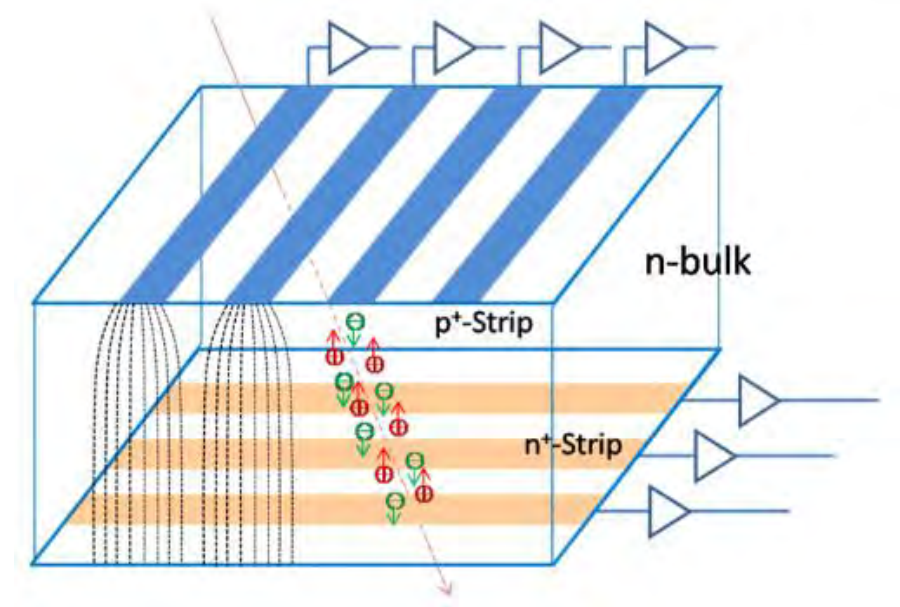

**(a)**

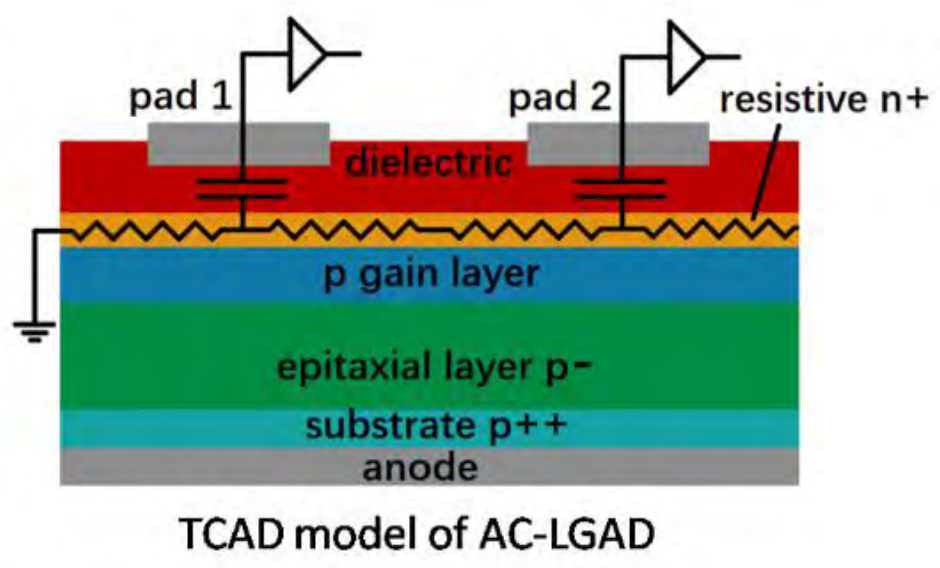

**(b)**

**Figure 3.20:** The Si-detector options are shown for (a) conventional double-sided detector with strips implemented on both sides of the wafer; (b) AC-LGAD detector with AC-coupling layers on top of the LGAD pads. A double AC layer configuration is proposed for 2D readout of the electron hits.

The Si-detectors are chosen for their low mass and high resolution in measuring electron impact positions. The options include the 2-dimensional (2D) Si-strip detectors shown in Figure 3.20a and the newly developed AC-LGAD depicted in Figure 3.20b. The AC-LGAD is also considered for the Silicon Tracking detector as described in Section 5.3.

The low-mass beam pipe minimizes the production of electromagnetic shower secondaries before the Si-wafer. The long LYSO array of 13.4 $X_0$ measures the beam electrons through energy. This requires that the primary electrons traverse the upstream materials—a short LYSO crystal (2 $X_0$), a flange, and a bellow (4.4 $X_0$)—totaling 6.3 $X_0$. The geometric acceptance of this system, illustrated in Figure 3.21a, is determined by scanning the detector response along the vertical ($y$-axis). The total energy measured in the long LYSO is plotted as a function of the incident angle in Figure 3.21b. The uniform acceptance plateau between 20 and 75 mrad corresponds to the LYSO front surface ranging from $|y|$=10 mm to 110 mm.

The $e/\gamma$ identification for radiative Bhabha events is conducted with electrons and photons generated by BHLUMI in conjunction with the LumiCal Geant4 simulation. The photons can be either ISR or FSR. The analysis targets scattered electrons entering the 2 $X_0$ LYSO

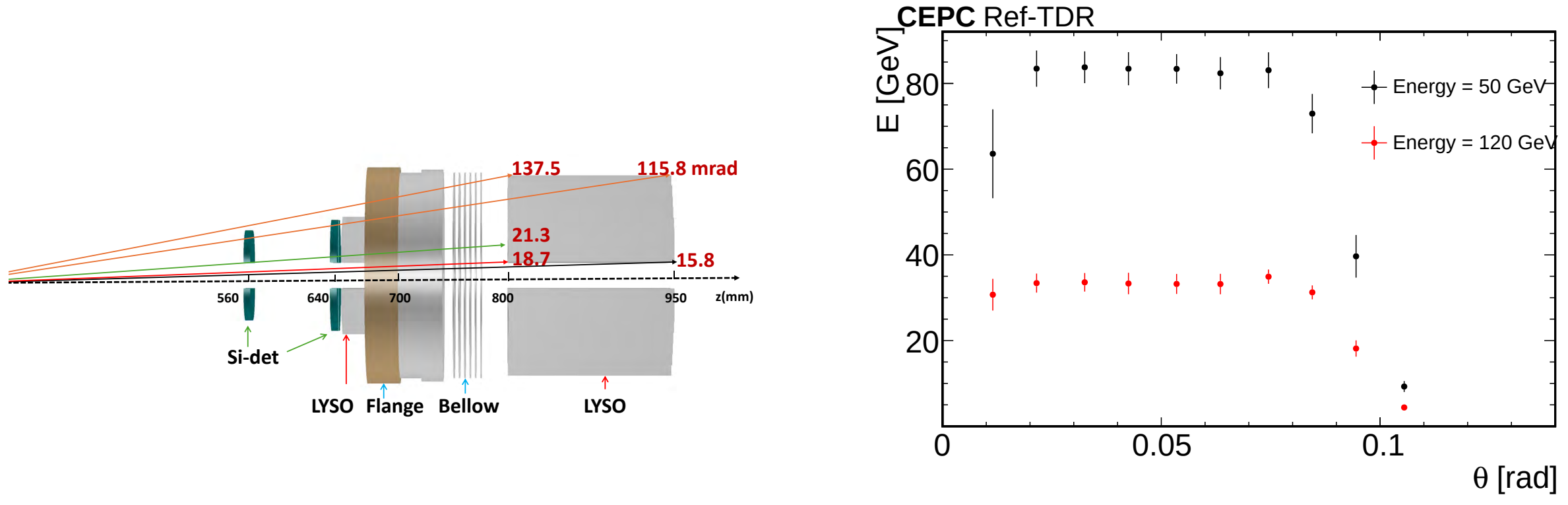

**(a)** **(b)**

**Figure 3.21:** The acceptance of the long LYSO detector is scanned along the vertical $y$-axis as a function of the electron $\theta$ angle, with the corresponding angles illustrated in (a), $y$-axis points to the up while $z$-axis points to the right. The total energy (E) deposited in the detector is plotted as a function of the electron incident angle ($\theta$) for 50 GeVand 120 GeVelectron beams in (b). The geometric acceptance of the detector is defined by the plateau in the deposited energy, which approximately ranges from 20 to 75 mrad, with sharp fall-offs to zero outside this range. The upstream materials (totaling 6.3 $X_0$ ) cause an energy loss that constitutes about 40% (30%) of the signal for 50 GeV(120 GeV) electrons within the acceptance.

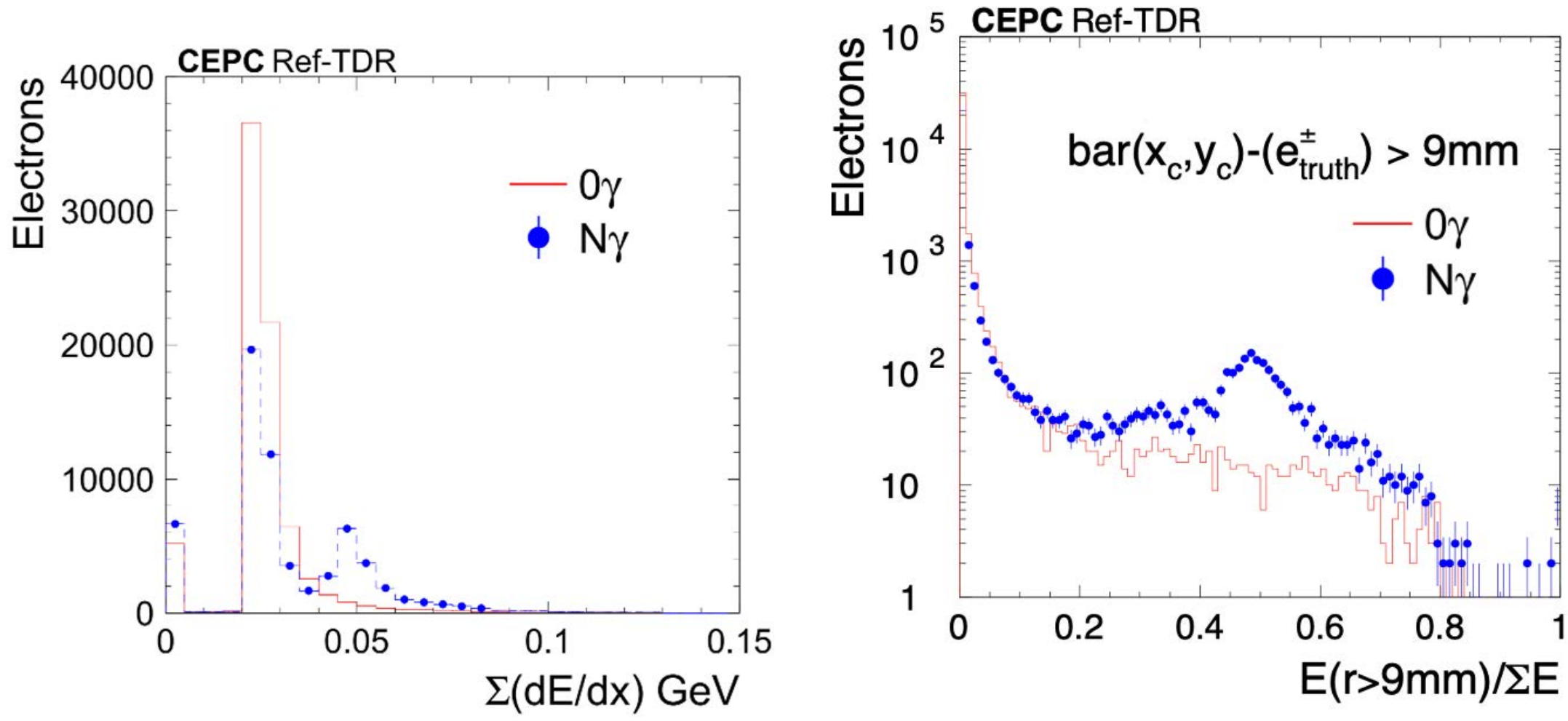

**(a)** **(b)**

**Figure 3.22:** The scattered electrons from BHLUMI with projected truth directions entering LYSO at $|z|$ = 647 mm and $|y|$ > 12 mm, are simulated with GEANT. The energy deposits of electrons in a half-circle module are plotted in (a). The double peak observed in the energy deposits for events with photon arises from the difference between ISR and FSR photons. FSR photons, which deposit energy close to the electron trajectory, contribute to the peak at higher energy. The ISR photons, on the contrary, typically deposite energy away from the electron and contribute more to the peak at lower energy. The fraction of energy deposited in LYSO bars 9 mm off the projected electron direction is plotted in (b), indicating the observability of a photon that can be identified through the LYSO bar segmentation.

arrays by examining the energy deposition pattern in the segmented $3 \times 3$ mm$^2$ LYSO bars. In Figure 3.22a, the energy deposits are compared between Bhabha events with and without incident photons that could be ISR or FSR. On a module of LYSO bars within the same $y$, $z$ hemisphere, electrons accompanied with an FSR photon exhibit energy deposits at twice the $\mathrm{d}E/\mathrm{d}x$ location. This accounts for approximately 1/4 of the events, resulting in the double-peak structure in the deposited energy distribution. The separation of FSR photons from the electron is demonstrated in Figure 3.22b, showing the energy deposited in cells centered 9 mm away from the projected electron position. The enhanced peak in this distribution indicates the presence of a photon hit.

#### 3.4.4.3 Systematic uncertainties on integral luminosity measurement

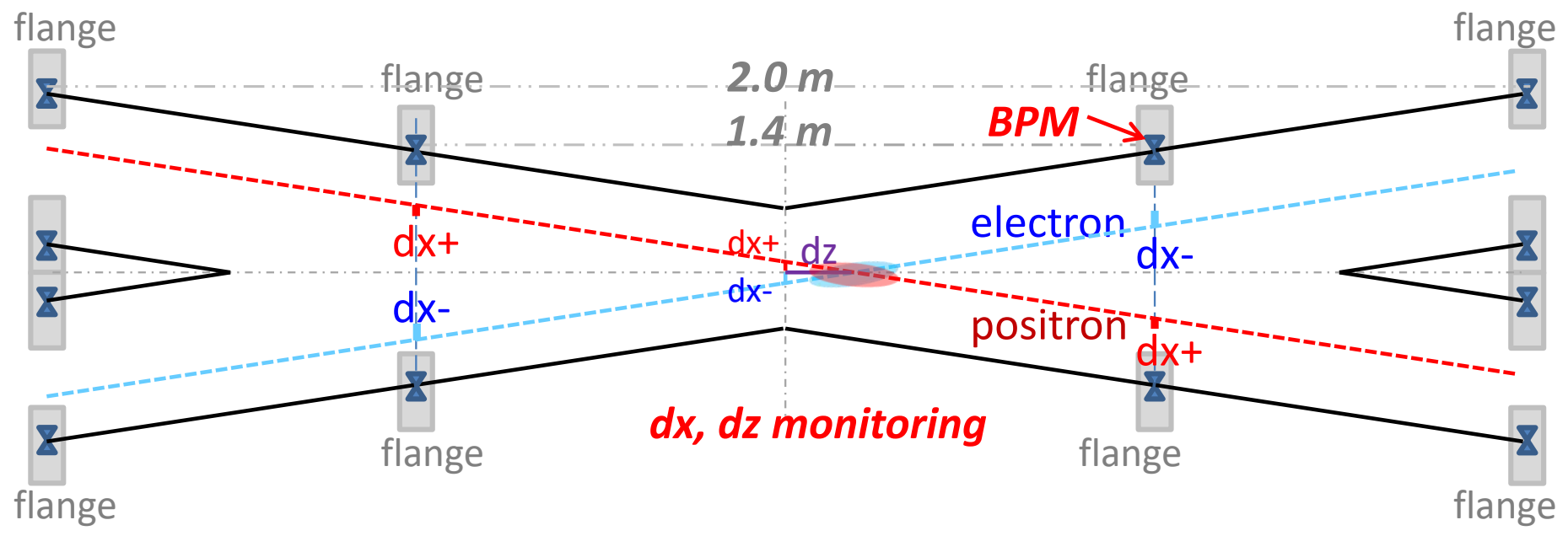

**Figure 3.23:** The beam currents in the interaction region are illustrated, showing the offsets to be measured with BPM.

The luminosity measurement is based on Bhabha events ($N_{\mathrm{Bh}}$) where both scattered electrons are detected in coincidence in the LumiCal modules on both $z$-sides. The luminosity is calculated using the formula: $\mathcal{L} = \frac{N_{\mathrm{acc}}}{\epsilon\sigma^{\mathrm{vis}}}$, where $\epsilon$ accounts for systematic uncertainties on instrumentation, and $\sigma^{\mathrm{vis}}$ on the theoretical calculation, depending on the precision on the fiducial acceptance. The fiducial acceptance is primarily determined by the lower $\theta$ edges, requiring a precision better than 1 μrad. The tolerance on the radius at $|z| = 560$ mm is 0.56 μm. The systematic uncertainties associated with the scattered electron $\theta$ can be categorized into four categories:

**a. Beam properties and interactions**

At the Z-pole, the beam energy spread is 0.13% Ref.[1]. Systematic uncertainties due to beam energy offsets can lead to deviations on the Bhabha cross section and counting uncertainties due to a longitudinal boost of the collision system caused by asymmetry in energy of the beams. The detailed study is presented in Ref.[35] and it is found that the beam energy can be determined with the 24 MeV variation for the beam-spread of 0.13%, using the central production of di-muon pairs, with the systematic contribution to the integrated luminosity uncertainty $\leqslant 10^{-4}$. As well, the longitudinal boost of the Bhabha center-of-mass frame with respect to the laboratory frame may be caused by the bias of one beam with respect to the other that should not be larger than 7 MeV as mentioned in Ref.[35].

Beam-beam interactions, as well as the interactions of incoming bunches with the Bhabha final states will occur impacting four-momenta of both initial and final states. Together with Ref.[36], there is ongoing study of these effects. Both effects are simultaneously present contributing to the loss of collinearity of Bhabha events that is coincidentally counted in left and right arms of the luminometer. If uncorrected, they will lead the relative loss of count in the luminometer of order of several permille at the Z-pole. However, the impact of electromagnetic deflection of the initial-state can be corrected from the measurement of the crossing angle using the s-channel di-muon production, where a crossing angle precision is of order of 1 μrad can be achieved for as little as 70 pb $^{-1}$ of integrated luminosity corresponding to less than 10 minutes run at the Z-pole. Electromagnetic deflection of the Bhabha final states leads to the relative luminosity loss below two permille and can be corrected in a simulation dependent way with the relative uncertainty of order of $10^{-4}$.

**b. Uncertainties on the IP position**

The position of IP is best measured by the BPM installed on the beam pipe. The closest BPM sets are located at $|z| = 700$ mm, inside the flanges, as illustrated in Figure 3.23. The single beam BPM positioned in front of the cryomodule can provide pulse measurements of the bunches corresponding to the $x$-$y$ positions of the beams.

The bunch size has the widths of $\sigma_x = 6$ μm and $\sigma_z = 10.6$ mm. The 33 mrad bunch crossing angle reduces the $z$-width to $\sigma_z = 380$ μm. Beam tuning will iterate on the timing of bunches arriving at IP, and perform the $x$-$y$ scans. At $\theta = 25$ mrad, the 1 μrad offset corresponds 40 μm at $|z| = 1000$ mm, where the BPM timing precision is required to be 1.3 ps. As discussed in Ref. [37], the timing precision better than three ps is sufficient to contribute to the relative systematic uncertainty of the integrated luminosity no more than $10^{-4}$. The BPM spacing precision in the $x$-$y$ plane is better than 1 μm. By scanning the beam currents in the $x$-direction, the crossing point is at $z = 0$, and the timing to the brightest luminosity shall yield the IP to the ideal beam pipe center.

**c. Uncertainties on measuring electron hits**

The impact of multiple scattering on the Bhabha cross-section, due to the beam-pipe material, was simulated using the BHLUMI-generated electron distribution processed through GEANT. Assuming that the LumiCal has the lower $\theta$ edge for $|y| > 20$ mm at $|z| = 1$ m, corresponding to a varying $\theta_{\min}$ starting from the lowest value of 20 mrad, the simulation includes multiple scattering effects. This scattering is modeled as a convolution of 100 μrad on the scattered electron angles $\theta$ and $\phi$.

Figure 3.24a shows the electron hits at the LumiCal $y$-edge, with the cut applied to the

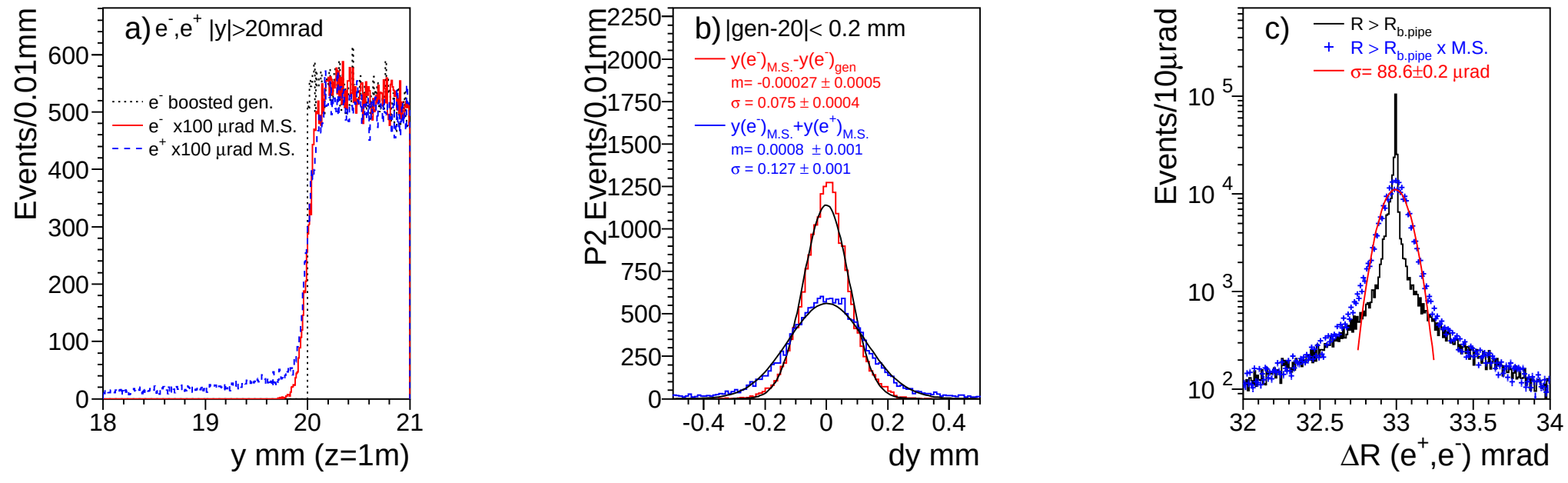

**(a)** **(b)** **(c)**

**Figure 3.24:** The scattered electrons (P2) and positrons (Q2) both that are accepted within the fiducial edges of the LumiCal at $|z| = 1$ m are plotted in (a) for their distribution in $|y|$. The dotted line represents the BHLUMI-generated P2 events cut at $|y| > 20$ mm. P2 Events smeared by multiple scattering (M.S.) are plotted in the red solid line. The Q2 events on the opposite $z$-side are plotted in blue dashed lines, with the tail representing events having radiative photons.

generated electrons. Multiple scattering deviates electrons in and out of the edge. The two Bhabha electrons are back-to-back symmetric with respect to the outgoing beam center, with $(x, y)$ coordinates of (16.5 mm,0) at $|z| = 1$ m. The $-y$ distribution of electrons on the opposite $z$-side is plotted to highlight the electrons from radiative events that have large $|y|$ beyond the cut (indicated by the blue line). The back-to-back tracking of the Bhabha electrons pairs will also provide calibration for the detector positions. The front silicon layer of the luminosity detector is designed to measure electron impact positions to a few microns. This sensor is a 2D fine-pitch strip detector, where the position is measured by strips that collect the ionization charges generated by a traversing electron.

**d. Uncertainties related to detector beam pipe, LumiCal survey and monitoring**

There are extensive studies performed on mechanical precision and alignment requirements in terms of relative positioning of the luminometer with respect to the IP and left and right arms of the luminometer with respect to each other, as well as the other mechanical effects like tilt and twist of the luminometer Ref. [37]. It is found that these effects will be no issue for the targeted luminosity precision at the Z-pole. The alignment precision of the front-layer Silicon detector is the most critical issue as well as the uncertainty of the inner aperture of the LumiCal of 1 µm to reach the targeted luminosity measurement precision of $10^{-4}$. An online survey system is required to monitor the fiducial edge position, ensuring that the uncertainty on the mean value is better than 1 µm. Other effect like the Gaussian spread of the measured radial shower position with respect to the true hit position and the maximal absolute uncertainty of the longitudinal distance between left and right halves of the luminometer were found to be of order of few hundreds of microns what should not be an issue with a tracking layer placed in front the luminometer and position monitoring system of the detector based on available FSI technologies.

## 3.5 Summary and future plans

The design of the entire interaction region and the luminosity measurement system was completed, and background level estimate was performed based on this design. Theoretical and simulation analysis results indicate that the background level achieved with the appropriately shielded design meets the requirements of the various detector systems for both the Higgs and low luminosity $Z$ modes. Furthermore, the luminosity measurement system can attain the target accuracy of $10^{-4}$ at the $Z$-pole. However, the current design faces several challenges. For instance, the implementation of shielding measures has resulted in cryogenic components that are relatively heavy. Additionally, some mitigation techniques, such as the collimators, could be further optimized. Critical technologies associated with the luminosity measurement detectors and detector beam pipe also require further validation. A prototype will be developed to verify the feasibility of these key technologies after further design optimization to verify the feasibility of these key technologies, thereby establishing a more robust foundation for future construction.

# Chapter 4 Vertex Detector

The identification of heavy-flavor ($b$- and $c$-) quarks and the $\tau$ lepton is essential for the CEPC physics program. It requires precise determination of the track parameters of charged particles in the vicinity of the IP, permitting the reconstruction of the displaced decay vertices of short-lived particles. This drives the need for a Vertex Detector (VTX) with low material budget and high spatial resolution. The detector is a cylindrical barrel with six silicon pixel sensor layers. It is optimized for the CEPC energy regime and uses modern sensors with integrated readout electronics.

The VTX design requirements, together with the overall detector layout, and estimates of the beam-induced background rate and radiation dose are outlined in Section 4.1. The sensor and readout technologies are discussed in detail in Section 4.2. The mechanical structure, cooling, and the service infrastructure are presented in Section 4.3, while the alignment and calibration schemes are discussed in Section 4.4. Past key technology R&D efforts supporting the VTX design are documented in Section 4.5. This includes the development of a full-size pixel chip and construction of a vertex detector prototype. Section 4.6 introduces an alternative layout concept. Detailed simulations of expected detector performance and the alignment strategy are analyzed in Section 4.7. The design summary and future development plans are provided in Section 4.8.

## 4.1 Detector overall design

### 4.1.1 Vertex detector requirements

As required for the precision physics program, the CEPC vertex detector is designed to achieve excellent impact parameter resolution, which, in the $r\phi$ plane can be parameterized by

$$\sigma_{r\phi} = a \oplus \frac{b}{p(\mathrm{GeV})\sin^{3/2}\theta} \tag{4.1}$$

where $\sigma_{r\phi}$ denotes the impact parameter resolution, $p$ the track momentum, and $\theta$ the polar track angle. The first term describes the intrinsic resolution of the vertex detector without multiple scattering and is independent of the track parameters, while the second term reflects the effects of multiple scattering. The parameters $a = 5$ µm and $b = 10$ µm GeV are taken as the design goals for the vertex detector. The main physics performance goals can be achieved with six concentric cylinders of pixelated vertex detector with the characteristics given in Table 4.1. The single-point resolution per layer should be better than 5 µm, while the full VTX material should not exceed 0.9% $X_0$, with particular care required in the innermost layers. The material of the beam pipe system, in front of the first layer, should be less than 0.5% $X_0$, while the first layer should be located as close as possible to the beam pipe.

**Table 4.1:** Physics Requirements for the Vertex Detector.

| **Parameter** | **Requirement** |
|---|---|
| Single-point resolution per layer | $\leq 5$ μm |
| Detector material budget | $\leq 0.9\% \ X_0$ |
| Angular coverage | $\lvert\cos\theta\rvert < 0.99$ |
| Beam pipe material budget | $< 0.5\%$ $X_0$ |
| Detector occupancy | $< 1\%$ |

To minimize the material budget, it is desirable to air cool the detector, which in turn requires the power consumption of the sensors and readout electronics to be kept below 40 $mW/cm^2$. The detector occupancy should not exceed 1% for efficient pattern recognition.

The radiation tolerance requirements, which are critical for the innermost detector layer, are driven by the beam-related backgrounds as described in Chapter 3.

### 4.1.2 Vertex detector design

The VTX is designed as six concentric cylindrical pixel layers covering a radius of 11.1 mm to 47.9 mm around the beam pipe, with a polar angle coverage of $\lvert\cos\theta\rvert < 0.99$. The layout of the VTX is shown in Figure 4.1 and Figure 4.2. Monolithic Active Pixel Sensor (MAPS) are used in all layers to meet the requirements of low power consumption, low material budget, and high spatial resolution listed in Table 4.1. In the inner four layers, stitched MAPS are gently curved to the required radius, so that each layer forms a seamless half-cylinder. These layers are collectively referred to as the Curved Vertex Layers (CVTXs). The outer two layers use traditional MAPS sensors named Planar Vertex Layers (PVTXs), of which the sensors are fixed inside and outside of a support structure to form a ladder. The ladders are arranged circularly to form a rough barrel, as shown in Figure 4.2. The pixel pitches are determined according to the requirement of 5 μm single-point resolution. All layers share a common MAPS front-end electronics that integrates amplification, discrimination and zero suppression in-pixel, achieving less than 5 μm single-point resolution with a timestamp precision of 100 ns. The 200 ns readout window for the VTX is chosen to balance high timing precision for resolving collision events at high frequencies (especially in both high and low-luminosity $Z$ modes) with low power consumption and efficient data processing. A stitched Repeated Sensor Unit (RSU) is the building block of the curved chips, maximizing wafer utilization. Average power density is required to be under 40 $mW/cm^2$, the limit of air-cooling.

The CEPC vertex detector is designed to operate at high collision frequencies, accommodating multiple collision modes. The collision frequency for the Higgs boson factory is approximately 1.7 MHz, while the $Z$ boson factory mode operates at about 14.5 MHz during the initial low-luminosity phase and increases to 43.3 MHz in the later high-luminosity phase. These frequencies pose challenges in maintaining low power consumption, which is a critical aspect of the VTX chip development for this project. A replacement of the VTX is expected after 10 years of operation, before the high-luminosity $Z$ boson factory mode.

Preliminary simulations and tests, based on the first VTX prototype, indicate that the TowerJazz 180 nm technology sensor power consumption at the low-luminosity $Z$ operation mode is larger than 60 mW/cm$^2$. This level of power dissipation exceeds the air cooling capacity of the VTX, resulting in sensor temperatures surpassing the operational upper limit of 30 °C, which may introduce higher noise levels and accelerate the aging of sensors, mechanical components, and other associated systems. To tackle this critical issue, the VTX will use Tower Partners Semiconductor Co. (TPSCo) 65 nm MAPS technology. In addition, Huali Microelectronics Corporation (HLMC) 55 nm MAPS technology is being explored as a potential alternative. This choice significantly reduces power consumption and also offers the potential for smaller pixel sizes, thereby enhancing spatial resolution. In summary, the VTX design parameters are listed in Table 4.2.

**Table 4.2:** Vertex Detector Design Parameters

| **Parameter** | **Design** |
|---|---|
| Spatial Resolution | ~ 5 µm |
| Detector material budget | ~ 0.8% $X_0$ |
| First layer radius | 11.1 mm |
| Power Consumption | < 40 mW/cm$^2$ (air cooling requirement) |
| Time stamp precision | 100 ns |
| Fluence | ~ $2 \times 10^{14}$ Neq/cm$^2$ |
| Operation temperature | ~ 5 °C to 30 °C |
| Readout electronics | Fast, low-noise, low-power |
| Mechanical Support | Ultralight structures |
| Angular Coverage | $\lvert\cos\theta\rvert < 0.99$ |

#### 4.1.2.1 Detector layout

**Inner layers with stitching design** For sensors constituting the curved layers of the VTX, the width of the curved sensor along the $\phi$-direction must correspond to the arc length of the semicircle with a radius of the CVTX. To achieve the required polar angle coverage, the length of the sensor's sensitive region along the beam direction ($z$-direction) is precisely defined. Due to limitations in current semiconductor manufacturing technology and processes, the sensors are arranged on 300 mm wafers, and the dimensions of a complete sensor are constrained by the wafer size. As a result, the number of sensors used in the z-direction varies for different CVTXs. Based on the radius and length requirements of the CVTXs, a $17.3 \times 20.0$ mm$^2$ RSU is defined as the basic unit (see Section 4.2.2.1). The repeating RSUs are stitched along the beam direction ($z$-direction). Three stitched sensors of different sizes are designed on a wafer, named type A, B and C as shown in Figure 4.3. Each type of sensor comprises modules arranged in multiple rows, where a module integrates several RSUs along with I/O pads and power supply positioned at both the left and right edges.

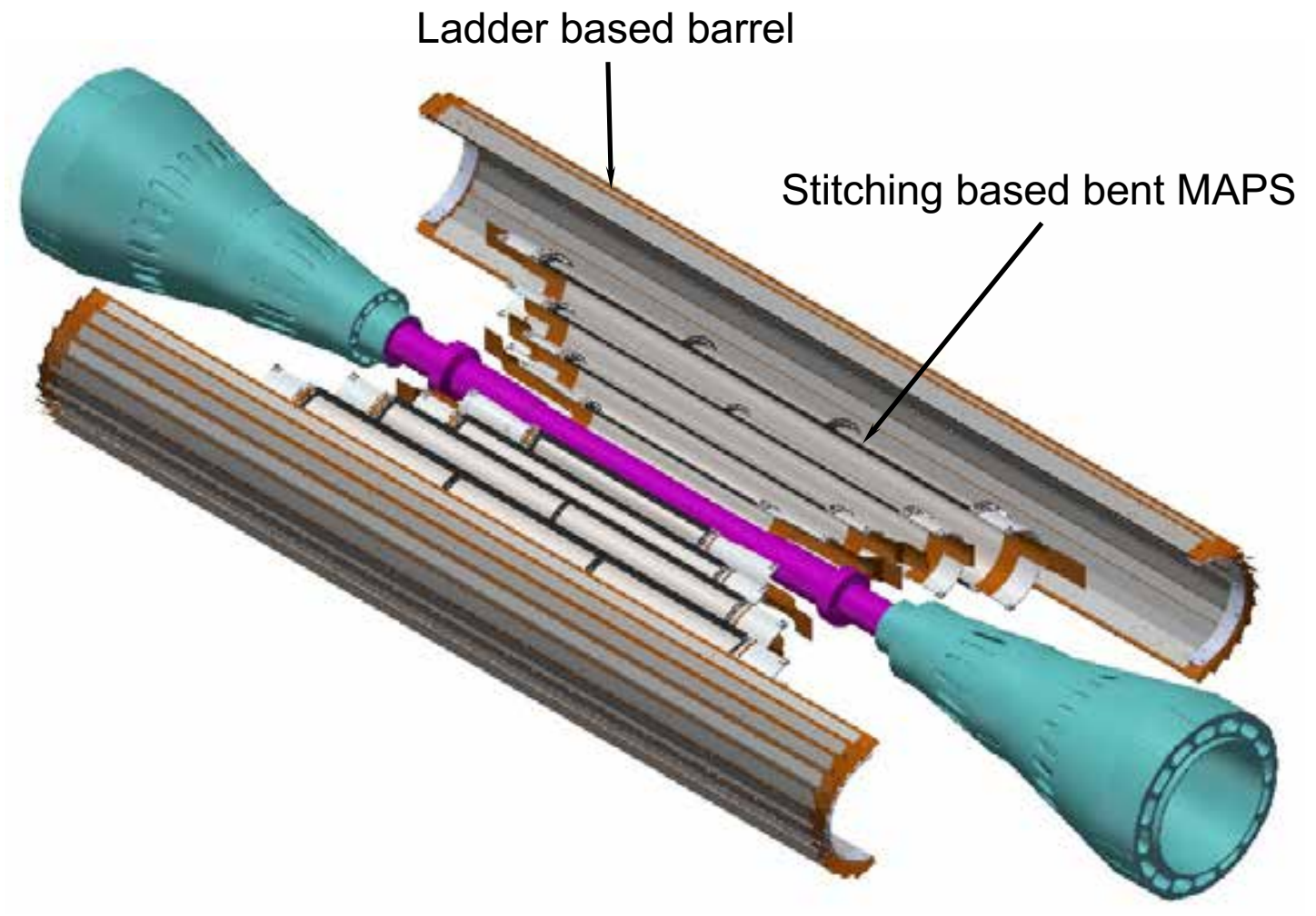

**Figure 4.1:** Diagram of the VTX. The first four layers are designed as single-layer structures (single-layer cylindrical design) using curved stitched sensors, while the last two layers are designed as double-layer ladder structures utilizing planar MAPS technology.

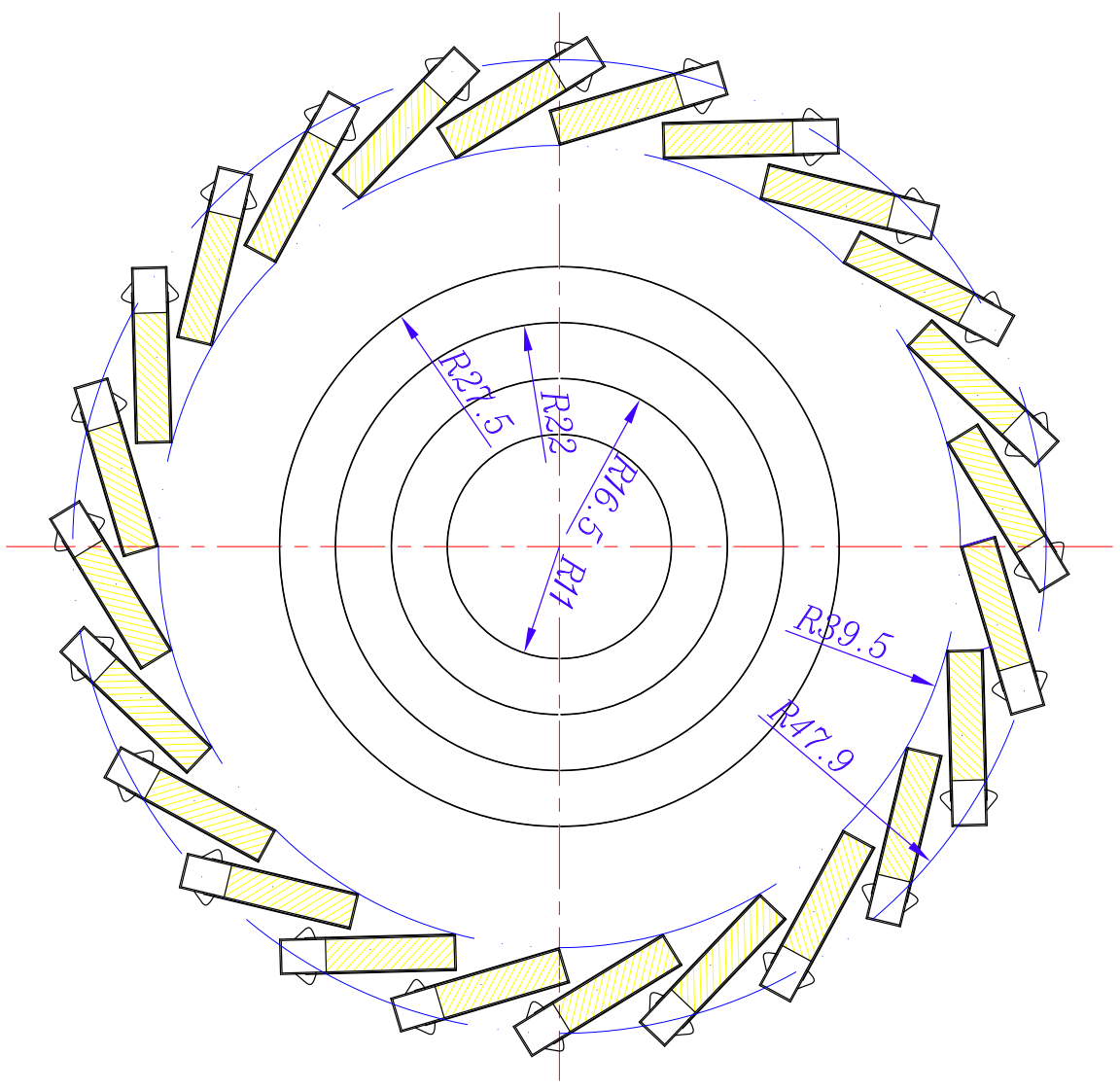

**Figure 4.2:** Sectional view of the VTX layout, with dimensions given in millimeters. The inner four layers are curved MAPS with semi-cylindrical structure. The fifth and sixth layers are planar MAPS sensors arranged inside and outside of a ladder structure.

CVTX 1 shares the Type C sensor with CVTX 3 and carries two Type C sensors (each containing two modules) arranged exclusively in the $\phi$-direction (see Figure 4.4).

As for CVTX 2, CVTX 3 and CVTX 4, all deploy four sensors: Type A sensors (three modules per sensor) for CVTX 2, Type C (four modules) for CVTX 3 and Type B (five modules) for CVTX 4; two sensors oriented along the z-direction and two along the $\phi$-direction. A 0.5 mm transverse seam between the two Right-end Power Blocks (RPBs) at $z = 0$ extends over the full

**Table 4.3:** Geometric configuration parameters. Radius and length, and material budget of the support structure for each layer of the VTX, as well as the arc length corresponding to CVTX and the height of ladder corresponding to PVTX.

| CVTX/ PVTX X | radius mm | length mm | arc length mm | height mm | support thickness μm |
|---|---|---|---|---|---|
| CVTX 1 | 11.1 | 161.4 | 69.1 | - | 45 |
| CVTX 2 | 16.6 | 242.2 | 103.7 | - | 32 |
| CVTX 3 | 22.1 | 323.0 | 138.2 | - | 31 |
| CVTX 4 | 27.6 | 403.8 | 172.8 | - | 29 |
| PVTX 5 | 39.5 | 682.0 | - | 3.3 | 300 |
| PVTX 6 | 47.9 | 682.0 | - | 3.3 | 300 |

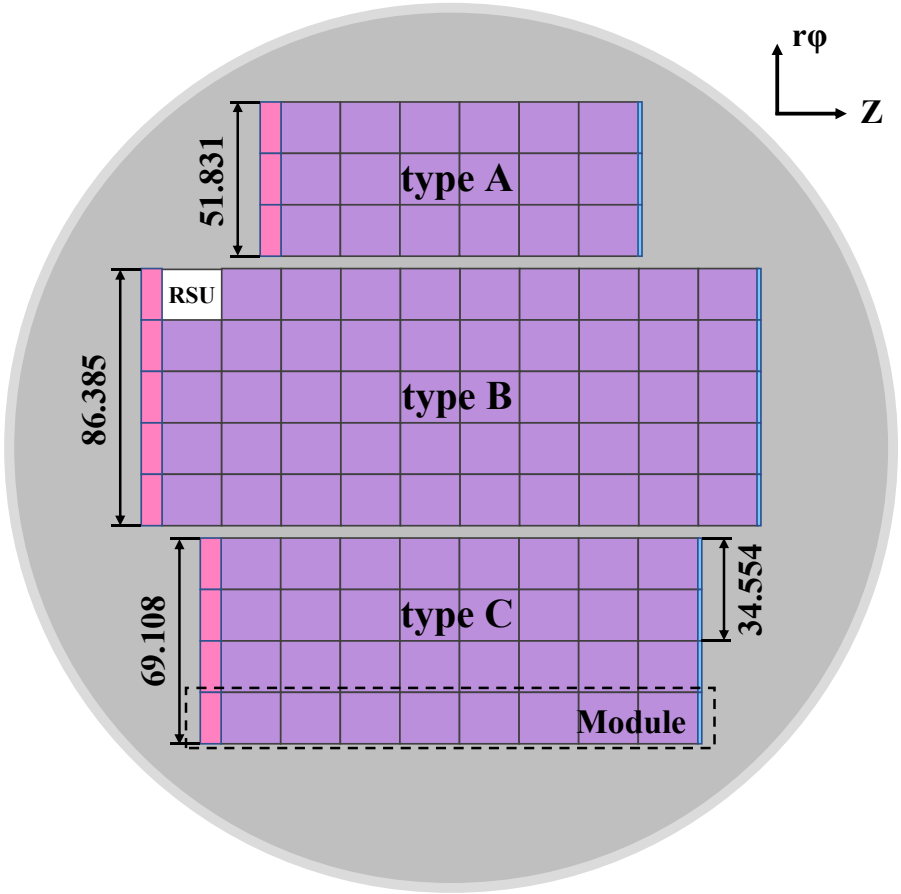

**Figure 4.3:** The arrangement of stitching frames on a 300 mm wafer. The units of the dimension marked in the figure are millimeters. The purple part represents the RSUs, the pink represents the Left-end Readout Block (LRB), and the blue represents the Right-end Power Block (RPB). The stitching technology is implemented exclusively in the $z$-direction. The short-dashed line area represents a module consisting of several RSUs, one LRB, and one RPB. To meet the geometric and coverage requirements of different VTX layers, Type A,B and C sensors are designed with varying lengths and widths.

$\phi$ circumference.

Considering the inefficient region in the $\phi$ direction of the stitched sensor, layers of the VTX are rotated by an angle when mounted. This angular adjustment reduces the likelihood of particles traversing through the overlapping inefficient regions, thereby mitigating tracking inefficiency. Each semi-cylindrical structure is supported by a material with a specific thickness (see Section 4.3.1.2).

**Outer layers with double-sided ladders design** The outer layers (layers 5 and 6) of the VTX do not use stitching technology due to wafer size limitations. These layers require larger sensors that exceed the practical stitching capabilities on standard 300 mm wafers. As a result, conventional double-sided planar MAPS ladders are adopted for simpler integration at larger

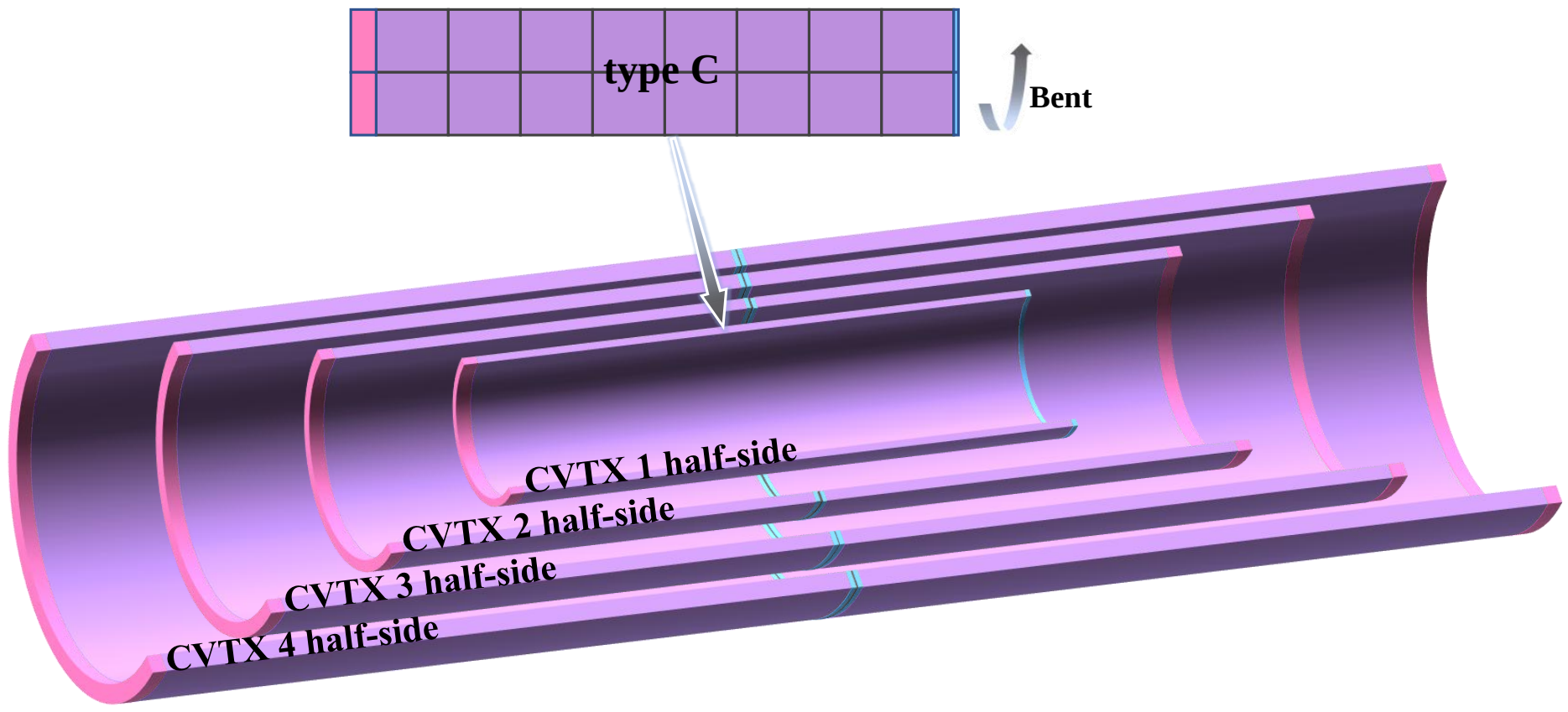

**Figure 4.4:** The Type C sensor is diced into a two-module sensor, which is bent along the $\phi$-direction to form a semi-cylindrical structure for the CVTX 1. For CVTX 2, two Type A sensors (each comprising three modules) are bent along the $\phi$-direction and arranged along the $z$-axis, forming a similar semi-cylinder, with LRBs at both ends and a maximum 0.5 mm seam between the RPBs (shown in gray). CVTX 3 and CVTX 4 are constructed in the same manner as CVTX 2, using two Type C and two Type B sensors, respectively.

radii.

Layer 5 and Layer 6 are constructed using planar MAPS technology with a double-sided ladder structure. Each ladder consists of 52 chips bonded to two FPCs on the front and back sides, with 26 chips on each side. Each chip sizes $15.9 \times 25.7\ \text{mm}^2$ with a thickness of 40 µm. The gap between adjacent chips is less than 0.1 mm to minimize inefficient regions.

The FPCs are glued to both sides of the support structure made of Carbon Fiber Reinforced Polymer (CFRP) to create a complete ladder cell as shown in Figure 4.5 and Figure 4.6, where the effective thickness of the support structure of CFRP hollow support pipe is 300 µm, as described in Section 4.3.1.3. The geometric center of the ladder unit rotates around the origin at a specific radius position from the $\phi = 0$ position, forming a barrel structure known as PVTX 5-6 to ensure that no particles leak out in the $\phi$-direction. A total of 24 ladder structures are utilized in the fifth and sixth layers. The number of sensors per layer, both of CVTX and PVTX, are summarized in Table 4.4.

**Table 4.4:** Total sensor counts for CVTX 1–4 (stitched sensors per RSU) and PVTX 5–6 (planar MAPS chips).

| CVTX / PVTX X | CVTX 1 | CVTX 2 | CVTX 3 | CVTX 4 | PVTX 5 | PVTX 6 |
|---|---|---|---|---|---|---|
| **Number of sensors** | 32 | 72 | 128 | 200 | 624 | 624 |

**Material Budget** Figure 4.7 presents the average material budget, expressed in units of radiation length ($X_0$), as a function of the polar angle $\theta$, averaged over the $\phi$ direction. The

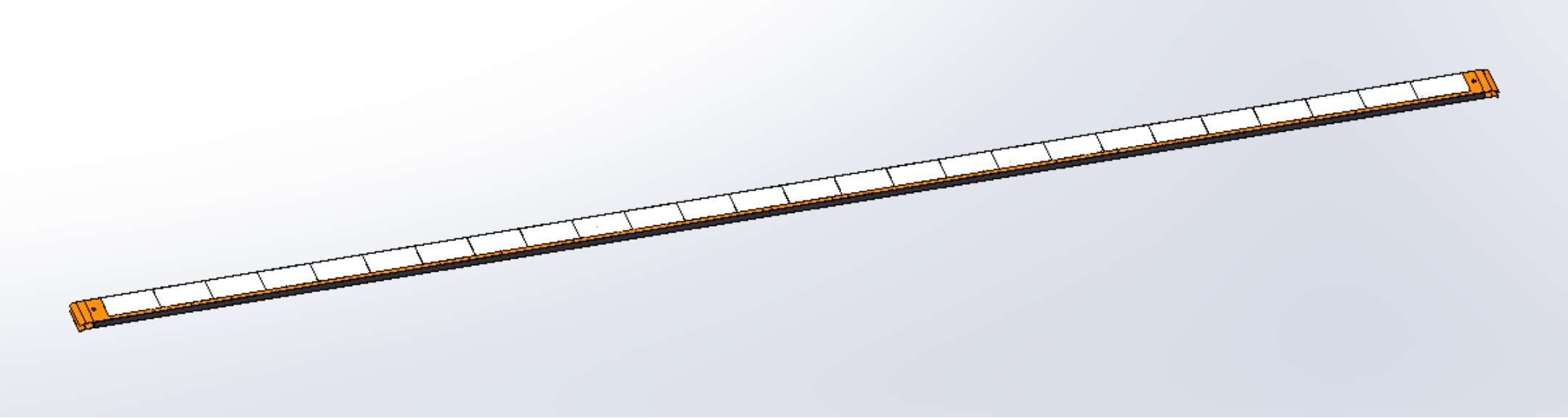

**(a)**

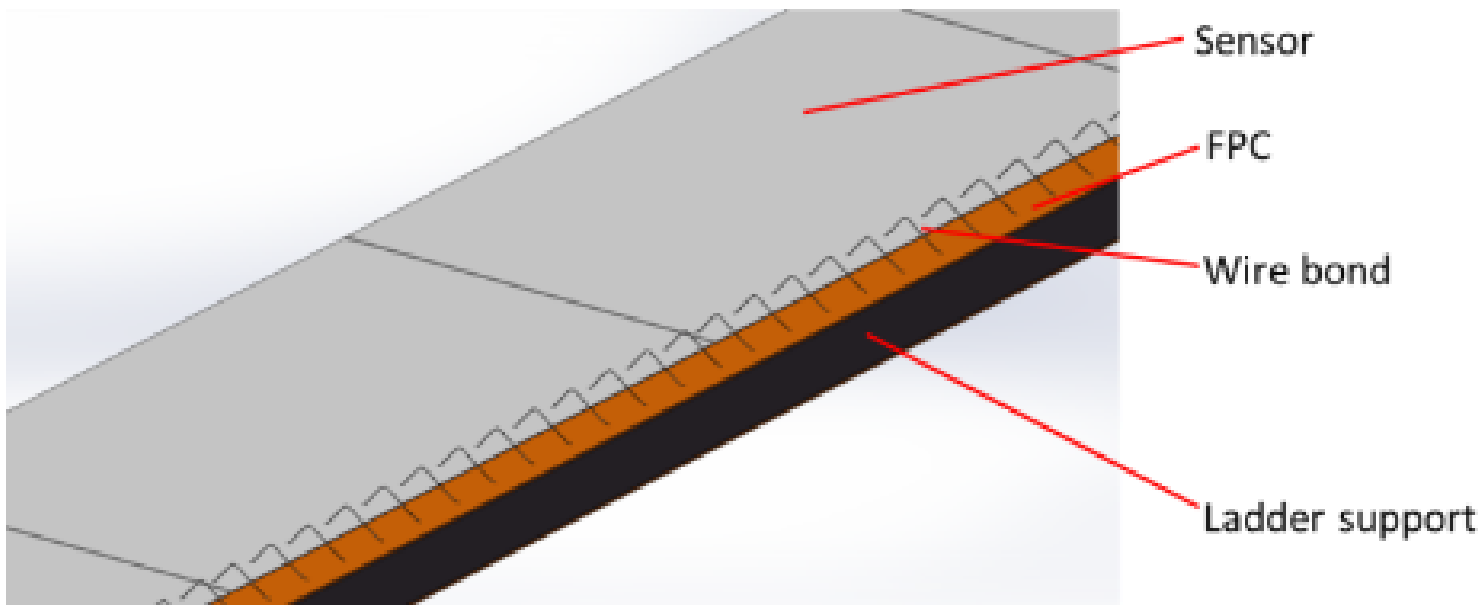

**(b)**

**Figure 4.5:** Layers 5 and 6 are constructed using planar MAPS technology with a double-sided ladder structure. Each ladder consists of 52 chips bonded to two flexible printed circuits (FPCs) on the front and back sides, with 26 chips on each side. (a) Full view of a ladder ; (b) Close-up view of a ladder showing sensors wire bonded to the FPC.

corresponding values for each VTX layer at $\theta = 90^\circ$ are detailed in Table 4.5.

**Table 4.5:** Average material budget of each VTX layer and the beam pipe, expressed in units of radiation length ($X_0$), evaluated at $\theta = 90^\circ$. The inner layers (Layer 1 ~ 4) benefit from the use of curved MAPS sensors, resulting in a significantly lower material budget.

| **Unit** | **Beam pipe** | **Layer 1** | **Layer 2** | **Layer 3** | **Layer 4** | **Layer 5** | **Layer 6** |
|---|---|---|---|---|---|---|---|
| $\bar{X}_0$ ($X_0$) | 0.454% | 0.067% | 0.059% | 0.058% | 0.061% | 0.280% | 0.280% |

### 4.1.3 Background estimation

The hit rate is a critical parameter influencing the design specifications of the sensors. While Chapter 3 provides a detailed description of individual components of beam-induced background, this section summarizes the total hit rate, data rate, occupancy, and other key performance indicators for each operating mode.

The hit rate is defined as the number of particles incident on a given RSU per unit time and unit area. The data rate is calculated as the product of the hit rate and the average cluster size, where the cluster size refers to the number of pixels activated by a single particle. The occupancy is assessed at the pixel level within the time window of a single bunch crossing.

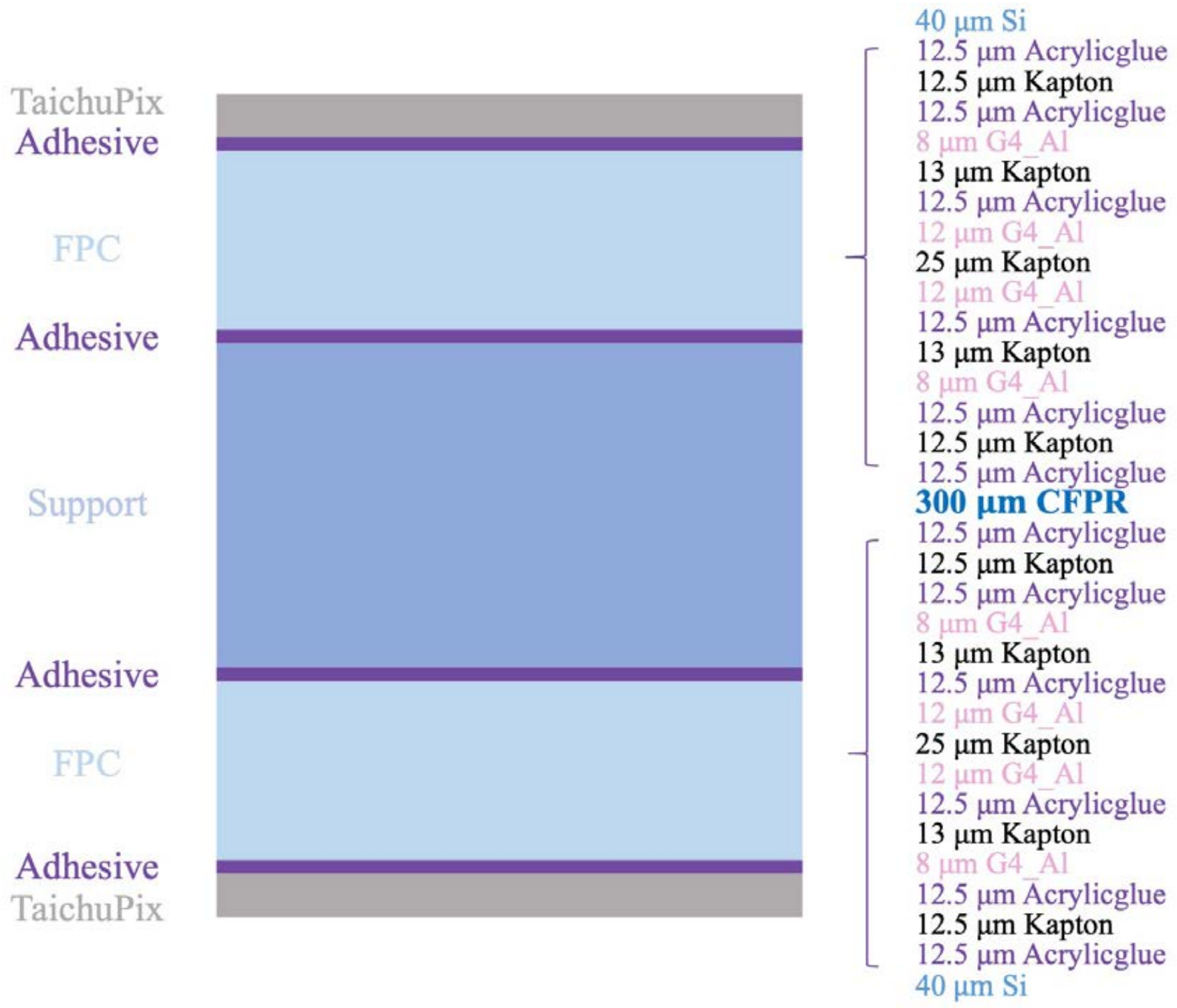

**Figure 4.6:** Schematic transverse cross-section of a double-sided ladder structure used in layers 5 and 6 of the VTX. The ladder consists of two TaichuPix sensors mounted on the outer surfaces, with a multi-layer internal stack comprising FPC, adhesives, kapton, aluminum, and a central CFRP support. The layout ensures mechanical stability, signal routing, and thermal management. All thicknesses are annotated on the right. This diagram is not to scale.

To ensure accurate data rate estimation, cluster size information must be incorporated. A 32-bit hit data format is assumed for each recorded hit pixel. Rather than assuming a fixed cluster size, an incidence-angle–dependent cluster-size model calibrated with TaichuPix-3 beam-test data [1] is adopted. This model takes into account the particle incident angle when computing cluster size. The data rate is computed using the following formula

$$\text{DataRate} = \text{HitRate} \times 32\ \text{bits/pixel} \times \text{ClusterSize}. \tag{4.2}$$

The resulting hit rates and corresponding data rates for each layer and running mode (50 MW for the Higgs mode and 12.1 MW for the low-luminosity $Z$ mode), including the applied safety factor, are summarized in Table 4.6. The various sources of beam-induced backgrounds, together with the applied safety factor, are discussed in detail in Chapter 3. Among them, pair production is estimated to be the dominant source for the beam-induced background in the VTX.

The spatial distributions of the hits from the beam-induced backgrounds under the Higgs boson factory mode and low-luminosity $Z$ mode operations are shown in Figures 4.8 and 4.9, respectively. The hit pattern of beam-induced backgrounds in the high-luminosity $Z$ mode is similar to that in the low-luminosity mode, but with hit rates increased by several times, which would affect the readout architecture design.

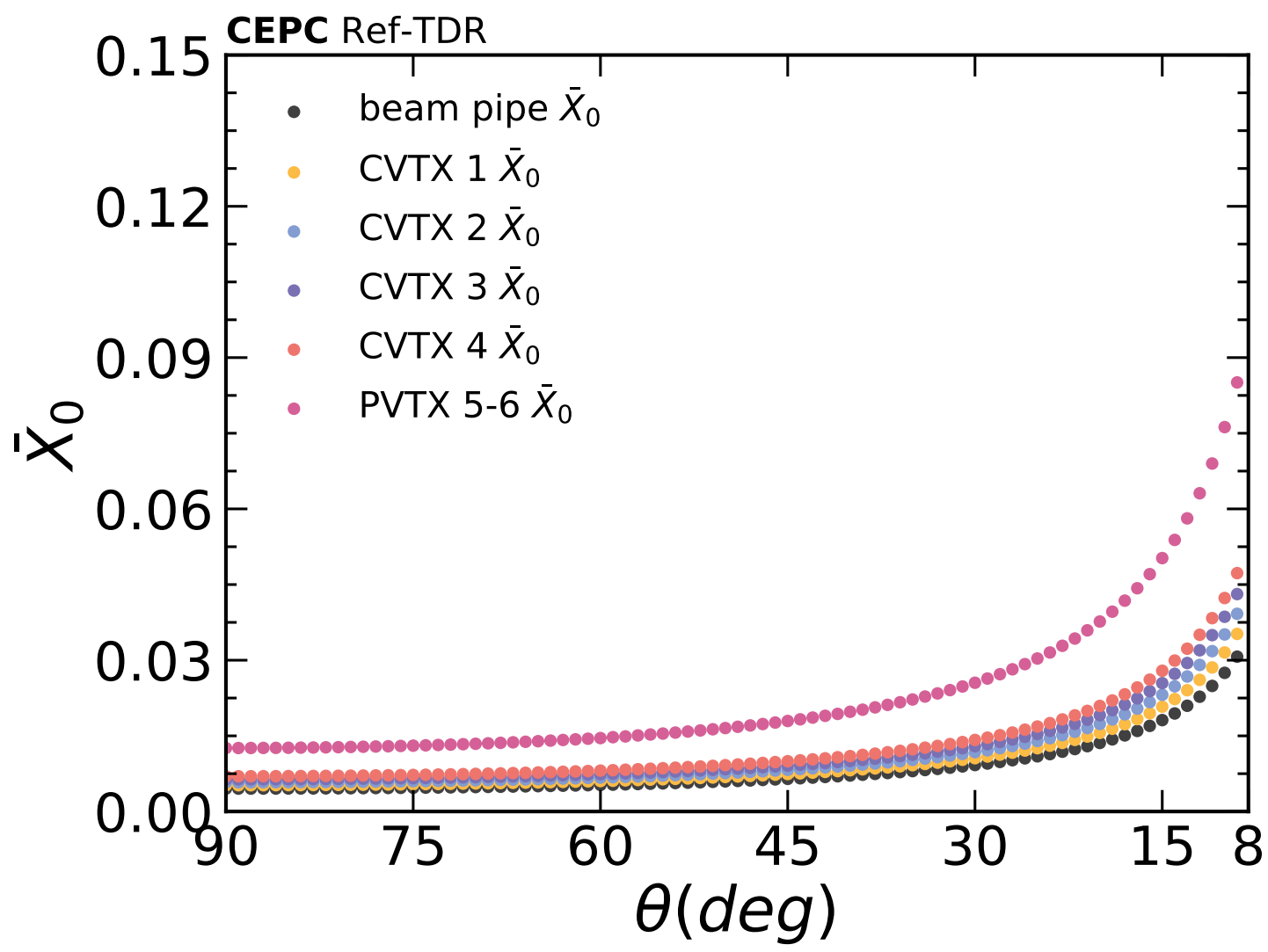

**Figure 4.7:** Average material budget $\bar{X}_0$ as a function of the polar angle $\theta$, averaged over the full azimuthal angle $\phi$, for each layer of the VTX and the beam pipe, is shown in a stacked plot. As $\theta$ decreases (i.e., moving toward the forward region), the effective path length through the detector increases, leading to a higher accumulated material budget. The four innermost curved MAPS layers (CVTX 1 ~ 4) exhibit significantly lower $\bar{X}_0$ values compared to the outer ladder-based layers (PVTX 5 ~ 6), reflecting the advantage of the low-mass inner design.

Occupancy values were calculated using one bunch spacing as the time window for each operational mode, and the results are summarized in Table 4.7. The results show that even in the innermost layers, the occupancy remains well below the typical upper limit of 1%, which is generally considered acceptable for efficient pattern recognition. This demonstrates that the VTX design is robust against beam-induced background. A more detailed evaluation of the VTX performance under beam-induced background conditions is provided in Section 4.7.

## 4.2 Sensors and electronics design

### 4.2.1 Sensor technology overview

The current design and development of the VTX are centered on utilizing MAPS technology for chip design and manufacturing. This technology integrates the sensor and readout electronics onto a single chip, significantly reducing both pixel size and power consumption while delivering high-performance detection capabilities. Such features meet the stringent requirements for high spatial resolution, low material budget, and fast readout speeds. The STAR experiment at RHIC successfully employs the MAPS technology in its VTX [2], and the STAR MAPS VTX features excellent spatial resolution [2, 3]. The Inner Tracker System-2 (ITS) of CERN's ALICE experiment is so far the largest scale MAPS system among high energy physics experiments [4]. The ITS-2 is a full-pixel silicon detector based on TowerJazz 180 nm technology. It boasts exceptional technical parameters, including approximately 5 μm high resolution, an extremely

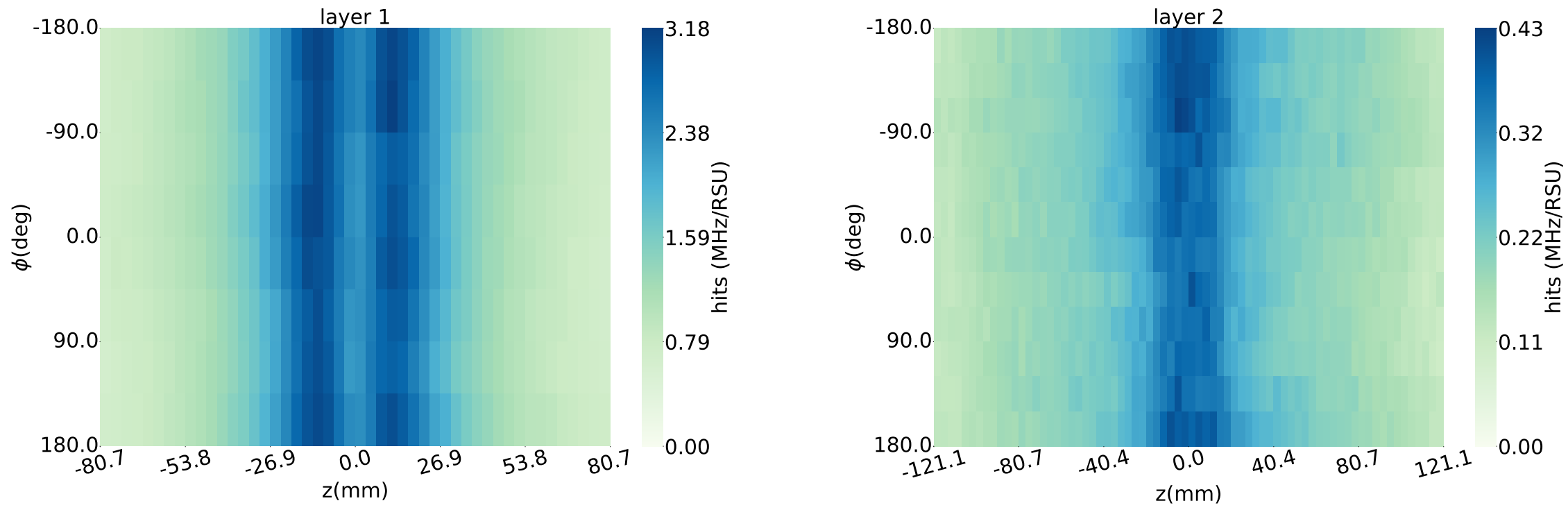

**(a)** **(b)**

**Figure 4.8:** Spatial distribution of beam-induced background hit rates in the VTX during Higgs mode operation. Each cell indicates a single RSU. The VTX geometry is approximately cylindrical, with the horizontal axis representing the global $z$-coordinate (along the beam pipe) and the vertical axis showing the azimuthal angle $\phi$. (a) Layer 1: The hit rate is relatively uniform across the layer, with a slight reduction near $z = 0$. (b) Layer 2: The distribution is also mostly uniform in $\phi$, but shows a concentration within $|z| < 40\,\text{mm}$. The remaining layers are not shown as their hit rates are significantly lower.

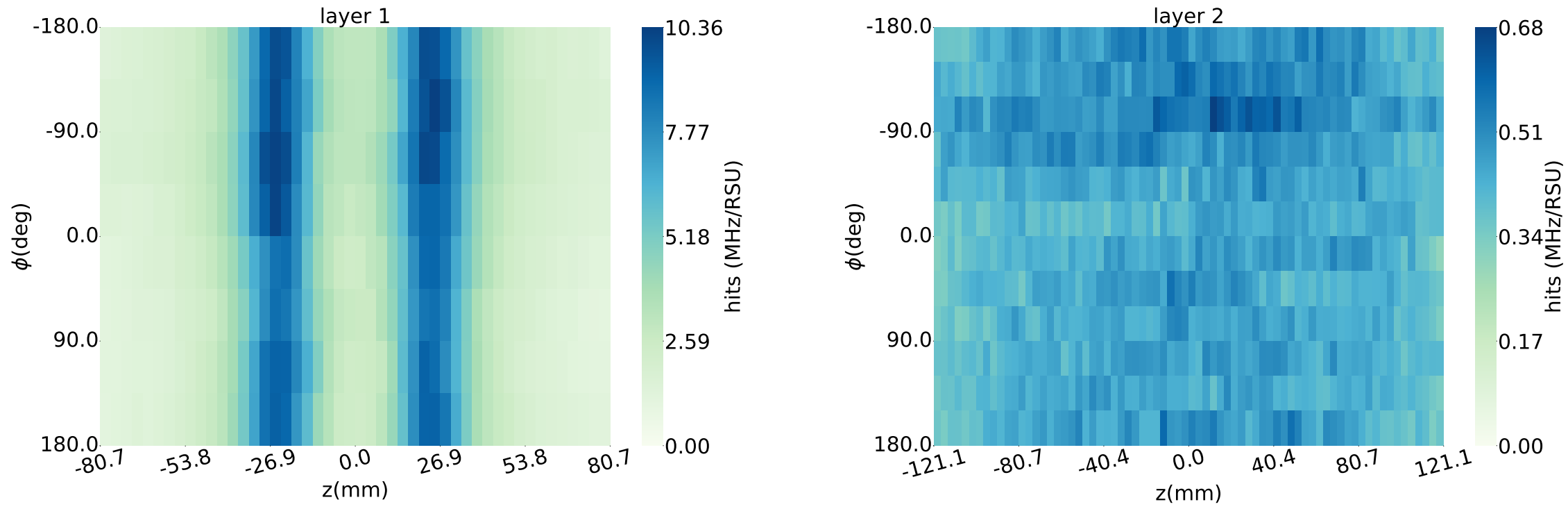

**(a)** **(b)**

**Figure 4.9:** Spatial distribution of beam-induced background hit rates in the VTX during low-luminosity $Z$ mode operation. Each cell indicates a single RSU. (a) Layer 1: The highest hit rates are observed in this innermost layer. (b) Layer 2: The hit rate pattern is more diffuse, with less pronounced structure.

low material budget of less than 0.3% $X_0$ per layer, stable readout support for collision frequencies up to 100 kHz, and power consumption as low as 40 mW·cm$^{-2}$ [4, 5]. The upcoming ITS3 upgrade for the ALICE experiment introduces the concept of "curved wafer-level chips", based on TPSCo 65 nm technology, aiming to achieve a self-supporting wafer structure [6, 7]. This innovation could potentially reduce the material budget of the VTX by a factor of five. The ITS3 project is currently in the research and development phase. The comparison of key parameters among different VTXs mentioned above is shown in Table 4.8.

**Table 4.6:** Estimated beam-induced background hit rates and data rates, including the applied safety factor, for the VTX layers in the Higgs mode (50 MW) and the low-luminosity $Z$ mode (12.1 MW) are summarized. Pair production is estimated to be the dominant source of beam-induced background. Details of the contributions from each background component to the VTX are discussed in Chapter 3. For both the Higgs and low-luminosity $Z$ modes, the highest hit rate is observed in the first layer of the VTX. Compared with the Higgs mode, the low-luminosity $Z$ mode exhibits significantly larger hit and data rates in the innermost layers, particularly between Layer 1 and Layer 2. This behavior is driven by the tighter bunch spacing and higher background particle density in the low-luminosity mode.

| **Layer** | **Ave. Hit Rate (MHz/**$\mathrm{cm}^2$**)** | **Max. Hit Rate (MHz/**$\mathrm{cm}^2$**)** | **Ave. Data Rate (Mbps/**$\mathrm{cm}^2$**)** | **Max. Data Rate (Mbps/**$\mathrm{cm}^2$**)** |
|---|---|---|---|---|
| Higgs mode: Bunch Spacing: 277 ns, 63% Gap | | | | |
| 1 | 6.2 | 12 | 760 | 1500 |
| 2 | 0.84 | 1.6 | 87 | 160 |
| 3 | 0.17 | 0.36 | 19 | 38 |
| 4 | 0.067 | 0.16 | 8.4 | 19 |
| 5 | 0.017 | 0.037 | 2.1 | 4.2 |
| 6 | 0.013 | 0.026 | 1.6 | 3.7 |
| Low-luminosity $Z$ mode: Bunch Spacing: 69 ns, 17% Gap | | | | |
| 1 | 15 | 39 | 2700 | 8100 |
| 2 | 1.7 | 2.6 | 240 | 400 |
| 3 | 0.72 | 1.2 | 110 | 240 |
| 4 | 0.43 | 0.94 | 70 | 210 |
| 5 | 0.10 | 0.19 | 14 | 31 |
| 6 | 0.078 | 0.15 | 11 | 23 |

### 4.2.2 Stitched sensor prototype design

The Complementary Metal Oxide Semiconductor (CMOS) stitching technology enables the fabrication of chips significantly larger than the size of a single lithographic reticle. In this process, the reticle layout is divided into multiple sub-regions (sub-frames), which are then sequentially exposed onto adjacent positions on the wafer according to a predefined stitching pattern. This technique allows the creation of large-area chips—potentially approaching the full wafer diameter—thus enhancing design flexibility and optimizing wafer utilization.

Considering the fabrication complexity and potential yield constraints associated with stitching, the baseline prototype adopts a one-dimensional (1D) stitching scheme aligned with the beam pipe direction. This choice simplifies manufacturing while allowing the desired elongated geometry for vertex detector applications. In parallel, the feasibility of two-dimensional (2D) stitching—aimed at achieving full wafer-scale integration—will be further investigated during the ongoing R&D phase.

**Table 4.7:** Occupancy estimation with safety factor for the VTX under Higgs (50 MW) and low-luminosity $Z$ mode (12.1 MW) operations, including the applied safety factor. The table reports both the average and maximum occupancy per pixel per Bunch Crossing (BX) for each detector layer. The results indicate that, even in the innermost layers, the occupancy remains well below the typical upper limit of 1% that is generally considered acceptable for efficient pattern recognition.

| **Layer** | **Mode** | **Ave. Occupancy** ($\times 10^{-5}$ **/ BX**) | **Max. Occupancy** ($\times 10^{-5}$ **/ BX**) |
|---|---|---|---|
| 1 | Higgs | 11 | 21 |
| 2 | | 1.2 | 2.3 |
| 3 | | 0.27 | 0.53 |
| 4 | | 0.12 | 0.27 |
| 5 | | 0.031 | 0.061 |
| 6 | | 0.023 | 0.054 |
| 1 | Low-luminosity $Z$ | 4.3 | 13 |
| 2 | | 0.38 | 0.62 |
| 3 | | 0.17 | 0.38 |
| 4 | | 0.11 | 0.33 |
| 5 | | 0.023 | 0.05 |
| 6 | | 0.017 | 0.037 |

**Table 4.8:** Comparison of sensor technologies used in vertex detectors across different experiments. The table presents key performance metrics for ALICE Pixel Detector (ALPIDE) (used in ALICE ITS2), MOSAIX (used in ALICE ITS3), TaichuPix-3 (the first VTX prototype), and the stitched sensor developed for the VTX (CEPC-Stitching). The listed metrics include technology node, power consumption, readout speed, and spatial resolution. The CEPC-Stitching sensor offers a favorable combination of low power consumption and high readout speed, enabled by its advanced 65 nm technology node.

| **Sensors** | **Technology Node** | **Power Consumption** | **Readout Speed** | **Spatial Resolution** |
|---|---|---|---|---|
| ALPIDE [4, 5] | 180 nm | 40 mW/cm$^2$ | Up to 100 kHz | 5 µm |
| MOSAIX [6] | 65 nm | 40 mW/cm$^2$ | 164 kHz | 5 µm |
| TaichuPix-3 [8, 9, 10] | 180 nm | 60 ~ 200 mW/cm$^2$ | 40 MHz | 5 µm |
| CEPC-Stitching | 65 nm | ≤ 40 mW/cm$^2$ | Up to 43.3 MHz | 5 µm |

#### 4.2.2.1 Sensor architecture and functional blocks

The concept of using a stitched sensor chip to construct half of a detector layer is inspired by the design of the ALICE ITS3 sensor [6, 7]. Encouraged by the promising results obtained from the MOnolithic Stitched Sensor (MOSS) and MOnolithic Stitched sensor with Timing (MOST) prototypes developed for the ALICE ITS3 project, the CEPC design adopts a similar approach. Detailed R&D on the stitching floorplan tailored to the VTX is currently underway. A preliminary layout of the stitched silicon wafer and the dimensions of each half-layer are shown in Figure 4.3.

Figure 4.10 presents a schematic floorplan of the sensor for Layer 1, which is realized by dicing out two adjacent modules. Each module operates independently. The sensitive area of a module consists of eight RSUs, each representing a repeated unit of the same design. The Left-end Readout Block (LRB), located on the left side of the module, provides signal routing to external systems and supplies power to the module. All power supplies, I/O pads, and services are connected exclusively through the LRB. The RPB, located on the right side, is dedicated solely to power distribution and contains only power transmission buses, without any direct connection to external services. This design minimizes the material budget in the central region of the detector.

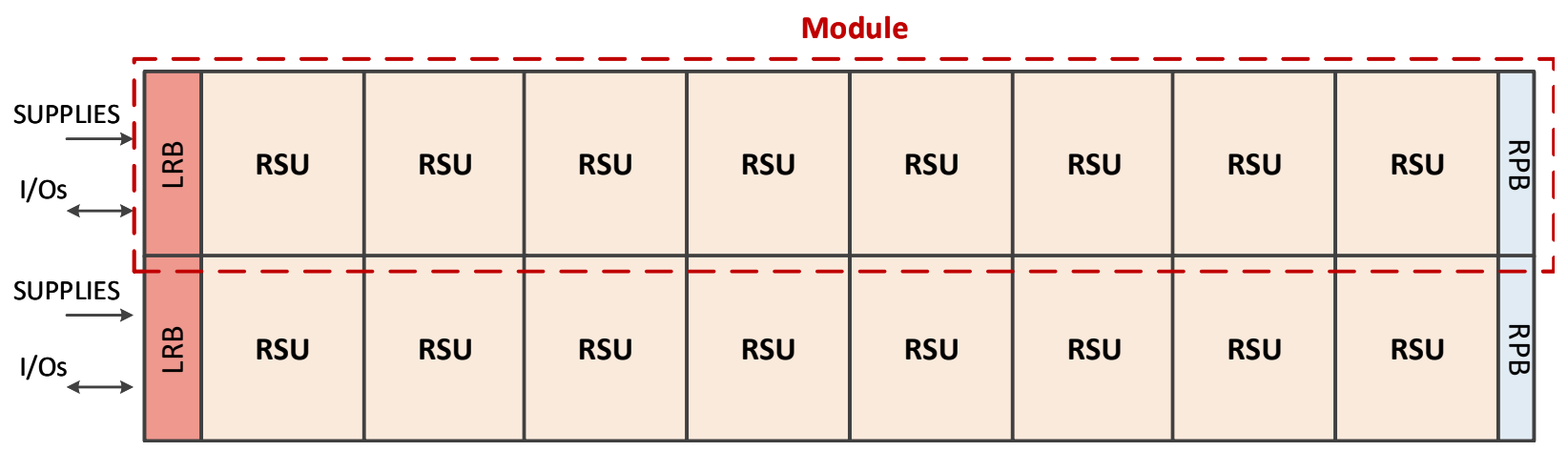

**Figure 4.10:** Schematic top-level floorplan of the sensor for Layer 1 (not to scale). The design consists of two identical modules, each comprising eight RSUs. One module is highlighted by the red dashed rectangle. The left side of each module includes a LRB for signal and power interfaces. The RPB, located on the right side, is dedicated solely to power distribution and contains only power transmission buses, without any direct connection to external services.

Figure 4.11 illustrates the preliminary floorplan of an RSU, which is divided into multiple identical sensor blocks. Each sensor block functions independently with its own biasing generator, slow control, and periphery readout circuits, as shown in Figure 4.12. Each sensor block includes local power switches on its right side, which can be individually activated by user control. This fine-grained power control improves fault tolerance: in the event of a supply short in one block, it can be powered down without affecting the functionality of the others. Since turning off a block introduces an inactive region in the sensitive area, the dimensions of each pixel matrix must be carefully chosen. In the baseline design, each power switch has a width of approximately 0.02 mm and each stitched interface has a width of approximately 0.06 mm, and they occupy only a small fraction of the total sensor block width (20 mm). This illustrates that the peripheral circuitry is extremely narrow relative to the overall sensor block, resulting in a very high fill factor for active detection. In future designs, the block size will be optimized to balance power granularity and data transmission performance.

In the current preliminary design, each RSU consists of 12 identical sensor blocks and four stitched data interface blocks arranged in two rows. Each sensor block connects directly to the LRB, and data are transmitted via dedicated point-to-point links. For sensor blocks located farthest from the LRB, where the transmission distance exceeds 20 cm, signal relay becomes necessary. This function is proposed to be integrated into the stitched data interface blocks, which handle both data transmission and control signal routing to the LRB.

Internally, the RSU contains repeated instances of multiple components: 12 sets of Pixel Matrices, Biasing Blocks, Power Switch Blocks and Matrix Readout Blocks, and four stitched data interface blocks.

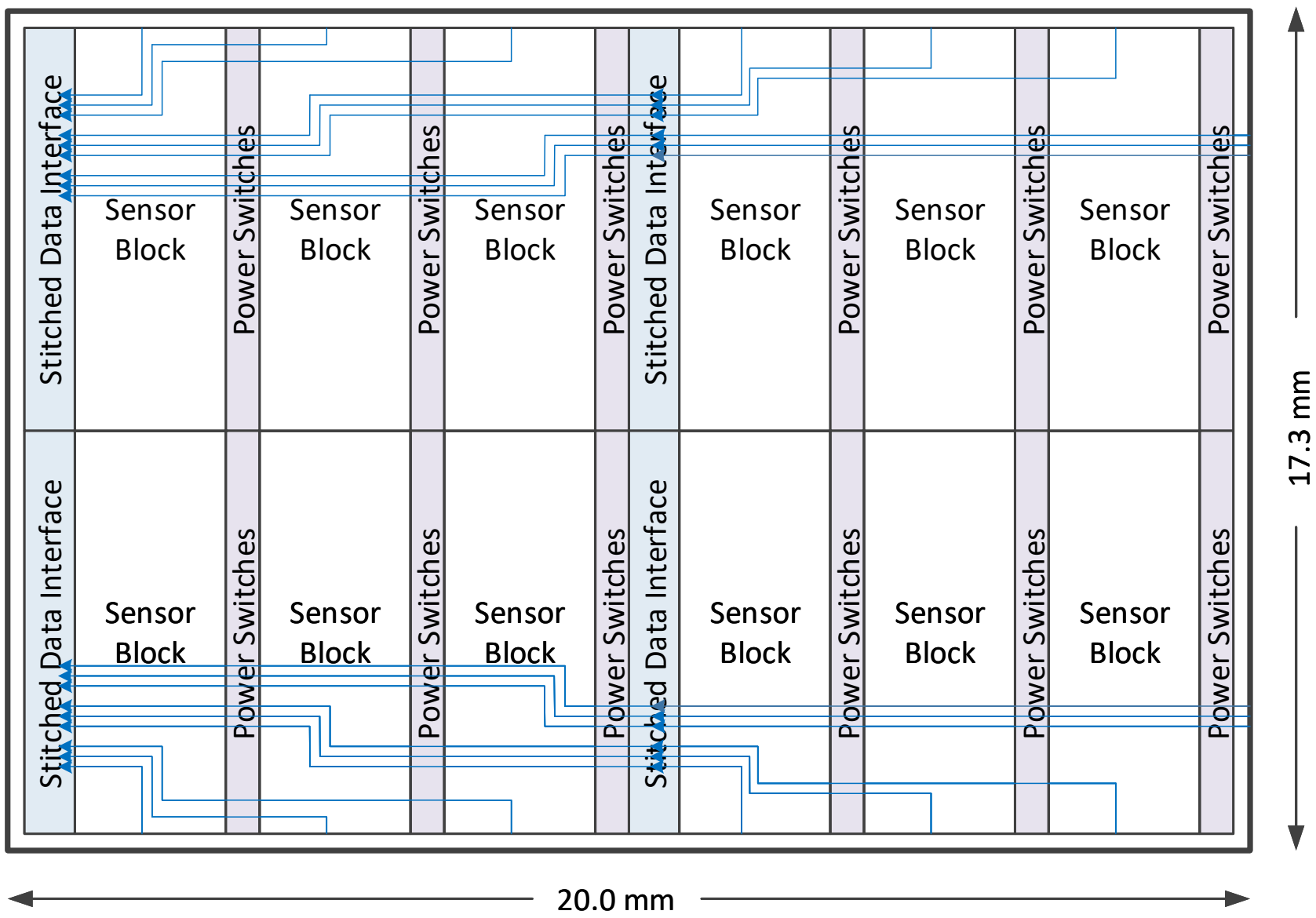

**Figure 4.11:** Proposed top-level floor plan of a RSU (not to scale). The blue lines represent the point-to-point data transmission lines traverse stitching boundaries of RSUs. The RSU is composed of multiple identical sensor blocks, each integrating a pixel matrix with its own biasing generator, slow control, and periphery readout circuitry. Each sensor block can be selectively powered on or off via dedicated power switch units, enhancing fault tolerance and enabling power granularity. Stitched data interface blocks are strategically placed to relay control signals and transmit readout data to the edge of the stitched sensor, ensuring reliable long-distance communication. In the baseline design, each power switch has a width of approximately 0.02 mm and each stitched interface has a width of approximately 0.06 mm, and they occupy only a small fraction of the total sensor block width (20 mm).

#### 4.2.2.2 Design of the repeated sensor unit

Each pixel cell integrates a sensing diode, a front-end amplifier, a discriminator and a digital pixel readout. The in-pixel digital readout benefited from the TaichuPix prototype[8]. Each pixel includes a hit storage register, as well as logic for configuring the pixel mask and test a pulse. Furthermore, each pixel can be individually masked and tested through charge injections.

**I. Sensing diode** A sensing diode is formed by a N-implant electrode in a P-type epitaxial layer. The sensing diode is designed at the center of each pixel, with a typical diameter of approximately 2 $\mu$m. The geometry of the sensing diode involves a trade-off among charge collection performance (i.e. efficiency, collection time, and radiation tolerance), area, and sensor capacitance. As discussed in Section 4.5.1, the sensor performance optimization was previously explored with the 180 nm TowerJazz MAPS technology. These studies provide foundational insights for enhancing sensor performance in the 65 nm TowerJazz MAPS process. A variety of pixel test structures are proposed to be designed in the initial phase of the stitched sensor R&D to further optimize the sensor geometry for compatibility with the 65 nm MAPS process. The structures include different electrode diameters, spacing between the electrode and the surrounding PWELL, PWELL geometries, and methods of applying reverse bias to the sensor diode. Additionally, the n-type implant modifications will be studied to optimize the performance trade-off between collection efficiency and position resolution.

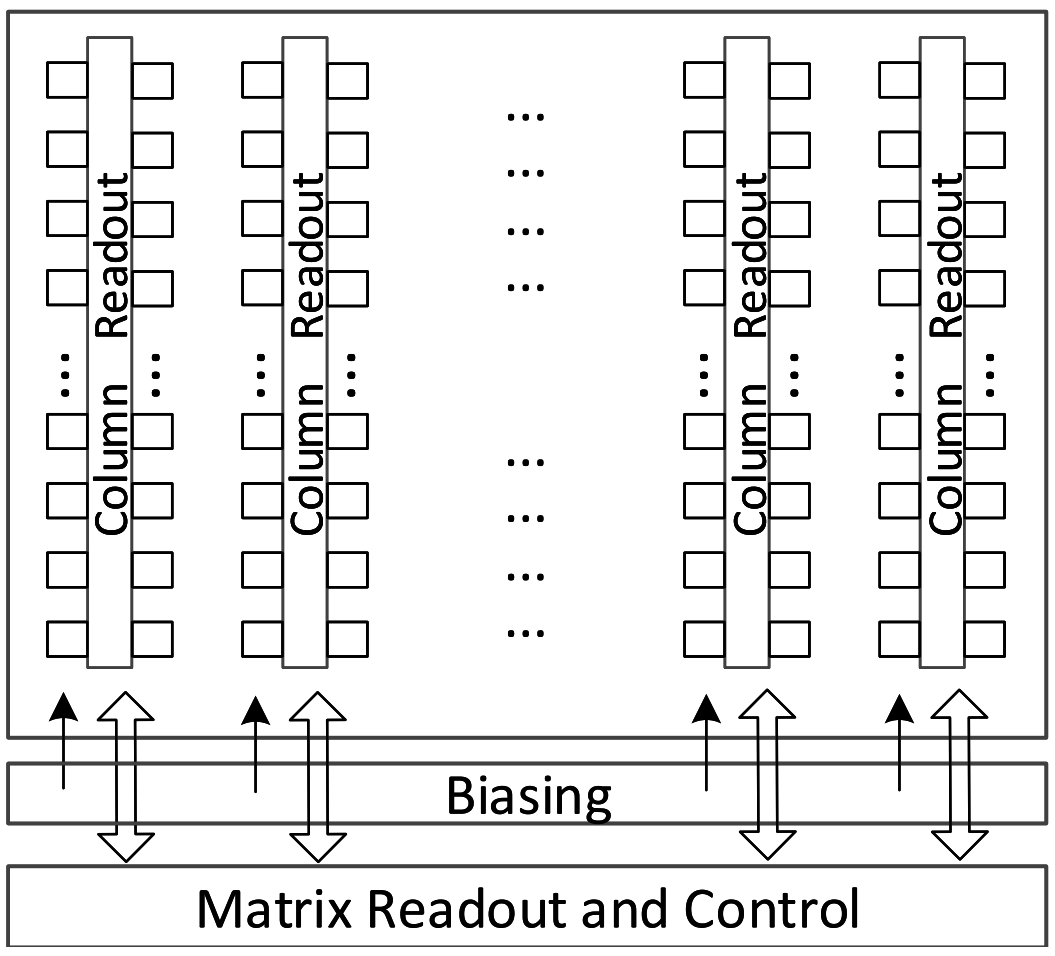

**Figure 4.12:** Architecture of the sensor block. It consists of a pixel matrix, a biasing generator and a periphery readout and control.

**II. Pixel front-end** The simplified schematic of the analog front-end is shown in Figure 4.13a, which was verified with the TaichuPix prototype (see Section 4.5.2). The topology of the circuit is derived from the ALPIDE sensor chip used in the ALICE ITS2 [5]. The analog front-end and the discriminator are continuously active.

**III. In-pixel digital logic** The in-pixel digital electronics, as shown in Figure 4.13b presents the diagram of the in-pixel logic. It features a hit latch set by a negative pulse from the output of the front-end. The pixel readout follows a Double Columns (Dcols) drain arrangement. The region for in-pixel digital readout logic is shared by two columns to minimize the crosstalk between analog signals and digital buses. Additionally, it saves space for the routing of the address encoder. The priority logic arbitrates the pixel readout, with the topmost pixel having the highest priority. The in-pixel logic also integrates configuration registers for calibrating the pixel front-end, for testing in-pixel readout logic, and for masking the faulty pixel. The configuration function is programmable through the setting of control bits, including MASK_EN, PULSE_EN, DPULSE and APULSE (as shown in Figure 4.13b).

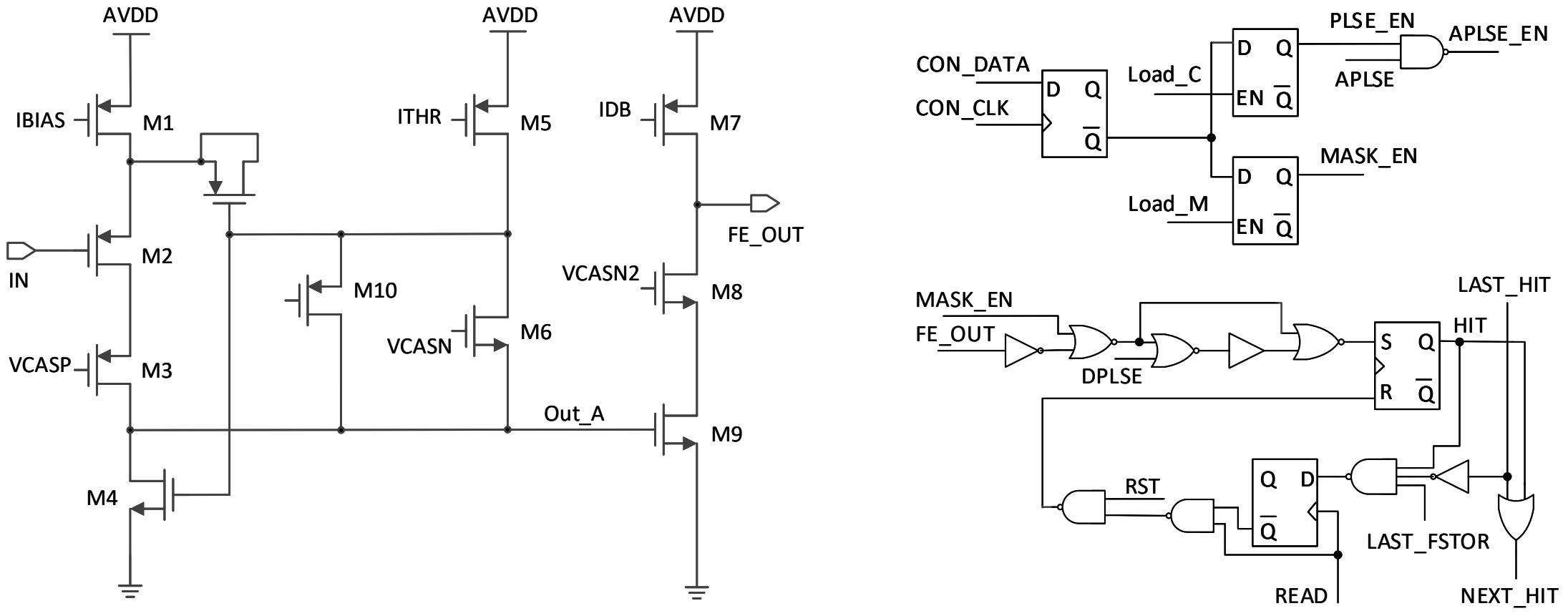

(a) (b)

**Figure 4.13:** Schematic and block diagram of the in-pixel circuitry. (a) Analog front-end, including a charge-sensitive amplifier and discriminator. (b) Diagram of the digital logic. It features a hit latch set by a negative pulse from the output of the front-end. The pixel readout follows a Dcols drain arrangement. The in-pixel logic also integrates configuration registers for calibrating the pixel front-end, for testing in-pixel readout logic, and for masking the faulty pixel.

**IV. Peripheral readout circuits on sensor** The main function of the on-chip peripheral readout circuits includes: sending control signals required by the pixel array and receiving data from the pixel array; buffering the data to smooth the output data rate; providing a slow control interface for chip configurations and tests.

The data from the pixel array are organized in Dcols. In order to achieve a detection efficiency close to 100%, all Dcols of pixel array are read out in parallel. A timestamp is recorded at the end of the Dcols for future data processing. Without a trigger signal, all the readout data are buffered and sent out. To accommodate higher data rates, a real-time data compression strategy can be implemented before sending data to First In First Outs (FIFOs). To implement the fast chip-level readout, a shareable architecture was proposed [11]. The architecture can accommodate the data rates in both the Higgs boson factory mode and the low-luminosity $Z$ boson factory mode during the first ten years of operation. A replacement of the VTX with upgraded technology is foreseen thereafter, with the readout architecture for the high-luminosity $Z$ boson factory mode discussed in Chapter 11.2.

#### 4.2.2.3 Design of the left-end readout block

All data from a row of RSUs of a module, see Figure 4.10, are collected and processed by the LRB located on the left side of the module. Primary functions of the LRB are illustrated in the block diagram of Figure 4.14. The RSU data are transmitted to the LRB via a large array of differential point-to-point on-chip serial data links. A group of receivers primarily compensates for phase differences in the incoming data streams from the RSUs and resamples them using a local fast clock. The data encoding blocks collect and encode the data with redundancy to correct errors during off-chip transmission. After encoding, the data are serialized at a rate of 11.09 Gbps and transmitted off-chip using a maximum of four serializers. The Clock Block supplies clocks required for the LRB and RSUs. Additionally, the LRB facilitates slow control and manages power switch control signals for the RSUs.

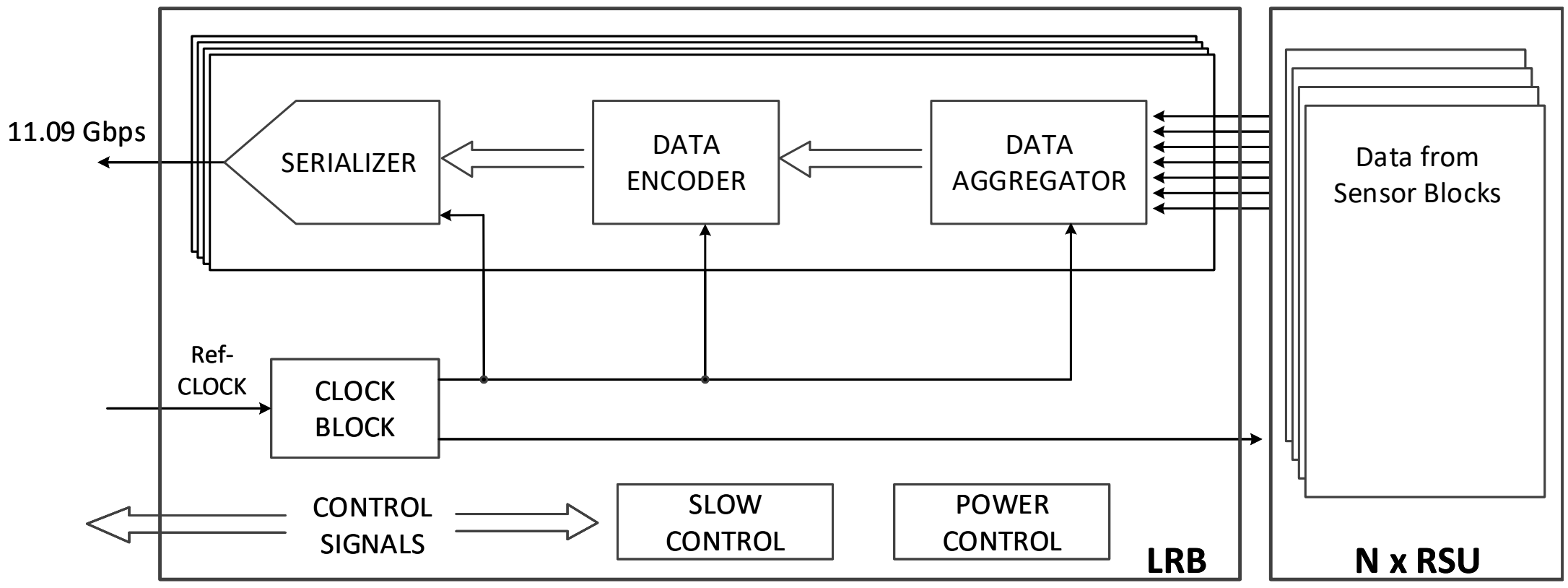

**Figure 4.14:** Proposed diagram of the Left-end Readout Block (LRB). All data from RSUs of one module must be transmitted to the LRB. 'N $\times$ RSU' labeled in the right part of the diagram represents that different layers in the detector contain a varying number of RSUs.

#### 4.2.2.4 Power consumption estimates

As illustrated in Figure 4.11 and 4.12, each RSU consists of pixel matrices, biasing blocks, matrix readouts, and data interface blocks. The expected power consumption of the main functional blocks of an RSU, as well as the total power consumption of a single RSU, are listed in Table 4.9. The entries for the pixel include the power of the pixel analog circuit and the pixel digital logic. In current calculations, it is assumed that the pixel size is $25 \times 25$ $\mu m^2$, which is consistent with the intrinsic position resolution value used in Section 4.1.2.1. The power of the pixel analog is calculated by the power density of the pixel front-end and the total area of 12 pixel matrices. The power consumption of the pixel digital in the table includes the dynamic power of the in-pixel digital logic and the Dcols address encoding logic (described in Section 4.2.2.2), as well as power consumption due to the leakage current. The contribution of the Biasing Blocks is estimated based on the experience of the design in the TaichuPix.

The value for the data interface blocks is obtained from the very preliminary considerations on the transmitting power of the control signals and the readout data of RSUs. Table 4.10 lists the power consumption estimates for the main functional blocks of the LRB (see Figure 4.14).

By combining the power estimates presented in Tables 4.9 and 4.10 with the surface areas of the RSU (3.46 $cm^2$) and the LRB (0.72 $cm^2$), the corresponding power dissipation densities are obtained, as summarized in Table 4.11. The expected power dissipation is sufficient to satisfy the requirements of both the Higgs mode and the low-luminosity $Z$ mode operations. For the Higgs mode operation alone, the total power consumption is expected to be reduced by more than 10% compared to the values listed in Table 4.11. All estimates assume an operation temperature of 27 °C. Further details on the cooling system and the expected operation temperature from thermal simulations are reported in Section 4.3.2. The impact of temperature on sensor performance, based on the TaichuPix-3 prototype chip, is shown in Figure 4.27.

**Table 4.9:** Estimates of power consumption of one RSU. All values are for 27 °C temperature and 1.2 V power supply voltage.

| Components | Pixel Analog | Pixel Digital | Biasing Block | Matrix Readout | Data Interface | RSU Total |
|---|---|---|---|---|---|---|
| Power [mW] | 36 | 30 | 8 | 40 | 17 | 131 |

**Table 4.10:** Estimates of power consumption of the LRB. All values are for 27 °C temperature and 1.2 V power supply voltage.

| Components | Clock Block | Data Aggregator | Data Encoder | Serializer | Slow & Power Control | LRB Total |
|---|---|---|---|---|---|---|
| Power [mW] | 36 | 120 | 80 | 32 | 80 | 348 |

**Table 4.11:** Estimates of average power dissipation per unit area over the main functional blocks composing the stitched chip.

| **Components** | **Power density** [mW/cm$^2$] |
|---|---|
| Repeated Sensor Unit | 38 |
| Left-end Readout Block | 485 |

### 4.2.3 Readout electronics

The readout electronics scheme of the VTX is illustrated in Figure 4.15. The system is designed to provide low-mass, high-bandwidth data transfer and stable power delivery to the stitched MAPS sensors, while ensuring compatibility with the mechanical constraints of the MDI region.

Taking Layer 1 ~ 4 as an example, each stitched sensor contains multiple RSUs connected to a LRB block. The LRB aggregates the data from the RSUs, distributes clock and control signals, and handles local power distribution. From the LRB, data are transmitted via low-mass FPCs to an Optical Array Tranceiver (OAT) module on the Front-End Electronics (FEE) board, which is installed in the service area outside the MDI region. This service area is located on the outer surface of the LYSO crystal in the FastLumi, which is illustrated in Figure 3.13. The OAT converts electrical signals into optical signals and transmits them over fibers to Back-End Electronics (BEE) in the counting room.

Power delivery follows a staged DC-DC conversion scheme (Power-at-Load (PAL)) to reduce resistive losses and cable mass. Electrical power from the power supply is sent to the front-end electronics board, where the PAL48 module steps down 48 V to 12 V, followed by the PAL12 module reducing it to 1.2 V for the stitched sensor. In addition, high voltage power supplies deliver the substrate bias voltage directly to the stitched sensors through dedicated cables, ensuring proper depletion and stable sensor operation during the lifetime of the detector.

The detailed design of the optical transceivers, power regulators, and overall readout architecture is discussed in Chapter 11. The routing of the FPCs from the sensors to the FEE boards is described in Section 4.3.3, while the cable routing from the counting rooms to the FEE boards is presented in Chapter 14.

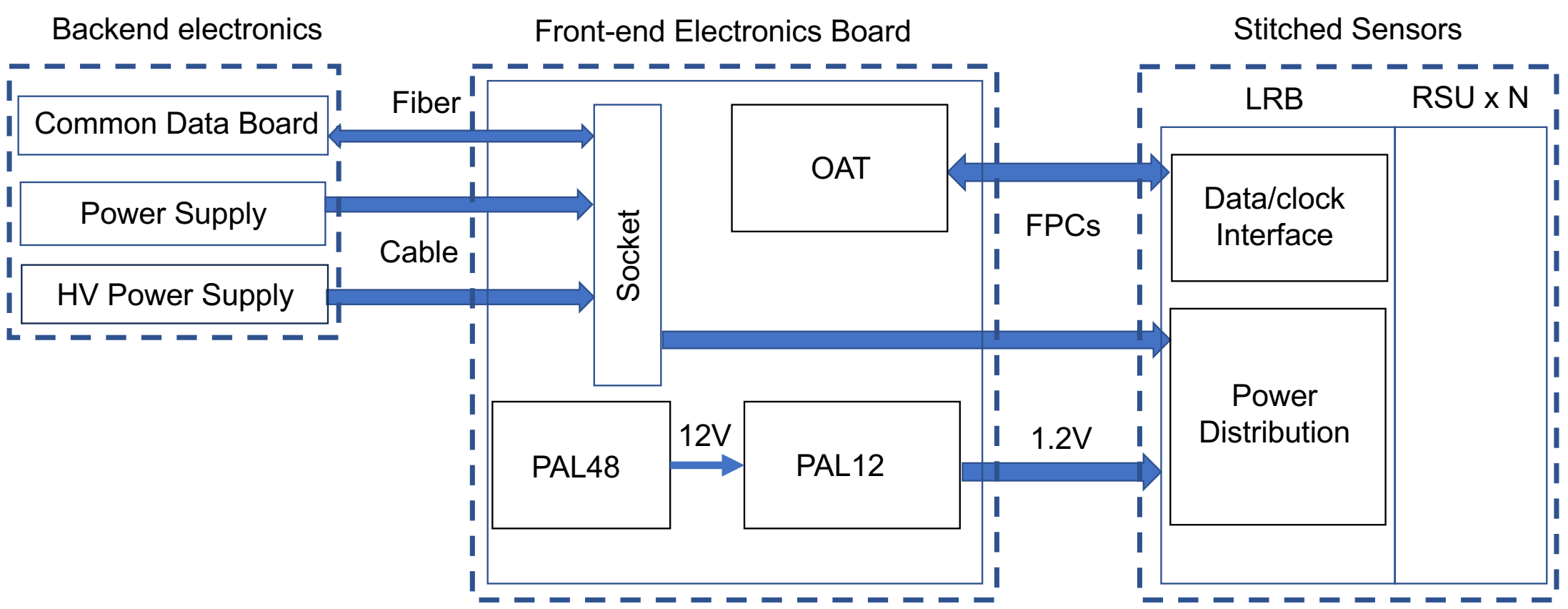

**Figure 4.15:** Diagram of the VTX readout electronics. The LRB in each stitched sensor collects data from multiple RSUs and transmits them via FPCs to the OAT module in the Front-end Electronics Board, which forwards the data to the BEE. Electrical power from the power supply is delivered to the stitched sensor through a staged DC-DC conversion chain (PAL), providing the required operating voltages. In addition, high voltage power supplies deliver the substrate bias voltage directly to the stitched sensors through dedicated cables, ensuring proper depletion and stable sensor operation during the lifetime of the detector.

## 4.3 Mechanics, cooling, and services

This section presents the mechanical design of the VTX, covering the support structure and its finite-element analysis, the detector's air-cooling concept, and the overall service layout.

## 4.3.1 Mechanics

### 4.3.1.1 General support structure

The baseline VTX design comprises four concentric cylindrical layers fabricated from curved MAPS sensors using stitching technology, followed by two outer layers assembled from double-sided planar ladders. The innermost cylindrical layer is positioned at a radius of 11 mm, providing a clearance of 0.3 mm from the beam pipe. Figure 4.1 shows the complete detector layout with its integrated mechanical support.

### 4.3.1.2 Assembly and support structure of the inner layers with curved MAPS

The inner region of the VTX comprises four CVTX layers, each with distinct radii and lengths as described in Section 4.1.2.1. Each CVTX layer consists of two half-cylindrical sensor assemblies that join to form a complete cylinder, as illustrated in Figure 4.16.

To preserve the curvature of the MAPS sensors, ultra-lightweight local supports made of Carbon Fiber Reinforced Polymer (CFRP) are employed. These supports are fabricated with slotted holes to minimize airflow resistance and reduce material budget.

The curved chips are wire-bonded along their lateral edge to the FPC, which integrates the readout electronics. To protect the wire bonding region, extended supports are attached to each cylinder, with the FPCs glued to these supports. The supports prevent deformation of the FPCs near the bonding joints, reducing the risk of damage. Each half-cylinder assembly therefore comprises the curved MAPS, the local support, and the extended support. Assemblies are mechanically independent and are mounted separately onto the beam-pipe assembly, with no physical contact between adjacent layers.

The design emphasizes the ease of assembly, reliable wire bonding, and independent installation of each half-cylinder. The innermost layer is mounted directly onto the beam pipe via its extended supports. The CVTX 2 – CVTX 4 are supported by intermediate structures located between their extended supports and the beam pipe.

The intermediate support for the second layer also incorporates slotted holes to promote airflow through the gap between the first and second layers. The third and fourth layers share a common frame-type intermediate support that provides mechanical rigidity while preserving airflow. The total cross-sectional area available for air cooling in the inner four layers is 19.8 $cm^2$, and the effective cross-sectional is 12.6 $cm^2$.

Finite-element analysis shows that, with both ends fixed, the structure undergoes minimal deformation and induces only very low stress on the curved sensors, ensuring safe operation.

### 4.3.1.3 Assembly and support structure of the outer PVTX layers

The double-sided ladder is the structural unit of Layers 5 and 6. It features pixel sensors mounted on inside and outside of a ladder support, where each sensor is glued onto a FPC,

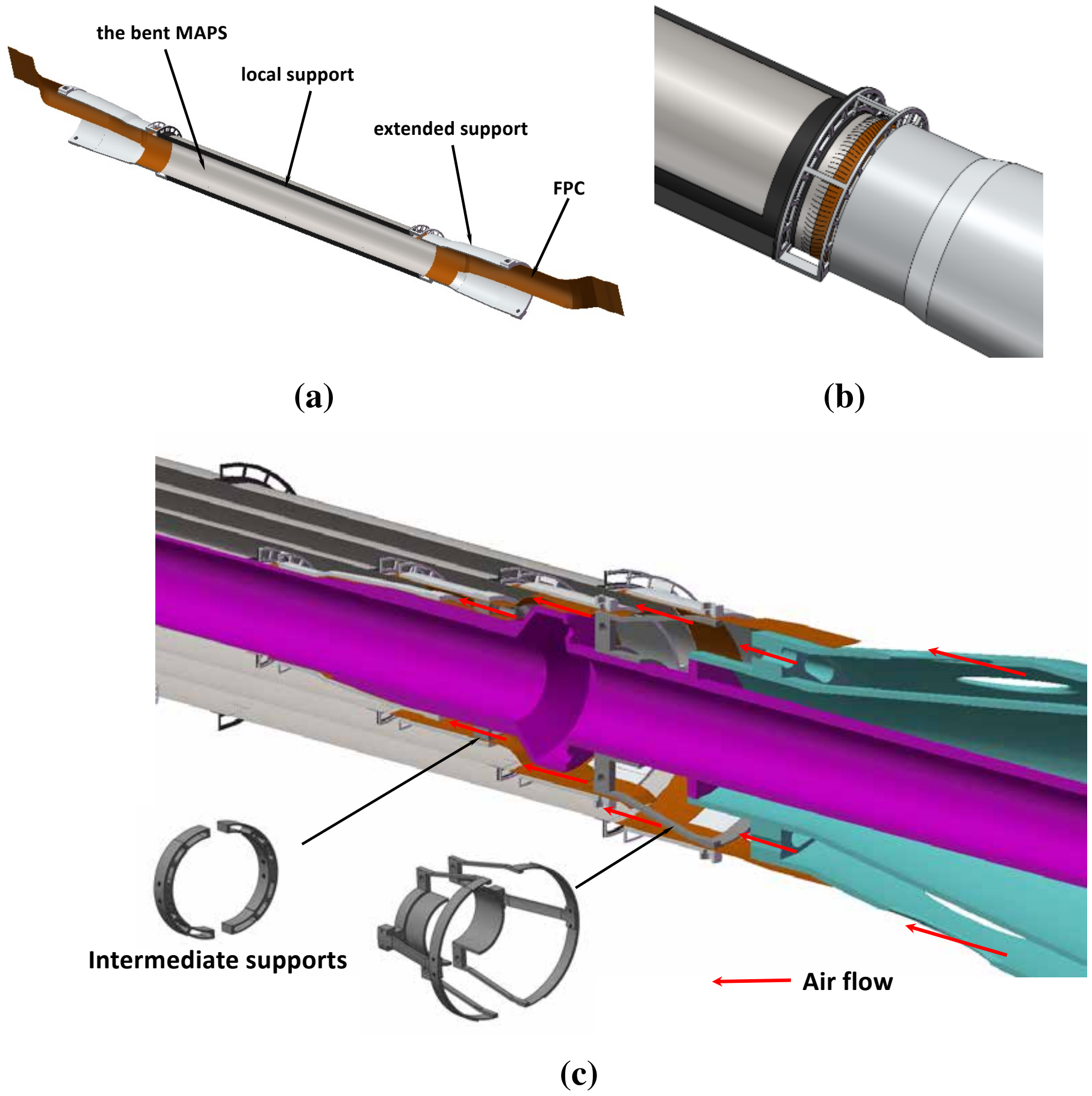

**Figure 4.16:** (a) The half-cylinder assembly of the innermost layer consists of the curved MAPS, the FPC, the local support and the extended support. (b) Extended support structure designed to prevent deformation of the FPC near the bonding region; (c) Installation of the cylinder layers and associated intermediate support structure. The intermediate support for the second layer includes slotted holes to allow airflow through the gap between the innermost and second layers. The third and fourth layers share common intermediate supports in the form of a frame-like structure, which provides mechanical strength while minimizing airflow obstruction. Airflow is indicated by red arrows in the figure.

which is in turn affixed to the support structure. The support is constructed from CFRP also.

The main body of the ladder support [12] adopts a hollow-shell structure [12] with overall dimensions of $682 \times 3.2 \times 17.5$ mm$^3$. The laminated shell is fabricated using ultra-thin CFRP plies. High-modulus CFRP (equivalent to M40) is used, and the total laminate thickness is 0.15 mm.

To conservatively estimate deformation, the weight of the sensors and FPCs is applied to the bare ladder support without accounting for their contribution to structural stiffness. Under this assumption, the expected deformation of the support from the finite element analysis is minimal and negligible. Previous evaluations further indicate that including the sensors and FPCs reduces the overall deformation by approximately 20% [12].

The outermost layer of the VTX, the ladder-based barrel, is not directly mounted on the beam pipe; it rests on and is secured to the intermediate conical parts that are mounted to the beam pipe, providing support for the vertex structure. The four CVTX layers are primarily attached to an extended section of the beam pipe.

The outer PVTX is constructed from overlapping double-sided ladders arranged circumferentially to form a continuous cylindrical detection surface. To simplify both assembly and integration, the PVTX is assembled by two half-barrels. Each half-barrel consists of two half side-rings located at either end, along with several ladders. The ladders are positioned and secured to the side-rings via tooth-like alignment surfaces. After pre-assembly on dedicated tooling, the two half-barrels are mounted onto the conical structure connected to the beam pipe, as illustrated in Figure 4.17.

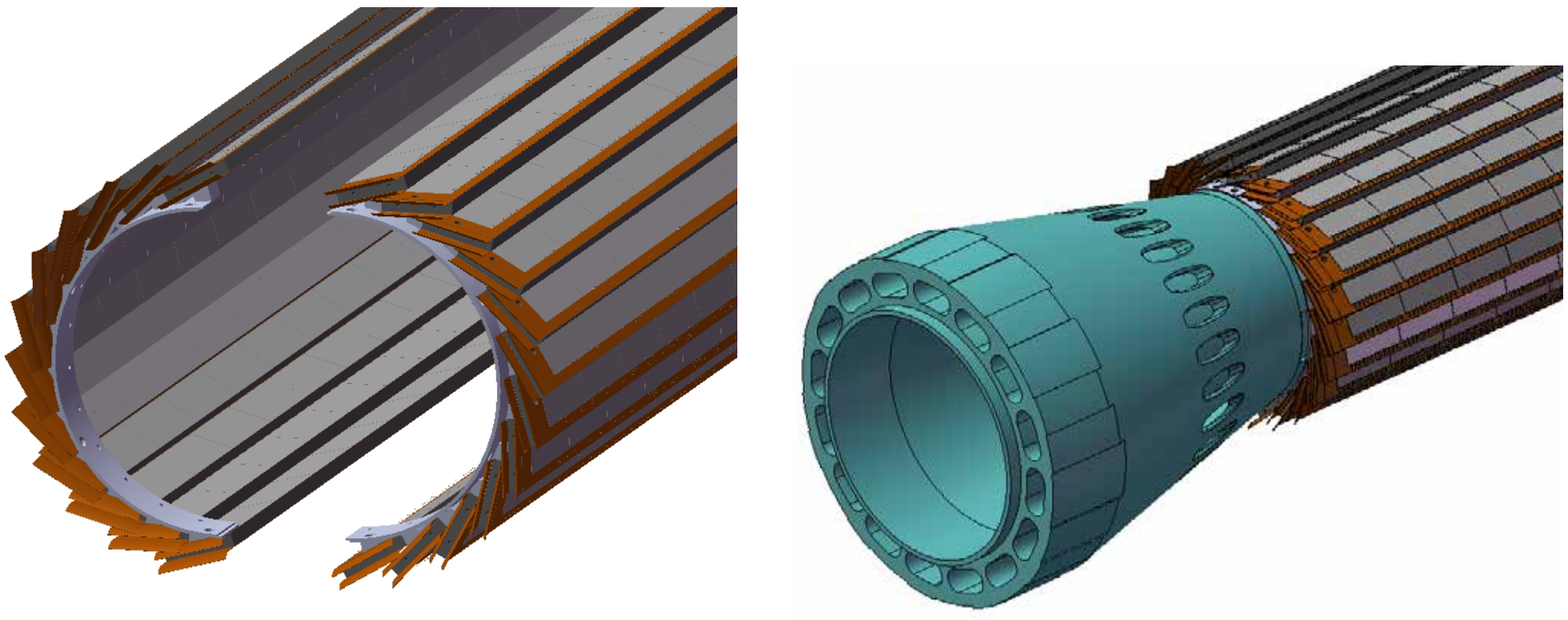

**(a)** **(b)**

**Figure 4.17:** (a) 3D model of the outer VTX layers composed of overlapping double-sided ladders mounted on two half side-rings. Each ladder is secured to the tooth-like structures on the ring. (b) Assembled view of the outer VTX layers mounted onto the conical intermediate support structure connected to the beam pipe.

### 4.3.2 Cooling

The heat dissipation of the VTX, including both the inner CVTX and outer PVTX layers, is estimated in Section 4.2.2.4. To ensure stable operation and prevent sensor performance degradation, the detector's operating temperature must be maintained below 30 °C.

Forced air convection was selected as the baseline cooling method due to the stringent material budget of the VTX.

The curved MAPS layers in the inner CVTX, particularly the innermost layer, present the most demanding thermal challenge because of their close proximity to the beam pipe and limited exposure to direct airflow.

To evaluate the performance of the cooling system, simulations were performed under various airflow speeds, as shown in Figure 4.18. The thermal simulations take into account the non-uniform power dissipation between the sensor's RSU and LRB, as described in Section 4.2.2.4. The results indicate a turbulent pattern under the current geometry and flow conditions, and demonstrate a clear trend of decreasing maximum temperature with increasing airflow speed, especially pronounced for the inner layers. At airflow speeds above 7 m/s, all four curved layers maintain temperatures below the required operational limit of 30 °C, with temperature gradients reduced to less than 10 °C. These results indicate that the proposed forced air cooling strategy can meet the thermal requirements of the full VTX detector. An airflow speed of 7 m/s, equivalent to a total flow rate of 3500 L/min, is selected as the baseline for the cooling design. Thermal simulations further demonstrate that this airflow is adequate to ensure the cooling performance of the outer PVTX layers.

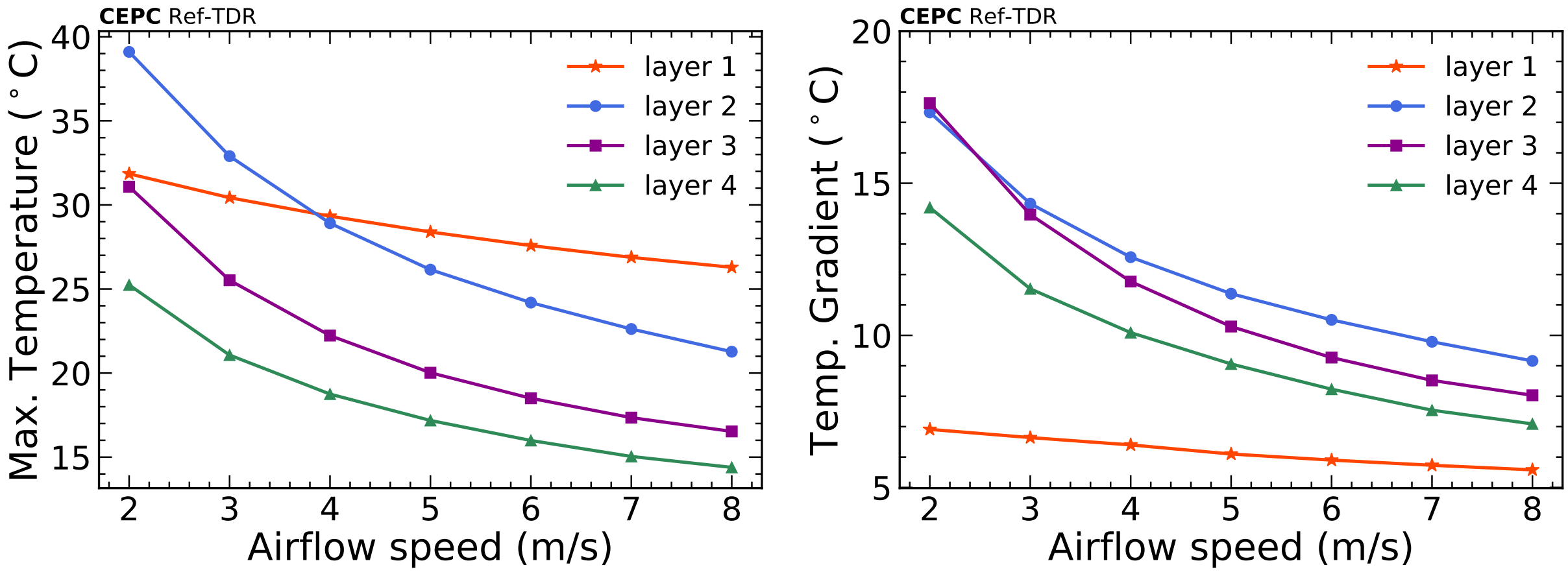

**(a)** **(b)**

**Figure 4.18:** Thermal simulation results for the curved MAPS cylinder layers of the VTX under different airflow speeds, with the beam-pipe surface temperature (see Chapter 3) included as a boundary condition. (a) The temperature response of each of the four curved layers (Layer 1 innermost to Layer 4 outermost) decreases markedly with increasing airflow; due to limited direct airflow and proximity to the beam pipe, Layer 1 remains the warmest across all cases. For airflow speeds ≥ 7 m/s, all layers stay within the 30 °C operational requirement. (b) The temperature gradient within each layer is largest for the innermost layer at low airflow and diminishes with increasing airflow, indicating improved in-layer thermal uniformity; For airflow speeds ≥ 7 m/s the inter-layer differences in gradients become less than 10 °C, consistent with the more uniform temperature field observed in (a).

### 4.3.3 Service infrastructure and cable routing

This section outlines the service infrastructure and integration considerations for the VTX. As the detector adopts forced air cooling, the primary service requirement is to ensure effective cooling across the full detector volume, which requires sufficient airflow through the gaps between adjacent layers. This requirement was addressed in the mechanical design. A key component in this design is the conical structure integrated into the beam pipe assembly. This hollow conical structure supports the outer PVTX and serves as a general-purpose air distributor. It receives air from one end and distributes it through dedicated ventilation holes directed toward various detector regions, as illustrated in Figure 4.19. In addition, both local and intermediate support structures are optimized to minimize airflow obstruction and ensure adequate ventilation across all layers.

Another requirement of the service is to route FPCs out of the extremely space-constrained beam pipe region. For the inner CVTX, each FPC is routed along the extended mechanical support of its corresponding curved sensor, converging past the outermost layer before being guided through slots in the intermediate conical structure.

For the outer PVTX ladders, FPCs from two adjacent ladders are grouped into a single conduit to reduce congestion. These conduits are guided by a dedicated FPC holder that preserves ventilation openings in the conical structure. The conduits are then routed through slots in the conical surface, as shown in Figure 4.19.

After routing out the MDI region, all FPCs are connected to front-end electronics boards via dedicated connectors. These front-end boards are equipped with optical transceivers and power regulators, as detailed in Section 4.2.3. To validate the feasibility of the proposed cable routing strategy, a half-dummy mechanical mockup was constructed. This mockup included curved MAPS cylinders and two short ladder structures. As shown in Figure 4.20, the test demonstrated the feasibility of the proposed service integration approach.

## 4.4 Alignment and calibration

The VTX is designed to precisely measure the trajectory parameters of charged particles close to the interaction point, enabling accurate reconstruction of decay vertices from short-lived particles. Effective alignment is crucial for achieving optimal resolution in the VTX.

The alignment strategy for the VTX faces significant challenges due to the implementation of the CVTX pixel sensors. The primary goals of this strategy include achieving high spatial resolution, reducing uncertainties of impact parameters, and ensuring stable alignment through efficient and automated processes.

### 4.4.1 Initial mechanical alignment and reference alignment system

Initial mechanical alignment will be conducted during detector assembly, utilizing precision optical survey instruments. Sensors will be aligned within a few micrometers of design specifications, and optical method will verify sensor curvature and cylindrical positioning. Fiducial markers placed strategically on sensor modules and supporting structures will facilitate optical tracking and ensure precise referencing. During the assembly phase of the CVTX layers, the curvature of the stitched sensors will be inspected. In later stages, the layer-to-layer coaxiality and $\phi$-stagger across layers will also be checked and recorded.

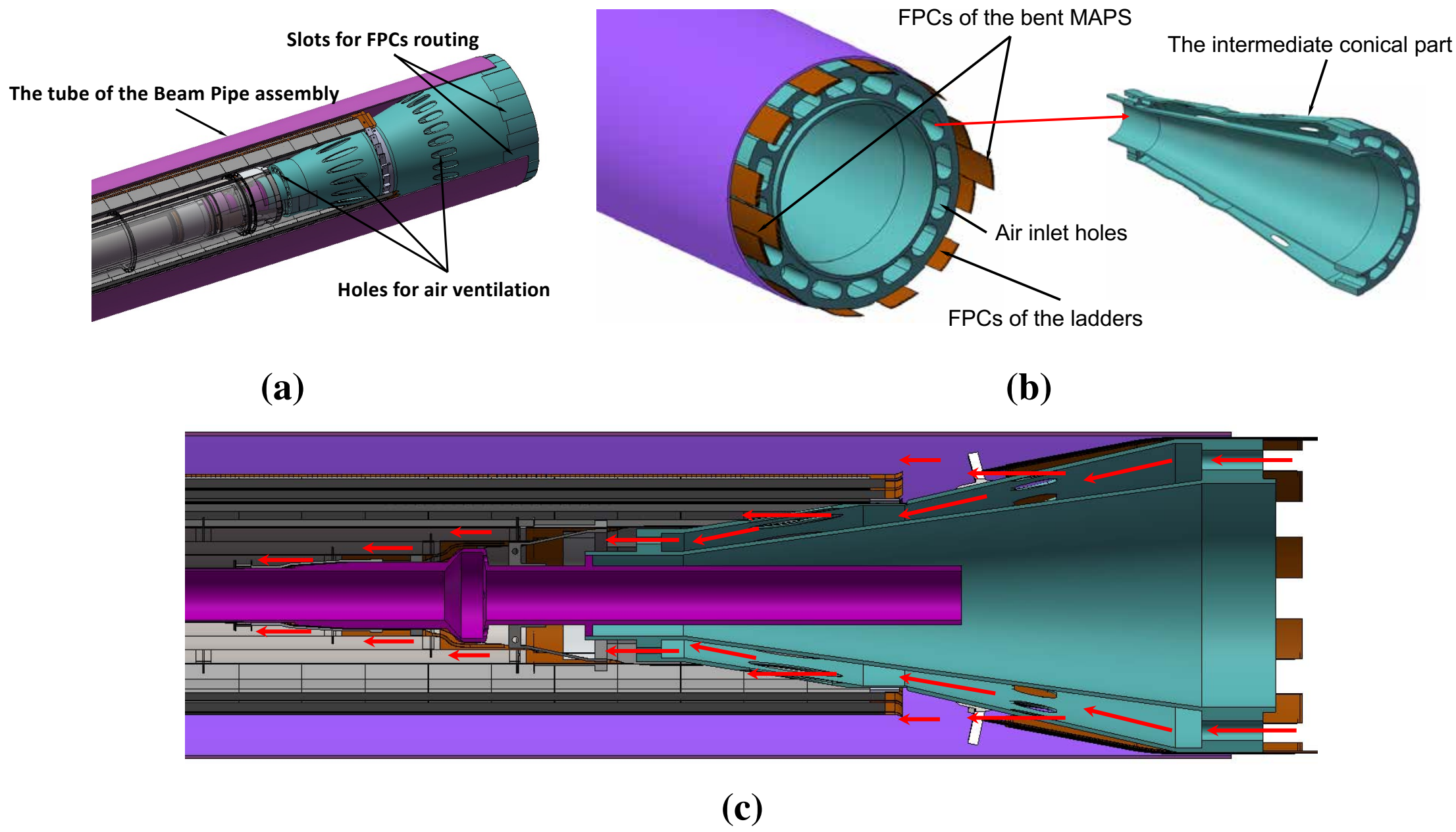

**Figure 4.19:** Airflow distribution and cable routing of the VTX. (a) Overview of the conical support structure, showing ventilation holes for airflow and dedicated slots for FPC routing. (b) 3D drawing showing the routing of FPCs from both the curved MAPS and ladders out through the conical structure on the side of the beam pipe assembly. (c) Cross-sectional view of the full assembly, illustrating the internal airflow paths (indicated by red arrows) through the conical air distribution structure.

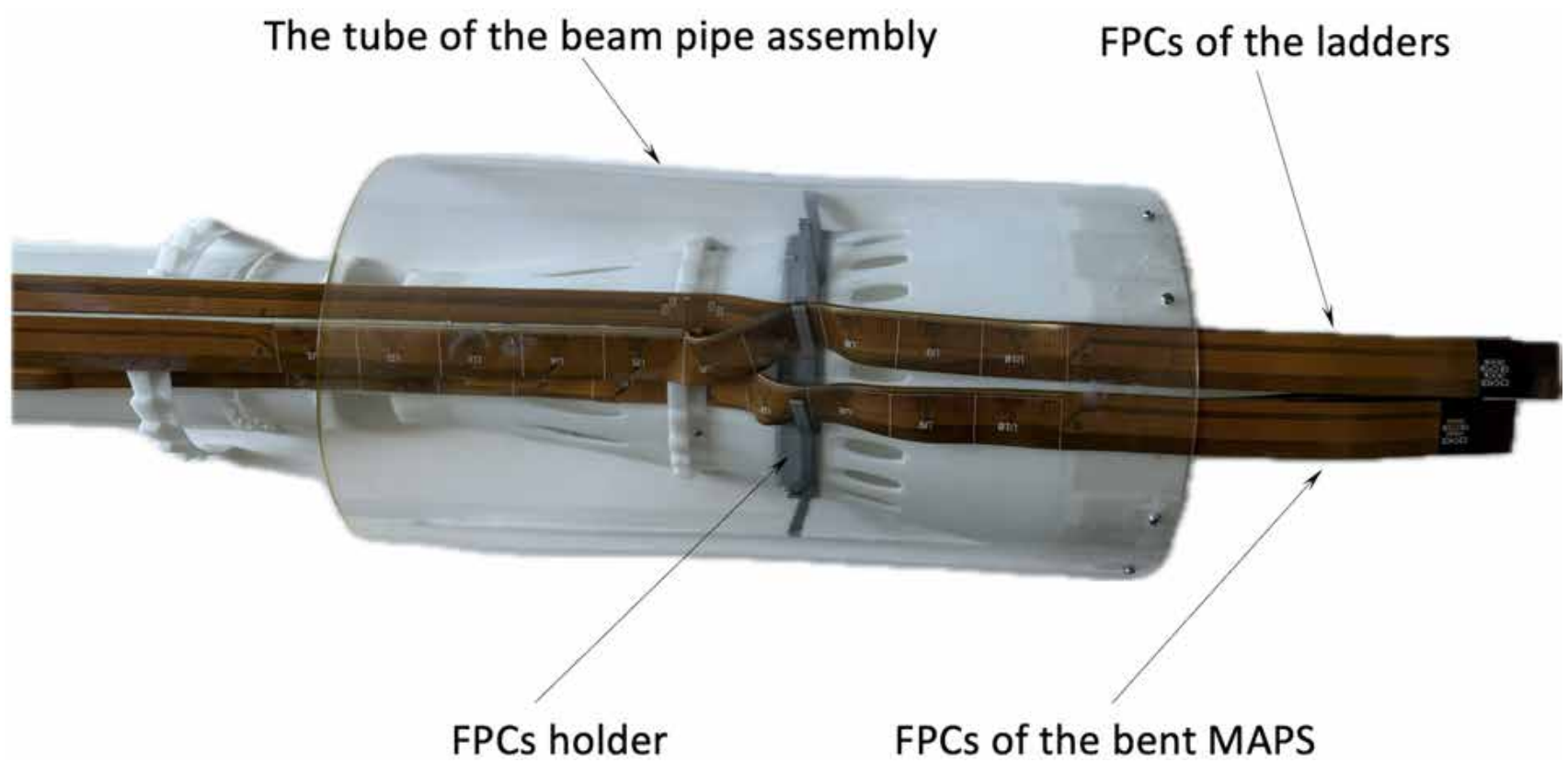

**Figure 4.20:** Half-dummy mockup of the vertex structure, illustrating key mechanical and service integration features. The setup includes curved MAPS cylinders, two short ladder modules, the intermediate conical support structure and a transparent half-tube representing the beam pipe enclosure. FPCs are merged using a dedicated FPCs holder mounted on the conical part. These bundled FPCs are then routed through designated slots and securely clamped by the outer beam pipe tube. This mockup validates the cable routing concept under realistic spatial constraints.

### 4.4.2 Track-based alignment

Mechanical alignment procedures during installation provide an initial level of precision in the VTX position. Typically this precision is significantly worse than the desired hit resolution.

Additional alignment (tracker-based alignment) is needed to account for the position, orientation, and surface deformations of the MAPS. Charged particle tracks obtained from collision data will facilitate iterative alignment corrections during both the commissioning and operational phases. Global and local $\chi^2$ minimization algorithms will refine sensor positioning, improving accuracy and ensuring optimal detector performance.

In the current detector design, simulations have demonstrated the ideal performance of the detector; however, in practical applications, it is necessary to consider typical deformation modes as shown in Figure 4.21. The deformation includes elliptical deformation; irregular, wavy distortion; circular distortion with uniform radial expansion.

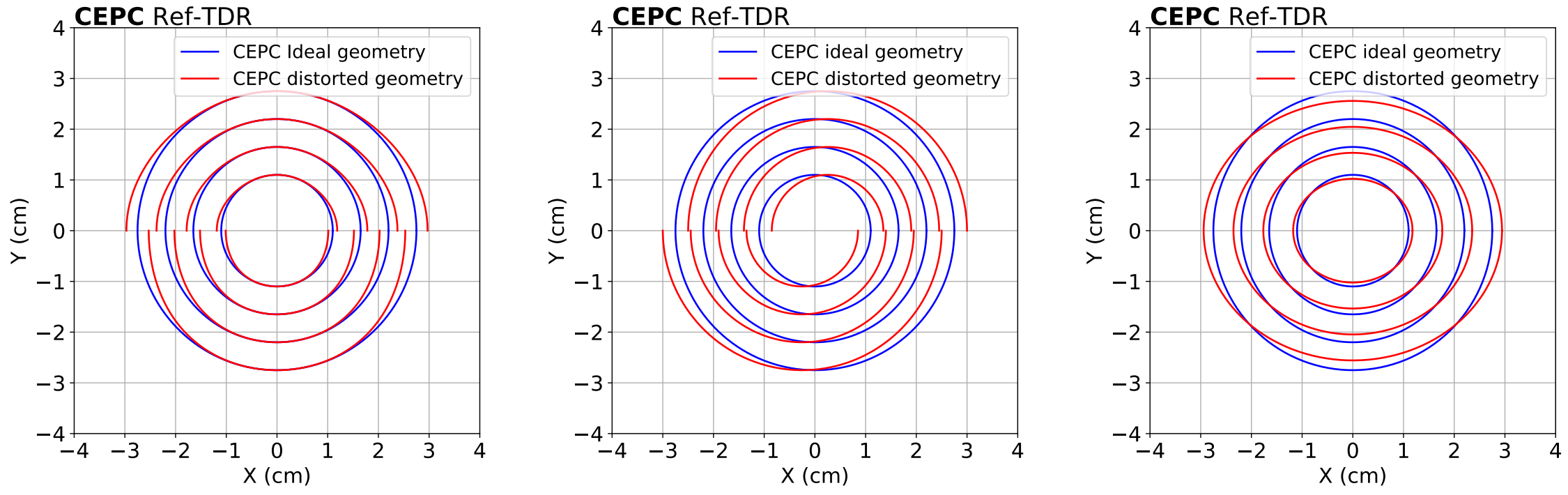

(a) (b) (c)

**Figure 4.21:** Illustration of hit positions in the transverse plane with ideal vertex geometry and three deformed geometry. The amount of deformation is amplified with respect to the expected one for visualisation. (a) Elliptical deformation. The red distorted geometry is elongated horizontally, demonstrating a deformation primarily along the $x$-axis. (b) Hit positions in the transverse plane exhibiting irregular, wavy distortion. The distorted geometry has noticeable undulations, indicating non-uniform deformation affecting both the $x$ and $y$ directions unevenly. (c) Hit positions in the transverse plane showing a circular distortion with uniform radial expansion. The distorted geometry presents a uniformly increased radius, representing an isotropic deformation compared to the ideal geometry.

These deformation modes can affect the detector's ability to determine the relative positions of collision products as they pass through the VTX, ultimately influencing the accuracy of the reconstructed collision parameters $d_0$ and $z_0$. The effect of each mode was studied in the simulation by modifying the position of hits according to the three modes of the deformation. In particular, for each simulated charged-particle trajectory, the intersection with the deformed geometry was computed to determine the corresponding position of the hit in the local detector coordinates. The track-reconstruction algorithm was then executed assuming an ideal geometry to quantify the effect of the detector deformations in the absence of a detector alignment procedure.

### 4.4.3 Real-time monitoring

A laser-based online alignment monitoring system will perform real-time checks, verifying alignment accuracy continuously.

Due to the air cooling design in the VTX, it is important to monitor the movement of the MAPS. Inspired by the laser alignment system of CMS experiment [13], it is proposed to install laser alignment system to keep track of these movement. As shown in Figure 4.22, laser sources are placed on the service portions of CVTX layers. There are two sources on both sides of inner three layers and one source at the right side of the Layer 4. This system utilizes near-infrared laser beams (1050 nm) directed through the VTX to detect potential movements or deformations of the mechanical structure. Laser sources located in service areas are triggered to send pulsed infrared light to CVTXs. By detecting their interaction with the silicon sensors, the system can infer movements of the CVTXs, ensuring that any deformations due to environmental factors like temperature, air flow, or magnetic fields are promptly identified.

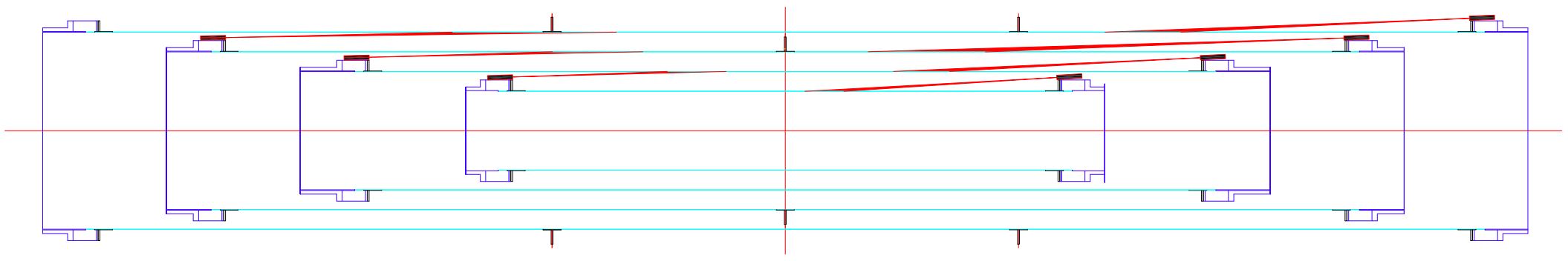

**Figure 4.22:** A laser-based online alignment monitoring system. Red lines diagonally connect with the aqua blue line (CVTX chip) are the lasers, and the laser sources are placed at both sides of the inner three layers and right side of the Layer 4 to monitor the movement of each layer.

To validate the laser calibration system, signals generated by the laser were simulated using Geant4 [14] within the CEPCSW framework. The laser was emitted as a point source from the position (13 mm, 0, −85 mm), directed at an angle of $\theta = 10^\circ$ and $\phi = 0$. A beam divergence of $\tan\alpha = 0.1$ is expected in the absence of a focusing system. The corresponding laser beam spot on Layer 2 of the VTX is shown in Figure 4.23.

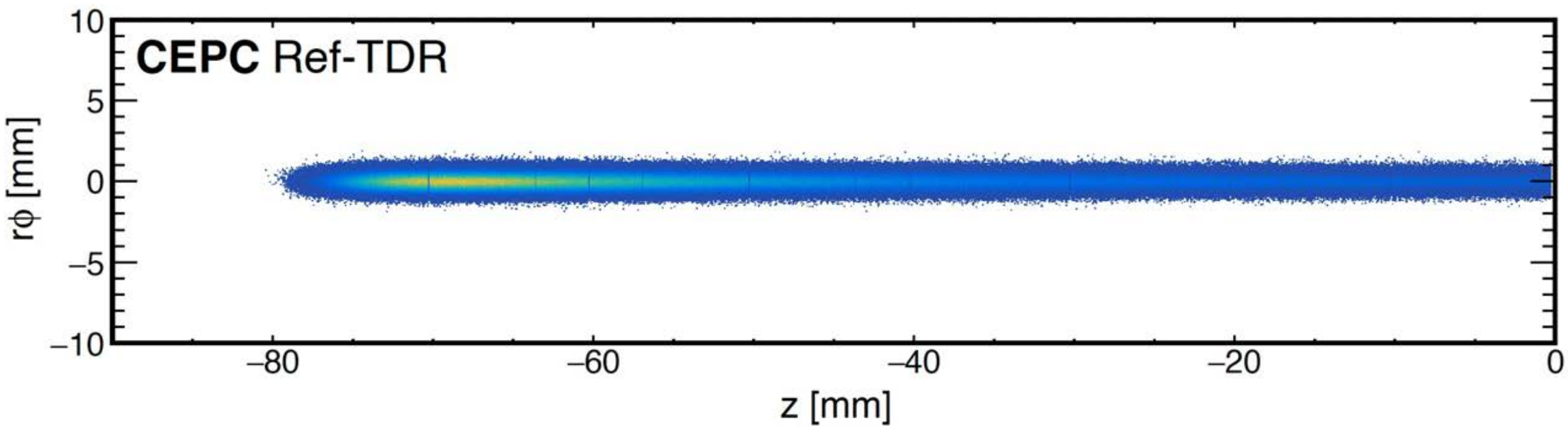

**Figure 4.23:** Laser beam spot on the Layer 2 from (13 mm, 0, −85 mm) in cylindrical coordinate system, where the horizontal axis is the $z$-coordinate, and the vertical axis is the $r\phi$ coordinate. Each region is divided by a blank or lower rate than its neighbor means one sensor.

When simulating deformation by radially displacing the second layer outward by $d = 10$ μm,

the observed difference corresponds to a distance of 50 μm, which is consistent with relationship $d/\tan\theta \approx 57$ μm. These findings indicate that the laser calibration system can infer positional changes through the distribution of pixel IDs, and the expected precision is at the micrometer level.

## 4.5 R&D efforts and results

To meet the stringent requirements of the VTX, significant efforts are devoted to advancing MAPS technologies, particularly through the TaichuPix and JadePix projects.

The JadePix project focuses on optimizing the charge collection diode design and the analog front-end circuitry, with particular emphasis on achieving low power consumption in the analog domain. In contrast, the TaichuPix project is dedicated to developing a fully functional digital architecture, incorporating in-pixel logic and a data-driven, column-drain readout scheme. This design is specifically tailored to meet the stringent timing and data throughput requirements of the high-luminosity CEPC environment, particularly during the low-luminosity $Z$-pole operation with high event rates.

### 4.5.1 JadePix series

Achieving a resolution of 3 μm requires a pixel pitch of less than 20 μm, which approaches the practical limits of current integrated circuit technologies. The JadePix project is based on a customized 180 nm MAPS process, modified and optimized from commercial CMOS imaging technology for ionizing radiation detection. In this process, the minimum achievable pixel pitch typically ranges between 25 μm and 30 μm. As a result, the central challenge in the JadePix development is to reduce the effective pixel size while maintaining low power consumption and supporting fast time stamping.

Since 2015, a total of five JadePix prototype chips were developed and iteratively improved. An overview of the chip generations and their main features is presented in Table 4.12.

**Table 4.12:** Overview of JadePix development

| Chip Name | Pixel Array | Pixel Pitch(μm$^2$) | Analog Front-end | Matrix Readout | Design Team |
|---|---|---|---|---|---|
| JadePix-1 | 128×192, 160×96 | 16×16, 33×33 | Source follower | Rolling Shutter | IHEP |
| JadePix-2 | 96×112 | 22×22 | Diff. Amp., CS Amp. | Rolling Shutter | IHEP |
| JadePix-3 | 512×192 | 16×23, 16×26 | ALPIDE | Rolling Shutter | IHEP, CCNU, SDU, DLNU |
| JadePix-4 | 356×489 | 20×29 | ALPIDE | AERD | CCNU, IHEP |
| JadePix-5 | 896×480 | 20×30 | ALPIDE | AERD | IHEP, CCNU |

JadePix-1[15] primarily focused on sensor optimization, comparing different geometric dimensions [16] and providing experimental evidence for the design of small charge collection electrodes [17]. As shown in Figure 4.24, the optimal spatial resolution was 2.7 μm [18], obtained through beam tests on an array of pixels with pixel pitches of 33 × 33 $\mu m^2$ pixel array.

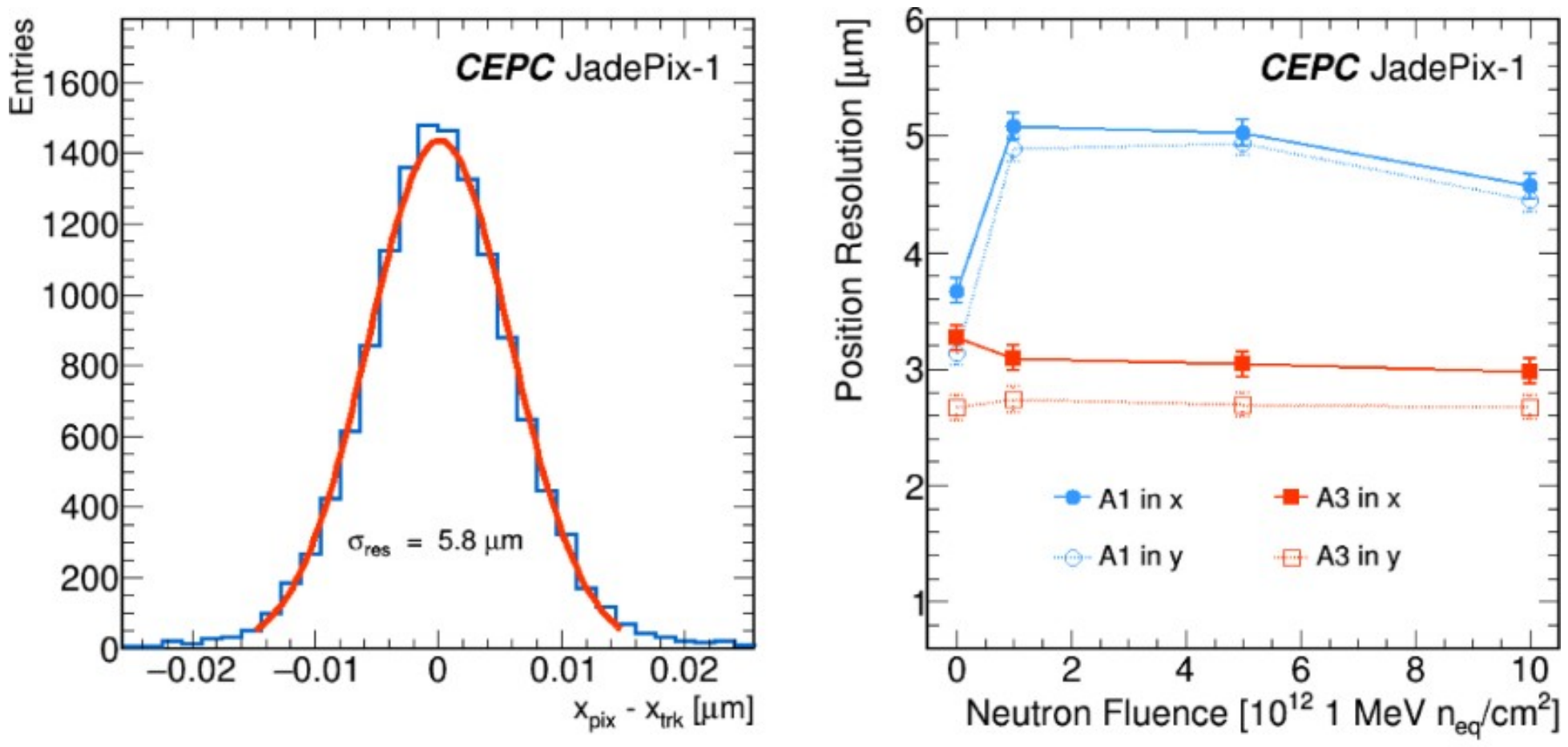

(a) (b)

**Figure 4.24:** (a) Residual distribution in the *x*-direction for the JadePix-1 sensor. (b) Position resolution as a function of neutron fluence for sensors with a pixel size of 33 × 33 $\mu m^2$, but different electrode geometries. A1 features a small electrode area of 4 $\mu m^2$, while A3 has a larger electrode of 15 $\mu m^2$. The resolution is measured in both the *x*- and *y*-directions, before and after neutron irradiation. Results indicate that smaller electrode size yields better intrinsic spatial resolution and is less sensitive to radiation-induced degradation.

JadePix-2 investigated AC coupling and different analog front-end amplifiers [19], attempting to optimize resolution, power consumption, and time stamping design using a 180 nm process based on the Minimum Ionising MOS Active pixel sensor (MIMOSA) architecture.

JadePix-3 utilized the low-power front-end amplifier of the ALICE Pixel Detector (ALPIDE)[5], combined with a Rolling Shutter readout architecture, achieving the smallest pixel design among similar chips at 16 × 23 $\mu m^2$. Infrared laser tests showed spatial resolutions of 3.4 μm (X-direction) and 2.7 μm (Y-direction) [20]. A five layers beam telescope system was constructed based on this chip, measuring spatial resolutions of 5.2 μm (X-direction) and 4.6 μm (Y-direction) on a 5.8 GeV electron beam [21]. The cluster size from beam tests was significantly larger than that from laser tests, adversely affecting the spatial resolution. The highest efficiency reached 99% at the optimum threshold of 160 electrons.

JadePix-4 and JadePix-5 employed ALPIDE analog front-ends and AERD readout architectures. The latter is an efficient sparse readout logic that can quickly read out the addresses of hit pixels at very low power consumption and apply time stamps around the periphery of the pixel array. The AERD readout architecture was adopted not only by various MAPS on the 180 nm process but also by the next-generation 65 nm Stitching process.

## 4.5.2 TaichuPix series

### 4.5.2.1 TaichuPix specification and performance

To cope with the high hit density at the CEPC, a dedicated MAPS sensor named *TaichuPix* was developed to deliver both high readout speed and excellent spatial resolution. TaichuPix-1[8] and TaichuPix-2[9], were fabricated as multi-project wafers, and TaichuPix-3, a full-scale prototype, was manufactured with an engineering run. The pixel matrix of TaichuPix-3 is 1024 × 512 with a pixel pitch of 25 μm, and a thickness of 150 μm. A VTX prototype was constructed, consisting of three layers of ladders, with double-sided mounted TaichuPix-3 sensors.

Each pixel of the TaichuPix-3 chip integrates a sensing diode, an analog front-end, and a digital readout. The analog front-end is designed based on the ALPIDE[5] chip, which is developed for the upgrade of the ALICE ITS[4]. In order to address the high hit rate of CEPC, the analog front-end of TaichuPix-3 was specifically optimized to ensure a quicker response. In addition, the digital readout includes a hit storage register, logic for pixel mask, and test pulse configuration. The digital readout follows the FE-I3[22] designed for the ATLAS pixel detector, but modified to adjust the pixel address generator and relocate the timestamp storage from within the pixel to the end of the column. This modification was necessary due to pixel size constraints. Furthermore, the Dcols drain peripheral readout architecture of the TaichuPix-3 chip employs an address encoder with a pull-up and pull-down network matrix. The design specifications of the TaichuPix-3 chip are summarized in Table 4.13.

**Table 4.13:** Design specifications of the TaichuPix-3 chip.

| **Specification** | **Index** |
|---|---|
| Pixel size | $25 \times 25\ \mu m^2$ |
| Dimension | $15.9 \times 25.7\ mm^2$ |
| Technology | TowerJazz 180 nm |
| Dead time | $< 500$ ns |
| Power density | $60 \sim 200\ mW \cdot cm^{-2}$ |

The performance of the pixel sensor was evaluated using an electron beam at Deutsches Elektronen-Synchrotron (DESY) II [10]. A dedicated six layers beam telescope based on the TaichuPix-3 chip was constructed to assess its tracking performance. Figure 4.25 shows the intrinsic spatial resolution as a function of the discriminator threshold. A resolution of 4.5 μm was achieved at the optimal threshold for TaichuPix-3. The detection efficiency was measured to be 99.7%, demonstrating that the sensor meets the stringent spatial resolution and efficiency requirements of the CEPC VTX.

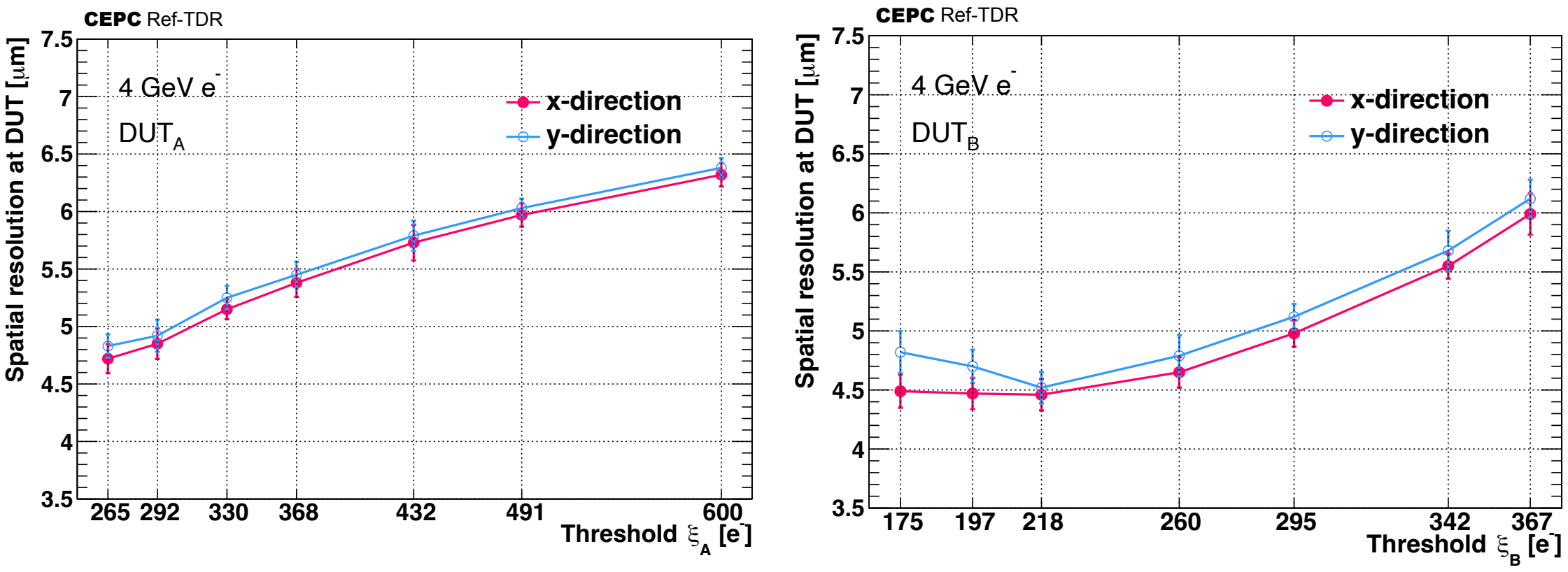

**(a)** **(b)**

**Figure 4.25:** A dedicated 6-layer beam telescope based on the TaichuPix-3 chip was constructed to assess its tracking performance at DESY II [10]. Spatial resolution as a function of threshold for DUT$_A$, a TaichuPix-3 chip fabricated using a modified process to enhance depletion (a), and DUT$_B$, a TaichuPix-3 chip fabricated using the standard process (b), evaluated in both the $x$-direction (shorter sensor side) and $y$-direction (longer sensor side). Error bars indicate the total systematic uncertainty.

#### 4.5.2.2 Prototype of a planar MAPS VTX

In order to meet the design requirements outlined in Table 4.13, a planar MAPS-based VTX prototype (shown in Figure 4.26) was developed and evaluated using an electron beam at DESY II [1].

The prototype comprises three concentric cylindrical layers positioned at radii of approximately 18.1 mm, 36.6 mm, and 59.9 mm. Each detector module, or ladder, features a double-sided structure capable of hosting pixel sensors on both sides. Each ladder integrates up to ten pixel sensor chips per side, along with a FPC and a carbon fiber mechanical support. For beam testing, two sensors were wire-bonded onto the FPC to cover the maximum active area permitted by the DESY II testbeam collimator. During the test, the electron beam traversed all six double-sided ladders, resulting in a total of 12 spatial measurement points.

The intrinsic spatial resolution was derived from an unbiased distribution of tracking residuals, excluding the Device Under Test (DUT) from the track fit. After alignment and threshold optimization, the spatial resolution reached 5.0 μm, with a detection efficiency of 99.6%.

**Air cooling** To maintain the detector at an appropriate operating temperature, air cooling must ensure that vibrations induced by forced airflow do not compromise spatial resolution. This aspect was investigated during the prototype R&D phase.

Test results show that the maximum vibration amplitude of the ladder support remains below 1.9 μm, even at an airflow rate of 4 m/s. Furthermore, during the beam test of the VTX prototype, air cooling using a fan reduced the chip temperature from 41 °C to 25 °C at a power dissipation of approximately 60 mW/cm$^2$, without degrading the spatial resolution. The strategy for controlling vibration in the baseline VTX design aligns with previous studies, relying on prototyping of critical structural components and subsequent performance validation through dedicated testing.

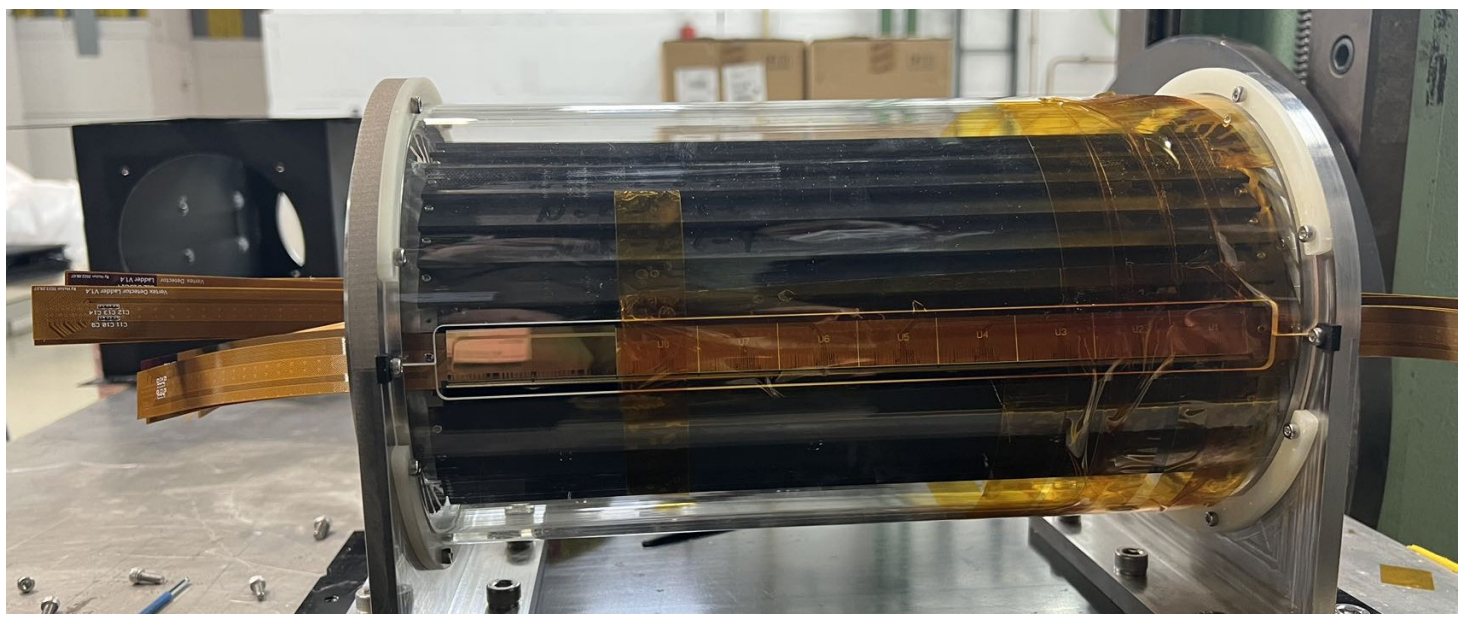

**Figure 4.26:** Photograph of the planar VTX prototype comprising six instrumented double-sided ladders (12 layers) assembled for beam testing. This full-scale prototype includes all mechanical ladders constructed from carbon fiber. Two TaichuPix chips were mounted on each side of every ladder. During the test, the electron beam passed through all six ladders, yielding a total of 12 spatial measurement points.

In addition to thermal management of the prototype, the dependence of the front-end noise on temperature was also characterized. As shown in Figure 4.27, the Equivalent Noise Charge (ENC) increases approximately linearly with temperature, from about 17 e$^-$ at 0 °C to about 22 e$^-$ at 70 °C. This trend is consistent with the expected contribution of thermal noise from the front-end electronics. Importantly, even at the highest tested temperature, the ENC remains below 25 e$^-$, which satisfies the design requirement for the VTX readout chip.

#### 4.5.2.3 Prototype of a stitched CMOS detector

A stitched MAPS prototype was developed using a 0.35 μm CMOS process, featuring a 14 μm-thick epitaxial layer and four metal routing layers. The process offers triple-well isolation; however, only Negative channel Metal-Oxide-Semiconductor (NMOS) transistors can be implemented within the pixel area. The complete pixel matrix consists of 644 rows and 3600 columns, formed by stitching basic arrays of 92 rows by 600 columns. The sensor operates in a rolling-shutter readout mode.

As a fully functional prototype, it integrates column-level discriminators, on-chip zero suppression, and interface circuits such as bias Digital-to-Analog Converter (DAC)s, analog buffers, Inter-Integrated Circuit (I2C) control, Phase-Locked Loop (PLL), Low-Voltage Differential Signaling (LVDS), and Low Dropout Regulator (LDO)s. The total chip area is approximately $11 \times 11$ cm$^2$. To study charge-collection efficiency and charge sharing, each basic pixel array is divided into six submatrices with different pixel sizes and diode configurations.

Preliminary test results of the prototype chip show that each row can be read out in 200 ns using an 80 MHz system clock. To adjust operating timing delays, dedicated buffer groups are inserted every 600 columns. The comparator threshold bias voltage ($V_{ref}/2$) is distributed across six groups to reduce load capacitance and maintain the comparator operating frequency. Each comparator dissipates 134 µW and exhibits a transient noise of 161 µV. Simulations indicate that the Fixed-Pattern Noise (FPN) across the 3600 comparators is about 1 mV, mainly due to long-distance timing distribution. Each reticle column incorporates an on-chip LDO, which can provide a stable clamping voltage for the pixels during the tests. Each LDO consumes approximately 10 mW, supports loads exceeding 10 nF, and can deliver transient currents greater than 1.3 mA, with a switching transient voltage variation below 200 µV.

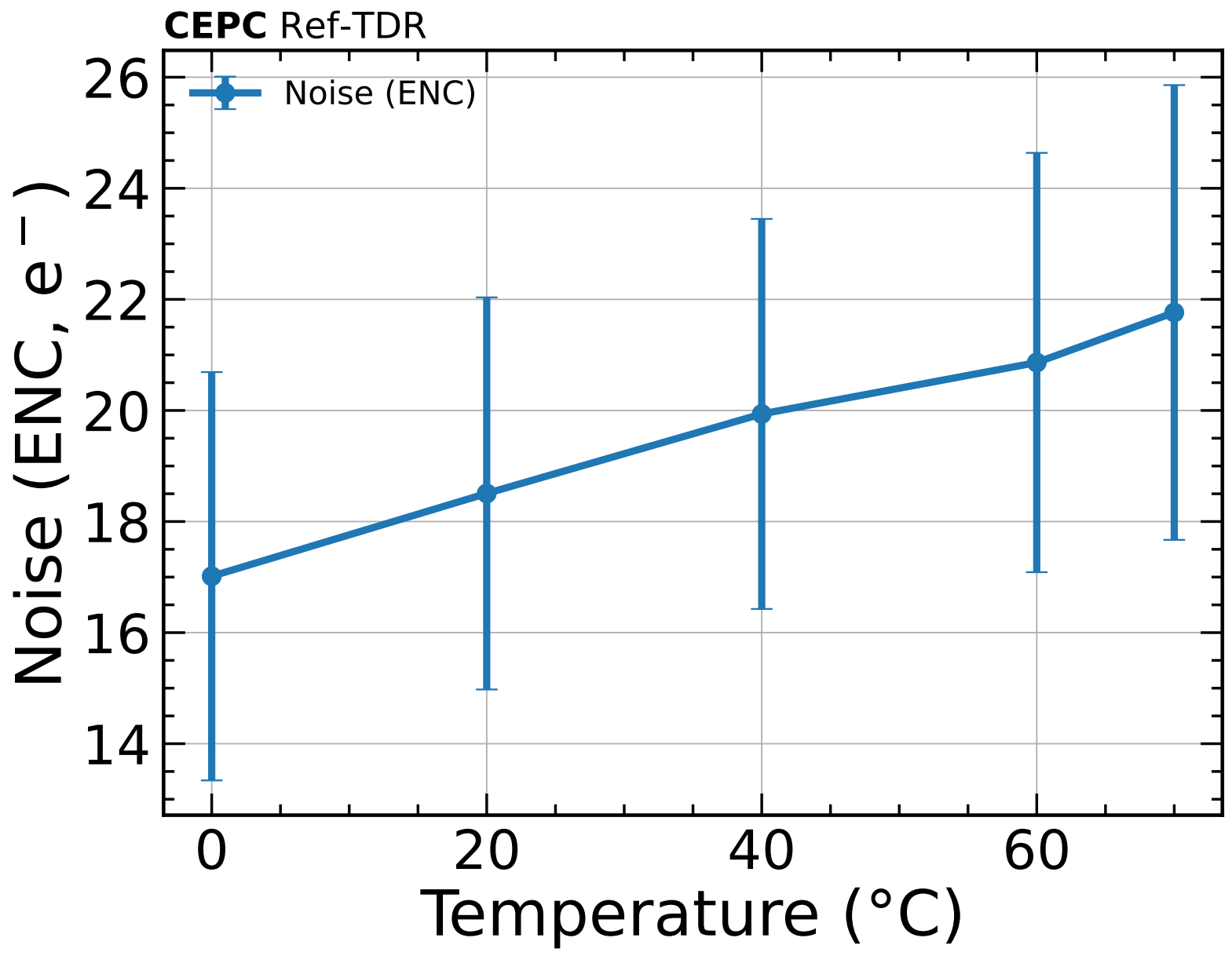

**Figure 4.27:** Measured noise performance of the TaichuPix-3 readout chip as a function of operating temperature. The Equivalent Noise Charge (ENC) is extracted at different temperatures ranging from 0 °C to 70 °C. The data show a linear increase of ENC with temperature, consistent with the expected thermal noise contribution from the front-end electronics. Error bars represent the statistical variation from repeated measurements.

The prototype is currently undergoing further testing, and its development is expected to provide valuable experience for designing stitched sensors in more advanced CMOS technologies.

**The radius bending test** Dummy wafers were used for bending tests, prototype manufacturing, and small-area MAPS to study the impacts of bending on the chip performance.

Silicon is a brittle material. It undergoes an irreversible brittle fracture when the stress exceeds its compressive strength. Wafers were thinned to thicknesses of 30 µm, 40 µm, and 50 µm, and bending tests—with and without a film—as well as fatigue tests at different radii and in various sequences, were successfully completed.

Currently, the bending limit has successfully achieved a minimum bending radius of 12

mm , as shown in Figure 4.28a. Additionally, more than 20 bending–recovery–bending cycles were performed on the same wafer. The prototype in Figure 4.28b was placed for over two years without damage.

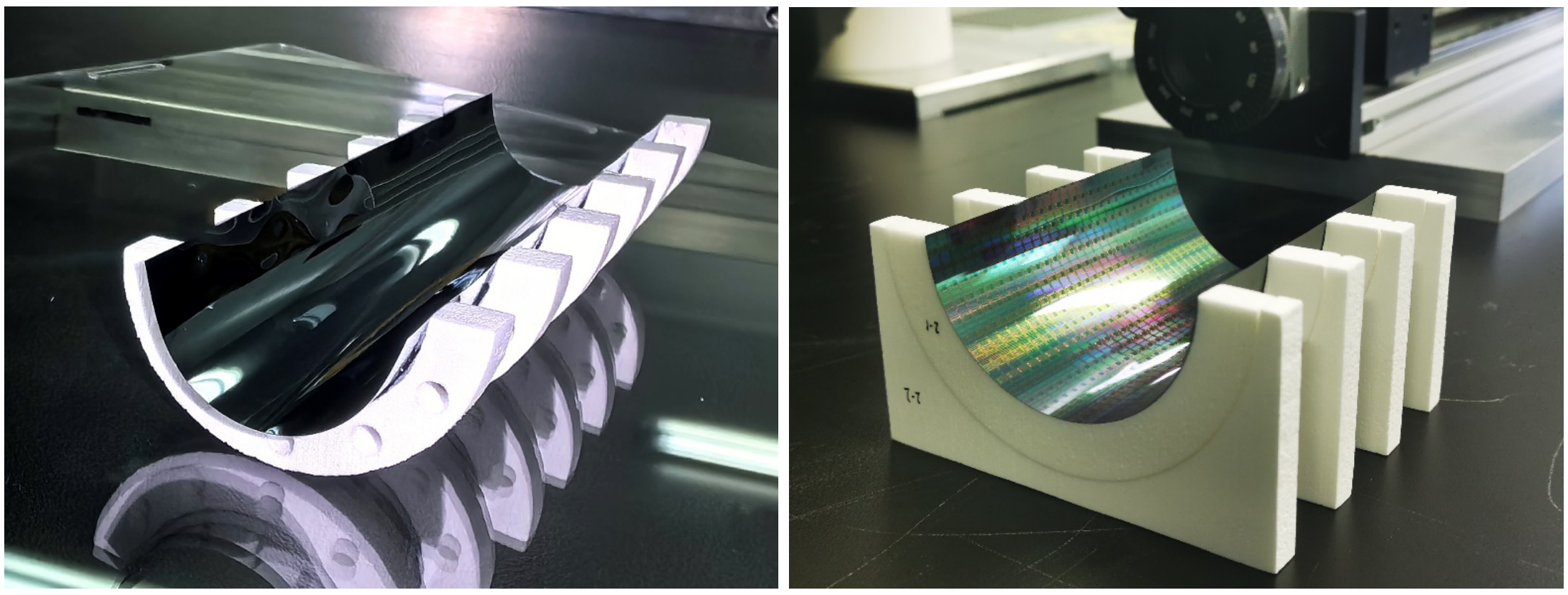

(a) (b)

**Figure 4.28:** Prototype used for wafer bending tests. (a) Evaluation of the mechanical bending limit, demonstrating that a minimum bending radius of 12 mm can be achieved without damage. (b) Long-term stability test of the bent wafer. The prototype was maintained in a bent configuration for over two years without exhibiting any visible damage, confirming the durability of the bending process.

## 4.6 Alternative design

An alternative VTX design featuring three layers of double-sided ladders with planar MAPS has also been considered, as illustrated in Figure 4.29. This design represents a conventional well-established approach and serves as an alternative fallback option should the baseline curved sensor technology encounter unforeseen challenges. Although it introduces additional material and mechanical complexity due to the need for structural support, ladder overlaps, and potentially thicker support components, the alternative design offers certain advantages: it is more robust and introduces fewer inefficient regions compared to the baseline.

Detailed parameters of the backup VTX layout are provided in Table 4.14. The intrinsic single-point resolution of the sensor is conservatively set to 5 μm, based on TaichuPix-3 beam test results.

## 4.7 Performance

### 4.7.1 Hit number and efficiency

There are two types of inefficient regions in the detector: one due to assembling pixel sensors into ladders or semi-cylinders, which form detector barrels; the other due to functional blocks in the sensor design, such as data interfaces and power switches, as shown in Figure 4.11. No hits will be produced when charged particles pass through these regions. During layout optimization, different deflections in the $\phi$ direction are applied to different layers to minimize the loss of the number of hits from a single trajectory. Furthermore, the width of the PVTX was increased to avoid empty sensor regions in the $\phi$ direction. However, when the width becomes too large, it could lead to a situation where both adjacent ladders at the same layer have signals, which increases the material volume of the overlapping area. To assess the reasonableness of the inefficient region and overlaps, simulations using charged Geantinos were performed to obtain the number of hits in the VTX.

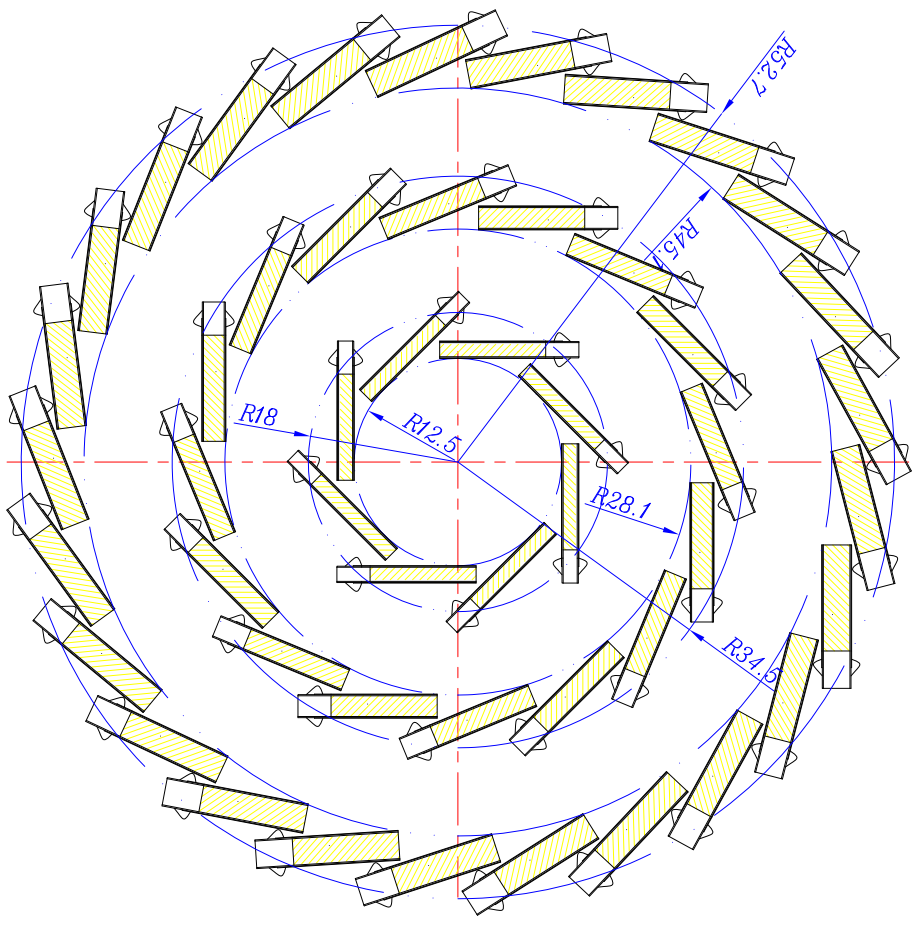

**Figure 4.29:** Sectional view of the VTX backup layout whose units are millimeters. All the three double-sided ladders (six layers) are composed of planar MAPS. PVTX 1-2, PVTX 3-4, and PVTX 5-6 include 8, 16, and 25 ladders, respectively.

**Table 4.14:** Geometric configuration parameters for the backup scheme. Radius and length, and material budget of the support structure for each layer of the VTX, as well as the height of ladder corresponding to PVTX.

| **PVTX X** | **radius** [ mm] | **length** [ mm] | **height** [ mm] | **support thickness** [μm] |
|---|---|---|---|---|
| PVTX 1-2 | 12.46 | 260.0 | 1.8 | 300 |
| PVTX 3-4 | 27.89 | 494.0 | 2.6 | 300 |
| PVTX 5-6 | 43.79 | 749.0 | 3.3 | 300 |

A metric known as Tracklet Efficiency is defined as the ratio of the total number of hits collected in the VTX by reconstructable trajectories to the total number of expected hits from all simulated tracks. It quantifies the detector's effectiveness in registering hits from traversing particles and is given by

$$\text{Tracklet Efficiency} = \frac{\sum_{i=1}^{N} \text{Hits}_i^{(\geq 4)}}{N \times L} \tag{4.3}$$

Here, $\text{Hits}_i^{(\geq 4)}$ denotes the number of hits produced by the $i$-th charged Geantino trajectory that leaves at least four hits in the VTX. $N$ is the total number of simulated events, and $L$ is the expected number of VTX layers a particle should traverse (six in this case).

As shown in Figure 4.30a, for charged Geantinos with a momentum of 20 GeV, the tracklet efficiency reaches 99.6% with an uncertainty of $10^{-4}$ at a polar angle of $\theta = 20°$.

Figure 4.30b presents the tracklet efficiency and its associated uncertainties for 20 GeV charged Geantinos passing through the VTX at different polar angles. The efficiency remains consistently near 100% across the full angular range, demonstrating the robust tracking capability of the detector.

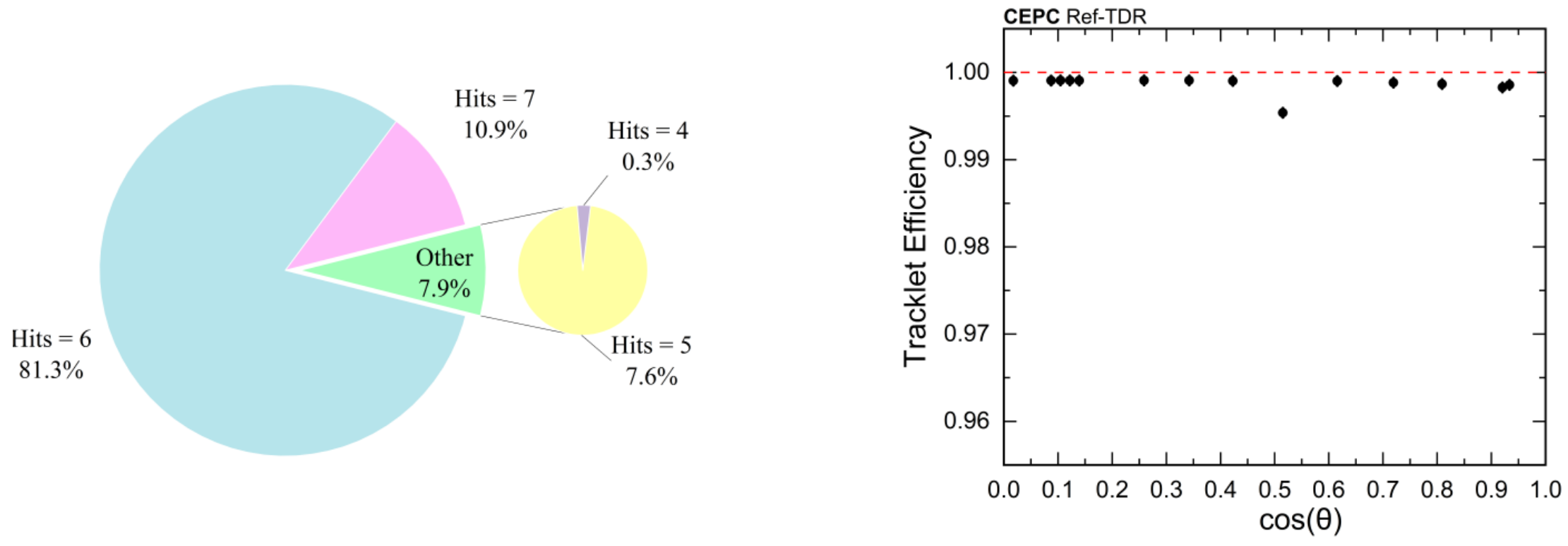

(a) (b)

**Figure 4.30:** Tracking performance of the VTX. (a) Distribution of the number of hits per track for 10,000 charged Geantinos with $p = 20$ GeVand $\theta = 20°$. Most tracks have six hits (81.3%), while 10.9% of the tracks have seven hits, and the rest are distributed among four, five, or other configurations. (b) Tracklet efficiency as a function of $\cos\theta$ for charged Geantinos at 20 GeV, originating from the interaction point and passing through the VTX. The efficiency remains above 99.6% across all polar angles, demonstrating robust tracking performance.

### 4.7.2 Resolution

The VTX configuration described in Section 4.1.2.1 serves as the baseline design for subsequent performance studies. An alternative (backup) scheme is also proposed and discussed in Section 4.6.

The baseline layout employs four inner layers of curved MAPS sensors (CVTX) mounted directly on the beam pipe, complemented by two outer layers composed of lightweight, double-sided ladders (PVTX). In contrast, the alternative scheme replaces the four curved inner layers with ladder-based structures using the same materials and sensor technologies as those in the baseline's outer layers. This replacement introduces additional material, particularly in the innermost region.

Simulation studies were performed using CEPCSW, with the full tracking system in place. The simulations assumed outgoing $\mu^-$ particles and a conservative single-point resolution of 5 µm for all pixel sensors. As shown in Figure 4.31, the baseline scheme demonstrates superior performance in $d_0$ and $z_0$ resolution, attributed to the reduced material budget in the inner layers and the closer proximity of the first active layer to the interaction point. While the alternative scheme features a more uniform and mechanically simpler structure, it yields a slightly degraded impact parameter resolution.

The observed dependence of the resolution on momentum reflects the dominant effect of multiple scattering at low momenta. Additionally, as the polar angle $\theta$ approaches the beam axis, the effective path length through material increases approximately as $1/\sin\theta$, which leads to enhanced multiple scattering and a corresponding degradation in resolution.

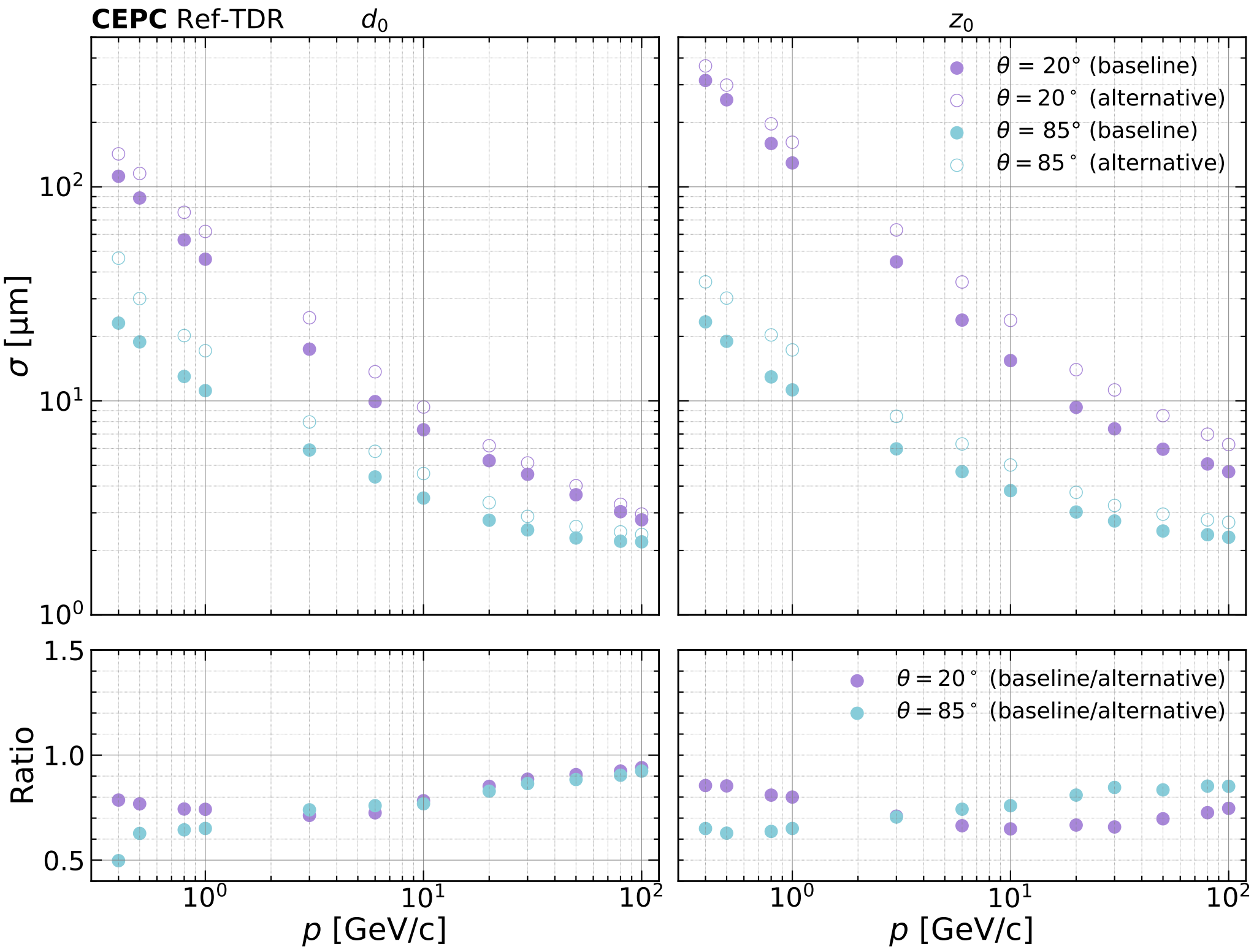

**Figure 4.31:** Impact parameter resolutions of $d_0$ and $z_0$ for the baseline and alternative vertex detector layouts, evaluated using full simulations within the CEPCSW framework. Results are shown for outgoing $\mu^-$ particles at two representative polar angles, $\theta = 20°$ and $\theta = 85°$, across a wide momentum range. The top panels show the absolute resolution values, while the bottom panels present the ratios of baseline to alternative configurations. Owing to its reduced material budget and more compact geometry, the baseline layout achieves superior resolution, outperforming the alternative scheme by approximately 10% - 40% in the momentum range from 0.4 GeV to 100 GeV.

### 4.7.3 Performance under sensor failure scenarios

During detector operation, individual VTX layers may fail or sustain damage due to various unforeseen factors. Such failures inevitably degrade the tracking performance. Therefore, the VTX design must be robust enough to ensure that even under the worst-case conditions, the performance remains within acceptable limits to support physics analyses.

To assess the robustness of the detector, simulation studies were carried out assuming the complete loss of signal from one layer at a time. The resulting degradation in impact parameter resolution is shown in Figure 4.32, comparing the baseline performance to scenarios where each of the six layers fails individually.

Results indicate that the loss of any single layer leads to a performance drop of no more than 25%, with the innermost layer contributing the most to the resolution. For partial or local failures, the overall performance impact can be approximated as a weighted average based on the proportion of damaged versus functional units. Given the extremely low probability of simultaneous multi-layer failures, the VTX is expected to maintain stable performance in realistic operational scenarios.

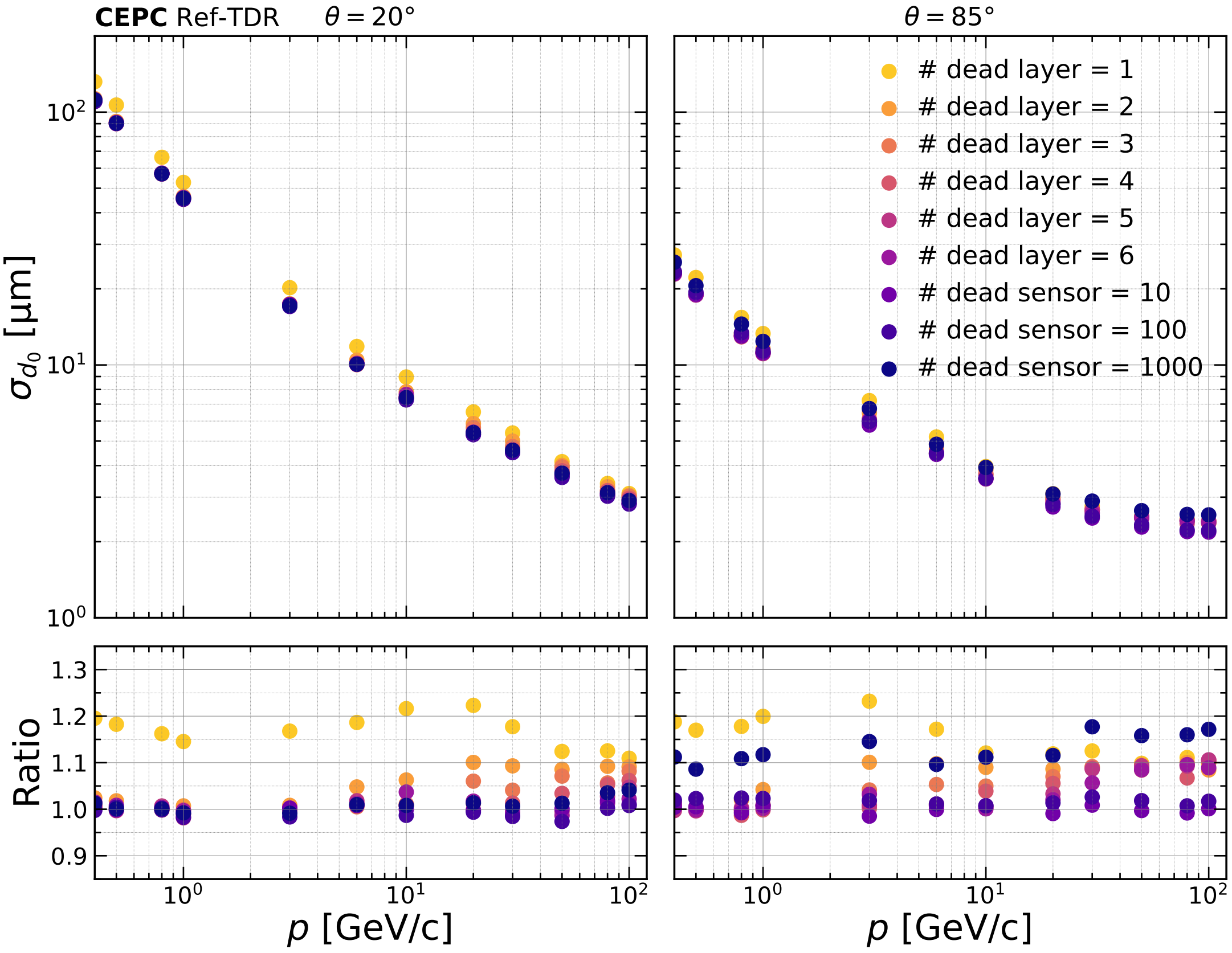

**Figure 4.32:** Impact parameter resolution $\sigma_{d_0}$ as a function of track momentum for the baseline VTX under various sensor failure scenarios. Results are shown for two polar angles, $\theta = 20°$ (left) and $\theta = 85°$ (right). The top plots display the absolute resolution, while the bottom plots show the resolution normalized to the baseline configuration. Failure scenarios include complete loss of individual layers (colored by layer index) and increasing numbers of random sensor failures (10, 100, and 1000 units). The plots demonstrate that the overall performance remains robust, with limited degradation even under partial sensor losses.

### 4.7.4 Performance with beam background

Given the close proximity of the VTX to the beam pipe, it is essential to assess the impact of beam-induced backgrounds on detector performance.

This assessment is conducted by overlaying realistic simulated beam backgrounds onto signal events, as described in Section 13.2.4. Figure 4.33 compares the $d_0$ and $z_0$ resolution with and without background overlay, showing no noticeable degradation. These results indicate that, under the background conditions predicted by current simulations, the VTX is expected to maintain excellent tracking performance.

However, the presence of beam background may increase the computational load and processing time for track reconstruction. Future efforts should therefore focus on optimizing reconstruction algorithms to maintain high efficiency in realistic experimental conditions.

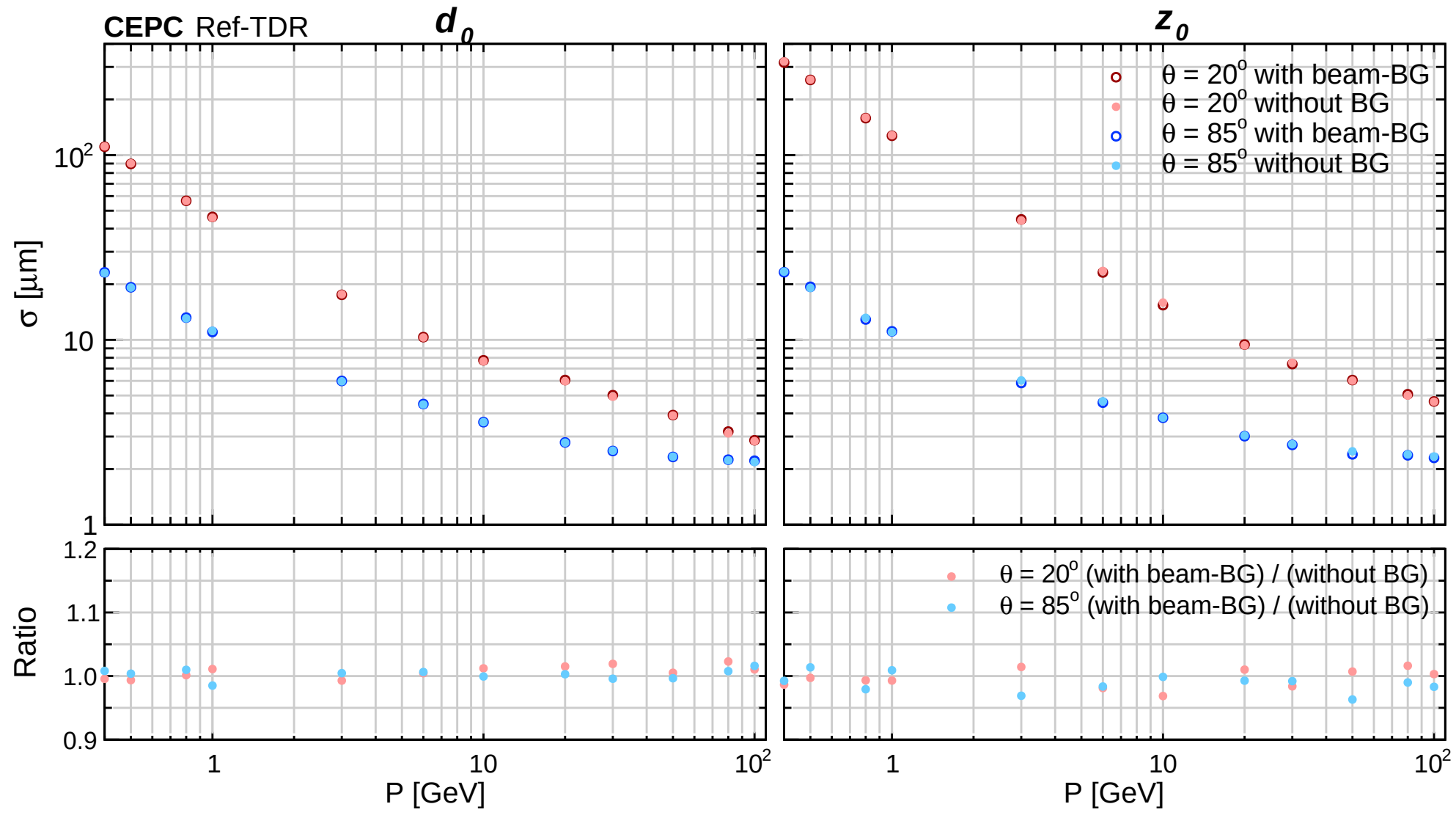

**Figure 4.33:** Comparison of resolution of the impact parameter $d_0$ and $z_0$ of tracks between clean single muon signal (without BG) and mixed beam background (with beam-BG) at polar angle of 85° and 20°.

## 4.8 Summary and future plan

The MAPS-based detector layout consists of six cylindrical layers of pixel sensors with pixel pitches in the order of 25 μm × 25 μm, enabling hit resolutions better than 5 μm. The inner four layers are curved MAPS cylinders. The outer two layers are based on double-sided ladder technology.

Initial VTX prototypes using double-sided ladder technology have demonstrated a spatial resolution better than 5 μm with air cooling in DESY testbeam. Geant4-based simulations demonstrate that the baseline design of VTX can achieve the target impact parameter resolution.

Future R&D priorities include:

1. Development of wafer-scale stitching pixel sensors
   - Initially, the stitching chip will leverage mature 180 nm technology (e.g., TowerJazz) to keep R&D risks and costs within reasonable bounds. The pixel cell and matrix design will build on the proven architecture of previous prototype chips, with focused efforts on stitching-related challenges and ensuring basic cell designs draw from verified experience.
   - The second-generation stitching chip will transition to 65 nm or 55 nm technology, with potential candidates including TPSCo's 65 nm technology and HLMC's 55 nm technology. Synergy with sensor development for future LHC upgrades is expected.
2. Ultra-thin mechanical supports and low-mass integration techniques.
   - The plan begins with exploring the construction of a mock-up featuring dummy heaters for thermal performance testing. Results will also validate thermal simulation models.
3. Construction of a full-scale VTX prototype to address challenges in mechanical precision, cooling performance, system-level integration and alignment.

# Chapter 5 Silicon Tracker

The Silicon Tracker (STK) is primarily designed to provide precise tracking for momentum measurements, as required by the CEPC physics programme across all data-taking modes. Located outside the Vertex Detector (VTX), the STK comprises the Inner Silicon Tracker (ITK) and the Outer Silicon Tracker (OTK), with the Time Projection Chamber (TPC) positioned between them.

The baseline design of the ITK employs monolithic High Voltage Complementary Metal-Oxide-Semiconductor (HV-CMOS) pixel sensors, achieving high spatial resolution and sufficient timing resolution for accurate bunch tagging. As the outermost tracking detector, the OTK — based on innovative microstrip AC-coupled Low Gain Avalanche Detector (AC-LGAD) technology — provides precise spatial measurement and extends the tracking lever arm, enhancing overall tracking performance, particularly in momentum resolution. In addition, it functions as a high-granularity Time-of-Flight (ToF) detector, offering precise timing measurement for particle identification.

This chapter begins with an overview of the Silicon Tracker in Section 5.1, followed by detailed descriptions of the ITK and OTK in Sections 5.2 and 5.3, respectively. Section 5.4 discusses the alignment strategies. Performance studies are presented in Section 5.5, and the chapter concludes with a summary and outlook in Section 5.6.

## 5.1 Overview

To support precise measurements of Higgs boson properties and other key physics objectives, the tracking system must deliver permille-level momentum resolution for charged particles with momenta below 100 GeV/c. In addition, ToF measurements are essential for particle identification in flavor physics studies and jet substructure analyses, particularly for hadron and jet ($b$ or $c$) tagging.

To meet these requirements, the baseline design of the STK includes three barrel layers and four endcap layers in the ITK, and one barrel layer and one endcap layer in the OTK, as shown in Figure 5.1. This layout enables precise determination of particle trajectories, with a bending lever arm of $\sim$ 1.6 m in the barrel region, spanning from the innermost ITK layer to the OTK. Additionally, the OTK provides precise timing measurement as a ToF detector for particle identification. The total surface area of the STK is $\sim$ 100 m$^2$.

Each detector layer is required to achieve a spatial resolution of 10 μm or better in the bending direction. At the same time, the detector material budget must be minimized to reduce multiple scattering, which is critical for maintaining good resolution at low momenta. The material thickness per ITK layer is constrained to be $< 1\%$ of a radiation length ($X_0$), while the OTK allows slightly more material, as it is the outermost tracking layer.

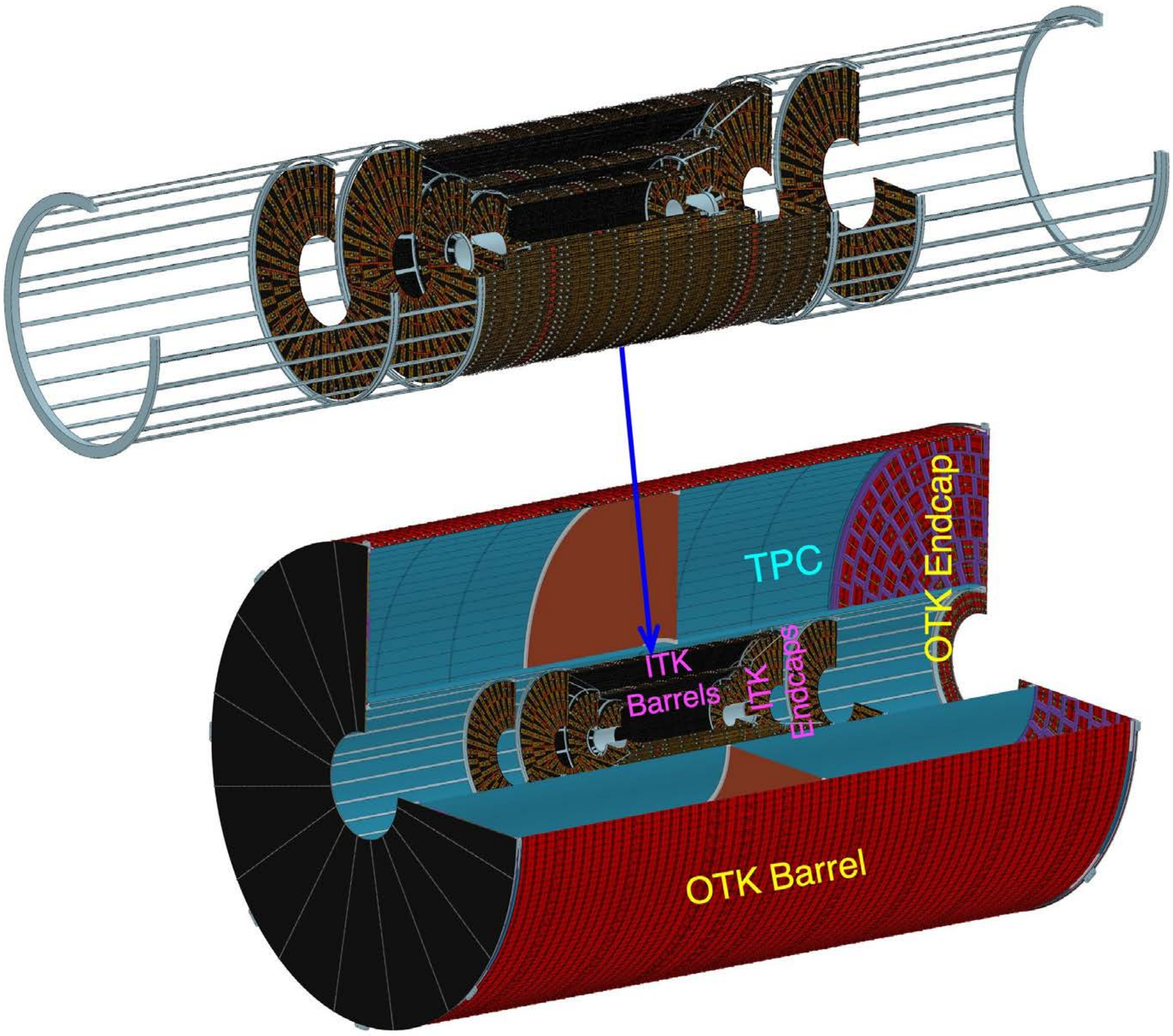

**Figure 5.1:** Layout of the Silicon Tracker (ITK and OTK). The ITK consists of three barrel layers and four endcap layers, together with two extended connection rings, forming the complete ITK assembly that is inserted into the inner barrel of the TPC. The OTK, as the outermost component of the tracker system, includes one barrel layer and one endcap layer, mounted outside of the TPC. It provides both high-precision spatial and timing measurements.

To meet these constraints, the ITK adopts a 150 μm-thick monolithic CMOS sensor, which integrates both the sensor and readout circuitry on a single chip. Under the high-luminosity $Z$-pole operation with 23 ns bunch spacing, a timing resolution of a few nanoseconds or better is required for bunch tagging. Accordingly, monolithic High Voltage Complementary Metal-Oxide-Semiconductor (HV-CMOS) pixel sensor technology has been selected for the ITK. For the CEPC specification, this sensor provides a spatial resolution of $\sim$ 8 μm in the bending direction and a timing resolution of $3 - 5$ ns, thanks to fast and large charge collection in a fully depleted sensor. This performance, achieved with moderate power consumption, significantly outperforms conventional partially depleted CMOS sensors. Common module designs are employed across the barrel and endcap regions to facilitate mass production. The mechanical structures are optimized for a low material budget and high stiffness, incorporating efficient embedded cooling, achieving a material budget of $\sim$ 0.7% $X_0$ per layer. Key ITK detector parameters are listed in Table 5.1.

For the OTK, the baseline technology is advanced microstrip AC-coupled Low Gain

Avalanche Detector (AC-LGAD), which provides both high-precision spatial resolution (~ 10 μm) for momentum measurement and high-precision timing (~ 50 ps) for particle identification. Unlike the monolithic ITK sensor, the OTK uses a hybrid sensor design, where the sensor and readout Application Specific Integrated Circuit (ASIC) are implemented as separate chips and then bonded together. The AC-LGAD sensor, combined with a dedicated readout ASIC chip currently under development, aims to deliver robust $K/\pi$/p separation below a few GeV/c, complementing the TPC's particle identification capability. The sensor modules, mechanical supports, and cooling structures of the OTK are specifically designed to meet the challenges of its large size, limited installation space, and strict material budget constraints (~ 1.5% $X_0$). Key OTK parameters are also listed in Table 5.1.

**Table 5.1:** Parameters and layout of the Silicon Tracker. The column labelled $\pm z$ shows the half-length of the barrel layers, and the $z$ position of the end-cap disks. The column labelled $\sigma_\phi$ and $\sigma_t$ represent the spatial resolution in the bending direction and time resolution, respectively.

| **Detector** | | **Radius $R$ [mm]** | | **$\pm z$ [mm]** | **Material budget [% $X_0$]** | **$\sigma_\phi$ [μm]** | **$\sigma_t$ [ns]** |
|---|---|---|---|---|---|---|---|
| **ITK Barrel** | Layer 1 (ITKB1) | 235.0 | | 493.3 | 0.68 | 8 | 3-5 |
| | Layer 2 (ITKB2) | 345.0 | | 704.8 | 0.68 | 8 | 3-5 |
| | Layer 3 (ITKB3) | 555.6 | | 986.6 | 0.68 | 8 | 3-5 |
| **OTK Barrel** | Layer 4 (OTKB) | 1,800 | | 2,840 | 1.6 | 10 | 0.05 |
| | | **$R_{in}$** | **$R_{out}$** | | | | |
| | Disk 1 (ITKE1) | 82.5 | 244.7 | 505.0 | 0.76 | 8 | 3-5 |
| | Disk 2 (ITKE2) | 110.5 | 353.7 | 718.5 | 0.76 | 8 | 3-5 |
| **ITK Endcap** | Disk 3 (ITKE3) | 160.5 | 564.0 | 1,000 | 0.76 | 8 | 3-5 |
| | Disk 4 (ITKE4) | 220.3 | 564.0 | 1,489 | 0.76 | 8 | 3-5 |
| **OTK Endcap** | Disk 5 (OTKE) | 406.0 | 1,816 | 2,910 | 1.4 | 10 | 0.05 |

As in the VTX, the STK hit rates are dominated by beam induced backgrounds, including pair production and single-beam backgrounds. Based on full simulations that incorporate all relevant background sources and the final detector layout, the average and maximum beam-induced background hit rates have been estimated for each STK layer across the three main operation modes, with a safety factor of 2 applied, as summarized in Table 5.2. The ITK and OTK readout systems are designed with sufficient safety margin to handle these rates under all operating conditions, including the most demanding high-luminosity Z-pole mode.

## 5.2 Inner silicon tracker (ITK)

The ITK is located between the VTX and the TPC. It consists of three barrel layers and four endcap layers, as detailed in Table 5.1, covering an area of ~ 20 $m^2$.

**Table 5.2:** Estimated average and maximum hit rates of the Silicon Tracker (safety factor 2 applied) [$10^3$ Hz/cm$^2$] for all layers across the three operation modes described in Chapter 1.

| | **Low Lumi.** $Z$ | | **Higgs** | | **High Lumi.** $Z$ | |
|---|---|---|---|---|---|---|
| $\mathcal{L}$ ($10^{34}$ cm$^{-2}$s$^{-1}$) | 26 | | 8.3 | | 192 | |
| | Average | Max | Average | Max | Average | Max |
| ITKB1 | 2.0 | 3.9 | 1.2 | 2.0 | 6.2 | 11.7 |
| ITKB2 | 2.0 | 6.6 | 1.1 | 2.6 | 5.9 | 19.8 |
| ITKB3 | 1.5 | 3.4 | 0.8 | 1.5 | 4.5 | 10.1 |
| OTKB | 1.3 | 2.2 | 0.7 | 1.2 | 3.9 | 6.6 |
| ITKE1 | 6.9 | 31.6 | 4.7 | 23.4 | 20.7 | 94.6 |
| ITKE2 | 8.0 | 49.0 | 4.1 | 19.6 | 23.9 | 146.8 |
| ITKE3 | 4.6 | 34.5 | 2.2 | 14.9 | 13.7 | 103.4 |
| ITKE4 | 5.1 | 17.1 | 2.5 | 7.4 | 15.2 | 51.2 |
| OTKE | 3.1 | 12.7 | 1.9 | 8.2 | 9.2 | 38.0 |

### 5.2.1 ITK design

The ITK design employs monolithic HV-CMOS pixel sensors fabricated using a 55 nm CMOS process to meet the requirements for performance, including power consumption. Key parameters are summarized in Table 5.3, and further details on the sensors are provided in Section 5.2.4.

**Table 5.3:** The HV-CMOS sensor key parameters

| **Parameter** | **Value** |
|---|---|
| Sensor size | 2 cm × 2 cm (active area: 1.74 cm × 1.92 cm) |
| Sensor thickness | 150 μm |
| Array size | 512 × 128 |
| Pixel size | 34 μm × 150 μm |
| Spatial resolution | 8 μm × 40 μm |
| Time resolution | 3-5 ns |
| Power consumption | 200 mW/cm$^2$ |
| Technology node | 55 nm |

#### 5.2.1.1 ITK barrel design

The ITK barrel design begins with the barrel module, which consists of 14 monolithic HV-CMOS pixel sensors arranged in two rows and seven columns, with their backside glued to a Flexible Printed Circuit (FPC) board that holds electronic components, including a radiation-tolerant Direct Current-Direct Current (DC-DC) converter and a data aggregation chip. The sensors are also electrically connected to the FPC via wire bonds, which transmit clock and command inputs, data output, and both low and high voltages for the sensors.

A schematic of a barrel pixel module is shown in Figure 5.2 (a). Each sensor has dimensions of 20 mm × 20 mm, with an active area of 17.4 mm × 19.2 mm, as listed in Table 5.3. The inactive area of each sensor mainly results from the region reserved for peripheral electronics along one edge, as well as from the guard rings designed to shape the electric field and protect against edge breakdown. Within the module, sensors are oriented such that most inactive regions are positioned along the two long edges of the module, as indicated by the grey areas in Figure 5.2 (a).

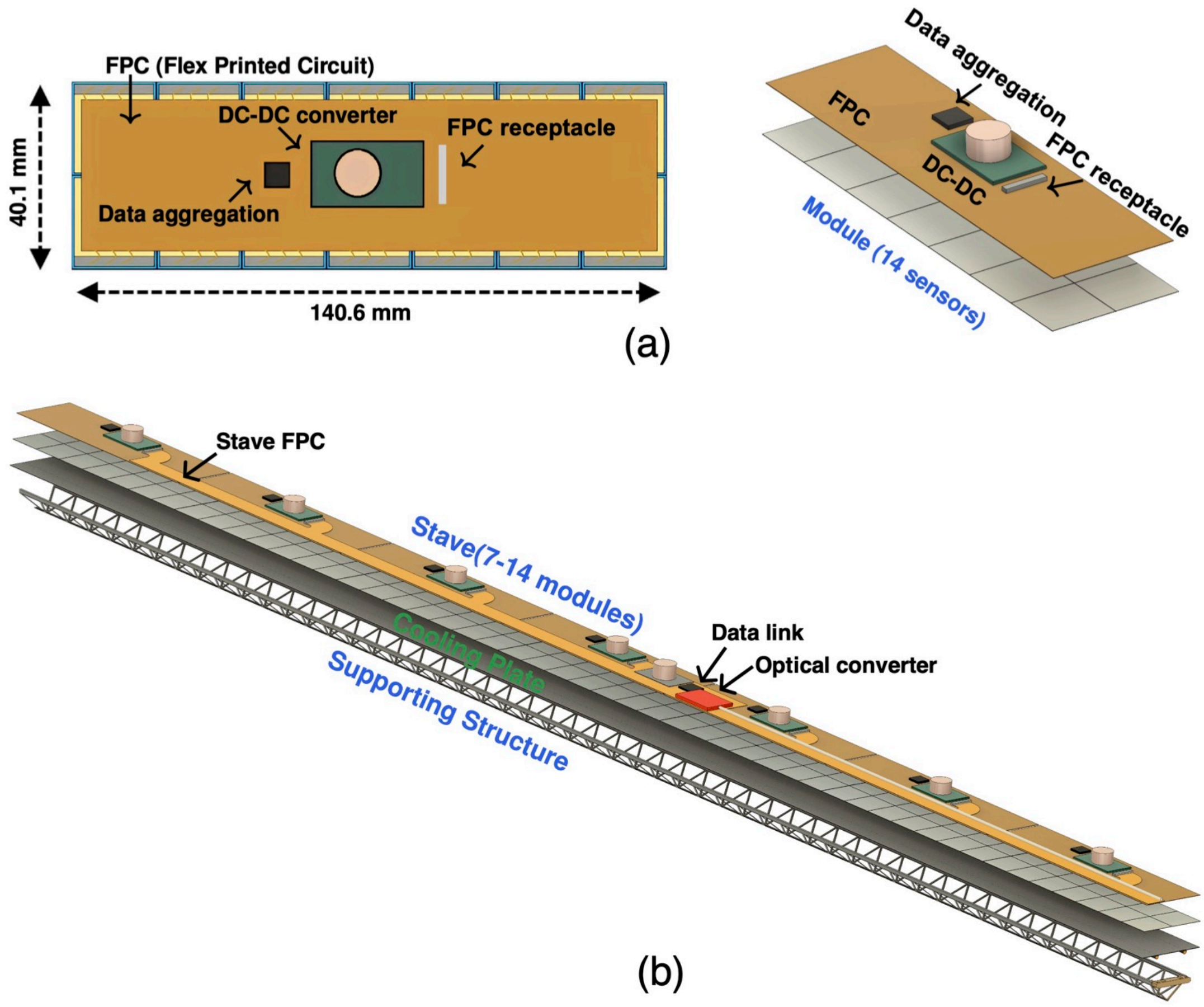

**Figure 5.2:** (a) ITK module and (b) ITK stave. The ITK module consists of 2 × 7 monolithic HV-CMOS pixel sensors, with their backside glued to a FPC board integrated with electronic components, including a DC-DC converter and a data aggregation chip. Each ITK stave is made up of 7, 10, or 14 modules, one or two long stave FPCs, a plate with embedded cooling tubes, and a truss supporting structure. A data link chip, an optical converter, and DC-DC converters are integrated at the center of each stave FPC.

Within each module, sensors are powered in parallel from a common low voltage input on the FPC, supplied by a DC-DC converter. A single downlink data line connects to the module, from which clock and command signals are extracted and distributed in parallel to each sensor. The uplink data streams from all sensors in a module are aggregated by a data aggregation chip and then transmitted off-module.

Every set of 7, 10, or 14 modules is assembled onto long supporting and cooling structures,

forming three types of staves with a width of 40.1 mm and lengths of 986.6 mm, 1,409.6 mm, and 1,973.2 mm, respectively, as illustrated by a stave in Figure 5.2 (b). These staves are used to construct the three ITK barrels (ITKB1, ITKB2, and ITKB3).

Modules for each stave are mounted on a carbon fiber supporting plate using a specialized assembly system with precision optical metrology and vacuum pick-up tools. The pick-up tools include adjustment mechanisms that allow fine-tuning of the module's position, achieving a placement accuracy better than 10 μm in the local coordinate system. A specific glue, characterized by its radiation resistance and high thermal conductivity, is used to establish robust mechanical and thermal contact between the modules' backside and the surface of the supporting structure.

After gluing all the modules onto a carbon fiber plate of a stave, one or two long FPCs (stave FPCs) are attached on top of the modules to transmit high voltage, low voltage, data, clock signals, and chip commands. For the innermost barrel layer (ITKB1), each stave uses a single FPC to connect all 7 modules, as shown in Figure 5.2 (b). For ITKB2 and ITKB3, each stave uses two stave FPCs, with each serving 5 (ITKB2) or 7 modules (ITKB3) from one end. Each stave FPC integrates a data link chip, an optical converter, and DC-DC converters, all located at its center.

The 150 V High Voltage (HV) and the original 48 V Low Voltage (LV) are both supplied from the end of the stave and transmitted along the stave FPC. The HV is delivered directly to individual modules for sensor biasing. The LV is stepped down to 12 V by the DC-DC converter at the center of the stave FPC, and then distributed to individual modules. Within each module, a local DC-DC converter further reduces the 12 V to 1.2 V to power the sensor circuits.

Uplink data from all sensors in a module are aggregated and sent through the stave FPC to the data link chip, which forwards the data to the optical converter. An optical fiber then transmits the data to the Back-End Electronics (BEE). The same optical fiber cable also carries downlink signals, enabling a bidirectional data link between the stave and the BEE. Details of the ITK electronics are further discussed in Section 5.2.2.

The overall stave, employing a truss structure and a plate with embedded cooling tubes, is designed for low mass (with a radiation length of 0.68% $X_0$) while providing high stiffness and effective heat dissipation, efficiently removing the heat generated by the modules. The details about the ITK mechanical and cooling design, as well as the performance, will be presented in Section 5.2.3.

The three ITK barrels have radii of 235.0 mm, 345.0 mm, and 555.6 mm. Since the inactive regions of the modules (see Figure 5.2 (a)) are concentrated along the two long edges of a stave, the staves in a barrel are mounted in a staggered structure. With this arrangement, each stave is aligned parallel to the $z$-axis and tilted by 55.85 mrad around it to allow sufficient overlap between neighboring staves, as illustrated in Figure 5.3 (a). To minimize dead space, there are 44, 64, and 102 staves required for the three individual barrels (see Figure 5.3 (b)). Table 5.4 summarizes the detailed information about the staves, modules, and sensors used for the construction of the 3 ITK barrels.

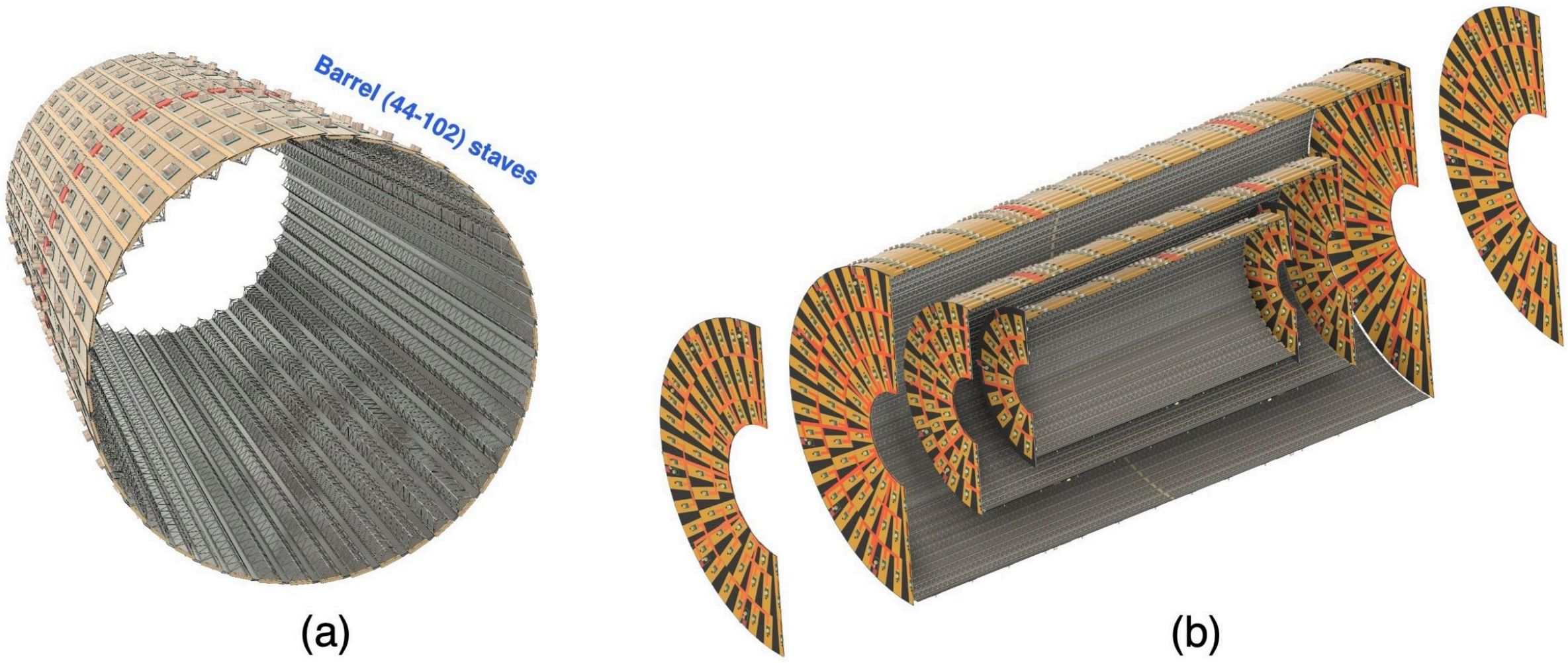

**Figure 5.3:** (a) ITK barrel and (b) ITK comprising three barrel layers and four endcap layers. Each ITK barrel contains 44, 64, or 102 staves arranged in a staggered structure to minimize dead area.

**Table 5.4:** Information about the staves, modules, and sensors used for the construction of the three ITK barrels.

| Barrel | Number of staves | Modules per stave | Sensors per module | Number of sensors | Sensor area [$m^2$] |
|---|---|---|---|---|---|
| ITKB1 | 44 | 7 | 14 | 4,312 | 1.72 |
| ITKB2 | 64 | 10 | 14 | 8,960 | 3.58 |
| ITKB3 | 102 | 14 | 14 | 19,992 | 8.00 |
| Total | 210 | | | 33,264 | 13.31 |

#### 5.2.1.2 ITK endcap design

The ITK endcap consists of four pairs of endcap disks — ITKE1, ITKE2, ITKE3, ITKE4 — with detailed layouts provided in Table 5.1. It uses the same monolithic HV-CMOS pixel sensors as the barrel and adopts similar module design concepts to streamline production. The endcap modules come in three variations containing 8, 12 or 14 sensors, with the 14-sensor module being identical to the one used in the barrel. The longer edge of modules are oriented along the radial ($r$) direction to optimize the position resolution in the azimuthal direction ($\phi$).

Modules are precisely positioned and glued to both sides of a carbon fiber honeycomb support disk for each endcap. Figure 5.4 (left panel) shows the module layout on the front side (facing the interaction point) and back side (facing away from the interaction point) of the fourth ITK endcap (ITKE4). As shown, there are three rings of modules on each side of the endcap. Each module in the outermost ring contains 12 sensors, those in the middle ring contain 14 sensors, and those in the innermost ring contain 8 sensors. To minimize dead detection area in the endcap, the modules on the back side are offset in the layout relative to those on the front side, creating overlapping coverage between the two surfaces. Figure 5.4 (right panel) provides a perspective view of the complete fourth ITK endcap, where the overlap regions are highlighted in dark green triangles.

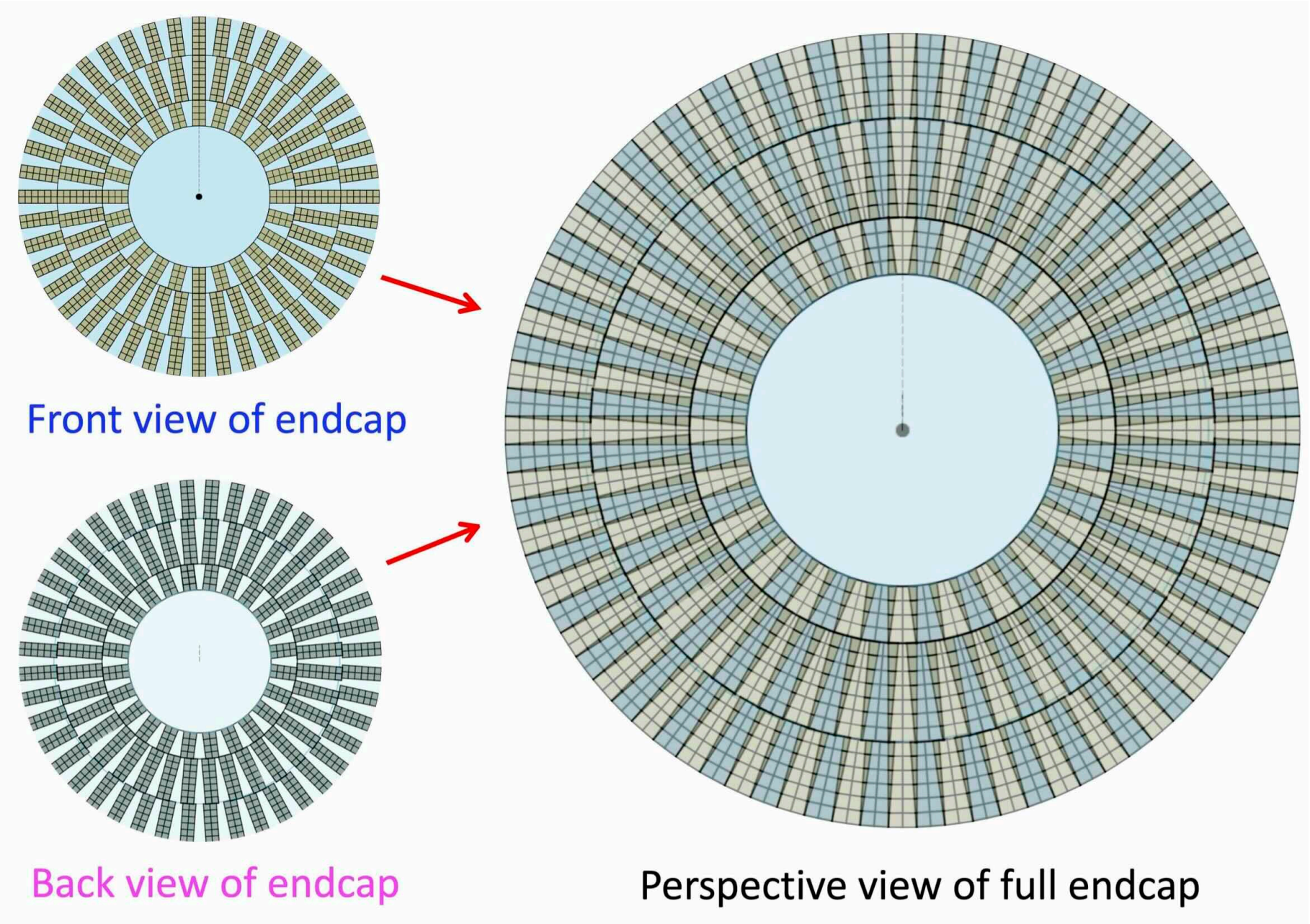

**Figure 5.4:** Perspective view of the sensor and module distribution for the fourth ITK endcap. The ITK endcap has double-sided detection surfaces, referred to as the "front view" (facing the interaction point) and the "back view" (facing away from the interaction point). The module layouts on the two sides are designed to complement each other to minimize dead detection areas. Overlapping detection regions between the two faces are highlighted as dark green triangles.

The layouts of the modules and their electronic components for the four ITK endcaps (ITKE1, ITKE2, ITKE3, and ITKE4) are displayed in Figure 5.5, with key details summarized in Table 5.5. In total, the four-disk pairs comprise 13,760 sensors and 1,236 modules used in the construction of the ITK endcaps.

**Table 5.5:** Information about the module and sensor specifications for the four pairs of ITK endcap disks

| Endcap | Number of rings per side | Number of modules per ring | Number of sensors per module | Total sensors |
|---|---|---|---|---|
| ITKE1 | 2 | 13,20 | 8,8 | 1,056 |
| ITKE2 | 3 | 16,24,28 | 8,8,8 | 2,176 |
| ITKE3 | 3 | 24,36,44 | 12,14,14 | 5,632 |
| ITKE4 | 3 | 24,36,44 | 8,14,12 | 4,896 |
| Total | | | | 13,760 |

After the modules are glued onto the carbon fiber disks, secondary data aggregation Printed Circuit Boards (PCBs) are mounted around the disk periphery, as shown in Figure 5.5 (e.g., four secondary boards on the front side of ITKE1). Each secondary board integrates a data link chip, an optical converter, DC-DC converters, and optionally a data aggregation chip. The modules connect to these secondary boards via sector FPC connectors (shown in orange-red in Figure 5.5), which transmit power, data, clock signals, and control commands between the modules and the secondary aggregation boards.

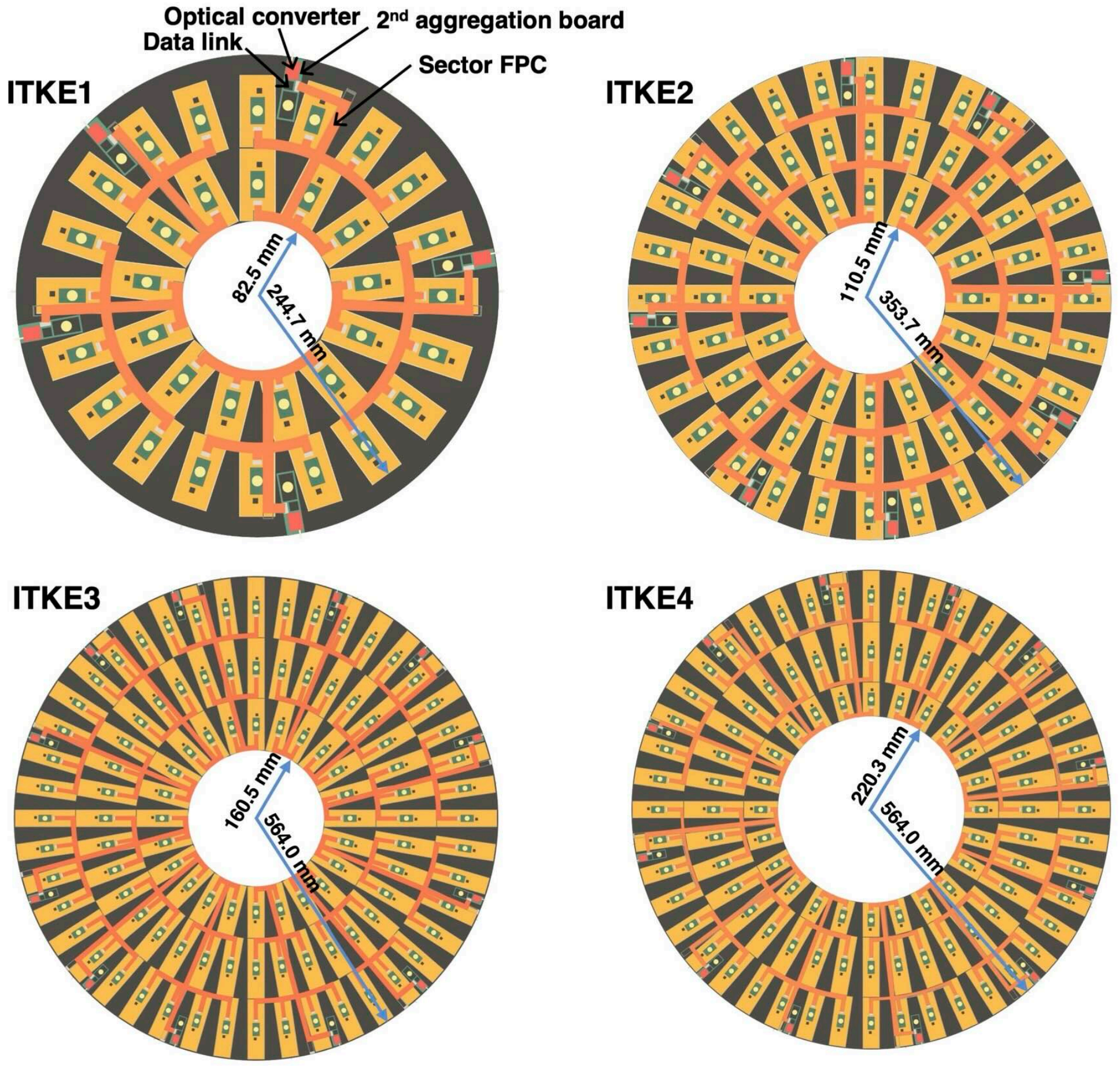

**Figure 5.5:** Module (yellow) layout of the four ITK endcaps: ITKE1, ITKE2, ITKE3, and ITKE4. Details are provided in Table 5.5. Secondary data aggregation boards — each integrating a data link chip, an optical converter, DC-DC converters, and optionally a data aggregation chip — are attached near the outermost rim of the endcap, with sector FPCs (shown in orange-red) connecting them to the modules.

The original 150 V HV supply, together with the 12 V LV output stepped down from the 48 V input by the DC-DC converter on the secondary board, provides power to the modules from the outermost rim of the ITK endcap. Readout data from each module is transferred through the sector FPC to the corresponding secondary board, where it is aggregated, converted into an optical signal, and sent via optical fiber to the BEE.

### 5.2.2 Readout electronics

In each ITK barrel module, 14 monolithic HV-CMOS pixel sensors are bonded to a common FPC, as shown in Figure 5.2. The module FPC transmits digital signals from the sensors to a data aggregation chip, and then via a long stave FPC to a data link chip, which subsequently sends the data out through the optical converter. To enable flexible data transmission, two ASICs are used for data aggregation: the data aggregation chip, called the Front-End Data Aggregator (FEDA), and the data link/interface chip, called the Front-End Data Interface (FEDI).

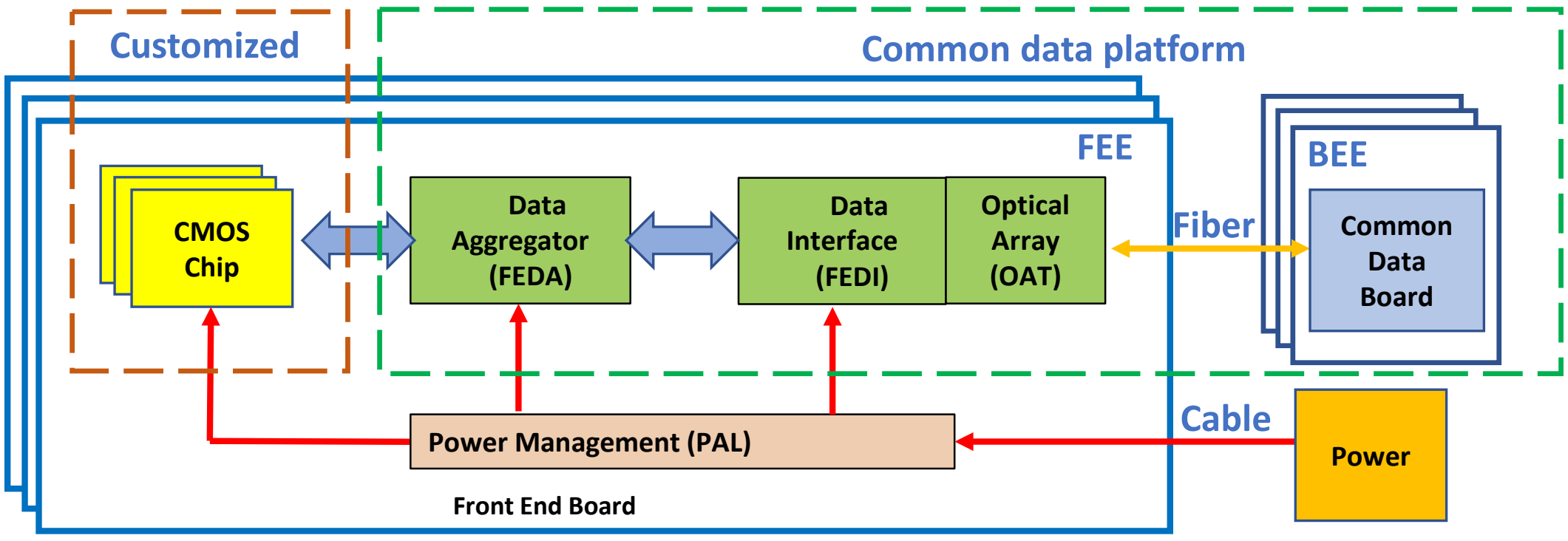

**Figure 5.6:** ITK readout and power supply scheme. The Front-End Data Aggregator (FEDA) chip collects data from multiple sensors and sends it to the Front-End Data Interface (FEDI) chip, which forwards it via the Optical Array Tranceiver (OAT) module to the Back-End Electronics (BEE). Power is supplied to both the sensors and readout chips through DC-DC converters (PAL).

As illustrated in Figure 5.6, the FEDA chip collects data from multiple sensors and transfers it serially. The FEDI chip [1] then gathers data from several FEDA chips and routes it to the Optical Array Tranceiver (OAT) module for optical readout. The hit rate varies across different radial regions as shown in Table 5.2 for different beam modes. To manage this variation, each aggregation ASIC includes a dedicated buffer. This buffer helps average the rate fluctuations and optimizes the speed of the e-link transceiver inputs. The aggregated data is transmitted via optical fiber from the detectors to the Back-End Electronics (BEE) and Trigger and Data Acquisition (TDAQ) system.

As shown in Figure 5.7, data from each module (up to 14 sensors) is first aggregated by a FEDA chip and then routed to the FEDI chip. The FEDI chip is designed to support a maximum valid data rate of 9.71 Gbps. It collects and aggregates data from seven uplinks, each operating at 1.39 Gbps (e-link), serializes it, and encodes it for high-speed transmission at 11.09 Gbps through an optical fiber.

Considering an average of $\sim$ 1.5 pixels firing per hit, with each fired pixel generating 42 bits of data, the designed per-sensor data bandwidth of 86.67 Mbps (applied to all barrel sensors and the inner radial endcap sensors) corresponds to a maximum hit rate of $4.1 \times 10^5$ Hz/cm$^2$ for a sensor with an active area of $1.74 \times 1.92$ cm$^2$. This is about 21 times higher in the barrel region and 3 times higher in the endcap region than the estimated peak background hit rates (after applying a safety factor of 2), under the most challenging high-luminosity $Z$-pole operation mode, as shown in Table 5.2.

For the synchronization of different detectors, a reference clock signal is distributed through optical fiber from the BEE to the FEDI chip. Clock recovery is managed by the FEDI's clock and data recovery circuit. The FEDI chip offers six clock output options: 43.33 MHz, 86.67 MHz, 173.33 MHz, 346.67 MHz, 693.33 MHz, and 1.39 GHz, which are used by the front-end electronics of the various detectors.

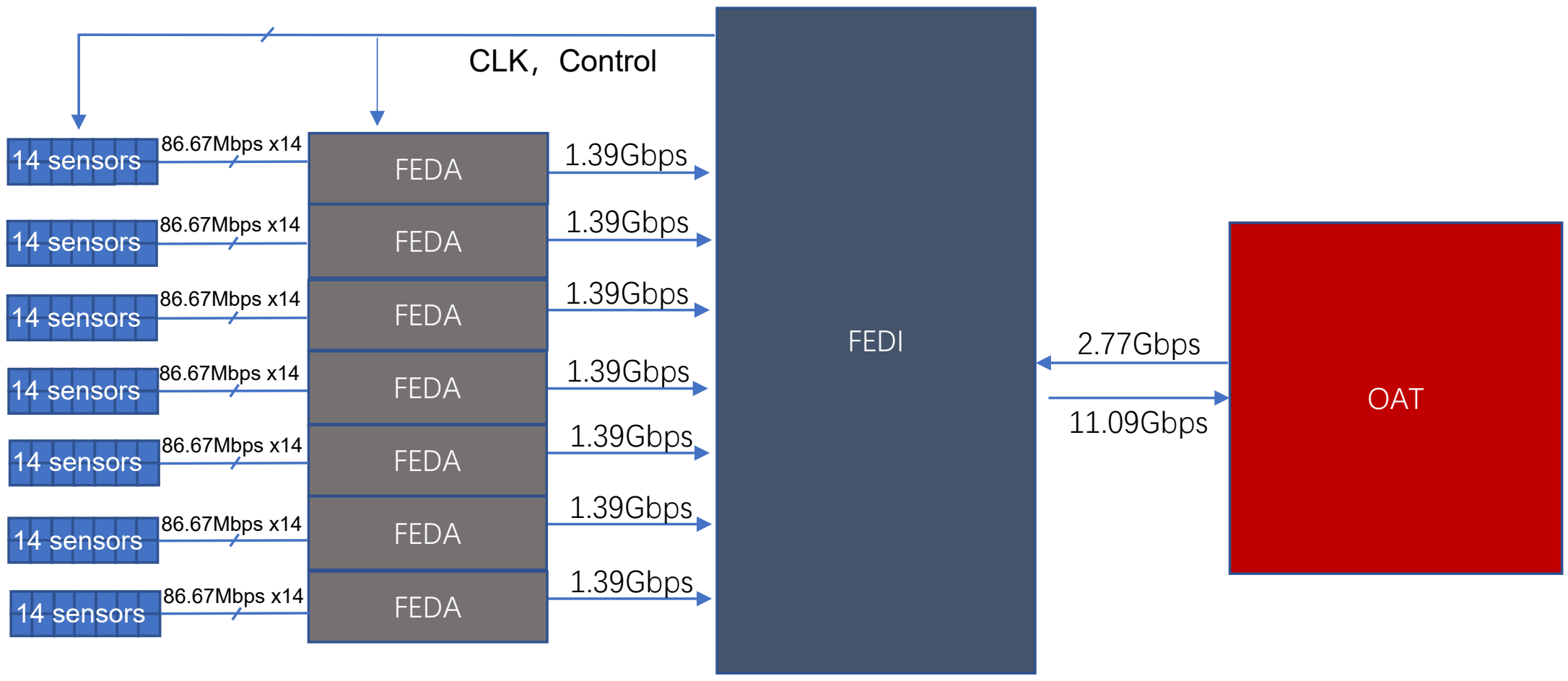

**Figure 5.7:** ITK readout electronics data flow. Data from 14 sensors are aggregated by a FEDA chip and transmitted via an e-link at 1.39 Gbps to the FEDI chip. Each FEDI chip collects and aggregates data from seven FEDA chips, serializes and encodes it, and forwards it to the OAT at an uplink rate of 11.09 Gbps. The OAT also provides a downlink of 2.77 Gbps for slow control, monitoring, and clock signals, which are distributed to the FEDI chip and then to the FEDA chips and sensors.

The FEDI chip also provides a range of slow control and monitoring features, including I2C master controllers, bidirectional Input/Output (I/O) ports, and a memory-like bus master controller for data and address management. The preliminary slow control implementation encompasses power control bits, configuration settings for the sensors, and monitoring functions such as DC-DC status and temperature measurements.

Considering a total power consumption of 200 mW/cm$^2$ and a sensor area of 2 cm $\times$ 2 cm, each module with 14 sensors requires $\sim$ 9 A at 1.2 V. A DC-DC converter (PAL12) is mounted on the module FPC to meet this current demand. The 48 V LV supplied from the power crate is first stepped down to 12 V using a DC-DC converter (PAL48), and further reduced to 1.2 V by the PAL12 converter, as illustrated in Figure 5.8.

As shown in Figure 5.2 (b) for a barrel stave and Figure 5.5 for endcap disks, the PAL48 converter is located at the center of the stave or the outer rim of the endcap. It steps down the 48 V LV input to 12 V, which is subsequently distributed to the individual modules via the long FPC. In each module, a dedicated PAL12 converter further reduce the 12 V LV to 1.2 V to supply power to the sensor circuits.

FPCs are used to transmit both low voltage and high voltage from the end of the barrel or endcap to the modules. These FPCs, together with optical fibers, are connected to composite cables from outside to deliver power and signals. In the ITK barrel, every three neighboring stave FPCs and their corresponding three fibers on the same side of the stave are grouped into one composite cable. Similarly, in the ITK endcap, every two neighboring sector FPCs and two fibers are connected to one composite cable. In total, the three ITK barrel layers use 128 composite cables, while the four pairs of ITK endcaps use 74.

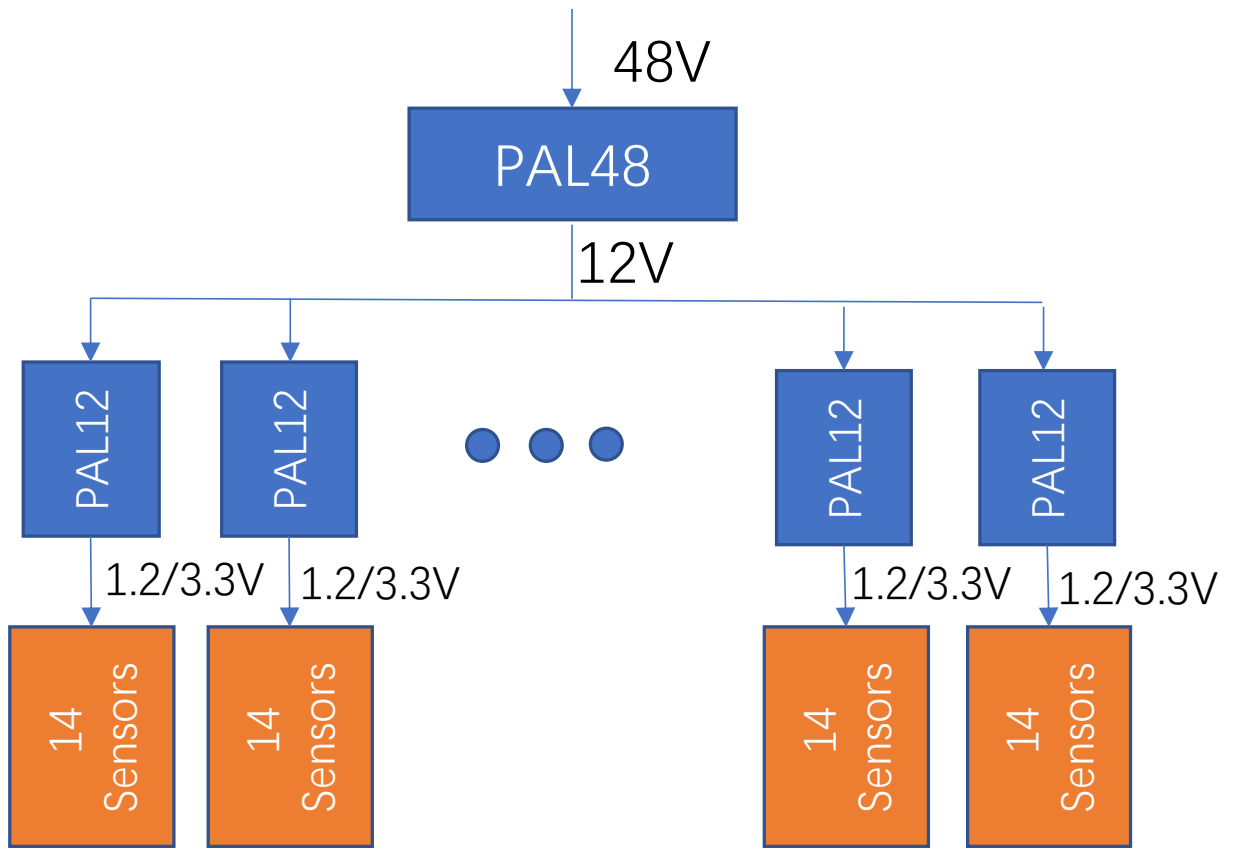

**Figure 5.8:** ITK readout electronics power distribution. The DC-DC converters operate in two stages: first, the PAL48 DC-DC converter reduces 48 V to 12 V; then, the PAL12 converter steps it down to 3.3 V for the Vertical-Cavity Surface-Emitting Laser (VCSEL) driver in the OAT and to 1.2 V for powering the sensor circuits and other readout electronics.

## 5.2.3 Mechanical and cooling design

The ITK barrel layers are azimuthally segmented into mechanically independent units called staves, which are mounted onto two end-wheel structures to form three barrel layers, as shown in Figure 5.1. Neighboring staves are partially overlapped to ensure the detector hermeticity, as illustrated in Figure 5.3. Four pairs of assembled ITK endcap disks are also mounted on the same end-wheel structures. Together with two extended connection rings, they form the complete ITK assembly. This entire ITK structure is inserted into the inner barrel of the TPC, with the extended connection rings fixed to the TPC end-wheels. The staves and endcap disks serve as the local support structures of the ITK. They are specifically designed to meet stringent requirements, including a low material budget, high structural rigidity, and efficient cooling.

### 5.2.3.1 Barrel local support

The ITK stave, a long structure spanning the full length of the ITK barrel, acts as the fundamental building block, integrating both structural and electrical components, as illustrated in Figures 5.2 (b) and 5.3. The design of the ITK stave is similar to that of the ITS2 used in ALICE [2]. The stave structure consists of three main units, as shown in Figure 5.9:

- Sensor Modules: These consist of sensors glued onto FPCs integrated with associated electronics.
- Cooling Plate: A carbon fiber plate embedded with cooling pipes that are in direct thermal contact with the sensors to efficiently dissipate heat. The cooling pipes are glued to the carbon plate, and the thermal contact is enhanced by a layer of graphite foil. To increase the structural rigidity, an additional thin layer of carbon fleece is added to both the top and bottom sides of the cooling plate.

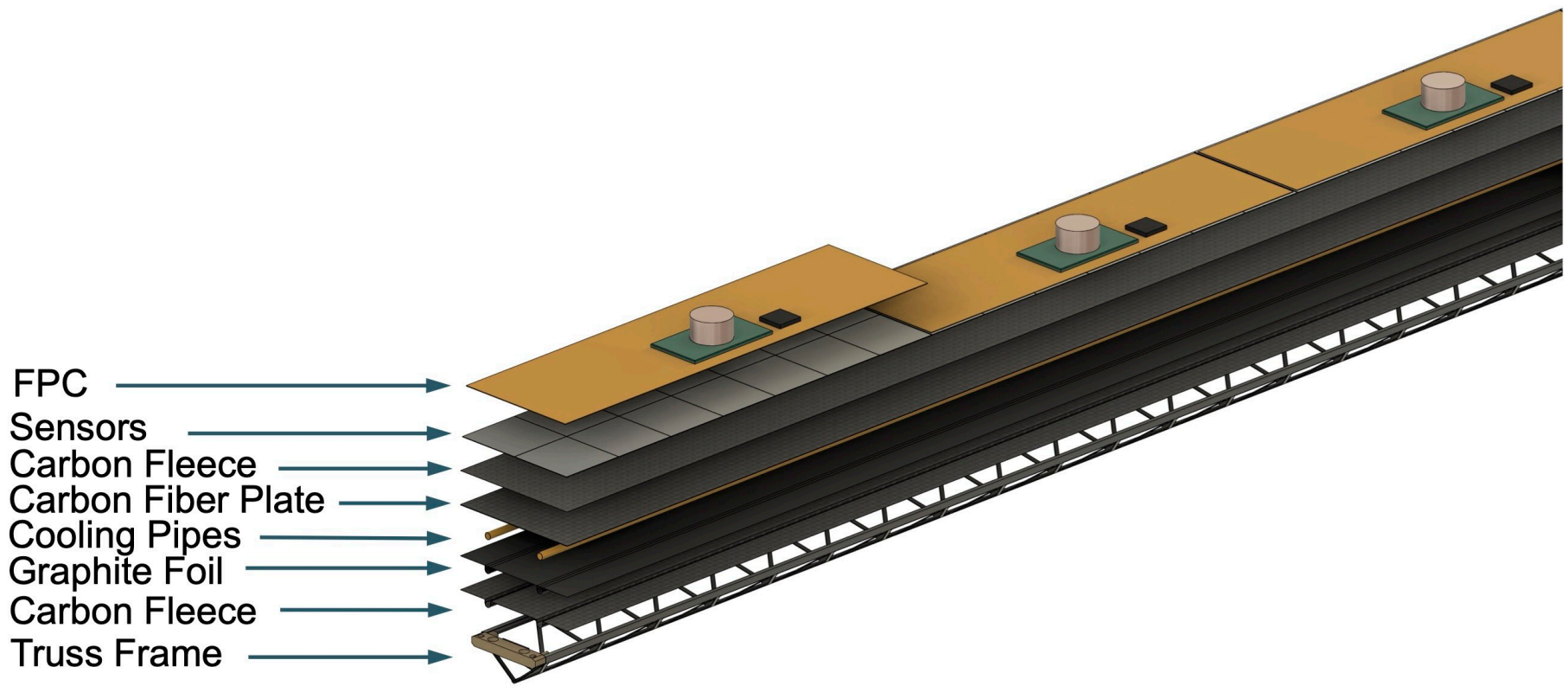

**Figure 5.9:** Structure of the ITK barrel stave. The stave consists of FPCs integrated with associated electronics and sensors, followed by a top carbon fleece layer, a carbon fiber plate, two cooling pipes, a graphite foil layer, a bottom carbon fleece layer, and a truss frame.

- Truss Frame: A carbon fiber support structure that provides mechanical support and enhances the stiffness of the stave.

**Materials** The ITK stave design must meet strict requirements to minimize the material budget. The Cooling Plate is stiffened by the Truss Frame, which has a triangular cross section made from carbon fiber with a high modulus, such as K13C2U [3], with a tensile modulus of up to 900 GPa. Other options, like M60J [4] with a tensile modulus of 588 GPa and M55J [5] with a tensile modulus of 540 GPa, may also be considered in the future. For the Cooling Plate, the carbon fiber used for the plate has high thermal conductivity, such as K13D2U, with a thermal conductivity of up to 800 W/(m·K). Additionally, the thermal performance is enhanced by a layer of graphite foil, which has high thermal conductivity in the horizontal direction, reaching over 1500 W/(m·K). This helps to maintain a more uniform temperature field across the plane.

Water with polyimide cooling pipes has been selected as the baseline cooling medium for the ITK stave. Polyimide possesses excellent properties such as high temperature resistance, corrosion resistance, radiation resistance, and high strength. Additionally, it has a low coefficient of friction and good sliding performance, which reduces frictional losses when transporting fluids or gases, thereby effectively improving the efficiency of the water-cooling system.

Table 5.6 lists the estimated contributions of the ITK stave to the material budget. The overall estimated material budget for a stave is 0.68% $X_0$. The overlap area between neighboring staves accounts for 8% of the stave area, corresponding to 6% of the stave materials.

**Structural characterization** The structural characterization is performed using finite element simulations. The two key structural parameters that define the achievable accuracy and stability in the position of the sensors are the stave sag under its own weight and the first natural frequency of the stave. The sag provides information on the deviation of the sensors' final positions relative to the nominal positions, while the first natural frequency indicates the frequency at which an external impulse can induce resonance phenomena in the structure, resulting in oscillations of the sensor positions.

**Table 5.6:** Estimation of the ITK stave material contributions. The wall thickness of the cooling tubes, averaged over the entire stave area and labeled with "†", is ∼ 16 μm.

| **Functional unit** | **Component** | **Material** | **Thickness [μm]** | **$X_0$ [cm]** | **Radiation Length [% $X_0$]** |
|---|---|---|---|---|---|
| Sensor Module | FPC metal layers | Aluminium | 100 | 8.896 | 0.112 |
| | FPC Insulating layers | Polyimide | 150 | 28.41 | 0.053 |
| | Sensor | Silicon | 150 | 9.369 | 0.160 |
| | Glue | | 100 | 44.37 | 0.023 |
| | Other electronics | | | | 0.050 |
| Cooling Plate | Carbon fleece layers | Carbon fleece | 40 | 106.80 | 0.004 |
| | Carbon fiber plate | Carbon fiber | 150 | 26.08 | 0.057 |
| | Cooling tube wall | Polyimide | 64† | 28.41 | 0.006 |
| | Cooling fluid | Water | | 35.76 | 0.028 |
| | Graphite foil | Graphite | 30 | 26.56 | 0.011 |
| | Glue | Cyanate ester resin | 100 | 44.37 | 0.023 |
| Truss Frame | Carbon fiber roving | | | | 0.080 |
| Power Bus FPC | | | | | 0.070 |
| Total | | | | | 0.677 |

Both the sag and the natural frequency for a given stave mass depend on the stave stiffness, which is mainly provided by the Truss Frame. As a conservative assumption, the contribution of the FPC and cooling pipes to the overall stiffness is not included in this analysis.

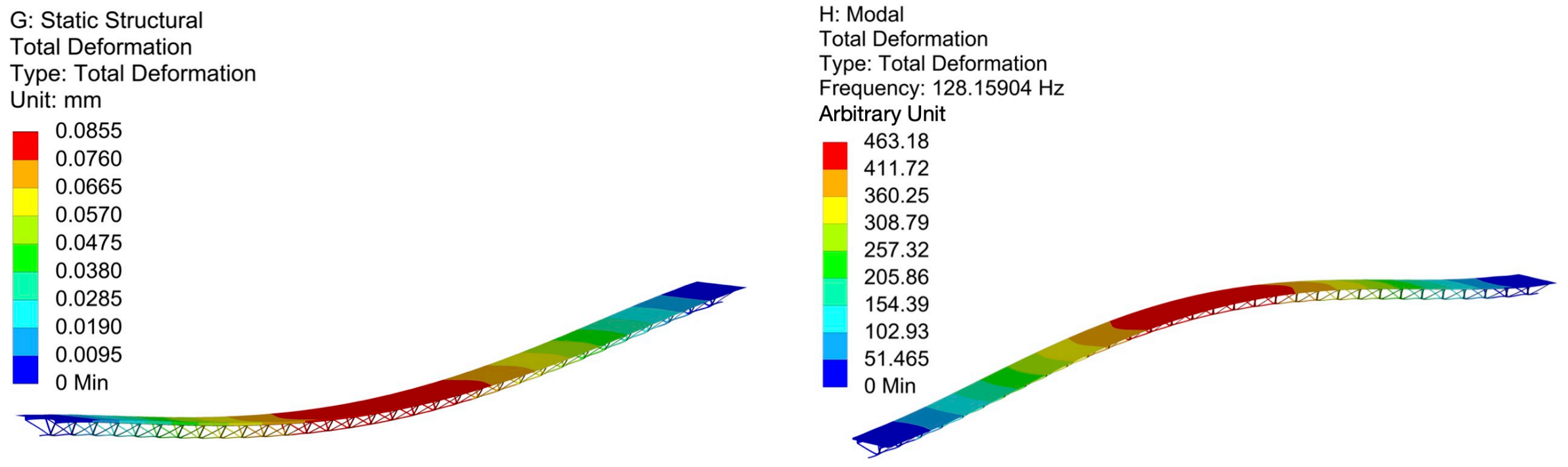

(a) Sag analysis (b) Modal analysis

**Figure 5.10:** Results of (a) gravitational sag analysis and (b) modal analysis for the stave of the first ITK barrel (ITKB1). Both ends of the stave were assumed to be fixed supports. The blue color indicates zero movement, while the red lines represent maximum movement.

Static and modal analyses were performed to evaluate the maximum deflections and the first natural frequencies, as shown in Figure 5.10 for the stave in the first ITK barrel (ITKB1). The results indicate sag values of 85 μm, 289 μm, and 896 μm for the staves in the three ITK barrels (ITKB1, ITKB2, and ITKB3), with corresponding first natural frequencies of 126 Hz, 69 Hz, and 34 Hz.

**Thermal characterization** Heat transfer simulation analysis is conducted to optimize the cooling design. The sensor specifications for ITK requires:

- The overall sensor operating temperature should not exceed 30 °C.
- The temperature uniformity across a single sensor should be maintained within 5 °C.

A water cooling fluid structure coupled finite element model was developed to study the temperature distribution along longitudinal length of the stave with the following configuration:

- The heat generated by sensors is uniformly distributed on the carbon fiber plate, with a magnitude of 200 mW/cm$^2$ (sensor heat flux).
- The cooling water enters the stave with a flow velocity of 2 m/s and a temperature of 5 °C.
- The two cooling polyimide pipes for the ITK stave have an inner diameter of 1.6 mm.
- Natural convection and radiative heat transfer are not considered.

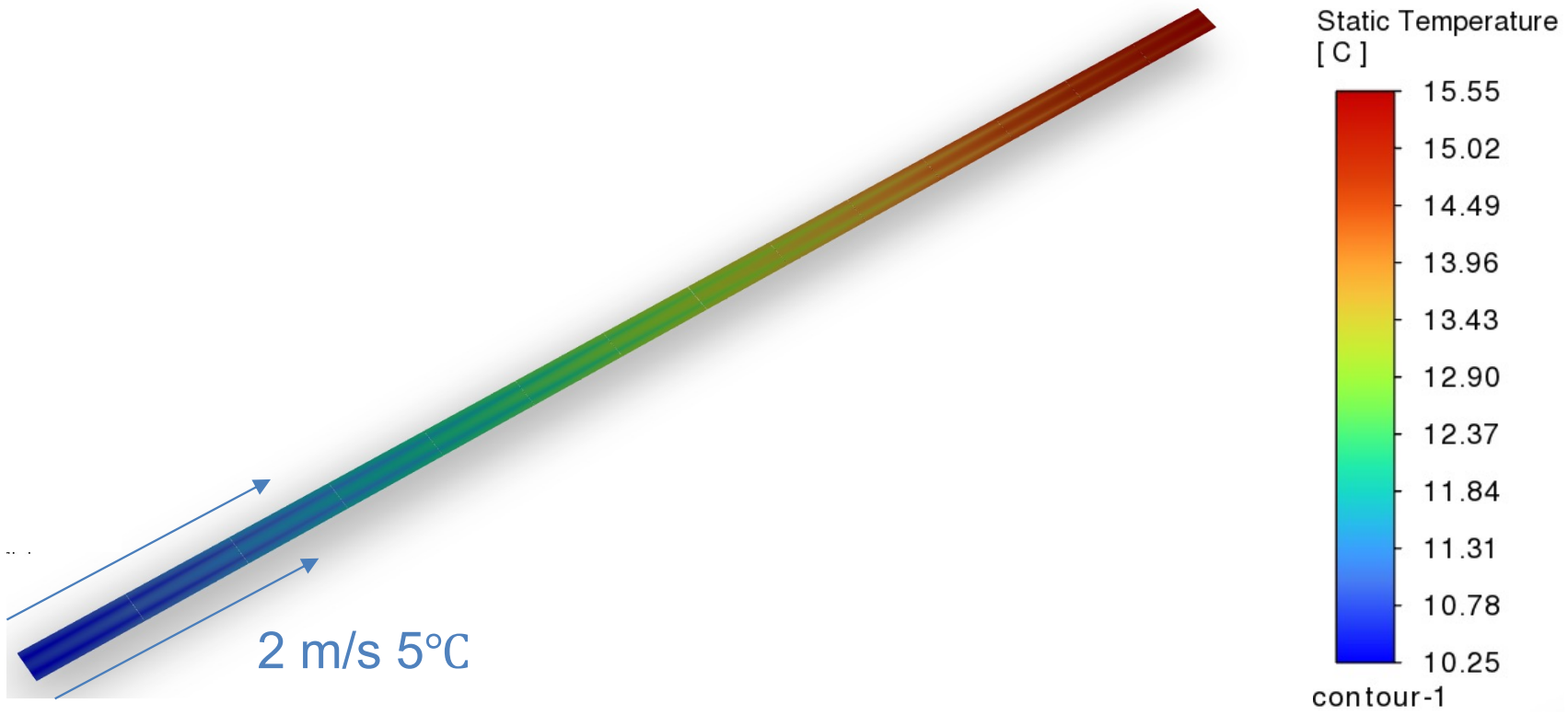

**Figure 5.11:** Simulation result of water cooling for the stave of the third ITK barrel (ITKB3), with a flow velocity of 2 m/s and a two-pipe inlet temperature of 5 °C. The temperature across the stave remains below 15.6 °C, and the temperature gradient along the stave is within 5 °C.

The simulation result, shown in the Figure 5.11, demonstrates that with both pipe inlets placed on the same side, the temperature gradient along the longest stave — ITKB3, with a length of 1,973 mm — can be controlled within 5 °C. The maximum temperature across the stave remains below 15.6 °C. With this configuration, the water cooling strategy meets the thermal requirements of the detector.

#### 5.2.3.2 Endcap local support

Figure 5.12 illustrates the structure of a single endcap layer (ITKE4). It consists of front and back layers of sensor modules — each glued to a carbon fiber plate (facesheet) — and a central opened cooling plate layer with mechanical and cooling structures, all sandwiched together. The sensor modules are attached to the front and back carbon fiber facesheets using thermal adhesive.

The main mechanical skeleton (frame) of the opened cooling plate consists of an inner ring and an outer ring, connected by eight radial ribs. In addition, carbon fiber honeycomb cores are embedded within the frame to enhance overall bending stiffness and maintain the flatness of the two facesheets. Serpentine grooves are machined into the honeycomb cores to accommodate cooling pipes. The outer ring interfaces with the inlets and outlets of the cooling loops. The gaps between the carbon fiber honeycomb and the cooling loops are sealed with low-mass, high thermal conductivity foam. Finally, the two carbon fiber facesheets (with the sensor modules) and the opened cooling plate are bonded using an adhesive infused with thermally conductive carbon particulates.

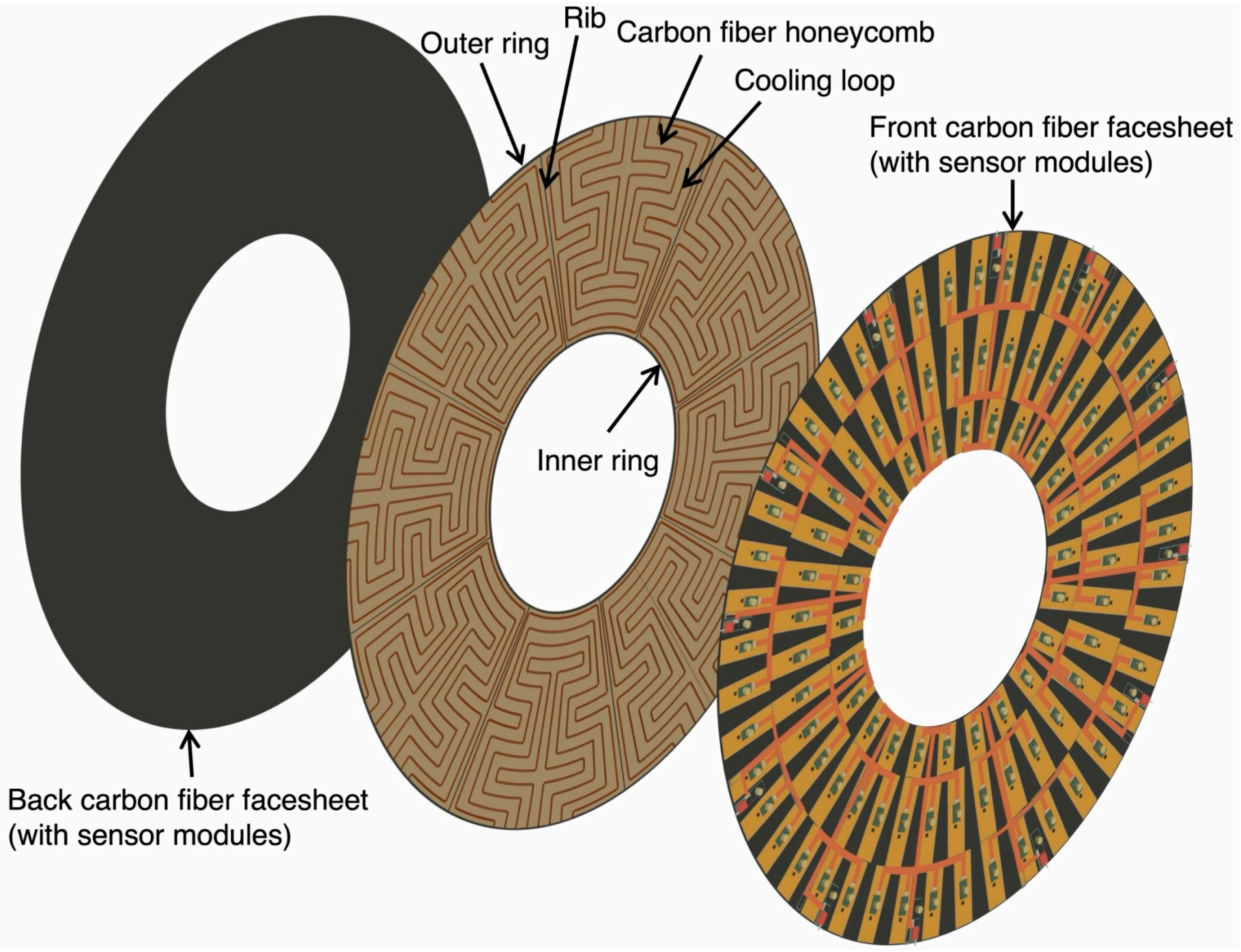

**Figure 5.12:** Structure of the ITK endcap (ITKE4). It consists of two facesheets with glued sensor modules, and a central supporting and cooling structure composed of an inner ring, an outer ring, eight connecting radial ribs, carbon fiber honeycomb, and integrated cooling loops.

**Materials** The two facesheets are constructed from layers of high-modulus unidirectional carbon fiber material combined with cyanate ester resins. The interface between the cooling tubes and the facesheets is formed using thermally conductive carbon foam, such as Allcomp K9. Water circulating through cooling loops of the cooling plate serves as cooling medium for the ITK endcap.

The frame, comprising the inner ring, outer ring, and eight radial ribs, is made from stiff materials. The primary candidate material for these components is carbon fiber. For the outer ring, high performance Polyether Ether Ketone (PEEK) polymer may also be considered as an alternative material.

Table 5.7 lists the estimated contributions of the ITK endcap to the material budget. The overall estimated material budget is 0.76% $X_0$.

**Structural characterization** The stiffness of the ITK endcap is primarily provided by the carbon fiber facesheets, honeycomb, eight radial ribs, inner ring, and outer ring. Finite element simulations, similar to those performed for the barrel, were also conducted for the endcap structural characterization. Analyses were carried out to evaluate the deformations for the ITK endcaps in both the nominal and flat-lying positions.

**Table 5.7:** Estimation of the ITK endcap material contributions. The wall thickness of the cooling tubes, averaged over the entire endcap area and labeled with "†", is ~ 24 μm.

| Functional unit | Component | Material | Thickness [μm] | $X_0$ [cm] | Radiation Length [% $X_0$] |
|---|---|---|---|---|---|
| Sensor Module | FPC metal layers | Aluminium | 100 | 8.896 | 0.112 |
| | FPC Insulating layers | Polyimide | 150 | 28.41 | 0.053 |
| | Sensor | Silicon | 150 | 9.369 | 0.160 |
| | Glue | | 100 | 44.37 | 0.023 |
| | Other electronics | | | | 0.050 |
| Structure | Carbon fiber facesheet | Carbon fiber | 150 | 26.08 | 0.057 |
| | Cooling tube wall | Titanium | 100† | 3.560 | 0.067 |
| | Cooling fluid | Water | | 35.76 | 0.027 |
| | Graphite foam+Honeycomb | Allcomp+Carbon fiber | 2000 | 186 | 0.108 |
| | Carbon fiber facesheet | Carbon fiber | 150 | 26.08 | 0.057 |
| | Glue | Cyanate ester resin | 200 | 44.37 | 0.045 |
| Total | | | | | 0.759 |

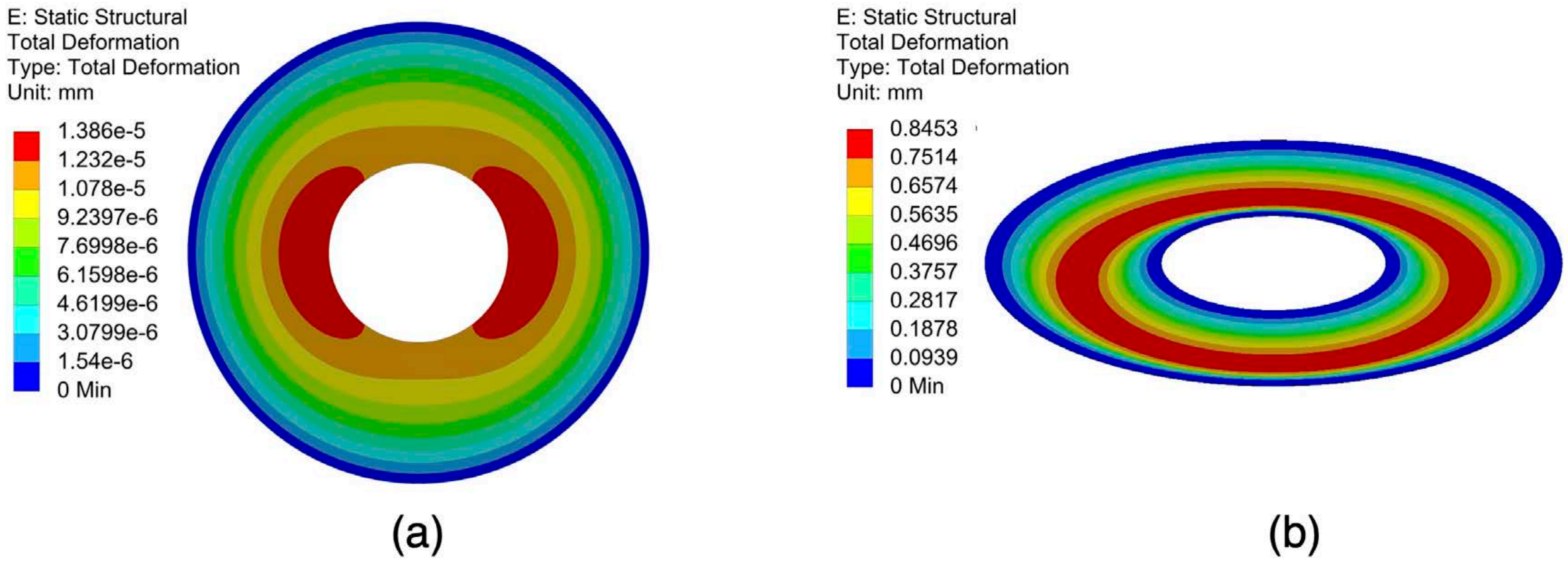

**Figure 5.13:** Deformations of the fourth ITK endcap in (a) the nominal position, with the outer ring of the endcap assumed to be a fixed support, and (b) the flat-lying position, with both the inner and outer rings of the endcap assumed to be fixed supports. The maximum sag is 0.8 mm in the flat-lying position (b), and negligible (< 1 μm) in the nominal position (a).

As an example, Figure 5.13 shows the deformation for the fourth endcap (ITKE4). A maximum sag of 0.8 mm is found for the ITKE4 in the flat-lying position. In the nominal position, it becomes negligible (< 1 μm). Overall, the stiffness of the ITK endcaps is adequate to ensure they can operate under various conditions.

**Thermal characterization** The operating temperature of the sensors in the ITK endcaps should meet the same requirements as those in the barrels, as previously discussed. Considering the compressive mechanical strength required for the detector on both faces and its ductility, titanium has been chosen as the material for the cooling pipes. Titanium offers high corrosion resistance, low weight, excellent radiation hardness, and high pressure resistance.

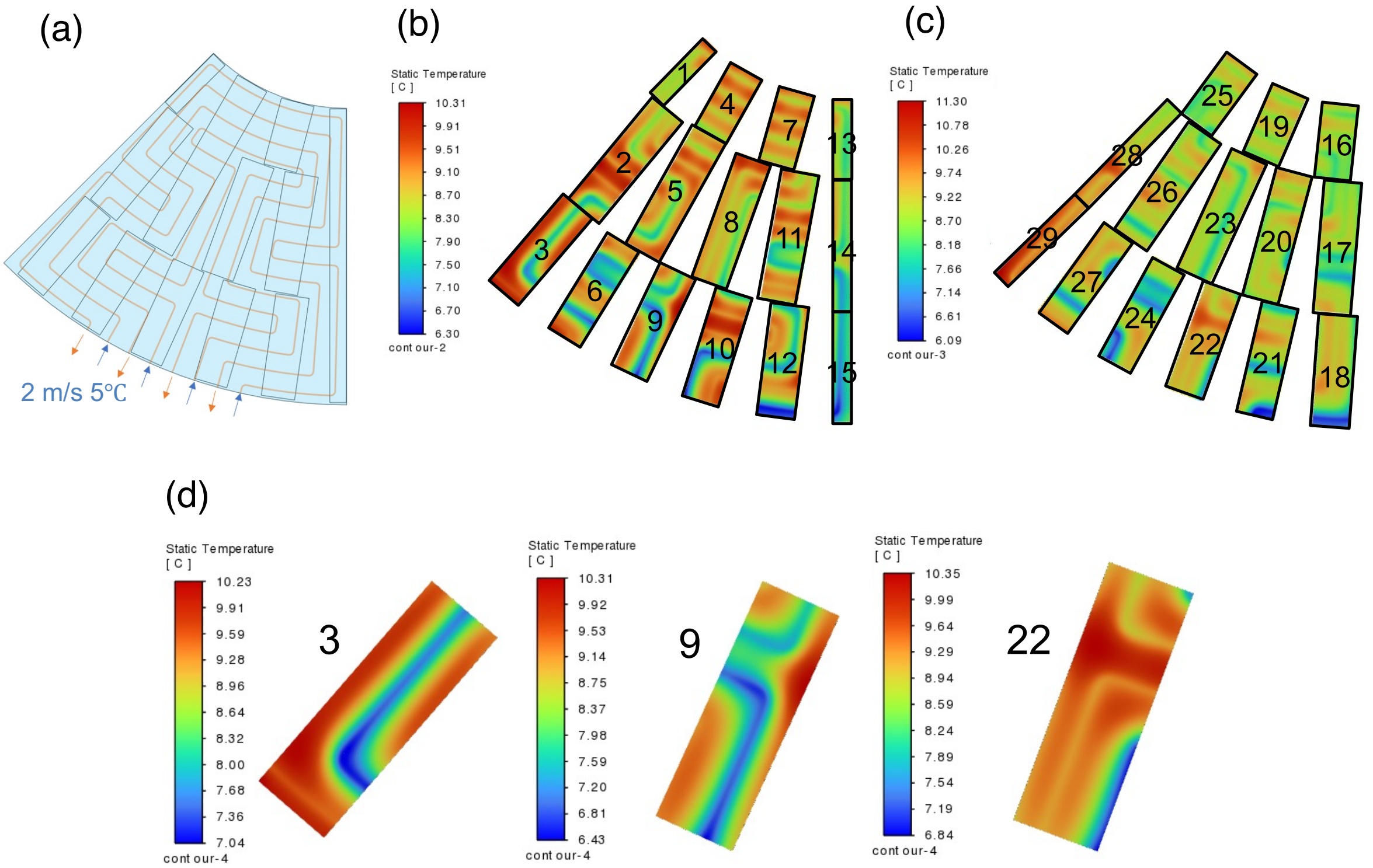

**Figure 5.14:** (a) Designed cooling loops for one eighth of the ITK endcap, featuring four closed loops arranged radially across 16 coil layers. Sensor temperature distributions for one eighth of the ITK endcap: (b) front-side sensors from 15 modules (numbered 1 to 15), (c) back-side sensors from 14 modules (numbered 16 to 29), and (d) selected temperature profiles from three modules (No. 3, 9, and 22).

The cooling loops for the ITK endcaps are custom-designed to achieve a small temperature gradient across the entire endcap disks. One eighth of the endcap is defined as a cooling unit. Figure 5.14 (a) shows the designed layout of the cooling loops for one eighth of the ITK endcap. This design employs four closed loops arranged radially across 16 coil layers. To ensure uniform temperature distribution across the endcap, the cooling loop design optimizations include radial layering, sector partitioning, and alternating inlet/outlet configurations. Titanium cooling pipes, adopted for the ITK endcaps, have an inner diameter of 1.6 mm and a wall thickness of 0.1 mm. The cooling water enters the ITK endcaps at 2 m/s and 5 °C, the same as in the barrel.

A finite element model was established to study the temperature distribution across the endcaps. Thermal simulations incorporate the thermal conductivity of cooling water, cooling pipes, graphite foam, carbon fiber facesheets, and detector sensors.

Figures 5.14 (b) and (c) show the simulation results for the temperature distributions across both faces of the one eighth of the fourth ITK endcap (ITKE4). The temperature of all sensors on both faces were found to be below 11.3 °C (well below the operation limit of 30 °C). The maximum temperature difference across the plane is below 5 °C, across one ITK module is below 4 °C (see Figure 5.14 (d)), and across one ITK sensor is below 2.5 °C (satisfying the requirement of $< 5$ °C). In all aspects, the effective cooling loop design with water cooling meets the detector's requirements in both thermal uniformity and operational functionality.

In addition to the water cooling, two-phase $CO_2$ flow cooling is also studied as an alternative solution, where the $CO_2$ undergoes a phase change from liquid to gas. This phase transition allows more heat to be carried away, improving temperature uniformity. Although water cooling has been selected as the baseline for the current design due to its overall system simplicity, ongoing R&D efforts are progressing in parallel for both cooling approaches.

### 5.2.4 HV-CMOS pixel sensor

Over the past two decades, monolithic CMOS sensor technologies have rapidly evolved for precise tracking by integrating sensing and readout electronics onto a single substrate. The introduction of deep n-well structures as charge collection electrodes in High Voltage Complementary Metal-Oxide-Semiconductor (HV-CMOS) sensors marked a significant advancement, enabling larger depletion volumes for efficient and fast charge collection as well as enhanced radiation hardness.

Most HV-CMOS sensors in high energy physics utilize 180 nm or 150 nm technology processes, such as those deployed in Mu3e [6] and planned for the LHCb upgrades [7]. However, HV-CMOS technology with a 55 nm feature size has been chosen as the baseline for the ITK for two key reasons. First, a smaller feature size offers the potential for better performance, including lower power consumption, while allowing more functionalities to be integrated within the same area, particularly benefiting digital circuits. Equally important is the reliability of the process access. Since the most advanced processes in industry often lead high energy physics applications by one or two decades, the processes used during the R&D phase may not be available during mass production. Therefore, starting with more advanced processes is more secure. This choice also aligns with the general trend in MAPS development using the CMOS Image Sensor (CIS) process, which is moving from 180 nm to 65 nm, such as the TaichuPix sensor developed by the CEPC team for the VTX detector, and front-end electronics are advancing even further to 28 nm processes.

For ITK R&D, a series of small-scale sensors (COFFEE1, COFFEE2) have been designed, submitted, and fabricated in Multi-Project Wafers (MPWs) using the 55 nm process. Notably, COFFEE2 was the first sensor prototype in the 55 nm HV-CMOS process worldwide. The performance of these small-scale sensor chips are evaluated to provide crucial input for the final detector design, as listed in Table 5.3. A small-scale prototype with full readout functionalities (COFFEE3) has also been designed, and further prototyping toward a full-size chip is planned, which will be covered in Section 5.2.5.

The COFFEE1 chip was fabricated in a 55 nm Low-Leakage process in an MPW in 2022. While this is not a High-Voltage process, it features a similar deep n-well structure underneath the electronics, which could serve as a large electrode, as illustrated in Figure 5.15 (a). The default wafers with low resistivity of a few $\Omega \cdot$ cm are used. This chip has an area of 3 mm $\times$ 2 mm, including arrays of passive diode sensors. The current-voltage curve shows a breakdown voltage at around −9 V, and a clear signal generated in the sensor in response to laser is observed. More results can be found in Ref. [8].

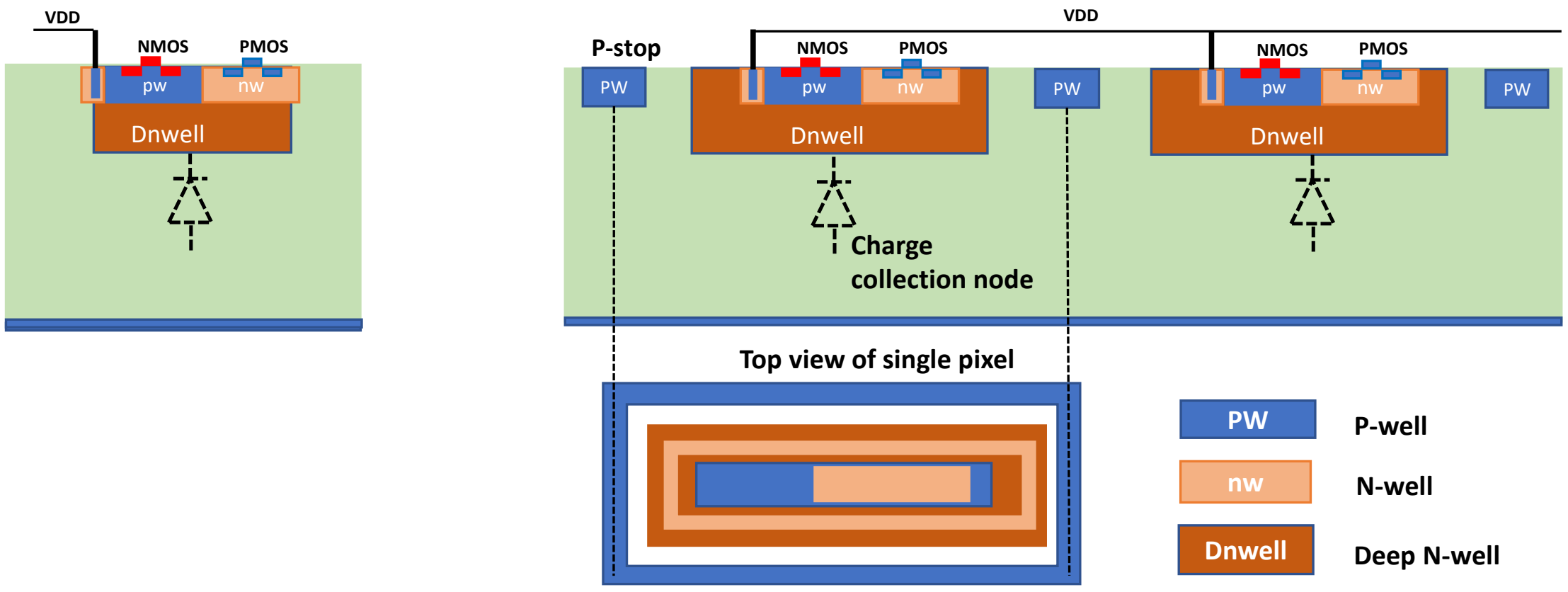

(a) Low-Leakage process (b) High-Voltage process

**Figure 5.15:** Cross-sections of the 55nm (a) Low-Leakage and (b) High-Voltage CMOS processes. The Low-Leakage process was used in the COFFEE1 chip, while the High-Voltage CMOS process was adopted in the COFFEE2 chip.

With the experience accumulated with COFFEE1, the first proof-of-principle sensor chip in 55 nm HV-CMOS process, COFFEE2, was developed. Figure 5.15 (b) shows the cross-section of this process, which is triple-well process including n-, p-, and deep n-wells. Up to ten metal layers can be used for fine pitch routing, including two thick metal layers for power. In this MPW, six metal layers are used, where one thick metal layer is used for power lines.

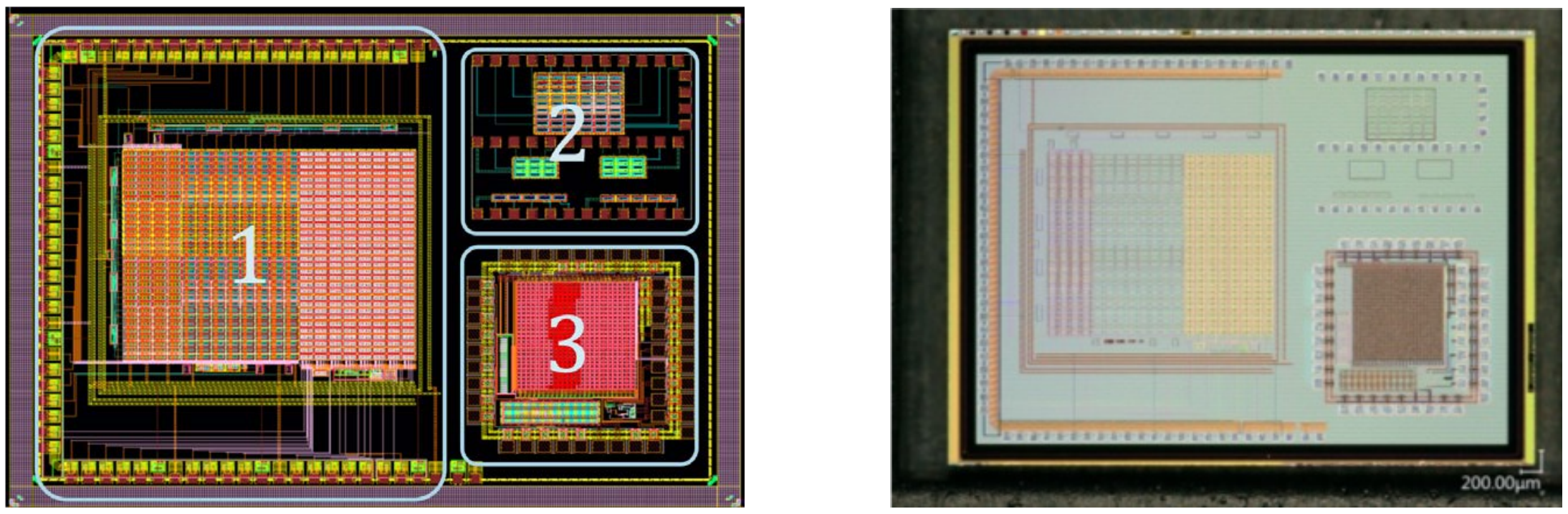

(a) (b)

**Figure 5.16:** (a) Floorplan and (b) photo of the COFFEE2 chip. The chip comprises three sections: Section "1" contains diode arrays and in-pixel circuits; Section "2" consists of passive diode arrays similar to those in COFFEE1; Section "3" is designed for imaging applications with small pixels and is not directly relevant to the CEPC use case.

The COFFEE2 chip has an area of 4 mm × 3 mm, with its floorplan and a photo shown in Figure 5.16. It is divided into three sections addressing different design purposes:

- Section "1" has a 32 × 20 pixel matrix with in-pixel circuits.
- Section "2" consists of passive diode arrays similar to COFFEE1 with the same pixel sizes as in the Section "1".
- Section "3" is designed for imaging application with small pixels for external applications outside of CEPC.

The pixel pitch was initially planned to be 25 × 150 μm$^2$ as in COFFEE1. However, this would be extremely challenging, as the narrower areas require a few micrometers of clearance to be kept from the edge of the deep n-well, leaving no more than 10 μm on one side for the circuit. Therefore, a less elongated pixel shape was adopted while keeping a similar pixel area, resulting in a pixel size of 40 × 80 μm$^2$.

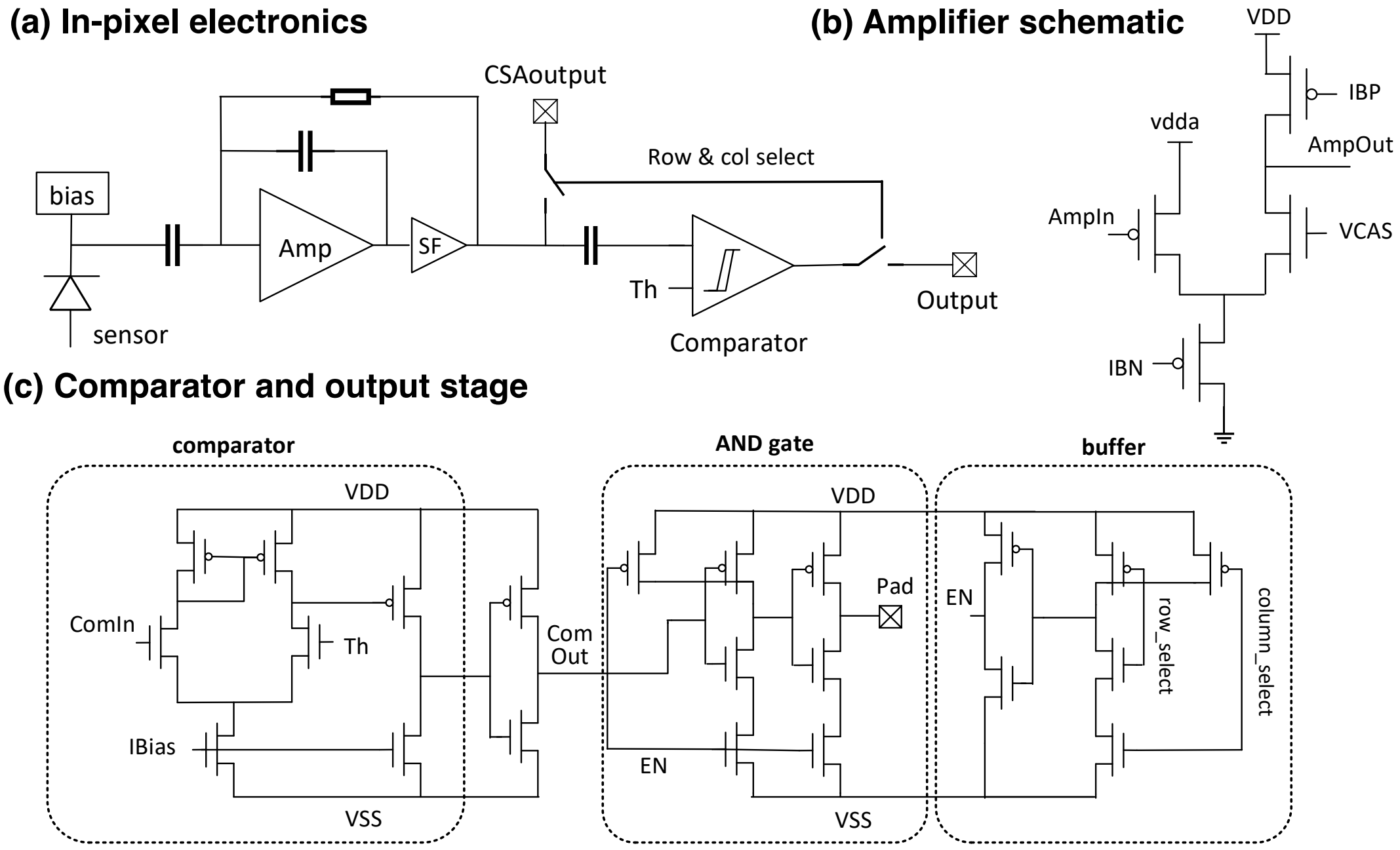

**Figure 5.17:** Schematic diagram of in-pixel electronics in Section "1" of COFFEE2. (a) The sensor generates a signal that is amplified by a charge-sensitive amplifier, consisting of a gain stage (Amp) and a source follower (SF). The amplified signal, labeled as "CSAoutput", is alternating current (AC)-coupled to a comparator, which converts the analog signal into a digital output. This output is routed to the exterior of the pixel via a shared output bus per column, with row and column select signals managing the pixel address. (b) Schematic of the amplifier (Amp), which operates using the current bias "IBN" and the voltage bias "VCAS". (c) Detailed circuit elements of the comparator and output stage. The comparator compares the input signal "ComIn" with the threshold reference "Th", using "IBias" as the current bias. The output signal "ComOut" passes through an AND gate and a buffer before reaching the external output pad. This configuration ensures robust digital signal readout, controlled by enable signals "EN", and facilitates row and column selection via "row_select" and "column_select". All bias signals are externally controlled, allowing fine-tuning of the amplifier during testing.

Figure 5.17 shows the schematic design of the in-pixel electronics in Section "1". A pixel consists of a Charge-Sensitive Preamplifier (CSA) connected to the front-end collection electrode through AC coupling. A comparator, via another AC coupling stage, is connected to the output stage of the CSA. This arrangement ensures the transmission of the analog signal while isolating the different stages electrically. Only one pixel in the array can be read out at a time when both the "column" and "row" are selected.

Several variations in diode or in-pixel electronics are implemented in Section "1", including:

- Gaps between neighboring deep n-wells are 10, 15, or 20 μm.
- With or without p-stop between pixels.
- Pixels with only amplifiers or with both amplifiers and comparators.
- By default, the comparators are implemented using both Positive channel Metal-Oxide-Semiconductor (PMOS) and Negative channel Metal-Oxide-Semiconductor (NMOS). However, in the left four columns, only NMOS-based comparators are designed. This NMOS-only design aims to mitigate potential crosstalk between the deep n-well and n-well.

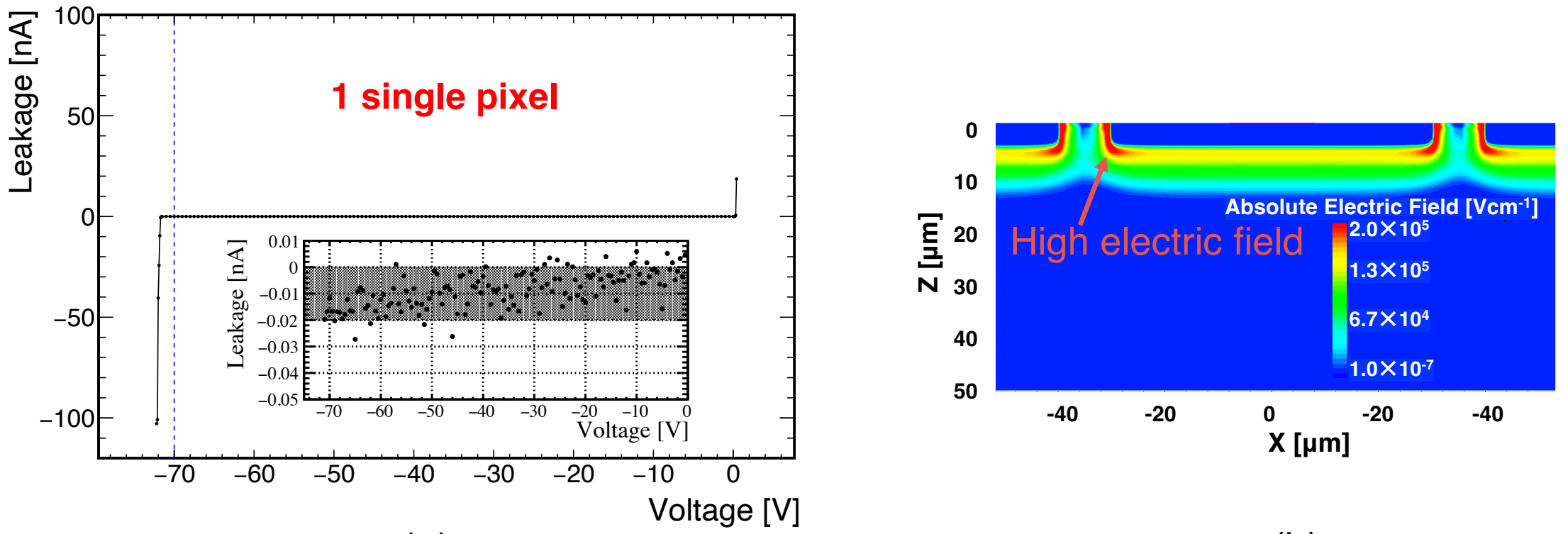

(a) (b)

**Figure 5.18:** (a) IV curve of a pixel in COFFEE2 chip; (b) Technology Computer Aided Design (TCAD) simulation of the passive diode under a reverse bias of −70 V. The breakdown voltage of the COFFEE2 chip is around −70 V, and the leakage current is at the pA level before breakdown.

The IV curve of a typical diode in Section "2" is shown in Figure 5.18 (a). The breakdown voltage can be as large as −70 V. The increase of breakdown voltage of COFFEE2 with respect to COFFEE1 is due to the different CMOS processes. The sudden increase in current beyond the breakdown voltage may indicate that the breakdown occurs at the edge of the deep n-well, which is consistent with the TCAD simulation shown in Figure 5.18 (b). The simulation also demonstrates that the breakdown voltage can be significantly increased by using higher resistivity wafers. The leakage current is at the pA level at low bias.

The capacitance as a function of bias voltage is shown in Figure 5.19 for (a) a single pixel and for (b) 8 connected pixels. Full depletion is not reached when the breakdown occurs.

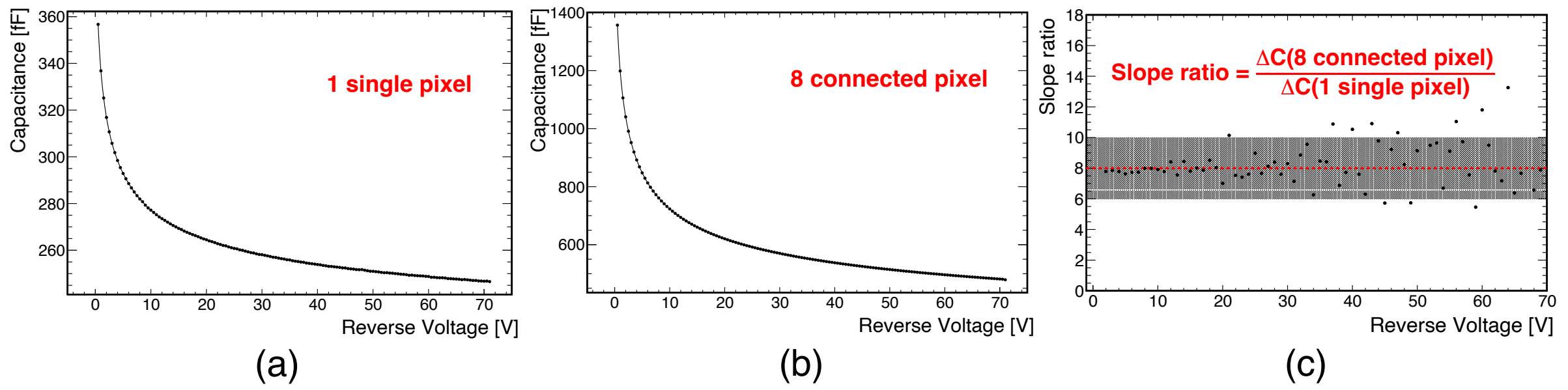

**Figure 5.19:** CV curves of (a) a pixel, (b) 8 connected pixels in the COFFEE2 chip, and (c) the ratio of capacitance of 8 pixels with respect to 1 pixel as a function of reverse bias. With the offset subtracted, the capacitance of a single pixel is $30 - 40$ fF at a bias of $-70$ V.

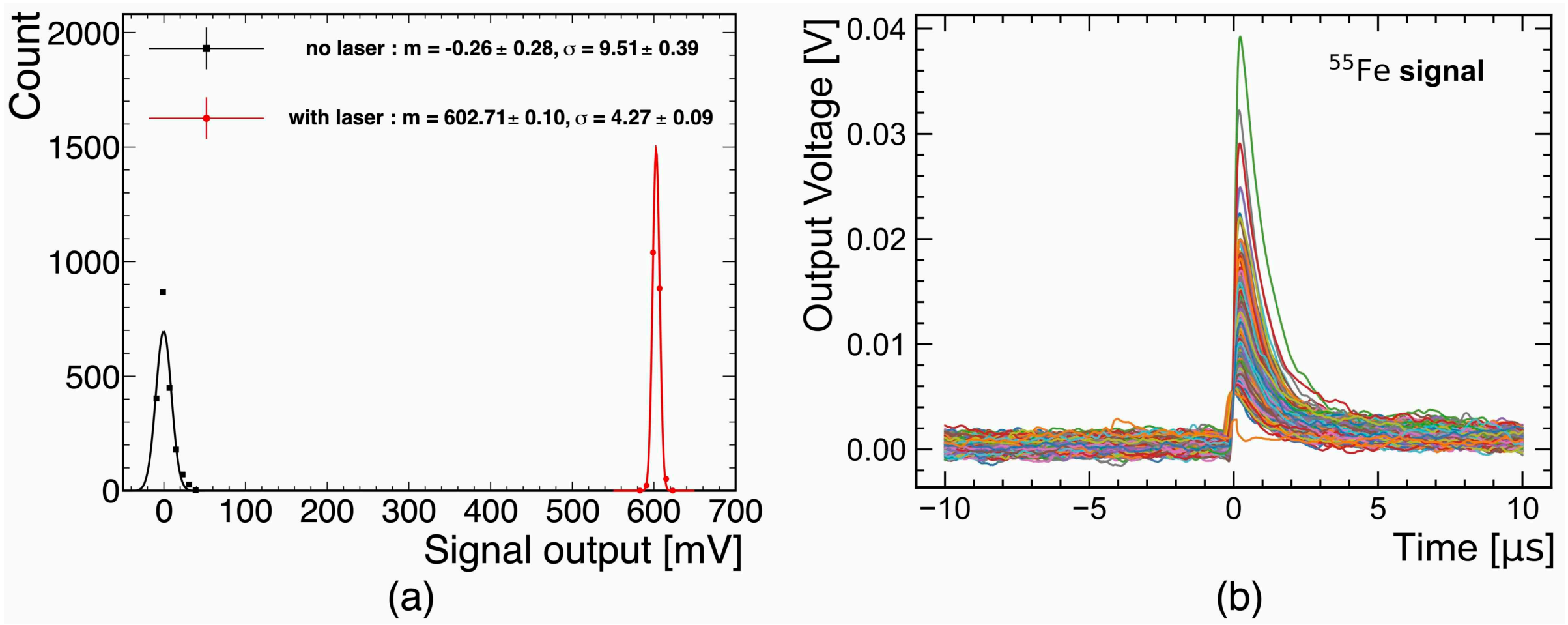

**Figure 5.20:** Responses to (a) red laser signals (red histogram) compared with no laser (black histogram), and (b) X-rays from $^{55}$Fe in a pixel of COFFEE2. Clear signal responses were observed for both the red laser and the $^{55}$Fe radioactive source.

The capacitance should be proportional to pixel area; however, an offset exists due to various reasons, such as the parasitic capacitance of metal routing. To eliminate the offset, the derivative of the CV curves is taken, and the ratio of the capacitance for 8 pixels versus a single pixel is extracted. The ratio agrees well with the ratio of their areas, as shown in Figure 5.19 (c). The raw capacitance of a single pixel at $-70$ V is about 200 fF, which becomes $30 - 40$ fF after subtracting the offset.

A dedicated carrier board for the test setup was designed and fabricated, which is connected to the general-purpose control and readout board [9], an Field-Programmable Gate Array (FPGA) board, and a personal computer. Clear responses to laser signals and radioactive sources, such as $^{55}$Fe, were observed from the amplifier output, as shown in Figure 5.20.

Following the successful verification of the sensor process and the in-pixel analog circuitry of COFFEE2, the subsequent COFFEE3 was designed and submitted for tape-out in early 2025, with fabricated devices received in May 2025. COFFEE3 incorporates an almost complete

ASIC readout framework, which can be further extended toward the final chip implementation. Comprehensive characterization of the chip is ongoing and is expected to be completed by the end of 2025. Further details on COFFEE3 and the HV-CMOS sensor R&D plan are provided in Section 5.2.5.

### 5.2.5 Future plan

In the coming years, substantial efforts will be dedicated to developing the full-size HV-CMOS sensor, evolving from small MPW productions to full reticle size. In parallel, module level and system level integration, including R&D on supporting electronics, mechanical structures, cooling systems, and detailed integration methodologies, will also be carried out.

Among the target sensor design specifications listed in Table 5.3, achieving a timing resolution of a few nanoseconds to tag the 23 ns bunch crossings is particularly challenging, especially when combined with the requirement for moderate power consumption and high hit density tolerance. To address this, a novel data-driven readout architecture featuring in-pixel Coarse-Fine Time-To-Digital Converters (TDCs) has been implemented in the latest sensor prototype, COFFEE3 (see Figure 5.21). This design fully exploits the 55 nm process's small feature size, enabling more functionality within the limited pixel area. The chip has been recently submitted and is currently undergoing comprehensive characterization.

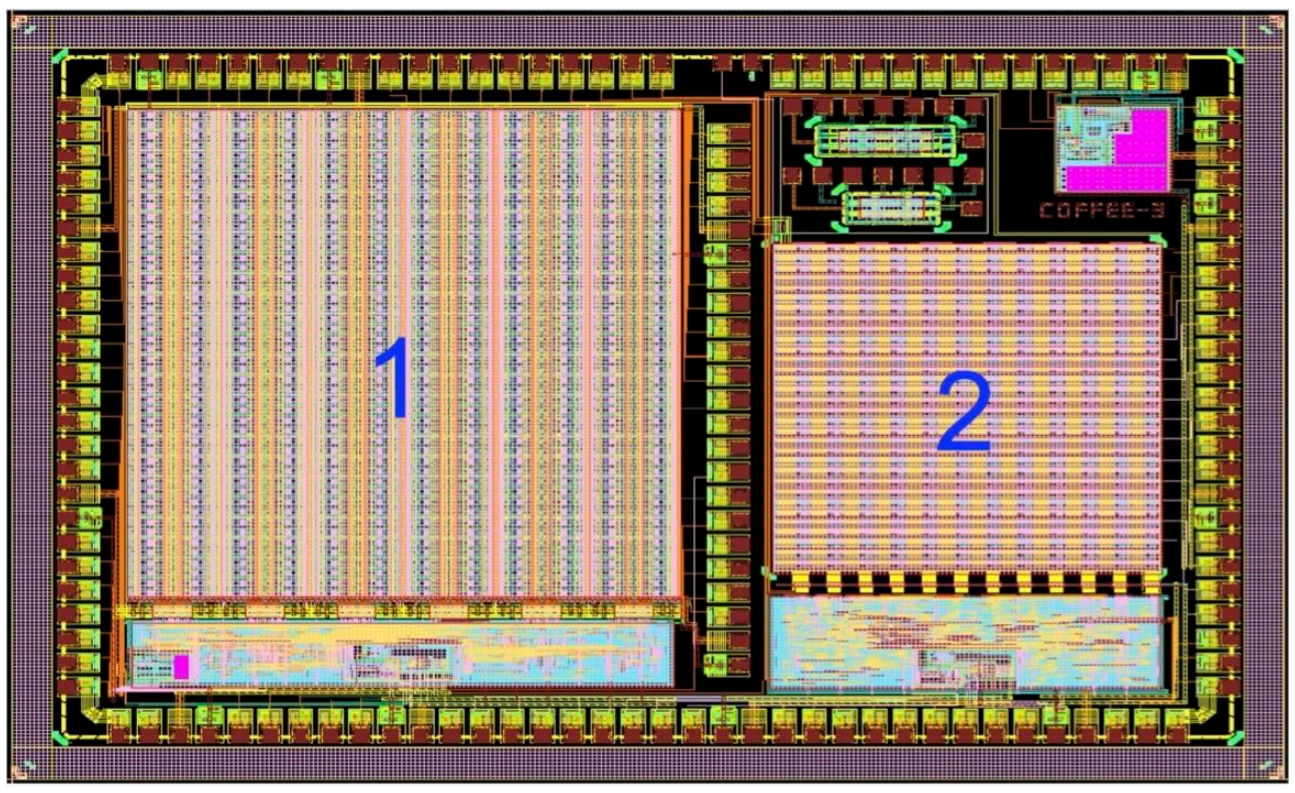

**Figure 5.21:** Layout of the COFFEE3 sensor chip, consisting of two distinct pixel array sections: Section "1" features a CMOS-based array, utilizing both PMOS and NMOS transistors, while Section "2" contains an NMOS-only pixel array, using exclusively NMOS transistors. Section "1" incorporates in-pixel Coarse-Fine TDCs, fully exploiting the small feature size of the 55 nm process.

Continuous efforts are also focus on improving the sensor design and fabrication processes. In the current HV-CMOS process used for the COFFEE series, the n-well of the PMOS transistor is directly in contact with the charge-collecting deep n-well (Figure 5.15). This configuration can induce voltage variations in the deep n-well due to PMOS transistor flipping, which are further amplified by the front-end electronics. To mitigate this cross-talk effect, a deep p-type layer has been proposed to separate the n-well from the charge collection electrode. Efforts are ongoing to enable such separation through a “quadruple-well process”. As an alternative to process modification, cross-talk can also be avoided by designing the pixel circuitry using only NMOS transistors. Both a CMOS-based array and an NMOS-only pixel array have been designed in COFFEE3, as shown in Figure 5.21. However, process modification remains the preferred solution, as it provides greater flexibility and enhanced pixel circuitry functionality.

Following COFFEE3 verification in late 2025, the focus will shift to process modification. In collaboration with the foundry, a new MPW submission using the “quadruple-well process” is planned for early 2026, followed by testing around mid-2026 to evaluate the modified process. In parallel, a quarter-size chip will be designed and submitted for tape-out after process verification. Testing of this chip, from late 2026 to mid-2027, will serve as an intermediate validation step before full-scale development.

From late 2026 through 2027, the project will advance to the full-size chip design, integrating all validated circuit and process improvements. The designed chip is scheduled for submission by the end of 2027. Final full-size chip testing is planned for 2028 to confirm complete functionality, performance, and production readiness. By the end of 2028, a fully validated, high-performance, and scalable HV-CMOS sensor is expected to be delivered for mass production.

Prototyping of ITK detector modules, along with their mechanical and cooling support structures for integration, will proceed in parallel with the development of the sensor chips. This includes the development of electronic components, methodologies, and tools for detector module assembly, studying support structures to optimize manufacturing procedures, and designing an efficient cooling system.

As the full-sized sensor prototype will only become available after several submissions, small-scale sensor chips together with the evolving off-chip electronic components, will serve as key intermediate steps for the system level R&D for both electrical and mechanical functionalities. Extensive prototyping studies — particularly on the mechanics, assembly and tooling — can be conducted using dummy silicon sensors. Specific traces can be added to these dummy sensors to define assembly procedures, especially when using an automatic gantry system. Dummy sensor modules, and other mechanical and cooling mockups, incorporating additional components that dissipate heat or generate external stress loads, will be useful for thermal and mechanical studies. Prototyping of the cooling system for the ITK and OTK using coolants such as $CO_2$ or water is currently underway.

## 5.3 Outer silicon tracker (OTK) with precision timing

Positioned close to the calorimeter, the OTK is the outermost tracking detector, covering an area of $\sim 85\,\mathrm{m}^2$. It is primarily designed to improve the momentum resolution by extending the tracking level arm with high spatial resolution, while also functioning as a precision ToF detector. This is achieved through the integration of novel microstrip AC-LGAD technology. Together with a high-resolution readout ASIC, the OTK is designed to provide a spatial resolution of ~ 10 μm and a time resolution of ~ 50 ps.

### 5.3.1 OTK design

The baseline design of the OTK consists of one barrel detector layer with a radius of ~ 1,800 mm and a length of 5,680 mm, along with a pair of endcaps positioned at $|z| = 2,910$ mm, covering a radial range of 406 mm – 1,816 mm. Both the barrel and endcap are constructed from AC-LGAD microstrip sensors with a strip pitch size of ~ 100 μm. These sensors are diced from 8-inch silicon wafers, having a rectangular shape for the barrel and a trapezoidal shape for the endcaps.

#### 5.3.1.1 OTK barrel design

The OTK barrel has two sensor types, with surface dimensions of 43.63 mm × 52.20 mm and 44.88 mm × 52.20 mm, which correspond to an active area of 43.03 mm × 51.60 mm and 44.28 mm × 51.60 mm, respectively. These two sensor types — differ only in length — are diced from a common 8" silicon wafer, as shown in Figure 5.22 (a). The thickness of the sensors is 300 μm. Each sensor contains 512 parallel strips with a strip pitch of 100 μm and strip lengths of 43.03 mm and 44.28 mm for the two sensor types, respectively.

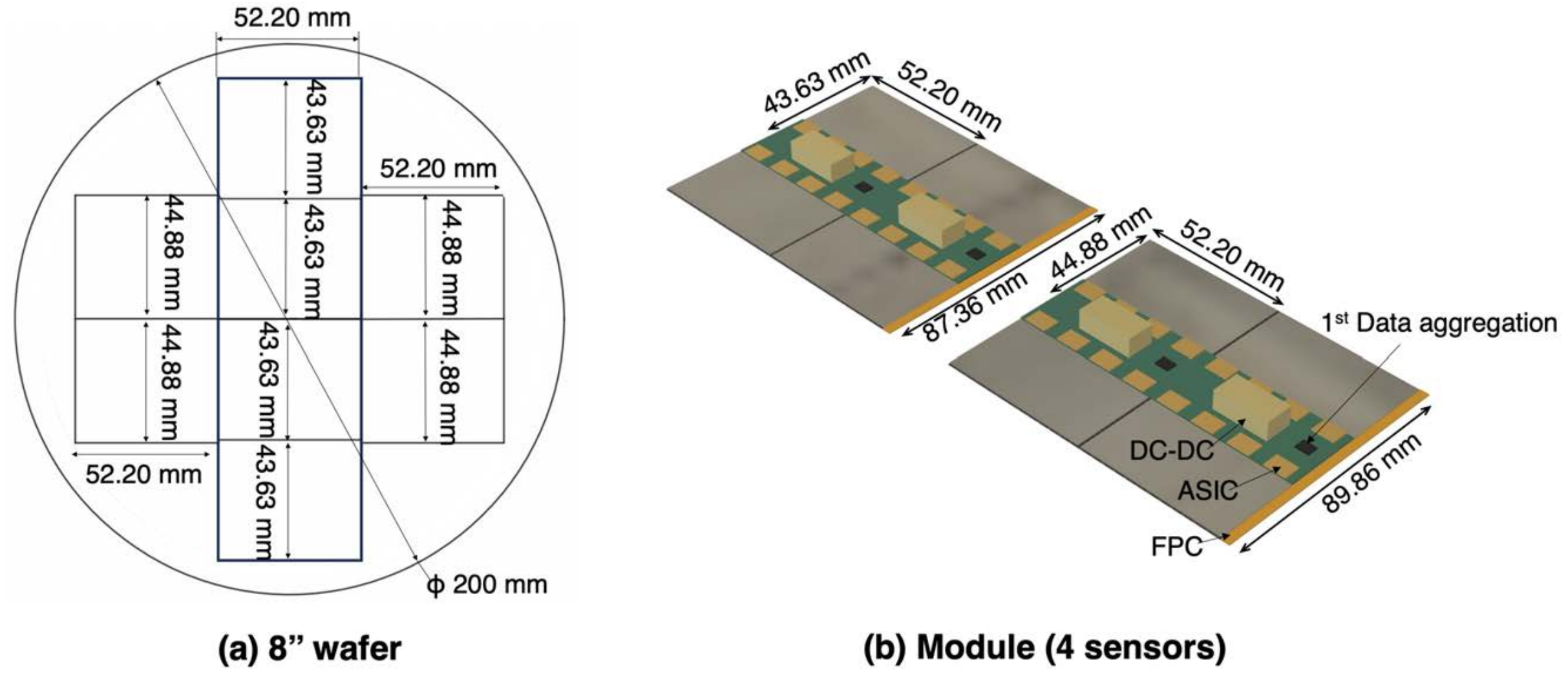

**Figure 5.22:** (a) Sensors and (b) modules for the OTK barrel. Eight sensors are diced from a single 8-inch silicon wafer, comprising four sensors with dimensions of 43.63 mm × 52.20 mm and four with dimensions of 44.88 mm × 52.20 mm. Each module consists of four sensors arranged in a 2 × 2 configuration, together with a low-mass PCB that integrates DC-DC converters, readout ASICs, and data aggregation chips. The two sensor lengths correspond to two types of modules, as illustrated in (b).

A barrel strip module consists of four sensors arranged in a 2 × 2 configuration and a low-mass PCB equipped with radiation-tolerant DC-DC converters, readout ASICs, and data aggregation chips, as illustrated in Figure 5.22 (b). The two sensor lengths correspond to two module types. A module is assembled by gluing the PCB to the silicon sensors using conductive epoxy. Thermal vias are included in the PCB design to improve vertical heat transfer from the ASICs to the sensor through the PCB. Each readout ASIC has 128 channels, connected to the silicon strips via wire bonding. The DC-DC converters provide low voltage (1.2 V) to two rows of readout ASICs (see Figure 5.22 (b)), with each row of ASICs wire-bonded to a set of strips on one half of a module. Each module contains 16 ASICs to read out 2,048 strips.

Every 8 modules (32 sensors) are glued to a common carbon fiber support plane with a thickness of 0.3 mm, forming a mechanical and electronic unit called a ladder, as shown in Figure 5.23 (a). There are two types of ladders, with lengths of 699.6 mm for short ladders and 719.6 mm for long ladders, assembled using the two types of sensors, respectively. A secondary data aggregation PCB, integrating an aggregation chip, a data link chip, an optical converter, and DC-DC converters, is glued near the middle of the ladder. At the edge of the ladder, a long kapton flexible printed circuit (ladder FPC) is connected (see Figure 5.23 (b)) to transmit high voltage (200 V) for sensor bias, low voltage for DC-DC, data, clock signals, and chip commands. Signals collected from the strips are digitized in the ASICs, aggregated within each module, transmitted through the FPC, and then reach the secondary data aggregation board. Data from all the modules in a ladder undergo secondary aggregation, passes through a data link chip, and is subsequently converted to optical signals via the optical module for readout.

To match the length of the OTK barrel, four long ladders and four short ladders are placed on the TPC outer barrel in a row, forming a long structure called a stave, with a total length of 5,680 mm, as shown in Figure 5.23 (b). In each stave, the central four ladders are short, with two long ladders positioned at each end.

After the ladders mounted onto the TPC outer barrel, two long power bus FPCs per stave are connected to the stave side, with each power bus FPC serving four ladders per half stave from one end. As shown in Figure 5.23 (c), each long power bus FPC is soldering to the four secondary data aggregation boards of the four ladders, and is placed just above the four shorter ladder FPCs. The high voltage (HV, 200 V) and original low voltage (LV, 48 V) are transmitted to the secondary data aggregation boards through the long power bus FPC. The DC-DC converters in the secondary data aggregation boards step down the 48 V LV input to 12 V. The 12 V LV, along with 200 V HV for sensor biasing, is distributed from the secondary data aggregation boards to the primary aggregation boards via shorter ladder FPCs. The DC-DC converters in the primary aggregation board of each module further drops the LV from 12 V to 1.2 V to supply power to the sensor readout ASICs.

Data readout from the sensor ASICs is aggregated on the primary aggregation board and transmitted to the secondary data aggregation board. The secondary data aggregation board processes the data, converts it to an optical signal, and transmits it through an optical fiber. The four optical fibers from four ladders per half stave are attached above the power bus FPC.

The diameter of the OTK barrel is 3,600 mm, requiring 110 staves to cover the full detector area of 65 m$^2$, as shown in Figure 5.23 (b). Neighboring staves are designed to overlap in the transverse direction to reduce the dead area. The detailed mechanical supporting designed for the OTK will be elaborated in Section 5.3.3.

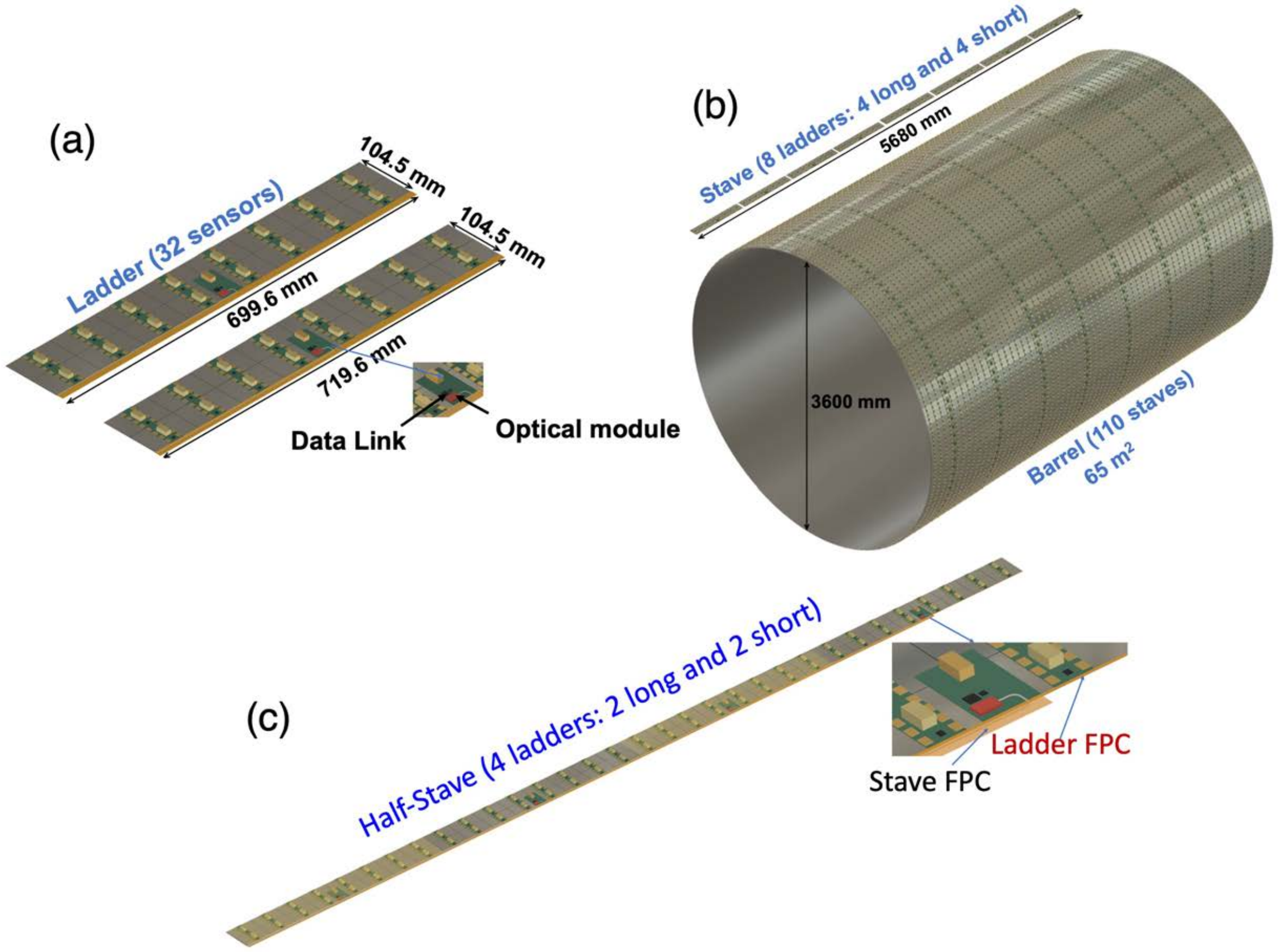

**Figure 5.23:** OTK (a) ladders, (b) barrel, and (c) half stave. Each ladder consists of eight modules (32 sensors) glued onto a carbon fiber support plane. A secondary data aggregation PCB — integrating an aggregation chip, data link chip, optical converter, and DC-DC converters — is attached near the ladder center, with a flexible printed circuit (ladder FPC) connected at the edge. Two types of ladders are used: short (699.6 mm) and long (719.6 mm). Four long ladders and four short ladders are arranged in a row on the TPC outer barrel to form a stave (5,680 mm). The OTK barrel contains 110 staves in total. The secondary data aggregation PCBs of the four ladders in each half stave are soldered to a power bus FPC (stave FPC) for power transmission from the stave end.

For the OTK barrel construction, the design of the OTK sensor mask and subsequent sensor dicing strategy from 8” silicon wafers, shown in Figure 5.22, have been optimized to maximize the usage efficiency of silicon wafers for the OTK barrel. For the entire OTK barrel, a total of 3,520 wafers are needed, with 15% greater silicon wafer utilization compared to conventional single-piece sensor dicing. Table 5.8 (a) summarizes the detailed information about the number of staves, modules, sensors, and readout ASICs used for the construction of the OTK barrel.

**Table 5.8:** Information about the modules, sensors, and readout ASICs used for the OTK construction

| **(a) Number of components used for the OTK barrel** | | | | |
|---|---|---|---|---|
| Staves | Ladders | Modules | Sensors | Readout ASICs |
| 110 | 880 | 7,040 | 28,160 | 112,640 |
| **(b) Number of components used for the OTK endcaps** | | | | |
| Sectors | Secondary aggregation boards | Modules | Sensors | Readout ASICs |
| 32 | 544 | 6,368 | 12,736 | 46,336 |

### 5.3.1.2 OTK endcap design

Each OTK endcap disk covers a surface area of $\sim 10\,\mathrm{m}^2$, corresponding to a radial range of 406 mm – 1,816 mm. Together with the barrel layer, the OTK provides full tracking coverage up to a polar angle of $|\cos(\theta)| < 0.99$. To facilitate sensor production and assembly, and to maximize silicon wafer utilization, the OTK endcap was specifically designed with trapezoid sensors.

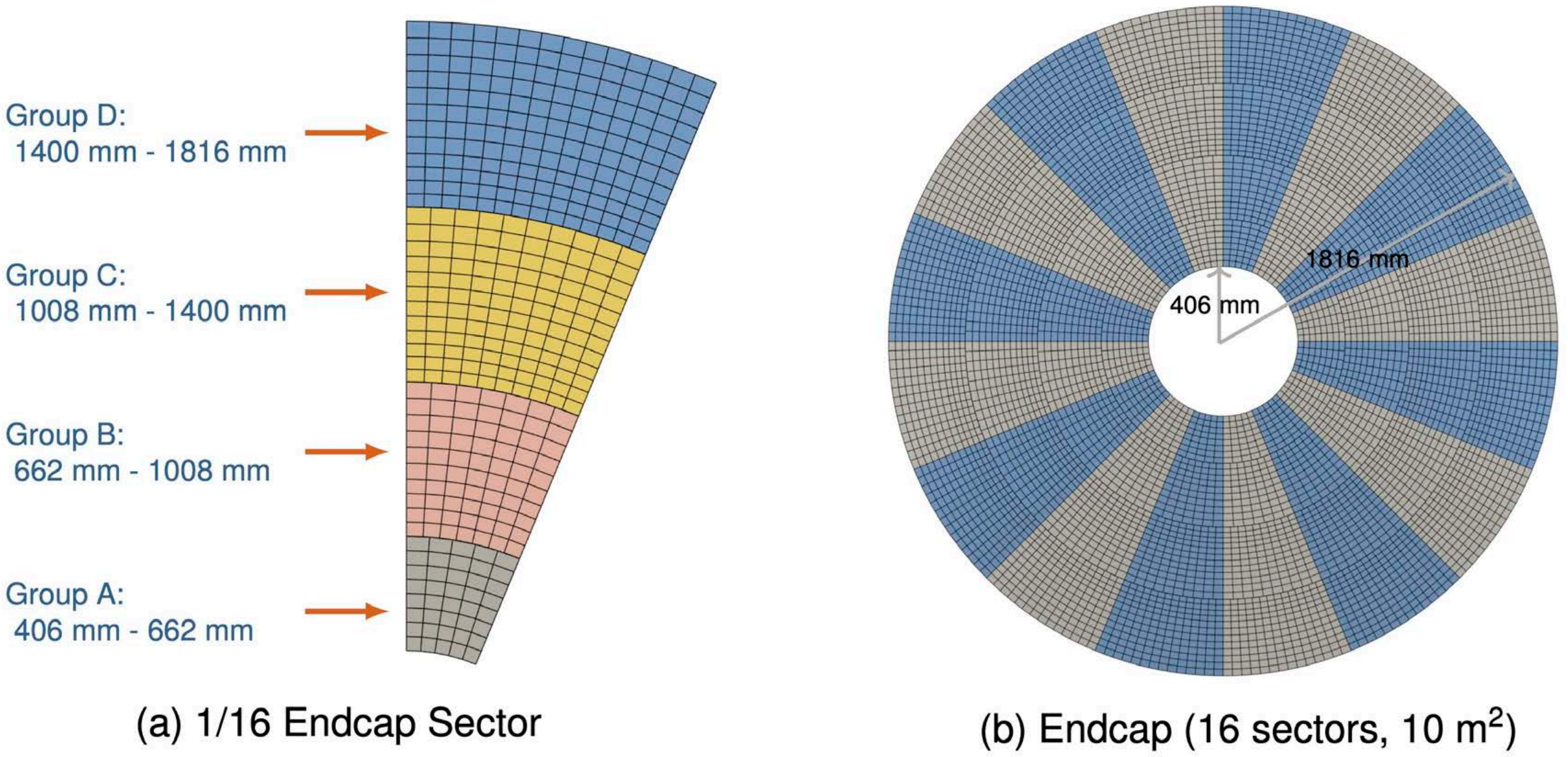

**Figure 5.24:** OTK sensor layout for (a) the 1/16 endcap sector and (b) the full endcap. The 1/16 sector of the OTK endcap consists of 42 rings of trapezoidal sensors, which are arranged into four groups: Group A, Group B, Group C, and Group D, each indicated by a different color.

The OTK sensors are fabricated from 8” silicon wafers. Figure 5.24 (a) shows a 1/16 sector of the OTK endcap, which consists of 42 rings arranged into four groups: Group A, Group B, Group C, and Group D, each indicated by a different color. Each group contains 2-4 subgroups of trapezoidal sensors, designed to fit to a common 8” silicon wafer. Figure 5.25 displays the wafer layouts for sensors from individual groups.

As shown in Figure 5.25, the sensor groups — specifically A1 and A2 in Group A, B1 and B3 in Group B, C1 and C2 in Group C, and D3 and D4 in Group D — each contain 4 sensors. Other sensor groups, namely B2 in Group B, C3 and C4 in Group C, and D1 and D2 in Group D, each contain 2 sensors. Each sensor has a set of implanted strips. Trapezoidal sensors with a middle width below 43 mm are classified as narrow sensors, while those above 43 mm are classified as wide sensors, with narrow and wide sensor containing 384 and 512 strips, respectively. The sensor geometric parameters are illustrated in Figure 5.25. For example, the surface dimensions of the top trapezoidal sensor of group A1 (A11) in Figure 5.25 (a) are 41.85 mm in top width, 39.34 mm in bottom width, and 31.90 mm in height. With a total of 384 strips, the radial strip pitch of the sensor A11 varies from a maximum of 107.4 µm at the top to a minimum of 100.9 µm at the bottom, while the strip length is ~ 31.2 mm.

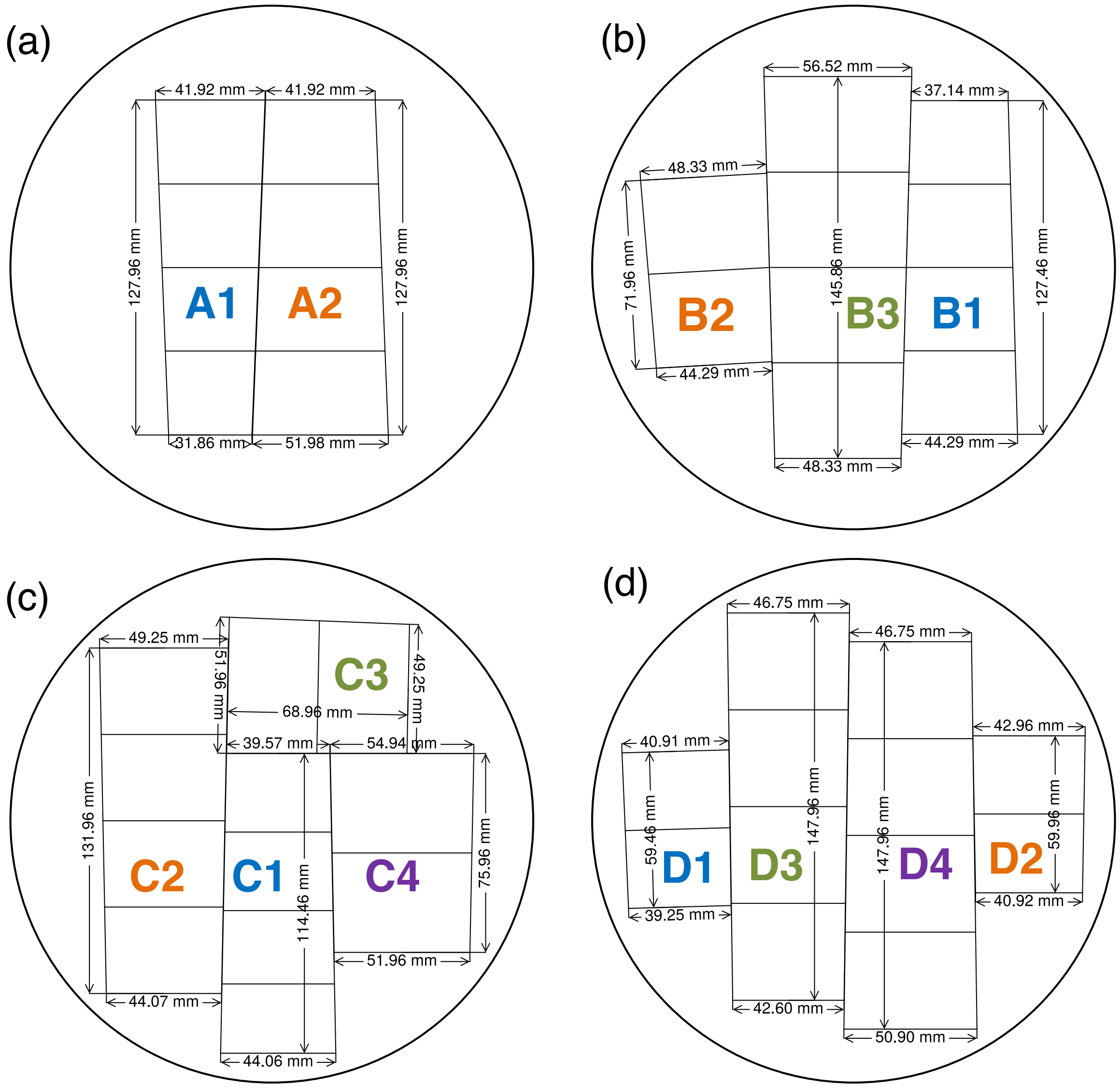

**Figure 5.25:** Four groups (A, B, C, and D) of OTK endcap sensors diced from 8” silicon wafer. Sensor subgroups B2, C3, C4, D1, and D2 each contain 2 trapezoidal sensors, while subgroups A1, A2, B1, B3, C1, C2, D3, and D4 each contain 4 sensors.

In the OTK endcap design (see Figure 5.24), each group of sensors is aligned to a 1/16

sector, with 16 sectors constituting the full endcap. This unique design requires only four masks for sensor fabrication and maximizes wafer usage efficiency. Table 5.9 summarizes the sensors and silicon wafer usage per OTK endcap. For each endcap, 6,368 silicon sensors will be fabricated from a total of 576 silicon wafers.

**Table 5.9:** Usage of silicon wafers, sensors, and ASICs per OTK endcap

| **Mask** | **A** | | **B** | | | **C** | | | | **D** | | | |
|---|---|---|---|---|---|---|---|---|---|---|---|---|---|
| Number of wafers | 80 | | 112 | | | 160 | | | | 224 | | | |
| Sensor subgroup | A1 | A2 | B1 | B2 | B3 | C1 | C2 | C3 | C4 | D1 | D2 | D3 | D4 |
| Sensors per subgroup | 4 | 4 | 4 | 2 | 4 | 4 | 4 | 2 | 2 | 2 | 2 | 4 | 4 |
| Number of sensors | 320 | 320 | 448 | 224 | 448 | 640 | 640 | 320 | 320 | 448 | 448 | 896 | 896 |
| ASICs per sensor | 3 | 4 | 3 | 4 | 4 | 3 | 4 | 4 | 4 | 3 | 3 | 4 | 4 |
| Number of ASICs | 960 | 1280 | 1344 | 896 | 1792 | 1920 | 2560 | 1280 | 1280 | 1344 | 1344 | 3584 | 3584 |

The basic assembly unit for the OTK endcap is the module. As shown in Figure 5.26 (a), each endcap module, similar to the barrel module, consists of two sensors and a low-mass PCB equipped with a DC-DC converter, readout ASICs, and a data aggregation chip. Each PCB, glued to the two sensors, serves as the primary data aggregation board, reading out two sets of strips. Each PCB has 6 ASICs for wide sensor modules and 8 for narrow sensor modules, with the ASICs wire-bonded to the strips.

For the OTK endcap, the 1/16 sector is designed as a complete mechanical and functional unit. To construct it, as shown in Figure 5.26 (b), 199 modules are glued on a 1/16 sector carbon fiber plane. Next, 17 secondary data aggregation boards are glued to the sensor surfaces of 1/16 sector, with each secondary data aggregation board connected to the two lateral rows of primary data aggregation boards. As illustrated in Figure 5.26 (a), each secondary data aggregation board includes a data aggregation chip, a data link chip, an optical module, and DC-DC converters. The primary aggregation board connects to the secondary aggregation board via a FPC connector. The lateral row of PCBs, or primary aggregation boards, in 1/16 sector are interconnected using compact connectors or wire soldering (the final method has yet to be decided). These connections enable the transmission of power, data, clock signals, and chip commands between the secondary aggregation board and any primary aggregation board.

As shown in Figure 5.26 (b) of a 1/16 OTK sector, 17 long power bus FPCs are arranged into three groups and fixed on three supporting frames. These FPCs extend from the outermost rim of the OTK sector to connect with the 17 secondary data aggregation boards. The high voltage (HV, 200 V) and original low voltage (LV, 48 V) are transmitted to the secondary data aggregation boards through these power bus FPCs. The DC-DC converter on the secondary data aggregation board steps down the 48 V LV input to 12 V and, along with the 200 V HV supply, distributes power to the primary aggregation boards. The 200 V is used for sensor biasing, while the DC-DC converter in the primary aggregation board of each module further drops the LV from 12 V to 1.2 V to supply power to the sensor readout ASICs.

Data readout from the sensor ASICs is aggregated on the primary aggregation board and transmitted to the secondary data aggregation board. The secondary data aggregation board processes the data, converts it to an optical signal, and transmits it through an optical fiber. The 17 optical fibers from the 17 secondary data aggregation boards in a sector are attached to the same mechanical frames as the power bus FPCs.

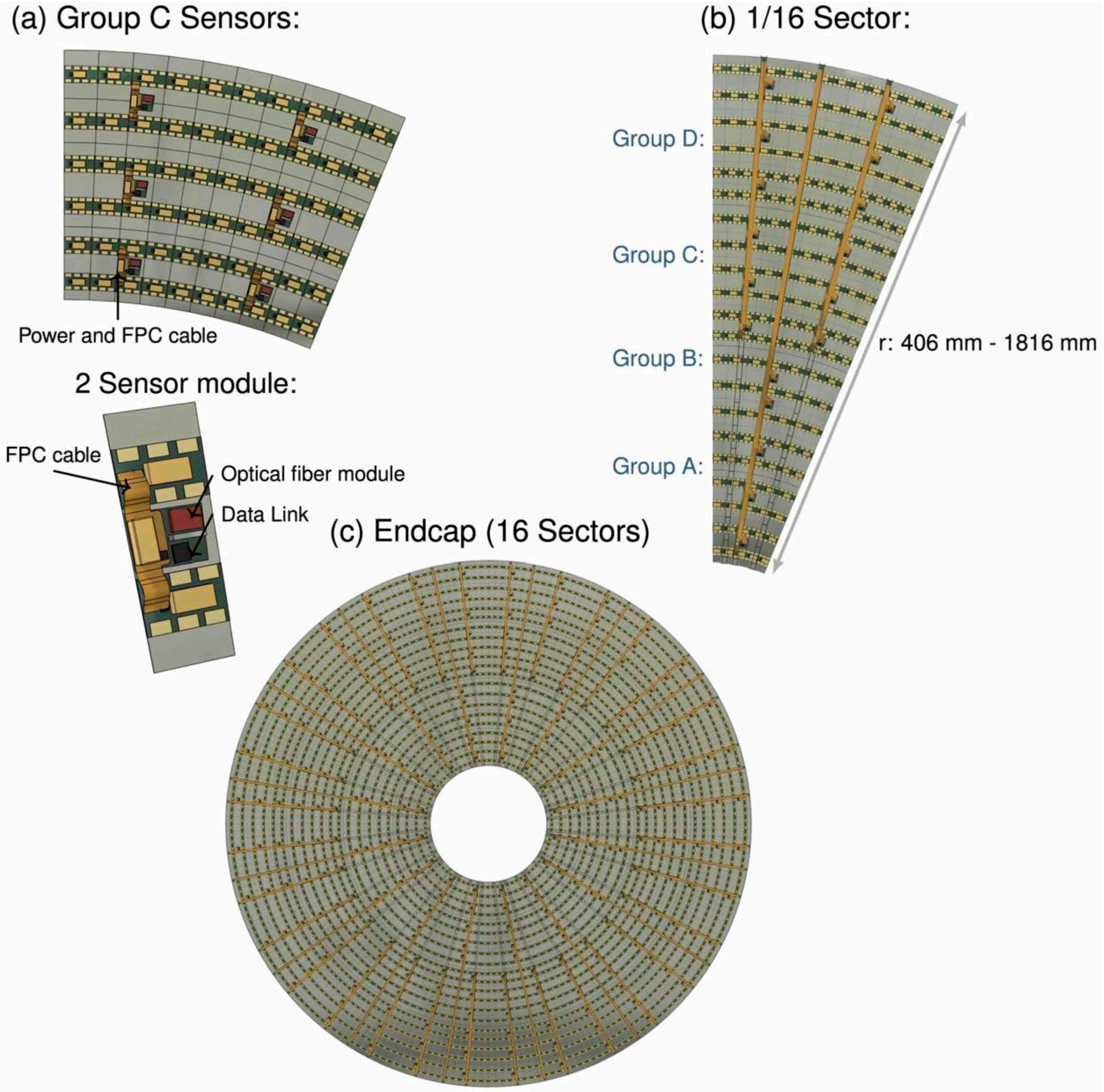

**Figure 5.26:** (a) The Group C sensors, (b) 1/16 sector of the OTK endcap, and (c) the full OTK endcap. Each endcap module consists of two sensors and a primary data aggregation board (a low-mass PCB equipped with a DC-DC converter, readout ASICs, and a data aggregation chip). Each secondary data aggregation board (a low-mass PCB integrated with DC-DC converters, a data aggregation chip, a data link chip, and an optical module) is glued to the sensors and connected to the two lateral rows of primary data aggregation boards via FPC connectors. The 1/16 OTK endcap sector contains 199 modules, 17 secondary data aggregation boards, and 17 long power bus FPCs arranged into three groups and fixed on three supporting frames for power transmission from the outermost rim of the endcap.

A complete OTK endcap consists of 16 sectors. Figure 5.26 (c) shows the OTK endcap with sensors and electronic components. Table 5.8 (b) summarizes the detailed information about the number of sectors, secondary data aggregation boards, sensors, and readout ASICs used for the construction of the OTK endcap pair. A detailed description of the OTK electronic design is provided in Section 5.3.2. The mechanical and cooling design for the OTK endcap will be presented in Section 5.3.3.

### 5.3.2 Readout electronics

Figure 5.27 shows the overall readout electronics scheme of the OTK. The AC-coupled signals from the Low Gain Avalanche Detector (LGAD) strips in the sensor are processed and digitized by the 128-channel LGAD Timing and Readout Integrated Chip (LATRIC) readout ASIC, on the Front-End (FE) board, also referred to as the primary data aggregation board (Figures 5.22 (b) and 5.26 (a)). The LATRIC ASIC, designed for high-precision timing and charge measurements, is described in detail in Section 5.3.5.

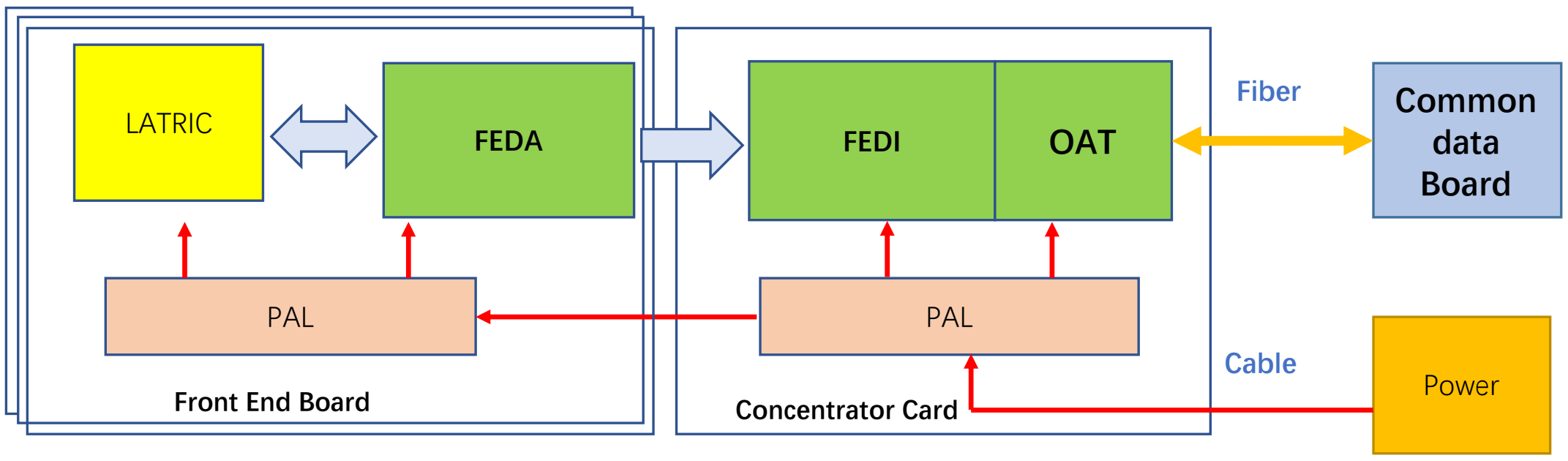

**Figure 5.27:** OTK readout electronics scheme. The Front-End Data Aggregator (FEDA) chip collects data from multiple LATRIC ASICs and transmits it to the Front-End Data Interface (FEDI) chip. The FEDI chip then forwards the data via the Optical Array Tranceiver (OAT) module to the Back-End Electronics (BEE). Power is delivered through DC-DC converters (PAL) to the LATRIC ASICs and other readout chips.

As illustrated in Figure 5.27, the data from the LATRIC ASICs is first aggregated by the FEDA chip on the FE board and the FEDI chip on the secondary data aggregation board (Concentrator Card), and then routed to the OAT module for fiber readout (Figures 5.23 (a) and 5.26 (a)). The fiber link transfers data between the backend electronics and the secondary data aggregation board.

Power is delivered from the crate via power cables and stepped down in two stages: from 48 V to 12 V, and then from 12 V to 1.2 V or 3.3 V by PAL DC-DC converters, which supply the LATRIC and other readout electronics.

#### 5.3.2.1 Front-end board

In the OTK module, two rows of LATRIC ASICs are integrated on the FE board (Figures 5.22 (b) and 5.26 (a)), with each row wire bonded to a set of LGAD strips. For the LGAD sensor, each fired strip generates 48 bits of data digitized by the LATRIC ASIC, with an average of about two strips firing per hit. The AC coupling design within the sensor [10] isolates the circuit from large leakage currents and protects the ASIC from potential damage caused by bias voltage in case of an LGAD failure.

The LATRIC ASIC operates on a 1.2 V power supply, which is regulated and filtered by the PAL12 (Power at Load for 12 V) ASIC — a radiation tolerant DC-DC converter capable of delivering up to 10 A of current, sufficient to power up to eight LATRIC ASICs.

On a barrel FE board (Figure 5.22 (b)), sixteen LATRIC ASICs are powered by two PAL12 converters, and their digitized data is aggregated by two FEDA chips. On an endcap FE board (Figure 5.26 (a)), six or eight LATRIC ASICs are powered by one PAL12, and their data is aggregated by one FEDA chip.

In the OTK baseline design, each FE board interfaces to the secondary data aggregation board, transferring the following signals:

- Clock: The FEDI provides 43.33 MHz e-clocks for the LATRIC ASICs.
- Data Readout:
  - Barrel FE board: Two uplink e-links (each at 346.67 Mbps) from two FEDA chips connect to the secondary data aggregation board, supporting a data rate of up to 43.33 Mbps LATRIC ASIC.
  - Endcap FE board: One uplink e-link (693.33 Mbps) from one FEDA chip connects to the secondary data aggregation board, supporting a data rate of up to 86.67 Mbps per LATRIC ASIC.
- Configuration/Sync/Reset:
  - Barrel FE board: Two downlink e-links (86.67 Mbps) are used for configuration, resynchronization, and reset, with each link shared by eight LATRIC ASICs.
  - Endcap FE board: One downlink is shared by six or eight LATRIC ASICs.
- Monitoring: Each FE board supports up to 16 temperature sensors, enabling real-time monitoring of both sensor and electronics temperatures.
- Power:
  - 200 V HV is used for sensor biasing.
  - 12 V LV is delivered from the secondary aggregation board to the PAL12 DC-DC converters on the FE board.

To enable multiple LATRIC ASICs to share a common configuration downlink, each ASIC is assigned a unique 4-bit address through dedicated Identification (ID) pins. The configuration protocol uses this address to configure each LATRIC chip individually via I2C. To ensure reliable reception of the downstream signals from the FEDI, the LATRIC chip can latch input signals on either the rising or falling edge of the clock.

#### 5.3.2.2 Concentrator Card

The Concentrator Card (CC), also referred to as the secondary data aggregation board, is designed to interface the system readout with multiple FE boards. For the barrel, each CC connects to 8 FE boards, with each FE board equipped with two FEDAs (see Figure 5.23 (a)). For the endcap, each CC connects to either 10 or 14 FE boards, with each FE board containing one FEDA (see Figure 5.26 (b)).

The FEDI on CC provides the interface from e-links to an OAT module. Each optical module includes a single channel receiver and a single channel transmitter for optical signals. The command downlink operates at 2.77 Gbps, while the uplink bandwidth supports up to 11.09 Gbps.

In the barrel, 16 FEDA chips on 8 FE boards are aggregated and linked to a single FEDI chip via 346.67 Mbps e-links. In the endcap, 10 to 14 FEDA chips are similarly connected to one FEDI chip, with each e-link operating at 693.33 Mbps. Considering each FEDA chip connecting to 8 readout ASICs in the barrel and 6 or 8 ASICs in the endcap, the resulting data rate per ASIC is up to 43.33 Mbps in the barrel and up to 86.67 Mbps in the endcap. These data rates correspond to a maximum supported hit rate of $8.0 \times 10^4$ Hz/cm$^2$ in the barrel (with an active detector area of $\sim$ 1.28 cm $\times$ 4.4 cm per ASIC) and $2.1 \times 10^5$ Hz/cm$^2$ in the endcap ($\sim$ 1.28 cm $\times$ 3.3 cm per ASIC). These supported hit rates are 12 times higher in the barrel and 6 times higher in the endcap than the estimated peak background hit rates (after applying a safety factor of 2), under the most challenging high-luminosity $Z$-pole operation mode, as shown in Table 5.2.

The distribution of a precise clock to the front-end is a critical requirement for the OTK system. The clocks in the backend system are recovered from the FEDI downlinks, synchronized to the 43.33 MHz frequency bunch crossing rate. An additional key feature of the FEDI is ensuring the precise clock distribution to the front-end system. This is achieved via a high-frequency clock noise filter in the Phase-Locked Loop (PLL) of the FEDI. However, low-frequency clock jitter and potential phase instability, particularly caused by temperature variations or the low-frequency response of the clock chain, will require special attention to ensure optimal system performance.

The FEDI also provides a set of slow control and monitoring features, including I2C master controllers, a JTAG master controller, programmable and bidirectional I/O ports, and a memory-like bus master controller with data and address management.

Power is distributed through PAL DC-DC converters. To power both the CC components and the FE boards, two types of DC-DC converters are used: PAL48 (Power at Load for 48 V) and PAL12. As shown in Figures 5.23 (c) and 5.26 (b), long power bus FPCs transmit both 200 V HV and 48 V LV from the end of the barrel or endcap to the CCs. The PAL48 converters on the CC step down the 48 V to 12 V, which, along with the 200 V HV, is then distributed to each individual FE board. Each CC also integrates a PAL12 converter that steps down the 12 V to either 1.2 V for FEDI and 3.3 V for the OAT.

The power bus FPCs, along with optical fibers, are connected to composite cables that deliver both power and signals from external sources. For each half stave in the barrel, four composite cables are connected to the FPC and four fibers. Similarly, each 1/16 endcap sector uses nine composite cables, with each cable connecting to two power bus FPCs and the corresponding two optical fibers. In total, the OTK barrel requires 880 composite cables, while the OTK endcaps use 288. To improve efficiency and compactness, serial powering is under consideration.

### 5.3.3 Mechanical and cooling design

Similar to the ITK, the OTK barrel is segmented in the azimuthal direction into structural units known as staves. Each stave is a long structure spanning the entire length (5,680 mm) of the OTK barrel. These staves are directly mounted on the TPC outer barrel — a carbon fiber cylinder integrally molded with a specifically designed support structure — to optimize support for the longer staves while minimizing material usage. The installation of the OTK barrel is illustrated in Figure 5.28 (a).

A pair of assembled OTK endcaps, together with the OTK barrel, form the complete OTK system, as shown in Figure 5.1 (b). The mechanical and cooling design of the OTK is driven by stringent requirements on material budget, structural robustness, and thermal efficiency.

#### 5.3.3.1 Barrel support

The cross section of the internal structure of an OTK barrel stave is shown Figure 5.28 (b). The OTK stave structure can also be divided into three functional units:

- Sensor Modules: Sensors are glued to low-mass PCBs equipped with DC-DC converters, readout ASICs, and other components.
- Support Structure: This structure consists of two layers of 0.3 mm thick carbon fiber facesheets sandwiching a carbon fiber honeycomb core. The transverse edges of the stave are enclosed by side closeouts, which are C-shaped channels made of carbon fiber. These side closeouts are important mechanical interfaces to the global support structures, increasing the stave's stiffness and reducing deformation.
- Cooling Structure: Sensors in the modules are attached to the carbon fiber facesheet using thermal adhesive. The area around the cooling pipes and the honeycomb core is filled with low-mass, high thermal conductivity foam. The carbon fiber facesheet and honeycomb are bonded using an adhesive infused with thermally conductive carbon particulates.

To achieve hermetic coverage, neighboring staves are partially overlapped via the stepped ramp ring mounting structure, as illustrated in Figure 5.28.

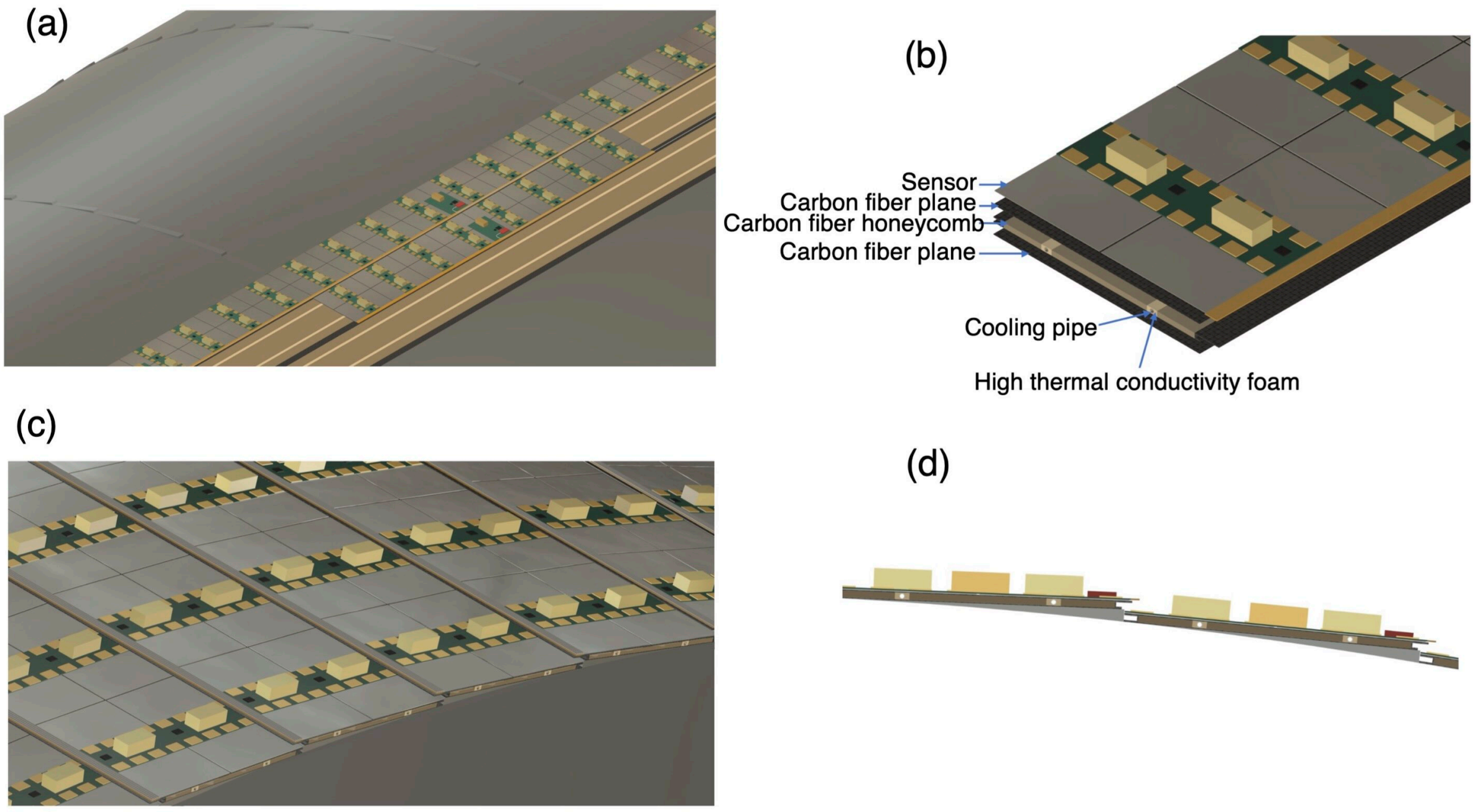

**Figure 5.28:** (a) Installation of the OTK barrel. Each stave is mounted onto the TPC outer barrel, a carbon fiber cylinder with stepped ramp rings. A lower carbon fiber facesheet (plane) with a honeycomb core is first attached, followed by the insertion of two ~ 6 m cooling pipes sealed with high-conductivity foam. Finally, eight ~ 0.7 m ladders are glued on top to complete the stave assembly. (b) Structure of the OTK barrel stave. Each stave contains sensor modules, top and bottom carbon fiber planes sandwiching a carbon fiber honeycomb core. Two cooling pipes are inserted into the honeycomb grooves, sealed with high-conductivity foam, and the transverse edges of each stave are enclosed by side closeouts. (c) Staggered structure of OTK staves and (d) cross section of two neighboring staves. Neighboring staves are staggered and partially superimposed to ensure the detector's hermeticity.

**Materials** The facesheets are constructed from layers of high-modulus unidirectional carbon fiber material combined with cyanate ester resins, designed to meet the overall stiffness requirements for local support and the thermal demands of conducting heat generated in the modules to the cooling structures embedded in the core.

The honeycomb core is made of lightweight, high-performance carbon fiber material. It maintains the separation of the facesheets to ensure good bending stiffness and provides a sufficiently flat and robust surface for adhesive attachment of the modules without requiring excessively thick glue layers.

The interface between the cooling tubes and the facesheets is formed using thermally conductive carbon foam, such as Allcomp K9. Water with titanium cooling pipes (tube wall thickness of 0.2 mm and a diameter of 5 mm) has been selected as the baseline cooling medium for the OTK.

Table 5.10 lists the estimated contributions of the OTK stave to the material budget. The overall estimated material budget for a stave is 1.6% $X_0$. The overlap area between neighboring staves accounts for 8% of the stave area, corresponding to 6% of the stave materials. With 110 staves, the total weight of the OTK barrel is ~ 300 kg.

**Table 5.10:** Estimation of the OTK stave material contributions. The wall thickness of the cooling tubes, averaged over the entire endcap area and labeled with "†", is ~ 60 μm.

| Functional unit | Component | Material | Thickness [μm] | $X_0$ [cm] | Radiation Length [% $X_0$] |
|---|---|---|---|---|---|
| Sensor Module | PCB metal layers | Cu | | 1.436 | 0.200 |
| | PCB Insulating layers | Polyimide | | 28.41 | 0.070 |
| | Sensor | Silicon | 300 | 9.369 | 0.320 |
| | Glue | | 100 | 44.37 | 0.023 |
| | Other electronics | | | | 0.100 |
| Structure | Carbon fiber facesheet | Carbon fiber | 300 | 26.08 | 0.115 |
| | Cooling tube wall | Titanium | 200† | 3.560 | 0.169 |
| | Cooling fluid | Water | | 35.76 | 0.105 |
| | Graphite foam+Honeycomb | Allcomp+Carbon fiber | 6000 | 186 | 0.322 |
| | Carbon fiber facesheet | Carbon fiber | 300 | 26.08 | 0.115 |
| | Glue | Cyanate ester resin | 200 | 44.37 | 0.045 |
| Total | | | | | 1.584 |

**Structural characterization** The stiffness of the stave is provided by the carbon fiber facesheets, honeycomb, and C-shaped carbon fiber channels. Similar to OTK stave, the maximum sag of the OTK stave under its own weight and the first natural frequency are used to estimate the displacement and stability of the sensors' positions.

A finite element model of an OTK stave for structural analysis was created, similar to the one used for the ITK stave. Based on this model, structural analyses were performed to evaluate the maximum deformation and the first natural frequency. The results indicate a sag of 135 μm for the OTK stave, with a corresponding first natural frequency of 76 Hz.

**Thermal characterization** According to the specifications for the OTK sensor, the cooling design must meet the following requirements:

- The overall sensor operating temperature should not exceed 30 °C.
- The temperature uniformity across a single sensor should be maintained within 5 °C.

Based on numerical simulations for different pipe diameters, and considering cooling system simplification, a pipe with an inner diameter of 5 mm using water cooling was selected as the baseline for the OTK stave cooling.

A water cooling fluid structure coupled finite element model was established to study the temperature distribution along the entire longitudinal length of the stave, considering a specific water cooling flow rate. The following configurations were used in the model:

- The detector dissipates heat at a flux of 300 mW/cm$^2$.
- Cooling water enters the stave with a flow velocity of 2 m/s and a temperature of 5 °C.
- Natural convection and radiative heat transfer were not considered.

Figure 5.29 shows the simulation result with both pipe inlets placed on the same side. The temperature gradient along the full stave length is controlled within 8.6 °C, and the maximum temperature remains below 14.6 °, satisfying the preliminary requirement. The temperature

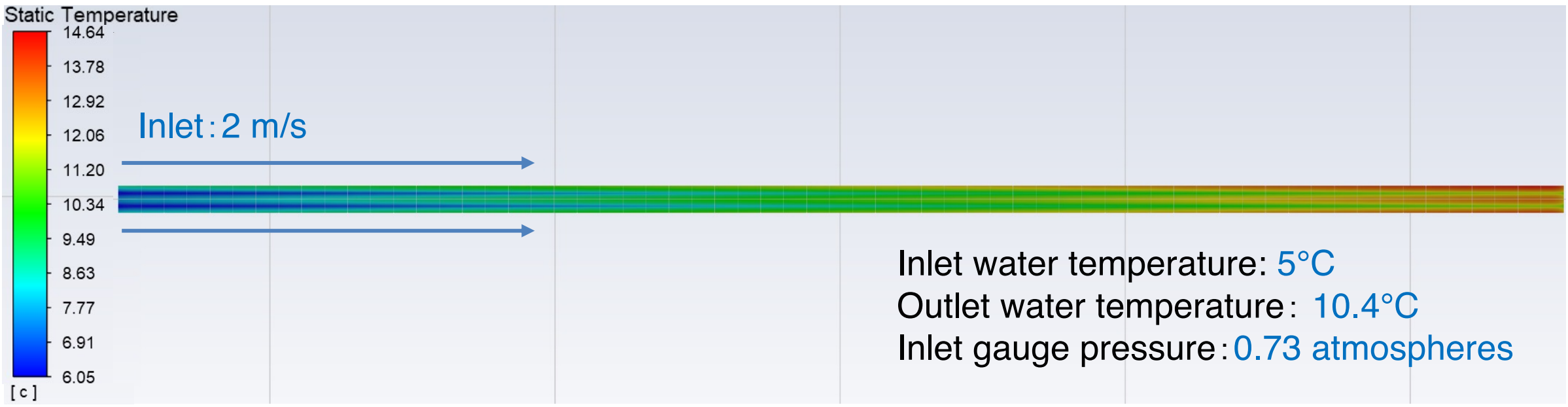

**Figure 5.29:** Simulation result of water cooling for the OTK stave, with a flow velocity of 2 m/s and a two-pipe inlet temperature of 5 °C. The temperature across the stave remains below 14.6 °C, and the temperature gradient along the stave is within 8.6 °C.

variation across a single sensor is found to be below 3 °C, which is within the specified limit.

#### 5.3.3.2 Endcap support

Figure 5.30 (a) illustrates the components of the OTK endcap, including a layer of sensor modules, a layer of cooling plates, connection rods, an inner structural ring, and an outer structural ring. To simplify the assembly and testing processes, the endcap is segmented into 16 sectors, each with its own dedicated detection, mechanical, and cooling structures. Figure 5.30 (b) shows a detailed view of one OTK sector, with main components summarized as follows:

- Sensor Modules: The sensors are glued to low-mass PCBs equipped with DC-DC converters, readout ASICs, and other components.
- Cooling Plate:
  - The cooling plate consists of two layers of high stiffness carbon fiber facesheets sandwiching a low-mass, carbon fiber honeycomb core with embedded serpentine cooling loops, sealed with high thermal conductivity foam. The composite pyrolytic graphite foam is the same as the one used in the OTK barrel stave.
  - The edges of the sector are enclosed by side closeouts, which interface with inlets and outlets of cooling loops. These closeouts, made of carbon fiber, serve as critical mechanical interfaces to the overall support structures, enhancing the sector's stiffness.
  - The sensors in the modules are attached to the cooling plate (more accurately, the carbon fiber facesheet) using thermal adhesive.

Sixteen composite rods are installed along the edges to join adjacent sectors, forming the complete OTK endcap disk. Figure 5.30 (c) illustrates the mechanical interface between neighboring sectors, where each composite rod extends from the inner to outer ring, connecting adjacent cooling plates. To reinforce the structural integrity of the assembly, inner and outer mounting rings are secured to the disk's perimeter, enhancing the connections between neighboring sectors. Together with the rods, these rings enhance the overall stiffness of the endcap and absorb induced stress. Additionally, the outer ring also serves as a routing structure for the cooling lines from the cooling plate.

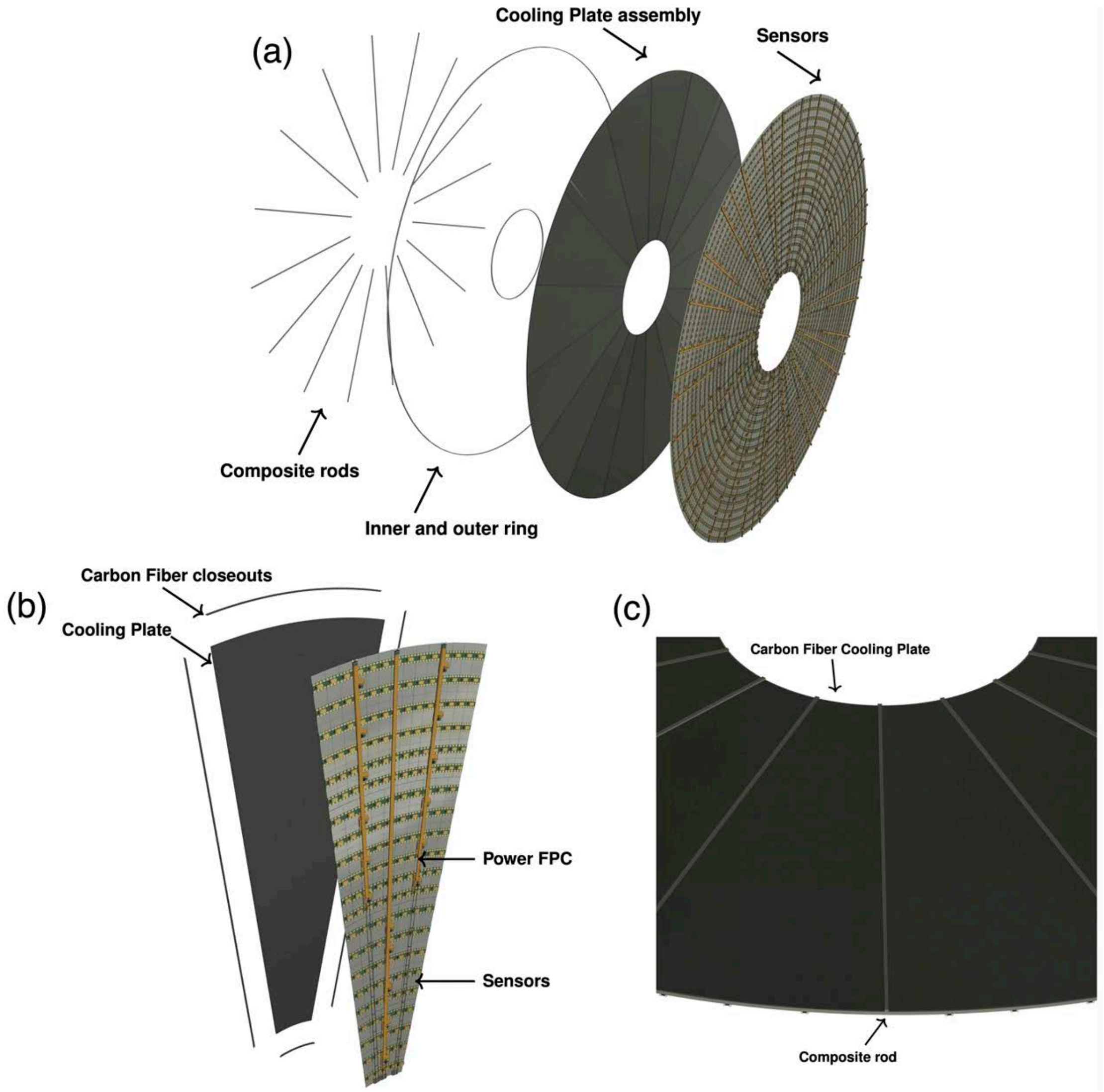

**Figure 5.30:** (a) Structure of the OTK endcap. It consists of a layer of sensor modules, a layer of cooling plates, connection rods, an inner structural ring, and an outer structural ring. (b) Structure of the OTK 1/16 sector, and (c) mechanical connection between sectors. Each sector consists of a sensor module layer and a cooling plate, enclosed by side closeouts. Sixteen composite rods are installed along the edges to join adjacent sectors, while inner and outer mounting rings are attached to the disk perimeter to reinforce inter-sector connections.

**Materials** The two facesheets of the cooling plate for each 1/16 sector are constructed from layers of high-modulus unidirectional carbon fiber material combined with cyanate ester resins. The interface between the cooling tubes and the facesheets is formed using thermally conductive carbon foam, such as Allcomp K9. Water circulating through cooling loops within the foam serves as the cooling medium for the OTK endcap.

The rods, inner ring, and outer ring, which connects sectors to an endcap disk, must be made from stiff materials. The rods are constructed from carbon fiber. For the inner ring and the outer ring, the primary candidate materials are the high-performance PEEK polymer and Torlon polyamide-imide technology.

Table 5.11 lists the estimated contributions of each OTK endcap disk to the material budget. The overall estimated material budget is 1.4% $X_0$.

**Table 5.11:** Estimation of the OTK endcap material contributions. The wall thickness of the cooling tubes, averaged over the entire endcap area and labeled with "†", is ~ 60 μm.

| Functional unit | Component | Material | Thickness [μm] | $X_0$ [cm] | Radiation Length [% $X_0$] |
|---|---|---|---|---|---|
| Sensor Module | PCB metal layers | Cu | | 1.436 | 0.200 |
| | PCB Insulating layers | Polyimide | | 28.41 | 0.070 |
| | Sensor | Silicon | 300 | 9.369 | 0.320 |
| | Glue | | 100 | 44.37 | 0.023 |
| | Other electronics | | | | 0.100 |
| Structure | Carbon fiber facesheet | Carbon fiber | 300 | 26.08 | 0.115 |
| | Cooling tube wall | Titanium | 200† | 3.560 | 0.168 |
| | Cooling fluid | Water | | 35.76 | 0.054 |
| | Graphite foam+Honeycomb | Allcomp+Carbon fiber | 3000 | 186 | 0.161 |
| | Carbon fiber facesheet | Carbon fiber | 300 | 26.08 | 0.115 |
| | Glue | Cyanate ester resin | 200 | 44.37 | 0.045 |
| Total | | | | | 1.371 |

**Structural characterization** The stiffness of the OTK endcap is primarily provided by the carbon fiber facesheets, honeycomb, and frames (including closeouts, rods, inner ring, and outer ring).

Structural analyses similar to those performed for the ITK endcap, were conducted to evaluate the maximum deformation for the OTK endcap in both the nominal and flat-lying positions. The results indicate a negligible sag (< 1 μm) in the nominal position, and a maximum of 1.4 mm in the flat-lying position. The OTK endcap is sufficiently stiff for both installation and operation.

**Thermal characterization** Similar cooling strategy to that of the OTK barrel is chosen as the baseline in the endcap, with the same water flow velocity and inlet water temperature. Similar to the ITK endcaps, the cooling loops for the OTK endcap are specifically designed to achieve a small temperature gradient across the entire surface of the endcap sensor plane. Figure 5.31 (a) shows the designed layout of the cooling loops for a 1/16 OTK endcap sector, consisting of five closed loops arranged across 40 coil layers. The inner diameter of the cooling coops is chosen to be 2.6 mm, with a wall thickness of 0.2 mm.

In the thermal simulations, the thermal conductivity of the cooling water, cooling pipes, graphite foam, carbon fiber face sheets, and detector sensors were taken into account.

Based on these configurations, Figure 5.31 (b) shows the simulation results for the temperature distributions of a 1/16 OTK endcap sector. As shown, the temperature of all OTK sensors remains below 15 °C. The maximum temperature difference across the sensor plane is below 8 °C, and the temperature difference across a single sensor is below 4 °C. Comparing the sensor operation limit of 30 °C and the temperature uniformity across a single sensor (< 5 °C), the OTK endcap cooling design meets the detector's requirements.

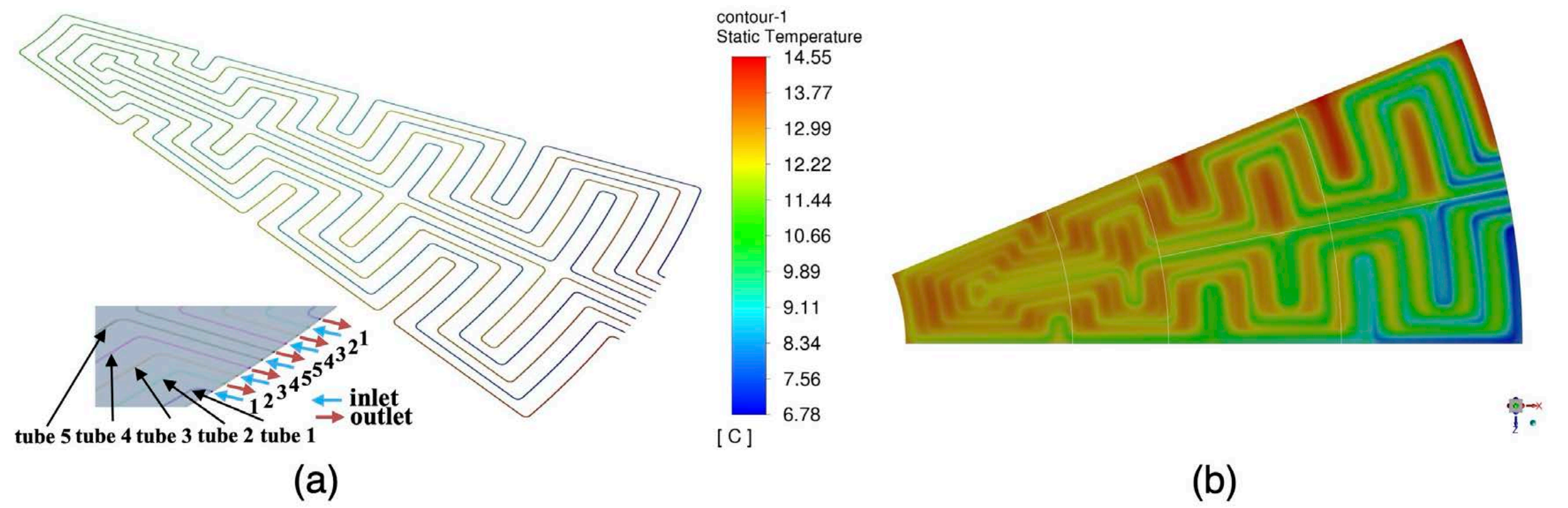

**Figure 5.31:** (a) Schematic of the 40 coil layer cooling loops designed for the 1/16 OTK endcap sector, featuring five closed loops arranged radially across the layers. (b) Simulated sensor temperature distributions for the 1/16 OTK endcap sector.

### 5.3.4 AC-LGAD sensor

Low Gain Avalanche Detector (LGAD) technology has been selected for use in the ATLAS High Granularity Timing Detector (HGTD) [11] and the CMS endcap timing layer [12] to mitigate pile-up effects in the HL-LHC experiment. LGAD sensors from several vendors have demonstrated timing resolutions of around 50 ps, both before and after irradiation. In particular, the LGAD developed by the Institute of High Energy Physics (IHEP) achieves a collected charge exceeding 15 fC and a timing resolution better than 35 ps before irradiation, and a charge of 4 fC with a timing resolution better than 50 ps after irradiation at a fluence of $2.5 \times 10^{15}\ n_{eq}/cm^2$ [13, 14].

However, despite their excellent timing capabilities, conventional LGADs face limitations in spatial resolution due to pixel size constraints. Furthermore, the presence of Junction Termination Extension (JTE) and P-stop structures between pixels introduces dead zones that reduce detection efficiency. To overcome these challenges and minimize the inactive area, AC-coupled LGADs (AC-LGADs) is being developed as an advanced evolution of LGAD technology.

In AC-LGADs, AC coupling electrodes and a dielectric layer are placed above the n+ layer of the LGAD, as illustrated in Figure 5.32. When a particle passes through the detector, the induced charge is distributed among electrodes near the impact position. By analyzing the signals collected from these adjacent electrodes, the particle's impact position can be precisely reconstructed. This enables AC-LGADs to deliver both high-precision timing and spatial measurements while achieving a nearly 100% fill factor, significantly enhancing their efficiency and performance in particle detection applications.

AC-LGADs have been extensively investigated by many research institutes and companies including National Institute for Nuclear Physics (INFN), Fondazione Bruno Kessler (FBK), Brookhaven National Laboratory (BNL), Hamamatsu (HPK), IHEP, and others [15, 16, 17, 18]. These detectors enable four-dimensional (4D) tracking by combining precise spatial and temporal measurements, and can also serve as ToF detectors [19]. Specifically, AC-LGADs with strip-shaped readout electrodes reduce the number of readout channels while maintaining exceptional spatial resolution along the bending direction [20], which has been adopted as the baseline design for the OTK.

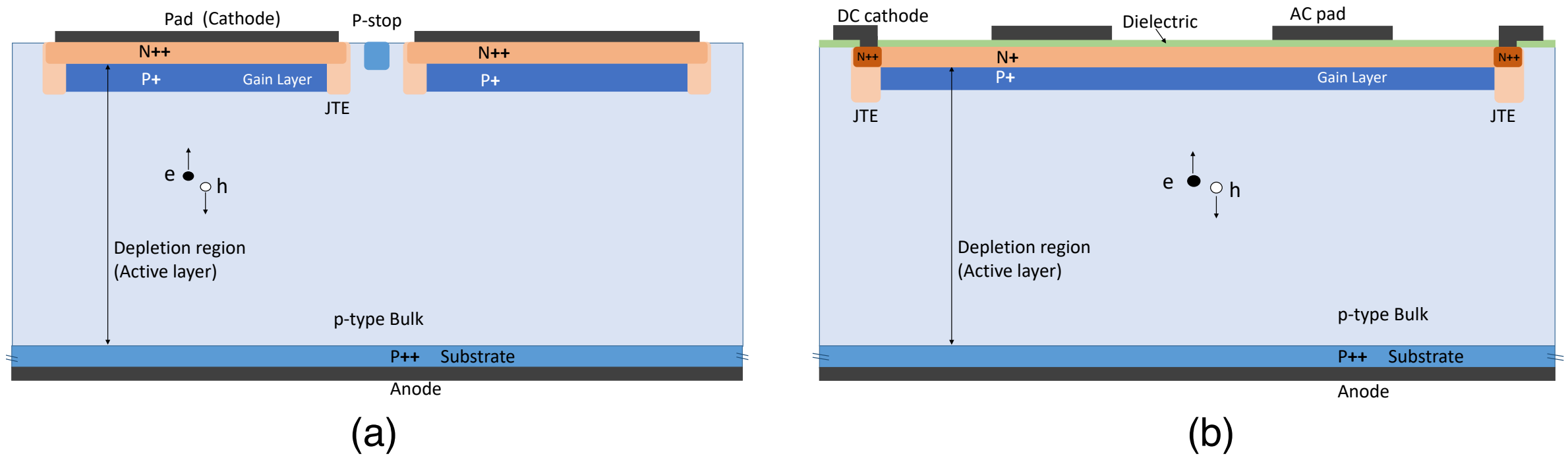

**Figure 5.32:** Structures of (a) standard LGAD and (b) AC-LGAD. In the standard LGAD (DC-LGAD), the readout electrode (pad) connects to the n++ layer, with Junction Termination Extension (JTE) and P-stop structures creating dead zones. In contrast, the AC-LGAD uses a thin dielectric layer to separate the metal AC-coupling electrodes from the n+ layer, achieving a nearly 100% fill factor.

Extensive R&D efforts have been undertaken at IHEP to advance the development of AC-LGADs. The process and structural parameters of these devices have been extensively simulated by using TCAD software, exploring the impact of various process parameters and structural configurations on detector performance.

#### 5.3.4.1 AC-LGAD simulation

Simulations of AC-LGAD devices were performed using the TCAD software, incorporating all key structural components such as the n+ resistance layer dose, AC coupling dielectric material and thickness, and metal pad ratio. The simulated model structure is illustrated in Figure 5.33. These simulations focus on AC-LGAD specific structures that differ from standard LGADs and have a significant impact on both timing and spatial resolution.

Key findings include:

- **Dielectric material and thickness:** Variations in the dielectric material and its thickness significantly influence spatial performance (Figure 5.34). Silicon nitride, with its higher dielectric constant compared to silicon oxide, enhances capacitive coupling when the thickness remains the same. As a result, AC-LGADs using high-dielectric-constant materials such as silicon nitride for the AC coupling layer can achieve improved spatial resolution.
- **n+ layer dose:** The n+ layer dose affects the charge sharing between electrodes, signal shape, and spatial resolution. As the n+ dose decreases, the resistivity of the layer increases, resulting in more signal charge being collected by the electrode closest to the particle impact position. The signal also changes more rapidly with the impact position, leading to improved spatial resolution [21].

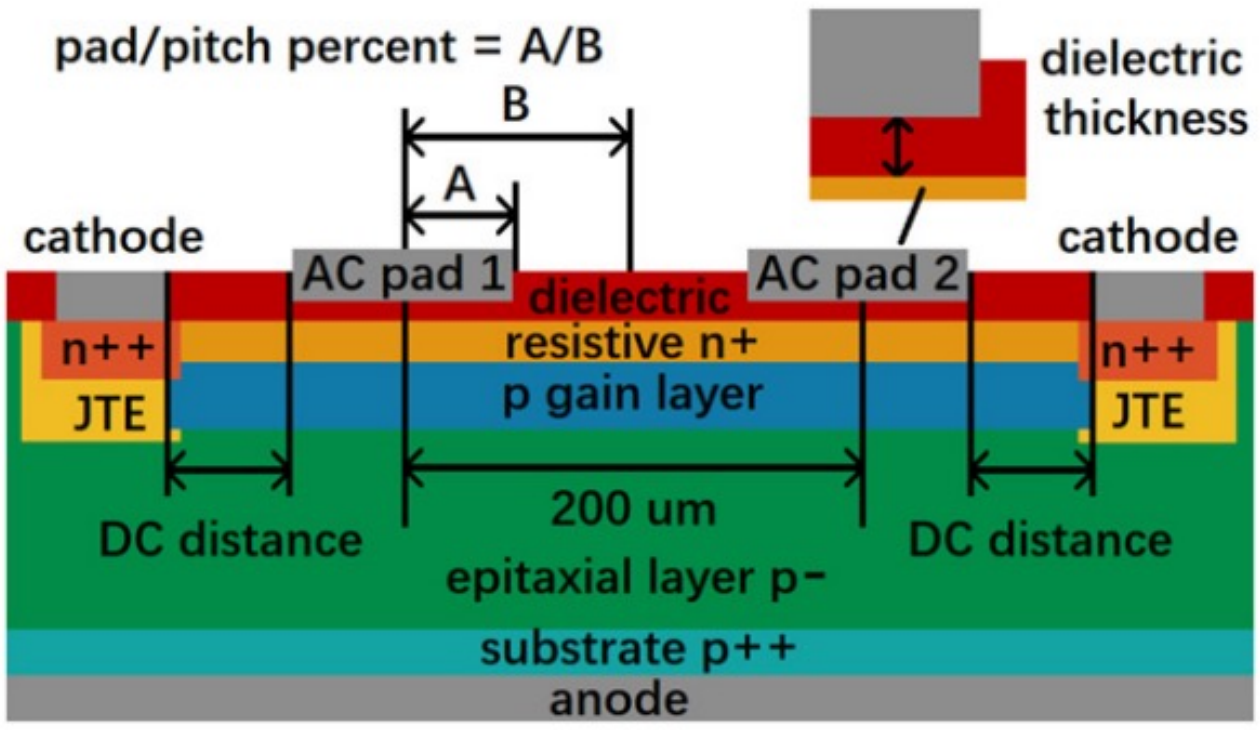

**Figure 5.33:** Sketch of the AC-LGAD simulation model with two AC pads, incorporating key structural components such as the n+ resistance layer dose, AC coupling dielectric material and thickness, and the metal pad ratio.

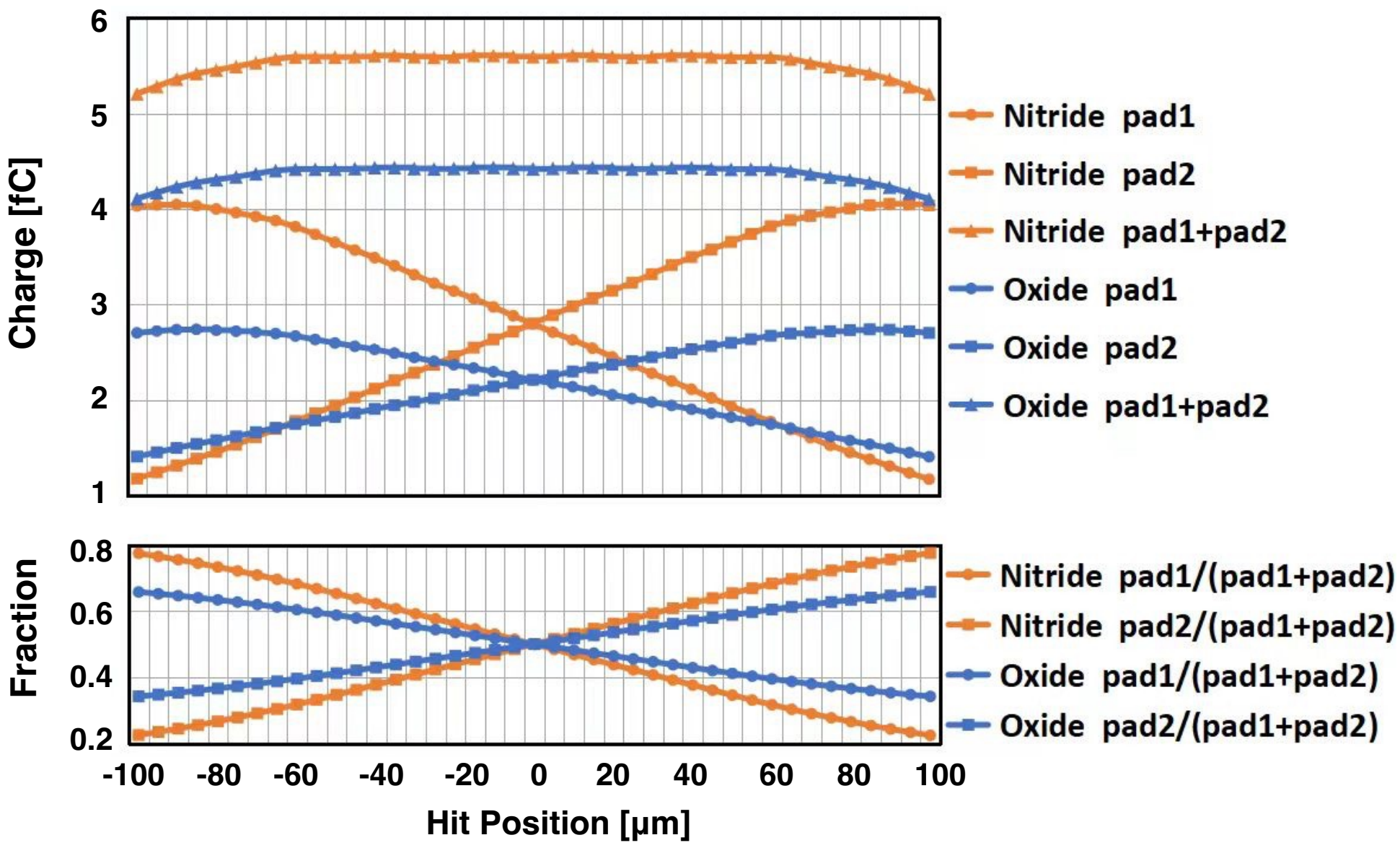

**Figure 5.34:** Charge collection of two neighboring pads (pad1 and pad2) as functions of hit position for two different dielectric materials: oxide (blue curve) and nitride (orange curve).

- **Metal pad pitch size and other structural parameters:** Additional structural parameters, such as the metal pad pitch size, were also simulated to evaluate their effects on overall detector performance. More details are provided in Ref. [21].

Based on these findings, various R&D prototypes were fabricated to further explore AC-LGAD performance. Dedicated testing systems were designed and implemented to study the timing and spatial resolutions of the developed devices.

### 5.3.4.2 Testing setup

The testing platforms, utilizing Transient Current Technique (TCT) laser scans and Beta tests, are illustrated in Figure 5.35. These setups comprise readout boards, amplifiers, oscilloscopes (Teledyne LeCroy HDO9204 with 2 GHz bandwidth and 20 G/s sampling rate), laser sources, and Beta sources. The AC-LGAD electrodes were wire-bonded to a four-channel readout board, which was designed according to the single-channel readout board developed by the University of California Santa Cruz [22]. Signals from the AC-LGAD electrodes were processed through a two-stage preamplifier system and subsequently recorded by the oscilloscope.

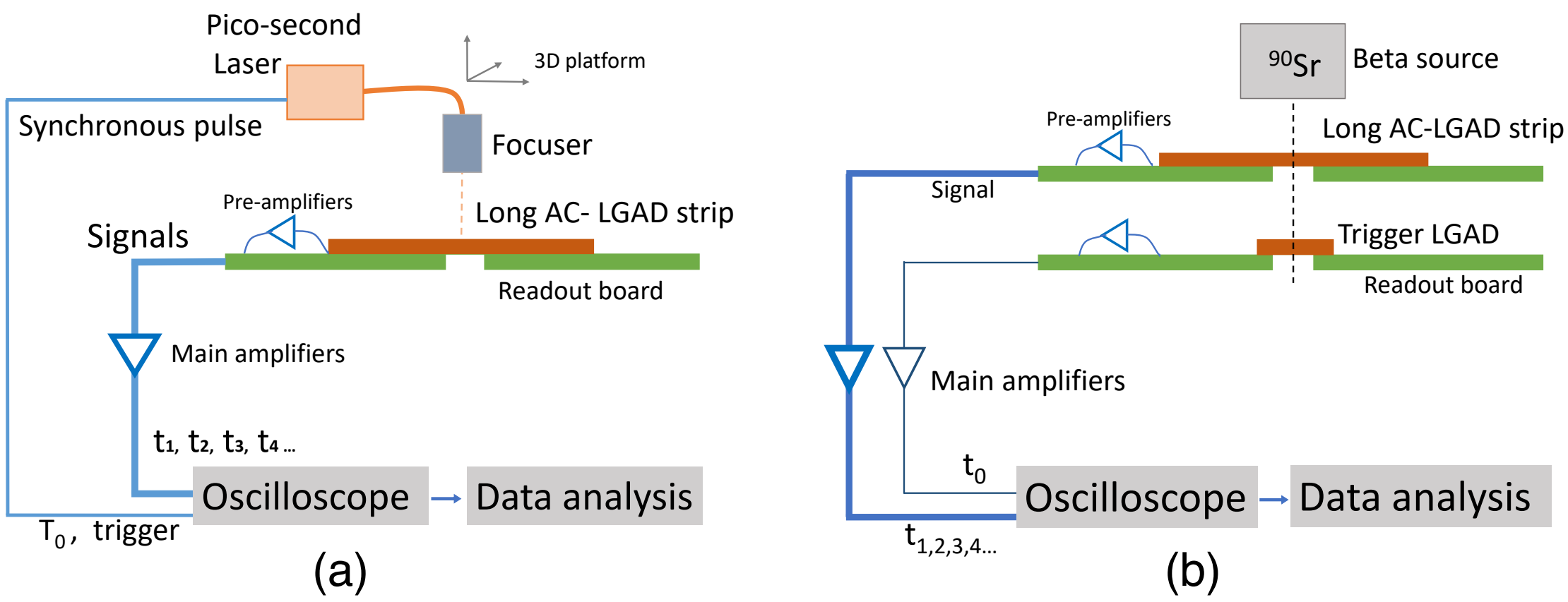

**Figure 5.35:** The schematics of the (a) TCT testing platform and (b) Beta testing platform. Each setup includes either laser or beta source, readout board, amplifiers, and oscilloscope. AC-LGAD electrodes are wire-bonded to a four-channel readout board, with signals processed through two-stage preamplifiers and recorded by the oscilloscope.

To evaluate the timing resolution of the DUT, a trigger LGAD with a timing performance ($\sigma_{\text{trigger}}$) of 28.5 ps was employed as the reference device. Signals from both the trigger LGAD and the AC-LGAD strip were recorded, and the time difference of flight for a Minimum Ionizing Particle (MIP) between the two devices was determined and defined as $\Delta T$. The standard deviation of $\Delta T$ ($\sigma_{\Delta T}$) was then calculated. Using this value, the timing resolution of the DUT sensor can be derived using the following equation:

$$\sigma_{\text{DUT}} = \sqrt{\sigma_{\Delta T}^2 - \sigma_{\text{trigger}}^2} \tag{5.1}$$

Transient Current Technique (TCT) laser scans were employed for highly accurate position resolution measurements. In this setup, a picosecond laser pulse with a wavelength of 1,064 nm strikes AC-LGAD sensor at a frequency of 20 MHz using a focuser. The focuser, mounted on a precision positioning table, ensures movement accuracy better than 1 µm. The laser systematically scans across the AC-LGAD sensor in µm steps. At each position, 1,000 waveforms are recorded before advancing to the next point. By analyzing the signals collected from electrodes near the laser hit position, the impact position is reconstructed. This enables the evaluation of spatial resolution through laser tests. The spatial resolution is defined as the standard deviation ($\sigma$) of the difference between the laser's actual position and the reconstructed position.

#### 5.3.4.3 Strip AC-LGAD prototype

A strip AC-LGAD prototype has been designed by IHEP [18]. It comprises strip electrodes, a dielectric layer, Direct Current (DC) electrodes, an n+ layer, a p+ layer, a p-type bulk, a p++ layer, and a backside aluminum anode. Figure 5.36 shows its detailed layout and schematics. It has a total thickness of 775 µm, includes a 50 µm p-type epitaxial layer (active layer). The length of the strip electrodes is 5.65 mm. Three pitch sizes of 250 µm, 200 µm, and 150 µm were designed, with the width of the strip metal electrodes being 100 µm.

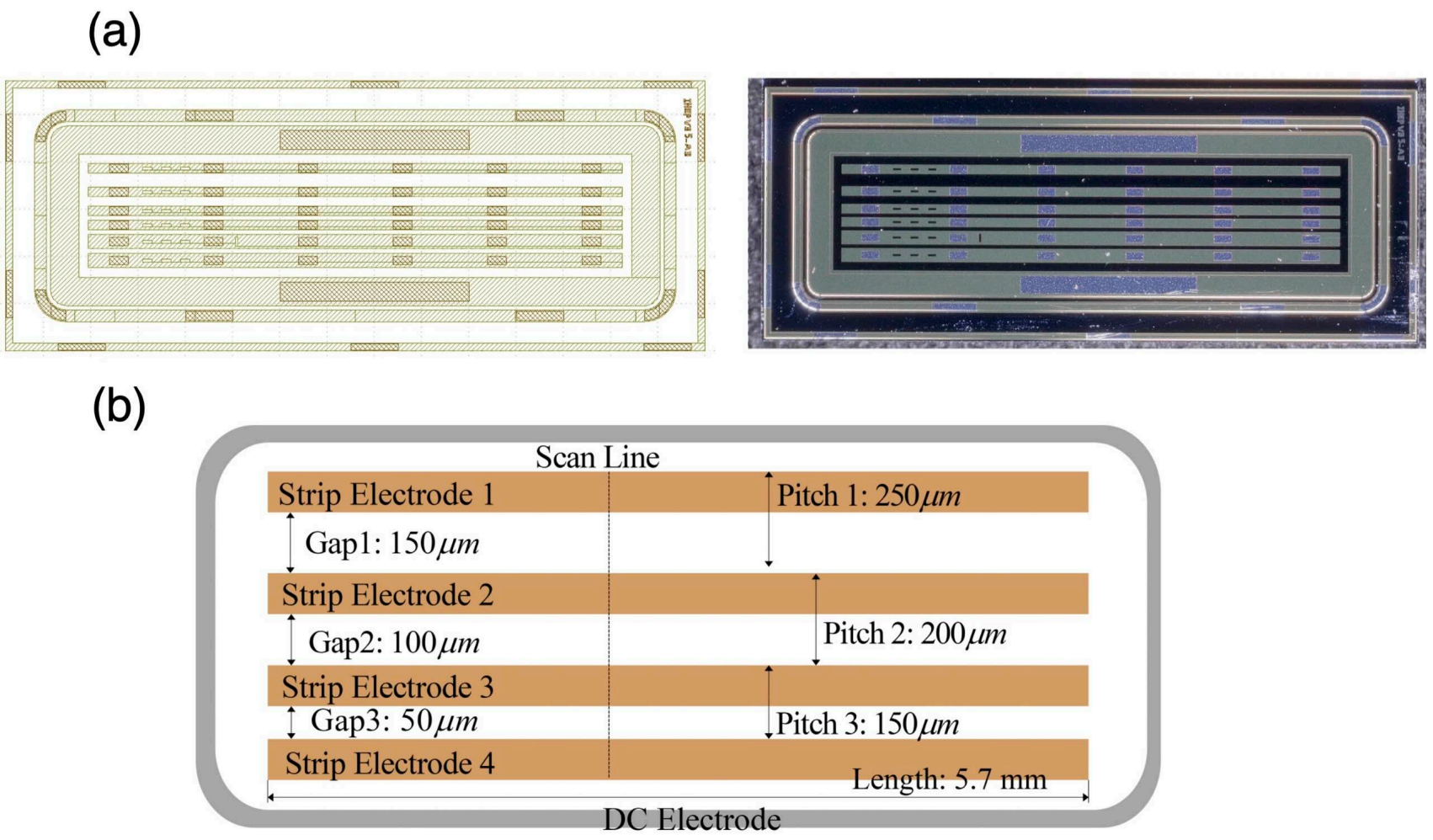

**Figure 5.36:** (a) The strip AC-LGAD layout and prototype, and (b) the schematic structure of the AC-LGAD prototype. The length of the strip electrodes is 5.65 mm, with three pitch sizes of 250 µm, 200 µm, and 150 µm. The width of the strip metal electrodes is 100 µm.

The breakdown voltage of the device is 385 V, with a leakage current on the order of 10 nA. The capacitance value is 28.9 pF when fully depleted. The gain layer depletion voltage ($V_{gl}$) is 21.5 V, while the full depletion voltage ($V_{fd}$) of the sensor is 34.8 V. The coupling capacitance between the strip electrodes and the DC electrode/n+ layer is measured to be 170 ± 0.2 pF.

**Timing performance from Beta testing** The timing resolution of the sensor was evaluated using a $^{90}$Sr beta source, where a calibrated LGAD (trigger LGAD) was placed above the AC-LGAD strip sensor and serves as the timing reference. The $^{90}$Sr beta source was positioned above the trigger LGAD to provide a stream of minimum ionizing particles.

Timing signals from the trigger LGAD and the AC-LGAD strip sensor were recorded. The time difference, denoted as $\Delta T$, was calculated as the difference between the signal arrival time of the trigger LGAD and that of the AC-LGAD strip sensor:

$$\Delta T = T_{\text{trigger}} - \frac{\sum_i a_i^2 T_i}{\sum_i a_i^2} \tag{5.2}$$

where $T_{\text{trigger}}$ is the arrival time measured by the trigger LGAD, while $T_i$ and $a_i$ are the arrival time and signal amplitude from strip electrode $i = 1, 2, 3$ of the AC-LGAD sensor, respectively. This amplitude weighting method enhances the contribution from the electrode closest to the particle hit position, effectively reducing noise and improving the timing resolution. The measured standard deviation of the time difference is $\sigma_{\Delta T} = 47.1$ ps. Considering the timing resolution of the trigger LGAD is $\sigma_{\text{trigger}} = 28.5$ ps, the timing resolution of the AC-LGAD strip sensor is calculated using Equation 5.1 and found to be $\sigma_{\text{AC-LGAD}} = 37.6$ ps.

**Spatial resolution** The TCT was used to study the spatial resolution of the prototype. When a focused laser pulse hits the sensor surface, the amplitude of the induced signal varies with the distance from the electrodes. This causes variations in the collected charge across the electrodes, which can be used for precise position reconstruction.

The position reconstruction method relies on the relative signal amplitudes recorded by the two strip electrodes nearest to the laser hit position. For example, when the laser or a particle hits the region between electrodes 1 and 2 (referred to as "gap 1" in Figure 5.36 (b)), the signal amplitudes on these two electrodes vary as a function of the hit location. The reconstructed hit position is characterized the amplitude ratio $R = a_2/(a_1 + a_2)$.

To obtain the spatial resolution, 1,000 waveforms were recorded for each laser hit position, resulting in 1,000 reconstructed positions per scan point. The difference between the reconstructed and the actual laser hit positions was computed for each event. For each gap region (e.g., gap 1), these differences were collected and plotted as a histogram. The standard deviation of the fitted distribution was used to estimate the device's spatial resolution.

Figure 5.37 shows the differences between the reconstructed and true hit positions for three pad-pitch sizes. The spatial resolutions for the AC-LGAD strip sensor with pitch sizes of 250 μm, 200 μm, and 150 μm are found to be 12.8 μm, 10.9 μm, and 8.3 μm, respectively.

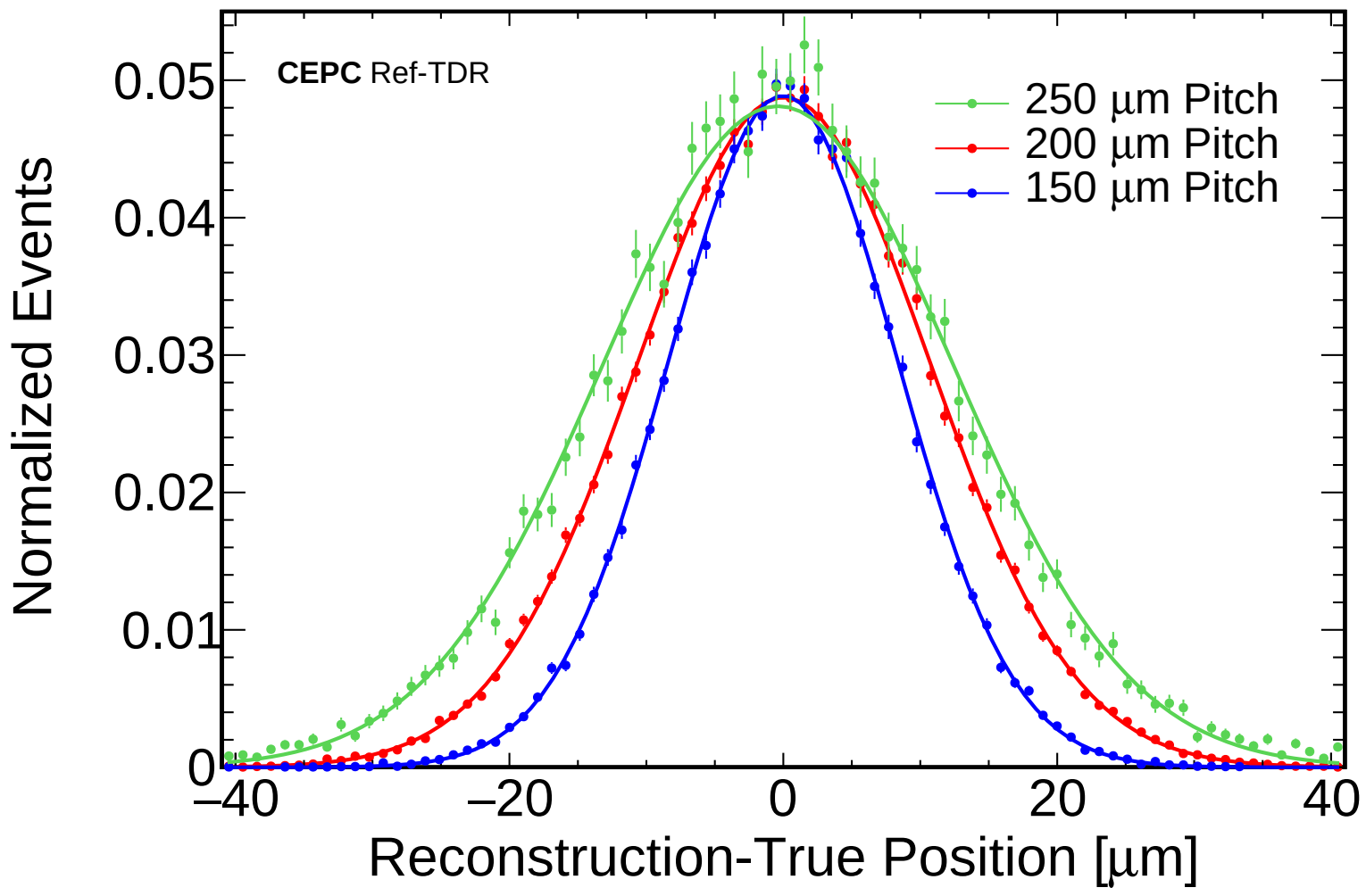

**Figure 5.37:** Distributions of difference in reconstructed and true hit positions for strip LGAD prototype with different pad-pitch sizes from TCT laser setup.

**Summary** Results from the 5.65 mm strip sensor prototype demonstrate a timing resolution of 37.6 ps and a spatial resolution of 8.3 μm (based on laser test with a strip pitch of 150 μm), fulfilling the performance requirements of the OTK system. This shows a great potential for the AC-LGAD strip as a promising candidate for 4D trackers in future particle physics experiments such as the CEPC. However, further studies are necessary to evaluate the timing and spatial performance of the longer AC-LGAD strip sensors as required by the OTK layout. To optimize the design, an isolated structure will be introduced into the AC-LGAD configuration to reduce sensor capacitance. Careful attention will be paid to the potential impact of this modification on spatial resolution. These refinements aim to ensure the sensor's compliance with the stringent requirements of the OTK system.

### 5.3.5 LGAD readout ASIC

The LGAD readout ASIC, LGAD Timing and Readout Integrated Chip (LATRIC), is designed to achieve both high time resolution for precise timing and good charge resolution for accurate position reconstruction, to fully exploit the sensor's outstanding performance. This section outlines the performance requirements, design considerations, and recent prototype testing of LATRIC.

#### 5.3.5.1 LATRIC architecture and requirements

LATRIC will feature 128 readout channels. The pitch between neighboring readout channels must be less than 100 μm to match the pitch of the LGAD strips. Each channel in LATRIC includes a preamplifier [23], a discriminator, and TDC for measuring both the Time-of-arrival (TOA) and Time-over-threshold (TOT), as illustrated in Figure 5.38. The ASIC also integrates common digital components, including a clock generator, data alignment logic, slow control configuration, and data transmission interfaces.

The requirements of the ASIC are driven by the electrical connections, targeted time resolution, and power constraint, etc. A summary of the final configuration for the LATRIC ASIC is presented in Table 5.12.

The time resolution of the OTK system depends on both the LGAD sensor and LATRIC. As the LGAD sensor exhibits a jitter of approximately 40 ps, to achieve an overall time resolution of 50 ps per hit, the contribution of LATRIC must be kept below 30 ps. Timing resolution is mostly limited by jitter and time walk. Given the small signal amplitude from the LGAD sensor, the analog front-end circuits — especially the preamplifier and discriminator — dominate the electronics' jitter contribution and are therefore critical for achieving OTK's precision timing performance. Time walk effects can be mitigated by applying corrections based on the correlation between TOA variations and TOT values.

The TOT serves as a charge indicator, as the signal amplitude is correlated with the duration of the signal above a specific threshold. Charge resolution depends on the intrinsic variations in the TOT width as well as time resolution. The Most Probable Value (MPV) of the charge collected from the LGAD sensor is approximately 16 fC. Considering the charge sharing with adjacent strips, the expected operating charge range is 8-50 fC.

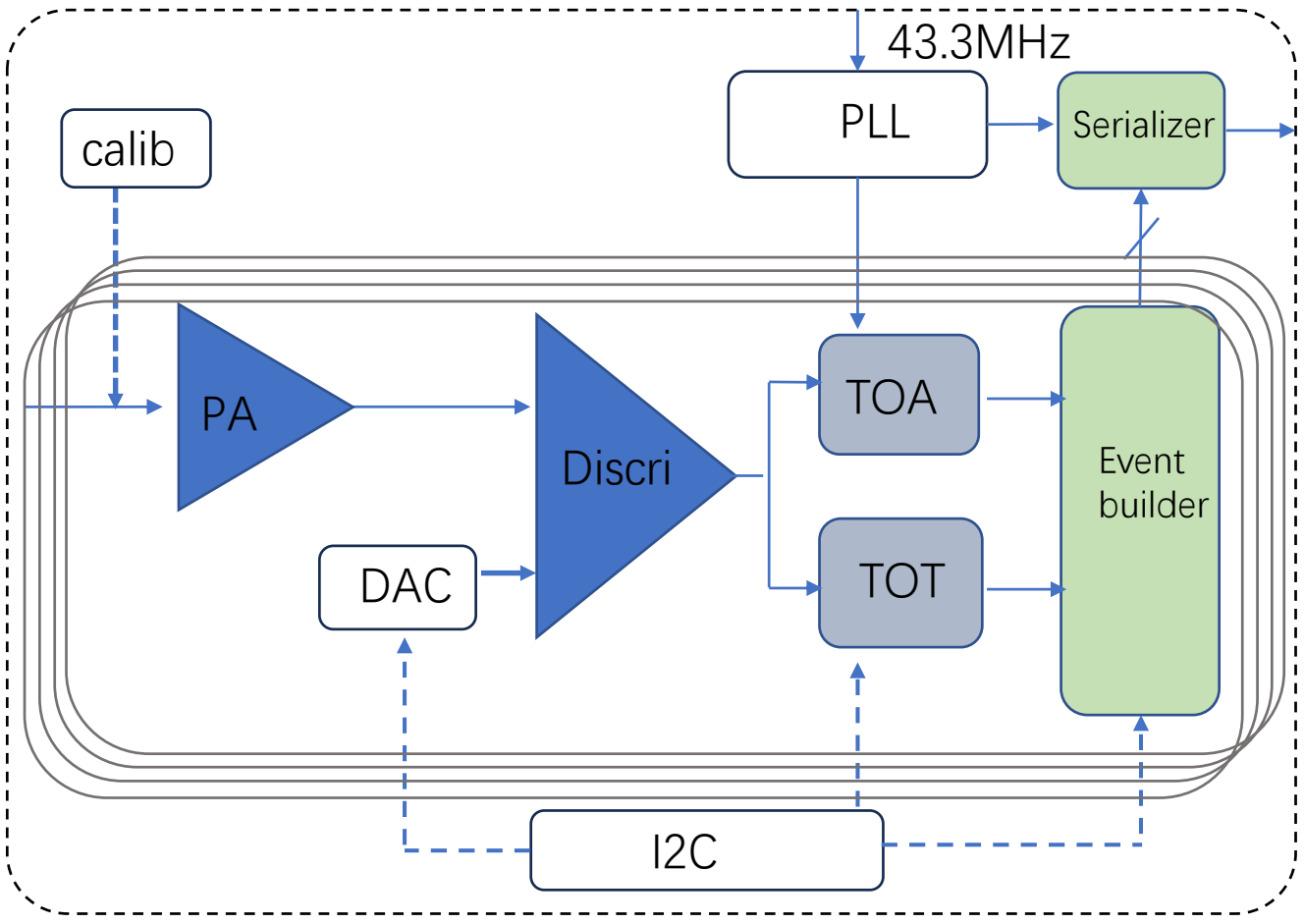

**Figure 5.38:** Global architecture of the LATRIC ASIC. The signal from the sensor is first amplified by the preamplifier (PA), and then discriminated through the discriminator (Discri) using a threshold provided by a DAC. The Time-To-Digital Converter (TDC) measures both the Time-of-arrival (TOA) and the Time-over-threshold (TOT). A Phase-Locked Loop (PLL) supplies the reference clock, and the data are transmitted via a serializer. An Inter-Integrated Circuit (I2C) interface is implemented to configure key parameters of the DAC, TDC, and the event builder.

**Table 5.12:** Configuration for the LATRIC ASIC (assuming a detector capacitance $C_d$ = 8 pF)

| Parameter | Value |
|---|---|
| Chip size | 1.2 cm × 0.5 cm |
| Voltage | 1.2 V |
| Number of channels | 128 |
| Channel pitch | < 100 μm |
| Single channel noise (ENC) | 0.8 fC |
| Cross-talk | < 10% |
| Maximum jitter | 30 ps at 16 fC |
| Minimum threshold | 4 fC |
| Dynamic range | 8 fC–50 fC |
| TDC conversion time | < 23 ns |
| Power dissipation per ASIC | 1.5 W (for occupancy < 1%) |
| Data size per fired channel | 48 bits |
| e-link driver bandwidth | 43.33 Mbps or 86.67 Mbps |
| Technology node | 55 nm |

#### 5.3.5.2 Single-channel readout electronics

This section presents key considerations in the designs of single-channel electronics — preamplifier, discriminator, and TDC — to achieve high performance in terms of compact area, low power consumption, and high precision.

**Preamplifier and Discriminator** As illustrated in Figure 5.39 (a), the preamplifier comprises two stages: a cascade amplifier (M1 and M2) as the first stage and a source follower (M3) as the second stage. The gain and bandwidth of the preamplifier depend on the transconductance (*Gm*) of transistor M1 and *Rf*.

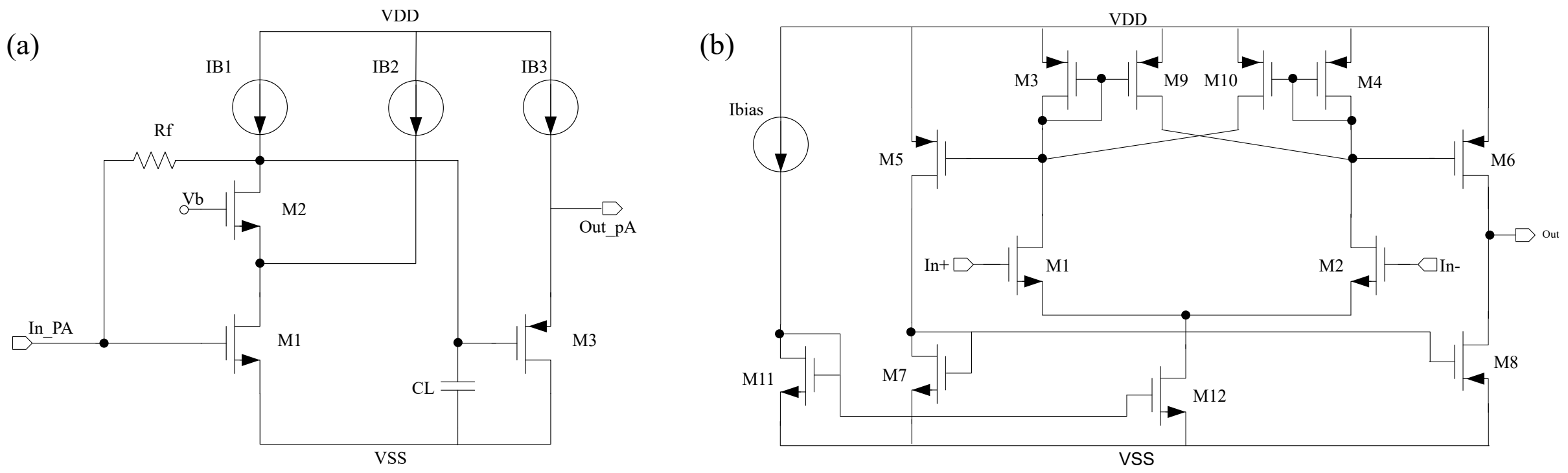

**Figure 5.39:** Schematic of the (a) preamplifier and (b) discriminator in LATRIC. The preamplifier is a transimpedance amplifier, where the feedback resistor *Rf* converts the input current to a voltage, followed by a buffer stage. The discriminator consists of two stages, with transistors M9 and M10 used for hysteresis regulation.

The discriminator comprises three stages of fully differential amplifiers, a comparator, and an internal buffer. The three amplifier stages receive small input pulses and amplify them to generate larger pulses suitable for the comparator. As illustrated in Figure 5.39 (b), the comparator consists of two stages. The first stage is a high-gain common-source differential amplifier using transistors M1 and M2. The second stage converts the differential output of the first stage into a single-ended signal. The discriminator threshold is connected to the inverting input and is set by an internal 10-bit DAC.

**TDC** The schematic of the TDC core [24] is illustrated in Figure 5.40. The LATRIC design faces two primary challenges: the limited area per channel and the need to achieve both time and charge measurements while maintaining low power consumption. Additionally, the pitch of the LGAD strips is only 100 μm, which necessitates that the height of the single-channel circuitry must also be less than 100 μm.

To address these constraints within a smaller area, a single delay line is employed to simultaneously measure the time of TOA and TOT, with each delay cell providing a delay of 30 ps. Flip-flops record the times of the signal's rising edge, falling edge, and the reference clock's rising edge, storing these values sequentially in registers. The chip employs a single delay line without a Delay-Locked Loop (DLL), using a cyclic structure to minimize the number of delay cells.

The delay of the delay line is affected by process variations, power supply voltage, and temperature. Therefore, a pulse self-calibration scheme is essential to compensate for these variations. This calibration is performed periodically using the system clock to measure and adjust the delay chain. The TDC output data width comprises 8 bits for TOT, 9 bits for TOA, 1 bit for hit flag, 8 bits for calibration, 7 bits for channel identification (128 channels), 8 bits for bunch ID, and 7 bits for chip ID, resulting in a total of up to 48 bits.

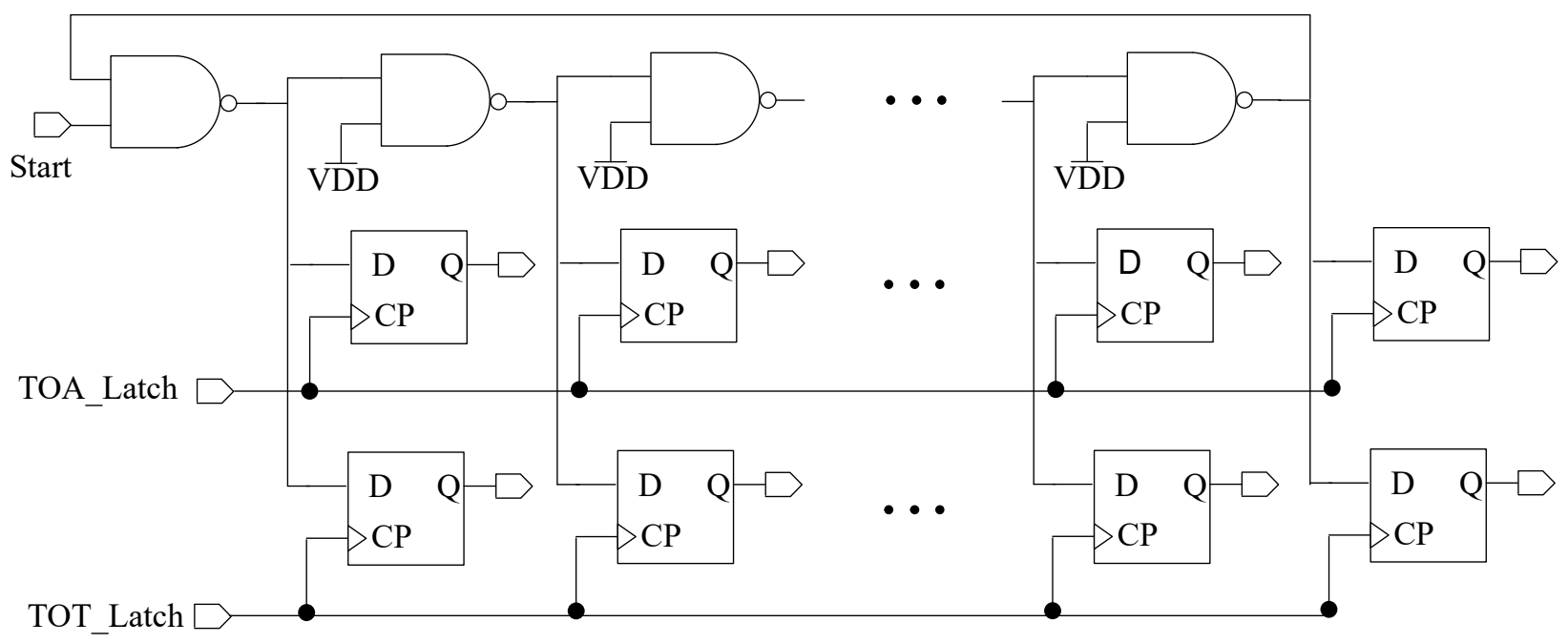

**Figure 5.40:** Basic structure of the TDC design based on a single multi-tapped delay line. The delay line consists of several NAND gates. The Time of Arrival (TOA) and Time Over Threshold (TOT) are measured independently. The states are latched upon the arrival of a rising or falling edge.

#### 5.3.5.3 Digital blocks and data handling

The digital blocks complement the analog front-end circuitry and operate at the full chip level to perform precise clock distribution, robust configuration, and ensure efficient data readout across all channels.

**Clock generation unit** The clock generator unit supplies clock signals to all functional blocks within the LATRIC chip. The oscillator in the PLL operates at 1.39 GHz, the highest frequency, and receives a 43.33 MHz clock as input. Effective clock distribution, particularly minimizing skew and jitter, is a critical design challenge. A binary tree, the most common and conservative clock distribution scheme, is planned for implementation. In this scheme, the clock signal branches from a central point to all destination nodes. The 7-stage tree structure ensures a balanced clock distribution network, with equal path lengths from the clock source to each channel. Buffers are added along the transmission path to maintain the quality of the clock signal.

**Data readout process** The TDC data from each channel is stored in a circular buffer. A scrambler is employed, and a Pseudo-Random Binary Sequence (PRBS) block is incorporated for testing purposes. The LATRIC chip supports three serial output data rates, depending on the anticipated occupancy of the chip.

The serial output data rate is configurable at either 43.33 Mbps or 86.67 Mbps for each LATRIC chip. Forward error correction schemes, such as 8b/10b or 64b/66b encoding, are applied to the serial output.

**Slow control** Slow control mechanisms are implemented to configure the registers within the chip. The chip includes various registers that manage different functions and operating modes, such as amplifier gain settings, test modes, data acquisition modes, and calibration controls. An I2C link is integrated into the ASIC and is configured through the I2C master in the FEDI chip.

**Monitoring** The ASIC monitoring primarily focuses on operational temperature, supply voltage, and the leakage current of the sensors. Although the ASIC itself is not highly sensitive to temperature, the performance of the LGAD sensors is temperature-dependent. Temperature monitoring via the ASIC channels also enables the detection of cooling failures. Voltage monitoring within the chip is used to assess functionality and can trigger the shutdown of malfunctioning modules when necessary.

#### 5.3.5.4 Prototype

A prototype chip, FPMROC, has been designed, incorporating eight complete readout channels. Each channel consists of a low-noise preamplifier, a discriminator, and a TDC for measuring both the TOA and TOT [24]. An integrated data event builder manages the data flow, while fast data serialization and a data driver handle the output. Peripheral circuits include DACs for calibration and threshold adjustment, a PLL to generate high-quality clocks, and a Serial Peripheral Interface (SPI) module for slow control. Additionally, the prototype features a charge injection circuit for testing and calibration purposes.

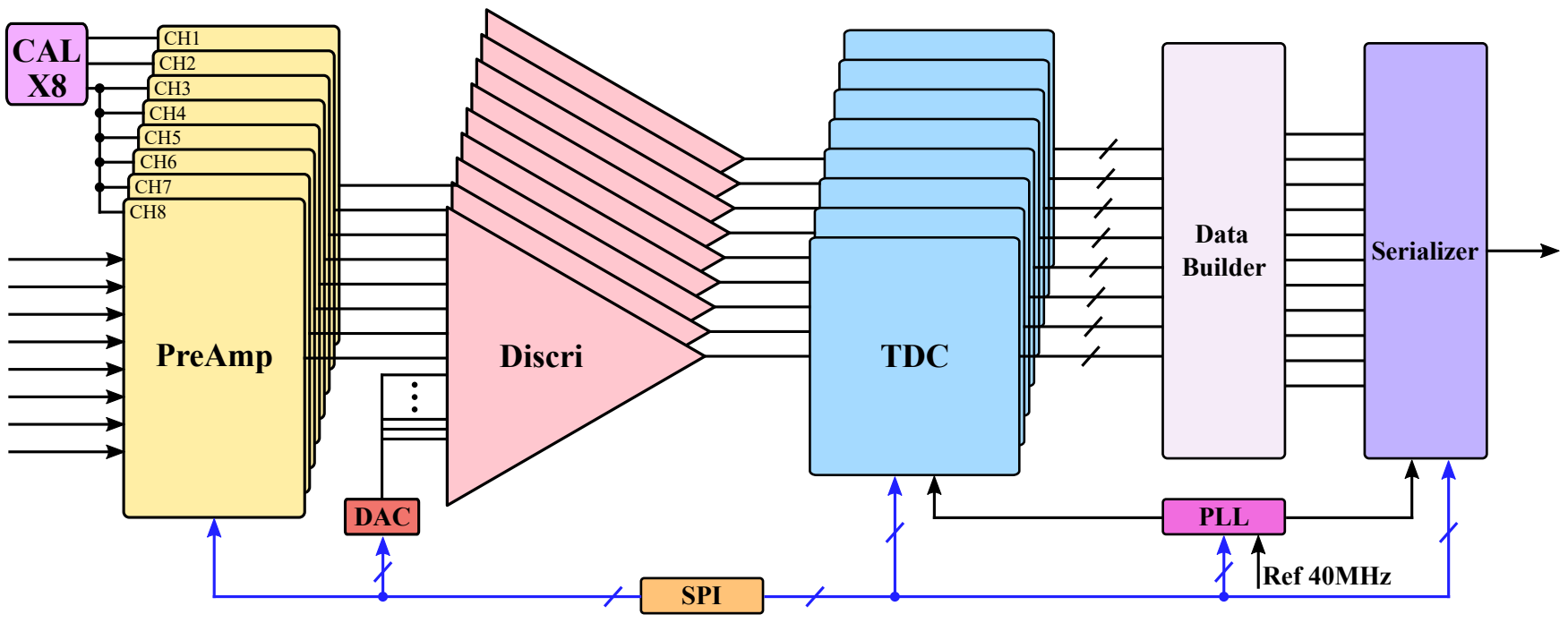

**Figure 5.41:** Block diagram of the FPMROC. It integrates eight channel circuits, including preamplifier, discriminator, TDC, event builder, and serializer. The TDC measures the Time of Arrival (TOA) and Time Over Threshold (TOT) of signals from the MCP-PMT. Chip configuration is realized via Serial Peripheral Interface (SPI).

Figure 5.41 illustrates the block diagram of the FPMROC ASIC. Eight channels collectively employ a data event builder for buffering, framing, scrambling, and encoding parallel data from multiple channels. The ASIC incorporates a serializer for off-chip data transmission at a rate of 10.24 Gbps, as well as a low-jitter LC-based PLL to generate 5.12 GHz and 40 MHz clocks for the serializer and TDCs, respectively. Additionally, an SPI interface is integrated to provide configurations of up to 200 bits.

Testing of this chip was successfully completed at the end of 2024. Further details on the design and performance of FPMROC can be found in Ref. [25]. In the second half of 2024, a new readout ASIC scheme was designed and submitted for wafer production in April 2025, along with the design of the new ASIC test system. This ASIC is intended to validate the performance of its key components — including the preamplifier, discriminator, and TDC modules — as well as to finalize the design of the ASIC test system. Performance testing, including radiation hardness testing for each component, is expected to be completed by the end of 2025. Further R&D plan are covered in Section 5.3.6.

### 5.3.6 Future plan

Significant R&D efforts are required to meet the design requirements of the OTK detector. These include major advancements in AC-LGAD sensor area scaling, substantial progress in ASIC development, and system-level prototyping of mechanical structures and cooling systems.

#### 5.3.6.1 Development of AC-LGAD strip sensor

The development of high-performance long strip sensors is still challenging as the impact of strip length on both timing and spatial resolution is not fully understood. It is one of the main focus of future R&D. Figure 5.42 shows a detailed tape-out plan for the LGAD strip sensors and their testing schedule, with key elements summarized below.

**Sensor Development and Testing** Using TCAD tools, models of AC-LGAD with various strip lengths and process parameters will be developed to study the impact of strip length on signal quality and to optimize spatial and timing performance of strip AC-LGAD. Prototypes with strip lengths of 1 cm, 2 cm, and 4 cm will be fabricated and tested. Meanwhile, sensors with one strip length will also have prototypes with different n+ dose and coupled capacitance. Testing will cover current-voltage, capacitance-voltage, spatial resolution, and timing performance, with subsequent radiation and beam tests.

**Design Adjustments** Based on test results, the strip length and pitch size may be modified to balance performance, yield, and readout requirements. One of our backup plans includes reducing the strip length from 4 cm to 2 cm and increasing the pitch size from 100 μm to 200 μm to maintain similar readout channels and power consumption. The final decision will be made based on the performance and yield.

#### 5.3.6.2 Development of LGAD readout ASIC

The development timeline of the LGAD readout ASIC (LATRIC) is closely aligned with that of the LGAD sensor, with the main focus on the design of a multi-channel ASIC, as shown in Figure 5.42.

Following the successful development and testing of the single-channel LATRIC V0, an 8-channel ASIC (LATRIC V1) will be developed in Q4 2025, incorporating a digital logic control section. This integrated ASIC design will then be submitted for wafer fabrication. By the first half of 2026, the test system design for this ASIC will be finalized. In mid-2026, the performance of the 8-channel ASIC will be evaluated, including crosstalk studies, followed by connection, debugging, and radiation-hardness testing in conjunction with the LGAD sensor.

By the end of 2026, the ASIC design will be refined, and the 64-channel ASIC (LATRIC V2) will be submitted for wafer fabrication. Simultaneously, a prototype of the LGAD readout frontend electronic system will be developed. In the mid-2027, performance tests of the 64-channel ASIC will be conducted, along with evaluations of the frontend prototype, ensuring seamless integration with the LGAD sensor.

By the end of 2027, the 128-channel LATRIC will be designed, integrated, and finalized for submission, paving the way for mass production.

Throughout the development process, the performance of the sensor–ASIC module will be thoroughly evaluated. A key focus will be on verifying whether the TOA and TOT metrics provide sufficient information to achieve the required timing and spatial resolutions. Additional efforts will aim to identify improvements in both the ASIC and the sensor to enhance the overall performance of the OTK detector.

Currently, wire bonding is the commonly used technique for connecting sensors and ASICs. This mature and reliable technology has been extensively applied in numerous experiments. However, alternative packaging and interconnection methods, such as solder-ball connections to the sensor or via an interposer board, will also be explored to further optimize the detector design and functionality.

In addition, to construct an OTK detector prototype, auxiliary chips — including the prototypes of data-link chip series (FEDI, FEDA, and OAT) and the power management DC-DC converters (PAL) — are scheduled for demonstration before 2027, while their final versions will be completed in the subsequent years.

#### 5.3.6.3 Summary

All the R&D mentioned above is part of a 3-year plan, as detailed in Figure 5.42. Considering the project timeline of more than 5 years from construction, the CEPC project is flexible enough to adjust its technical approach and has sufficient backup plans for the engineering phase.

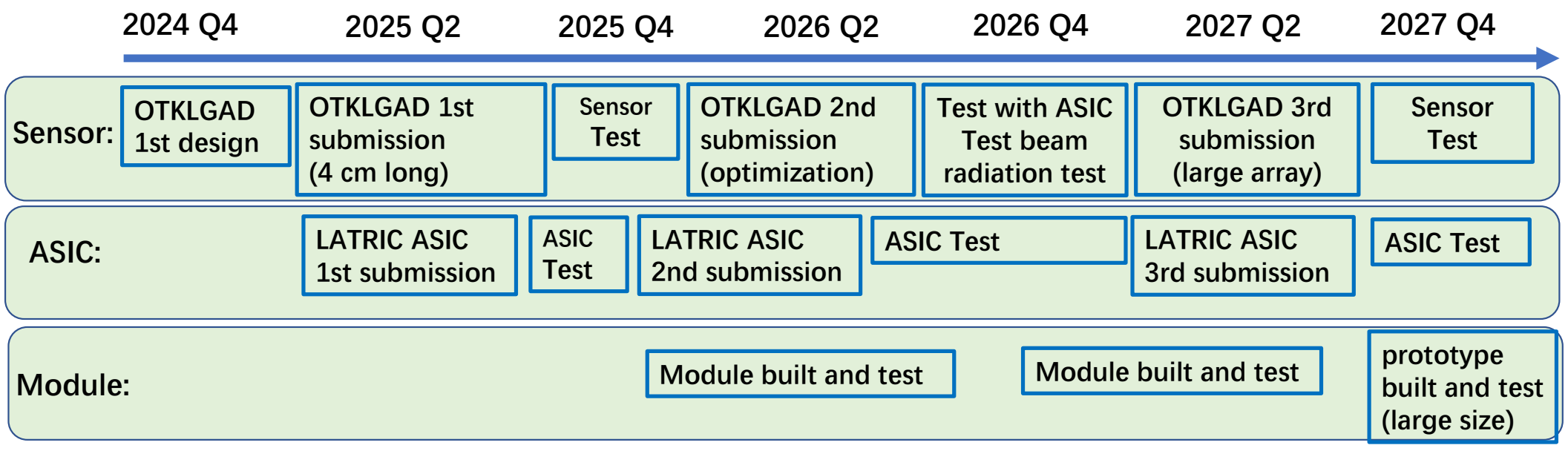

**Figure 5.42:** R&D timeline for the OTK, including sensor, ASIC, module, and mechanical and cooling components

For instance, if the performance of the AC-LGAD strip sensor degrades significantly for large-dimension strip sensor, we may consider using a smaller sensor, or a conventional large-dimension sensor with or without an external ToF detector. Additionally, if monolithic AC-LGAD technology reaches maturity by the engineering phase, the AC-LGAD sensor could be applied not only to the OTK layer but also to the outermost ITK layer. In addition to precise spatial measurement, this would enable accurate timing measurement, enhancing PID for very low-momentum particles that cannot penetrate the TPC, thereby extending PID coverage from $> 0.7$ GeV/c down to $> 0.3$ GeV/c.

## 5.4 Survey and alignment

Precise knowledge of the silicon sensor positions is critical for achieving micron-level track position resolution for the Silicon Tracker. The tracker alignment process consists of two key steps: optical survey (mechanical assembly precision) and track-based alignment.

### 5.4.1 Mechanical assembly and optical survey

The assembly of both the ITK and OTK follows a hierarchical structure, starting from small modules and progressively building up to larger components. Sensors are first integrated into modules, which are then mounted onto ladders, staves, or endcap plates. These mechanical units are subsequently installed onto their respective docking structures, forming the barrels and endcaps of the Silicon Tracker.

The electrical and mechanical sensor modules will be assembled using manual alignment jigs or pick-and-place machines. Each sensor in a module has fiducial reference marks precisely engraved and aligned with the pixel or strip array. A three-dimensional Coordinate Measuring Machine (CMM), equipped with an optical probe, will measure the relative positions of the sensors within a module and verify their alignment with the next supporting structure, such as a ladder, stave, or endcap plate serving as the module's local support. These measurements will be compared against external reference markers on the FPC or module local support as part of an optical survey.

#### 5.4.1.1 ITK installation and survey

For the ITK, the next step involves the precise mounting and measurement of the relative positions of staves and endcaps on their common docking structure (end-wheels). The positions of staves and endcaps will be measured and referenced against designated reference points on the end-wheels, using reference markers on the staves or endcap plates, including FPCs and sensor markers.

After each ITK layer is installed, a comprehensive survey of all targets on the end-wheels, staves, and endcaps within that layer will be conducted using a tracking system.

Once the ITK construction is complete, the entire detector will be inserted into the TPC inner barrel (see Figure 5.1). Due to the limited viewing angles for accessing the ITK through the TPC inner barrel holes, a laser tracking system will be used to reach the reference markers on the outermost ITK endcaps. This system will establish the reference between the ITK local coordinate system and the global coordinate system.

#### 5.4.1.2 OTK installation and survey

For the OTK barrel, the relative positions of the ladders will be measured against the stepped ramp rings on the TPC outer barrel (see Figure 5.28). Reference markers on the ladders will be compared against targets on the stepped ramp rings. Once the entire OTK barrel is assembled, a complete optical survey of all ladders will be performed relative to the reference points on the TPC end-wheels.

For the OTK endcap, the assembly involves joining 16 sectors, which are then mounted onto the common endcap support frame shared with the OTK and the ECAL. The docking points among sectors, sector connection rods, inner ring, and outer ring will be carefully monitored to ensure precise mounting. During installation, the OTK endcap will be referenced against the OTK&ECAL common endcap support frame.

By using reference targets on the TPC end-wheels and the OTK&ECAL common endcap support frame, the local coordinates of both the OTK barrel and endcap will be referenced to the global coordinate system.

#### 5.4.1.3 Target optical survey precision

Table 5.13 summarizes the targeted optical survey precision for the Silicon Tracker. These metrology and survey data serve as starting points for the final track-based alignment.

**Table 5.13:** Target optical survey precision for ITK and OTK

| Component | Sensor in module | Module on plate | Plate to end-wheel | End-wheel to global |
|---|---|---|---|---|
| ITK Barrel | ~ 5 μm | ~ 10 μm | ~ 100 μm | ~ 100 μm |
| ITK Endcap | ~ 5 μm | ~ 10 μm | ~ 100 μm | ~ 100 μm |
| OTK Barrel | ~ 5 μm | ~ 10 μm | ~ 100 μm | ~ 100 μm |
| OTK Endcap | ~ 5 μm | ~ 10 μm | ~ 100 μm | ~ 100 μm |

### 5.4.2 Track-based alignment

Track-based alignment is a method for achieving high-precision detector positioning, based on reconstructed particle tracks. The global composite alignment approach [26] will be applied to the track-based alignment.

#### 5.4.2.1 Global composite alignment

The tracker components are assembled in a hierarchical mechanical support structure. For example, in the ITK barrel, sensors are first assembled into modules, modules are then glued onto staves, and staves are mounted onto the tracker end-wheels. Each component is positioned relative to its immediate support structure with six degrees of freedom: three translations and three rotations.

Since all subcomponents within the same supporting structure are subject to common shifts and highly correlated displacements, applying alignment directly at the sensor level would neglect these correlations and could distort the detector geometry.

The global composite alignment approach addresses this by defining alignment parameters at each hierarchical level relative to its immediate supporting structure. A global alignment procedure is then applied to simultaneously determine the displacement parameters of all detector components across different mechanical hierarchical levels, ensuring all correlations are properly accounted for. In this approach, both the global detector alignment parameters and the local track parameters are determined simultaneously through a vast $\chi^2$ minimization.

#### 5.4.2.2 Alignment plan and datasets

The planned track-based alignment consists of two stages: (a) alignment using cosmic muons without a magnetic field, and (b) alignment during $e^+ e^-$ collisions.

Cosmic muon data is crucial for the alignment process. Unlike beam-induced tracks, cosmic muons traverse the detector along various paths, making them valuable for constraining and correcting detector deformations. This helps resolve the so-called $\chi^2$ invariance issue in beam-induced track alignment for $e^+ e^-$ collisions.

A dedicated data collection phase using straight muon tracks (with the magnetic field off) is planned after the completion of ITK and OTK detector construction. The detector will be installed $\sim$ 100 m underground. Due to significant energy loss in the ground materials such as rock and reinforced concrete, only a small fraction of high-energy muons (with $E > 100$ GeV) can penetrate and reach the detector. Therefore, a cosmic muon data-taking period of several months is planned to achieve sufficient statistics for precise alignment.

In addition to cosmic muon data, potential deformations of the tracking detectors after the magnetic field is turned on must be considered. During beam operation, the process $e^+e^- \to \mu^+\mu^-$ is a "golden channel" for the track alignment. Continuous data acquisition allows for real-time alignment of the tracker system, correcting potential displacements or deformations caused by external factors such as mechanical vibrations, thermoelastic movement, and differential settling of detector components. This alignment process must be periodically repeated to maintain optimal tracking performance.

## 5.5 Performance

This section highlights key tracking performance of the tracking system, focusing on the tracking momentum resolution and PID capabilities.

The total momentum resolution of a tracking system is mainly determined by two main factors: (a) the intrinsic spatial resolution of the position measurements and (b) the effects of multiple scattering from particle interactions with detector materials, which cause track deflections. For low-momenta tracks, the impact of multiple scattering dominates due to its inverse dependence on momentum, whereas for high-momenta tracks, the intrinsic spatial resolution of the tracker becomes the dominant factor. The CEPC tracking system benefits significantly from the integration of a gaseous detector TPC with a very low material budget and silicon detectors (VTX and STK) with high intrinsic spatial precision.

The tracking system also provides good PID performance, particularly in $K/\pi/$p separation, which is critical for CEPC flavor physics, searches for long-lived particles, and studies of jet substructure composition. This is primarily achieved by combining precise time measurement from the OTK as the ToF and the $\mathrm{d}N_p/\mathrm{d}x$ measurement (primary cluster counting) from the gaseous detector TPC.

### 5.5.1 Momentum resolution of the barrel region

Figure 5.43 shows the transverse momentum resolution ($\sigma_{p_\mathrm{T}}/p_\mathrm{T}$) as a function of transverse momentum at a polar angle of 85°, for three configurations: the silicon tracker including the VTX (in blue), the TPC alone (in red), and the full tracking system (in black). As expected, in the low momentum (< 10 GeV/c) region , the TPC provides better resolution. At high momenta, the silicon tracker significantly outperforms the TPC. The combination of both detectors provides the best performance across the entire momentum range.

Figure 5.44 presents the overall momentum resolution of the combined tracking system in the barrel region for three representative polar angles, satisfying the CEPC tracking requirement. As shown in Figure 5.44, the momentum resolution of the combined tracking system at a given polar angle $\theta$ in the barrel can be parameterized as:

$$\begin{aligned} \left(\frac{\sigma_{p_\mathrm{T}}}{p_\mathrm{T}}\right)_\mathrm{Si} &= a p_\mathrm{T} \oplus \frac{b}{\beta\sqrt{\sin\theta}} \\ \left(\frac{\sigma_{p_\mathrm{T}}}{p_\mathrm{T}}\right)_\mathrm{TPC} &= a s_1 p_\mathrm{T} \oplus \frac{b s_2}{\beta\sqrt{\sin\theta}} \\ \left(\frac{\sigma_{p_\mathrm{T}}}{p_\mathrm{T}}\right)_\mathrm{Combined} &= \frac{1}{\sqrt{\left(\frac{\sigma_{p_\mathrm{T}}}{p_\mathrm{T}}\right)_\mathrm{Si}^{-2} + \left(\frac{\sigma_{p_\mathrm{T}}}{p_\mathrm{T}}\right)_\mathrm{TPC}^{-2}}} \end{aligned} \tag{5.3}$$

where $a = 2.1 \times 10^{-5}$, $b = 2.2 \times 10^{-3}$, $s_1 \approx 4$, and $s_2 \approx 0.6$. In this parameterization, contributions from both the silicon tracker and the TPC are taken into account.

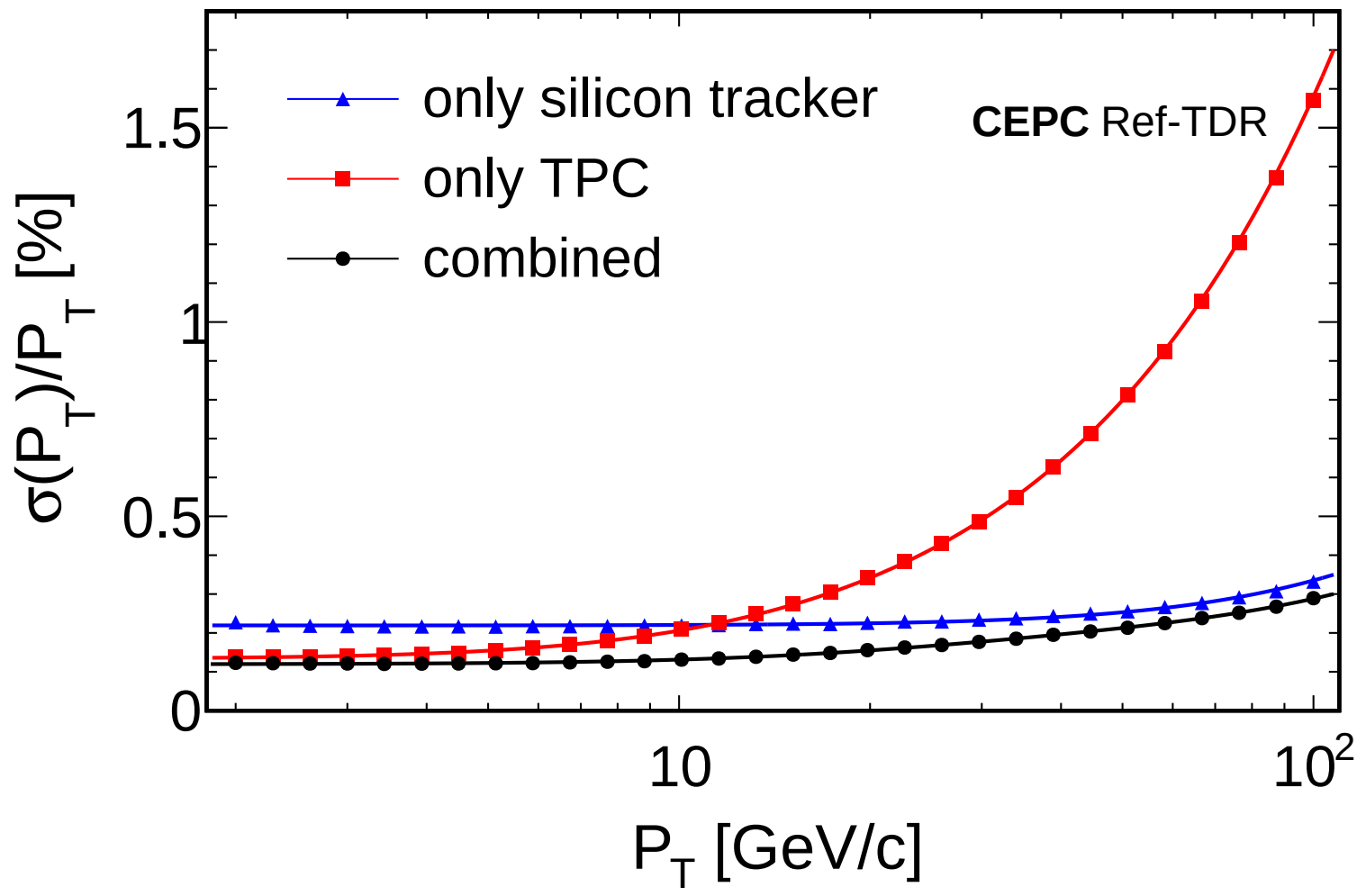

**Figure 5.43:** Transverse momentum resolution as a function of transverse momentum for the silicon tracker only (including the VTX, blue dots), TPC only (red dots), and the complete tracking system (black dots) at a polar angle of 85°.

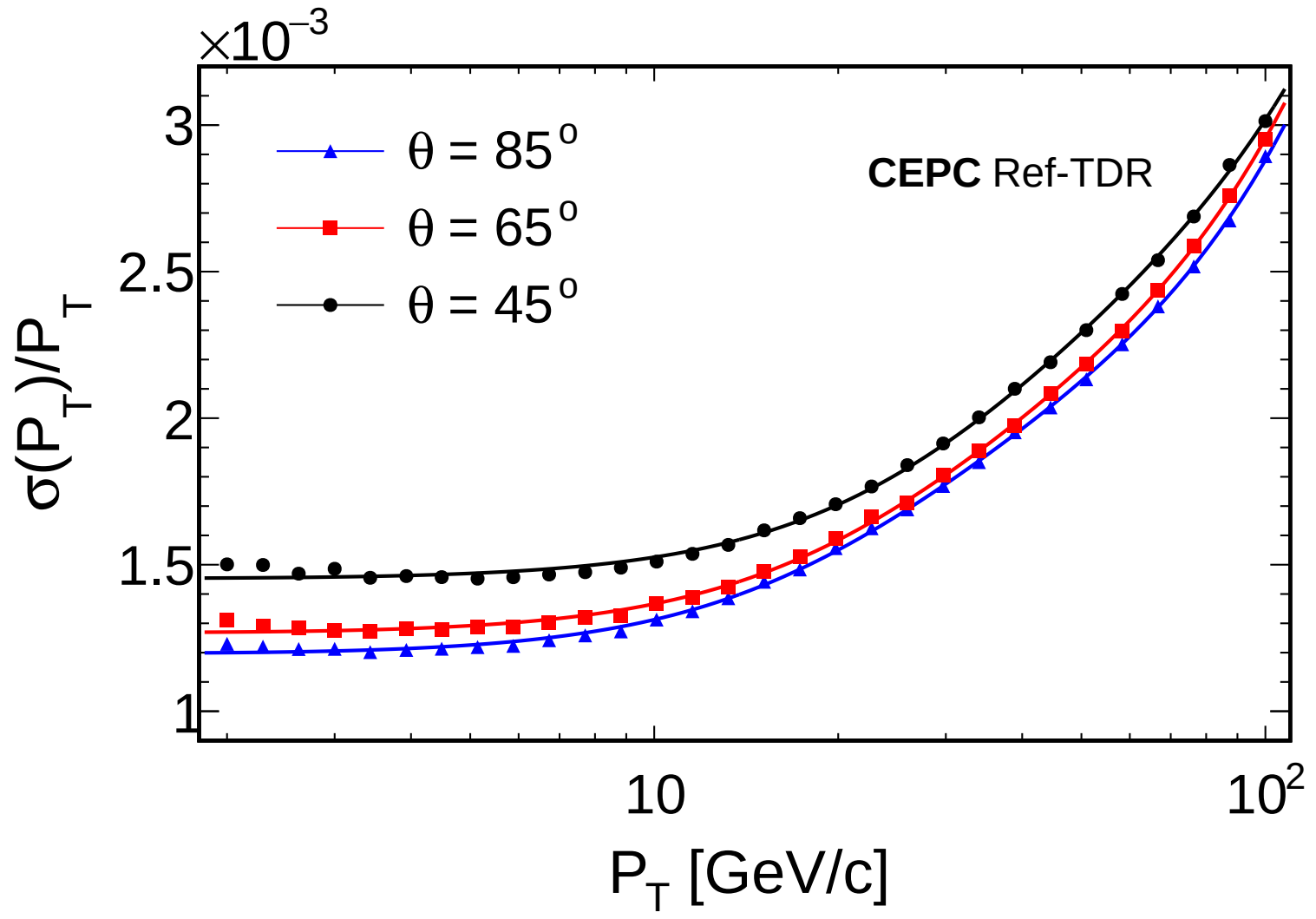

**Figure 5.44:** Transverse momentum resolution as a function of transverse momentum for the complete tracking system at polar angles of 85° (blue dots), 65° (red dots), and 45° (blue dots), where the curves are the fitting results with Equation 5.3.

### 5.5.2 Momentum resolution of the forward region (endcap)

The endcap region of the tracking system covers polar angle $8^\circ < \theta < 30^\circ$, including the silicon tracker and TPC overlap region ($11.5^\circ < \theta < 30^\circ$) and the very forward region ($8^\circ < \theta < 11.5^\circ$), which is exclusively covered by the silicon tracker. In this region, the tracking performance is primarily limited by the relatively short bending lever arm. This effect is partially mitigated by the extended barrel design of the CEPC tracking system.

Figure 5.45 illustrates the transverse momentum resolution as a function of transverse momentum for the endcap region, using the silicon tracker alone. The momentum resolution of the silicon tracker in the endcap region can be parameterized as:

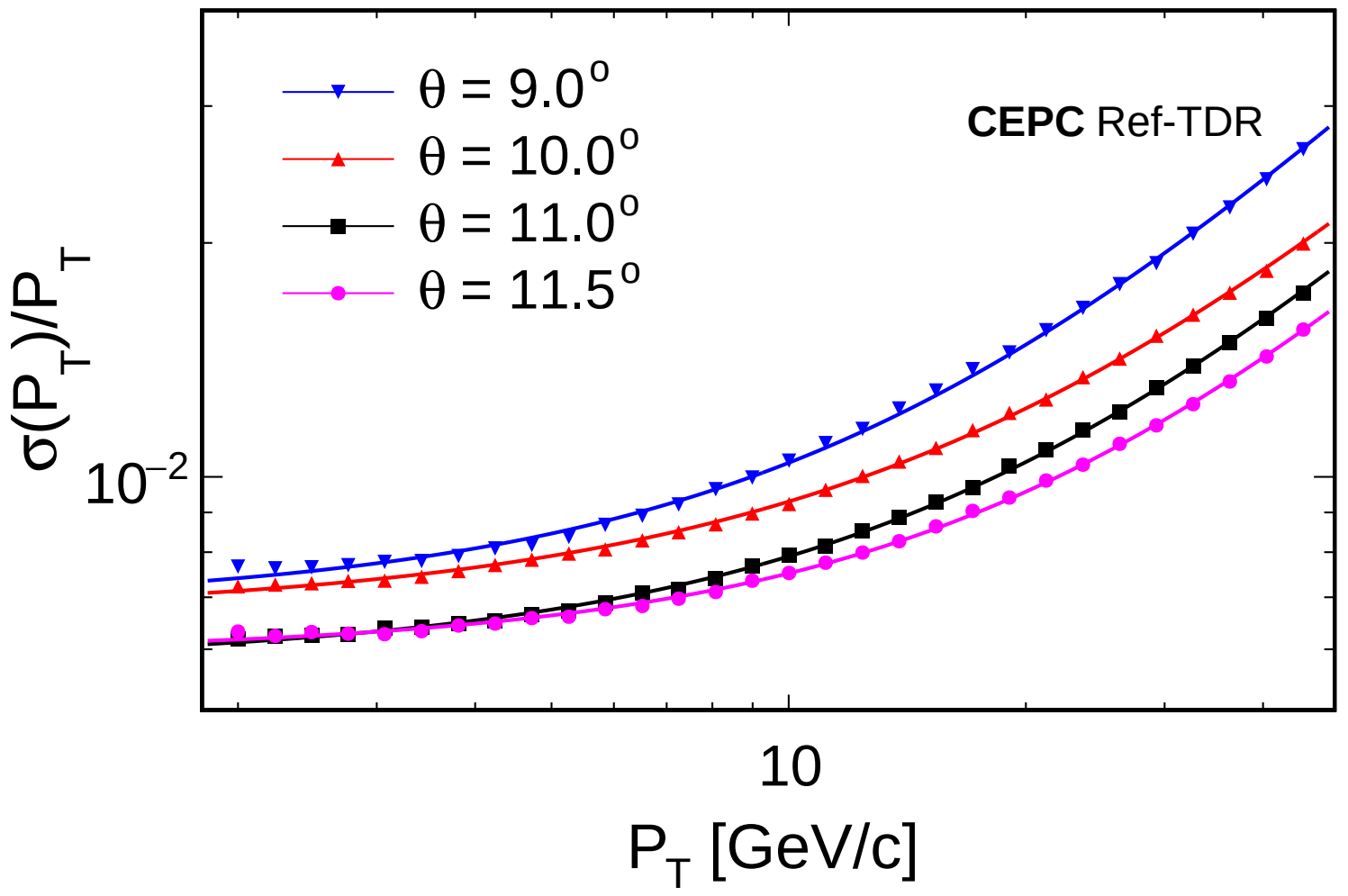

**Figure 5.45:** Transverse momentum resolution as a function of the transverse momentum for different polar angles at endcap region. The curves are fitting results with Equation (5.4).

$$\frac{\sigma_{p_\mathrm{T}}}{p_\mathrm{T}} = \frac{a' p_\mathrm{T}}{(\tan\theta)^2} \oplus \frac{b'}{\beta \tan\theta \sqrt{\cos\theta}} \oplus \frac{c' \sqrt{p_\mathrm{T}}}{\sqrt{\beta}(\tan\theta)^{\frac{3}{2}}(\cos\theta)^{\frac{1}{4}}} \tag{5.4}$$

where $a' \approx 1.1 \times 10^{-5}$, $b' \approx 1.1 \times 10^{-3}$, and $c' = 1.2 \times 10^{-4}$.

### 5.5.3 Particle identification performance

The separation power, $\eta_{AB}$, for distinguishing two particle types, A and B, can be defined as:

$$\eta_{AB} = \frac{|\mu_A - \mu_B|}{\frac{1}{2}(\sigma_A + \sigma_B)} \tag{5.5}$$

where $\mu_A$ and $\mu_B$ are the mean values of the PID estimator for particles A and B, and $\sigma_A$ and $\sigma_B$ are the standard deviations of the PID estimator.

Figure 5.46 shows the separation power between $K$ and $\pi$ as a function of momentum, using the ToF (OTK) with a 50 ps timing resolution. The blue curve corresponds to the barrel region at a polar angle of $\theta = 85°$, while the red curve corresponds to the endcap region at $\theta = 25°$. As seen, compared with the the barrel region, the ToF in the endcap region exhibits better PID performance due to the longer flight path to the OTK endcaps. Overall, the ToF provides good PID performance below 2 – 3 GeV/c.

The momentum cutoff of the ToF below ~ 0.9 GeV/c is caused by the particle's inability to penetrate the OTK detector, which results from the curvature bending in the magnetic field.

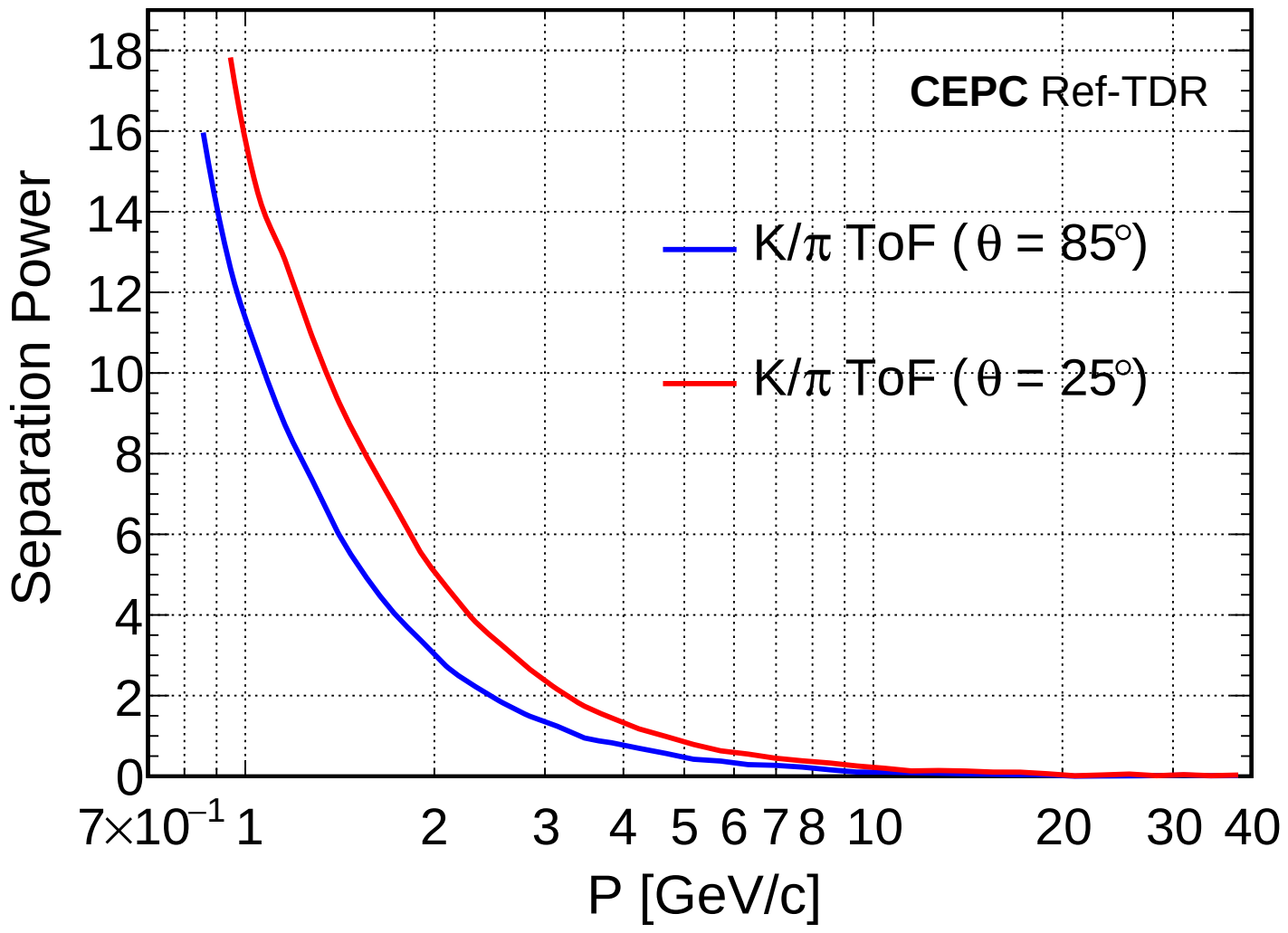

**Figure 5.46:** Separation power between $K$ and $\pi$ as a function of momentum at the polar angles $\theta = 85°$ (blue curve) in the barrel region and $\theta = 25°$ (red curve) in the endcap region, using the ToF (OTK) with a 50 ps timing resolution. As shown, the ToF provides good $K/\pi$ separation below 2 − 3 GeV/c.

## 5.6 Summary

The Silicon Tracker is a vital detector system for the CEPC, essential for precise particle trajectory determination and particle identification. As the system evolves through continuous technological improvements and optimizations, it will be steadily transitioning into the engineering phase in the next few years. Continuous innovations in sensor technology, readout electronics, mechanical design, and the development of the cooling system will ensure that the Silicon Tracker meets the high demanding requirements of CEPC physics.

# Chapter 6 Time Projection Chamber

The charged particle tracking system of CEPC includes a TPC embedded in the silicon tracking system described in Chapter 5. The TPC provides measurements of numerous space points, which is not only beneficial for tracking, but also enables ionization measurements for charged PID.

In this chapter, the physics requirements and the overall concept of the TPC detector are introduced in Section 6.1. Section 6.2.5 details the TPC design, including the chamber geometry, field cage, endplates, readout modules, high-granularity readout electronics, and auxiliary subsystems. Section 6.3.4 presents the completed R&D efforts and describes the key technologies that require further development to address challenges such as beam-induced background, calibration, and alignment. The simulation and performance of TPC are described in Section 6.4.4.2. Section 6.5 evaluates a drift-chamber-based design and its PID performance, as a possible alternative to TPC. Finally, the chapter ends with a summary and future plans in Section 6.6.

## 6.1 Overview

TPCs have been extensively studied and used in different fields, especially in particle physics experiments such as ALEPH [1], DELPHI [2], STAR [3], and ALICE [4]. Furthermore, the LC-TPC Collaboration [5] has conducted extensive studies of a large TPC for a linear $e^+e^-$ collider detector, which serves as the foundational reference for this report. The volume of the gaseous detector results in a low material budget, while the high density of spatial measurement points enables excellent pattern recognition capability. However, attention must be paid to the space charge distortion resulting from the accumulation of positive ions in the drift volume [6]. This effect is especially important in high-rate conditions. To mitigate this, the CEPC TPC design leverages high-granularity readout technology, building upon foundational R&D from the high-energy physics community, including TPC concepts developed for the ILD [7] and Electron Ion Collider (EIC) [8] projects.

The CEPC tracker system, comprising the ITK, the TPC, and the OTK, is designed to achieve $2\times10^{-5}$ $(\mathrm{GeV}/c)^{-1}$ transverse momentum resolution for charged particles with momenta below 100 $\mathrm{GeV}/c$. As such, the TPC is required to provide a standalone momentum resolution of $10^{-4}$ $(\mathrm{GeV}/c)^{-1}$ with a magnetic field of 3 T. To reduce multiple Coulomb scattering effects that degrade momentum resolution, the material budget of the TPC must be minimized. Furthermore, the TPC should provide excellent PID capabilities, which are critical for flavor physics studies and jet substructure analysis. This is achieved by leveraging ionization energy loss ($\mathrm{d}E/\mathrm{d}x$) [9] or number of activated pads ($\mathrm{d}N_p/\mathrm{d}x$), in combination with ToF information from the OTK, enabling a $3\sigma$ separation between $\pi$, $K$, and protons with momenta up to 20 $\mathrm{GeV}/c$.

To meet these stringent requirements, the TPC is designed to consist of a large-volume cylindrical chamber and readout detector modules located on two endplates, as shown in Figure 6.1, featuring:

- An inner cylinder (radius: 0.6 m) and an outer cylinder (radius: 1.8 m) with a length of 5.8 m, constructed from ultra-lightweight, low-$Z$ materials to minimize material budget to about 1.14% $X_0$ and provide sufficient rigidity.
- Two hermetically sealed endplates that integrate readout modules with high-granularity readout pads for precise 3D track reconstruction.
- A working gas mixture characterized by low transverse diffusion and fast drift velocity, identical to the T2K gas ($Ar/CF_4/iC_4H_{10}$ = 95/3/2), to reduce the impact on positional measurement precision.

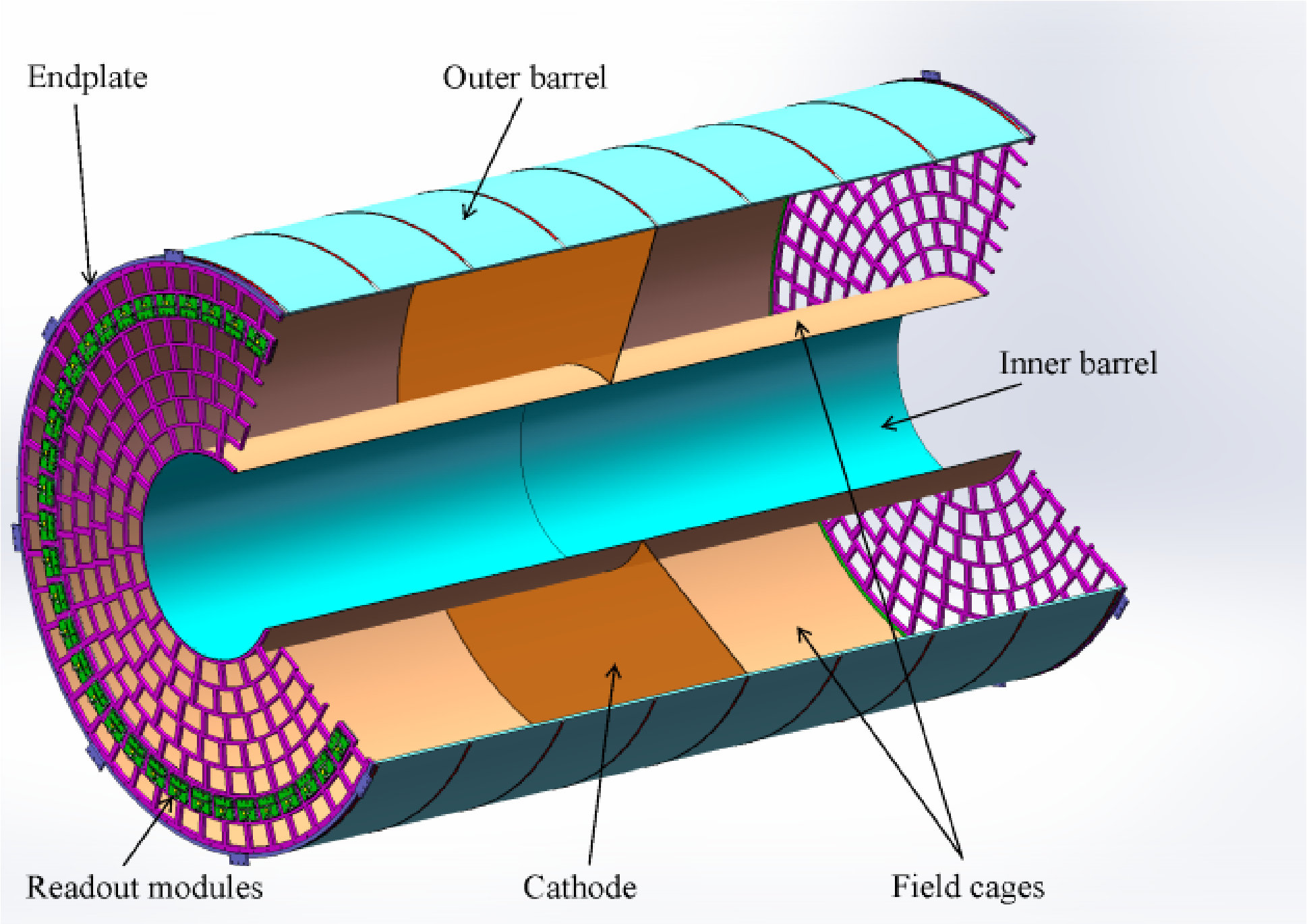

**Figure 6.1:** The TPC is a cylindrical, gas-filled tracking device whose axis coincides with the nominal beam direction. The barrel encloses a field cage assembly that produces a uniform electric drift field along the $z$-axis. Two readout endplates mechanically support the barrels, while the central cathode plane is positioned at the barrel's mid-plane.

To reach the tracking performance goals, advanced Micro Pattern Gaseous Detector (MPGD) [10] is used for gas amplification devices in the TPC, such as Micro-Mesh Gaseous Structure (Micromegas)[11]. Compared to traditional wire chambers, Micromegas offers readout segmentation that is more than an order of magnitude smaller. This high granularity enables a superior spatial resolution of approximately 100 µm. In addition, the spatial and temporal spreads of the signals at the endplates are significantly reduced, allowing for better two-particle separation power. The use of a Micromegas detector equipped with an additional mesh can significantly suppress the Ion Backflow (IBF), thereby reducing the space charge effect and electric field distortion.

Transverse charge diffusion is mitigated through combined strategies. Beyond the use of gas mixtures with low diffusion and fast drift velocity, a strong detector solenoid magnetic field further suppresses diffusion by constraining the electron drift paths. This magnetic compression is particularly crucial given the TPC's long drift distance of approximately 2.9 m, as it directly improves $\sigma_{\text{point}}$ and enhances the ability to separate two tracks.

To meet the high-hit-rate conditions at the CEPC, high-granularity readout with a pad size of 500 μm × 500 μm is adopted in the readout detector module design. Small readout pads can significantly reduce the voxel occupancy (the ratio of fired to total 3D readout cells) of the TPC. In addition, the high-granularity readout enables two distinct measurement methods: the traditional $\mathrm{d}E/\mathrm{d}x$ (energy deposition per unit length of the track) method and $\mathrm{d}N_p/\mathrm{d}x$ (the number of activated pads along the track segments) method. The $\mathrm{d}N_p/\mathrm{d}x$ method has the potential to significantly improve the PID performance due to a lower sensitivity to Landau tails than the $\mathrm{d}E/\mathrm{d}x$ method.

**Table 6.1:** Critical parameters of the TPC.

| **Parameters** | **Values** |
|---|---|
| Length (outer dimensions) | 5800 mm |
| Radial extension (outer dimensions) | 600–1800 mm |
| Cathode potential | -63,000 V |
| Gas mixture | T2K: $Ar/CF_4/iC_4H_{10}$ = 95/3/2 |
| Drift velocity | $\sim$ 8 $cm/\mu$s |
| Maximum electron drift time | $\sim$ 34 μs |
| Readout detector | Double-mesh Micromegas |
| Pad size | 500 μm × 500 μm |
| Gas gain | $\sim$ 2000 |
| Readout modules per endplate | 244 |
| Cooling | Water cooling circulation system |

The critical design parameters of the TPC are shown in Table 6.1. The potential of the cathode located at the center of the chamber is -63 kV, generating a uniform drift electric field of approximately 230 V/cm between the cathode and the readout detectors. Under this drift electric field, the maximum drift time of the ionized electrons around the cathode in the T2K gas is about 34 μs.

Each endplate is equipped with 244 readout detector modules (488 in total), and each module contains a double-mesh Micromegas detector with a high-granularity readout anode to meet the requirement of high-count-rate conditions and provide excellent PID. The high-density readout scheme based on high-granularity anodes presents challenges in the number of readout channels and power consumption. Consequently, dedicated readout ASIC chips have been designed, and a water cooling system has been implemented to handle the electronics heat dissipation.

## 6.2 Detailed design

### 6.2.1 Chamber and field cage

Figure 6.1 illustrates the detailed design of the chamber. The central high-voltage electrode divides the TPC chamber into two symmetric halves. Two field cages are tightly attached to the inner surface of the outer cylinder and the outer surface of the inner cylinder. Faraday cages covering the exterior surface of the TPC are used for electromagnetic shielding.

To achieve a low material budget while maintaining structural rigidity and minimizing deformation, both cylinders employ a composite sandwich construction. The outer cylinder consists of a Nomex honeycomb core (approximately 20 mm thick) sandwiched between a 100-μm-thick aramid paper and an outer skin of 200-μm-thick carbon fiber sheet. The inner cylinder features a similar structure, with a thinner Nomex honeycomb core of about 5 mm. In addition, a 100-μm-thick polyimide insulating layer is glued to the aramid paper on both cylinders to enhance electrical insulation.

The structure of the cylinders is shown in Figure 6.2. With such design, both the inner and outer cylinders are lightweight, rigid, and gas-tight, and there is sufficient electrical insulation between the field cage and the Faraday cage.

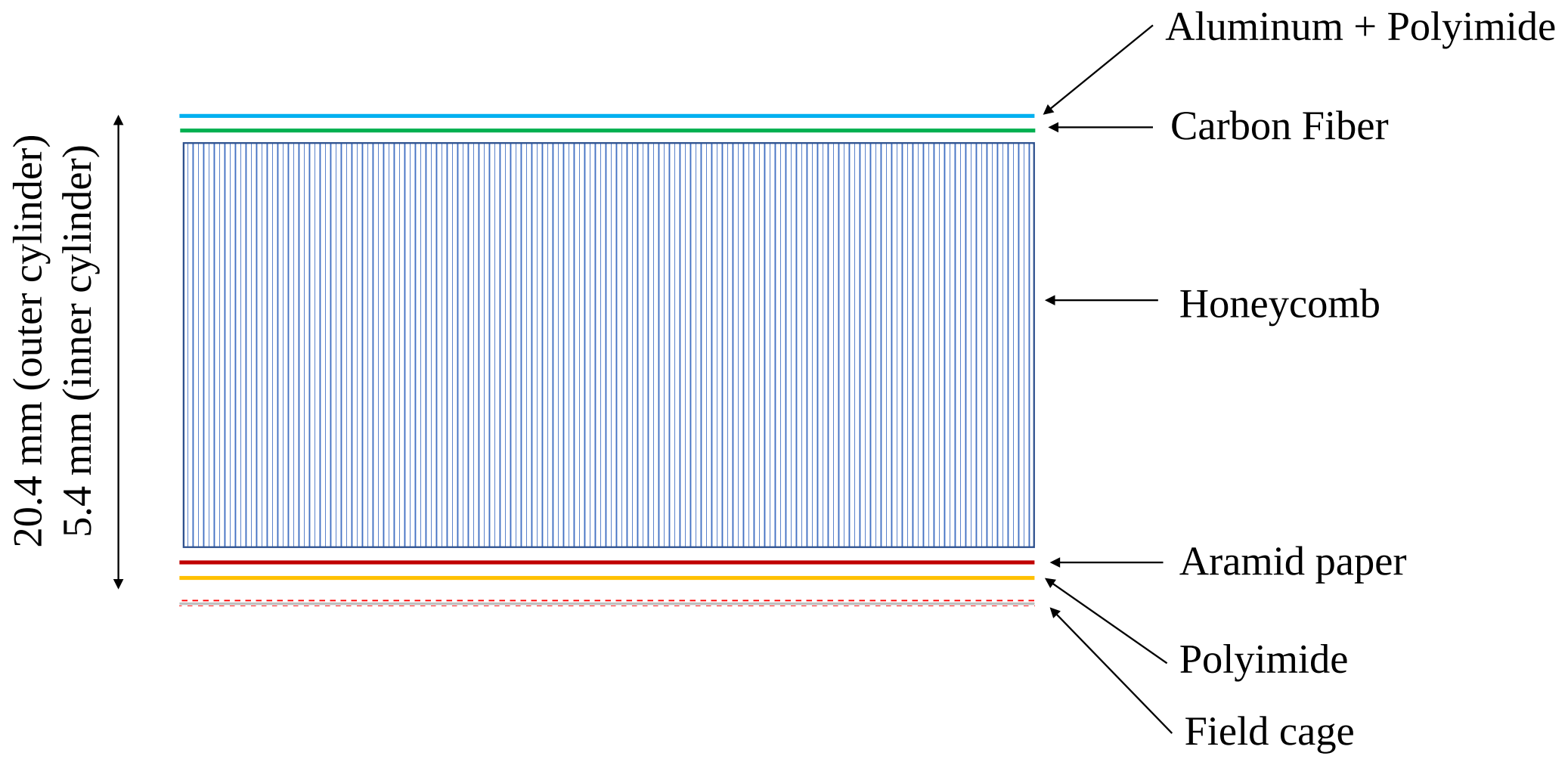

**Figure 6.2:** Structure of the TPC inner and outer cylinders. The Nomex honeycomb core is embedded between the carbon fiber layer and aramid paper layer, forming a light, rigid and gas-tight structure, with sufficient insulation between the field cage and the Faraday cage.

The material composition and the material budget estimation of the inner and outer cylinders are summarized in Table 6.2. The total thickness of the outer wall is 20.4 mm, with a material budget of 0.69% $X_0$, while the total thickness of the inner wall is 5.4 mm, with a material budget of 0.45% $X_0$. A high modulus QM55 carbon fiber will be used in the construction of the inner and outer cylinders.

**Table 6.2:** Material composition and material budget estimation of the TPC barrel, in layers from outer radius to inner radius. The thickness ($X$), the radiation length ($X_0$), and material budget ($X/X_0$) are provided. The total thickness of the outer wall is 20.4 mm, with a material budget of 0.69% $X_0$. The inner wall has a total thickness of 5.4 mm and a material budget of 0.45% $X_0$.

| Component | Material of each layer | $X$ [mm] | $X_0$ [cm] | $X/X_0$ [%] |
|---|---|---|---|---|
| | | **TPC outer wall** | | |
| Faraday cage shield | Aluminum | 0.05 | 8.9 | 0.06 |
| Faraday cage substrate | Polyimide | 0.05 | 28.6 | 0.02 |
| Outer wall support cylinder | QM55 carbon fiber | 0.2 | 25.28 | 0.08 |
| Outer wall support cylinder | Nomex honeycomb | 19.6 | 800 | 0.25 |
| Outer wall support cylinder | Aramid paper | 0.1 | 35 | 0.03 |
| Insulating layer | Polyimide | 0.1 | 28.6 | 0.03 |
| Mirror strips layer | Copper | $0.012 \times 0.95$ | 1.44 | 0.08 |
| Field cage substrate | Polyimide | 0.05 | 28.6 | 0.02 |
| Field strips layer | Copper | $0.012 \times 0.95$ | 1.44 | 0.08 |
| Glue | Epoxy | 0.2 | 35.3 | 0.06 |
| **Total** | | **20.4** | | **0.69** |
| | | **TPC inner wall** | | |
| Field strips layer | Copper layer | $0.012 \times 0.95$ | 1.44 | 0.08 |
| Field cage substrate | Polyimide | 0.05 | 28.6 | 0.02 |
| Mirror strips layer | Copper | $0.012 \times 0.95$ | 1.44 | 0.08 |
| Insulating layer | Polyimide | 0.1 | 28.6 | 0.03 |
| Inner wall support cylinder | Aramid paper | 0.1 | 35 | 0.03 |
| Inner wall support cylinder | Nomex honeycomb | 4.6 | 800 | 0.06 |
| Inner wall support cylinder | QM55 carbon fiber | 0.2 | 25.28 | 0.08 |
| Faraday cage substrate | Polyimide | 0.05 | 28.6 | 0.02 |
| Faraday cage shield | Aluminum | 0.05 | 8.9 | 0.06 |
| Glue | Epoxy | 0.2 | 35.3 | 0.06 |
| **Total** | | **5.4** | | **0.45** |

The TPC field cage, an assembly comprising precisely spaced and electrically isolated conductive rings or strips, works together with the cathode and anode planes to establish and maintain a highly uniform electric field throughout the drift volume. The overall design of the field cage involves the geometry of the potential rings, the resistor chains, the central HV membrane, the gas container, and a radiation source calibration system. It must withstand the high voltage on the central HV membrane with the minimum possible material budget.

While theoretical optimization calls for minimizing the strip gap and width to maximize field uniformity, the practical design is constrained by comprehensive considerations. These include the resistor chain design, soldering feasibility, and the precision limits of flexible PCB manufac-turing. Excessively narrow strips, for instance, introduce significant soldering challenges such as solder bridging between the strips and require tighter tolerances in PCB fabrication.

As shown in Figure 6.3(a), the final dimensions were set to a strip gap of 0.1 mm and a strip width of 2 mm, representing a balance between performance goals and manufacturing constraints. The effectiveness of this design is validated by Finite Element Analysis (FEA), as shown in Figuree 6.3(b). The FEA results of the drift electric field ($E_z$), indicates that a highly uniform field across the entire TPC chamber is achieved beyond a distance of approximately 5 mm from the copper strip.

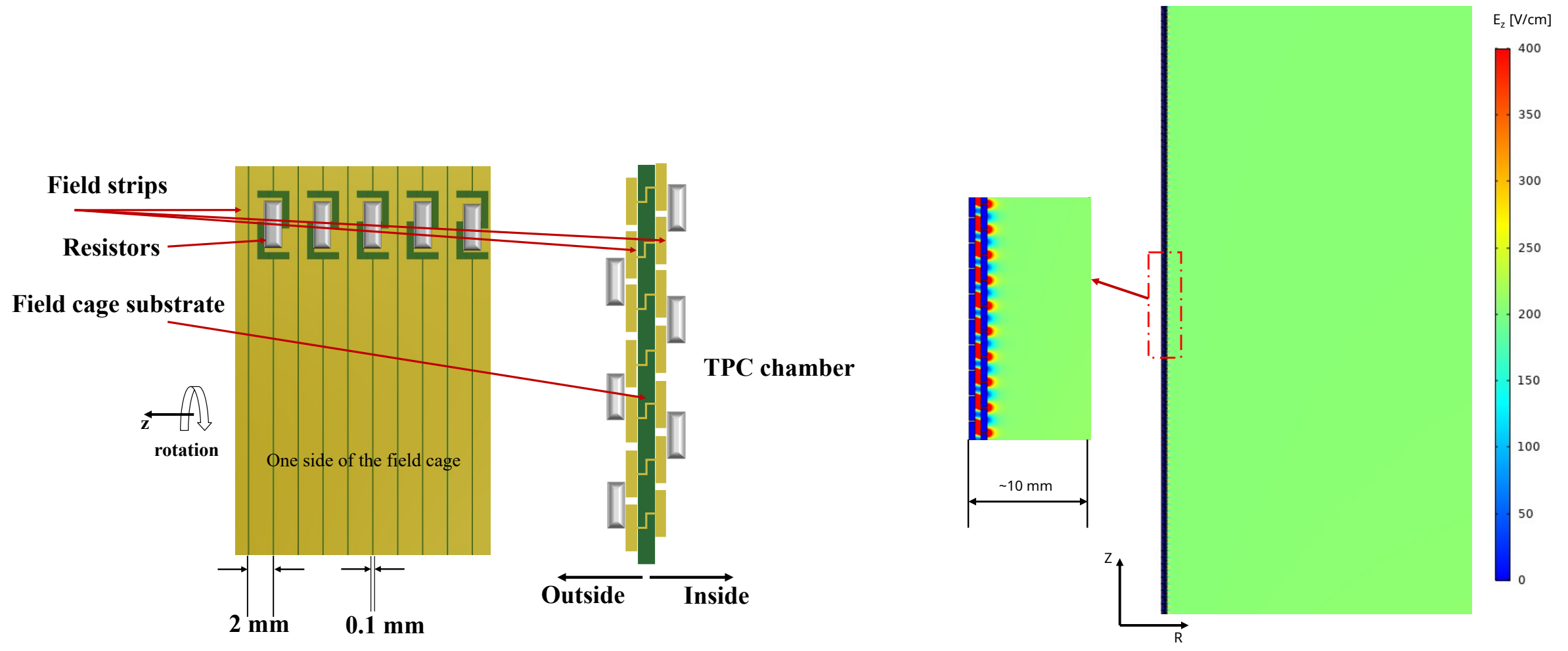

**(a)** Detailed design

**(b)** FEA results of drift electric field

**Figure 6.3:** Design of the field cage. (a) Layout of double-layer copper strips with 2 mm width and 0.1 mm gap. (b) Finite element simulation results of drift electric field $E_z$ of the double-layer field cage.

In this design, the strips on the back of the field cage are offset by 1 mm from those on the front, positioning them precisely at the center between two strips on the opposite side. This arrangement effectively compensates for the electric field non-uniformity caused by the voltage-free state of the strip gaps. The strips on the front and back of the field cage are designed with an inter-layer separation. The soldering positions of the resistors on the front and back of the field cage are staggered to prevent heat transfer during soldering from damaging solders on the opposite side. The strips on the top and bottom surfaces of the field cage are connected through vias, simplifying the voltage loading process. During operation, high voltage is applied to only one side of the field cage and is transmitted to the opposite side through the via holes, thus achieving a uniform voltage distribution.

The cage ensures that ionization electrons drift along straight, predictable trajectories toward the readout plane, minimizing spatial distortions and enabling accurate three-dimensional reconstruction of particle tracks.

### 6.2.2 Endplate and readout modules

Each endplate of the TPC has an area of approximately 10 $m^2$, consisting of a 25-mm-thick aluminum plate with seven rings of non-uniformly arranged windows along the radial direction for the installation of the readout detector modules, as shown in Figure 6.4. The aluminum endplate offers a combination of lightweight and high strength, making it suitable for maintaining structural integrity while minimizing the material budget. The readout detector modules and the connection to the electronics, as well as the cooling of the electronics, are highly correlated to the endplates. To minimize the impact on ECAL particle flow measurement in the forward direction, the total material budget of the endplate is kept below 15% $X_0$. Figure 6.5 shows the detailed connection between the outer barrel and the endplate. An aluminum ring will be glued to the outer cylinder at each end, and each endplate will be bolted to the corresponding aluminum ring and sealed with an O-ring.

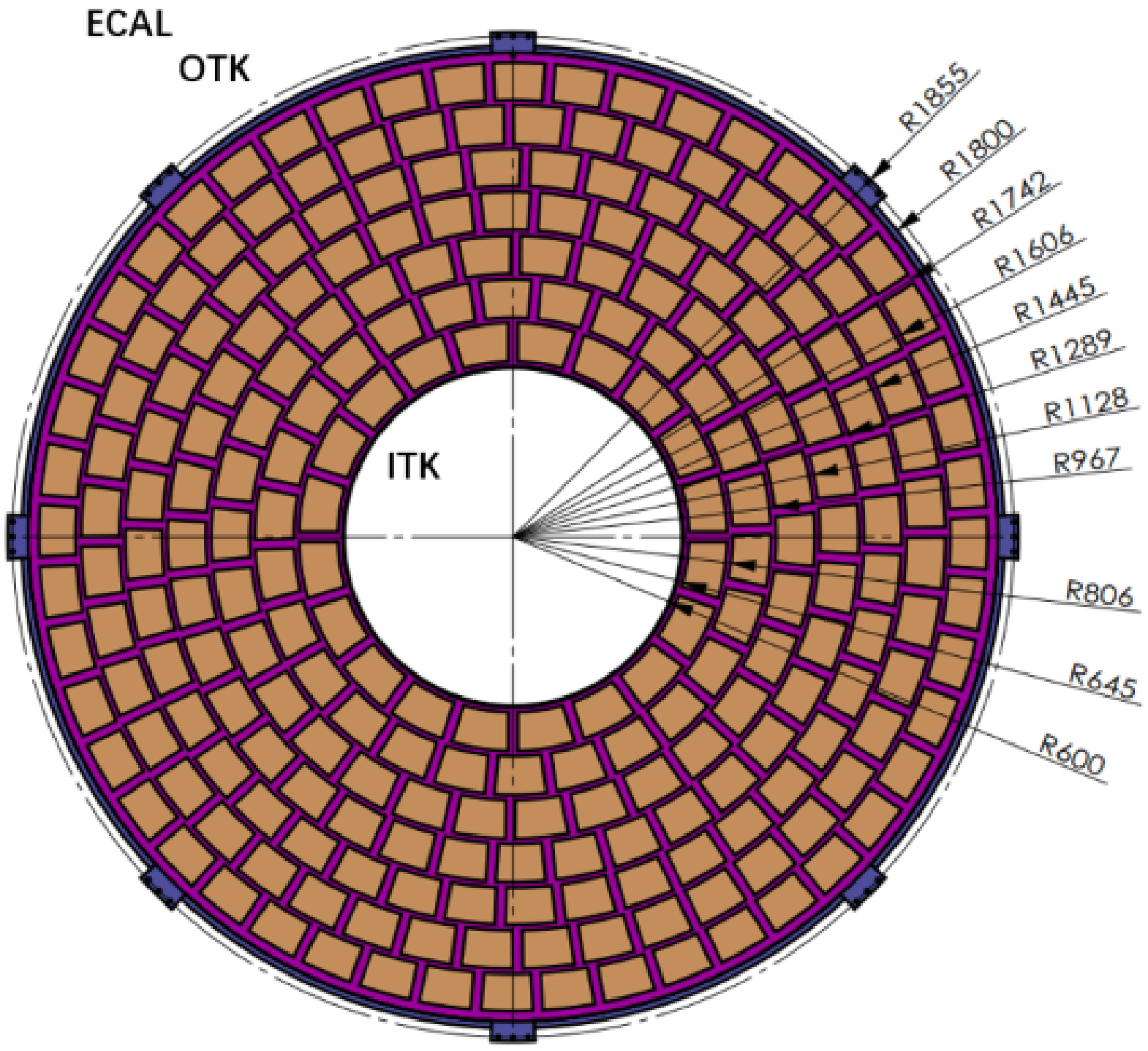

**Figure 6.4:** Two-dimensional layout of the endplate of the TPC (unit: mm). Seven rings of windows are arranged radially for installing the readout detector modules. The inner radius and outer radius of the endplate are 600 mm and 1800 mm, respectively.

The readout modules with an aluminum alloy support frame will be installed in the separated windows of the endplates, as shown in Figure 6.6. An O-ring will be used for gas sealing between the aluminum support frame and the inner surface of the endplates. Figure 6.7 shows the design of the readout detector modules. To maximize the sensitive area, seven slightly different module sizes are used for the seven rings along the radial direction. The maximum external envelope dimensions of the modules are listed in Table 6.3.

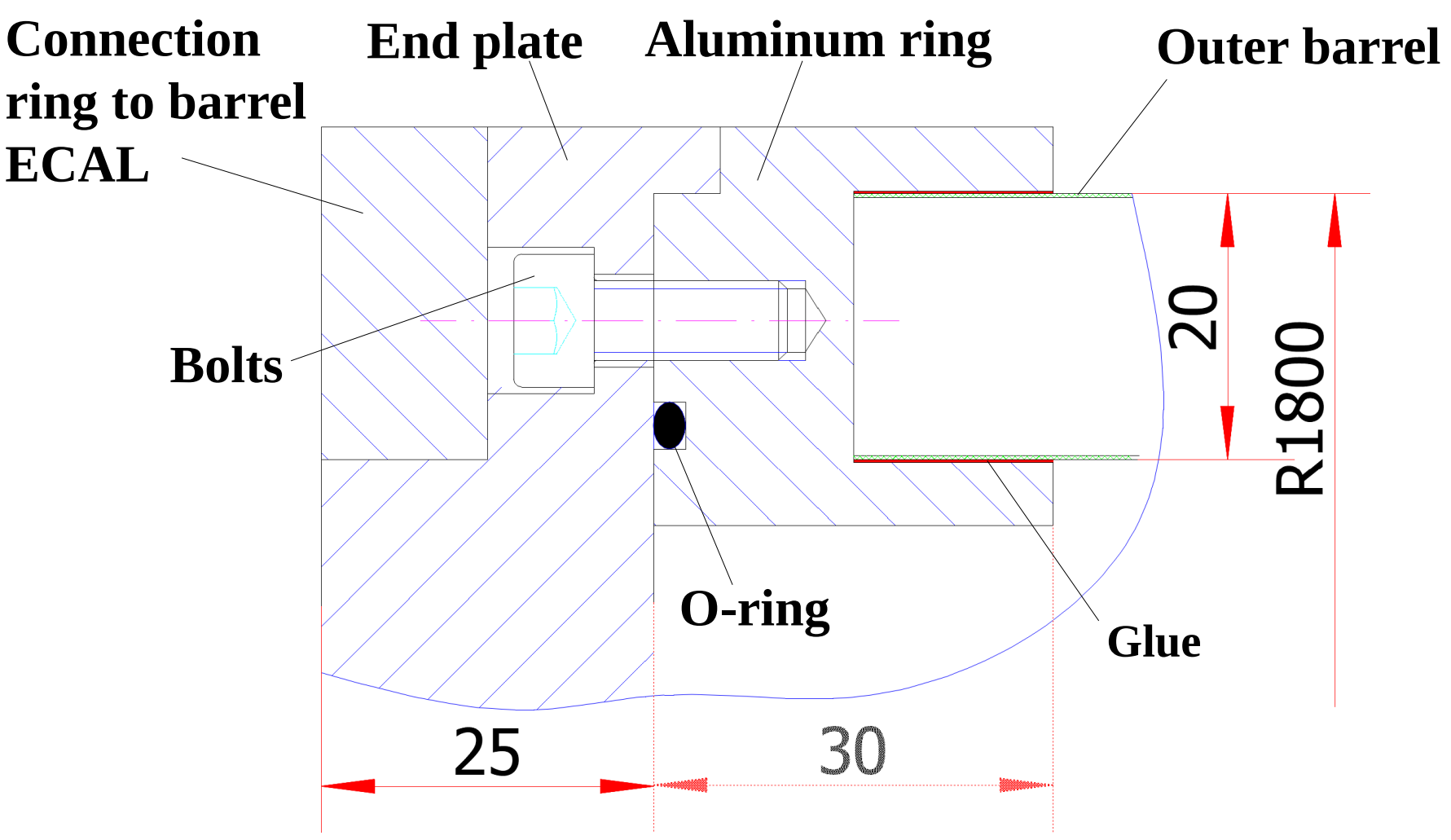

**Figure 6.5:** Connection between the outer cylinder and the endplates. There is an aluminum ring glued to the outer cylinder at each end. The endplate will be bolted with the aluminum ring and sealed with an O-ring.

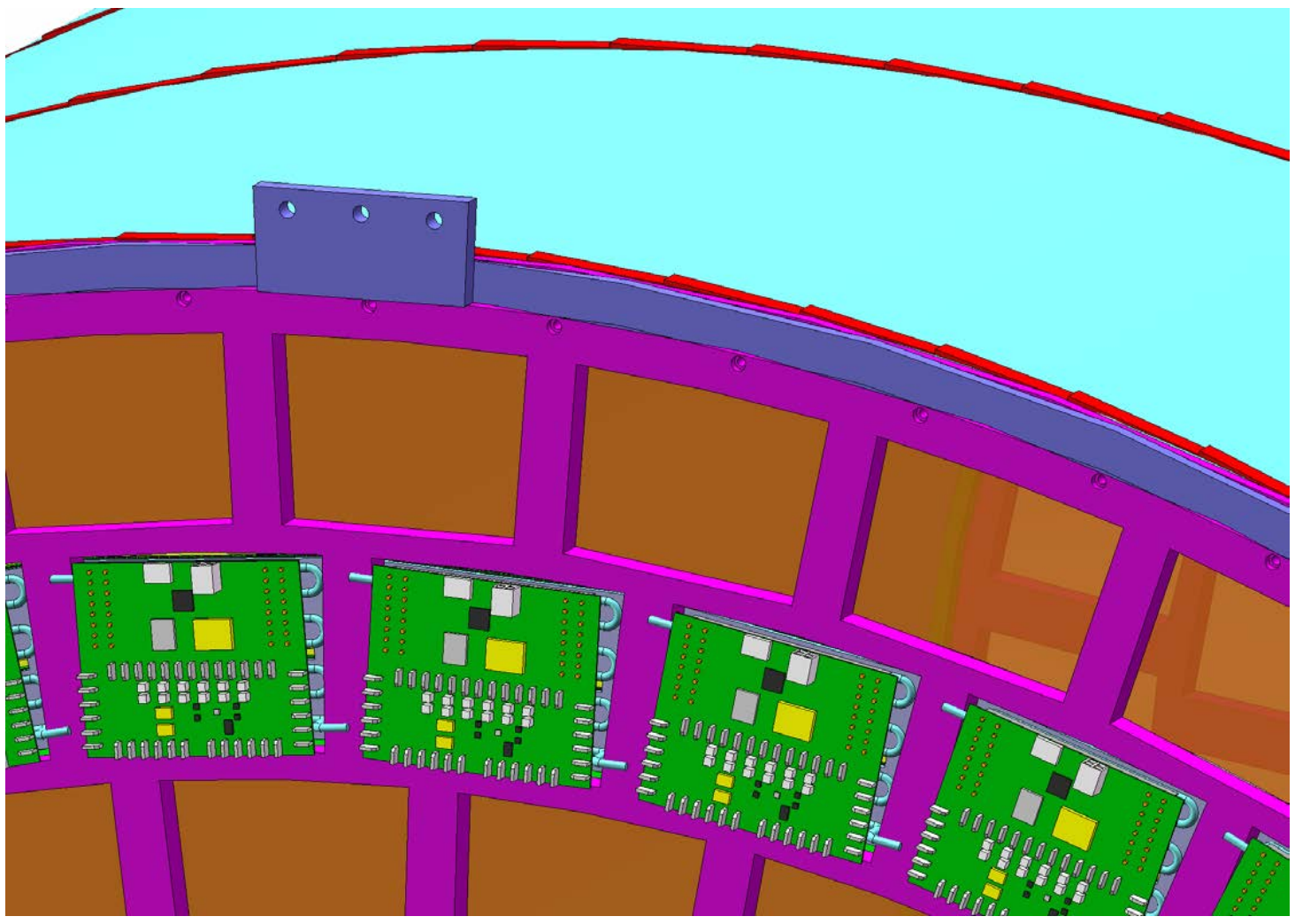

**Figure 6.6:** Examples of the readout modules installed on the endplate. Each endplate will be equipped with 244 individual detector modules.

Each module is mainly composed of a double-mesh Micromegas detector with a high-granularity readout anode, a readout ASIC chip array mounted on the back of the anode board (FEE), and a readout board connected to the FEE board. The readout board is responsible for delivering power, sending configuration commands to the ASIC chips, and receiving and packaging data from the chips. After packaging, the data is transferred to the back-end electronics through optical fibers.

Reducing the size of the readout pads is critical to minimize the TPC occupancy. Based on preliminary simulation results, a readout pad size of 500 μm × 500 μm including the pad gap was adopted, resulting in high-granularity readout on the sensitive endplate of the TPC. The Micromegas detector equipped with an additional mesh serves as the first option to reduce ion

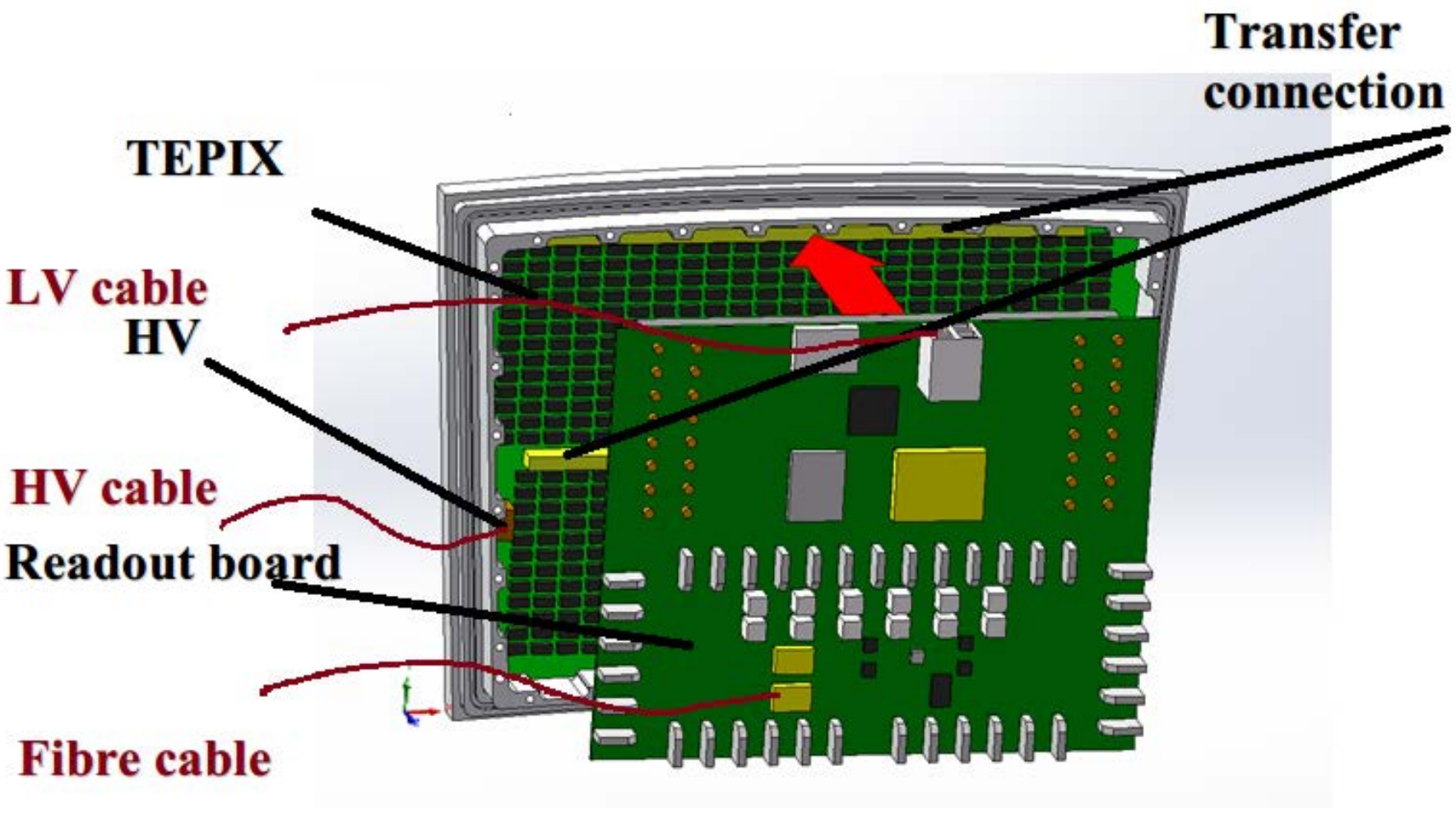

**Figure 6.7:** Design of the TPC endplate readout modules. Each module is mainly composed of a double-mesh Micromegas detector with high-granularity readout anode, a readout ASIC chip array mounted on the back of the anode board (FEE), and a readout board connected to the FEE board.

backflow.

In the mechanical design, the dimensions of the FEE board are consistent with those of the aluminum alloy support frame. To accommodate the installation requirements of the detector modules, the FEE board incorporates a curved edge and two precision positioning holes, ensuring accurate alignment and reliable connection between the FEE board and the aluminum alloy support frame. The FEE board also integrates HV connectors to provide HV (300–500 V) for the double-mesh Micromegas detector.

**Table 6.3:** Geometric specifications of the readout modules. In the maximum external envelope dimensions, the third dimension represents the combined thickness of the readout detector, the FEE board, the readout board, the cooling system, and the shielding cover.

| Layer ID | External envelope dimensions (mm$^3$) | Modules/layer | Channels/module |
|---|---|---|---|
| 1 | $228.1 \times 155.0 \times 80.0$ | 20 | 131617 |
| 2 | $226.7 \times 155.0 \times 80.0$ | 26 | 130774 |
| 3 | $224.9 \times 160.0 \times 80.0$ | 30 | 133944 |
| 4 | $221.9 \times 160.0 \times 80.0$ | 36 | 132140 |
| 5 | $232.6 \times 160.0 \times 80.0$ | 40 | 138554 |
| 6 | $229.6 \times 160.0 \times 80.0$ | 44 | 136726 |
| 7 | $247.9 \times 160.0 \times 80.0$ | 48 | 147649 |

In Figure 6.7, the HV, LV, ground, and optical-fiber signal cables of each readout module are first bundled into a single composite cable. This composite cable passes through the detector's outer envelope and is routed toward the FEE board and the readout board, where it splits into individual HV, LV, ground and fiber lines for final connection to the module. On each endplate, there are 244 composite cables routed sector-wise to six external bulkhead connectors, while the cables pass through the terminals, enter each sector, and then spread along the module rails to the corresponding readout modules.

### 6.2.3 Structural finite element analysis

In order to evaluate the stability and reliability of the overall structure design of the TPC, FEA has been performed. In the FEA, the loads include self-weight of the TPC and the weight of the OTK, which is mounted on the TPC and supported by the outer cylinder and two endplates. The weights of the TPC components, including the inner and outer cylinders, the endplates, and the readout detector modules, are listed in Table 6.4. The total weight of the OTK (detailed design described in Chapter 5) is 300 kg, including 180 kg support structures fixed to the outer cylinder of the TPC at nine support points along its length. The constraints were applied to the eight lifting ears installed on each endplate, consistent with the actual installation where the TPC is fixed to the ECAL via these points. 200 Pa positive gas pressure inside the chamber is also taken into account in the FEA.

**Table 6.4:** The weights of the TPC detector components.

| **TPC components** | **Weight (kg)** |
|---|---|
| Inner cylinder | 10 |
| Outer cylinder | 67 |
| Endplates | $295 \times 2 = 590$ |
| Readout modules, cables and cooling | $500 \times 2 = 1000$ |

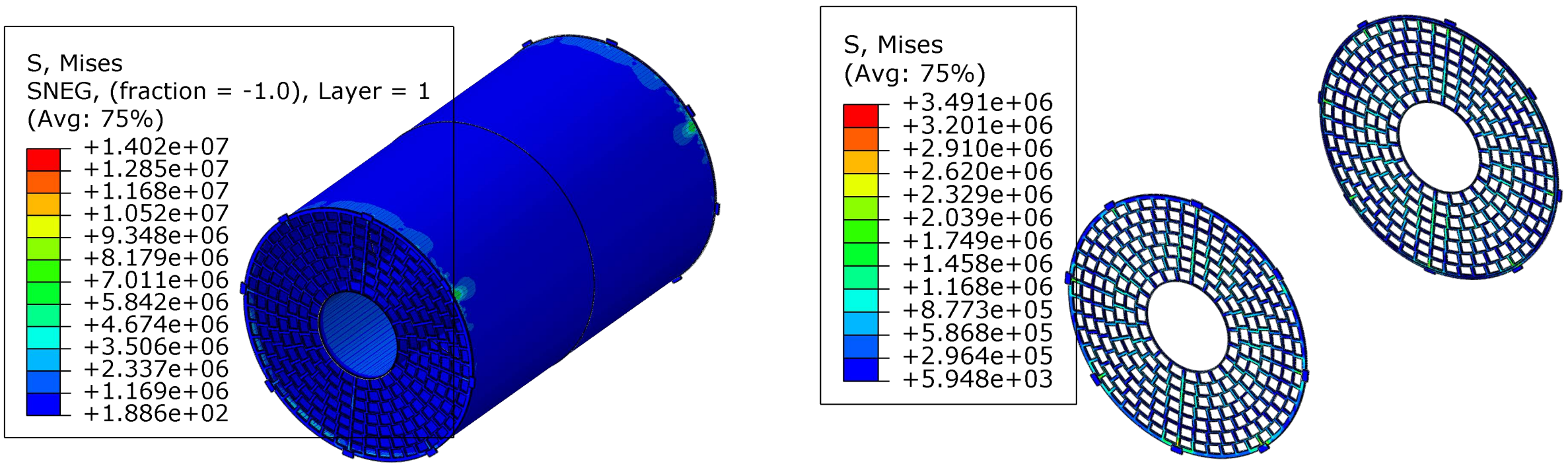

**(a)** Stress in the barrel

**(b)** Stress in the Endplates

**Figure 6.8:** The TPC von Mises stress contour plots based on FEA. Units are in Pa. (a) Stress in the barrel, (b) Stress in the Endplates. The maximum stress in the endplate is 3.5 MPa, while the maximum stress in the barrel is 14 MPa at the connection between the barrel and the endplates, which is mainly caused by stress singularity, and the actual stress should be less than this value.

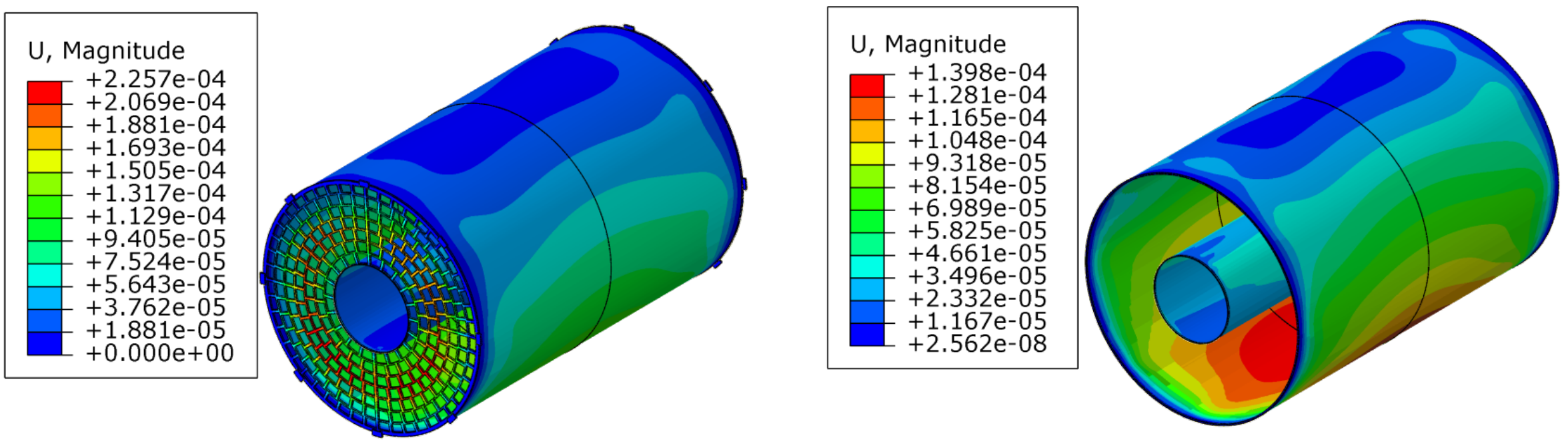

**(a)** Deformation of the whole structure **(b)** Deformation of the barrel

**Figure 6.9:** The TPC deformation contour plots based on FEA. Units are in meter. (a) Deformation of the whole structure, (b) Deformation of the barrel. The maximum deformation of the whole structure is 226 μm, located at the endplates, and the maximum deformation of the barrel is 140 μm, located at the bottom of the outer cylinder.

The results of the FEA are shown in Figure 6.8 and Figure 6.9. Figure 6.8 presents the von Mises stress distribution. The maximum stress in the endplates is 3.5 MPa, which is located at the position of the support block of the endplates. However, the maximum stress in the barrel is 14 MPa at the connection between the barrel and the endplates, which is caused by stress singularity. More detailed analysis will be conducted in the following studies. Figure 6.9 shows the deformation distributions. The maximum deformation of the barrel is 140 μm, located at the bottom of the outer cylinder, while the maximum deformation of the endplates is 226 μm. This is primarily caused by the load of the OTK, inducing an outward deformation of the endplates under the effect of the internal gas pressure. Regarding the impact on the electric field, this degree of the deformation is acceptable. Additionally, the impact of the deformation on the OTK can be corrected through laser measurement and alignment algorithms.

In conclusion, the mechanical design of the TPC using ultra-light sandwich structure of the cylinders can ensure the stability of the TPC with acceptable deformation, indicating that the design is reasonable. Future work will focus on further design optimization based on carbon fiber production processes.

### 6.2.4 High-granularity readout and electronics

The block diagram of the TPC readout electronics is shown in Figure 6.10. Each readout module comprises a double-mesh Micromegas detector, a FEE board, and an associated readout board. The readout module also includes high-voltage and low-voltage supply cables for system operation, as well as a cooling system to dissipate heat generated by the electronics.

In order to achieve high spatial resolution, high rate capability, and excellent PID performance, high-granularity readout with a pad size of 500 μm × 500 μm is selected based on preliminary simulations, which is also driven by power consumption considerations. The high-granularity readout can provide the $\mathrm{d}N_p/\mathrm{d}x$ (the number of activated pads along the track segments) measurement. Compared to the d$E$/d$x$ method, the d$N_p$/d$x$ method has a lower sensitivity to Landau tails. Therefore, it has the potential to significantly improve the PID performance. Using the d$N_p$/d$x$ method, the $K/\pi$ separation power of the TPC reaches approximately $3\sigma$ at 20 GeV/c (details are described in Section 6.4.4). Although technologies like the GridPix detector have achieved a finer granularity of 55 μm × 55 μm for future experiments like the EIC [12, 13], their high power consumption and cooling requirements make them unsuitable for the CEPC TPC.

This high-granularity readout scheme of the TPC results in over 30 million readout channels per endplate, presenting major challenges in terms of electronics density and power consumption[14].

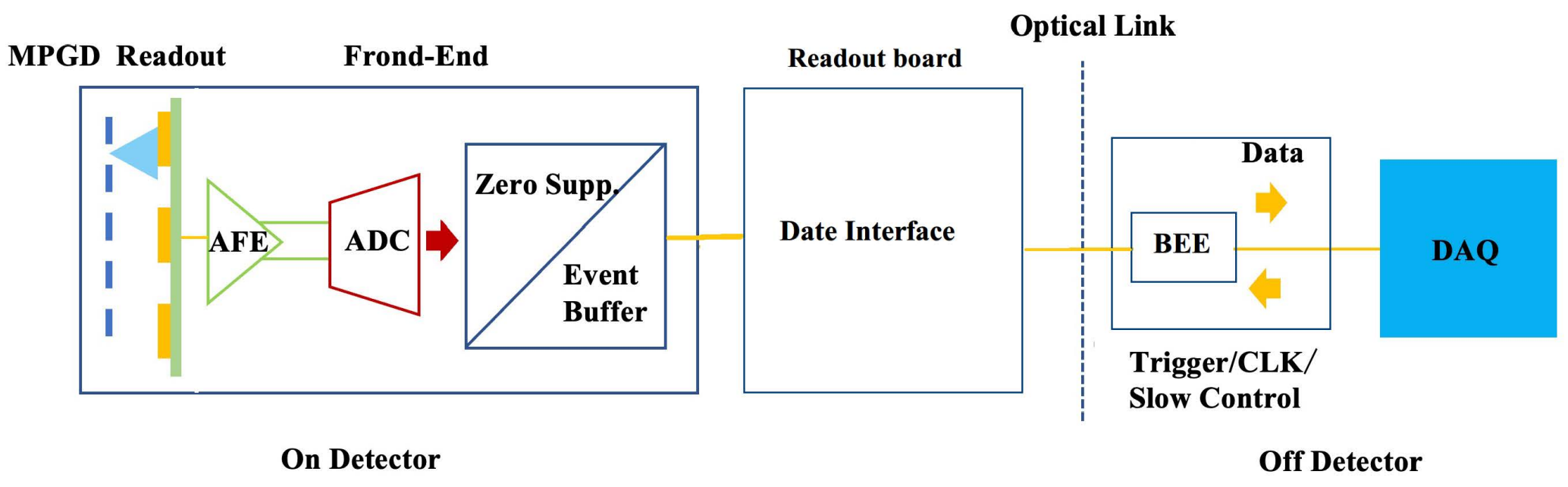

**Figure 6.10:** Architecture of the TPC readout electronics. The FEE will be mounted directly on the MPGD readout board. It includes an Analog Front-End (AFE), comprising a preamplifier and a shaper, followed by an Analog-to-Digital Converter (ADC), and a Digital Signal Processors (DSP) unit. Signals are sent through high-speed optical links to the off-detector BEE and Data Acquisition (DAQ) system.

#### 6.2.4.1 Low power consumption FEE

A low-power readout ASIC named TEPIX [15] is designed to measure the charge and arrival time of each signal. To achieve the connection between the high-density readout pads of the Micromegas detector and the multi-channel TEPIX chips, an organic interposer will be developed using advanced multi-chips packaging technology, instead of the traditional PCB. The key specifications of the FEE and the TEPIX chips are listed in Table 6.5.

Figure 6.11 shows the schematic diagram of the TEPIX chip. Each channel contains a CSA with two gain settings (40 mV/fC and 6 mV/fC), a Correlated Double Sampling (CDS) shaper, a discriminator, 4-channel sample/hold circuits, and a Wilkinson ADC, enabling simultaneous energy (8-10 bits) and time-of-arrival measurements. The chip is designed to operate at 10 kHz frame rate. The pipeline maintains an effective integration time greater than 90%, allowing the detection of up to four events per pad per frame (theoretically 40 kcps/pad).

The TEPIX ASIC is designed to operate in various trigger modes, encompassing both global trigger mode and self-trigger mode. It is capable of storing signal amplitudes and timing information for up to four events before processing. Upon receiving a trigger signal, the analog amplitudes are converted in parallel by Wilkinson-type ADCs.

**Table 6.5:** The key specifications of the FEE (500 μm × 500 μm pad) and the TEPIX chips using 65 nm process.

| Items | Specifications |
|---|---|
| Total number of channels | >30 Million per endplate |
| Pad size | 500 μm × 500 μm |
| Equivalent Noise Charge (ENC) | 100 $e^-$ |
| Dynamic range | 0.1–20 fC |
| Charge buffer dynamic range | 8-10 bits |
| Event rate | 12 kHz/cm$^2$ max. in low-lumi. $Z$ mode |
| Event size | 64 bits |
| Data rate | <10 Gbps per module with compression |
| Power consumption | <100 mW/cm$^2$ |

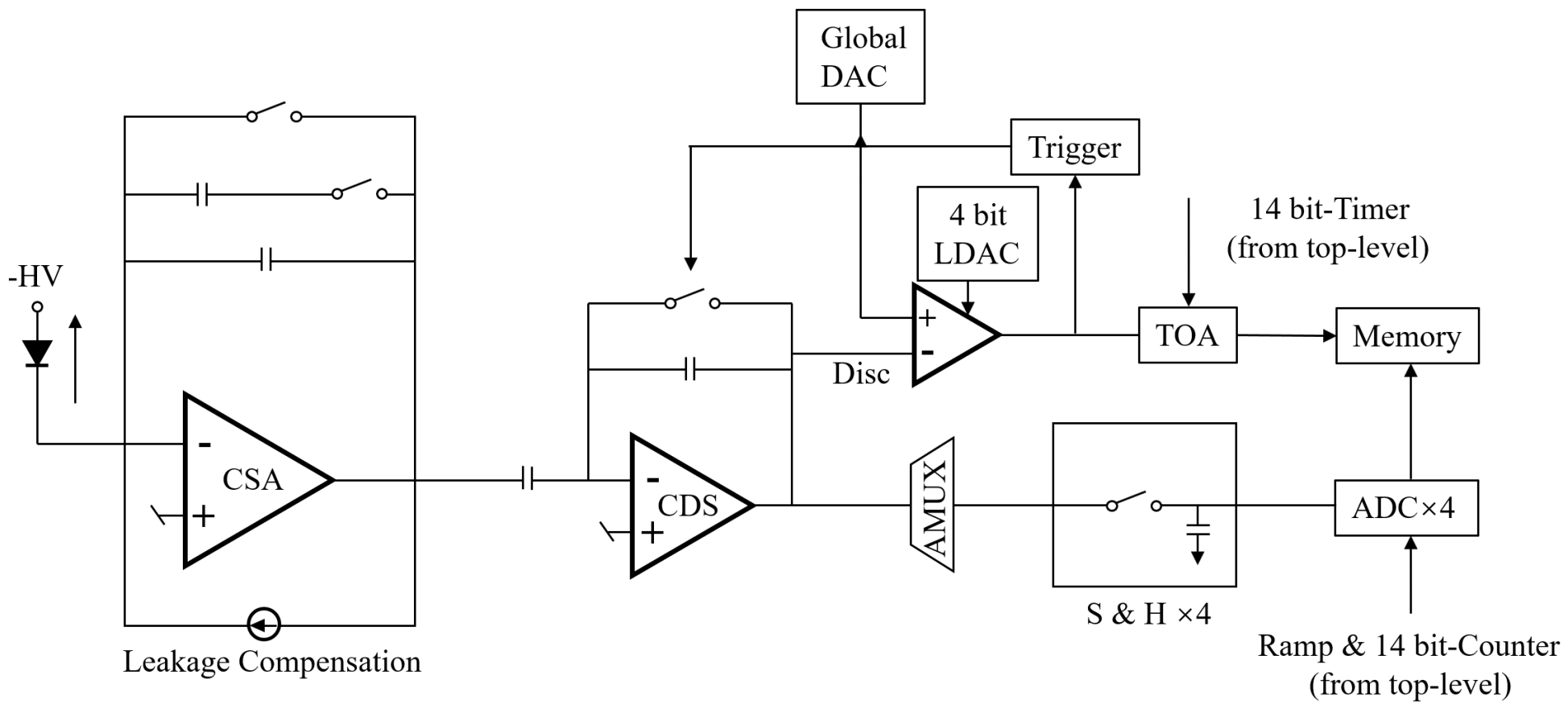

**Figure 6.11:** Schematics of the TEPIX chip, with each channel integrating a charge-sensitive preamplifier, a correlated double sampling shaper, a discriminator, a time counter and analog memory circuits.

The architecture of the TEPIX ASIC demonstrates significant advantages over conventional peak-detection and ToT methods in count rate capability and linearity. Key innovations include pad-level storage cells for high count rates, global summation triggering for charge-sharing events, and three operating modes (Local/Global Self-Trigger, Global External Trigger).

#### 6.2.4.2 Radiation hardness

At the CEPC, the TPC readout electronics are positioned several meters away from the collision point, where the radiation dose is relatively low (less than 1 krad, as discussed in Chapter 3). This considerable distance from the high-radiation environment reduces the demand for specialized radiation-hardened technologies for the readout electronics. However, energetic particles can occasionally induce instantaneous failures, such as Single Event Upset (SEU) or Single Event Latch-Up (SEL). These rare events can potentially impact system performance or even cause irreversible damage. To mitigate these risks, it is essential to apply radiation-tolerant design principles to the TPC readout electronics. These principles aim to ensure that the overall system performance remains robust and unaffected by such rare but potentially disruptive events. By integrating these design strategies, the TPC readout system can maintain its operational integrity and reliability, thereby supporting the high-precision requirements of the CEPC physics studies.

The majority of Application-Specific Integrated Circuits (ASICs) employed in some detector systems at the high energy physics are fabricated in commercial grade CMOS technologies spanning nodes from 180 nm down to 65 nm. Thinner oxides are expected to provide an increased tolerance to ionizing radiation effects. Studies have shown that 65 nm CMOS technology can tolerate high TID levels well beyond the CEPC requirements [16, 17]. In the following Section 6.3.1.3, the developed ASICs have also been validated by X-ray irradiation.

#### 6.2.4.3 Multi-chip interposer

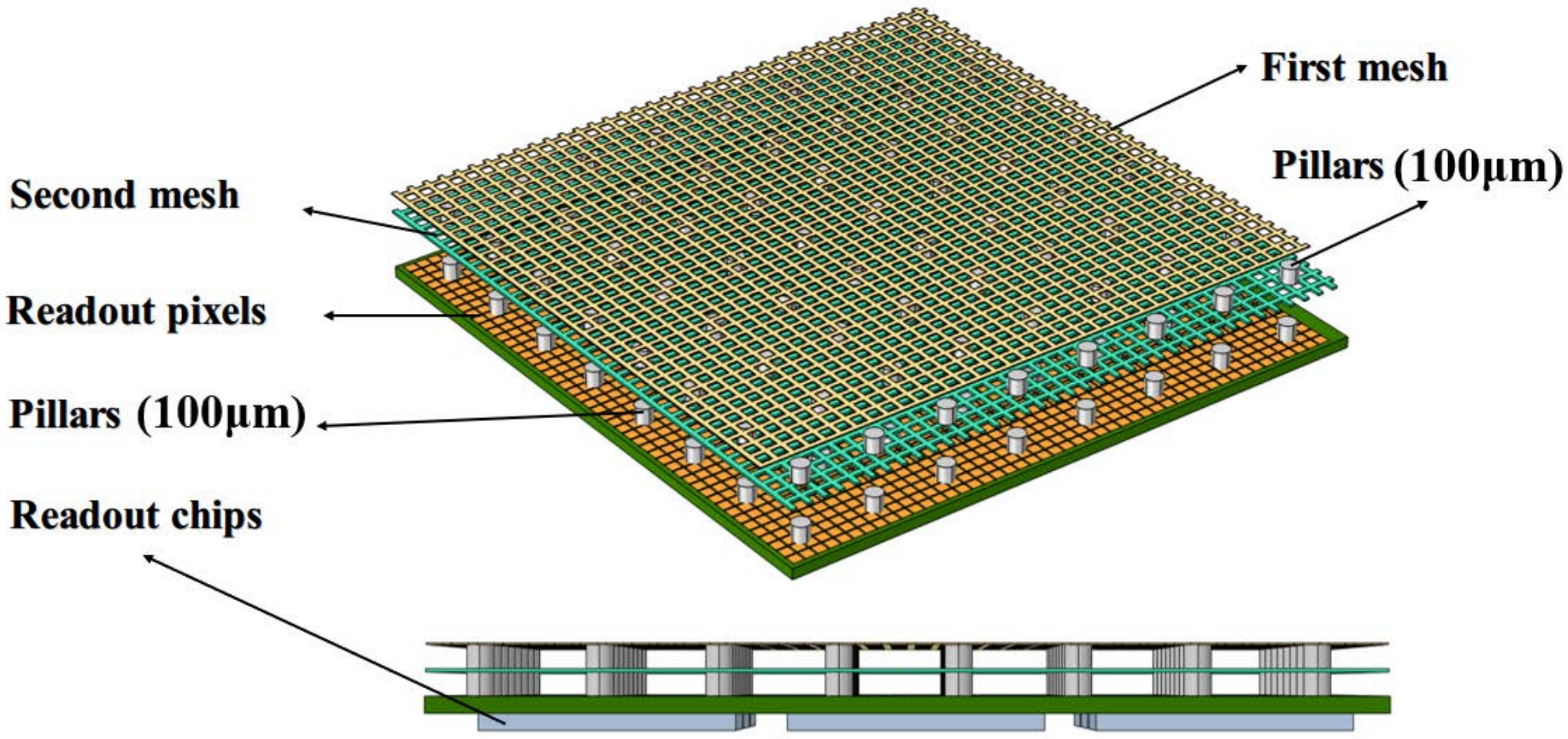

**Figure 6.12:** Conceptual drawing of the readout panel as viewed from the chamber side. The first mesh serves as the IBF barrier to suppress positive ions. The second mesh is positioned over the pillars on the multi-chip interposer, serving as the main amplication mesh of the Micromegas detector. The interposer routes the charge collected on the small pads to the readout ASICs on the back of the interposer. One ASIC connects to 256 small pads.

A multi-chip interposer will be employed to route the charge-collecting pads to the inputs of the readout ASICs, as depicted in Figure 6.12. This approach offers the flexibility to implement different pad sizes using the same readout chip, thereby facilitating detector optimization. To achieve a pad size of 500 µm × 500 µm, the interposer will be fabricated using organic substrate technology, for its fine routing and drilling, as well as cost-effectiveness compared to other alternatives such as silicon and glass.

The minimum Line/Space (L/S) of the interposer is 15 μm and the minimum hole size is 50 μm. The routing from the pads to the chip input is minimized to reduce parasitic capacitance and, consequently, electronic noise. The power and other I/O pins are routed to a connector on one side of the interposer, which also allows connection to three sides of the ASIC chip. The interposer also provides critical electrical isolation between the high-voltage domain of the detector pads and the low-voltage ground of the readout electronics. The fabrication process involves via formation, sidewall insulation, barrier/seed deposition, copper electroplating to fill the vias, and back-grinding to expose the opposite surface.

This design, leveraging techniques from advanced multi-chip modules, meets all mechanical and electrical requirements for the CEPC TPC, enabling a compact and high-performance interconnection system.

### 6.2.5 Cooling

Cooling is critical for the high-granularity readout TPC. The dominant heat source is the ASIC array on each readout module, with an estimated total power consumption of approximately 10 kW per endplate. An efficient thermal-management system is therefore required to remove the heat and maintain the readout ASICs within their specified operating temperature range. In Figure 6.13, a schematic diagram of the cooling scheme for a single readout module is presented. A serpentine water-cooled tube is intimately bonded to the cooling plate, efficiently extracting heat from every ASIC and maintaining the ASIC chip temperature below approximately 28 °C to ensure the proper functioning of the chips. Above the cooling plate, the signal aggregation readout board is mounted via interface connectors.

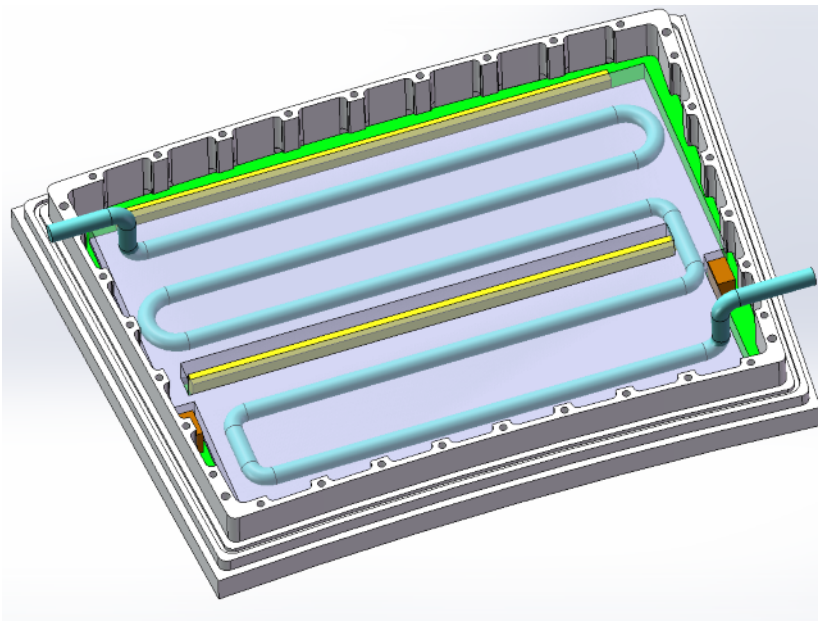

**Figure 6.13:** Schematic diagram of the cooling scheme for a single readout module: a water-cooled plate with one inlet and one outlet tube is thermally coupled to the ASIC side. Two yellow interface pieces establish the mechanical and electrical connection to the readout board.

Figure 6.14 illustrates the cooling pipe routing scheme for a quarter of the endplate. This quarter incorporates five independent cooling loops, with each loop connected in series to approximately 12 readout detector modules. In order to simplify the routing of the cooling pipe, the 12 readout modules connected to each circuit are distributed on the same layer or adjacent two layers. Scaling up to the full endplate, one endplate will comprise 21 parallel cooling loops.

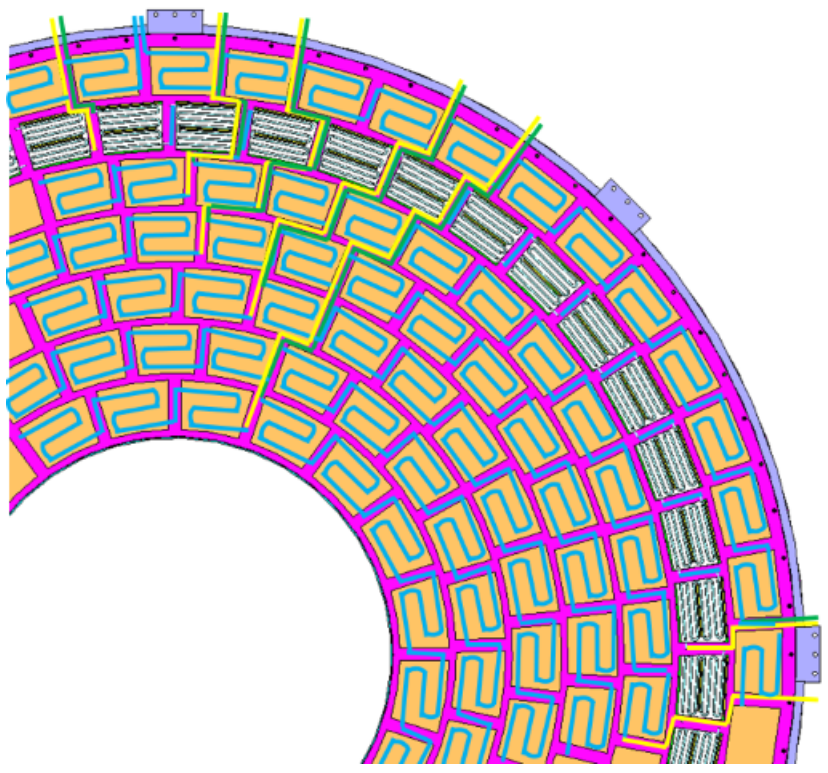

**Figure 6.14:** Schematic diagram of cooling pipe routing scheme for a quarter of the endplate. The quarter contains five independent cooling loops, and each loop is connected in series to about 12 readout detector modules.

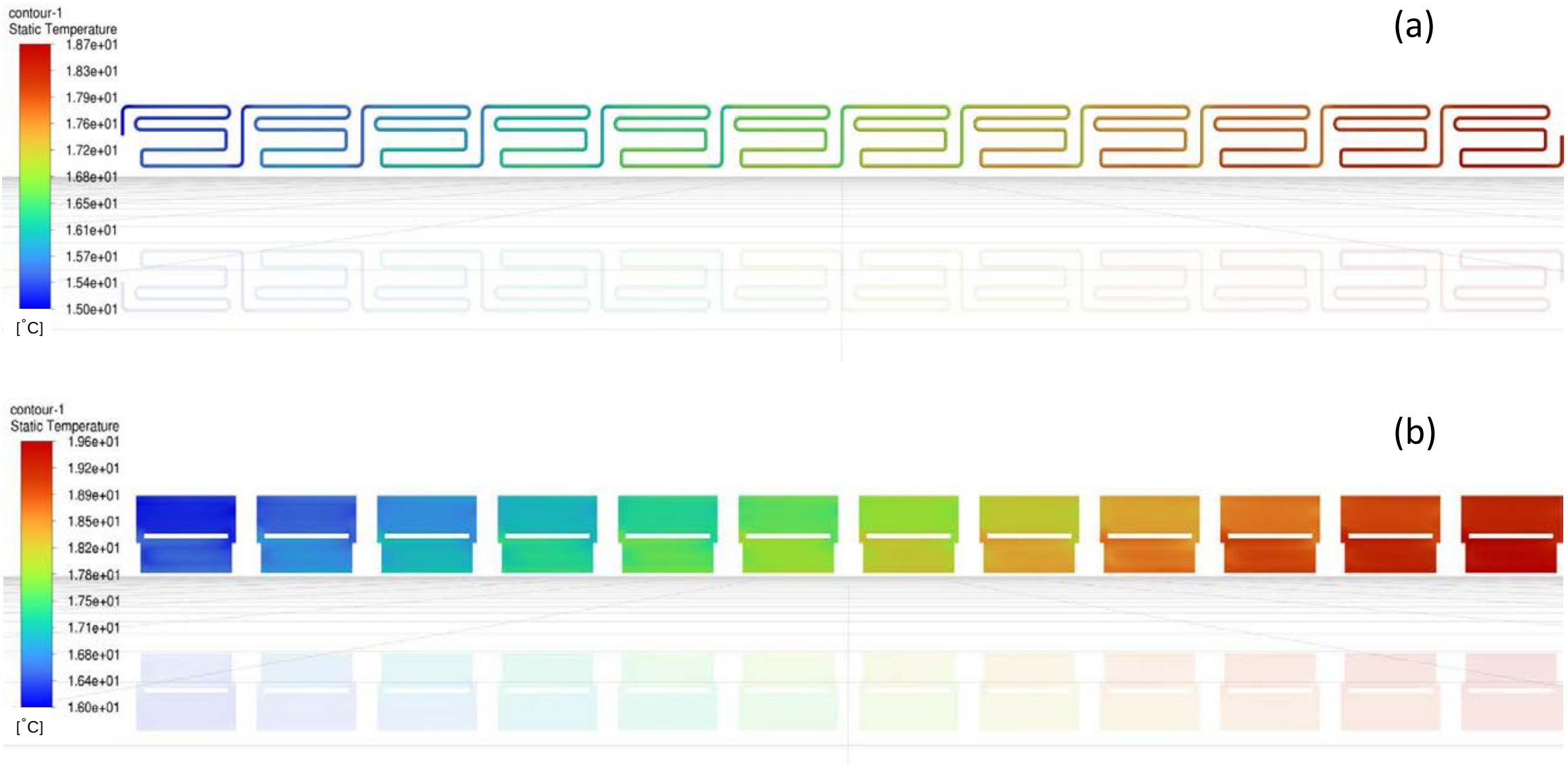

**Figure 6.15:** Thermodynamic FEA results of a cooling loop with 12 readout modules connected in series. (a) The temperature distribution of the cooling loop water pipes. (b) The temperature distribution of ASIC chips on the cooling loop. The temperature at the inlet is 15 °C. After the cooling water flows through the 12 readout modules, the outlet temperature is approximately 18.7 °C, indicating a water temperature increase of 3.7 °C. The corresponding maximum temperature of the ASIC is about 19.6 °C, which meets the heat dissipation requirements for stable operation of the ASICs.

Based on the specifications of the ASIC chips and the design of the readout modules, each readout module dissipates approximately 40.5 W. Assuming a cooling flow velocity of 1.5 m/s through a 6-mm-diameter pipe, the temperature rise of a single cooling loop based on thermodynamic FEA is 3.7 °C, as shown in Figure 6.15, which confirms the effectiveness of the proposed cooling scheme.

With an optimized inlet water temperature of 15 °C, the maximum outlet temperature is maintained at 18.7 °C, and the corresponding maximum temperature of the ASICs is about 19.6 °C (Figure 6.15), ensuring stable thermal conditions for the ASIC operation. Further design optimization of the cooling pipes and cooling plates will be guided by thermodynamic FEA.

## 6.3 R&D Efforts and key technologies

Based on the foregoing design studies and the critical technical issues identified, the dedicated R&D efforts have been under way for several years. Experimental tests have already yielded some results. This section summarizes the progress achieved to date and discusses the key technical challenges that remain to be addressed.

### 6.3.1 R&D efforts and results

This subsection describes the experimental studies carried out with the TPC readout modules and prototypes, including the generation and application of Ultraviolet (UV) laser tracks, the front-end electronics for the TPC module, the radiation tests of key functional blocks within the ASICs, and the development of the high-granularity readout module prototype.

#### 6.3.1.1 Study on TPC module and prototype

A TPC prototype testing system, as shown in Figure 6.16, was set up for the key technology studies. The TPC prototype featured with a drift length of 500 mm and an effective area of 200 mm × 200 mm with double-Gas Electron Multiplier (GEM) readout structure. It operated with the T2K gas at a gain of approximately 3400 in the relative low-gain mode [18]. To facilitate the performance testing, the TPC prototype was integrated with a 266 nm ultraviolet (UV) laser system, which could provide six horizontal laser tracks in the prototype chamber. Over the past few years, various performance metrics and testing results of the TPC prototype, integrated with UV laser tracks, have been published [19], including measurements of hit resolution, $\mathrm{d}E/\mathrm{d}x$ resolution, and drift velocity.

The $\mathrm{d}E/\mathrm{d}x$ resolution of the prototype was measured to be (8.9 ± 0.4)%. Extrapolating it to the substantial track length, the $\mathrm{d}E/\mathrm{d}x$ resolution of the TPC with a readout pad size of 1 mm × 6 mm (220 layers) is estimated to be (3.36 ± 0.26)%, assuming 100% available hit points [20]. The resolution of the TPC only decreases by 0.6% when considering 10% invalid hit points.

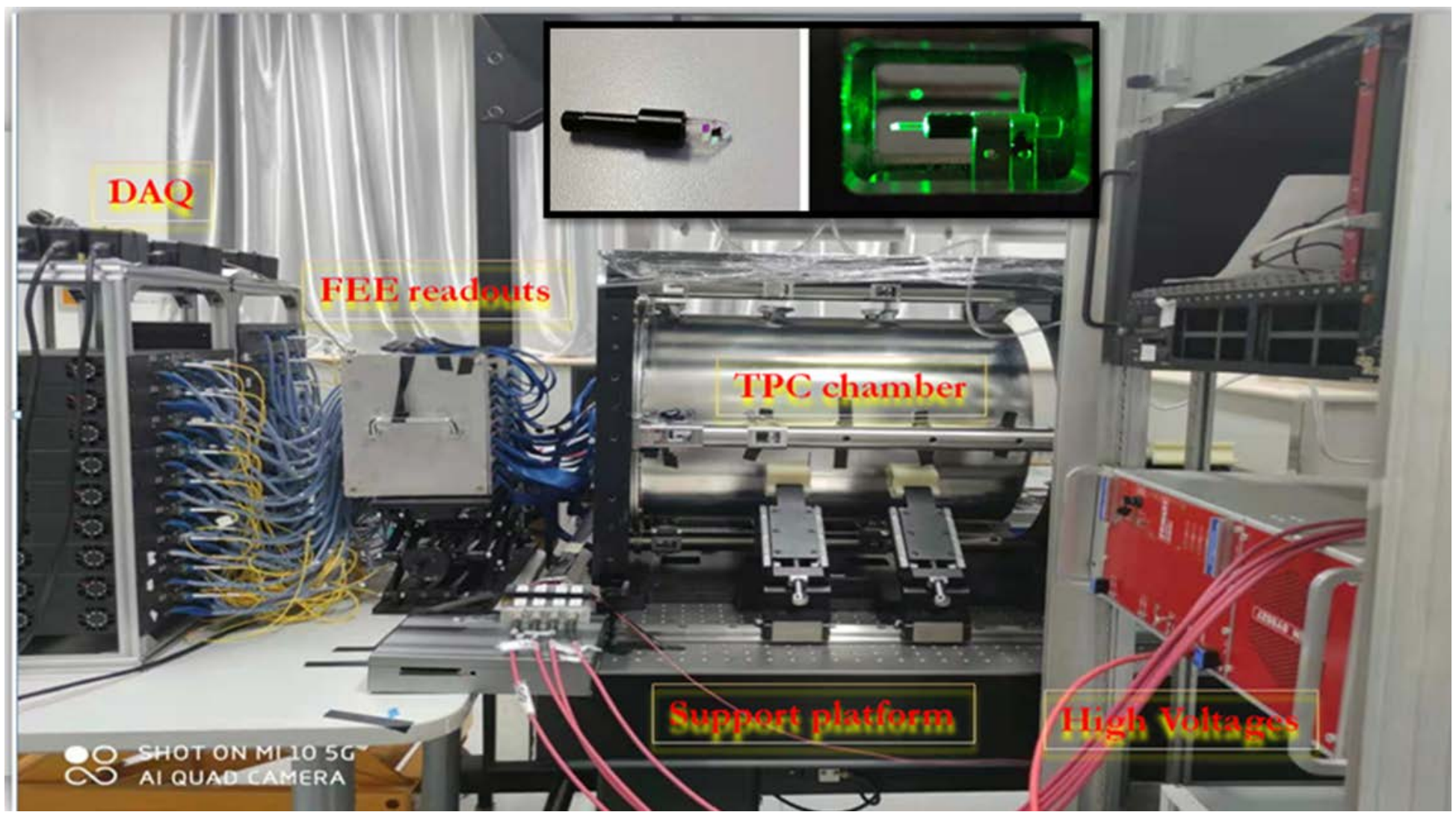

**Figure 6.16:** A TPC prototype integrated with six horizontal UV laser beams, with a drift length of 500 mm and an effective readout area of 200 mm × 200 mm.

#### 6.3.1.2 Front-end electronics R&D

A TEPIX ASIC chip prototype has been developed to demonstrate the design and the performance. To address the critical challenge of power consumption, this prototype employs a 180 nm CMOS process and a circuit structure featuring simple analog circuits. The chip integrates 128 channels in a 2.2 mm × 5.6 mm area, as shown in Figure 6.17. The testing results of the prototype are listed in Table 6.6. The noise and power consumption exceed the specifications, which will be improved with advanced 65 nm process in future work.

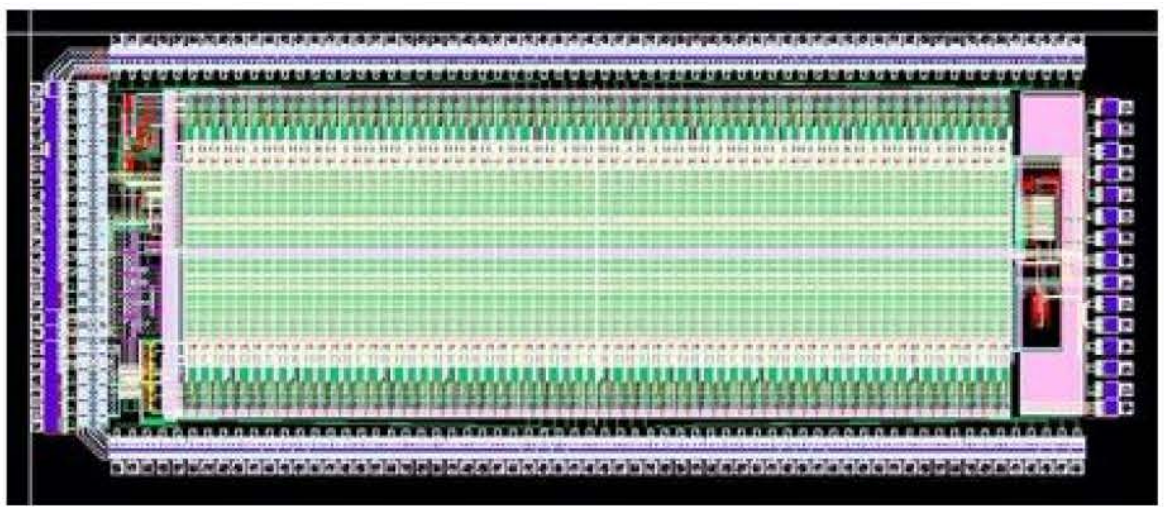

**Figure 6.17:** Layout of the TEPIX prototype with 128 channels using 180 nm process. The size of ASIC chip is 2.2 mm × 5.6 mm.

In the tests, the master bias current of the AFE was adjusted using an external resistor. The total current of the test PCB increases linearly with the master bias current. To estimate the power consumption of the AFE, the total current was modeled as the sum of the AFE current (assumed to be proportional to the master bias current) and a constant current from other components. Therefore, the current consumed by all other components on the PCB was estimated from the y-intercept of the linear fit, a value determined to be 9.73 mA. Based on this model, the power consumption of the AFE was estimated to be 2.0 mW per channel at the nominal master bias current of 25 µA, matching the simulation result of 1.98 mW. The testing results of the AFE and the Successive Approximation Register (SAR) ADC demonstrate excellent agreement with the specified performance metrics, thereby fulfilling the required performance criteria. The power consumption of the entire AFE and ADC core circuits has been optimized to approximately 3.0 mW per channel, representing a significant reduction of at least 50% compared to other contemporary ASICs with similar architectures.

**Table 6.6:** Testing results of the 128-channel TEPIX chip using 180 nm process.

| Parameters | Testing results |
|---|---|
| ENC | 300 $e^-$ |
| Dynamic range | 0.24–25 fC |
| Integral Non-linearity (INL) | <1% |
| Time resolution | 10 ns |
| ADC | 10 bits |
| Total power consumption | 0.3 mW/channel |

#### 6.3.1.3 Radiation tests of ASIC prototypes

To validate the radiation tolerance of the critical functional blocks within the readout ASICs, dedicated irradiation tests were performed. Three types of ASIC prototypes with different functional blocks were designed and fabricated, including a five-channel AFE chip, a SAR ADC chip, and a mixed-signal chip with both AFE and ADC.

The TID tests were performed using a $^{60}$Co source at Xi'an Northwest Institute of Nuclear Technology[21]. Three AFE and SAR ADC chips were exposed to the radiation source. An FPGA board shielded with lead bricks was used to provide the sampling clock for the SAR ADC. The power supply and signal generator were located in the control room and interconnected with the chips through high-speed coaxial cables. These chips were irradiated at a dose rate of 50 rad (Si)/s at room temperature with the total dose up to 1 Mrad (Si). The components were then annealed at room temperature for 168 hours (7 days). The critical functional blocks of the chips were studied, and all three devices exhibited stable performance with no significant change after the 1 Mrad (Si) exposure, thereby validating their radiation tolerance.

#### 6.3.1.4 Readout module prototype

Using the prototype of the TEPIX chip, a $32 \times 32$ pad-array interposer prototype has been developed to integrate the TEPIX chips to the charge-collecting pads (pad size: 500 µm $\times$ 500 µm).

Figure 6.18 illustrates the prototype of the readout detector module after integration of the Micromegas detector and the FEE.

The readout prototype study is designed to evaluate two configurations: a single and a double-mesh Micromegas. The single Micromegas module serves to validate detector long-term stability and the spatial resolution, while the double-mesh Micromegas configuration is employed to quantify IBF suppression and the spatial resolution. The readout detector prototype will be tested using a test beam.

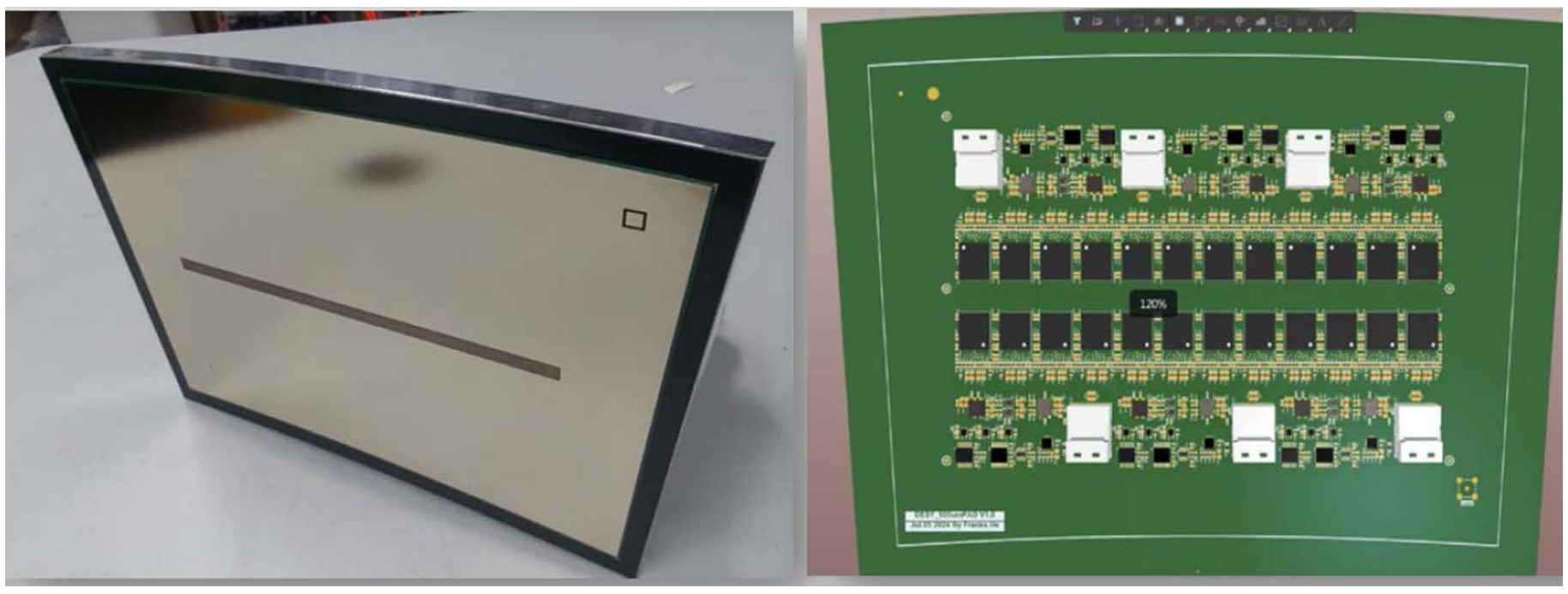

**Figure 6.18:** Prototype of the TPC readout detector module.

### 6.3.2 Ion backflow suppression

The spatial resolution of the TPC is ultimately limited by the space-charge effect, which primarily arises from two sources: positive ions produced by the ionization of the working gas itself, and, more critically, ions backflowing from the amplification region of the readout detectors. This IBF occurs in the following way: primary electrons drifting to the TPC endplate are multiplied in the readout detectors, generating signals; an equal number of ions (corresponding to multiplied electrons) are created and some of them drift back into the TPC volume at roughly 1% of the electron drift velocity. Although the readout structures are designed to confine ions to the multiplication region, a small fraction inevitably leaks back. These ions distort the electric field, perturb the primary electron drift, and degrade the TPC performance.

The IBF ratio is defined as $\epsilon_{\text{IBF}} = n_{\text{IB}}/n$, where $n_{\text{IB}}$ is the number of ions entering the TPC volume and $n$ is the total number of generated ions. Minimizing the IBF is essential for achieving optimal TPC performance.

Primary ion production can be mitigated by optimizing the MDI region, while continuous suppression of the IBF can be achieved by introducing a second mesh for the readout detector or applying a graphene coating to the amplification device.

MPGDs like Micromegas offer an intrinsic advantage for suppressing this IBF, by acting as a gate for electron transmission and a blocker for ions. Adding additional mesh can further enhance the IBF suppression capability of the Micromegas detectors. Double-mesh Micromegas detectors were setup and tested. The testing results are shown in Figure 6.19. At a gain of about 2000, the IBF ratio × gain can be reduced to below 1. This means that, on average, fewer than one ion backflows into the drift volume for each primary ionization electron, effectively mitigating the space charge effect.

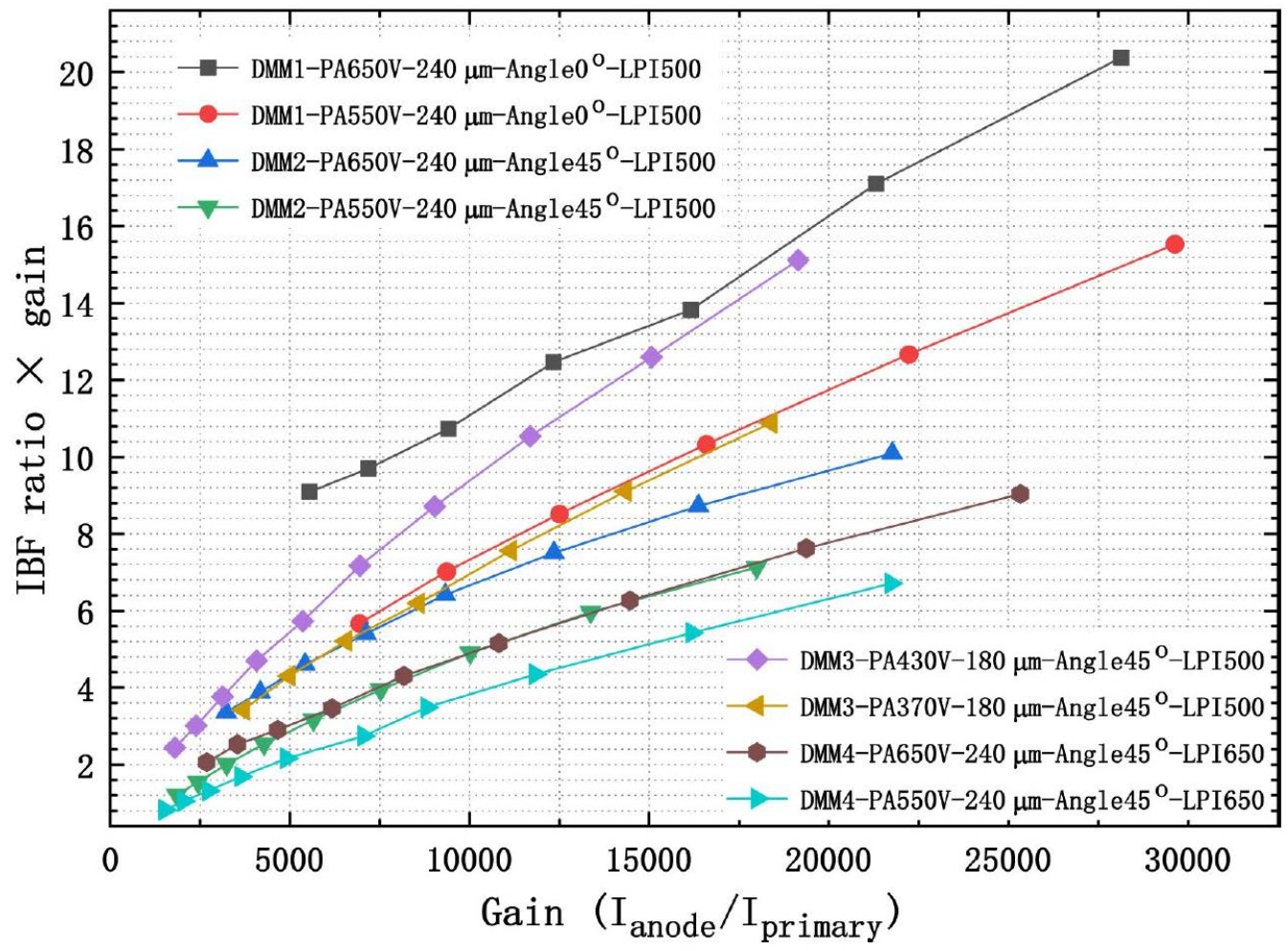

**Figure 6.19:** Results of the ion backflow suppression by using a double-mesh structure detector (the pre-amplification mesh integrated with a Micromegas detector). The $x$-axis represents the gain of the Micromegas detector, and the $y$-axis indicates the number of ions flowing back into the TPC volume after electron multiplication in the Micromegas detector. The curves in different colors correspond the testing results obtained with different detector configuration parameters. When using 500 or 650 LPI meshes, with a high voltage of 550 V on the pre-amplification mesh, a crossing angle of 45 °, and a gap of 240 μm between the two mesh layers, the IBF ratio × gain achieves a value of less than 1 at a gain of about 2000 (two green curves) [22].

Recently, significant progress has been made in the study of graphene and mutil-mesh structure for positive ion backflow suppression [23]. Graphene, the thinnest known material, consists of a single carbon atom layer forming a microporous structure, enabling it to block ions while allowing electrons to pass through. Experiment results have shown that monolayer graphene membranes exhibit exceptional mechanical properties, allowing them to remain suspended independently over micron scale pores. Therefore, they have become a promising candidate for future detector optimizations.

### 6.3.3 Beam-induced background estimation

The beam-induced background, primarily from sources such as $e^+e^-$ pairs and beam-halo muons detailed in Chapter 3, remains a key challenge for the TPC operation at the CEPC. Although the electron-positron collision environment in the CEPC is cleaner than that of hadron colliders, beam-induced background is unavoidable.

In general, the momenta of beam-induced background particles are much lower than that of particles originating from physics events. These low momentum particles typically do not enter the TPC directly but may interact with the beam pipe or other upstream materials, producing secondary recoil particles. Some of these secondaries can enter the TPC and ionize the working gas, contributing to readout occupancy. Furthermore, the accumulation of slowly drifting ions in the TPC can reduce the uniformity of the drift electric field and thus degrade the spatial resolution. Simulations based on the MDI model presented in Chapter 3 have been performed to evaluate the beam-induced background impact, focusing on high-granularity readout TPC voxel occupancy and space charge distortion [24].

#### 6.3.3.1 Simulation flow of beam-induced background in the TPC

As shown in Figure 6.20, a complete simulation flow has been established to study the effect of beam-induced background on the TPC performance. The beam-induced background generation is described in Chapter 3. The output from the different generators is exported to the CEPCSW bunch by bunch for full detector simulation. The CEPCSW framework has integrated DD4hep toolkit [25], which can provide a full detector description, including detector geometry, materials, visualization attributes and readout information.

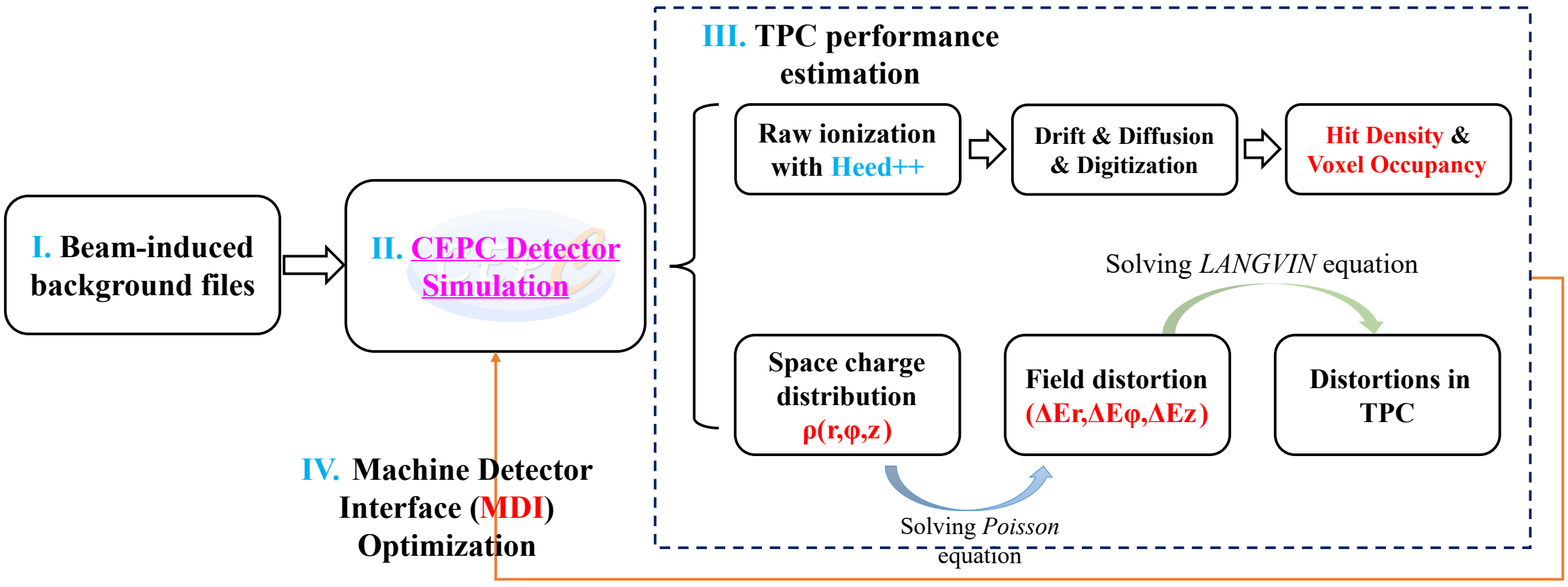

**Figure 6.20:** Simulation flow of beam-induced background that consists of four steps. Step I involves generating beam-background seeds. Step II is the full-scale detector simulation. Step III focuses on the performance estimation. The final step is the MDI region optimization.

Two studies have been performed: hit density and voxel occupancy estimation, and space distortions caused by the space charge effect. For the occupancy calculation, it is important to obtain the number of fired pads and the signal time. A new tool, called *TrackHeedSimTool* invokes *TrackHeed* [26] in Garfield++ to perform ionization simulation at the molecular level and record each primary electron along the track. Each primary electron undergoes a process of drift, diffusion, and amplification to generate signals on the Micromegas anode. Each fired pad is then mapped to a 3D voxel to calculate the voxel occupancy. The estimation of space distortions involves calculations of the space charge distribution, the electric field distortion, and its impact on the motion of electrons, which are discussed in Section 6.3.3.4. The TPC simulation results are used as an important reference for the optimization of the CEPC MDI region.

#### 6.3.3.2 High-granularity readout TPC occupancy

The voxel occupancy is a critical parameter for the TPC's track-finding performance. The voxel occupancy of the high-granularity readout TPC is the number of 3D readout cells (voxels) that see a signal, divided by all voxels in the TPC. For the CEPC TPC, the time window is determined by the maximum electron drift time of ~34 μs, which corresponds to integrating at least 122 bunch crossings in the $H$ operation (277 ns bunch spacing time). Given a typical drift velocity ($V_d$) of ~8 cm/μs in the T2K gas and the 43 MHz DAQ sampling rate, the effective voxel size is defined as 0.25 mm$^2$ in the $x - y$ plane (dictated by the pad size) and 1.86 mm in the $z$-direction (calculated as $V_d \times 23.25$ ns). Therefore, within a maximum drift time of 34 μs, the total voxels in the TPC reach $9 \times 10^{10}$. The values of the average and maximum hit density in different running modes of the CEPC are provided in Chapter 3. The maximum voxel occupancy is calculated to be less than $7.5 \times 10^{-4}$ in the innermost layer of the TPC. This level is safe for the TPC operation and has no impact on track finding.

#### 6.3.3.3 Main energy deposition contribution

The particle type and energy distributions in the $H$ and low-luminosity $Z$ mode after 10 BX are shown in Figure 6.21 (a). Photons and electrons are main components, and other types of particles, including positrons, protons, and neutrons, are negligible. Figure 6.21 (b) shows the energy distributions of the photons that interact with the TPC working gas, showing a majority of photons below 10 MeV and a prominent peak at 0.511 MeV from positron annihilation. In the $H$ mode, approximately $7.2\times10^4$ photons cross the TPC per 10 BX, of which about 1800 photons interact with the working gas. The low-luminosity $Z$ mode presents slightly larger numbers of photons and higher energy deposition than the $H$ mode. Photons interact with matter via mechanisms such as the photoelectric effect, Compton scattering, and pair production. At energies in the range of 0.1 MeV to 5 MeV, Compton scattering is dominant in the gas. Many secondary electrons with extremely low momentum ($\sim$ MeV$/c$) will be produced in the TPC, contributing to more than 98% energy deposition in the CEPC $H$ and low-luminosity $Z$ modes. It is important to analyze the properties, e.g. initial position distribution and energy distribution, of these low-energy photons.

#### 6.3.3.4 Estimation of space distortions

Maintaining the TPC's spatial resolution requires mitigating the distortions caused by accumulated positive ions, which originate from beam-induced backgrounds and accumulate due to their slow drift velocity. The basic flow of estimating the distortions caused by the space charge effect is shown in Figure 6.20. The first step is to calculate the space charge density $\rho_{\mathrm{sc}}(r, \varphi, z)$. The number of primary ions produced in each hit is determined by the ratio of energy deposition to the effective ionization potential of the gas. The space charge density is

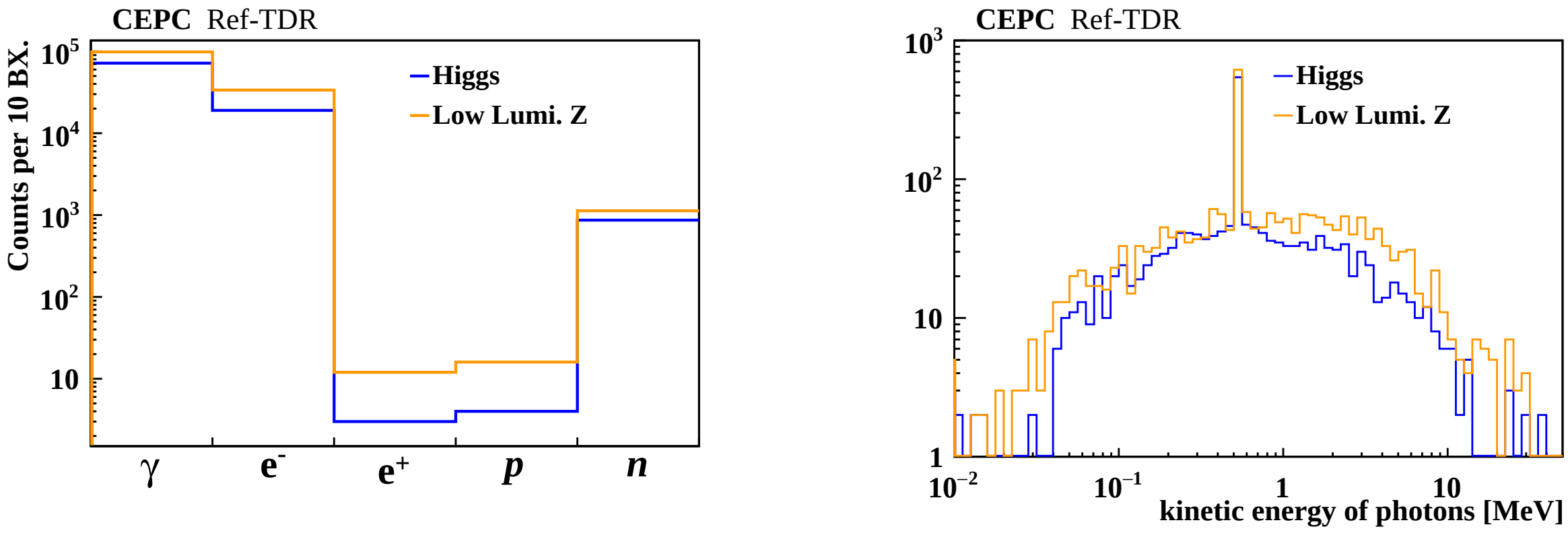

**(a)** Composition of secondary particles **(b)** Energy distribution of photons

**Figure 6.21:** (a) Composition of secondary particles in the TPC arising from beam-induced background. (b) The energy distribution of photons interacting with the working gas. Blue line: $H$ mode; Yellow line: low-luminosity $Z$ modes.

given by Equation 6.1:

$$\rho_{\text{sc}} = e\frac{\text{d}E_{\text{dep}}/\text{d}V}{I_{\text{gas}}} \cdot \frac{L_{\text{max drift length}}}{v_{\text{ion}}} \cdot BX_{\text{freq.}} \tag{6.1}$$

where $dE_{\text{dep}}$ is the energy deposited in the volume element $dV$, $I_{\text{gas}} = 26$ eV is the average ionization energy of the gas, the drift velocity of ions $v_{\text{ion}}$ is taken as 5 m/s [27], and $BX_{\text{freq.}}$ is the bunch crossing frequency varying in different running modes. The term $L_{\text{max drift length}}/v_{\text{ion}}$ defines the maximum time an ion remains in the sensitive volume. Multiplying this by bunch crossing frequency $BX_{\text{freq.}}$ gives the number of bunch crossings during this period, which acts as an accumulation factor for the ion charge.

Assuming that those photons with energy less than 10 MeV can be shielded, the steady space charge density $\rho_{sc}$ distributions in the TPC as a function of radius are shown in Figure 6.22, which only includes primary ions. An empirical function (Figure 6.2) is used to fit the data to get a continuous and smooth curve for subsequent calculations:

$$\rho(r) = \frac{a}{r^k} + \rho_0 \tag{6.2}$$

where $a$, $k$, $\rho_0$ are fitted parameters. From the fitting results, it can be concluded that the space charge density in the TPC due to the beam-induced background decreases in the radial direction for both the CEPC $H$ and low-luminosity $Z$ modes, with the power exponents $k$ close to 3. As shown in Figure 6.22 (a), if only the contribution of primary ionization is included (without photons with energy less than 10 MeV), the maximum space charge density is only 0.04 nC/m$^3$ at the TPC inner radius in the $H$ mode. The maximum space charge density increases by a factor of about 10 due to larger bunch frequency and beam lost rate, approaching 0.62 nC/m$^3$ in the low-luminosity $Z$ mode.

And then, the field distorted components ($E'_r$,$E'_\varphi$,$E'_z$) are calculated by solving the *Poisson* equation based on Green's function method, as mentioned in Ref. [28]. The electric field due to

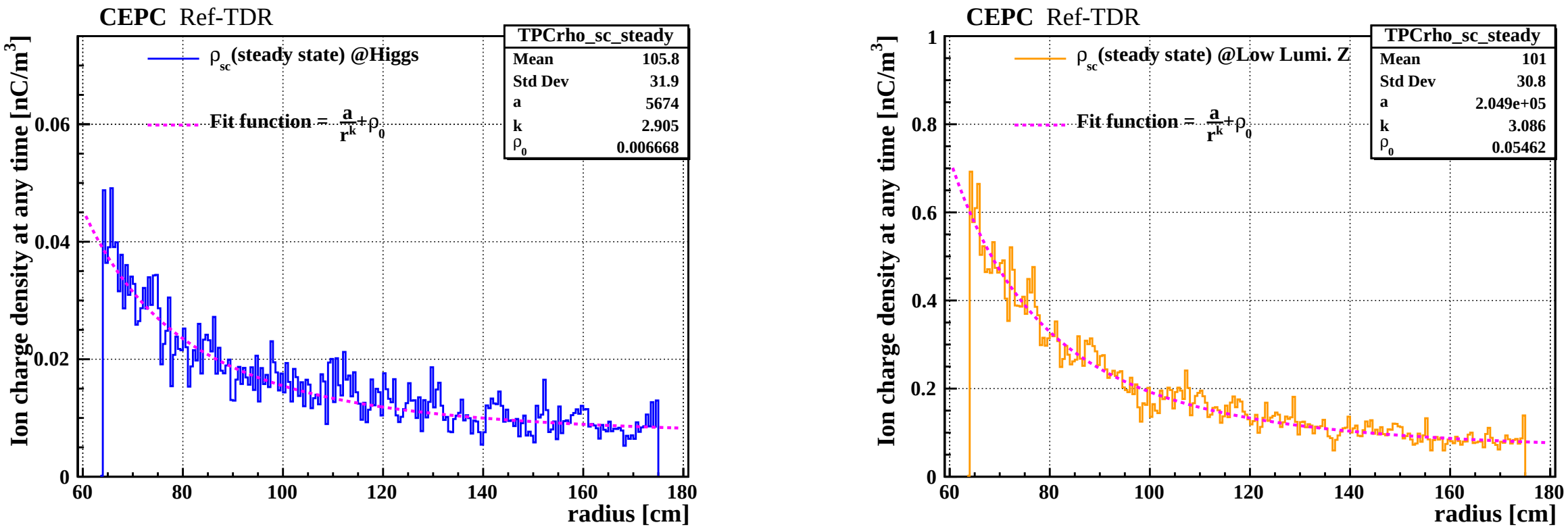

**(a)** $H$ mode **(b)** Low-luminosity $Z$ mode

**Figure 6.22:** Charge density in the TPC as a function of radius (after averaging over $\varphi$ and $z$ directions) and fitted by Equation 6.2 (Magenta dashed line in the plots). (a): $H$ mode; (b): Low-luminosity $Z$ mode.

arbitrary space charge density $\rho_{\text{sc}}(r, \varphi, z)$ can be evaluated using Equation 6.3:

$$\mathbf{E}(r, \varphi, z) = -\nabla\Phi(r, \varphi, z) = -\frac{1}{\epsilon_0}\int r'dr' \int d\varphi' \int dz' \rho_{\text{sc}}(r', \varphi', z') \cdot \nabla G(r, \varphi, z, r', \varphi', z') \tag{6.3}$$

where $\Phi(r, \varphi, z)$ is the potential, $\epsilon_0$ is the permittivity of the free space. The geometry of the TPC can be approximated by two concentric cylinders with radii $R_{\text{inner}} = a$ and $R_{\text{outer}} = b$, where $a < b$. The cylinders are closed at $z = 0$ and $z = L$. The Green's function $G(r, \varphi, z, r', \varphi', z')$ under such a geometry have been derived, and useful representations of the solutions are given in Ref. [29].

The motion of electrons, under the influence of electric and magnetic fields, is described by the *Langevin* Equation [27]. When the electrons drift to the anode under a non-uniform field, distortions occur both in the radial and azimuthal directions. The distortions ($\Delta_r$, $\Delta_{r\varphi}$) at different drift lengths $L$ are given by Equation 6.4:

$$\begin{aligned} \Delta_r &= \int_0^L \frac{1}{1+\omega^2\tau^2} \times \frac{E_r'}{E_z'} dl \\ \Delta_{r\varphi} &= \int_0^L \frac{-\omega\tau}{1+\omega^2\tau^2} \times \frac{E_r'}{E_z'} dl \end{aligned} \tag{6.4}$$

where the Hall parameter $\omega\tau$ is about 10 under 3.0 T magnetic field and 230 V/cm electric field in the T2K gas, simulated by Garfield++ [30]. As Equation 6.4 shows, the azimuthal distortion ($\Delta_{r\varphi}$) is larger than the radial distortion ($\Delta_r$) by the factor of $\omega\tau$. Consequently, the radial distortion is negligible for reconstructing most high $p_{\text{T}}$ tracks. In contrast, the azimuthal distortion has a more significant impact on both high and low $p_{\text{T}}$ tracks. When the electrons drift towards the anode, the electrons close to the inner cylinder of the TPC and close to the outer cylinder of the TPC are deflected in two opposite directions under the action of the distorted electric field. Using the computed $E_r/E_z$ ratio maps, the $r\varphi$ distortion under different drift lengths can be calculated using Equation 6.4.

Azimuthal distortion as functions of drift distances have been calculated, and the results are shown in Figure 6.23. The distortion get larger with the increase of the drift length, and the most severe distortion appears at the smallest radius in both modes. When only the primary ion contribution is considered, the maximum distortions after the 2.75 m drift in the CEPC *H* mode and low-luminosity *Z* mode are measured to be approximately 10 µm and 150 µm (azimuthal), respectively. However, if low energy photons with an energy below 10 MeV cannot be shielded and enter the TPC, and the contribution of the IBF is also included (IBF ratio × gain = 1), the maximum space charge density in the TPC will increase to 1.7–3.3 nC/m$^3$ in the CEPC two operating modes (Figure 6.24). This leads to severe track distortion on the order of 1 mm in the low-luminosity *Z* mode, as shown in Figure 6.25.

The estimation of the TPC space charge distortion and mitigation strategies in *H* and low-luminosity *Z* modes are listed in Table 6.7. The space charge density with the contribution of low-energy photons in the TPC is about 1/36 of the ALICE TPC [31] if the IBF can be suppressed to the primary ion level.

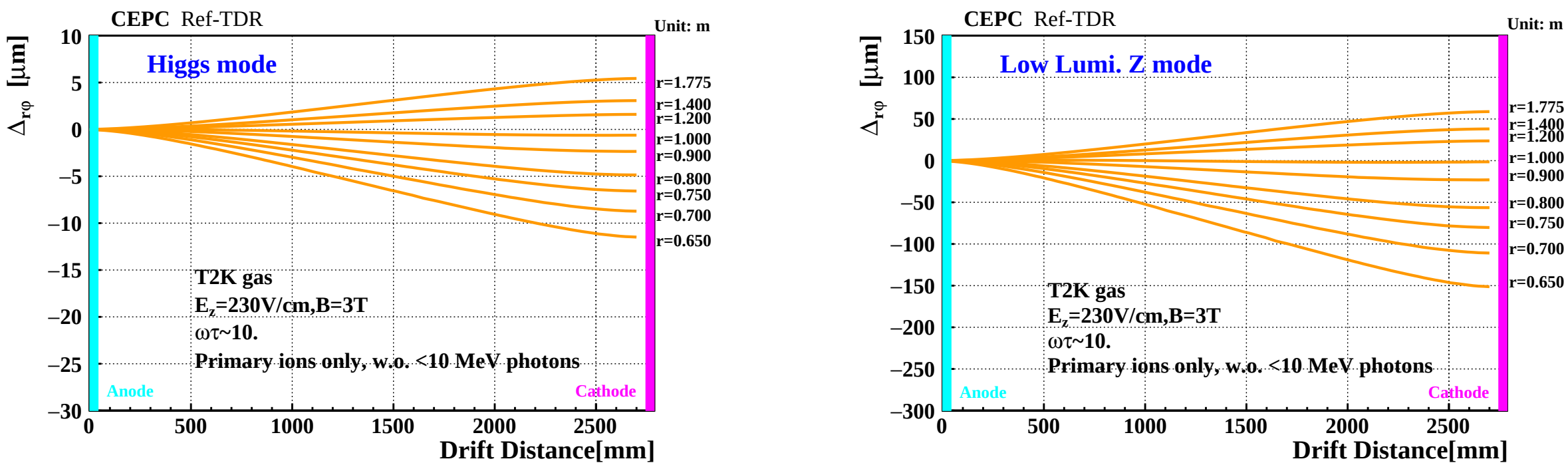

(a) *H* mode (b) Low-luminosity *Z* mode

**Figure 6.23:** Azimuthal distortion as functions of drift distances without the contributions of ≤ 10 MeV low energy photons and the IBF in CEPC different operating modes. (a) *H* mode, (b) Low-luminosity *Z* mode. Each orange solid line shows the azimuthal distortion at a different radius.

### 6.3.4 Calibration and alignment

During the data taking, many factors have negative impacts on the TPC performance, including space charge distortion, changes in gas densities caused by temperature and pressure fluctuations, inhomogeneity of the magnetic and electric fields, mechanical imperfection in the detector installation. Space charge distortion and inhomogeneity of the magnetic and electric fields can significantly affect the electron drift process, which can degrade the precision of track reconstruction. The drift velocity of electrons in the gas is influenced by both the electric field and the properties of the gas itself. Calibration and alignment are crucial for mitigating these effects and achieving good spatial resolution and momentum resolution for the TPC. Both on-detector and track-based calibration methods are applied.

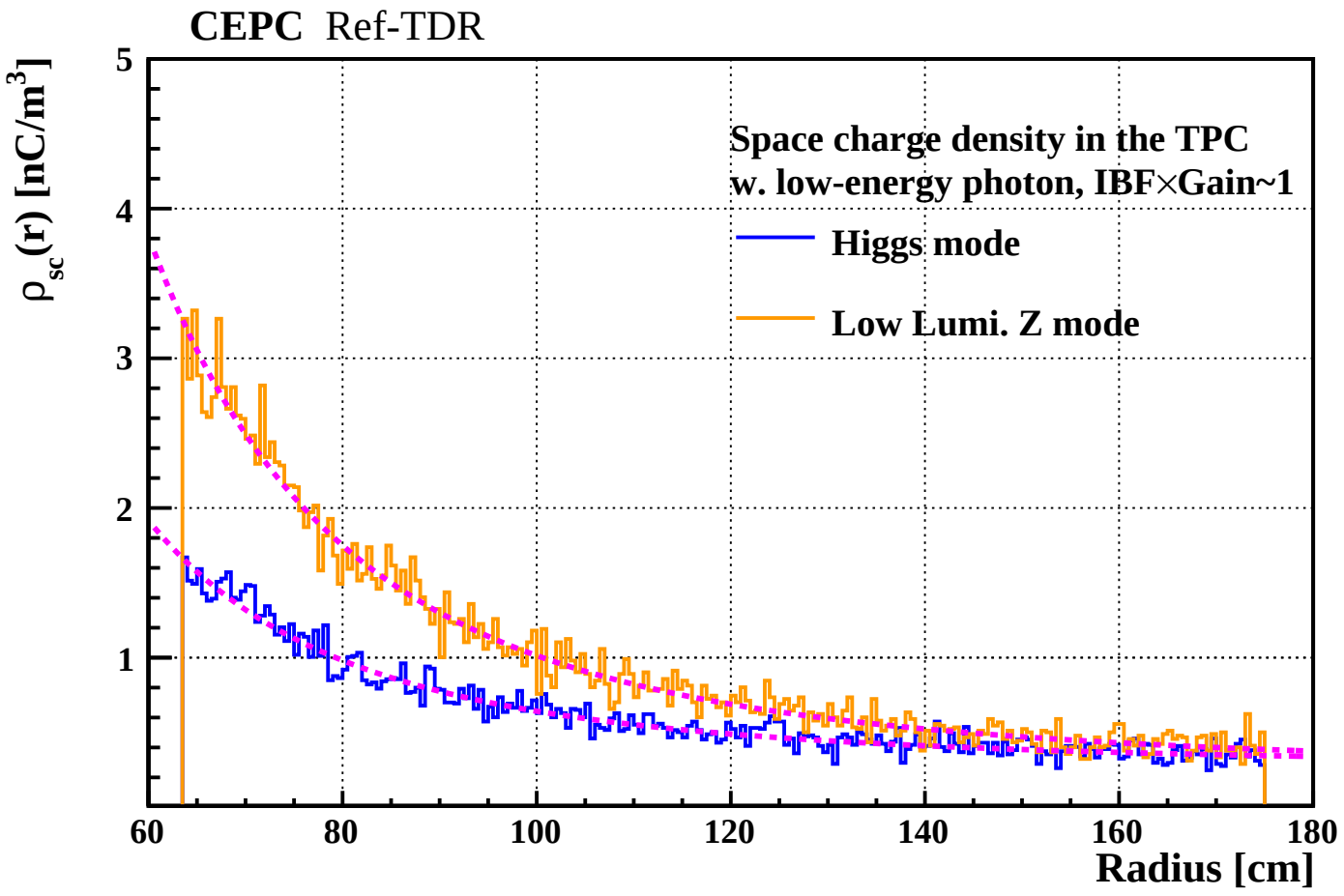

**Figure 6.24:** Charge density, including the contributions of ≤ 10 MeV low energy photons and the IBF, as a function of radius and fitted by Equation 6.2 (Magenta dashed line in the plots). The blue and orange lines represent the $H$ mode and Low-luminosity $Z$ mode, respectively.

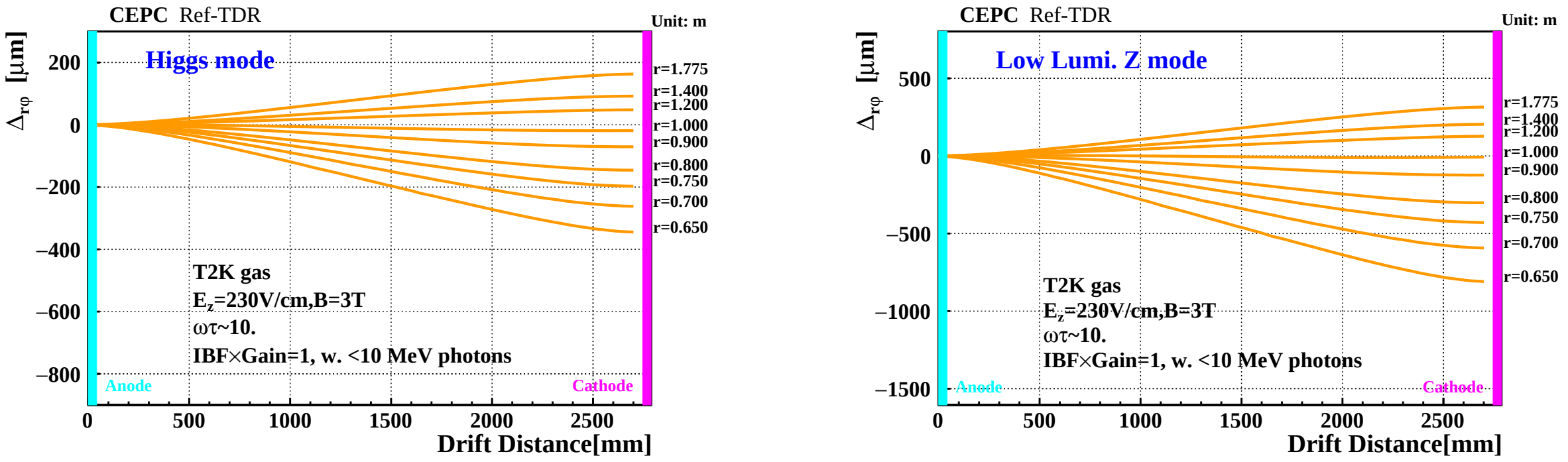

**(a)** $H$ mode  **(b)** Low-luminosity $Z$ mode

**Figure 6.25:** Azimuthal distortion as functions of drift distances in CEPC different operating modes, including the contributions of ≤ 10 MeV low energy photons and the IBF. (a) $H$ mode, (b) Low-luminosity $Z$ mode. Each orange solid line shows the azimuthal distortion at a different radius.

A primary goal of on-detector calibration is to characterize the electron drift velocity as a function of electric field strength. This is achieved using a 266 nm ultraviolet laser system to emulate particle tracks for performance testing [32]. The laser beams are designed to enter the TPC chamber obliquely through the endplate. The ultra-violet laser beams offer several advantages: negligible ionization-density fluctuations, precise beam steering, and the superior spatial resolution for long-track and double-track measurements. Moreover, the laser is neither perturbed by nor perturbs the electromagnetic field, making it ideal for investigating field-induced track distortions.

The track-based calibration is essential for achieving high-precision calibration of drift velocity and correction of space charge distortion. With the CEPC tracking system, charged tracks can be reconstructed with very high precision by combining both the ITK and the OTK, providing significant advantages for the TPC calibration. Bhabha events can be used for the TPC calibration. Based on the full track reconstruction, charged tracks can be refitted excluding the measured TPC clusters to provide a reference for the spatial position correction of the TPC clusters, following the approach used in the ALICE TPC calibration [33]. Through analysis of a large statistics of events, a three-dimensional function between the drift time and space point can be derived. The function will be saved in the calibration database and applied in the final data reconstruction. Monitoring of data quality will be performed throughout the data taking period to facilitate prompt calibration updates.

**Table 6.7:** Estimation of the TPC detector space charge distortion in the CEPC $H$ mode and low-luminosity $Z$ mode.

| Mode | Maximum space charge density | Remarks | Mitigation strategies |
|---|---|---|---|
| $H$ | $\rho_{sc0}$~0.04 nC/m$^3$ | w.o. low-energy photons (<10 MeV) and IBF contributions | Acceptable |
| $H$ | $\rho_{sc1}$~1.67 nC/m$^3$ | w. low-energy photons (<10 MeV) and IBF contributions (IBF ratio × gain = 1) | Optimizing shielding and developing dedicated distortion calibration and correction algorithm |
| low lumi. $Z$ | $\rho_{sc2}$~0.62 nC/m$^3$ | w.o. low energy photons (<10 MeV) and IBF contributions | Acceptable |
| low lumi. $Z$ | $\rho_{sc3}$~3.32 nC/m$^3$ | w. low-energy photons (<10 MeV) and IBF contributions (IBF ratio × gain = 1) | Optimizing shielding and developing dedicated distortion calibration and correction algorithm |

During TPC installation, mechanical tooling and the laser-based alignment systems position the sub-components of the detector with accuracy ranging from 0.1 to 1 mm, depending on the specific component. Offline track-based alignment will be applied to achieve high-precision position correction. As a sub-detector of the CEPC tracking system, the TPC will undergo track-based alignment simultaneously with the silicon trackers. The method and procedure can be found in Section 5.4.

## 6.4 Simulation and performance

### 6.4.1 TPC simulation framework

Full simulation and reconstruction software has been implemented in CEPCSW according to the CEPC TPC design parameters, which enables the study of tracking performance and PID capabilities. The workflow is shown in Figure 6.26. The software framework provides an interface for the incident charged particles generated from the generator. When the charged particle enters the TPC volume, both the ionization process and the transport of drifting electrons are simulated. A digitization algorithm has been implemented to simulate the multiplication of drifting electrons and signal readout at the end of the chamber. The information from the simulation and digitization is written into the MC event file as input for the event reconstruction. At the reconstruction level, the tracks of charged particles are reconstructed based on the clustering from readout pads. Based on tracking results, an ionization reconstruction is performed for PID. Both tracking and PID performance studies are performed within this simulation framework.

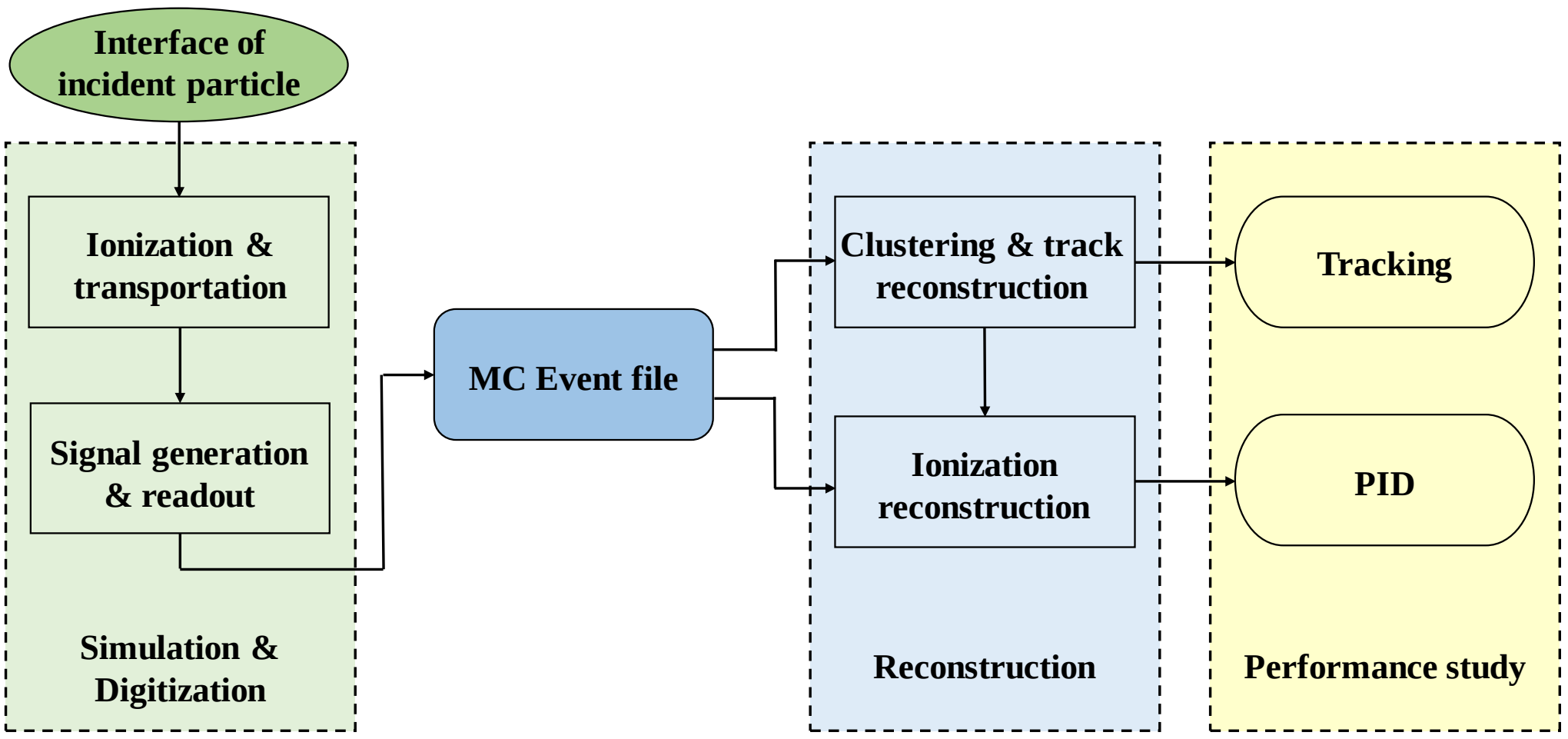

**Figure 6.26:** Workflow of the MC study for charged particle detection and reconstruction. The information from the simulation and digitization, including the ionization process, the transport of drifting electrons within the TPC volume, and the signal generation and readout, is written into the MC event file. At the reconstruction level, the clustering algorithm, the track and ionization reconstruction are implemented. The performance study including tracking and PID is based on the full simulation and reconstruction.

### 6.4.2 Simulation and digitization of TPC

As shown in Figure 6.27, the TPC simulation and digitization workflow includes the ionization process, the transport of drifting electrons, the amplification and the signal readout at the end of the chamber. The information of ionization electrons is recorded as MC Truth for use in subsequent drift and amplification parameterized models. Electrons that have undergone amplification are collected at the readout detectors. The electronic noise simulation uses parameters based on ASIC testing data, implemented via a parameterized model.

The simulation of the ionization process within the CEPC software framework is implemented using the Heed package in Garfield++ program, which can accurately simulate the ionization process in gases [34]. The ionization cluster information, including the cluster size, positions, and times, is recorded as MC Truth for subsequent simulation processes in the software framework. Additional technical details can be found in Reference [35].

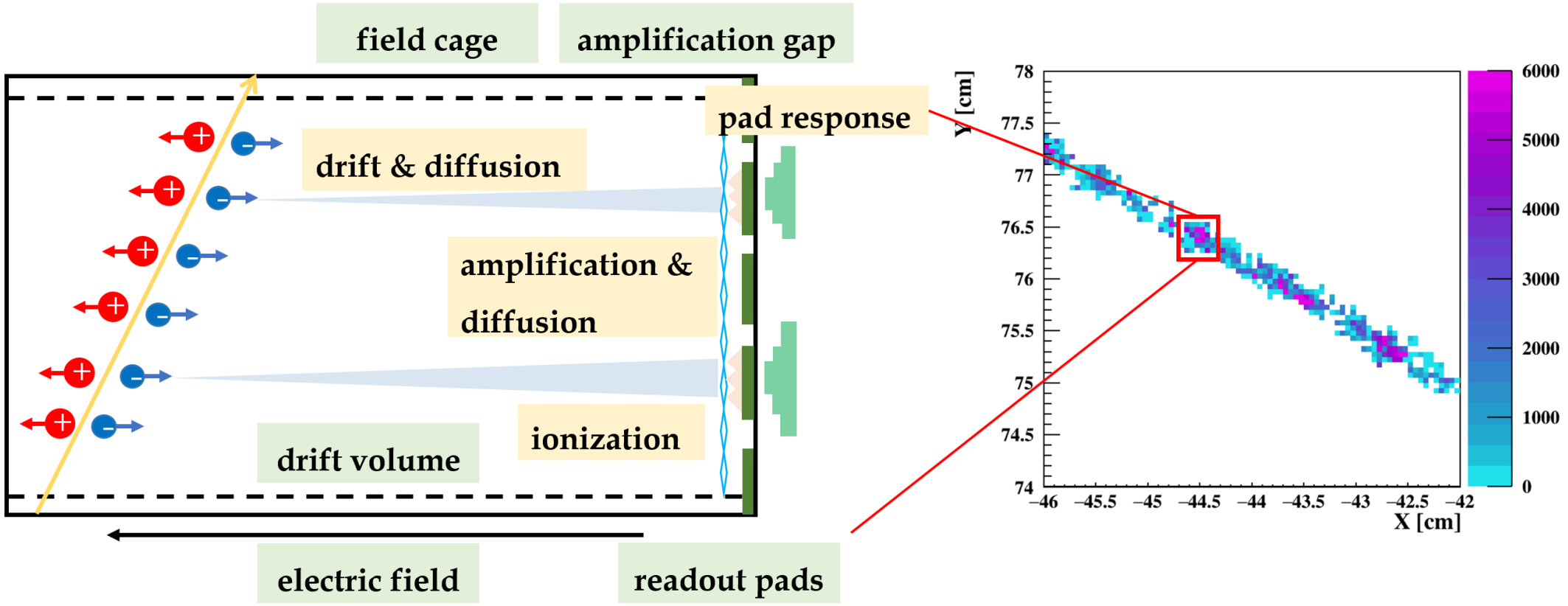

**Figure 6.27:** Schematic diagram illustrating the operational principle of the high-granularity readout TPC. The drift process of ionization electrons in the TPC volume is simulated, accounting for diffusion effect. The amplification process and signal generation at the TPC end region with high-granularity readout are implemented.

A parameterized model based on the Garfield++ simulation results is applied to simulate the drift process since the full simulation using Garfield++ is extremely time-consuming. The drift velocity and the parameters of diffusion effect can be obtained from Garfield++ simulation. The drift time and position uncertainties as functions of drift distance ($L$) with fit to Equation 6.5 are shown in Figure 6.28. The results are used to perform time and position sampling of drifting electrons, modeling the diffusion effect including both longitudinal and transverse diffusions.

$$\sigma_i = C_0 + C_1\sqrt{L} \tag{6.5}$$

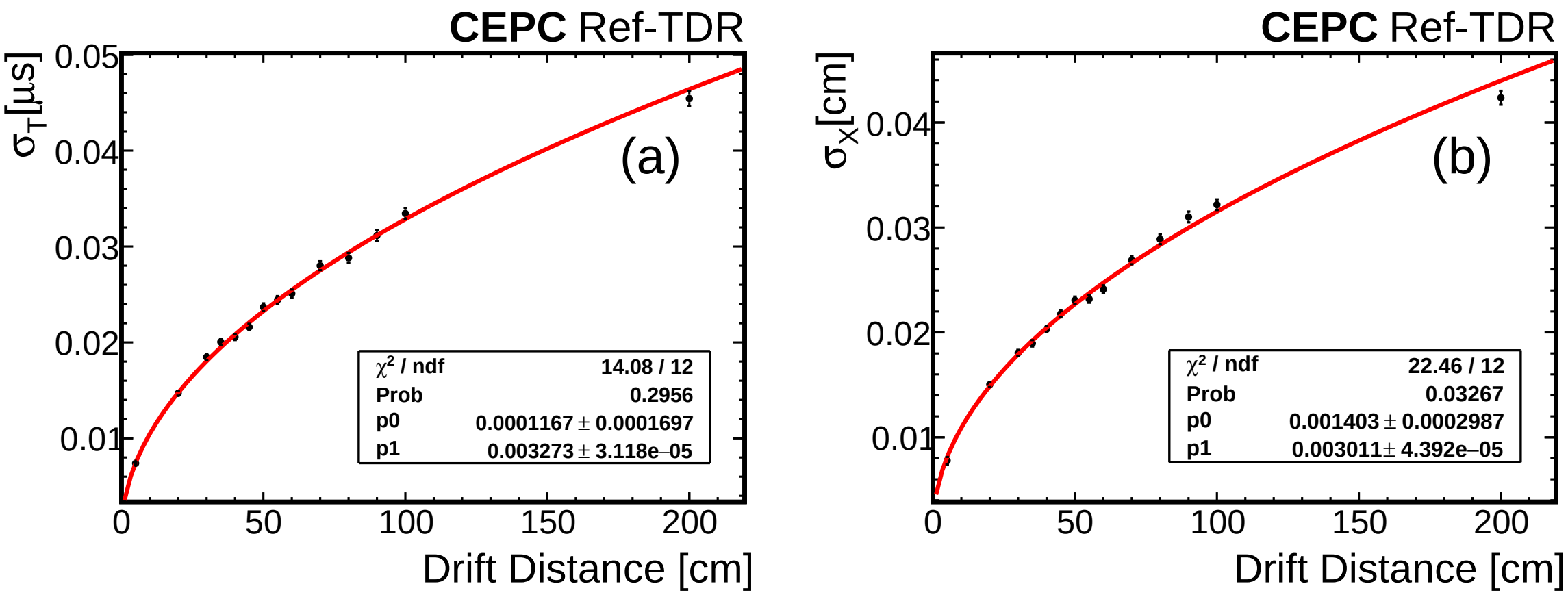

**Figure 6.28:** (a) Longitudinal diffusion parameters $\sigma_T$ and (b) transverse diffusion parameters $\sigma_X$, as functions of the drift distance under a 3.0 T magnetic field. The transverse diffusion coefficient $C_1$ obtained by fitting the plot of $\sigma_X$ versus the drift distance is 30 $\mu$m/$\sqrt{\text{cm}}$.

The Micromegas amplification is simulated by modeling the smallest repetitive unit of its periodic micro-mesh structure. Figure 6.29(a) shows the COMSOL model with a standard 400-mesh stainless steel wire (diameter: 25 μm, pitch: 45 μm). The 100 μm avalanche region achieves a gain of 2000, meeting the readout requirements. The single-electron gain distribution is fitted with Equation 6.6 [36] to extract $C_0$, $\langle G \rangle$, and $\theta$, as shown in Figure 6.29(b).

$$P(G) = C_0 \frac{(1+\theta)^{(1+\theta)}}{\Gamma(1+\theta)} \left( \frac{G}{\langle G \rangle} \right)^{\theta} exp \left[ -(1+\theta) \frac{G}{\langle G \rangle} \right] \tag{6.6}$$

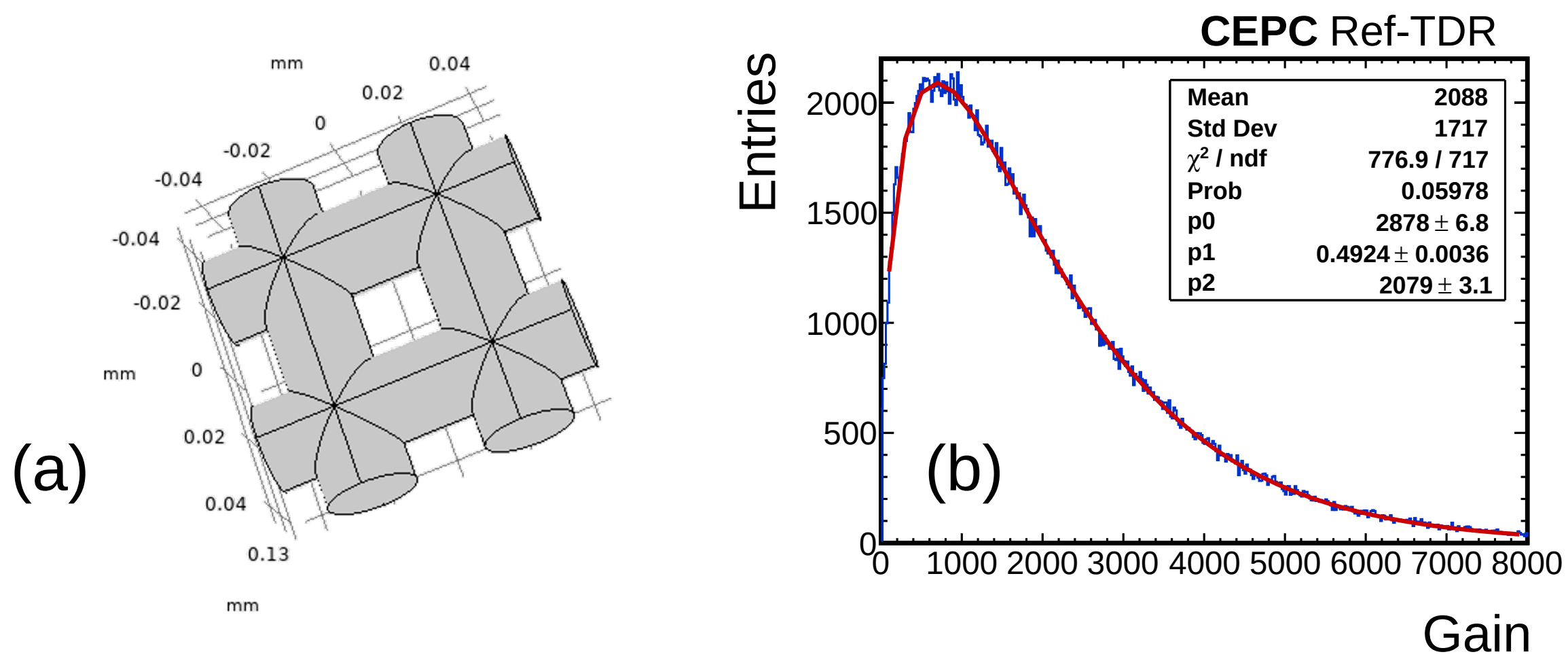

**Figure 6.29:** (a) Minimum periodic unit of the Micromegas micro-mesh structure obtained from COMSOL simulations; (b) Simulation results of the single-electron gain distribution and Polya fitting.

After amplification, the signal is read out and mapped to a readout unit based on its position. The drift time and distance are recorded in a sparse matrix for reconstruction. Electronic noise is applied to all pads, using a Gaussian parameterization model, with 100 $e^-$ per channel.

### 6.4.3 Spatial resolution and momentum resolution

In the TPC, the spatial resolution is primarily limited by the diffusion effect due to the long-distance drift of the ionization electrons. Figure 6.30 shows the comparison of the transverse spatial resolution as a function of drift distance under different conditions. The curves are calculated using Equation 6.7, where $\sigma_{r\phi_0}^{\text{pad}}$ is the intrinsic resolution of the geometry, $D_{r\phi}$ is the transverse diffusion coefficient of the gas, and $L$ is the drift distance. The magenta curve represents the transverse spatial resolution for the 1 mm × 6 mm pad readout. For the 500 μm × 500 μm high-granularity readout, the red curve shows the transverse spatial fluctuation of a single electron, reflecting the combined effects of diffusion and readout granularity. In contrast, the practical spatial resolution is represented by the green curve, which signifies the effective resolution achieved by grouping signals from adjacent layers of high-granularity pads to form a virtual pad with an area comparable to that of a traditional one. This combination, which incorporates the average number of electrons derived from the simulation, suppresses single-layer uncertainties. The results demonstrate that the high-granularity pad readout achieves a better spatial resolution compared to the traditional pad readout.

$$\sigma_{r\phi}^{\text{pad}} = \sqrt{(\sigma_{r\phi_0}^{\text{pad}})^2 + LD_{r\phi}^2} \tag{6.7}$$

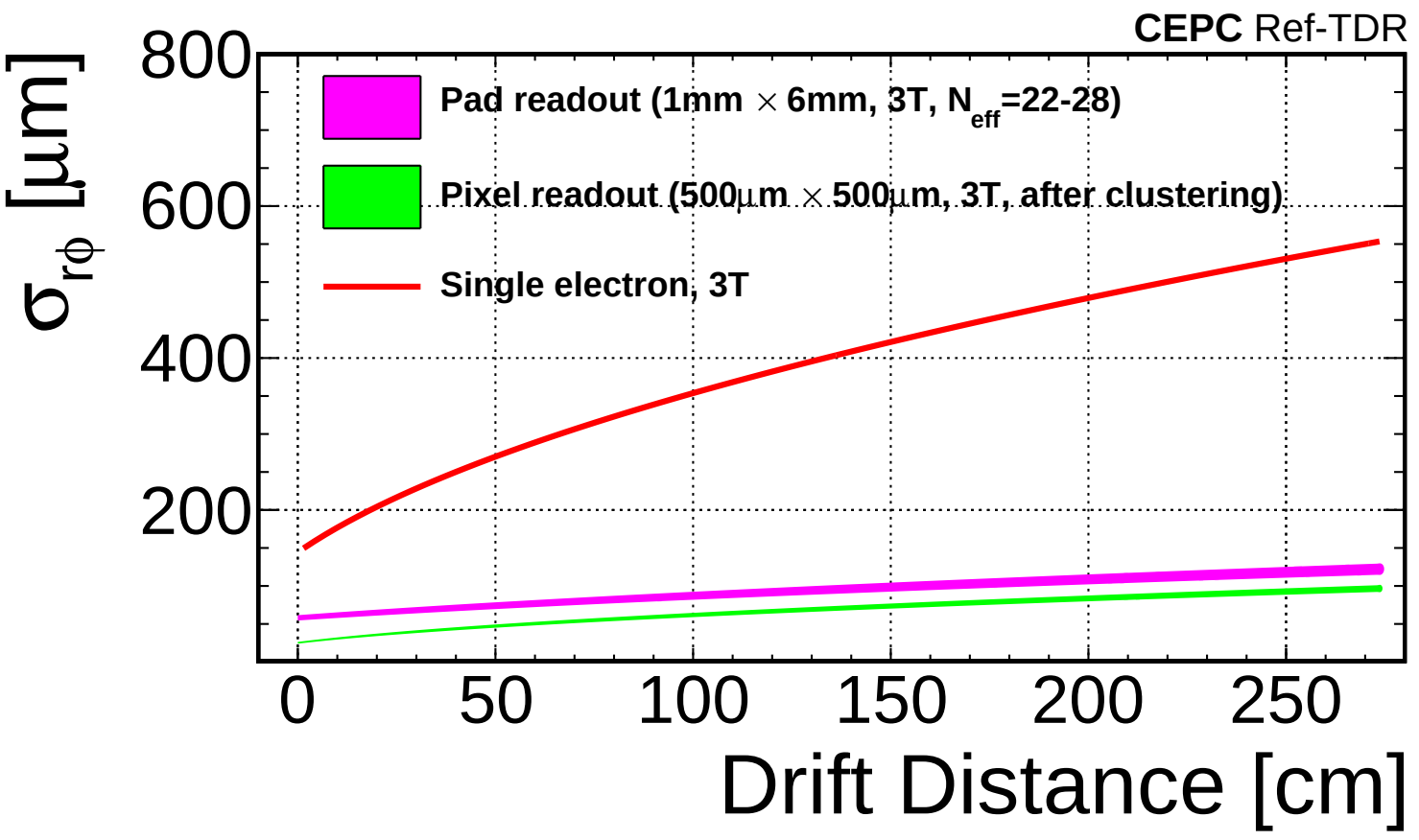

**Figure 6.30:** Spatial resolution as a function of drift distance for 1 mm × 6 mm pad readout at 3.0 T (magenta), 500 μm × 500 μm pad readout (single electron) at 3.0 T (red), and 500 μm × 500 μm pad readout after clustering at 3.0 T (green).

Figure 6.31 shows the spatial resolution and the clustering algorithm that provides hit clusters for track reconstruction. Figure 6.31 (a) presents the dependence of cluster spatial resolution ($r$-$\phi$ plane) on the drift distance, using hits from 10 small-pad layers. The blue curve represents the ideal scenario without considering the space-point distortions. In practice, the electron drift process is affected by inhomogeneity of the electric field due to space charge distortion and the mechanical deformation, and non-uniformity of the magnetic field, which cause the space-point distortion effect and degrade the spatial resolution. As mentioned in Section 6.3.4, the data-driven track-based calibration can be applied to mitigate the impact of the space-point effect on the spatial resolution. The calibrated spatial resolutions in the $H$ and low-luminosity $Z$ modes through a preliminary estimation are shown in figure 6.31 (a) (green and red curves, respectively). The resolutions reach their optimums at 300 mm drift distance. Below 300 mm, the resolutions degrade as the drift distance decreases. This is because a short drift distance results in a narrow electron cloud. When the electron cloud width ($\sigma_d$) is less than 0.2 × pad width ($w$), only two pads collect charge. As shown in figure 6.31(b), the Center Of Gravity (COG) algorithm, which requires charge spread over at least three pads to accurately reconstruct the position, becomes biased and fails in this case. For 500 μm × 500 μm small pads in the T2K gas at 3.0 T, this effect becomes negligible when the drift distance exceeds 400 mm. Above 300 mm, the resolutions degrade gradually with the drift distance due to the dominant effect of electron diffusion. This behavior is incorporated into the TPC digitization algorithm.

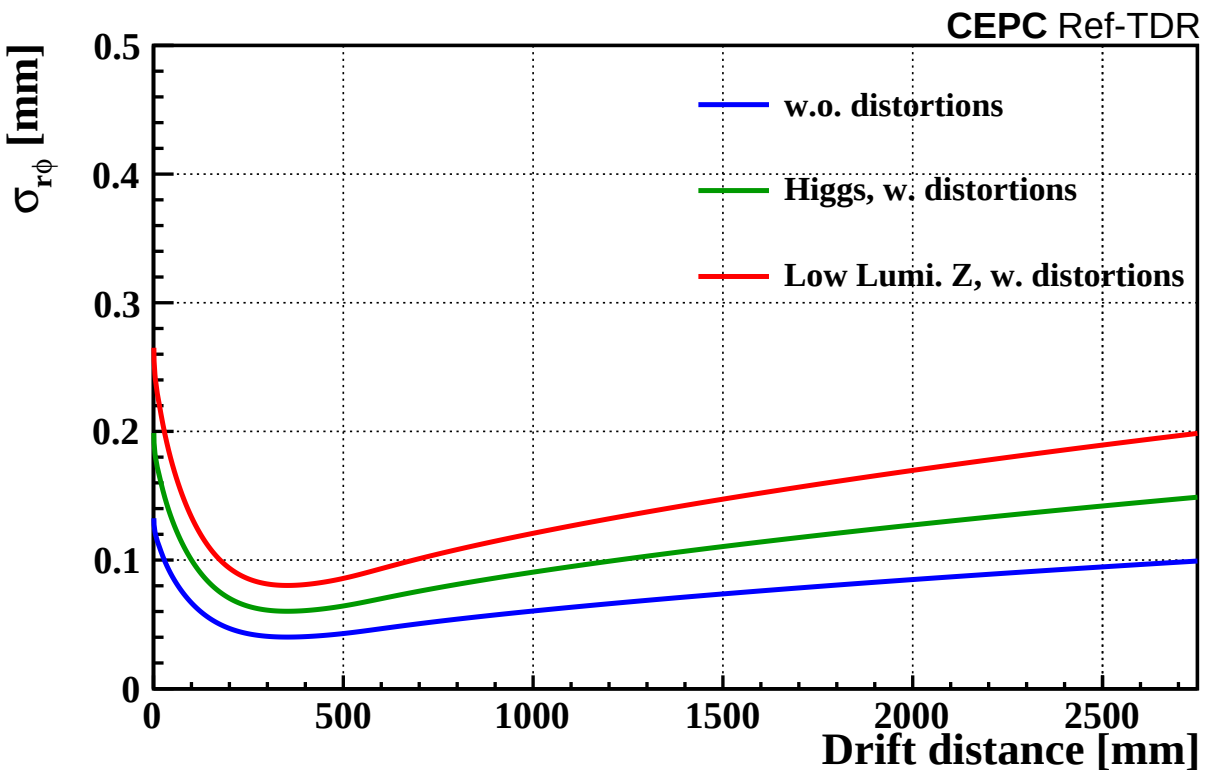

**(a)** Spatial resolution as a function of the drift distance

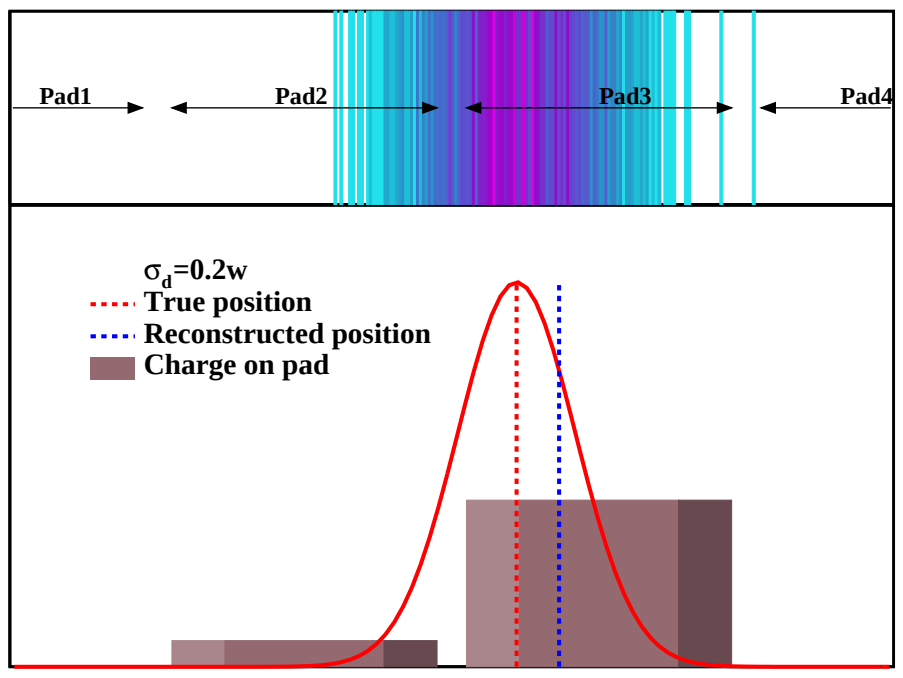

**(b)** Schematic of the COG

**Figure 6.31:** (a) Spatial resolution of the TPC in $r - \phi$ plane with a clustering of ten layers of pads as a function of the drift distance. The blue, green and red curves represent the ideal scenario without considering space-point distortions, the $H$ and low-luminosity $Z$ modes with considering space-point distortions and calibration, respectively. (b) Schematic of the Center of Gravity method, the COG is the point where the total weight of an object is considered to be concentrated.

With the full simulation and reconstruction, the transverse momentum resolution curves with only the TPC at different incident angles are obtained, as shown in Figure 6.32 (the space-point distortions are not included). The crossover points between 7 and 10 GeV/c highlight the competition between the multiple scattering effects and the electron diffusion effect. In the region below the crossing points, the resolution improves with the increase of $\theta$, since the momentum resolution is dominated by the multiple scattering. Tracks with $\theta = 85°$ that are almost perpendicular to the $z$-axis, have the shortest track lengths, and are minimally influenced by the multiple scattering effects. Tracks with smaller $\theta$ suffer from worse resolution due to their longer path lengths and increased scattering. While in high momentum region, the momentum resolution, which is governed by the single-point spatial resolution of the TPC, degrades with the increase of $\theta$. In the case of tracks with $\theta = 85°$, the spatial resolution is not good due to the impact of the diffusion effect, since all the ionization electrons on the track have the longest drift distances. As the incident angle $\theta$ decreases, the drift distance for electrons on the outer portion of the track is reduced, thereby improving the spatial resolution and leading to superior momentum resolution at high momentum.

Figure 6.33 presents the transverse momentum resolution as a function of momentum at the incident angle of 85° when using only the TPC and the combined information of the TPC, the ITK, and the OTK. For the case of using only the TPC, compared to the idea scenario without space-point distortions, the transverse momentum resolution of the high momentum tracks significantly degrades in both the $H$ and low-luminosity $Z$ modes if space-point distortions with calibration are taken into account. Notably, this performance loss is largely mitigated by combining the TPC information with that of the ITK and OTK. The high-precision hits from the external silicon trackers provide effective compensation for the TPC's impaired spatial resolution. This ensures a more robust final momentum resolution, with the maximum degradation being only approximately 11% for 100 GeV/c tracks in the low-luminosity $Z$ mode.

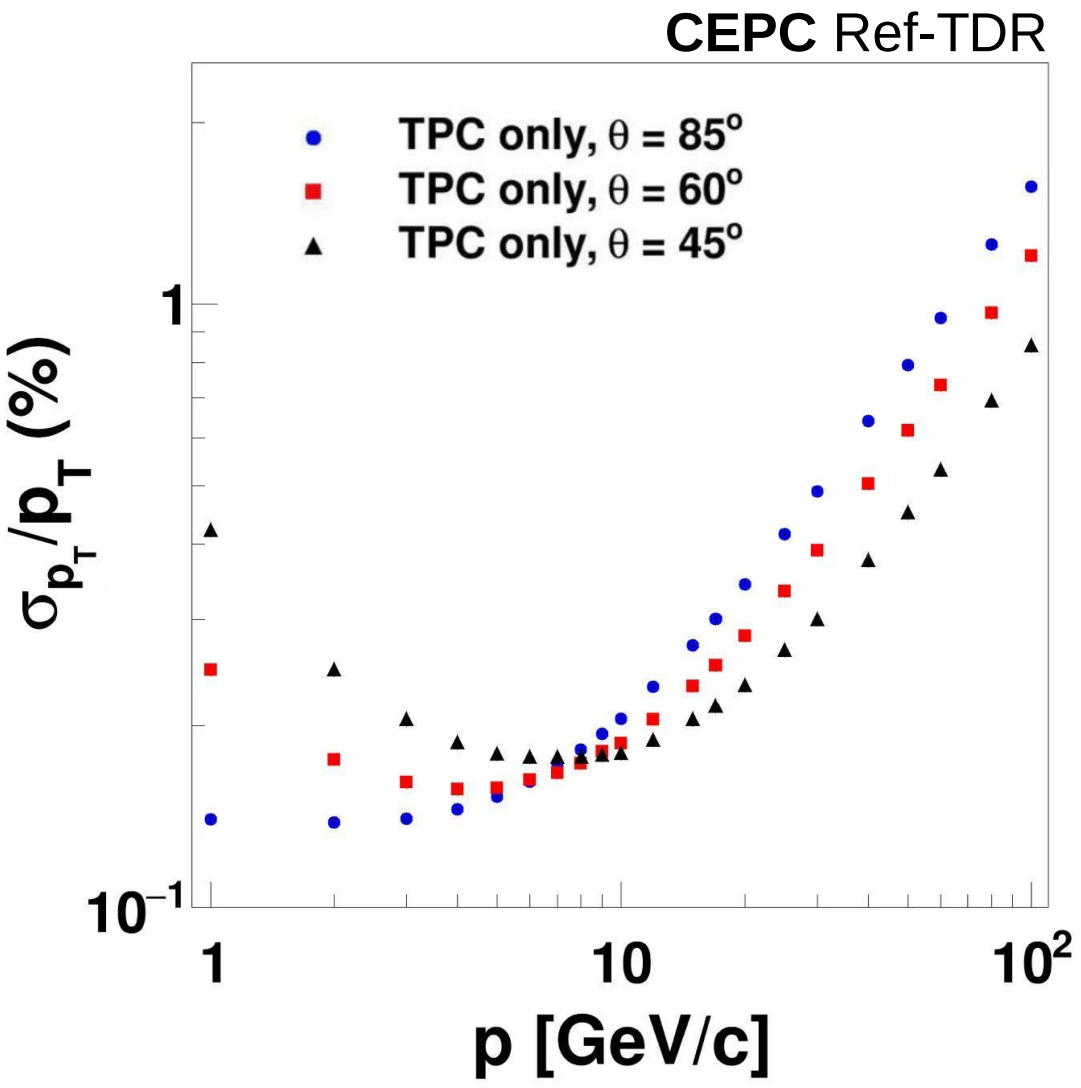

**Figure 6.32:** Transverse momentum resolution as a function of momentum using only the TPC at incident angles of 85° (blue solid circles), 60° (red solid squares), and 45° (black solid triangles), obtained from full simulation without space-point distortions in CEPCSW.

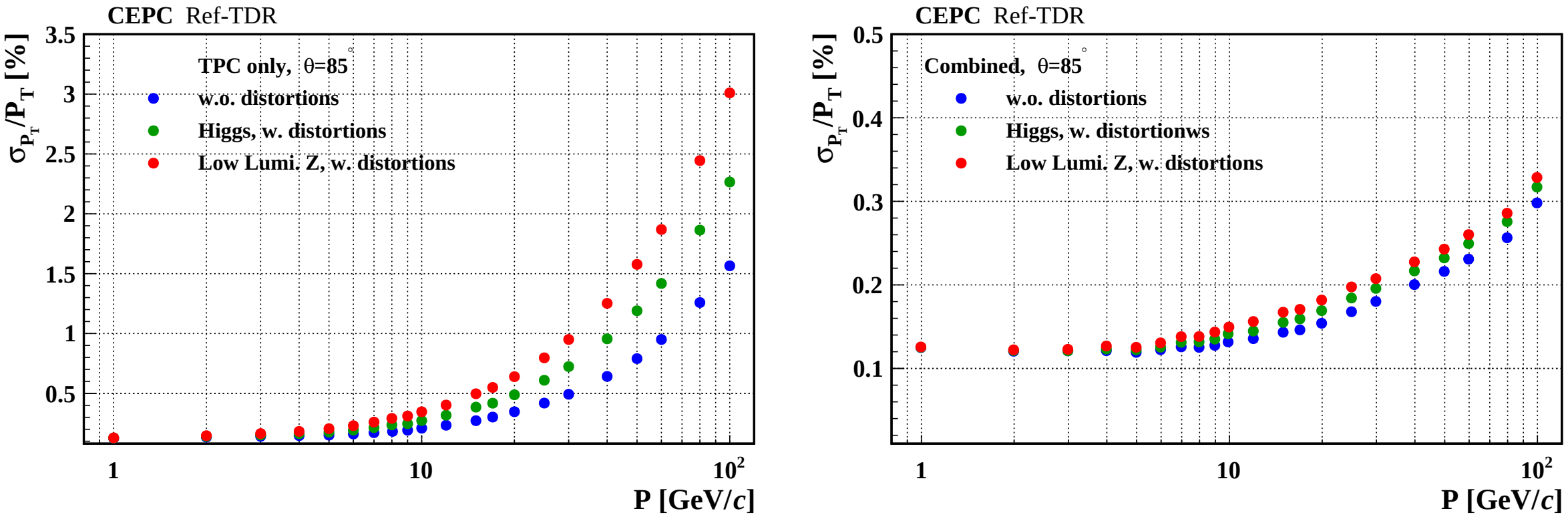

**(a)** Only TPC **(b)** TPC + ITK + OTK

**Figure 6.33:** Transverse momentum resolution as a function of momentum at the incident angle of 85°, obtained from full simulation in CEPCSW. The blue, green and red solid circles represent the scenario without considering space-point distortions, the $H$ and low-luminosity $Z$ modes with considering space-point distortions and calibration, respectively. (a) Using only the TPC information. (b) Using combined information of the TPC, the ITK, and the OTK.

### 6.4.4 Particle identification

Particle identification is vital for CEPC physics studies, including jet flavor tagging and flavor-physics studies. The detector should distinguish charged hadrons over a broad momentum range, spanning tens of GeV/c. The high-granularity readout TPC, which provides more than 2000 hits per track, supports both the traditional $\mathrm{d}E/\mathrm{d}x$ measurement of total energy deposition and the novel $\mathrm{d}N_p/\mathrm{d}x$ technique, which counts the number of activated pads along the track segments. By mitigating the Landau fluctuations inherent in $\mathrm{d}E/\mathrm{d}x$ measurement, the $\mathrm{d}N_p/\mathrm{d}x$ method can significantly enhance the PID performance.

In this section, the reconstruction methods for $\mathrm{d}E/\mathrm{d}x$ and $\mathrm{d}N_p/\mathrm{d}x$ and the evaluation of their respective PID capabilities are presented.

#### 6.4.4.1 Reconstruction Algorithms

Reconstruction algorithms for both $\mathrm{d}N_p/\mathrm{d}x$ and $\mathrm{d}E/\mathrm{d}x$ have been developed for the high-granularity readout TPC. The $\mathrm{d}N_p/\mathrm{d}x$ method utilizes the proportional relationship between ionization events and the number of activated pads. Secondary ionizations, however, introduce a Landau distribution in the activated pad counts across different TPC layers. To mitigate this effect, a truncated-mean approach is applied. In practice, the $\mathrm{d}N_p/\mathrm{d}x$ method counts threshold-crossing pads per layer, or over multiple layers, and the final $\mathrm{d}N_p/\mathrm{d}x$ value ($N_p$) is obtained as the mean of the truncated samples, representing the average number of threshold-crossing pads per unit track length. The $\mathrm{d}E/\mathrm{d}x$ method follows a parallel structure, determining the total energy deposition per layer, or over multiple layers, and computing the average of the truncated data to estimate the energy loss per unit length, denoted by $E$. The truncated-mean algorithms have two adjustable parameters: the number of combined layers ($1 \sim 10$) and the truncation ratio ($0 \sim 40\%$), which are independently optimized, as shown in Figure 6.34.

#### 6.4.4.2 Performance of the particle identification

With full simulations and reconstruction, the PID performance was evaluated. Figure 6.35 shows the resolution of $\mathrm{d}N_p/\mathrm{d}x$ and $\mathrm{d}E/\mathrm{d}x$ as functions of particle momentum. The red dotted line with stars represents the 500 µm × 500 µm high-granularity readout, while the blue dashed line with triangles corresponds to the 1 mm × 6 mm pad readout. The $\mathrm{d}N_p/\mathrm{d}x$ method achieves a resolution of about 2%, more than twice as good as that of $\mathrm{d}E/\mathrm{d}x$.

Separation power between two charged particle species, defined in Equation 6.8 [37, 38], is a key metric for the PID. Figure 6.36 presents the $K/\pi$, $K/p$, and $\pi/p$ separation power using $\mathrm{d}N_p/\mathrm{d}x$ reconstruction as a function of momentum at various polar angles. The $K/\pi$ separation power exceeds $3\sigma$ at 20 GeV/c and remains above $2\sigma$ at 40 GeV/c. As $\theta$ decreases, the separation power improves due to the longer track length. All curves reach their minimum at the low momentum area, creating a "blind area" that can be addressed by incorporating ToF information from the OTK.

$$\eta_{A,B} = \frac{|\mu_A - \mu_B|}{\frac{1}{2}(\sigma_A + \sigma_B)} \tag{6.8}$$

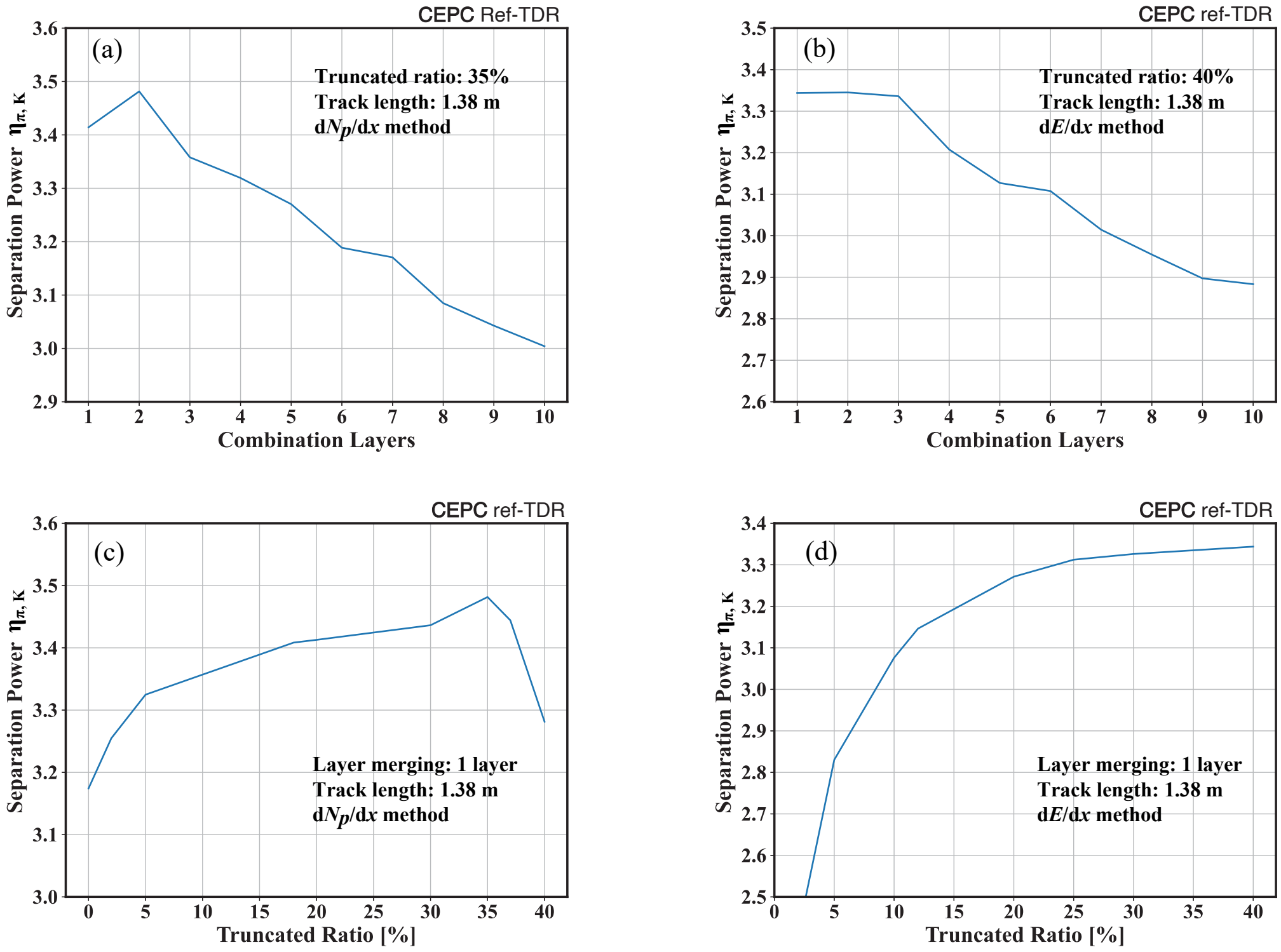

**Figure 6.34:** Parameter optimization of $\mathrm{d}N_p/\mathrm{d}x$ and $\mathrm{d}E/\mathrm{d}x$ algorithms. The figure of merit is the $K/\pi$ separation power defined in Eq. 6.8. The truncated ratio or the number of combination layers are optimized while keeping the other fixed. (a) $\mathrm{d}N_p/\mathrm{d}x$ with 35% truncation achieves optimal $\pi/K$ discrimination at two layers; (b) $\mathrm{d}E/\mathrm{d}x$ with 40% truncation performs best at one layer; (c) $\mathrm{d}N_p/\mathrm{d}x$ shows optimal 35% truncation at fixed one layer; (d) $\mathrm{d}E/\mathrm{d}x$ confirms best performance at 40% truncation (one layer).

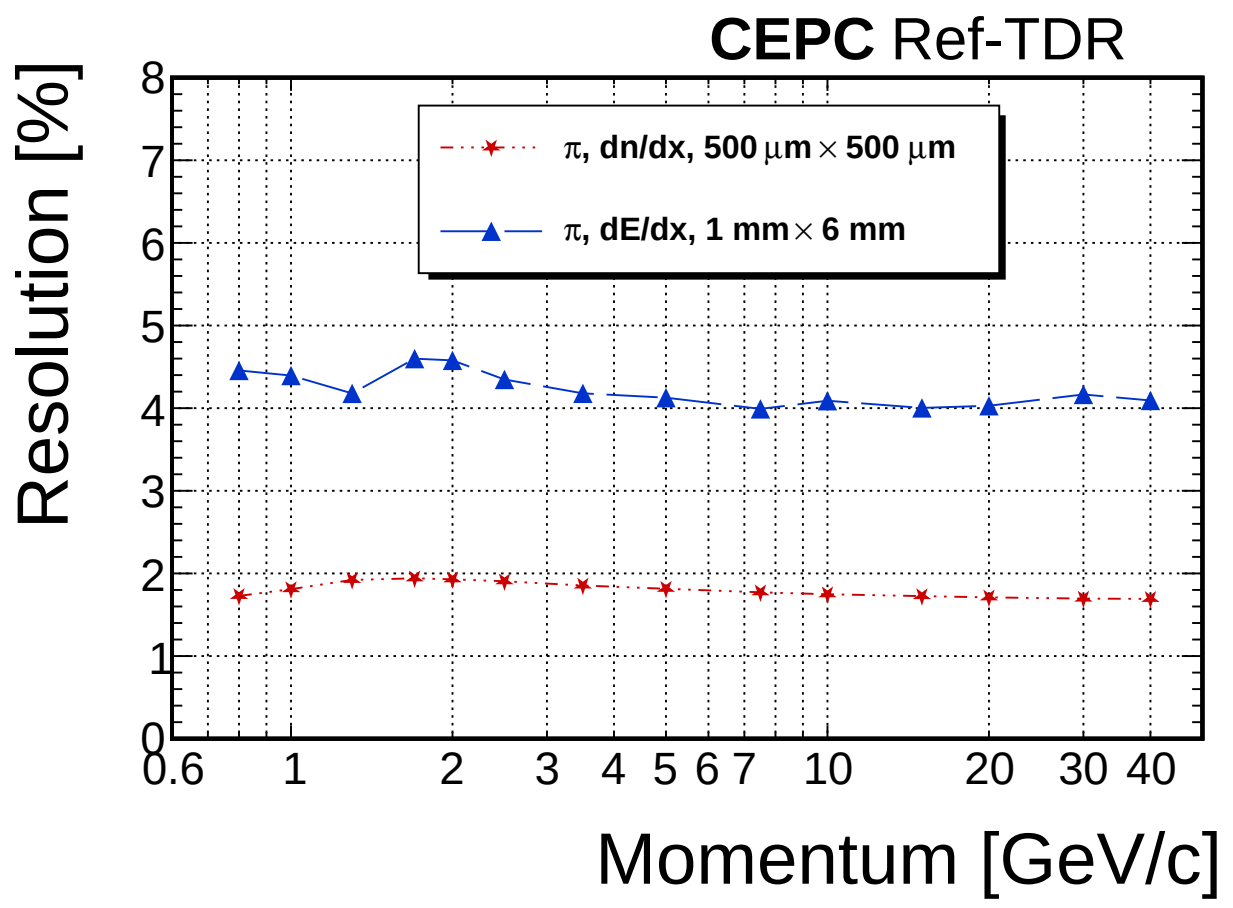

**Figure 6.35:** Resolutions of $\mathrm{d}N_p/\mathrm{d}x$ with 500 µm × 500 µm high-granularity readout (red dotted line with stars) and $\mathrm{d}E/\mathrm{d}x$ with 1 mm × 6 mm pad readout (blue dashed line with triangles) as functions of momentum for charged pions with a polar angle of 60°.

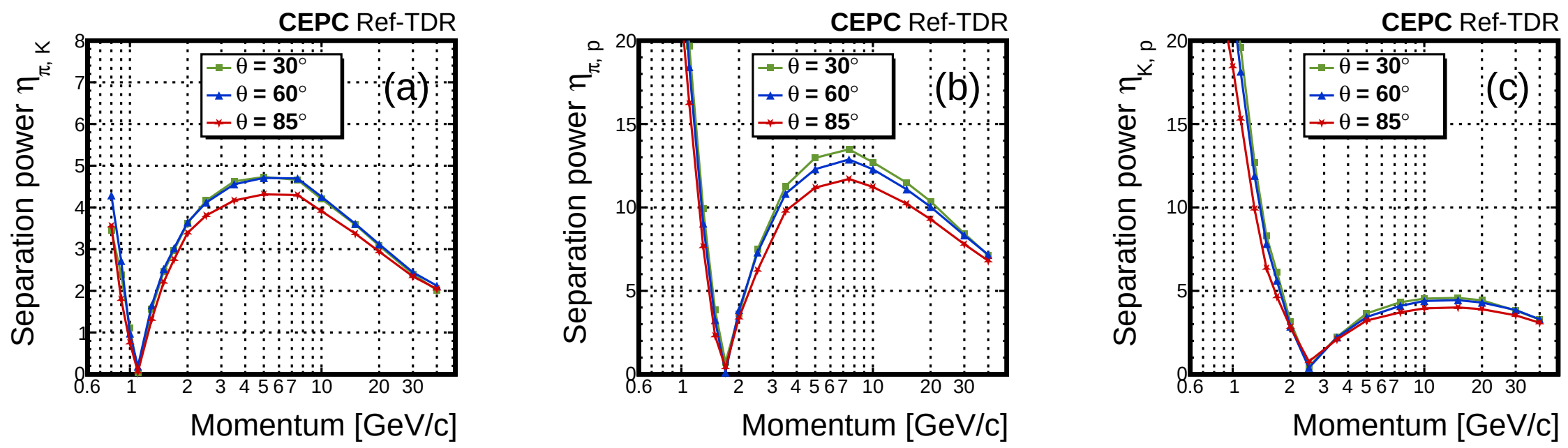

**Figure 6.36:** Separation power of (a) $\pi$ and $K$, (b) $\pi$ and $p$, (c) $K$ and $p$, as functions of momentum at incident angles of 30° (green solid curve with solid squares), 60° (blue solid curve with solid triangles), and 85° (red dotted curve with stars), obtained with full simulation.

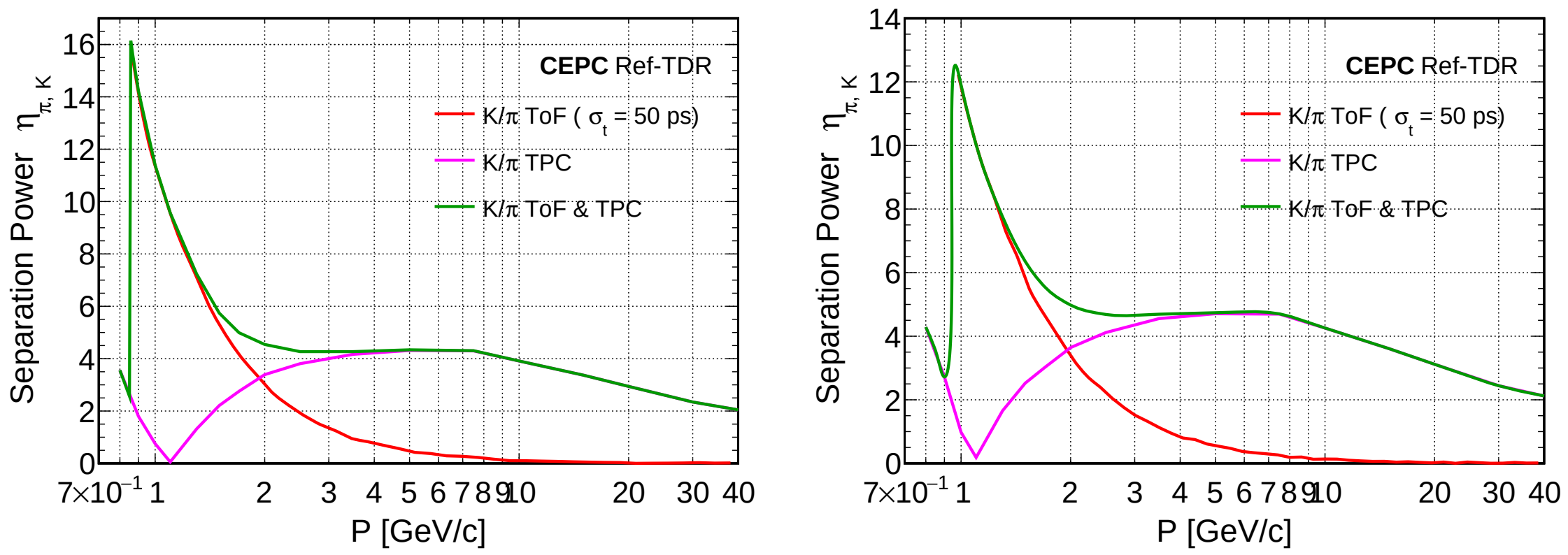

(a) $\theta = 85°$ (b) $\theta = 60°$

**Figure 6.37:** Separation power between $K$ and $\pi$ as a function of momentum, at the polar angles of (a) $\theta = 85°$ and (b) $\theta = 60°$. The red curves, pink curves, and green curves indicate the separation power using only the ToF measurement from the OTK, only the TPC, and the combination of both, respectively.

Figure 6.37 shows the $K/\pi$ separation power when ToF measurements from the OTK (using LGAD with 50 ps timing resolution) is included. Above 2 GeV/c, the TPC alone provides better PID separation power than ToF measurement, while below 2 GeV/c, ToF measurement outperforms the TPC. The ToF cutoff around 0.9 GeV/c arises because low-momentum particles cannot penetrate the OTK due to magnetic bending as detailed in Section 5.5. By combining the TPC with ToF information from the OTK, the "blind area" at low momentum is effectively eliminated, resulting in strong separation power across the entire momentum spectrum.

## 6.5 Alternative Solution: Drift Chamber

A Drift Chamber (DC) has been designed as an alternative gaseous detector [39]. Its preliminary design specifies a 5800 mm total length and a radial span from 600 mm to 1800 mm. The inner wall is formed by a carbon-fiber cylinder, while the outer support comprises a carbon-fiber frame with eight longitudinal hollow beams and eight rings, all enclosed within a gas-tight envelope. To withstand the cumulative wire tension with minimal deformation, the aluminum end plates adopt a multi-stepped, tilted profile. In total, the chamber holds about 64 wire layers. To fulfill both PID and momentum-measurement requirements, full simulations dictated an 18 mm × 18 mm cell size. Each cell encloses a central 20 µm gold-plated tungsten sense wire, surrounded by eight 80 µm gold-plated aluminum field wires arranged in a approximately square lattice. A 90% helium (He) plus 10% isobutane ($iC_4H_{10}$) gas mixture achieves the target primary-ionization density for cluster counting ($dN/dx$) measurements.

A waveform based simulation and digitization framework was developed to study $dN/dx$ based PID. The simulation uses a parameterized Garfield++ model, incorporating geometry, gas mixture, and high-voltage settings that replicate the test beam setup [40]. The ionization patterns produced by relativistic charged particles are modeled using Heed. To enhance computational efficiency, the full Garfield++ avalanche and induced-current calculations were placed with parameterized modules that aggregate each pulse's timing, amplitude, and shape into an analog current waveform per cell.

Extracting $dN/dx$ demands isolating individual electron arrival peaks from waveforms heavily affected by pileup and electronic noise. Traditional methods, involving a two-step process of derivative-based peak finding followed by timing-based cluster merging, depend on incomplete raw hits and expert heuristics, limiting their accuracy. In contrast, supervised Machine Learning (ML) algorithms learn complex waveform features directly from labeled data. An ML based cluster counting algorithm for $dN/dx$ reconstruction was therefore implemented [41, 42].

**Table 6.8:** $K/\pi$ separation powers for ML based and traditional algorithms. The ML based algorithm has a roughly 10% improvement for momentum range from 5.0 to 20.0 GeV/c.

| | **Momentum** ( GeV/**c**) | | | | | | |
|---|---|---|---|---|---|---|---|
| | 5.0 | 7.5 | 10.0 | 12.5 | 15.0 | 17.5 | 20.0 |
| ML alg. | 4.6$\sigma$ | 4.6$\sigma$ | 4.4$\sigma$ | 4.2$\sigma$ | 3.8$\sigma$ | 3.5$\sigma$ | 3.2$\sigma$ |
| Traditional alg. | 4.2$\sigma$ | 4.3$\sigma$ | 4.1$\sigma$ | 3.8$\sigma$ | 3.6$\sigma$ | 3.3$\sigma$ | 2.9$\sigma$ |

The PID performance via $K/\pi$ separation power was quantified over a 1.2 m track. Table 6.8 compares two algorithms across momenta. The ML cluster counting approach consistently outperforms traditional methods, achieving a 10% greater separation power. According to the square root of track length scaling law, this improvement is equivalent to the gain from a 20% increase in detector radial coverage, which offers substantial cost saving. Preliminary CEPC criteria call for 3$\sigma$ $K/\pi$ separation up to 20 GeV/c. With ML reconstruction, the current DC design (r = 600 ~ 1800 mm) comfortably meets these PID requirements.

To demonstrate the $dN/dx$ method, fast readout electronics were developed and tested together with a detector prototype [43]. The preliminary test results have basically verified the performance of the readout electronics and the feasibility of the $dN/dx$ method. In the next research, the DC design and the ML-based algorithms will be further optimized. In addition, a multilayer drift chamber prototype and a full-length prototype will be developed to study and optimize the resolution of $\mathrm{d}N/\mathrm{d}x$.

## 6.6 Summary and future plan

The CEPC TPC with high-granularity readout is designed to deliver good momentum resolution and powerful PID capabilities for charged particles across a broad momentum range. Its design, spanning a 0.6 m to 1.8 m radius and 5.8 m length, integrates low-mass construction, high-granularity readout modules, advanced front-end electronics, and optimized cooling system. Extensive simulations and R&D efforts have addressed key challenges in spatial resolution, momentum measurement, PID capability, beam-induced background suppression, and voxel occupancy, confirming the detector's ability to meet stringent tracking and PID performance requirements.

Forthcoming IBF suppression studies will utilize TPC prototypes in which the standard mesh is intergrated with a monolayer of graphene. This graphene-coated mesh will be evaluated for its ability to suppress IBF, leveraging graphene's high thermal stability and capability to withstand a one-atmosphere pressure differential. These studies will be conducted in collaboration with CEA-Saclay, Lyon University and NIKHEF.

To advance the key technologies, further R&D efforts are essential, particularly in developing the readout detector modules and a full-scale drift-length prototype. As illustrated in Figure 6.38, two TPC prototypes have been designed: the first integrates seven readout modules, while the second features the full 2.9 m drift length. To evaluate the performance of the TPC prototypes and provide experimental validation for the high-granularity readout approach, beam tests have been planned. Within the LC-TPC framework and CERN R&D Collaboration on gaseous detector (DRD1)[44] international collaboration framework, our group has scheduled a beam test at DESY in Hamburg, Germany [45].

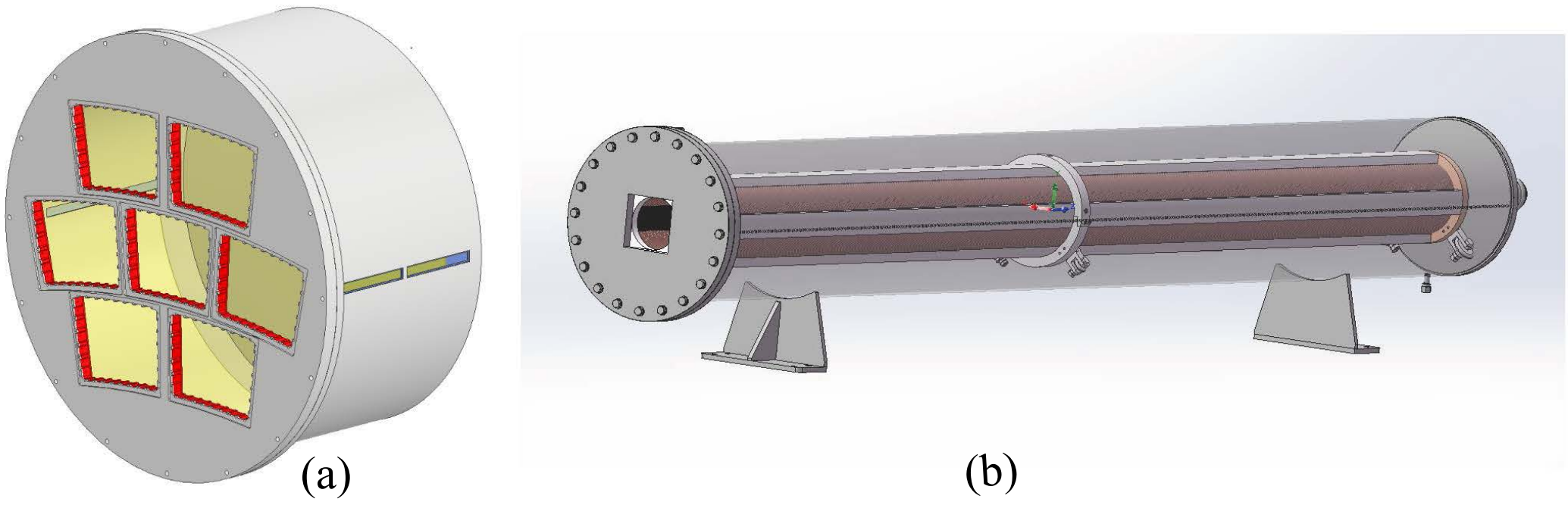

(a) (b)

**Figure 6.38:** Design of TPC prototypes. (a) A prototype assembled with seven readout modules to study the module performance. (b) A prototype features the full 2.9 m drift length to study the electric field and the electron drift.

Moreover, further studies on the precise simulation including the beam-induced background, space-point distortions, and inhomogeneity of the magnetic field will be performed in CEPCSW. The optimization of the readout pad size and the study of calibration methods and procedure will also be explored through simulations and prototype tests.

# Chapter 7 Electromagnetic Calorimeter

The primary objective of the ECAL is to identify electrons, positrons and photons and to measure their energies and positions with a high precision in a wide range, from tens of MeV to around 100 GeV, while also serving as the first section for detecting hadronic showers. It should also provide $\pi^0$-photon discrimination. Based on a Particle Flow Algorithm (PFA), the ECAL is highly segmented both longitudinally and transversely at the level of $X_0$ (radiation length) and $R_M$ (Moliere radius), respectively, to ensure superior separation of nearby particle showers. In conjunction with the tracking system, matching of charged clusters in the ECAL to charged tracks measured in the tracking system and determination of neutral clusters are indispensable for the particle-flow reconstruction.

State-of-the-art calorimeters optimized for PFA are expected to achieve an unprecedented jet energy resolution, but their EM energy resolution is inherently limited by the sampling structures. On the other hand, traditional crystal-based calorimeters, known for their excellent EM energy resolution, are incompatible with PFA, limiting their ability to deliver a good jet energy resolution. This challenge has driven the need for an innovative calorimeter design that bridges this gap. Therefore, a high-granularity homogeneous calorimeter using dense scintillating crystals was chosen, aiming to achieve an unprecedented jet energy resolution and an excellent Electromagnetic (EM) energy resolution. The ECAL design should emphasize maximum possible hermeticity, modularity, and scalability.

This chapter is organized as follows: Section 7.1 first introduces the ECAL, followed by the ECAL design in Section 7.2 and key technologies in Section 7.3. Dedicated R&D activities and performance studies will be discussed in Section 7.4 and Section 7.5, respectively. ECAL alternative options will be covered in Section 7.6 and a summary will be presented in Section 7.7.

## 7.1 Overview

The ECAL is designed to be an imaging homogeneous calorimeter with finely segmented crystals and SiPMs based on the particle flow design [1, 2, 3]. It consists of one barrel and two endcaps with crystal modules and covers 99% of the detector's solid angle. Scintillating crystals have been selected as the ECAL sensitive volume due to its excellent intrinsic energy resolution ($\sigma_E/E \leq 3\,\%/\sqrt{\mathrm{E(GeV)}}$), high density, fast decay time, linearity and feasibility to achieve high granularity. Each ECAL module consists of multiple longitudinal layers. Long crystal bars (with a typical length of 40 cm) are arranged to be orthogonal to each other in every two neighboring layers. Each crystal bar is read out by two SiPMs at both ends. Insensitive materials in between crystals are minimized to guarantee precision energy measurements.

Among the various inorganic scintillating crystals widely used in calorimetry, such as Lead Tungstate ($PbWO_4$), BGO ($Bi_4Ge_3O_{12}$), lutetium yttrium oxyorthosilicate ($Lu_{2(1-x)}Y_{2x}SiO_5$ or LYSO), and cesium iodide (CsI), BGO has been selected as the baseline option, which is driven by its exceptional material properties, including the high density and high light yield, which enable a compact ECAL design while achieving an excellent energy resolution. The BGO calorimeter in the L3 experiment [4] at the Large Electron-Positron Collider (LEP) was a pioneering application of BGO crystals in high-energy physics. The technology for growing long BGO crystals and mass production has been well established, making it a reliable and scalable choice for large-scale applications. Compared to LYSO, BGO offers a more cost-effective solution, particularly for the substantial volume required. Additionally, BGO exhibits favorable mechanical properties, allowing for efficient processing, including cutting and surface treatments, which are critical for precise detector assembly. On the other hand, Bismuth Silicate (BSO) has been identified as a promising alternative option due to its potential for further cost savings, primarily due to the absence of germanium, a relatively expensive material. BSO also offers faster scintillation compared to BGO, which could enhance the timing performance. A major technical challenge to be addressed for BSO lies in developing and demonstrating the capability to produce long crystals with excellent optical quality and uniformity, which is essential for mass production.

The schematics of an ECAL module in typical dimensions of $40 \times 40 \times 27$ cm$^3$, as shown in Figure 7.1, contain over 400 long crystal bars and 800 readout channels. A typical ECAL module (also referred to as a 'supercell') consists of 18 longitudinal layers and, on average, 27 crystal bars per layer. The cross section of each crystal bar is $1.5 \times 1.5$ cm$^2$. All readout components of SiPMs and ASICs are embedded in the readout boards, which are integrated in a light-weighted and robust mechanics structure along with cooling pipes and cables. The modular structure allows for scalability and system integration. It should be noted that the module schematics here is only for illustration and the actual modular design is presented in subsequent sections.

The barrel ECAL is cylindrical, with a total length of 5800 mm along the $z$-axis (namely the beamline direction), an inner radius of 1830 mm from the IP and a total thickness of less than 300 mm. There are 15 segments along the $z$-axis and 32 segments in the azimuthal direction. The module dimensions are mainly determined by the maximum possible length of crystal bars and minimum gaps in between modules, so as to minimize energy loss in insensitive areas reserved for services such as readout electronics, cooling, and supporting structures.

Two endcap discs are designed to guarantee the hermeticity up to 99 % of the solid angle. The space for the MDI is reserved in each disc (700 mm wide) with an outer radius aligned with the barrel (i.e. 2130 mm). The endcap module design is optimized to minimize the number of crystal types. The detailed designs is presented in Section 7.2.5.1 and 7.2.5.2.

Figure 7.3 shows azimuthal and polar angle distributions of ECAL materials budgets including crystals and insensitive materials for readout, mechanics, and cooling. An energy correction algorithm has been developed to compensate for the energy loss of showers in crack regions, which will be presented in detail in Section 7.3.

To meet the requirements and fully exploit the potential of the homogeneous PFA-oriented

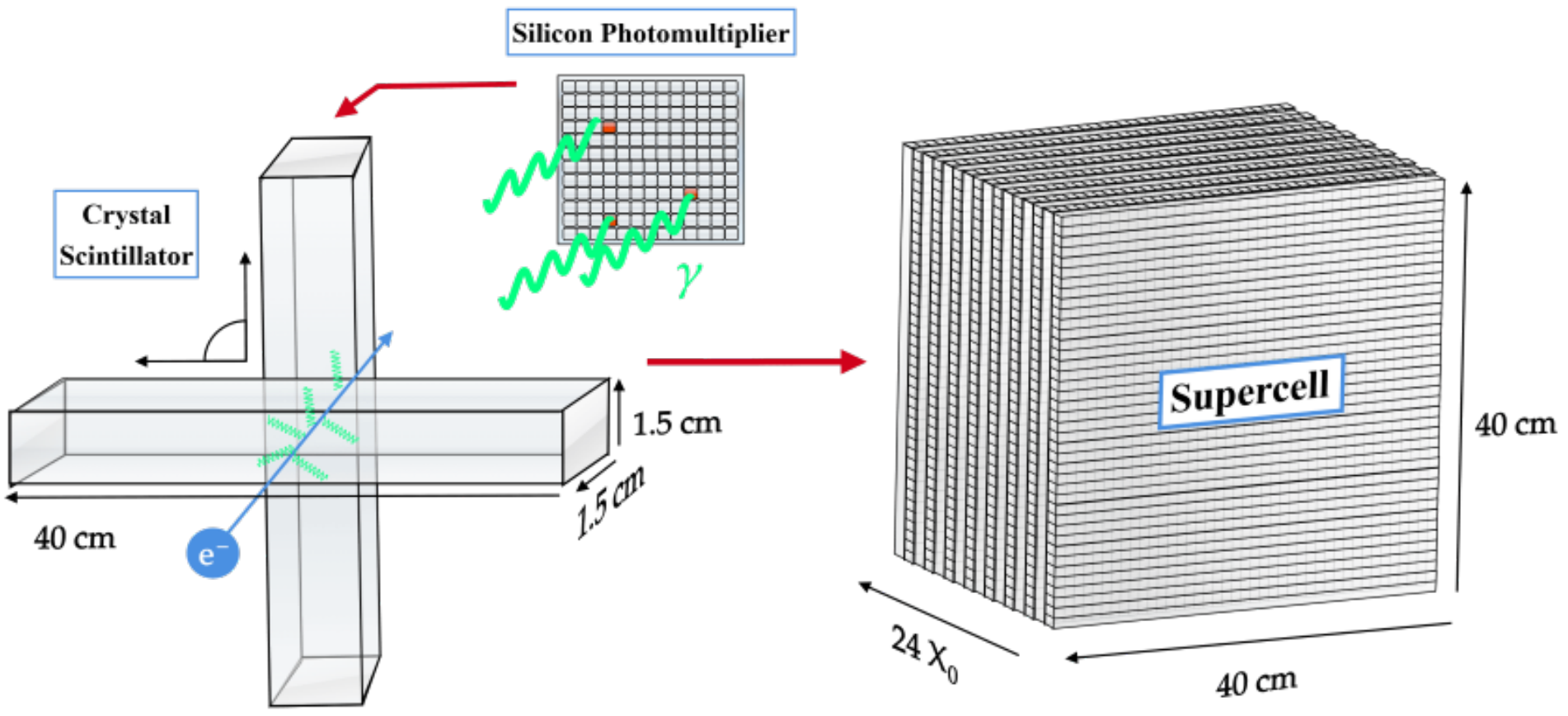

**Figure 7.1:** The schematics of an ECAL module (i.e. a supercell) with the orthogonal arrangement of long crystal bars to achieve fine segmentation in the longitudinal direction and transverse planes. Each crystal bar is individually wrapped with a reflective foil to ensure a high light collection efficiency and a good response uniformity and avoid any optical crosstalk between crystals. The scintillation light of each crystal is read out by two SiPMs attached at each crystal end.

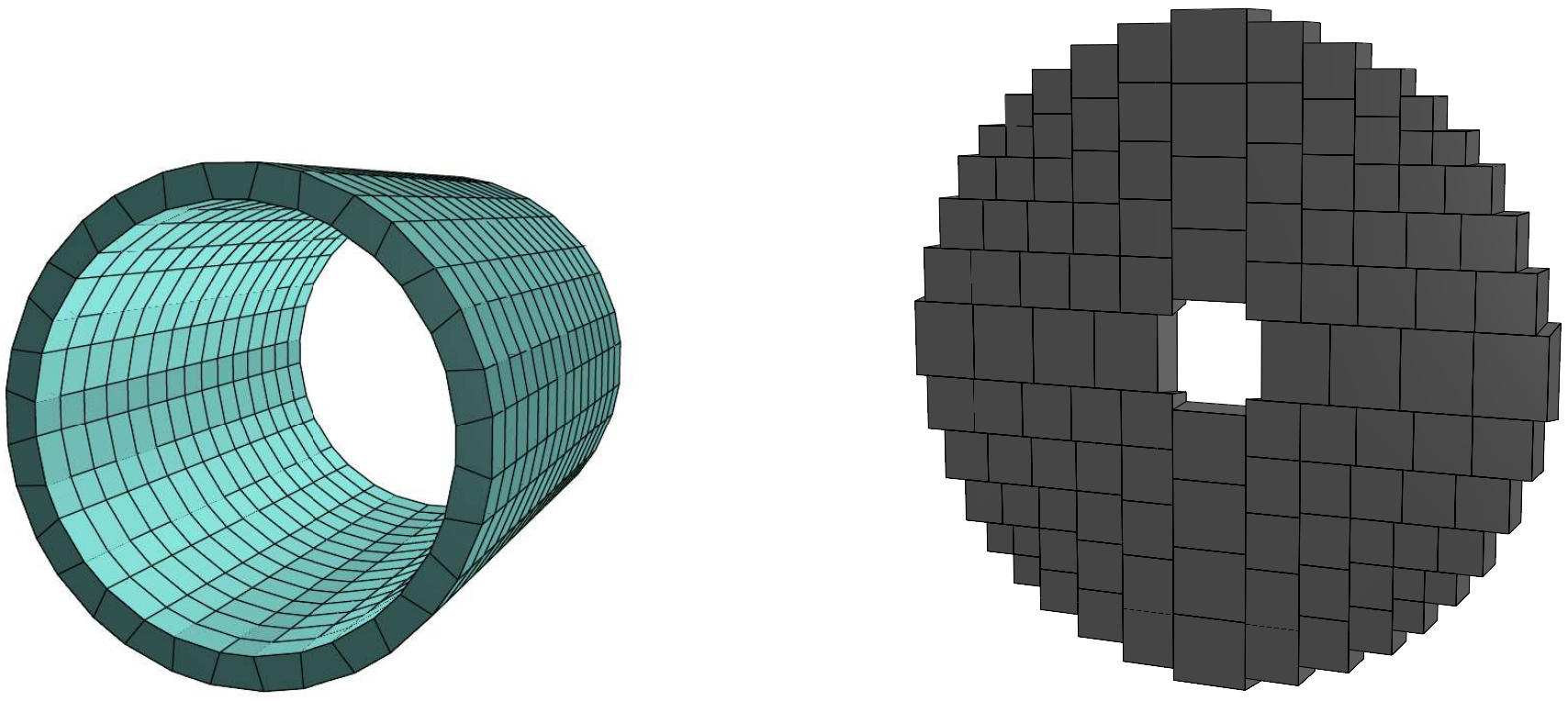

**(a)** **(b)**

**Figure 7.2:** The ECAL design schematics: the barrel in a cylinder of 480 modules (a) and one of two endcap discs with over 200 modules (b). The ECAL modular designs have been implemented in the CEPCSW.

ECAL, there are several issues to be addressed. The orthogonal arrangement of long crystal bars would result in challenges regarding pattern recognition of particle showers, necessitating advanced reconstruction software development. The algorithm is designed to disentangle overlapping showers efficiently and reconstruct particle clusters accurately. The placement of readout electronics and cooling systems around modules also needs to avoid interference with the sensitive detection elements. Effective thermal management is crucial for ensuring the stable operation of the SiPMs. Moreover, the integration of long crystal bars introduces potential issues related to mechanical stability and alignment.

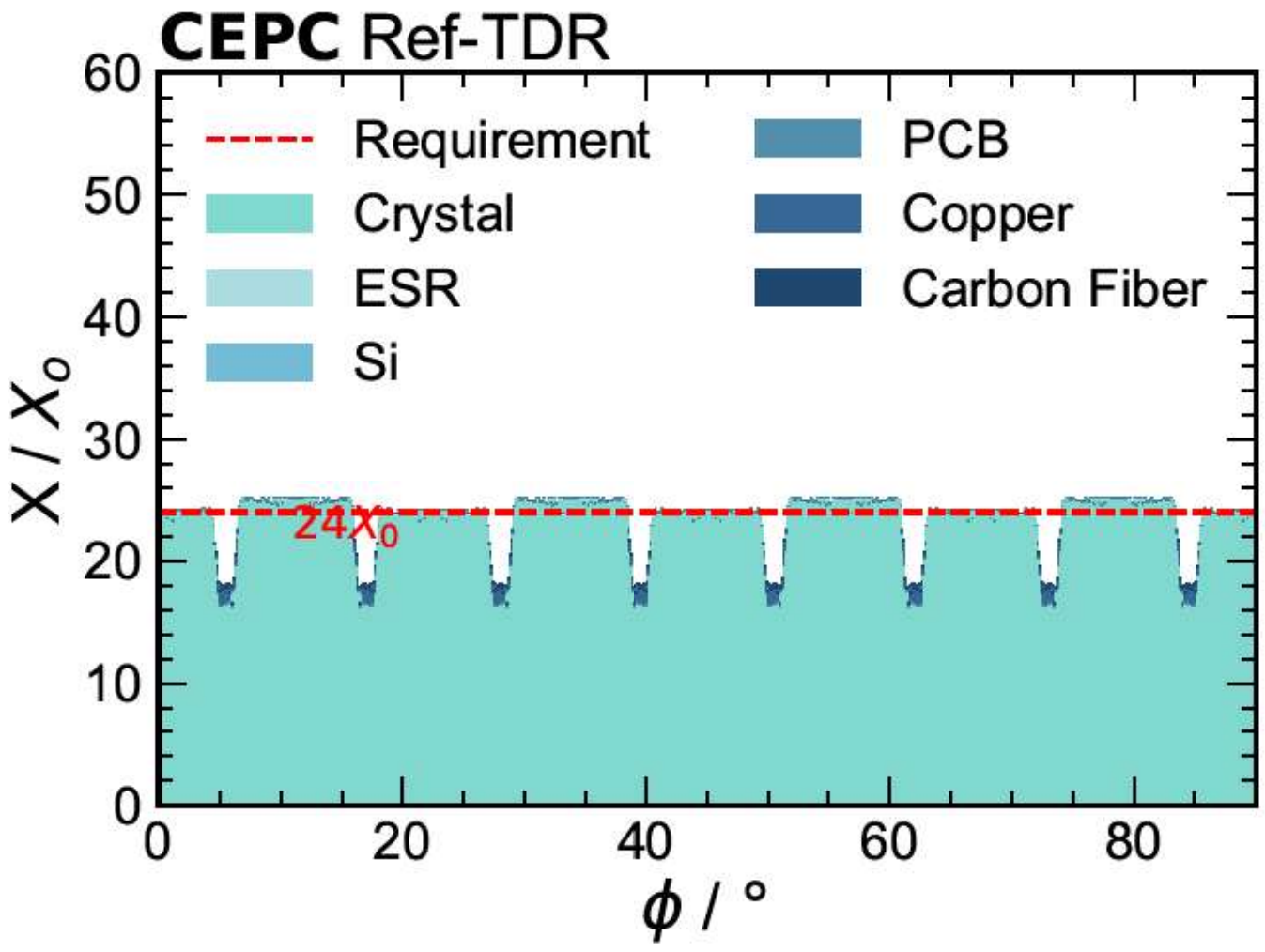

**(a)**

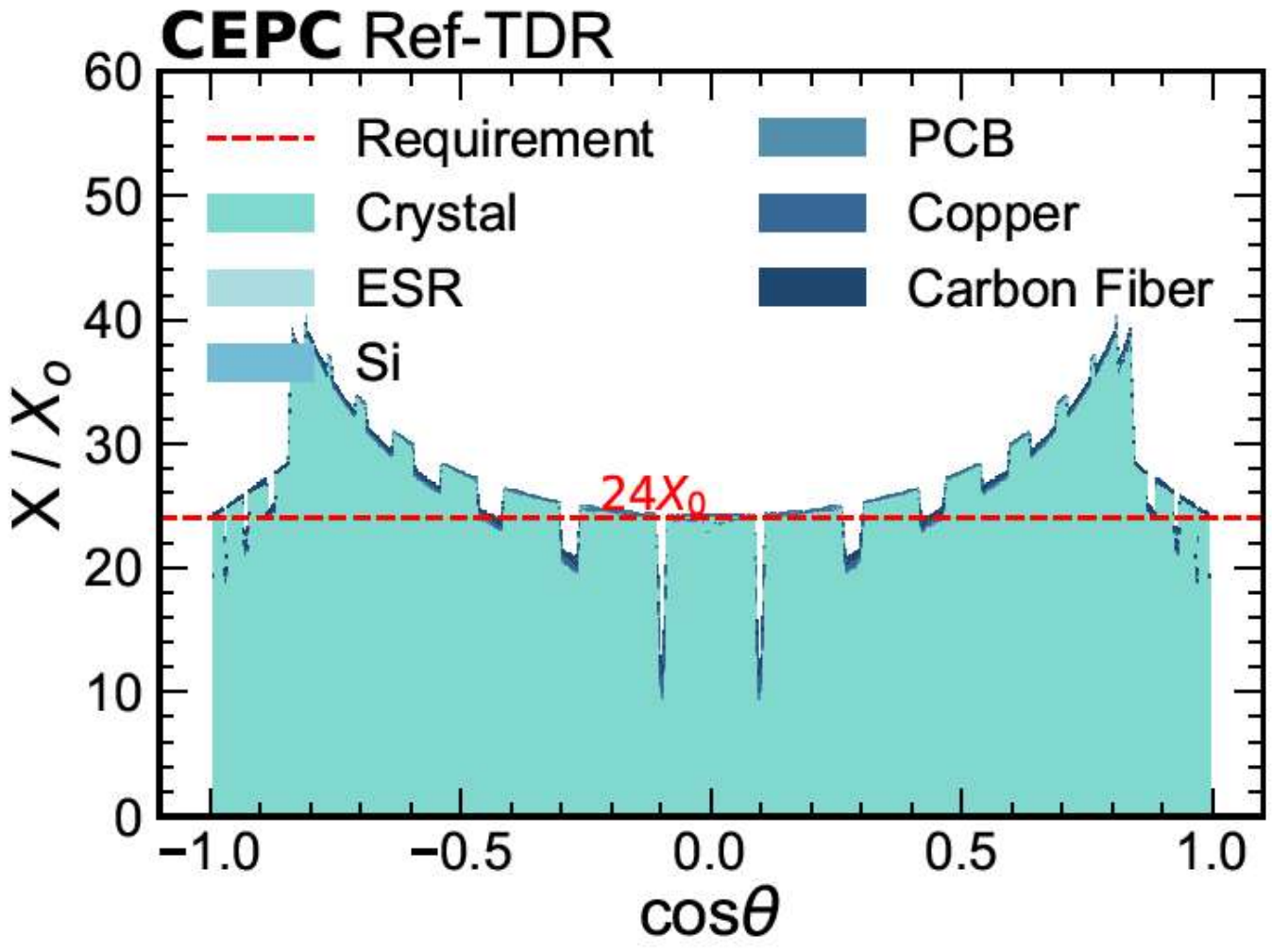

**(b)**

**Figure 7.3:** ECAL material budget scans along with the azimuthal angle (a) and the polar angle (b). Both sensitive and passive materials are including, with the color coding representing crystals, SiPMs, readout electronics, reflective foil, carbon-fiber-based mechanics and copper-based cooling.

## 7.2 ECAL design

### 7.2.1 Detector specifications

The ECAL granularity has been optimized to balance the mass-production feasibility of long crystal bars, the number of readout channels, and PFA performance. The ECAL hardware specifications primarily focus on performance, such as the MIP response, the crystal response uniformity, the signal dynamic range, timing resolution, and the temperature gradient. Their specifications, as summarized in Table 7.1, have been studied based on the Geant4 full simulation to quantify the requirements.

**Table 7.1:** Specifications for crystals and their properties

| Parameters | Value | Remarks |
|---|---|---|
| MIP response | 300 p.e/MIP | For EM energy resolution < 3 % |
| Energy Threshold | 0.1 MIP | To balance EM resolution and SiPM noise suppression |
| Crystal response uniformity | < 1% | For consistent responses in crystals |
| Dynamic Range | 0.1 MIP to 3000 MIP | The range of the maximum and minimum responses per unit |
| Time Resolution | 0.5 ns | For MIP signals |
| Temperature Gradient | ≤ 6 K | For crystal response uniformity and SiPM noise control |

**MIP response and energy threshold** The EM energy resolution predominantly depends on the MIP response in terms of the number of photons detected by SiPMs and the energy threshold, especially in the low energy region. Therefore, scintillation photon statistics, the SiPM response and non-linearity effects, and modeling of readout electronics have been included in a Geant4 simulation of crystal ECAL modules to study their influence. The light output in terms of MIP response for each BGO crystal with the thickness of 15 mm corresponds to an MIP energy deposition of 13.3 MeV (as the most probable value). Figure 7.4a shows the EM energy resolution with the MIP response of 300 p.e at different energy thresholds. It demonstrates that the EM energy resolution of better than $3\,\%/\sqrt{\mathrm{E(GeV)}}$ can be achieved. Figure 7.4b shows that a higher MIP response per crystal can further improve the stochastic term of the ECAL. A MIP response of 300 p.e. appears sufficient for the target energy resolution.

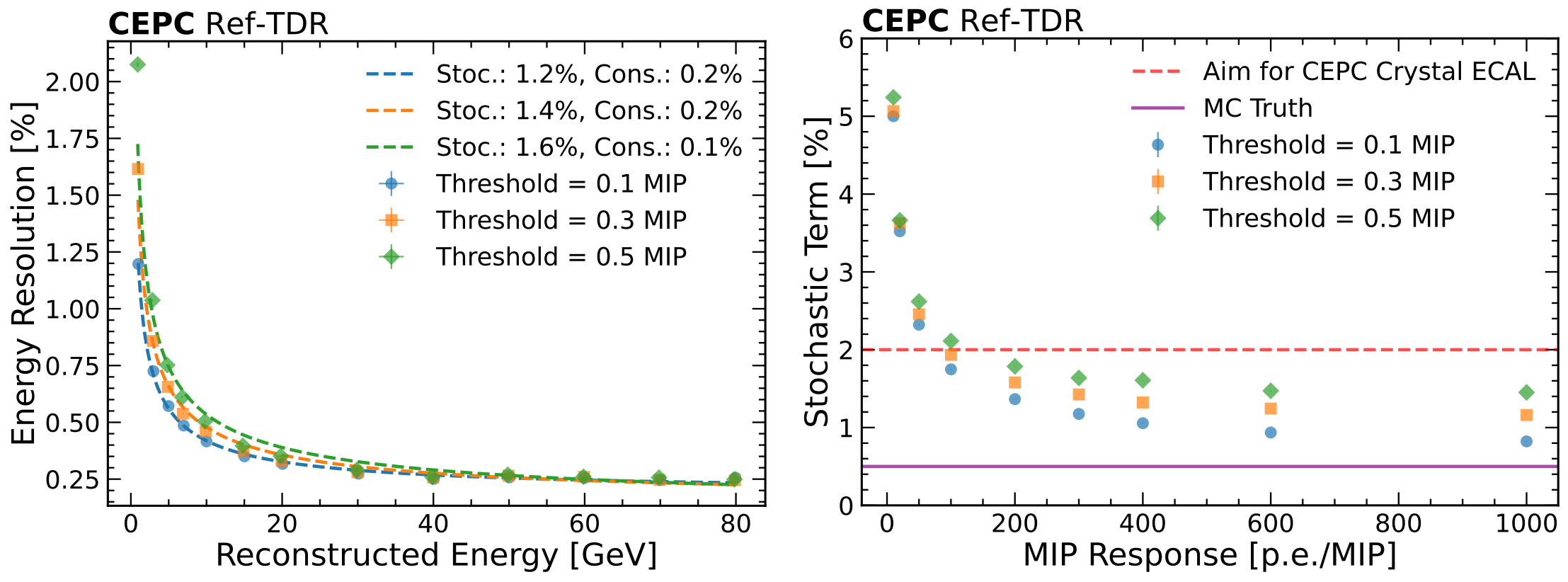

**(a)** **(b)**

**Figure 7.4:** (a) EM energy resolution with different energy thresholds per channel, where the MIP response is set to 300 p.e/crystal (b) the stochastic term extracted from the EM energy resolution with the varying MIP response and the trigger energy threshold for each channel.

**Dynamic range** The ECAL is required to maintain an excellent linearity over a large dynamic range for photons from sub-GeV (i.e. $\pi^0 \to \gamma\gamma$) to 115 GeV from Higgs decays. EM showers of up to 180 GeV are expected for a potential upgrade near the top-quark threshold at 360 GeV. A readout ASIC with automatic switching between several gain stages is a key component to enable the required dynamic range and provide an excellent response linearity. The upper dynamic range is sufficient to handle high-energy processes without compromising low-energy performance. Different gain stages will be inter-calibrated using physics events (e.g. Bhabha events, $Z \longrightarrow e^+e^-$ decays) as well as on-board calibration schemes.

Based on a full detector simulation of 180 GeV electrons from Bhabha scattering at $\sqrt{s}$ = 360 GeV, the maximum energy deposition per crystal bar is expected to be below 45 GeV. Photons from $H \longrightarrow \gamma\gamma$ are expected to be less than 25 GeV. The distributions are shown in Figure 7.5.

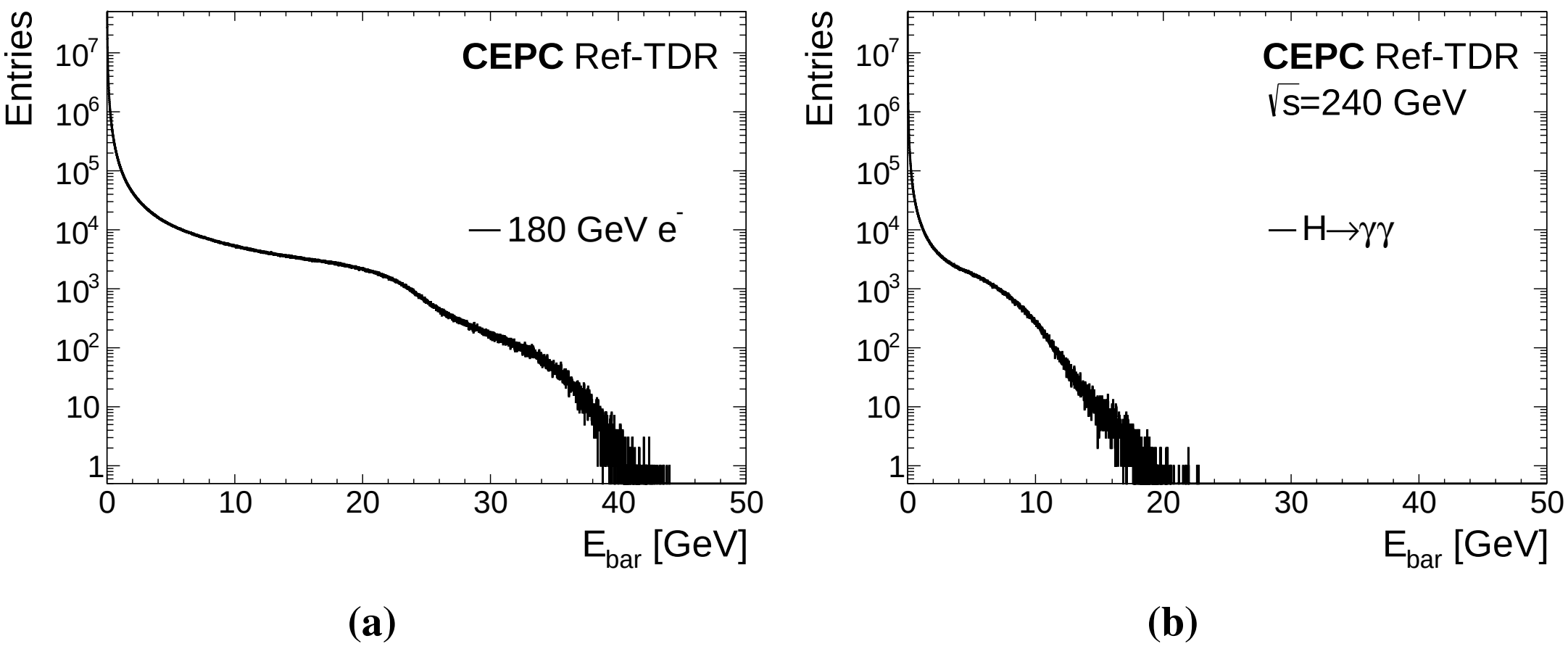

**Figure 7.5:** The energy deposition of a single crystal bar in different physics processes: (a) 180 GeV electrons for the CEPC operated at 360 GeV, with a maximum energy deposition of approximately 45 GeV; (b) $H \longrightarrow \gamma\gamma$ for the CEPC at 240 GeV, where the maximum deposited energy reaches up to 25 GeV.

**Crystal response uniformity** The crystal response needs to be uniform when particles hit at different positions of a single long crystal bar. Its non-uniformity is defined as the standard deviation of the mean response. Based on the simulation studies, the non-uniformity of the crystal response is required to be less than 1 % in order to ensure the target EM energy resolution.

**Timing resolution** Simulation studies and beam tests have been performed to evaluate the timing resolution of BGO crystal bars. For single-MIP signals, the timing resolution per readout channel is around 0.5 ns, due to the long crystal geometry and the relatively slow scintillation time of BGO (around 300 ns). For particle shower signals large than a single MIP, the timing resolution better than 0.2 ns can be achieved. The application of timing information in calorimetry is expected to improve clustering performance, with its full potential to be explored using PFA.

**Temperature gradient** Temperature gradient has to be well controlled as it will affect the crystal light yield, the SiPM operation voltage and the SiPM noise level. Simulation studies indicate that a temperature gradient controlled within 6 degrees can have a negligible contribution to the ECAL energy resolution.

### 7.2.2 Design with orthogonal long crystals

In traditional crystal calorimetry design, crystal bars were arranged to project toward the IP, but the lack of longitudinal segmentation made them incompatible with PFA. This new design introduces orthogonally oriented long crystal bars with 3D segmentation to enable PFA compatibility. However, this design must address two critical issues: overlapping showers and hit ambiguity, as discussed below.

**ECAL reconstruction and pattern recognition** The ECAL reconstruction faces two major challenges: overlapping showers in the same crystals and positioning ambiguity caused by the crystal bar geometry. This ambuguity occurs when multiple particles enter a single module, resulting in spurious hits (or 'ghost hits') as shown in Figure 7.6.

A new particle-flow algorithm named Crystal bar ECAL reconstruction PFA (CyberPFA) [5] has been developed to address the issues of overlapping showers and ambiguity. The method employed by CyberPFA is briefly outlined as follows. Firstly, track positions are extrapolated from the tracker to the ECAL. Adjacent hits are clustered, and local maximum in each layer are used as 'seeds' to define the core region of the shower. Topological merging is applied to improve hadronic shower reconstruction, and overlapping showers are split using expected shower profiles. Ambiguity removal is achieved by an algorithm that incorporates longitudinal development patterns, module geometry constraints, and timing information. Details are described in Chapter 13.

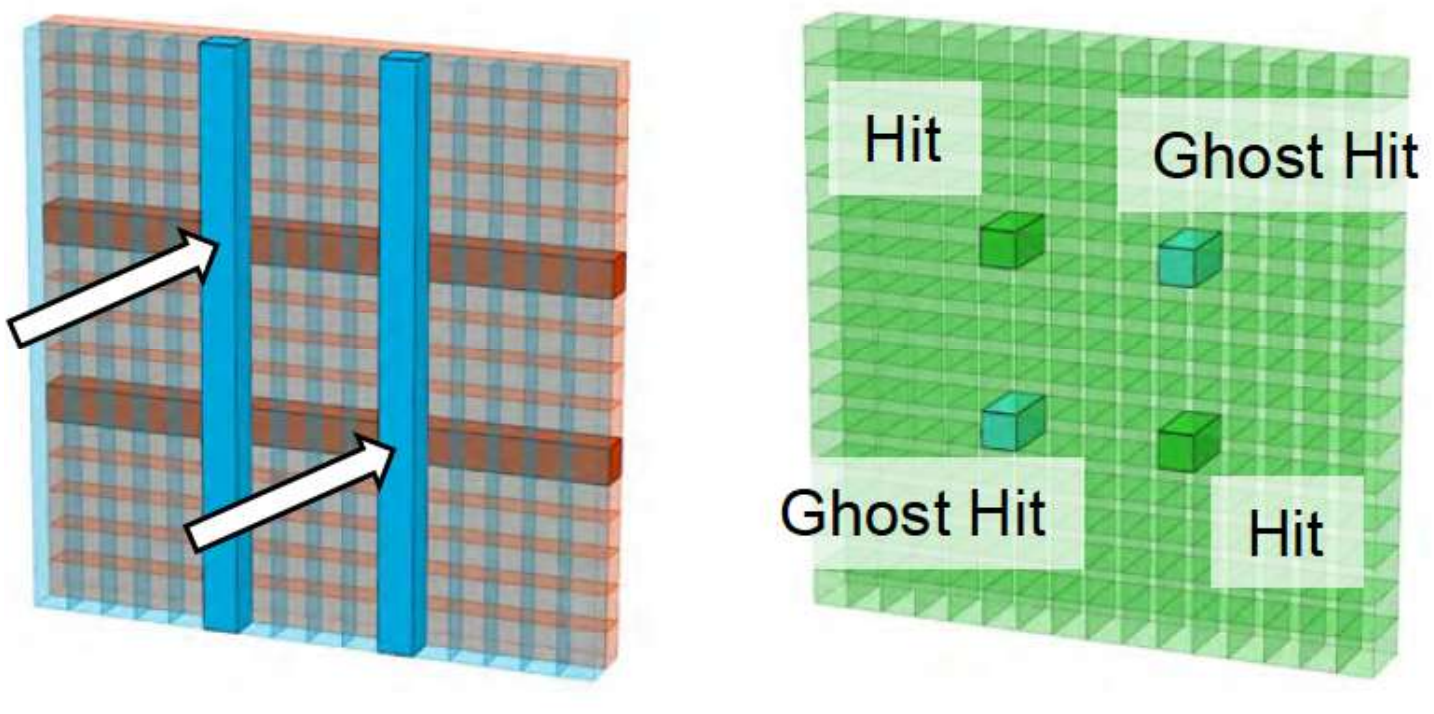

**Figure 7.6:** Schematics of a possible ambiguity scenario of two photons (both with an energy of 5 GeV) with an incident angle perpendicular to the transverse plane of an ECAL module.

**Separation of two close-by particles** The two-particle separation efficiency is defined as the ratio of events successfully reconstructed with two Particle Flow Objects (PFOs) over the total number of events with two incident particles. This efficiency is a critical figure of merit for CyberPFA, particularly for reconstructing events with near-by showers, such as photon pairs produced in $\pi^0 \to \gamma\gamma$ decays. Figure 7.7 presents the separation efficiency for two 5 GeV photon showers in the normal incidence to the ECAL module transverse plane as a function of their lateral distance. At a distance of 3 cm, the efficiency reaches 100 %.

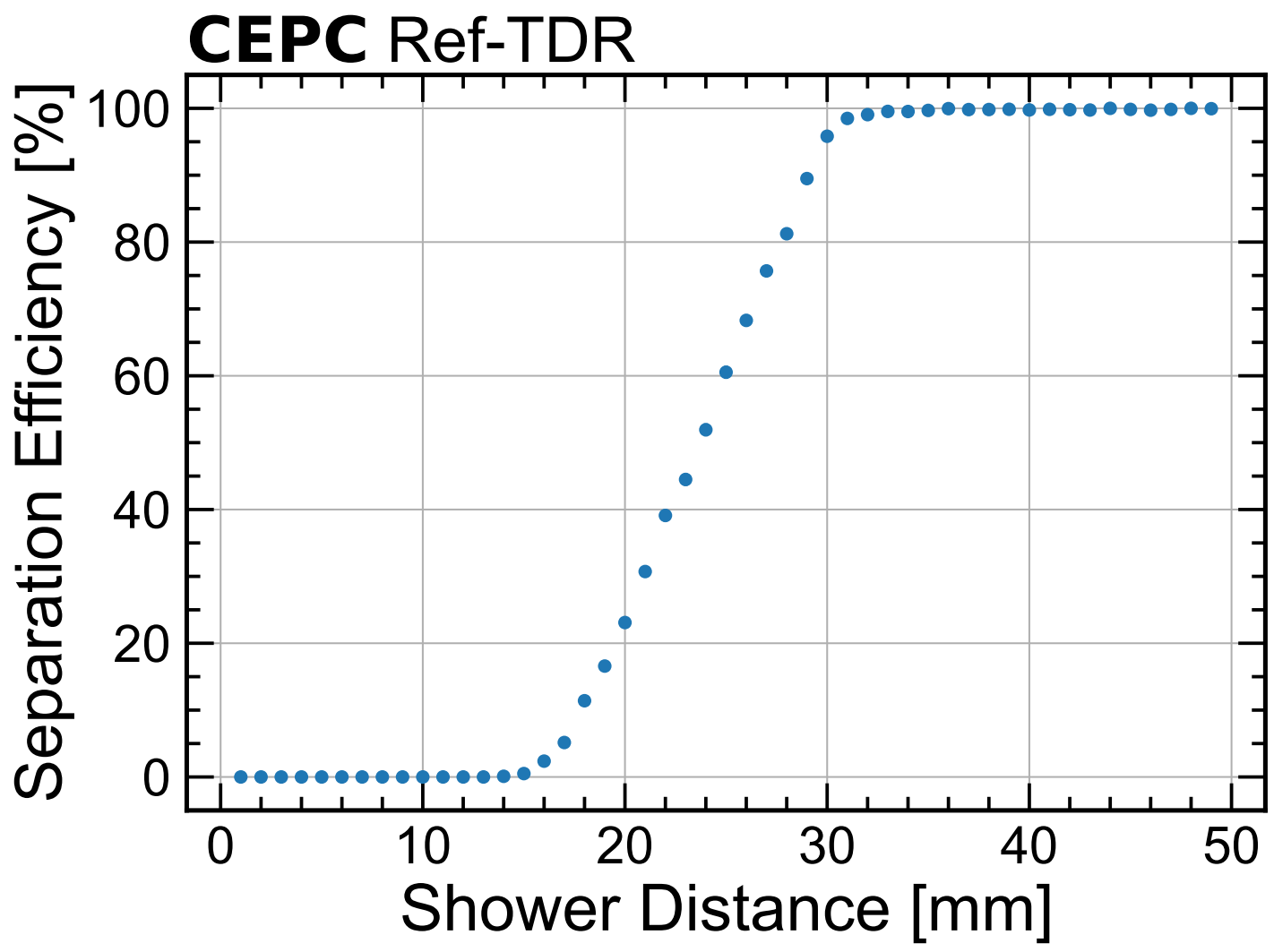

**Figure 7.7:** Separation efficiency of two close-by photons with varying incident positions in between. The separation is successful when two neutral particle-flow objects (PFOs) are reconstructed, each with $E_{\text{PFO}}/E_{truth}$ > 75% and PFO-truth angle < 0.25°

## 7.2.3 Detector components

### 7.2.3.1 Scintillating crystals

As the baseline scintillating crystal option, BGO offers a density of 7.13 g/cm$^3$ and a radiation length of $X_0$ = 1.13 cm [6], which enables a compact ECAL design with a total depth of 27 cm, corresponding to $24X_0$, for efficient containment of electromagnetic showers. BGO also exhibits a high intrinsic light yield in the range of 8000 to 10000 photons/MeV, enabling the use of relatively small SiPMs with a typical sensitive area of 3 × 3 mm$^2$. This configuration achieves an excellent signal-to-noise ratio, critical for precise energy measurements. While the scintillation decay time of BGO is relatively slow at 300 ns, it is already sufficient for operations in Higgs and Low-Luminosity Z modes, where pile-up effects from beam-induced backgrounds have minimal impacts on performance, as discussed in detail in Section 7.4.5.

For BSO as an alternative option, its density (6.80 g/cm$^3$) and radiation length ($X_0$ = 1.15 cm) are very close to those of BGO, ensuring that the ECAL depth and total crystal volume would remain nearly identical. BSO has an intrinsic light yield approximately four times lower than that of BGO, but compensated by a significantly faster scintillation decay time of around 100 ns [7]. It makes BSO an intriguing option for improved timing performance, particularly in scenarios where precise timing is critical. However, the lower light yield poses challenges for achieving the comparable energy resolution, and further developments are needed to optimize BSO for better optical quality and the possibility of large-scale mass production.

#### 7.2.3.2 Silicon photomultiplier

Silicon Photomultiplier (SiPM) [8, 9], a compact solid-state photosensor with the sensitivity to detect single photons, is selected as the baseline option of photosensor for the ECAL. Major considerations for the selection of SiPM candidates for the crystal ECAL are discussed as below.

**SiPM requirements and specifications** The SiPM for the crystal ECAL is required to have a high dynamic range, a moderate Photon Detection Efficiency (PDE) to match the BGO crystal scintillation light (maximum emission around 480 nm) and an acceptable Dark Count Rate (DCR). A typical SiPM sensitive area of $3 \times 3$ mm$^2$ is chosen as it can satisfy the MIP response and the total terminal capacitance requirements. SiPM candidates equipped with pixel arrays of typical pixel pitches of 6 µm and 10 µm are selected to mitigate non-linearity effects with large light intensities. Additionally, the inter-pixel crosstalk probability should be as low as possible (e.g. $\leq 1$ %) to ensure minimum impact to the response linearity, which is also critical for maintaining accuracy at a high light intensity. In addition, fast SiPM pixel recovery time is extremely important to further extend the SiPM linear response range for large scintillation light signals, as more pixels can recover in time to further detect more photons from BGO crystals with the relatively slow scintillation time.

Major SiPM options with a focus on the dynamic range and their typical parameters are summarized in the Table 7.2 [10, 11, 12]. Three SiPM candidates are equipped with small pixel pitches of 6 µm and 10 µm offer a large dynamic range and a good signal-to-noise ratio. NDL-SiPMs use the epitaxial layer as the quenching resistor and avoid the use of polysilicon to ensure a higher PDE for small pixel pitches. The HPK SiPM employs optical trenches around each pixel to isolate crosstalk photons from neighboring pixels, leading to a substantially low level of inter-pixel crosstalk. Based on a newly developed simulation model of SiPM non-linearity effects in the BGO crystal readout, all three SiPM options can satisfy the dynamic range requirement. Dedicated test-stands and further measurements are needed to validate the model and evaluate the SiPM non-linearity effects with experimental data.

### 7.2.4 Electronics

The electronics system, which measures the deposited energy in crystals and the arrival time of the particle, consists of three parts: (1) the front-end ASICs responsible for converting the SiPM charge and the time stamp to digital signals; (2) the data aggregation from the front-end ASICs and data transmission via fibers to the back-end electronics; (3) the power distribution and management to ensure the proper operation by components such as SiPMs and ASICs.

**Table 7.2:** Parameters of SiPM candidates selected for the crystal ECAL

| **SiPM Type** | **NDL EQR06** | **NDL EQR10** | **HPK S14160-3010PS** |
|---|---|---|---|
| Pixel Pitch µm | 6 | 10 | 10 |
| Pixel Quantity in $3 \times 3\,\mathrm{mm}^2$ | 244,719 | 90,000 | 89,984 |
| Pixel Gain | $8 \times 10^4$ | $1.7 \times 10^5$ | $1.8 \times 10^5$ |
| Typical peak PDE | 30 % (at 420 nm) | 36 % (at 420 nm) | 18 % (at 460 nm) |
| Typical DCR (20 °C) | 2.5 MHz | 3.6 MHz | 700 kHz |
| Inter-pixel Crosstalk | 12 % | N/A | < 1 % |
| Terminal Capacitance (pF) | 45.9 pF | 31.5 pF | 530 pF |

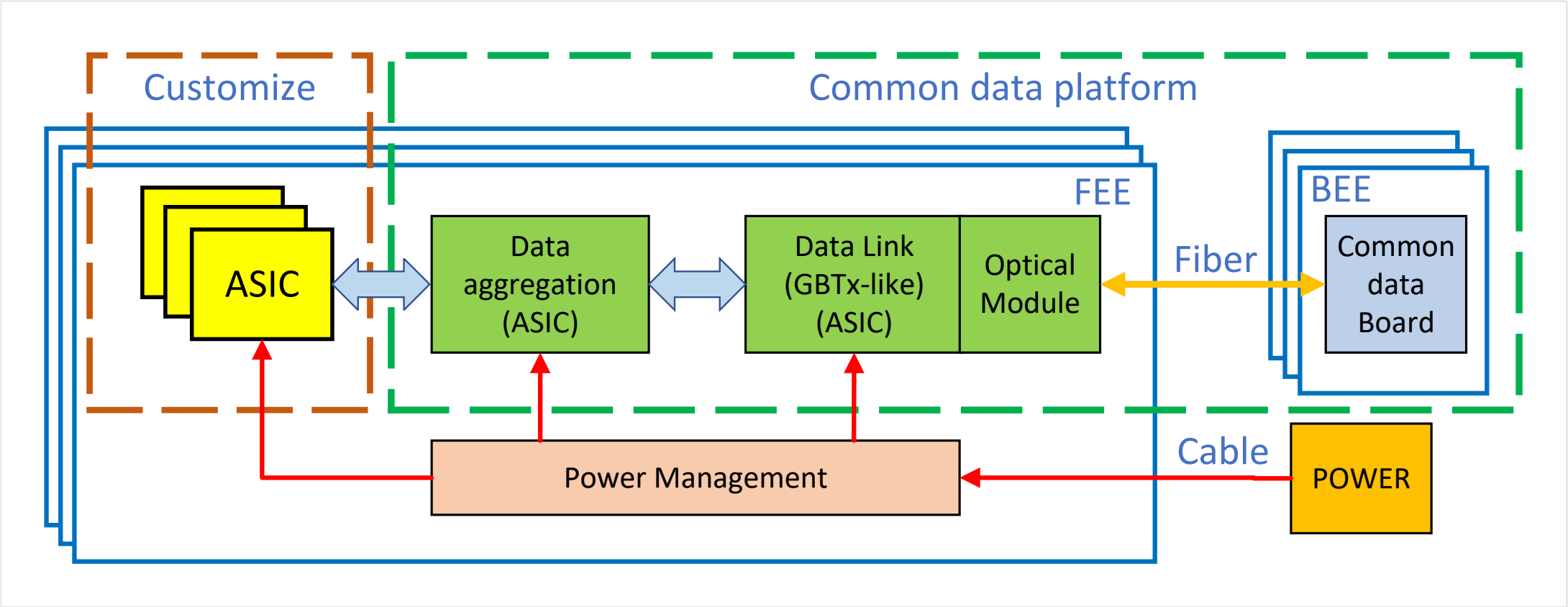

**Figure 7.8:** Schematics of the ECAL readout electronics with the modular design for the SiPM readout: the Front-End Electronics ASICs, connected to data aggregation ASICs that transmit digital data via optical data links and data aggregator ASICs to the Back-End Electronics boards.

**Table 7.3:** Requirements of ECAL front-end electronics

| **Parameters** | **Requirements** |
|---|---|
| Charge Range | 0.128 pC~ 3.84 nC (0.1~3000 MIP) |
| Charge Resolution | 10 % (1.0 MIP), 1 % (100 MIP) |
| Timing Resolution | 200 ps (1 MIP), 100 ps (12 MIP) |
| Integral Non-linearity | < 1 % |
| Average Event Rate/channel | 13 kHz |
| Max Event Rate/channel | 230 kHz |
| Typical Signal Rising Edge | 40 ns |
| Typical Signal Width | ~ 1 µs |

In fact, HCAL, and the Muon Detector will use similar SiPMs for the scintillator readout. A new ASIC, named "SiPM ASIC for Calorimeter (SIPAC)", will be developed as a common front-end electronics component for all scintillator-based sub-detectors. The design of SiPM ASIC for Calorimeter (SIPAC) will follow state-of-the-art ASIC technologies, which have been successfully demonstrated [13, 14, 15, 16]. The ECAL electronics requirements are summarized in Table 7.3, with the target SiPM capacitance of 50 pF and the typical SiPM pixel gain of $8\times10^4$.

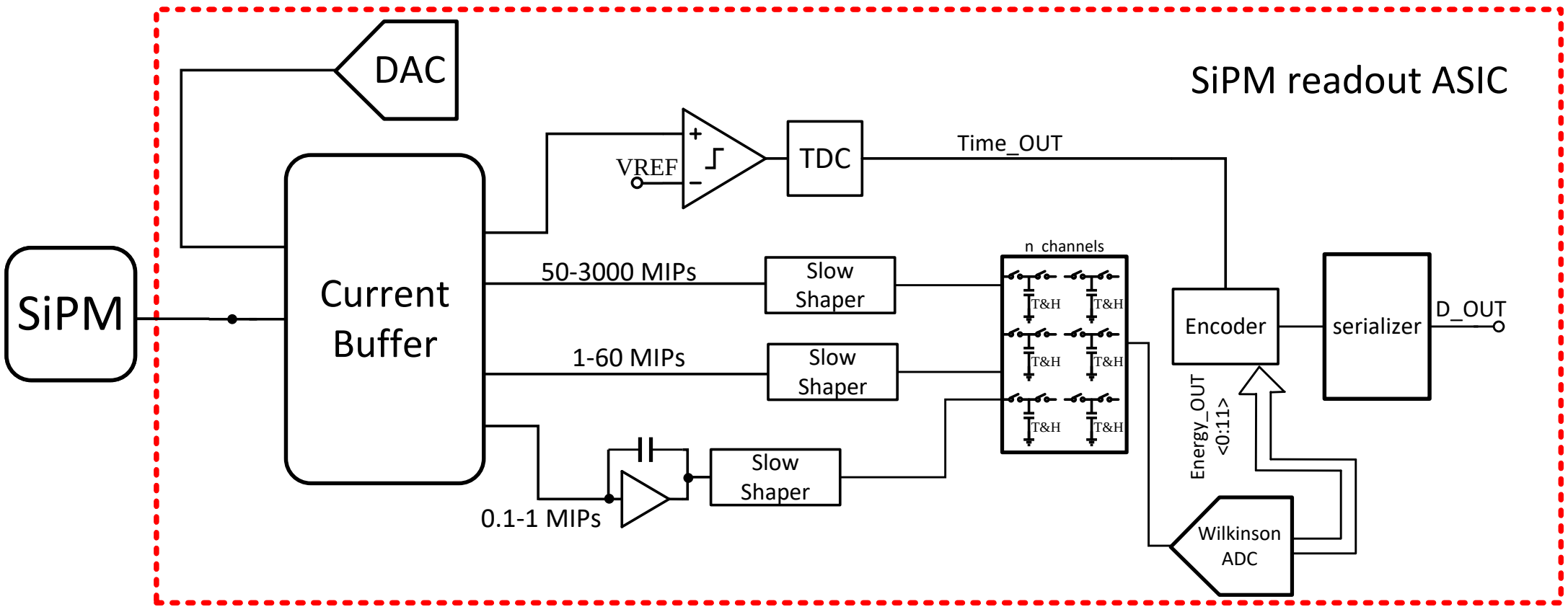

**Figure 7.9:** Block diagram of one channel of SIPAC ASIC. The single-channel employs a current buffer, followed by two slow shapers and an ADC for charge measurement, and a discriminator with a TDC for time measurement. Signals are sent to the back-end electronics system through high-speed serial links. A DAC is used for tuning the bias voltage of the SiPM.

The SIPAC chip is designed with multiple channels and the single-channel readout schematic is shown in Figure 7.9. Each channel has a current buffer, a discriminator, and a TDC for the TOA measurement. Two-stage gain modes are designed for the charge measurement in the required dynamic range, followed by two slow shapers and then shared with a Switched-Capacitor-Array block for signal sampling.

Front-end readout circuits employ multiple gain paths to achieve measurements across a wide dynamic range. The energy path utilizes a slow shaper with two stages of low-pass filters, while the timing path employs a fast shaper with a band-pass filter.

When passing the trigger threshold, signals after shapers are then digitized by a 12-bit ADC and a TDC with a timing resolution of 100 ps. The digital codes generated by the ADC and TDC are encoded and serialized by the digital module for data transmission. The TDC is designed to measure the TOA and it employs a hybrid measurement structure combining both coarse and fine counting. The coarse counting is derived from a counter, while the fine counting is determined by a delay line.

Precise tuning of the SiPM bias voltage can be achieved using an on-chip DAC. AC coupling is implemented for signal transmission which isolates the adjustment effects of the DAC from the readout circuit.

### 7.2.5 Mechanics and cooling

The ECAL mechanics structure is based on CFRP featuring high strength and light weight, as the primary option to minimize the material budget while enabling integration of detector components. Such a large-scale CFRP-based mechanical system for the ECAL with a total volume of 24 $m^3$ BGO crystals shall take into account the structural integrity, thermal management, and alignment precision. This subsection will describe the ECAL mechanics design for both the barrel and endcaps.

The general mechanical design of the ECAL system is shown in Figure 7.10. The barrel ECAL is affixed to the barrel HCAL via dedicated structural interfaces while the OTK layer are mounted on the endcap ECAL discs. FEA is used to evaluate the stress and deformation distributions as well as temperature gradients.

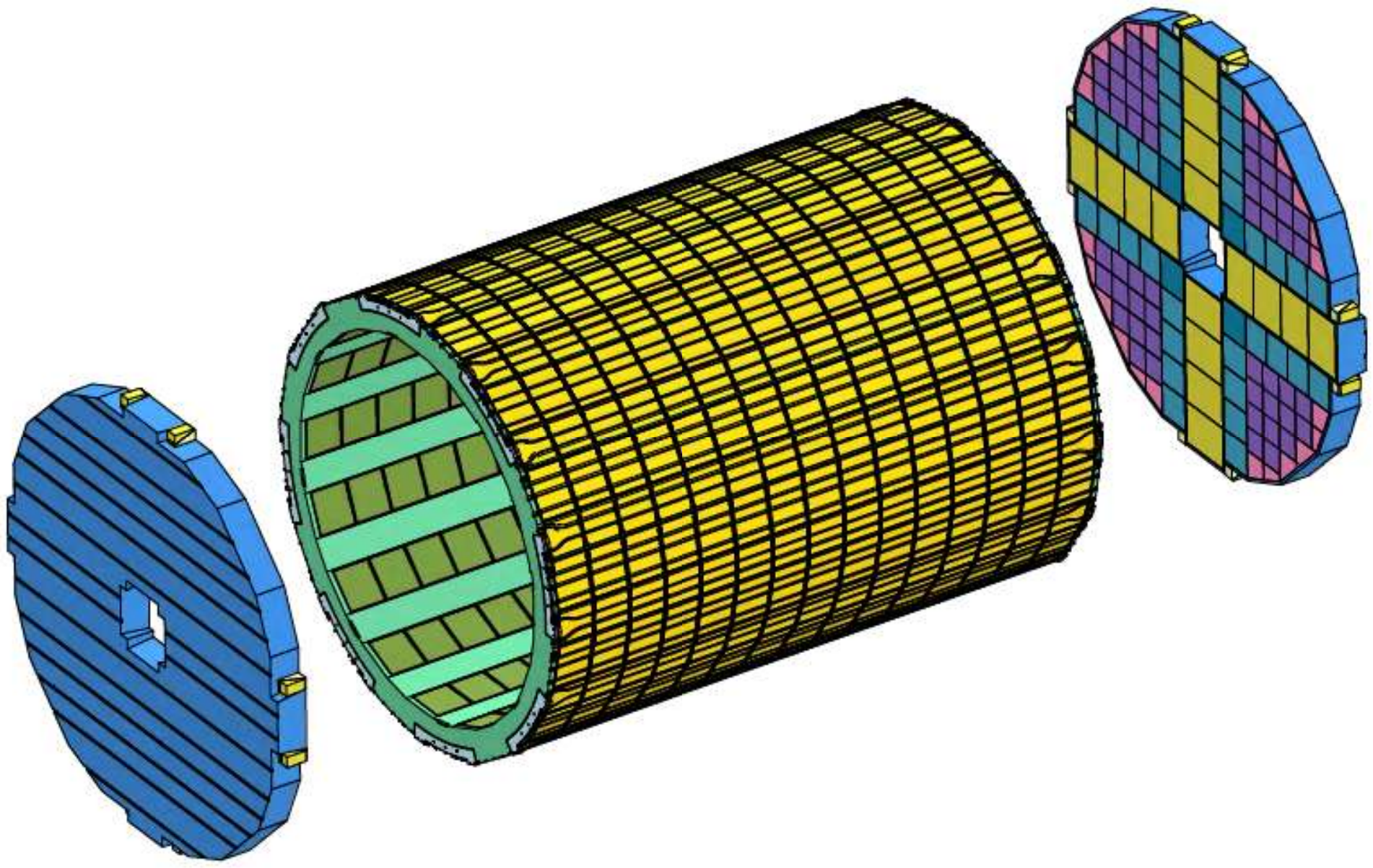

**Figure 7.10:** The ECAL mechanical design with one barrel and two endcaps. The barrel ECAL has 480 modules, which are integrated into a carbon-fiber frame (in lighter green), and is affixed to the barrel HCAL via 8 interconnection points. Cooling plates and cooling pipes are colored in light yellow. Two ECAL endcap discs along with endcap OTK layers are mounted onto the HCAL from each side. Endcap ECAL modules, indicated in various colors for different dimensions and designs, are integrated into one of the discs of a carbon-fiber frame, indicated in blue.

#### 7.2.5.1 Barrel ECAL

The barrel ECAL is 5800 mm long with an inner and an outer diameter of 3660 mm and 4260 mm, respectively. To ensure a full angular coverage and avoid detection inefficiencies due to cracks between adjacent modules, the ECAL is segmented into 15 equal partitions along the beam axis. The cylindrical part of the barrel is divided into 32 sectors with two types of trapezoids alternatively arranged. This design avoids module projective cracks directly pointing to the IP and therefore guarantees that particles emerging from the IP will not go through the cracks completely, as shown in Figure 7.11.

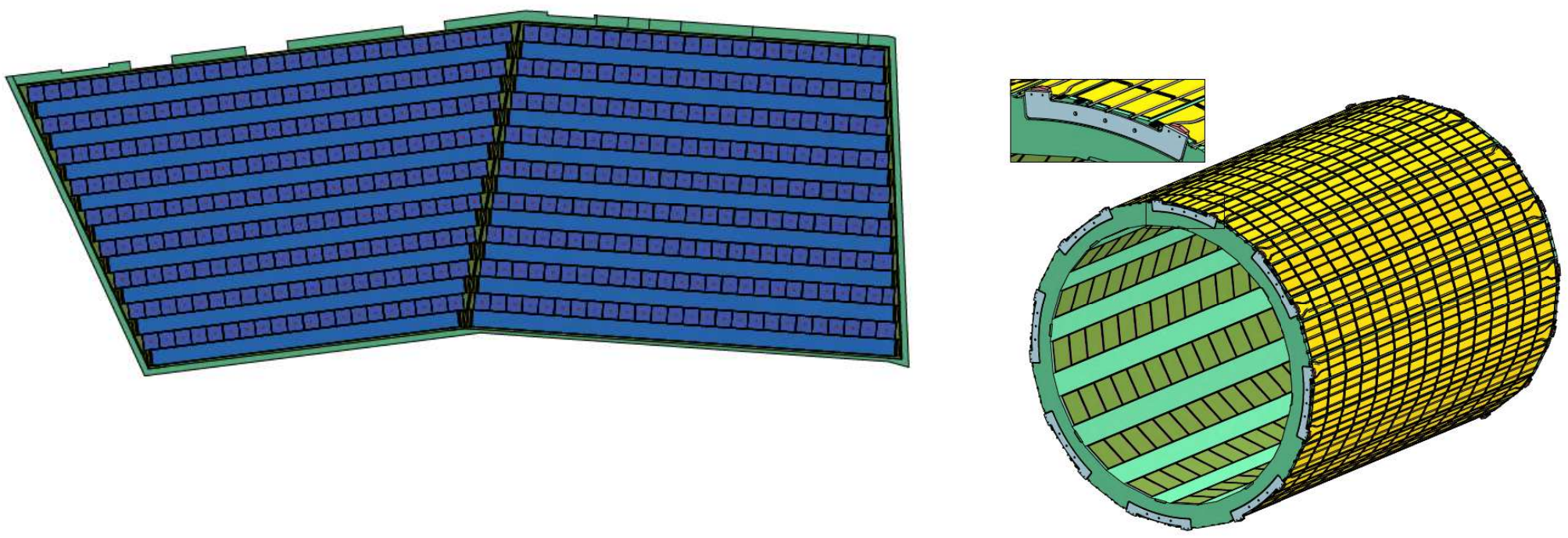

**(a)** **(b)**

**Figure 7.11:** The side view of two types of barrel ECAL modules (a); the stainless steel structure for inter-connecting the ECAL barrel to the barrel HCAL, which are embedded in the main carbon-fiber frame (b). The two module types are named ”Type-A” for the trapezoid shape with its shorter side at the bottom and ”Type-B” for the trapezoid shape with its longer side at the bottom. Detailed FEA results of the ECAL-HCAL inter-connections are described in Chapter 14 of Mechanics and Integration.

**Barrel mechanical structure design** The barrel mechanical structure is illustrated in Figure 7.10, including a cylindrical CFRP frame to fix 480 modules. Each module is mounted in a compartment of the frame and consists of 18 longitudinal layers of crystal bars. As shown in Figure 7.12, crystals bars, readout boards, and cooling services are integrated in a carbon-fiber frame for each module. Crystal bars in one layer are grouped and housed within extra CFRP honeycomb boxes for structural integrity. Signal cables and cooling pipes are routed at the module backend, which are integrated to minimize interference with sensitive detection areas. The requirements include: (1) tolerances of the outer and inner contour within [−5 mm, 0 mm] and [0 mm, 5 mm], respectively; (2) modules aligned within 2 mm accuracy; (3) positioning tolerances controlled within ±0.5 mm.

**Barrel structure FEA results** FEA has been performed to assess the structural stability of the carbon-fiber frame and evaluate mechanical stress to crystal bars. The structural response was simulated under conditions with a total weight of 130 t, when 480 modules are fully integrated. The stress distributions in three different directions (i.e. the stress in parallel to the carbon-fiber direction, the stress perpendicular to the carbon-fiber direction and the in-plane shear stress) show that they stay well within the specifications of the cost-effective CFRP option named "T700". A maximum deformation of the main frame is 3.4 mm. For individual modules, the maximum deformation of 0.23 mm occurs in the central module of the top row. Given the 1 mm gap reserved for module assembly tolerance, this deformation can be accommodated within the available space. This indicates that the lateral stress resulting from the cumulative forces exerted by neighboring modules remains well within tolerance and satisfies the design requirements.

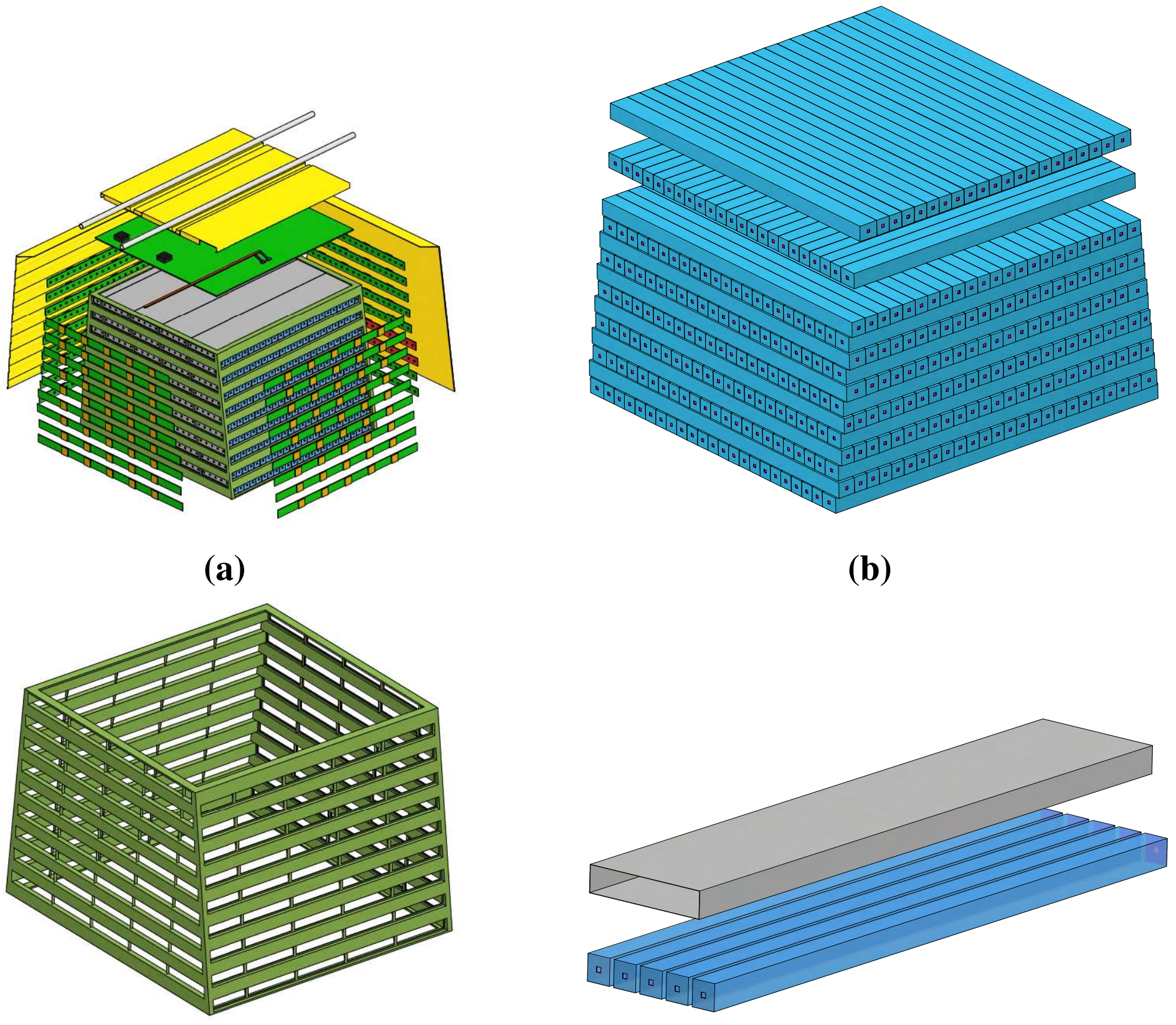

(a) (b)

(c) (d)

**Figure 7.12:** The schematics of a barrel "Type-B" module with crystals, readout boards and the cooling service. Crystal bars (in blue) are grouped and fixed in a light-weighted box (in gray) and mounted in a carbon-fiber frame (in dark green). Readout boards (in light green) and cooling plates and pipes (in yellow) are also integrated within the frame.

Since the design of barrel modules is similar to that of endcaps, FEA results of a single ECAL module are presented in Section 7.2.5.2.

**Cabling and cooling pipe layout** A 10 mm space in addition to the ECAL outer radius is reserved for routing both signal cables and cooling pipes, which is also shared by back-end data concentration boards located at the rear of each module. Optical fibers connect the back-end data concentration boards to the external data acquisition system. For each azimuthal slice, there are 15 optical fibers for digital signals, where 7 fibers are routed toward one end and 8 towards

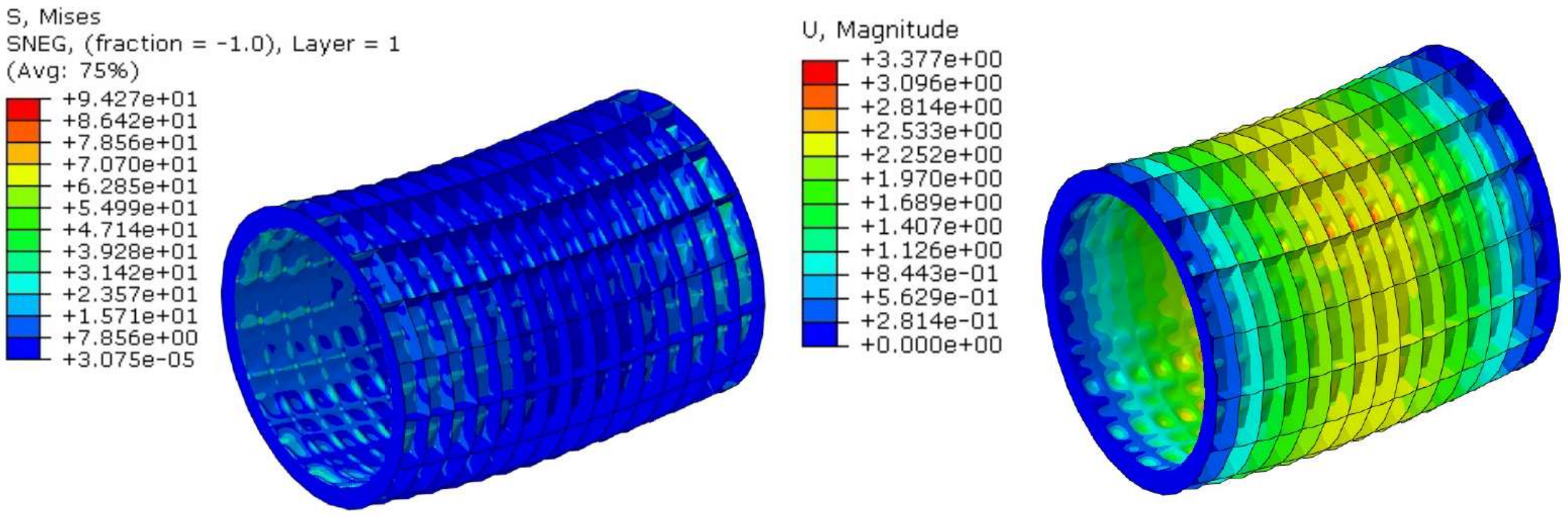

**(a)** **(b)**

**Figure 7.13:** Distributions of the von Mises stress in MPa (a) and the deformation in mm (b) of the barrel CFRP frame with the full load of all ECAL barrel modules in the FEA simulation. The maximum stress is 94.3 MPa and the maximum deformation is 3.4 mm.

the other end.

The cooling system dissipates heat by 1 mm thick copper cooling plates attached to the readout boards and copper cooling pipes installed in the module rear part. The cooling pipes, each with a round shape and a diameter of 10 mm, are grouped in parallel and brazed or soldered onto the cooling plate. Copper spacers are added to support the pipes and maintain a reliable thermal contact with the cooling plate. Three cooling pipes are arranged in each ”Type-A” module, and two cooling pipes in each ”Type-B” module, as illustrated in Figure 7.11.

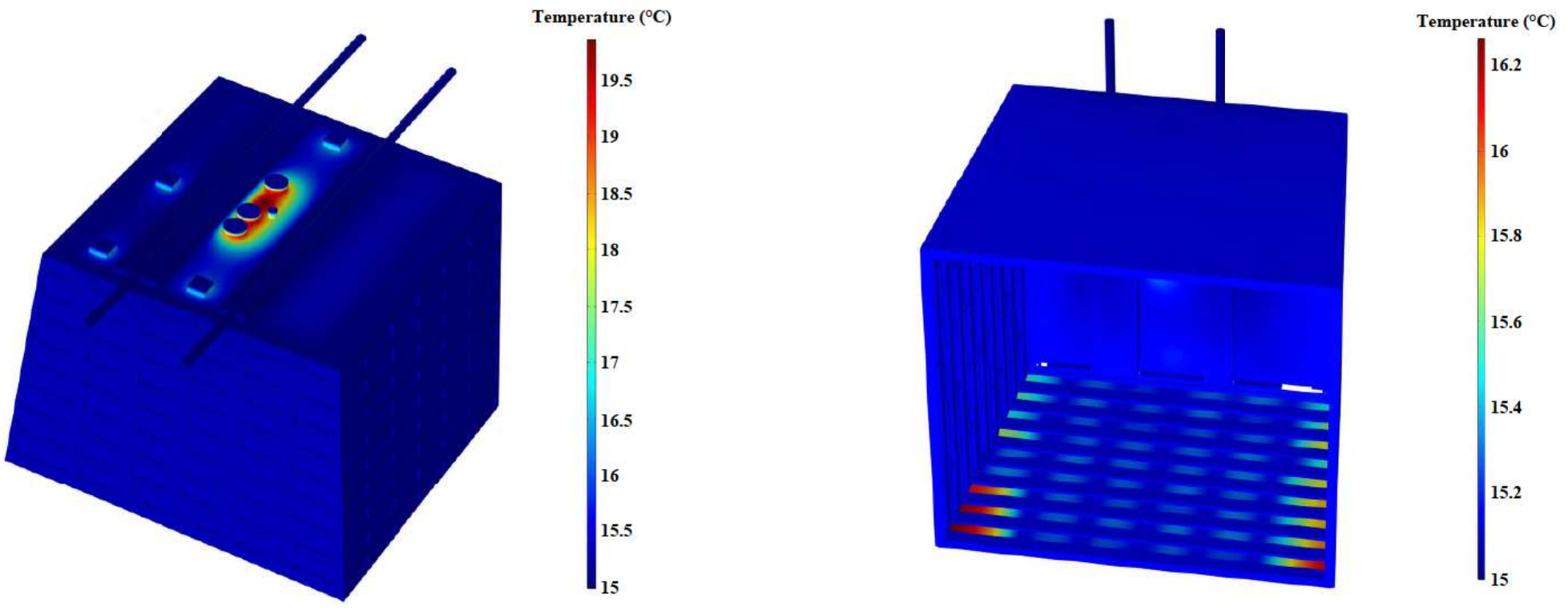

**(a)** **(b)**

**Figure 7.14:** The FEA results on temperature distributions of a barrel module integrated with readout boards and cooling pipes: the temperature distributions including all components (a) and excluding the data concentration chips (b). The simulation assumes an inlet cooling water temperature of 15 °C, an environmental temperature of 17 °C, and the nominal power dissipation of the ASICs. A coefficient of 0.5 $\mathrm{W/m^2 \cdot K}$ is used to simulate the limited convective heat transfer with the confined environment of ECAL modules. The module perspective in (b) is rotated approximately 90 degrees relative to that in (a) to enable a detailed visualization of the temperature distribution within the module.

**Cooling Performance** Thermal simulations were performed to evaluate the temperature distribution of barrel modules, accounting for the power dissipation of all readout chips and data concentration boards. Distilled water is used as the coolant, with a flow rate of 10 cm/s. Figure 7.14 shows the resulting temperature distribution across a single module and its back-end PCB area. The temperature rise in a module is less than 4.5 °C, mainly localized near the data concentration chips. Additionally, the maximum temperature variation within the module is less than 1.2 °C, which can satisfy requirements.

#### 7.2.5.2 Endcap ECAL

The ECAL endcap is divided into several zones, as shown in Figure 7.15, to minimize types of scintillating bars. In total, there are a total of 6 module types. To ensure hermetic coverage, the modules indicated by yellow and blue are designed with a trapezoidal structure.

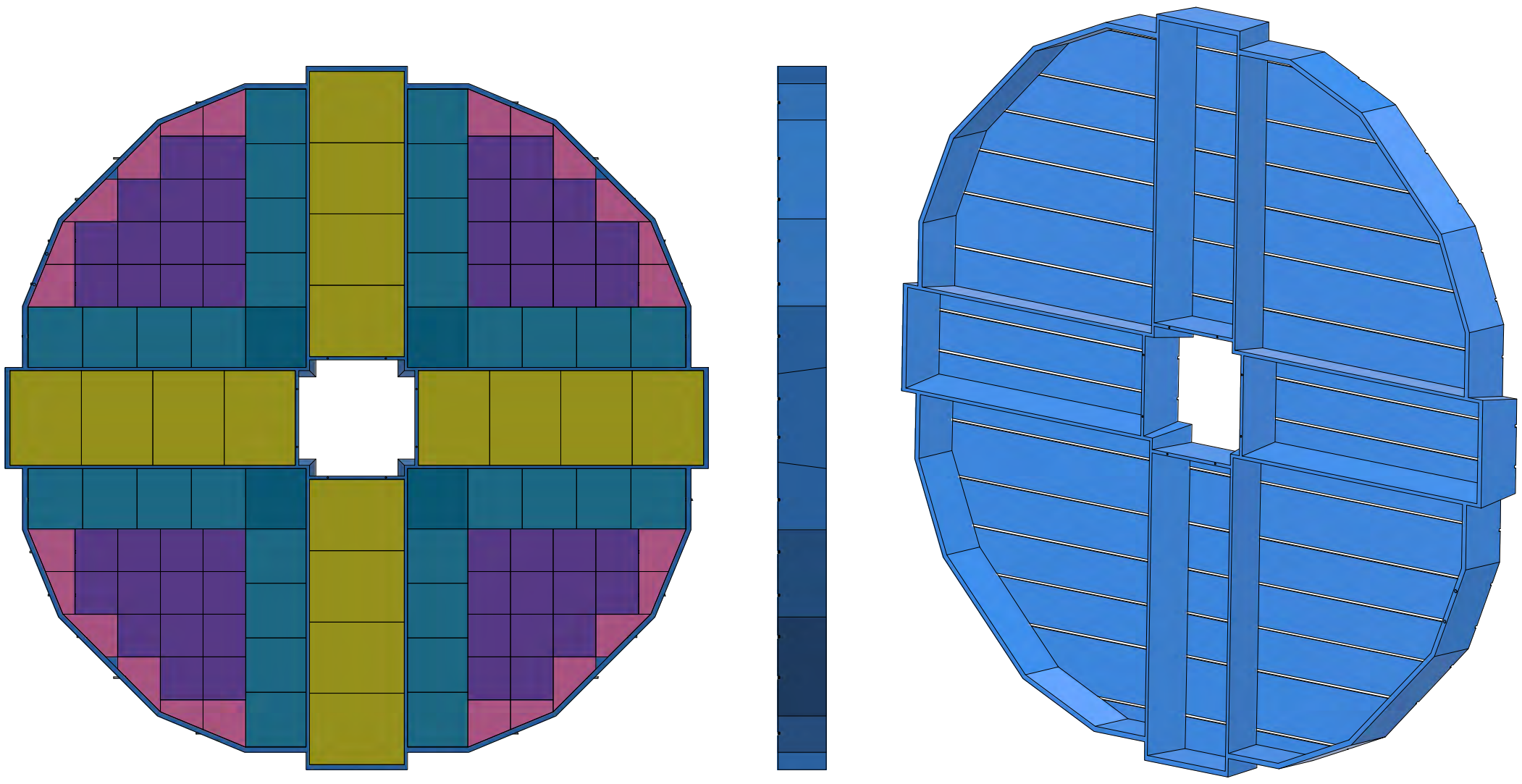

**(a)** **(b)**

**Figure 7.15:** The overall ECAL endcap design with modules in different dimensions (a) and a CFRP frame of an ECAL endcap disc with modules mounted (b). For the modules on the same horizontal plane as colliding beam lines, module gaps are optimized to avoid projectile cracks directly collinear with the IP. The color coding represents different module types with various shapes and dimensions.

The framework structure of the endcap is shown in Figure 7.15b. Cross-shaped reinforcements are placed in the middle of the framework and square grooves are introduced into the backplate to facilitate the installation of cooling pipes. To minimize gaps between modules, no intermediate structural layers are placed to separate adjacent modules. Therefore each module frame has to sustain its own module weight and also the total weight of all modules above it, which is a major difference from the barrel module design. The endcap module with the largest size and the maximum load capacity is selected for the detailed mechanical design, and the specific structure is shown in Figure 7.16.

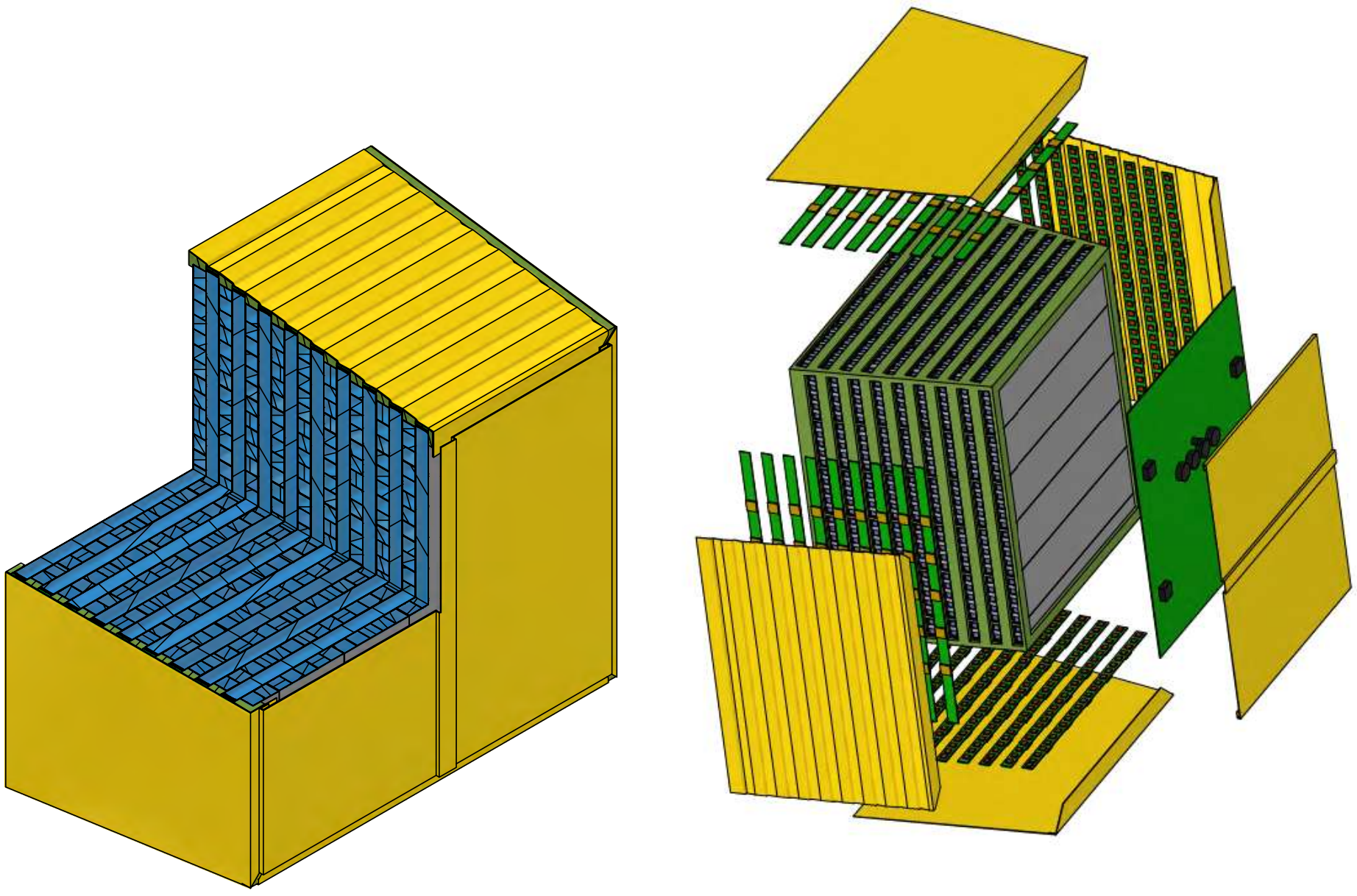

**(a)** **(b)**

**Figure 7.16:** The endcap ECAL module structure and components placed in the vertical direction: the cutaway view of a single endcap ECAL (a) and the explosion view of an endcap ECAL module (b). Crystal bars (blue) and readout boards (dark green) are fully integrated in the carbon-fiber frame (light green) and covered by copper cooling sheets (yellow) on the four lateral sides. A data concentration board (the square in dark green) is attached and covered by a cooling plate, where cooling pipes are also integrated.

As shown in Figure 7.16, a typical endcap module weighs around 0.4 t, including a carbon fiber frame, crystal bars (a total of 1,204 crystal bars for this module), electronics readout boards, a data concentration board, power and signal cables, copper sheets and pipes for cooling.

**Table 7.4:** Deformation and stress in various directions of the CFRP frame including the weight of endcap modules

| Parameter | Unit | Values for T700 | Reference Value/Range |
|---|---|---|---|
| Deformation | mm | 0.099 | |
| Fiber direction stress | MPa | 98.9/-126.7 | 1600/-900 |
| Lateral stress | MPa | 4.1/-3.0 | 22/-120 |
| In-plane shear stress | MPa | 2.0 | 20 |

As with the barrel, an FEA was performed to study the structural strength of endcap modules. The deformation of the CFRP frame and the stress in three directions are shown in Table 7.4 when the frame bears a full gravity load of endcap modules. All stresses and deformations are well within requirements. The cooling design of an endcap ECAL disc is mostly similar to that of barrel modules, as shown in Figure 7.17. Therefore, similar temperature distributions are expected for the endcap modules.

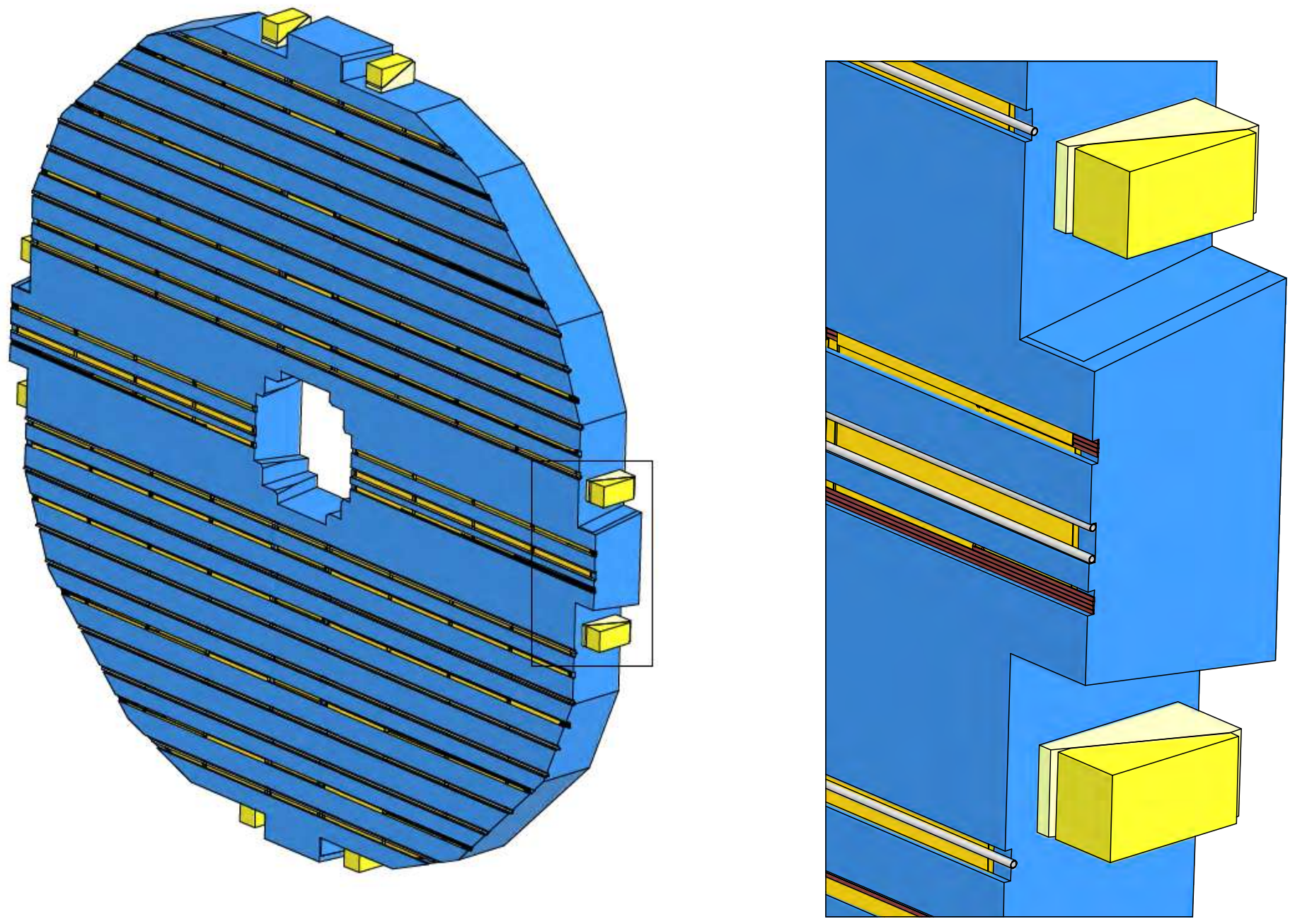

**(a)** **(b)**

**Figure 7.17:** The cooling pipes (light gray), signal and power cables (red) and copper plates (yellow) integrated in the endcap ECAL disc (a) and detailed zoom-in of cooling pipes and cabling (b).

### 7.2.6 Calibration and monitoring

Effective calibration is essential to maintain ECAL performance throughout its operational lifespan. This can be achieved through a combination of on-detector calibration and calibration using physics events.

Before collision data taking, pre-calibration is done by extracting key parameters of crystals and SiPMs from the quality control database during the module assembly and commissioning, some using cosmic-ray muons. Critical measurements include the crystal light yield uniformity and SiPM parameters, such as the breakdown voltage, gain, dark noise rate, and photon detection efficiency. All these parameters will be used to equalise the ECAL response per channel and to correct SiPM non-linearity effects.

The reconstructed energy within a cluster can be described as:

$$E = \sum_i C_i(t) \cdot (A_i - P_i) \cdot R_i \cdot G_i \cdot S_i, \tag{7.1}$$

where $E$ is the energy sum over crystals with signals above the energy threshold and $i$ is the channel index, $A_i$ indicates the signal amplitude and $P_i$ the combined pedestal and noise amplitude, $R_i$ is the inter-calibration gain ratio of an ASIC channel, $G_i$ is the ADC-GeV calibration factor, $S_i$ describes a SiPM non-linearity correction factor and $C_i(t)$ a radiation-damage related correction factor.

In the following, each term of Eq. (7.1) will be discussed in detail. Following this, the calibration precision and the monitoring system will be addressed.

#### 7.2.6.1 On-detector calibration

The on-detector calibration relies on datasets from charge injections and random triggers. This enables the extraction of pedestals, noises and the multi-stage gain of each channel.

The TID and NIEL levels are considered safe for most of crystals and SiPMs except in the regions closest to the beam pipe. The expected levels of irradiation are so low that an optical monitoring system for transparency is considered unnecessary, and the calibration will be performed using collision data.

**Pedestal and noise calibration** The electronics pedestal and SiPM noise for each channel, given by $A_i$, shall be measured periodically. Random triggers without beam collisions will be used for the following purposes: firstly, the definition of an adequate trigger threshold on a channel-by-channel basis while retaining sensitivity to low-energy signals; secondly, the identification of problematic channels with excessive noise levels so that they can be suppressed by other trigger schemes; lastly, providing an input to a time-dependent pedestal and noise calibration.

**SiPM non-linearity calibration** The characterization and calibration of SiPM non-linearity follow a multi-stage strategy. The effects are first quantitatively studied by accounting for pixel recovery during the relatively slow BGO scintillation time (typically 300 ns), which increases the effective number of pixels since individual pixels can fire multiple times per scintillation event. The SiPM non-linearity curve can be measured in the lab or testbeam with a dedicated setup. Key parameters that determine the curve can be extracted for a small batch of SiPMs, such as the breakdown voltage and the inter-pixel crosstalk probability. Correlations of these parameters to the Quality Assurance, Quality Control (QA/QC) database can be established so that the different non-linearity curve of each SiPM is monitored and recorded in the database. Therefore, an initial off-detector calibration is performed by extracting key SiPM parameters from the QA/QC database, which are then applied to all SiPMs in the ECAL and used during commissioning. Finally, in situ calibration is carried out using Bhabha events to monitor and correct for residual non-linearity effects throughout the ECAL operation.

**ASIC calibration** Charge injections will be used to calibrate the channel response within an ASIC and multiple gains of each channel. Each ASIC channel receives periodic injections of known electrical charges to map the digitized output response in ADC codes over the full dynamic range. The overlapping range between the high and low gain modes is covered to determine gain transitions (namely inter-calibration factors) . Calibration results will be cross-validated using collision data, such as Bhabha events.

**Calibration for beam-induced backgrounds** Calibration data sets of beam-induced background can be acquired using random triggering or other dedicated trigger schemes during beam collisions and will be used for tuning the threshold, masking noisy channels, and performing time-dependent corrections.

#### 7.2.6.2 Collision data calibration

Collision data during the operation can be used to ensure the calibration precision over a wide energy range with well-constrained systematic uncertainties, which include electrons or positrons from Bhabha scattering, leptons from $W$ and $Z$ bosons, or photons from the decay of a neutral pion.

**Bhabha scattering calibration** Bhabha scattering events ($e^+e^- \rightarrow e^+e^-$) are used as a calibration source due to its large cross-section and clean event topology [17]. The energy of an electron or positron ($E$) measured in crystals can be compared to the momentum ($p$) measured in the tracking system. The $E/p$ method can be used for the inter-calibration of ECAL modules. The expected Bhabha events collected per hour in the barrel and endcap regions shown in Figure 7.18. The TID rate (per hour) accumulated by each ECAL module was also studied in BIB simulation under these conditions. The simulation results indicate that the TID in the barrel region is much lower than in endcaps. High-statistics Bhabha events especially in endcaps will be used for in situ calibration of crystal transparency changes due to TID. The relationship between TID and BGO crystal transparency loss exhibits two distinct regimes [18]. For TID $\leq$ 200 rad, a dose of around 10 rad would induce a loss of $<$ 1% of the BGO transparency loss based on an approximate linear interpolation. For TID above 200 rad, the BGO transparency decreases much more slowly along with the TID. Under nominal operating conditions, the BIB simulation studies show that the TID level in most ECAL regions remains well below 10 rad per hour. Therefore, the high statistics of Bhabha events collected per hour, particularly in the endcaps, are sufficient to enable in situ calibration, thereby compensating for transparency changes induced by these relatively low radiation levels.

$Z \rightarrow e^+e^-$ **calibration** $Z \rightarrow e^+e^-$ events present another source of calibration. Electron and positron candidates can be selected by requiring good tracks and good matching between the track and energy cluster in the calorimeter. Non-resonant $e^+e^-$ production or $\tau$ lepton decays can be excluded via the selection around the peak of the $Z$ invariant mass. Any differences between the invariant mass distribution of the $e^+e^-$ pair and the known $Z$ mass indicate offsets to be calibrated.

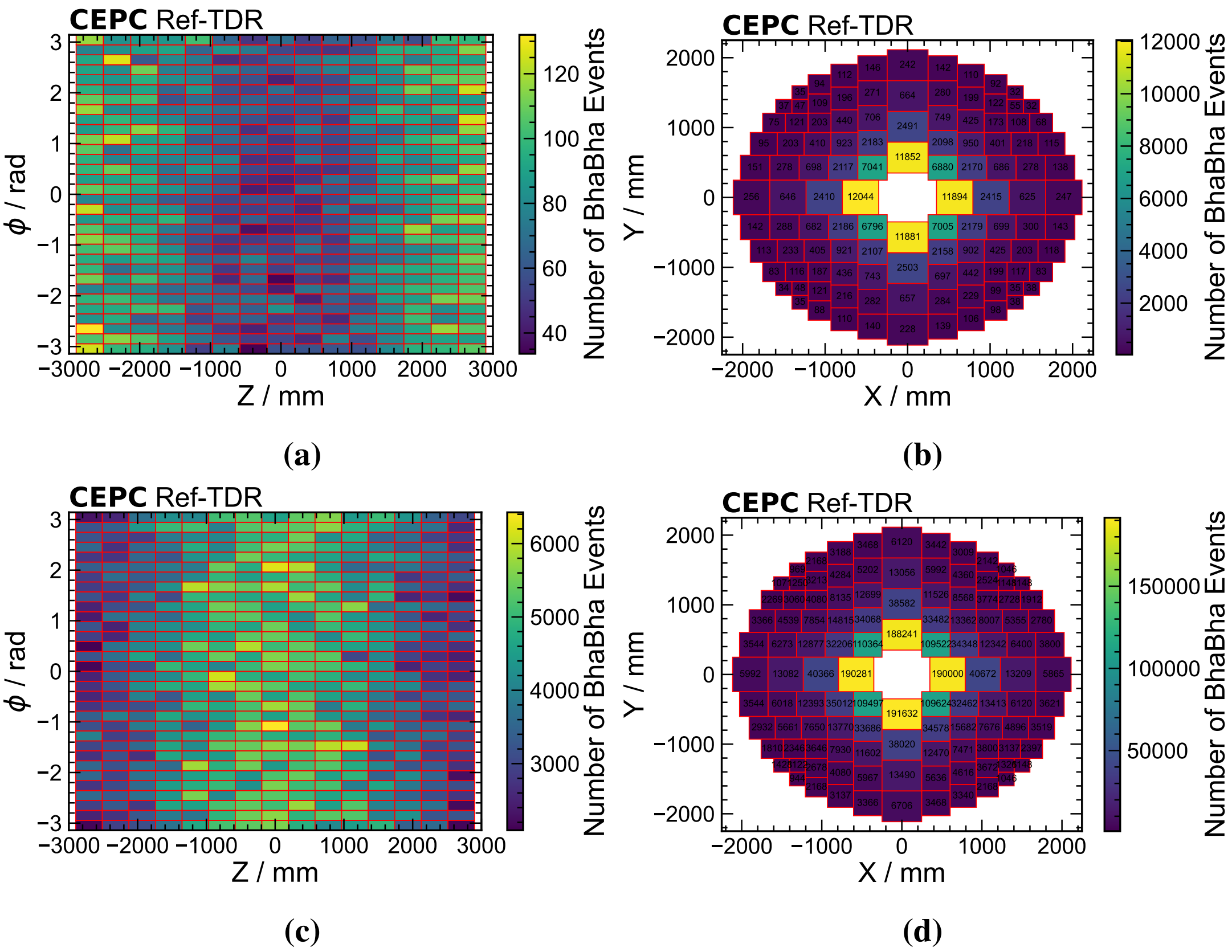

**Figure 7.18:** Bhabha events collected per hour for each module in the barrel (a) and endcaps (b) in the Higgs mode, and in the barrel (c) and endcaps (d) in the Low-Luminosity Z mode.

**$\pi^0$ calibration** $\pi^0 \to \gamma\gamma$ from hadronic events can be used for low-energy calibration to ensure uniformity of response across different ECAL channels, and to provide an absolute energy scale based on the $\pi^0$ invariant mass. A $\pi^0$ production rate of 40 kHz is expected from $e^+e^- \to q\bar{q}$ in the Low-Lumi Z mode. Isolated photons can be selected using electromagnetic shower characteristics. Clustering algorithms can be applied to reconstruct the electromagnetic shower and to remove the hits affected by dead or noisy channels. The calibration of the ECAL energy scale is achieved by comparing the measured and true invariant mass of $\pi^0$, iteratively applying correction factors to individual crystals until convergence, and determining a global correction factor for the absolute energy scale.

**MIP calibration** MIPs can be used for channel-wise energy calibration. The high granularity allows to develop and implement a tracking algorithm to select MIP hits in consecutive longitudinal layers. The absolute energy scale depends on the path length of the MIP traversing the crystal bar. MIPs are commonly produced in $e^+e^- \to \mu^+\mu^-$ and $e^+e^- \to q\bar{q}$ interactions, where 'punch-through' hadrons from jets can be selected. The expected rate is O(10 Hz) in the Higgs mode and O(10 kHz) in the Low-Luminosity Z mode.

**Other calibration schemes** Other physics processes such as $J/\psi \to e^+e^-$ and $W \to e\nu$ will be complementary to extend the calibration to specific energy regimes: $J/\psi$ can be used for low-energy corrections and $W$ decays for the ECAL response to high-energy isolated electrons with the missing energy from neutrinos. In addition, events of $Z$ bosons decaying into a pair of muons where one muon radiates a photon ($Z \to \mu\mu\gamma$) can also be used to cross-check the energy calibration of photons.

#### 7.2.6.3 Calibration precision

The precision of the calibration plays a key role in the calorimeter resolution. The constant term of the energy resolution will be degraded by non-uniformity in the calibration. By contrast, precise calibration ensures a uniform energy response across channels and helps to minimize this effect. To evaluate the calibration precision, a simulation study was performed to quantify the effect of inter-channel uncertainties on the EM resolution after calibration. Figure 7.19 shows the ECAL energy resolution for 5 different precisions ranging from 1 % to 10 %. This result indicates that the calibration precision should target for 1 %, yielding around a 0.3 % degradation on the constant term.

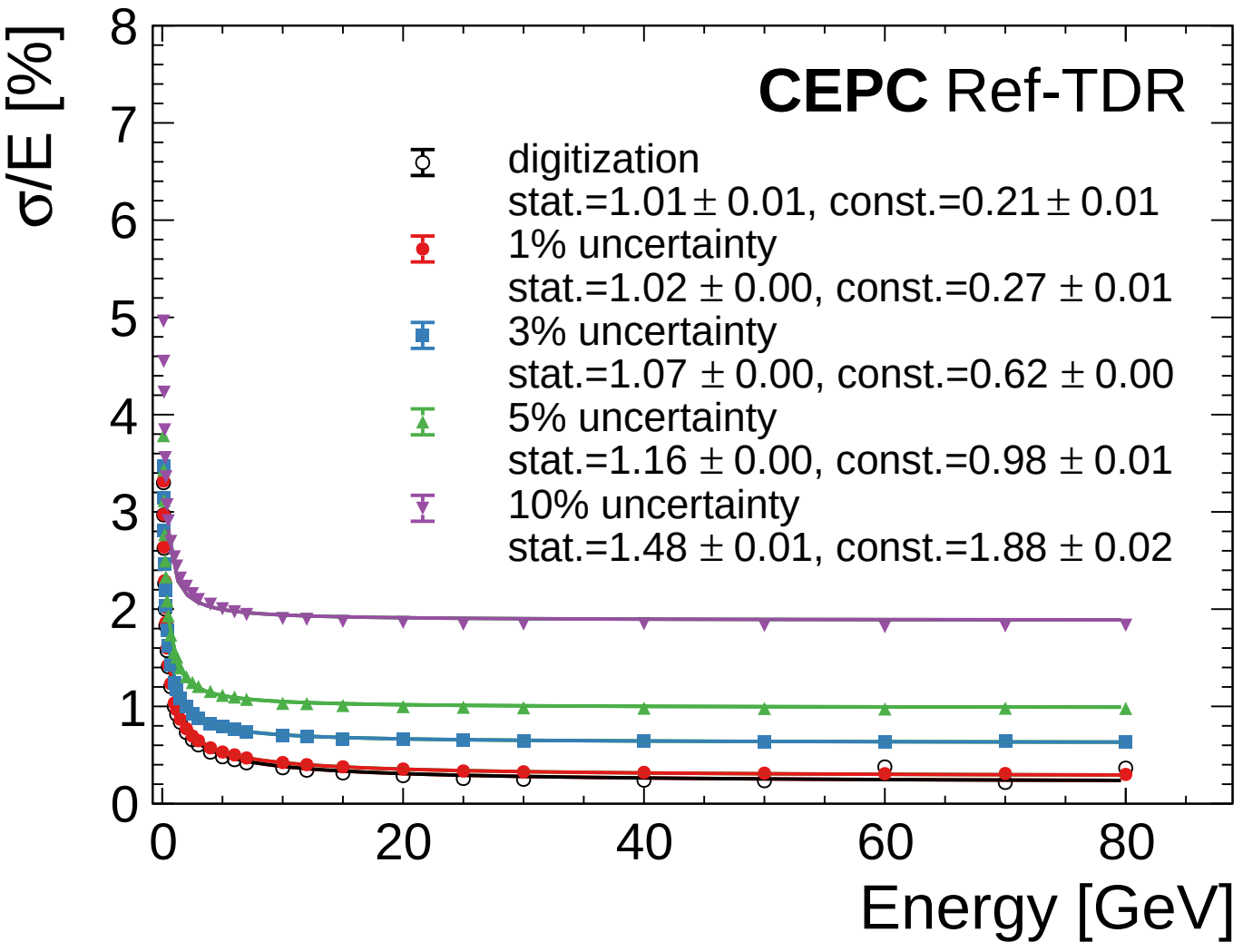

**Figure 7.19:** ECAL energy resolution as a function of calibration precision. The resolution is measured with single photons with energy in the range 0.1 GeV to 80 GeV.

Precise calibration requires a high signal-to-noise ratio in each detector channel. Additionally, sufficient calibration events are necessary to minimize statistical fluctuations, and efficient signal selection and fitting algorithms are helpful.

#### 7.2.6.4 Monitoring system

The temperature of crystals and SiPMs will be monitored to correct temperature variation offsets during data-taking. The temperature monitoring precision aims for 0.1 °C, which is required to accurately measure and control the temperature gradient of 0.3 °C per layer (Section 7.2.5.1, 7.3.5) The monitoring system is used to regulate the ECAL environmental and operational parameters, integrating the safety system, temperature, humidity, high voltages for SiPM biasing, and low voltages for electronics respectively. It is designed to ensure the thermal stability of the crystals and SiPMs. Monitoring of leakage current tracks the SiPM degradation caused by radiation damage.

## 7.3 Optimization and key performance

The ECAL design has been optimized to address key challenges such as pattern recognition, energy resolution, and temperature effects. A new particle-flow algorithm was developed to resolve ambiguities in long crystal arrangements and improve the separation of nearby particle clusters. The longitudinal depth of the ECAL was optimized for efficient EM shower containment, while the crystal granularity affects the shower separation, the total number of readout channels and crystal mass production. The module boundaries were adjusted to minimize gaps and cracks. Additionally, temperature effects on SiPM noise and crystal light yield were studied to quantify the impact on energy resolution. The digitization of Monte Carlo simulation to detector response was developed to ensure realistic descriptions of energy and timing measurements in the simulation.

### 7.3.1 Module optimization

**Longitudinal depth** Optimization of the ECAL longitudinal depth was achieved by studying longitudinal shower profiles in simulation. Photons in the energy range of 0.1 GeV to 100 GeV were used to evaluate the effect of the total ECAL depth on the energy reconstruction. Different depths in the range of 210 mm to 350 mm were studied. Figure 7.20a shows the contained fraction of energy as a function of crystal thickness. Depths greater than 24 $X_0$ were found to offer diminishing returns in terms of containment. Similarly, Figure 7.20b shows the most probable containment fraction of the shower energy reaches 99 % at 24 $X_0$, presented in terms of $H \to \gamma\gamma$ process with the maximum photon energy of around 115 GeV. The ECAL depth is thus chosen to be 24 $X_0$.

**Module structure optimization** The barrel module has a trapezoid shape chosen to avoid projectile cracks from the IP, shown in Figure 7.21 (a). The inclined angle was scanned using photons, as shown in Figure 7.21 (b) and was chosen to be 12° to balance between effective radiation length and influenced angle range. The impacts of cracks were studied by full simulation and effects to physics measurements are negligible.

The ECAL endcap design also maximizes the coverage and avoids possible projectile cracks pointing to the IP at the horizontal plane of the beamline. An inclined angle of 12° is designed

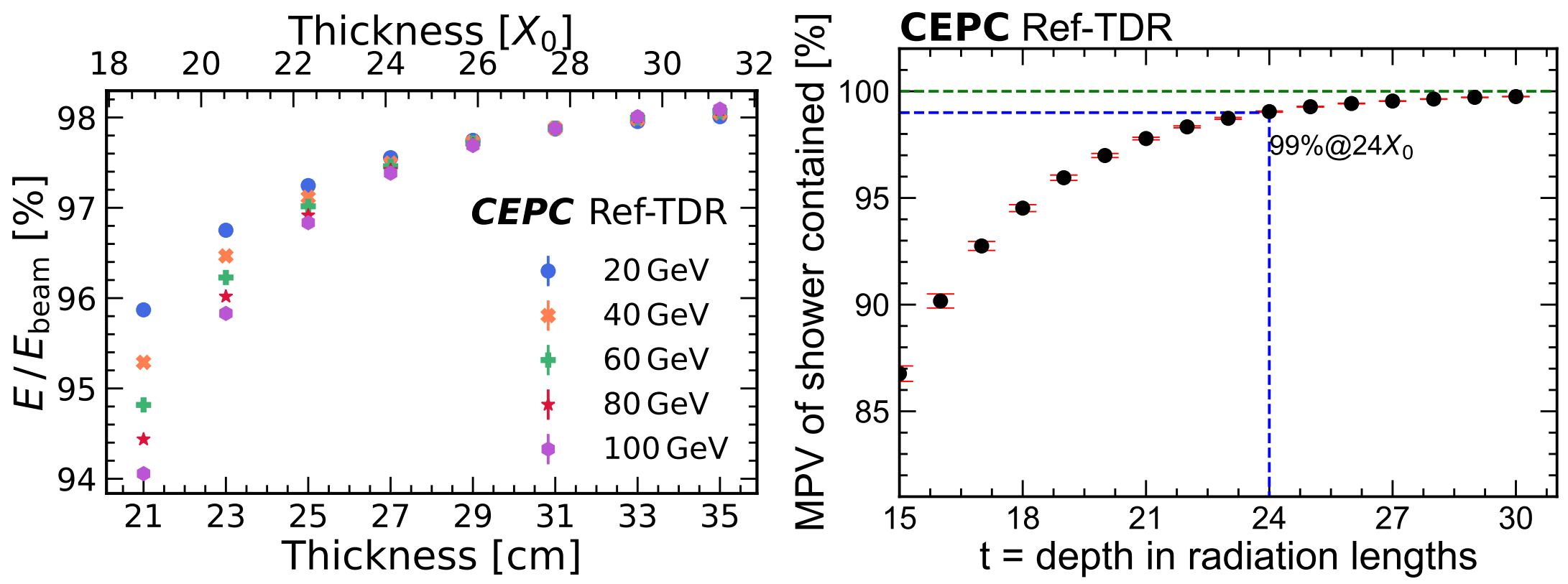

**(a)** **(b)**

**Figure 7.20:** Energy containment versus the total depth of the ECAL based on BGO crystals ($X_0$ = 1.12 cm): (a) using single photons in the energy range of 20 - 100 GeV; (b) using photons with the maximum possible energy in the channel of $H \rightarrow \gamma\gamma$

for each of 16 modules in a cross-shaped mechanical frame. Modules are arranged from the innermost module outward, filling each sector and the circular area.

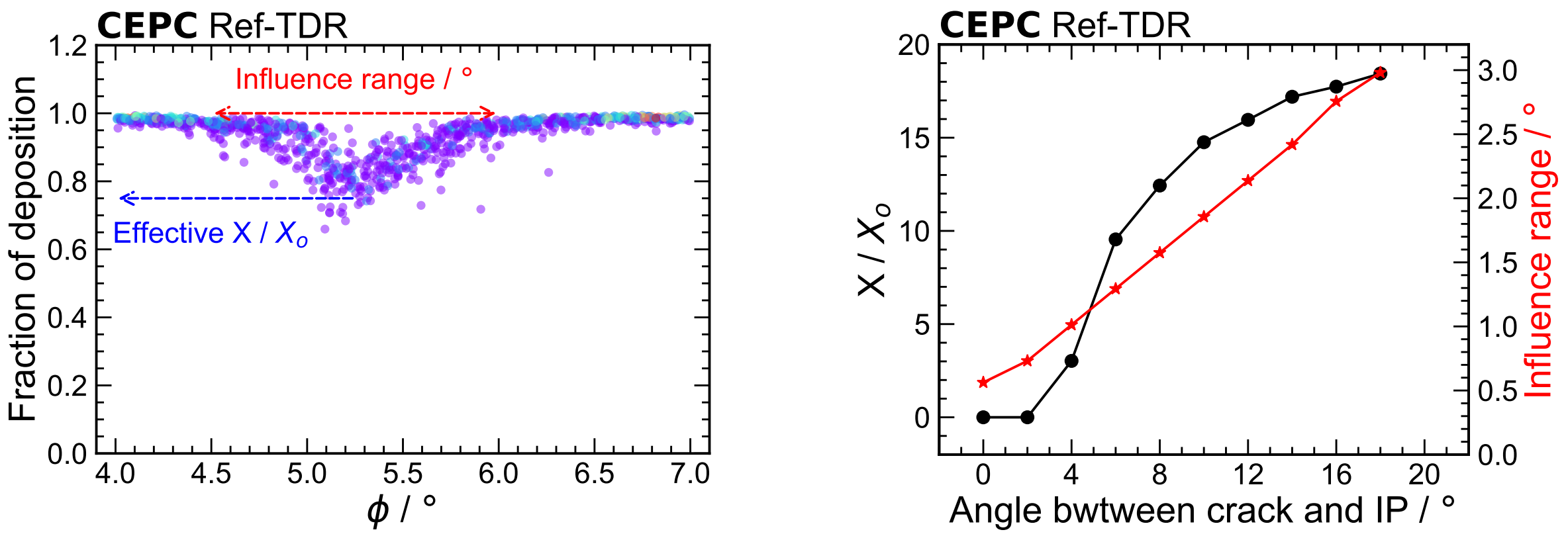

**(a)** **(b)**

**Figure 7.21:** The optimization of the inclined angle between barrel modules, which involves a trade-off between the influenced range (a) and the effective radiation length (b). An inclined angle of 12° was selected to ensure the total radiation length and to reduce unnecessary material interactions within module boundaries.

## 7.3.2 Granularity optimizations

Choosing the dimensions of the crystal involves a trade-off between PFA performance (pattern recognition, separation performance) and feasibility in terms of crystal production and the number of readout channels required. The ECAL granularity has been optimized to balance the performance, cost, and power dissipation using simulation.

Crystal bars with cross-sections of 10×10 mm$^2$, 10×20 mm$^2$, 15×15 mm$^2$, and 20×20 mm$^2$ were studied to assess the performance. The photon reconstruction of CyberPFA employs the

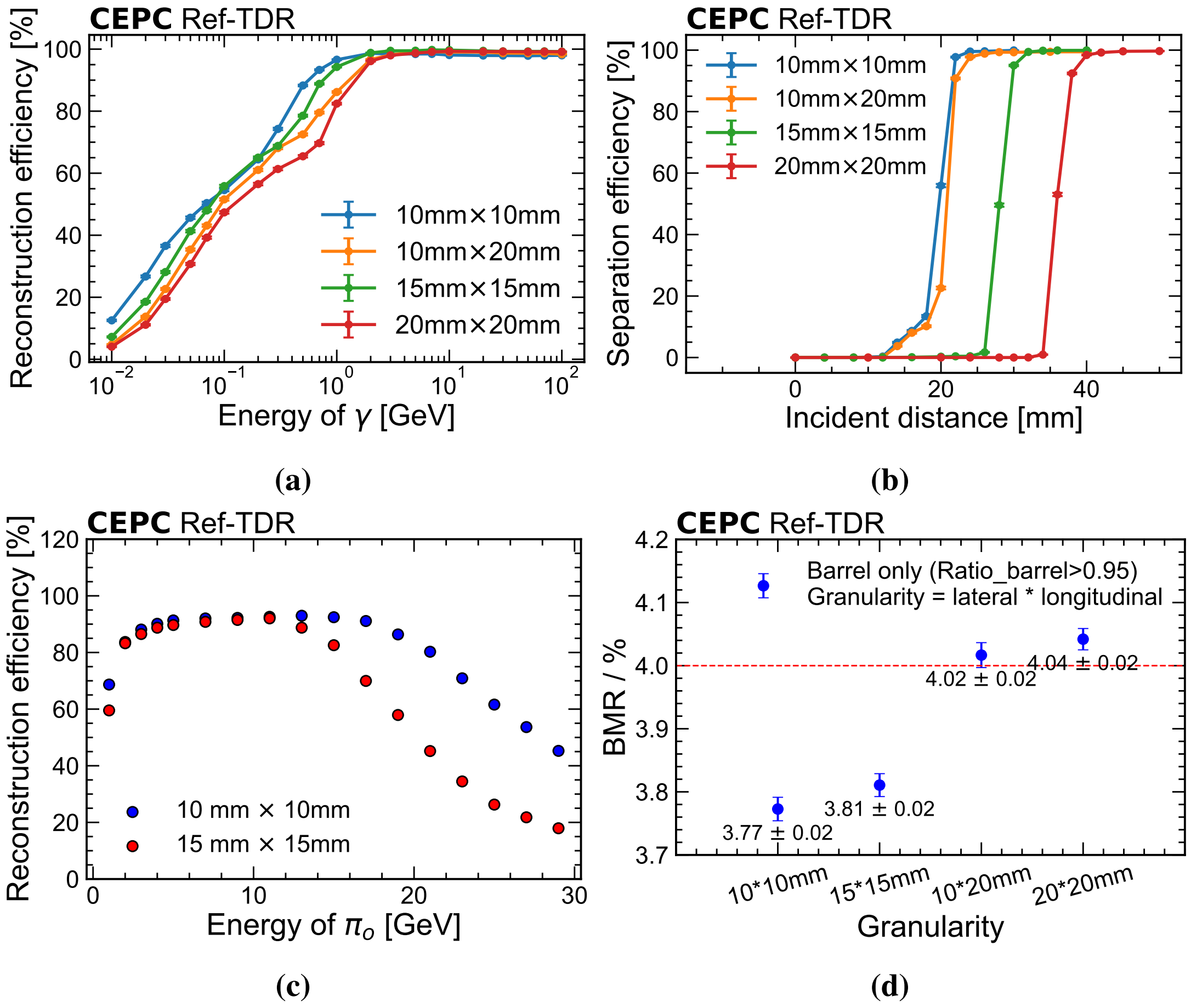

**Figure 7.22:** The crystal ECAL performance in four different crystal granularity configurations: the single photon reconstruction efficiency along with the photon energy (a), the two-photon separation efficiency along with the incident distance (b), the $\pi^0$ reconstruction efficiency versus the $\pi^0$ kinetic energy (c) and the BMR of Higgs decaying into two gluon jets (d).

Hough Transform, referring to the longitudinal shower axis. Figure 7.22a shows that a 20 mm thick crystal considerably degrades the reconstruction efficiency of the ECAL, compared to a 10 mm thickness. Figure 7.22b shows that the two-photon separation efficiency depends on the transverse granularity. Figure 7.22c shows the $\pi^0$ reconstruction efficiency degrades with transverse granularity for energies above 15 GeV. Further improvements can be expected by using electromagnetic shower profiles to discriminate a single photon from a $\pi^0$. Figure 7.22d shows that the ECAL granularity configurations of $10 \times 10$ mm$^2$ and $15 \times 15$ mm$^2$ can meet the requirement of the BMR better than 4% for jets in Higgs hadronic decays. Here results did not yet into account experimental effects that may further degrade the performance.

The major conclusion from this study is that the crystal granularity of $15 \times 15$ mm$^2$ is optimized in terms of the physics performance, crystal production, cost and design constraints. Subsequently, results presented in the following sections will focus on the ECAL design of $15 \times 15$ mm$^2$ if not specified otherwise.

### 7.3.3 Simulation: digitization

Digitization is necessary to simulate the realistic detector response of the ECAL as part of the simulation and is mandatory to model the physical aspects for measuring the energy deposition by particles and timing information, including the scintillation light emission, decay, and propagation processes, the SiPMS response and non-linearity, and readout electronics. The ECAL digitization also includes extra models to simulate temperature effects and SiPM noises. Digitization parameters used in CEPCSW have been validated based on the dedicated R&D works and summarized in Table 7.5.

**Table 7.5:** Table of ECAL digitization parameters in CEPCSW.

| Process | Parameters | Unit | Value |
|---|---|---|---|
| **Scintillator (BGO)** | | | |
| | Intrinsic light yield | ph/MeV | 8200 |
| | MIP energy | MeV/MIP | 13.4 |
| | Effective light yield | p.e/MIP | 300 |
| **SiPM (NDL EQR06 11-3030D-S)** | | | |
| | Active area | $mm^2$ | $3 \times 3$ |
| | Pixel pitch | μm | 6 |
| | Pixel counts | | 244719 |
| | Gain | | $8 \times 10^4$ |
| | PDE | % | 18 |
| | Crosstalk | % | 12 |
| | Gain fluctuation | % | 8 |
| | DCR | MHz | 2.5 |
| **Electronics** | | | |
| | ADC precision | bit | 12 |
| | Time gate | ns | 300 |
| | Gain | | Multi-gain modes |
| | Pedestal (mean) | ADC | 50 |

**Scintillation photon statistics** In the digitization process of scintillation photons, the energy deposition is converted to the number of photons detected by a SiPM for one crystal (denoted as $N_{p.e.}$). A series of parameters are modeled, including the light attenuation of crystals, the light collection efficiency, and the SiPM PDE. The number of detected photons is based on a Gaussian distribution for a given mean response in terms of $N_{p.e.}$.

**SiPM non-linearity effects** The SiPM response becomes non-linear when the number of detected photons ($N_{p.e.}$) approaches the total number of SiPM pixels. A dedicated model has been developed to describe this non-linearity, incorporating factors such as the SiPM pixel density, the inter-pixel crosstalk probability, the pixel recovery effect and PDE. For BGO and BSO crystals, the SiPM pixel recovery effect is particularly significant, as the relatively slow scintillation process allows many pixels to be fired multiple times, effectively increasing the number of available pixels.

**Readout electronics** The charge of the SiPM response is determined by $N_{p.e.}$ and the SiPM pixel gain, which is converted to ADC counts based on the SiPM-readout ASIC design. Multiple-gain stages are modeled to cover the full dynamic range while maintaining the required integral non-linearity within 1%. In each gain mode, a 12-bit resolution is applied for the charge-to-digital conversion. The digitization process incorporates Gaussian-distributed noises, modeled on actual measurements of the readout electronics, where a decrease in the gain results in an increase in the equivalent noise energy.

### 7.3.4 Linearity and energy resolution

The detector geometry is implemented with DD4hep [19] in CEPCSW based on the ECAL mechanical design. The insensitive materials of each ECAL module include the carbon-fiber structures, copper cooling plates, and readout boards. An algorithm has been developed to correct the energy for clusters near the module gaps. A mapping of the correction factors for different incident angles in $\theta$ and $\phi$ is derived and can be used for the energy reconstruction. Figure 7.23 illustrates the energy distribution of 6 GeV and 60 GeV photons before and after the energy correction. The algorithm not only adjusts the MPV of the energy distribution to be consistent with the true value but also eliminates the long tail due to module cracks.

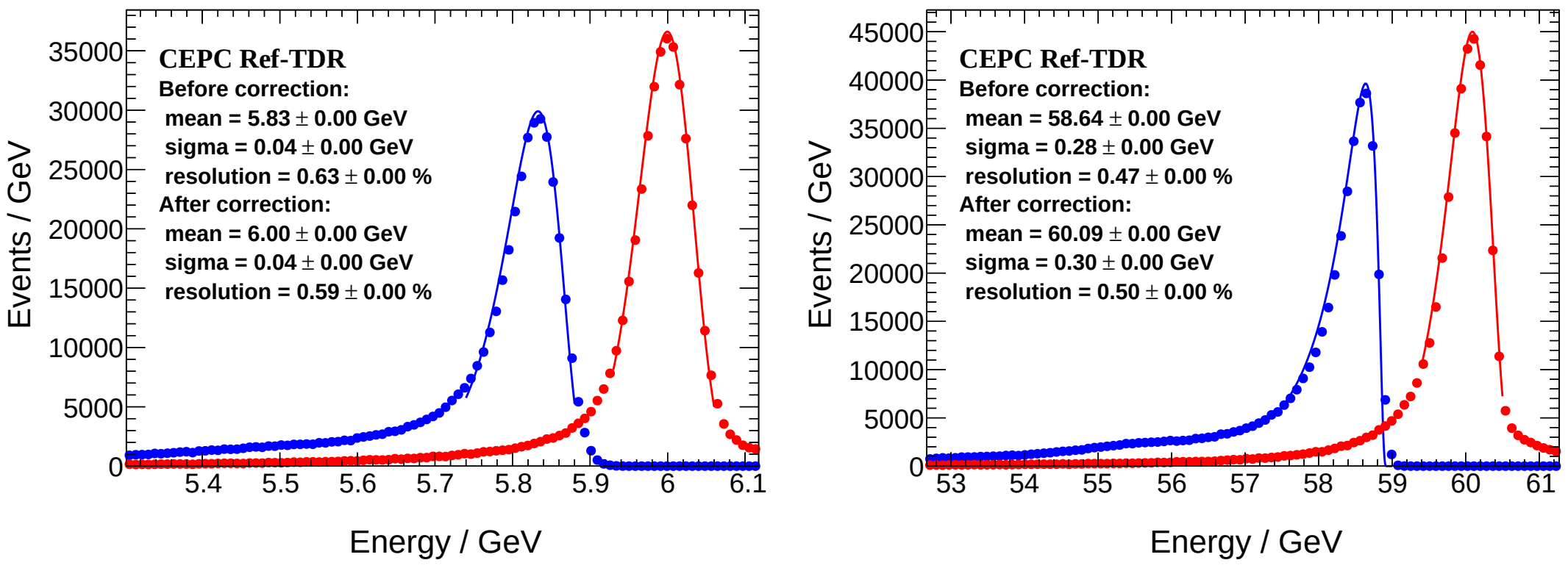

**(a)** 6 GeV **(b)** 60 GeV

**Figure 7.23:** Energy deposition of 6 GeV and 60 GeV photon with and without the energy correction. The directions of the photons are $\theta \in [40°, 140°]$ and $\varphi \in [0°, 360°]$.

The energy linearity and resolution of single-energy photons with various incident angles are studied using Geant4 simulation with digitization in Section 7.3.3. The resulting EM energy linearity is well below 1% as shown in Figure 7.24a. The EM energy resolution of the ECAL alone (without upstream materials from the tracking system) is

$$\frac{\sigma_E}{E} = 1.14\,\%/\sqrt{\text{GeV}} \oplus 0.44\,\%\,, \tag{7.2}$$

as shown in Figure 7.24b. It should be noted that the ECAL shower leakage will become gradually more prominent along with the photon energy and therefore degrade the EM resolution. Further corrections using longitudinal shower profiles in the ECAL as well as the shower tail information from HCAL can be applied to further improve the energy resolution in the high energy region.

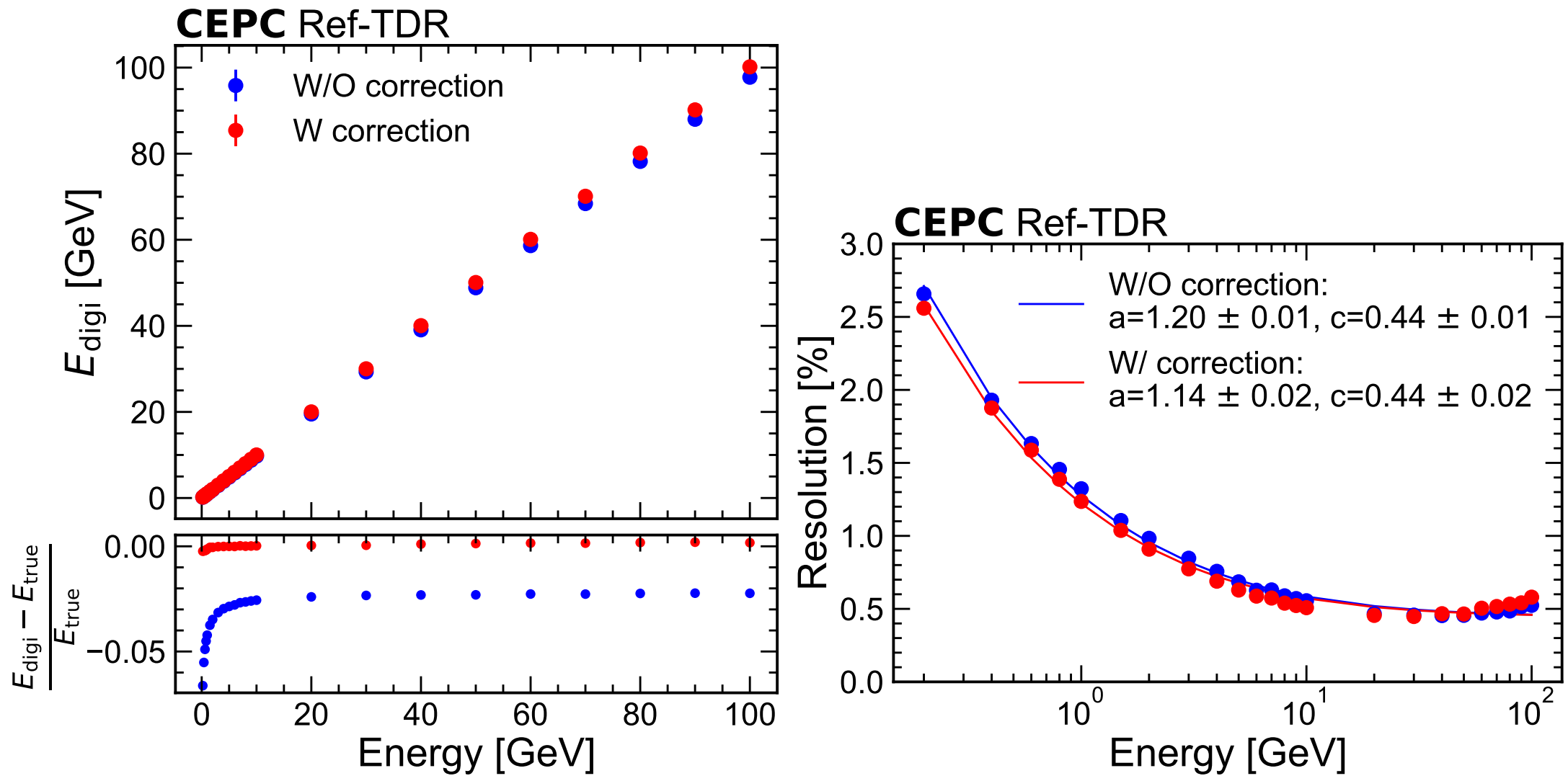

**(a)** Energy linearity **(b)** Energy resolution

**Figure 7.24:** Single-photon performance of the ECAL, after applying the digitisation model, with and without the energy correction.

The impact of module gaps on energy resolution has been investigated by simulating single photons at various energy points across the module gaps along the $\phi$ direction. The dependence of energy resolution on the azimuthal angle $\phi$ is shown in Figure 7.25a. Meanwhile, Figure 7.25b depicts the thicknesses of major components, including crystals, readout PCB, module gaps, copper plates, and carbon fiber structures along the $\phi$ direction. The results reveal that in the module gap regions, energy resolution for photons deteriorates slightly due to the reduced effective thickness of crystals and the increased proportion of insensitive materials.

### 7.3.5 Temperature effect studies

The SiPM noise is sensitive to the temperature in an exponential way, while the intrinsic light yield of BGO and BSO has a negative linear dependence on the temperature. This section discusses the temperature impacts on the ECAL performance.

The primary source of temperature variation in the detector is heat generated by the electronics. To account for temperature effects in the simulation, a temperature field is constructed to model the spatially non-uniform thermal response across ECAL regions. For simplicity, the CEPCSW framework uses an idealized linear temperature gradient, decreasing from the inner to the outer layers of each module, with a uniform temperature assumed within each layer. Detector parameters in the outermost layer follow the values listed in Table 7.5. In the simulation, three

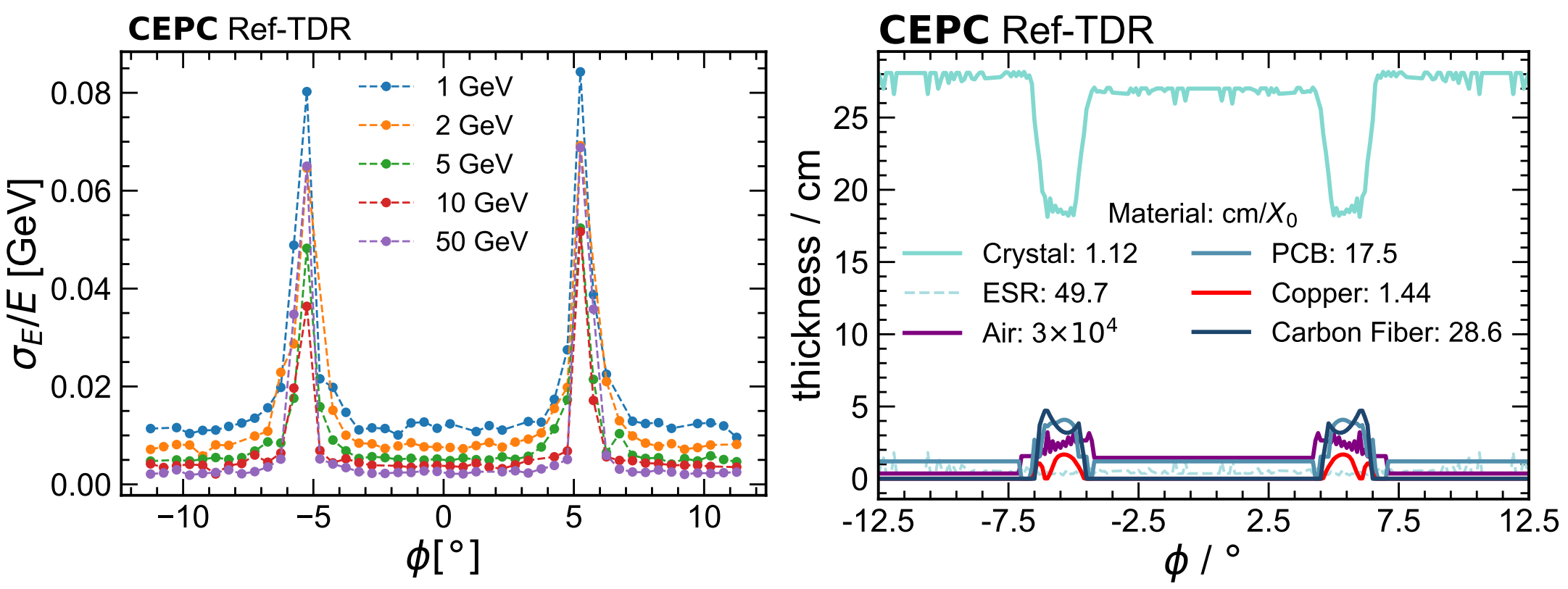

**(a)** **(b)**

**Figure 7.25:** Quantitative studies on impacts of module gaps on the energy resolution: (a) energy resolutions corresponding to incident photons at different azimuthal angles (denoted as $\phi$) and (b) thicknesses of major components at varying azimuthal angles. Copper cooling plates dominate the total radiation length, while carbon-fiber structures and air gaps contribute mostly to the total thickness. Radiation lengths of the component materials are also added.

temperature-dependent parameters are modeled: the crystal intrinsic light yield, the SiPM gain, and the SiPM DCR, with their corresponding temperature coefficients provided in Table 7.6. Real-time compensation for light yield and gain is expected to be achieved using temperature sensor readings and calibration coefficients.

**Table 7.6:** Temperature coefficients in simulation, which are all relative to 25 °C [20, 21]

| Parameter | Description | Value |
|---|---|---|
| $C_{T-LY}$ | Temperature-dependent variation of the BGO intrinsic light yield | $-1.38\,\%\,°\mathrm{C}^{-1}$ |
| $C_{T-Gain}$ | Temperature-dependent variation of the SiPM gain | $-3\,\%\,°\mathrm{C}^{-1}$ |
| $C_{T-DCR}$ | Temperature-dependent variation of the SiPM DCR | $10^{0.042\cdot\Delta T(°\mathrm{C})}\,\%$ |

An increase in temperature reduces the crystal light yield and the SiPM gain while increasing the SiPM noise. Elevated noise levels can cause non-hit channels to exceed the trigger threshold. For the entire ECAL, the impact of noise grows with number contributing to the readout. Figure 7.26 shows the dependence of the stochastic, constant, and noise terms of the energy resolution on the temperature difference within an ECAL module, including results with varying numbers of modules. The results show that the stochastic term increases with the temperature variation, while the constant term remains largely unaffected. The noise term remains stable at small temperature variations, suggesting that the energy threshold effectively suppresses noise under these conditions. However, with larger temperature variations, the noise term begins to rise, and its impact becomes more pronounced as more modules contribute noises. Overall, when the temperature variation is below 6 °C, its impact on the energy resolution is minimal.

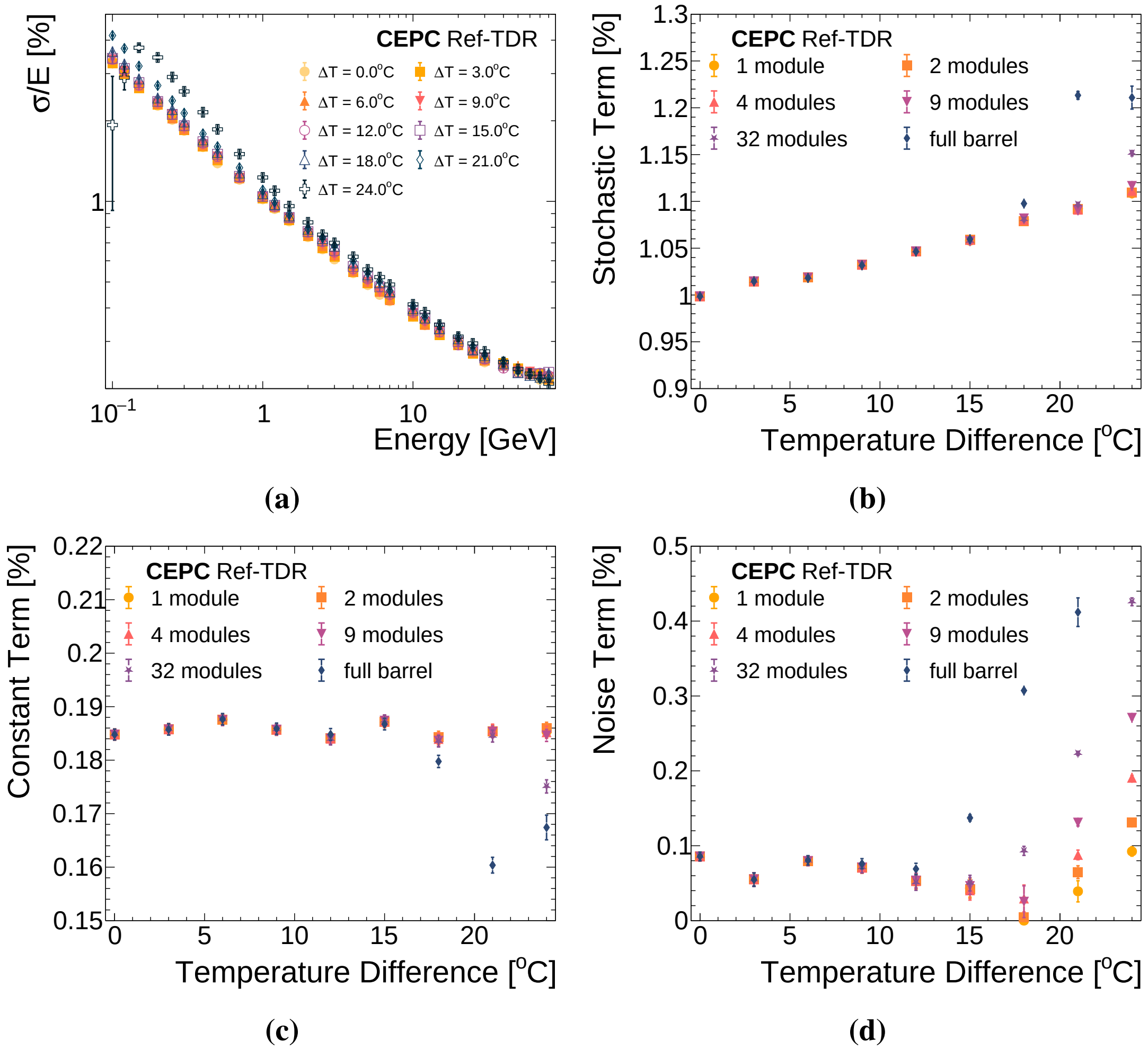

**Figure 7.26:** Energy resolution under different temperature differences and the dependence of its stochastic, constant, and noise terms on the resolution. The temperature difference refers to the difference between the first and the last layers of ECAL module, assuming a uniform temperature gradient. Points in different colors represent the inclusion of noise from varying numbers of modules. Increasing temperature difference leads to a larger stochastic term and noise term in the energy resolution.

## 7.4 ECAL prototyping

ECAL prototyping activities focus on evaluating key detector performance and addressing technical challenges. The MIP response and timing resolution of detector components with crystals and SiPMs were tested with beam particles. A crystal calorimeter prototype integrated with ASICs was successfully developed and exposed with electron beams to study the EM performance and validate the simulation and the digitization.

### 7.4.1 Crystals

BGO crystals as an essential ECAL component have been extensively tested to assess the MIP response and uniformity. BGO bars wrapped with Enhanced Specular Reflector (ESR) films, as shown in Figure 7.27 (a), are in the transverse dimensions of $1\times1$ cm$^2$ and $1.5\times1.5$ cm$^2$. The lengths of different BGO crystal bars are in the range of 4 cm to 60 cm.

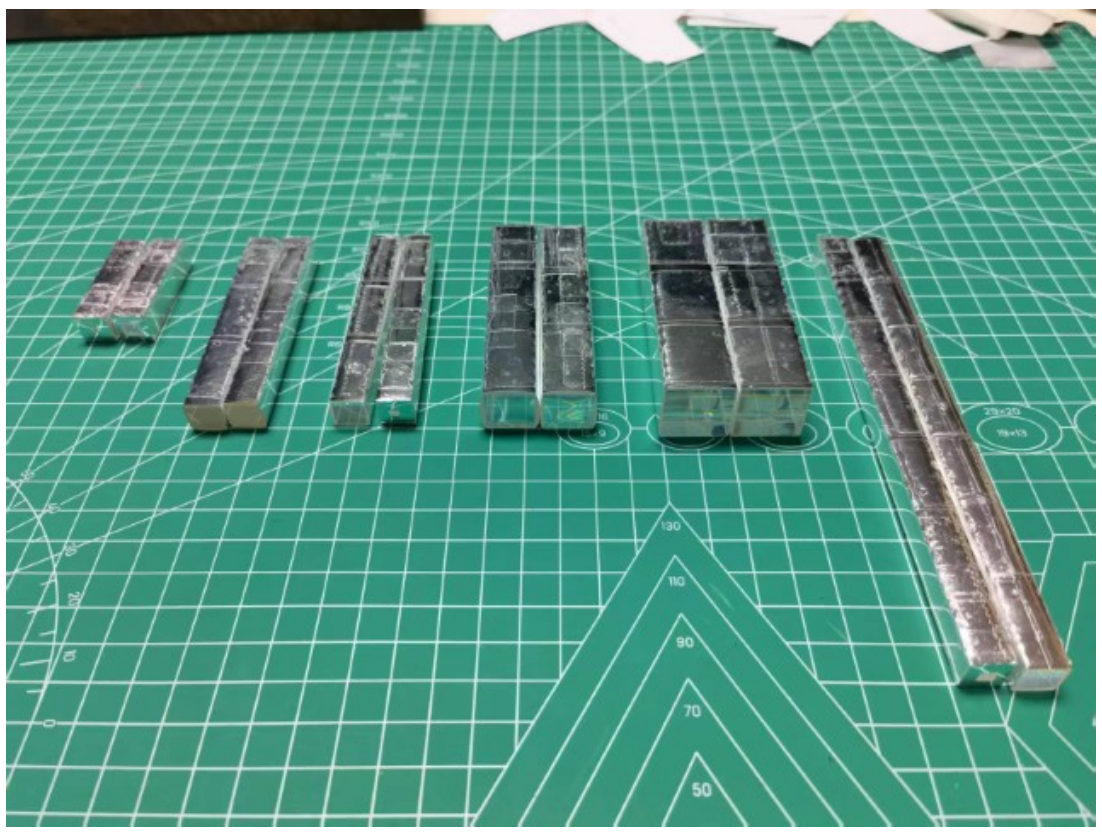

**(a)**

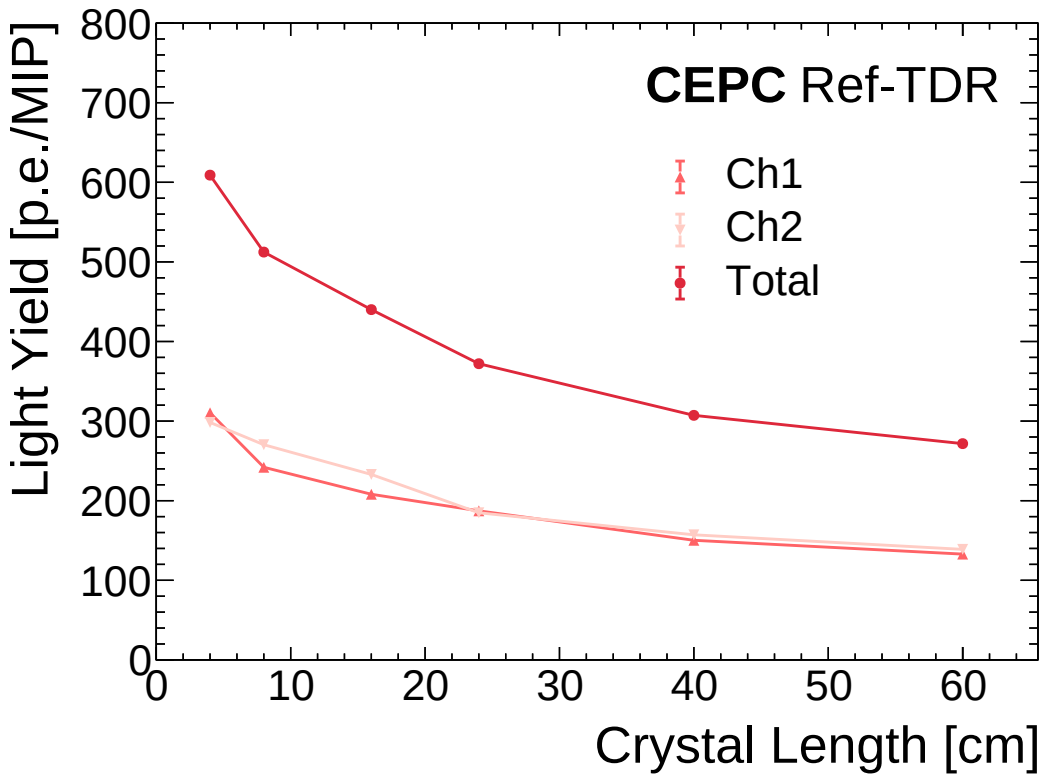

**(b)**

**Figure 7.27:** BGO crystals were individually wrapped with ESR foil (a); MIP responses of BGO crystals with the same cross section of $10 \times 10$ mm$^2$ but different lengths, including individual SiPM signals from two ends of each crystal bar and the signal sum (b), which were measured using 10 GeV$\pi^-$ beams with the incident position in the geometric center of each BGO crystal and the normal incident angle to the crystal length direction.

**MIP response** The MIP response of BGO crystals was tested with 10 GeV pions at CERN, and scaled to SiPMs of NDL EQR06-11-3030D-S with a sensitive area of $3 \times 3$ mm$^2$. The measurement results of 1-MIP signals for BGO crystals with different lengths are presented in Figure 7.27b, with the beam particles incident at the center of each crystal bar. The MIP response decreases with increasing crystal bar length due to the crystal self-absorption of a longer path of light propagation. The effective light yield was approximately 307 p.e/MIP for the $1 \times 1 \times 40$ cm$^3$ BGO with a signal integration time window of 300 ns, which leaves around 50% headroom for the MIP response requirement.

**Response uniformity** The response uniformity of a crystal-SiPM detector unit, defined as the consistency of signal amplitude across different incident positions, is critical for accurate energy measurement. Position-dependent variations may arise from crystal inhomogeneities, imperfections, and the SiPM-crystal coupling, which could be more prominent for long crystals. To quantify this effect, the MIP response uniformity was measured using the 10 GeV $\pi^-$ beam at CERN.

The results of the uniformity scan for $1 \times 1 \times 40$ cm$^3$ and $1.5 \times 1.5 \times 60$ cm$^3$ BGO crystals are shown in Figure 7.28. For a $1\times1\times40$ cm$^3$ BGO crystal bar, a good MIP response uniformity within 1.5 % was achieved, where the uniformity is defined as the ratio of the maximum response minus the minimum response over the average response. However, the MIP response uniformity with a $1.5 \times 1.5 \times 60\,\mathrm{cm}^3$ BGO crystal bar is around 10 % due to the defects in the long crystal bar. Detailed test results can be found in [22].

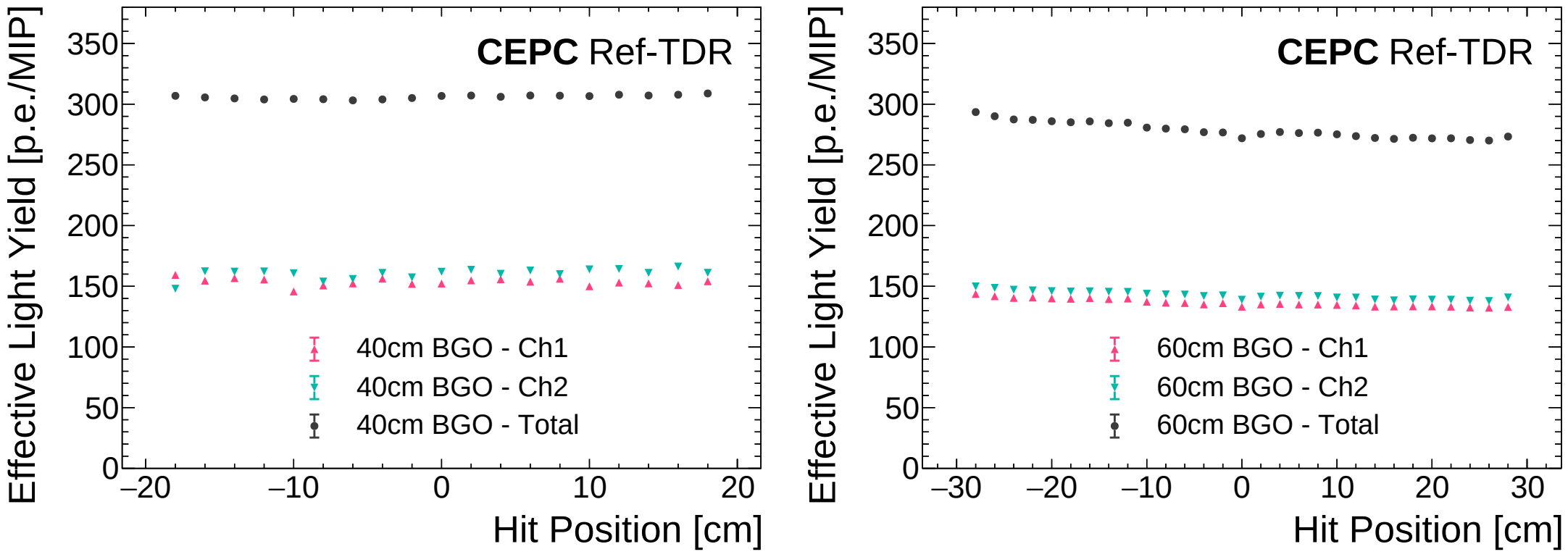

(a) (b)

**Figure 7.28:** The response of BGO-SiPM units to 10 GeV $\pi^-$ beams incident at different positions on the crystal bar. The $1 \times 1 \times 40\,\mathrm{cm}^3$ BGO unit demonstrates good response uniformity (a), while the $1.5 \times 1.5 \times 60\,\mathrm{cm}^3$ crystal shows slightly worse uniformity, with the MIP light yield gradually decreasing from one end to the other end (b). It should be noted that the responses have been properly scaled to the target SiPM sensitive area of $3 \times 3\ \mathrm{mm}^2$ using a signal integration time window of 300 ns. The uniformity of the 60 cm BGO crystal is generally worse than the 40 cm one, which is largely due to the degraded optical quality of crystal growth and this effect is more pronounced when the crystal length is more than 40 cm.

### 7.4.2 SiPM

The R&D studies on SiPMs focus on noise impacts to the EM performance and the dynamic range for scintillation signals.

#### 7.4.2.1 Noise studies

SiPM noise is significantly affected by temperature, bias voltage, inter-pixel crosstalk and after-pulsing. Measurements were carried out for 3 types of SiPM, each of which was wrapped in black tape to eliminate interference from ambient light.

The dark noise rate measured at different thresholds are shown in Figure 7.29. At the threshold of 0.5 p.e, DCR values were consistent with the product data sheets. As the threshold increases, the DCR gradually decreases. However, this reduction was less pronounced for the NDL SiPM than that of the two HPK SiPMs due to a higher level of inter-pixel crosstalk. Even at high thresholds, the NDL SiPM exhibited a relatively high noise rate. The crosstalk rate of the NDL SiPM was measured to be about 12 %. These DCR values have been applied to the ECAL simulation.

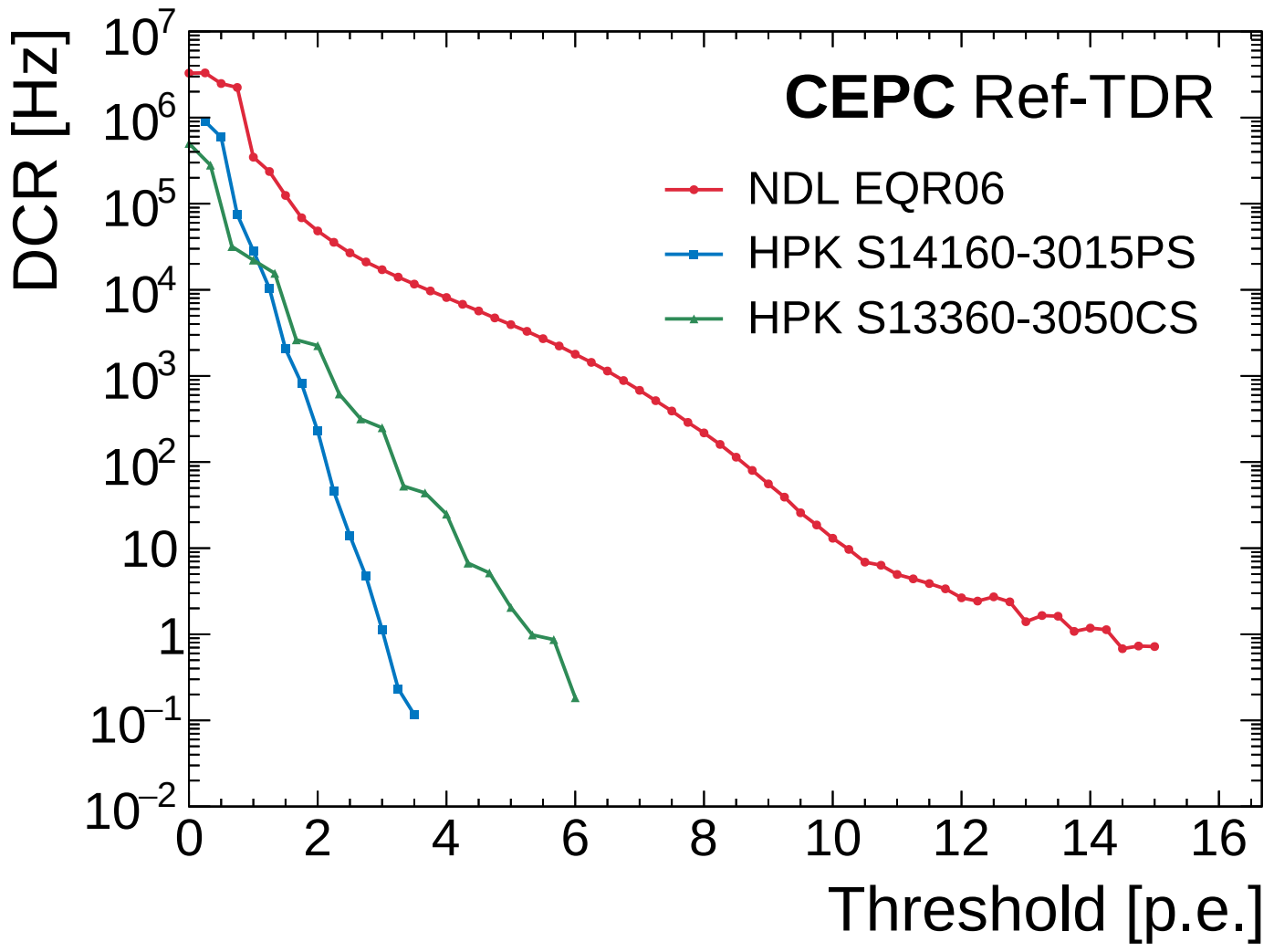

**Figure 7.29:** Dark Count Rates measured by different SiPMs, including NDL EQR06 11-3030D-S, HPK S14160-3015PS and HPK S13360-3050CS. The noise rate decreases with the increasing threshold.

Considering the 0.1 MIP energy threshold (which is equivalent to 15 p.e. per SiPM), the noise level is below 1 Hz, which has marginal effects to the ECAL performance.

#### 7.4.2.2 SiPM dynamic range

The maximum energy deposition in a crystal is expected to be at the shower core of 180 GeV electrons, or high-energy photons (up to 115 GeV) from Higgs decays. A scan of the ECAL barrel region was performed using 180 GeV electrons. Figure 7.5(a) shows the energy deposition per crystal in the barrel ECAL. The maximum energy in an individual crystal reaches around 45 GeV. Given the required MIP response of 300 p.e/MIP, this maximum energy corresponds to around $5 \times 10^5$ p.e/channel, imposing stringent requirements on the dynamic range of the SiPMs.

The SiPM dynamic range depends on the pixel density and the pixel recovery time. A comprehensive SiPM response model is developed to examine the response characteristics of SiPMs with a large number of pixels and a fast recovery time for relatively slow BGO scintillation light, described in Ref. [23].

The simulated SiPM response as a function of incident light intensity is shown in Figure 7.30. Two SiPMs with very small pixel pitches of 10 µm and 6 µm are investigated. The relevant parameters for these devices are summarized in Table 7.2. In Figure 7.30, the horizontal axis represents the effective photon count that the SiPM would detect in the absence of saturation—i.e., the product of the SiPM PDE and the incident photon count—while the vertical axis shows the relative residual of the number of detected photons with respect to the ideally linear response. As the incident light intensity increases, the SiPM initially displays a linear response but gradually deviates from linearity as saturation effects become significant.

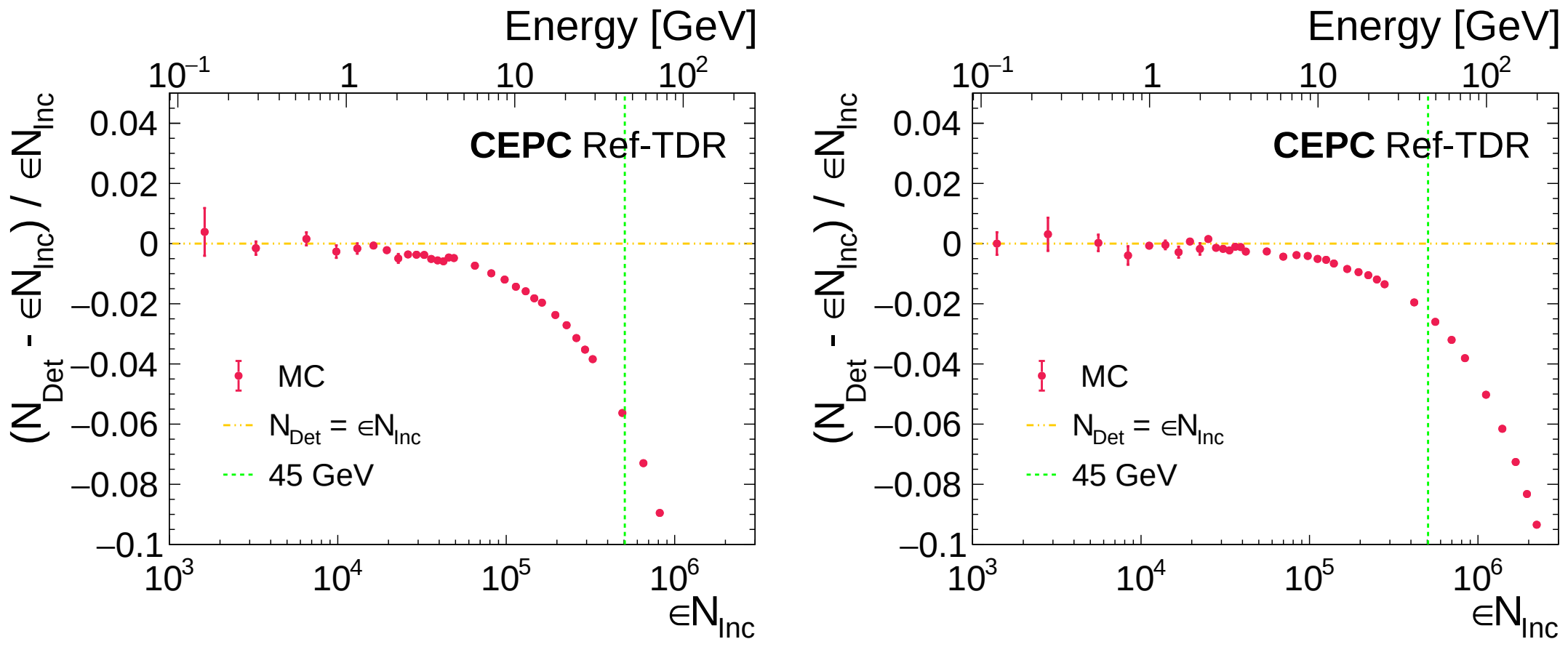

**(a)** **(b)**

**Figure 7.30:** Simulated SiPM responses to BGO ($1 \times 1 \times 40\,\mathrm{cm}^3$) scintillation light: (a) NDL EQR10 11-3030D-S with a 10 µm pixel pitch, and (b) NDL EQR06 11-3030D-S with a 6 µm pixel pitch. In the figures, the red points represent the simulation results, the blue dashed line corresponds to the expected non-linearity SiPM response using a function described in [24].

According to the simulation results in Figure 7.5(a), for a maximum energy deposition of 45 GeV in a single crystal bar, corresponding to roughly $5 \times 10^5$ p.e./channel, the SiPM with a 6 µm pixel pitch exhibits a non-linearity of only −2.36 %, indicating negligible saturation effects. And the SiPM with 10 µm pixel pitch from NDL shows non-linearity values of approximately -5.8 %, demonstrating more pronounced saturation.

On the other hand, after characterizing the non-linearity of the SiPM, a correction can be applied. The SiPM non-linear response is influenced by factors such as the bias voltage, the inter-pixel crosstalk probability and the pixel recovery time. Proper quality control of the SiPMs in mass production is essential to ensure the non-linearity calibration accuracy.

A dedicated testbeam setup was developed to measure the SiPM non-linearity curve. A 40 cm long BGO crystal bar, read out by two SiPMs, was exposed to high-energy electron beams up to 300 GeV at CERN. As shown in Figure 7.31, the effective number of SiPM pixels are greatly increased due to pixel recovery effects, which substantially extends the SiPM dynamic range. As discussed in the SiPM calibration strategy in Section 7.2.6.1, this curve can be applied to correct the SiPM non-linearity in the whole dynamic range. The target maximum response of $5 \times 10^5$ p.e. remains far from the SiPM saturation region and thus maintains headroom before significant saturation, which ensures the correction with a high precision. The SiPM simulation can generally well reproduce the non-linearity tendency of data, while remaining discrepancies between the data and simulation remain to be further studied.

### 7.4.3 Timing performance

The timing resolution of a crystal-SiPM unit is defined as the standard deviation of differences between time stamps of SiPM signals from two ends of a crystal for a fixed impact-point position. It is expected to further enhance shower clustering performance and to suppress contributions from random noises.

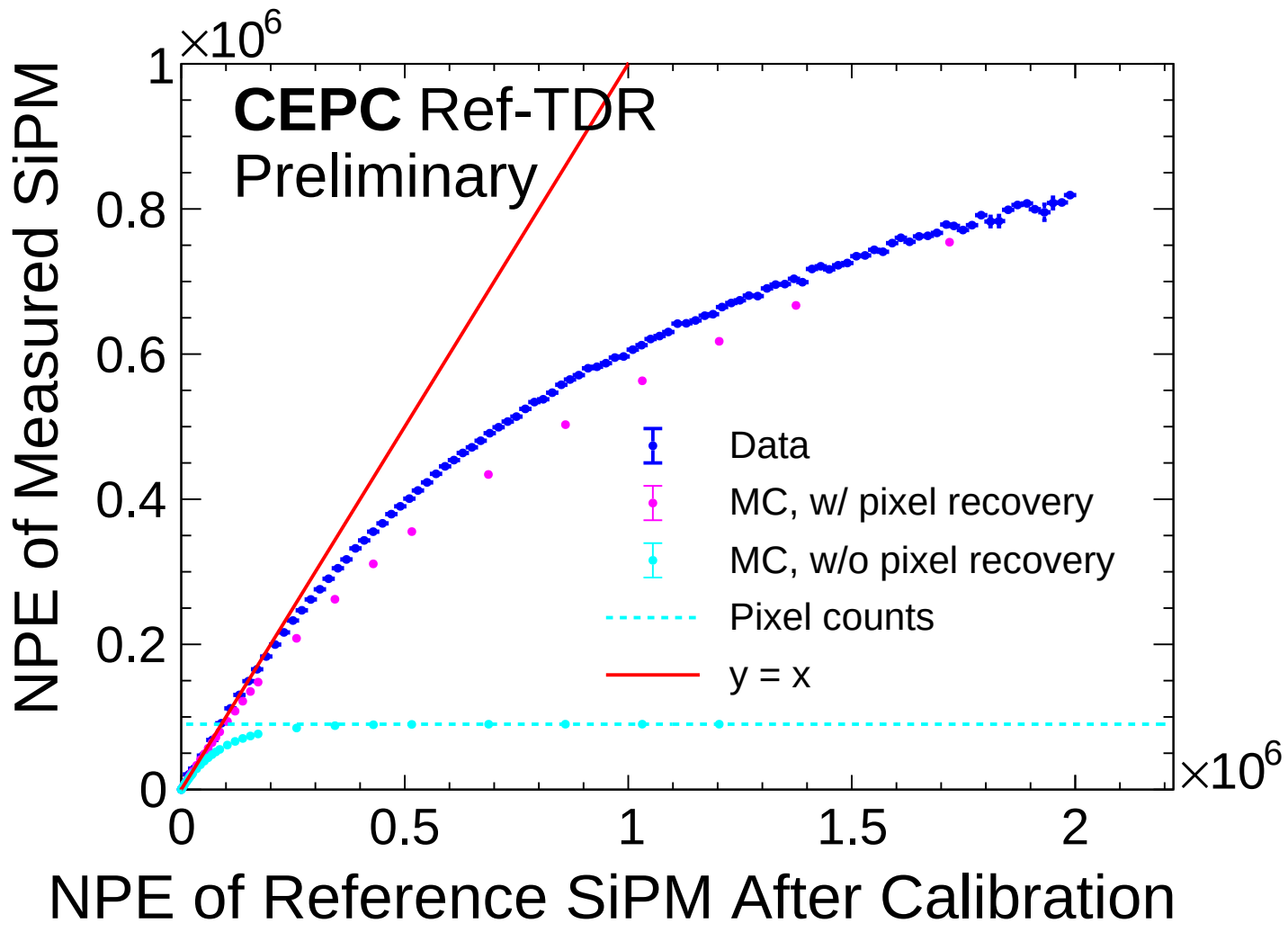

**Figure 7.31:** The SiPM non-linearity measured with high-energy electron beams using a BGO crystal and two SiPMs (HPK S14160-3010PS). Results of testbeam data and MC simulation with and without pixel recovery effects are superimposed.

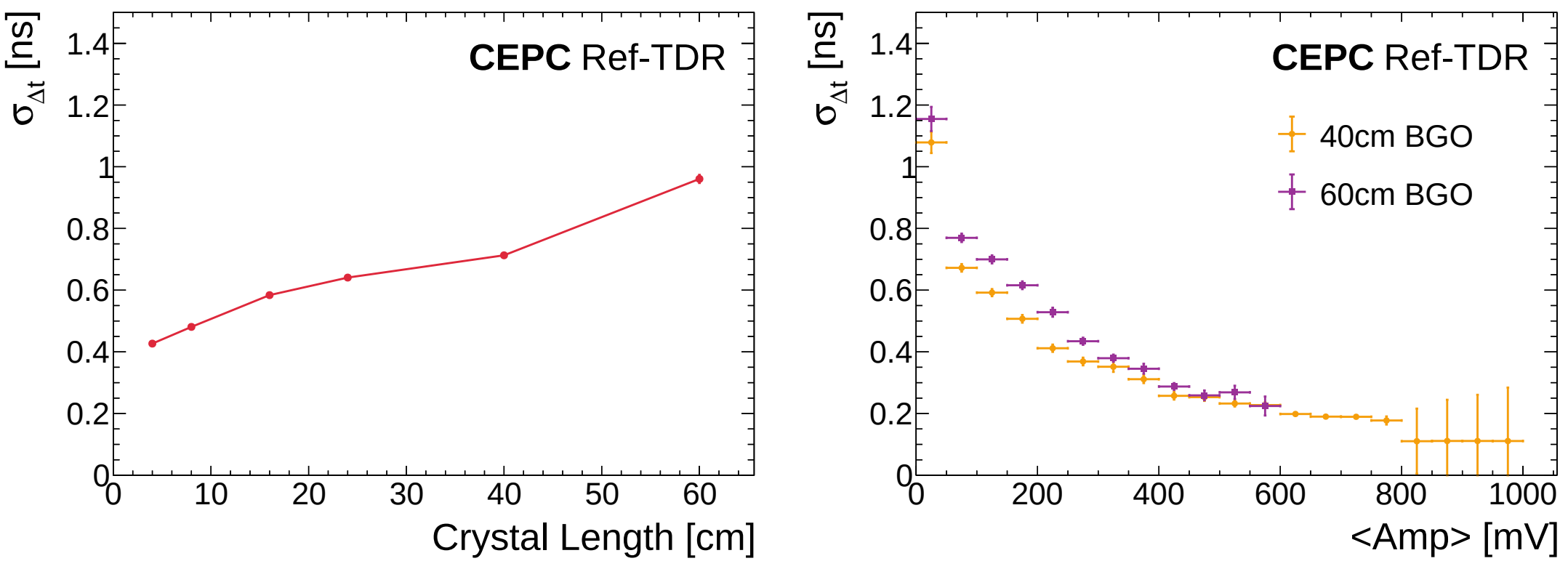

**(a)** **(b)**

**Figure 7.32:** Time resolution of BGO crystal-SiPM units at 1 MIP energy level, which degrades with increasing crystal length (a); (b) Time resolution of the $1\times1\times40\,\text{cm}^3$ and $1.5\times1.5\times60\,\text{cm}^3$ BGO crystal units under electromagnetic showers induced by 5 GeV electrons, as a function of the average signal amplitude. Data from varying thickness of crystals as the pre-shower are combined.

The timing resolution of BGO crystals with the 1-MIP response was measured using 10 GeV $\pi^-$ beams at CERN, as shown in Figure 7.32a. Different lengths of BGO crystals were tested with incident beam particles perpendicular to the center of each crystal bar. A degradation in timing resolution was observed with increasing crystal length. For the $1 \times 1 \times 40\,\text{cm}^3$ BGO bar, the MIP-level timing resolution was around 700 ps for two ends, corresponding to around 500 ps per end. The timing resolution of the $1 \times 1 \times 40$ cm$^3$ BGO crystal unit shows a uniform distribution along with different incident positions. Similar timing performance has been achieved by using an alternative definition of the timing resolution by summing the time stamps from two SiPMs in order to completely eliminate any dependence on the incident particle positions. The timing resolution includes contributions from components of the test stand, including the BGO crystal, the SiPM, the amplifier, and the oscilloscope, where the electronics contribute less than 120 ps.

Timing resolutions of BGO crystal bars to EM showers with lengths of 40 cm and 60 cm were further investigated using an electron beam of 5 GeV at DESY. Extra crystals of varying thicknesses were placed upstream of the crystals under study as a passive pre-shower. The dependence of timing resolution on signal amplitude was studied, as shown in Figure 7.32 (b), based on data of different signal amplitudes from EM showers. For the 40 cm BGO crystal, the time resolution achieved better than 200 ps when the average signal is larger than 12 MIPs. The 60 cm BGO crystal shows slightly inferior timing performance compared to that of the 40 cm crystal but still reaches a timing resolution of 200 ps [22].

### 7.4.4 Prototype and beam tests

A prototype of the ECAL with SiPMs and readout ASICs has been successfully developed and tested with high-energy beam particles. Main aims are in the following: (1) to address critical issues at the system level, including the integration of SiPMs, ASICs and readout boards with the mechanics and cooling system; (2) to evaluate the key EM performance with electron beams and (3) to validate the digitization model developed and implemented in CEPCSW.

#### 7.4.4.1 Development of an ECAL prototype

Considering typical dimensions of EM shower profiles, the first ECAL prototype is designed with a transverse size of $12 \times 12\,\mathrm{cm}^2$ and a total thickness of 24 cm for containment of EM showers. The performance of the prototype has been simulated in Geant4 using electrons in the range of 1 GeV to 60 GeV. The simulation shows that the EM resolution of the prototype can achieve a result of better than $3\%/\sqrt{\mathrm{E(GeV)}}$.

The prototype consists of 72 BGO crystal bars (each in $12 \times 2 \times 2\,\mathrm{cm}^3$) and has a total of 12 longitudinal layers, with a total thickness of 24 cm ($21.4\,\mathrm{X}_0$). Each longitudinal layer is equipped with 6 crystal bars. Each crystal bar, wrapped with ESR and aluminum foils, is read out with one SiPM on each side. The electronics readout is based on the CITIROC-1A chips with a total of 144 readout channels. A digitization model has been adapted from the one implemented in CEPCSW by taking into account different dimensions of crystal bars, properties of SiPMs and ASICs.

#### 7.4.4.2 Beam-test results of the prototype

The prototype was successfully tested at CERN using high-energy beams of muons and electrons. All components of beam instrumentation were removed to ensure the minimum possible upstream material and the lowest possible beam momentum spread. Beam test results are shown in Figure 7.33. The energy linearity is within 1% for the beam momentum above 1 GeV. The EM energy resolution is $1.96\,\%/\sqrt{\mathrm{E(GeV)}} \oplus 0.54\,\%$ using the beamline collimation setting of ±1.5 mm, which is similar to that with a wider collimation of ±3.0 mm. The prototype demonstrates that the EM energy resolution of better than $2\%/\sqrt{\mathrm{E(GeV)}} \oplus 1\%$ is achieved.

Based on a dedicated beamline simulation model developed for the CERN PS-T9, preliminary results indicate that a beam momentum spread of around 1% at 1 GeV. After subtracting this contribute to the EM energy resolution, the stochastic term can be further reduced to around 1.7%.

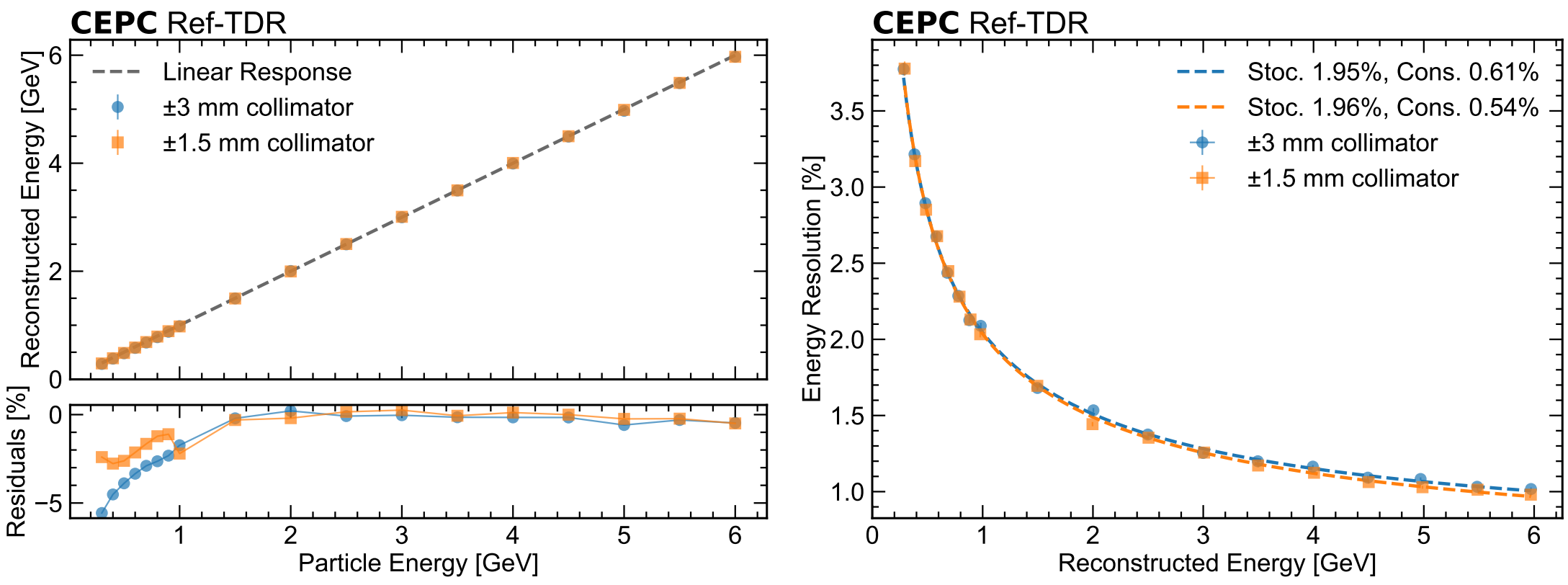

**(a)** **(b)**

**Figure 7.33:** EM performance of the crystal ECAL prototype with electron data in the momentum range of 0.3 GeV to 6 GeV: the response linearity (a) and the energy resolution (b). Various collimator settings were tested to study their impact on the beam momentum spread and intensity.

### 7.4.5 Beam-induced backgrounds

BIB is simulated to assess its influence on ECAL performance and its mitigation parameters. The nominal design (operated at the power of 50 MW) from the CEPC accelerator TDR [25] is used with a bunch spacing of 277 ns for the Higgs mode and 69 ns for the Low-Luminosity Z mode, respectively. MC samples including the pair production and single beam backgrounds are used for ECAL studies, where no safety factor are applied. Extra hit rates and distributions of TID and NIEL in the barrel and endcap regions are estimated. The influence on the ECAL energy resolution in different time windows of signal integration is also studied.

#### 7.4.5.1 Estimates of hit rates and data throughput

The time stamps of BIB hits in ECAL include bunch crossings, and the flight time of the background particles up to each crystal. The time distribution of BIB hits in the ECAL within a certain range of bunch crossings is shown in Figure 7.34.

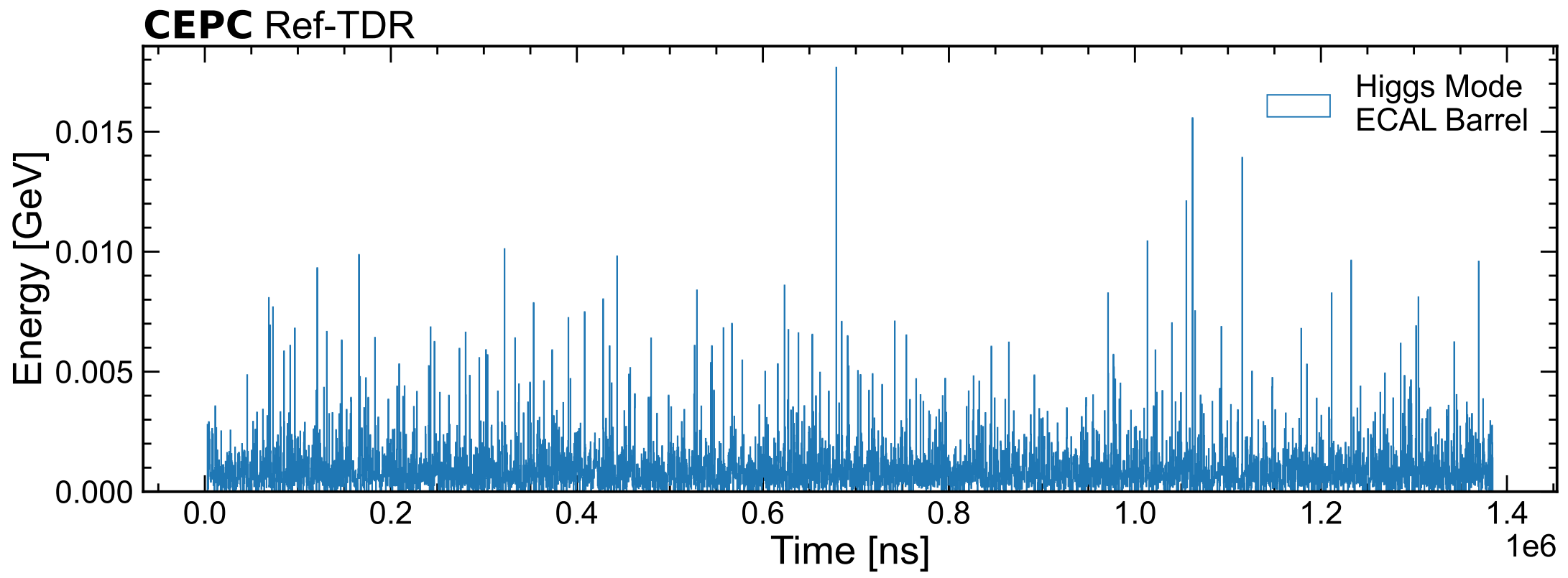

**Figure 7.34:** Time stamp distribution of signals from beam-induced backgrounds.

Based on the hit time distribution of each single crystal, a time window of signal integration in the ECAL is applied to determine the detected photons per channel. If this quantity exceeds the trigger threshold of 0.1 MIP, this hit is regarded as an effective one. The counting rate is defined as the number of hits above the trigger threshold in a given crystal within the signal integration window.

BIB hit rates in the barrel and endcap regions are shown in Figure 7.35. The endcap exhibits a higher BIB hit rate than that of the barrel for obvious reasons. BIB hit rates in the Z mode are higher than that of the Higgs mode due to the higher luminosity and a shorter bunch spacing. The present BIB hit rate measurements indicate that the ECAL can maintain sufficient data throughput with a safety factor of two.

On average, 1300 electrons per collision, which are mostly low-energy in the forward direction, are produced in pair production. The BIB hit rate is generally low for the barrel ECAL except the first few layers in modules near the endcaps, as shown in Figure 7.36. In the endcap regions, crystals with a high BIB rate are mostly near the beam pipe as shown in Figure 7.36. Similar to the hit patterns in the barrel region, the first endcap layers have considerably higher rates.

In the endcap regions, crystals with a high BIB rate are mostly near the beam pipe as shown in Figure 7.36. Similar to the hit patterns in the barrel region, the first endcap layers have considerably higher count rates.

#### 7.4.5.2 Simulation results of TID

The TID received by the ECAL will influence the performance and longevity of the detector components. BIB is the primary contribution to TID, interacting with sensitive elements, leading

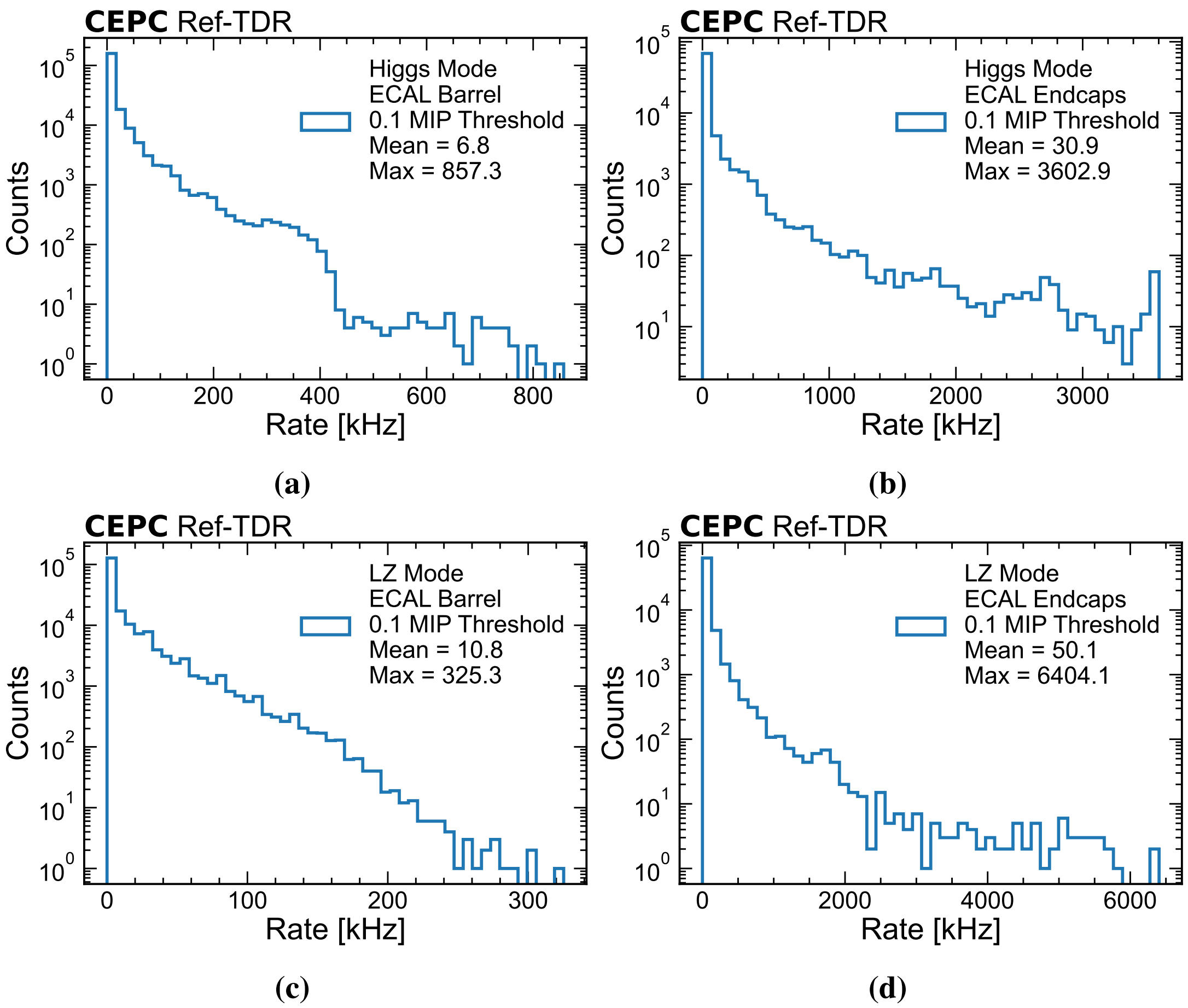

**Figure 7.35:** BIB hit rate distributions of crystals in the barrel (a) and endcaps (b) in the Higgs mode, and in the barrel (c) and endcaps (d) in the Low-Luminosity Z mode.

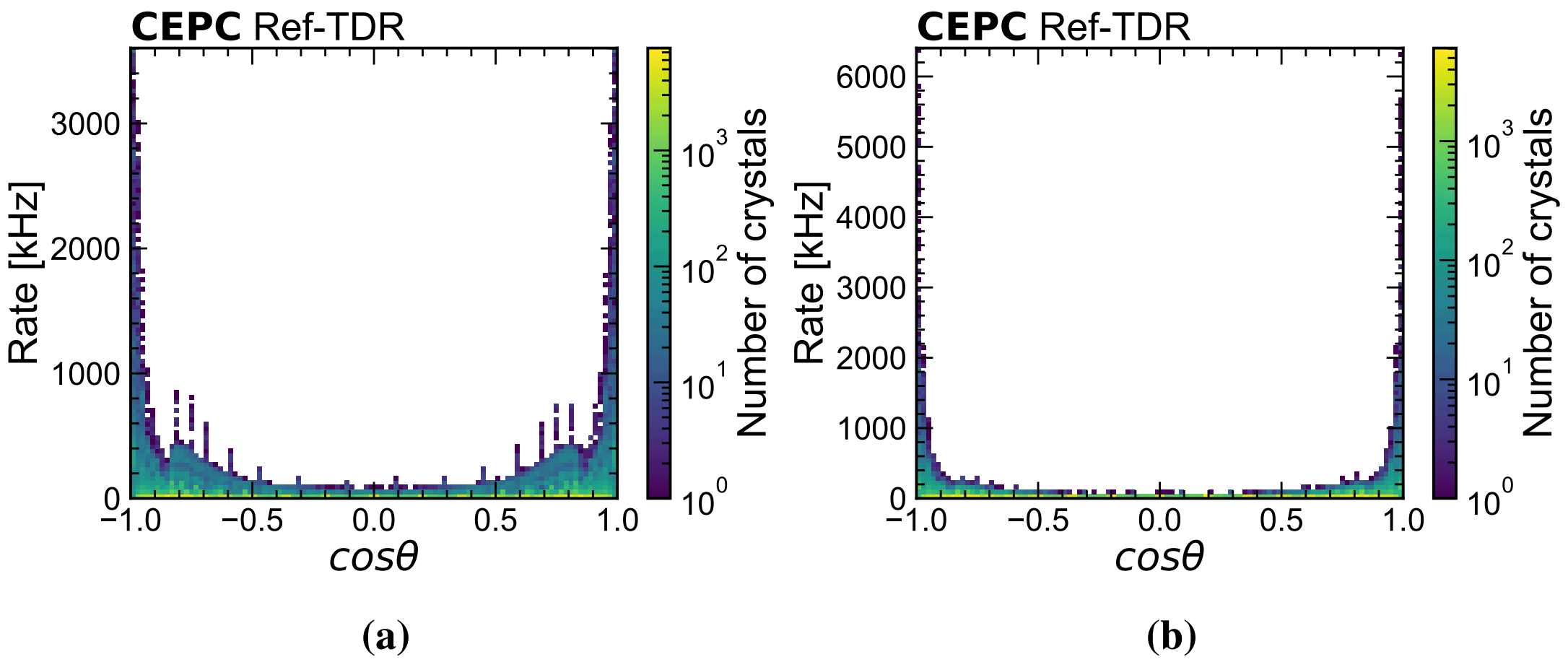

**Figure 7.36:** BIB hit rate distributions in the Higgs mode (a) and the Low-Luminosity Z mode (b) along with $\cos\theta$, where $\theta$ denotes the polar angle. The color scale represents the number of crystals associated with a specific BIB rate at a given polar angle.

to radiation damages to the sensors, electronics, and crystals.

Figure 7.37 shows that for a crystal of volume $1.5 \times 1.5 \times 40\,\mathrm{cm}^3$ and mass of 641.7 g, the maximum annual dose would be 323 Gy in Higgs mode if the annual operation time is 7000 hours. Gamma-induced radiation damage studies [18] have shown that BGO can withstand a TID level up to $10^5$ Gy with a relative light output drop less than 50%. This indicates that BGO crystals remain safe and reliable over the full operational lifespan, even when applying a safety factor of two to the beam-induced background estimates.

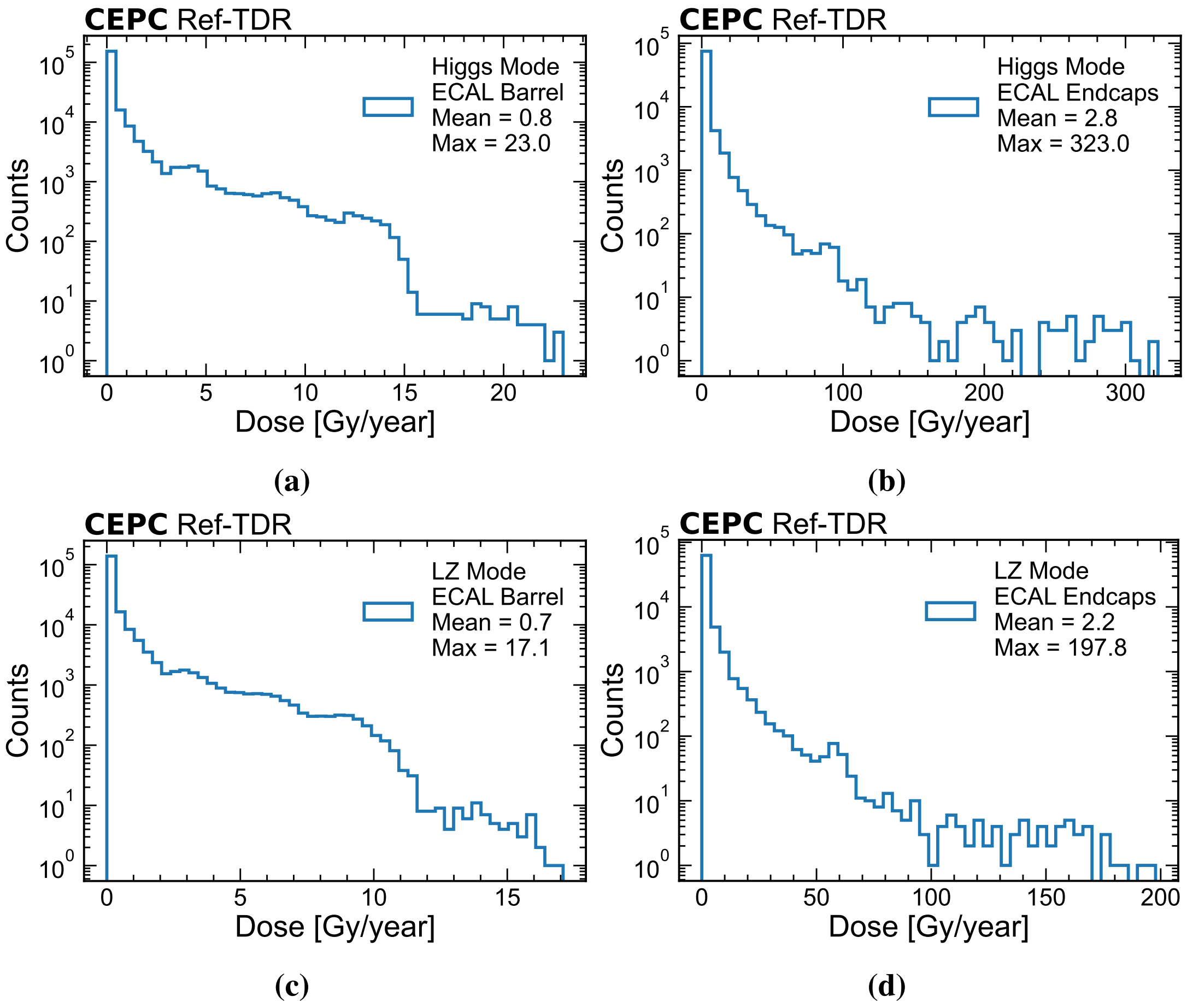

**Figure 7.37:** TID distributions of crystals in the ECAL: in the barrel (a) and endcaps (b) in the Higgs mode, and in the barrel (c) and endcaps (d) in the Low-Luminosity Z mode.

#### 7.4.5.3 Occupancy

Occupancy is defined as the fraction of crystals with energy depositions over the energy threshold of 0.1 MIP during the bunch spacing of 277 ns in the Higgs mode or 69 ns in the Low-Luminosity Z mode. The occupancy will influence the trigger system, the signal-to-noise ratio and the energy measurement precision. Figure 7.38 shows the distribution of ECAL occupancy in Higgs and Low-Luminosity Z modes. It can be seen that the occupancy of the endcap is larger than that of the barrel because the dominant pair production process is mainly in the forward direction. The occupancy in the Higgs mode is higher than that of the low luminosity Z mode because there are more energetic particles travelling deeper into the ECAL.

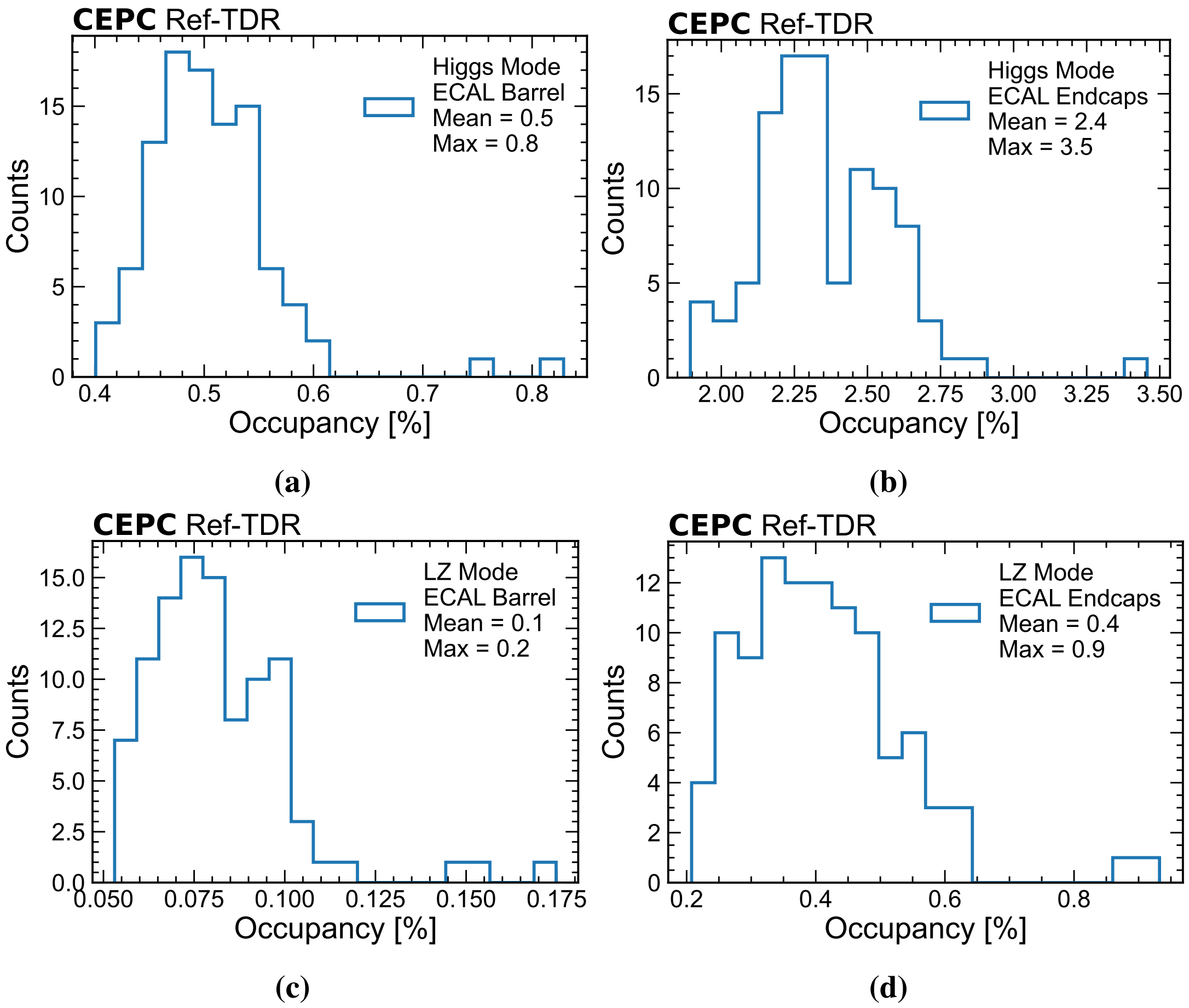

**Figure 7.38:** Occupancy distributions of BIB hits in the ECAL: in the barrel (a) and endcaps (b) in the Higgs mode, and in the barrel (c) and endcaps (d) in the Low-Luminosity Z mode.

Figure 7.39 shows that the occupancy tends to decrease as the depth of the detector increases. Removing noise from the first layer during the reconstruction is a potential solution to mitigate beam-induced backgrounds. The barrel occupancy asymptotically approaches zero in deeper longitudinal layers, whereas the endcap displays distinct behavior: the first layer maintains high occupancy, intermediate layers sustain non-zero values due to fired crystals near the beam pipe, and the outermost layer exhibits a moderate occupancy increase from single beam background contributions.

#### 7.4.5.4 Impacts to energy resolution

The energy deposition caused by beam-induced background will deteriorate the results of energy resolution and linearity. The energy deposition of the EM shower induced by a single photon incident at the center of the module, combined with the energy deposition of the beam-induced background in the same module, constitutes the final photon energy. As

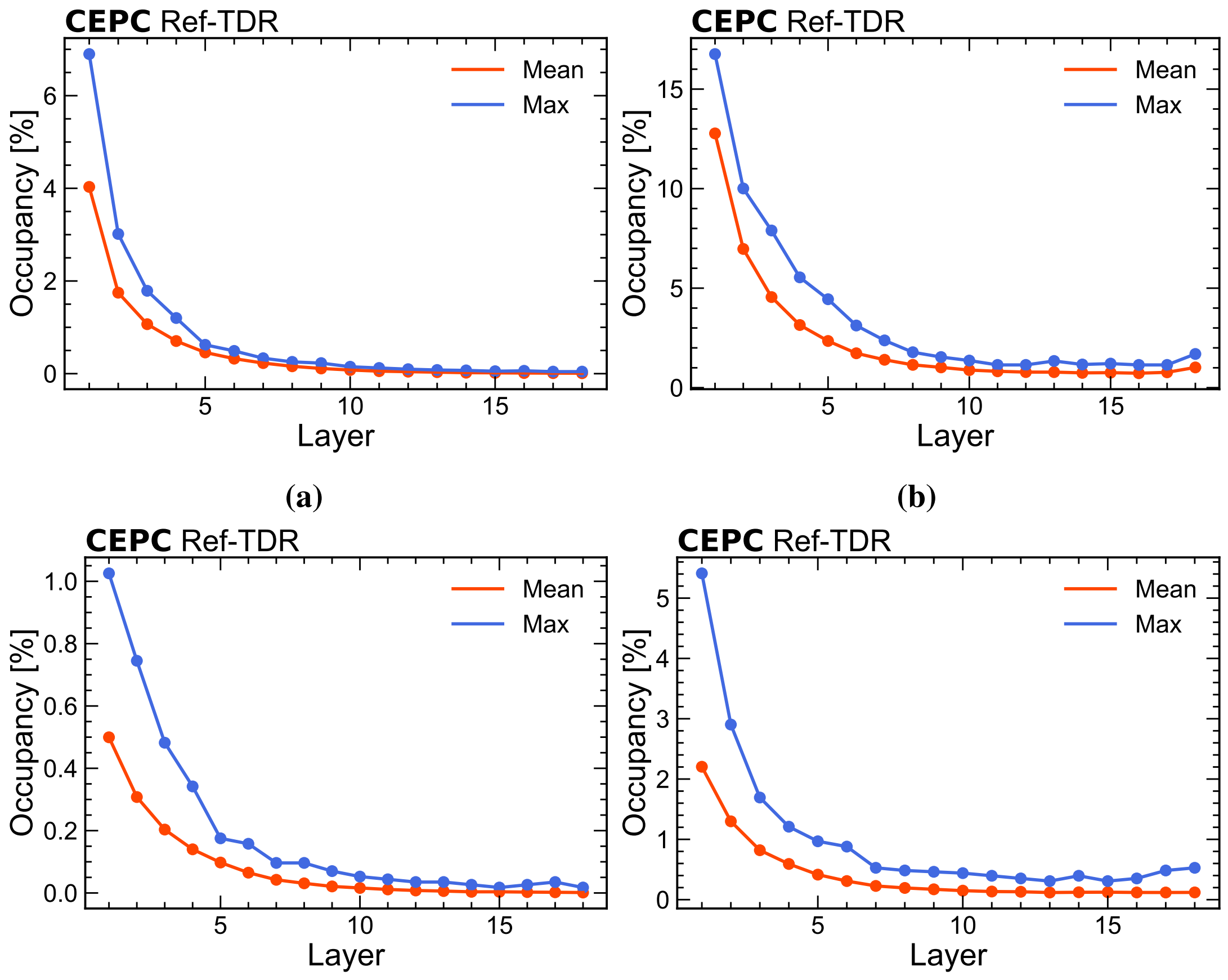

(a) (b) (c) (d)

**Figure 7.39:** Occupancy distributions of BIB hits in different ECAL longitudinal layers: in the barrel (a) and endcaps (b) in the Higgs mode, and in the barrel (c) and endcaps (d) in the Low-Luminosity Z mode.

Figure 7.40 shows, the effect is more pronounced, especially at the low energy with a significant contribution to the noise term. A longer time window will introduce more beam-induced background, resulting in a larger energy range affected. A conclusion from this study is that the time window for signal integration should be within 300 ns.

#### 7.4.5.5 Radiation damage of SiPMs: impact to performance

The effect of dark noise subtracting on calorimeter response was studied using electron shower and SiPM dark noise simulations in Ref. [26]. The resolution and the linearity of the calorimeter were measured as a function of the DCR, and associated to an approximate relation between fluence and temperature obtained for a SiPM in Ref. [27], at 20 °C. Figure 7.41a shows that the linearity suffers heavily due to the radiation damage at energies less than 5 GeV. At the highest fluence studied, a deviation of around 42 % is observed at the neutron fluence of $1 \times 10^{10}\ n_{eq}(1\ \text{MeV})\text{cm}^{-2}$. This result can be explained by the biasing of measured energy due to charges from dark counts exceeding the ECAL energy threshold. The diminishing influence of the dark count rate with increasing energy arises because the distribution of dark noise is energy-independent (i.e. it depends not on the number of active SiPMs during the event but on the spread of the charge distribution measured by the SiPM). This distribution is skewed toward higher energies due to the threshold cut criterion. Figure 7.41b shows that an offset to the resolution is small but non-negligible. The stochastic term and the noise term degrades by approximately 0.2 % and 1 %, respectively, in the fluence range of $1\times10^{7}$ cm$^{-2}$ to $1\times10^{10}$ cm$^{-2}$.

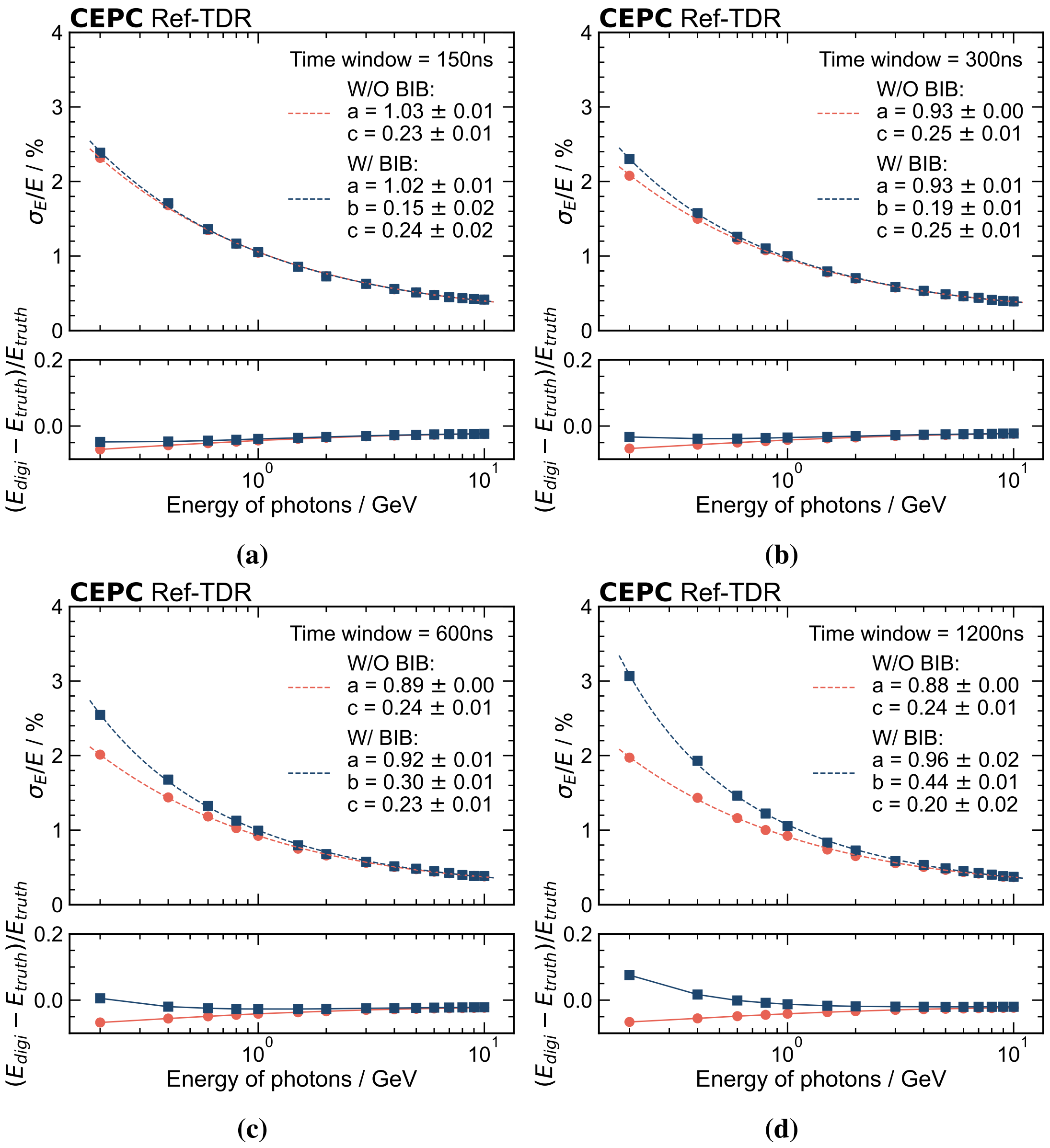

**Figure 7.40:** BIB impacts on energy resolution and linearity in the Higgs mode using different signal integration windows of: 150 ns (a), 300 ns (b), 600 ns (c) and 1200 ns (d).

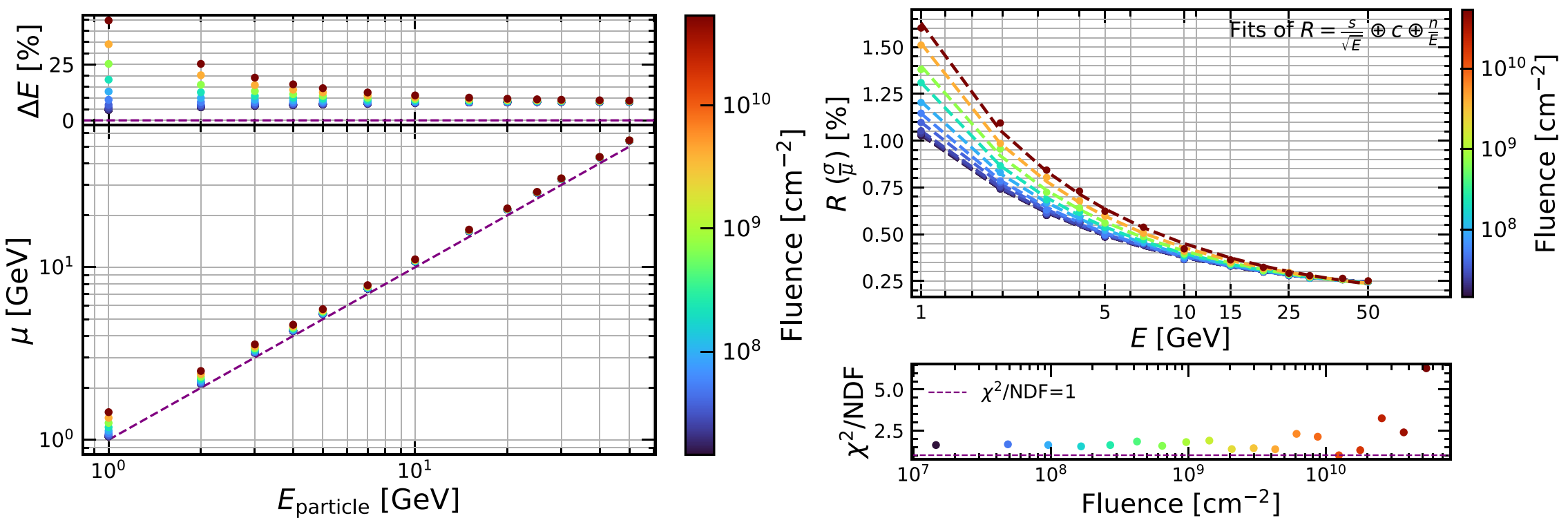

**(a)** **(b)**

**Figure 7.41:** ECAL linearity (a) and energy resolution (b) versus SiPM DCR, expressed in terms of fluence. The main graph of (a) shows the mean from a Crystal Ball fitting function versus the particle energy. The top inset shows the deviation from linearity. The dashed purple line represents the ideal linearity. The main graph of (b) depicts the energy resolution versus particle energy in inverse square-root scaling, with dashed lines indicating the fit of the resolution. The bottom inset shows reduced $\chi^2$ of the fits versus fluence. The color scale for both figures indicates fluence from low (blue) to high (red).

## 7.5 Performance studies

### 7.5.1 $\pi^0 \to \gamma\gamma$ process

$\pi^0$ meson reconstruction performance is crucial for precision measurements of final states containing neutral pions [28, 29, 30]. The typical $\pi^0$ kinematics of $\mathrm{B}_s^0 \to \pi^0\pi^0$ from $Z$ boson decays is used in CEPCSW to study the $\pi^0$ reconstruction efficiency and the $\pi^0$ mass resolution, as shown in Figure 7.42, where $\pi^0$ mesons are in the kinetic energy range of 0.5 to 30 GeV.

Figure 7.42a shows that the ECAL achieves a good $\pi^0$ reconstruction efficiency for the $\pi^0$ kinetic energy below 15 GeV. As the $\pi^0$ energy increases above 15 GeV, the narrowing angle of two photons decaying from $\pi^0$ is a major limiting factor for the $\pi^0$ reconstruction efficiency. The $\pi^0$ invariant mass resolution, as shown in Figure 7.42b, can be better than 10% for $\pi^0$ below 15 GeV. The ECAL angular resolution in the higher energy is a dominating factor that contributes to the $\pi^0$ invariant mass resolution.

Furthermore, the shower shape information extracted from longitudinal and lateral segmentations can be used to further enhance the capability to distinguish a high energy $\pi^0$ from a single photon. Preliminary results, shown in Figure 7.43, indicate that combining CyberPFA reconstruction with a new method based on the second moment of the lateral shower shape achieves a $\pi^0$ reconstruction efficiency above 80 %, with a photon-to-$\pi^0$ misidentification rate below 20 %. Further improvements are expected by incorporating additional shower shape variables and advanced machine learning techniques.

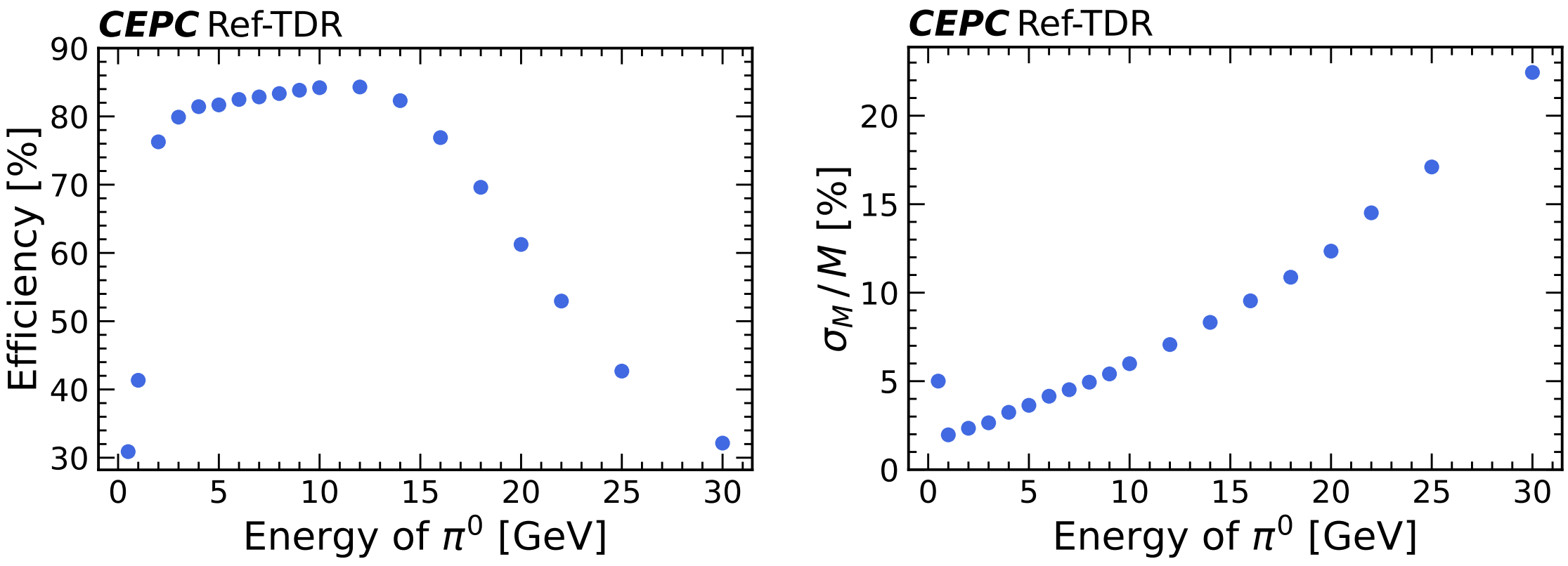

(a) (b)

**Figure 7.42:** The CyberPFA reconstruction performance of $\pi^0 \to \gamma\gamma$ in the $\pi^0$ kinetic energy range of 0.5 to 30 GeV: the $\pi^0$ reconstruction efficiency (a) and the reconstructed $\pi^0$ invariant mass resolution (b) along with the $\pi^0$ kinetic energy. The $\pi^0$ reconstruction efficiency is defined as the fraction of $\pi^0$ decay events reconstructed in the invariant mass range of 50 to 200 MeV.

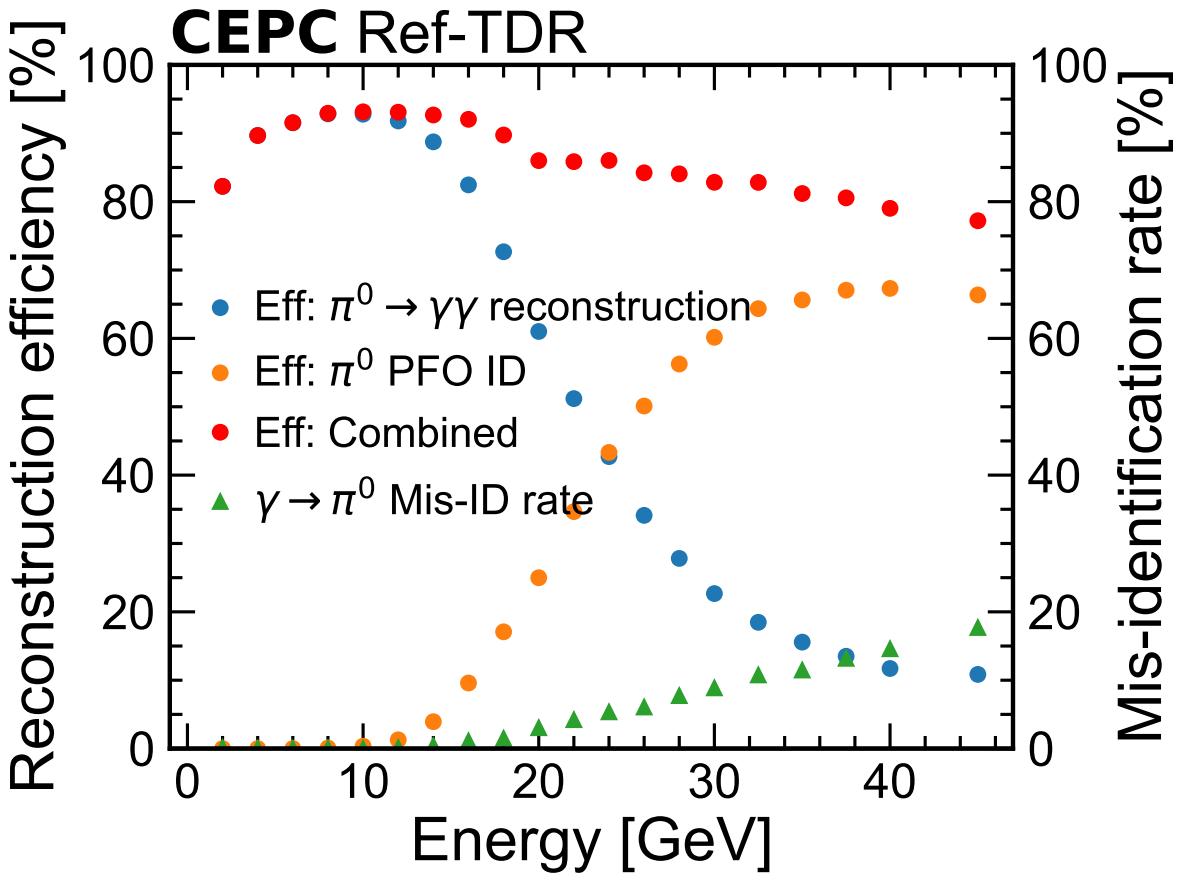

**Figure 7.43:** The ECAL reconstruction performance of for $\pi^0 \to \gamma\gamma$ in the $\pi^0$ kinetic energy range of 2 to 45 GeV. The $\pi^0$ reconstruction efficiency as a function of $\pi^0$ energy is presented for three different scenarios: (1) using CyberPFA alone (blue points), (2) applying the new $\pi^0$ identification technique based on the second moment of shower profiles (orange points) and (3) combining CyberPFA with the new technique (red points). The green points indicate the misidentification rate, defined as the fraction of a single photon incorrectly identified as a $\pi^0$ by the second moment method.

### 7.5.2 $H \to \gamma\gamma$ process

A major physics benchmark process, $H \to \gamma\gamma$, is crucial to precision measurements of the Higgs properties. As no hadrons in the final state are involved, ECAL plays a pivotal role in its reconstruction. Using Whizard[31], electrons and positrons collide at $\sqrt{s} = 240\,\text{GeV}$ and the Higgs bosons are produced via $e^+e^- \to ZH$, and the two particles decay via $Z \to \nu\bar{\nu}$ and $H \to \gamma\gamma$ respectively. A typical event display is shown in Figure 7.44a.

The reconstructed invariant mass of Higgs bosons fitted with a double-sided Crystal Ball

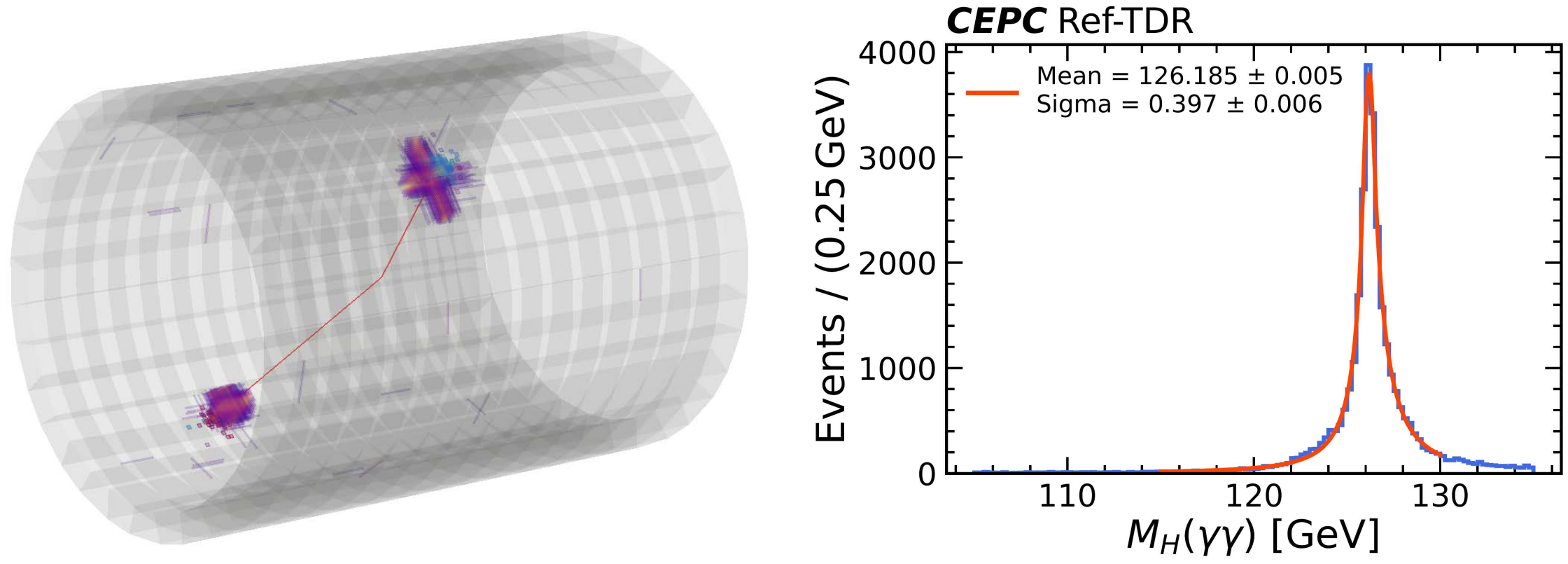

**(a)** **(b)**

**Figure 7.44:** Results of the reconstruction for the $H \to \gamma\gamma$ process: the display of a typical event (a) and the distribution of the Higgs invariant mass reconstructed from two photons (b).

function is shown in Figure 7.44b. It shows that the $\sigma$ is $(0.397 \pm 0.006)$ GeV. The constant term in Equation 7.2 dominates in this process, due to the high energy of the two photons.

## 7.6 ECAL alternative options

Alternative ECAL options include Silicon Tungsten (SiW) and Scintillator-Tungsten (ScW) sampling calorimeters, the stereo crystal ECAL and the dual-readout crystal ECAL. The SiW option, as part of the ILD calorimetry system, is discussed in Chapter 18.7. The crystal EM calorimeter with the dual-readout capability is covered by the IDEA detector concept in Chapter 19. The ScW and stereo crystal options are presented as follows.

### 7.6.1 Scintillator-tungsten sampling calorimeter

Scintillator-tungsten electromagnetic calorimeter (ScW-ECAL) provides a cost-effective PFA-calorimetry option to achieve moderate EM and jet performance. ScW-ECAL uses plastic scintillator for the sensitive elements interleaved with tungsten plates to significantly reduce the cost and facilitate assembly (e.g. without complex silicon sensor alignment and wire bonding). A sensitive layer consists of plastic scintillator strips in dimensions of $45\times45\times2$ mm$^3$, which are positioned orthogonal to each other in two adjacent layers to achieve a fine effective transverse granularity of around $5\times5$ mm$^2$. Each strip is wrapped with ESR foil and read out individually by one SiPM. The mechanical structure is similar to the SiW-ECAL option, with a major exception that the spacing between absorber plates is increased to accommodate scintillator strips.

A first ScW-ECAL physics prototype was developed with 26 layers of tungsten absorber interleaved with plastic scintillator strips and SiPMs. The EM energy resolution has a stochastic term in the range of 13 % to 14 % (consistent with Geant4 simulations) and a constant term in the range of 3 % to 4.5 % in the various configurations and regions [32]. A complete ScW-ECAL physics prototype [33] was followed with 30 alternating layers with a depth of 21.5 $X_0$ (266 mm thick). The EM energy resolution achieved $12.5\,\%/\sqrt{\mathrm{E(GeV)}} \oplus 1.2\,\%$ and linearity within 1.1 % tested with electrons with energies in the range 2 GeV to 32 GeV. A ScW-ECAL technological prototype of 30 sampling layers (with a depth of $22X_0$ and a total of 6,320 readout channels) has been successfully developed [34] and tested with high-energy beam particles [35].

### 7.6.2 Stereo crystal calorimeter design

A conceptual crystal ECAL design has been proposed [36] using a stereo structure within the transverse plane to achieve segmentations in the longitudinal direction. Instead of aligning the crystal bars toward the interaction point, they are oriented toward the beam axis with a specific pointing angle relative to the detector's radial direction. To improve position resolution in the transverse direction and resolve the bias introduced along the azimuthal angle due to the pointing angle, even and odd layers along the $z$-axis are configured with opposite pointing angles. The readout of the crystal bars using SiPMs is positioned at the outer end of each crystal. The front-end electronics, along with the cooling system, are placed behind the ECAL shielding. All crystal bars are designed with a consistent trapezoidal shape, ensuring a homogeneous structure within the barrel and endcaps.

## 7.7 Summary

The ECAL is a novel imaging homogeneous calorimeter based on finely segmented scintillating crystals read out by SiPMs, to achieve an EM energy resolution of $\sigma_E/E \leq 3\,\%/\sqrt{\mathrm{E(GeV)}} \oplus 1\,\%$ and to enable particle-flow reconstruction for precision studies on Higgs, electroweak and flavor physics and searches for new physics. The high-granularity homogeneous ECAL represents a significant advance over legendary calorimeters in the LEP and LHC-era, leveraging breakthroughs in scintillating crystal quality, SiPM dynamic range, and low-power ASIC design. The detector provides a maximum coverage (hermeticity) and operates over a dynamic range of tens of MeV up to 100 GeV, accommodating particle reconstruction from $\pi^0$ mesons, $Z/W$/Higgs bosons and potentially the top quark. The mechanical structure uses lightweight CFRP frames and is optimized for compact modular integration. A dedicated cooling system is integrated to stabilize temperatures for crystals and SiPMs. The ECAL granularity derives from $\sim$ 285,500 crystals with transverse sizes of $15 \times 15\,\mathrm{mm}^2$, each of which is coupled individually with SiPMs at both sides, corresponding to a total of 571,000 SiPMs. Front-end readout employs low-power multi-channel ASICs to digitize signals.

The Geant4 simulation and the target EM performance have been validated by prototype beam tests with electron beam data. Test-beam results for a crystal calorimeter prototype, using electron beams in the energy range of 0.3 GeV to 6 GeV, demonstrate that an EM energy resolution of $\sigma_E/E < 2\,\%/\sqrt{\mathrm{E(GeV)}} \oplus 1\,\%$ has been achieved. Physics benchmarks including $\pi^0 \to \gamma\gamma$ reconstruction and $H \to \gamma\gamma$ invariant mass resolution of 0.33 % are studied with CEPCSW full simulation.

# Chapter 8 Hadronic Calorimeter

The HCAL serves as the key detector component through its essential role in jet energy reconstruction within the PFA paradigm. It provides essential discrimination power between charged and neutral hadrons through complementary measurements, matching charged hadron showers to tracker information while independently reconstructing neutral hadron showers. The Glass Scintillator HCAL (GS-HCAL) is proposed as the baseline design for its crystal-like density and comparable performance, while maintaining excellent moldability at substantially reduced cost.

We acknowledge that the proposed GS-HCAL system is currently in its early stage of R&D and requires further development over the coming years to achieve full technical readiness for construction. This chapter demonstrates its fundamental feasibility by detailing the design, development, and performance evaluation of the system and its potential to meet the CEPC physics requirements. An introduction of the GS-HCAL concept is given in Section 8.1, followed by the geometrical and mechanical design in Section 8.2. Key challenges and solutions appear in Section 8.3, with component R&D and prototype validation in Section 8.4. Performance simulations and alternative technologies are presented in Section 8.5 and Section 8.6, respectively, which complete the technical discussion before the roadmap in Section 8.7.

## 8.1 Overview

Within the PFA paradigm, the HCAL is uniquely responsible for measuring the dominant hadronic components of jets, comprising approximately 62–65% charged hadrons and 10% neutral hadrons, which collectively represent over 70% of the total jet energy (see, e.g., Ref [1]). This capability is indispensable for precision measurements of dijet invariant masses, particularly in $W/Z$/Higgs boson decays where the target boson mass resolution of 3–4% [2] directly depends on the HCAL performance.

The calorimeter design requirements are consequently driven by the need to optimize the reconstruction of these hadronic components, which present distinct technical challenges for charged versus neutral particle detection. For charged hadrons, the HCAL primarily provides cluster shape information, which is then matched to corresponding tracks in the central tracker, with the final energy measurement derived from the track momentum reconstruction. In this case, the absolute energy resolution of the HCAL itself is less critical. However, for neutral hadrons, which leave no tracks in the tracking system, the HCAL must independently provide both a precise spatial reconstruction of shower clusters and an accurate energy measurement. In addition to neutral hadrons, good energy resolution is also well-suited for detecting long-lived particles (LLPs) that may decay within the HCAL volume. This dual requirement necessitates an HCAL with exceptionally fine three-dimensional granularity for optimal shower topology reconstruction, combined with the best possible energy resolution for neutral hadrons.

The baseline GS-HCAL design addresses the physics requirements through its high-density Glass Scintillator (GS) cells and SiPM readout, having both the necessary spatial segmentation and energy sampling fraction to enable precise reconstruction of both neutral and charged hadrons, fulfilling the core requirement for particle-flow calorimetry.

The geometric architecture of the GS-HCAL comprises a barrel section and two endcaps as illustrated in Figure 8.1, with a total thickness of 6 nuclear interaction lengths ($\lambda_I$) in steel and GS that ensures hermetic containment of hadronic showers. The barrel consists of 16 trapezoidal sectors arranged in an exadecagonal cylinder with inner and outer radii of 2140 mm and 3455 mm, respectively, spanning 6460 mm along the beam axis. Each sector contains 48 longitudinal layers with total thickness 1315 mm, progressively increases from a front width of 851 mm (facing the interaction point) to a back width of 1375 mm. The endcaps mirror this segmentation with 16 disk-shaped sectors per endcap, featuring inner and outer radii of 450 mm and 3455 mm.

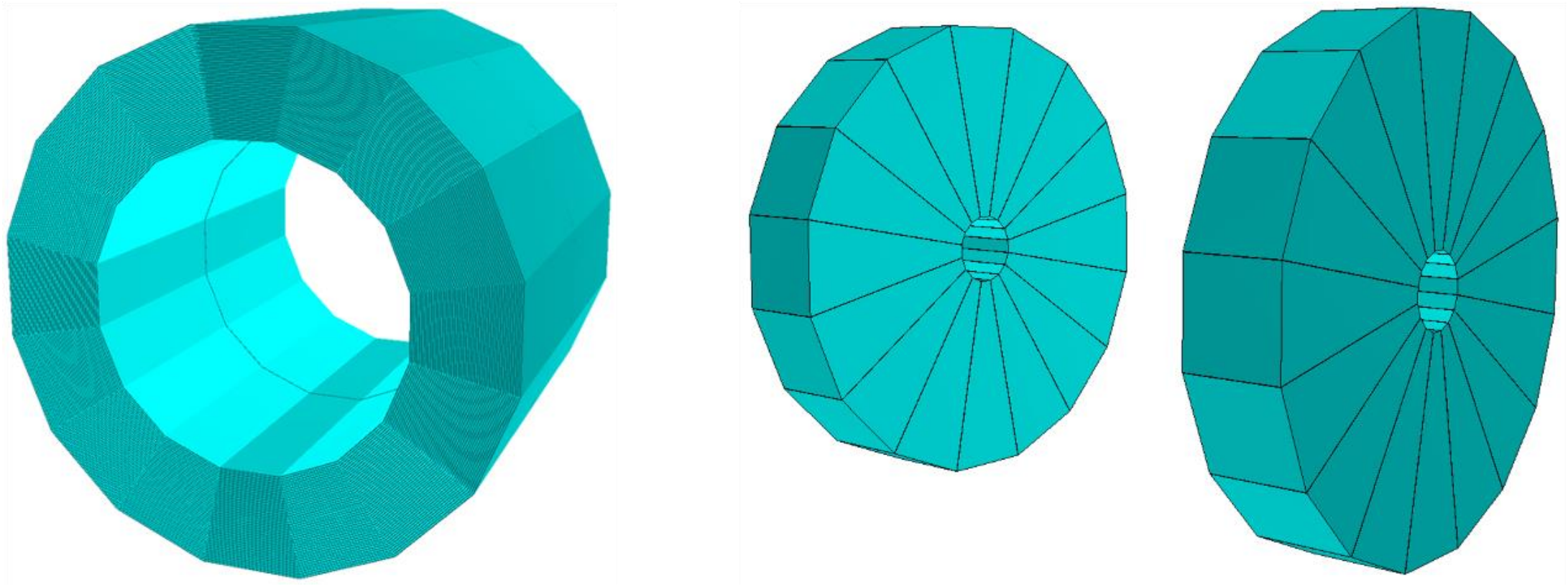

**Figure 8.1:** Geometric configuration of the GS-HCAL system showing (left) barrel with 16 trapezoidal sectors forming a cylindrical structure and (right) endcap disks with matching sector segmentation. The barrel inner diameter measures 4280 mm with outer diameter 6910 mm, while endcaps complete the hermetic coverage at both beam ends.

The design features 5.22 million identical GS cells, each measuring 40 mm $\times$ 40 mm $\times$ 10 mm. These cells are arranged in 48 longitudinal layers, providing a total thickness of $6\lambda_I$. Each layer consists of steel absorber plates with embedded cooling channels, supporting modular detection units that combine GS cells with readout electronics. The barrel contains approximately 3.2 million cells (955 tons), while each endcap holds about one million cells (367 tons). Each cell provides 60–100 Photoelectron (p.e.)/MIP light output, read out by 3 mm $\times$ 3 mm SiPMs. This performance is enabled by the close match between the GS emission spectrum, which peaks at 400 nm, and the peak sensitivity of the SiPMs. The 30% sampling fraction between steel absorbers and GS represents a distinct design choice compared with traditional HCAL, so as to reduce the sampling term in the energy resolution and thereby enhance the sensitivity to non-standard signatures such as LLPs and other new physics phenomena. The readout system processes signals from 5.22 million channels through low-noise electronics capable of resolving 0.1–100 MIPs. Calibration combines physics processes like $Z \to \mu\mu$ decays with dedicated Light Emitting Diode (LED) monitoring to maintain long-term stability. Cooling systems manage the 15 mW/channel power dissipation through integrated water channels.

Simulations confirm the design meets CEPC requirements, with projected energy resolution of $30\%/\sqrt{E} \oplus 6.5\%$ before calibration. The $6\lambda_I$ thickness provides sufficient shower containment, while reconstruction algorithms further improve performance. Mechanical integration uses modular boxes for assembly and maintenance, with alignment precision maintained at the sub-millimetre level. A prototype system will undergo beam tests in 2025–2027 to validate the design before full production. Ongoing developments focus on improving the constant term through advanced reconstruction techniques and material optimizations.

The following sections present the complete technical implementation of the GS-HCAL system. The mechanical design and geometrical configuration of the calorimeter modules are described first, covering both the conceptual framework and practical realization. Subsequent documentation of research and development achievements demonstrates validated technical solutions, including the glass scintillator production and readout system integration. The discussion also identifies remaining challenges requiring further development, particularly in radiation-hard SiPM optimization and large-scale assembly techniques. Together these serve as both a record of demonstrated capabilities and a roadmap for future prototyping efforts needed to finalize the production design.

## 8.2 HCAL design

Following the overview, this section details the engineering implementation of the GS-HCAL system. The design transforms physics requirements into mechanical solutions through modular construction principles that enable the precise assembly of 48 longitudinal layers. Key features include an optimized material composition that achieves the 30% sampling fraction while minimizing the proportion of non-sensitive material.

The 16 trapezoidal sectors in the barrel and in the matching endcap disks implement the same layer structures while accommodating distinct geometric constraints. Each 27.2 mm-thick layer combines 9.8 mm steel absorbers with 10.2 mm GS active elements, maintaining uniform 40 mm $\times$ 40 mm $\times$ 10 mm cell granularity throughout. The following subsections document the standardized layer architecture, its adaptation to barrel and endcap geometries, the mechanical interfaces, and structural validation through finite element analysis.

### 8.2.1 Single layer structure

A single-layer design of the GS-HCAL is illustrated in Figure 8.2(a). The GS-HCAL consists of alternating layers of GS cells as the active medium and steel plates as the absorber material. This configuration provides high calorimeter density while maintaining fine segmentation for precise shower imaging.

The total thickness of an individual GS-HCAL layer is 27.2 mm corresponding to about 0.125 $\lambda_{\mathrm{I}}$. The top of the layer features a 2 mm-thick upper cover. Below this lies the 10 mm-thick GS active layer with a cell size of 40 mm × 40 mm as shown in Figure 8.2(b), which produces scintillation light when traversed by particles. This is followed by a 3.2 mm layer containing PCB and ASIC chips that handle signal processing and readout. A 2 mm-thick bottom cover provides structural support beneath the PCB. The steel absorber forms the base of the structure with a thickness of 9.8 mm. The cooling pipes with an outer radius of 3 mm are embedded in the absorber. The SiPMs are mounted on a PCB that spans multiple GS cells, with one SiPM positioned at the center of each GS cell. The material contributions are 0.0805 $\lambda_{\mathrm{I}}$, 0.0425 $\lambda_{\mathrm{I}}$, and 0.0024 $\lambda_I$ (1 : 0.53 : 0.03) for the steel absorber, GS, and PCB, respectively. The calorimeter's sampling fraction is approximately 31%. This design optimizes the balance between material interaction length and scintillation efficiency, enabling effective particle detection within a compact cell structure.

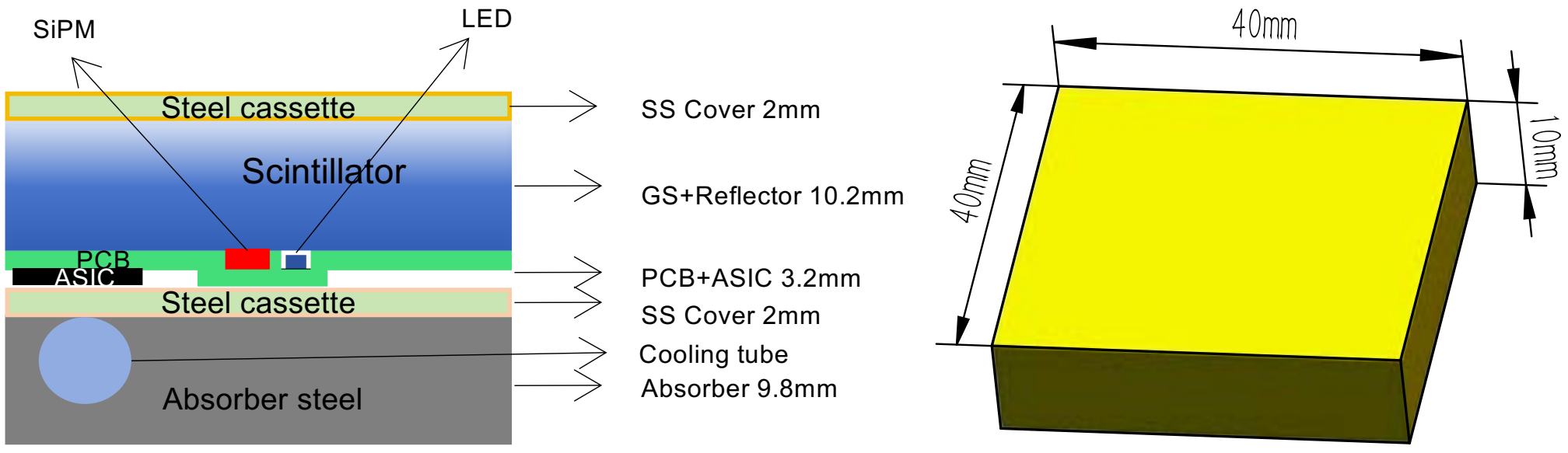

(a) (b)

**Figure 8.2:** (a) Single layer structure of GS-HCAL. (b) One GS cell.

### 8.2.2 Barrel

The barrel section is divided into 16 trapezoidal sectors, providing a precise and symmetric segmentation, as shown in Figure 8.3. The inner and outer radii of the barrel measure 2140 mm and 3455 mm, respectively, while the length of the barrel along the beam direction extends to 6460 mm. The trapezoidal shape sector is characterized by a front width of 851 mm, corresponding to the narrower side facing the interaction point, and a back width of 1375 mm, corresponding to the wider side facing outward. The total thickness of the barrel sector is 1315 mm, corresponding to $6\lambda_{\mathrm{I}}$. The longitudinal segmentation is set to 48 layers based on PFA optimization studies conducted with the Plastic Scintillator Hadronic Calorimeter (PS-HCAL) prototype [3]. This configuration remains optimal for the GS-HCAL design since both calorimeters share the same total thickness of $6\lambda_I$, ensuring comparable shower development characteristics despite their different active media. Each sector gradually expands in width from the front face to the back face while maintaining a uniform height. The total weight of the barrel is 955 tons.

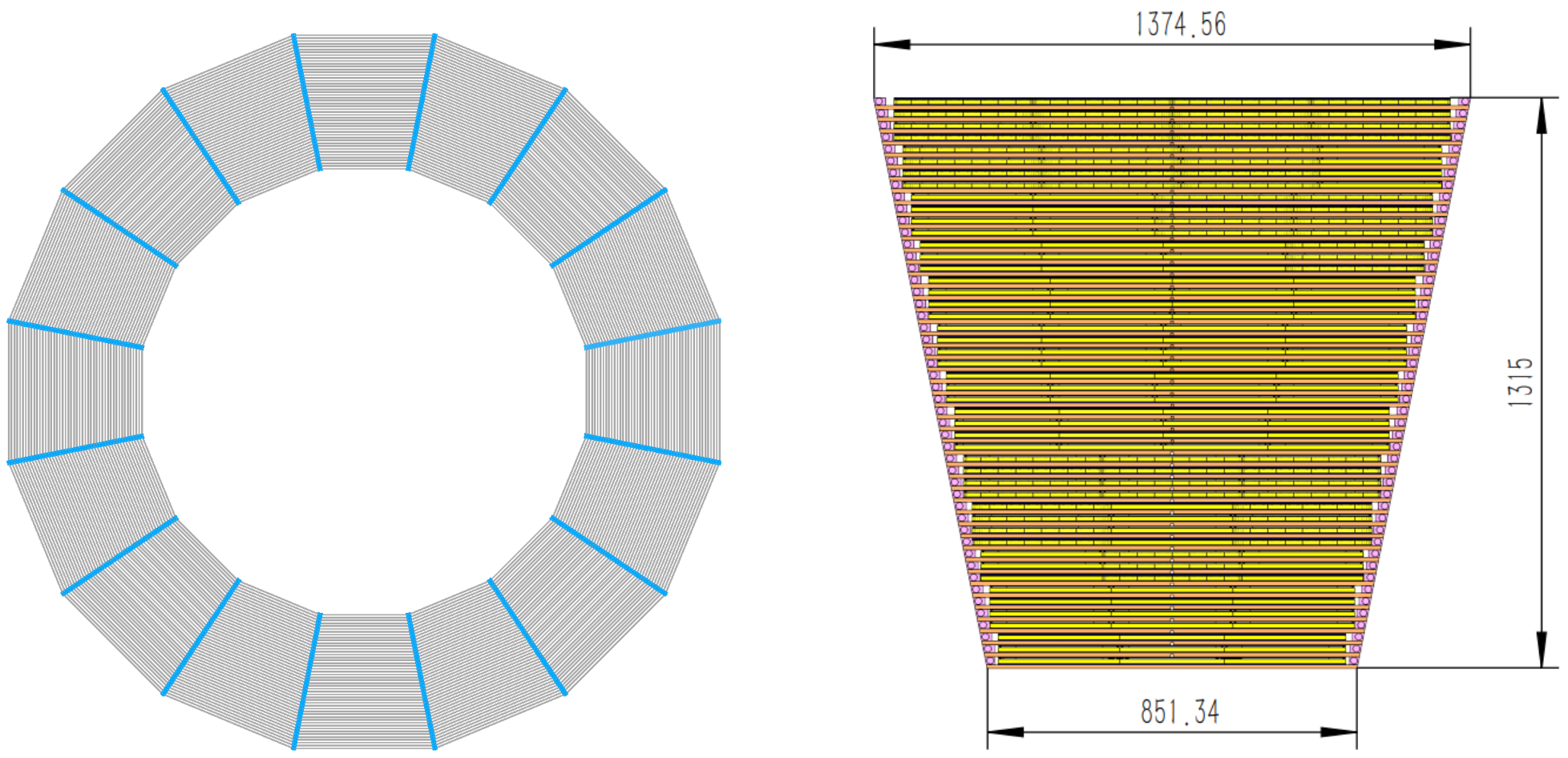

**(a)** **(b)**

**Figure 8.3:** Dimensions of the GS-HCAL barrel. (a) HCAL Barrel cross-sectional view showing 16 trapezoidal sectors arranged in a ring with an inner diameter of 4280 mm and an outer diameter of 6910 mm. (b) Side view of a single trapezoidal module along the beam direction, with a front width of 851 mm, a back width of 1375 mm, and a height (depth) of 1315 mm.

Inherent to the sector design, this geometry has 16 radial cracks but maintains effective coverage through compensating effects. Hadronic showers predominantly initiate in the preceding BGO ECAL with its 1.2 nuclear interaction lengths, causing lateral spread that bridges sector boundaries. The solenoidal magnetic field induces sufficient muon curvature guarantees detection even if trajectories initially align with sector edges.

As shown in Figure 8.4, each layer of a barrel sector consists of a steel absorber interleaved with active detection elements described in Section 8.2.1. The steel absorber plate forms the structural base of each layer, extending continuously along the full 6460 mm length of the barrel. This long single-piece steel construction is well within modern manufacturing capabilities, representing a routine engineering solution for large-scale structures. This monolithic absorber serves both as the primary hadronic interaction medium and as the load-bearing support structure for the entire layer. Upon this rigid base, there is 30 or 40 modular boxes precisely positioned in each layer. Each box contains a complete detection unit comprising GS, SiPM, PCB, ASIC chips, cables, cover plates etc., as mentioned already.

Taking the 48th layer of a sector as an example, the 40 modular boxes are positioned as shown in Figure 8.4. The 48th layer consists of 4 rows along the $\phi$ direction, with 10 boxes in each row to cover the full barrel length along the beam direction. Each box contains 8 GS cells along its width and 16 cells along its length, totaling 128 cells. Across the entire layer, this configuration provides 5120 cells. Each box is first aligned using registration pins, then secured with 4 bolts through corners. Each corner features 3 holes: one for box assembly, one for precision alignment pins, and one for final bolts that structurally connect adjacent absorber layers. Adjacent boxes are separated by a 2 mm gap to accommodate thermal expansion, while their cover plates are designed to ensure mechanical stability across the full sector width.

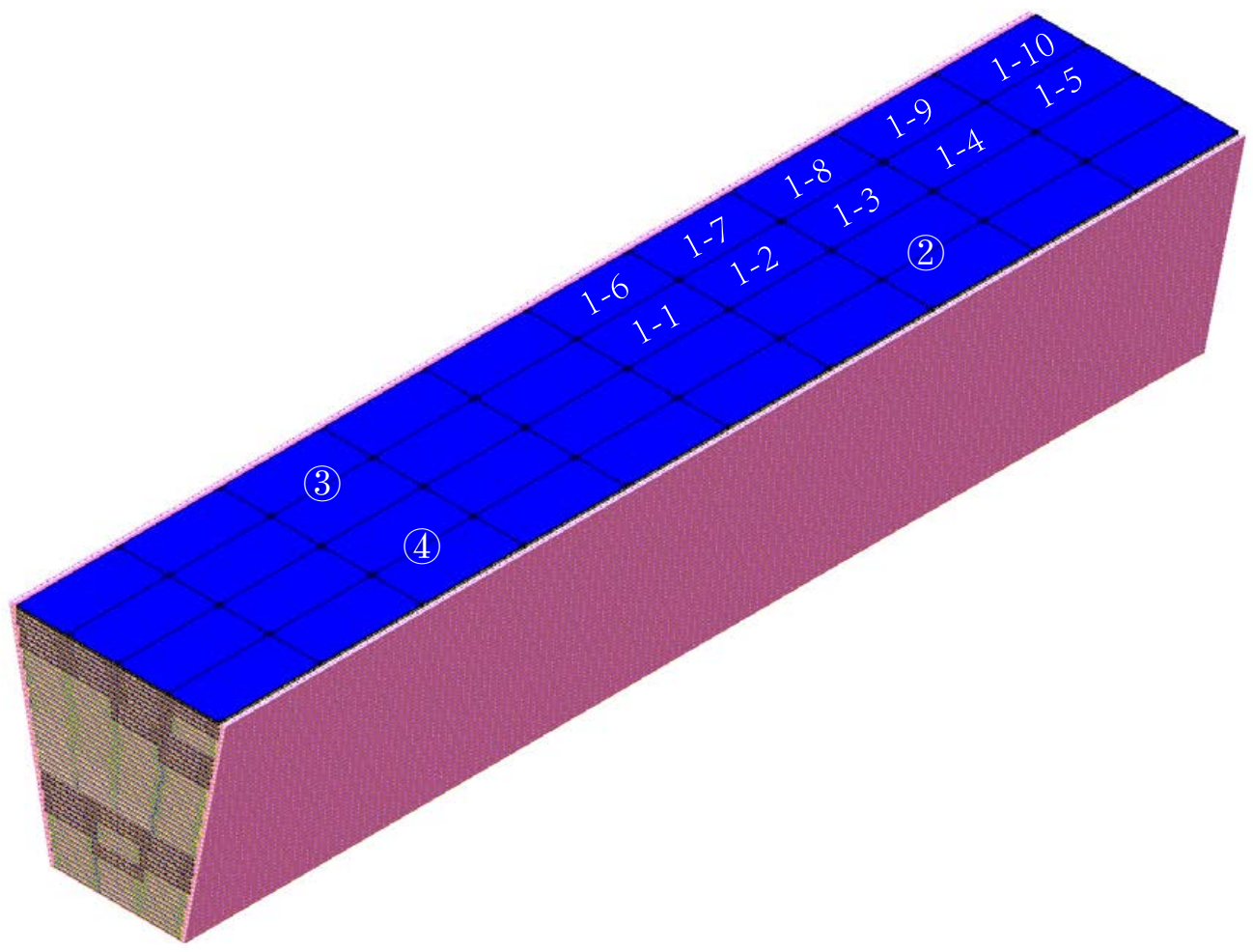

**Figure 8.4:** A sector of the GS-HCAL barrel, also shown is the arrangement of 40 boxes in the outermost (48th) layer. The 4 quadrants from (1) to (4) contain 10 boxes numbered as $n$-1 to $n$-10, where $n$ is the quadrant number.

The inner box structure is shown in Figure 8.5. The trapezoidal sector geometry requires adaptive box arrangements across its 48 layers. As the width of each layer decreases, three box widths (320 mm, 280 mm, and 240 mm) are deployed in combinations to maintain full coverage. The outermost layers employ four 320 mm boxes, while progressively inward layers use mixed combinations (e.g., three 320 mm plus one 280 mm box) to accommodate the narrowing profile. The innermost layers transition to three 280 mm boxes, ensuring surface coverage throughout the sector's radial gradient with minimized gaps. This scheme preserves the 40 mm $\times$ 40 mm $\times$ 10 mm cell granularity while accommodating the sector's geometric constraints.

A water cooling system is specifically designed to manage thermal loads mostly from ASIC chips. Taking the 48th layer as an example, a heat load of 76.8 W of this layer (5120 channels $\times$ 15 mW/ch) is generated by the front-end electronics. Each of the 48 layers in a sector incorporates 4 water cooling pipes embedded within the steel absorber plates, positioned in close proximity to the PCB-mounted ASIC chips. As shown in Figure 8.6, the system is organized hierarchically with shared coolant distribution, every eight consecutive layers are grouped together, sharing common input and output manifolds. This configuration results in six independent cooling loops per sector (48 layers / 8 layers per group), each with dedicated supply and return piping. Round water pipes with an outer radius of 3 mm are used. The localized four-pipe arrangement per layer ensures efficient heat extraction from the ASICs, while the grouped manifold system simplifies the overall cooling infrastructure. While the current design provides sufficient cooling capacity, future optimizations may explore more cost-effective thermal management solutions that could reduce complexity without compromising performance.

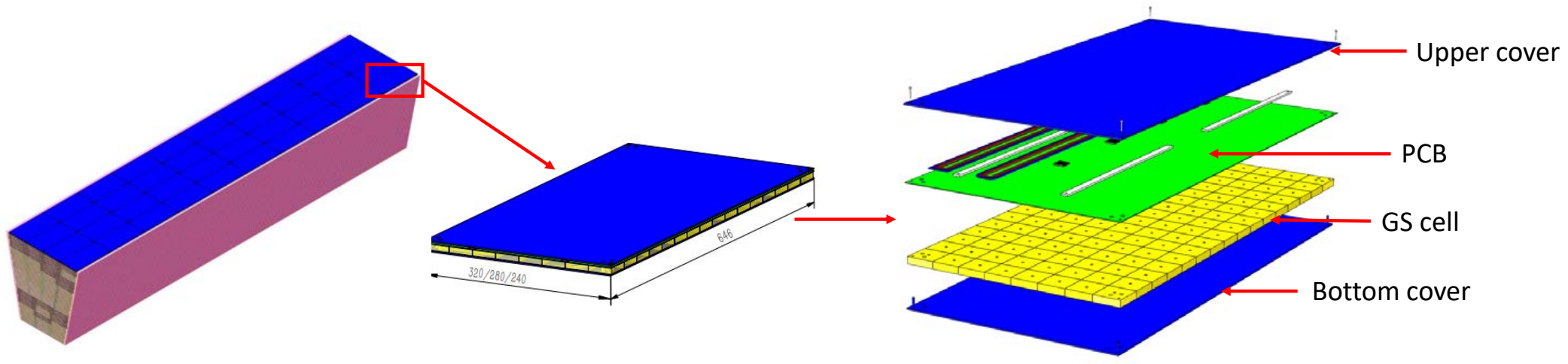

(a) (b) (c)

**Figure 8.5:** Illustration of the GS-HCAL sector structure. (a) A full barrel sector composed of 48 layers, (b) the dimensions of an individual box with varying widths (240 mm/280 mm/320 mm) and a fixed length of 646 mm, (c) the internal structure of a single box, including the cover plates, GS cells, PCB with ASIC chips, cables, and cooling water pipes.

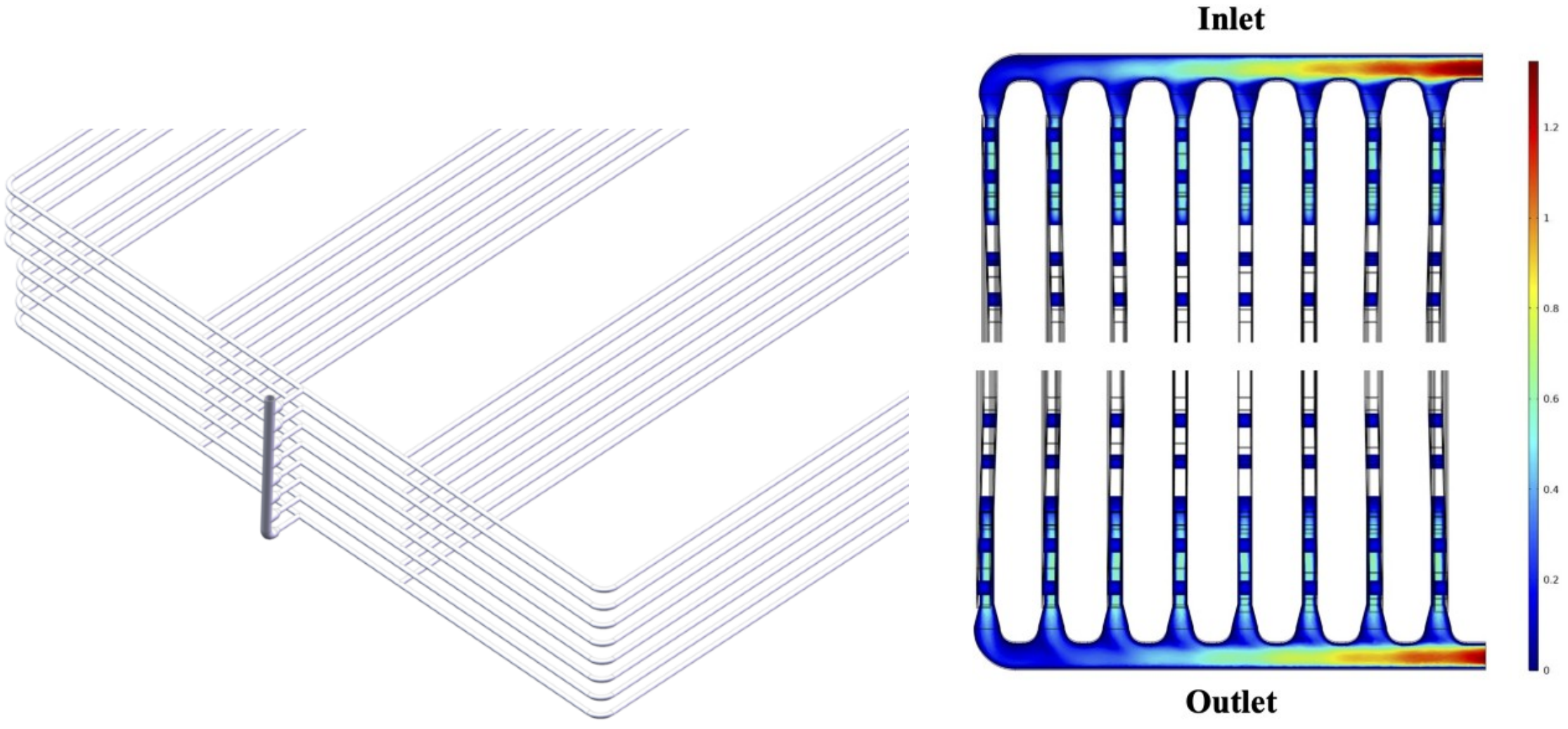

(a) (b)

**Figure 8.6:** Cooling pipe routing for a barrel sector of the GS-HCAL. (a) Schematic of the eight-in-one pipeline design. (b) Simulated flow velocity (m/s) distribution of the coolant along the pipeline, with inlet and outlet positions indicated.

For the barrel, the total number of cells, boxes, layers and sectors are summarized in Table 8.2 and their total weights are listed in Table 8.1.

### 8.2.3 Endcap

The endcap section complements the barrel to provide full geometric coverage at both ends along the beam direction. Figure 8.7(a) and (b) show the front and side view of one of the endcap disks. The dimension and structure match with the barrel design of 16 sectors, 450 mm inner radius and 3455 mm outer radius. The thickness of the sector is 1315 mm divided among 48 layers, the same numbers as the barrel. The shape in Figure 8.7(c) corresponds to one sector of the GS-HCAL endcap. It has a top width of 1307 mm and a total height of 2862 mm measured along the radial direction from the beam axis. The sector features a tapered geometry, with a wider outer edge and a narrower inner width. It is divided into 6 boxes along the radial direction. The top 4 boxes are arranged symmetrically in two pairs, and the bottom 2 boxes are single. The endcap shares the same single layer structure as that of the barrel, as detailed in Section 8.2.1. This segmentation supports both mechanical integration and longitudinal sampling for hadronic shower reconstruction. Each endcap weighs 362 tons.

**Table 8.1:** Weight data of different components of barrel HCAL.

| Items | Weight/sector (kg) | Weight/barrel (tn) |
|---|---|---|
| Cover plate | 10287 | 164.6 |
| PCB plate | 390 | 6.2 |
| GS | 19277 | 308.4 |
| Absorber layer | 27515 | 440.2 |
| Support structure | 1883 | 30.1 |
| Auxiliaries | 350 | 5.6 |
| Total | 59702 | 955.2 |

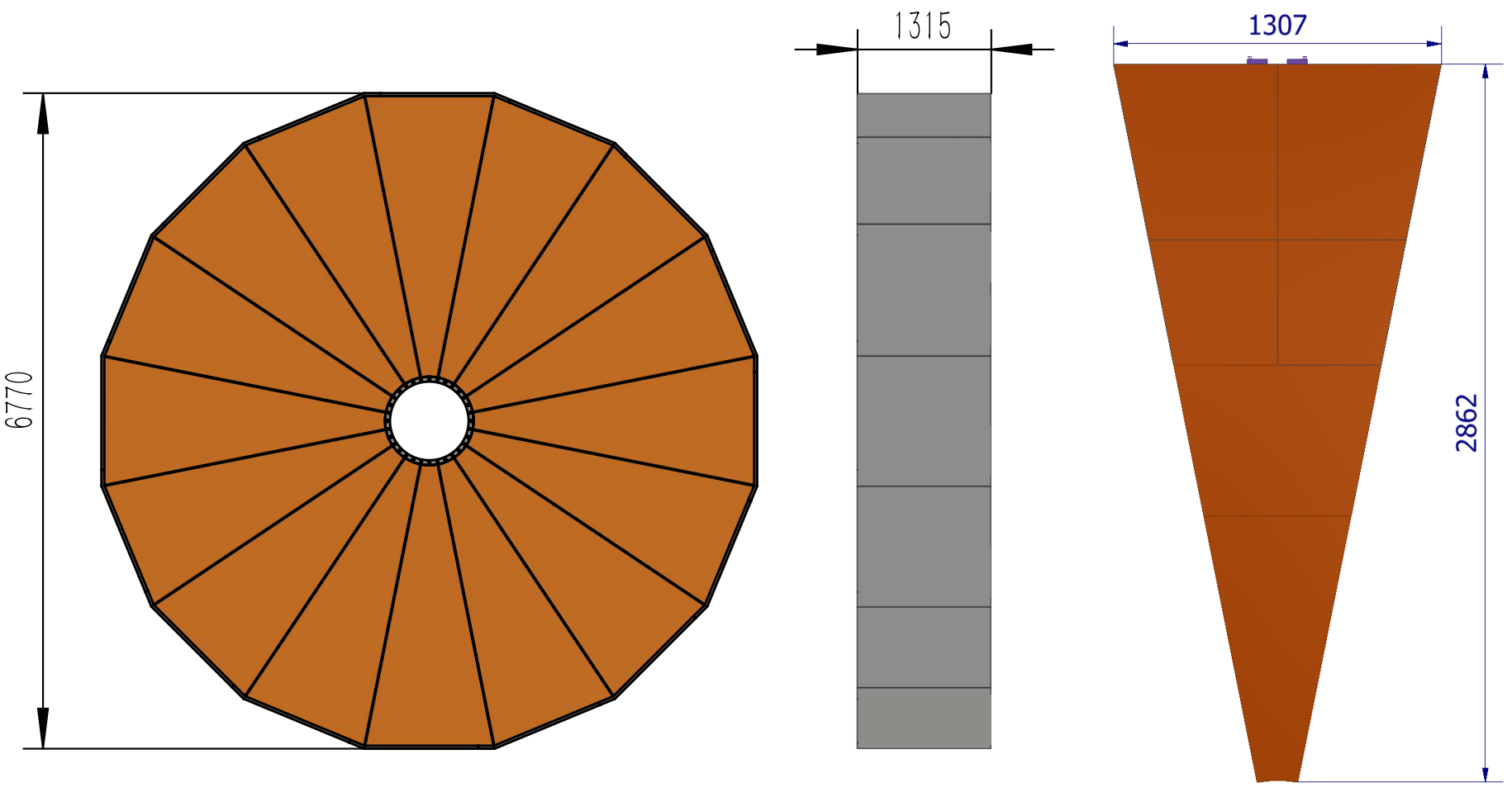

**(a)** **(b)** **(c)**

**Figure 8.7:** Dimensions of the GS-HCAL endcaps. (a) Front view of the endcap along the beam axis, showing the full diameter of 6670 mm, (b) Side view revealing the endcap thickness of 1315 mm; (c) Detailed view of a single sector, the complete endcap consists of 16 such identical sectors.

Similar to the barrel design, the 16 radial sectors in endcaps introduce projective cracks whose effects are mitigated by comparable compensation mechanisms. The preceding ECAL ($1.2\lambda_I$) ensures lateral spread across sector boundaries, while the solenoidal field curves muon trajectories away from crack alignments. Additional energy corrections can be applied during reconstruction to further compensate for any residual effects.

The construction of the endcap begins with 1405 individual detector cells assembled into a single layer sector, with 16 of these sector layers combined to form one complete disc-shaped layer. A total of 48 such layers are sequentially stacked to complete the module. The assembly process starts with mounting a 50 mm-thick outer structural plate (non-absorbing) to provide rigidity and serve as the installation reference. Active layers and steel absorber plates are then alternately installed horizontally. The assembled components are carefully rotated into their final vertical orientation during installation, with particular attention to minimizing mechanical stress on the GS cells to preserve their structural and optical integrity.

The cooling system of the endcap is designed in such a way that each detector layer has 16 separate cooling regions as shown in Figure 8.8. Each region contains one dedicated cooling water channel. This design creates 16 directional cooling paths per layer, which connect vertically when all 48 layers are assembled, matching channels connect vertically, forming parallel water circuits through the stack. Same round water pipes with a 12 mm inner diameter is used also for the endcaps. In total, there are 48 parallel cooling pipes in the complete endcap structure.

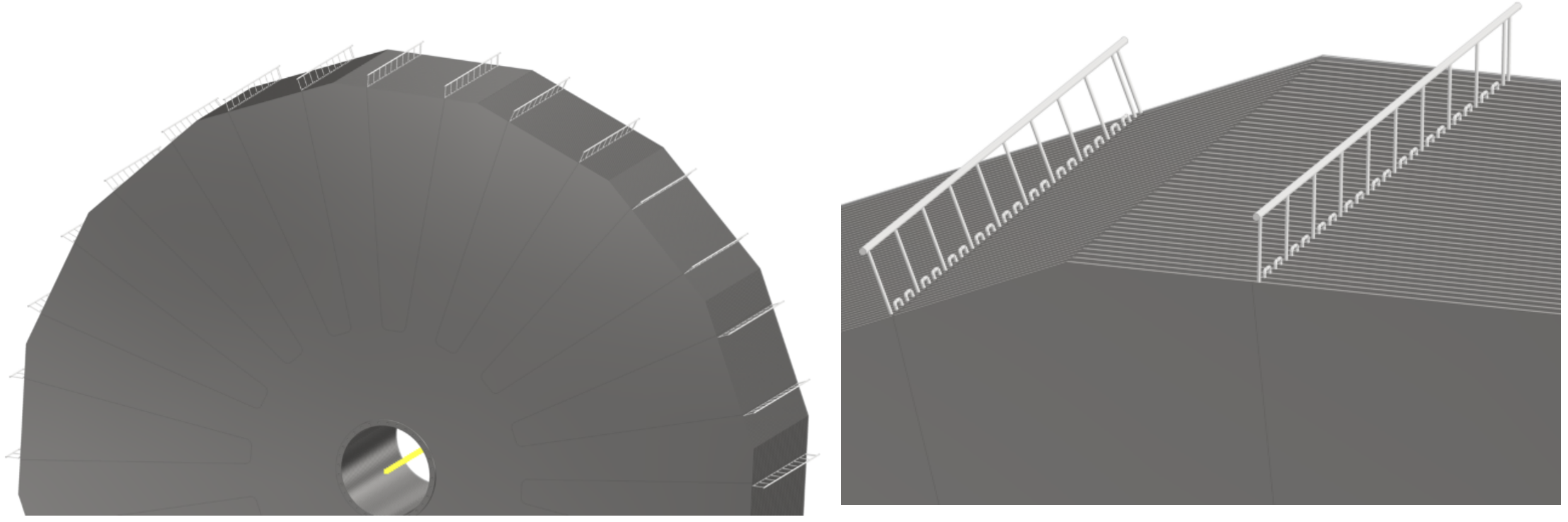

(a) (b)

**Figure 8.8:** Cooling pipe structure for the endcap, where (a) is showing a half of the endcap, and (b) is showing a zoomed in corner of the endcap.

### 8.2.4 Mechanical feasibility

The mechanical feasibility of the GS-HCAL design was validated through structural analysis. The design demonstrates sufficient strength to withstand assembly stresses while maintaining detector performance through controlled deformation. Particular attention was given to the barrel's capacity to support the ECAL load, as the GS-HCAL structure bears the full 135-ton weight of the ECAL system while preserving alignment precision. Figure 8.9, shows the deformation distribution of the GS-HCAL without and with ECAL loading.

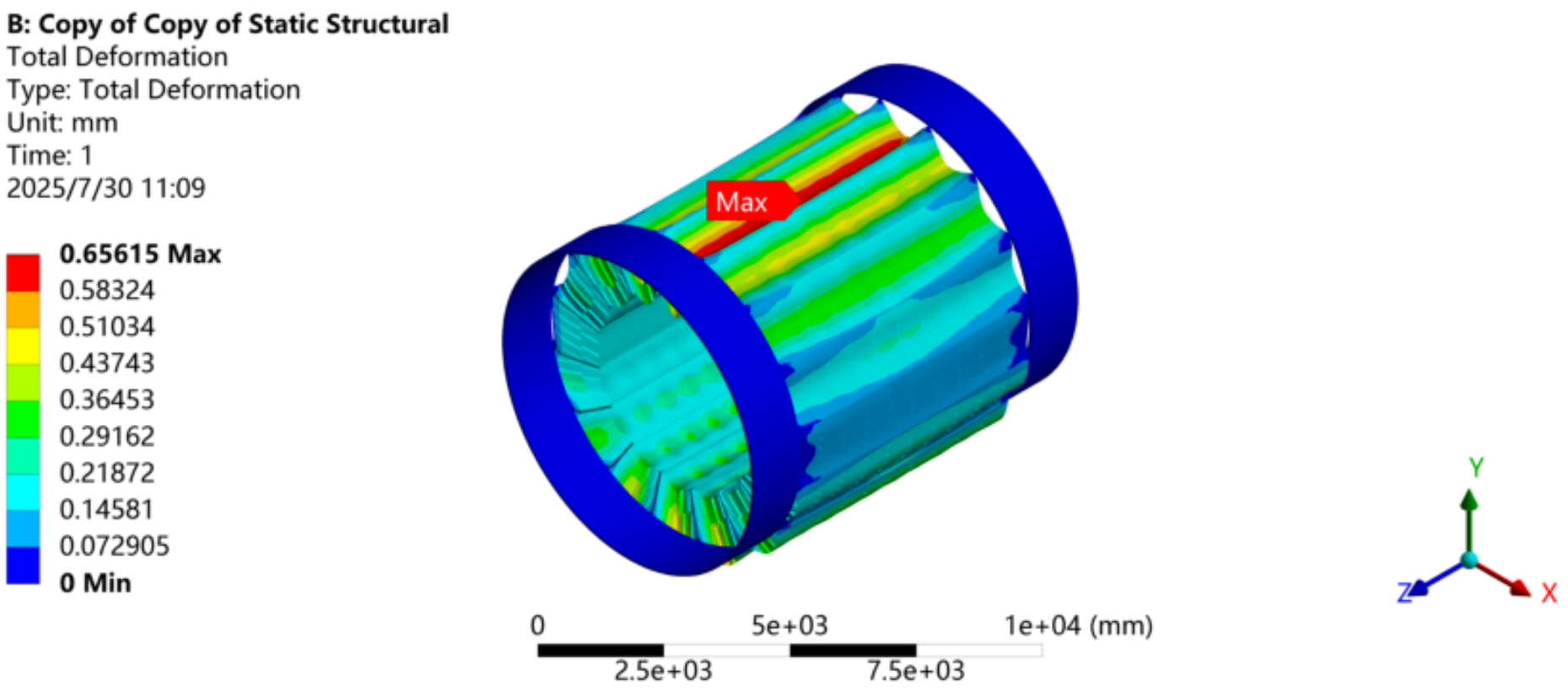

**(a)**

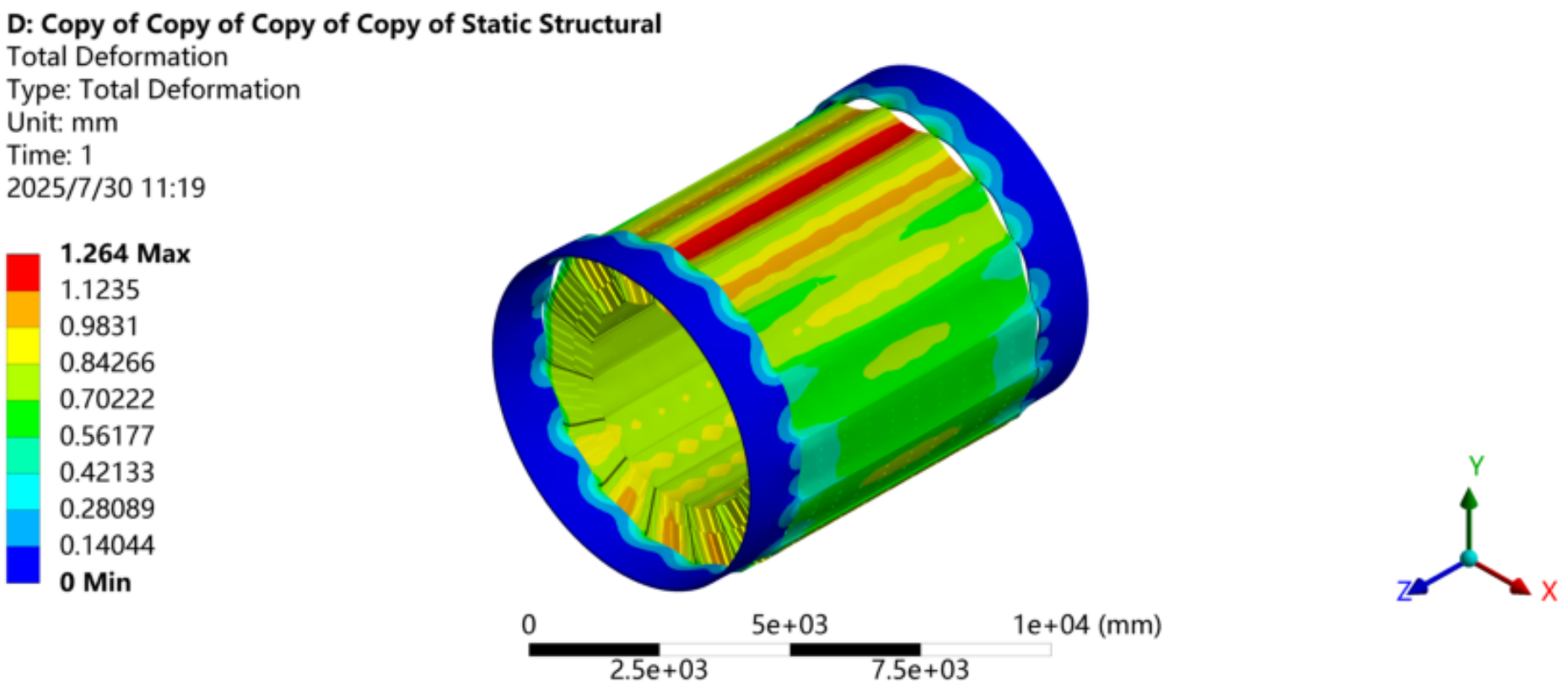

**(b)**

**Figure 8.9:** Deformation distribution of the GS-HCAL (a) without and (b) with ECAL loading.

The analysis demonstrates that the design successfully handles all mechanical loads with substantial safety margins. During assembly, maximum deformations remain below 0.66 mm with localized stresses of 190 MPa, well below the yield strength of structural materials. The most demanding scenario occurs when supporting the 135-ton ECAL, where peak stresses reach 371 MPa at specific connection points. These critical areas will utilize titanium alloy (TC4) components to maintain safety factors above 2.2, while the majority of the structure employs standard stainless steel.

Key performance metrics comprehensively validate the design's mechanical integrity. The system maintains layer-to-layer deformation differences below 0.2 mm, well within the accommodation range of 2.4 mm gaskets, while intra-layer deformations are constrained to less than 0.5 mm, resulting in merely 20 MPa stress within the glass scintillators. Crucially, the scintillator boxes demonstrate robust performance by withstanding 0.5 mm bending conditions while maintaining a safety factor exceeding three times the allowable stress limits. Furthermore, all manual handling scenarios generate stress levels fully compliant with the design specifications, confirming practical feasibility during installation and maintenance operations.

The continuous steel absorber design provides exceptional rigidity along 6.46 m barrel length, while the modular box system maintains precise alignment of active components. These results collectively verify that the GS-HCAL mechanical design satisfies all structural requirements for successful implementation in the CEPC detector system.

### 8.2.5 Summary

Table 8.2 gives the numbers of cells, boxes, layers, and sectors for the GS-HCAL barrel and endcap. The barrel consists of 16 sectors, each composed of 48 layers, totaling 768 layers, with 27,840 boxes and 3,212,800 cells. Each endcap is built with 16 sectors and 48 layers, corresponding to 3,072 boxes and 1,006,080 cells per endcap. Since there are two endcaps and one barrel, the total number of cells is about 5.22 millions.

**Table 8.2:** Numbers of cells, boxes, layers, and sectors for the barrel and endcap.

| Part | Cell | Box | Layer | Sectors |
|---|---|---|---|---|
| Barrel | 3,212,800 | 27,840 | $48 \times 16$ | 16 |
| Endcap (2 sides) | $1,006,080 \times 2$ | $3,072 \times 2$ | $48 \times 16 \times 2$ | $16 \times 2$ |

## 8.3 Key technologies and major challenges

The CEPC GS-HCAL, leverages several advanced technologies and strategies to address critical challenges posed by its stringent physics requirements. These technologies encompass enhancing glass scintillation properties, optimizing photon detection efficiency, addressing glass weight and mechanical integration, mitigating SiPM performance and managing detector cost.

One critical technology aspect is the optimization of GS properties, particularly regarding the light output and attenuation length. The GS-HCAL employs high-density GS (~ 6 g/cm$^3$) with moderate intrinsic light yields around 1000–2000 photons per MeV. Although GS intrinsically exhibit lower photon yields compared with crystals, careful optimization, including surface polishing and reflective coatings, can achieve substantial effective light output of 60–100 p.e./MIP. The current attenuation length of 6.0 cm represents significant progress, further improvements remain desirable to optimize performance.

The significant weight of GS materials introduces a major mechanical engineering challenge. Each endcap of the GS-HCAL alone weighs 362 tons, while the barrel weighs 955 tons, mandating robust, yet precisely aligned mechanical structures. To manage the considerable loads, a modular design approach featuring rigid frames and carefully engineered cassette-style modules has to be adopted, facilitating installation, maintenance, and ensuring mechanical stability and alignment precision throughout the detector's lifetime.

To validate and refine the proposed design solutions, extensive prototyping and testing of the GS-HCAL have to be undertaken. The GS-HCAL prototype should include alternating layers of high-density GS and steel absorber, read out using advanced SiPMs. Beam test studies have to be perfomed to confirm the feasibility of achieving the targeted energy resolution around $30\%/\sqrt{E}$, validating the design's potential in meeting stringent physics requirements.

The SiPMs are critical components for photon detection, being selected due to compactness, and high PDE, which can exceed 50% at the relevant wavelengths (420–460 nm). Achieving high PDE is crucial for compensating the moderate intrinsic photon yield from the GS. Despite these advantages, SiPMs present challenges such as dark noise, with typical rates of tens to hundreds kHz per $mm^2$. Further work is required to implement effective strategies such as operational cooling, development of advanced electronics for noise filtering, and design of sophisticated data acquisition algorithms, in order to suppress dark noise and ensure high detector sensitivity.

Cost considerations significantly influence the technological strategies adopted for the GS-HCAL. With millions of readout channels envisioned, the cost-per-channel of SiPMs and scintillator production will substantially impact overall project economics. To address this, collaborative procurement strategies and industry partnerships should be further developed to manage SiPM costs, and advancements in scalable manufacturing processes for GS need to be pursued to reduce costs, with the goal of making GS economically competitive compared with traditional plastic alternatives.

Addressing mechanical challenges, the structural design has to ensure mechanical rigidity, precision alignment, and ease of assembly and maintenance. Innovations such as modular cassette-based designs are under development to enable manageable assembly and efficient replacement or servicing, which will be critical for maintaining detector performance over extended operational periods.

In summary, the GS-HCAL for CEPC aims to address multiple technological challenges through focused R&D efforts in scintillator optimization, photon detection technologies, cost-effective production methods, noise reduction techniques, and advanced mechanical design solutions. These ongoing strategic developments are essential to enable the realization of a highly granular, cost-effective, mechanically robust, and high-performance calorimeter required to realise ambitious physics goals.

## 8.4 Technology development and prototyping

The pursuit of high-performance hadron calorimetry has always been driven by the need to balance cost and energy resolution. Among the various scintillator materials explored, GS have emerged as a compelling candidate, offering a unique combination of high density, moderate light yield, and scalability—qualities essential for next-generation collider experiments. The evolution of GS in calorimetry reflects a series of iterative advances, from early exploratory designs to modern granular systems enabled by breakthroughs in photodetection and signal reconstruction. This section reviews the key milestones in glass-scintillator R&D, with particular emphasis on recent developments that have transformed their feasibility for large-scale HCAL.

### 8.4.1 Introduction

The conceptual foundation for GS in calorimetry dates back to the 1980s, spurred by the Crystal Clear Collaboration [4] at CERN, which pioneered investigations into high-Z glass matrices for ECAL. Among the earliest prototypes, the HED-1 GS calorimeter [4] demonstrated the material's potential despite limitations in radiation length and light output.

Subsequent efforts, such as CERN's exploration of Cerium-Doped Heavy Metal Fluoride Glass (HFG:Ce), further highlighted these challenges.

Its susceptibility to radiation damage rendered it impractical for high-luminosity experiments. These early attempts revealed a critical gap, the need for a material that could simultaneously deliver high density, radiation resistance, and efficient light transmission, a gap that remained unresolved for decades due to technological constraints.

A persistent limitation of traditional GS was their excessive self-absorption, which severely attenuated light output in long-bar geometries. This drawback stifled numerous proposals for scintillating glass-based calorimeters, as the resulting light yield did not meet the requirements for precise energy reconstruction. However, the maturation of SiPMs and the adoption of PFA in the 2010s redefined the feasibility of GS, which triggered IHEP to propose GS-HCAL [5, 6, 7]. The mass production of SiPMs drastically reduced the cost of high-efficiency photodetection, enabling the use of granular scintillator blocks where light collection no longer depended on long attenuation lengths. Concurrently, PFA-based calorimetry eliminated the need for longitudinal light transmission by leveraging fine-grained, cubic type of spatial segmentation and topological clustering, shifting the focus to compact, optically isolated cells. These innovations relaxed the stringent light yield requirements for GS, provided they could achieve a minimum threshold of 1000 ph/MeV and couple with SiPMs exhibiting $\geq$40% photon detection efficiency. This paradigm shift rendered previously impractical designs viable, reigniting interest in GS as a cost-effective alternative to crystals.

Recent advances in GS technology (post-2020) initiated by IHEP have focused on optimizing the density-light yield balance while ensuring scalability for large detectors. The GS-HCAL initiative, benefiting from the PS-HCAL design of the Calorimeter for the Linear Collider Experiment (CALICE) collaboration, employs GS cells with densities of approximately 6 g/cm$^3$ readout by SiPM arrays to achieve high-precision energy sampling.

Modern glass variants at a density of 6 g/cm$^3$ now achieve light yields of 1000–2000 ph/MeV despite inherent self-absorption, meeting stringent resolution requirements for particle-flow calorimetry [8, 9, 10]. These advancements incorporate activator doping techniques (e.g., $Ce^{3+}$) and matrix composition refinements (e.g., fluorides), producing glasses with decay times of about 500 ns. The technology benefits from parallel progress in mass production techniques across the field, including continuous melting and precision molding methods developed through academic-industrial partnerships.

The GS-HCAL project represents one important pathway in this global development effort, demonstrating the maturity of glass scintillator technology through systematic validation. International collaborations including the worldwide CERN Detector R&D Collaboration (DRD) program are addressing key challenges such as inter-cell uniformity, SiPM integration, and radiation tolerance. Recent milestones include successful testing of 40 mm × 40 mm × 10 mm cells and larger-scale validation, enable hadronic energy resolution of $\sigma/E \sim 30\%/\sqrt{E}$, advancing beyond conventional calorimetry limits. The following subsections detail the technical foundations supporting this progress, from material development to system integration.

## 8.4.2 Glass scintillator

The development of high-performance scintillation glass is fundamental to the success of the GS-HCAL project. Over the last seven years, the Glass Scintillator collaboration led by IHEP, consisting of universities, research institutes, and industries have thoroughly studied and developed various glasses [8, 9, 10]. Figure 8.10 shows the light yields versus density characteristics of the various GS samples invented in the past years by the Glass Scintillator collaboration and other groups. The Gadolinium Fluoro-Oxide (GFO) glass was selected ultimately as the baseline material due to its optimal balance between performance characteristics and manufacturability for large-scale production [11, 12, 13]. As detailed in Table 8.3, GFO glass achieves a density of 6.0 g/cm$^3$ with favorable radiation length (1.59 cm) and nuclear interaction length (24.2 cm), making it particularly suitable for hadronic shower measurements. While BGO crystals show superior absolute performance in some parameters, GFO's competitive light yield (~1500 ph/MeV), decay time (~500 ns), and lower production temperature (1250 ℃) present significant advantages.

The glass composition features $Ce^{3+}$ as the luminescent center, chosen for its strong emission, fast decay, and efficient energy transfer from $Gd^{3+}$ modifiers [14]. The borosilicate matrix ($B_2O_3$-$SiO_2$) provides excellent thermal and optical properties, with $Al_2O_3$ additives enhancing chemical stability. Gadolinium serves as the optimal heavy element modifier, offering high density while maintaining low radioactivity and cost-effectiveness compared with lanthanum or lutetium alternatives.

### 8.4.2.1 Light yield

Figure 8.11 shows the pictures of GFO glasses with a size of 40 mm × 40 mm × 10 mm under natural (such as regular room light) and ultraviolet light. Figure 8.12(a) shows the X-Ray Excited Luminescence (XEL) spectrum and transmission spectrum of a GFO glass. The transmittance in the visible light range exceeds 80%. Combined with the XEL spectrum, it can be observed that there is a certain self-absorption effect in the glass [17]. Additionally, as the glass thickness increases, its transmittance in the visible spectrum decreases from approximately 80% to around 75%. The measurable light output of GFO glass is a critical factor for HCAL applications. Extensive $\gamma$-ray testing has demonstrated that the light yield of GFO glasses can consistently exceed 1000 ph/MeV. When measured with an XP2020 (2 inch) Photomultiplier Tube (PMT), the detected photo-electron number reaches 1/3 that of BGO crystals with identical dimensions [18], as shown in Figure 8.12(b). Figure 8.12(c) presents the scintillation decay profile of GFO glass under $\gamma$-ray excitation. While maintaining a light yield of 1000 ph/MeV, the decay time of the slow component can be controlled to below 500 ns. This scintillation timescale is fully compatible with the high granularity of the GS-HCAL design, where the small cell size ensures excellent spatial resolution without requiring fast timing properties from the scintillator material itself.

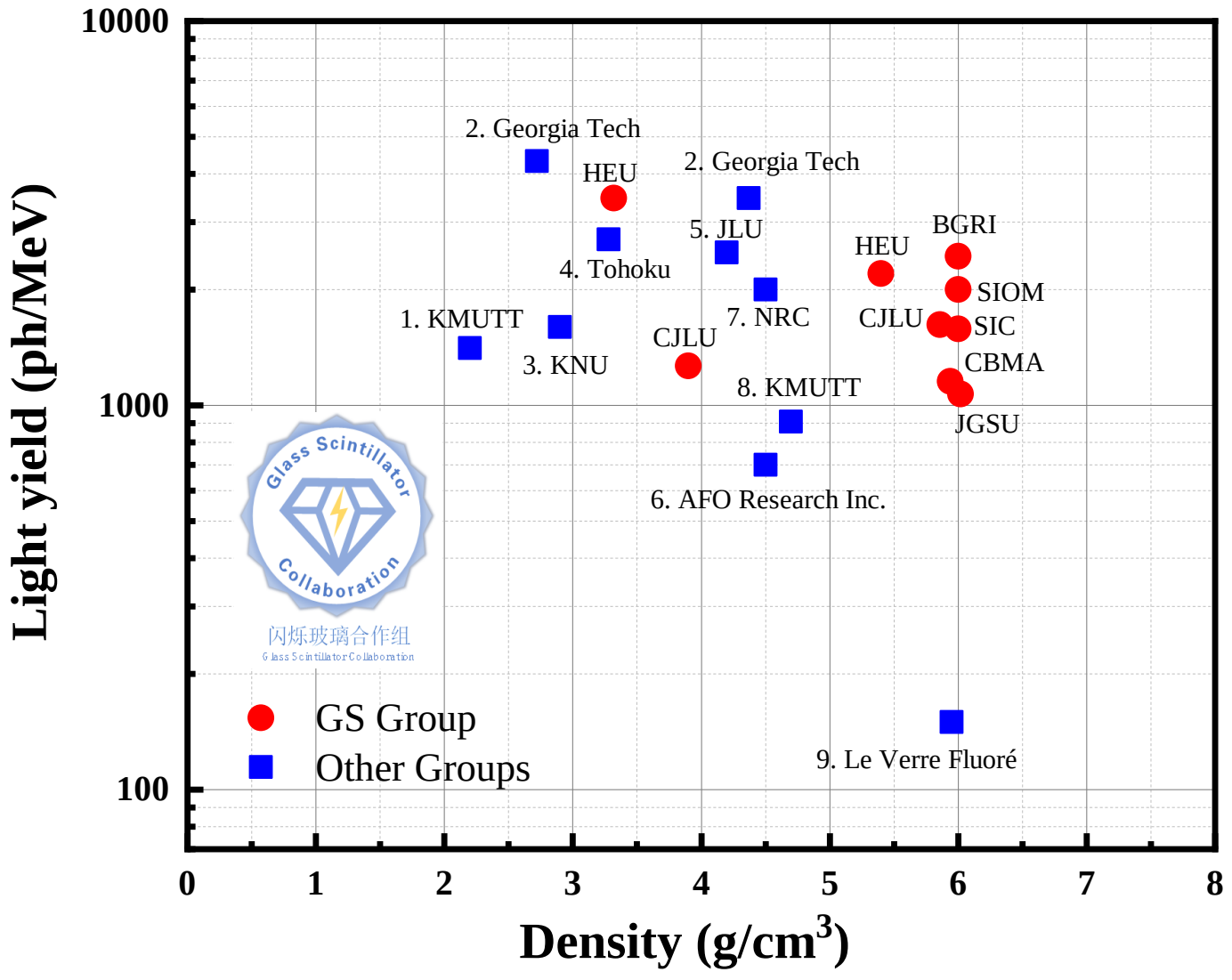

**Figure 8.10:** The light yields versus density characteristics of the various GS samples invented in the past years by the Glass Scintillator collaboration and other groups.

**Table 8.3:** Key parameters of the GFO GS with a size of 5 mm × 5 mm × 5 mm, compared with the BGO [15] and Barium Di-Silicate BaO-2SiO$_2$ (DSB) glass [16] of the same size.

| Key parameters | GFO glass | BGO | DSB Glass |
|---|---|---|---|
| Density (g/cm$^3$) | 6.0 | 7.13 | 4.2 |
| Melting point (°C) | 1250 | 1050 | 1550 |
| Radiation length (cm) | 1.59 | 1.12 | 2.62 |
| Molière radius (cm) | 2.49 | 2.23 | 3.33 |
| Nuclear interaction length (cm) | 24.2 | 22.7 | 31.8 |
| $Z_{eff}$ | 56.6 | 71.5 | 49.7 |
| $dE/dx$ (MeV/cm) | 8.0 | 8.99 | 5.9 |
| Emission peak (nm) | 400 | 480 | 430 |
| Refractive index | 1.74 | 2.15 | |
| Light yield (ph/MeV) | ~ 1500 | 7500 | 2500 |
| Energy resolution (% at 662 keV) | ~ 23 | 9.5 | |
| Scintillation decay time (ns) | ~ 60, 500 | 60, 300 | 90, 400 |

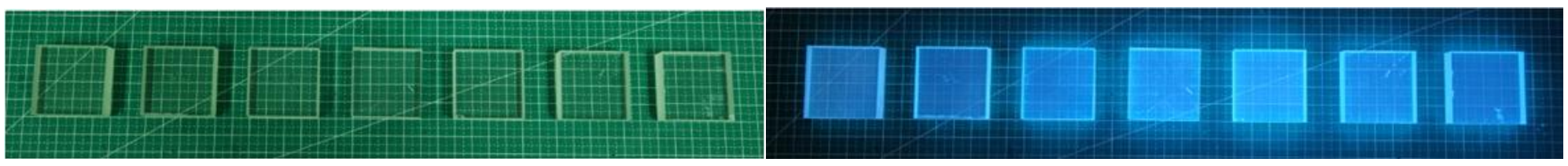

(a) (b)

**Figure 8.11:** GFO glass cells with the size of $40 \times 40 \times 10$ mm$^3$ under (a) natural light and (b) ultraviolet light.

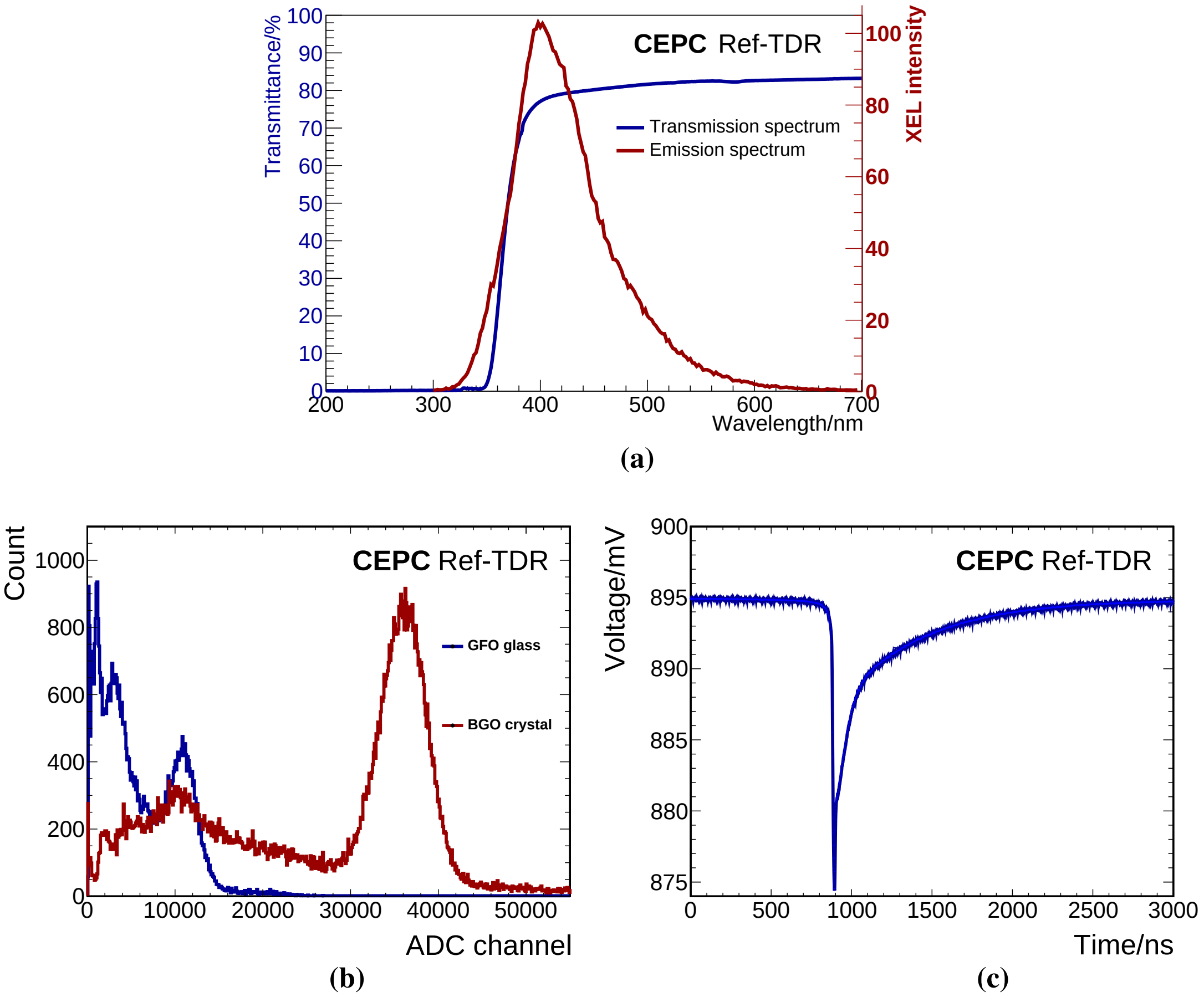

**Figure 8.12:** (a) Transmission (%) and XEL (%) spectra of large GFO glass, (b) Energy spectra of GFO GS and BGO crystal with same dimension. (c) Scintillation decay curve of the GFO glass.

### 8.4.2.2 Light attenuation length

The light attenuation length (LAL, $L_0$) serves as a fundamental parameter for assessing light transmission performance of scintillators. It can be determined through thickness-dependent transmittance measurements using the relation $L_0 = (L_2 - L_1)/\ln(T_{L_1}/T_{L_2})$ [19], where $L_1$ and $L_2$ denote two distinct sample thicknesses satisfying $L_2 > L_1$, $T_{L_1}$ and $T_{L_2}$ represent the transmittance values measured at thicknesses of $L_1$ and $L_2$, respectively. The variable $T_L$ is defined as the ratio of light intensity at thickness $L$ to the initial light intensity. This method accounts for full-volume light penetration while eliminating position uncertainty.

As shown in Figure 8.13, the best-performing GFO GS sample achieves a light attenuation length of approximately 6 cm at 400 nm. This wavelength corresponds to the emission peak of the GS sample. Across the visible spectrum beyond 400 nm, the light attenuation length remains stable, consistently ranging between 6 cm and 6.8 cm, ensuring efficient light collection. This result serves as the standard reference in the TDR for all subsequent performance evaluations.

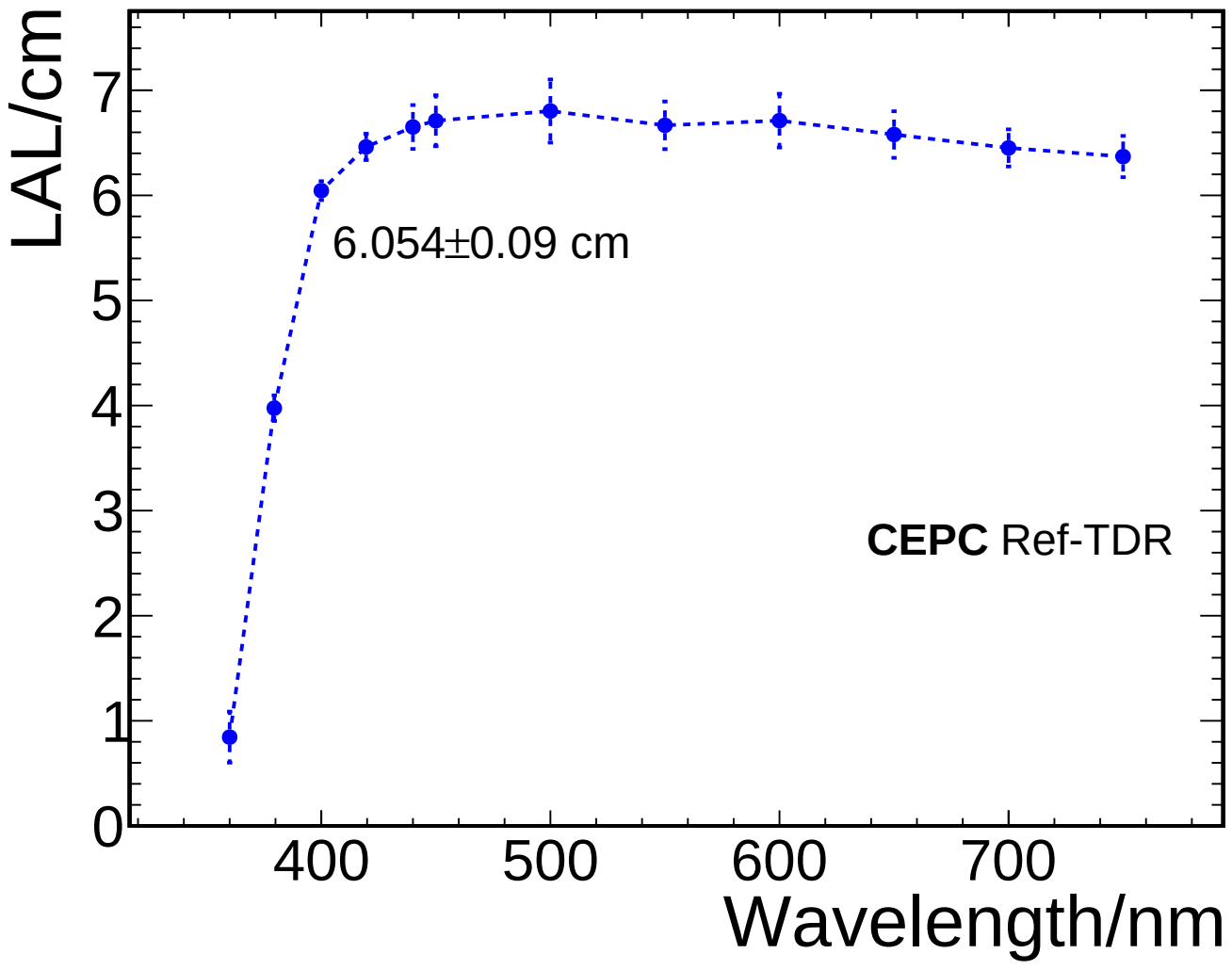

**Figure 8.13:** Attenuation length spectrum measured using transmittance method.

#### 8.4.2.3 Radiation tolerance

According to the current running plan, during the 10 year Higgs run, with a luminosity of $8 \times 10^{34}$ cm$^{-2}$s$^{-1}$, the irradiation dose reaches 40 Gy. For the Z run, the luminosity is $192 \times 10^{34}$ cm$^{-2}$s$^{-1}$, so a rough estimate suggests that one year of Z running would result in an irradiation dose of 96 Gy. As shown on Figure 8.14, the light yield can keep more than 70% compared with the value before any irradiation for a radiation dose below 100 Gy. This observation is confirmed by irradiation tests [20] using proton beams at the Associated Proton Beam Experiment Platform (APEP) [21].

#### 8.4.2.4 Quality control

The production of GS tiles with reproducible and controlled quality is an essential component of the technology validation plan. Approximately 290 samples with size of 40 mm × 40 mm × 10 mm were produced recently. The initial quality assessment indicates that about 60% of these tiles produced meet the target light yield specification exceeding 1000 ph/MeV. Variations in light yield between individual tiles are observed. These variations will be corrected through dedicated calibration procedures, including the LED monitoring system and MIP inter-calibration using muon beams or cosmic rays, to minimize their impact on the constant term of the energy resolution. This manufacturing experience provides a crucial foundation for scaling up production with better quality in the coming years.

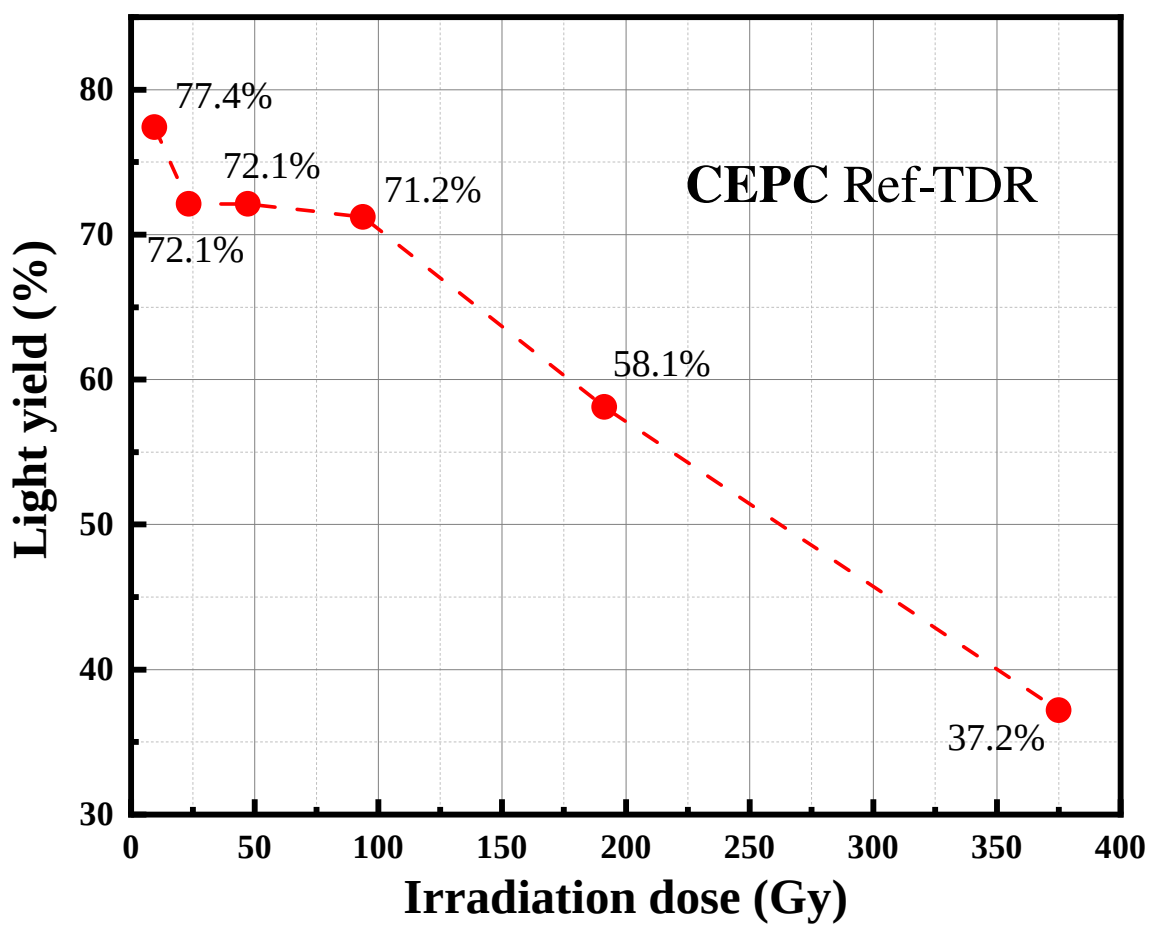

**Figure 8.14:** Gamma irradiation results showing the relative light yield of the scintillator as a function of irradiation dose. The light yield decreases gradually with increasing dose, reaching 71.2% at 100 Gy.

A major production run is scheduled for 2025-2026, aiming to manufacture approximately 10,000 GS tiles required for the full-size prototype. This mass production will serve as the platform for developing and refining the mass production techniques and quality control procedures necessary to ensure consistent performance throughout the GS-HCAL system. An automatic batch testing platform will be developed to quickly measure the properties of the GS tiles and to control their quality. The establishment of rigorous quality control metrics, including light yield uniformity, attenuation length verification, and dimensional tolerances, will be paramount to achieving the required energy resolution for both single hadron and jets.

### 8.4.3 Photon detector

SiPMs have gained significant importance in calorimetry applications due to their compact form factor, high gain, low operating voltage, excellent radiation hardness, and insensitivity to magnetic fields. The Compact Muon Solenoid (CMS) experiment at the LHC, for instance, is upgrading its endcap calorimeter with the High Granularity Calorimeter (HGCAL) [22]. The hadronic part (CE-H) of the HGCAL employs SiPMs to read out scintillator cells. Designed to withstand high radiation fluxes while delivering superior spatial resolution for particle showers, the HGCAL benefits from the inherent radiation tolerance and fine granularity of SiPMs. Similarly, the CALICE collaboration [23] has developed a highly granular calorimeter prototype for future collider experiments, utilizing SiPMs coupled to plastic scintillator cells in electromagnetic calorimeters. Their scalability and compactness make SiPMs ideal for systems requiring PFA.

The CMS experiment has further replaced its Hybrid Photodiodes (HPDS) with SiPMs in the barrel hadronic calorimeter [22], capitalizing on their improved timing resolution, higher PDE, and enhanced radiation resilience. SiPMs are also being investigated for dual-readout calorimeters, where simultaneous measurement of scintillation and Cherenkov light from particle showers could improve energy resolution.

Commercially available SiPMs are offered by several manufacturers, including HPK, Onsemi (SensL), and Chinese producers such as Novel Device Laboratory (NDL) and Rayquant (JBT), as summarized in Table 8.4. Key SiPM parameters include pixel pitch size, pixel count, and active area, which directly influence performance metrics like dark current and gain.

The PDE of SiPMs is wavelength-dependent, typically peaking around 420 nm. For optimal performance, the PDE needs to align with the emission spectrum of the scintillator material. For example, the GFO GS emits at 400 nm, necessitating high PDE at this wavelength. Dark count rates, typically on the order of 100 kHz/mm$^2$, introduce noise and scale with pixel size, requiring careful mitigation in high-precision systems.

For the GS-HCAL application, the primary selection criteria include high PDE at 400 nm, high gain, low noise, and cost-effectiveness, with less stringent dynamic range and timing resolution requirements compared with the ECAL. Unlike the ECAL design that utilizes precise timing for PFA clustering, the 40 mm $\times$ 40 mm $\times$ 10 mm cell segmentation of the GS-HCAL inherently provides 3D spatial information for clustering without requiring stringent timing resolution. Based on current performance metrics, the HPK S14160-3050 and NDL EQR20-11-3030 are promising candidates for prototype testing. However, rapid advancements in semiconductor technology are expected to yield even more optimized SiPMs tailored for hadron calorimetry, warranting continued evaluation of emerging devices.

**Table 8.4:** The parameters of some typical commercial SiPMs.

| **Supplier** | **HPK** | | **NDL** | **JBT** |
|---|---|---|---|---|
| Type | S14160-3015PS | S14160-3050HS | EQR20-11-3030-S | JSP-TP3050-SMT |
| Pitch [$\mu$m] | 15 | 50 | 20 | 50 |
| Number of pixels | 39984 | 3531 | 22500 | 3364 |
| Terminal capacitance [pF] | 530 | 500 | 157.5 | 170 (fF) |
| Breakdown voltage ($V_B$) [V] | 38 ± 3 | 38 ± 3 | 27.2 ± 1 | 24.6 ± 0.2 |
| Recommended operation voltage [V] | $V_B + 3$ | $V_B + 2.7$ | $V_B + 5$ | $V_B + 2$ |
| Peak sensitive wavelength [nm] | 450 | 450 | 420 | 420 |
| Peak PDE at PSW [%] | 32 | 50 | 47.8 | 35 |
| PDE at 400 nm [%] | 27 | 47 | 45 | 33 |
| Gain | $3.6 \times 10^5$ | $2.5 \times 10^6$ | $8.0 \times 10^5$ | $2.1 \times 10^6$ |
| DCR [kHz/mm$^2$] | 700–2100 | 10–100 | 150–450 | 120–270 |

The high DCR of SiPMs represents a critical consideration in applications involving weak signals. For GS-HCAL, the relatively low light yield of GS presents a challenge in achieving the desired 0.1 MIP detection threshold. Several methods exist for suppressing SiPM dark noise, including temperature reduction and electronic threshold adjustment. However, since the HCAL will operate at room temperature (approximately 21 °C), which provides negligible noise reduction, threshold optimization emerges as the most practical approach for noise suppression.

Figure 8.15 illustrates the DCR characteristics as a function of threshold level (0–10 p.e.). Measurements indicate a baseline DCR of 255 kHz/mm$^2$ at 0 p.e. threshold for the NDL EQR20 SiPM, compared with 112 kHz/mm$^2$ for the HPK S14160-3050 SiPM. Given the GS light output specification of 60 p.e./MIP, a 0.1 MIP threshold corresponds to approximately 6 p.e.. The data demonstrate significant DCR suppression at higher thresholds: at 5 p.e., the DCR reduces to approximately 28 Hz/mm$^2$ and 12 Hz/mm$^2$ for the NDL and HPK devices, respectively. Extrapolating to the 10 p.e. operational threshold suggests even greater noise reduction for the standard 3 mm × 3 mm SiPMs used in the HCAL design.

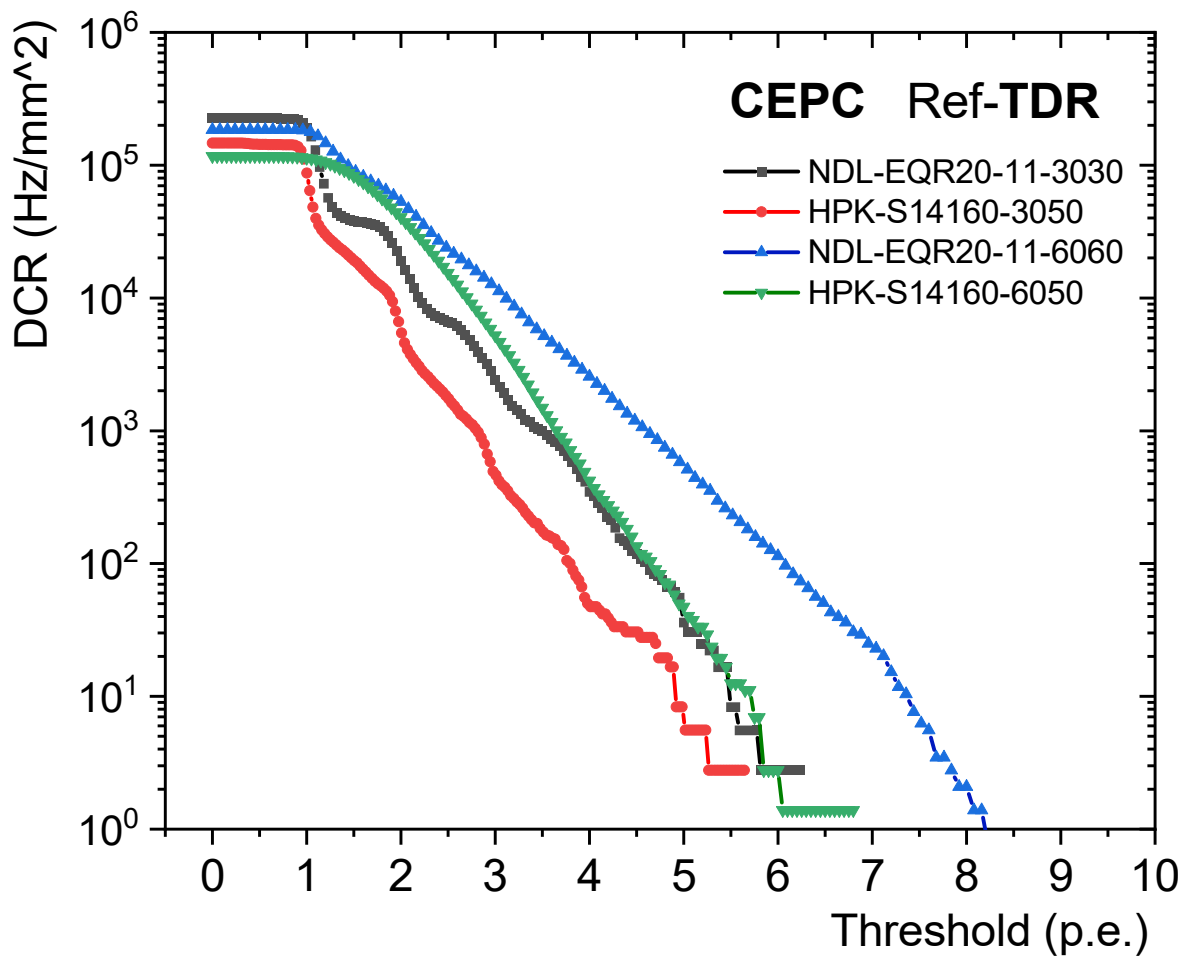

**Figure 8.15:** Dark counting rates of several SiPMs as a function of threshold.

### 8.4.4 Simulation study of attenuation length

We conduct GEANT4-based optical simulations of the GS coupled to SiPMs to characterize performance and identify optimization pathways. The standalone simulation framework enables detailed photon tracking, with particular emphasis on light output and attenuation length, which are key parameters for scintillator performance. The study focuses on how the GS attenuation length affects photon detection efficiency at the SiPMs, providing crucial data for material optimization. When ionizing particles deposit energy in the GS, a portion is converted to detectable visible light, similar to conventional scintillators. The simulations fully model this process from energy deposition to photon detection.

The detector geometry implements the baseline design of the 40 mm × 40 mm × 10 mm GS cell with a single 3 mm × 3 mm SiPM positioned at the cell center inside the readout PCB. The entire periphery is wrapped in Teflon reflective film. All characteristics listed in Table 8.3 are incorporated with particular attention to attenuation length and emission peak. The physics processes include a physics list ('QGSP-BERT') and optical processes. The physics list accounts for ionization, bremsstrahlung, multiple scattering, pair production, Compton scattering, and photoelectric effects. The optical processes include scintillation and Cherenkov light generation, Rayleigh scattering, bulk absorption, and boundary processes. Both the GS and SiPM are designated as sensitive detectors, with the GS interacting with muons while the SiPM detects emitted photon

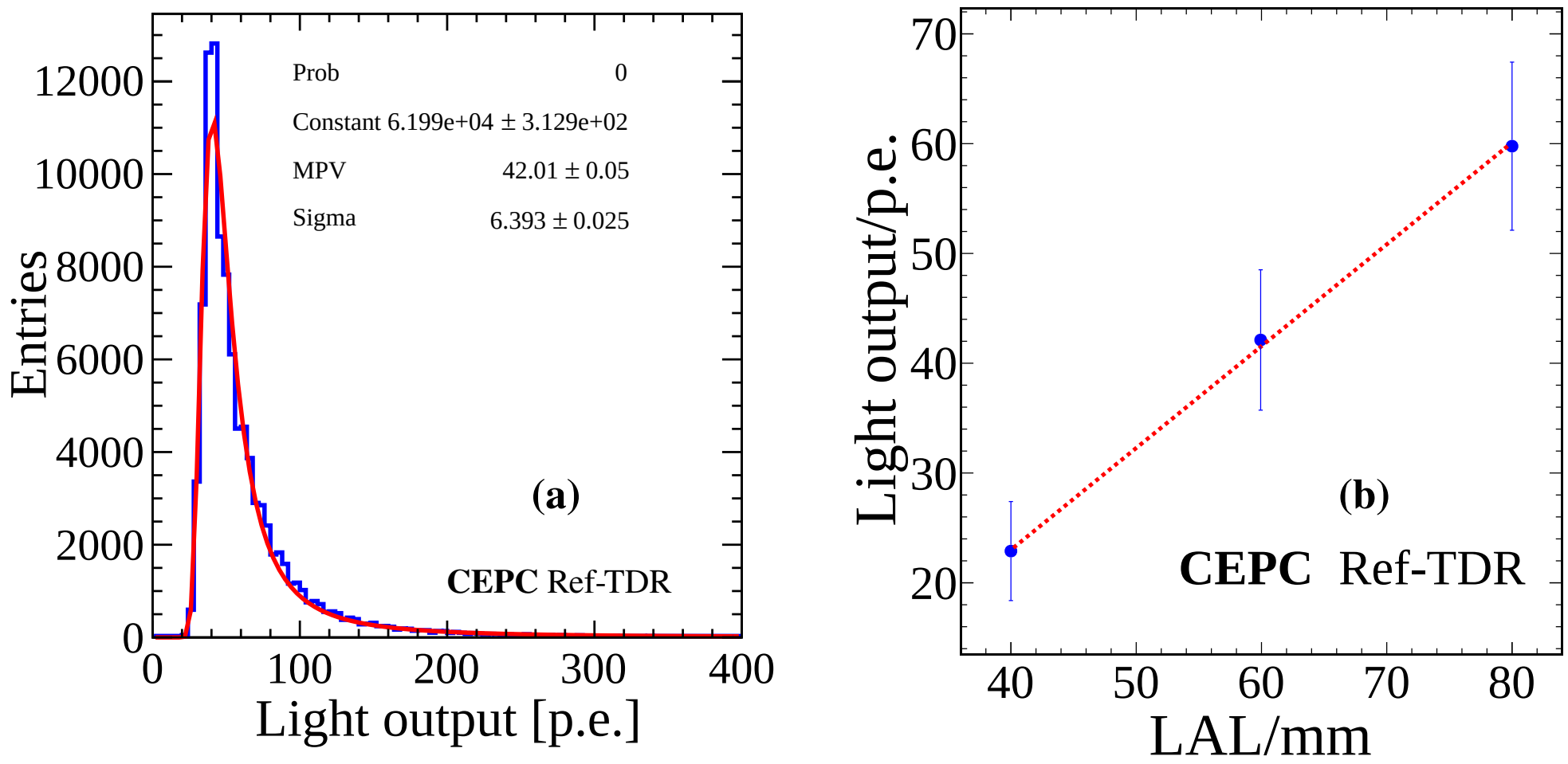

**Figure 8.16:** (a) The resulting light output (p.e.) distribution of simulated GS cell assuming an attenuation length of 60 mm. (b) The fitted light output MPV results as a function of different attenuation lengths such as 40, 60, and 80 mm for simulated GS cells.

Figure 8.16(a) shows the light output distribution for simulated GS cells with 60 mm attenuation length, yielding a fitted effective light output of 42.05±0.01 p.e./MIP. Figure 8.16(b) demonstrates the expected increase in light output MPV with longer attenuation lengths, as the reduced photon absorption naturally preserves more scintillation light. The reflective film wrapping ensures efficient photon collection by the SiPM, with the 60 mm attenuation length close to that of the measurements.

Validation tests revealed differences between simulation and experimental measurements. Figure 8.17(a) shows the simulated photon arrival time distribution with fast and slow components of 96 ns and 1445 ns, respectively, comparably longer than the reference values of 60 ns and 500 ns from Table 8.3. Figure 8.17(b) shows the deposited energy distribution for 5 GeV muons, and the most probable value from the Landau fit is approximately 7.3 MeV/MIP, in good agreement with theoretical expectations.

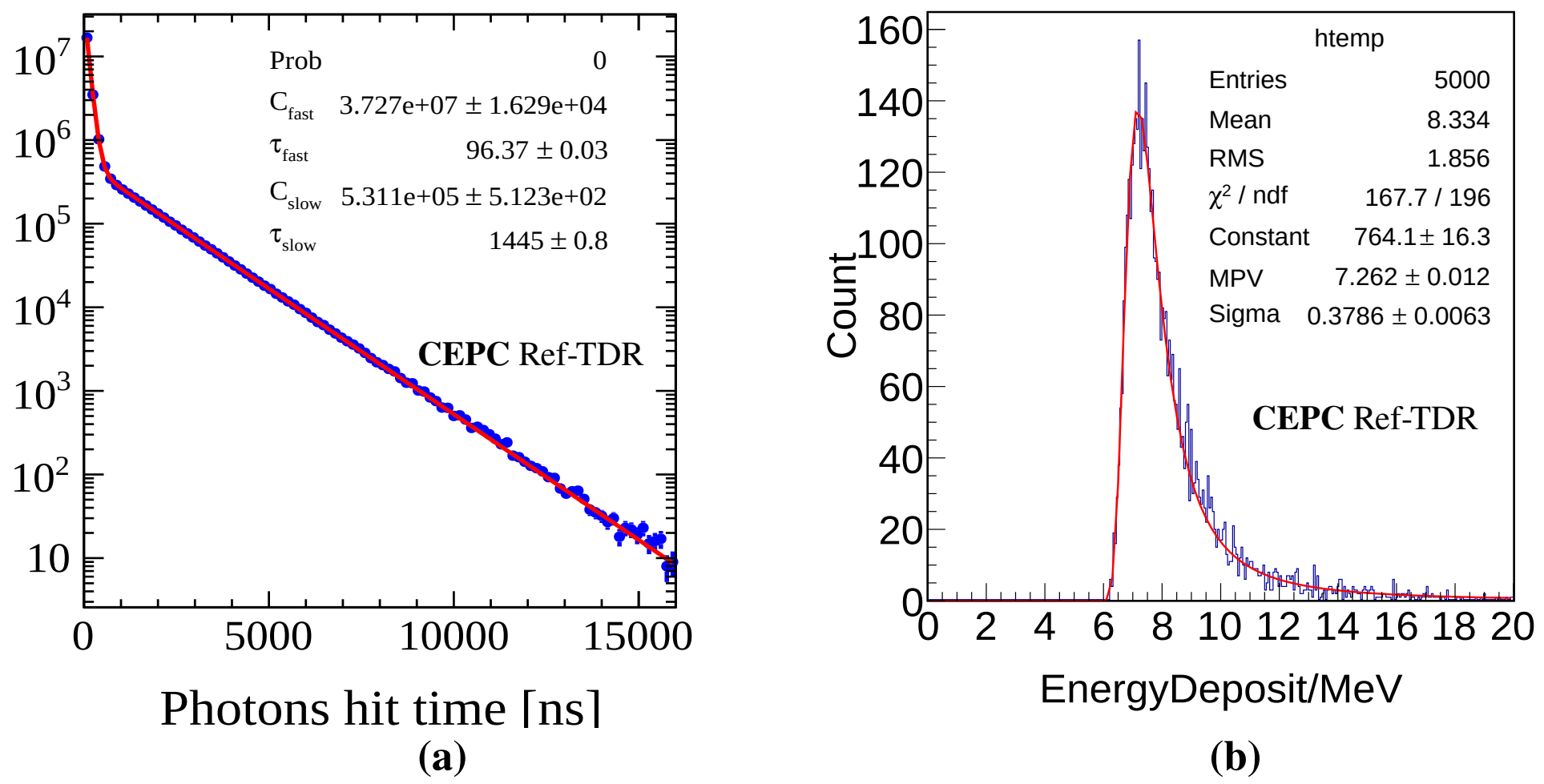

**(a)** **(b)**

**Figure 8.17:** The results checked to validate the physical process in simulation, where (a) is the signal decay time (fast and slow components), and (b) is the muon deposited energy in GS cell (10 mm thick).

### 8.4.5 Measurements

Several measurements were performed using GS samples of 40 mm × 40 mm × 10 mm to evaluate the effective light output under different configurations. Three distinct measurement approaches were implemented: one utilizing 662 keV gamma rays from a $^{137}$Cs radiation source, another employing cosmic ray muons with expected energy deposition of 7.7 MeV/MIP, and a third using 5 GeV electron beams at KEK. In all cases, the measurements were conducted using SiPM sensors mounted at the center of a GS cell on readout boards equipped with pre-amplifiers, with different sensor models optimized for each specific measurement scenario.

For the $^{137}$Cs measurement, the HPK S14160-3050HS SiPM is selected, offering a weighted PDE of 44.5%. The measurements of cosmic rays were conducted using HPK S13360-3050PE SiPM with a weighted PDE of 32.5%. The electron beam test employed an NDL EQR20 11-3030 SiPM with a weighted PDE of 42.5%. The weighted mean value of PDE refers to that shown in Figure 8.12(a).

Prior to these measurements, comparative tests of ESR and Teflon reflective films demonstrated superior performance of Teflon, showing approximately 20% higher reflectivity than ESR films. This enhancement is due to the diffused reflectivity and the softer composition of Teflon enabling better optical contact with the glass surface. ESR films may introduce slight air gaps and exhibit particularly poor reflectivity in the 380–400 nm range [24], which is around the peak of the GS scintillation light spectrum. All measurements employ Teflon films as reflective material.

The $^{137}$Cs source is positioned at the center of the opposite face of the GS cell, resulting in a maximal 10 mm propagation distance between source and sensor. The measured spectrum is shown in Figure 8.18(a). With a calibrated conversion factor of 208 ADC/p.e. and the known 662 keV gamma energy of $^{137}$Cs, the derived light output reaches 12.9 p.e./MeV, corresponding to about 99 p.e./MIP.

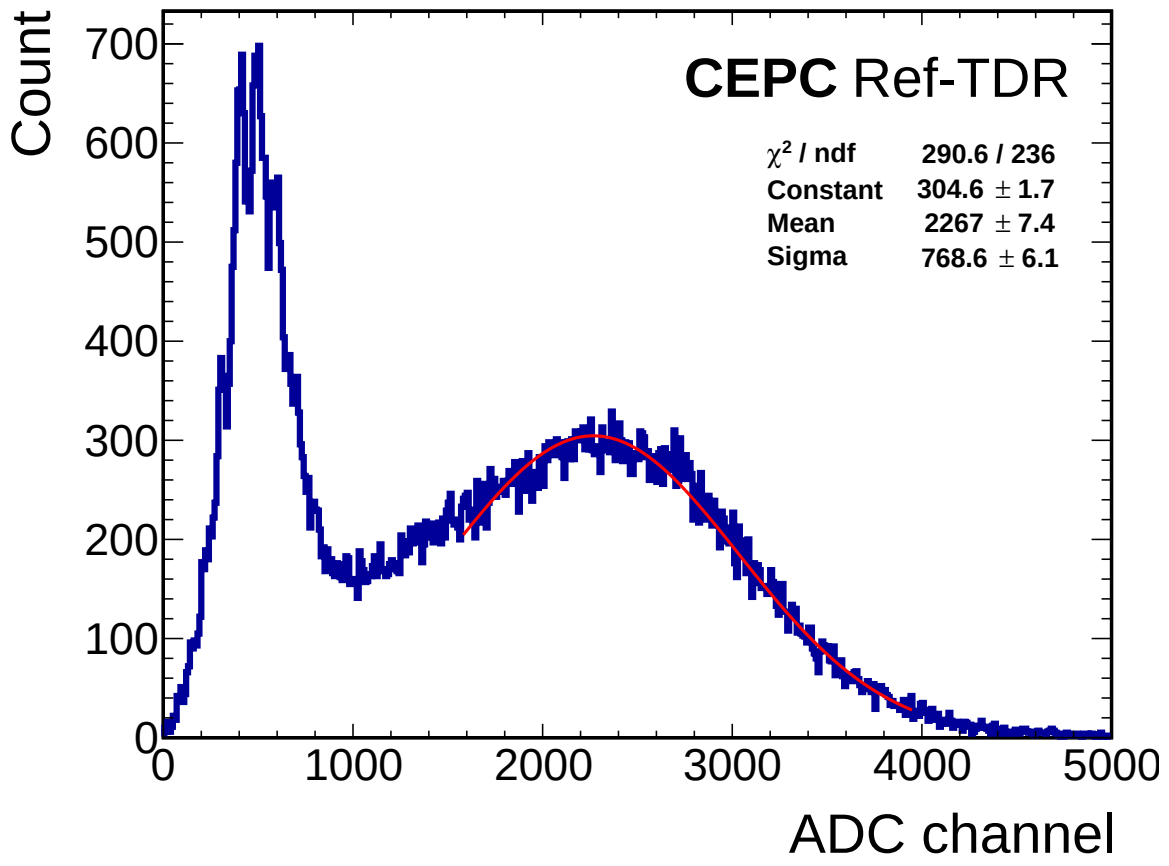

(a)

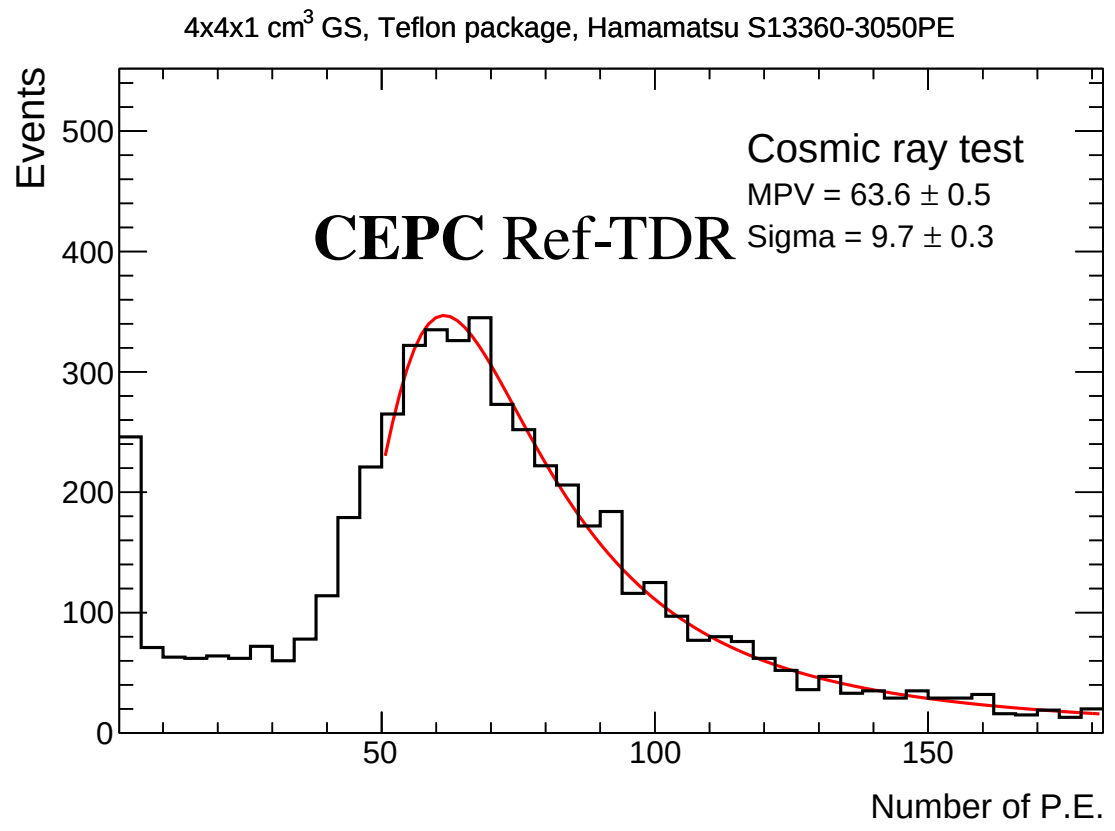

(b)

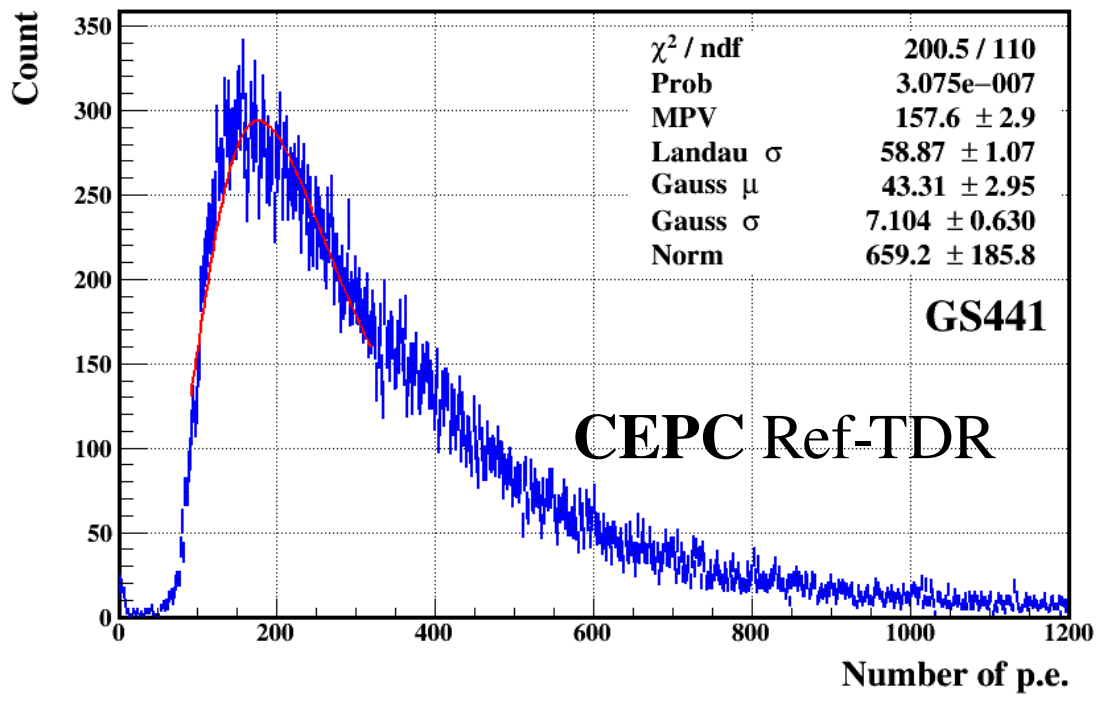

(c)

**Figure 8.18:** Measurement results of the GS cell with SiPM readout, the same GS cell size of 40 mm × 40 mm × 10 mm is used. (a) The measured ADC spectrum of GS cell with a SiPM (HPK S14160-3050HS) using $^{137}$Cs radiation source. (b) Cosmic ray test results measured using HPK S13360-3050PE SiPM, resulting a light output about of 64 p.e./MIP. (c) Response to 5 GeV electrons measured at the KEK beam test facility using an NDL EQR20-11-3030 SiPM. A Landau fit yields a light output of 157.6 ± 2.9 p.e./MIP.

The cosmic ray measurement configuration employs two plastic scintillator trigger panels surrounding the GS cell, with an upper panel sized 35 cm × 5 cm × 1 cm and a lower panel sized 10 cm × 5 cm ×1 cm. Coincident signals from both plastic scintillator panels trigger the readout of GS cell. As shown in Figure 8.18 (b), the cosmic ray test results of GS+SiPM gives a light output of 63.6 ± 0.5 p.e./MIP.

For the electron beam test, 5 GeV electrons were directed onto the center of the GS cell. The resulting spectrum, measured using the NDL SiPM, is shown in Figure 8.18(c). A Landau fit yields a most probable value (MPV) of 157.6 ± 2.9 p.e./MIP for the light output.

The higher light yield from $^{137}$Cs measurements compared with cosmic rays arises from different scintillation light propagation paths. The $^{137}$Cs setup has a shorter light propagation path of about 10 mm while cosmic muons generate light across the full 40 mm × 40 mm GS cell, with an average propagation path of about 20 mm to reach the SiPM.

The significantly higher light output observed with 5 GeV electrons, compared with both the $^{137}$Cs source and cosmic muons, is expected and stems from the fundamental difference in energy deposition. A relativistic muon (~MIP) deposits energy primarily through ionization along a straight track. In contrast, a high-energy electron initiates an electromagnetic shower within the scintillator volume. Although the 10 mm thickness of the GS cell is insufficient to fully contain a 5 GeV shower, leading to significant energy leakage, the localized energy density deposited within the cell is vastly greater than that of a single MIP. This results in a much higher scintillation light yield, despite attenuation effects, consistent with the observed measurement.

When comparing with the simulation results in Section 8.4.4, the measured cosmic ray light output of 64 p.e./MIP exceeds the simulated value of 42 p.e./MIP. This discrepancy may originate from longer simulated decay times and lower MIP response as described before. Simulation result may also be affected by uncalibrated parameters for energy deposition and optical photon transportation, since experimental data from the newly developed GS have never been incorporated into the GEANT4 fine-tuning.

As this represents early stage research of the GS-HCAL, substantial work remains to improve the simulation accuracy, including refining the GEANT4 physics list for GS and incorporating more experimental data to better match observed performance.

## 8.4.6 Calibration

### 8.4.6.1 Calibration using physics processes

Several physics processes can produce isolated hadrons, such as hadronic $\tau$ decays in $Z \to \tau^+\tau^-$ events or isolated hadrons within jets. By leveraging particle flow reconstruction, the tracker can precisely determine the energy of these hadrons. Applying a veto on energy deposits in the ECAL allows for the selection of pure hadronic showers in the HCAL. The $E/p$ ratio of these showers can then be used for calibration purposes. This method is particularly suitable for time-dependent monitoring of the HCAL response, though not for absolute channel-by-channel calibration.

The global calibration of the GS-HCAL can be performed using MIP signals from prompt muons produced in collisions. These muons, identified by the inner tracker and muon chamber systems, predominantly lose energy through ionization, with a response that scales nearly linearly with their path length through the HCAL cells. The $e^+e^- \to Z \to \mu^+\mu^-$ process provides an ideal source of such high-momentum isolated muons. Figure 8.19 shows the kinematic distributions of muons at $\sqrt{s} = 91.2$ GeV.

During Z-pole running, this process occurs at a rate of 584 Hz at the design instantaneous luminosity. A conservative estimate yields 47 events per hour within a $\Delta\phi \times \Delta\cos\theta = 0.018 \times 0.018$ acceptance window, corresponding to one 4 cm × 4 cm GS cell at the central barrel

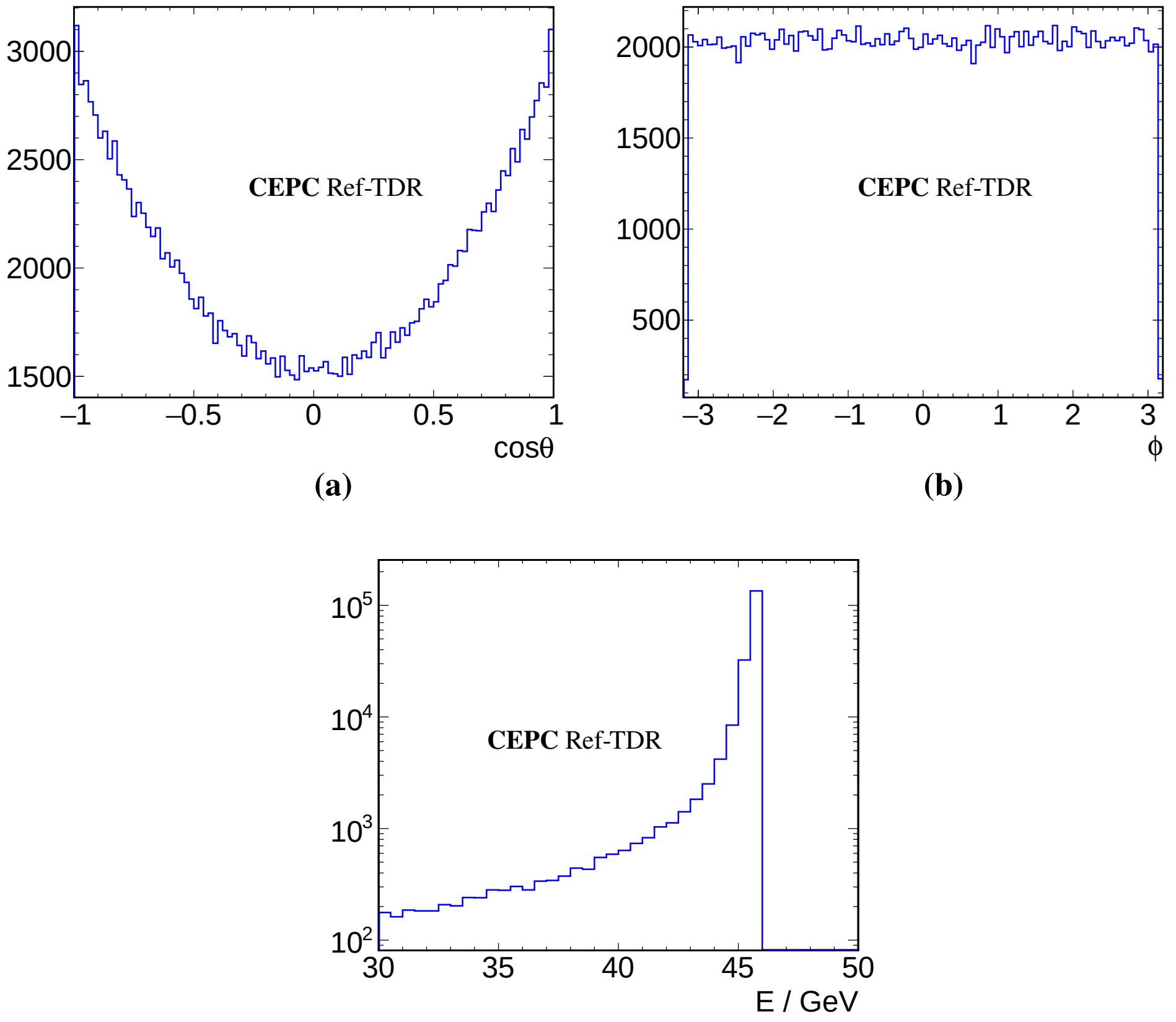

(a) (b) (c)

**Figure 8.19:** Generator-level distributions of (a) $\cos\theta$, (b) $\phi$ and (c) energy of the muons in $e^+e^- \to \mu^+\mu^-$ process at $\sqrt{s} = 91.2$ GeV.

HCAL surface. The event rate increases significantly with $\cos\theta$. Assuming regular two-week calibration runs in Z mode (16 hours per day), we can collect $4.71 \times 10^8$ di-muon events for HCAL calibration, achieving a statistical precision of about 1‰.

The dominant process in Z-pole operation, $e^+e^- \to Z \to q\bar{q}$, provides resonant jets that enable combined ECAL and HCAL calibration. Figure 8.20 shows the invariant mass distribution of the dijet system of the $Z \to q\bar{q}$ process. The invariant mass $m_{jj}$ calculated from the reconstructed jet energies ($E_{\text{jet1}}$, $E_{\text{jet2}}$) should match the known $Z$ boson mass, at least for light-quark jets. Here $E_{\text{jet}} = a \times E_{\text{ECAL}} + b \times E_{\text{HCAL}}$ represents the energy reconstruction for individual jets. This method serves as an important cross-check of detector performance across the full acceptance by examining objects in different phase space regions.

#### 8.4.6.2 Hardware calibration

The GS-HCAL requires precise performance monitoring to compensate for potential radiation damage and aging effects in its GS and SiPM components. A critical aspect is the correction of inter-tile light yield variations, which directly impact the constant term of the energy resolution. The proposed calibration system employs two complementary approaches, a MIP inter-calibration using cosmic rays or muon beams, and an online LED-based monitoring

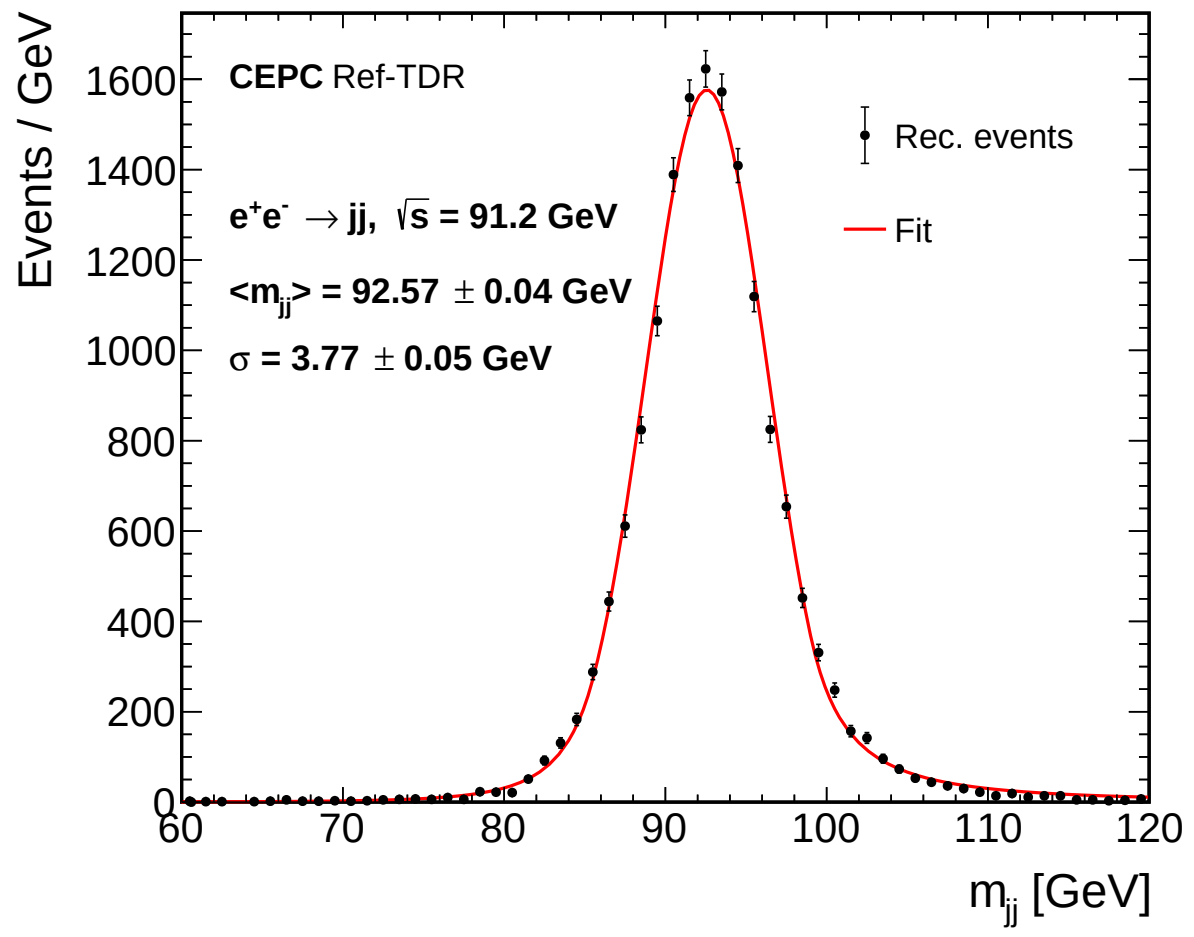

**Figure 8.20:** Distribution of dijet invariant mass of $Z \longrightarrow q\bar{q}$ process at $\sqrt{s}$ = 91.2 GeV.

system.

The MIP inter-calibration utilizes the well-defined energy deposition of minimum ionizing particles from cosmic rays or dedicated muon beams. This provides an absolute energy scale reference that allows for inter-calibration of response variations between different GS tiles, ensuring uniform energy measurement across the entire calorimeter system.

The LED monitoring system uses a centrally located LED with optical fiber distribution to multiple scintillator cells, enabling tracking of pedestal levels, gain characteristics, and response linearity across the detector. Figure 8.21 illustrates the system architecture where a programmable pulse generator drives the LED with adjustable voltage and pulse width settings. Photons propagate through fibers into each GS, traversing the scintillator material before detection by individual SiPM components. This configuration simultaneously characterizes GS optical properties and SiPM performance while maintaining a compact geometry with minimal cabling requirements.

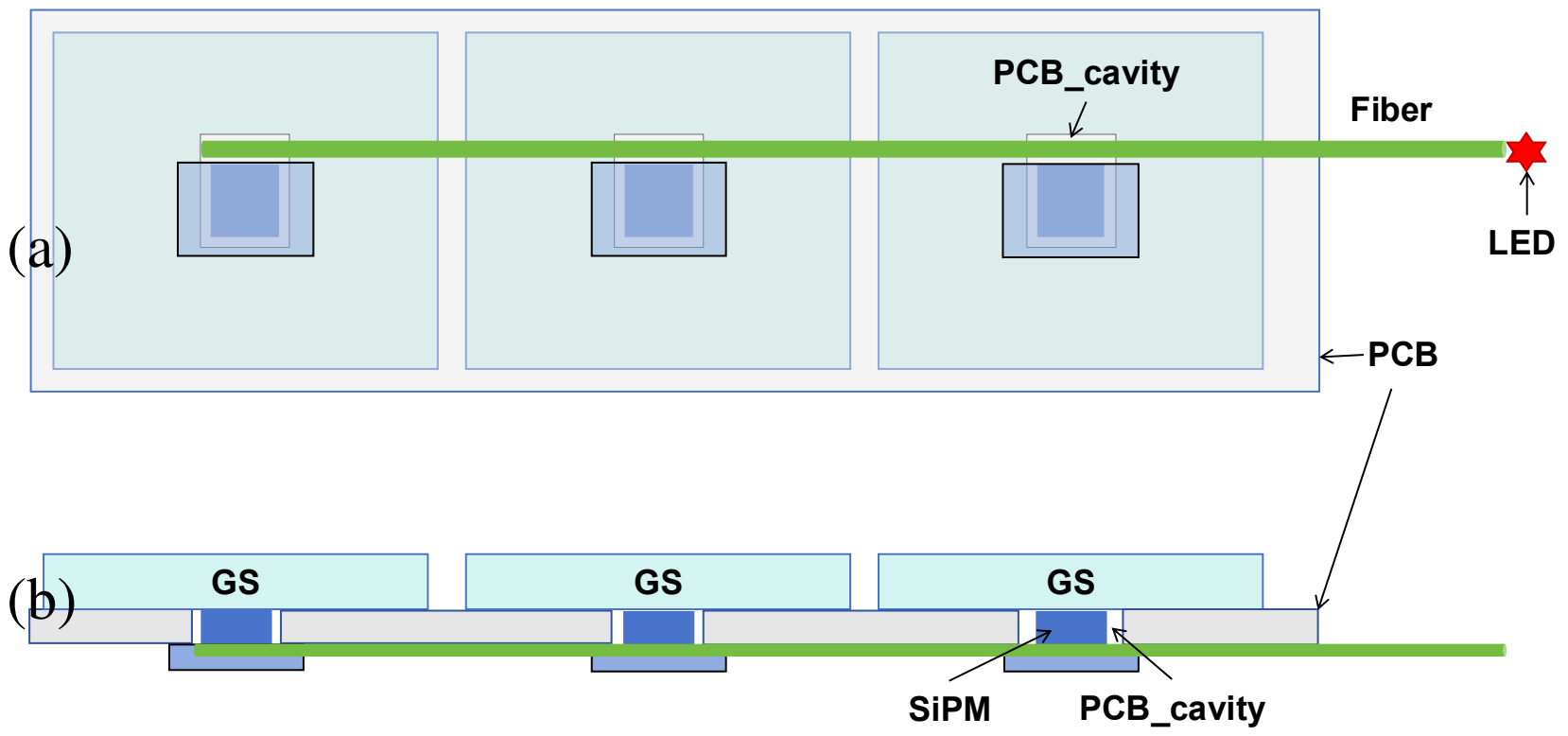

**Figure 8.21:** Schematic diagram of the GS-HCAL calibration system showing the optical fiber connections between adjacent cells, with (a) displaying the top view and (b) showing the side view of the LED light distribution mechanism.

The programmable pulse generator controls the LED light output and triggers the DAQ

acquisition system. During operation, the DAQ interface scans through voltage and pulse width settings while recording SiPM responses to determine light yields, attenuation characteristics, and gain stability. The photon path through fibers and scintillator material provides sensitivity to both component degradation and long-term performance variations, with ongoing optimization of calibration procedures for aging effects.

### 8.4.7 Readout electronics for R&D

The HCAL design incorporates more than five million channels of SiPMs, requiring front-end electronics for signal processing and readout. The readout system consists of several key components: a stable, low-noise power supply to provide SiPM bias voltage; a charge-sensitive pre-amplifier for the SiPMs; and a shaping stage to optimize the signal-to-noise ratio and timing resolution through analog to digital converter. Subsequent processing includes discriminators for photon counting and timing measurements. The system also incorporates ADCs to digitize pulse amplitudes for energy measurements. The design of each component emphasizes stability, low noise, and high channel density to meet the requirements.

**Table 8.5:** SiPM characteristics that drive the requirements for the readout electronics design.

| **Item** | **Requirement** |
|---|---|
| Charge dynamic range | 0.8 pC~800 pC (0.1~100 MIPs) |
| Charge resolution | 10% of 1.0 MIP |
| SiPM capacitance | $\leq$ 100 pF |
| SiPM gain | $\geq 5\times10^5$ |
| Average event rate/channel | 2 kHz |
| Max event rate/channel | 50 kHz |
| Typical signal | 2~3 mV/p.e. |

To enable immediate SiPM characterization and HCAL prototype testing, an interim FEE solution [25] was developed as commercial options proved inadequate. This customized system supports various commercial SiPM configurations during research and development while providing design references for the final GS-HCAL electronics. The key specifications for this solution are detailed in Table 8.5.

Figure 8.22 shows the pre-amplifier developed for the GS-HCAL prototype. The pre-amplifier is implemented on a compact PCB (Figure 8.22(a)), designed to couple directly to the GS cell, and its schematic diagram is shown in Figure 8.22(b). The circuit features a gain of $\pm$20 V/V, a bandwidth of 400 MHz, and a baseline noise of 300 μV RMS, with an input impedance of 50 Ω. The performance of the pre-amplifier is demonstrated in Figure 8.22(c) and 8.22(d), showing clean multi-photoelectron signals. Figure 8.22(e) displays a typical waveform from a 3 mm $\times$ 3 mm SiPM (EQR20-11-3030D-S) coupled with the pre-amplifier, while Figure 8.22(f) presents the ADC spectrum of the SiPM dark noise, demonstrating well-separated photoelectron peaks. The implemented solution achieves high-speed circuitry that preserves signal integrity across the full dynamic range.

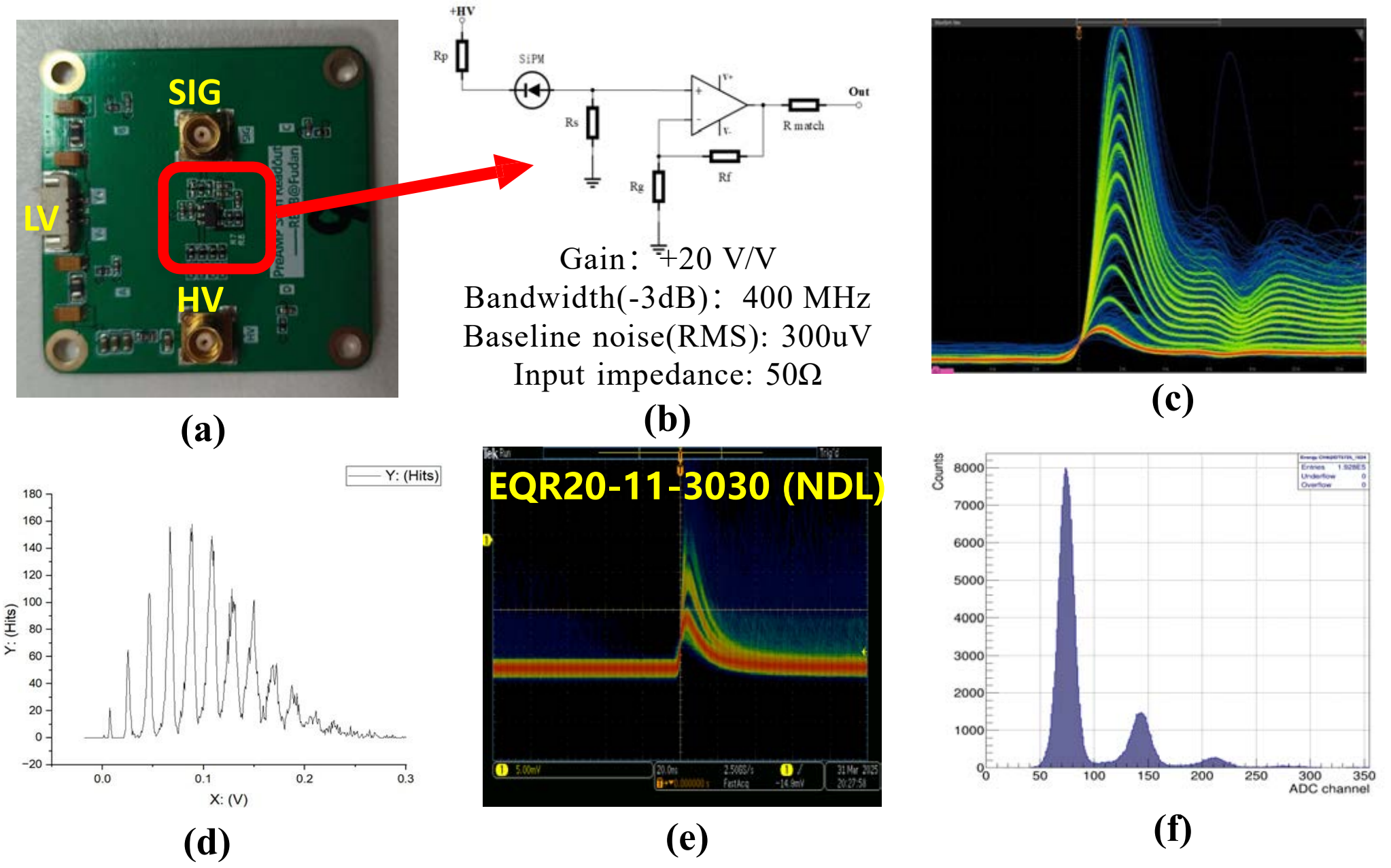

**Figure 8.22:** The pre-amplifier for HCAL SiPM study. (a) one pre-amplifier channel PCB coupling to GS cell. (b) the schematic diagram of the pre-amplifier. (c) and (d) the performance of the pre-amplifier. (e) Signal of one 3 mm × 3 mm SiPM (EQR20 11-3030D-S) coupling with a readout PCB. (f) ADC distribution of dark noise of a 3 mm × 3 mm NDL SiPM.

### 8.4.8 Prototype

A prototype of the GS-HCAL was proposed to be developed following the single-layer design as illustrated in Figure 8.2 in Section 8.2.1. This prototype implements a steel-scintillator sandwich structure in a compact 0.6 $m^3$ volume (52.0 cm × 52.0 cm × 130.6 cm). The primary objectives are to validate the novel GS technology at larger scale and optimization of the PFA for the GS-HCAL. This prototype contains 8112 GS cells arranged in 48 layers with total thickness of about $6\lambda_I$, serving as a critical testbed for the full GS-HCAL systems. Each scintillator layer consists of 169 cells (13×13 array) with individual cell dimensions of 40 mm × 40 mm × 10 mm, optically coupled to 3 mm ×3 mm SiPMs for readout. The steel absorber plates use S304 stainless steel (GB/T 4237-2015, equivalent to AISI 304) with 9.8 mm thickness. The system employs SiPM photodetectors with more than 30% photon detection efficiency and a gain greater than $10^5$, selected based on their performance in previous experiments like GECAM [26] and JUNO-TAO [27, 28].

Simulation studies using 80 GeV proton beams guided the prototype design optimization. As shown in Figure 8.23, simulations determined that a 13×13×48 cell configuration achieves 95% energy containment while balancing cost and performance. Despite this optimization, energy leakage remains significant due to the prototype's reduced size, leading to expected resolution degradation of the constant term (Figure 8.24). The deposited energy distribution demonstrates that most energy deposits concentrate along the beam center, with the selected prototype size approaching full energy deposition levels.

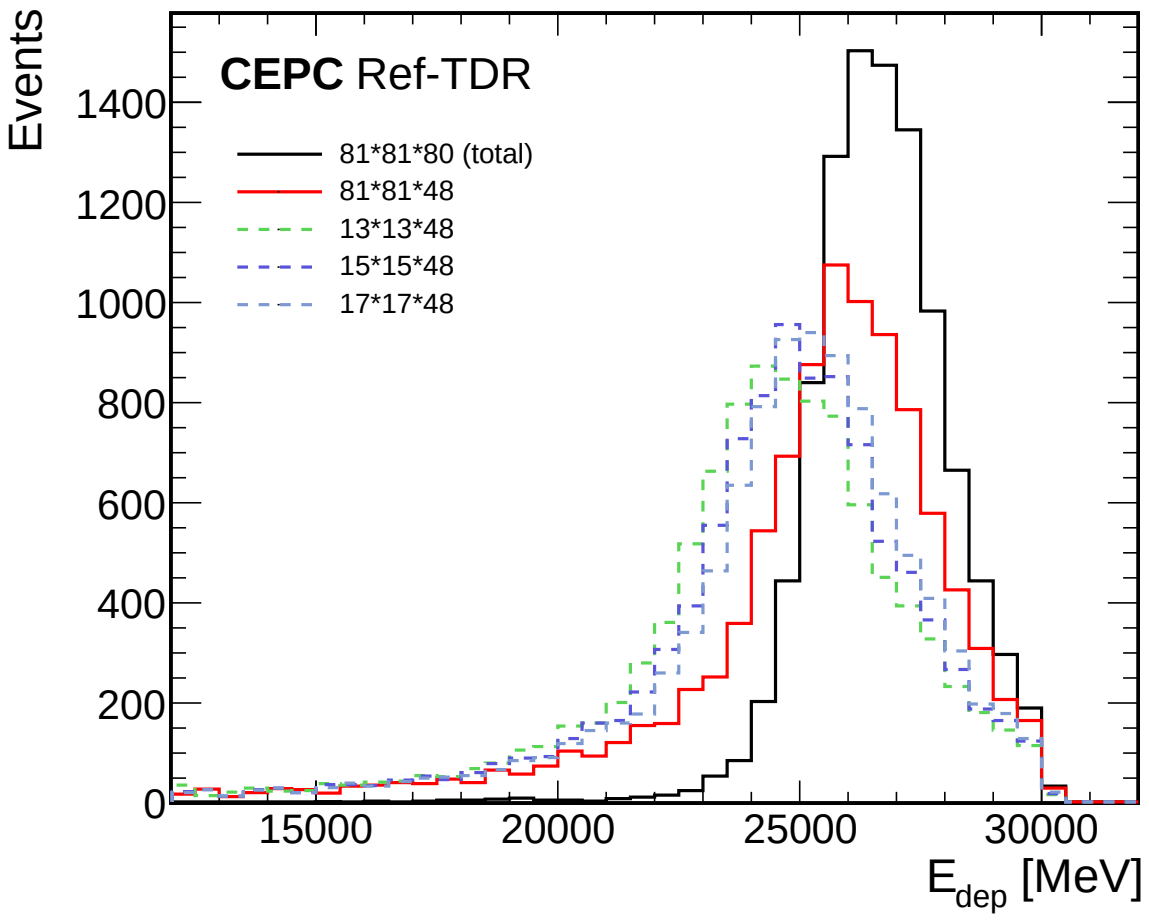

**Figure 8.23:** Distribution of deposited energy in prototype with different size using 80 GeV $\pi$ from simulation.

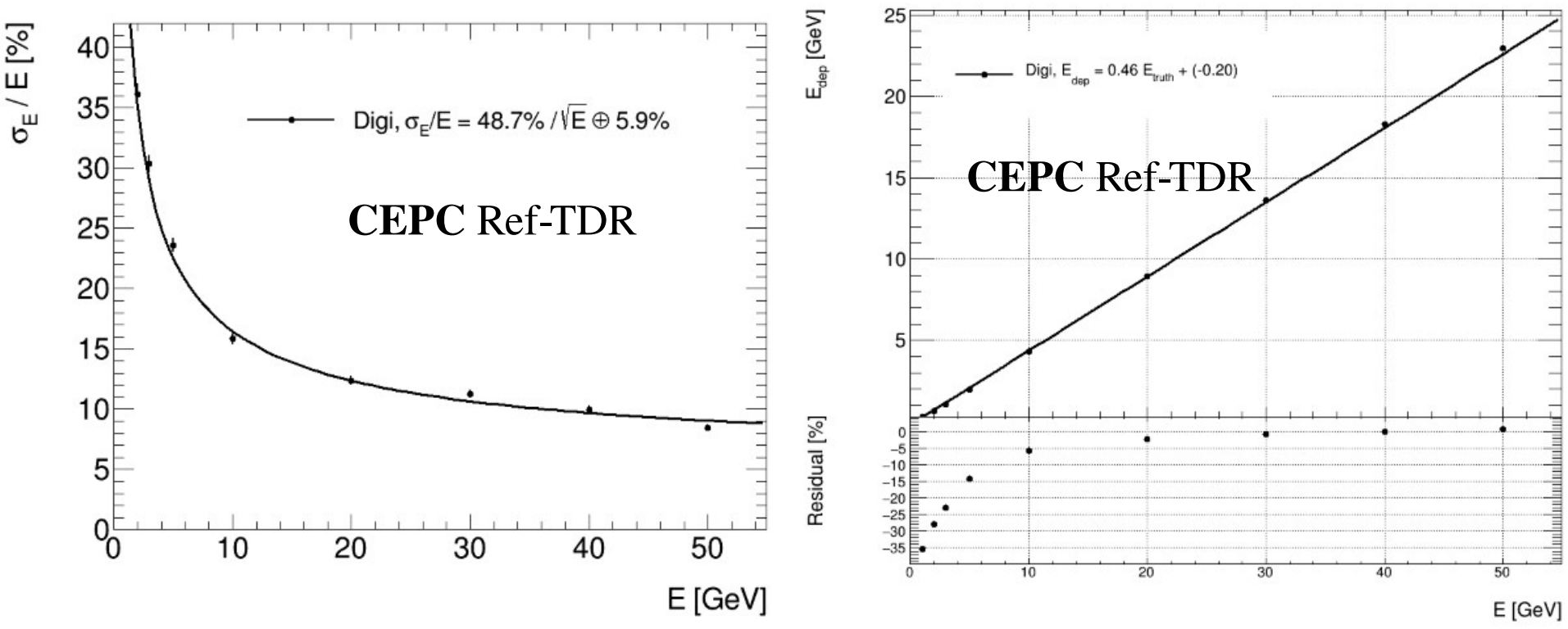

**(a)** **(b)**

**Figure 8.24:** (a) Energy resolution and (b) energy linearity of GS-HCAL prototype based on simulation.

The construction and testing program proceeds in phases. A 3×3×7 mini-prototype is currently under construction for initial beam tests at CERN in October 2025. The full 8112-channel prototype is scheduled for completion by end-2026, with cosmic ray tests planned at IHEP and subsequent beam tests preferably in 2027. The beam test program will utilize a dedicated platform with ±25 cm positioning accuracy (<1 mm precision), ±30° rotation capability, and 5-ton capacity to systematically study shower profiles across the detector volume. Alternative test sites including the China Spallation Neutron Source (CSNS) (1.6 GeV proton beam) and the facility at University of Tokyo were identified if CERN beamtime become unavailable.

Several critical subsystems were developed specifically for the prototype. The calibration system employs blue LEDs to monitor SiPM gain and response characteristics across the full dynamic range. For electronics readout, the prototype will utilize the new ASIC under development, featuring low power consumption of 15 mW per channel. Reflective coating uses titanium oxide and Teflon for higher light collection efficiency. Cooling is implemented via a water-based system with precisely regulated flow rates and temperatures.

## 8.5 Simulation and performance

The primary objective of this section is to validate the GS-HCAL design, including hadronic energy resolution, linear response across the energy spectrum, and mitigation of beam-induced background effects on signal reconstruction. Leveraging extensive simulation studies in the absence of prototype data, we demonstrate the design's compliance with these requirements.

This section describes the GS-HCAL's global performance, focusing on energy linearity, resolution parameterization (with emphasis on the constant term), and BMR studies for benchmark physics process. We then analyze beam-induced backgrounds to quantify their impact on the calorimeter and outline mitigation strategies.

### 8.5.1 Simulation setup

The energy linearity and resolution of the GS-HCAL are estimated using GEANT4 simulation within the CEPCSW framework. To study the intrinsic energy response of the GS-HCAL, all inner sub-detectors are removed. A digitization model is constructed, taking into account the following factors:

- Birks constant. Currently a preliminary value of $C_{Birks} = 0.01$ is used. A measurement using heavy ion beam is being planned for more accurate determination.
- Light yield. Based on measurements of GS samples, the intrinsic MIP light yield can reach 1500 ph/MeV. Consequently, the detected effective light output is expected to be at least 60 p.e./MIP that delivers a descent signal to noise ratio above a 0.1 MIP threshold.
- Attenuation length. It is expected to be 6 cm according to current measurements. More in-depth studies are needed.
- Threshold. Through scanning the impact on energy resolution, a 0.1 MIP threshold is determined.
- SiPM response. The response of the SiPM is incorporated into the model.
- Electronic system. The characteristics of the electronic system are considered.
- After calibration. Other factors are considered as having negligible effects.

## 8.5.2 Hadron energy resolution

For hadron calorimetry, the energy resolution is typically characterized by the formula,

$$\frac{\sigma_E}{E} = \frac{a}{\sqrt{E}} \oplus b,$$

where $a$ represents the stochastic term and $b$ is the constant term. The GS-HCAL design yields an estimated energy resolution of $\sigma_E/E = 29.8\%/\sqrt{E} \oplus 6.5\%$ and an energy non-linearity within 2% before calibration, as shown in Figure 8.25. The default hadron package used is 'QGSP-BERT', and others were tested. This result outperforms the stochastic term of traditional HCALs, which typically have an energy resolution of $\sigma_E/E = 60\%/\sqrt{E} \oplus 3\%$. The relatively large constant term in the energy resolution is attributed to the longitudinal leakage in this 6 $\lambda_I$ design, which is constrained by the total detector volume and cost. And this constant term can be further brought down by software compensation such as corrections in PFA.

In the CEPC collision environment, as depicted in Figure 8.26, more than 98% of hadronic objects in the final state have energies of less than 60 GeV. This enables the GS-HCAL to achieve optimal energy resolution for the physics of interest.

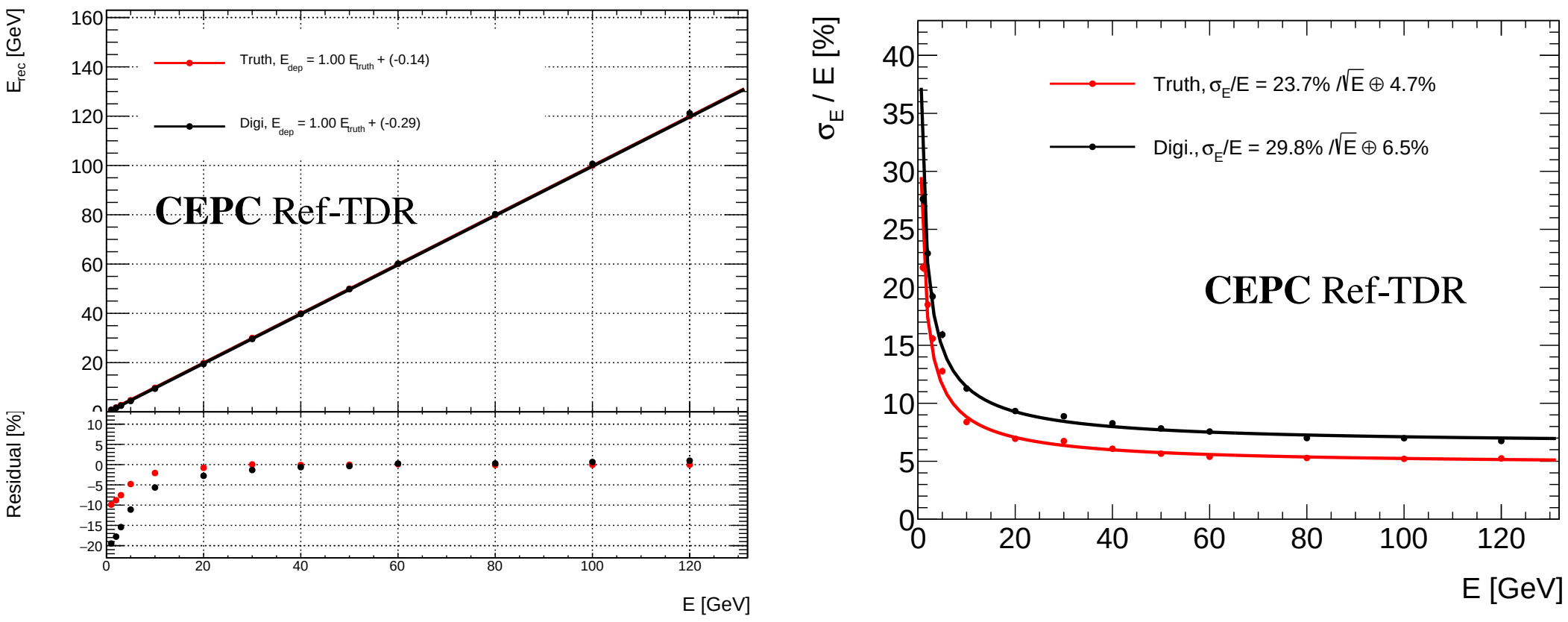

**(a)** **(b)**

**Figure 8.25:** (a) Energy linearity and (b) energy resolution of GS-HCAL with the digitization model.

Although the current digitization model provides a conservative estimate of 6.5% for the constant term in the energy resolution, it is crucial to recognize that several contributing factors can be mitigated through calibration and reconstruction improvements. The preliminary simulation incorporates non-optimized parameters, such as a short light attenuation length and longitudinal leakage effects. Future calibration procedures are expected to partially compensate for these issues.

A detailed simulation study was performed to evaluate the impact of various effects on the single hadron energy resolution of the GS-HCAL. As shown in Figure 8.27, three key factors were investigated, the light output and energy threshold, the light attenuation length of the GS, and the Birk's constant used in the modeling of scintillation quenching. These parameters were found to significantly affect the calorimeter performance, particularly in the low-energy regime, highlighting the importance of optimizing the scintillator properties and calibration methods for precise energy measurement.

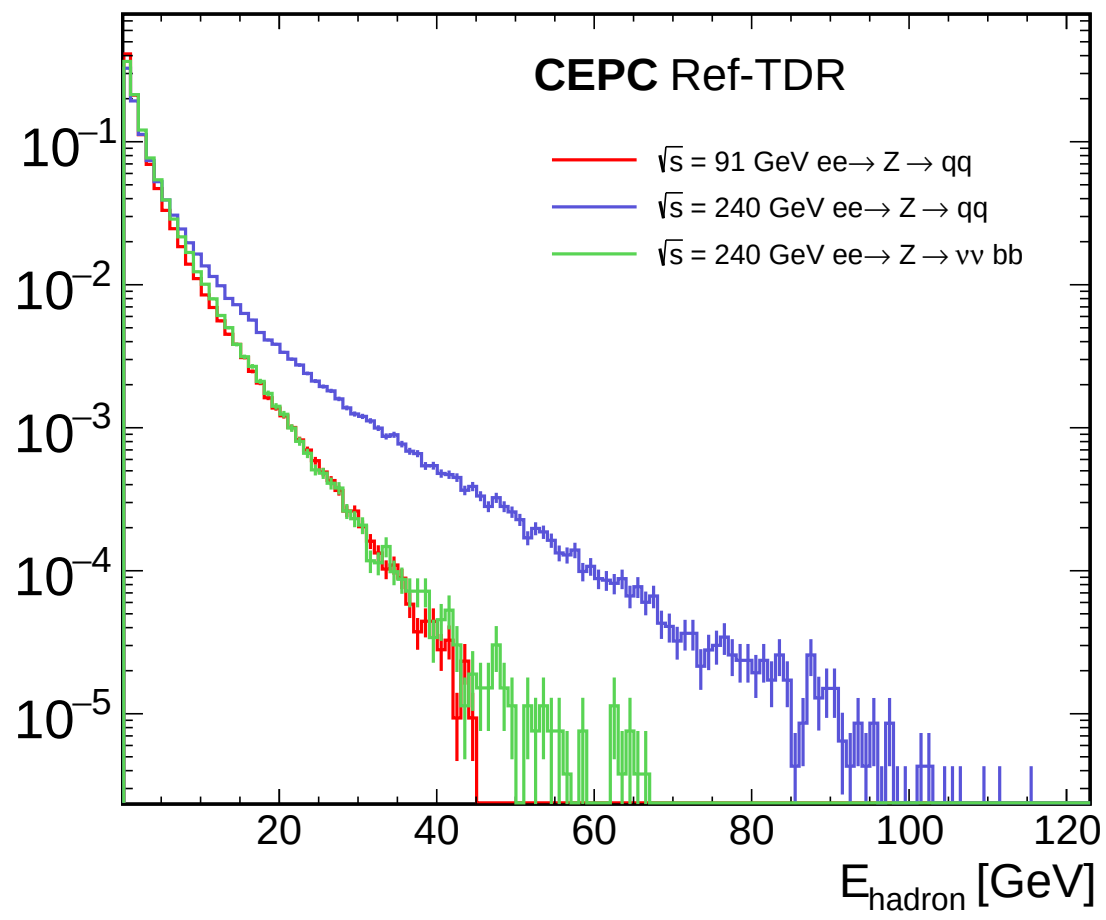

**Figure 8.26:** Hadronic objects energy distribution in the most energetic processes in $e^+e^-$ collision.

In initial studies of the GS-HCAL, the energy resolution yielded a constant term ($b$) of approximately 5–6% based on GEANT4 full simulations with all inner sub-detectors (e.g., ECAL) removed. It was uncertain whether this value was typical for hadronic calorimeters, as it appeared higher than expected. This raised concerns about potential underlying issues, such as non-uniformities or calibration errors. Subsequent investigation shows that in the GS-HCAL, the depth was sometimes insufficient to contain hadronic showers, leading to energy leakage.

Specifically, for hadronic showers that initiate late or develop with great depth, the GS-HCAL configuration could permit a substantial portion of the energy to escape. To verify the hypothesis that longitudinal leakage was the cause of the observed large constant term, we increased the depth of the GS-HCAL from 48 layers (6 $\lambda_I$) to 80 layers (10 $\lambda_I$), the constant term dropped from 4.7% to around 2.9%.

If using only lower energy $\pi$s or requiring hadrons to initiate their first interaction within the first two layers of the GS-HCAL (corresponding to the first $0.25\lambda_\text{I}$), the constant term can reduce to about 3%. This result agrees with previous measurements from PS-HCAL prototypes [3, 29], where similarly small constant terms were achieved using comparable selection criteria, specifically requiring ≥4 hits in the first 5 layers to suppress late-interacting hadrons.

In the full detector configuration, the ECAL with about $1.2\lambda_\text{I}$ will cause approximately 70% of hadrons to initiate showers before reaching the GS-HCAL. This upstream shower development is expected to significantly reduce leakage effects and consequently lower the constant term in the combined ECAL and HCAL system.

As the study for the GS-HCAL progresses, a multifaceted approach is being deployed to further reduce the constant term. Advanced energy compensation techniques [30, 31, 32] are being integrated with improved containment strategies, while machine learning methods correlate 3D shower topologies to enhance reconstruction accuracy. Furthermore, as shown in Ref. [29], advanced PFA can reduce this constant term even further.

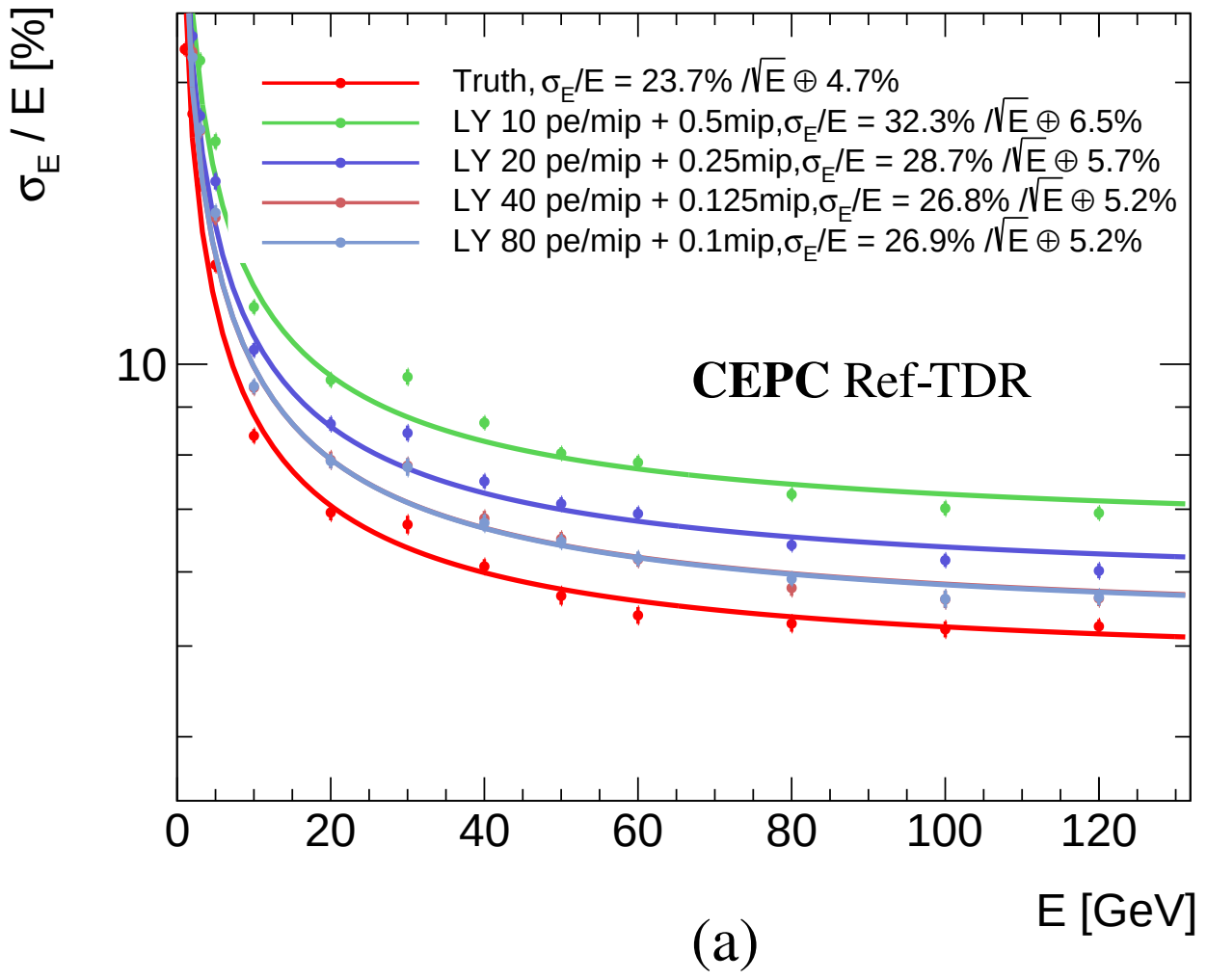

(a)

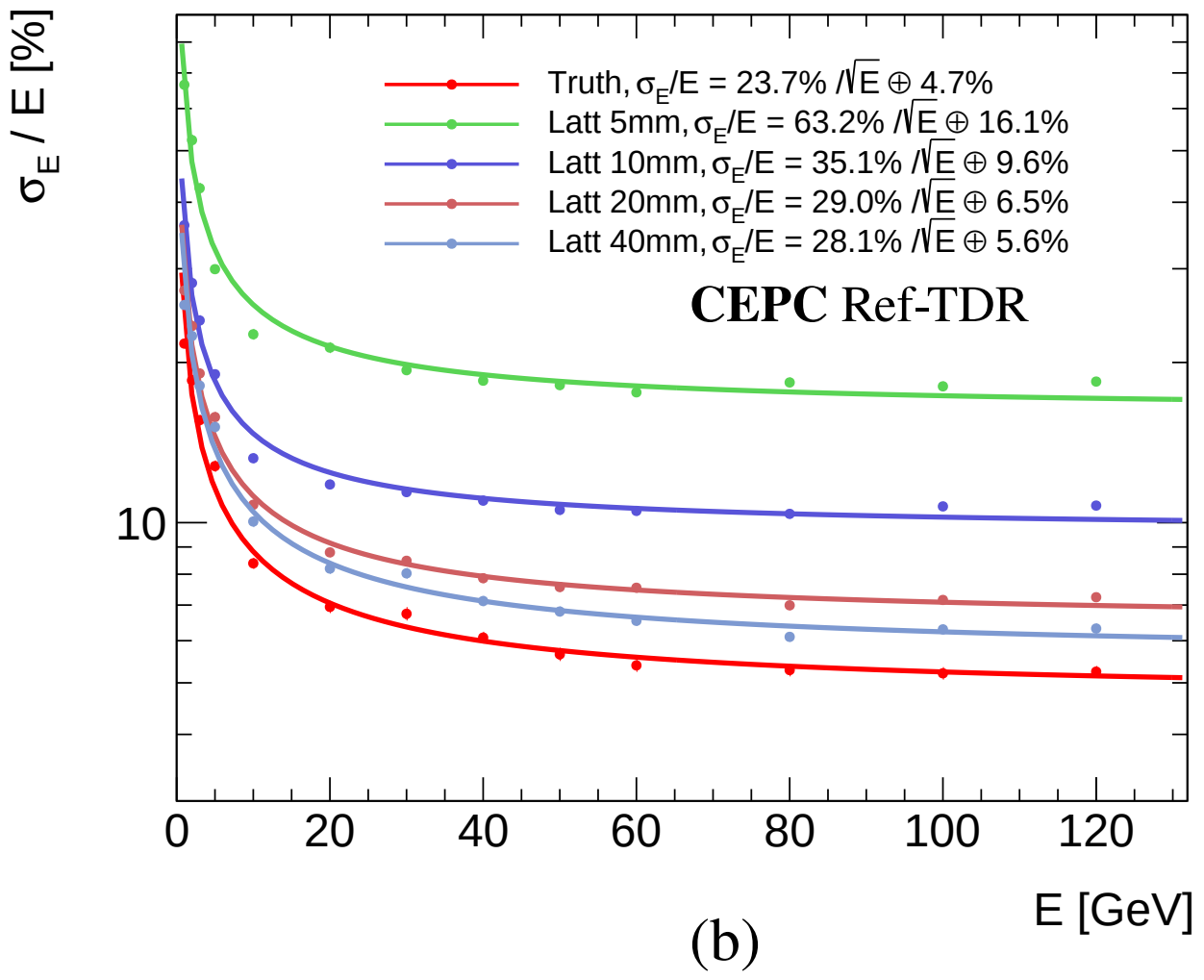

(b)

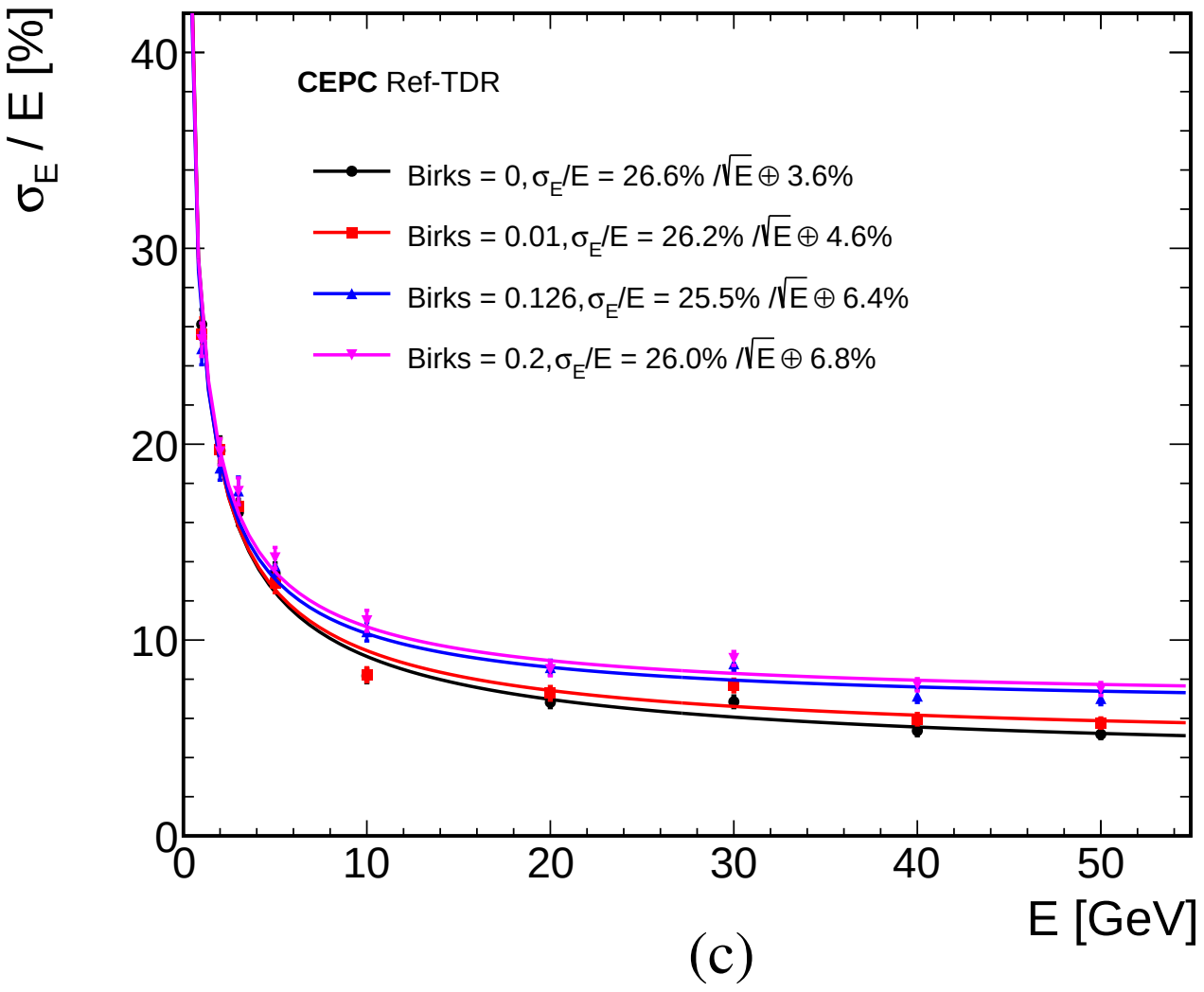

(c)

**Figure 8.27:** The simulated energy resolution with varies of (a) light output and energy threshold, (b) GS attenuation lengths, (c) Birk's constants.

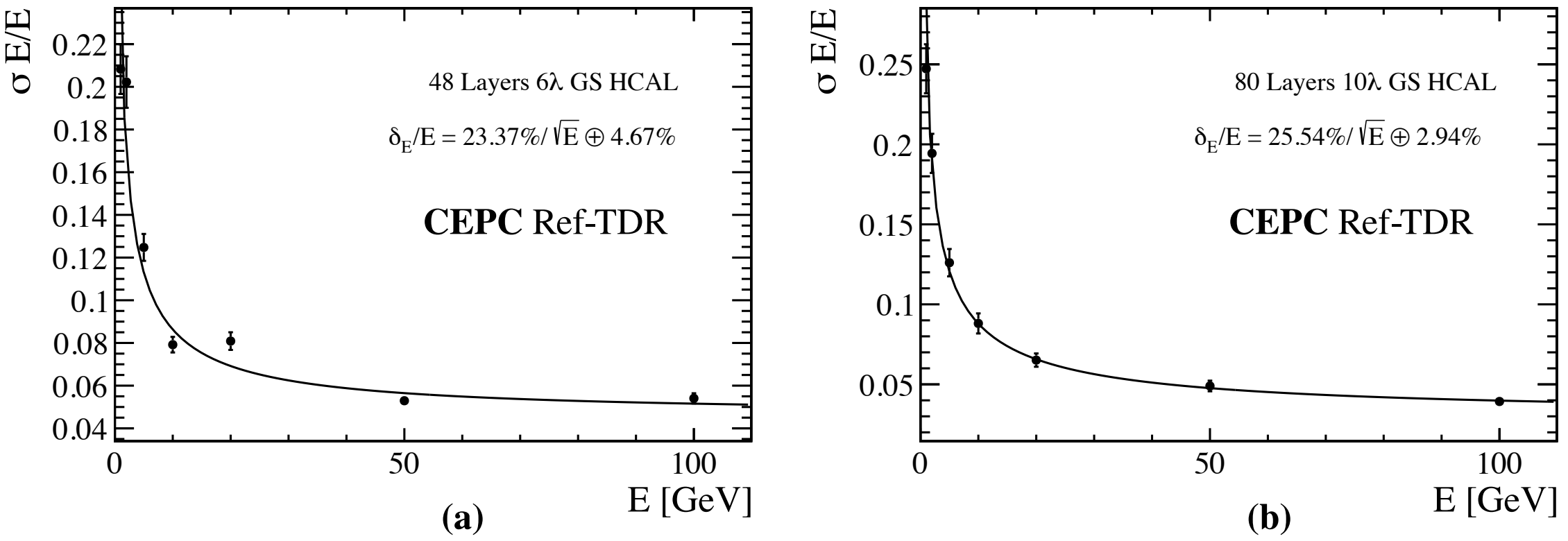

**Figure 8.28:** Comparison of two GS-HCAL configurations with different depths. A shallower setup with 6 $\lambda_I$ (a) yields a pronounced high tail in the energy resolution and a large constant term. A deeper setup with 10 $\lambda_I$ (b) effectively contains the showers, reducing the tail and lowering the constant term to a more typical value.

### 8.5.3 Energy response studies

We have performed a calibration study on GS-HCAL response to electrons and various hadrons based on full Geant4 simulations, considering only the GS-HCAL without the presence of an ECAL or other inner subdetectors. The calibration was carried out by fitting the true energy $E_{\text{true}}$ against the raw calorimeter response $S_{\text{HCAL}}$ using a linear model $E_{\text{true}} = C \times S_{\text{HCAL}}$, thereby extracting the calibration constant $C$ for each particle species. The ratio $C_{\text{hadron}}/C_e$ is the inverse of the mean detector response for an electron relative to that of a hadron. In essence, a value greater than unity indicates that a hadron of a given energy produces a smaller signal than an electron of the same energy.

This empirically measured quantity, often called the electron-hadron response ratio, should not be confused with the intrinsic compensation ratio $e/h$ defined in the context of hadronic shower development. The latter specifically characterizes the calorimeter's response to the electromagnetic component (e.g., from $\pi^0$ decays) versus the non-electromagnetic component (e.g., from nuclear breakup) within a single hadronic shower.

The results are summarized in Table 8.6. The measured response ratios indicate a non-compensating behavior at the level of 9–14%, meaning that hadrons produce less signal per unit energy than electrons. This is consistent with typical sampling calorimeters that are not optimized for compensation. It is important to note that these values represent an overall response difference between electrons and hadrons, and do not directly reflect the $e/h$ value as defined above. The $e/h$ parameter requires disentangling the electromagnetic fraction ($f_{\text{em}}$) and the non-electromagnetic component of the shower, which was not attempted in this preliminary study.

| Particle | $C$ (slope) | $C_{\text{hadron}}/C_e$ |
|---|---|---|
| Electron | $2.94 \pm 0.01$ | – |
| $\pi^-$ | $3.20 \pm 0.05$ | $1.09 \pm 0.02$ |
| $K^+$ | $3.28 \pm 0.05$ | $1.12 \pm 0.02$ |
| Neutron | $3.35 \pm 0.07$ | $1.14 \pm 0.02$ |
| Proton | $3.36 \pm 0.07$ | $1.14 \pm 0.02$ |

**Table 8.6:** Calibration slopes $C$ for the GS-HCAL and the corresponding electron-hadron response ratios $C_{\text{hadron}}/C_e$, with the uncertainty reprsent the limited MC statistics.

The necessity for corrections based on these measured electron-hadron response ratios stems from the fundamental challenge in hadron calorimetry. As detailed in Ref. [33], the energy resolution for hadrons is dominated by fluctuations in the electromagnetic fraction of the shower. The intrinsic compensation ratio $e/h$ serves as the true figure of merit, where a value of unity signifies perfect compensation and elimination of the degrading effects of $f_{\text{em}}$ fluctuations on resolution. Specifically, the constant term in the energy resolution is strongly dependent on the degree of non-compensation, $|1 - h/e|$. Therefore, precise knowledge of $e/h$ is crucial for optimizing the calorimeter's performance. Future studies will focus on extracting the true $e/h$ value through detailed GEANT4 simulations that separate the electromagnetic and non-electromagnetic components of hadronic showers. By analyzing the origin of energy deposits, the average electromagnetic fraction $\langle f_{\text{em}} \rangle$ and its fluctuations can be determined. A precise calibration based on this intrinsic $e/h$ parameter is expected to reduce the constant term in the energy resolution.

## 8.5.4 Physics performance

The performance of the detector on physical events is evaluated through PFA, which synthesizes information from tracker, calorimeters, and other sub-detector systems. The Higgs boson decay $H \to gg$ serves as the benchmark to quantify the detector performance.

Event generation is performed using WHIZARD v1.9.5 [34, 35] with next-to-leading-order matrix elements, followed by full detector response using GEANT4 v11.2 within the CEPCSW framework. Digitized signals from trackers and calorimetric energy deposits are reconstructed into charged-particle trajectories and electromagnetic/hadronic clusters, respectively, with reconstruction algorithms detailed in Section 13.3. By aggregating the four-momenta of all reconstructed particles, the fitted Higgs boson mass from gluon pairs is $126.32 \pm 0.04$ GeV$/c^2$, $\sigma(m_{jj}) = 4.90 \pm 0.04$ GeV$/c^2$. The Higgs boson invariant mass resolution (BMR) is 3.88%, as shown in Figure 8.29.

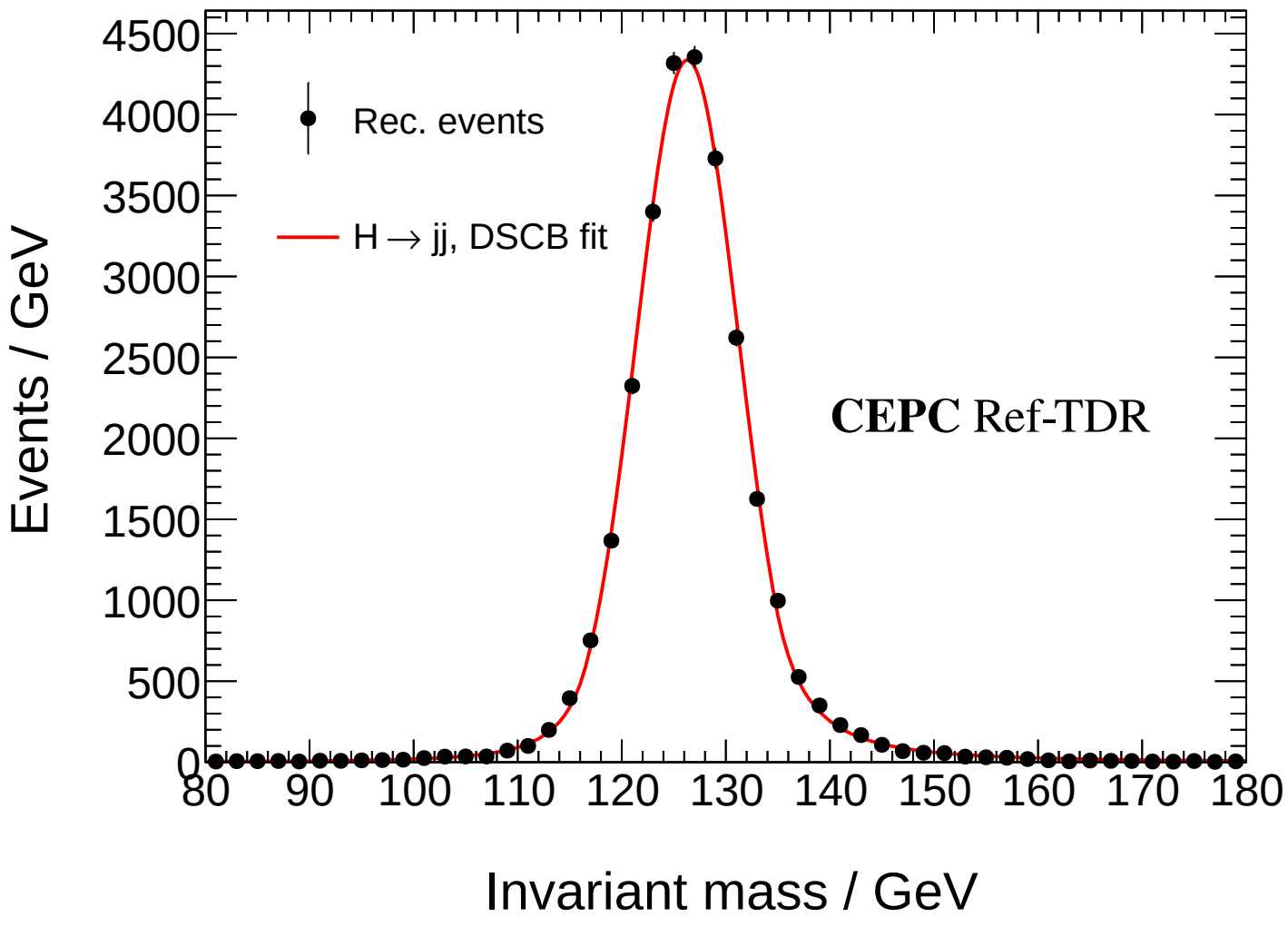

**Figure 8.29:** Reconstructed invariant mass of two jets using $H \rightarrow gg$ process. The fitted Higgs boson mass is $126.32 \pm 0.04$ GeV/$c^2$, $\sigma(m_{jj}) = 4.90 \pm 0.04$ GeV/$c^2$.

## 8.6 Alternative HCAL option

Multiple technological approaches of the CEPC calorimeters were investigated to achieve the required jet energy resolution. While the baseline PFA-based GS-HCAL using GS offers superior performance, extensive R&D efforts within the CALICE collaboration were made and alternative solutions were systematically evaluated over the past two decades. These include gaseous detector-based approaches like the Digital HCAL (DHCAL) [36, 37] and Semi-Digital Hadronic CALorimeter (SDHCAL) [38, 39], as well as plastic scintillator cell designs (Analog HCAL (AHCAL)) [3]. These alternatives represent different trade-offs between spatial granularity, energy response linearity, and system complexity.

The SDHCAL approach utilizes glass Resistive Plate Chamber (RPC) technology with digital readout, achieving high granularity through small pad sizes while maintaining robust operation [38, 40, 41, 42, 43]. Complementary to this, the collaboration has developed analog PS-HCAL prototypes using plastic scintillator cells with SiPM readout, offering improved energy resolution through analog signal processing. Furthermore, MPGD-based solutions including resistive Micromegas and Resistive Plate WELL (RPWELL) technologies were explored as potential alternatives [39], demonstrating the versatility of gaseous detectors in high-radiation environments. Beam tests at CERN have shown these technologies typically achieve single hadron energy resolution around $\sim 60\%/\sqrt{E}$.

Alongside the PFA approach, the dual-readout concept was investigated as an alternative methodology [44, 45, 46, 47, 48], though the current baseline remains firmly rooted in the PFA paradigm. The following subsections provide summaries of two key PFA-compatible alternatives that have reached advanced prototype stages: the SDHCAL and AHCAL designs, both of which have demonstrated promising performance in extensive test beam campaigns.

### 8.6.1 RPC-SDHCAL

Motivated by the excellent efficiency, high homogeneity, and fine lateral segmentation that gaseous detectors can provide, the concept of SDHCAL is proposed aiming for future colliders. In contrast to glass scintillators, the lateral segmentation of the AHCAL with RPCs is determined by the highly granular digital readout pads (1 $cm^2$ each) [36, 37], rather than the detector segmentation itself. In addition, the minimum thickness of the active layers is also a key issue to be addressed, since the HCAL is to be placed inside the solenoid. Such reconstruction can resolve hadronic showers from different particles and identify delayed neutrons with higher efficiency, which ultimately improves the energy response.

To achieve excellent resolution in the energy measurement of hadronic showers, a binary readout of the gaseous detector is the simplest and most effective approach. However, a lateral segmentation of a few millimetres is needed to ensure good linearity and to resolve energy deposition. Such a lateral segmentation leads to a huge number of electronic channels resulting in a complicated read-out system design. A cell size of 1 cm × 1 cm is found to be a good compromise that still provides a good resolution at moderate energy. However, simulation studies show that saturation effects are expected to show up at higher energy (>40 GeV). This happens when many particles cross a single cell at the center of a hadronic shower. To reduce these effects, multithreshold electronics (semi-digital) readout is chosen to improve the energy resolution by utilizing the particle density information. These elements were behind the development of an SDHCAL.

The large number of electronic channels has two important consequences: on one hand, the high power consumption results in increasing temperature, which affects the performance of the active layers; on the other hand, the number of service cables needed to power and read out these channels also increases. These two aspects can degrade the performance of the HCAL if they are not addressed properly.[2]

The SDHCAL is a sampling calorimeter that uses stainless steel as the absorber and Glass Resistive Plate Chambers (GRPC) as the sensitive medium, and is designed to be as compact as possible with its mechanical structure being part of the absorber. The GRPC and the read-out electronics are conceived to achieve minimal dead regions [38]. This design renders the SDHCAL optimal for the application of the PFA techniques [1, 49, 50]. A prototype was designed and commissioned, and test beam experiments were conducted at the PS and SPS facilities of CERN, which enabled detailed performance studies of the SDHCAL.

The prototype SDHCAL consists of 48 active layers. Each layer is equipped with a 1 m × 1 m GRPC and an Active Sensor Unit (ASU) of the same size hosting on the side in contact with the GRPC, pick-up pads of 1 cm × 1 cm size each and 144 HARDROC2 ASICs [51] on the other side. The GRPC and the ASU are assembled within a cassette made of two stainless steel plates, both of which have a thickness of 2.5 mm. The 48 cassettes are inserted into a self-supporting mechanical structure made of 49 plates, each 15 mm thick, of the same material as the cassettes, bringing the total absorber thickness to 20 mm/layer. The gap between two consecutive plates is 13 mm, which allows for the insertion of one cassette with a thickness of 11 mm. In total, the SDHCAL represents approximately 6 interaction lengths ($\lambda_I$). The HARDROC2 ASIC has 64 channels to read out 64 pick-up pads. Each channel has three parallel digital circuits. The parameters of which can be configured to provide 2-bit encoded information per channel. This indicates whether the charge seen by each pad has passed any of the three different thresholds associated with each digital circuit. This multi-threshold readout is used to improve the energy reconstruction of hadronic showers at high energies ($>$ 30 GeV), compared with the simple binary readout mode as explained in Ref. [42].

In 2012, using the beams provided by the SPS facility at CERN, the performance of the SDHCAL prototype was studied in the energy range above 10 GeV [42]. To study its performance under different beam conditions and at lower energy points, the SDHCAL prototype was exposed to hadrons at both the SPS and the PS facilities in 2015. It was first exposed to $\pi^-$ beams of (3, 4, 5, . . . , 11) GeV at the PS facility and then to positively charged hadrons of (10, 20, 30, . . . , 80) GeV at the SPS facility. At both beamlines, about 10, 000 events were collected at each energy point.

The total energy of the hadronic shower is reconstructed using the number of hits registered at three different thresholds (NHit1, NHit2, NHit3), following a weighted sum approach as described in Ref. [42]. The weights $\alpha$, $\beta$, and $\gamma$ are parameterized as second-order polynomials of the total hit count (NHit = NHit1 + NHit2 + NHit3), with their coefficients determined by minimizing a $\chi^2$ function that compares reconstructed energy ($E_{\text{rec}}$) to beam energy ($E_{\text{beam}}$). The $\chi^2$ incorporates an expected stochastic resolution term $\sigma/E_{\text{beam}} \sim 1/\sqrt{E_{\text{beam}}}$, ensuring optimal calibration across different energy points.

Based on the $\pi$ test beam data collected at CERN SPS and PS with energy ranging from 3 to 80 GeV, energy linearity and resolution of the prototype are obtained, as shown in Figure 8.30. Apparently, the energy linearity is within about 4%, and the energy resolution for the $\pi$ beam is $65\%/\sqrt{E} \oplus 2.5\%$.[41]

The SDHCAL prototype demonstrates the feasibility of a highly granular gaseous calorimeter with semi-digital readout for future colliders. Its compact design and multi-threshold capability effectively address the challenges of shower saturation and channel complexity, achieving performance metrics that validate its potential for PFA applications.

### 8.6.2 PS-AHCAL

Considering the performance, maturity and cost, a viable approach is to develop AHCAL based on plastic scintillators. Two AHCAL prototypes are built and extensively studied within the CALICE collaboration [3, 52, 29].

As the baseline HCAL design in CEPC CDR, a large number of simulation studies were carried out to address optimal design such as absorber and scintillator thickness, sampling layer

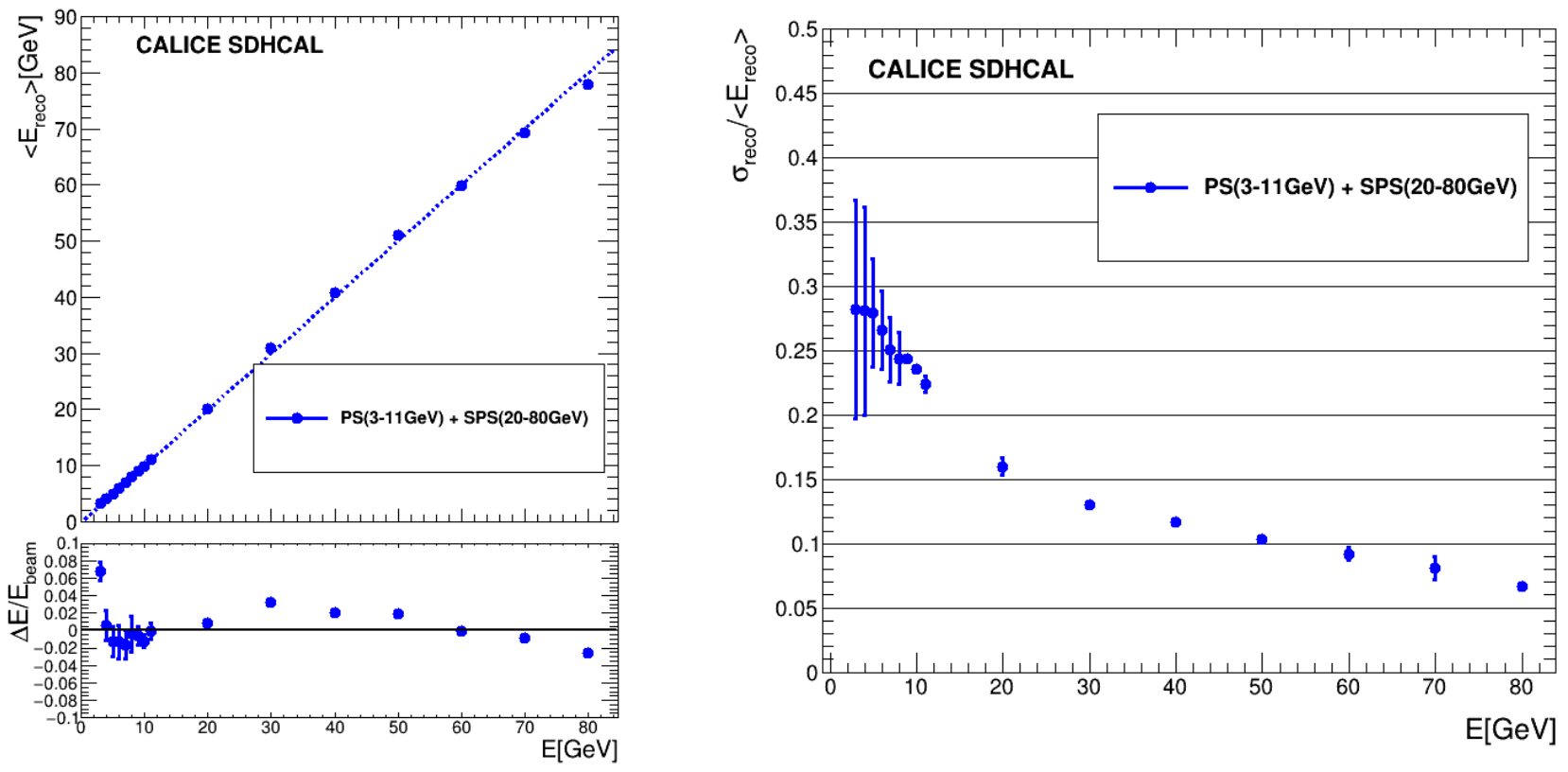

**(a)** **(b)**

**Figure 8.30:** (a) Energy linearity and (b) energy resolution using all the test beam data collected at CERN with PS and SPS [41].

and cell size with respect to physics performance [52, 29]. An AHCAL technological prototype with a size of 72 cm × 72 cm × 120 cm was developed by University of Science and Technology of China, Shanghai Jiao Tong University and IHEP. It has 40 layers, corresponding to $4.7\lambda_I$ in total. Each layer has 324 (18 × 18) plastic scintillators; each scintillator couples with a SiPM mounted on PCB. The prototype has a total of 12960 readout channels and was successfully tested using CERN SPS and PS beams in 2022 and 2023.

A single sensitive unit is composed of a scintillator cell and a SiPM. The 40 mm × 40 mm × 10 mm scintillator cell is designed with a 5.5 mm × 5.5 mm × 1.1 mm groove at the bottom to accommodate SiPM, together with a LED in the groove adjacent to the SiPM for calibration. The scintillator cells are produced using a cost-effective injection molding technique. Subsequently, the scintillator cell is wrapped with ESR. An automated batch test platform was designed with 144 SiPM (HPK S13360-3025PE) mounted on PCB. About 15000 plastic scintillator cells were tested using $^{90}Sr$ radiative source. The measured light output, corrected for the set-up response non-uniformity, is around 12.9 p.e./MIP. About 91.6% of scintillators are qualified within 10% of light output window that assures the uniform light output of scintillators to be used for AHCAL prototype [53]. For the AHCAL prototype, a SiPM (HPK S14160-1315PS) with higher PDE is placed in the groove to collect the fluorescence light. The light output of this unit was approximately 17 p.e./MIP. The non-uniformity was approximately 6.7%.

The electronic chip we used is SPIROC2E [54, 55, 56], each contains 36 channels, corresponding to 36 sensitive units. Each channel employs two preamplifiers with different gains to enhance the dynamic range. Following the high gain preamplifier, a fast shaper and a discriminator are used to provide self-trigger. Once triggered, the signals from the high gain preamplifier and the low gain preamplifier are recorded in the analog memory after the slow shaping. Subsequently, the signals stored in the analog memory are converted into digital signals by a 12-bit ADC.

Two calibration systems were designed to monitor the status of electronics and SiPMs. The charge signal could be injected into the SPIROC chip to probe the response of all channels. By varying the amount of injection charge, the gain ratio of high gain and low gain could be calibrated. The other calibration method utilizes an LED, placed adjacent to the SiPM to calibrate and monitor it. The SiPM response to the LED exhibits good single photon separation. The intervals between photon peaks are approximately 20 ADC, representing the gain of the SiPM.

The HCAL Base Unit (HBU) board serves as the structural and functional platform for mounting the sensitive components and performing analog-to-digital signal conversion. Plastic scintillator cells, each wrapped with ESR film and measuring 3.5 mm in thickness, are fixed to one side of the HBU, which itself consists of a 2.5 mm-thick PCB. These assemblies are housed within a mechanically robust steel cassette structure designed to provide optimal support for the sensitive layer while maintaining handling convenience. The structural integrity of the cassettes is ensured by 2 mm-thick steel sheets at both the top and bottom surfaces, with a total internal accommodation for the 3.5 mm scintillator-ESR composite, 2.5 mm PCB, and a 4 mm clearance space for electronic components (including a 1 mm tolerance margin), resulting in an overall cassette thickness of 14 mm.

To study the performance of the AHCAL prototype, three beam tests were carried out at the CERN SPS H4, H2, and PS beamlines. About 65 millions events were collected, including muons, electrons and negative pions in a wide energy range, from a few GeV to hundreds GeV, to study response of AHCAL prototype to high-energy particles. Detailed studies were performed using test beam samples, advanced machine learning methods (e.g., residual neural network, graph neural netowork) are introduced to improve particle identification, and to obtain clean $\pi$ hadron samples with high purity [57]. The energy response linearity and energy resolution of the AHCAL prototype were studied using high-energy $\pi$ events, as shown in Figure 8.31a and 8.31b respectively. The measured energy linearity is better than 1.5%, and the fitted energy resolution (shown in red curve) is about $56.2\%/\sqrt{E} \pm 2.5\%$ [29].

In the initial studies, a standard energy reconstruction technique based on calibrated sub-detector energy sums was employed. However, this method has limitations in accurately resolving hadronic energies due to the complex nature of hadronic showers. Hadronic showers in calorimeters are characterized by a significant amount of energy loss through processes like nuclear interactions, which are difficult to precisely account for with a simple summation approach.

This reconstruction takes advantage of the high granularity of the detectors, using local energy density information. Studies, such as those using the combined electromagnetic and hadronic calorimeter systems of the CALICE collaboration (e.g., the Silicon-Tungsten Electromagnetic Calorimeter (Si-W ECAL) and the scintillator-SiPM based AHCAL), have shown that the software compensation-based algorithm can improve the hadronic energy resolution by up to 30% compared with the standard reconstruction [58, 30]. When operated in hadron beams at CERN and Fermilab, the combined system data demonstrated comparable energy resolutions to those achieved for data with showers starting only in the AHCAL. This success indicates the effectiveness of the algorithm in handling the complex energy deposition patterns in hadronic showers.

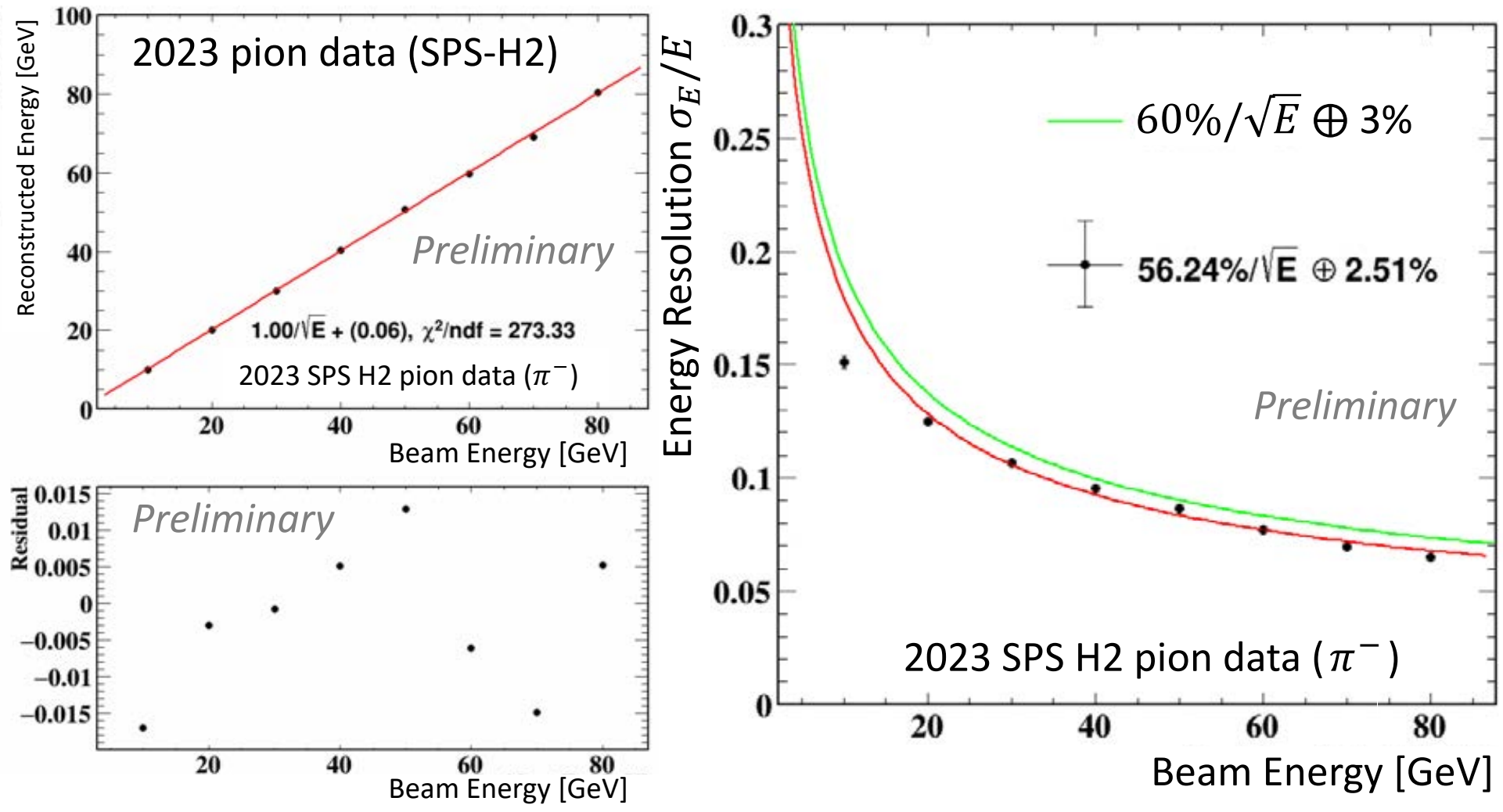

(a) (b)

**Figure 8.31:** (a) Measured energy linearity and (b) energy resolution of the AHCAL prototype using high energy negative pions, where red curve is function fitted to data, and green curve is a reference.

## 8.7 Summary and future plan

The CEPC HCAL aims to precisely measure hadronic jet energy with high resolution and hermetic coverage, which is crucial for Higgs, electroweak and flavor physics studies. The baseline design, GS-HCAL, uses high density glass scintillator and SiPMs. It has a barrel and two endcaps, with a total of about 5.22 million cells, and is designed for PFA. The single layer structure consists of alternating GS cells and steel absorber plates. The barrel and endcap geometries are carefully designed for mechanical stability and efficient particle detection.

Past R&D efforts have focused on various aspects. High performance GFO glass was developed, with studies on light output and attenuation length. SiPMs were selected as photon detectors, and efforts are ongoing to suppress their dark noise. Simulation studies have characterized the performance of the GS-HCAL, and a prototype is being developed for testing. Alternative solutions such as (S)DHCAL and AHCAL have also been explored, providing different and feasible approaches to hadronic calorimetry.

Looking ahead, several key steps are needed to meet the detector design requirements. For the GS-HCAL, although the constant term of about 5% in the energy resolution is sufficient for a 3–4% BMR, it would be better if it was further reduced below 3%. This may be achieved through calibration and reconstruction improvements, such as developing more sophisticated energy weighting and compensation algorithms, implementing position dependent light col-lection efficiency corrections, and applying machine learning techniques to enhance energy reconstruction.

The GS-HCAL development follows an aggressive yet realistic timeline:

- **2024–2025**: Finalize 40 mm × 40 mm × 10 mm glass tile design and initiate mass production of 10,000 prototype tiles;
- **2026**: Complete 0.6 $m^3$ prototype assembly (8,112 tiles) and begin beam tests at CSNS or CERN;
- **2027**: Integrate a full-scale module and validate its performance using cosmic-ray and test beam.

In terms of components, continuous research on GS should aim to further optimize their performance, especially in increasing the light output and improving the attenuation length. For SiPMs, with the advancement of technology, more optimized devices are expected to emerge. Collaboration with domestic partners is essential to reduce costs while maintaining high performance.

The development of the readout electronics is crucial. A customized front-end electronics solution is needed to meet the requirements of SiPMs and the HCAL prototype. This should consider factors like charge dynamic range, timing resolution, and power consumption. As the R&D progresses, the HCAL electronics will be integrated into the whole CEPC electronics system.

Finally, extensive tests and studies of the prototypes, both the mini-prototype and the full-scale one, is necessary. Beam tests at CERN or alternative CSNS proton beam facility will help validate the design and performance. These tests will provide valuable data for further improvements, ensuring that the HCAL can meet the stringent requirements of the CEPC physics program and contribute to significant discoveries in high energy physics.

# Chapter 9 Muon Detector

In the CEPC Reference Detector, muons are identified primarily in the PFA-oriented calorimeters and their momenta are measured in the tracking system. An outermost muon detector system is envisioned to provide redundancy, aid muon identification in busy environments, and reduce the pion fake rate. Embedded in the solenoid flux return yoke, the outermost muon detector is designed to identify muons with high standalone efficiency ($> 95\%$) for those with momenta $> 5$ GeV/$c$ over the largest possible solid angle. While the identification requirement drives the design, the muon detector also provides the opportunity for standalone measurements of muon momenta.

The muon detector will significantly enhance the identification of muons produced within jets, such as those from $b$-hadron decays. Moreover, the detector can compensate for leaking energetic showers and late showering pions from the calorimeters, which could help to improve the jet energy resolution. It can also aid in searches for Long-Lived Particles (LLPs) that decay far from the IP, but still inside the detector.

This chapter outlines the design and development of the muon detector. Section 9.1 introduces the performance requirements for the detector and summarizes the design. Section 9.2 elaborates on the detailed design, including its geometry, baseline technology, and electronics system. Section 9.3 discusses the R&D program for advancing the detector technology. Section 9.4 presents performance evaluation from simulation studies. Section 9.5 addresses challenges and considerations related to mass manufacturing, assembly, and detector integration. Finally, Section 9.6 concludes the chapter by exploring alternative design options.

## 9.1 Performance Requirements

To achieve high precision in measurements of final states involving muons and to broaden the scope of searches for new physics, the performance requirements for the muon detector are defined. The detector design is based on Plastic Scintillator (PS) materials, Wavelength-Shifting (WLS) optical fibers, and SiPM sensor technologies, chosen for their high-performance capabilities and cost effectiveness. The detector layout is optimized through extensive simulation studies, with realistic input parameters obtained from R&D efforts.

High muon identification ($\mu$ID) efficiency and minimal background contamination are essential for advanced precision physics, including measurements of the Higgs boson, electroweak interactions, and rare decay processes. In particular, the muon detector plays a critical role in identifying Higgs boson production via the $e^+e^- \to ZH$ channel, where the Higgs boson is determined from the recoil of $Z \to \mu^+\mu^-$. Its placement outside calorimeters enhances the sensitivity to muons originating from $b$-hadron and $c$-hadron decays, and signatures of physics beyond the SM. While $\mu$ID performance drives the detector design, its potential standalone capability to measure muon momentum independently adds significant value. This dual functionality enables exploration of exotic phenomena, such as LLPs that produces detectable decay products outside the tracking system but before the muon detector.

Furthermore, the muon detector will significantly enhance the identification of muons generated within high-energy jets and mitigate the effects of energetic showers leaking into the detector. It also addresses late-showering pions or muons originating from calorimeters, as pion decays could produce low-momentum muons. Collectively, these functionalities are expected to enhance the precision of jet energy reconstruction and improve overall resolution.

The detector shall achieve a time resolution of approximately 1 ns per channel. By leveraging multiple hits from a single muon track across the layers of the muon detector, the time resolution can be significantly better than 1 ns. This improvement is further enhanced when multiple muon tracks are produced in a single $e^+e^-$ collision. The muon detector's superior time resolution in reconstructing muon tracks can serve as input for the trigger system and provide an independent determination of the event T0. The integration of muon detector information as general trigger signals into the trigger system is discussed in Section 12.4.2.2 of Chapter 12.

The muon detector offers significant advantages in the search for signals of new physics, such as LLPs. An LLP with a displaced vertex may produce no detectable signal in the inner subdetectors but generate a few charged-particle tracks within the muon detector. In such scenarios, the muon detector can provide a dedicated trigger signal specifically designed to identify LLPs. The timing capability of muon detector also enhances searches for LLPs by reducing background and noise outside the signal time windows.

Based on the defined physics requirements, comprehensive analysis of the selected technologies, and detailed detector simulation, the muon detector is expected to achieve the following performance specifications:

- Solid angle coverage: $> 98\% \times 4\pi$
- $\mu$ID efficiency: $> 95\%$
- $\pi \to \mu$ fake rate: $< 1\%$
- Spatial resolution: $\sim$ 1 cm
- Time resolution: $\sim$ 1 ns
- Rate capability: 50 Hz/ cm$^2$

The detector system is constructed from fundamental detection units known as Plastic Scintillator Units (PSUs). Each PSU integrates three core components: a PS strip, a WLS fiber embedded inside the strip, and a SiPM optically coupled to one end of the WLS fiber. When a charged-particle track traverses the PSU, scintillation photons are generated in the PS strip and absorbed by the WLS fiber, which shifts their wavelength and guides them toward the SiPM. The SiPM then converts the photons into electronics signals, enabling precise readout and determination of the particle's hit position. The detailed structure of a PSU will be described in Section 9.2.3.

The muon detector is embedded within the flux-return yoke of the solenoid magnet, po-sitioned outside the main solenoid coil. As illustrated in Figure 9.1, the system comprises a twelve-sided (dodecagonal) barrel structure, with two endcaps at either end, each divided into four sectors. The barrel covers the polar angles from 46.2° to 133.8°, while the endcaps extend the range to 7.8° to 172.2°, ensuring near-complete azimuthal acceptance, covering approximately $0.98 \times 4\pi$ solid angle.

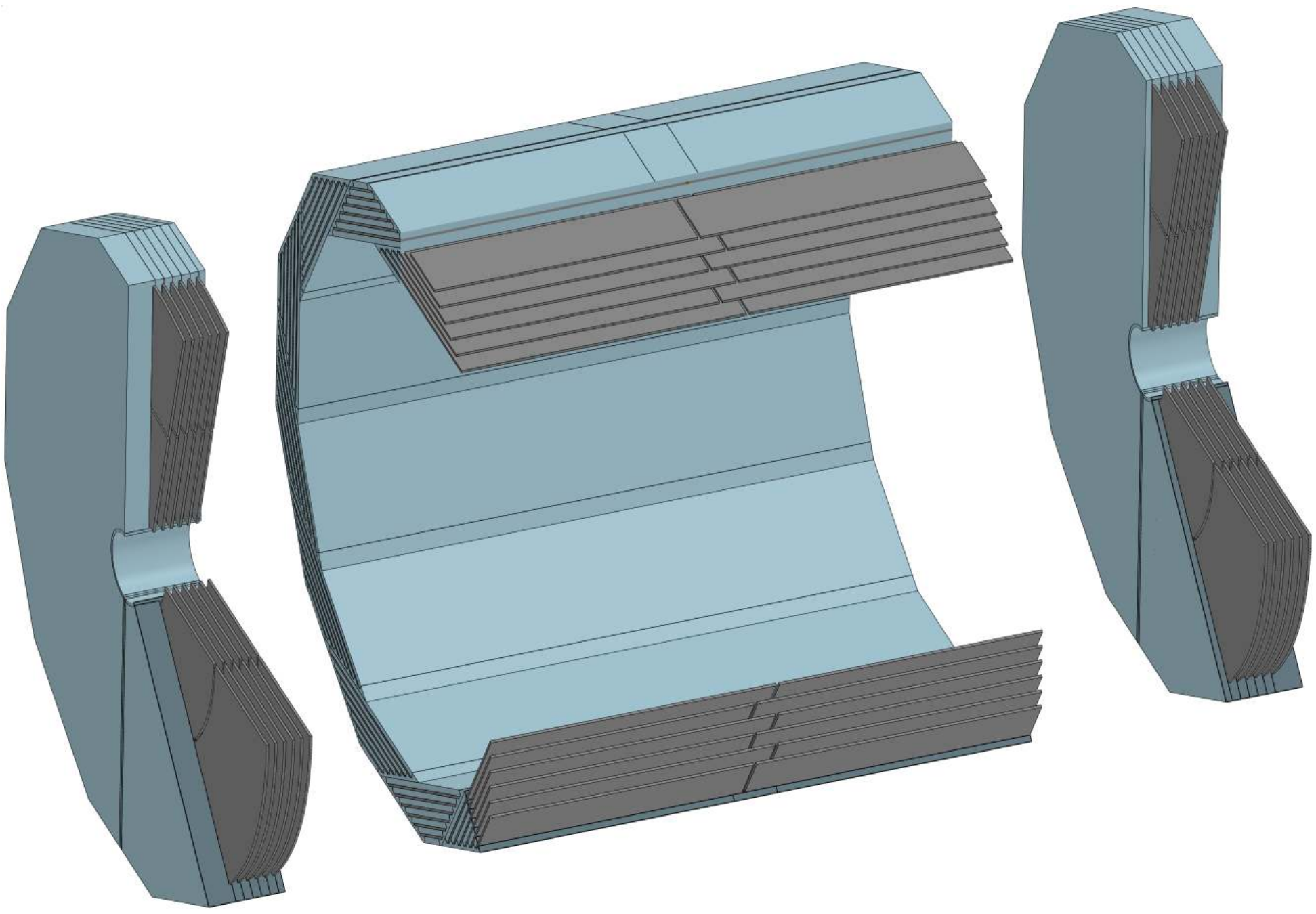

**Figure 9.1:** Overall view of the muon detector. It is composed of a dodecagonal barrel and two endcaps. The muon chambers (dark gray) are incorporated into six layers inside the iron yoke (sky blue).

Each sector (a dodecant of the barrel, or a quadrant of the endcaps) includes six detector layers interleaved with seven iron plates. The iron plates provide 4 cm gaps to house muon detector modules. Each iron plate has a thickness of 10 cm, except for the outermost layer of the endcaps, which has a thickness of 40 cm to meet the requirements of the magnetic field system. At each endcap, an additional outer layer is used to hold paraffin for shielding against neutron backgrounds from the accelerator tunnel. The paraffin layer is discussed in Chapter 14 and will not be addressed further in this chapter. There is a reinforcing stiffener between the plates of a gap, partitioning the gap into two volumes. Thus, there are a total of twelve detector modules in each sector.

A detector module consists of two perpendicular layers of PSUs, as explained in Section 9.2.3. The PS strip of a PSU has a rectangular cross section measuring 4 cm × 1 cm and various lengths. The two-layer configuration enables three-dimensional position reconstruction with a performance equivalent to that of 4 cm × 4 cm × 2 cm cubic cells, delivering a spatial resolution of approximately 1 cm for a single charged-particle track. Ghost tracks can emerge in sectors that have multiple tracks. The detector channel's time resolution of 1 ns provides an additional measurement of the position with a spatial resolution of approximately 17 cm. By leveraging the time information from multiple hits in the detector, these ghost tracks can be effectively removed.

In a 3T magnetic field, a muon with momentum above 5 GeV/$c$ generated at the IP traverses the muon detector along a nearly straight trajectory before escaping or decaying. However, low-momentum muons produced indirectly, such as those from hadronic showers or jets in calorimeters, can also penetrate the detector if created in proximity. Optimized for performance, the detector can achieve a standalone detection efficiency exceeding 95% while maintaining high purity, a critical feature for effectively distinguishing muons from background charged pions in high-collision environments.

Recent advancements in PS materials, WLS optical fibers, and SiPM technologies have enabled the development of extruded PS-based systems as a viable solution for muon detector. The integration of PSUs as active detector elements offers significant advantages: structural simplicity, high-rate capability, low operational voltage enabled by SiPMs, and excellent time resolution. These attributes align closely with the stringent performance requirements of modern muon detectors.

Cost-benefit analyses further indicate that the PS-based design is economically competitive with an RPC system. After evaluating the technical merits and cost efficiency, the extruded PS technology, enhanced by WLS fiber readout and SiPM sensors, is selected as the baseline design for the muon detector. In contrast, RPC technology is kept as a backup option.

## 9.2 Overall Detector Design

The design of the muon detector has undergone substantial evolution since the publication of the CDR [1], driven by advancements in technological R&D and simulation-based optimization. This section provides the refined detector architecture, integrating insights from material studies, spatial resolution improvements, and signal processing. Key aspects of the system, including the geometric configuration, the electronics readout system, the structures of PSU and detector modules, and the cable arrangement, are detailed in the following subsections.

### 9.2.1 Geometry configuration

The detector consists of a central barrel and two endcaps, with the key parameters detailed in Table 9.1. Six layers of modules made of PS strips are embedded in each yoke sector. Significant efforts have been made to minimize dead zones, resulting in a total ineffective area of approximately 0.34%. They originate from two primary sources: two magnet chimney holes (~ 0.04%, discussed in Chapter 10) and the gaps between the endcap sectors (~ 0.3%).

Figure 9.2 illustrates the design of the barrel, characterized by a spiral dodecagonal geometry constructed from trapezoidal-shaped sectors. These dodecants are precisely arranged to form a cylindrical detection apparatus, with each sector oriented such that its longer edge faces inward and shorter edge radiates outward. The trapezoidal configuration enables adjacent sectors to interlock seamlessly along the dodecagonal perimeter: the longer edge of a sector aligns with the slanted edge of its neighbor, creating a continuous structural pattern. These interlocking connections are integrated into the iron yoke layers, ensuring mechanical stability, robust alignment, and uniform distribution of structural forces across the assembly.

**Table 9.1:** Design parameters of the muon detector.

| Parameter | Value |
|---|---|
| Solid angle coverage | $> 0.98 \times 4\pi$ |
| Total detection area | $4.8 \times 10^3$ m$^2$ |
| Number of layers per sector | 6 |
| Thickness of detector module | 35 mm |
| Barrel inner radius | 4245 mm |
| Barrel outer radius | 5185 mm |
| Barrel length | 8950 mm |
| Barrel physical thickness | 940 mm |
| Endcap inner radius | 550 mm |
| Endcap middle radius | 2800 mm |
| Endcap outer radius | 5085 mm |
| Endcap physical thickness | 1240 mm |

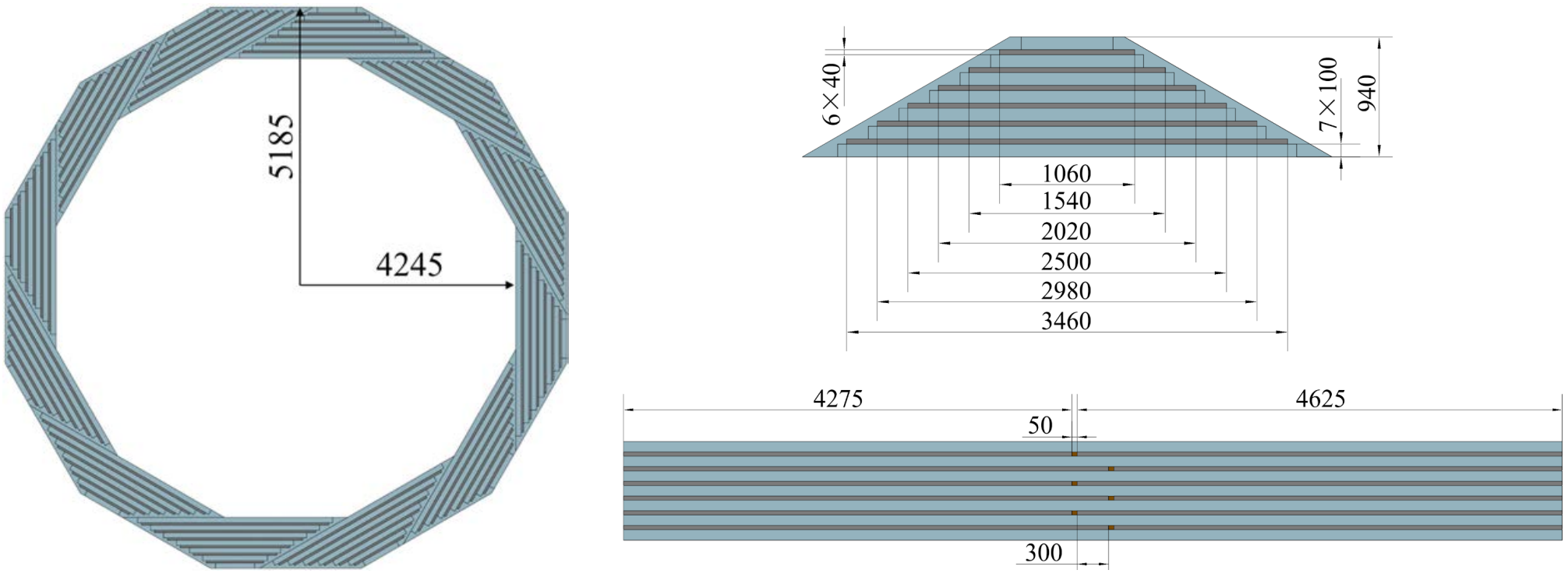

**(a)** **(b)**

**Figure 9.2:** (a) Barrel yoke structure with six layers of integrated muon detectors viewed axially. The full yoke structure features a regular dodecagonal configuration composed of long elements with an isosceles trapezoidal transverse shape. (b) Transversal and longitudinal views of the yoke elements containing the muon detectors. The units of the geometry labels are in millimeters.

The barrel has a total length of 8,950 mm, segmented by 50 mm-wide structural stiffeners. To eliminate potential dead zones, the stiffeners are not centrally positioned but offset by 300 mm between successive layers, creating a staggered arrangement. This design produces two distinct modules per layer: one spanning 4,275 mm and the other 4,625 mm in length. Furthermore, the width of each dodecant tapers progressively across layers, decreasing from 3,460 mm at the innermost layer to 1,060 mm at the outermost layer.

Figure 9.3 depicts the structural design of an endcap, featuring a radial coverage extending from 650 mm to 5,085 mm. As highlighted in Section 9.4.3, background levels near the beam pipe are significantly higher than those in other areas. To address this, the detector layers are segmented into inner and outer modules by 50 mm wide stiffeners. The inner modules, spanning 650 mm to 2,800 mm radially, are designed to withstand higher hit rates due to their proximity to the beam pipe region. In contrast, the outer modules spanning 2,850 mm to 5,085 mm detect particles at larger angles. Both module types are positioned to facilitate efficient cable management, with dedicated pass-throughs integrated in their designs. Lengths of the PS strips within each module are optimized, with the longest measuring 2.7 m for inner modules and 4.2 m for outer modules, ensuring consistent signal strength across the detection area.

The muon detector consists of 240 modules in total: 144 in the barrel, and 48 in each of the two endcaps. The barrel configuration features twelve distinct module types differentiated by variations in length and width dimensions. The endcap modules are divided into two specialized types corresponding to the geometric requirements of their inner and outer zone locations. These 240 detector modules constitute $4.3 \times 10^4$ PSUs, a total sensitive length of 119 km, and a total sensitive area of $4.8 \times 10^3$ m$^2$.

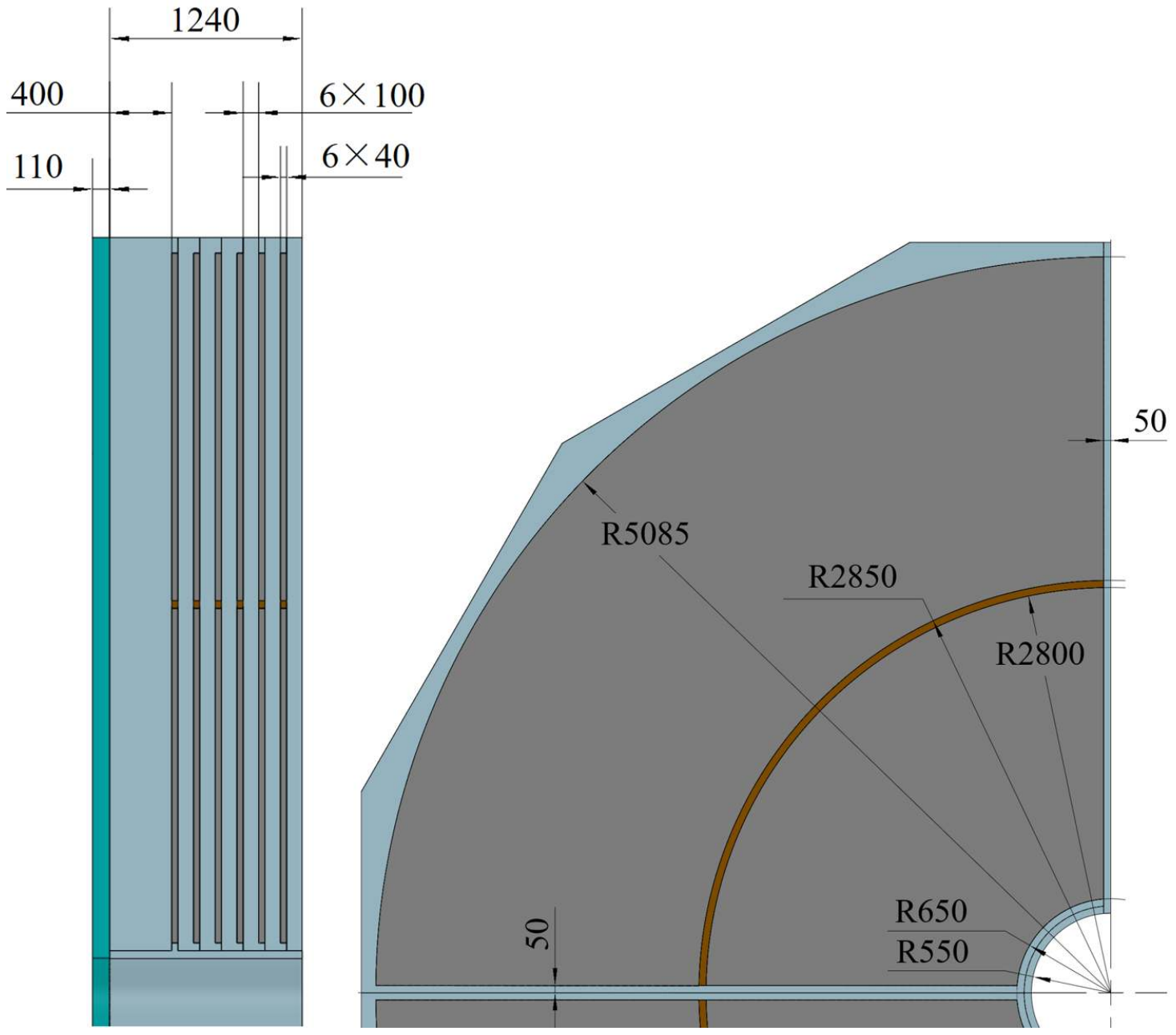

**Figure 9.3:** Geometric configuration of an endcap muon detector, showing side and front views of the module (gray), including the arrangement of inner ($R \in$ [650 mm, 2800 mm]) and outer ($R \in$ [2850 mm, 5085 mm]) components divided by a 50 mm wide stiffener shown in orange. Outside the outermost iron plate is a paraffin layer. The units of the geometry labels are in millimeters.

### 9.2.2 Electronics readout system

Figure 9.4 illustrates the architecture of the on-detector electronics system for the muon detector. The design is organized into three primary functional stages. The core components are SiPM tile, Front-End Electronics Board (FEB), and Management Board (MB). Detailed technical specifications, performance benchmarks, and operational validation results are provided in Chapter 11.

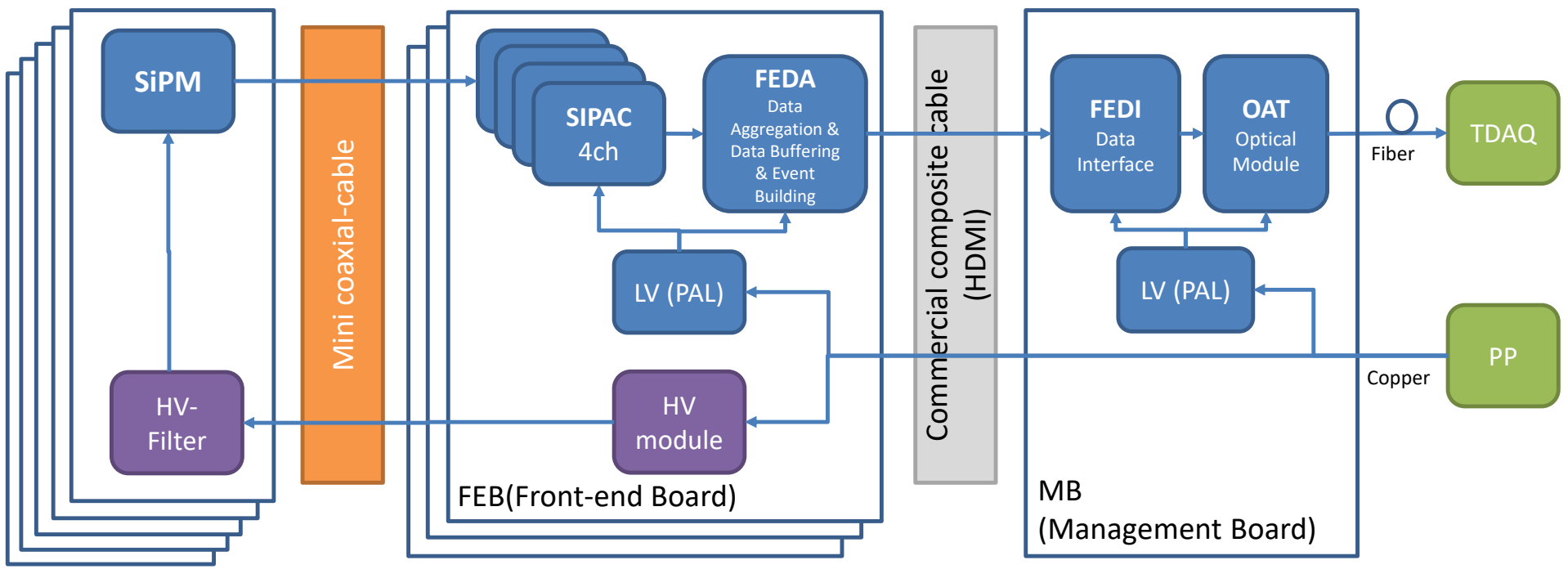

**(a)**

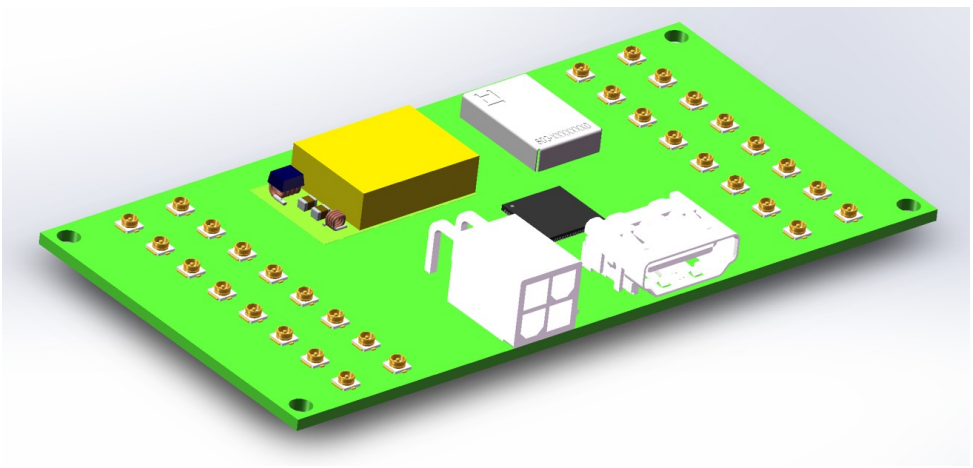

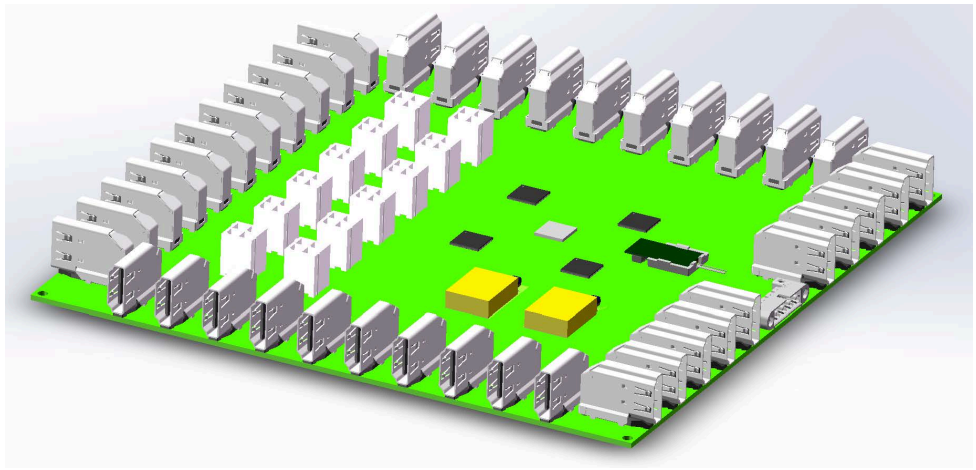

**(b)** **(c)**

**Figure 9.4:** The electronics readout system of muon detector. (a) Block diagram for the on-detector electronics. (b) An FEB contains a mini HV generator to drive the SiPMs and connectors for LV power cables, data cable, and 32 readout channels. (c) An MB contains connectors for data cables, LV power cables, and fiber cables.

1. **SiPM Tile**: A SiPM tile is installed at one end of each PS strip and serves as a photon detector. It converts photon signals generated by charged-particle tracks in the PS strip into electrical signals.
2. **FEB**: Having an area of 5 cm × 10 cm and thickness less than 1.5 cm, located inside a detector module, the FEB features a design that integrates 32 readout channels and a mini power generator. The FEB incorporates a custom-designed front-end ASIC chip, which performs charge integration of the SiPM output and records precise signal arrival time. The digitized data is then forwarded to the subsequent data aggregation and transmission stage.

   The mini HV power generator, designed using the MAX5026 module, provides an output voltage range of 0 – 71 V and a current output of up to 2mA. It features a ripple noise of 2 mV/Vpp and a maximum power consumption of 200mW. The FEB also includes an on-board DC-DC converter to regulate power supply levels appropriate for both the ASICs and the SiPMs.

3. **MB**: Located outside a detector sector, the MB serves as the central node for multiple functions, including data aggregation, clock distribution and synchronization, fast control command handling, and power management. It receives digitized data from the FEBs, consolidates information, and converts them into high-speed optical signals, which are transmitted via optical fibers to the off-detector (back-end) electronics in the electronics room. The MBs also receive clock and fast control signals from the back-end system and distribute them to the FEBs, ensuring precise timing coordination across the entire detector. The muon detector system comprises 40 MBs, each connected to the back-end electronics through dedicated power cables and optical fiber links.

A control and monitoring system utilizes hardware measurements in conjunction with rapid data analysis to ensure high data quality. Since the SiPM's response is highly sensitive to ambient temperature fluctuations, temperature sensors are placed on the FEBs inside the detector modules, and the temperatures are continuously recorded. The feasibility of adjusting the operation of the SiPMs at various voltages to compensate for temperature drifts is currently under investigation.

The currents of the HV system integrated on the FEBs for the SiPMs are also recorded. During experimental operation, the currents show sensitivity to background levels. Zero-current readings identify channels that are dead or disconnected. A hot channel can also be identified by its current. SiPM currents from calibration runs help monitor the long-term stability and are linked to the accumulated radiation dose.

A hundred thousand random triggers per calibration run ensure reliable pedestal measurements for each channel. The pedestal width effectively indicates the levels of SiPM and electronics noise. These data help determine the spectrum of photoelectron peaks, enabling the calculation of gain. However, this accuracy may decline over several years due to an increase in SiPM DCR resulting from radiation damage. Muon detection efficiency for each channel can be monitored online. The muon detector's Data Quality Monitor (DQM) histograms show current efficiency, muon trigger rate, and angular distribution for each channel.

### 9.2.3 Detector module

As illustrated in Figure 9.5, a PSU consists of an elongated PS strip with a cross section of 4 cm × 1 cm, a WLS fiber embedded within the PS strip, and a SiPM mounted on a compact PCB. The PS strip is coated by a 98% reflective layer made of $TiO_2$ or Teflon. The thickness of the reflective layer is ≤ 0.1 mm. Using a coupling component, the WLS fiber with a diameter of 1.2 mm (D12) or 2.0 mm (D20) is optically coupled to a SiPM, which has an effective photosensitive area of 3 mm × 3 mm.

Efficient photon collection relies critically on optimal coupling between the WLS fiber and

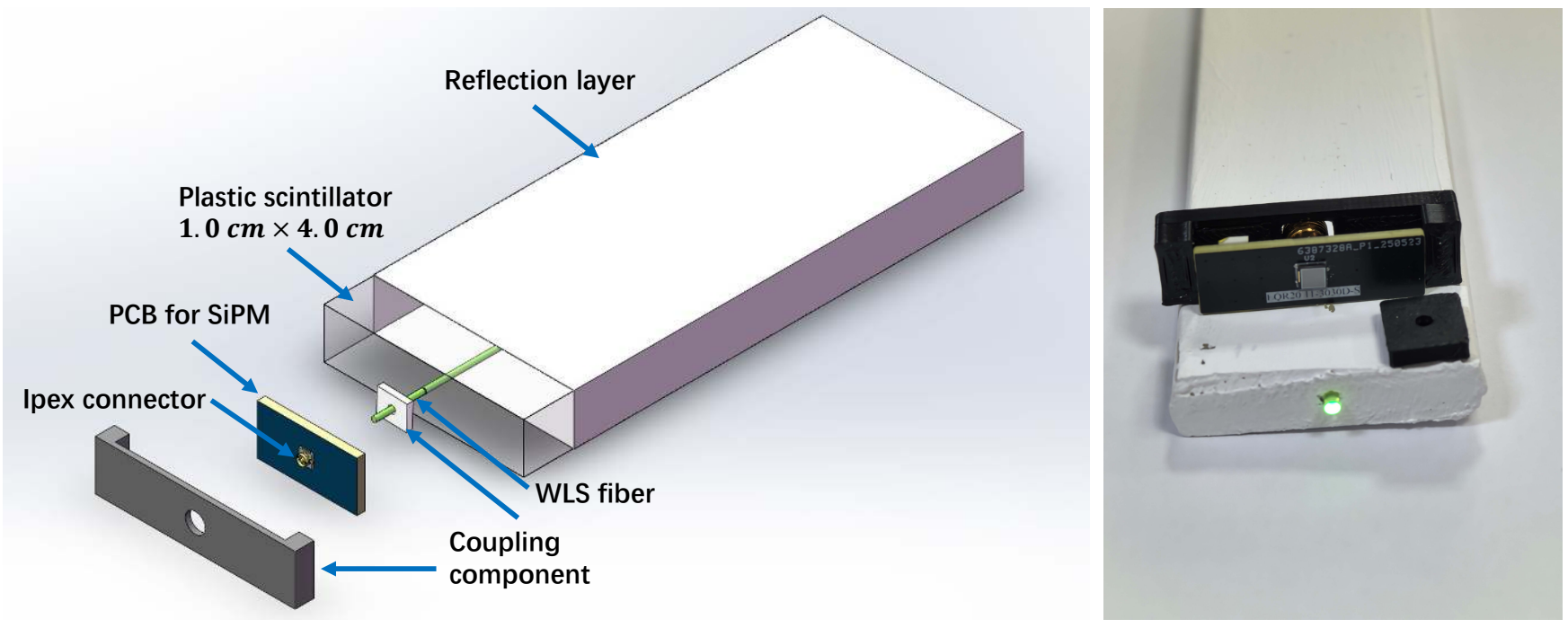

**(a)** **(b)**

**Figure 9.5:** (a) A schematic view of a PSU, featuring an elongated PS strip with a cross section of 4 cm × 1 cm, a WLS fiber fitted inside the PS, a SiPM mounted on a compact PCB, and a coupling component to optically connect the terminal of WLS fiber and the SiPM active surface. (b) A photo of the components for a prototype PSU, which shows the SiPM.

the SiPM surface. Such a coupling is achieved by integrating a precisely machined rectangular slot to immobilize the SiPM and a dedicated alignment hole to fix the fiber in position. This design ensures rigorous coaxial alignment between the SiPM's active surface and the fiber's terminal, minimizing photon loss and enhancing the signal integrity.

The 4 cm width of the PS strips is primarily determined by the requirement of the spatial resolution of the muon tracks. As a muon travels from the IP to the muon detector, multiple scattering introduces positional uncertainty along its trajectory. For a muon track with a momentum above the threshold of about 5 GeV$/c$, the scattering-induced spatial uncertainty is below 1.3 cm. The selected 4 cm strip width provides a spatial resolution of $\sim$ 1 cm through hit reconstruction across multiple detector layers.

A module consists of two orthogonally arranged layers of PSUs, enabling multidimensional signal detection. Figure 9.6 depicts the internal structures of a barrel detector module in (a), an inner endcap module in (b), and an outer endcap module in (c). More details of the internal structure of a barrel module are shown in Figure 9.7. Mechanical support of the module is provided by an aluminum frame composed of two 1 mm-thick surface plates and four perimeter borders with a thickness of 4 mm.

As shown in Figure 9.7, positioned between an aluminum plate and a PSU layer is a 6.5 mm-thick high-strength plastic sheet. The bottom sheet covers the interior area of the frame. The top sheet is recessed by 200 mm along the long edge and 250 mm along the short edge of the frame. This reduction creates clearance for routing electronics connections while maintaining mechanical integrity. The bottom sheet serves as a substrate for the adhesion of the bottom-layer PSU strips. The top-layer PSU strips are bonded to the bottom PSU layer. Intra-layer strips of both layers are interconnected for structural cohesion. The top sheet is adhered to the top PSU layer after installation, using glue to secure the module structure.

A dedicated 100 mm-long window is created at a corner of one module to accommodate power cables and data cables. As shown in Figure 9.6, the corner is along the long border and close to the short border in a barrel module, or along the vertical border and close to the outer arc border in an endcap module. The window has a width of 15 mm in a barrel module, and a width of 6.5 mm in an endcap module. More details are described in Subsection 9.2.4.

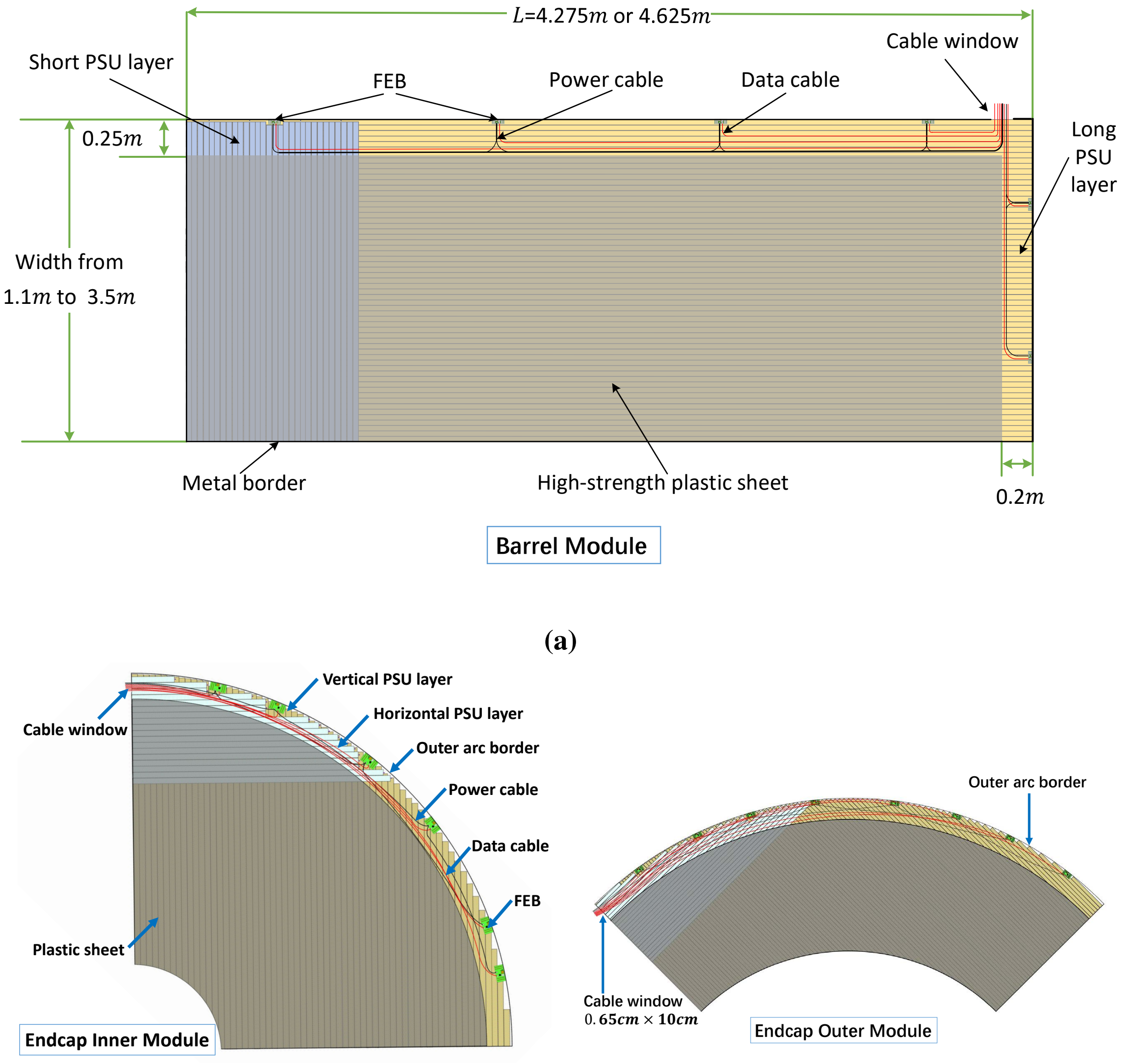

(b) (c)

**Figure 9.6:** (a) Schematic layout of a barrel detector module illustrating the arrangement of FEBs, routing for LV power and data cables, and integrated cable management pathways; (b) Detailed view of an inner endcap module structure; (c) Outer module design showcasing components analogous to those in a barrel module.

The PSUs of the bottom layer are aligned along the $z$ direction. Each fiber is coupled at one end to a SiPM oriented toward the endcap. The PSUs of the top layer are aligned along the $R-\phi$ direction. The SiPM-equipped sides are oriented towards the detector's outer edge to mitigate radiation-induced degradation. The end of a WLS fiber contacting no SiPM features a

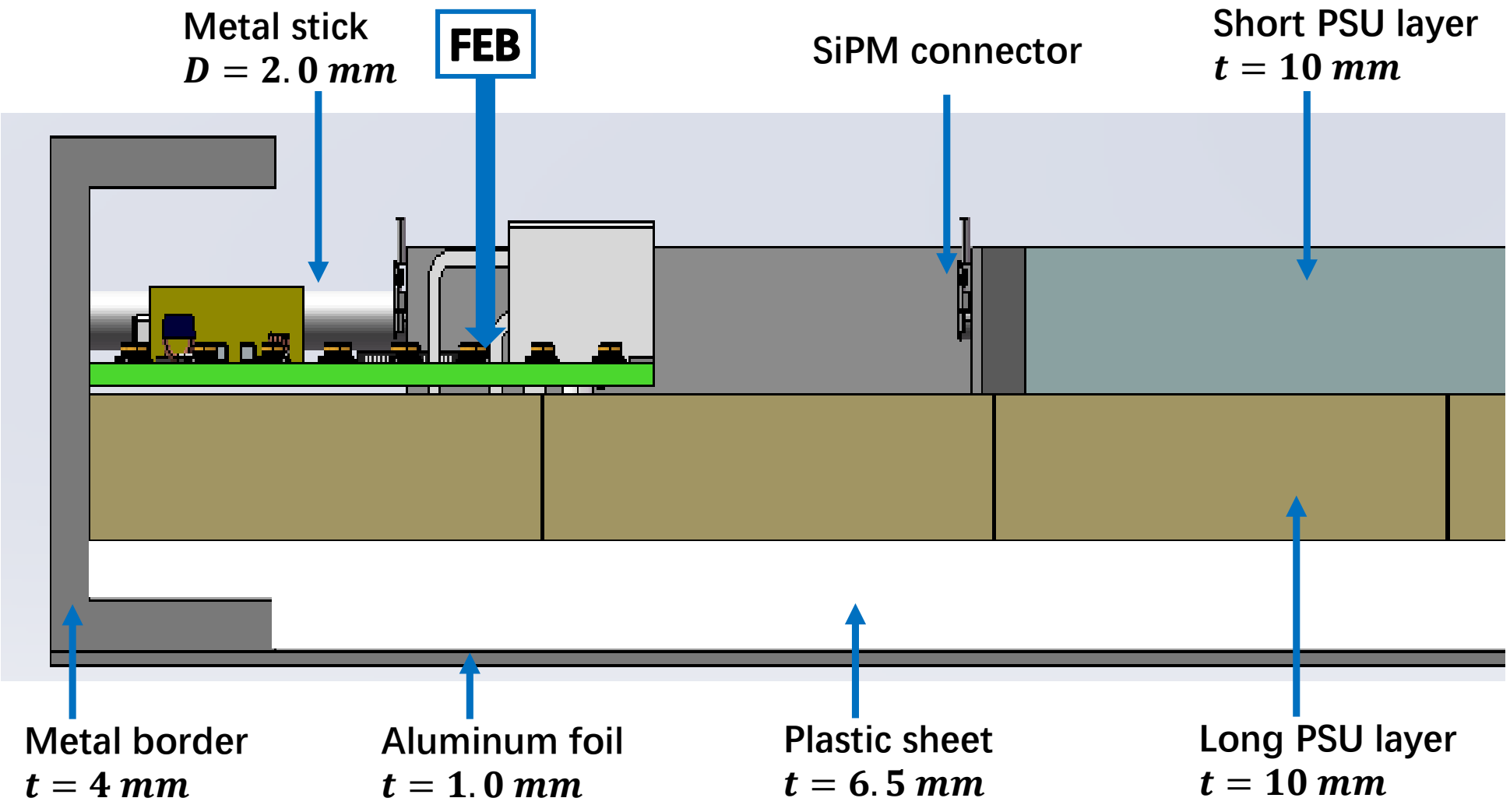

**(a)**

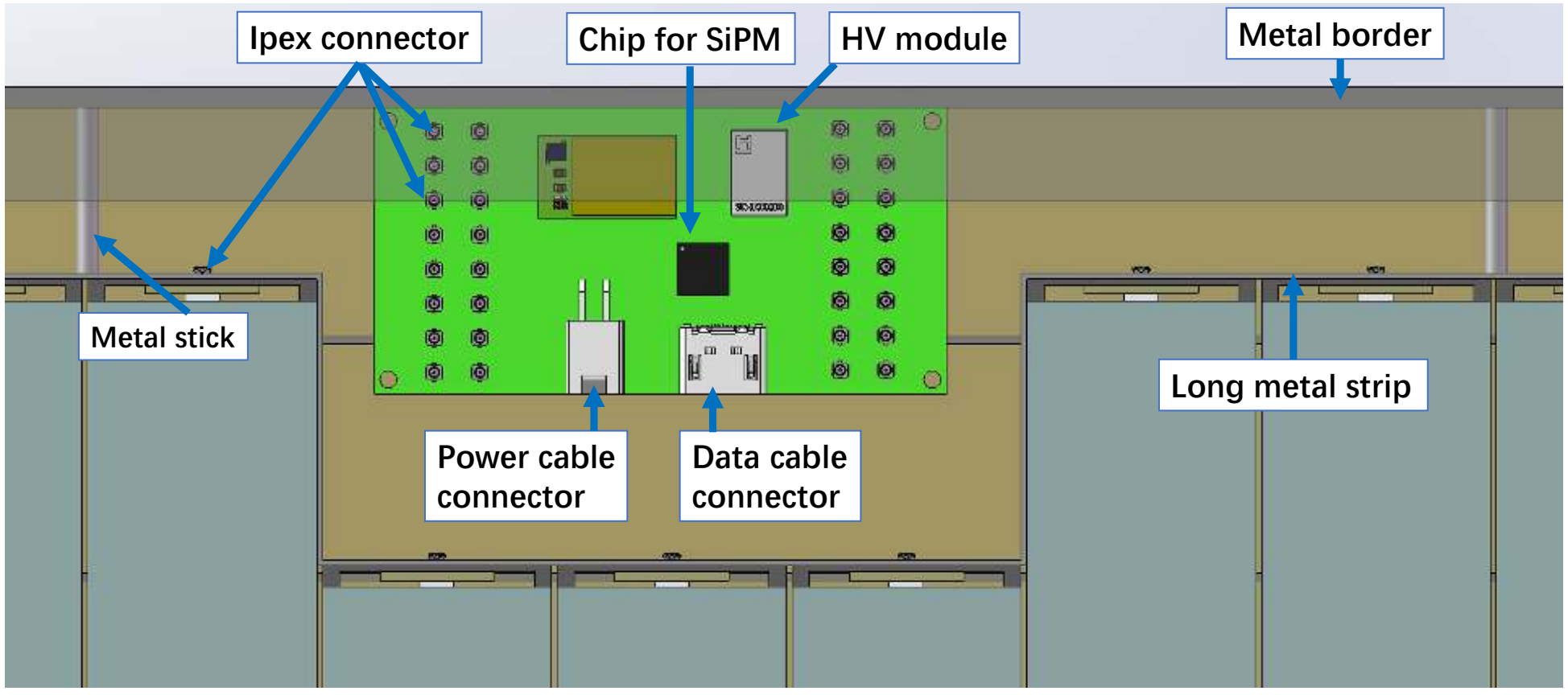

**(b)**

**Figure 9.7:** Detailed close-up views of a barrel module in the electronics connection region: (a) Cross-sectional perspective depicting the mechanical support of an aluminum frame and surface plates, plastic sheets, arrangement of the two PSU layers, SiPM connection, and FEBs. (b) Top-down view highlighting a designated 3 – 6 cm space for FEBs, cable management between FEBs and SiPMs.

silver-coated tip to enhance light reflection.

Figure 9.8 illustrates the arrangement of FEBs with PSUs across the six layers of modules in one side of a barrel detector sector. Each barrel sector, as well as each endcap sector, contains 12 modules.

The endcap modules have slightly different shapes but similar structures. A similar procedure is followed for arranging the bottom layer of PSUs (horizontal and facing the IP), the FEBs, and the top layer of PSUs (vertical), as shown in Figures 9.6b and 9.6c for the inner and outer modules, respectively. The FEBs are mounted on the outer arc border of the sector frame. In case of overlap occurring between FEBs for the bottom and the top PSU layers, their positions are slightly shifted but remain in proximity. This design ensures that the inner area adjacent to the beam pipes is fully covered by PS strips integrated with WLS fibers of the inner modules.

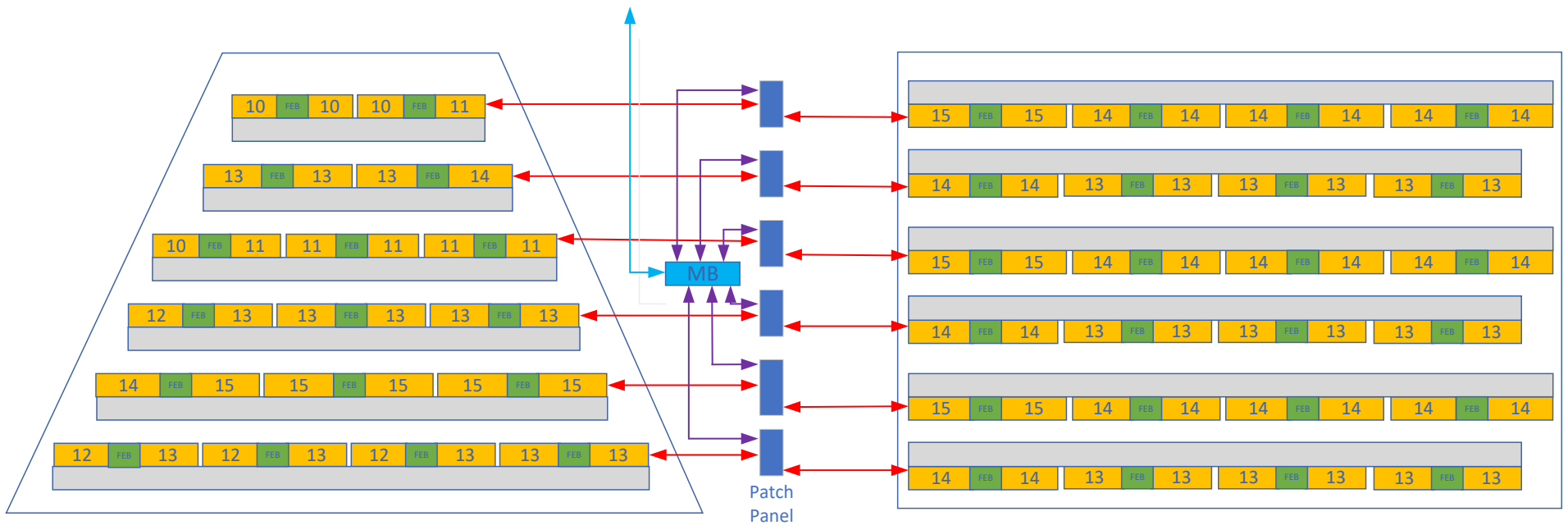

**Figure 9.8:** Configuration of FEBs and PSUs groups in the six modules on one side of a barrel sector. (Left) FEBs mounted on the short frame edges, with corresponding numbers of long PSUs. (Right) FEBs mounted on the long frame edges, with corresponding numbers of short PSUs. The MBs are positioned outside the detector modules, and their specific locations are illustrated in Figure 9.10.

FEBs are integrated into the detector modules, necessitating careful design of their corners to ensure proper functionality. Figure 9.7 provides a detailed view of the area surrounding an FEB within a barrel module. The FEB is positioned above the bottom PSU layer and mounted along the long straight border in a barrel module or the outer arc border in an endcap module. To accommodate the installation of the FEB, the lengths of the three adjacent PS strips are reduced by 2 cm.

Some PSU strips are oriented downward in both the barrel and endcaps. Such a PSU strip has a weight up to 1.4kg in the barrel and up to 1.7kg in the endcaps. To enhance mechanical stability and prevent gravitational displacement within the barrel module, a long metal strip is installed parallel to the module's long edge. This strip, with a thickness of 2 mm and a width of 1 cm, features holes of 4 mm diameter to accommodate the cabling for the SiPMs. Additionally, two metal rods, each with a diameter of 2 mm, are positioned around each FEB to maintain the spacing between the PSU strips and the long edge, as illustrated in Figure 9.7. A similar approach is applied in the endcap modules, with the exception that the long metal strip, aligned parallel to the outer arc border, is shaped into a stepped configuration to accommodate the module's geometry.

### 9.2.4 Arrangement of cables

Figure 9.9 shows the connection between PSUs and an FEB established using mini coaxial cables with a diameter of 1 mm and two Ipex connectors with a size less than 2 mm, ensuring both signal integrity and mechanical adaptability. The connection between the FEB and the MB utilizes HDMI cables to ensure the robust transmission of data and control signals. The diameter of an HDMI cable is less than 4 mm. Meanwhile, each MB distributes power to the FEBs inside one detector module in a daisy chain through LV cables. An FEB has two groups of power connectors, one for the input from the MB or a previous FEB, another for output to the next FEB, until the last FEB. This creates a series connection for power distribution, reducing the number of power cables from the MB. There are two groups of power cables from an MB, one for the bottom PSU layer and the other for the top PSU layer, respectively.

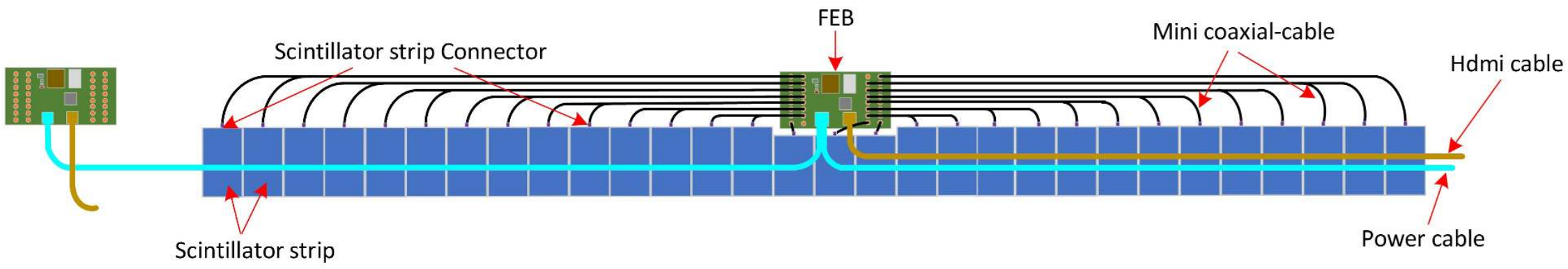

**Figure 9.9:** Diagram illustrating an FEB interfacing with its associated PSUs and MB. The FEB connects to the PSUs via 1 mm thin coaxial cables. The LV power is supplied to the FEBs from the MB through a daisy-chained LV cable network. Data communication between the FEB and MB is facilitated by dedicated HDMI cables.

Figures 9.10a and 9.10b show the cable windows in a barrel module and an endcap module, respectively. Each window is designed to accommodate the cabling for two groups of power cables and 6 – 8 data cables. Figures 9.10c and 9.10d illustrate the positions of two MBs on a sector, which serve to connect the 12 modules installed within the sector.

In a barrel sector, the cable windows are positioned near the endcaps and the outer edge of the muon detector. Six channels, each with a cross section of 3 cm $\times$ 15 cm, are constructed on a single yoke cover plate. The power and data cables exiting the cable windows are routed through these channels and connected to the two MBs.

In an endcap sector, a vertical channel with a cross section of 4 cm $\times$ 5 cm is created within the gap between iron plates after the installation of one inner module and one outer module, with the designed geometry of the detector modules and the stiffener (see Figure 9.3). The cables exiting the cable windows of these two modules are directed outward from the yoke. The two MBs are positioned on the top or bottom of the yoke. The connection from FEBs to MBs in an endcap sector follows a similar configuration to that in a barrel sector.

## 9.3 Key technology R&D

Muon detectors utilizing PS technology have demonstrated exceptional performance in large-scale experiments, such as the Belle II KLM detector [2]. Recently, similar designs have been adopted by the Chinese experiments Large High Altitude Air Shower Observatory (LHAASO) [3] and Jiangmen Underground Neutrino Observatory-Taishan Antineutrino Observatory (JUNO-TAO) [4]. However, the systems employ significantly shorter PS strips compared to those in our muon detector, where strips exceed 4 m in length, with some approaching 5 m. The extended strip length in our design introduces challenges.

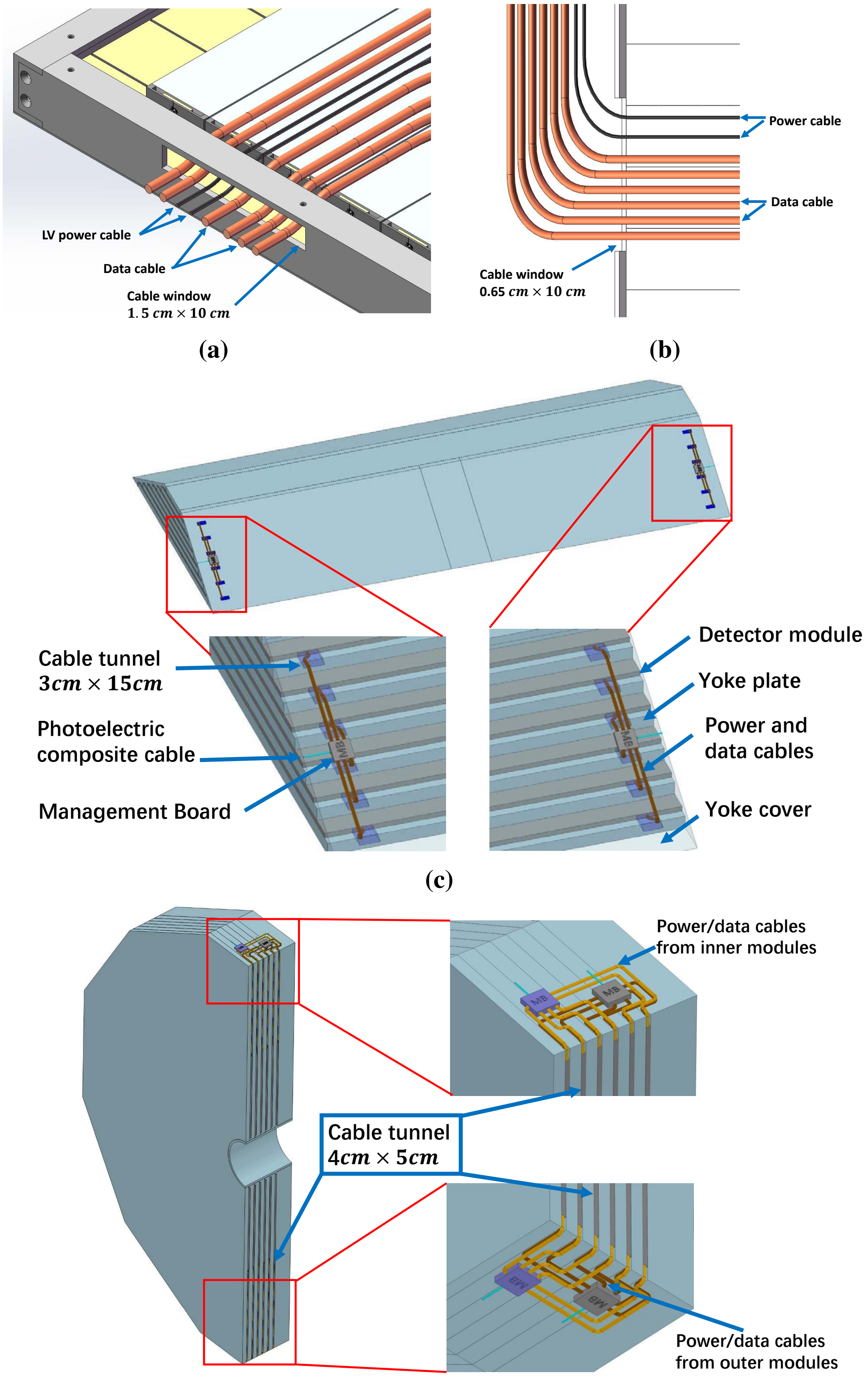

**Figure 9.10:** Schematic diagram illustrating the connection of 12 modules within a barrel or endcap sector to MBs. Two MBs are mounted on the detector yoke, with cables routed through module windows and yoke holes in the barrel cover plate or through dedicated channels in the endcaps. The two colors in (d) indicate cable bunches from inner and outer modules of the endcaps.

Specialized preamplifiers [5] were designed to process SiPM signals and evaluate the performance of multiple SiPMs models. During the early stages of R&D, the properties of 1.5 m long PS strips were investigated and the methods to enhance photon collection were explored [6]. Building on these findings, the PS strip design was optimized to improve photon production efficiency through three key advancements: enhanced integration of WLS fibers, increased fluorescence efficiency within the scintillator material, and refined fabrication techniques. The development of technology for manufacturing long PS strips with a length of 5 m was finally completed, and their performance characteristics were evaluated.

### 9.3.1 Investigation of SiPMs

SiPM offers various advantages in photon detection, such as high quantum efficiency, compact size, and low operation voltage. The limitations of SiPM's DCR can be mitigated by adjusting the threshold of the number of photoelectrons (nPE). Given the substantial luminosity at CEPC, PS presents significant potential with high-rate performance. The decay times of PS and WLS fiber can be as short as a few nanoseconds, so that the primary constraint lies within the readout system. In the inner module of an endcap near the beam pipe, a long strip of about 2.7 m could achieve an average rate capability of up to 18 kHz/ cm$^2$, far exceeding the maximal expected rate of 50 Hz/ cm$^2$, which was based on background simulations described in Section 9.4.3.

To achieve excellent performance from the SiPM, a compact, high-speed, low-noise preamplifier was developed [5], which offers a stable gain of about 21, a bandwidth of 426 MHz, a baseline noise level of approximately 0.6mV, and a time resolution of 20 ps, making it suitable for use with various SiPMs. The photon collection is assessed by examining pulse heights or ADC counts, and the precise timing information is derived using the leading-edge or constant-fraction discrimination method.

As discussed in Ref. [6], SiPMs from HPK and NDL were tested. The HPK S13360 series Multi-Pixel Photon Counters (MPPCs), with a pixel size of 50 μm, offers a sensitive area of 1.3 mm × 1.3 mm. The Epitaxial Quenching Resistor (EQR)-15 11-3030D-S series, with a pixel size of 15 μm, features a larger sensitive area of 3.0 mm × 3.0 mm, significantly exceeding the cross sections of all the WLS fibers tested in the R&D. Experimental results show that the S13360 MPPC achieves a gain of $1.1 \times 10^5$ at an operational voltage of 57.0 V. In comparison, the EQR-15 11-3030D-S has a gain of $0.7 \times 10^5$ at 27 V.

Given the extensive use of D20 WLS fibers, MPPCs with a sensitive area of 1.3 mm×1.3 mm are unsuitable for optimal matching. To address this, HPK has introduced new MPPCs with a larger sensitive area of 3 mm × 3 mm, such as the S14160-3050 with a pixel size of 50 μm. Additionally, the new NDL EQR-20 SiPM, featuring a pixel size of 20 μm, demonstrates significantly enhanced gain and reduced DCR. The two types of SiPM have a similar PDE of 50%. The test results for the MPPC S14160-3050 and the NDL EQR20-11-3030 are presented in Figure 9.11.

Figure 9.11b shows that the cross talk of the EQR20-11-3030 is less than twice that of the S14160-3050, with one S14160-3050 sample exhibiting cross talk levels comparable to the EQR20-11-3030. As illustrated in Figure 9.11c, the DCR decreases rapidly for both devices as the nPE threshold increases. By raising the nPE threshold by one photoelectron, the EQR20-11-3030 achieves a DCR similar to that of the S14160-3050. At an nPE threshold of six, both SiPM types demonstrate a DCR less than $10\,\mathrm{Hz/\,cm^2}$. Figure 9.11d presents the breakdown voltages for 10 samples of each SiPM type, with deviations within 0.5 V. The EQR20-11-3030 exhibits an average breakdown voltage 10 V lower than that of the S14160-3050.

As detailed in Section 9.3.3, the nPE of the SiPM signals significantly exceeds 15, with an average value of approximately 30 at the far end of the 5 m long PSU prototypes. The NDL EQR20 SiPM meets the requirements of the muon detector, as the nPE threshold can be tuned within the range [6, 15], achieving a significantly lower DCR while maintaining a detection efficiency close to 100%. Although the NDL SiPMs exhibit lower gain compared to the HPK MPPCs, this limitation does not compromise their effectiveness in the PSU application. According to the product datasheets, the EQR20-11-3030 has a terminal capacitance of 57.5pF, significantly lower than the 500pF of the S14160-3050, resulting in a much shorter FWHM of 7 ns for the single photoelectron pulse. Given the additional advantages of a 10 V lower breakdown voltage and reduced cost, the NDL EQR is currently the preferred SiPM option.

### 9.3.2 R&D of PS strips

Extruded PS strips produced with a polystyrene extrusion method are used in the muon detector, as illustrated in Figure 9.5. Initial extensive R&D work is done to study these PS strips [6] and is summarized here. The work used 1.5 m strips with a groove created for embedding a WLS fiber. The study compared the performance of WLS fibers, SiPMs, and reflection coatings. The improvements of photon collection were investigated carefully, including the use of optical glue.

The 1.0 mm-diameter BCF-92 fiber from Saint-Gobain and the 1.2 mm-diameter Y11(200) fiber from Kuraray were tested. The results show that the light yield from the Kuraray fiber is about 2.3 times higher than that of the Saint-Gobain fiber [6].

The study shows that the connection between the WLS fiber, PS strip, and SiPM is crucial for the detector performance, as it affects both detection efficiency and time resolution. Methods include creating a coupling component to secure the WLS fiber and SiPM, polishing the end of the WLS fiber, applying a reflective layer to the scintillator in addition to the manufacturer's coating, and using optical glue between PS and WLS fiber. The coupling component provides a strong connection between the WLS fiber's end and the SiPM's sensitive area.

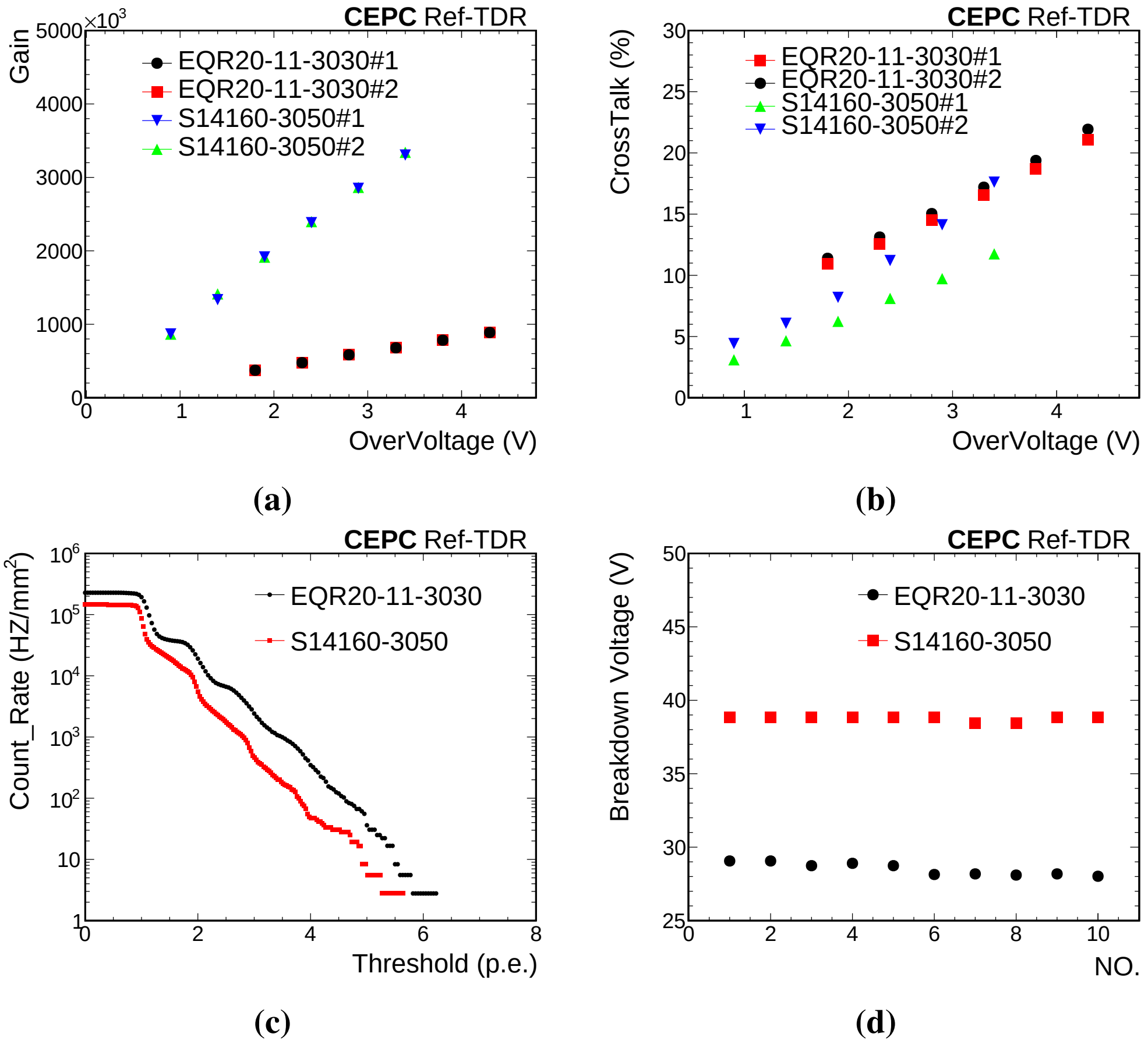

**Figure 9.11:** Test results of SiPMs from NDL and HPK. (a)-(b): Overvoltage dependence of gain and cross talk. (c): DCR vs. nPE threshold. (d): Breakdown voltages for 10 samples of each SiPM type.

Tests comparing reflective coatings indicate that $TiO_2$ slightly outperforms Teflon. However, Teflon offers production simplicity and cost-effectiveness, suggesting that further optimization is possible using thicker Teflon layers. SYLGARD-184 optical glue was used to improve light transfer between the fiber and the scintillator. This commercially available glue has a refractive index of approximately 1.4, matching that of the extruded scintillator. Tests with $TiO_2$-coated strips showed a 63% increase in nPE at the far end of the strip. Teflon-coated strips also saw a 30% improvement, highlighting the effectiveness of optical glue in boosting photon collection efficiency.

The early R&D phase involved assembling six detector channels and performing cosmic ray tests [6]. Both NDL SiPMs and HPK MPPCs showed comparable photon collection efficiencies, despite varying cross talk levels. Most strips maintained efficiencies above 90% at a threshold of 8P.E. The time resolution is closely linked to the nPE. Achieving a time resolution below 1 ns requires nPE $>$ 35 for NDL SiPMs and nPE $>$ 30 for MPPCs. Enhancing the scintillator's light yield by using larger SiPMs or employing multiple or larger diameter fibers can improve the time resolution. Time resolutions along the long strips were approximately 1.4 ns for NDL

SiPMs and 1.2 ns for MPPCs.

### 9.3.3 Performance of full-size PSU prototypes

The early R&D findings show that the technology using extruded PS strips has impressive efficiency and time resolution [6]. However, the tests used 1.5 m strips with a groove. These strips are too short and brutal for embedding WLS fibers for the muon detector. A new method that creates a hole in the PS strip during the production process is studied. Simulations indicate that changing from a groove to a hole can increase the nPE by a factor of 1.4, and increasing the fiber diameter from 1.2 mm to 2.0 mm results in an additional 2.8 times increase. Since there are many different channels in the muon detector, two fiber diameters can be used: D12 for PS strips shorter than 3 m, and D20 for longer strips.

A novel PS strip featuring an integrated WLS fiber hole is successfully developed, demonstrating 5 m length and twice the fluorescence yield. Four PS strips coated with $TiO_2$ and equipped with D20 WLS fibers were evaluated. Due to the substantial length of the PSUs, neither optical glue nor aluminum foil was applied. The test results for one PS strip with dual-end readouts are presented in Figure 9.12. At an nPE threshold of 15, the efficiency at the far end exceeds 95%. As illustrated in Figure 9.12c, the time resolution at the far end of a single readout channel is measured to be approximately 1 ns. By integrating time data from both ends, the 5 m PSU prototype achieves a time resolution of 0.6 – 0.7 ns. This improvement arises from averaging the synchronized signals from the two ends, which mitigates both statistical and systematic uncertainties in the time measurement, thereby enhancing precision.

## 9.4 Simulation and detector performance

This section presents the results of a GEANT4-based simulation integrated with the CEPCSW framework, evaluating the simulated detector performance and analyzing hit rates with background simulation from MDI.

### 9.4.1 Detector simulation and signal reconstruction

The detector simulation begins with a standalone run using GEANT4, which is later integrated into the CEPCSW. Charged particles, including muons and pions, are generated to interact with the detector, facilitating an evaluation of its performance. Following this, reconstruction algorithms have been developed for charged-particle tracks, based on the energies deposited in the PS strips of the detector.

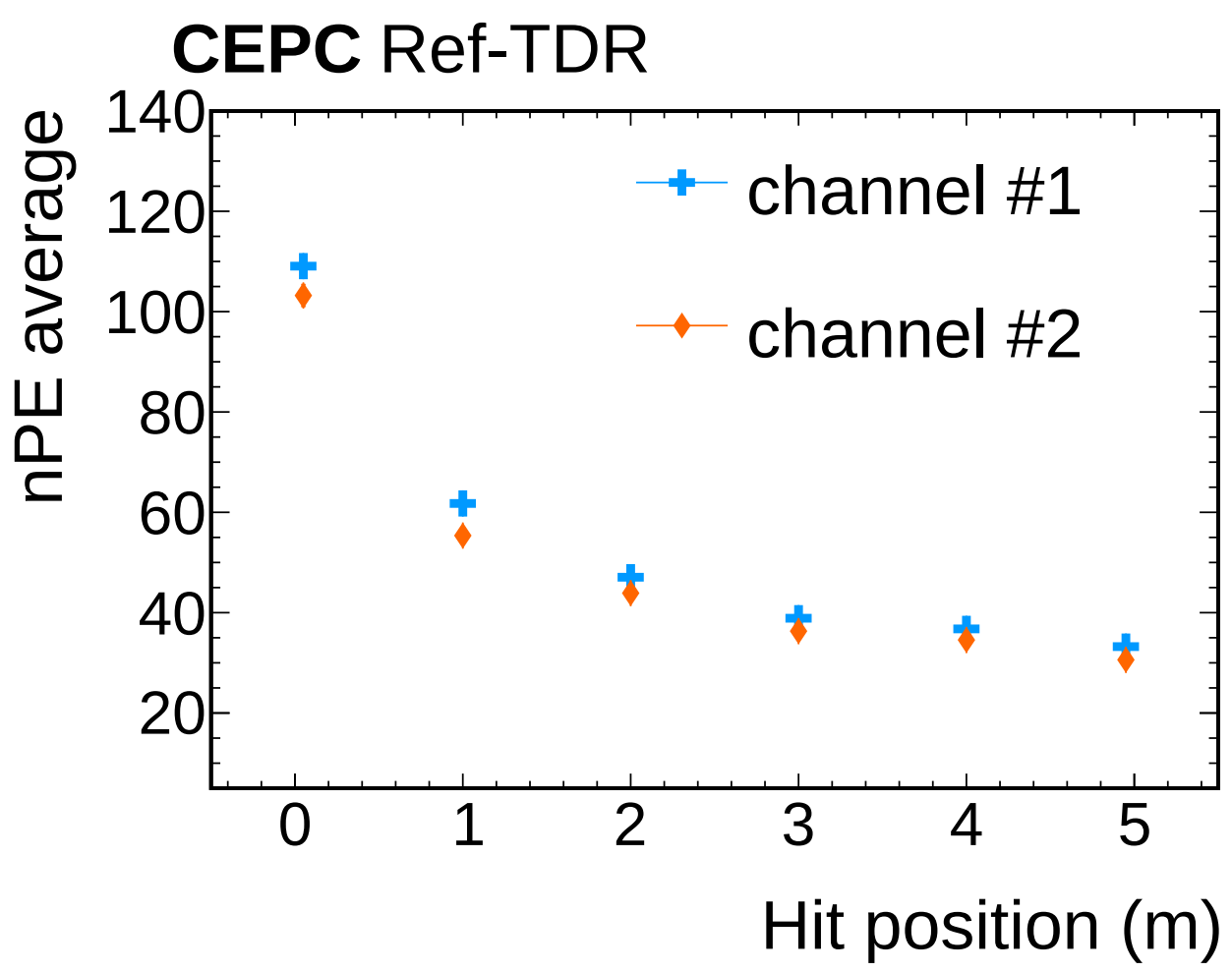

**(a)**

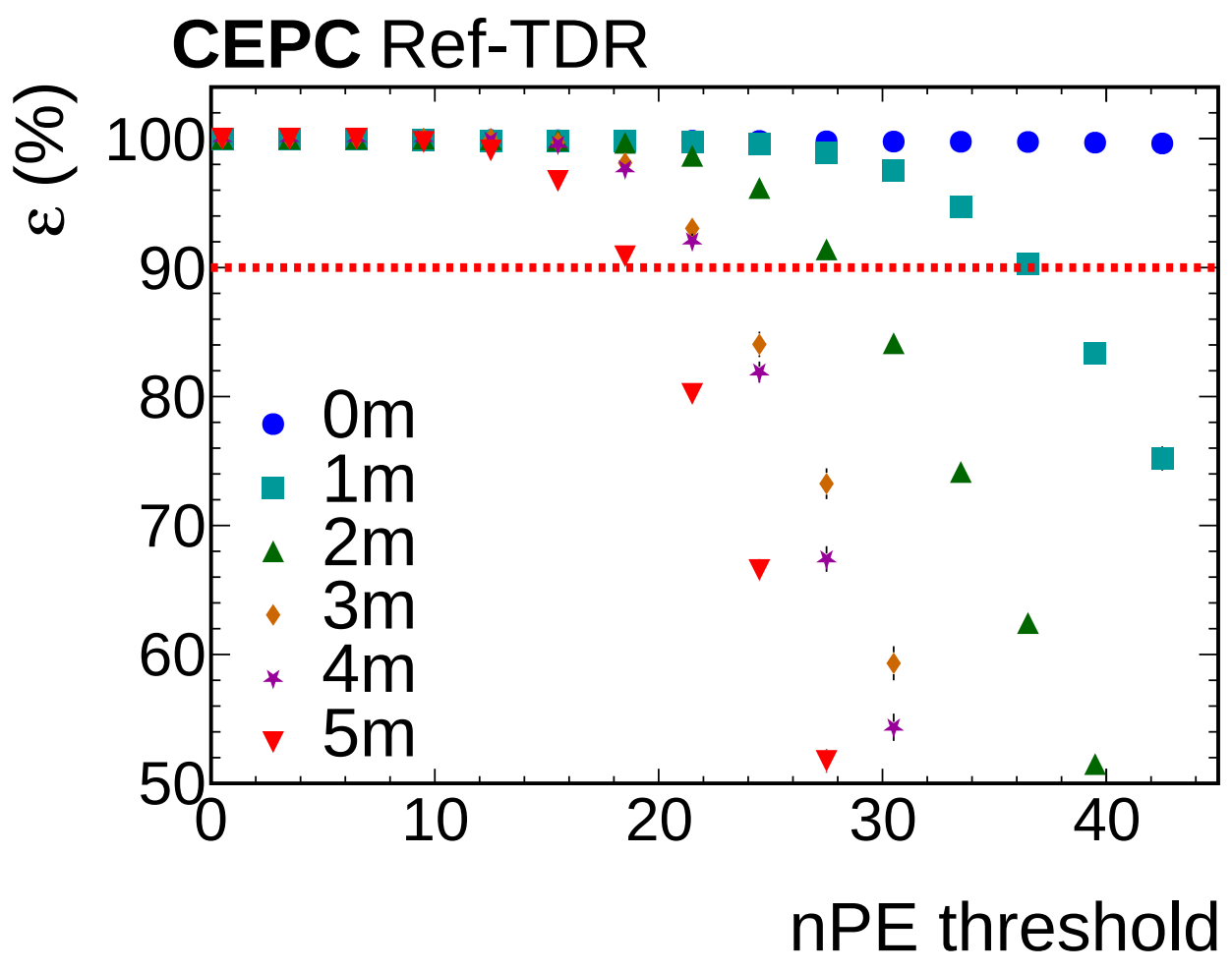

**(b)**

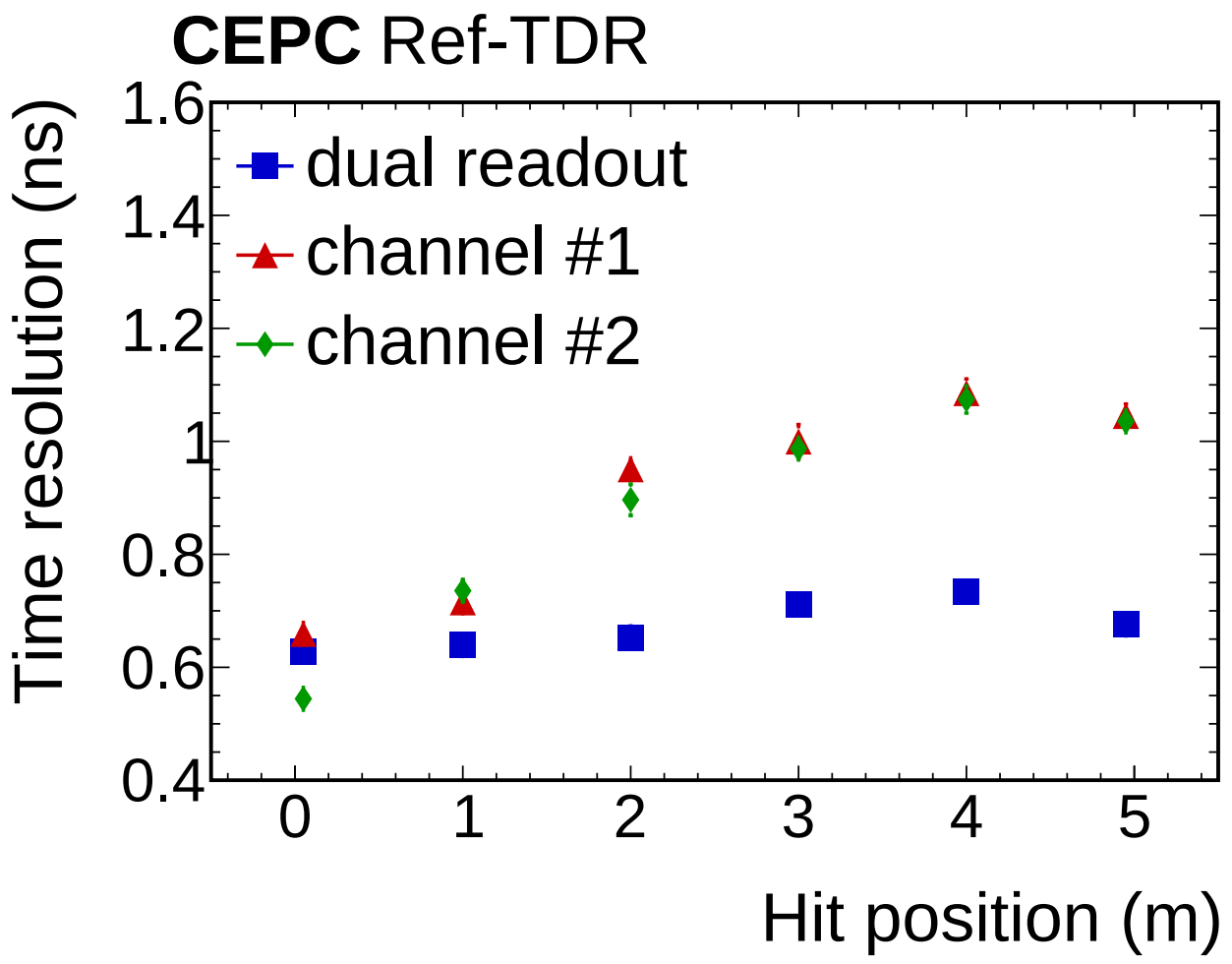

**(c)**

**Figure 9.12:** Performance evaluation of the 5 m PSU prototype with dual readout in a cosmic ray test: (a) average nPE along the strip, (b) detection efficiency of a single readout at varying nPE thresholds, and (c) time resolution measurements along the strip, including results from dual-end readout configurations. Here, channels #1 and #2 correspond to the readout channels at the two ends of the PS strip.

#### 9.4.1.1 Simulation framework and expected performance

Independent simulations are conducted using the standalone GEANT4 framework to model the scintillator detector and simulate optical processes following muon hits. The simulated detector is then integrated into the DD4hep framework for modular design and management. The standalone simulation employs the FTFP_BERT physics list to model particle interactions with PS materials, accurately simulating secondary particle generation and propagation. Additionally, G4EmStandardPhysics_option4 is used to model electromagnetic processes. Since the PS material for the muon detector is not included in the GEANT4 material library, it is defined based on experimental data, with parameters such as the refractive index fine-tuned to match the experimental results.

The muon detector is integrated into the CEPCSW framework using DD4hep, enabling efficient and modular simulation and management. Geometry parameters are separated from their definitions for easy updates. Approximately $1\times10^5$ fluorescent photons are produced when a muon passes through a PS strip. Although the OpticalPhysics library is available for simulating scintillation and photon collection, it is not used due to high computational costs. Instead, the number of photons received by the SiPMs is estimated directly based on the simulated energy deposition in the scintillator. Experimental measurements during the R&D phase verify the performance of photon collection.

When a muon traverses a PS strip, interaction details such as position, time, momentum, energy deposition, and location within the strip are recorded. A sample of 10 GeV/$c$ muon tracks with random directions within the solid angle is simulated. Figure 9.13a displays the energy deposition distribution in PS strips, showing a broad peak around 1.8 MeV, slightly higher than that of a MIP muon passing through 1 cm in PS. This shift is attributed to the random entry angles of muons at different positions in the detector, which increase the path length of the muon track within a PS strip. Additionally, backgrounds with low deposited energy are observed.

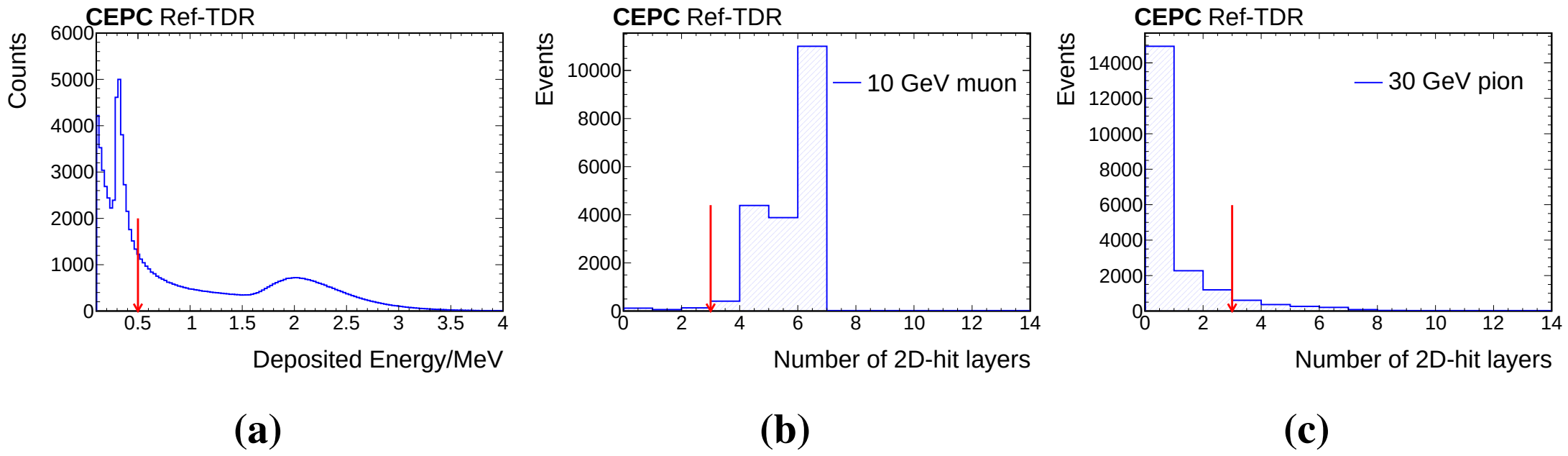

**Figure 9.13:** Simulation of energy deposition and the number of 2D-hit layers in the muon detector. (a) Energy deposition in PS strips. Distributions of the number of 2D-hit layers for (b) 10 GeV/$c$ muon tracks and (c) 30 GeV/$c$ pion tracks, with a hit threshold of 0.5 MeV energy deposition.

Based on the distributions presented in Figures 9.12a and 9.13a, and considering a reason-able nPE requirement (~ 8), a deposited energy threshold of 0.5 MeV is used to define a **hit**. In each detector layer, the two hits from the two PSU layers are combined to determine a **2D-hit**. This criterion is applied to evaluate the number of 2D-hit layers for muon tracks in the muon detector.

Figure 9.13b shows the distribution of the number of 2D-hit layers for the sample of 10 GeV/$c$ muon tracks. Most tracks exhibit 2D-hits in three or more layers, while a small fraction register hits in seven layers across different sectors. Additionally, a sample of 30 GeV/$c$ pion tracks is simulated, with the corresponding 2D-hit layer distribution shown in Figure 9.13c. A requirement of $\geq 3$ layers of 2D-hits effectively vetoes the majority of pions.

#### 9.4.1.2 Reconstruction

In experimental operations, the available data consist of the ADC signals from the SiPMs and the positions of the scintillator strips. To estimate the SiPM ADC based on energy deposition in the simulation, the nPE is first calculated using the energy deposition and the distance between the hit position and the SiPM, following the relationship illustrated in Figure 9.12a. Subsequently, a random number is drawn from the ADC distribution obtained from cosmic ray tests.

A signal from a PSU is counted as a hit if the nPE exceeds a predefined threshold. According to the R&D results shown in Figure 9.12b, a wide range of $6 < \text{nPE} < 15$ can be utilized for threshold optimization. The reconstruction process involves determining the hit positions from these signals, ensuring accurate and reliable data interpretation.

A muon passing through a detector module usually produces ADC signals in the two PSU layers. In barrel, the bottom PSU layer provides accurate $x$ and $y$ coordinates, while the top PSU layer contributes an accurate $z$ coordinate. In the endcaps, one PSU layer provides accurate $y$ and $z$ coordinates, while another one provides accurate $x$ and $z$ coordinates. Two-PSU-layer structure of a detector module and its position provide a complete 3D trajectory point for a muon track.

### 9.4.2 Performance in $\mu$ID and pion fake rate

The muon detector provides two key performance metrics: spatial and time resolutions. Each contributes differently to the standalone $\mu$ID, pion veto (pion fake rate), track reconstruction, and background rejection. This section focuses on the standalone $\mu$ID and pion fake rate provided by the muon detector.

Comprehensive cosmic ray tests on prototype PS strips have demonstrated a time resolution of 1 ns. For short PS strips, the time resolution improves to approximately 0.8 ns, corresponding to a spatial resolution of about 14 cm along the strip, calculated based on the speed of light propagation in the WLS fiber. This enhanced timing capability is particularly advantageous for resolving ambiguities in long PS strips and eliminating ghost hits arising from multiple particle crossings. Additionally, precise time information significantly contributes to reducing the pion fake rate.

The muon detector's contribution to the momentum resolution of charged particles is constrained by the 4 cm width of its PS strips. Yet, it remains sufficient for $\mu$ID and background rejection. Penetrating particles can be identified through multiple approaches: first, by verifying the consistency between measured hits and tracks extrapolated from the inner detectors, and second, by leveraging penetration depth information derived from the muon detector layers to improve discrimination capabilities.

#### 9.4.2.1 Standalone $\mu$ID

Two MC samples of muons ($\mu^-$ and $\mu^+$) with a momentum of 10 GeV/$c$, uniformly distributed in solid angle covering the barrel and the endcaps, are simulated to evaluate the performance of the standalone $\mu$ID. Based on the distribution presented in Figure 9.13b, a requirement of the number of 2D-hit layers $\geq 3$ is applied for the standalone $\mu$ID. Figure 9.14 displays the maps for the obtained standalone $\mu$ID efficiency.

The standalone $\mu$ID demonstrates high efficiency across most regions, except for areas around the 10 cm gaps between the endcap sectors, which account for 0.3% of the solid angle and exhibit reduced efficiency. The magnetic field causes the positions of these low-efficiency regions to differ for the $\mu^-$ and $\mu^+$ samples. With a requirement of $\mu$ID efficiency $< 50\%$, the fraction of events is 0.35%, consistent with the geometric calculations of the muon detector. Additionally, regions with slightly reduced efficiency are observed at the interface between the barrel and endcap, as well as around the center of the barrel due to the 50 mm wide stiffeners.

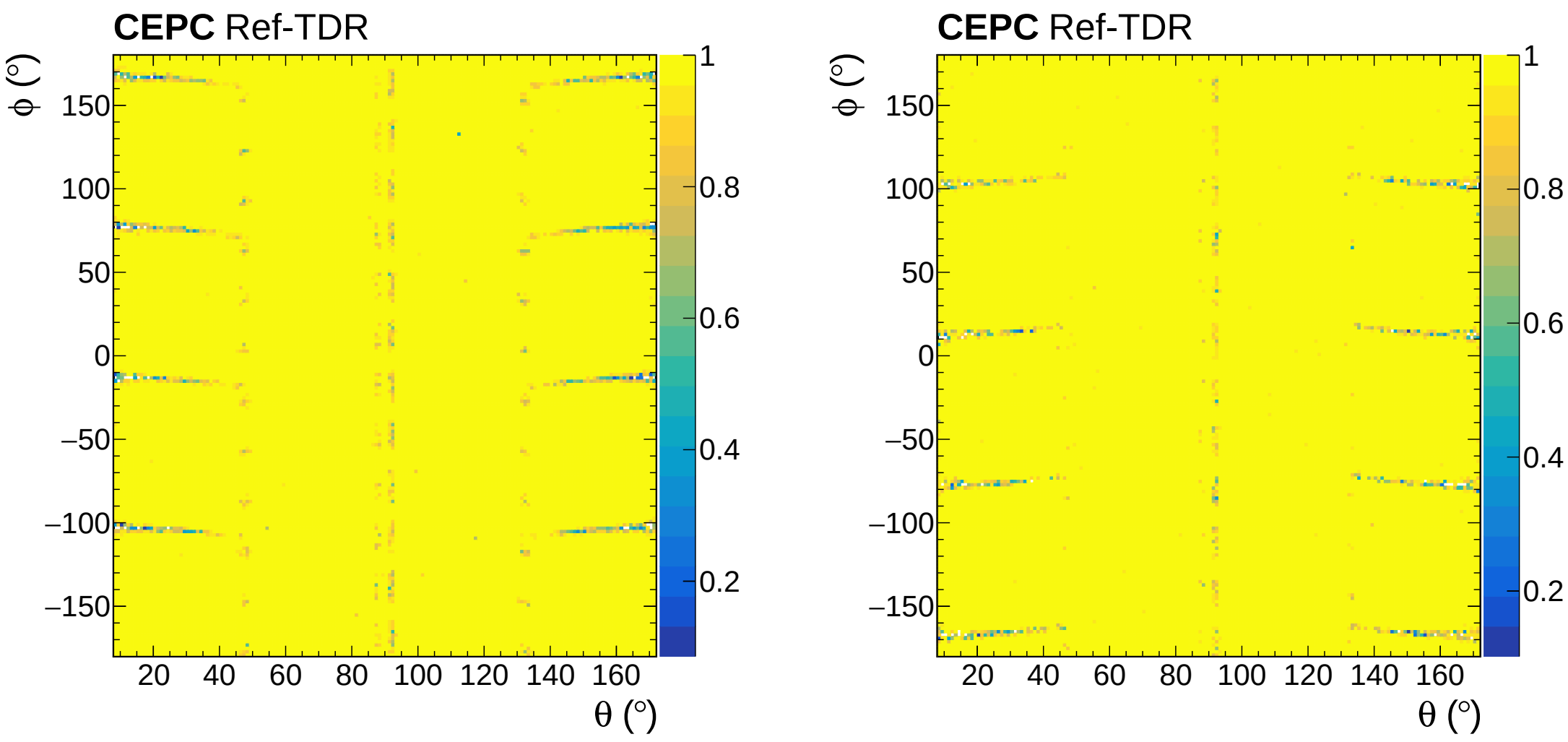

(a) (b)

**Figure 9.14:** Efficiency map of the standalone $\mu$ID based on muon detector as a function of polar ($\theta$) and azimuthal ($\phi$) angles for (a) $\mu^-$ tracks and (b) $\mu^+$ tracks with a momentum of 10 GeV/$c$. The muon detector covers the region $7.8° < \theta < 172.2°$.

MC samples of muons with varying momenta are used to study the muon detector standalone $\mu$ID efficiency in the momentum range from 1 GeV/$c$ to 10 GeV/$c$. Each sample has $2 \times 10^5$ events. When considering the acceptance within the muon detector, the muon detector achieves a standalone efficiency of > 95% (mostly 98%) for muons above the threshold of approximately 4.5 GeV/$c$, as shown in Figure 9.15. Here, the low efficiency zones of 10 cm gaps in the endcaps are included. The momentum threshold is primarily determined by muon energy deposition in the calorimeters. Muons with momenta above 3 GeV/$c$ exhibit a stable peak at 2.4 GeV of the deposited energy, as shown in Figure 9.15b for momenta ranging from 3 GeV/$c$ to 10 GeV/$c$. This energy loss reduces the muon's momentum, making it difficult for muons with initial momenta below 4 GeV/$c$ to reach the muon detector.

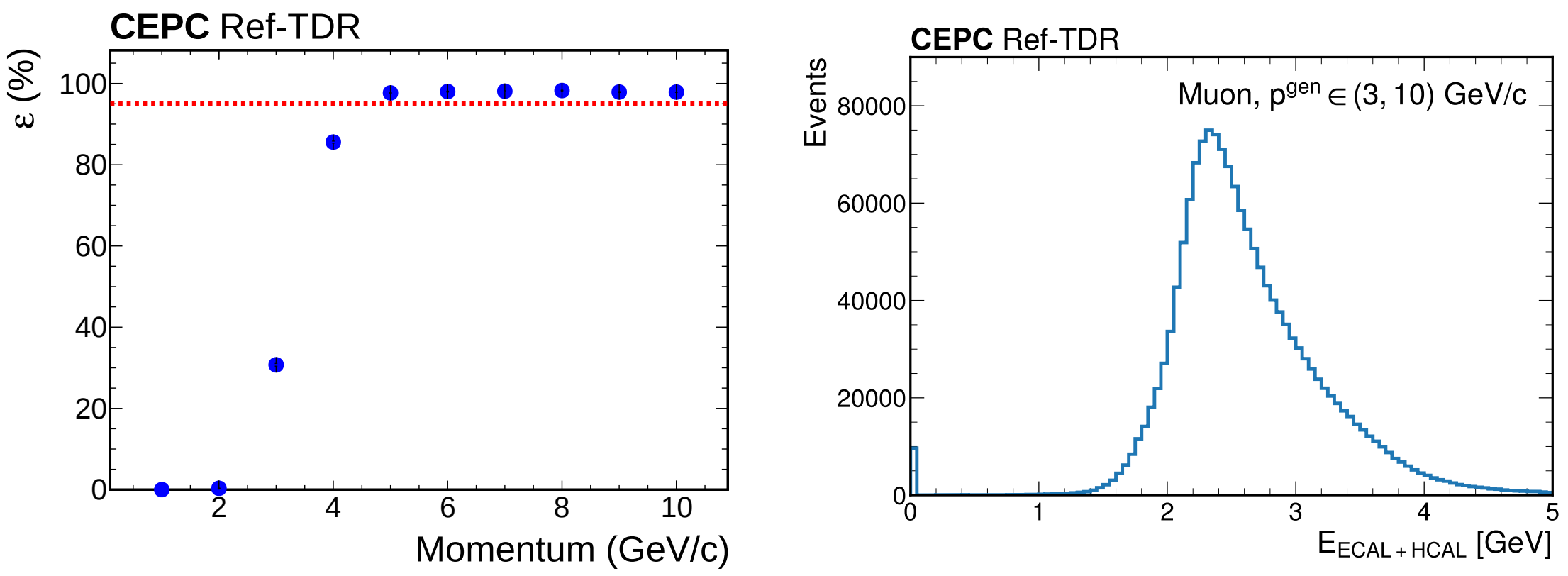

(a) (b)

**Figure 9.15:** (a) Muon detector standalone $\mu$ID efficiency versus momentum of the charged particles. The dashed line represents an efficiency of 95%. (b) The energy deposited in the calorimeters by a muon with an initial momentum ranging from 3 GeV/$c$ to 10 GeV/$c$.

### 9.4.2.2 Standalone pion fake rate

The muon detector's multi-layer design enables effective discrimination between muons and pions. High-momentum muons typically penetrate the muon detector and produce clear signals across multiple detector layers. However, pion tracks, particularly from the 30 GeV/$c$ sample, often register diffuse hits in the muon detector due to leakage from the HCAL. Even with a requirement of the number of 2D-hit layers $\geq 3$ for $\mu$ID, 8.0% of pion tracks remain unvetoed. This fraction varies to 4.5% and 11.5% for 20 GeV/$c$ and 40 GeV/$c$ pions, respectively, as listed in Table 9.2.

The muon detector's excellent time resolution provides a new tool for pion veto. Muons traverse the detector at velocities close to $c$, while hadronic showers from high-momentum pions produce complex signals, including slower components. Figure 9.16 illustrates hit times for muon and pion samples across various momenta. Each sample has $2 \times 10^5$ events. For muons (Figure 9.16a), tracks typically arrive at the detector 15 ns post-collision and exit within 10 ns, with 4 GeV/$c$ muons delayed due to the significant energy loss in the calorimeters. In contrast, pions (Figure 9.16b) exhibit broader hit time distributions, with many hits occurring $> 30$ ns post-collision. As pion momentum increases, so does the number of 2D-hits, including those with hit time exceeding 30 ns.

As listed in Table 9.2, by imposing the requirements of hit time $< 30$ ns and $\mu$ID, the pion fake rate is reduced to 0.62%, 0.71%, and 0.66% for 20 GeV/$c$, 30 GeV/$c$, and 40 GeV/$c$ pions, respectively. While tighter time cuts further suppress pion fakes, they also slightly reduce $\mu$ID efficiency. Achieving a 1 ns time resolution is essential to enhance muon-pion separation in the detector.

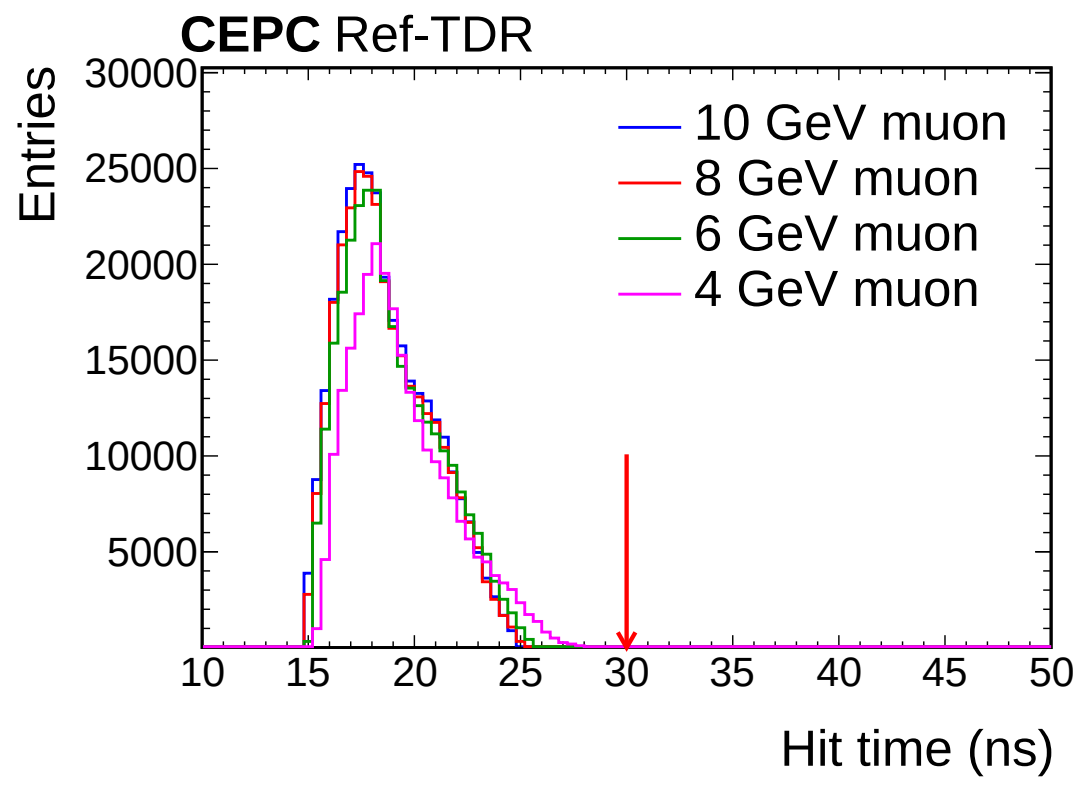

(a)

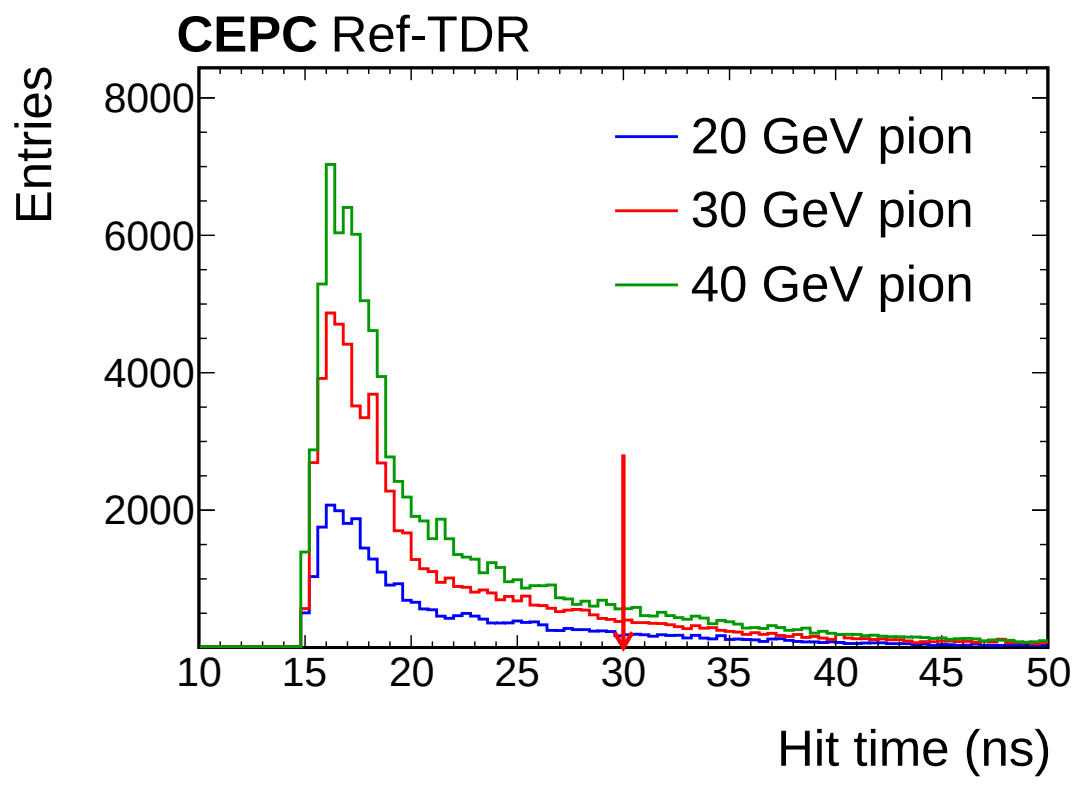

(b)

**Figure 9.16:** Distributions of hit time for (a) muon samples with momenta of 4 – 10 GeV/$c$ and (b) pion samples with momenta of 20 – 40 GeV/$c$. Each sample has $2 \times 10^5$ events.

Both software and simulation are still under development, alongside detector R&D and optimization. More sophisticated algorithms for the $\mu$ID, pion fake rate, and improved simulation, such as better time resolution, hit time distribution, and hit position distribution, are being implemented. These advancements will help reduce efficiency drops in the standalone $\mu$ID around problematic regions. These regions include the gaps between endcap sectors, gaps between the barrel and endcaps, and the stiffeners around the center of the barrel. Future iterations will enable a more accurate and robust analysis of the MC simulation data.

### 9.4.3 Beam-induced background estimation

During operations, various sources contribute to the background noise, including the operational environment, cosmic rays, and beam-induced background, with the latter being the most hazardous. One of the key objectives of the muon detector is the search for long-lived particles, with a particular focus on decays into a $\mu^+\mu^-$ pair. However, background events frequently resemble genuine signals, resulting in a reduction in the signal-to-noise ratio and an increased rate of misidentification. In extreme scenarios, excessively high background rates can inundate the electronics, making it challenging to recover signals.

**Table 9.2:** The standalone detection efficiencies of $\mu^{\pm}$ and $\pi^{\pm}$ in the muon detector, evaluated under the $\mu$ID requirements (number of 2D-hit layers $\geq 3$) and a hit time constraint of $< 30\,\text{ns}$, as a function of momentum.

| **Particle** | **Momentum** ( GeV/$c$) | **Efficiency** (%) | |
|---|---|---|---|
| | | $\mu$ID requirement | $\mu$ID and hit time $< 30\,\text{ns}$ |
| $\mu^{\pm}$ | 2 | $1.0 \pm 0.1$ | $0.3 \pm 0.1$ |
| | 3 | $34.0 \pm 0.3$ | $30.7 \pm 0.3$ |
| | 4 | $90.8 \pm 0.2$ | $85.6 \pm 0.2$ |
| | 5 | $98.0 \pm 0.1$ | $97.7 \pm 0.1$ |
| | 6 | $98.2 \pm 0.1$ | $98.0 \pm 0.1$ |
| | 7 | $98.3 \pm 0.1$ | $98.1 \pm 0.1$ |
| | 8 | $98.5 \pm 0.1$ | $98.3 \pm 0.1$ |
| | 9 | $98.2 \pm 0.1$ | $97.9 \pm 0.1$ |
| | 10 | $98.3 \pm 0.1$ | $97.9 \pm 0.1$ |
| $\pi^{\pm}$ | 20 | $4.54 \pm 0.15$ | $0.62 \pm 0.06$ |
| | 30 | $7.98 \pm 0.19$ | $0.71 \pm 0.06$ |
| | 40 | $11.54 \pm 0.23$ | $0.66 \pm 0.06$ |

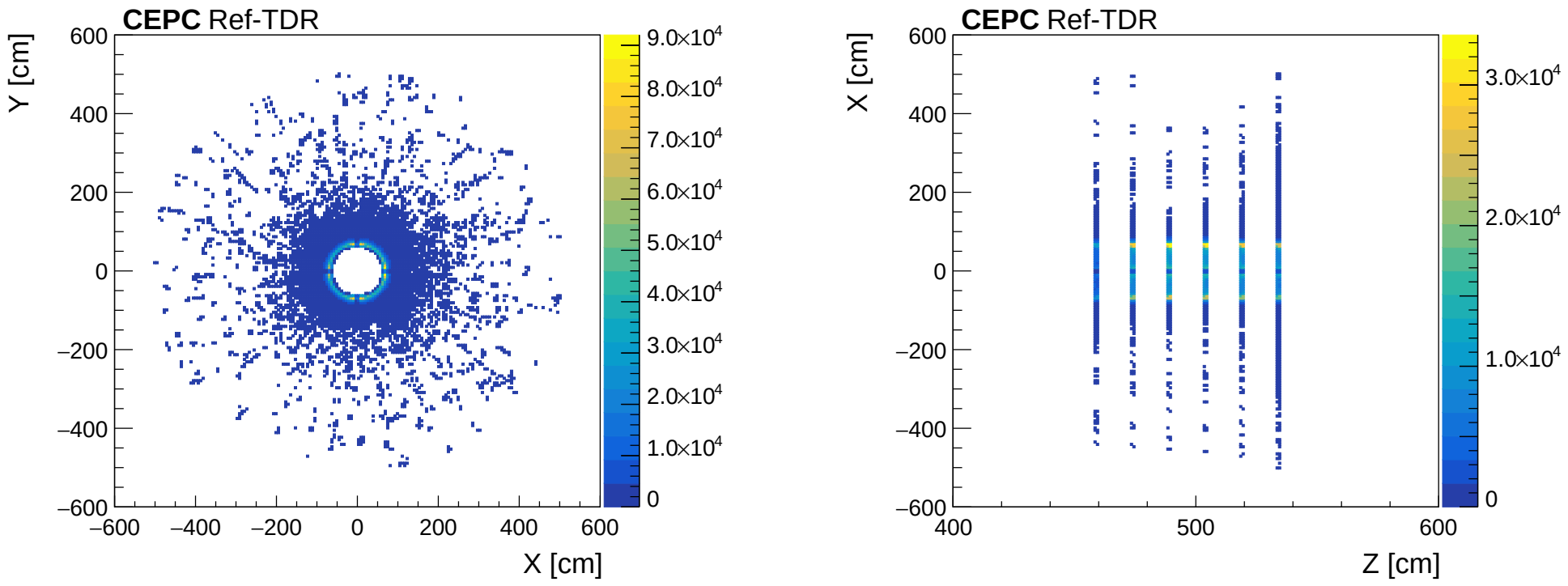

**(a)** **(b)**

**Figure 9.17:** The number of hits in (a) the $X - Y$ panel and (b) the $X - Z$ panel, obtained from $2 \times 10^{6}$ simulated beam-induced background events in one endcap.

The beam-induced backgrounds in the muon detector are simulated with Geant4 using the CEPCSW software package, along with a detailed analysis of the simulation samples. Figure 9.17 displays the general hit maps from the analysis, showing that most hits are in the forward and backward endcaps, with nearly no hits in the barrel. Table 9.3 presents the average and maximum hit rates per layer along with the occupancy.

A maximum hit rate of $13.7\,\text{Hz}/\,\text{cm}^2$ is achieved, far below the rate capability of the muon detector based on the scintillator. According to a test on SiPM readout [5], the detector could offer a capability of about $18\,\text{kHz}/\,\text{cm}^2$. According to the background levels, a rate capability of $50\,\text{Hz}/\,\text{cm}^2$ is required for the muon detector, which primarily drives the design of the electronics

**Table 9.3:** Average hit rate in the PSUs of the endcaps due to backgrounds at the $Z$ pole.

| **Layer#** — | **Maximum hit density** ($\times 10^{-7}$/ cm$^2$/$BX$) | **Maximum hit rate** (Hz/ cm$^2$) | **Average hit rate** (Hz/ cm$^2$) | **Occupancy** ($\times 10^{-4}$) |
|---|---|---|---|---|
| 1 | 7.86 | 10.24 | 0.80 | 7.6 |
| 2 | 7.86 | 10.24 | 0.96 | 7.6 |
| 3 | 10.48 | 13.66 | 1.22 | 10.0 |
| 4 | 9.16 | 11.96 | 1.16 | 8.8 |
| 5 | 6.58 | 8.60 | 0.88 | 6.2 |
| 6 | 5.24 | 6.82 | 0.28 | 5.0 |

system.

## 9.5 Mass production and detector installation

This section describes the production of the PS strips, the assembly of the PSUs, the construction of the detector modules, and the integration of the yoke-detector sectors. The PS strips will be produced at the company and subsequently delivered to the detector construction hall located at the ground level above the experiment hall. The construction of detector modules will proceed in parallel with the assembly of the yoke sectors. The two will be integrated into the yoke-detector sectors, which will be transported to the experimental hall for final installation.

### 9.5.1 Manufacturing of PS strips

PS strips will be fabricated using extrusion technology by the manufacturer, which has actively participated in the R&D outlined in Section 9.3. The technological advancements and production expertise gained during the R&D phase will establish a benchmark for the new production of high-quality PS strips.

The production process yields continuous PS strips with lengths of up to 5 m, each featuring a central hole with a diameter of 2.5 mm (2.0 mm) to accommodate D20 (D12) WLS fibers. As depicted in Figure 9.6, the PSUs in the muon detector are designed with varying lengths. The manufacturer will receive detailed specifications, including the required lengths, quantities, and diameters of the PS strips. During production, the continuous strips will be cut to the specified lengths, ranging from approximately 1 m to 4.6 m. Additional spare strips of the longest length will be provided to ensure operational readiness, as they can be further cut to any required lengths during the construction of detector modules.

A reflective coating will be applied to the strips during the manufacturing process. Periodic quality control tests will be conducted to verify the uniformity of the coating thickness, light output, and attenuation length before the construction of detector modules.

### 9.5.2 Assembly of PSUs

The PSU structure is shown in Figure 9.5. Upon arrival at the construction hall, the holes in the PS strips, intended for embedding WLS fibers, will be thoroughly cleaned. The strips will then be organized and labeled based on their lengths and the diameters of the holes. The WLS fibers will be cut to match the lengths of the corresponding PS strips.

One end of the fiber will be inserted into the coupling component of the PSU and aligned with the sensitive area of the SiPM. This end should be carefully cut to ensure optimal surface flatness at the terminal. The fiber will be secured to the coupling component using glue. The opposite end of the fiber will be coated with a silver layer to improve light collection efficiency.

The WLS fiber will then be inserted into the hole of the PS strip, and the coupling component will be affixed to the terminal of the PS strip. Optical glue may be applied within the hole to enhance light collection further.

Before assembling the detector modules, cosmic ray tests will be conducted on the PSUs to ensure proper functionality and performance.

### 9.5.3 Construction of detector modules

The structures of detector modules are shown in Figures 9.6 and 9.7. Before constructing detector modules, the necessary aluminum materials for the frames must be prepared, including straight or arc-shaped borders with a thickness of 4 mm, long support metal strips with a width of 1 cm, and surface plates with a thickness of 1 mm. The metal rods, with a diameter of 2 mm, should be made of steel. Additionally, a sufficient quantity of high-strength plastic sheets should be provided.

The module assembly begins with the connection of the borders and the fixation of the bottom layer plastic sheet to these borders. The plastic sheet fills the entire area within the borders and is adhered to the aluminum surface plate. Above the plastic sheet lies the bottom PSU layer, which corresponds to the long PSU layer in a barrel module or the horizontal PSU layer in an endcap module. The PSUs of the bottom layer are affixed to the plastic sheet using adhesive, leaving a gap of approximately 3 cm from the border to accommodate cabling between the SiPMs and FEBs. The second layer of PSUs is then installed, constituting the short PSU layer in a barrel module or the vertical PSU layer in an endcap module. The pre-designed metal support strip is mounted on the SiPM side of the PSUs and secured to the border using 2 mm-diameter metal rods.

The FEBs are positioned and mounted on the 4 mm-thick borders, and the SiPMs are connected to the FEBs via mini coaxial cables. Following this, the LV power cables and data cables are connected to the FEBs and routed above the top PSU layer, with connectors leading to the MBs through the cable window.

The smaller high-strength plastic sheet is then placed over and adhered to the top PSU

layer, as illustrated in Figure 9.6. The final step involves covering the assembly with the second aluminum surface plate, which is attached to the plastic sheet and the frame borders, forming a sealed box structure that completes the module assembly. After the construction of a module, the performance of each PSU should be checked.

### 9.5.4 Detector installation

The assembly process of the yoke structure is detailed in Chapter 14. Detector module fabrication will occur concurrently with the assembly of the yoke sector. Due to the extensive surface area coverage required by these modules, transporting them over long distances presents substantial challenges and potential risks. Thus, the detector construction and storage hall will be strategically located adjacent to the site designated for final integration of the yoke-detector sectors.

The yoke sector will be positioned horizontally during the integration process. The detector modules are inserted into their preassigned slots using specialized adjustable tables engineered to enable precise rotation control, lateral movement, and vertical height alignment relative to the yoke structure. Following precise alignment, each module will be guided into the gap between the iron plates. Upon reaching its designated coordinates, the module will be permanently secured to prevent shifts. This is critical to maintain positional integrity during subsequent phases, including transportation to the experimental hall and final installation.

Each module will be electronically certified before the final integration step. For a barrel sector, the cover plate will be installed, with the power and data cables from the FEBs passing through the cable windows on the cover plate. The cables outside the windows, particularly their connectors, will be well-protected and securely fastened to the surface of the cover plate. For an endcap sector, the cables exiting the module cable windows are routed as illustrated in Figures 9.6b and 9.6c, protected within the cable channel. The MBs are installed at the positions shown in Figures 9.10c and 9.10d.

The fully assembled sector will undergo extensive testing with cosmic rays to validate its functionality and integration with all modules. A systematic procedure will be developed to assess performance, identify and address potential issues, evaluate quality assurance metrics, and ensure compliance with the QA/QC standards. All findings, corrective actions, and outcomes will be thoroughly documented and recorded.

The transportation of the yoke-detector sectors to the experimental hall and their subsequent installation into the spectrometer are detailed in Chapter 14. Strict protocols will be enforced throughout these operations to protect the cables and MBs. After the installation, initial electronics verification tests will be conducted to confirm system functionality, followed by extended cosmic ray tests to assess long-term performance.

## 9.6 An alternative solution based on the RPC technology

RPCs are being considered for the muon detectors due to their effectiveness in high energy physics experiments, such as those in LHC experiments [7, 8, 9, 10], for their fast response and high efficiency. They are essential for muon detection, triggering, and precise time measurements. A prominent example is the CMS experiment, where RPCs play a vital role in identifying and reconstructing muons in both the barrel and endcap regions. They are part of the L1 trigger system, operating in avalanche mode to ensure better timing resolution and durability, facilitating rapid muon identification.

RPCs are gaseous detectors that function through avalanche multiplication in a strong electric field. They consist of two parallel resistive plates, typically made of Bakelite or glass, separated by a gas gap. High voltage is applied across these plates, creating a strong electric field within the gas. When a charged particle, like a muon, passes through, it ionizes the gas molecules. The electric field accelerates the resulting electrons, causing further ionization and creating an electron avalanche. This avalanche induces a detectable signal on the readout strips located on the outer surfaces of the resistive plates. An RPC detector usually includes resistive plates, gas gaps, a HV system, readout strips, and FEEs.

1. Resistive Plates: These plates, typically made of high-resistivity materials such as Bakelite, form the boundaries of the gas volume. The high resistivity prevents signal propagation along the plates, ensuring localized signal readout.
2. Gas Gap: The space between the resistive plates is filled with a precisely selected gas mixture, usually comprising gases such as Argon, Freon, and Isobutane. This combination is optimized for efficient ionization and avalanche formation.
3. HV system: Typically, a few kilovolts are applied across the plates to generate the strong electric field needed for avalanche multiplication.
4. Readout Strips: Metallic strips on the outer surfaces of the resistive plates capture the induced signals from the avalanche. These strips are linked to readout electronics that amplify and digitize the signals.
5. Front-End Electronics: These electronics amplify, shape, and filter signals from the readout strips, provide timing information, and transmit data for further processing.

RPC offers good time and spatial resolutions, as well as two operational modes: avalanche and streamer. The avalanche mode provides superior time and spatial resolution, as well as higher rate capability, while the streamer mode features simpler readout electronics and lower cost. RPCs are known for their simple structure, cost-effectiveness, large area modules, and robustness. However, challenges such as rate capability, aging effects, and noise reduction are associated with their operation.

### 9.6.1 Performance requirement for the RPC option

The performance parameters for the RPC option align with Table 9.1. Coverage is affected by the iron yoke structure, especially the spiral partition in the barrel. The readout strip width significantly impacts position resolution, while digitized signals capture charge and time information in the FEEs. RPCs can achieve 95% detection efficiency at the high voltage efficiency plateau, although efficiency is influenced by material resistivity, gas gap, and gas mixture, all requiring optimization.

Long-term stability is a challenge, as observed efficiency degradation is noted in some experiments after years of operation. Rate capability is also a focus for RPC technology, emphasizing thin-gap designs and low-resistivity materials. Initial estimates indicate a minimum configuration of six layers of barrel RPCs and four layers of endcap RPCs is necessary for adequate solid angle coverage and tracking efficiency.

### 9.6.2 Electrode materials and gas mixture

Materials with high volume resistivity, ranging from $10^9$ to $10^{12}\Omega$ cm, are selected as electrodes for RPCs, with bakelite and glass being the most common. Bakelite, with a lower resistivity of $10^9$ and $10^{10}\Omega$ cm, supports higher event counting rates and greater mechanical rigidity than glass. In contrast, glass, which has one or two orders of magnitude higher resistivity, does not require oiling of internal surfaces during construction and is easier to maintain.

Both bakelite and glass RPCs typically operate in avalanche mode. At the LHC, the high luminosity of hadron collisions presents significant challenges due to elevated event rates. For example, the A Toroidal LHC ApparatuS (ATLAS) RPC system maintains a trigger rate of 100 Hz/ $cm^2$ to 100 kHz/ $cm^2$ [7]. Thin-gap bakelite RPCs provide quick information on muon tracks, allowing trigger logic to determine their multiplicity and approximate energy range.

The baseline material is float glass, while phenolic glass, a type of HPL consisting of glass fibers embedded in phenolic resin, is an alternative being explored for its low and adjustable electrical resistivity and enhanced mechanical properties.

The gap is filled with a suitable gas mixture at atmospheric pressure, which serves as the target for ionizing radiation. The gas mixture interacts with incoming particles to produce ionization and charges, which are then collected by readout panels outside the chambers. A commonly used gas mixture, also employed by the ATLAS RPC system, consists of Tetrafluoroethane ($C_2H_2F_4$)/Iso-Butane (Iso-$C_4H_{10}$)/Sulphur hexafluoride ($SF_6$) in proportions of 94.7%, 5.0%, and 0.3%, respectively. This mixture offers a relatively low operating voltage (due to the low $SF_6$ concentration), non-flammability, and low cost, while ensuring a stable plateau for safe avalanche operation.

The current RPC gas mixture has a high global warming potential. It is the only known gas suitable for operating the ATLAS RPCs with the present front-end electronics in avalanche mode and at full efficiency. To reduce dependence on these gases, which new regulations may ban, discontinue on the market, or make prohibitively expensive, an ongoing R&D program is searching for alternatives.

### 9.6.3 Preliminary test results

A prototype of a 1 m × 1 m RPC module from the ATLAS upgrade is tested. The module features a 1.2 mm gas gap using R134a. A preamplifier with a gain of 16 is applied to the signal. Figure 9.18 shows the efficiency with high voltage applied and the time difference between the RPC and the trigger. When the high voltage is increased to 6.8kV, an efficiency above 90% is achieved. The trigger has a time resolution better than 100 ps, indicating the RPC's time resolution is approximately 0.5 ns.

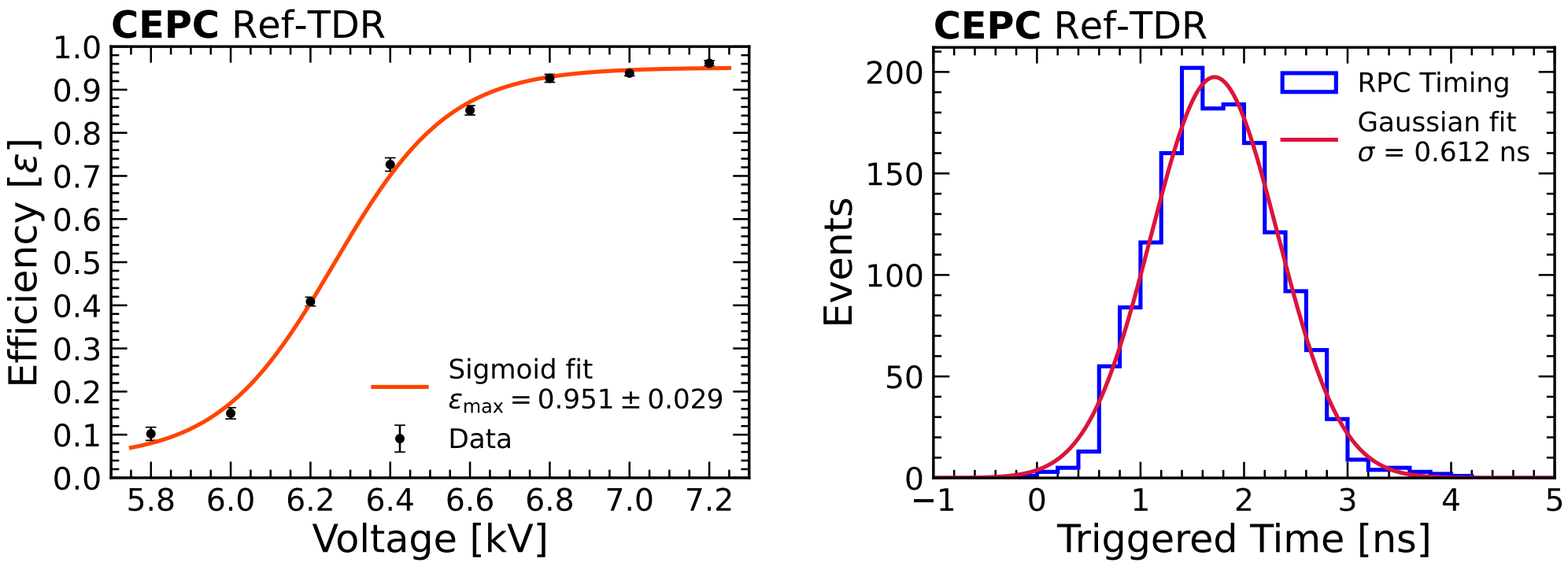

(a) (b)

**Figure 9.18:** Performance of the RPC prototype in the cosmic ray test, where (a) shows the efficiency with the HV applied, and (b) the time difference distribution between the RPC signals and the trigger signals.

## 9.7 Summary and further research plan

The muon detector is designed to achieve a $\mu$ID higher than 95%, a pion fake rate below 1%, a spatial resolution of approximately 1 cm, and a time resolution of about 1 ns. Utilizing extruded scintillator, WLS fiber, and SiPM technology, the detector comprises 240 modules: 144 for the barrel and 96 for the endcaps. Covering a total detection area of $4,800\ m^2$, it features 43,000 channels, 119 km of sensitive WLS fiber, and requires $48\ m^3$ of extruded scintillator. RPC is considered a backup option.

The R&D of the muon detector, utilizing PS strips, WLS fibers, and SiPMs, has shown exceptional performance. However, further investigation is necessary to fully explore the potential of this technology in various physics applications, particularly in the search for LLPs. Our strategic plan includes the following key initiatives:

- Study with new WLS fibers.
- Enhance the light yield and photon collection efficiency of the 5 m PS strips.
- Construct multiple detector modules for comprehensive system testing, with a focus on stable long-term operation.
- Conduct radiation hardness assessments on the PS strips, WLS fibers, and SiPMs.
- Refine track reconstruction algorithms and improve the measurement of low-momentum muons.
- Expand the exploration of the physics potential, specifically in the search for LLPs.

# Chapter 10 Detector Magnet System

The CEPC detector magnet is a large superconducting solenoid with an iron yoke, designed to provide a 3 T axial magnetic field at the interaction point [1]. The solenoid has a large bore to house the barrel hadronic calorimeter. This Chapter presents the technical design of the detector magnet. Following a brief overview of the magnet system requirements in Section 10.1, detailed discussions of the magnetic field distribution, solenoid coil, conductor structure, magnet configuration, and yoke design are provided in Section 10.2. Section 10.3 covers R&D on conductors and coil prototypes. Section 10.4 summarizes simulation results, including magnetic field uniformity, coil stress, and quench analysis. Section 10.5 describes the cryogenic system and cryostat design. Alternative conductor and magnet designs are discussed in Section 10.6, and Section 10.7 concludes with a summary and future plans. Compensating magnets (anti-solenoids and quadrupoles) are briefly mentioned here; detailed discussions are provided in Section 3 and the accelerator TDR volume, Section 4.3.4 [2].

## 10.1 Requirements for the magnet

The CEPC detector magnet follows the design concepts of the BES III magnet [3], and CMS [4] and ILD [5]. The superconducting magnet is an iron-yoke-based solenoid designed to provide an axial magnetic field of 3 T at the interaction point. The solenoid is supported by an aluminum alloy cylinder and cooled indirectly by liquid helium to an operating temperature of 4.5 K. The magnet is located outside the hadronic calorimeter detector, as shown in Figure 10.1. A room temperature bore is required with 7.07 m in diameter and 9.05 m in length. Table 10.1 lists some of the magnet dimensions. A conventional Low Temperature Superconductor (LTS) solenoid is employed, featuring aluminum stabilized NbTi/Cu Rutherford cable, reinforced with an aluminum alloy structure, and is under development. LTS technology was chosen because of its maturity, reliability, accumulated experience, and low cost [3, 6]. The requirement for the non-uniformity of the magnetic field along $z$-axis in the TPC region is less than 7%, $\left|\frac{B-B_{\mathrm{ave}}}{B_{\mathrm{ave}}}\right| < 0.07$.

**Table 10.1:** Dimensions of the detector magnet

| Parameter | Value |
|---|---|
| Inner diameter | 7070 mm |
| Thickness | 700 mm |
| Outer diameter | 8470 mm |
| Length | 9050 mm |

The magnet assembly weighs approximately 250 tons, excluding the yoke. To minimize dead regions in the muon detector embedded within the yoke, the barrel yoke is designed as a twelve-fold rotationally symmetric structure. The total yoke weigth is approximately 3010 tons, comprising a 1610 ton barrel yoke and two 700 ton endcap yokes. Thermosiphon cooling is selected due to its uniform temperature distribution and superior heat-transfer efficiency compared with forced-flow methods. A small-scale prototype is in development to validate pipe-weld reliability and cooling performance.

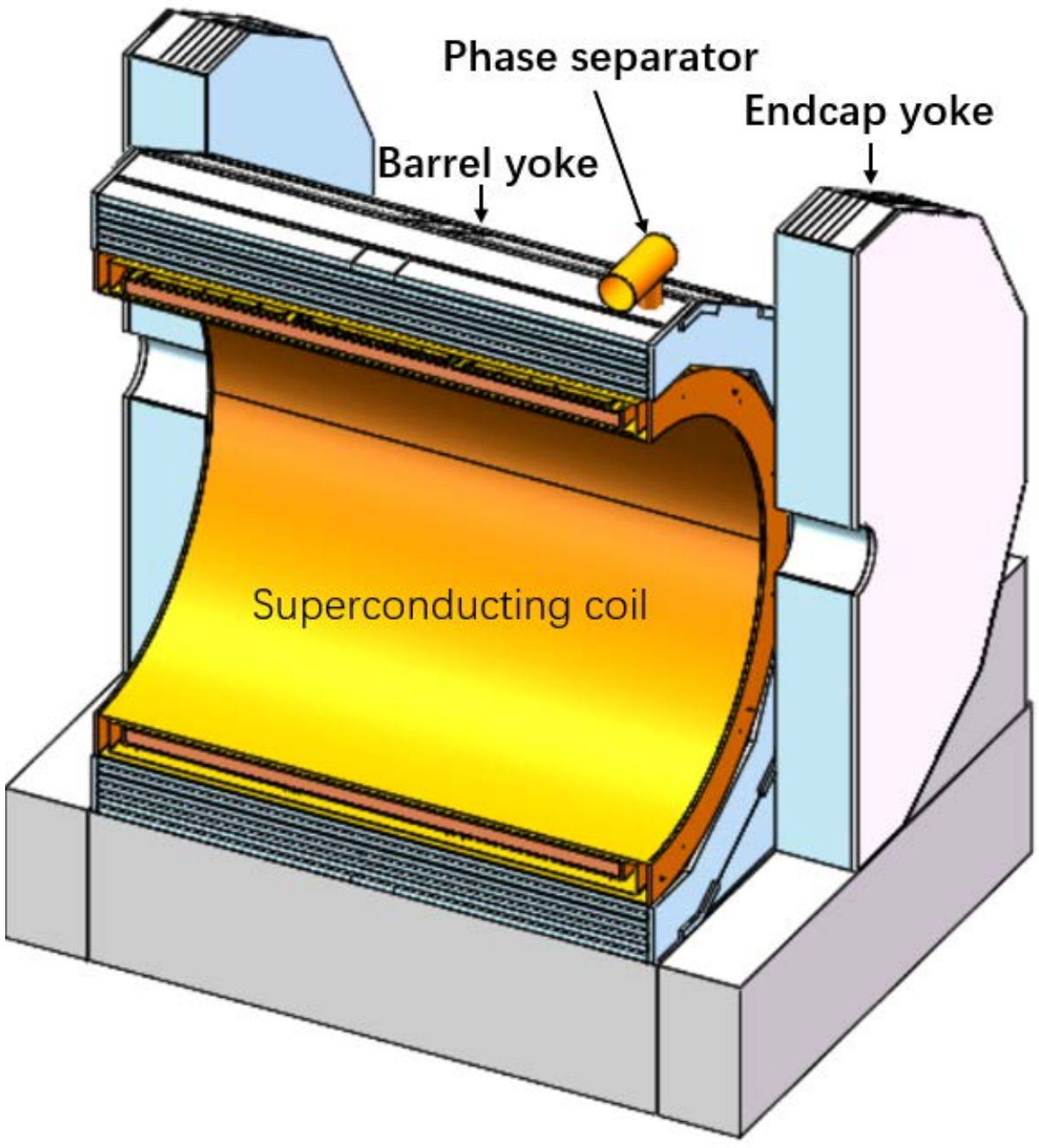

**Figure 10.1:** Schematic diagram of the magnet. The magnet is located between the HCal and Muon detector. The yoke has a dodecahedral structure. The magnet has two chimneys, one of them is connected to the phase separator as shown in the diagram.

## 10.2 Detailed design of the detector magnet

The superconducting magnet consists of several critical components, including the superconducting coil, cryogenic system, quench protection system, power supply, control system, and so on. These elements work in unison to ensure the magnet operates stably at the required superconducting temperature and generates the necessary magnetic field strength.

### 10.2.1 Magnet parameters

The cross section of the magnet is shown in Figure 10.2. The inner diameter of the coil is 7.3 m. The coil is divided into three modules axially and consists of four layers radially, with 120 turns per layer at each end module and 110 turns per layer in the middle module. The axial length of the coil is 8.15 m. Detailed specifications are presented in Table 10.2.

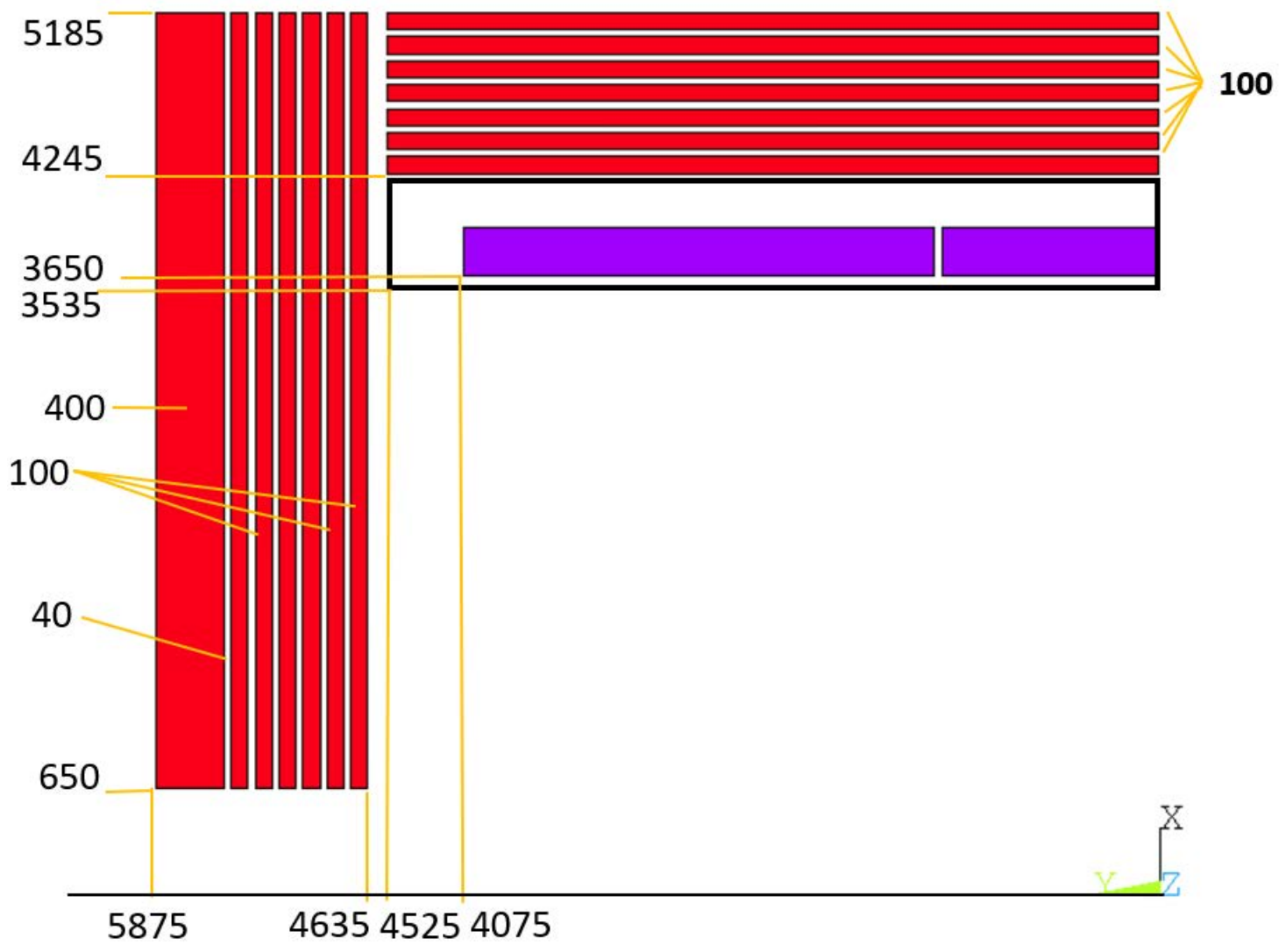

**Figure 10.2:** Cross-section of the magnet (dimensions in millimeters). There are seven layers of barrel yoke and endcap yoke. The gap thickness between the yoke layers is 40 mm.

The coil operates under two main conditions: cooling to 4.5 K and excitation to 17 kA. Figure 10.3 shows the 2D simulation model of the coil. The 4-layer coil is arranged in an orderly manner, with an external fiber glass insulation thickness of 0.5 mm for the cables, and an additional 1 mm insulation made of glass fiber epoxy between the coil and the support. The support structure has a thickness of 30 mm and is made of 5083 aluminum alloy.

**Table 10.2:** Main parameters of the coil

| Parameter | Value |
|---|---|
| Coil Inner diameter | 7300 mm |
| Coil Outer diameter | 7916 mm |
| Coil length | 8150 mm |
| Central magnetic field | 3 T |
| Operating current | 17 kA |
| Coil Module gap | 50 mm |
| Inductance | 11 H |
| Stored Energy | 1.54 GJ |

### 10.2.2 Conductor structure

The design of the conductor must meet both mechanical and Residual Resistivity Ratio (RRR) requirements simultaneously. This consideration was integral to the project since its inception and has led to the development of the conductor structure depicted in Figure 10.4.

The conductor is a box-structured aluminum-stabilized cable consisting of three components: the Rutherford type superconducting cable, a high purity aluminum stabilizer, and an aluminum alloy reinforcement. This structure is manufactured through a dual extrusion process, and its overall characteristics are detailed in Table 10.3.

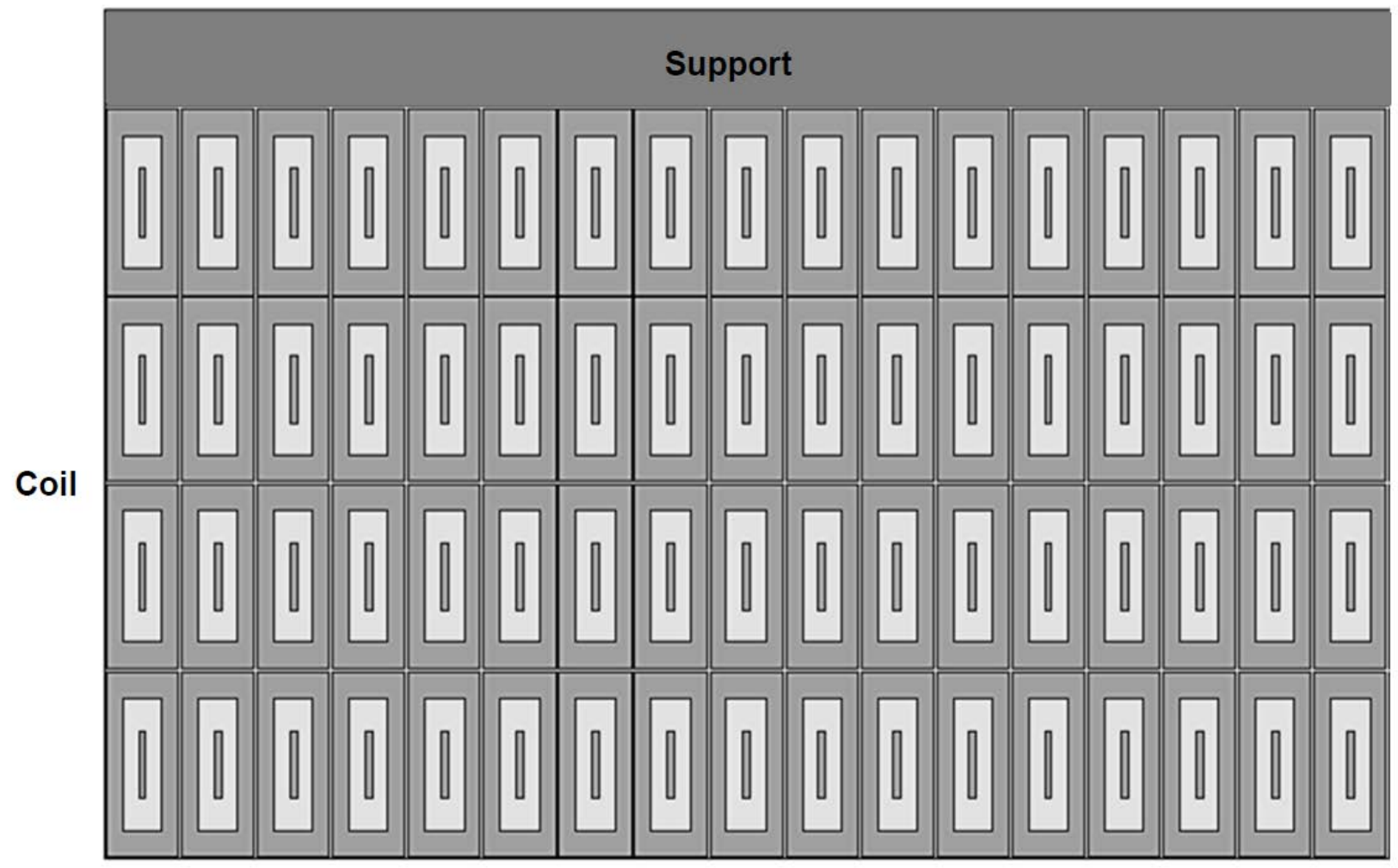

**Figure 10.3:** Structure of the coil. There are four lays Box-type conductor, winding inside the support cylinder.

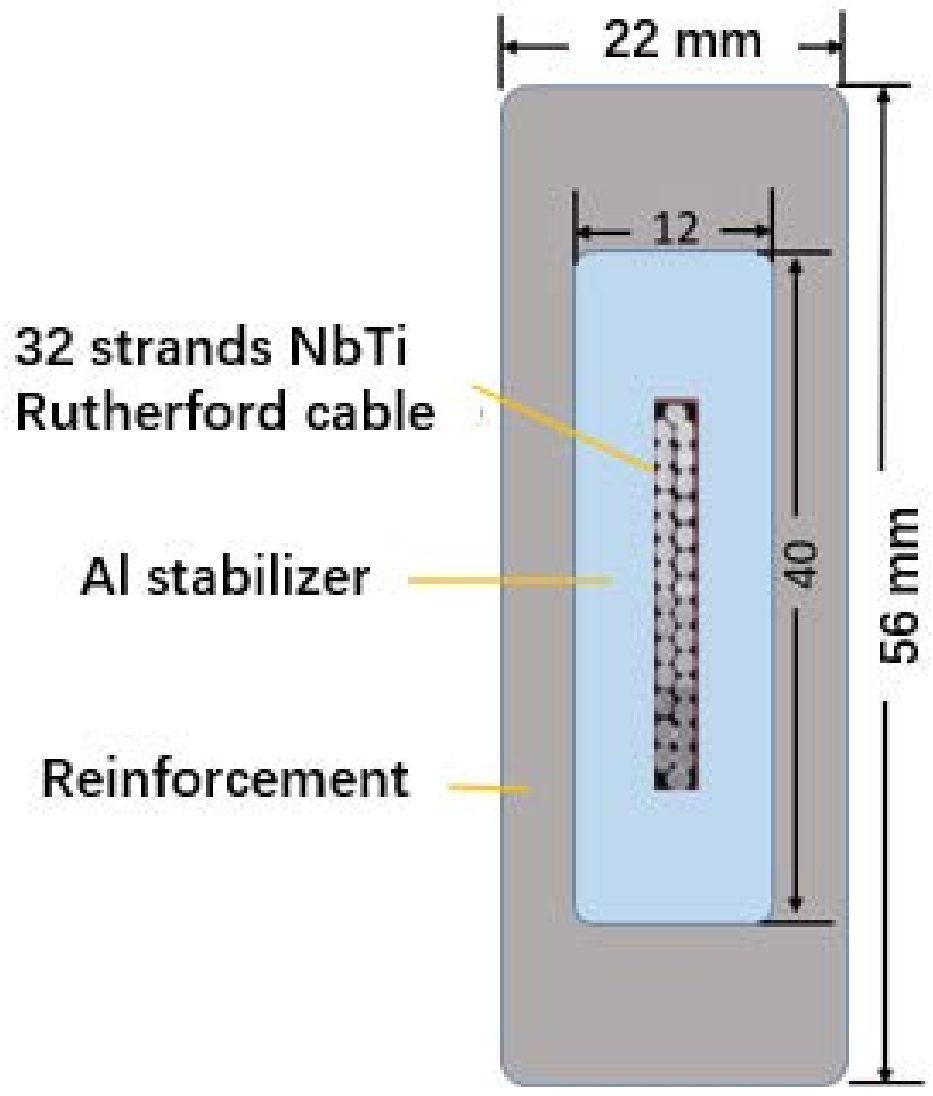

**Figure 10.4:** Cross section of aluminum stabilized superconductor used in detector magnet. The conductor has a box-type geometry and comprises three elements: NbTi Rutherford cable, aluminum stabilizer, and aluminum-alloy reinforcement. It is obtained by dual extrusion process.

### 10.2.3 Magnet structure

Figure 10.5 shows the structural design of the detector magnet, comprising from inner to outer layers: the dewar (stainless steel 304), thermal shield (aluminum alloy 5083), superconducting coil, support cylinder (aluminum alloy 5083), liquid-helium pipes, and suspension structure (titanium alloy). Multilayer insulation is applied to minimize radiative heat load.

**Table 10.3:** Conductor overall characteristics

| Parameters | Value |
|---|---|
| Nominal design current | 17 kA |
| Critical Current (Ic) at 4.2 K and 5 Tesla | >51 kA |
| Total length of conductor | 33 km |
| Overall dimension (Without Insulation) | 56 mm × 22 mm |
| Superconducting cable cross section area | 2.2 mm × 20.5 mm |
| Al stabilizer cross section area | 12 mm × 40 mm |
| Al stabilizer RRR value | >700 |

Section 10.5.4 describes the mechanical design in detail.

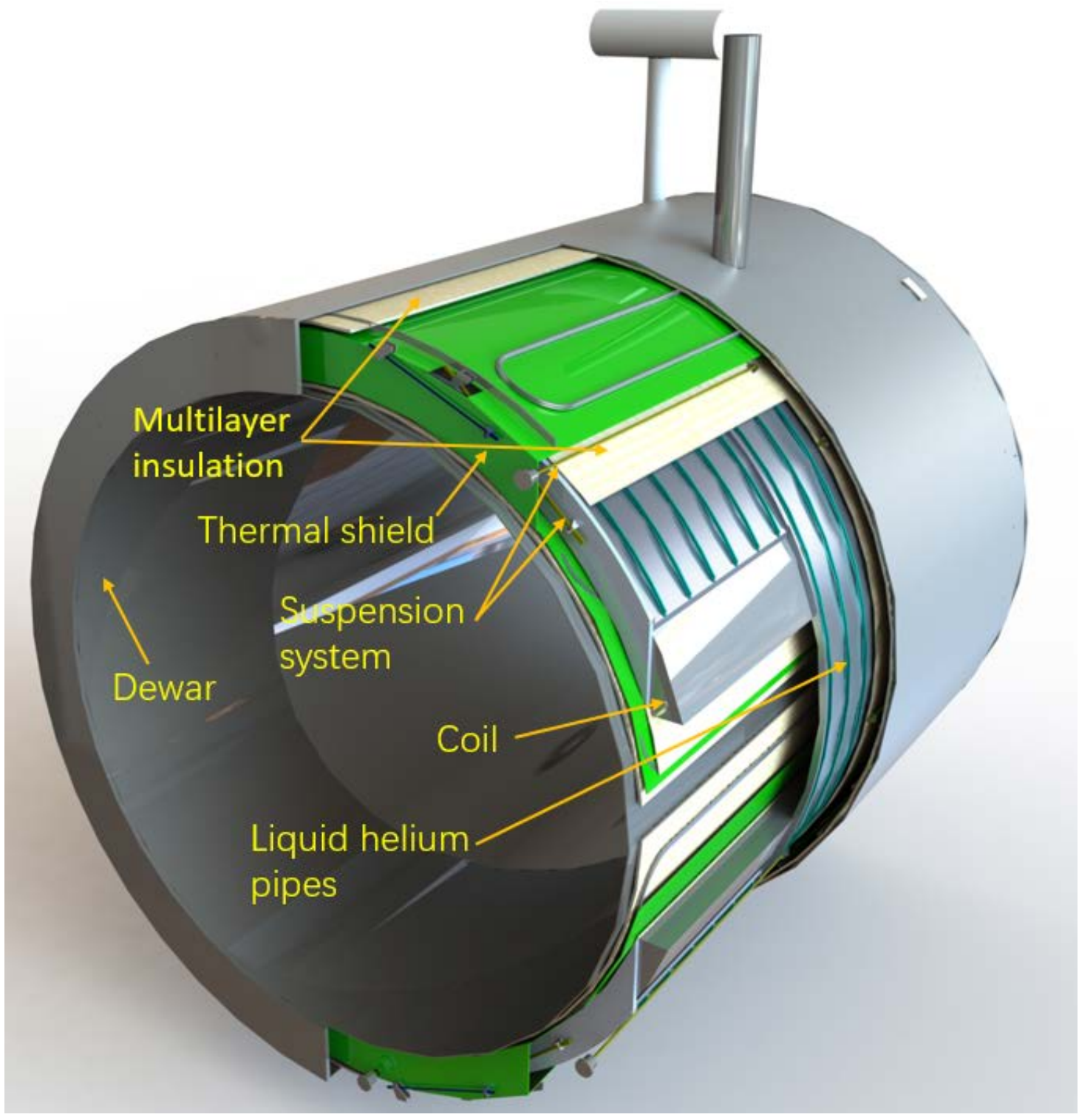

**Figure 10.5:** Mechanical design of the detector magnet. The internal structure of the magnet can be seen from this picture, including coils, thermosiphon cooling pipes, thermal shield and cold helium gas pipes, coil suspension system, Dewar and multi-layer insulation.

### 10.2.4 Yoke

The yoke provides support, fixation for internal sub-detectors, a magnetic flux return for magnets, particle absorption (excluding muons), and installation space for Muon detector. Yoke structure consists of a fixed twelve-set barrel yoke and two movable endcap yokes. The structural design and detailed parameters of the yoke is described in Chapter 14.2.1.

## 10.3 R&D and prototypes

The magnet features a substantial solenoid structure, with the main solenoid utilizing an internal winding technique. This technique involves four layers of superconducting cables wound on the inner wall of the support cylinder. The coil uses aluminum stabilized NbTi/Cu Rutherford cables for this purpose [7]. This section will introduce the R&D of the conductor and the development of dummy coil, including the progress of NbTi Rutherford cable development, the box-structured aluminum stabilized superconductor and the dummy coil development.

### 10.3.1 Aluminum stabilized NbTi Rutherford cable

The NbTi strand was designed based on the experience acquired in the development of conductors for previous Al-stabilized solenoids. Considering the electrical properties degradation due to the cabling process, the critical current density of NbTi strands 2700 A/ $mm^2$ at 5 T, 4.2 K are required for the final conductor. The strand parameters are shown in Table 10.4.

**Table 10.4:** NbTi superconducting strands parameters.

| **Parameters for wires** | **Value** |
|---|---|
| Filament diameter | 34 μm |
| Number of filaments | 708 |
| Strand diameter | 1.30±0.01 mm |
| Cu/SC ratio | 1.07±10% |
| Ic at 5 T, 4.2 K | >1750 A |
| RRR of copper | >100 |

Rutherford cables are assembled from 32 superconducting strands.The cabling process includes twisting, shaping, and forming. The strands are fed through a stranding machine under controlled tension, where both the machine and the wire spool rotate synchronously around the axis. This rotation causes the wire to spiral around a core rod, resulting in a helical configuration. Later, the wire is passed through a rolling mill for shaping.

**Table 10.5:** Rutherford cable parameters.

| **Parameter** | **Value** |
|---|---|
| Number of strands | 32 |
| Cable cross section | $21 \times 2.38\,mm^2$ |
| Transposition pitch | ~ 200 mm |
| Filling ratio | >83% |
| Cable Ic at 4.2 K, 5 T | >51 kA |

The aluminum coating production line for the cable primarily includes the cleaning of Rutherford cable and aluminum rods, co-extrusion forming, water cooling, and cable rewinding.

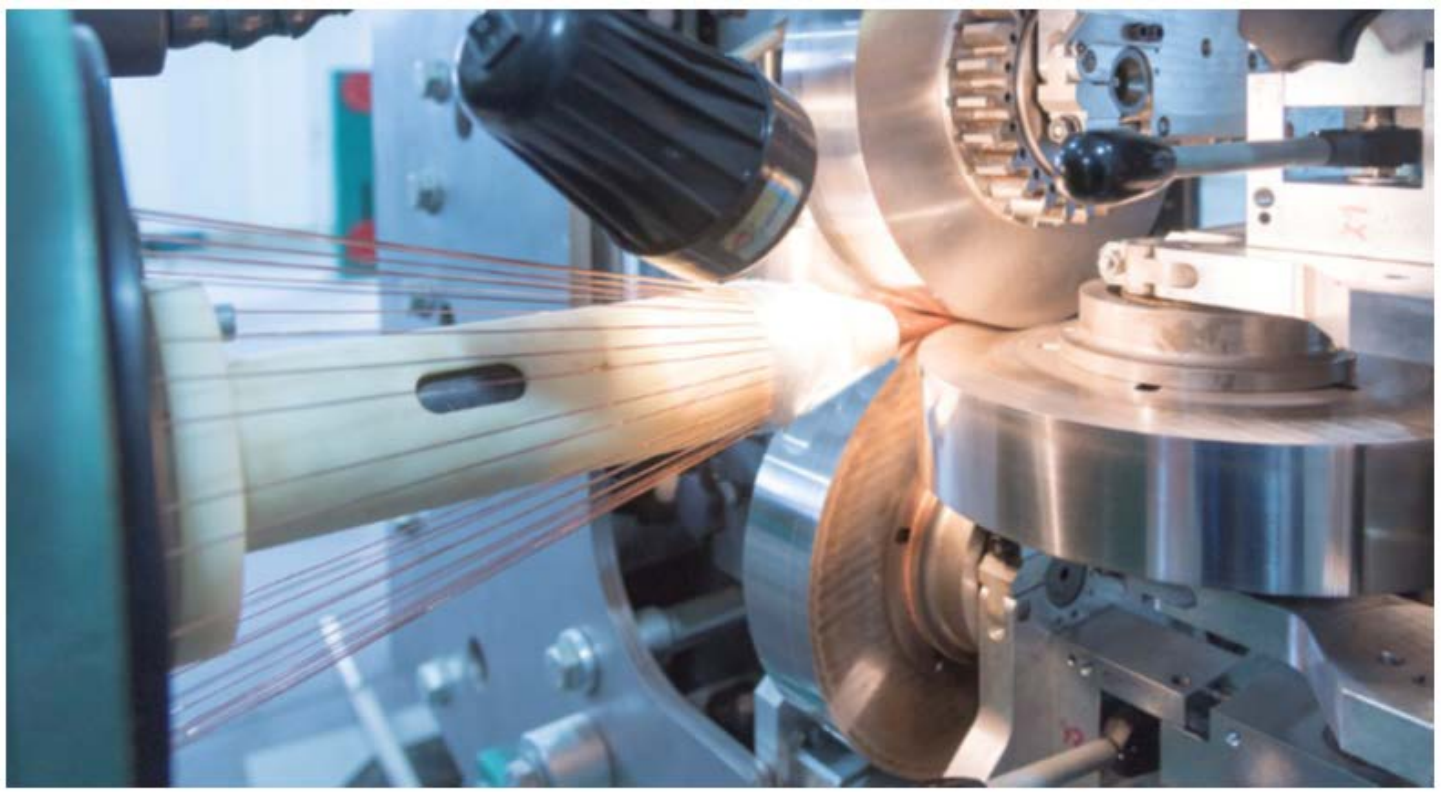

**Figure 10.6:** Manufacturing of Rutherford cable

After ultrasonic cleaning and drying, the Rutherford cable is passed through the forming mold that was preheated to 450 °C. The aluminum rods, after the oxidation layer was removed and cleaned, enter the extrude wheel groove. It is drawn into the mold's forming cavity by the frictional force of the chamfered wall, wrapping around the Rutherford cable to create a rectangular aluminum-stabilized conductor that is later extruded.

### 10.3.2 Developing progress of Box-Structured aluminum stabilized superconductor

The box-shaped conductor configuration is fabricated via a secondary co-extrusion process, using the aluminum-stabilized superconducting conductor as an insert. To reduce costs, preliminary process validation employed a 16-strand configuration. A 16-strand aluminum-stabilized superconductor was first developed, with the Rutherford cable supplied by Wuxi Toly Co. Following cabling, the NbTi wire's critical current degraded by less than 5%, while the RRR of the copper matrix decreased by approximately one-third.

Bond quality between the Rutherford cable and aluminum stabilizer constitutes a critical performance parameter for the conductor. The measured interfacial shear strength reaches 35 MPa. Figure 10.7 shows the shear-test configuration, performed using an AET-20K microcomputer-controlled universal testing machine, with four test specimens depicted.

Figure 10.8 presents a micro graph of the aluminum-copper interface obtained via electron microscopy. The absence of a clear diffusion zone suggests that the Rutherford cable surface was not adequately cleaned before co-extrusion, and surface oxides may have hindered metal-to-metal diffusion. To address this, future processes will incorporate surface cleaning and preheating of the cable before extrusion.

The 16-strand aluminum stabilized conductor was used as an insert and then through a second co-extrusion process with high-strength aluminum alloy to form reinforce, thereby enhancing the yield strength of the cable. Different grades of aluminum alloy materials and mold preheating processes were tried. A five-meter-long cable sample was obtained with box

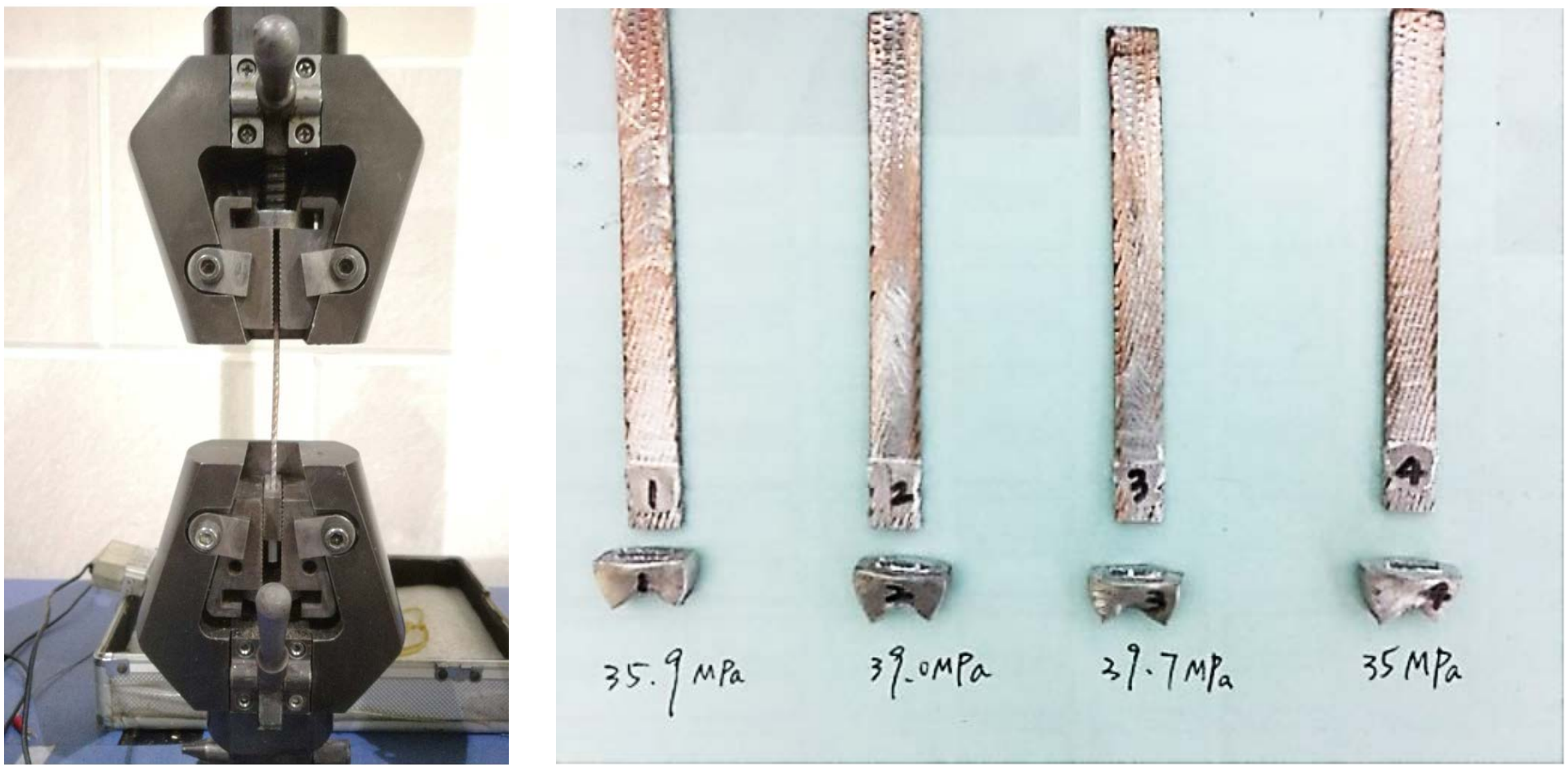

**(a)** **(b)**

**Figure 10.7:** (a) Test of shearing strength. AET-20K machine adopts microcomputer control, electronic sensors, and digital display instruments to achieve the measurement and control of parameters such as force and deformation, as well as data processing functions; (b) Four samples of shearing test. The magnitude of the bonding force directly reflects the quality of adhesion between the cable and the aluminum stabilizer. The shear strength is larger than the required (20 MPa).

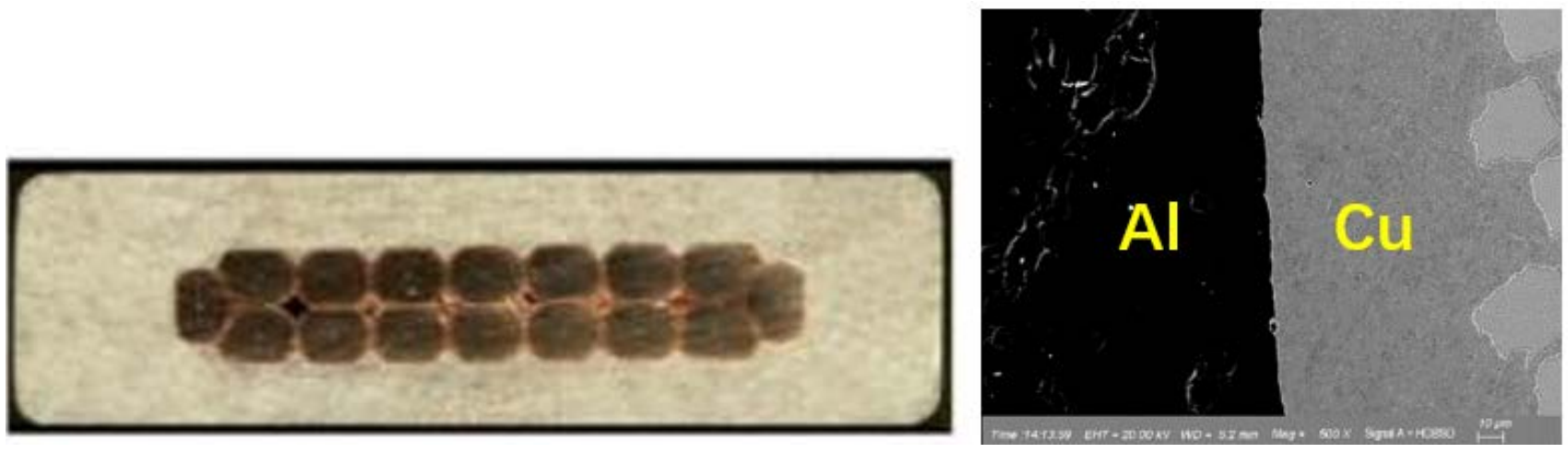

**Figure 10.8:** Cross-section of the superconducting conductor and SEM image of bonding surface

configuration superconductor from the pure aluminum and insert by a second co-extrusion process. The cable manufacturing process was interrupted due to the breakage of the core wire. The sample was developed and tested with about 10% critical current degradation in the second co-extrusion process. Figure 10.9 shows the second co-extrusion process and the cable structure. The team developed the 32 strands NbTi Rutherford cable and finished the first co-extrusion process. The last step secondary co-extrusion is been developing, seen in Figure 10.10.

### 10.3.3 Development of dummy coil

The dummy coil development evaluates manufacturing feasibility for large-diameter coils using inner-winding techniques and Vacuum Pressure Impregnation (VPI). Collaborating with

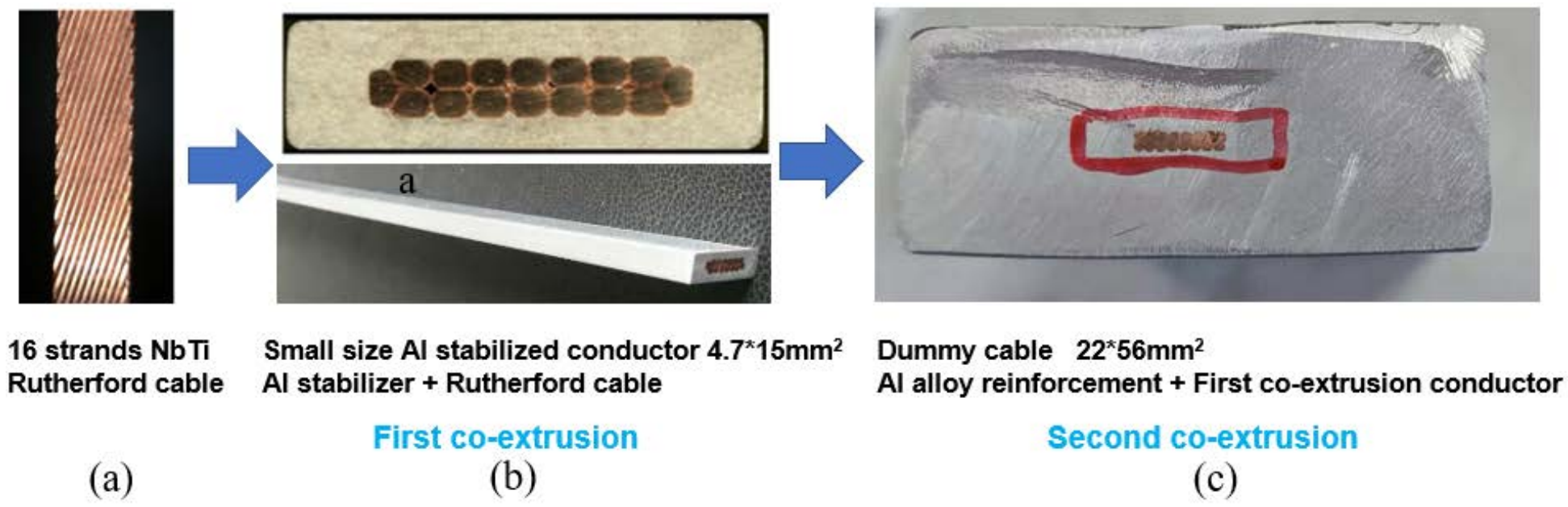

**Figure 10.9:** Second co-extrusion process. (a) The 15 strands NbTi Rutherford cable; (b) aluminum stabilized cable 4.7 mm × 15 mm as the insert ; (c) Box configuration type conductor with pure aluminum and aluminum stabilized cable 56 mm × 22 mm.

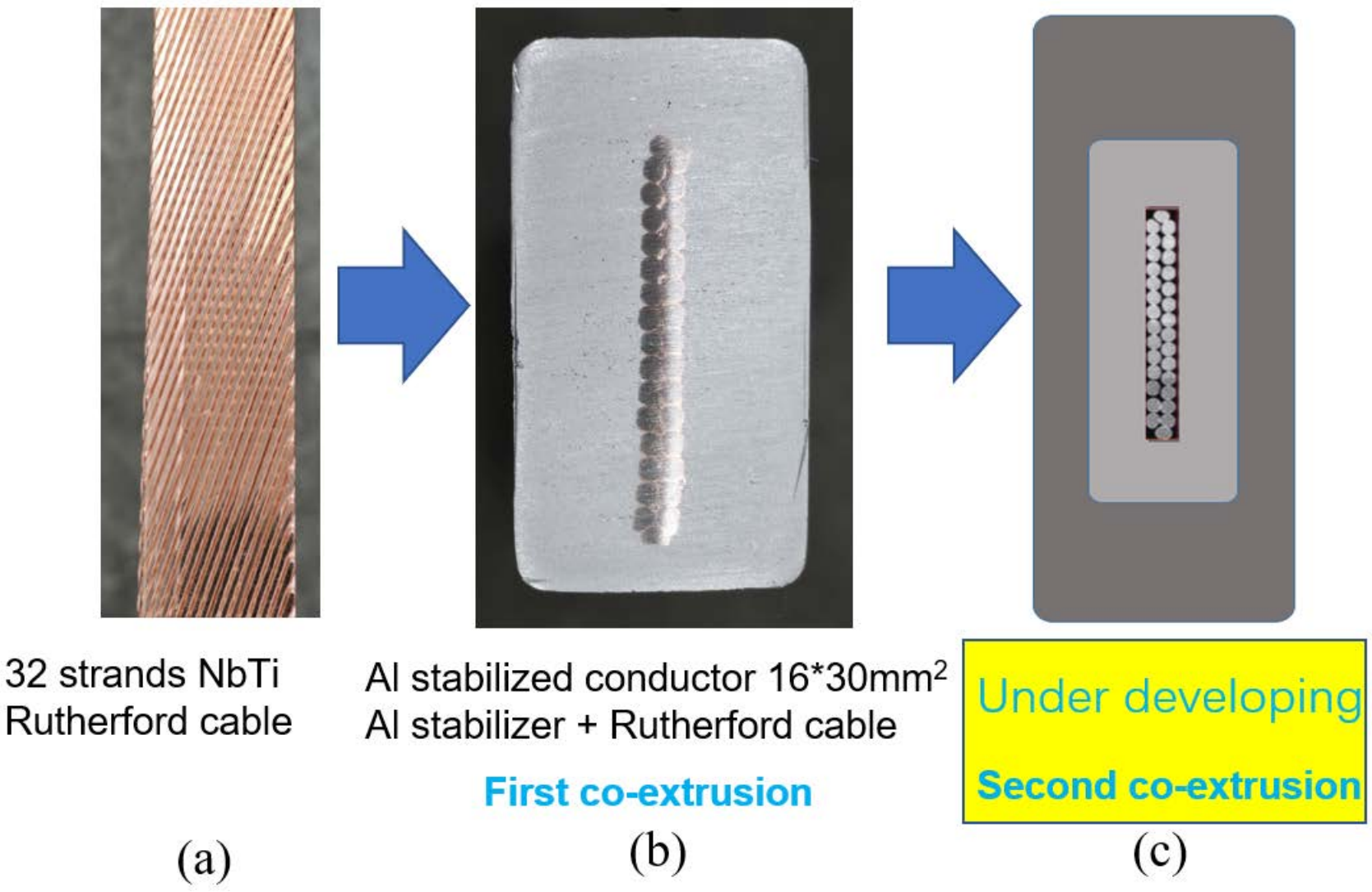

**Figure 10.10:** Full size Box-type superconductor. (a) The 32 strands NbTi Rutherford cable; (b) aluminum stabilized cable as the insert; (c) Box configuration type conductor 56 mm × 22 mm.

Hefei Keye Company, a dedicated winding platform for large solenoid coils was established. The coil comprises an aluminum-alloy support cylinder with four helical winding layers (10 turns/layer), each formed from a single aluminum-alloy conductor. Figure 10.11 shows the production line, including cable drum, straightening machine, buffering support, winding former, rotary platform, and mold.

Dummy cables were fabricated via co-extrusion of heat-treated 6061 aluminum (420 ℃, 8 h) and copper, producing four 250-m conductors with 56 mm × 22 mm cross-sections. A schematic diagram of coil VPI mold is shown in Figure 10.12. Post-VPI (Figure 10.13), the coil achieved:

- ±1 mm dimensional accuracy;
- > 500 V turn-to-turn insulation strength;
- > 2000 V layer-to-cylinder ground insulation.

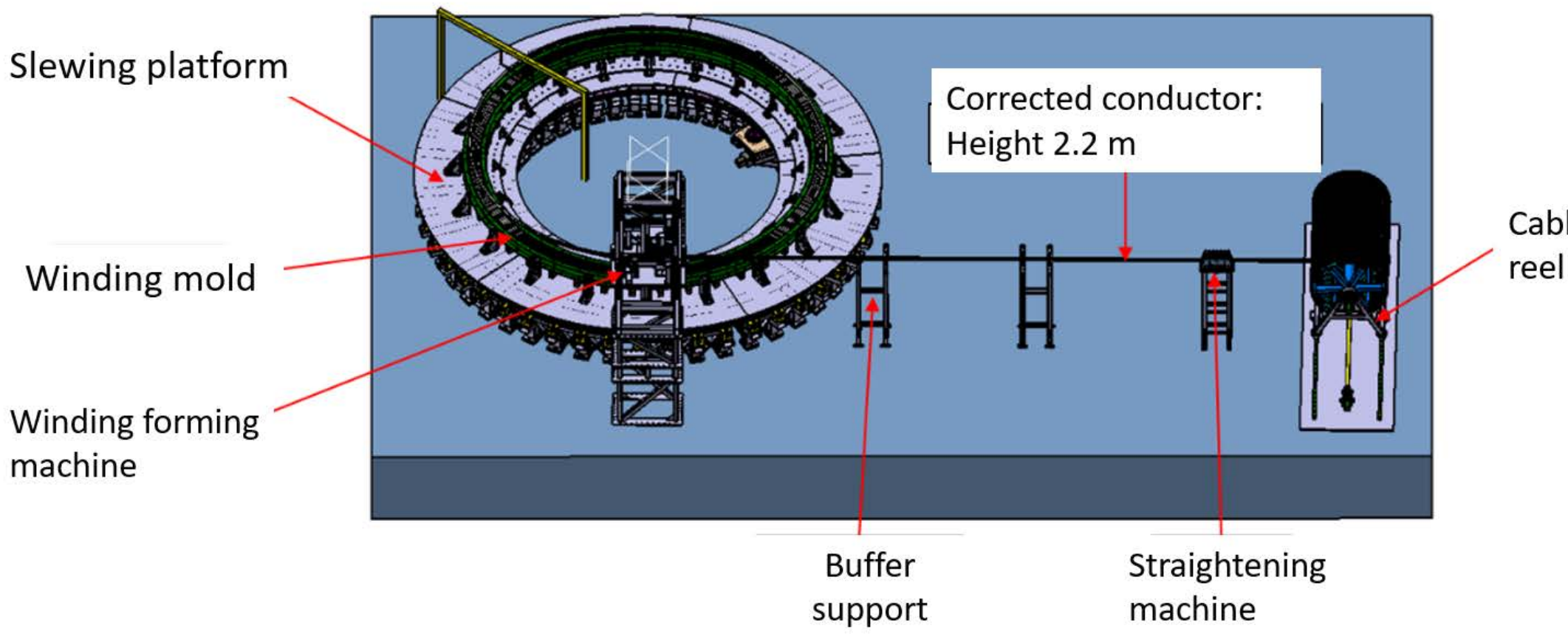

**Figure 10.11:** Diagram of the large-scale superconducting coil winding production line. It mainly includes the following components: cable reel, straightening machine, buffering support frame, winding forming machine, rotary platform.

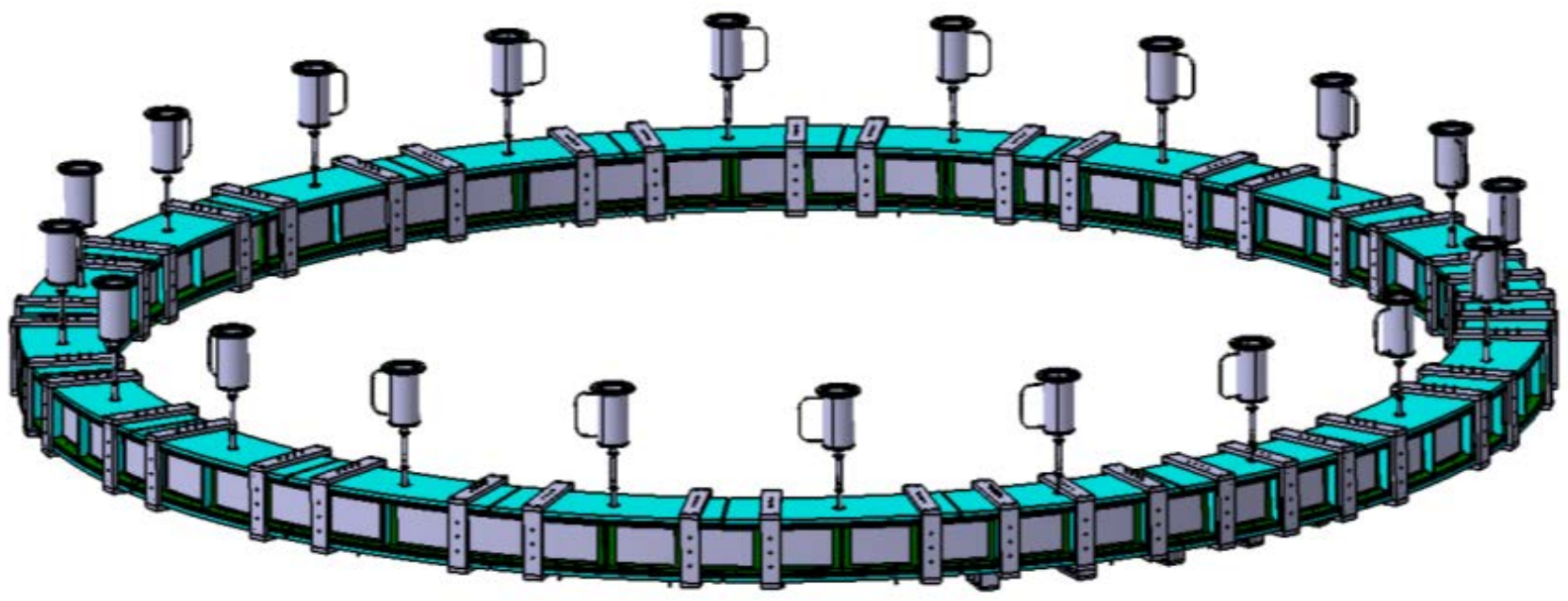

**Figure 10.12:** Schematic diagram of VPI mold. Above is the epoxy-filled vessel; the epoxy is drawn into the coil under vacuum.

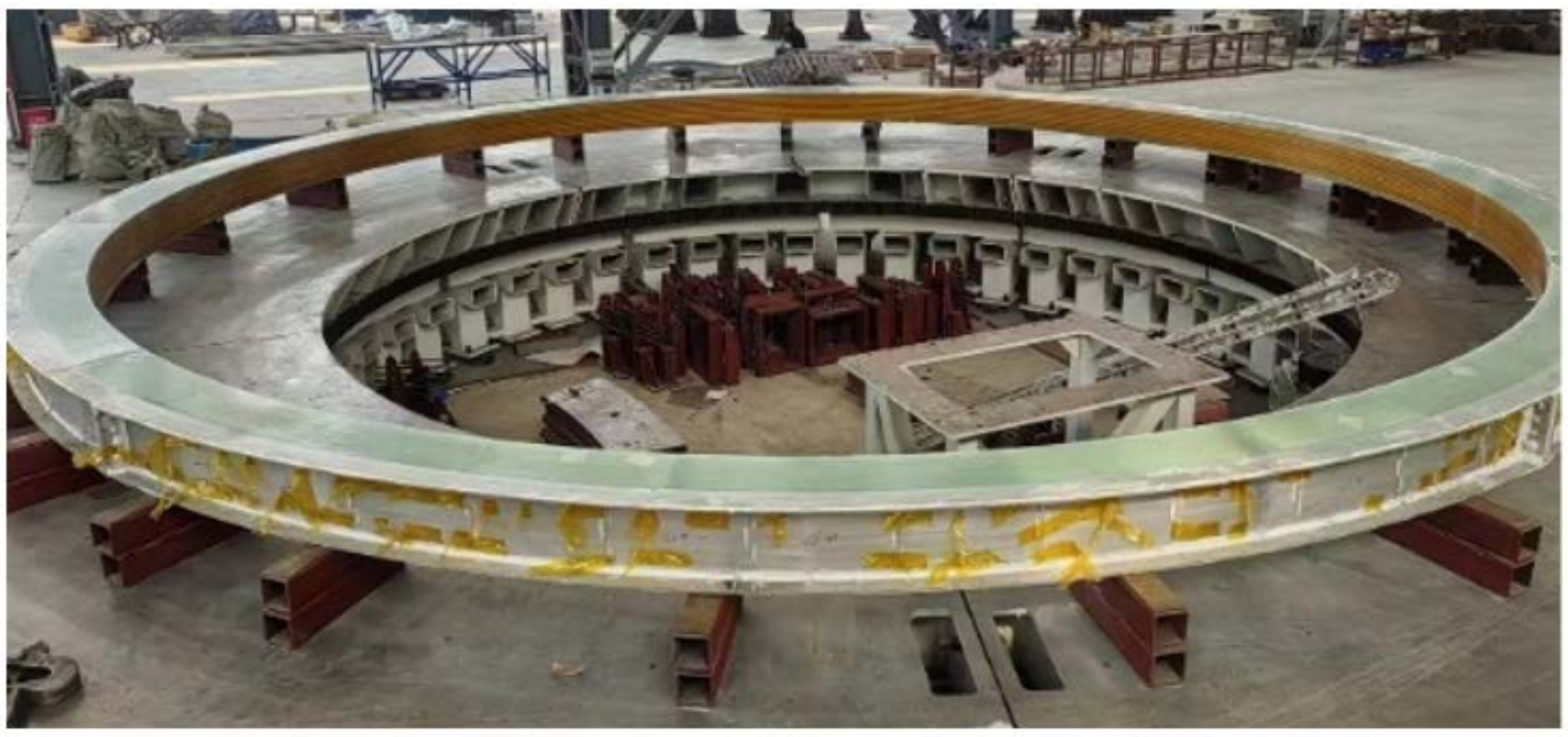

**Figure 10.13:** Dummy coil after coil winding and VPI. The inner diameter of the coil is 7.2 m. There are four layer and ten turns each layer. The manufacturing accuracy is controlled within ± 1 mm. The entire coil fabrication process was validated as feasible.

## 10.4 Simulation and performance

This section introduce the simulation results of magnetic field profiles, uniformity, coil stress, joint design with anti-solenoid, quench and stability analysis.

### 10.4.1 Field profiles and uniformity

The magnetic field distributions are calculated using Finite Element Method (FEM). Figure 10.14 shows the two-dimensional magnetic field distribution. Since the barrel yoke adopts a spiral structure rather than an axisymmetric structure, a three-dimensional magnetic field simulation is conducted. The corresponding results are shown in Figure 10.15. The magnetic field strength at the IP is 3 T, while the peak field on the coil is 3.3 T, and on the yoke is 3.8 T. Figure 10.16 displays the magnetic field distribution along the axis of the coil.

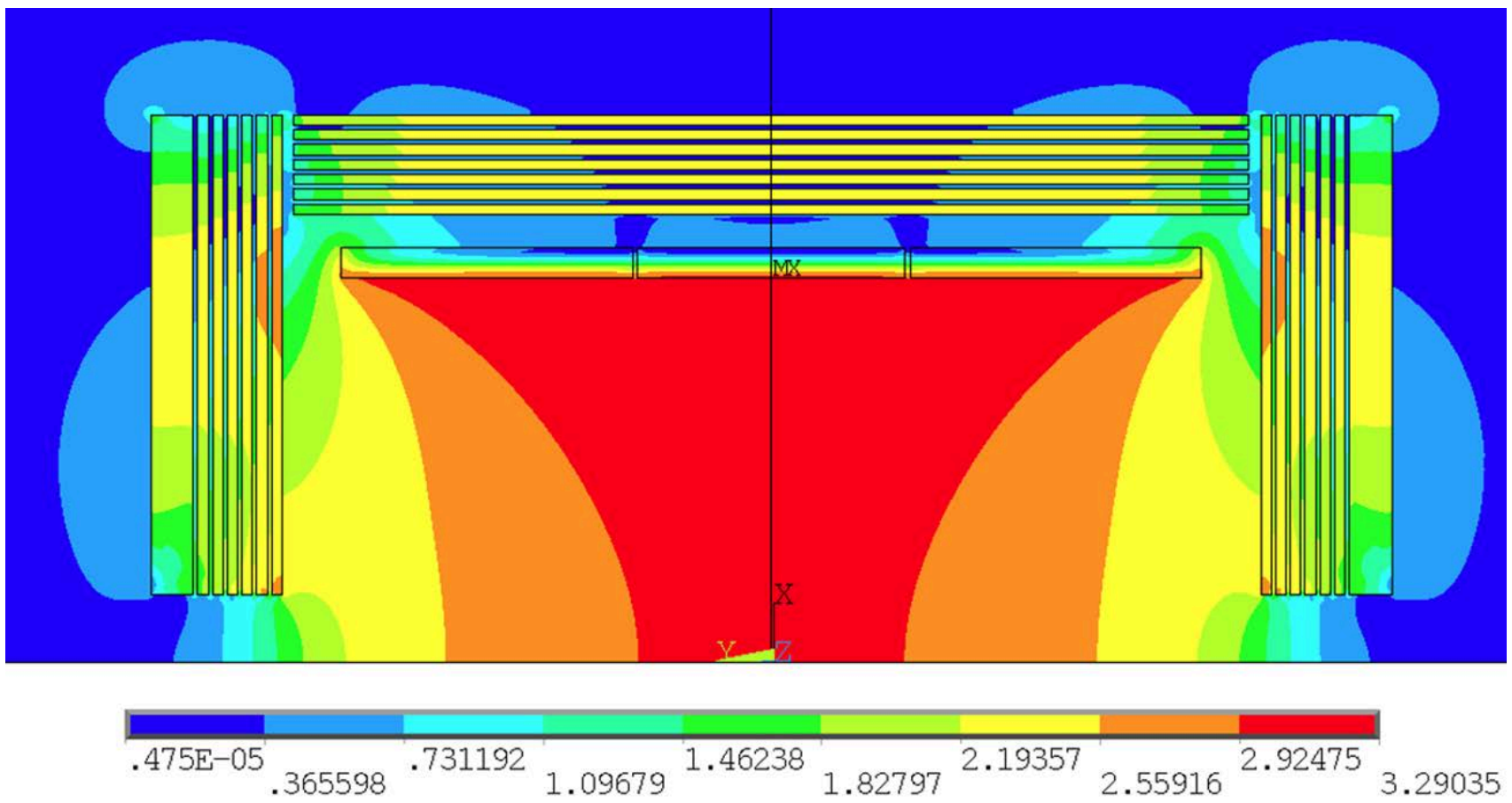

**Figure 10.14:** Two-dimensional Magnetic field distribution of the solenoid. The barrel yoke adopts a symmetrical structure. The magnetic field strength at the IP is 3 T, while the peak field on the coil reaches 3.3 T.

The magnetic stray field outside the yoke of the detector need to meet the requirements of the electronics, accelerator components and maintenance interventions. Additionally, the booster, located in the tunnel 25 m from the interaction region, requires the magnetic field to remain below 50 Gauss, potentially necessitating extra shielding. The stray field distribution of the detector magnet is illustrated in Figure 10.17 and Table 10.6. the 50 Gauss surface is located approximately 32.3 m axially and 25.4 m radially, while the 100 Gauss surface is at 25.6 m axially and 20.2 m radially.

When the coil is perfectly centered, its symmetrical structure and balanced electromagnetic forces result in a net force of zero. In this ideal scenario, the suspension structure only needs to support the coil's weight and the tension caused by structural deformation due to thermal contraction. However, during actual installation, deviations from the design position are inevitable due to manufacturing and measurement tolerances. These misalignments, combined with the

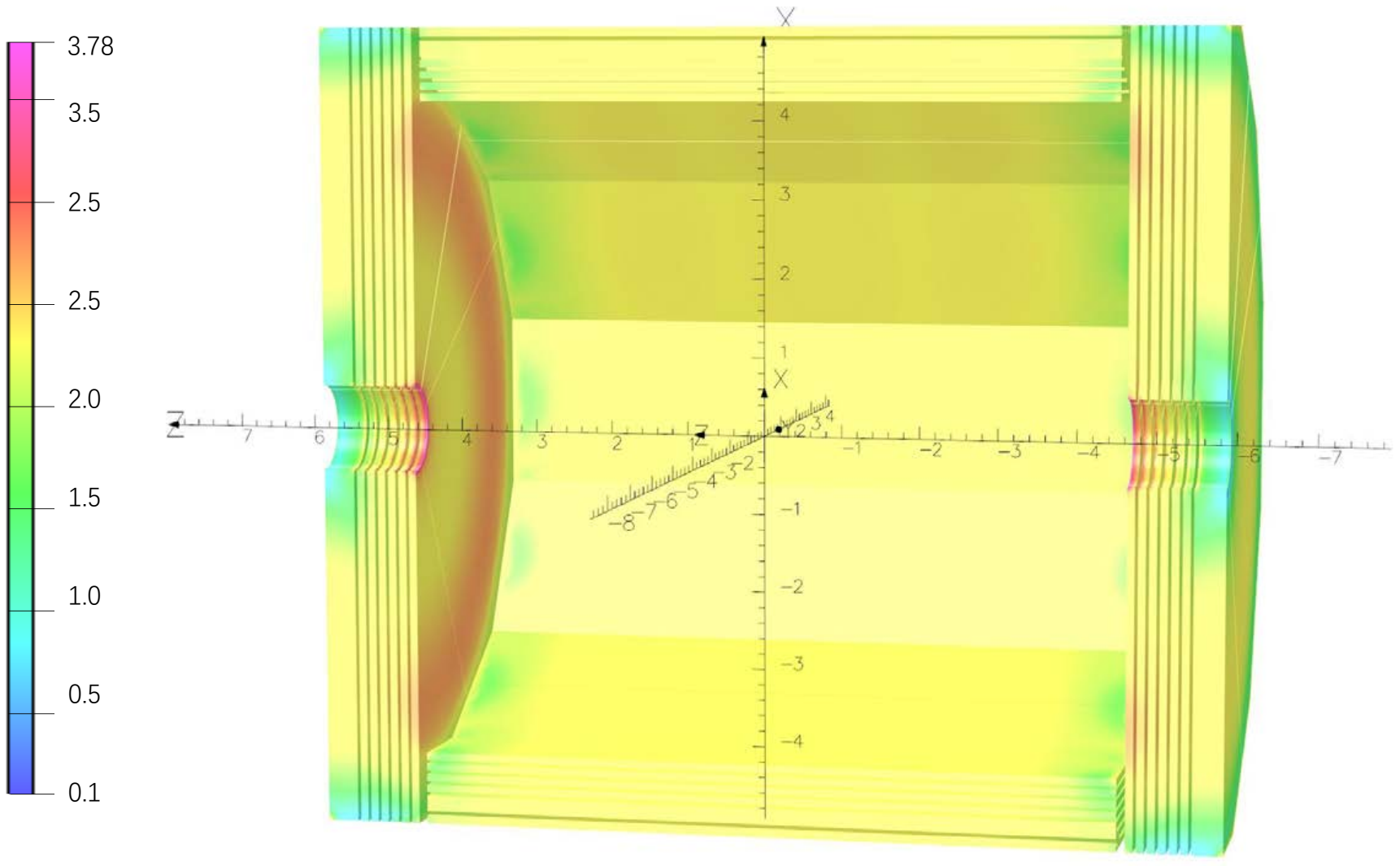

**(a)**

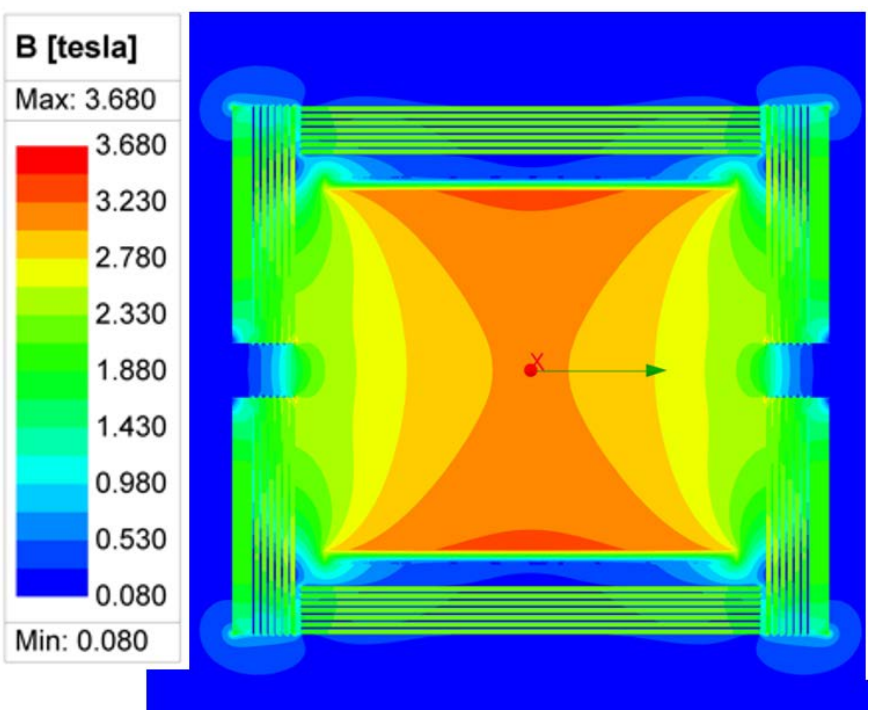

**(b)**

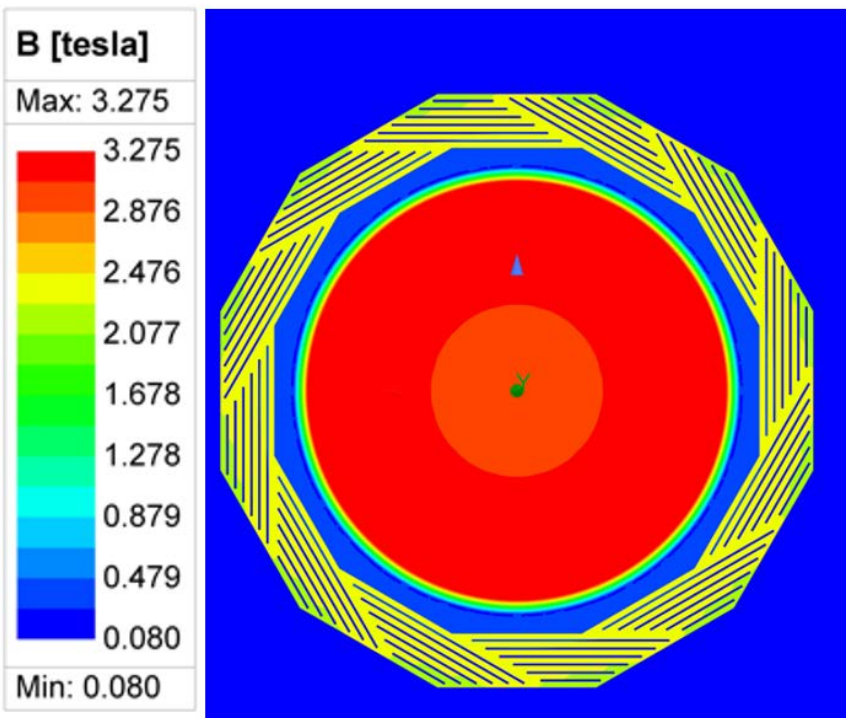

**(c)**

**Figure 10.15:** Three-dimension magnetic field distribution of the solenoid. (a) Magnetic field distribution of the yoke with a peak value of 3.78 T, (b) $xz$ plane through IP and (c) $xy$ plane through IP .

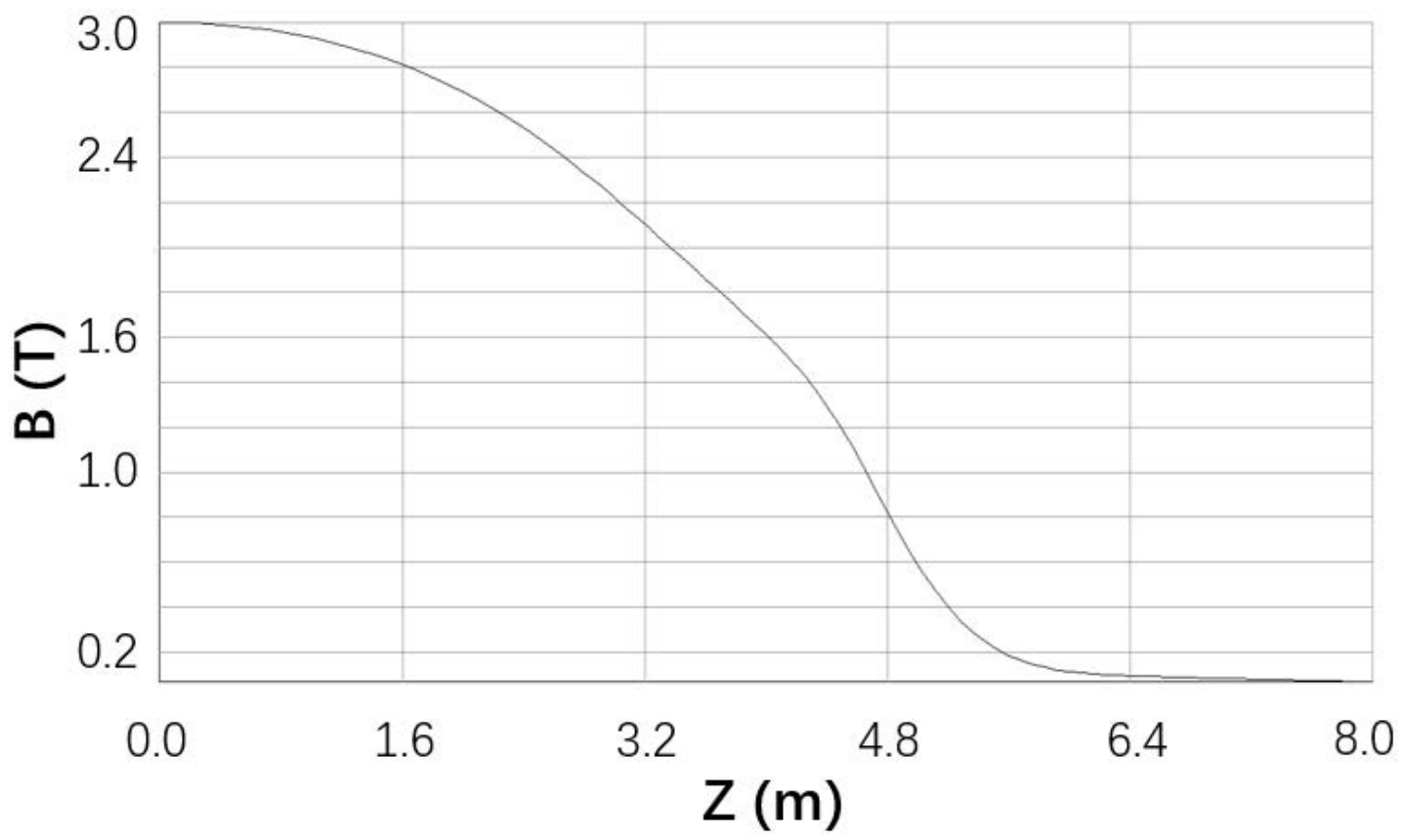

**Figure 10.16:** Magnetic field distribution in axial direction. The horizontal axis represents the axial position, ranging from the collision point to 8 m. The vertical axis represents the magnetic field strength, with the magnetic field strength at the IP point being 3 T.

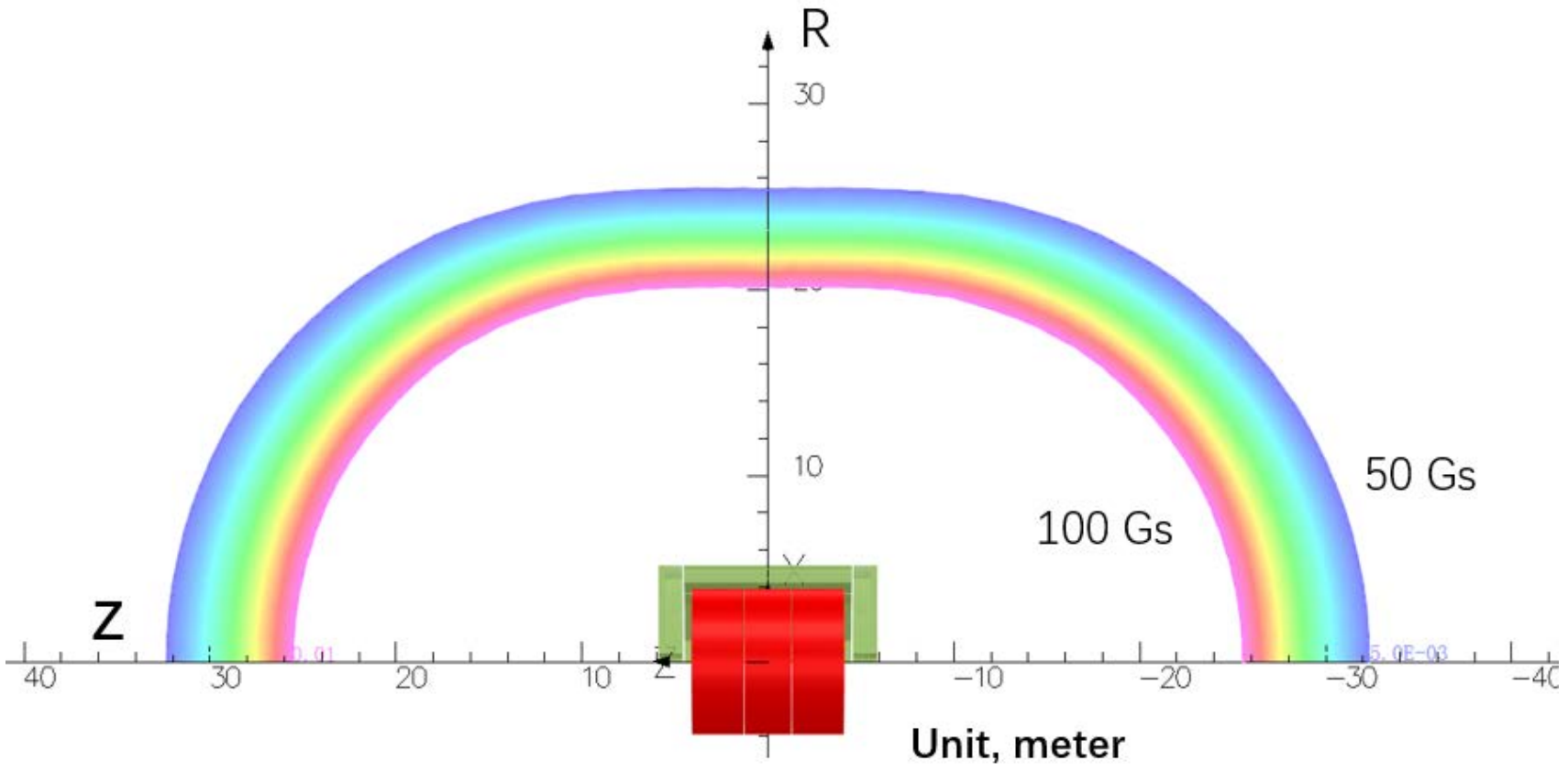

**Figure 10.17:** Stray field distribution. The horizontal axis represents the axial position, while the vertical axis represents the radial position. The position of the 50 and 100 Gauss lines. The inner part is 100 Gauss, the outer blue part is 50 Gauss.

**Table 10.6:** Stray field positions and distances.

| Stray field | Direction | Position |
|---|---|---|
| 50 Gs | R direction | 25.4 m |
| | Z direction | 32.3 m |
| 100 Gs | R direction | 20.2 m |
| | Z direction | 25.6 m |

presence of yokes, can disrupt the electromagnetic force balance, resulting in additional loads on the nearest yoke. Table 10.7 summarizes the forces and torques acting on the coil under various misalignment scenarios. These electromagnetic forces must be carefully considered in the design of the coil's suspension structure.

**Table 10.7:** Forces and torques of coil offset

| Axial displacement (cm) | 1 | 3 | 10 |
|---|---|---|---|
| **Axial force (kN)** | 391 | 1247 | 5276 |
| **Radial displacement (cm)** | 1 | 3 | 10 |
| **Radial force (kN)** | 49.3 | 258 | 708 |
| **Angular tilt (degree)** | 1/407.5 = 0.14° | 3/407.5 = 0.42° | 10/407.5 = 1.41° |
| **Torque (kN· m)** | 1678 | 7020 | 17647 |

The magnetic field distribution in the TPC region is crucial for particle identification. The uniformity and gradient of the magnetic field have an impact on detection efficiency.

The magnetic field distribution was simulated without anti-solenoids and focusing quadrupole magnets. Figure 10.18 shows the magnetic field distribution in the TPC, which is relatively uniform with a difference of 0.44 T between the maximum and minimum values.

The dimensions of the ITK region are illustrated in Figure 10.19. For the magnetic field calculation, this region is simplified and approximated as a rectangular-like area. Figure 10.19 shows the magnetic field distribution in the ITK region, with a variation of 0.11 T between the maximum and minimum field strengths.

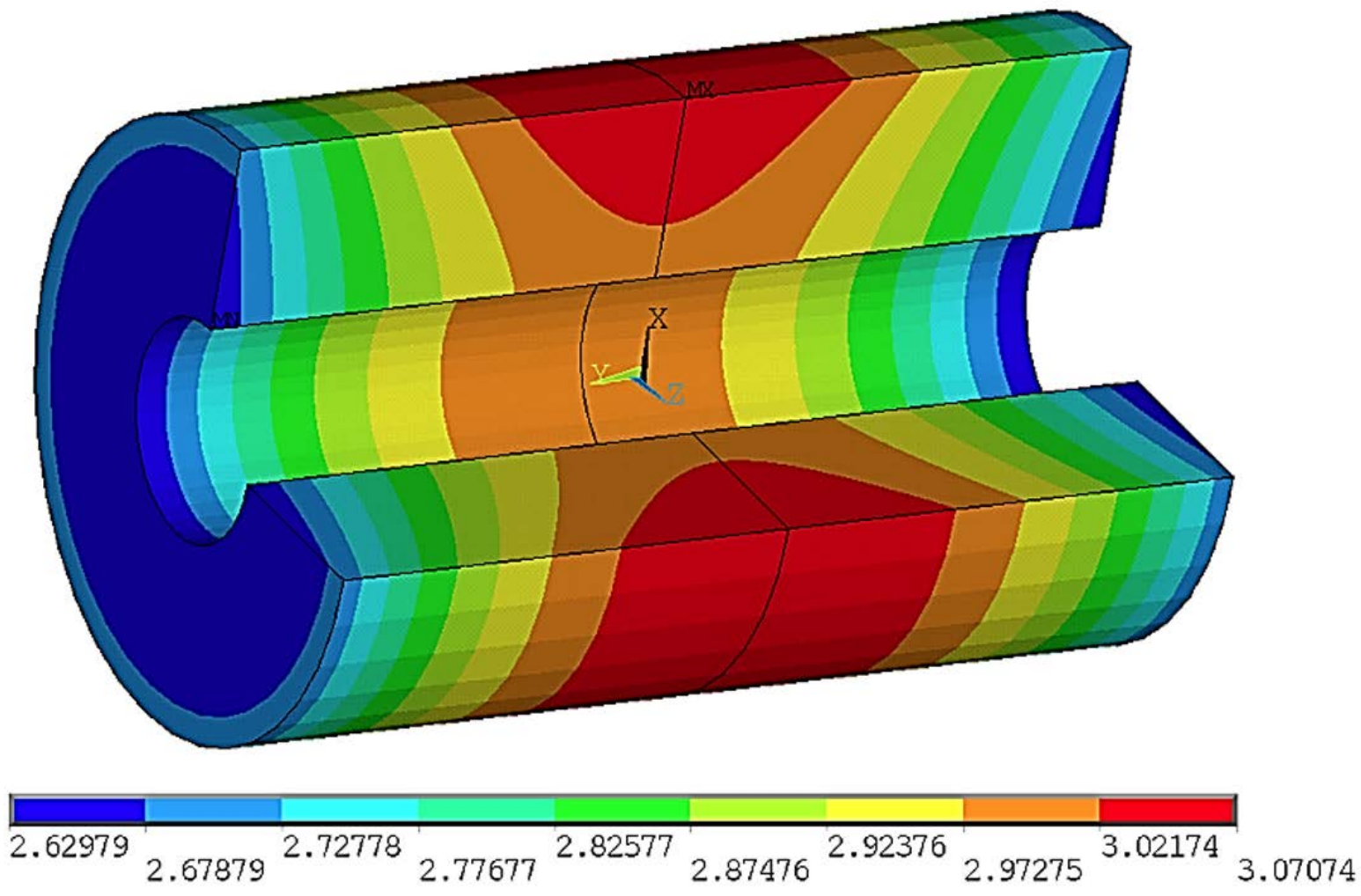

**Figure 10.18:** Magnetic field distribution in TPC region without compensation solenoid. The difference is 0.44 T between the maximum and minimum values.

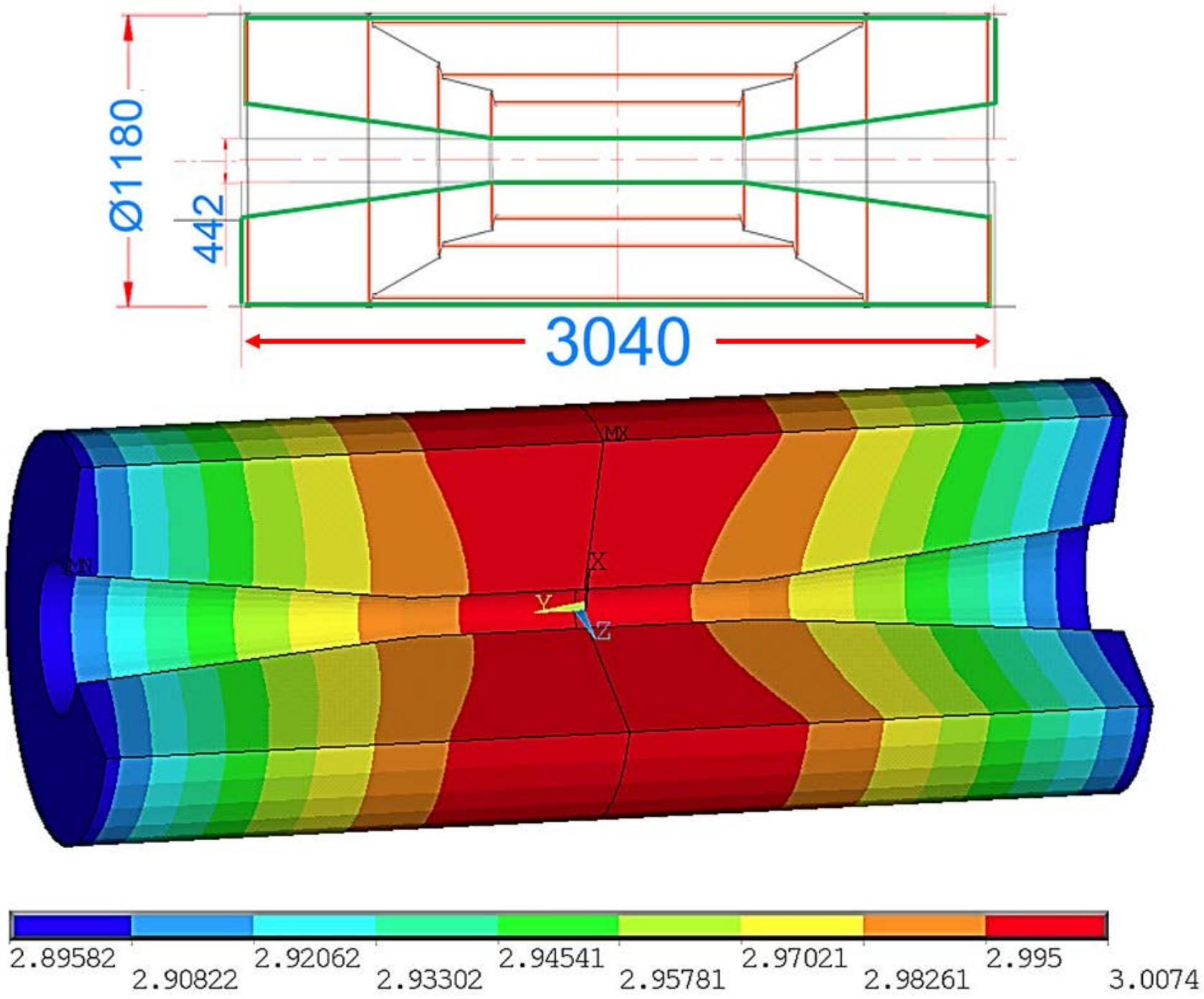

**Figure 10.19:** ITK dimensions (units is mm) and magnetic field distribution without compensation solenoid. It is an almost uniform magnetic field distribution with only a difference of 0.11 T. For simplicity, it was approximated as the shape of the green area

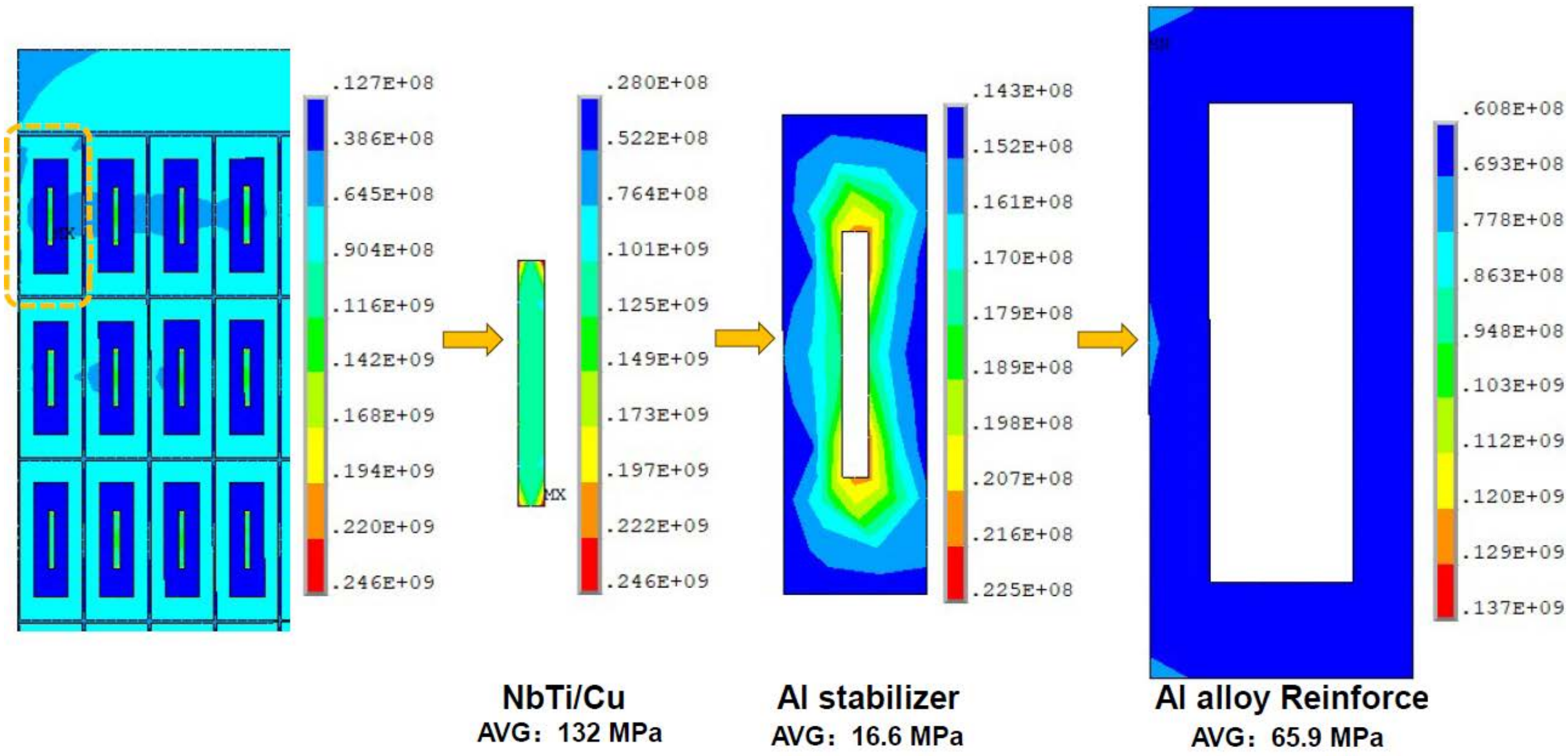

**Figure 10.20:** Von Mises stress distribution of the coil was analyzed after cooling to 4.5 K and energizing to 3 T. The maximum stress on the coil is 246 MPa, occurring in the NbTi/Cu cable. The maximum stress on the high-purity aluminum stabilizer is 22.5 MPa, while the aluminum alloy and aluminum alloy support experience maximum stresses of 137 MPa and 127 MPa, respectively. The maximum shear stress on the epoxy glass fiber is 10.6 MPa.

## 10.4.2 Coil stress analysis

Stress evaluation of the coil was performed using FEA simulation. The materials and parameters for each coil component are listed in Table 10.8. The aluminum stabilizer is made of high-purity aluminum.

**Table 10.8:** Material parameters used in the model

| Materials | Young modulus | Poisson's ratio | Thermal expansivity | Rp0.2 at 4.2 K |
|---|---|---|---|---|
| NbTi/Cu | 147 GPa | 0.31 | $8.79 \times 10^{-6}$/K | >500 MPa |
| Al | 69 GPa | 0.34 | $14.23 \times 10^{-6}$/K | 34 MPa |
| Reinforcement | 77.7 GPa | 0.327 | $14.16 \times 10^{-6}$/K | >350 MPa |
| Fiber glass‖ | 20 GPa | 0.21 | $8.45 \times 10^{-6}$/K | - |
| Fiber glass⊥ | 12.5 GPa | 0.21 | $25.5 \times 10^{-6}$/K | - |

Due to significant differences in thermal expansion coefficients between NbTi/Cu and aluminum, localized high stresses occur near joint surfaces after cooling. Upon coil excitation, stress gradually decreases from the center toward the ends under Lorentz force influence. Table 10.9 summarizes the Von Mises stress distribution under these conditions.

Figure 10.20 illustrates the Von Mises stress distribution in the coil after cooling to 4.2 K and energizing to 3 T, showing stresses on the NbTi/Cu Rutherford cable, aluminum stabilizer, and aluminum alloy reinforcement. The maximum stress on the coil is 246 MPa, located in the NbTi/Cu cable. The high-purity aluminum stabilizer experiences a maximum stress of 22.5 MPa, while the aluminum alloy and aluminum alloy support reach maximum stresses of 137 MPa and 127 MPa, respectively. The maximum shear stress in the epoxy glass fiber is 10.6 MPa. Focusing on the outermost conductor turn, where peak stress occurs, nodal solutions were extracted to calculate average material stresses. The average Von Mises stresses in this turn are 132 MPa for NbTi/Cu, 16.6 MPa for the aluminum stabilizer, and 65.9 MPa for the reinforcement. These are listed in Table 10.10. Maximum stresses in all coil materials remain within allowable limits, providing enough safety margins for stable coil operation.

**Table 10.9:** Von Mises stress and strain on coil

| **Material** | **Von Mises stress (MPa)** | **Von Mises strain (‰)** |
|---|---|---|
| | **Coil at 4.5 K** | |
| Aluminum stabilizer | 7.7-20.0 | 0.1-3.01 |
| NbTi/Cu cable | 177-279 | 1.36-2.15 |
| Al alloy | 7.0-43.3 | 0.05-0.1 |
| Al alloy support | 6.3-36.3 | 0.1-0.44 |
| | **Coil at 4.5 K, energized to 3 T** | |
| Aluminum stabilizer | 14.3-22.5 | 0.85-4.09 |
| NbTi/Cu cable | 28-246 | 0.2-1.9 |
| Al alloy | 60.8-137 | 0.7-1.7 |
| Al alloy support | 62.2-127 | 0.7-1.57 |

**Table 10.10:** The maximum and average Von Mises stresses of each material of the conductor in the maximum stress turn. Maximum stress in each coil material remains within allowable limit, providing enough safety margins for stable coil operation.

| **Materials** | **Max von Mises (MPa)** | **Avg von Mises (MPa)** | **$Rp_{0.2}$@4.2 K (MPa)** | **Allowable stress @4.2 K (MPa)** |
|---|---|---|---|---|
| NbTi/Cu | 246 | 132 | > 500 | 300 |
| 4N8 Al stabilizer | 22.5 | 16.6 | ~ 34 | 22.7 |
| Aluminum alloy | 137 | 65.9 | > 350 | 233 |

### 10.4.3 Joint design with anti-solenoid and focusing magnets

The anti-solenoid coils and focusing quadrupole magnets are the critical components of the accelerator, designed to counteract the magnetic field effects of the solenoid coil on the beam, thereby directly influencing accelerator performance.

Figure 10.21 shows the simulation model, the outer part is the yoke, the inner parts are the coils, the outer is the detector solenoid coil, the small parts are the anti-solenoid coils, the focusing quadrupole coils locate in the anti-solenoid coils.

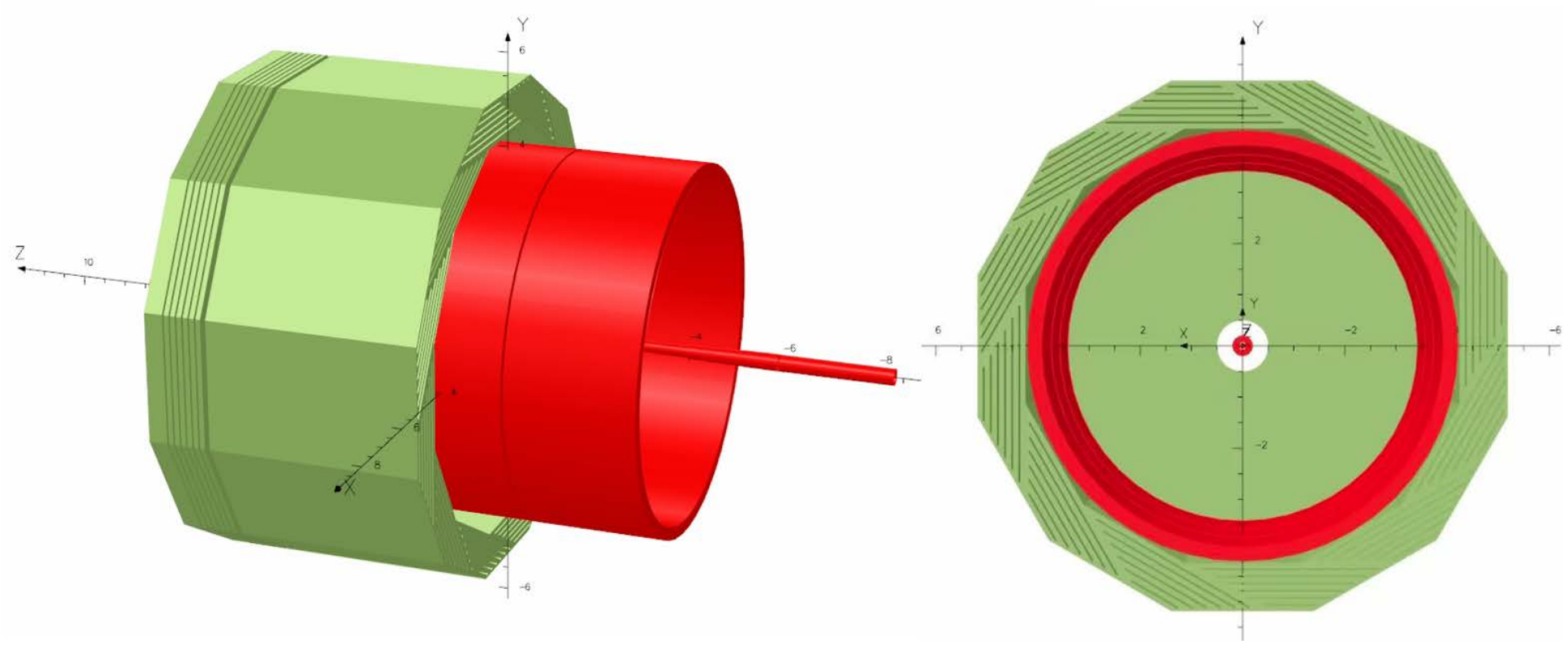

**Figure 10.21:** Joint design of detector solenoid coil, anti-solenoid coils and focusing quadrupole coils. The outer part is the yoke, the inner parts are the coils, the out is the detector solenoid coil, the small parts are the anti-solenoid coils, the focusing quadrupole coils locate in the anti-solenoid coils.

Figure 10.22 illustrates the magnetic field distribution of TPC region combined solenoid, anti-solenoid and quadrupole magnets. The magnetic field difference is 0.438 T.

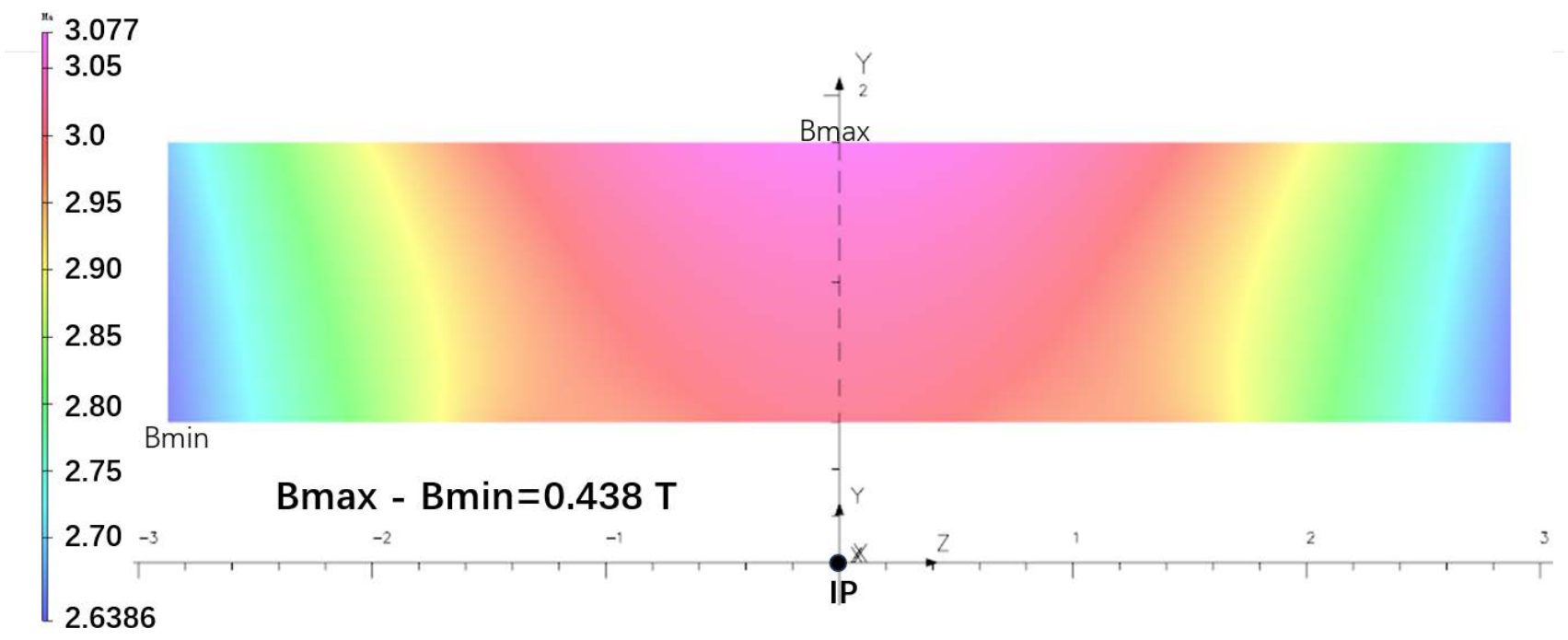

**Figure 10.22:** Joint magnetic field distribution of TPC region. The magnetic field difference is 0.438 T.

The dimensions of the TPC region and the magnetic field distributions along paths are shown in Figure 10.23a. The non-uniformity on path is less than 7%, which meet the requirement of TPC: the non-uniformity of magnetic field along $z$-axis in TPC region less than 7%.

## 10.4.4 Quench and stability analysis

### Quench detection and protection

The quench protection scheme for the coil is based on a dump resistor. After a quench occurs, the current rapidly decreases, depositing most of the energy into the dump resistor outside the Dewar. The external support cylinder, which has good thermal contact with the coil, will also generate eddy current that result in quench back effects, heating the coil and speeding up the quench process. Figure 10.24 shows the equivalent electrical circuit of the superconducting coil.

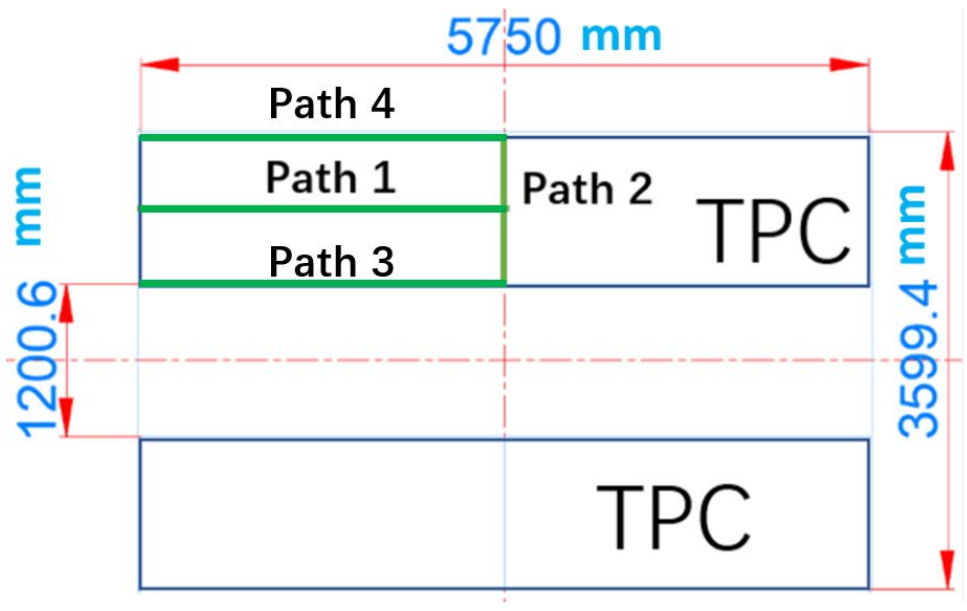

**(a)**

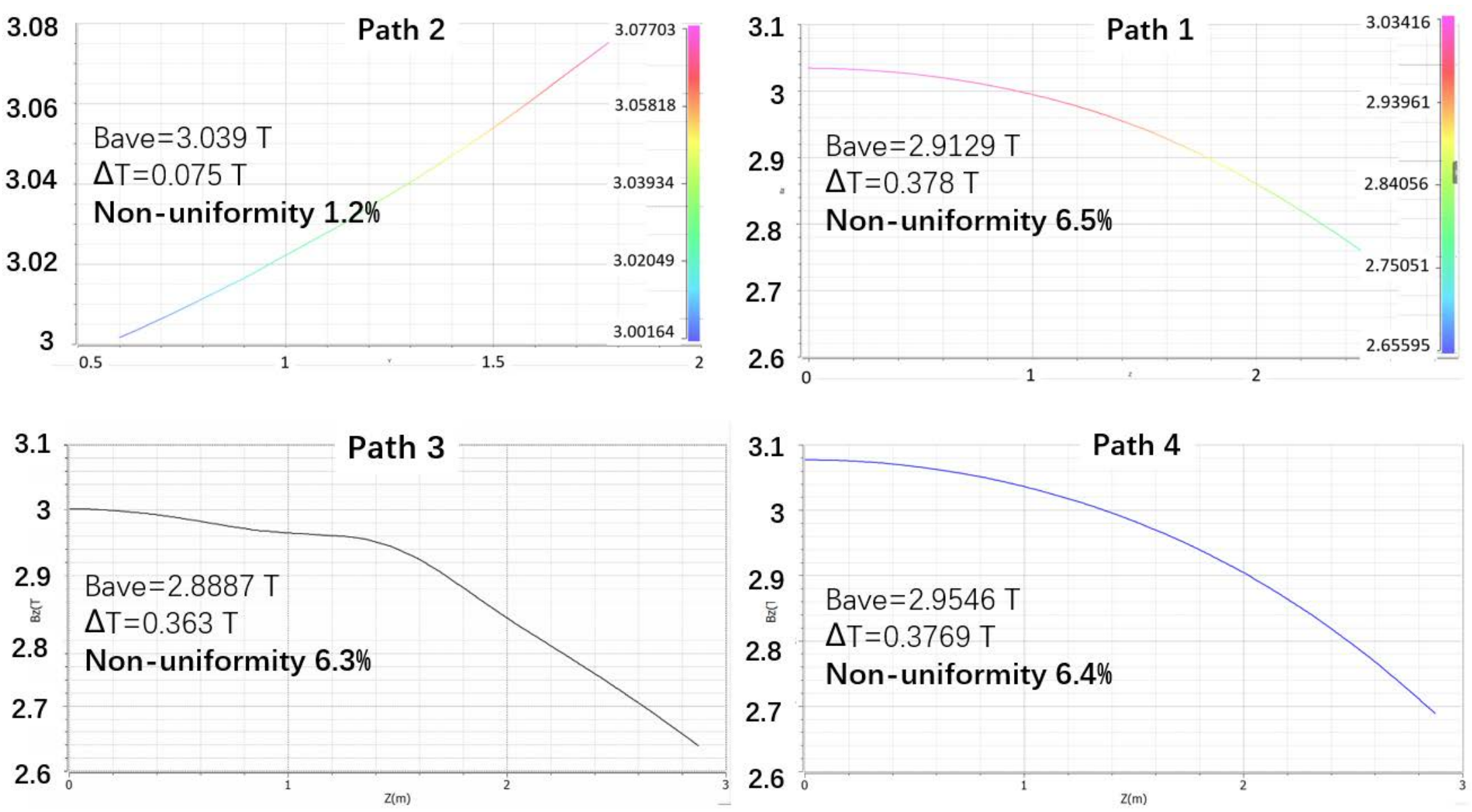

**(b)**

**Figure 10.23:** (a) The dimension of TPC region (b) magnetic field distribution on path 1, path 2, path 3 and path 4. The non-uniformity on path along $z$-axis is less than 7%.

Quench detection relies primarily on monitoring the coil's voltage signal. The time required to confirm a quench and to open the circuit breaker introduces a power-shutdown delay of a few seconds. As the breaker delay increases, the hot spot temperature rises accordingly, indicating that the delay should be kept . An additional approximate 30 ms is needed to connect the dump resistor.

Using the Multi-Integral of Inductance and Time (MIITs) fast quench calculation method, a parameter scan was performed for quench energy resistance, quench delay, and stabilizer RRR. Under the constraint of an end voltage below 1000 V, the highest energy extraction efficiency is obtained with a dump resistance of 50 $m\Omega$ and a stabilizer RRR above 600. To provide a stability margin, the CEPC conductor stabilizer is required to have an RRR larger than 700.

Figure 10.25 presents the coil current decay, temperature rise and voltage variation based on a 50 $m\Omega$ dump resistor and an RRR of 700 at different delay times. Table 10.11 lists key quench parameters based on different delay times, 0.05 s, 2 s and 10 s. The average post quench coil temperature is 59.2 K to 62.25 K, and 74.23% to 70.68% of the stored energy is extracted.

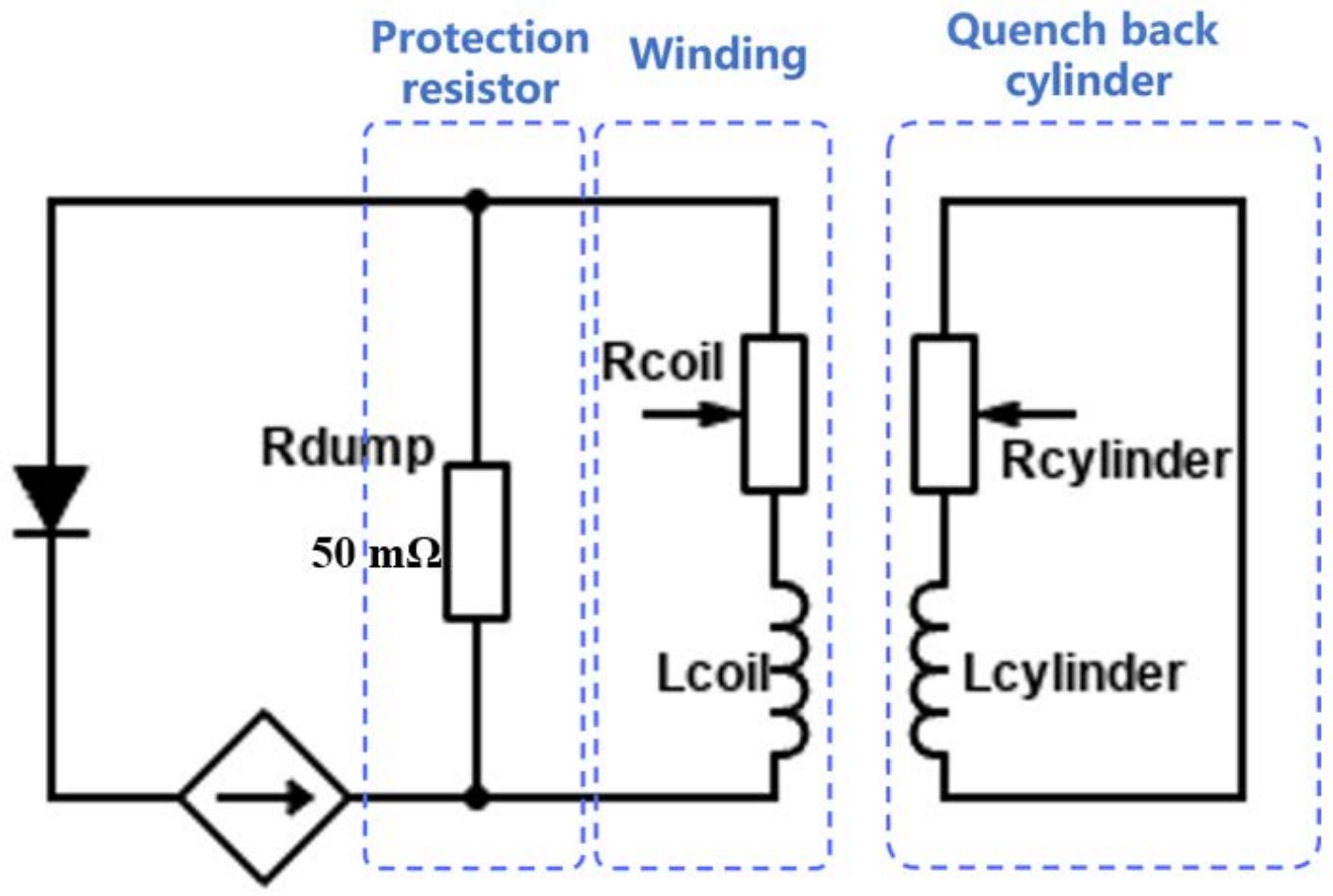

**Figure 10.24:** Equivalent electrical circuit of the winding configuration. The quench protection scheme for the coil is based on a 50 mΩ dump resistor. The external support cylinder will also generate eddy current that result in quench back effects, heating the coil and speeding up the quench process.

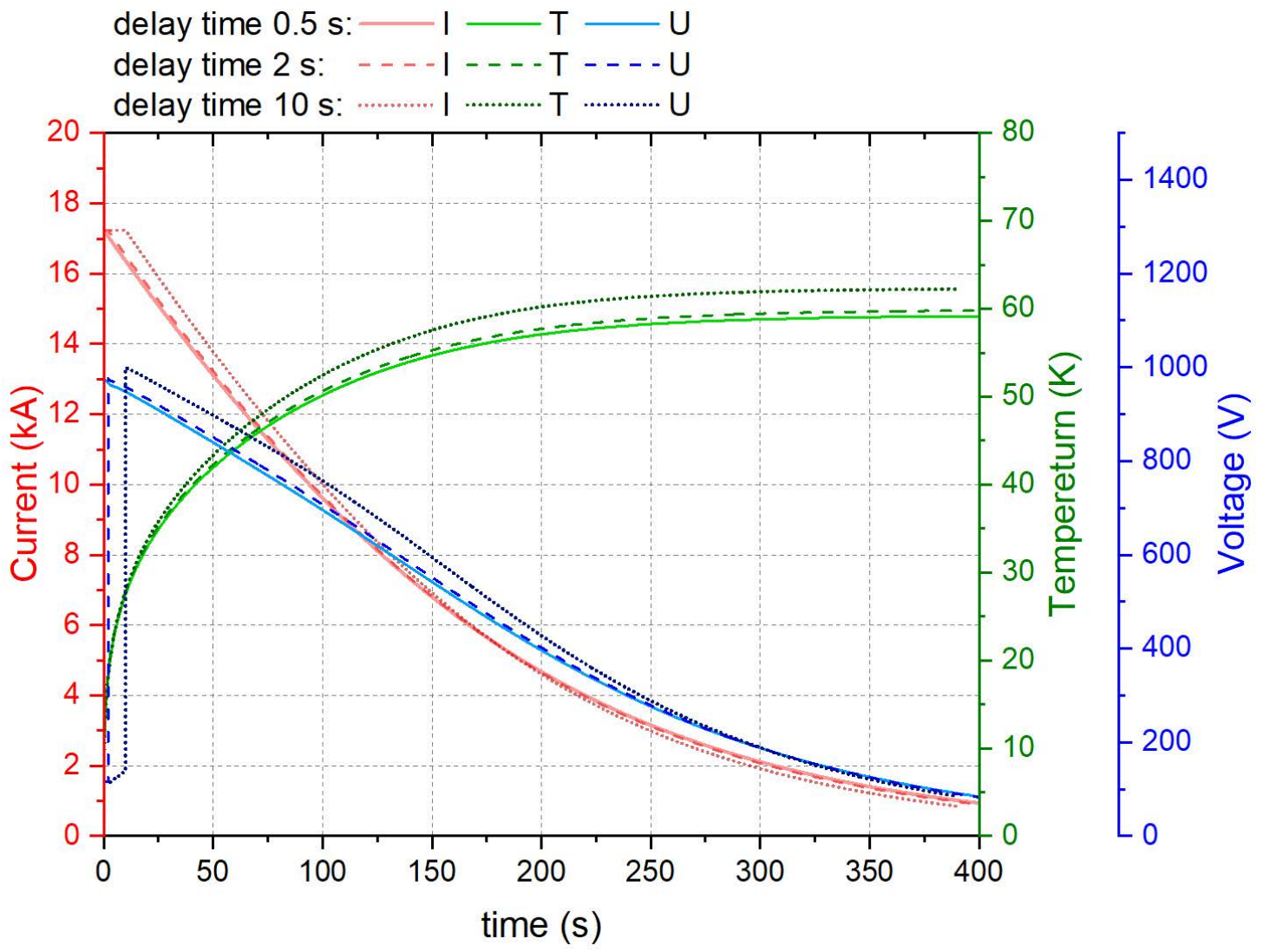

**Figure 10.25:** Coil current decay, temperature rise and voltage variation based on a 50 $m\Omega$ dump resistor and an RRR of 700 at different delay times.

### Quench characteristic

With a stabilizer RRR of 700 and the corresponding thermal conductivity and specific heat of the cable materials, the quench propagation velocities are determined to be 2.514 m/s along the cable, 0.089 m/s axially, and 0.139 m/s radially.

Assuming a thermal disturbance is applied to the innermost two-turn cable of the coil,

**Table 10.11:** Quench parameters for different detection delay times.

| Parameter | Delay time | | |
|---|---|---|---|
| | 0.05 s | 2 s | 10 s |
| Average coil final temperature | 59.2 K | 59.8 K | 62.25 K |
| Effective time constant* | 159.55 s | 159.9 s | 161.2 s |
| Extracted energy ratio | 74.23 % | 73.66 % | 70.68 % |
| Magnet final resistance | 0.040 Ω | 0.040 Ω | 0.049 Ω |
| Terminal voltage | 976 V | 976 V | 999 V |

**time when the current is* $I = I_0/e = 6340\,A$.

the Minimum Quench Energy (MQE) is calculated to be approximately 23.68 J, as shown in Figure 10.26.

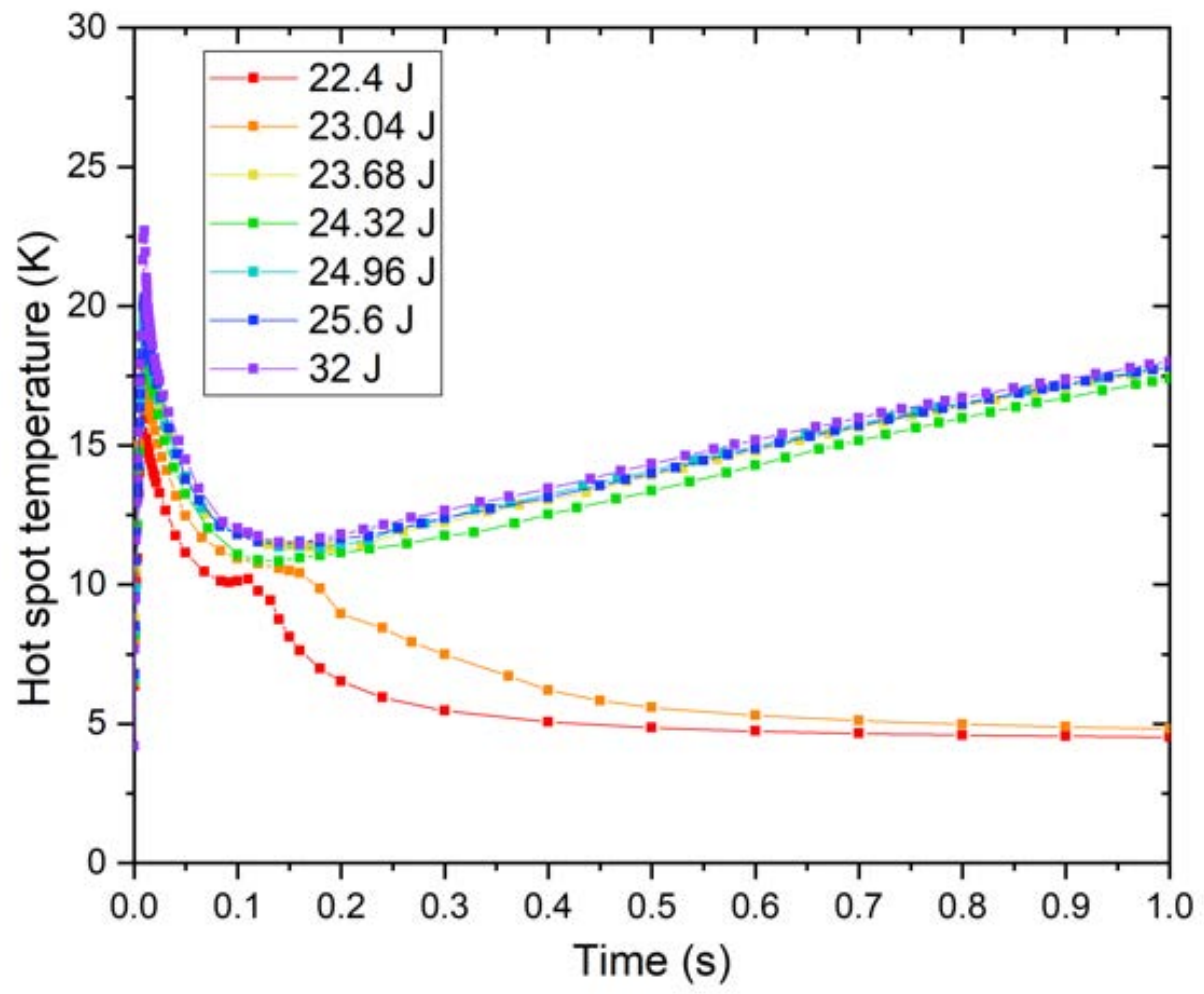

**Figure 10.26:** Coil hot spot temperature with different thermal disturbance energy. The MQE of the coil is about 23.68 J

The quench and quench-back behavior under a localized thermal disturbance are also simulated. The quench is initiated at the innermost turn at the coil center. The resulting hot spot temperature reaches 124.6 K, with a maximum temperature difference across the coil of about 29.9 K, as illustrated in Figure 10.27. This temperature gradient is not expected to cause significant localized thermal stress.

Due to the dump resistor and the quench-back effect from the support cylinder, the coil current decays more rapidly. The entire coil is fully quenched within 10 s, and the hot spot temperature stabilizes within an acceptable range after approximately 115 s.

The worst condition is also simulated, the quench protection is failed and all the stored energy releases on the coil. Figure 10.28 shows the coil current decay, temperature rise and voltage variation without and with the quench protection. Table 10.12 shows the temperature, voltage parameters of all energy, 50% energy and 30% energy release on the coil. The stored energy density of CEPC is 10.6 $J/g$. The magnet can also survive at the worst condition.

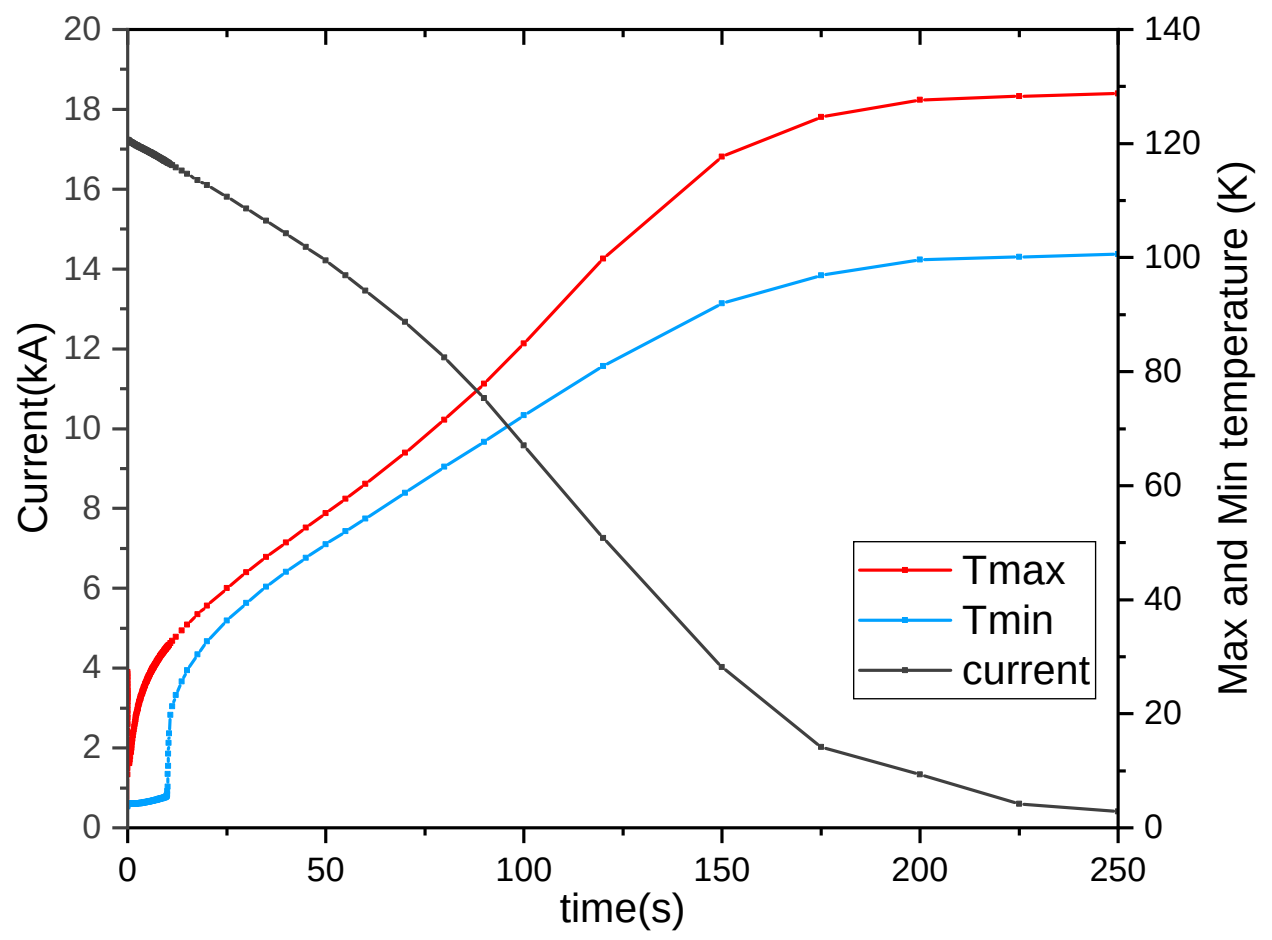

**Figure 10.27:** Maximum and minimum temperature distributions of the coil under the MQE 23.68 J thermal perturbation energy. The quench is initiated at the innermost turn at the coil center. The hot spot temperature reaches 124.6 K.

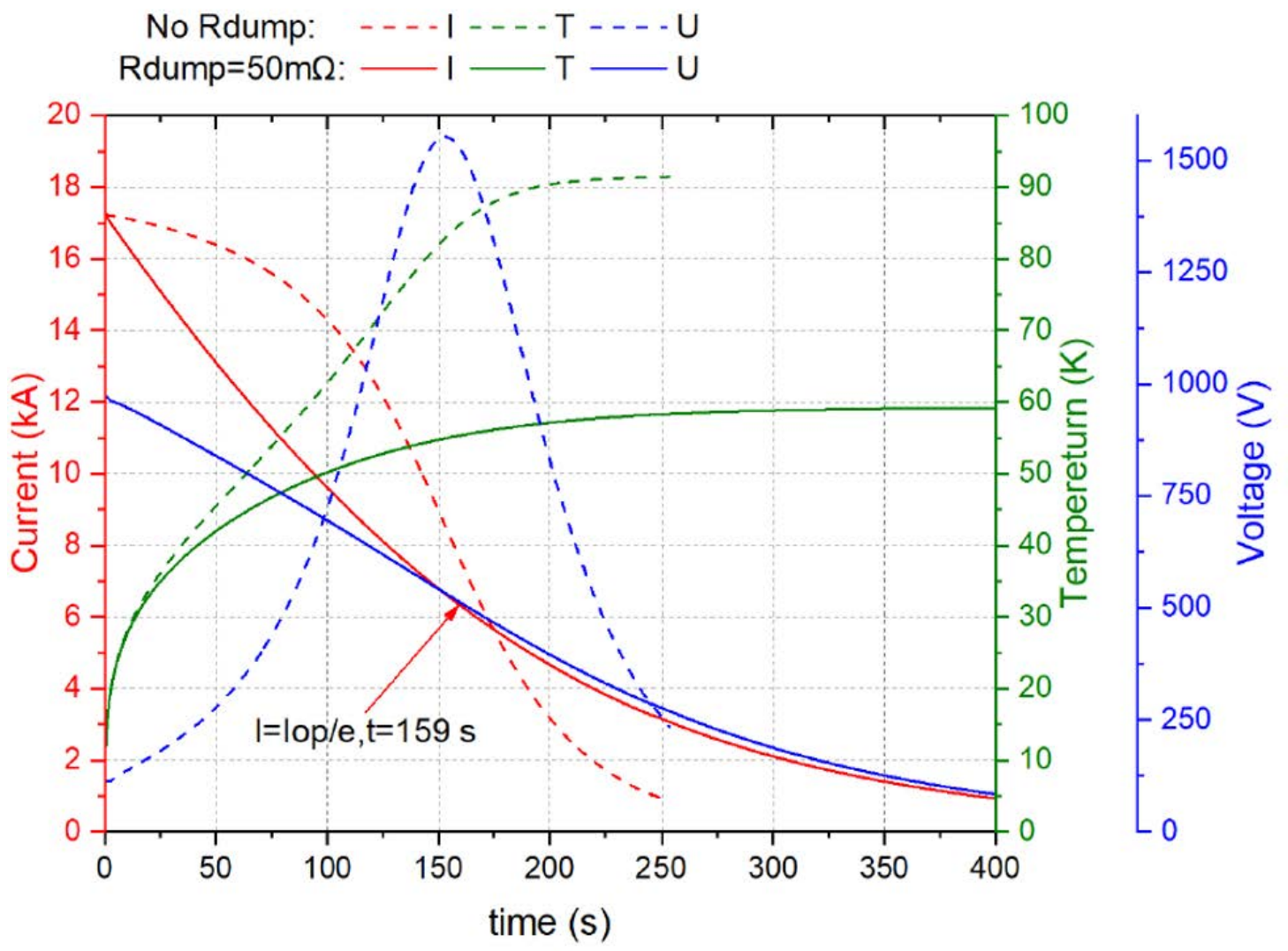

**Figure 10.28:** Coil current decay, temperature rise and voltage variation without the dump resistor and with the dump resistor.

**Table 10.12:** Quench parameters for different dump resistances.

| **Parameter** | **0** | **30 mΩ** | **50 mΩ** |
|---|---|---|---|
| Average coil final temperature | 91.42 K | 70.1 K | 59.22 K |
| Effective time constant | 169 s | 186 s | 159 s |
| Extracted energy ratio | 0 % | 54.7 % | 74.22 % |
| Terminal voltage | 1552 V | 666 V | 976 V |

## 10.5 Cryogenic system and cryostat

The cryogenic system is responsible for cooling the superconducting coil, ensuring it operates within its superconducting temperature range and minimizing the risk of quench.

It is a 4 K liquid helium system, divided into four main parts: (1) the coil cryostat inside the magnet vacuum vessel, (2) the compressor and refrigerator plant (including the liquid helium Dewar), (3) the cryogenic transfer and distribution system (valve box and piping), and (4) the helium recovery and purification system.

The core is the coil cryostat, which contains: a 145-ton superconducting coil indirect cooled by a 4 K liquid helium system, a 13-ton thermal shield cooled by 40 K helium gas, and a room temperature vacuum vessel. Two chimneys penetrate the yoke to house current leads and cryogenic lines, and a phase separator above the magnet enables stable thermosiphon circulation.

The cryogenic scheme for the detector magnet completed the overall thermal load estimation and process design, as well as the layout design of the main equipment. Small-scale experimental studies were conducted on certain key technologies, such as the integration welding technology for large-size magnets and cooling channel, and the efficient cooling structure for current leads.

### 10.5.1 Heat loads calculation

The heat loads on the superconducting magnet cryogenic system consists of static and dynamic components. Radiation dominates the static load and can be greatly reduced by Multilayer Insulation (MLI). However, because MLI layer adds thickness, the number of layers must be optimized to balance insulation performance with spatial constraints. After evaluating these requirements, 10 MLI layers are installed between the coil and the cold shield, and 40 layers are installed between the cold shield and the 300 K Dewar. The resulting radiation heat loads are listed in Table 10.13.

**Table 10.13:** Radiation heat loads and MLI parameters. Different layers of MLI relates to different radiation heat flux.

| | Surface area ($m^2$) | Number of layers | Heat flux ($W/m^2$) | Heat load (W) |
|---|---|---|---|---|
| 4 K | 400 | 10 | 0.082 | 32.6 |
| 60 K | 424 | 40 | 0.886 | 373 |

#### 4 K heat load

Saturated liquid helium for the superconducting coil is supplied from the phase separator. Under the design operating pressure of about 1.2 bar, the helium saturated temperature is 4.4 K. All heat loads are evaluated at this temperature; the resulting steady state heat input to the coil at the nominal current of 17 kA is summarized in Table 10.14.

**Table 10.14:** Steady state heat input at nominal coil current

| Component | Heat Load (W) |
|---|---|
| Radiation heat load | 32 |
| Support heat load | 7 |
| Phase separator / Pipes / Valves | 30 |
| Conductor junctions | 3.76 |
| Current leads | 100 |
| **Total static heat loads at nominal current** | **173** |

During magnetic field ramps, eddy currents induced in the coil introduce an additional dynamic heat load. FEA gives a peak value of about 193 W for this component. The resulting total heat load at 4 K is listed in Table 10.15.

**Table 10.15:** Heat load of 4 K

| Component | Heat Load (W) |
|---|---|
| Static heat loads | 173 |
| Peak value of dynamic heat load | 193 |
| **Total** | **366** |

#### 60 K thermal shield heat load

The heat load on the thermal shield consists mainly of radiation and conduction through the support structures. This load is absorbed by the cold shield and is referred to as the 60 K heat load; the detailed values are given in Table 10.16.

**Table 10.16:** Heat load of 60 K thermal shield

| Component | Heat Load (W) |
|---|---|
| Radiation heat load | 373 |
| Support heat load | 54 |
| Phase separator / Pipes / Valves | 160 |
| Current leads | 2000 |
| Dynamic heat loads | 65 |
| **Total** | **2652** |

### 10.5.2 Design of coil cooling system

Figure 10.29 shows the coil cryogenic system, which comprises two largely independent circuits: a saturated liquid helium/helium gas loop for the 4 K region and a cold shield cooling loop. The 4 K loop operates as a thermosiphon. Cooling channels welded around the coil are connected to a phase separator mounted above the magnet, forming a closed circulation path; the welded cable joints ensure efficient heat transfer. A fraction of the saturated liquid helium is diverted to cool the current leads. Both loops are rated for a maximum pressure of 18 bar to accommodate possible quench events.

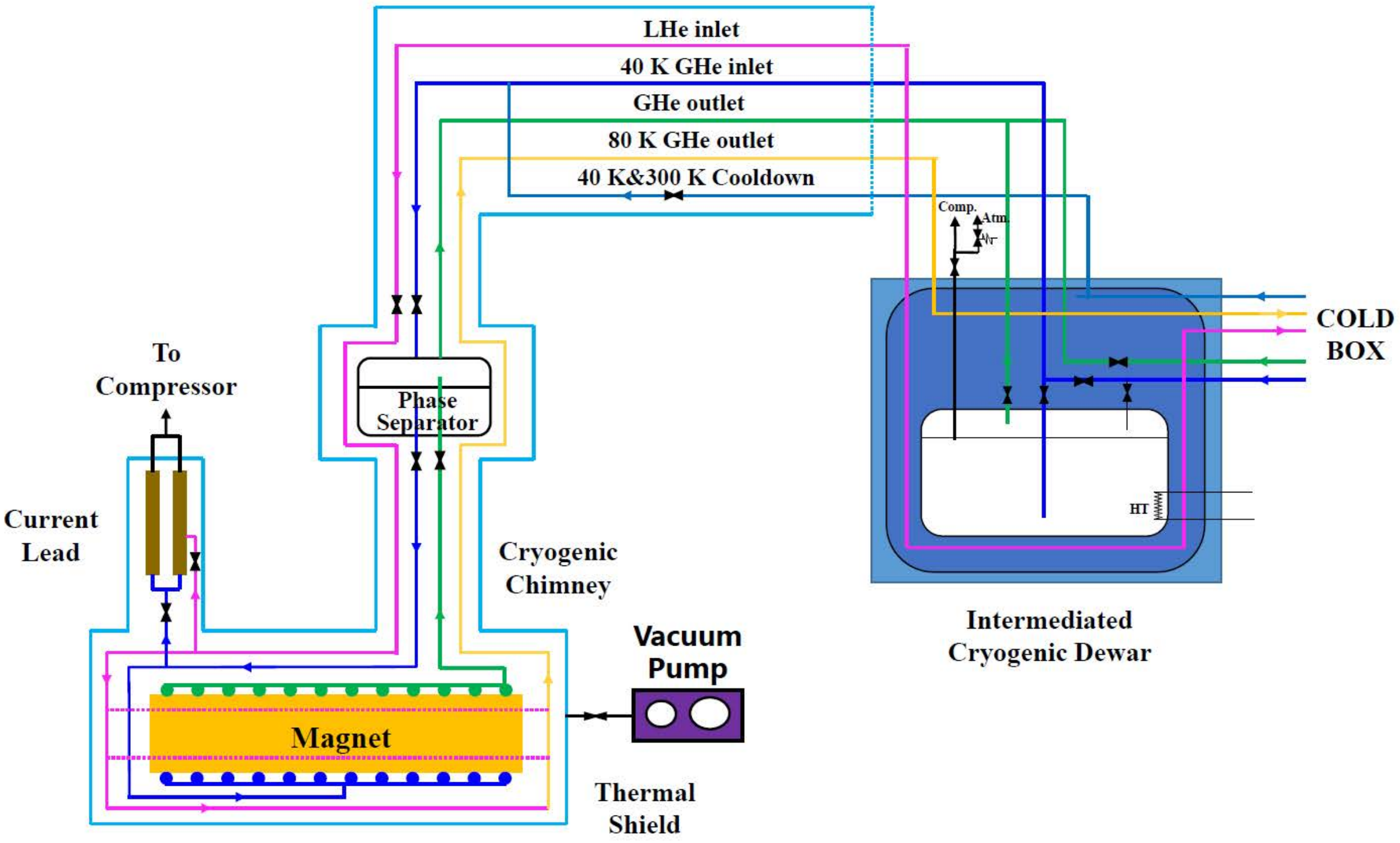

**Figure 10.29:** Coil cryogenic circuits. The cryogenic system features two independent circuits: a liquid-helium loop that cools the 4 K coil and current leads, and a separate line that maintains the cold shield at 40–80 K.

### Main circuit

The structural parameters and thermal hydraulic parameters of the main line are shown in Table 10.17. Two operating modes were considered: steady-state (normal) and current ramp up and down; the latter imposes the dynamic heat load on the coil.

**Table 10.17:** The structural parameters of the liquid helium main circuit.

| Parameter | Normal | Current ramp up and down |
|---|---|---|
| Average temperature | 4.2037 K | 4.4008 K |
| Max temperature | 4.2042 K | 4.45 K |
| Tube spacing | 250 mm | |
| Tube inner diameter | 15 mm | |
| Work pressure | 1.05 bar | |

The main cooling loop operates on the principle of thermosiphon, a method distinct from the helium bath approach and commonly used in low loss superconducting magnets. This design greatly reduces the size and complexity of the cryogenic system, but it also places stricter demands on flow and heat transfer. Any deficiencies in these areas could lead to uneven temperature distribution within the coil.

Thermosiphon cooling uses gravity and density differences to drive circulation. This passive, moving-part-free scheme was successfully validated in the CMS [8] and ALEPH [9] solenoids, offering high reliability and allowing the saturated liquid helium to self-adapt to local heat loads. The proposed cooling circuit consists of 33 parallel semicircular pipes arranged in a $3 \times 11$ configuration, each with an inner diameter of 15 mm, as shown in Figure 10.30.

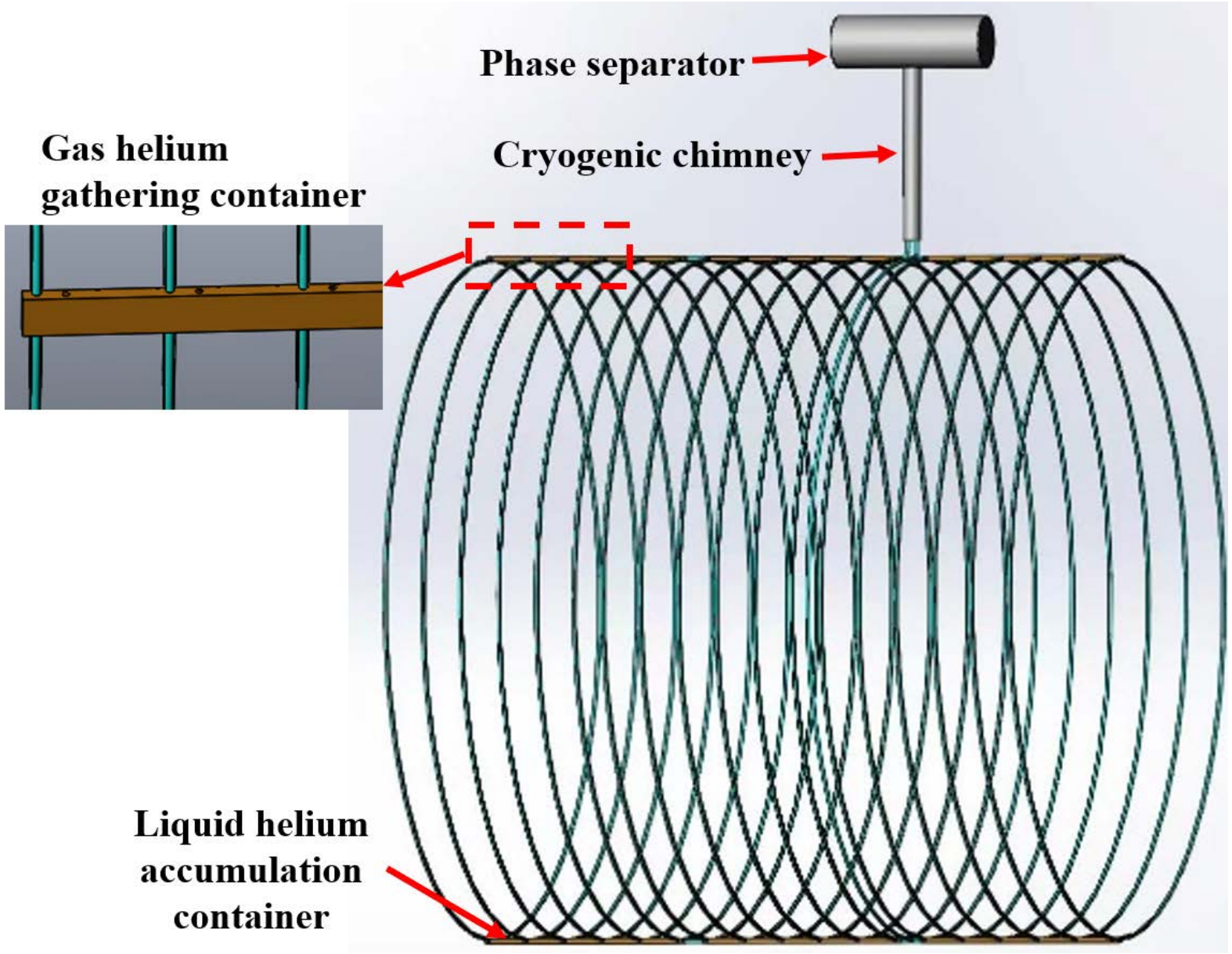

**Figure 10.30:** Layout of the thermosiphon cooling channel. It is planned to use 33 parallel semicircular pipes as the main cooling channel, with an inner diameter of 15 mm for each pipe.

### Thermal shield circuit

The cold shield is fabricated from two curved aluminum plates joined to form a cylindrical enclosure, with the superconducting coil positioned at its center. A serpentine cooling tube, with an inner diameter of 15 mm, is welded to the surface of the shield. Detailed specifications for the tube and the shield's thermal hydraulic parameters are provided in Table 10.18 and Table 10.19.

**Table 10.18:** Cold Shield Structural Parameters

| Parameter | Value |
|---|---|
| Length | 90 m $\times$ 2 |
| Inner diameter | 15 mm |
| Work pressure | 3 bar |
| Pressure drops | 0.27 bar |

**Table 10.19:** Cold Shield thermal hydraulic parameters

| **Parameter** | **Value** |
|---|---|
| Inlet temperature | 40 K |
| Outlet temperature | 65 K |
| Average temperature | 54 K |
| Flow rate | 8 g/s |

### Temperature distribution

To estimate the coil's temperature distribution under steady-state conditions, the boundary conditions are shown in Figure 10.31a. The coil experiences a radiative heat flux of 0.08 W/ $m^2$ and dynamic heat load 193 W. Conductive heat flows through structural components are as follows: 0.095 W per transverse tie rod, 0.35 W through radial tie rods, and 0.52 W via the main tie rod. The inlet supplies saturated liquid helium at 4.2 K with 0% vapor fraction. Under these conditions, the resulting temperature distribution across the coil is presented in Figure 10.31b, exhibiting a maximum temperature difference of 0.25 K.

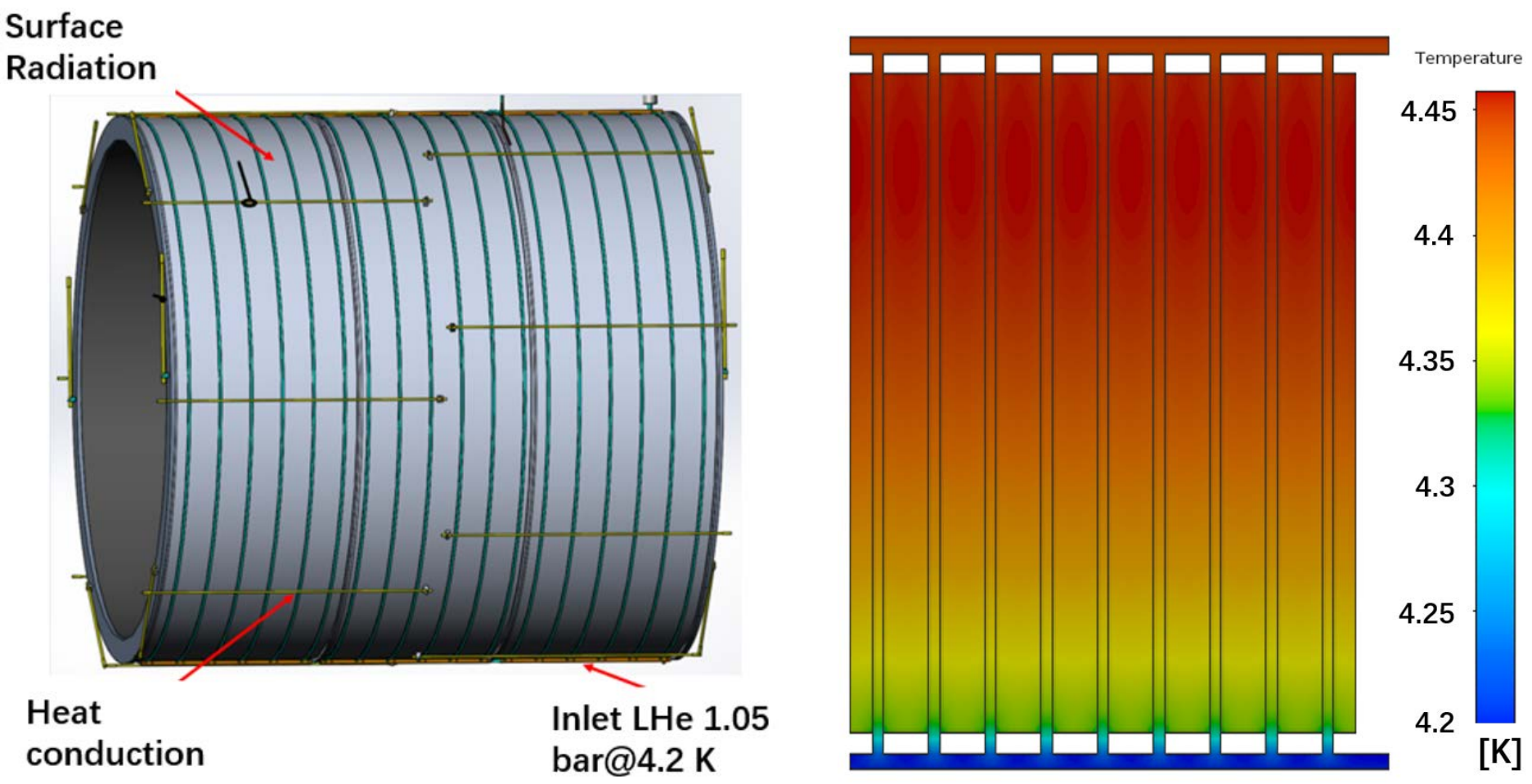

**(a)** **(b)**

**Figure 10.31:** Boundary Conditions and Temperature Contour of the coil. (a) The simulated boundary conditions, surface heat radiation 0.082 W/ $m^2$, dynamic heat load 193 W; (b) The temperature distribution of the liquid helium pipes.

A realistic welded structure to simulate the temperature distribution on the conductor was considered, assuming the worst-case scenario: the contact area between the liquid-helium pipe and the support cylinder consists only of the welds, which are single-sided and spaced apart, with each weld having a cross-section of 5 mm × 100 mm, as shown in Figure 10.32a. Figure 10.32b shows the temperature distribution of the coil in the current ramp up and down work condition. The maximum temperature on the conductor is less than 4.5 K, which ensures the magnet work safely.

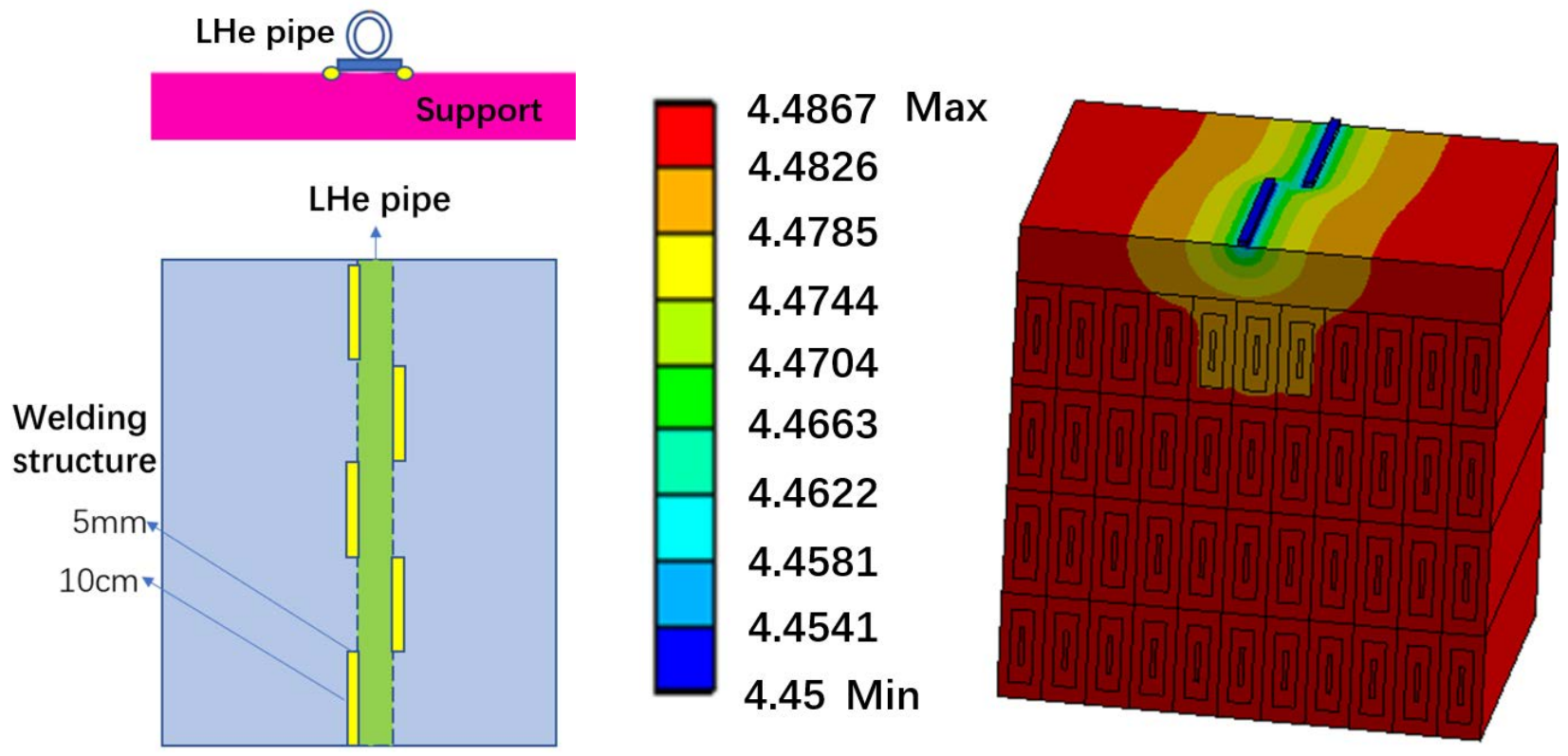

(a) (b)

**Figure 10.32:** (a) Welding structure of liquid helium pipe and the support; (b) The temperature distribution of the coil. The maximum temperature on the conductor is less than 4.5 K.

To estimate the temperature distribution of the thermal shield, the boundary conditions shown in Figure 10.33a are applied. The external radiative heat flux on the cold shield is set to 0.89 W/m$^2$. Under these conditions, the calculated fluid temperatures inside the thermal shield pipes are 40 K at the inlet and 65 K at the outlet. The resulting average temperature of the thermal shield is 54.26 K, as illustrated in Figure 10.33b.

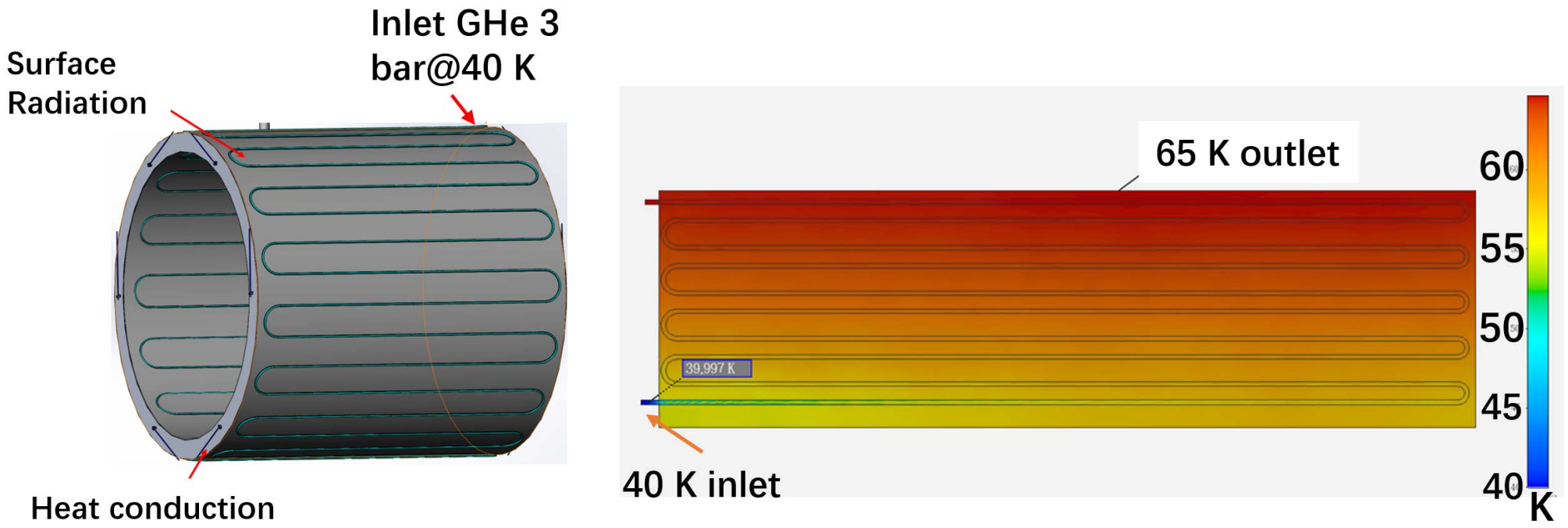

(a) (b)

**Figure 10.33:** Boundary Conditions and Temperature Contour of the thermal shield. (a) The simulated boundary conditions, surface heat radiation 0.89 W/m$^2$; (b) The temperature distribution of the coil. The calculated fluid temperatures inside the thermal shield pipes are 40 K at the inlet and 65 K at the outlet. The resulting average temperature of the thermal shield is 54.26 K.

### Cryogenic sequences

The operation of the superconducting magnet cryogenic system is divided into five stages: pre-cooling, normal operation, slow discharge during power failure, re-cooling after quench, and warm-up.

During pre-cooling, cold helium gas is simultaneously supplied to both the thermal shield and the main line. The maximum temperature difference between the coil and the thermal shield is continuously maintained below 50 K until both reach temperatures below 60 K. At this point, the main line switches to saturated liquid helium supplied from a 5000-L dewar to continue cooling and accumulation.

The total helium gas inlet pressure during cooling is approximately 10 bar, with a flow rate estimated at 150 g/s. It is expected to take about 3 weeks to cool 145 tons of cold mass.

### 10.5.3 Cryogenic plant

#### Compressor and Helium storage tank

The helium compressors for the detector superconducting magnet's cryogenic system primarily consist of two screw compressors, along with a standby compressor. All these compressors are located within the cryogenic hall. The total rated mass flow of the compressors is approximately 180 g/s, and the maximum outlet pressure is 18 bar.

In addition to the main compressors, a slightly smaller screw compressor is installed within the same building to serve as a recovery compressor. This unit is used to recover helium gas from the cryogenic system after the main compressors and refrigerators are shut down.

Clean helium gas is primarily stored in two 250 $m^3$/20 bar gas tanks, which are set up in an outdoor area. This storage capacity is slightly more than twice the amount of helium required for the normal operation of the entire superconducting magnet cryogenic system. Additionally, the helium storage tanks of the superconducting magnet cryogenic system can be connected to the helium storage tanks of the accelerator cryogenic system to provide an extra supply of helium.

#### Cold box, pipeline and 5000 L Dewar

Based on the estimated total heat load, a helium refrigerator with a rated power of 1.5 kW at 4.2 K is selected. The cold head of the refrigerator and the associated 5000 L liquid helium dewar are located above the detector magnet tunnel and are connected to the superconducting magnet via a local valve box (phase separator), minimizing heat leak along the cryogenic transfer lines. The cold head does not use liquid nitrogen pre-cooling, instead, it directly employs a three-stage helium turbine to construct a Clausius cycle for the production of liquid helium.

#### Cryogenic process design

The cryogenic plant comprises the compressor, refrigerator (cold box), liquid helium dewar, main distribution valve box, superconducting magnet distribution valve box, and coil cooling equipment. As illustrated in Figure 10.34.

The compressor raises the pressure of high-purity helium gas to 13 bar before sending it to the high-pressure side of the refrigerator. The high-pressure helium gas is then cooled to 5 K through a three-stage turbine and a seven-stage heat exchanger. After that, it passes through a J-T valve for throttling, resulting in liquid helium, which is stored in the liquid helium dewar.

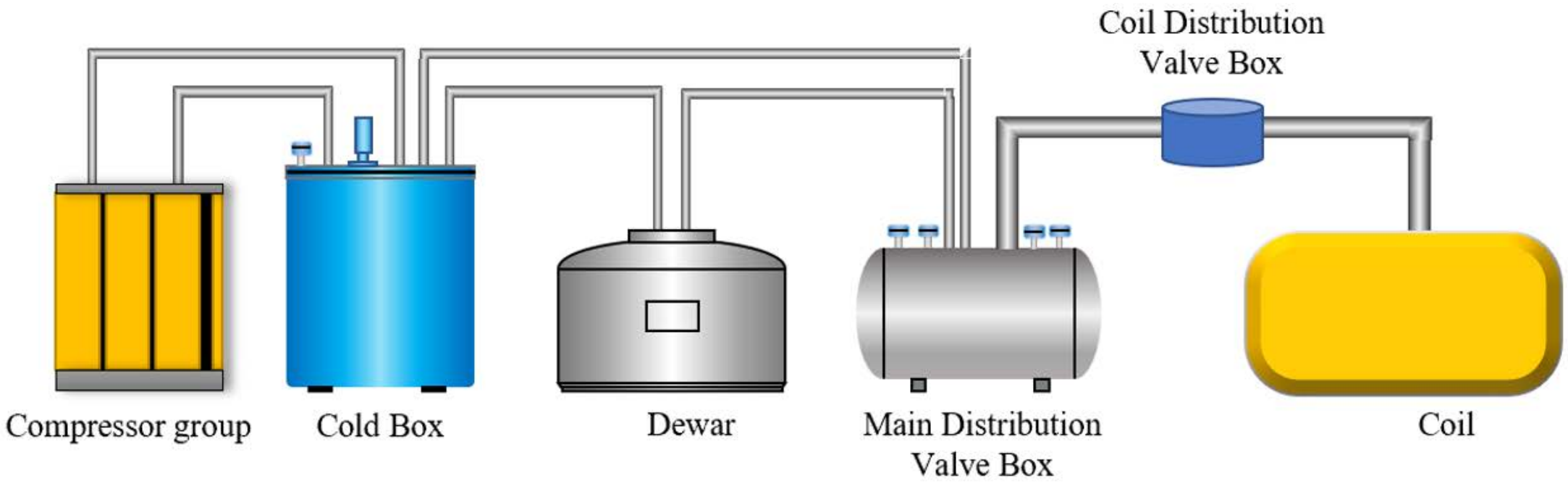

**Figure 10.34:** Cryogenic equipment and flow. The cryogenic plant comprises the compressor, refrigerator (cold box), liquid helium dewar, main distribution valve box, superconducting magnet distribution valve box, and coil cooling equipment.

The cold helium gas from the refrigerator and the liquid helium from the dewar are distributed to various usage points through the main distribution valve box. The portion allocated to the superconducting magnet cryogenic system is routed through the superconducting coil cryogenic distribution valve box and then sent to the superconducting coils, cold shield, and current leads. The corresponding return gas flows back to the superconducting coil cryogenic distribution valve box and is then delivered back to the refrigerator.

**Helium recovery and purification system**

It is essential to have a set of helium recovery and purification systems in place. These systems allow operators to clean the system before start-up, decontaminate it during shutdown periods, and decontaminate bulk helium gas supplies before entering the active system. Additionally, the system enables the recovery of helium gas during unplanned power outages and short maintenance periods, where a full system warm-up is not feasible, or when the total gaseous helium inventory of the system exceeds the available helium gas storage capacity.

A gas management panel for the recovery and purifier compressor system is used to regulate the recovery compressor suction and discharge independently of the primary gas management system of the cold box compressor system. It also allows for the purification of individual gas helium storage tanks and bulk helium gas supplies. The recovery compressor system is interconnected with the cold box primary high-pressure supply and low-pressure return. When combined with the primary gas management system, the entire system can be scrubbed and purified.

## 10.5.4 Cryostat

The detector magnet's cryostat comprises a vacuum dewar, a thermal shield, and cold-mass supports. As shown in Figure 10.35, the outermost vacuum dewar is fabricated from 304 stainless steel and withstands an external atmospheric pressure of 1 bar in addition to the weight of internal components such as the coils and thermal shields. The coil is suspended inside the dewar by titanium-alloy rods. To ensure stable operation, heat leakage must be minimized. A thermal shield made of 5083 aluminum alloy surrounds the superconducting coil within the vacuum dewar.

Two chimneys are located on the top of the dewar: i) One houses the cryogenic lines, maintaining the solenoid below 5 K during operation. ii) Another chimney accommodates the current leads and serves as the main pumping and venting line.

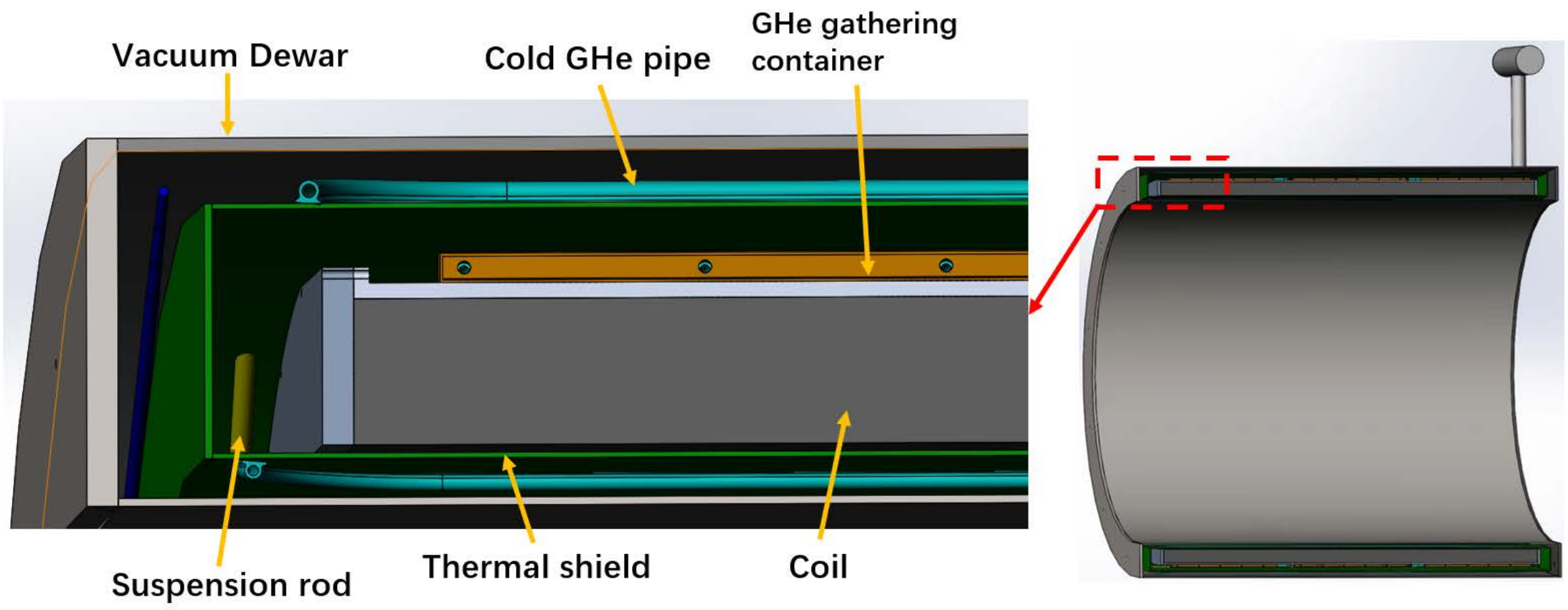

**Figure 10.35:** Structure of the detector magnet. The magnet comprises components such as a vacuum dewar, thermal shield, coil, liquid helium and cold gas helium pipes and suspension rods.

### Vacuum dewar

The vacuum dewar comprises two concentric cylindrical shells joined by two end flanges via structural welds. It must withstand large electromagnetic forces arising from cylindrical imperfections and misalignment between the solenoid and the return yoke during magnet excitation, as well as the pressure differential across the vessel wall after evacuation and the weight of the magnet's cold mass. Structural calculations include accelerations of 1.5 g in the axial (x and z) directions and 2.5 g in the vertical (y) direction during magnet transport. Shell thicknesses are 15 mm for the inner cylinder, 25 mm for the outer cylinder, and 60 mm for the end flanges.

### Thermal shield

Fabricated from 5083 aluminum alloy, the thermal shield also consists of two concentric cylindrical shells connected by two end flanges and integrated with associated pipework.

### Coil suspension system

The coil suspension system ensures precise and stable support of the cold mass within the vacuum vessel. It must withstand the cold mass's self-weight and magnetic forces arising from coil misalignment relative to the yoke. The suspension system comprises titanium alloy rods, with a total of 28 radial and axial rods. The rod parameters are listed in Table 10.20.

**Table 10.20:** Parameters of the titanium alloy rods. There are three types of rods used in the coil suspension system. Different rods have different functions.

| **Rod Type** | **Diameter** ( mm) | **Length** ( mm) | **Screw** |
|---|---|---|---|
| Axial rods | 30 | 3530 | M36×3 |
| Radial rods | 40 | 1700 | M48×4 |
| Main pull rods | 48 | 1650 | M56×4 |

## 10.6 Alternative solutions

This section presents two alternative conductor designs, high strength aluminum stabilized superconductor and welded-structure superconductor. High temperature superconducting magnet option is also studied and make some progress.

### 10.6.1 Alternative solutions for superconducting conductors

**Development of high strength aluminum stabilizer superconductor**

Figure 10.36 shows a Rutherford cable co-extruded with an Al-0.025%Ni-0.025%Be stabilizer. This alloy is under R&D to achieve required mechanical/electrical properties. Quench simulations and coupled electromagnetic-mechanical analyses require stabilizer RRR larger than 350.

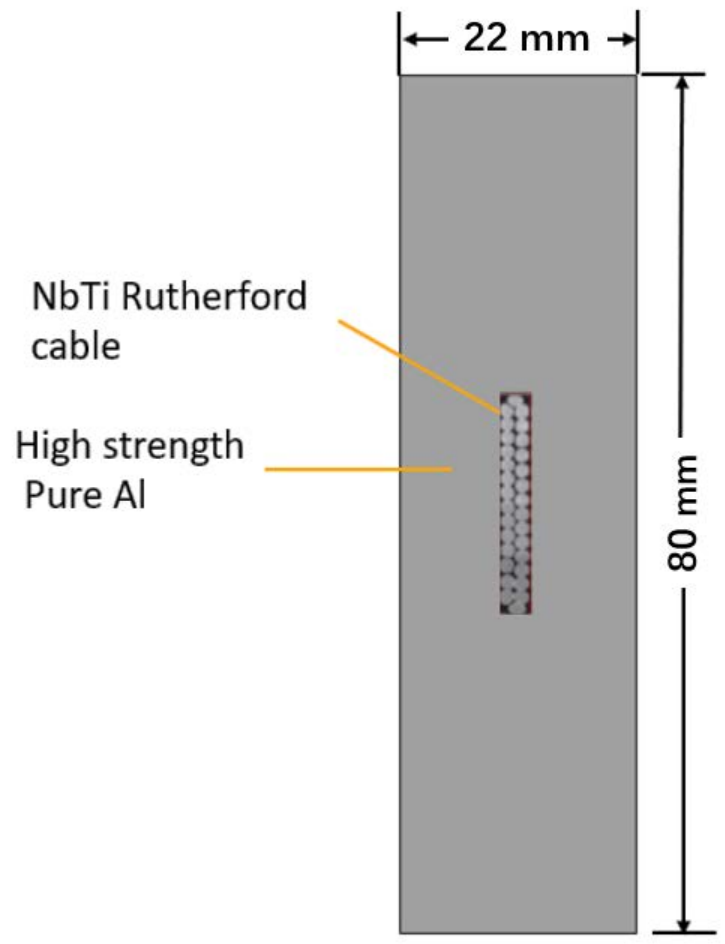

**Figure 10.36:** High strength and high RRR aluminum stabilized superconducting conductor. The center is NbTi Rutherford cable, surrounded by high strength aluminum stabilizer.

Doped aluminum stabilizer R&D conducted with Xinjiang Zhonghe Co. selected eight intermediate alloys with minimal resistivity impact. Ni, Be, and Er were prioritized for doping-ratio tests. RRR measurements used a GM-cryocooler fixed-point method. Four-probe resistance measurements at 293 K and 4.2 K determined RRR and yield strength for various doping ratios (Figure 10.37).

Figure 10.38 shows an Al-Ni-Be stabilized superconducting cable[10], where co-extrusion embeds a 32-strand Rutherford cable into a 56 mm × 22 mm cross-section. Critical current measurements of the strands before and after extrusion show no significant degradation. Cold-working and performance tests continue.

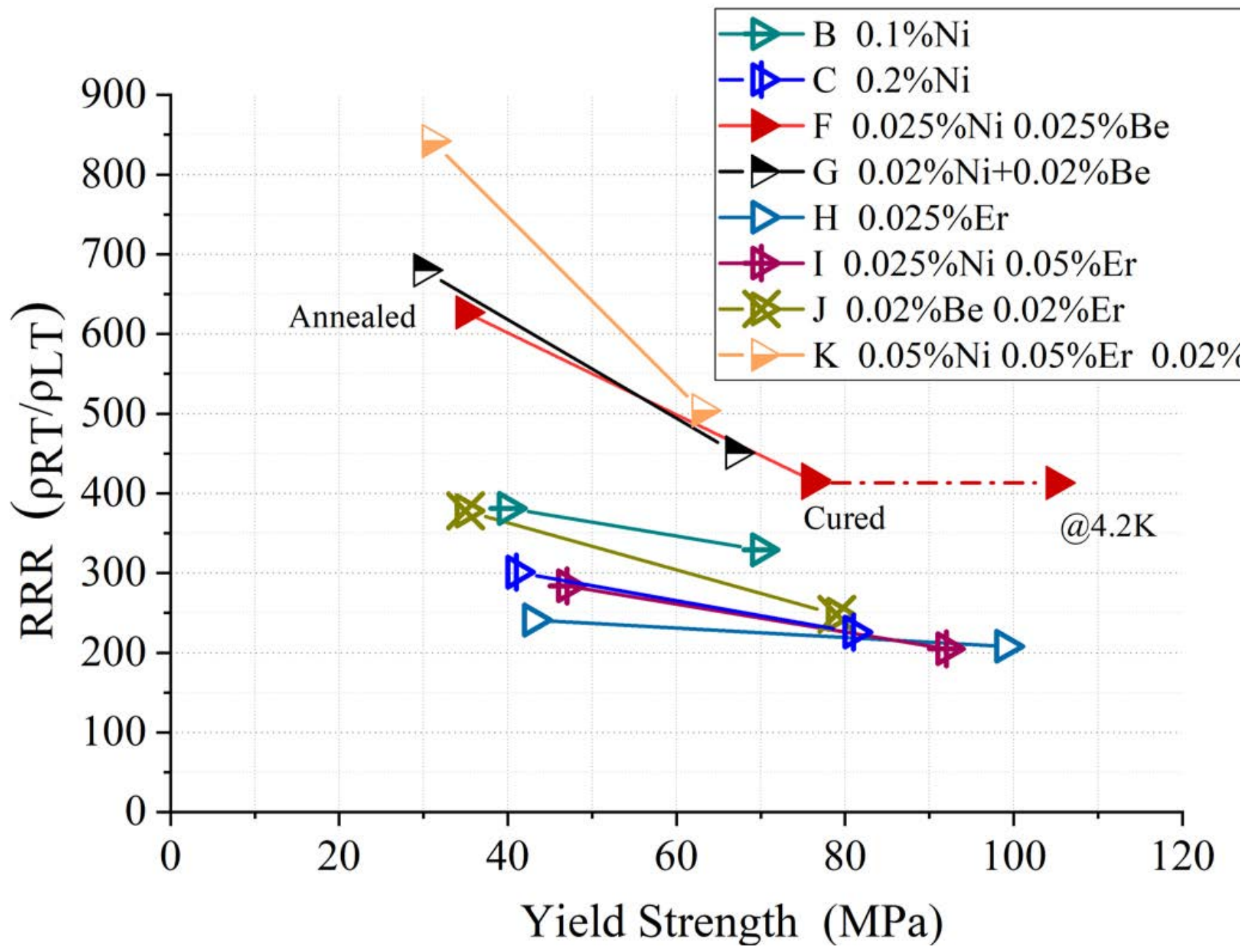

**Figure 10.37:** RRR values and yield strengths of eight different aluminum samples doped with combinations of Nickel, Beryllium, Erbium, Boron, and Cadmium.

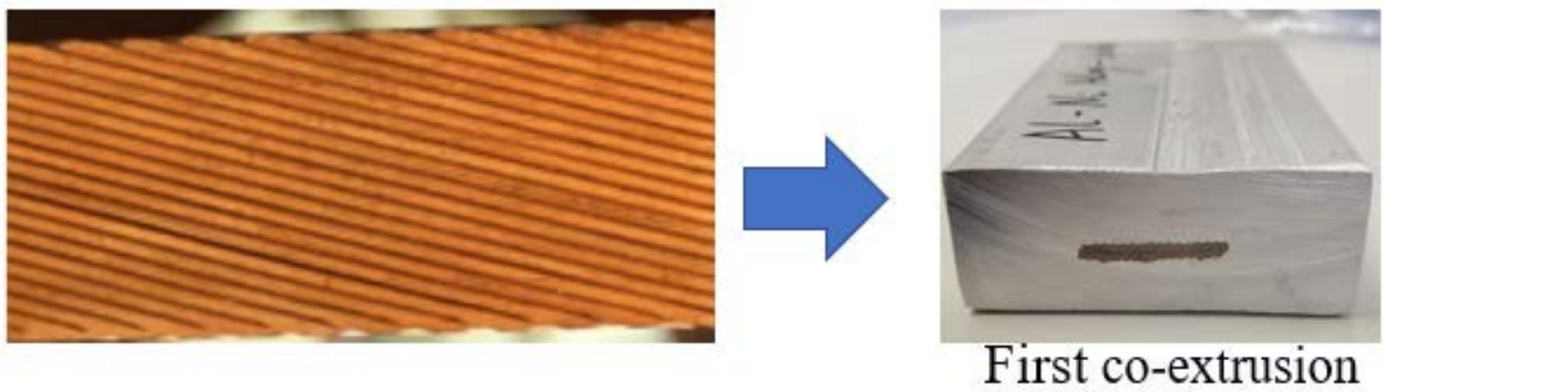

**Figure 10.38:** The Al-Ni-Be stabilizer conductor sample

### Welded-structure aluminum stabilized conductor

This cable design enhances mechanical strength by welding aluminum alloy onto both sides. Its structure, illustrated in Figure 10.39, closely resembles the conductor configuration used in the CMS experiment [11].

The welded structure is relatively straightforward to manufacture. The main challenge is the requirement for a large vacuum chamber for electron beam welding, which significantly increases costs. We have already partnered with manufacturers to produce short cable samples. If the development of the aforementioned two types of cables encounters difficulties, the plan for this type of cable will be activated.

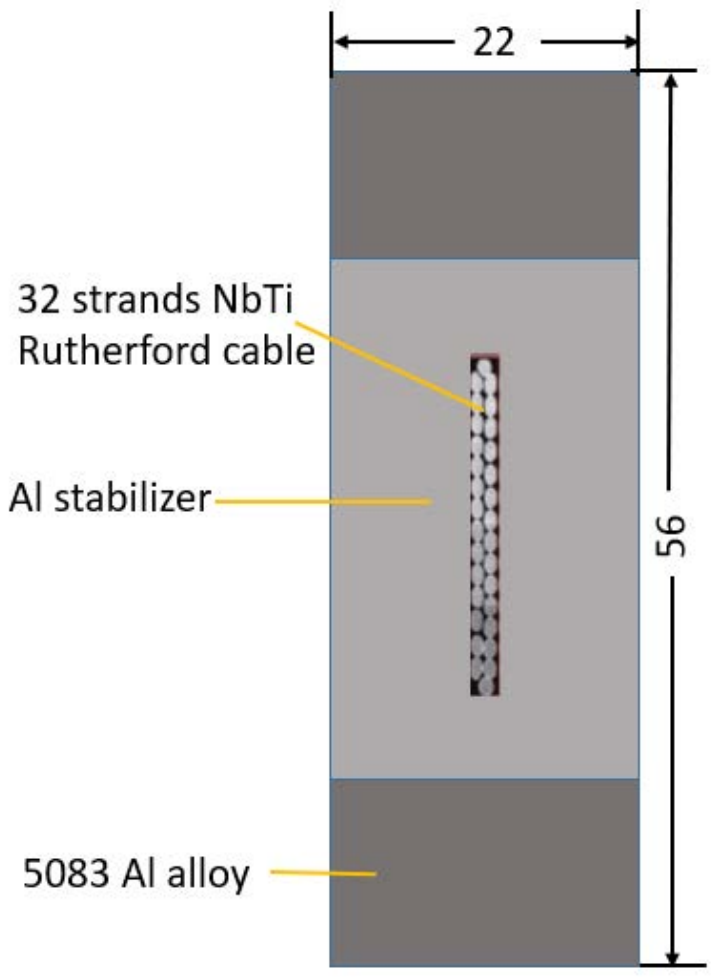

**Figure 10.39:** The welded-structure superconducting conductor (units is mm). Enhance the cable mechanical strength by welding aluminum alloy onto both sides.

### 10.6.2 HTS option for detector magnet

High Temperature Superconductor (HTS) represent the future direction for high-field or elevated-temperature applications, offering clear advantages over low-temperature superconductors. Consequently, we are investigating a HTS option for the CEPC detector magnet. In this scheme the coils locate inside the HCAL, imposing stringent demands on coil, support, and cryostat materials and requiring an ultra-thin, low-radiation-length magnet.

Due to rapid progress in second-generation HTS conductors, the HTS design offers unique benefits. REBCO tapes provide very high tensile strength (typically $>$ 400 MPa) and a thin, 50 μm Hastelloy substrate, while have a broad temperature margin. IHEP is developing an Aluminum stabilized Stacked REBCO Tape Conductor (ASTC) to advance this HTS solution.

**Table 10.21:** Size of the HTS Detector Magnet

| **Parameter** | **Value** |
|---|---|
| Central field (T) | 3 |
| Length of magnet ( mm) | 8000 |
| Outer diameter ( mm) | 4960 |
| Inner diameter ( mm) | 4660 |
| Iron yoke mass (tons) | 1143 |

### Physical design of the HTS magnet

The CEPC detector magnet system requires a 3 T central field. The schematic diagram of superconducting coil and cable structure refers to Figure 10.40. The operating current is 29 kA. The main parameters of the HTS magnet are given in Table 10.22.

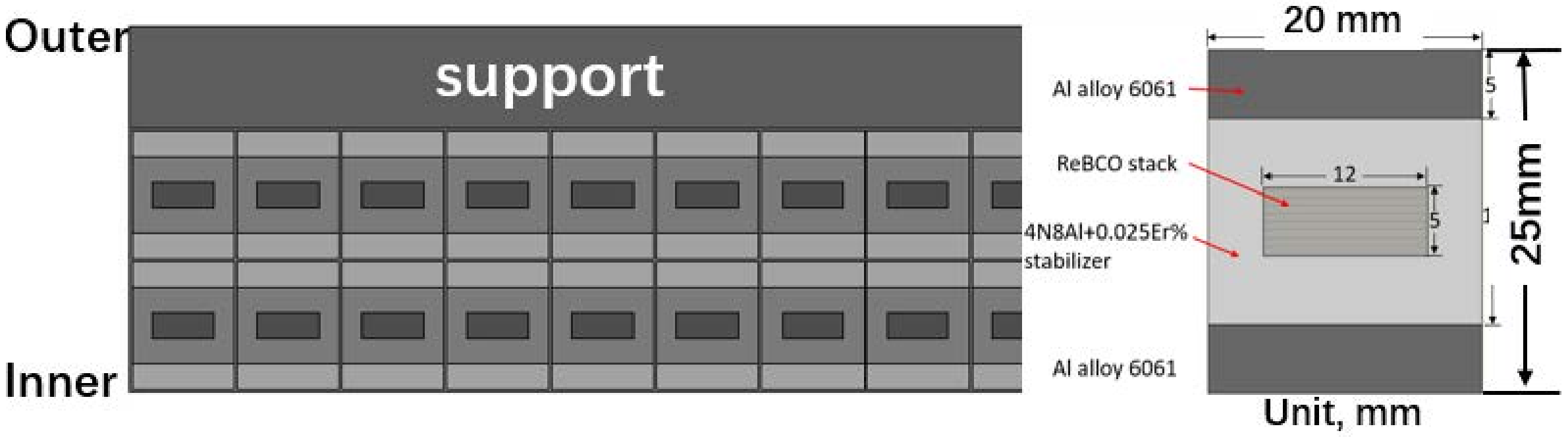

**Figure 10.40:** HTS coil and cable structure. The superconducting coil is designed with two layers. There are 300 turns each layer, and total 600 turns.

**Table 10.22:** HTS Magnet Parameters

| Parameter | Value |
|---|---|
| Current | 28000 A |
| Inductance | 1.27 H |
| Stored energy | 500 MJ |
| HTS cable length | 10.7 km |
| ASTC mass | 16.6 tons |
| Total weight | 48 tons |

The maximum perpendicular magnetic field on the cable is 2 T. Multiple REBCO tapes are stacked to improve the overall current of the cable. The aluminum stabilizer is needed to protect an HTS cable during a quench by temporarily carrying the current when HTS loses its superconductivity, as well as carrying part of the Lorentz forces. The traditional choice of cable stabilizers is high-purity aluminum or doped aluminum to increase the mechanical strength.

### Developing progress of ASTC

A novel aluminum stabilized HTS cable is proposed in this scheme. The conceptual design of ASTC can be seen in Figure 10.41. There are mainly two steps in the design: first, multilayer REBCO tapes are stacked together to form a stacked cable. Then, the stacked cable is co-extruded with pure aluminum or high strength and high RRR value aluminum alloy. The ASTC prototype cable is under development and have achieved some progress [12]. Figure 10.42 shows the ASTC cable processing process and the cable structure.

Significant progress was made in the development of small size ASTC cable. The cross section is $15\times10$ mm$^2$, tape width: 4 mm, tape thickness: 80 µm. 14-layer REBCO tapes ASTC cable. By optimizing the process parameters, such as preheating temperature, extrusion speed, Al outer layer thickness, and improving the internal structure of the cable, the critical current of the ASTC prototype cable is increased to 905 A at 77 K, self-field. The Ic degradation of the cable is less than 20%. Two 100-meter-long cables were produced for subsequent research.

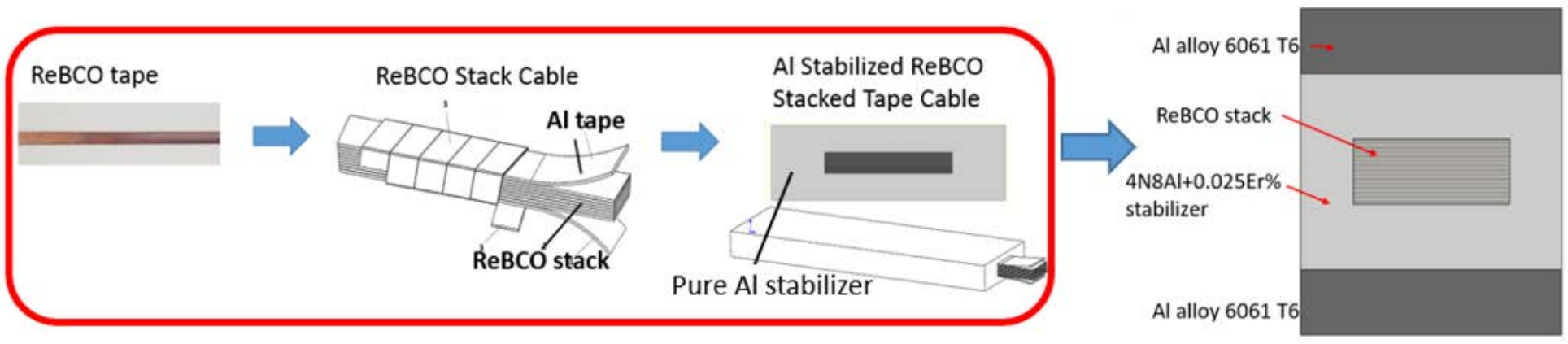

**Figure 10.41:** Conceptual design of ASTC.

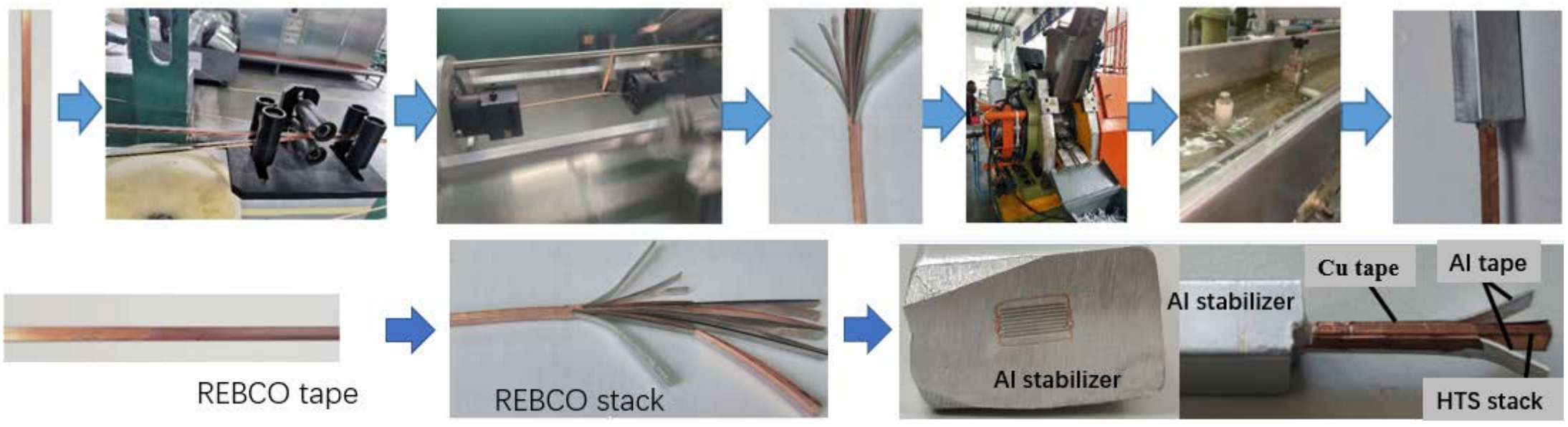

**Figure 10.42:** The ASTC cable processing process.

### Ongoing work and plan

Large size ASTC cable development. Previous ASTC cables were based on 4 mm wide tapes, and future cable development is using 12 mm wide tapes to achieve a higher current density.

A 100-meter-long ASTC is used to fabricate small-sized HTS experimental coils. The coil diameter is approximately 600 mm based on the research of the bending radius of the cable [13]. The following process researches are carried out : (1) Winding process: Study the Non-insulated winding method for cable winding coils; (2) Quench detecting methods. Comparison of different quench detection methods such as voltage signal, temperature signal, and optical fiber detection.

## 10.7 Summary and future plan

The detector magnet is a large solenoid. It uses the LTS scheme. It is located outside the HCAL, surrounded by Muon detector and yoke. Its inner diameter is 7070 mm, with a thickness of 700 mm and a length of 9050 mm The central magnetic field strength is 3 T. the team have conducted the physical design, the cryogenic system design, the mechanical system design, and have also carried out the development of box-type superconducting cables.

**R&D in the future**

The production of detector superconducting magnets relies on key enabling technologies: aluminum-stabilized cables, coil winding platforms, cryogenic systems, mechanical supports, and electrical controls. Preliminary R&D on these subsystems is essential.

**Cable Development** Full-scale 100 m Al-stabilized NbTi/Cu Rutherford cables will be completed by the end of 2026. Small-scale cables and Al-Cu prototypes are already fabricated. Key tasks include:

- Enhancing mechanical strength via cold-working, alloy reinforce welding, and high-strength/high-RRR Al development
- Improving cable-stabilizer bonding through removing surface impurities and oxides, and preheating
- Tests of the cable, including Ic and mechanical tests, tension of the cable, stretching and compression both transverse and longitudinal

**Coil Winding Process** A four-layer model coil with an inner diameter of 7.2 m was wound on an industrial helical platform, achieving ±1 mm accuracy after leak testing and epoxy impregnation.

**Cryogenic System** Thermosiphon cooling is implemented leveraging proven expertise in low-temperature simulation, assembly, and system integration to ensure reliable cryogenic performance. A key technical focus involves achieving robust leak-tightness for the extensive aluminum tube welding joints. Dedicated process development is underway to establish optimized welding techniques for the multi-channel tubing system with integrated gas collection and liquid accumulation tanks, with an estimated research and validation period of approximately 12 months.

**Mechanical System** The 145-ton coil requires suspension with $\leq$ 1 mm concentricity. The Support rod design will undergo 12 months of R&D validation through fabrication and mechanical testing of titanium-alloy and carbon-fiber solutions.

**Scaled Model Coil** We have finalized a comprehensive model-coil plan: a 1.2 km, full-size superconductor will be wound into a four-layer, thirteen-turn-per-layer segment, seen in Figure 10.43. The program also covers fabrication, testing and validation of the winding platform, impregnation procedure, cable performance and cryogenic circuit. Building on our earlier joint work, the activity will be executed in close collaboration with industrial partners.

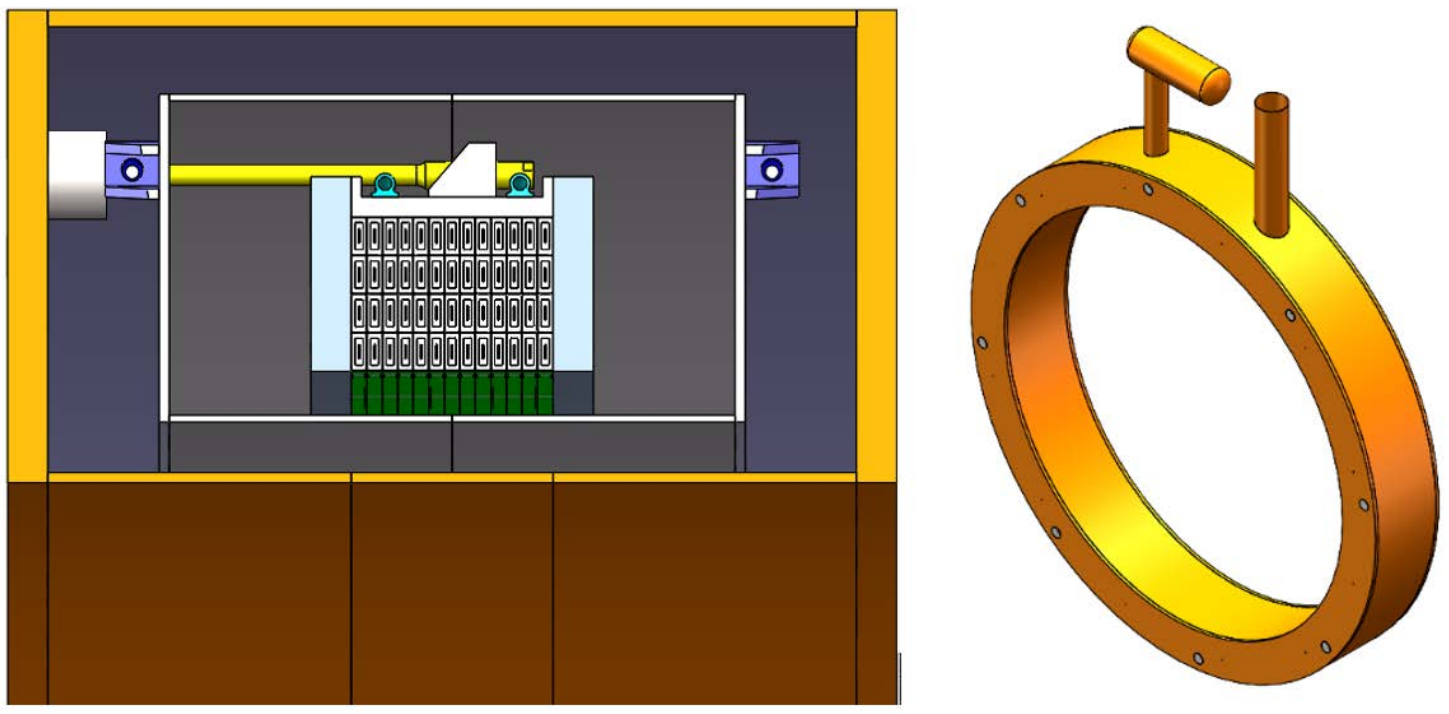

**Figure 10.43:** Scaled model coil design

# Chapter 11 Readout Electronics

The readout electronic system involves signal amplification, analog-to-digital conversion, processing signals from millions of detector channels, and buffering these signals and interfacing with the trigger system. Additionally, the system ensures precise synchronization of the clocks across all sub-detectors with the accelerator's collision bunch crossing and control of all electronics components.

This chapter focuses mainly on the common readout electronics design, including the global readout strategy, common interfaces such as data transmission, power distribution, and clocking. However, the detailed designs of the FEE of each sub-detector are covered in the corresponding detector chapters (Chapter 4–9).

The chapter is organized as follows. Section 11.1 introduces the requirements for each sub-detector's readout electronics. Section 11.2 analyses the overall requirements and proposes a unified baseline solution and framework structure for the readout electronics. The detailed designs of the key components are then discussed: Section 11.3 presents the design of the data interfaces, while Section 11.4 discusses wireless-transmission-based data readout as an alternative. Sections 11.5, 11.6, and 11.7 cover detector clock synchronization, front-end power distribution, and the BEE and its interface with the trigger system, respectively. Section 11.8 summarizes the total requirements for fibres, cables, crates, racks, and counting room space. Finally, Sections 11.9 and 11.10 describe the design team and the long-term development schedule of the electronic system.

## 11.1 Detector readout requirements

For the readout electronic system, two things are crucial: first, to meet the readout requirements of all sub-detectors, and second, to create a comprehensive common electronics framework for the overall spectrometer design. This approach ensures that all sub-detector readout electronics are designed with common interfaces and standards, which promotes modularization, enhances design independence, and enables plug-and-play capability and upgradeability. At the system level, this strategy also helps with more efficient scheduling and communication among the various systems. In the sub-detector chapters, preliminary readout electronics schemes are discussed. Related parameters, which are essential for designing the common electronics framework, are summarized in Table 11.1.

Table 11.1 first summarizes the chip channels and data width for each sub-detector and the module size, on which a data link is transmitted the front-end data to the BEE. Module data rates for Higgs and Low Luminosity $Z$ mode are calculated in the following rows. The values take into account the MDI beam-induced background estimation (see in Chapter 3) and module size for each data link. Now the beam-induced background rate and shielding for the High

**Table 11.1:** Critical inputs of the readout electronic system. Beam-induced background rates are estimated by the MDI study (in Chapter 3), then the corresponding data rate for each module is calculated. For each sub-detector, only the endcap modules are summarized in this table, as they usually has higher requirement on data readout than barrel. In the table, ”stch. chip” is the short form of ”Stitched chip”, ”GS” for the ”Glass Scintillator”, ”Agg Brd” for the ”Aggregation Board”, ”LowZ” for the ”Low Luminosity $Z$ ”. For the VTX, the background rate of the innermost layer is taken to estimate the highest module data rate. For ECAL and HCAL, the unit of the background rate is MHz per bar/glass.

| | **VTX** | **ITK** | **OTK** | **TPC** | **ECAL** | **HCAL** | **Muon** |
|---|---|---|---|---|---|---|---|
| Chn/ chip | $512 \times 1024$ | $512 \times 128$ | 128 | 128 | 4–16 (common SiPM ASIC) | | |
| Data Width | 32 bit / hit | 42 bit / hit | 48 bit / hit | 48 bit / hit | 48 bit / hit | | |
| Cluster size | 3 pixel | 1.5 pixel | 2 strip | N/A | N/A | | |
| Module Size/ Link | 32.7 cm$^2$ / stch chip | 423.3 cm$^2$ / stave | 365.7 cm$^2$ / stave | 461 cm$^2$ / module | 856 chn w/ 1.5 cm bar | 600 GS / Agg Brd | Agg Brd |
| Beam-induced background rate (MHz/cm$^2$) | | | | | | | |
| Higgs | 6.4 | $3.0 \times 10^{-3}$ | $1.9 \times 10^{-3}$ | $2.4 \times 10^{-3}$ | $6.2 \times 10^{-2}$ | $2.4 \times 10^{-4}$ | $1.4 \times 10^{-6}$ |
| LowZ | 15.0 | $5.4 \times 10^{-3}$ | $3.1 \times 10^{-3}$ | $5.2 \times 10^{-3}$ | $1.0 \times 10^{-1}$ | $2.4 \times 10^{-4}$ | $9.2 \times 10^{-7}$ |
| Data rate per Link (Mbps) | | | | | | | |
| Higgs | $2.00 \times 10^4$ | 80.0 | 66.7 | 53.1 | $2.55 \times 10^3$ | 6.91 | $< 10$ |
| LowZ | $4.71 \times 10^4$ | 144 | 109 | 115 | $4.11 \times 10^3$ | 6.91 | $< 10$ |

Luminosity $Z$ mode are still being optimized. It is expected the rates to be about five times higher than those in the Low Luminosity $Z$ mode.

## 11.2 Global architecture

Designing all FEE systems based on a common platform is an essential way to meet the challenges of the next-generation large collider experiment. This ensures uniformity in data and power interfaces and enables BEE to receive data, perform slow control, and configure FEE through a unified interface. This modular approach enables for scalability by increasing the number of generic modules based on the size of the sub-detectors. To start it is necessary to decide whether the electronics and TDAQ systems adopt a FEE trigger scheme or a full data transmission scheme.

### 11.2.1 Readout strategy

Table 11.2 provides a general comparison of the two typical trigger readout schemes. Generally speaking, both the full data transmission and the FEE trigger scheme have their own advantages in different aspects. In typical applications, these approaches are respectively supported by various large particle physics experiments, such as the CMS [1, 2], the Large Hadron Collider beauty (LHCb) [3, 4], the ATLAS [5, 6], the Belle II experiment (Belle II) [7, 8], and the BESIII [9, 10]. In China's collider spectrometer experiments, such as those represented by BESIII, the front-end trigger-based approach is widely adopted. Non-collider experiments, such as the Jiangmen Underground Neutrino Observatory (JUNO) [11] and the LHAASO [12], generally explore the front-end waveform sampling schemes. However, overall, the implementation of trigger algorithms still follows relatively traditional approaches, such as data compression and detector information extraction.

**Table 11.2:** Comparison of the full data transmission and FEE trigger readout strategy.

| Characteristics | Full Data Transmission | FEE-Trigger | Which is better |
|---|---|---|---|
| Trigger info. generation | On BEE | On FEE | Full Data Transmission |
| Possible trigger latency allowed | Medium-to-long | Short | |
| Compatibility on Trigger Strategy | Hardware/ software | Hardware only | |
| FEE-ASIC complexity on Trigger | Simple | Complex on algorithm | |
| Upgrade possibility on new trigger | High | Limited | |
| FEE data throughput | Large | Small | FEE-Trigger |
| Maturity | Mature but relatively new | Very mature | |
| Resources needed for algorithm | High | Low | |
| Representative experiments | CMS, LHCb, ... | ATLAS, Belle II, BESIII, ... | |

Based on the project goals and timeline, the electronics-TDAQ framework with full data transmission is chosen as the primary strategy for the Reference Detector's electronic system. The traditional FEE trigger readout framework will serve as a backup. The decision is based on these key points:

**i)** Physics Exploration and Flexibility: Full data transmission retains maximum potential for discovering new physics by reading out all raw detector information. Front-end triggers, based on known physics, might miss unknown processes. Full data transmission also simplifies future upgrades, requiring only enhancements to signal processing and readout capabilities, without redesigning the electronics.

**ii)** ASIC Design and Iteration: Full data transmission accelerates ASIC development by focusing on detector signals and beam-induced backgrounds, establishing key interface specifications early. Front-end trigger strategies complicate ASIC design with additional requirements like on-chip triggers and readout designs, increasing iteration risks and timelines.

**iii)** System Versatility and Simplification: The full data transmission framework enables for a unified interface between FEE and BEE, simplifying the overall design. Trigger algorithms are implemented in back-end FPGAs, which can be hot-programmed as needed, achieving detector independence and reducing the need for customized designs.

**iv)** Data Rate Capability: The FEE can handle the required data rates, including beam-induced backgrounds and critical detector information. Even with encoding overhead, the data rates are within typical interface capabilities, ensuring that the system can manage the detector's first operation phase before the High Luminosity $Z$ Mode.

**v)** Future High Luminosity $Z$ Mode: For this second operation phase, optimizations are possible. MDI shielding can reduce beam-induced background rates, and advanced data transmission technologies are expected for the VTX. Dynamic thresholds for ECAL can also help reduce the data rate.

## 11.2.2 Baseline architecture of the electronics-TDAQ system

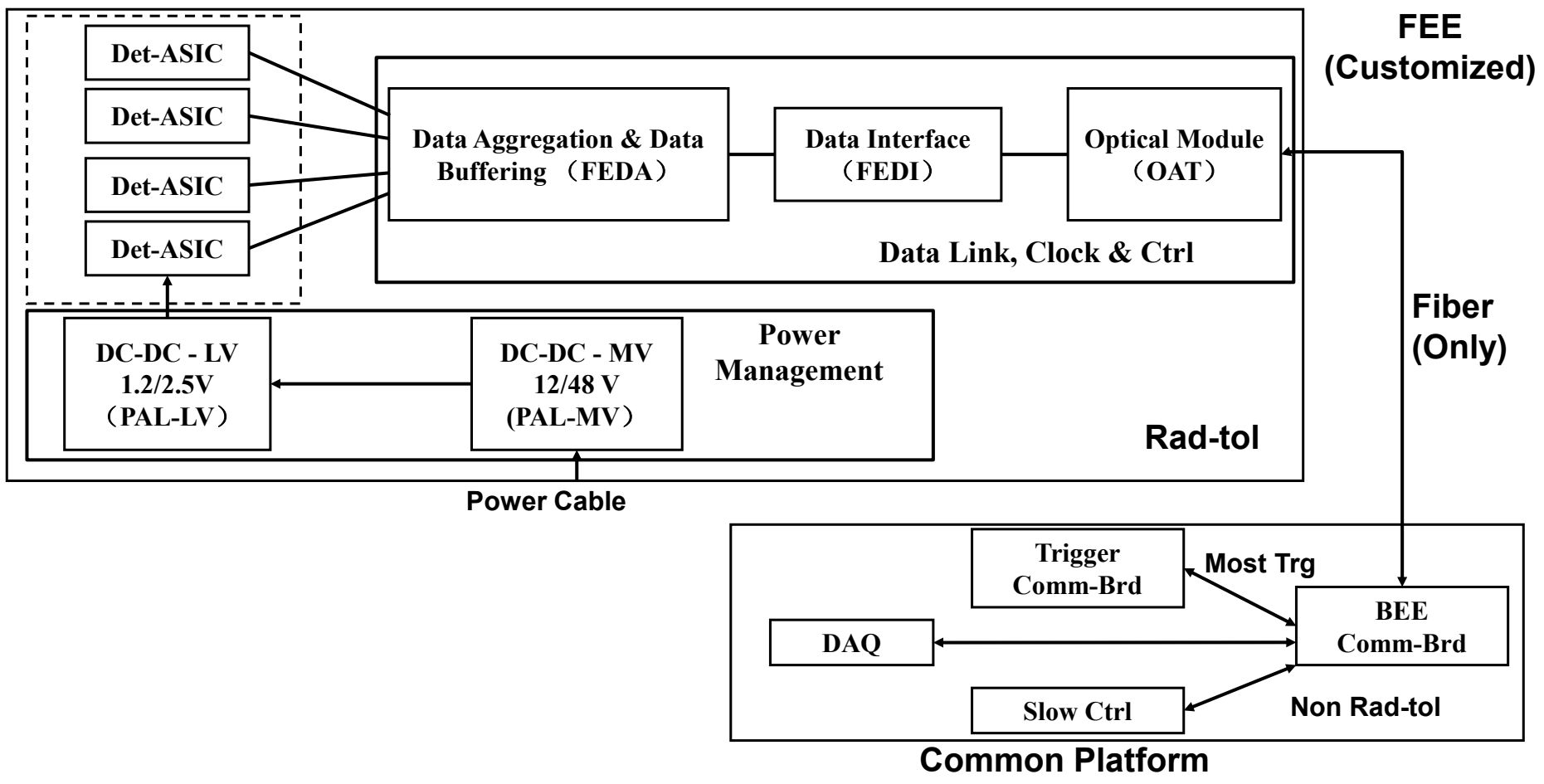

**Figure 11.1:** Global framework of the electronics-TDAQ System of the Reference Detector. It includes key blocks of the electronic system, such as the front-end part and the back-end part. The proposed design is based on a common platform approach, where the designs are shared by all sub-detector readouts.

Figure 11.1 shows the electronic system architecture based on these considerations. It uses a full data transmission scheme and consists of two main parts: customized FEE and a common platform. The customized FEE are designed for each sub-detector system's specific requirements and layout. However, only the front-end readout ASIC, which handles full data transmission of detector signals and beam-induced backgrounds, is fully customized. The data interface and power supply systems are still based on the common platform. The data interface platform receives low-speed data from the front-end ASIC via the FEDA, aggregates it into high-speed serial data, encodes it through the FEDI, and converts it into optical signals using the OAT module. These optical signals are transmitted via optical fibers to the BEE, completing the data transmission chain. The front-end PAL module converts external high-voltage power supplies into the required voltages for the front-end chips. The BEE and trigger system use a common PCB approach, scalable based on the data volume of different sub-detector systems. All trigger signals are transmitted between the common back-end board and the trigger board, allowing flexible algorithms based on physical objectives.

Dividing the electronic system into customized front-end and common platform parts allows the FEE to be fully ASIC-based, achieving radiation tolerance. The BEE can be placed in shielded environments away from the collision point, enabling the use of high-performance commercial devices without concerns for radiation tolerance.

As a backup for full data transmission, the front-end trigger readout scheme is a reasonable option. If the final detector beam-induced background exceeds the transmission capability of the common data interface, the FEE can be adjusted to revert to this backup plan during the High Luminosity $Z$ operation period in the second operation phase. The specific steps are:

**i)** When the detector beam-induced background data rate exceeds the transmission capability of the data interface, additional fiber channels can be added, each with effective data rate of 9.71 Gbps, to achieve a maximum data interface transmission capacity of 38.84 Gbps.

**ii)** When the data rate levels of certain detector modules still exceed the data rate limits, advanced data compression algorithms can be deployed in the front-end ASIC chip, such as extracting key information about hit clusters in the track and reducing redundant time stamp information output, thereby reducing the data rate levels at the front end.

**iii)** If these optimizations are insufficient, a fast trigger strategy can be further considered for detectors with transmission bottlenecks (e.g., the VTX). This is considered only for the High Luminosity $Z$ mode in the second operation phase when a new VTX can be replaced. The process involves: buffering detector signals and background data on the front-end chip; the trigger system generating fast trigger arbitration information quickly and sending it to the front-end chip via a dedicated channel; the front-end chip comparing buffered data with the fast trigger information and transmitting preliminarily arbitrated data to the BEE. This approach reduces the front-end data rate but requires a fast trigger algorithm based on detector optimization and physical objectives. Further details are discussed in Chapter 12.

## 11.3 Common electronics interface

Besides the ASIC design, the common FEE blocks, including the Data Link and the FEE Power Module, also play key roles in the global architecture of the Electronics System.

### 11.3.1 Overall architecture of the data interface

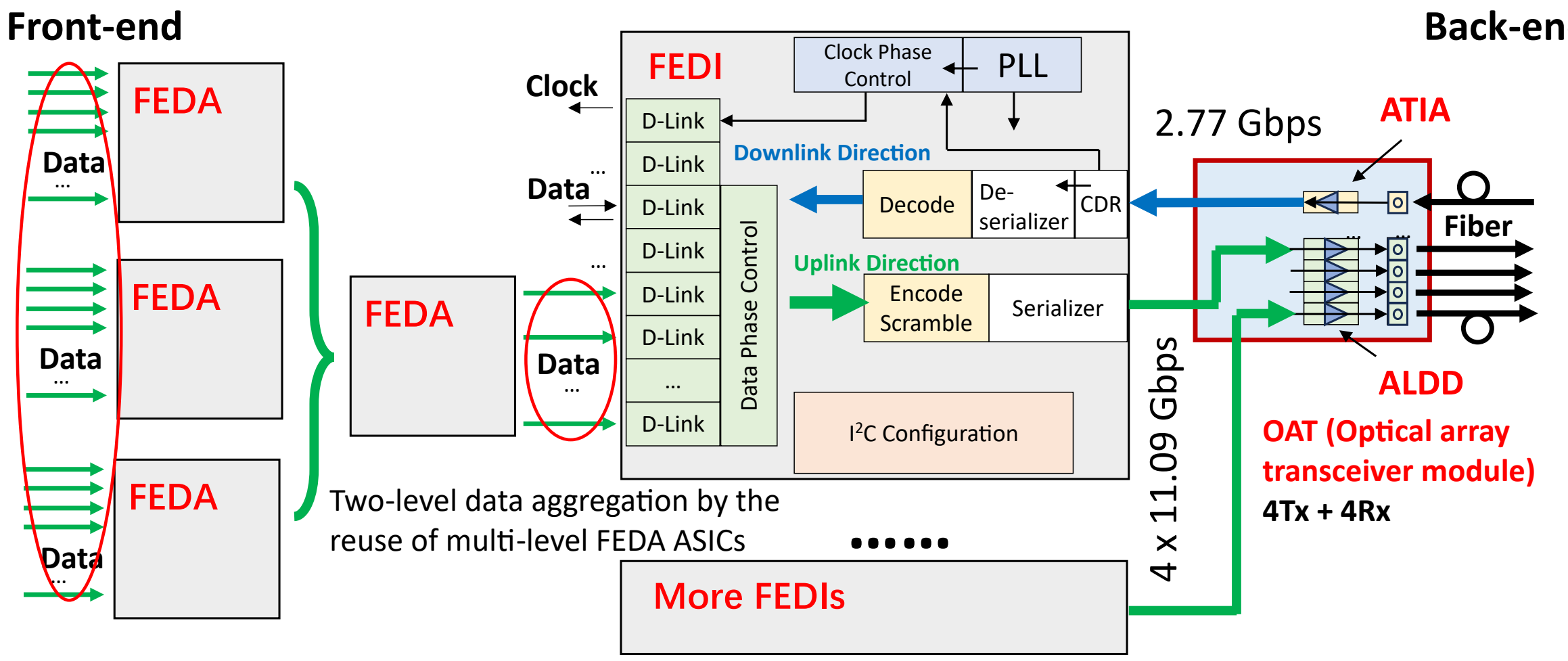

**Figure 11.2:** The data interface has a series of ASICs and a custom Optical Array Tranceiver (OAT). The ASICs include the Front-End Data Aggregator (FEDA), Front-End Data Interface (FEDI), Array Laser Diode Driver (ALDD), and Array Transimpedance Amplifier (ATIA).

Figure 11.2 illustrates the overall architecture of the data link system designed for the Reference Detector. It aims to build a general purpose, high-speed, bi-directional optical transmission system that can be used in the front-end readout electronics of various sub-detectors. The overall system structure follows the CERN Versatile Link Project and the GBT series ASICs [13].

The data link system has a series of ASICs and a customized OAT. The ASICs include the FEDA, FEDI, ALDD, and ATIA. FEDA serves as a data pre-aggregator for multiple readout channels across various sub-detectors, with single-channel data rates ranging from 43.33 Mbps/ch to 346.67 Mbps/ch. FEDI functions as a bi-directional data interface. It receives multi-channel serial data, each at 1.39 Gbps/ch from FEDA, combines all input data together to form an 11.09 Gbps high-speed serial output for the uplink. Additionally, FEDI receives single channel 2.77 Gbps command or configuration data coming from the BEE, recovers and distributes the clock, trigger, and multi-channel data signals to the FEE, serving as the downlink function. ALDD is a four-channel $4 \times 11.09$ Gbps/ch laser array driver that amplifies the high-speed serial data from FEDI to drive the VCSEL. ATIA is a four-channel $4 \times 2.77$ Gbps/ch transimpedance amplifier, which receives high-speed current signals from the pin-diode and amplifies them to regular high-speed differential signals. Both ALDD and ATIA are integrated within the OAT module, which is a custom array optical module with four Transmitters (Tx) and four Receivers (Rx). More detailed functions and interface of these ASICs are presented in the following sections.

### 11.3.2 FEDA: Front-End Data Aggregator ASIC

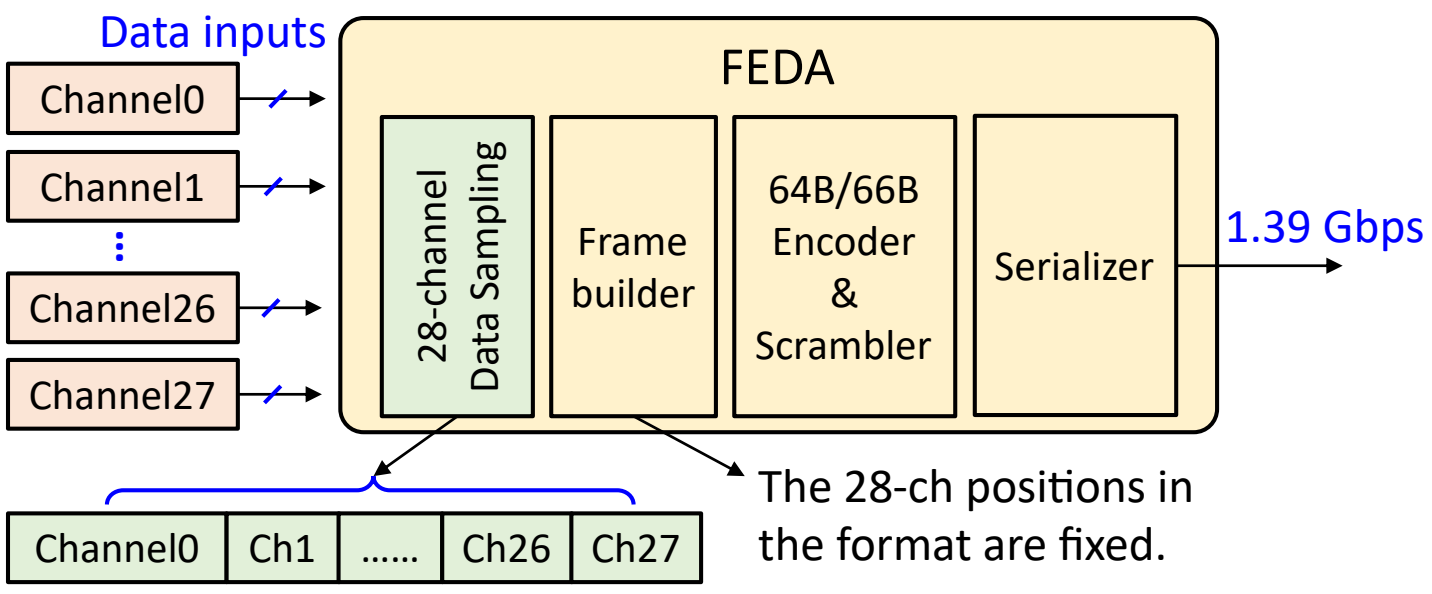

**Figure 11.3:** The FEDA processes multiple data channels and rates from front-end sub-detectors, supporting maximum 28 channels at the data rate of 43.33 Mbps. It consists of a channel-data sampling, frame builder, 64B/66B encoder and a serializer at a fixed output data rate of 1.39 Gbps.

Figure 11.3 shows the block diagram. The FEDA processes multiple data channels and rates from front-end sub-detectors, serializing them into a single data stream. The output data rate must align with the FEDI requirement of 1.39 Gbps, based on a 43.33 MHz detector system clock.

The data sampling module samples each channel per clock cycle, transmitting the 28-bit data to the frame builder. Since channel positions in the frame format are fixed, no additional address bits are needed. The data is then encoded and scrambled using a 64B/66B scheme before being serialized into a 1.39 Gbps output stream. The input data rate ranges from a minimum of 43.33 Mbps (supporting up to 28 channels) to a maximum of 346.67 Mbps (supporting 3 channels at 346.67 Mbps and 1 channel at 173.33 Mbps). Additional operation modes are shown in Table 11.3.

### 11.3.3 FEDI: Front-End Data Interface ASIC

The FEDI architecture integrates a high-precision clock system, high-speed Serializer/Deserializer (SerDes), data builder, configuration interface, and monitoring circuitry, as shown in Figure 11.2. It processes a 7-channel data stream at 1.39 Gbps received from the FEDA through dedicated D-links. The data processing pipeline includes phase alignment, encoding, DC-balancing via the data builder, and serialization to 11.09 Gbps before transmission to the OAT.

The clock system has a PLL, Clock Data Recovery (CDR) and Clock Phase Control (CPC). Prototype circuits are developed to evaluate their functionality and performance. Both the PLL and CDR circuits support system clock frequencies of 40 MHz and 43.33 MHz, exhibiting a random jitter below 1 ps. The phase noise values at a 1 MHz offset are measured at -113 dBc/Hz and -108 dBc/Hz for the PLL (5.12 GHz) and CDR (2.56 Gbps), respectively [14]. Additionally,

**Table 11.3:** Interface information of the FEDA. The operation modes describe different combinations of the available channels along with the data rate. While the sum of the data rate keeps unchanged, the combinations of the channel numbers and data rate are flexible.

<table>
<tr><th>Operation modes</th><th>Output data rate</th><th>Description</th></tr>
<tr><td>28 ch × 43.33 Mbps</td><td rowspan="11">1.39 Gbps</td><td rowspan="11">Channels 0 to 7 serve as an example. When CH0 operates at 346.64 MHz, CH1–CH7 cannot receive input. Similarly, when CH0 operates at 173.32 MHz, CH1–CH3 are disabled, and when CH0 operates at 86.67 MHz, CH1 is disabled. The remaining channels are configured in the same manner.</td></tr>
<tr><td>14 ch × 86.67 Mbps</td></tr>
<tr><td>7ch × 173.33 Mbps</td></tr>
<tr><td>3 ch × 346.67 Mbps + 1 ch × 173.33 Mbps</td></tr>
<tr><td>3 ch × 346.67 Mbps + 2 ch × 86.67 Mbps</td></tr>
<tr><td>3 ch × 346.67 Mbps + 4 ch × 43.33 Mbps</td></tr>
<tr><td>2 ch × 346.67 Mbps + 3 ch × 173.33 Mbps</td></tr>
<tr><td>2 ch × 346.67 Mbps + 2 ch × 173.33 Mbps + 2 ch × 86.67 Mbps</td></tr>
<tr><td>2 ch × 346.67 Mbps + 2 ch × 173.33 Mbps + 1 ch × 86.67 Mbps + 2 ch × 43.33 Mbps</td></tr>
<tr><td>1 ch × 346.67 Mbps + 2 ch × 173.33 Mbps + 4 ch × 86.67 Mbps + 4 ch × 43.33 Mbps</td></tr>
<tr><td>. . . . . .</td></tr>
</table>

the SerDes block operates correctly, with the total jitter of the serializer at 10.24 Gbps maintained below 43 ps.

#### 11.3.3.1 FEDI input/output channels

In the uplink direction, FEDI receives multi-channel data from FEDA. It performs data reception, alignment, scrambling, encoding, frame building, serialization, and outputs high-speed serial data to OAT. In the downlink direction, FEDI receives high-speed serial data from OAT. It performs data alignment, clock recovery, decoding, descrambling, outputs parallel data to the front-end, and provides clocks and slow control data to FEE.

The bi-directional data transmission capabilities of FEDI are summarized in Table 11.4. The FEDI operates at a serial data rate of 2.77 Gbps for downlink, and at 11.09 Gbps for uplink. It provides 16 channels of clock signals to the front-end with configurable frequencies and phases. The input and output electrical signal specifications for FEDI are detailed in Table 11.5 and Table 11.6.

#### 11.3.3.2 Protocol

**Digital processing** Figure 11.4 illustrates the digital processing architecture of FEDI. In the downlink, a serial data stream (2.77 Gbps) is first deserialized to parallel format (IFRAMED). The IFRAMED data is then de-interleaved, de-framed and de-coded with error correction, and de-scrambled to get the Header, Bidirectional Control (BDC), Unidirectional Control (UDC), and payload data. In the uplink transmitter path, data is first scrambled for DC balance, then encoded using Forward Error Correction (FEC) [15]. The encoded bits undergo interleaving to form IFRAMEU, which is then serialized for transmission.

**Table 11.4:** Data transmission capability of the FEDI. It supports a maximum of 7-channel data inputs, a single 11.09-Gbps serial output, eight output clock channels with selectable frequencies, a maximum of 16-channel downlink payload, and an external control channel.

| Capability | Channel Number | Data/Clock Rate |
|---|---|---|
| Uplink payload data channels (IN) | maximum 7 | 1.39 Gbps/ch |
| Uplink external control (EC) channel (IN) | 1 | 86.67 Mbps |
| Downlink payload data channels (OUT) | maximum 16 | 86.67 - 346.67 Mbps/ch |
| Downlink EC channel (OUT) | 1 | 86.67 Mbps |
| Ouput clock channels | 16 | 43.33 /86.67 /173.33 /346.67 /693.33 MHz or 1.39 GHz |

**Table 11.5:** Input electrical signal specification of FEDI. The input signal of 1.39 Gbps complies with the differential Current Mode Logic (CML) standard.

| Input information | Description |
|---|---|
| Signal type | Differential |
| Differential termination | 100 Ω |
| Differential peak-to-peak amplitude | Minimum 140 mV |
| Reception of the common mode voltage (Vcm) | 0 V ˜ 1.2 V |
| Coupling mode | Configurable AC or DC coupling |
| Internal Vcm for AC coupling | 600 mV |
| Data rate for the uplink | 1.39 Gbps |
| Data rate for the EC channel | 86.67 Mbps |

**Table 11.6:** Output electrical signal specification of FEDI. The output signal of 11.09 Gbps complies with the CML standard.

| Output information | Description |
|---|---|
| Signal type | Differential |
| Differential termination | 100 Ω |
| Differential peak-to-peak amplitude | 200 – 800 mV (programmable) |
| Common mode voltage | 600 mV |
| Coupling mode | AC or DC coupling |
| Data rate for the downlink | 86.67 /173.33 /346.67 Mbps per channel |
| Data rate for the EC channel | 86.67 Mbps |
| Clock frequency | 43.33 /86.67 /173.33 /346.67 /693.33 MHz or 1.39 GHz |

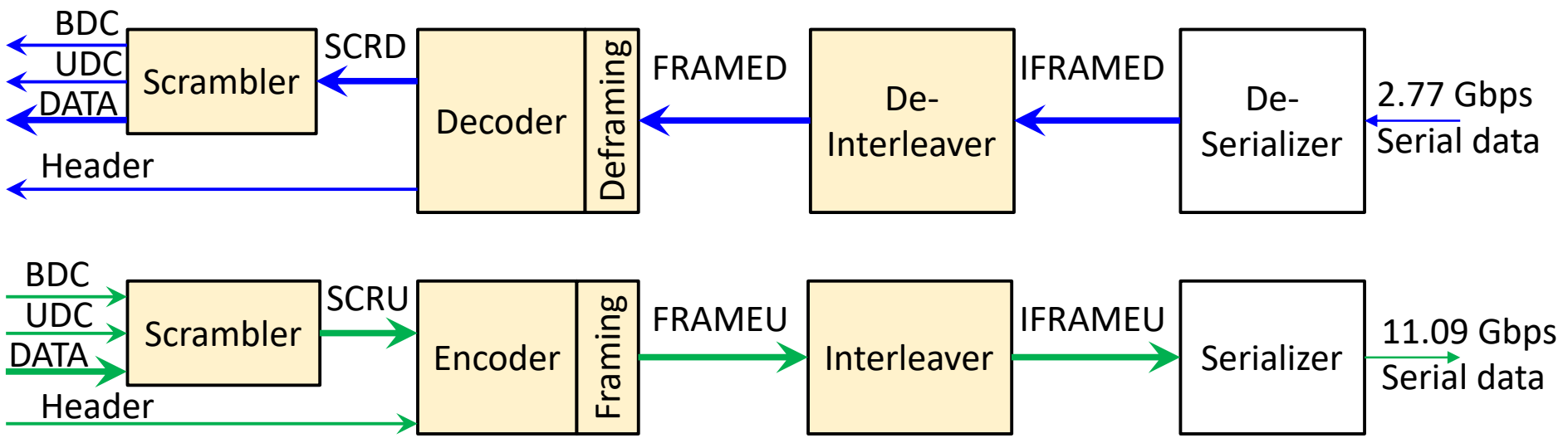

**Figure 11.4:** The encoding/decoding block diagram of FEDI consists of a 2.77-Gbps downlink and 11.09-Gbps uplink.

**Interleaving/de-interleaving** FEDI performs de-interleaving for the downlink transmission. The downlink FEC code supports single-symbol error correction (t = 1) with 3-bit symbols. To cover the entire frame (36 data bits), four encoders are employed, and the interleaving of their codes enables error correction for up to four consecutive symbols, equivalent to 12 consecutive erroneous bits. For the uplink transmission, FEDI handles interleaving, allowing for the correction of a consecutive 24-bit error burst.

**Encode/decode** Due to factors such as noise, inter-symbol interference, or SEU, data can be corrupted during transmission. FEC [15] addresses this issue by enabling error correction without requiring data retransmission. This is achieved by transmitting additional ”redundant” bits alongside the original data. While this improves transmission reliability, it reduces the available data bandwidth. The FEDI employs Reed-Solomon Codes [16], a class of FEC codes that operate on “non-binary” symbols formed of m bits, providing robust error correction for noisy channels.

**Scramble** Both uplink and downlink paths use scrambling to mitigate DC imbalance and help for a reliable CDR. FEDI uses a parallel implementation for its self-synchronizing scrambler and descrambler. This design offers three key benefits: it matches the ASIC N-parallel input, enables high data throughput with a lower internal clock frequency, and minimizes latency. For testing purposes, the scrambler / de-scrambler can be bypassed.

**Data frame** In optical communication systems, data transmission is susceptible to burst noise interference, which can cause consecutive bit errors. To address this issue, a robust error correction scheme combining FEC with interleaving is implemented. Figure 11.5 presents the data frame of the downlink and uplink, both operating at a frame rate of 43.33 MHz. The downlink frame has 64 bits resulting in a data rate of 2.77 Gbps. The uplink frame length is 256-bit for 11.09 Gbps transmission, which is compatible with lpGBT [17].

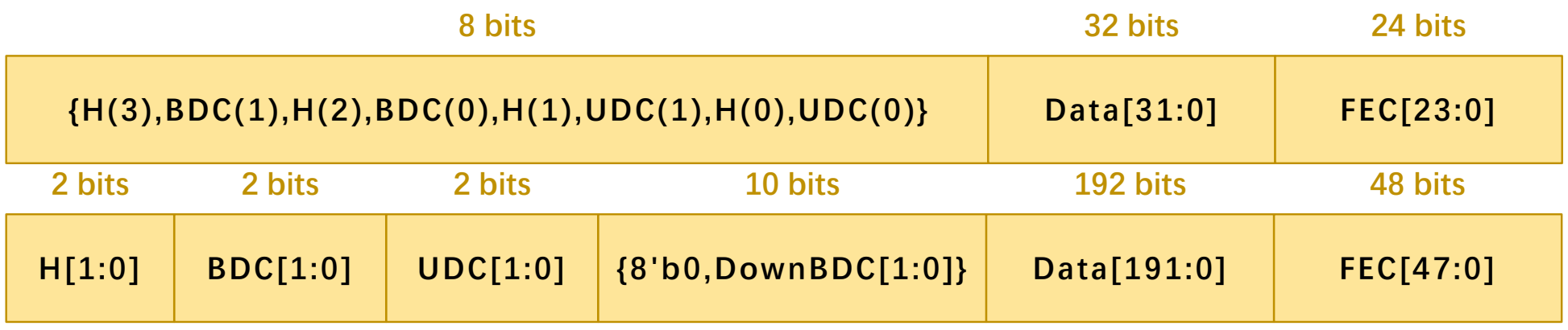

**Figure 11.5:** (a) The downlink frame has 64 bits resulting in a data rate of 2.77 Gbps. (b) The uplink frame length is 256-bit for 11.09 Gbps.

### 11.3.3.3 Design of the core circuits

**On-chip clock system** Figure 11.6 shows the architecture of the clock system. The PLL receives a 43.33 MHz system clock and generates the required clocks for FEDI, while also offering up to 16-channel differential clock outputs externally. These outputs support configurable frequencies for all 16 channels and configurable phases for two of the 16 channels. Each output clock channel can operate at six frequencies: 43.33, 86.67, 173.33, 346.67, 693.33, and 1386.67 MHz. The CPC regulates the phases, providing a resolution of 45.08 ps across all frequencies, with a full range of 23.07 ns. The CDR consists of a differential Alexander Phase Detector (PD), a Frequency Detector (FD), a differential charge pump (CPc) and a low pass filter (LPFc), sharing a Voltage-Controlled Oscillator (VCO) with the PLL. The clock system can operate in two modes. When in the CDR mode, the 43.33 MHz reference clock is inactive. The CDR recieves a 2.77 Gbps data stream and recovers a synchronized data stream and a 5.55 GHz clock. When in the PLL mode, the system requires the reference clock to synthesize a 5.55 GHz clock.

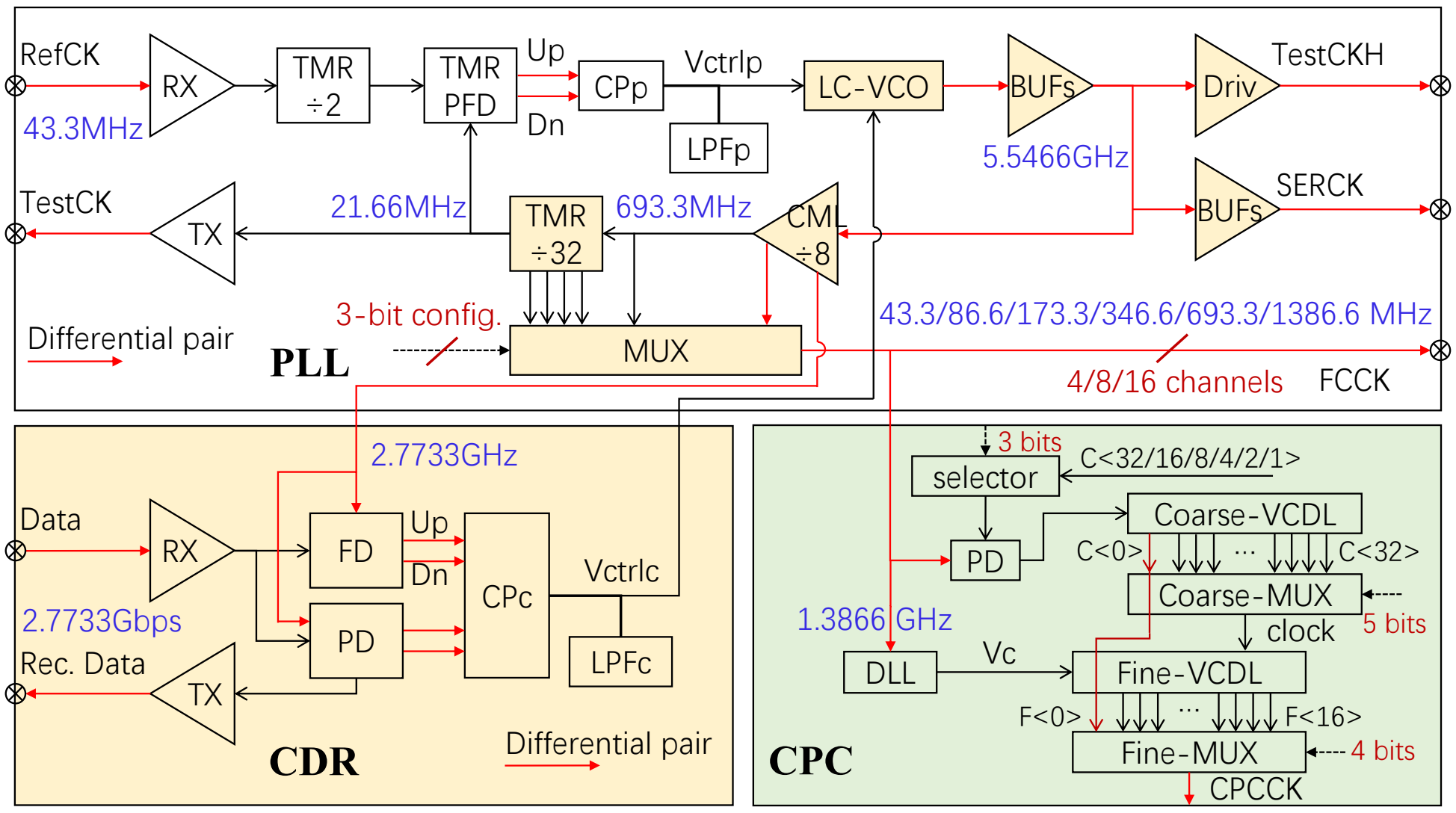

**Figure 11.6:** The on-chip clock system mainly comprises a Phase-Locked Loop (PLL), a Clock Data Recovery (CDR) and a Clock Phase Control (CPC).

**Serializer** Figure 11.7a illustrates the overall block diagram of the serializer, which employs a two-stage 4:1 Multiplexer (MUX) circuit in a cascaded configuration to serialize 16 channels of 693.33 Mbps parallel input data into a single 11.09 Gbps high-speed serial data output. The design has four low-speed 4:1 MUXs, a high-speed 4:1 MUX, a clock distribution, 16 Receivers (RXs), a high-speed Transmitter (TX) and a PRBS self-test circuit.

**De-serializer** Figure 11.7b shows the architecture of the de-serializer, which has a Data Receiver Circuit (DRX), a Clock Receiver Circuit (CRX), two dividers (by 4), four 1:4 Demultiplexers (DEMUXs), several Single-Ended-To-Differential (S2D) circuits, and LVDS drivers. The input data rate and clock frequency are 2.77 Gbps and 2.77 GHz, respectively. The DRX circuit first converts the input data into single-ended 2.77 Gbps CMOS signals. These signals are then processed by the first-stage 1:4 DEMUX, generating four parallel channels of 693.33 Mbps data. Each of these four channels is further demultiplexed by four parallel 1:4 DEMUXs, achieving a 4:16 demultiplexing operation and producing 16 channels of single-ended 173.33 Mbps data, and then output in parallel by the LVDS drivers.

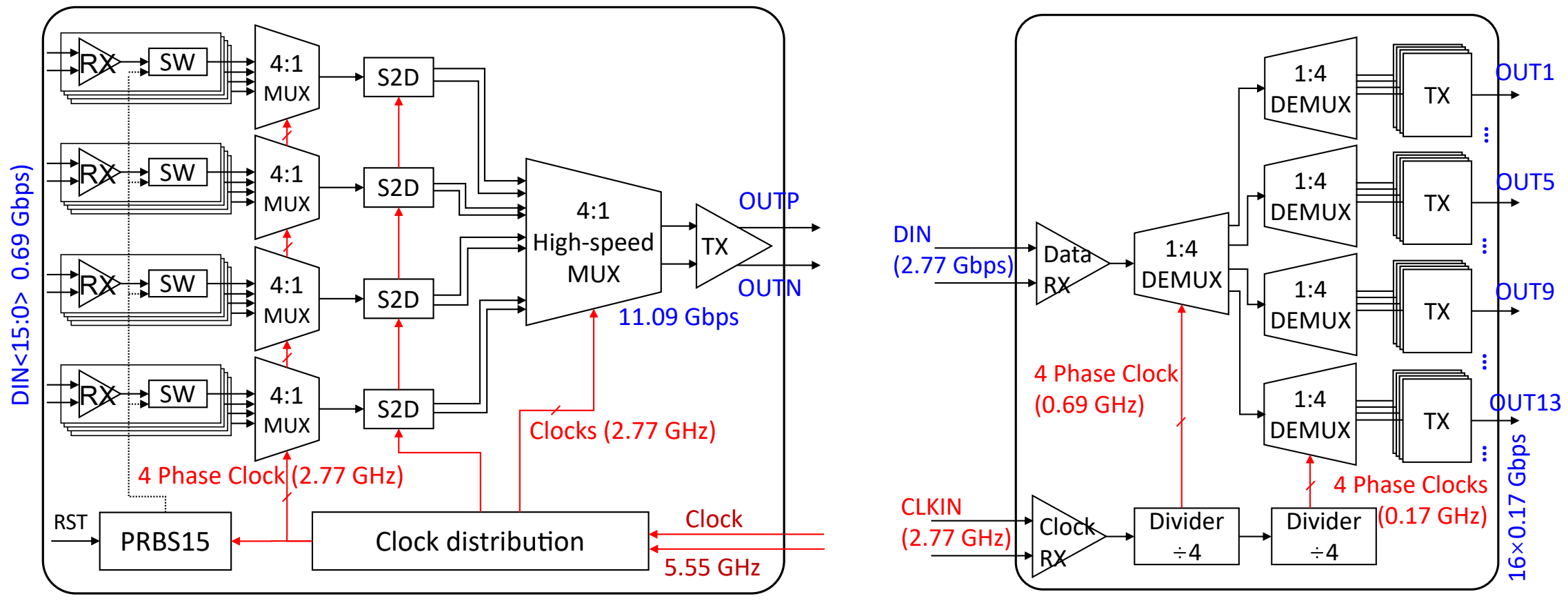

**(a)** **(b)**

**Figure 11.7:** (a) The serializer employs a two-stage 4:1 MUX circuit in a cascaded configuration to serialize 16 channels of 693.33 Mbps parallel input data into a serial 11.09 Gbps output. (b) The deserializer consists of a data receiver (DRX), five 1:4 Demultiplexers (DEMUX), an LVDS output circuit, a Clock Receiver (CRX), and a divide-by-4 circuit (Divider/4).

**Data Phase Control (DPC)** Figure 11.8 presents the architecture of the DPC designed for automatic phase/delay adjustment of multi-channel data inputs to the FEDI chip. The DPC comprises three key components: a DLL, a Voltage-Controlled Delay Line (VCDL), and a digital phase-selection logic. It can operate in two modes: auto-tracking and training. In auto-tracking mode, all delay units and taps operate continuously, with the digital logic performing real-time optimal phase selection. In training mode, calibration uses prepared data. The digital logic locks the initial tap selection and maintains this configuration without further adjustments.

The current design can support automatic phase alignment for data rates of 173.33, 346.67 and 693.33 Mbps, and 1.39 Gbps.

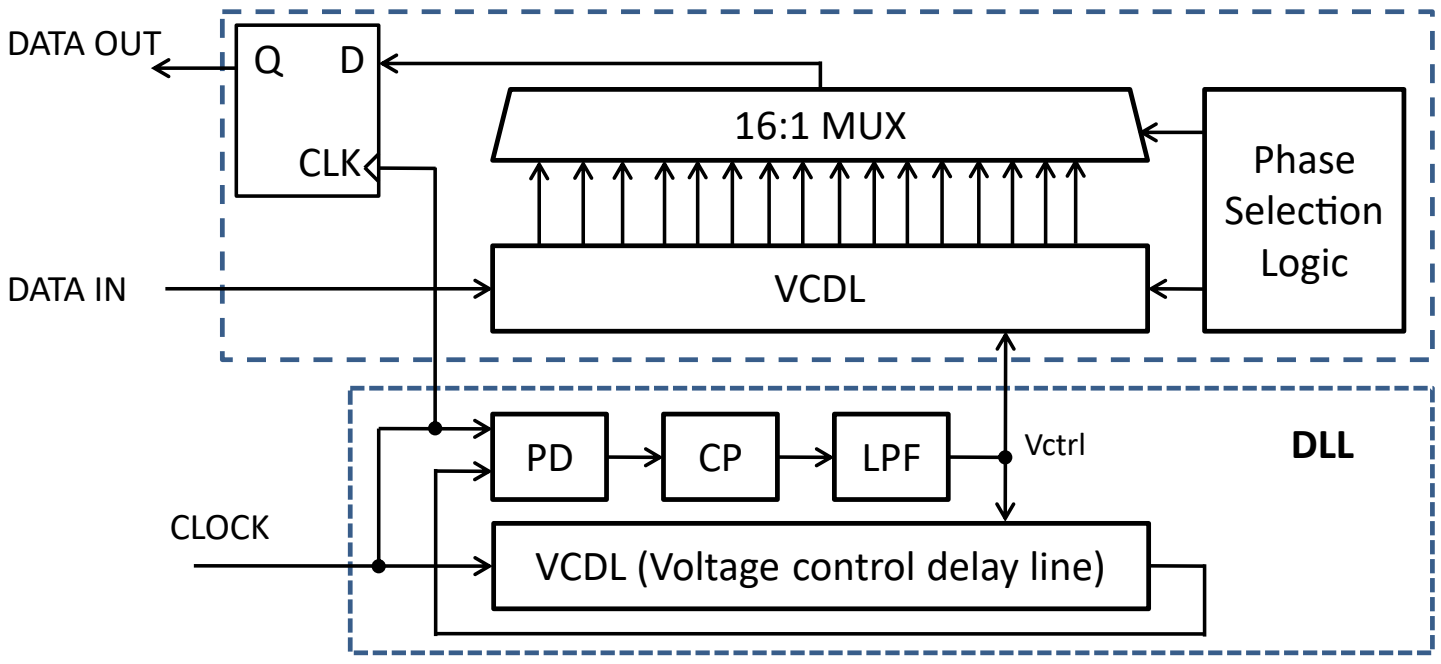

**Figure 11.8:** DPC module comprises a Delay-Locked Loop (DLL), a Voltage-Controlled Delay Line (VCDL), and digital selection logic.

## 11.3.4 ALDD: VCSEL driver

ALDD is a four-channel VCSEL array driver. Each channel has an input equalizer, a limiting amplifier, and a novel output-driving stage, as shown in Figure 11.9 [18]. The equalizer receives a pair of 11.09 Gbps/ch differential signals. The limiting amplifier receives the signals from the equalizer and further amplifies the differential signals with sufficient gain and bandwidth for the output driver. The output driver converts the amplified differential voltage signals from the limiting amplifier to a single-ended high-speed current signal and drives the external VCSEL [18]. The specification of this VCSEL driving ASIC is presented in Table 11.7.

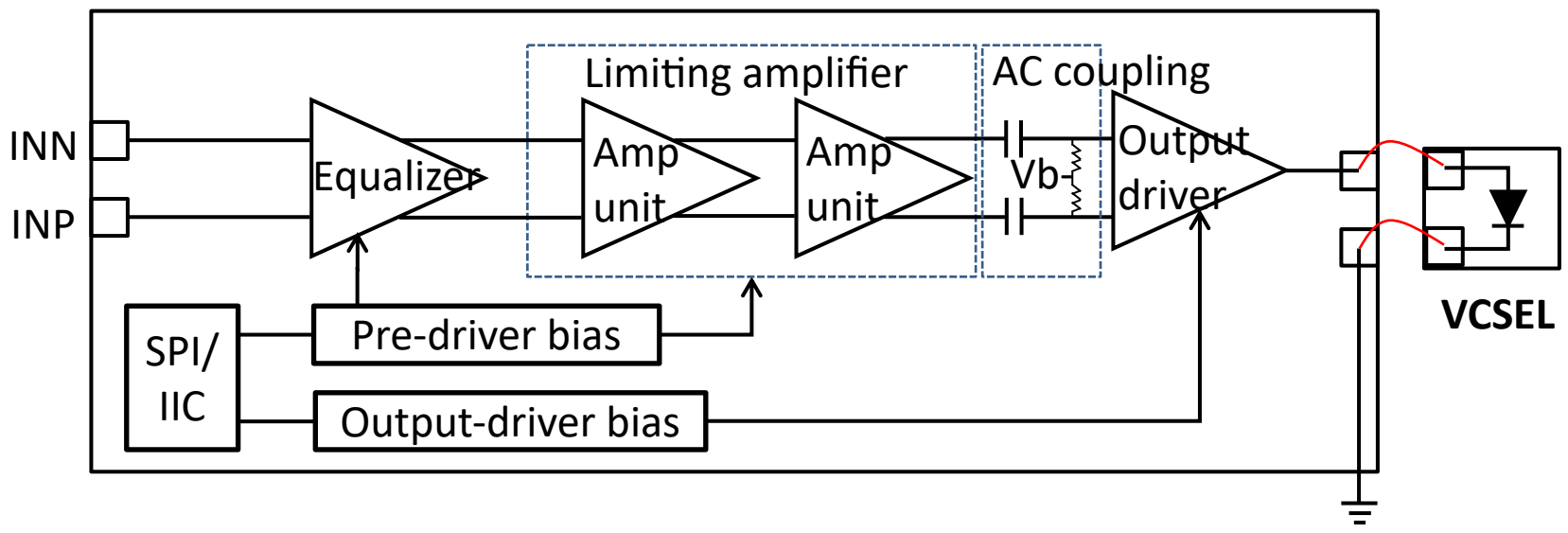

**Figure 11.9:** The VCSEL driver consists of an equalizer stage, a limiting amplifier stage, and a novel output driver stage.

## 11.3.5 ATIA: Trans-impedance receiver

The ATIA is a four-channel Photodiode (PD) array receiver with a data rate capability exceeding 2.77 Gbps per channel. Figure 11.10 shows the block diagram of the single-channel design. Each channel has a Transimpedance Amplifier (TIA), a Limiting Amplifier (LA), and a driver stage. A fully differential cascade TIA with programmable feedback resistance is designed to achieve low noise and adjustable bandwidth. The DC Offset Cancellation (DCOC) circuit controls the DC current in the input stage, allowing the cancellation of the offset DC along the amplifying chain. The photodiode signal is AC coupled to the TIA using on-chip capacitors. A photodiode biasing circuit is designed and integrated in the chip to ensure the proper biasing of the photodiode for high irradiation levels. The driver stage has a programmable pre-emphasis circuit and an output driver.

**Table 11.7:** Specifications of the ALDD. It comprises four channels with two power supply voltages, where the 3.3-V power supply is dedicated to the driving stage.

| Parameter | Design indicators |
|---|---|
| Power supply voltage | 1.2 V and 3.3 V |
| Power consumption | 50 mW/ch (typical) |
| | 200 $mw$ |
| | 4×11.09 Gbps/ch |
| Bit rate | 11.09 Gbps/ch |
| Differential input signal amplitude | 200 mV p-p min. |
| Differential input impedance | 100 Ω |
| Maximum equalizer equilibrium strength | > 7 dB |
| Limiting amplification stage gain | > 12 dB |
| Limiting amplification level bandwidth | > 9.8 GHz |
| Output current amplitude | 5 mA |
| Maximum pre-emphasis strength | > 2.5 dB |
| Simulate ISI jitter | < 15 ps |

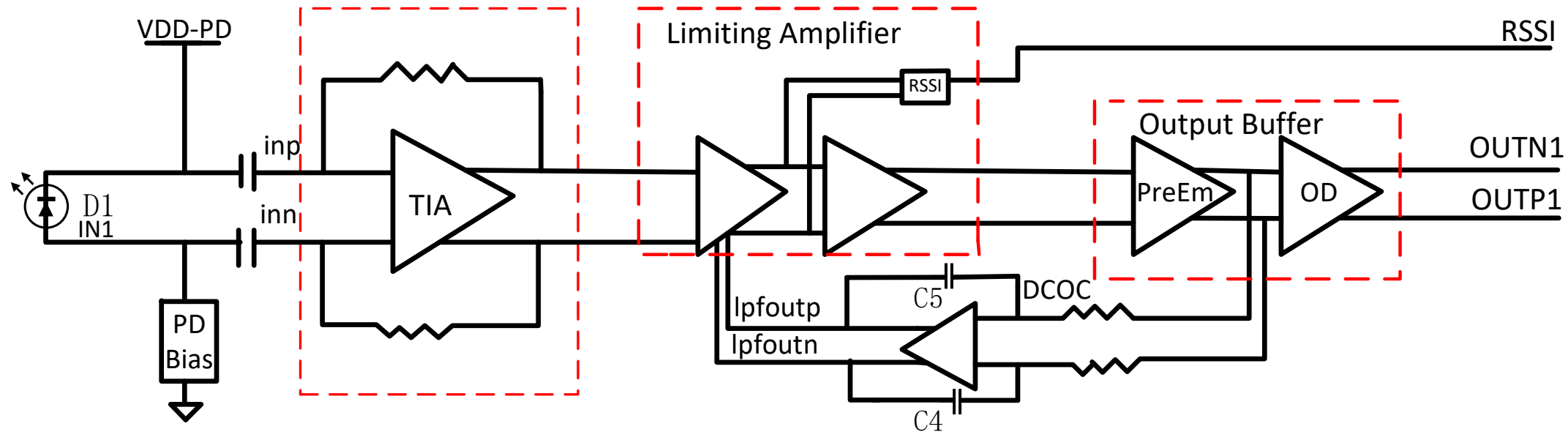

**Figure 11.10:** Block diagram of the single channel in ATIA receiver. Each channel consisting of a Transimpedance Amplifier (TIA), a Limiting Amplifier (LA), and a driver stage called Output Buffer.

The main design specifications are shown in Table 11.8. The TIA is designed to achieve a high transimpedance gain (> 20 kΩ), high bandwidth (> 2 GHz), and low input referred noise (<2 µA RMS).

**Table 11.8:** Specifications of the ATIA. It comprises four channels with two power supply voltages, where the high power supply is for the photodiode.

| Parameter | Design indicators |
|---|---|
| Bit rate | 2.77 Gbps |
| Photodiode capacitance | 200 fF to 2 pF |
| High cut-off frequency | 2 GHz |
| Low cut-off frequency | < 1 MHz |
| Sensitivity for BER=$10^{-12}$ | 20 µA p-p (-17 dBm) |
| Transimpedance gain | 20 kΩ |
| Output differential voltage | 400 mV p-p (50Ω) |
| PD bias supply voltage | 2.5 V / 3.3 V |
| Circuit supply voltage | 1.2 V |
| Power consumption | < 100 mW/ch |
| Total jitter | 0.085 UI (< 34 ps at 2.77 Gbps) |

### 11.3.6 OAT: Optical Array Transceiver module

OAT is a custom 4 Tx + 4 Rx array optical module designed for front-end readout electronics. It integrates a ALDD, a ATIA, a commercial VCSEL array, and a PD array, along with optimized electrical and optical interface designs. The electrical socket uses the Firefly connector from Samtec [19]. Figure 11.11a illustrates the optical link with Chip-On-Board (COB) assembly of transceivers, which is suitable for detector readout of a hundred meter. The COB modules can be compact for housing a total of up to 12 channels of VCSEL and PD, and their driver and TIA chips, respectively. A prototype transceiver equipped with 4-transmitter and 4-receiver channels in dimension of $10 \times 20$ mm$^2$ is shown in Figure 11.11b. It is utilizing a prism of 2.75 mm in height for light coupling to a Mechanical Transfer (MT) type fiber connector. The plug-in to electrical connector on mother board will require further consideration on electric connector and heat dissipation [20].

Radiation resistance of optical link components is required in collider environments. The opto-electronics of VCSELs and PDs are studied for degradation caused by NIEL [21, 22]. The ionizing doses on ASICs can have effects on currents and noises due to build-up on Si / $SiO_2$ interfaces [23]. Fibers in radiation can have color-centers generated. The radiation induced attenuation Radiation Induced Attenuation (RIA) as a function of TID is given by RIA = 10 × $\log_{10}$ $(P_0/P_T)$/Length, where $p_T$is the optical power transmitted after receiving a total ionizing dose T. The known rad-hard fiber of F-doped is graded for 1 Gbps speed. The RIA is reported for 0.02 dB/m at 100 kGy(Si) [24]. The telecom grades of high speed (at 10 Gbps level) fibers are mostly Ge-doped with radiation resistance depending on the doping complex and fabrication methods. The responses in ionizing doses were tested in $^{60}$Co gamma-ray. A 10-Gbps multi-mode Ge-dope fiber is found durable in radiation, with the RIA of 0.05 dB/m at

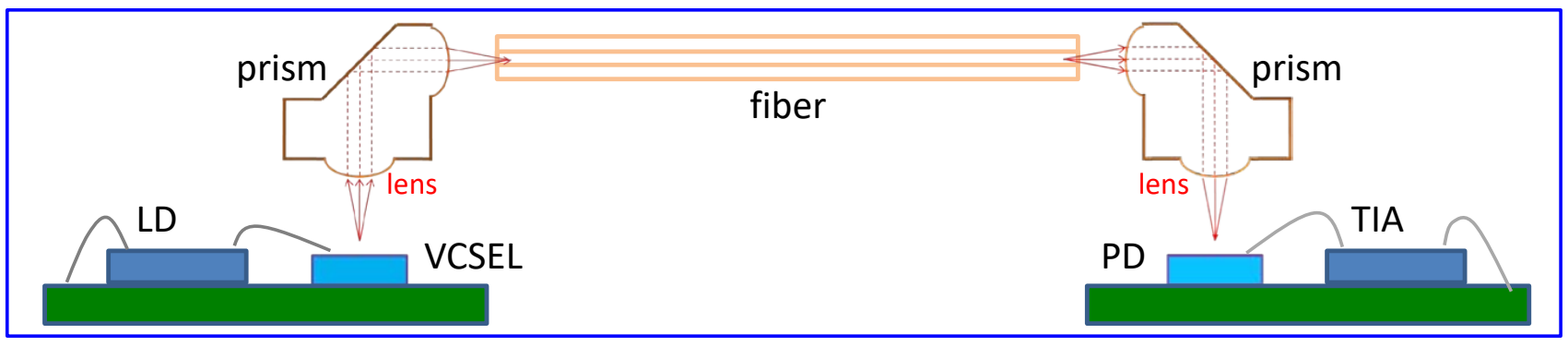

**(a)**

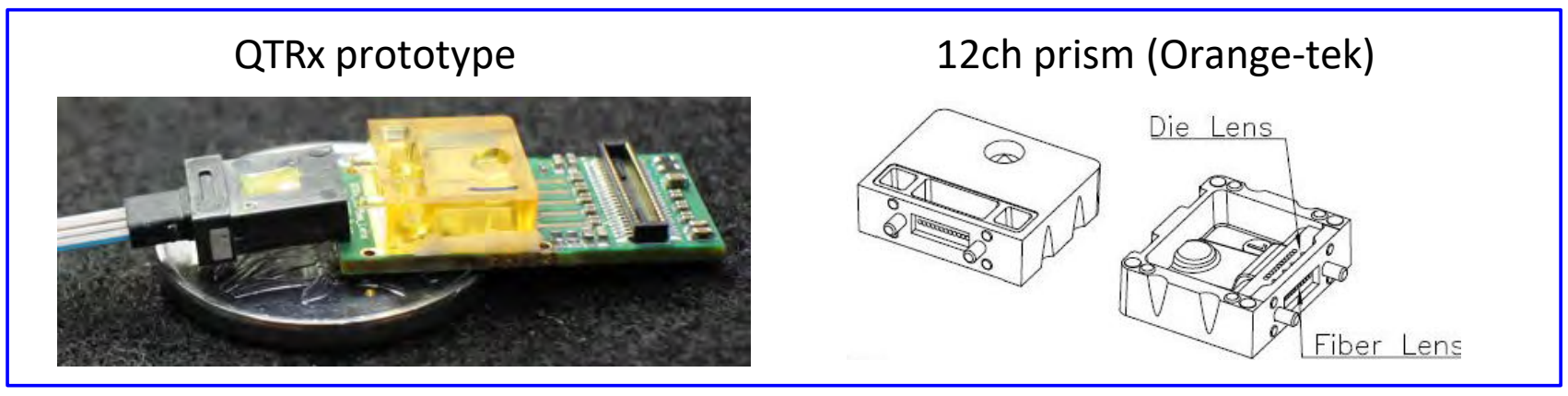

**(b)**

**Figure 11.11:** (a) illustrates the optical link with Chip-on-Board assembly of transceivers. (b) shows a prototype transceiver (QTRx [20]) using a Hirose electric connector. The prism is made in PEI plastic with the dimension of 9.5 mm in width and 2.75 mm in height.

300 kGy(Si) [25].

The RIA characteristics of the rad-hard fiber were tested in $^{60}$Co gamma-rays. In Figure 11.12a the measurements of a sample tested at a dose rate of 39.4 Gy/hr are plotted. The sample was maintained at 13 °C in a water tank chilled by a compressor plate. The annealing (dashed line) has recovered the defeats generated in irradiation (solid line) by up to 50% at the beginning. At higher dose rate the file-up of defects is higher, and at lower temperature the recovery is stagnant. However, after annealing the RIA depends only on TID. Plotted in 11.12b is a compiled RIAs of five samples irradiated at different dose rates, at 32 °C. The dashed line joints the RIAs in irradiation at the daily doses. The points are RIAs after 10 hours annealing.

### 11.3.7 Development plan

The data link system is the core block in the readout electronic system. Its development status not only proves the performance of the chip series but also validates the full scheme of the common readout electronics. The first prototype of the chip series is scheduled to be demonstrated before 2027, which will also support some sub-detector R&Ds. The final version will be ready before 2029. By then, a real standalone FEDI chip with bi-directional communication capabilities, and the OAT module with a tiny size, will be showcased as a platform.

## 11.4 Alternative scheme based on wireless communication

As previously described, the baseline readout architecture of the Reference Detector electronic system employs conventional front-end electrical links and back-end fiber optic links.

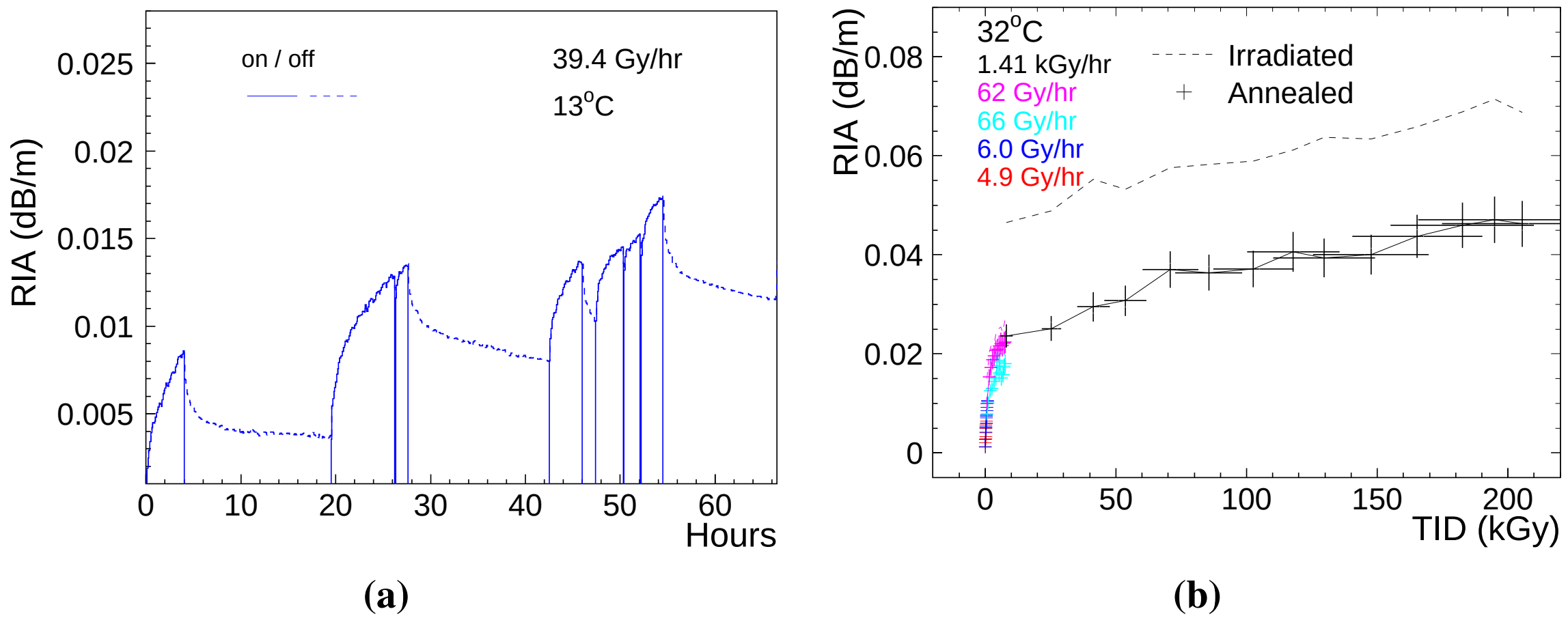

**Figure 11.12:** (a) The Radiation Induced Attenuation (RIA) of a rad-hard Ge-dope fiber sample is plotted in irradiation test at dose rate of 39.4 Gy/hr at 13 °C, in irradiation (solid line) and in annealing (dashed line). (b) The daily RIAs of five fiber samples tested at 32 °C at various dose rates are plotted in irradiation (dashed line) and after 10 hours annealing (points).

To further reduce transmission material, research and development efforts are undertaken to implement wireless communication-based data links on the detector as a next-generation readout scheme. This section introduces the concept of this alternative scheme and its current development status.

### 11.4.1 Motivation

High-energy physics experiments require the transmission of large volumes of detector signals, traditionally achieved through cable and optical fiber connections. However, extensive reliance on cables and fibers presents several critical challenges. First, it increases detector dead zones, reducing detection efficiency. Additionally, the increased material budget from extensive wiring systems exacerbates these issues, significantly raising system errors and the risk of particle misidentification. Such misidentification can profoundly impact experimental results and theoretical predictions [26].

Beyond technical challenges, cable costs can be substantial. Limited routing space in experimental setups complicates cabling, increasing installation time and labor. These factors underscore the need for alternative solutions to address these limitations.

Adopting wireless communication systems presents a viable alternative with several advantages:

1. Reduced Cable and Fiber Usage: Minimizing physical cabling enables more flexible system installation, leading to cost savings and easier modifications or expansions to accommodate evolving research needs.
2. Improved System-Level Crosstalk Control: Wireless systems enhance interference and signal overlap management, improving data transmission quality, which is critical for precise measurements in high-energy physics experiments.

3. Enhanced Reliability and Radiation Resistance: Wireless communication exhibits greater resilience to environmental factors, including radiation, ensuring robust data integrity in harsh conditions typical of high-energy experiments.
4. Minimized System Space and Material Mass: Reducing or eliminating cables and fibers decreases the experimental apparatus's physical footprint and material mass, which is crucial for particle detector designs where size and weight impact performance.
5. Efficient Broadcast Control and Clock Distribution: Wireless systems enable broadcast control and clock distribution without additional cables, facilitating seamless synchronization across multiple detectors and enhancing operational efficiency.
6. Flexible Data Acquisition and Triggering Configuration: Wireless solutions offer greater adaptability in configuring data acquisition and triggering systems, allowing researchers to reconfigure data channels and settings without physical wiring changes, facilitating rapid testing and optimization.

Given these significant advantages, wireless communication emerges as a compelling alternative for high-energy physics experiments. By addressing traditional cabling system limitations, this approach enhances operational efficiency and data integrity while fostering innovation in experimental design and execution. Adopting wireless technology could significantly advance high-energy physics research capabilities, facilitating discoveries that expand our understanding of fundamental particles and forces.

### 11.4.2 Wireless application proposal

As shown in Figure 11.13, a wireless solution offers a more streamlined design through alternative data pathways, as illustrated. In this proposal, data is transmitted radially between layers rather than relying solely on fixed fiber connections. Once collected in the outermost layer, data is transmitted axially toward the endcap. This strategy reduces internal cabling volume and simplifies the overall architecture by minimizing interference risks inherent in highly wired environments.

The wireless system implementation uses a transmission node to relay data from the inner to the intermediate layer [27]. A repeater in the intermediate layer efficiently receives inner layer data and forwards it to the outer layer. The outermost layer hosts the maximum number of nodes, with each node achieving a maximum bandwidth of 6.25 Gbps according to current design estimates. This configuration requires approximately 15,000 links with transmission distances of 10 cm to 25 cm. Millimeter-wave transmission technology is proposed to enhance data transfer rates and overall system performance.

Axial transmission offers an alternative method, enabling high-bandwidth data transfer from the barrel section to the endcap or from the endcap to external systems. This method serves as a direct optical fiber replacement, employing free-space optical transmission with optimized collimation positions and Dense Wavelength Division Multiplexing (DWDM) technology to enhance optical path data rates. This design enables data aggregation from the barrel section ladder and wireless optical transmission to the endcap. The narrow optical beam width significantly minimizes interference with other optical pathways. Then a high-density optical component arrangement without performance compromise is feasible. The high optical transmission bandwidth helps high-bandwidth data transfer post-aggregation. This design enhances data throughput while optimizing overall system communication efficiency.

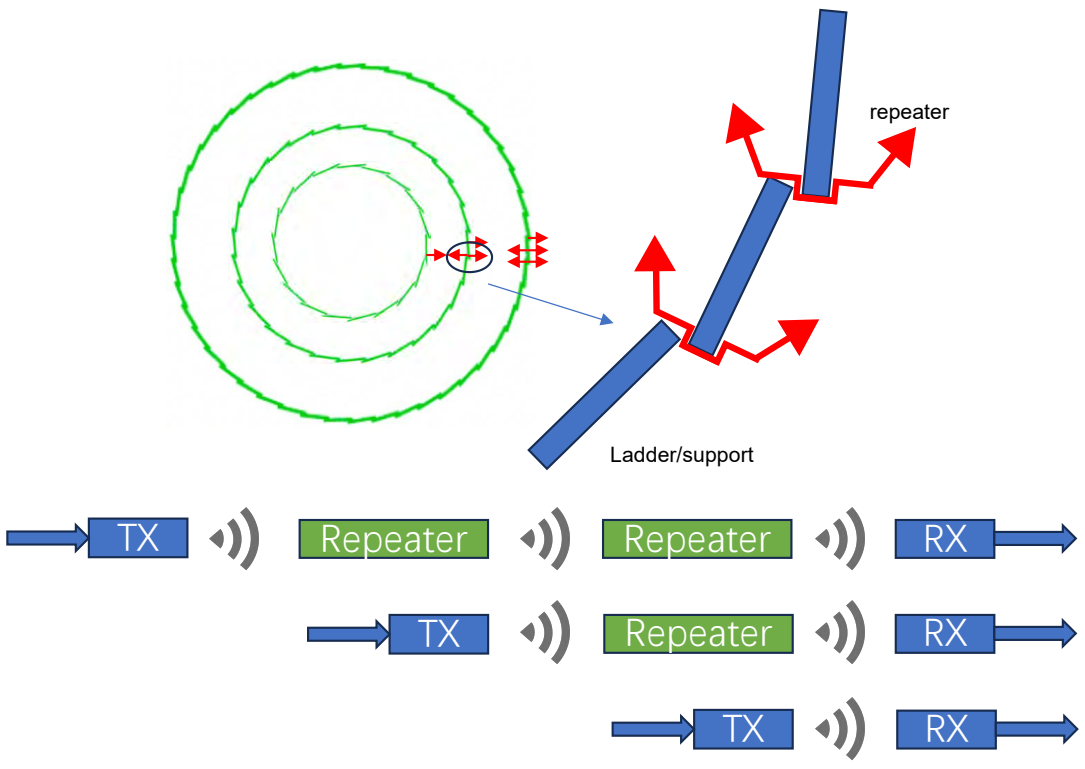

**Figure 11.13:** Data transmission between layers with repeater. The repeater is used to receives inner layer data and forward it to the outer layer.

## 11.4.3 Technology study

### 11.4.3.1 Traditional Wi-Fi solutions

Traditional Wi-Fi solutions primarily operate within the 2.4 GHz and 5 GHz frequency bands. Their key advantages include technological maturity and widespread adoption, with Institute of Electrical and Electronics Engineers (IEEE) 802.11 series standards defining the most universal wireless local area network specifications—commonly known as Wi-Fi.

Various commercial modules with standard interfaces are readily available. These modules typically include readily available software and drivers, simplifying installation. This eliminates the need for complex Wi-Fi client chip designs.

For testing, it uses the Raspberry Pi Zero 2 W, a mature System-On-Chip (SoC) supporting 2.4 GHz 802.11b/g/n Wi-Fi. The module measures 6.5 cm$\times$3 cm and consumes approximately 2 W. The module communicates with an FPGA board via dual SPI buses for receive and transmit functions. PC testing demonstrates wireless uplink and downlink bandwidth of up to 22.03 Mbps. Additional multi-channel testing is necessary. With multiple channels, bandwidth is shared among endpoints connected to a single switch. However, independent endpoints spaced over 2 m apart typically do not interfere.

Test results reveal significant limitations of these solutions. Transmission density is constrained by the low frequency range and limited channel availability. Furthermore, high-power antenna miniaturization challenges further reduce transmission density. Despite these limitations, traditional Wi-Fi's ease of development makes it suitable for low-bandwidth applications like slow control data and small system testing.

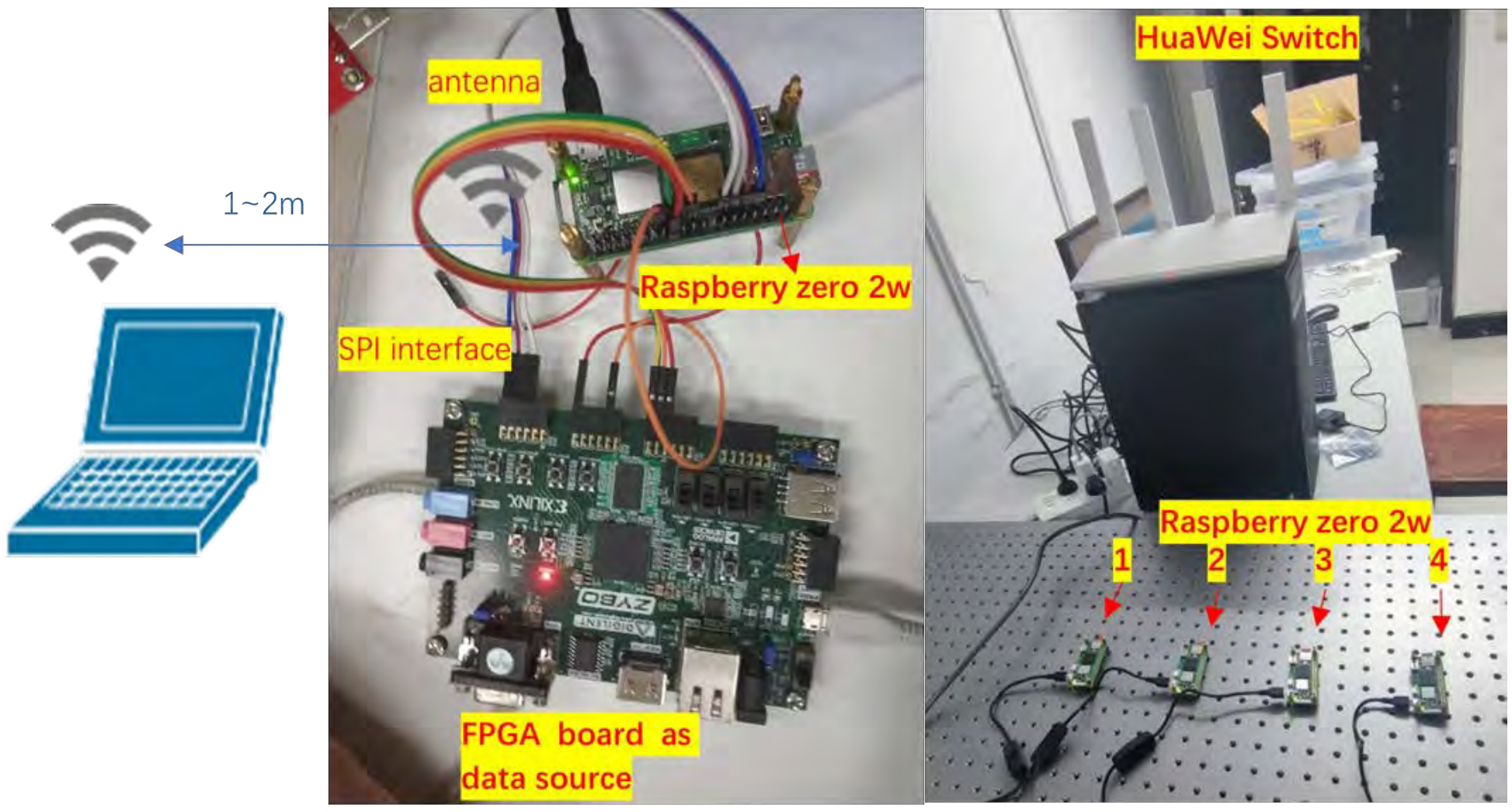

**Figure 11.14:** Test setup based on Raspberry Pi Xero 2W. The standalone and multi-channel testing results indicate the limitation in the bandwidth and power consumption.

#### 11.4.3.2 Millimeter wave

Frequency ranges are currently categorized into three bands: microwave (0.3–30 GHz), millimeter wave (30–300 GHz), and terahertz (0.1–10 THz). Future 6G and Wi-Fi 7 transmission technologies are anticipated to use these higher frequency bands. These frequency ranges represent a significant advancement over traditional Wi-Fi, providing theoretically robust solutions to bandwidth limitations. However, higher bandwidth capabilities come with increased power consumption and implementation costs.

As these technologies remain under development, commercial availability of relevant chips is now limited. Significant technical challenges and stringent foreign regulations pose substantial manufacturing barriers and cost constraints. This project will focus on collaborating with domestic research institutes and commercial entities to explore system development using existing or emerging millimeter wave and terahertz chips.

Testing uses the SK202 evaluation board, featuring ST-Microelectronics' commercial 60 GHz RF transceiver chip, the ST60A2. This chip supports data transmission rates up to 6.25 Gbit/s, making it ideal for high-speed wireless communication applications. The chip demonstrates efficient power consumption, measured at 44 mW (Tx) and 27 mW (Rx), while maintaining high performance.

1. Transmission distance and bandwidth test:

   The test setup involved positioning two test boards face-to-face, both connected to a com-puter. The inter-board distance was systematically varied to assess network connectivity, transmission bandwidth, and packet loss rate. The test results are detailed in Table 11.9. At distances below 5 cm, transmission speeds of up to 900 Mbps were achieved, demonstrating the chip's capability to sustain high data rates over short ranges. However, link establishment failed at distances exceeding 6 cm, highlighting the technology's limited effective transmission range.

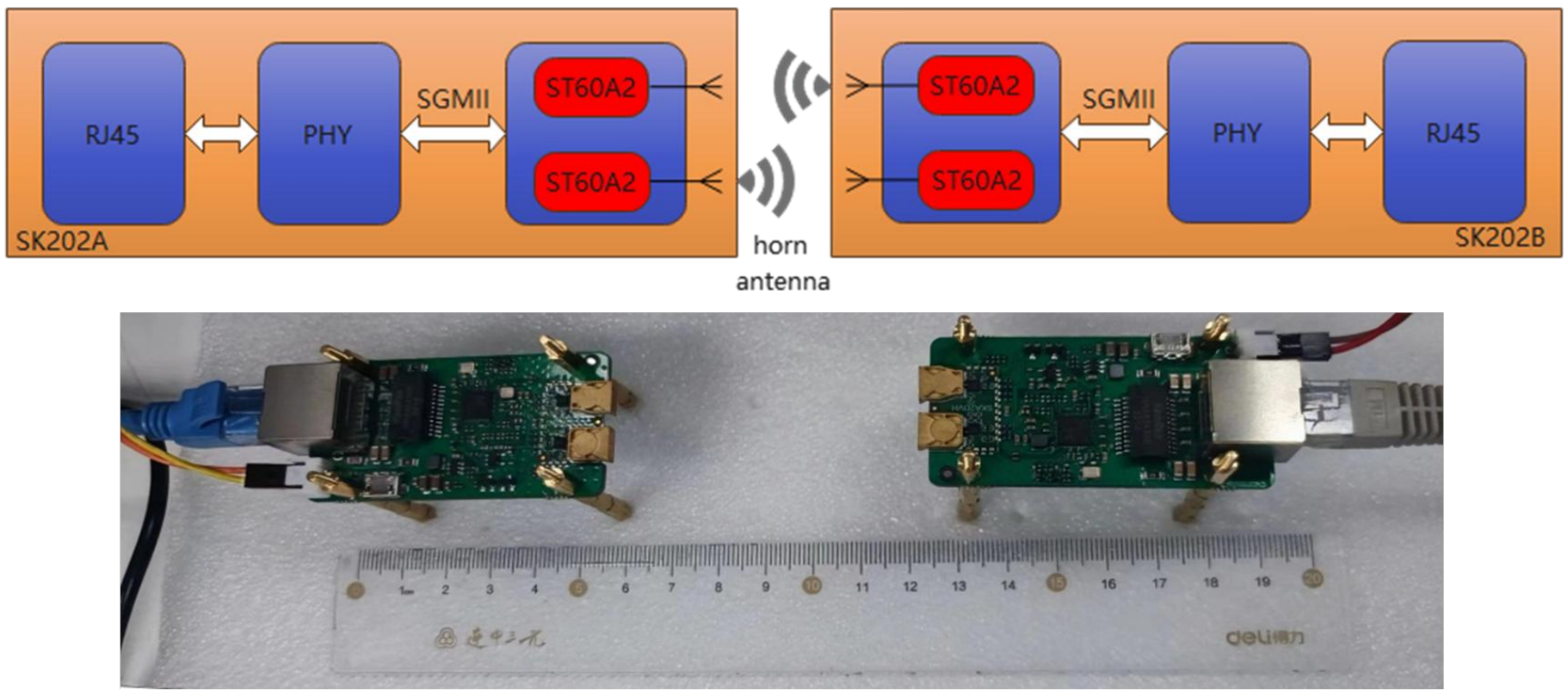

**Figure 11.15:** The mini-meter wave test with commercial Radio Frequency (RF) transceiver chip. The test setup involved positioning two test boards face-to-face, both connected to a computer.

**Table 11.9:** Transmission distance and bandwidth.

| Distance (cm) | Bandwidth (Mbps) | Packet loss rate |
|---|---|---|
| 1 | 914 | 0.031% |
| 3 | 917 | 0.061% |
| 5 | 915 | 0.05% |
| 6 | 913 | 0.13% |
| $> 6$ | No link | No link |

Test results indicate that commercial devices' transmission distance and speed fall short of required specifications, necessitating further research and development. Nevertheless, this commercial module can serve as a foundation for additional testing and analysis.

2. Penetration test:

The penetration capabilities of millimeter waves were evaluated through testing with various materials. The test results are summarized in Table 11.10. Testing demonstrated that metallic materials show strong millimeter wave blocking capabilities. This result is consistent with the fundamental principle of electromagnetic wave reflection by metals, which prevents wave penetration. Consequently, millimeter wave communication effectiveness is reduced by metallic obstacles.

In contrast, non-metallic materials like plastics, glass, and wood show varying degrees of millimeter wave permeability, enabling partial transmission. These materials help for the millimeter wave propagation, making them suitable for applications requiring unobstructed transmission.

**Table 11.10:** Penetration test at 3 cm distance.

| Material | Thickness | Penetration Ability |
|---|---|---|
| Paper | 2 mm | Yes |
| Plastic | 2 mm | Yes |
| FR4 PCB | 1.6 mm | No |
| Flex | 0.2 mm | No |

#### 11.4.3.3 Optical wireless communication

Free-Space Optical Communication (FSO), or optical wireless communication, transmits data using light waves across the visible, ultraviolet, and infrared spectra. This technology uses light properties to transmit information over distances without physical cables. Unlike traditional wired communication that relies on physical media such as fiber optics or copper, wireless optical communication employs air or vacuum as the transmission medium. FSO requires a clear line of sight between transmitter and receiver, as obstacles can disrupt the optical signal. The detector's optical path requires careful consideration. Optical frequencies provide significantly greater bandwidth than conventional radio frequencies, enabling higher data rates [28].

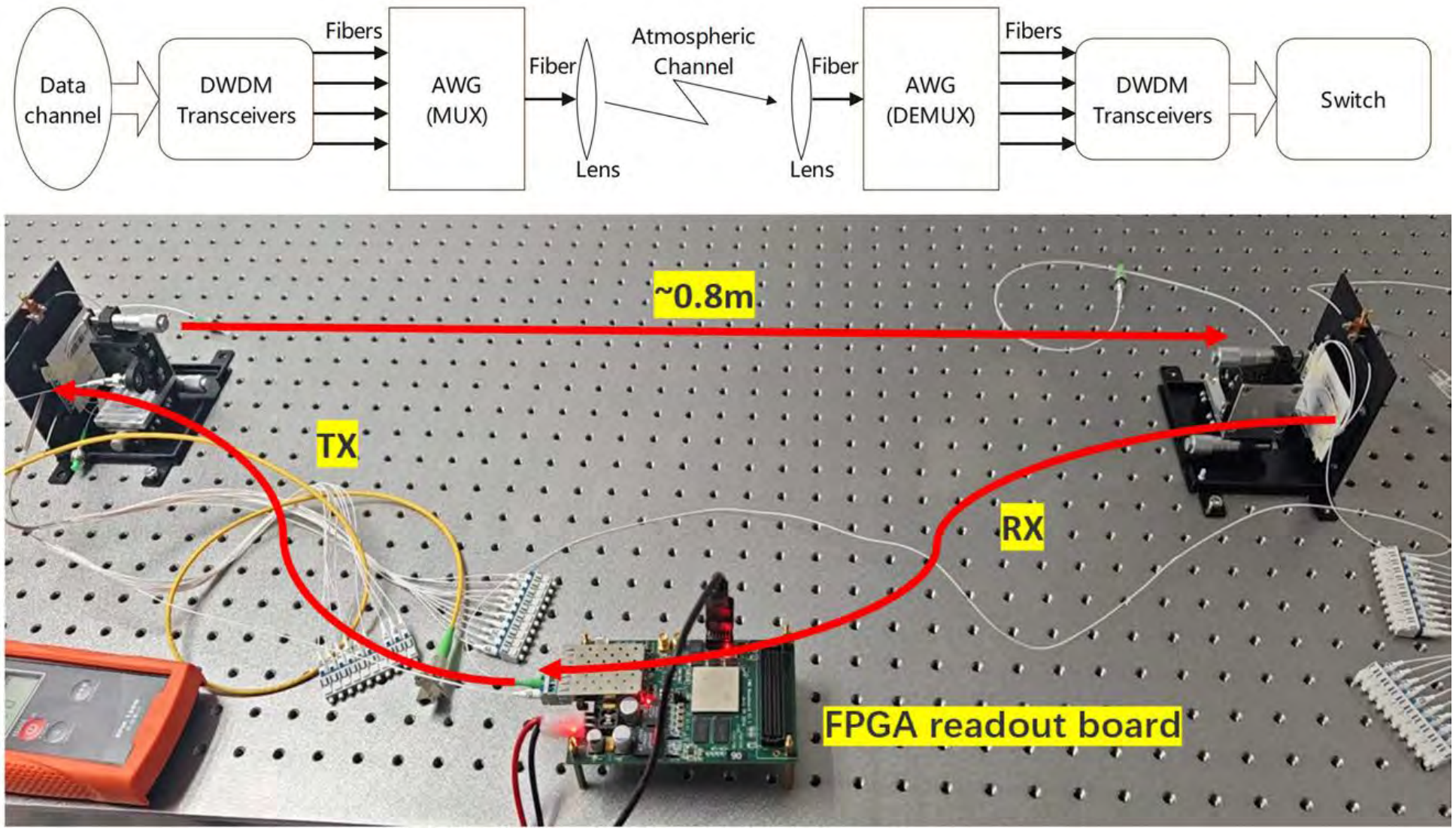

**Figure 11.16:** Optical wireless test setup. Data generation is handled by Xilinx's Gigabit transceiver (GTX) module, operating at 10 Gbps per channel. By integrating 12-channel DWDM onto a single optical link, the system achieves total data rates exceeding 100 Gbps.

A demonstration setup is established to validate wireless optical transmission. Figure 11.16 illustrates the setup's structure and physical components. The optical transceiver employs standard Small Form-factor Pluggable (SFP) modules. Data generation uses Xilinx's GTX module, operating at a single-channel line rate of 10 Gbps. Our configuration integrates 12-channel DWDM optical waves on a single optical link. This integration achieves data transmission rates exceeding 100 Gbps.

The demonstration setup successfully demonstrates the feasibility of high-capacity wireless optical data transmission. Achieving over 100 Gbps data transmission highlights the potential of advanced modulation and multiplexing techniques in wireless optical communication, enabling future high-speed data transmission applications across various environments.

## 11.5 Clocking systems

In large-scale detector systems, the clocking system is a critical component that significantly influences electronics performance. Furthermore, certain sub-detectors, particularly the OTK, aim to achieve high-precision timing measurements of up to 30 ps. This requires careful consideration of the clocking system design. This section presents a comprehensive analysis of clock distribution, encompassing both BEE-to-FEE and TDAQ-to-BEE directions.

### 11.5.1 Basic structure

The clock system must perform two primary functions:

1. Provide a reference clock for FEE: This clock establishes the baseline frequency for sampling, time measurement, and energy measurement across all sub-detectors. In the accelerator design, the clock frequency is 43.33 MHz, which sets the base clock frequency for the entire clocking system. Moreover, for some sub-detectors such as the OTK, the timing measurement aims for a precision at the level of 30 ps. This consequently leads to the requirement for the timing distribution network to achieve picosecond-level stability.
2. Provide absolute time stamping for detectors: This ensures the trigger system receives temporally correlated physical data. The baseline implementation uses optical links for Timing, Trigger, and Clock (TTC) information distribution, as illustrated in Figure 11.17. The timing distribution network's back-end consists of FPGAs with multiple transceivers.

Figure 11.17 illustrates the TTC distribution structure. Each subsystem features a TTC distribution crate that functions as a central communication hub with the trigger system. Within each crate, a top-level TTCD board receives TTC information from the trigger and clock system. The TTCD board distributes this information to all Level-2 TTCDs within the crate through the back-plane. Subsystems with fewer than 10 BEE crates do not require Level-2 TTCDs.

Level-2 TTCDs transfer TTC information to BEE crates via front-panel optical fiber connections. Each BEE crate contains a TTC distributor that receives information from Level-2 TTCDs and distributes it to all BEE boards via the backplane. This design ensures low-latency communication, crucial for maintaining precise time stamping across the system.

On the FEE, FEDI recovers the synchronized clock from BEE board data with a fixed latency. The latency between the TTCD and the FEE is determined by the cable length and the fixed delay introduced by the FEDI chip and GTX transceiver. This fixed delay can be calibrated using a physical calibration method.

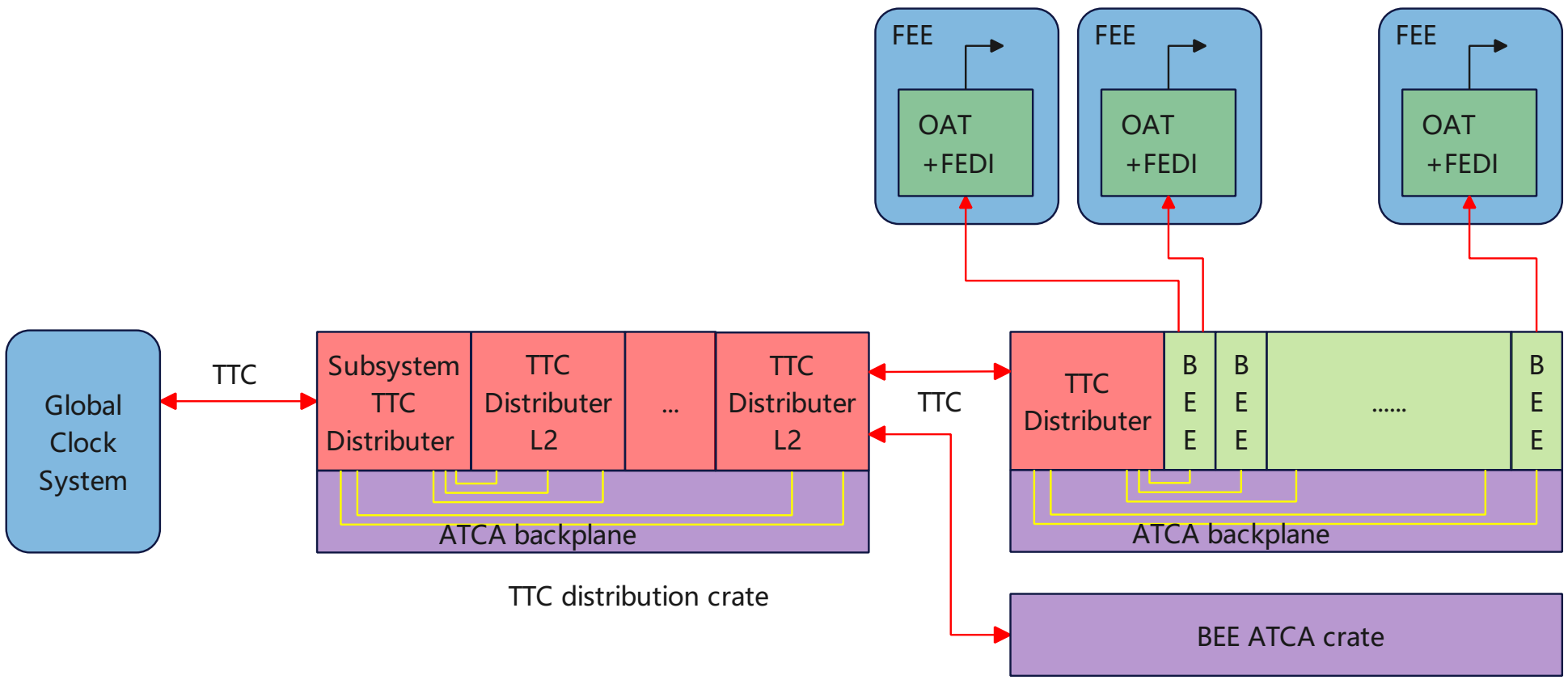

**Figure 11.17:** TTC distribution structure. Each subsystem has a TTC distribution crate containing a TTC Distributor (TTCD) board, which receives TTC signals. The TTCD board then distributes these signals via the crate's backplane to all Level-2 TTCDs.

### 11.5.2 TTC distributor

The TTC distributor module distributes TTC information through fan-out functionality [29]. It connects master and slave devices via a single optical fiber, simultaneously transmitting high-precision synchronous clock and timestamp data. It synchronizes the sender and receiver clocks.

Our clock synchronization system is based on the Precision Time Protocol (PTP), specifically the IEEE 1588 standard. This protocol enables precise clock synchronization across networked systems, ensuring highly accurate device synchronization [30]. The skew between the master clock and the final slave clock is maintained below 200 ps. According to the recent test report [31], the jitter of skew between master clock and the slave clock can achieve 12 ps. Figure 11.18 illustrates the timing compensation scheme concept.

Our approach centers on phase difference measurement using the DDMTD technique [32]. This technique employs advanced digital signal processing to precisely measure time offsets between the reference clock and incoming timing signals. Phase relationship analysis yields precise measurements essential for effective synchronization.

The measurement accuracy of DDMTD depends on the amplification factor. Specifically, a higher amplification factor results in greater measurement accuracy but also increases the time required for a single measurement. Additionally, a glitch phenomena impact DDMTD accuracy. When the amplification factor exceeds a certain threshold, the increase in glitches due to metastability issues suppresses further improvement in DDMTD accuracy. Under these conditions, the measurement accuracy reaches approximately 10 ps.

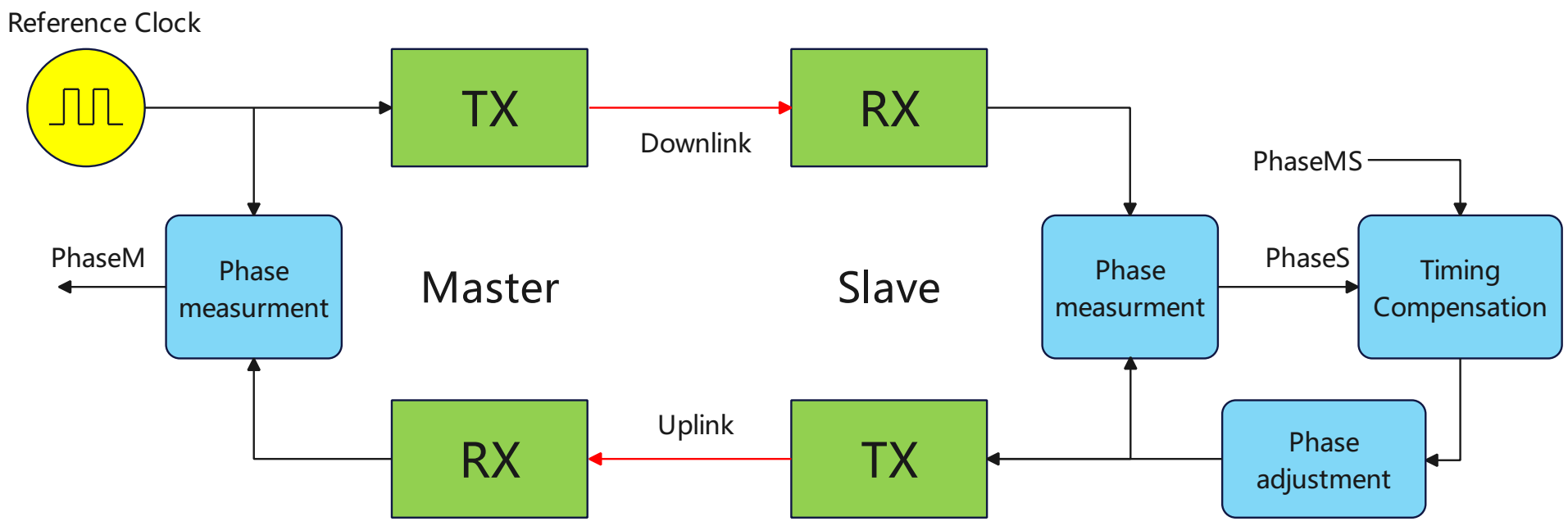

**Figure 11.18:** Concept of a timing compensation scheme. The proposed scheme employs the Digital Dual Mixer Time Difference (DDMTD) technique for precise time offset detection, implements dynamic phase compensation through IEEE 1588 PTP, and incorporates a jitter cleaner to effectively suppress high-frequency timing fluctuations.

Timing compensation is crucial in this process. Following PTP 1588 protocol principles, the timing compensation values is calculated based on phase measurement discrepancies. These values determine the required local clock phase adjustments for accurate reference clock alignment.

The phase adjustment process uses a high-performance jitter cleaner Si5345 to mitigate short-term jitter, reducing the Time Interval Error (TIE) jitter of output clock to less than 1 ps. The phase adjustment precision can achieve the 1 ps level. Effective jitter reduction ensures stable and reliable front-end timing signal delivery. This improves overall synchronization accuracy and enhances time-sensitive application performance.

During system operation, several factors affect clock synchronization accuracy. First, temperature variations cause changes in the electronic and fiber delays, making the coefficients in the adjustment scheme deviate from their ideal fixed values. This requires either calibration or the development of a new temperature compensation solution. Second, during system reboots, the uncertainty within the FPGA transceiver now reaches up to 200 ps. A new approach is developing to reduce this uncertainty, with a target of achieving 30 ps. Alternatively, physical calibration can be employed to correct this uncertainty, as it represents a fixed offset without power cycle.

## 11.6 Low voltage power system

The power supply is a crucial element in electronics systems, as it has a direct impact on their overall performance and functionality. Figure 11.19 provides the power distribution scheme for the low-voltage supply system of the Reference Detector. In the counting room, where there is no radiation and strong magnetic field, both the Commercial Off-The-Shelf (COTS) AC-DC power supply and the custom-designed DC-DC converter are housed. The AC-DC power supply converts 220 V to 380 V AC to 110 V DC. The DC-DC converter steps down the voltage from 110 V to 48 V.

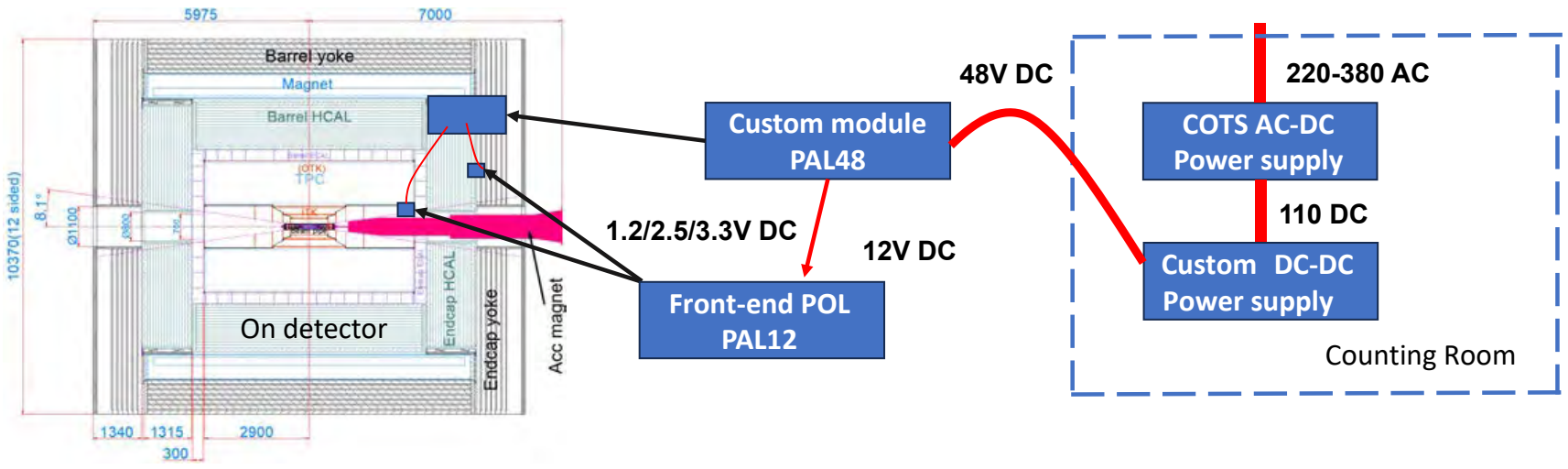

**Figure 11.19:** Low-voltage power distribution. The system uses a two-stage DC-DC conversion architecture: the PAL48 module reduces 48 V input to 12 V near the endcap to minimize high voltage transmission losses, while the PAL12 module on the front-end board further converts 12 V to the required 1.2 V, 2.5 V or 3.3 V.

### 11.6.1 Power supply distribution

Two main schemes for the FEE power distribution are Serial Powering [33, 34] and Parallel Powering [35, 36, 37]. While Parallel Powering is a more conventional and mature scheme, Serial Powering shows various advantages in many aspects, including less material, less cabling, and easier installation. However, for some sub-detectors like the VTX, stitched sensors will be used for the inner layers. This results in a common substrate covering a large area. Furthermore, there is also a possibility that a negative bias voltage are used for the substrate. Both of these factors introduce issues with adopting the Serial Powering scheme. Given that a common power distribution scheme is provided to all the sub-detectors, Parallel Powering is chosen as the baseline scheme for the FEE power distribution. Meanwhile, the field of Serial Powering in R&D is to be explored as an alternative scheme.

Since the AC-DC power supply is very mature as a COTS, the main design task for the power distribution system is focused to develop an efficient FEE power module under the Parallel Powering scheme. A two-stage DC-DC conversion architecture is implemented, based on PAL power modules. In the first stage, the PAL48 module receives 48 V input from the long cable extending from the counting room. This module steps down the voltage to 12 V. The PAL48 module is strategically installed near the endcap, reducing high-voltage cable length to minimize voltage drop and ensure efficient power delivery.

In the second stage, the PAL48 module's 12 V output is delivered to the front-end board via finer, more manageable cables. On the front-end board, the PAL12 module converts the 12 V supply into specific lower voltages: 1.2 V for analog readout chips and 2.5 / 3.3 V for digital transmission chips.

This two-stage conversion approach optimizes power distribution efficiency and enables a compact design that minimizes radiation exposure to front-end sensitive electronic components. It ensures precise voltage regulation, maintaining detector readout electronics performance integrity in radiation-prone areas' challenging environmental conditions.

## 11.6.2 BUCK DC-DC converter

As critical components in detector power systems, buck converters reduce the power loss from high-voltage transmission to the lower voltages required by FEE [38]. High-voltage, low-current power transmission reduces cable count, minimizes voltage loss, and enhances detector system power efficiency. However, key buck converter modules—reference, pre-buck, and on-chip LDO—are susceptible to radiation effects, potentially causing level drift and circuit failures. Therefore, implementing module-specific radiation-hardening measures is crucial.

The buck-type DC-DC converter's location within the detector, coupled with high radiation and magnetic field exposure, precludes the use of magnetic-core inductors [39]. Area and space limitations further restrict energy storage inductance values to relatively low levels. However, FEE demands stringent power noise performance, requiring exceptionally low ripple voltage. Traditional methods of increasing inductance values to reduce ripple voltage are no longer viable. Instead, ripple voltage minimization is achieved through increased switching frequency. The proposed DC-DC controller targets switching frequencies in the MHz range. Moreover, detector operation radiation environments can cause soft errors or hard failures, including burnout, in silicon-based circuits. Therefore, the DC-DC controller design must withstand both ionizing and non-ionizing radiation effects.

Based on research and literature review of radiation-hardened buck-type DC-DC converters, the controller system structure will be investigated. The investigation will focus on parallel output current capabilities, high switching frequency, efficient circuit structures, and specific implementation methods. System-level modeling and simulation will optimize the controller's system architecture.

### 11.6.2.1 Design specification

From the FEE requirements, most of the front-end chips operate at 1.2 V. Some components may require 2.5 V or 3.3 V power. Given the short on-time for high conversion ratios, the proposed DC-DC conversion employs a two-stage approach.

**Table 11.11:** DC-DC converter specification.

| Parameter | Specification |
|---|---|
| Input Voltage Range (V) | 36 ~ 48 |
| DC Output Voltage (V) | 12 and 1.2 (or 1.8, 3.3 optional) |
| Maximum Output Current (A) | 1, 10 |
| Efficiency (%) | 80 ~ 85 |
| Output Ripple Voltage (mV) | < 5 |
| Protection Features | Over voltage, over current, short circuit protection |
| Operating Temperature Range (°C) | -10 ~ +65 |
| Dimensions | 50 mm × 20 mm × 6.7 mm (including shielding) |

The specifications of the DC-DC converter are listed in Table 11.11. Considering the most critical requirement from the VTX, the power chips must withstand an estimated radiation dose of 5 Mrad (Si) per year. Additionally, they must be resistant to a strong magnetic field of 3 T. This leads to a specific design for the inductor components, while a high efficiency of the DC-DC converter must be preserved.

The floorplan of the PAL module, which includes a customized DC-DC converter, is shown in Figure 11.20. The air-core inductor occupies a significant area, while the size of the capacitors also increases to accommodate high voltage.

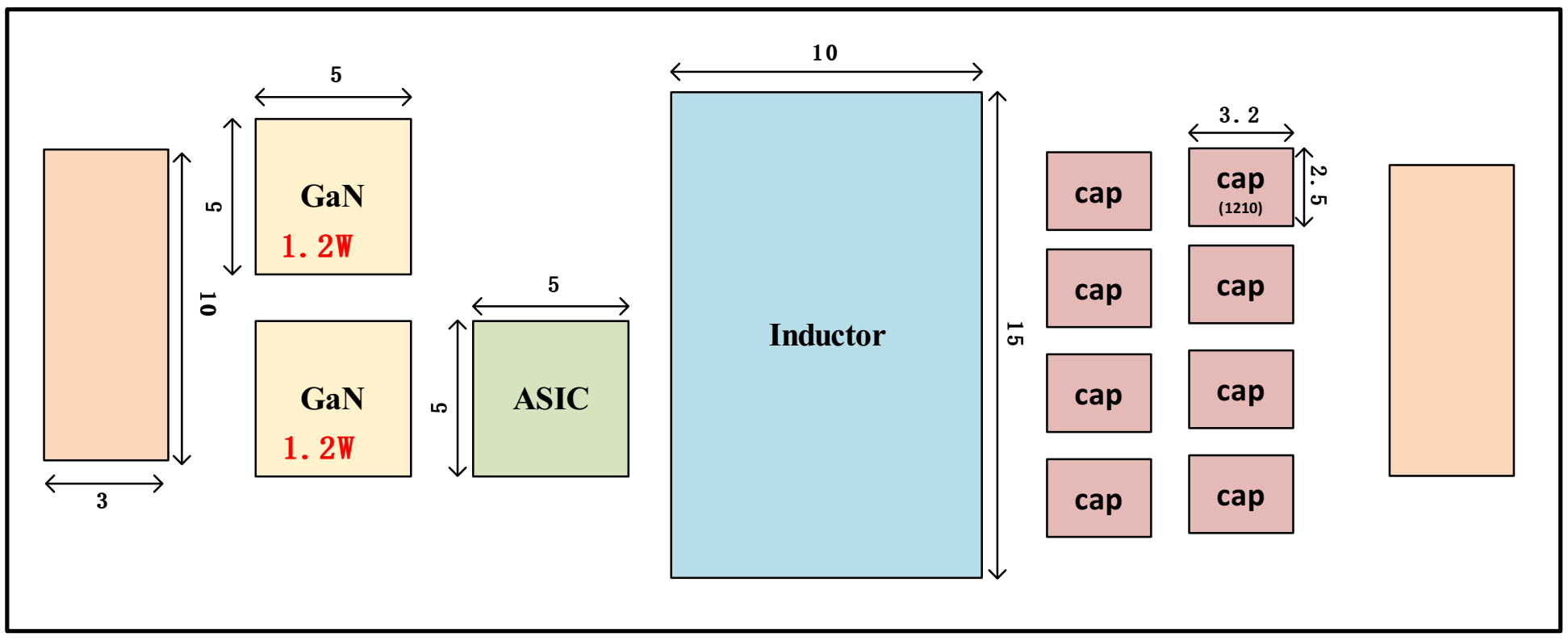

**Figure 11.20:** Floorplan of PAL DC-DC module for heat simulation. ASIC is the DC-DC controller, which can switch two GaN transistors to adjust the output voltage.

#### 11.6.2.2 Structure of DC-DC converter

The proposed DC-DC converter structure, shown in Figure 11.21, operates in voltage-mode Pulse Width Modulation (PWM). The output voltage and current are sampled and compared with a reference voltage, generating PWM pulses to adjust the switch on-time and form a negative feedback loop. The PWM controller includes an error amplifier, a comparator, a driver, a dead-time controller, a ramp generator, an oscillator, and soft-start circuitry. However, this mode faces challenges such as loop frequency compensation and stability issues due to parasitic elements, and suboptimal transient performance. To address these, current-mode PWM is considered as a backup structure.

GaN power transistors have seen rapid development in recent years, characterized by a wide band gap. Compared with traditional power transistors, GaN offers low on-resistance, high switching frequency, and high current density [40]. Their low switching loss enhances power efficiency. Given these advantages, GaN transistors hold great promise for use in switching power applications. The designed DC-DC converters will therefore adopt GaN transistors.

Since the detectors are surrounded by a solenoid magnet, conventional inductors with magnetic cores cannot function in such a strong magnetic field. Air-core inductors appear to be the only viable option, although achieving high inductance with a compact size is challenging [39]. The space available for the power module is severely limited by each detector. Commercial air-core solenoids are available and relatively small, but their open magnetic field can cause unexpected Electro-Magnetic Interference (EMI). The air-core toroid is proven to be a good candidate. Therefore, customized toroids are studied. Moreover, shielding methods should also be studied considering the size, material budget, and magnetic field constraints.

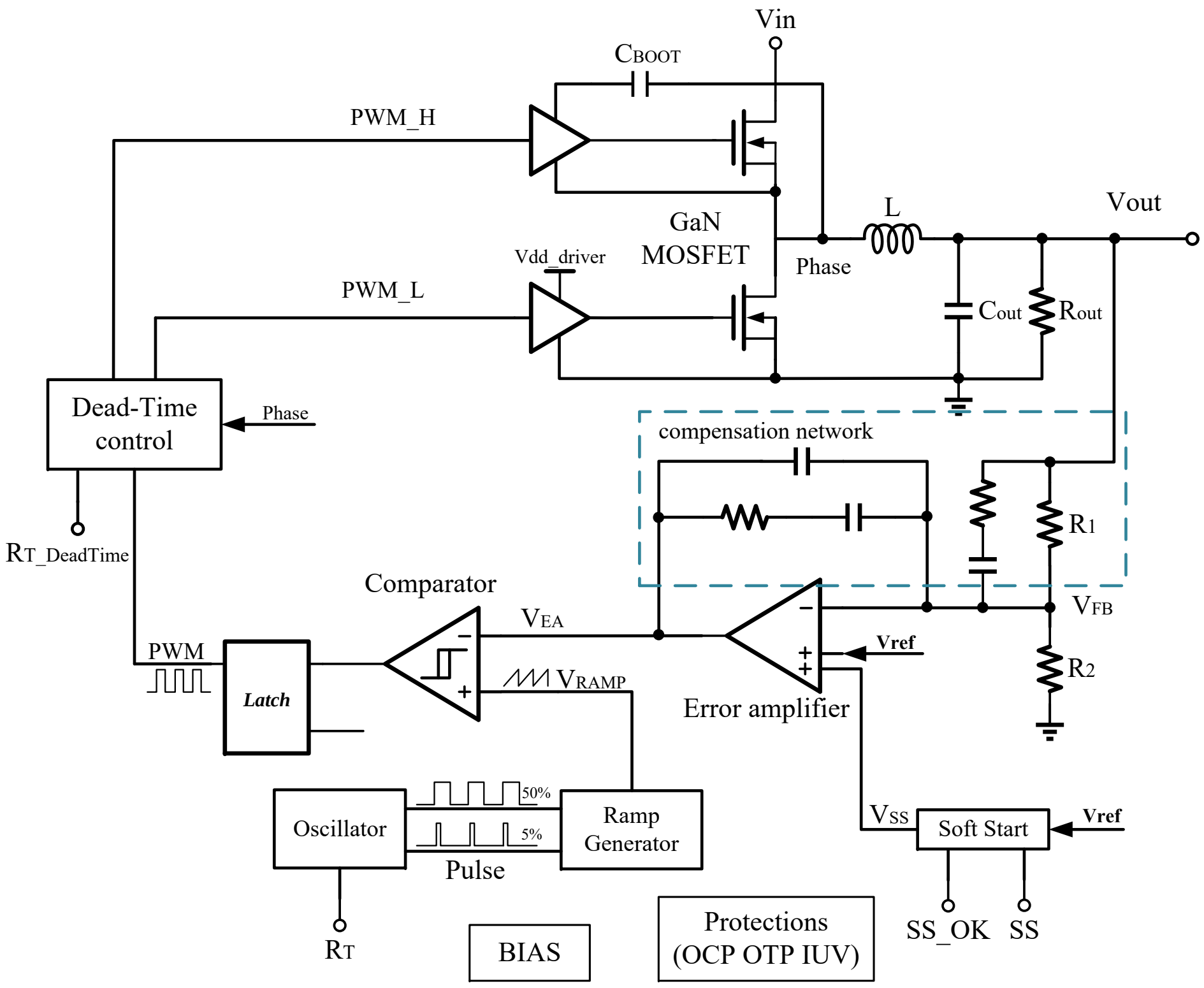

**Figure 11.21:** Structure of DC-DC converter with voltage mode of PWM. Error voltage is compared with a fixed-frequency ramp voltage to generate a voltage pulse, whose width is adjusted by the feedback voltage. GaN transistors are planed to be used to improve power efficiency.

#### 11.6.2.3 Development plan

The development plan for the low-power system involves two parallel R&D approaches. First, the front-end power module, which features a compact size and magnet-proof design, will be demonstrated before 2027. This module's DC-DC controller is based on a commercial chip, whose interface definition will serve as a reference for the self-developed PAL controller. The small-sized module will be provided to the sub-detectors for prototype validation. Second, the core DC-DC controller will be designed and integrated into the module before 2029, making the front-end module radiation-tolerant as well.

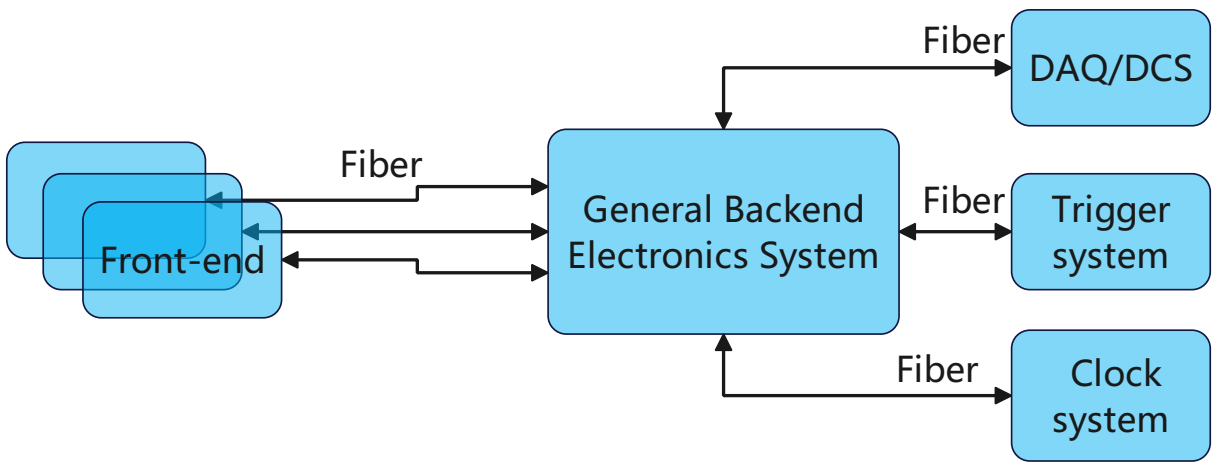

**Figure 11.22:** General BEE system connection. The BEE receives raw data from front-end, dynamically generates and dispatches trigger signals to the trigger system, receives trigger decisions, and transmits validated data to the DAQ system. It maintains strict synchronization with the global clock distribution network and coordinates operational parameters via the Detector Control System (DCS).

## 11.7 Back-end electronics

The BEE system serves as the bridge between the FEE and higher-level data processing systems. Upon receiving raw data from the FEE, the BEE performs several critical functions to ensure effective data acquisition. The relationships between the BEE system and other systems are shown in Figure 11.22. The key functions of the system are:

1. Data Reception: Receives raw data from the FEE.
2. Initial Data Processing and Compression: Processes the incoming data, filters out noise, enhances signal quality, and performs data compression.
3. Trigger Signal Generation: The BEE gathers detector hit and time information required by the trigger system. These signals are crucial for initiating specific operations and ensuring that data acquisition timing aligns with relevant events. Compared with the FEE-Trigger scheme, the back-end trigger scheme concentrates data on the BEE board, benefiting from its large buffer space, more powerful FPGA, and higher communication speed with the TDAQ system. This approach also makes the system more flexible in terms of trigger strategy due to the FPGA's hot-programming capability.
4. Data Caching: The back-end system temporarily caches raw data for a defined duration, ensuring access to this information while waiting for the trigger system's decision. This feature is vital for maintaining data integrity and accessing relevant data during decision processing.
5. Final Trigger Decision Processing: After a predetermined delay, the trigger system sends a final trigger decision back to the BEE. Based on this decision, the back-end filters the buffered data to select information that meets the final trigger criteria.
6. Data Packaging and Transmission: Once the relevant data is identified, it is packaged and sent to the DAQ system for comprehensive analysis and long-term storage.
7. Communication with the DCS: The BEE system exchanges status monitoring and control signals with the DCS, ensuring effective oversight and management of both the front-end detectors and the BEE. This communication is essential for maintaining operational integrity and system performance.

8. Synchronization Clock Distribution: The BEE communicates with the synchronization clock distribution system to ensure precise timing across all components. This synchronization is critical for accurate data acquisition and processing, especially in high-speed applications.

Overall, the BEE system is responsible for critical data management processes that enhance the efficiency and effectiveness of the data acquisition system. Its ability to generate trigger signals, perform initial data processing, maintain synchronization, and ensure smooth communication between various components reinforces its pivotal role in high-performance data acquisition environments.

## 11.7.1 BEE hardware design

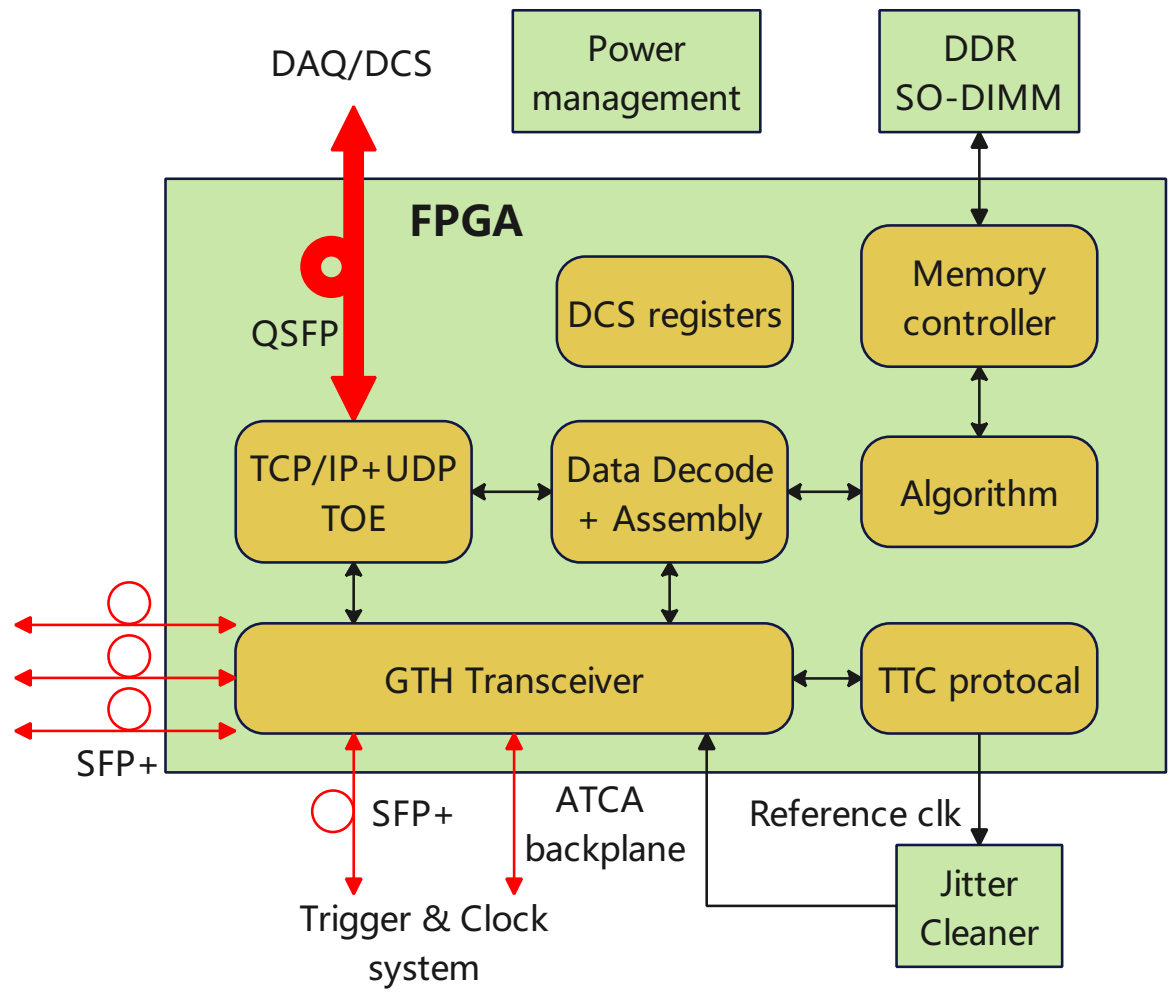

**Figure 11.23:** General BEE board structure.

The hardware components of the custom BEE are illustrated in Figure 11.23 and primarily consist of the following key sections:

1. FPGA: Serving as the core processing unit of the BEE board, the FPGA controls and executes the majority of the BEE functions.
2. Clock Jitter-Cleaner: This component ensures that timing signals meet the requirements for high-speed data connections.
3. Dynamic Random Access (DDR) Memory: DDR provides necessary memory resources for data buffering and temporary storage during processing.
4. Power Management: This section manages power distribution to ensure that all hardware components receive stable and adequate power.

From preliminary evaluations conducted with the FEE, it is estimated that the fiber optic line rate for data transmission to the BEE is about 11 Gbps [41]. To accommodate this high data throughput, a custom communication protocol is implemented. Additionally, connections to the trigger system and the clock distribution system also use fiber optic links operating at the same 11 Gbps line rate. In contrast, communication with the DAQ and DCS uses 40 Gbps Quad Small Form-factor Pluggable Plus (QSFP+) optical modules to enable robust and high-speed connectivity.

These requirements place stringent demands on both the speed and the number of high-speed serial transceivers integrated into the FPGA. Furthermore, the FPGA is tasked with performing initial processing on the raw data, necessitating a considerable level of processing capability.

To ensure that our system meets these operational criteria, a comparative analysis is conducted of the performance specifications of several widely used FPGA models currently available on the market. The results of this comparison are summarized in Table 11.12, it leads to the initially selection of the XC7VX690T-2FFG1158C as our preferred FPGA model. This choice is based on its capability to satisfy the high-speed communication requirements and processing demands of the BEE system, ensuring efficient operation and data integrity in a high-throughput environment.

**Table 11.12:** Comparison of common FPGA performance parameters.

| **FPGA** | **XC7K325 -2FFG900C** | **XCKU040 -2FFVA1156E** | **XC7VX690T -2FFG1158C** | **XCKU115 -2FLVF1924I** | **KU060 –** |
|---|---|---|---|---|---|
| **Logic Cells (k)** | 326 | 530 | 693 | 1,451 | 725 |
| **DSP Slices** | 840 | 1,920 | 3,600 | 5,520 | 2,760 |
| **Memory (Kbits)** | 16,020 | 21,100 | 52,920 | 75,900 | – |
| **Transceivers** | 16 (12.5 Gb/s) | 20 (16.3 Gb/s) | 48 (13.1 Gb/s) | 64 (16.3 Gb/s) | 32 (16.3 Gb/s) |
| **I/O Pins** | 500 | 520 | 350 | 832 | 624 |

## 11.7.2 Prototype performance

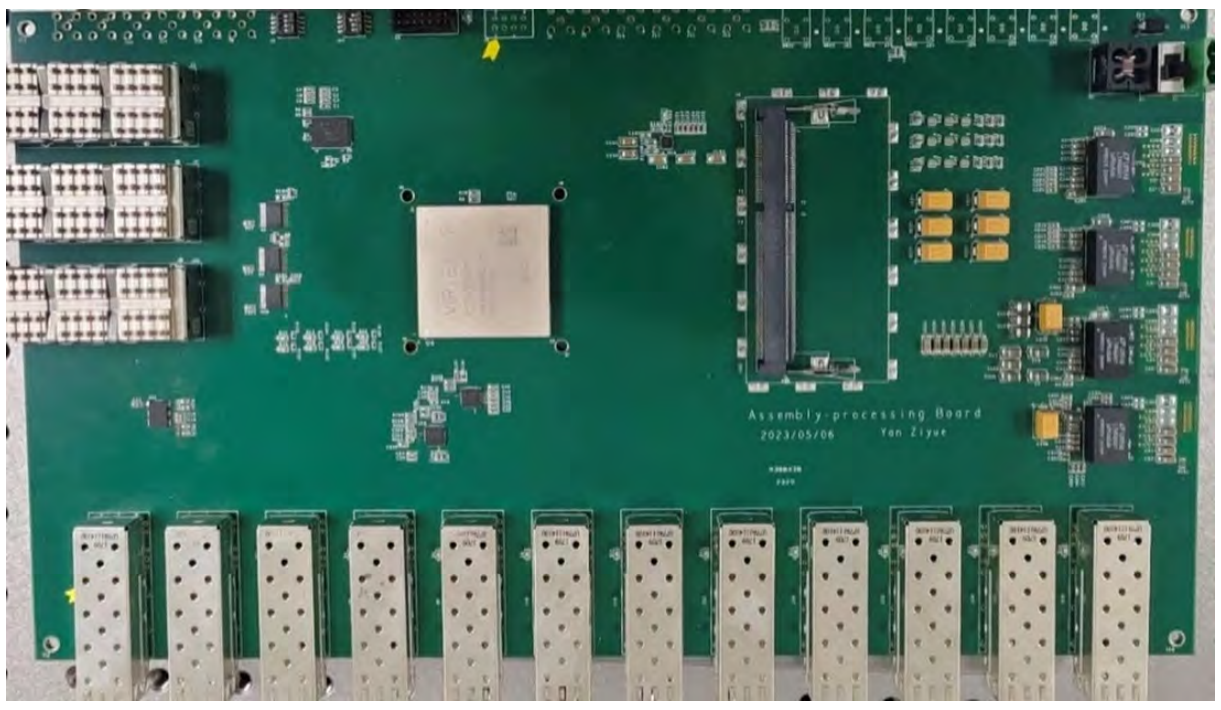

**Figure 11.24:** BEE board prototype. The board is equipped with twelve SFP+ interfaces and two QSFP+ interfaces.

The physical overview of the data aggregation processing board is presented in Figure 11.24. The board is equipped with twelve SFP+ interfaces and two QSFP+ interfaces, which are utilized for connecting the GTH transceiver modules of the FPGA. These interfaces serve as the communication pathways for data transmission between the data aggregation processing board and the upper computer, and the readout system that encompasses multiple readout channels. The total power consumption is approximately 40 W.

The GTH transceivers offer higher data rates and improved performance compared with GTX transceivers, making them suitable for applications that demand higher data transmission bandwidth. Additionally, the board is equipped with a 204-pin DDR3-SODIMM module, which is intended for large data buffering during the algorithm processing phase. This configuration ensures that the board can efficiently manage the substantial data flow necessary for effective data aggregation and processing.

Loopback tests were conducted to assess the transmission quality of 12 GTH links. The loopback tests used a PRBS pattern at a rate of 120 Gbps (single channel at 10 Gbps, PRBS $2^{31}-1$). The tests were carried out until the error rate reached a Bit Error Rate (BER) of $10^{-15}$ (the Ethernet transmission standard for 10 Gbps requires an error rate of below BER $10^{-12}$).

## 11.8 Back-end crates, racks and cabling

The requirements for crates and racks related to electronics systems primarily encompass data transmission, low-voltage power supply, and high-voltage power supply. After investigating the currently available industry standards, it is decided to adopt the Advanced Telecommunications Computing Architecture (ATCA) standard with a rack height of 42U. This standard offers a relatively advanced and mature design that meets both current and future needs.

Since the counting room is outside the radiation environment, there is no further concern about the tolerance of the crates and racks to radiation. It is natural to choose COTS instruments for all data crates, low-voltage AC–DC power supplies, and high-voltage power supplies. Considering that the performance of COTS equipment will undoubtedly be enhanced year by year, only the constraint of interface with the mechanical systems is analyzed in this section. The specific products can be selected from mainstream COTS offerings during the construction phase, ensuring that their performance will fully meet present requirements in future.

### 11.8.1 Crates and racks

As previously discussed, the data rate of a single optical fiber channel is 11.09 Gbps. Considering the ATCA standard for designing back-end data crates, a 9U ATCA board can handle 16 input data channels, while a single crate can accommodate a maximum of 14 plugins. By using two plugins for control, one for clock synchronization (as TTC), and one reserved for future upgrades, each crate can accommodate up to 10 back-end boards.

To fully utilize the ATCA rack space, 2U of height is allocated to the cooling space for each crate. Therefore each rack can accommodate three back-end crates, with the remaining space reserved for data exchange switches for the DAQ system.

The power to the FEE provides by 110 V commercial DC power supply crates. Based on the counts of the FEE modules, it is decided to divide the crates into two categories to save the total volume and, consequently, the overall cost. The first category is a high-power crate with a single-channel power limit of 100 W and a capacity of 48 channels per crate. The second category is a low-power crate with a channel limit of 40 W and a capacity of 96 channels per crate. Both types of crates have a height of 3U, and each crate requires 1U of cooling space. Therefore, a 42U rack can accommodate 10 crates. Furthermore, the power supply for these crates can be provided by a dedicated crate with a height of 6U and a capacity of 10 output power channels, with a total power capacity of 60–70 kW per crate. A 42U rack can accommodate five of these power supply crates.

The high voltage for most sub-detectors ranges from 10 to 200 V. However, the power consumption of these high-voltage power supplies is relatively low. To meet the requirements of the detector high-voltage power supplies, the high-voltage supply crates are designed with a maximized number of channels. Each crate provides 224 channels, each with individually adjustable voltage values. With a height of 8U and an additional 2U for cooling, each rack can accommodate four crates. For detectors that require even higher voltages (such as the TPC), the crate height increases to 10–12U, allowing three crates per rack.

### 11.8.2 Overall interface with FEE

For sub-detector front-end modules, the ”one optical, one power” approach is adopted for data and power connections. Data, including uplink, downlink, slow control, and clock, is transmitted through a fiber bundle based on the front-end data interface. The number of fiber channels is determined by the front-end module’s data rate, with a maximum of four uplink fibers (at 44 Gbps). The FEE power consumption includes the electronics chips and data transmission chipsets. Considering the DC-DC module efficiency of 85%, an additional 18% is added to the total power consumption of the FEE based on the front-end chips’ power usage.

Furthermore, to simplify and unify global wiring, all cables and fibres are routed via photoelectric composite cables to the counting room. Each composite cable comprises a pair of a low voltage power cable with its respective ground wire, a pair of a high-voltage cable with its ground, and one fibre bundle, as shown in Figure 11.25. The cable diameter is constrained to 5 mm, driven by the most challenging space requirements of detectors, such as the OTK and HCAL. Due to this constraint, two types of low-voltage power cables are selected to accommodate varying sub-detector requirements. Specifically, AWG-12 (American Wire Gauge (AWG)) cable (diameter: 2 mm; maximum current: 15 A) is used for high-current applications, while AWG-18 cable (diameter: 1 mm; maximum current: 3 A) is employed for standard cases. Following the ”one optical, one power” modular design principle, composite cables are allocated to individual detector modules: fibres connect directly to OAT modules, while power terminates at PAL

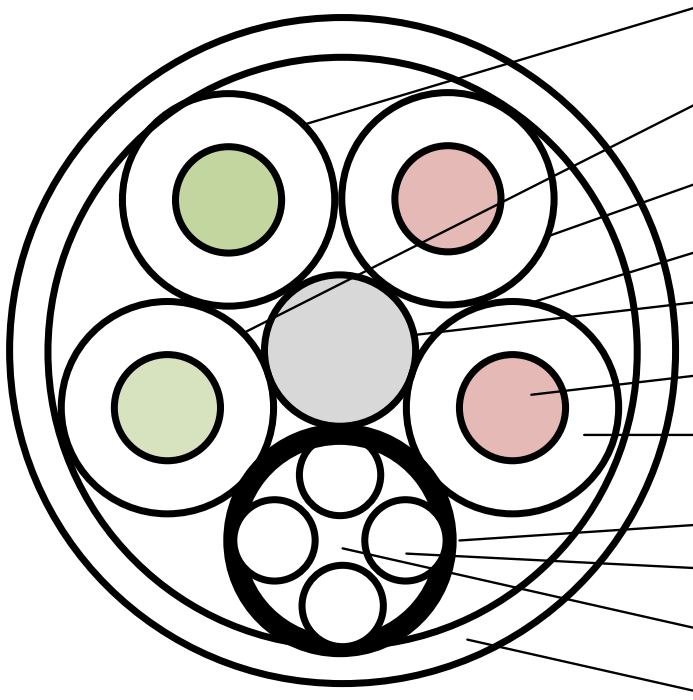

**Figure 11.25:** Sketch of a photoelectric composite cable. It comprises a pair of low-voltage power conductor together with its ground wire, a pair of high-voltage conductor with its ground wire, and a fiber bundle. All conductors and fibers are enclosed within an outer jacket to form a single cable.

modules. Inter-detector cabling follows specific design implementations.

Regarding electronics connectivity to the counting room, composite cables may extend up to 100 m in length. To help detector installation, power sockets and fibre reshuffles serve as patch panels, separating cables and fibres into on-detector and off-detector segments. These panels are positioned at the outer radius of the yoke to ensure adequate clearance. Detailed mechanical designs are discussed in Chapter 14.

### 11.8.3 Consideration of the counting room

The analysis of cables and crates for sub-detector electronics is summarized in Tables 11.13 and 11.14. The electronic system requires 83 data crates, 175 power supply crates, and 62 detector high-voltage crates, corresponding to 27 data racks, 24 power racks, and 21 detector high-voltage racks, respectively. Based on the raw data volume and estimated trigger rate, the TDAQ system needs about 100 general trigger boards (see in Chapter 12), equivalent to about 20 ATCA crates or 10 racks. For the power supply of the 92 racks themselves, ten high-voltage AC crates are needed, corresponding to two high-voltage power racks. Thus, the minimum total number of racks required is 94. Given the limited and non-expandable underground space, potential data volume fluctuations, and possible upgrades in the second operation phase, preliminary considerations suggest doubling the number of crates for redundancy, which results in a total capacity of approximately 200 crates. Assuming a crate size of 0.5 m × 0.5 m, with cooling space between crates, a center-to-center distance of 2 m (1.5 m adjacent spacing), and 2.5 m front-to-back, the electronics area can be organized across two floors, each with an area of 500 m$^2$(total 1000 m$^2$). Each floor can accommodate 100 crates arranged in a 10 × 10 grid, resulting in a floor size of 25 m × 20 m.

**Table 11.13:** Electronic system cabling and crates on data. The usage of fibers, common back-end boards, crate usage, and so on are summarized in this table. The detailed numbers are derived from the sub-detector design related to the modules and readout architecture. The fibers summarized here is the fiber channels to the BEE, while the "Fibers per Module" indicate the detailed fiber cores in the fiber bundle of the corresponding composite cable. More specifically, for the VTX, ITK, and OTK, the number of fibers used per module may vary depending on the location and the estimated beam-induced background. For the ECAL, since it cannot be replaced during the entire operational period, the maximum possible data transmission capability is ensured from the very beginning. Therefore, four fibers are installed for each module. However, only two of them are used for data transmission during the Higgs and Low Luminosity $Z$ modes, based on the expected data rate. After switching to the High Luminosity $Z$ mode, all four fibers are enabled to achieve the maximum data transmission capability. The numbers in brackets refer to the second operational phase, while the regular numbers refer to the baseline case.

| **Detector** | **Max Data Rate/ Fiber (Gbps)** | **Fibers/ Module** | **Fibers** | **BEEs** | **Crates** |
|---|---|---|---|---|---|
| VTX | 8 | 1–2 | 96 | 6 | 1 |
| TPC | 0.1 | 1 | 496 | 32 | 4 |
| ITK-Barrel | 2.88 | 2–3 | 376 | 24 | 2 |
| ITK-EndCap | 4.4 | 2 | 148 | 6 | 1 |
| OTK-Barrel | 1.4 | 1 | 880 | 55 | 4 |
| OTK-EndCap | 1.4 | 1–2 | 544 | 34 | 4 |
| ECAL-Barrel | 4.8 | 2 (4) | 960 (1,920) | 60 (120) | 6 (12) |
| ECAL-EndCap | 4.8 | 2 (4) | 448 (896) | 28 (56) | 4 (8) |
| HCAL-Barrel | 0.14 | 1 | 5,568 | 348 | 36 |
| HCAL-EndCap | 1.75 | 1 | 3,072 | 192 | 20 |
| Muon-Barrel | 0.01 | 1 | 24 | 2 | 1 |
| Muon-EndCap | 0.01 | 1 | 16 | 1 | - |
| **Total** (Upgraded) | - | - | 12,628 (14,036) | 788 (876) | 83 (93) |

## 11.9 Previous electronic system R&D for particle physics experiments

The electronics research team comprises China's leading experts in high-energy physics, covering a comprehensive spectrum of research directions, including front-end ASIC design, back-end high-speed readout, system-level electronics, and advanced electronics technologies.

Building on its foundational work with BESIII, the team has successfully spearheaded and executed the development of electronics systems for several major high-energy physics facilities in China, such as the BESIII experiment, the Daya Bay Neutrino Experiment, LHAASO, and JUNO. These systems have demonstrated long-term operational stability, providing a robust foundation for the acquisition of high-quality physical data. Furthermore, the team has made significant contributions to international collaborations, including ATLAS, CMS, and next-generation Enriched Xenon Observatory (nEXO).

**Table 11.14:** Electronic system cabling and crates on power. The usage of cables, front-end low voltage power, high voltage usage, crate usage, and so on are summarized in this table. The detailed numbers are derived from the sub-detector design related to the modules and readout architecture. The ”Power Channels” are identical for both low-voltage and high-voltage, due to the strategy of utilizing the composite cable. Consequently, this numbers are also the composite cable numbers for each sub-detector. Only the global cables to the counting room are considered in this table, while the inter-detector cables after splitted from the composite cable are discussed in the corresponding sections of each sub-detector. For the high voltage, most of the sub-detectors suffer from radiation damage, and the high voltage must be increased to compensate for the resulting loss in efficiency. The table summarizes the maximum values rather than the nominal ones, to illustrate the requirements for the high-voltage power supplies.

| Detector | Power Channels (Composite Cables) | Total LV Power (kW) | LV Crates | Max. HV (V) | HV Crates |
|---|---|---|---|---|---|
| VTX | 96 | 0.45 | 2 | -10 | 1 |
| TPC | 496 | 16 | 6 | -500 | 4 |
| ITK-Barrel | 128 | 26.6 | 6 | 300 | 2 |
| ITK-EndCap | 74 | 13.8 | 4 | 300 | 2 |
| OTK-Barrel | 880 | 195 | 42 | 500 | 4 |
| OTK-EndCap | 288 | 60 | 14 | 500 | 2 |
| ECAL-Barrel | 480 | 17.5 | 5 | 60 | 4 |
| ECAL-EndCap | 224 | 9.5 | 4 | 60 | 2 |
| HCAL-Barrel | 5,568 | 66.1 | 58 | 60 | 26 |
| HCAL-EndCap | 3,072 | 41.3 | 32 | 60 | 14 |
| Muon-Barrel | 24 | 0.76 | 1 | 60 | 1 |
| Muon-EndCap | 16 | 0.45 | 1 | 60 | - |
| **Total** | 11,346 | 447.46 | 175 | - | 62 |

Now, the team is actively expanding its international engagement. Through participation in CERN R&D Collaboration on electronics and on-detector processing (DRD7), more international research groups are welcome to join this effort and collaborate on the future development of advanced electronics systems.

## 11.10 Main schedule for future works beyond the Reference Detector TDR

Developing all the ASICs in a very limited time, including front-end ASICs and common ASICs, is undoubtedly a challenging task. Table 11.15 summarizes the current status of all the target ASICs for the electronic system. Fortunately, in previous R&D efforts, most of the critical blocks of these ASICs are verified. Some even have prototype systems that can be subjected to beam tests. The remaining tasks are mainly focused on integrating all necessary blocks, defining the main functionality of the digital circuits, and tuning the performance to meet the final requirements, with no show stoppers identified. Moreover, ”sister chips” can be found for almost all the ASICs. Only a representative of each is listed. These chips have similar functionality to the final goal, although some aspects of their performance may not be fully satisfied. However, they can be very helpful during the early stages of detector prototyping. They enable the R&D of the sub-detectors and ASICs to progress in parallel.

**Table 11.15:** Current status of all the ASICs. According to their readiness, the maturity can be divided into nine levels: 1: basic principles observed; 2: technology concept formulated; 3: experimental proof of concept; 4: technology validated in lab; 5: technology validated in relevant environment; 6, 7: R&D of key technologies finished and validated through prototyping, the performance meet requirements; 8: System design finalized, technologies for key components validated in experiments, have capability of small batch production; 9: technology is mature, prototype system produced and tested; the team has successful record in similar projects and can start mass production or construction immediately.

| **Name** | **Application** | **Functionality** | **Maturity** | **Similar chips** |
|---|---|---|---|---|
| FEDI | Common Elec | Data Link | 5˜6 | lpGBT [17] |
| OAT | Common Elec | Optical | 7 | VTRx+ [42] |
| FEDA | Common Elec | Data Aggregation | 5 | lpGBT [17] |
| PAL | Common Elec | DC-DC | 5 | bPol [43] |
| Taichu | VTX | VTX-Stitched | 5˜6 | MOSAIX [44] |
| TEPIX | TPC | Pixel TPC | 6˜7 | Gridpix [45] |
| COFFEE | ITK | HVCMOS | 5 | MightyPix [46] |
| LATRIC | OTK | LGAD-TOF | 5 | ALTIROC [47] |
| SIPAC | SiPM ASIC | ECAL, HCAL, Muon | 6 | HGCROC [48] |

Regarding the TDR, the primary objective of the current electronic system is to establish a preliminary feasible baseline scheme. Accordingly, Figure 11.26 outlines the research plan for the next five years. The electronic system development will proceed in stages, with the release of the Ref-TDR serving as the first milestone. This milestone will provide the final design scheme for the electronic system, informed by detector requirements derived from beam-induced backgrounds simulations. The final design will include critical parameters such as data volume and power consumption estimates, which will serve as inputs to the trigger and mechanical systems.

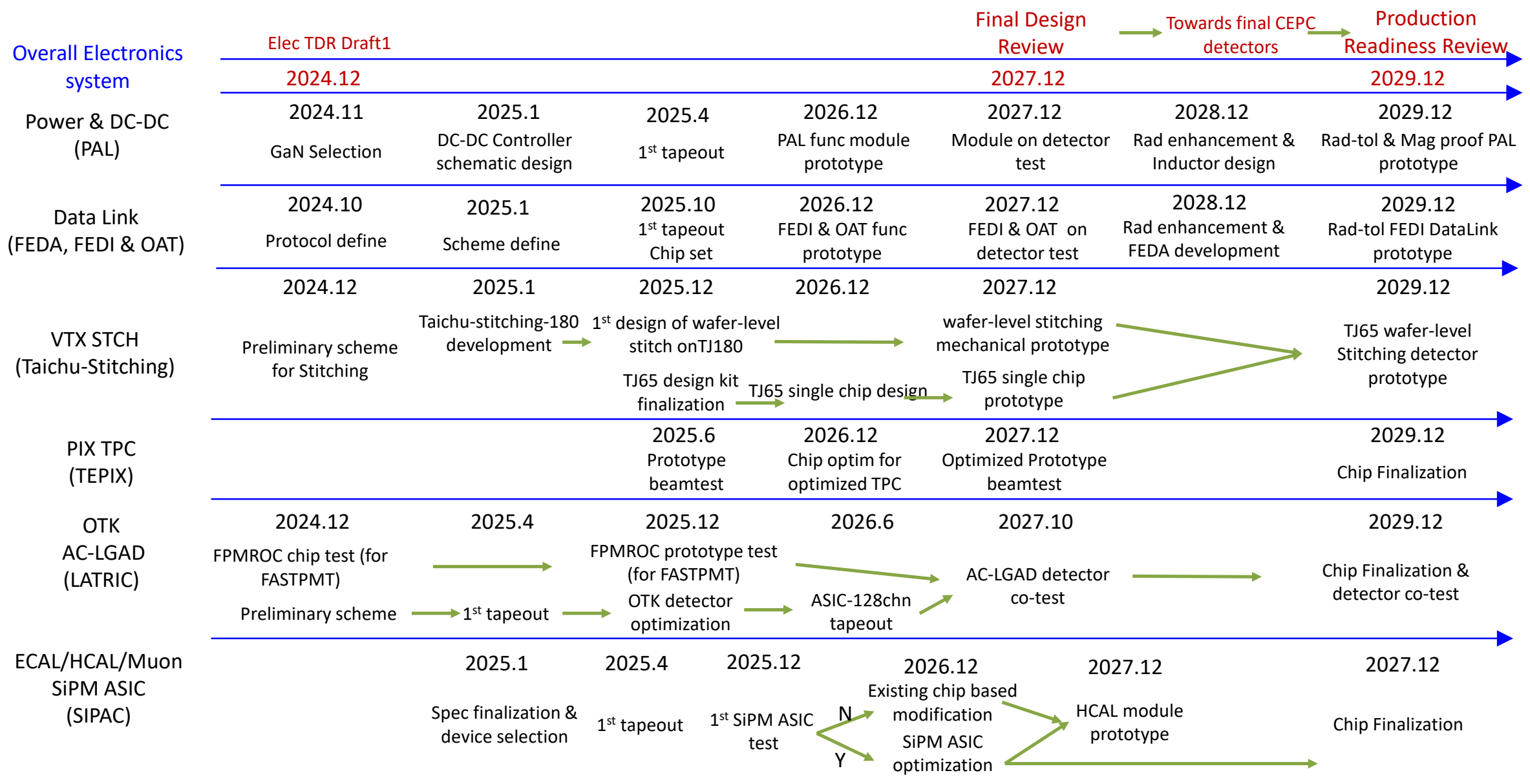

**Figure 11.26:** Main schedule for the next five years. It outlines the key milestones for the development of individual ASICs. Following the first major milestone—the release of the Ref-TDR—the plan sets 3-year near-term targets to finalize the design of all ASICs and 5-year long-term targets to deliver a prototype system.

Subsequently, 3-year (short-term) and 5-year (long-term) goals are set to complete prototype verification for every subsystem and to finalize both the front-end chips and the common electronics interface. Achievement of the short-term goal is gated by the ASIC Final Design Review (FDR), at which point the full chip design must be frozen; achievement of the long-term goal is gated by the Production Readiness Review (PRR), after that the chips must be successfully co-tested with detectors in the prototype systems. Detailed road-maps for each front-end ASIC are given in the respective subsystem chapters. This section summarizes the key electronics R&D efforts—ASIC development in particular—to provide a single reference for unified management during the ensuing engineering implementation phase.

## 11.11 Summary

This chapter outlines the overall framework of the electronic system, which adopts a full data transmission approach based on counting rate estimates from beam-induced backgrounds simulations. With a back-end triggering strategy, the system is designed to meet the requirements for exploring new physics. The electronic system framework comprises customized front-end components, common data interfaces, power modules, and BEE. Early stage R&D results for each component demonstrate the technical feasibility of the scheme, with no show stopper identified for the electronics framework or sub-detector readout. Additionally, conservative backup schemes based on conventional triggers and innovative approaches utilizing wireless communication are considered to accommodate varying levels of risk and ambition. Beyond the Ref-TDR releasing plan, the electronic system will establish short-term (3-year) and long-term (5-year) goals to achieve the overarching objectives of electronic system development.

# Chapter 12 Trigger and data acquisition

Collision rates in modern particle colliders reach few to tens of megahertz. Yet, the Higgs production rate remains exceedingly small. This requires the TDAQ to achieve real-time event selection of rare physical events among the beam background noise.

This chapter presents the technical design of the Reference Detector's TDAQ system. Subsequent sections address critical topics including system requirements and overall structure design as Section 12.1, First Level Hardware Trigger (L1) as Section 12.2, software-based High Level Trigger (HLT) as Section 12.3, trigger simulations and algorithms as Section 12.4, DAQ as Section 12.5, and integrated detector and Experiment Control System (ECS) as Sections 12.6 and 12.7.

## 12.1 Overview

The trigger system reconstructs particle tracks, energy clusters, and signatures (e.g., electrons, photons, muons, jets), which are used to suppress background while retaining target physics events. Physical observables like momentum and polar angle are analyzed to distinguish processes of interest. Events failing predefined trigger criteria are permanently discarded. An optimal trigger system maximizes retention of relevant physics events while minimizing background acceptance, with performance quantified by trigger efficiency and data purity metrics.

The system must operate on collision events occurring at frequencies up to tens of MHz. Trigger decisions must be executed within the limited data buffering window of electronic systems (typically microseconds), after that signals initiate event packaging. The data acquisition system reads out triggered events from detector electronics, transfers them to computing systems for aggregation and event reconstruction. Rapid full-data reconstruction of physical objects enables secondary filtering, retaining high-quality events and reducing data storage volumes.

This chapter details the architecture and requirements for the TDAQ and online systems, supporting a non-empty bunch crossing rate of 1.34–40 MHz and background data throughput from 100 GB/s (Higgs mode) to 1 TB/s (Z-boson pole). Simulations show the trigger system reduces rates to the range of few to tens of GB/s efficiently and event rate to 30 kHz (Higgs) and 120 kHz (low lumi-Z). The Level-1 trigger employs a three-stage hardware system: the first stage generates module-level triggers from backend electronics; subsequent stages integrate sub-detector triggers and execute global selection. The DAQ system manages tens GB/s post-trigger data via network transport, using parallel processing for HLT tasks (event assembly, classification, analysis) to reduce event rates/sizes before storage. The ECS provides AI-enhanced unified monitoring and control. The Detector Control System interfaces directly with hardware for monitoring, feeding data to the ECS, which also supplies centralized IT infrastructure, databases, and platform support for all systems.

### 12.1.1 Requirements

As shown in Table 3.3, the CEPC's operational complexity arises from its multi-phase design and different distinct operation modes. Each phase introduces distinct luminosity profiles, collision frequencies, and background characteristics. The TDAQ system must therefore adapt to dynamic operating conditions, balancing aggressive background suppression with robust signal retention across diverse physics programs.

According to the experimental roadmap, the CEPC will prioritize Higgs-mode operation for approximately one decade, with periodic low-luminosity $Z$ mode (12.1 MW) runs interspersed throughout this phase. Consequently, the TDAQ system shall be designed to:

- Support all operation modes through a unified technical framework
- Ensure scalable architecture accommodating luminosity progression from initial low-intensity to ultimate high-luminosity regimes
- Implement modular upgrade capabilities for future high-luminosity $Z$ mode and $t\bar{t}$ mode requirements
- Maintain algorithm adaptability for novel physics trigger development and subsequent adjustments

#### 12.1.1.1 Physics event rate

According to Table 3.3, the non-empty bunch crossing rate operates at 1.34 MHz in Higgs mode and 12 MHz (12.1 MW) in $Z$ mode. The TDAQ system must implement event selection at these beam collision frequencies, preserving all physical process events while suppressing background rates to physics-comparable levels ($\mathcal{R}_{\mathrm{bkg/phys}} < 1$).

At the peak luminosity, the expected event rates of the trigger system for the Standard Model processes at the Higgs and $Z$ modes are summarized in Table 12.1 and 12.2. The cross-section is estimated by BesTwogam [1] for the di-photon events, and by the Whizard generator [2, 3, 4] for the others. The Higgs event rate is expected to be 0.017 Hz for the Higgs mode and the $Z$ resonance decaying into $q\bar{q}$ event rate is expected to be 8 kHz for the low luminosity $Z$ mode. The trigger system is optimized to maximize the collection efficiency for Higgs production events, two and four fermions processes for the Higgs mode. For the $Z$ mode, the system aims to collect the two fermions processes with maximal efficiency. The di-photon processes are considered as the optional signals, with lower priority for the trigger algorithm design. The rate of the Bhabha events within the detector region($|\cos\theta|$ <0.99) is expected to be about 80 Hz for Higgs mode and 1.7 kHz for the low lumi. $Z$ mode. The total event rate of all the physical processes produced by electron-positron collisions in the Higgs mode is about 500 Hz. For the low luminosity $Z$ mode, the total event rate of all detectable physical processes produced is about 10 kHz. At a depth of 100 meters underground, the cosmic ray event rate [5, 6] is estimated to be about 0.463 $\mathrm{m}^{-2}\mathrm{s}^{-1} \times 11\ \mathrm{m} \times 11\ \mathrm{m} = 56$ Hz. The exotic processes, such as long-lived particles, are under consideration. Comparing the non-empty bunch crossing rate with the desired data storage rate, a reduction of about three orders of magnitude is required for both the Higgs and $Z$ modes.

**Table 12.1:** Expected Standard Model processes event rate at the Higgs mode for 50 MW.

| Higgs mode processes | Cross section (fb) | Event rate (Hz) |
|---|---|---|
| **Physical events (top priority)** | | |
| Higgs production | 203.7 | 0.017 |
| Two Fermions processes (exclude Bhabha) | $6.4 \times 10^4$ | 5.3 |
| Four Fermions processes | $1.9 \times 10^4$ | 1.6 |
| Bhabha | $1.0 \times 10^6$ | 80 |
| **Diphoton process (low priority)** | | |
| $\gamma\gamma \to b\bar{b}$ | $1.6 \times 10^6$ | 136 |
| $\gamma\gamma \to c\bar{c}$ | $2.1 \times 10^6$ | 173 |
| $\gamma\gamma \to q\bar{q}$ | $6.0 \times 10^7$ | 4963 |
| $\gamma\gamma \to \mu^+\mu^-$ | $2.1 \times 10^8$ | 17210 |
| $\gamma\gamma \to \tau^+\tau^-$ | $2.3 \times 10^6$ | 193 |

**Table 12.2:** Expected Standard Model processes event rate at the $Z$ mode for 12.1 MW.

| $Z$ mode processes | Cross section (fb) | Event rate (Hz) |
|---|---|---|
| **Physical events (top priority)** | | |
| $q\bar{q}$ | $3.1 \times 10^7$ | 7970 |
| $\mu^+\mu^-$ | $1.5 \times 10^6$ | 400 |
| $\tau^+\tau^-$ | $1.5 \times 10^6$ | 396 |
| Bhabha | $6.6 \times 10^6$ | 1714 |
| **Diphoton process (low priority)** | | |
| $\gamma\gamma \to b\bar{b}$ | $2.7 \times 10^5$ | 71 |
| $\gamma\gamma \to c\bar{c}$ | $5.1 \times 10^5$ | 132 |
| $\gamma\gamma \to q\bar{q}$ | $3.5 \times 10^7$ | 9014 |
| $\gamma\gamma \to \mu^+\mu^-$ | $1.3 \times 10^8$ | 33696 |
| $\gamma\gamma \to \tau^+\tau^-$ | $6.3 \times 10^5$ | 163 |

#### 12.1.1.2 Detector response

The Higgs boson, produced in electron-positron beam collisions, is unstable and decays into various final states. The physical objects resulting from the final states include electrons, photons, muons, jets, etc. Charged particles, such as electrons and muons, traverse the tracker, leaving hits on each subdetector. These hits are reconstructed into tracks using tracking reconstruction algorithms. Muons, which penetrate the ECAL and HCAL, leave hits in the Muon Detector, enabling their reconstruction via dedicated muon reconstruction algorithms. Photons, which rarely interact with the tracker, deposit most of their energy in the ECAL, where they are fully absorbed. Similarly, hadrons deposit energy in both the ECAL and HCAL, where they are eventually absorbed.

On the other hand, the main background processes, the beam-induced background, including both photon backgrounds and beam loss backgrounds, will generate many hits in the tracker and ECAL detector. The energy deposition on both ECAL and HCAL for the beam-induced backgrounds is smaller than other physical processes, while in the ECAL endcap region, a few GeV energy deposition is observed within the crystal modules near the beam pipe. And generally, the number of hits in the Muon detector from the beam-induced background is negligible in the barrel region, but in the endcap region, a very high background level appears at the endcaps closest to the beam pipe, as mentioned in Section 3.3.

The development of trigger algorithms involves several key considerations, including the detector response characteristics for different particle types, the distinction between signal and background in each subdetector, the energy deposition patterns in the calorimeters, and the identification of relevant information from each subdetector for triggering purposes. Background processes include beam-induced backgrounds, as discussed in Section 3, as well as detector and electronic noise, which will be investigated in future studies.

The VTX and ITK, located close to the beam pipe, experience relatively high background levels. The trigger system design of the Reference Detector must account for the real-time readout capabilities and buffering constraints of the electronics.

The readout time windows for the VTX, ITK, OTK, ECAL and Muon detector are 69 ns, and HCAL is 1000 ns. Additionally, the TPC exhibits a long drift time for charged particles, initially estimated at 34 μs [7]. When correlated with the bunch spacing in different operational modes, this implies that a single TPC readout window will contain background data from hundreds to thousands of collisions. It needs special readout and trigger treatment for pileup tracks and events from multiple collisions. When the event rate is sufficiently high, all TPC data must be read out by DAQ.

#### 12.1.1.3 Electronics requirements

The current electronics system design transfers all detector data with over threshold and zero compression from front-end electronics(called FEE) to back-end systems (called BEE). In this architecture, the trigger system acquires feature information exclusively from BEE, which substantially reduces the design requirements and implementation complexity for both the electronics and trigger hardware subsystems. However, this centralized approach inherently increases trigger latency more than 10 μs. Since the BEE are designed with larger DDRs capable of caching more than a millisecond of raw data, trigger latency won't be much of a requirement.

During high-luminosity operations, the data volume generated by VTX may exceed the transmission capacity between FEE and BEE, resulting in incomplete data transfer. To address this limitation, the trigger system design requires incorporation of a rapid preliminary triggering stage. This stage would utilize partial detector data to perform fast event decision with latency less than 6 μs, thereby allowing the reduction in data rate to manageable levels for complete transmission. And MDI will continue to do shielding optimization to reduce the background. Further, it is planned to replace the VTX with a new one, which has higher bandwidth from FEE to BEE.

#### 12.1.1.4 Raw data rate and event size estimation

Table 12.3 details the timing windows for subdetectors and event size computations in both Higgs and $Z$ modes. The "Full Data" metric is derived as:

$$\text{Full Data} = (\text{Ave. Hit Rate}) \times (\text{Detector Area}) \times (\text{Bit Width}) \qquad (12.1)$$

where hit rates are extracted from Tables 3.6 and bit widths from Table 11.1. Per-bunch data sizes are calculated by normalizing Full Data against non-empty bunch crossing (1.34 MHz for Higgs mode, 12 MHz for $Z$ mode). Event-level data sizes accumulates data from multiple collisions within the time window plus the one clock trigger uncertainty (23 ns).

These elevated data rates stem from continuous data streaming before trigger validation. Current projections indicate $Z$ mode background rates will reach 1 TB/s due to enhanced luminosity. In the high luminosity Z mode, the data volume of the detector will be significantly increased. The TDAQ system should have scalability to cope with such situations.

**Table 12.3:** Time windows for each subdetector and the event size calculation for both Higgs and $Z$ mode. The "Full Data" is calculated by multiplying the Ave. Hit Rate, from Table 3.6, by the detector area, and the bit width from Table 11.1. The data size per bunch is derived by dividing the "Full Data" by the non-empty bunch crossing rate, with the Higgs mode rate being 1.34 MHz and the $Z$ mode 12 MHz. The data size per event is obtained after accounting for the impact of the time windows of each sub-detector plus the trigger uncertainty (23 ns) on individual events.

| | **VTX** | **ITK** | **OTK** | **TPC** | **ECAL** | **HCAL** | **Muon** | **Total** |
|---|---|---|---|---|---|---|---|---|
| Time windows (ns) | 69 | 69 | 69 | 34000 | 69 | 1000 | 69 | |
| 50 MW Higgs mode Full Data (Gbps) | 130 | 21.2 | 82.7 | 26.4 | 752 | 26.6 | <1 | 1040 |
| Data size / bunch (kB) | 12.1 | 1.98 | 7.71 | 2.46 | 70.1 | 2.48 | <0.1 | 96.9 |
| Data size / event (kB) | 12.1 | 1.98 | 7.71 | 303 | 70.1 | 9.92 | <0.1 | 405 |
| 12.1 MW $Z$ mode Full Data (Gbps) | 307 | 37.8 | 139 | 57.1 | 1202 | 27.2 | <1 | 1771 |
| Data size / bunch (kB) | 3.20 | 0.394 | 1.45 | 0.595 | 12.5 | 0.283 | <0.1 | 18.4 |
| Data size / event (kB) | 6.40 | 0.788 | 2.90 | 293 | 25.0 | 4.53 | <0.1 | 333 |

### 12.1.2 Overall design

Generally there are three types of trigger modes: all-hardware trigger, a combination of hardware and software triggers, and all-software trigger. All-software trigger has the advantages of simplifying the design and scale of the electronics system, facilitating the deployment of complex algorithms, detection and alteration. But the overall resource requirements, compared with hardware trigger, are not cost-effective enough, since the physical event rates of the Reference Detector are sufficiently low relative to the bunch crossing rate. Therefore, the current basic trigger schema of Reference Detector is the combination of software trigger and hardware trigger. And the all-software trigger scheme will be the alternative trigger scheme for the Reference Detector and easier to upgrade with current electronics design schema.

Due to the very large amount of data from the VTX during high luminosity $Z$ operation mode, it is possible that the transmission of full data cannot be realized within the on and BEE transmission capability. Therefore, a Fast Trigger (FT) process is considered to be added within the upper level trigger of the basic scheme, and a FT is should be generated by a small number of detectors to limit the data rates for completing the data transmission.

With this consideration, the overall structure of TDAQ and online system is shown as Figure 12.1. The baseline trigger system contains two level triggers, the level 1 trigger(L1) and high level trigger(HLT).

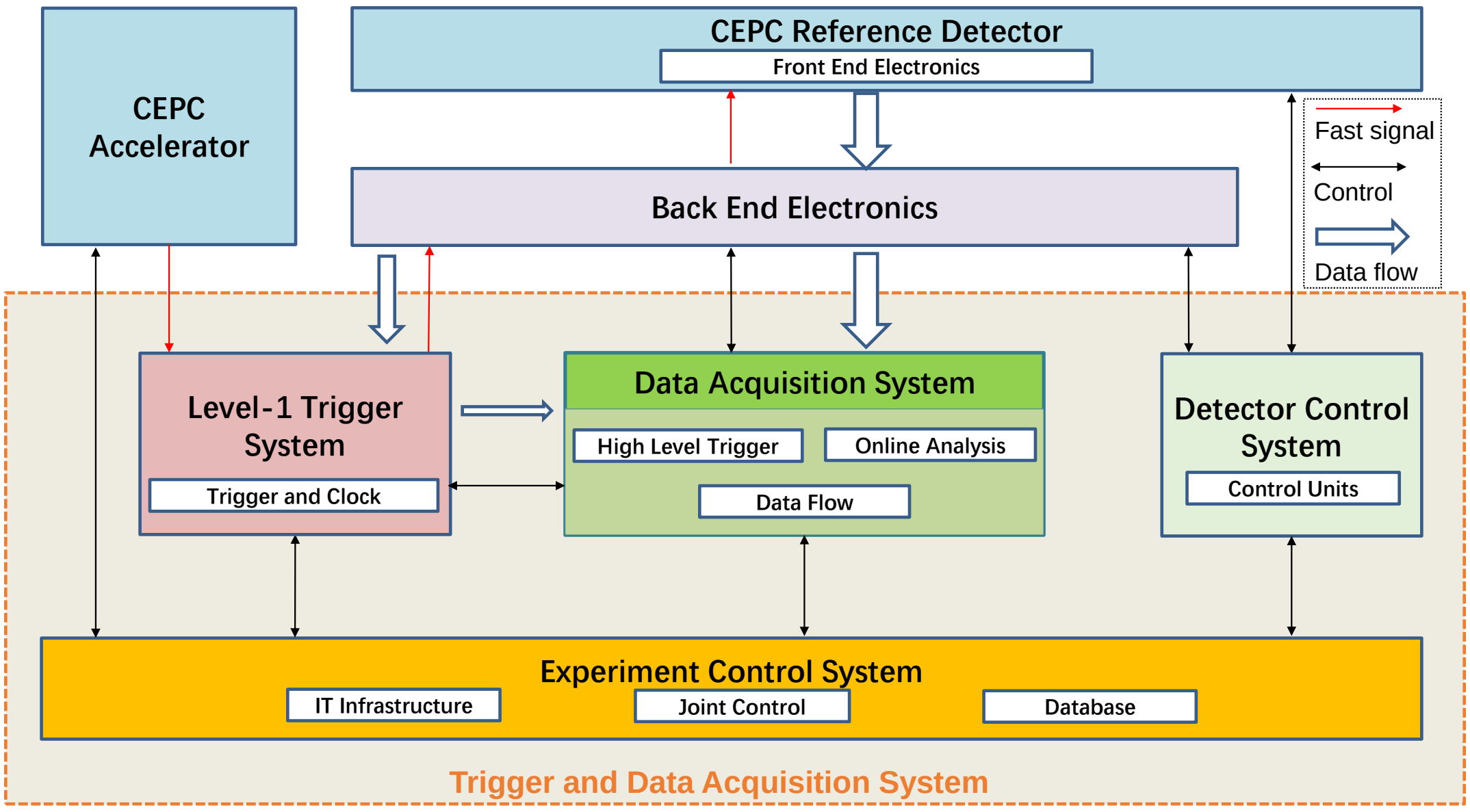

**Figure 12.1:** Overall structure of TDAQ and Online system for Reference Detector. Hardware L1 trigger system generate L1 and send to electronics system. DAQ readout electronics data passed L1 and execute HLT to filter background events and compress data rate. DCS control operation of detectors and electronics, and ECS provide infrastructure service and joint control and monitoring for whole TDAQ system. .

The baseline design for the L1 trigger uses the information from ECAL, HCAL, and Muon detector, while the information from VTX, ITK, TPC and OTK is under consideration for use. Events are selected through a trigger menu consisting of a series of specific trigger algorithms that are used to check whether each event meets specific requirements. When an event meets any of the requirements in the trigger menu, the complete detector information for that event is read and transferred to the next step in the HLT system. Therefore, the trigger menu needs to have the ability to collect the physical processes of interest while being able to exclude a large number of background processes. The L1 trigger will be designed to suppress most of the beam background for both Higgs and $Z$ mode. Most of the L1 trigger algorithm will be running on an FPGA farm, which has the advantages of low latency and real time. As the principle of the trigger is to keep all the good events and reduce the background events to acceptable level of DAQ readout, it is challenging to maintain the data purity.

The TPC is a special case, since the readout time is 34 microseconds, and so once the event rate is above about 30kHz, there are multiple overlapping events in the drift volume at the same time. This requires the TPC to be read out continuously by DAQ, and data from one clock cycle is generally part of several distinct events and so must be distributed to multiple L1 accept event buffers for further processing. However, the TPC data readout itself does not require an L1 trigger, it can still utilize the L1 trigger information to facilitate the integration and subsequent processing of the data. When an L1 trigger is present, the back-end electronics boards send the trigger information and the corresponding 34 μs data to the DAQ. If the 34 μs data also contains information for the next trigger, the packet will be truncated and the subsequent data will be included in the next triggered packet. Detector data are packed based on the trigger ID and bunch crossing ID. Data with trigger IDs within a certain range (for example, 0–9) are packed into a single block. The number of triggers included in each can be configured according to the number of subsequent processing nodes, ensuring that enough data is retained on each node for parallel processing. Subsequent processing is based on this data block structure. When processing TPC data, the corresponding 34 μs data segment can easily be located by referencing the trigger ID. Furthermore, it is straightforward to process the data from other detectors that share the same trigger ID as the TPC data.

After an event passes the L1 trigger screening, it is further screened by the HLT. The HLT can use the information from the complete set of detectors. Therefore some more complex offline reconstruction algorithms can also be applied to the HLT system. In the L1 trigger, the complete track cannot be reconstructed due to the lack of information from the time-projection chamber, whereas in the HLT, the track can be reconstructed. By analyzing the intersection points of the reconstructed track, the vertex can be reconstructed. Since most of the tracks from the beam background do not originate from the vicinity of the collision point, reconstructing the vertex effectively eliminates most of the tracks from the beam background. The particle flow object can be reconstructed using an offline algorithm by means of a calorimter and a track detector, and the specific physical objects can be distinguished to reconstruct different physical objects, such as photons, electrons, kaon, pion, etc. The combination of these physical objects can be used to reconstruct the particle flow object and further suppress the background process.

On top of the L1 trigger, the HLT will further suppress of the beam background for both Higgs and $Z$ mode. The expected target event rates for the trigger system in the different modes of the Reference Detector are listed in Table 12.4. The physical event rate is introduced in Section 12.1.1.1. The detail calculation for the DAQ readout rate is shown in Table 12.12. According to the beam background event size shown in Table 12.3, the raw event size is estimated to be about 0.3-0.4 MB. The total DAQ storage rates are calculated based on the corresponding raw event size that could be compressed with full reconstruction and detailed Regions Of Interest

(RoI) analysis at HLT.

**Table 12.4:** Trigger rate estimation table for different run conditions. The physical event rate is introduced in Section 12.1.1.1. The detailed readout rates of each sub-detector are shown in Table 12.12. The total DAQ storage rates are calculated based on the corresponding raw event size that could be compressed with full reconstruction and detailed RoI analysis at HLT.

| **Running mode**<br>**SR power** | **Higgs**<br>**50 MW** | ***Z***<br>**12.1 MW** |
|---|---|---|
| Non-empty bunch crossing rate(MHz) | 1.34 | 12 |
| Luminosity ($10^{34}/\text{cm}^2/\text{s}$) | 8.3 | 26 |
| Physical event rate (kHz) | 0.5 | 10 |
| L1 triger rate (kHz) | 20 | 120 |
| DAQ readout rate (Gbyte/s) | 5.34 | 11.9 |
| HLT rate (kHz) | 1 | 20 |
| Raw event size (kbyte) | 405 | 333 |
| DAQ storage rate (Gbyte/s) | 0.405 | 6.66 |

Aiming at the high throughput and high bandwidth characteristics of the experiment, it is necessary to design and develop a new online data processing framework, focusing on TB/s scale data streaming and online real-time data processing challenges. In the streaming readout data acquisition system, a high-performance data acquisition framework and online processing platform are crucial, while online software trigger and data compression are the core aspects of the system. It is necessary to study all aspects of data processing and related technologies, including software trigger algorithms, online instance reconstruction and categorization, zero data compression and lossless compression techniques. According to the requirements of the experiment, it is necessary to explore the balance between the deployment of algorithms and the use of resources by combining the compression characteristics of each algorithm, and strive to reduce the cost of hardware and maximize the efficiency. It is necessary to carry out in-depth exploration and pre-study in this direction, master the relevant technical means, and deploy a prototype system to accumulate implementation experience, so as to provide an important reference for the eventual large-scale application of future experiments.

Due to the extensive use of ASICs in the FEE of the detector leading to an increase in integration, the data acquisition link and slow control link are fused together, allowing the front-end and back-end interactions of data acquisition and full control to be combined and realized together. Therefore as Figure 12.1 shows, the new online system design considers to extract the electronics control from the traditional data acquisition system, combine it with part of the advanced functions of the detector control, and unite with other related systems to build a unified experimental control system, complete the unified monitoring and joint control of all the facilities of the whole experiment, and apply artificial intelligence methods to enrich and simplify the manipulation of the whole system.

ECS is crucial for the efficient operation of experiment, with its automated control platform achieving coordinated control and real-time monitoring of subsystems, simplifying complex operational procedures, accelerating the response to control commands, and ensuring the accuracy of control processes. The auxiliary monitoring platform, utilizing LoraWAN technology, easily deploys a large number of widely distributed monitoring nodes, significantly enhancing the safety of the experiment and data analysis capabilities. The intelligent operations and maintenance platform employs advanced ML and large language model algorithms to accurately pinpoint faults and facilitate rapid resolution. The intelligent shift assistant reduces the workload of shift personnel and maintenance staff, improving operational efficiency and ensuring the stable operation of experimental facilities.

## 12.2 Level 1 trigger

The baseline scheme of the L1 trigger system is designed for Higgs boson and low-lumi $Z$ mode with the hardware trigger structure being compatible with the other modes. In the case of Higgs boson and low-luminosity $Z$ modes, the background event rate is relatively low and the number of data events is relatively small, and the high-speed link between FEE and BEE can transmit all the digitized data to the BEE. When all the front-end data are transmitted, the data can be buffered in the BEE, and the buffering time is improved considerably to the order of milliseconds due to the DDR buffering technology used on BEE and common trigger board. In this case, the trigger latency requirement is no longer particularly strict, and it also makes it possible for TPC to participate in the L1 hardware trigger. For the high-lumi $Z$ mode case, the amount of VTX data increases significantly, and there is a risk that the high-speed link will not be able to transmit all of the front-end data. In this case, the generated FT signal can be sent to the FEE via the Trigger Control and Distribution System (TCDS) and BEE for filtered transmission of the FT trigger of the VTX FEE data. In this case, the FEE need to control the data transfer delay between the FEE and BEE involved in fast-trigger logic based on the buffering capacity of the VTX FEE.

### 12.2.1 System architecture

A proposed architectural scheme for the L1 trigger system is presented in Figure 12.2. Functionally, the L1 trigger system is divided into two parts. One is trigger logic part, and the other one is TCDS part. According to the study at this stage, the L1 trigger of the Reference Detector is mainly generated by information from the ECAL, HCAL, and Muon detector. If the information of the first three detectors cannot satisfy the trigger system requirements, the track information of the ITK, OTK and TPC and part of VTX detectors can be used as a joint consideration.

The trigger logic is divided into three levels: in the first level, the trigger subsystem of each detector receives the pre-processing trigger information from the BEE and generates the trigger primitives information, such as hit counts, cluster information, track segments, track segment counts and energy information in sector area; in the second level, the global trigger subsystem of each detector receives the sector-level trigger primitive information and generates the global detector hit or cluster information, track counts and position information, energy and position information, and other trigger primitives; in the third level, global trigger logic is applied.

The global trigger combines two or more detector trigger primitives based on the characteristics of the final state particles of interest to do particle identification and filter interesting events to generate L1 signal (and FT if needed). The L1 (and FT if needed) signals are sent to BEE of each detector for data readout via the TCDS system.

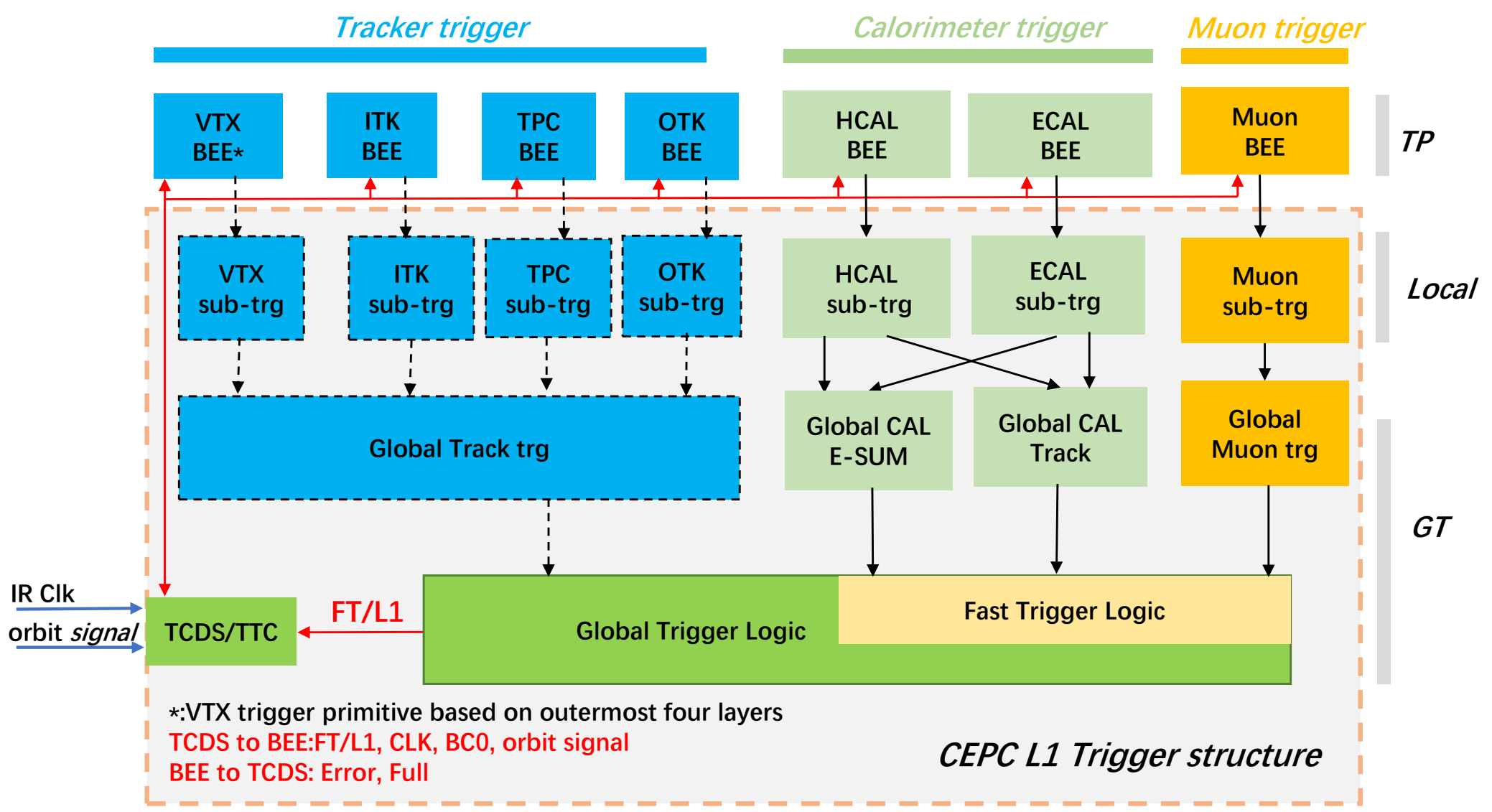

**Figure 12.2:** Trigger architecture diagram. The trigger has three levels. Solid line section is the baseline of trigger structure. Dotted line section is the tracker trigger section if needed.

L1 Trigger system employs positional, energy, and temporal correlations in detector signals to select valid events. This requires synchronized real-time data transmission and processing between trigger subsystems and plug-ins aligned with the system clock. Implementing trigger high-speed real-time transmission protocols—a critical technical component—ensures this temporal coherence. The proposed protocol leverages high-speed serial interfaces and multiplexed channel bonding to standardize data transmission from back-end electronics to the trigger system, as well as inter-plugin communication within the trigger layer. Key specifications include data rates, clock synchronization with spectrometer detectors, and standardized packet formats optimized for signal timing, energy profiles, spatial clustering, and detector hit rates—collectively fulfilling the trigger system's low-latency requirements. Trigger information transmission pathways—from back-end electronics to trigger layers and internal trigger-layer interconnects—are designed for compatibility across luminosity regimes, supporting both high-luminosity $Z$ operation and low-luminosity $Z$/Higgs boson modes.

### 12.2.2 Trigger menu

Since events failing to meet trigger criteria are permanently discarded, physical process identification accuracy must approach 100%. To achieve this, a trigger logic menu containing all physics events of interest must be designed. In systems like Belle II, over 160 distinct trigger primitives are implemented, with events retained when any single condition is met.

### 12.2.3 Common trigger board

Figure 12.3 shows the design structure of a common trigger prototype board. The common trigger board is the core component of the L1 trigger system. The prototype of the trigger system building and trigger algorithm performance testing and data acquisition bandwidth testing are all based on this component. The trigger electronics board possesses high-speed inter-connectivity, high-performance data processing, high-bandwidth data readout, and a large capacity for data buffering. In the design of the trigger electronics board, it is proposed to use the high-performance FPGA chip of AMD Xilinx Company as the core data parallel real-time processing chip; for high-speed data interconnection capability, it is proposed to use the high-density fiber optic transceiver and the high-speed serial core (RocketIO) inside the FPGA; for high data readout bandwidth, it is proposed to use the hardware high-speed link combined with the TCP protocol or the Remote Direct Memory Access (RDMA) protocol; RAM is combined with DDR buffering technology to realize the buffering of trigger data. The SoM is proposed for the management of the trigger board, and the configuration of slow-control parameters. A network switch is used for the FPGA, the SoM, the power supply control module, and the board network interface to provide a network data switch path. Based on RocketIO and optical technology, the trigger electronics board is proposed to support 40 fiber-optic channels, and the data transmission line rate of a single channel is better than 16Gbps.

### 12.2.4 Trigger control and distribution

The TCDS, shown in Figure 12.4, plays a core role in TDAQ, with its main functions being:

1) To receive the clock signal and the bunch crossing gap signal (orbit signal) from the accelerator, as well as to receive the FT and L1 signal generated by the L1 trigger system;

2) To send the clock signal, bunch crossing gap signal, and the decided L1 signal to the trigger and the electronics subsystems of each detector. Each subsystem then performs internal fan-out distribution.

3) To receive and provide feedback on status information from various BEE systems, such as buffer full signals (or BUSY signals), link error states, etc. When the full (BUSY) signal or error signal is fed back to TCDS, TCDS will mask the L1 signal until full or error signals are cleared to avoid data overwriting in the data packing buffer in DAQ readout.

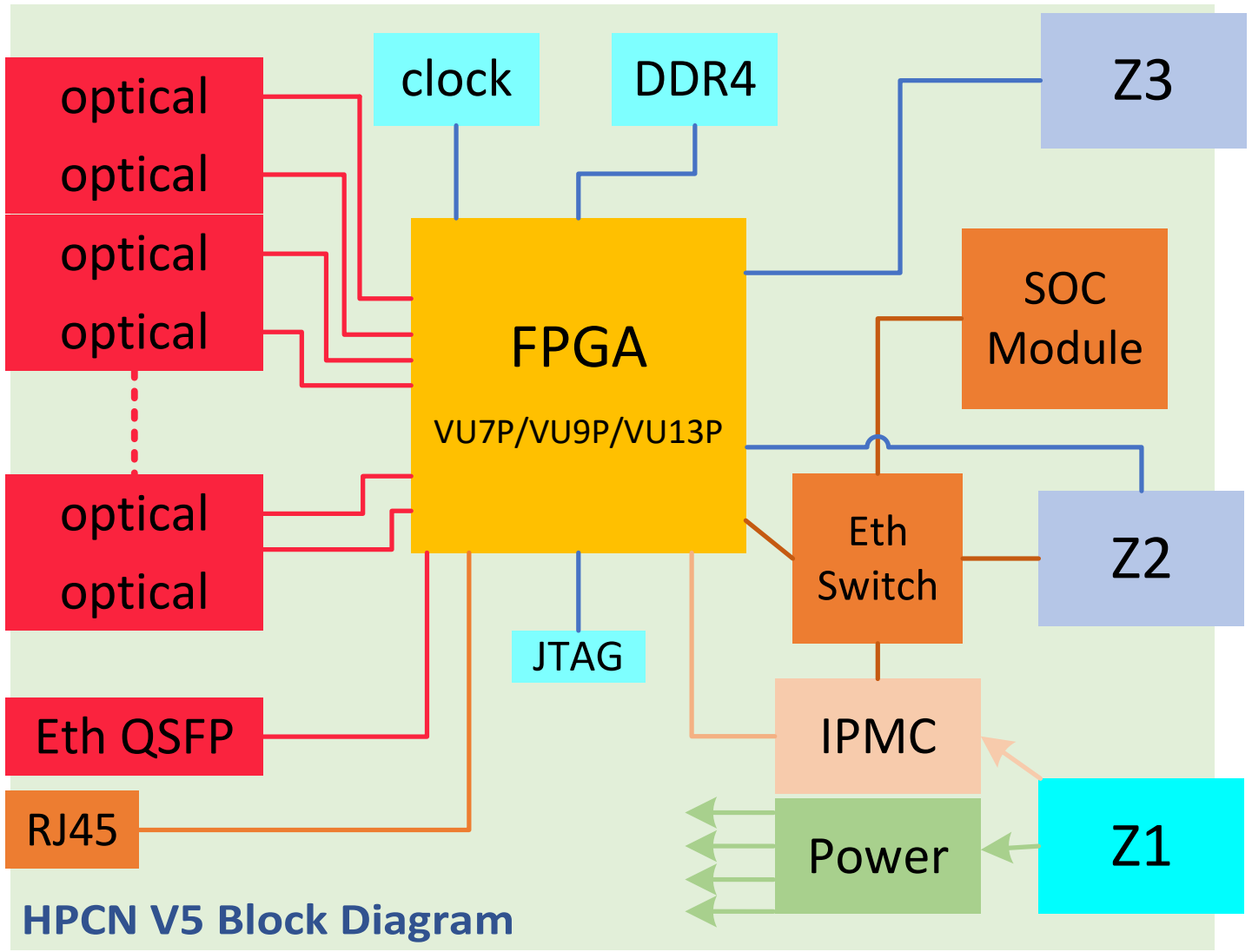

**Figure 12.3:** Diagram of the proposed structure of the trigger electronics plug-in principle prototype. It is based on ATCA standard. AMD Xillinx FPGA will be the first option for high speed data data transmission and high performance data processing.

In Figure 12.4, the TCDS/TTC module is the central control module that receives system clock and orbit signal from the accelerator. It also receives L1/FT from L1 trigger logic and fans-out these signals to L1 trigger and BEE system, and collects feedback status signals from the BEE system. The Data Concentrate and Timing Distribution (DCTD) board is the sub-system level control module in the ATCA crate. It has a master and slave mode. In master mode, the DCTD serves as the interface for each BEE system. It fans-out clock, trigger, and fast control signals to slave DCTD, and collects ATCA crate level status feedback from the slave DCTD. The slave DCTD acts as the central control module within the ATCA crate. It fans-out clock, trigger, and fast control signals to the ATCA back-end board and collects ATCA back-end board status, which it then feeds back to the master DTCD boards.

In the TCDS, the interfaced BEE subsystems include the VTX, ITK, TPC, OTK, ECAL, HCAL and Muon systems, Luminosity electronics subsystem, and the trigger system.

The connection between the trigger control and distribution system and each subsystem uses high-speed serial links, with the system clock being recovered from the high-speed serial link using CRD technology at the BEE. In the downstream direction, (from TCDS to each off detector electronics subsystem), the trigger signals such as L1 and Bunch Crossing Zero (BC0), reset signals, etc., are distributed. In the upstream direction, (from each BEE subsystem to TCDS), status information from each BEE subsystem is fed back, including buffer full states, system error states, and BEE card identification information, etc.

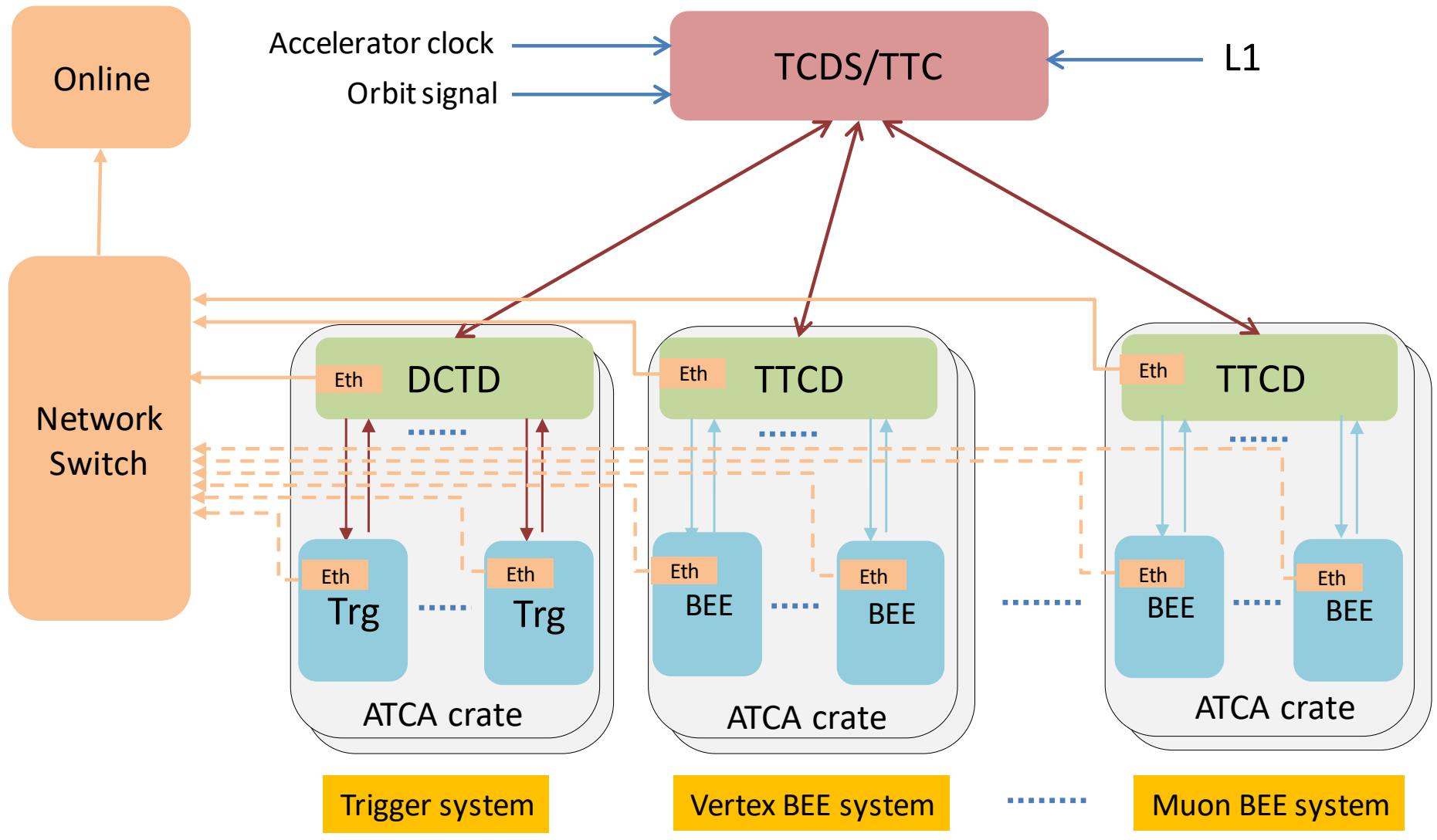

**Figure 12.4:** TCDS/TTC proposed structure. DCTD is Data Concentrate and Timing Distribution board. TTCD is Trigger Timing and Control Distribution for each sub-detector BEE system. Details of TTCD please refer Figure 11.17.

## 12.2.5 Resource estimation

In the L1 trigger strategy, the baseline trigger configuration utilizes the ECAL, HCAL, and Muon detectors, while track triggers incorporate the VTX, ITK, OTK, and TPC detectors when required. This resource estimation focuses on Higgs boson studies and low-luminosity Z boson operation modes.

Within this framework, trigger primitives for tracking detectors (VTX, ITK, OTK) contain cluster coordinates, dimensions, and timing data. Calorimeter systems (ECAL, HCAL) provide cluster positions, spatial extents, energy measurements, and temporal information. The muon detector supplies hit counting metrics comprising hit count, spatial distribution, and synchronization data. For the TPC detector's potential trigger involvement, further investigation is required—current preliminary estimates derive solely from its data throughput analysis. Global trigger resource allocations for each subsystem employ sector-based segmentation corresponding to the polar angle $\phi$ as the foundational calculation method. For global track trigger, Calorimeter track finding, and Global trigger, a sector will be 60° wide. For the energy trigger of the calorimeter, a sector will be 120° wide.

All trigger primitives information is acquired from back-end electronics. Transmission of trigger primitives to the trigger system occurs at a 10 Gbps line rate, delivering 8 Gbps effective bandwidth after accounting for 8B/10B encoding overhead. The L1 trigger system employs 16 optical channels for connectivity with back-end electronics boards or upstream trigger boards. By analyzing trigger information data volumes derived from detector background simulations, it can estimate both the required optical links per detector sub-trigger system and the corresponding number of trigger boards. For system installation, each ATCA crate supports up to 10 trigger boards plus a minimum of one DCTD board. This methodology yields the preliminary trigger system scale estimates presented in Table 12.5.

**Table 12.5:** L1 Trigger resource estimation table. The number in parentheses represents the resources required for track trigger additional, in which VTX, ITK, OTK and TPC will be used. Muon Barrel and Edncap share one trigger board.

| Detector | BEE Board Num | TP Bits | Fiber num to trg/BEE | Trigger Board | ATCA Crate | DCTD Boards |
|---|---|---|---|---|---|---|
| VTX | 6 | – | – | (6) | (1) | (1) |
| TPC | 32 | – | 1 | (2) | (1) | (1) |
| ITK-Barrel | 24 | 64 | 1 | (2) | (0) | (0) |
| ITK-EndCap | 6 | 64 | 1 | (2) | (1) | (1) |
| OTK-Barrel | 55 | 64 | 1 | (4) | (0) | (0) |
| OTK-EndCap | 34 | 64 | 1 | (3) | (1) | (1) |
| ECAL-Barrel | 60 | 48 | 3 | 12 | 1 | 1 |
| ECAL-EndCap | 28 | 48 | 3 | 6 | 1 | 1 |
| HCAL-Barrel | 348 | 48 | 1 | 22 | 3 | 3 |
| HCAL-EndCap | 192 | 48 | 1 | 12 | 2 | 2 |
| Muon-Barrel | 2 | – | 1 | — | — | — |
| Muon-EndCap | 1 | – | 1 | 1 | 1 | 1 |
| Global Trigger | – | – | – | 18+(12) | 2+(2) | 2+(2) |
| TCDS | – | – | – | 1 | 1 | 1 |
| Sum | – | – | – | 72+(31) | 11+(6) | 11+(6) |

## 12.3 High level trigger system

### 12.3.1 Schema design

The HLT system is a software-based filtering system that processes detector data to identify physics events of interest. This system performs two primary functions: effectively suppress background events through precision online filtering, and generate RoI information to enable targeted data compression. While the HLT accesses more comprehensive event data and operates with longer decision latency compared to L1 triggers, it employs more sophisticated algorithms for precise event identification. Unlike offline reconstruction systems, which prioritize reconstruction accuracy as their primary metric, the HLT must balance two competing priorities in its algorithms: sufficient precision for background suppression and execution speed compatible with available computing resources.

The Belle II experiment directly uses the offline reconstruction algorithm in the HLT system to construct the RoI for the silicon pixel detector, which reduces the data size by 60% [8]. And the latency is 1-2 seconds per track running on a Central Processing Unit (CPU) core for Belle II HLT [9]. The LHCb experiment has upgraded the trigger system to software-only trigger schema from the RUN3 operation period. Its HLT1 is operating with Graphics Processing Unit (GPU) based system, which is now the flagship of the online trigger system [10]. Using only the CPU is feasible for current data rate, but GPU acceleration delivers superior performance and greater computational efficiency, though it requires more R&D investment.

For the Higgs operation mode, where the L1 trigger rate is relatively low, the HLT system prioritizes optimized reconstruction algorithm design. For the high-luminosity $Z$ operation mode, the focus shifts to developing heterogeneous computing acceleration technologies alongside algorithmic improvements to handle increased throughput.

The key features of the HLT system in general include the design of customizable algorithms, which utilizes particle tracking, calorimeter clustering, and detector-specific signatures to filter events based on physics objectives. Since the HLT system is implemented between L1 and the DAQ system, it is necessary to design the integration with these two systems accordingly. The technical design for HLT system needs to consider the following items:

- Distributed Computing: Use a network of processors to handle the massive computational workload that requires hundreds of milliseconds to process an event.
- Advanced Software Tools: Leverage ML, pattern recognition, and simulation-based matching to enhance the accuracy of event classification.
- Data Volume: Even after the L1 trigger system, still tens of GB/s of data generated in the detector need to be handled by HLT system for low and high luminosity $Z$ operation modes.
- Real-Time Constraints: Ensuring rapid decision-making without sacrificing the accuracy of event selection.
- Detector Limitations: Dealing with noise, pileup, or incomplete data due to detector inefficiencies.

Due to the demands of anomaly detection for some new physical events, it is desirable to have direct access to the raw data so that loss of these signals by coarse L1 triggering algorithms can be avoided, but such signals are usually unknown and not clearly regular, requiring frequent modifications and adjustments to the triggering algorithms, which is more difficult to achieve at L1 triggering system, although new ML algorithms may help to ameliorate this problem.

Software triggering can be accelerated by GPUs and FPGAs in addition to CPUs. Figure 12.5 shows the hetorogeneous computing platform for the HLT system. Different functions can be implemented onto CPU, GPU or FPGA based on that algorithm is best suited to that processor. The following features need to be implemented in HLT system.

- Event building: Raw data from different readout units need to be constructed as a complete event or a full event for a specific sub-detector.
- Data Preprocessing: Refine clusters from L1 output using advanced algorithms. Perform time alignment and coordinate transformations.
- Feature Extraction: Compute cluster shapes, energy sums, and time spreads.
- Track Seed Finding: Implement Hough transform or template matching for rapid seed identification.
- Track reconstruction: Reconstruction of tracks based on multiple tracking sub-detectors, especially, for the TPC detector, which it has 34 $\mu$s signal window.
- Lightweight PIDs: Run simplified ML inference models to identify particle types.

- Calibration: Apply the loose calibration constants for alignment and calibration.
- Physics Skim: Application of the HLT filter to keep dedicated physics skim.
- Data Reduction: Suppress irrelevant regions and forward RoI data to the dedicated sub-detector.

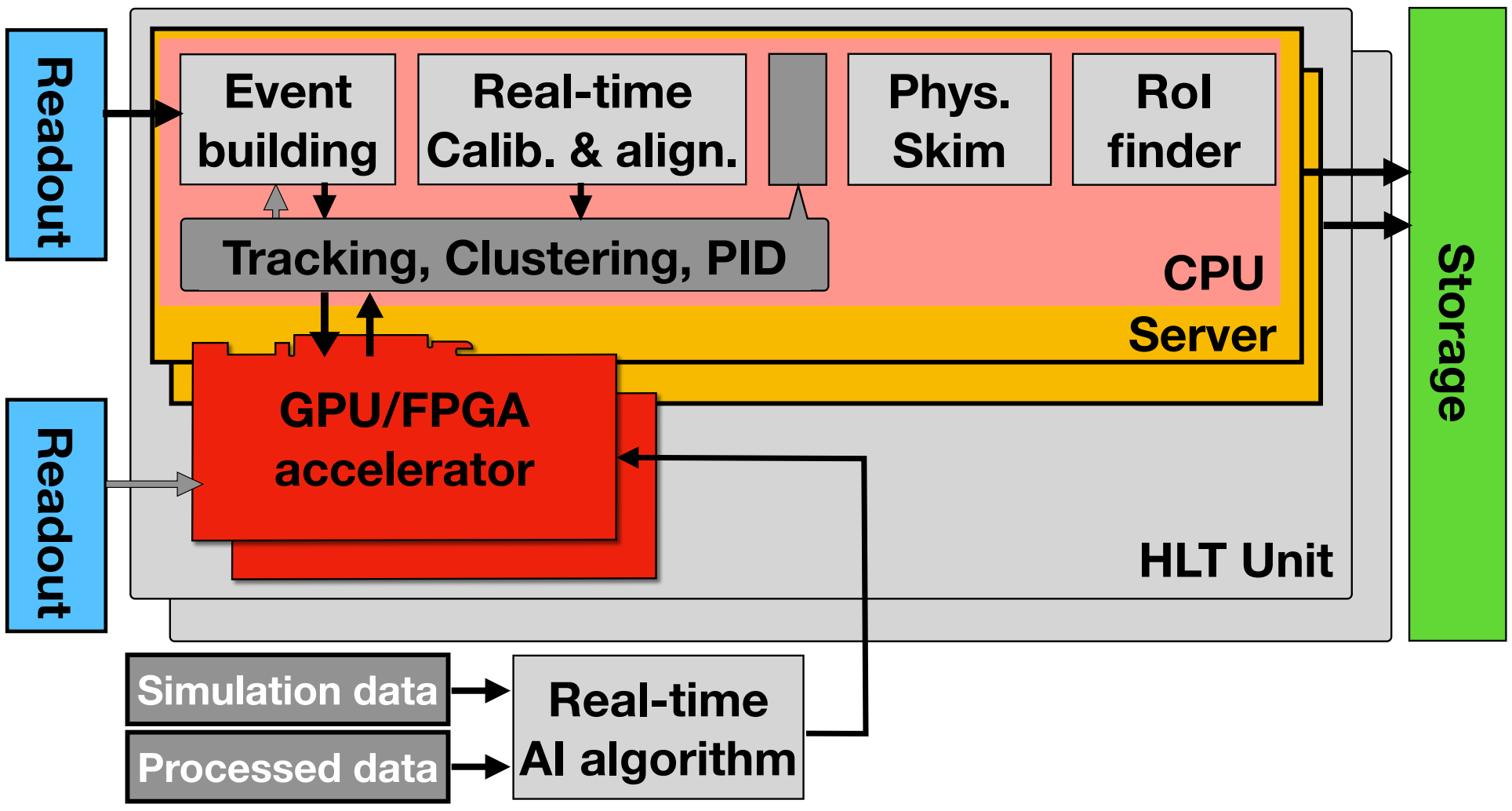

**Figure 12.5:** Image of heterogeneous computing platform for HLT system. HLT is required to complete the tracking, Clustering and PIDs by fast reconstruction, which can be accelerated by GPU/FPGA. Some other tasks such as Event building and real-time calibration can be done on the CPU.

### 12.3.2 HLT algorithms and software

Offline software tracking reconstruction algorithms, as mentioned in Chapter 13, are tested for the design estimation of HLT system. The offline tracking algorithms contain five steps, SiliconTracking, ForwardTracking, Tracksubset, Clupatra, and FUllDCTracking. For the beam background, no tracks are reconstructed in more than 80% of the events, and on average it needs one CPU to run about one second at the Higgs mode and about 1.2 second at the $Z$ mode for the full tracking reconstruction algorithm. For the physical events, it depends on the physical topology: for a simple and clean process, such as $Z(e^+e^-)H(\mu^+\mu^-)$, it needs 456 milliseconds to finish the full tracking reconstruction algorithm, while for a more complicated process, such as $Z(e^+e^-)H(ZZ)$, it needs about 6.5 seconds. The ML technique has been widely developed to build the HLT algorithms at different experiments.

The drift chamber noise suppression algorithms based on Graph Neural Network (GNN) edge classification and interaction networks, developed for the BESIII and STCF experiments, can significantly improve track reconstruction efficiency [11]. The Exa.TrkX collaboration, led by CERN, has developed a GNN track reconstruction framework supporting highly parallelized and pipeline-compatible operations [12, 13, 14], which has been successfully deployed in High-Luminosity LHC (HL-LHC) track reconstruction systems for ATLAS, CMS, and LHCb experiments. The ML technique should also be utilized in the general design and development of HLT algorithms to improve the latency and performance.

Instead of explicitly splitting software into online and offline part, both could be unified into a single software framework supporting reconstruction and selection both for online and offline processing. Having an unified software framework would mainly assure data processing consistency between online and offline stage limiting possible effect of software on physics exploitation. Additionally, this would help with general development, as the same code could be deployed online and offline, and also with the software maintenance as the total amount of code would decrease.

### 12.3.3 GPU acceleration

The use of GPUs is currently the prevailing way of parallel computing, occupying a place in image processing, Artificial Intelligence (AI) acceleration and other aspects, and has been widely and deeply applied. In high-energy physics collider experiments, track reconstruction, particle identification and so on are very similar to these tasks, the use of GPU acceleration of these physical processes is consistent with the characteristics of the task itself.

The development of GPUs in recent years has facilitated the development of various industries, especially the AI industry, and the development of AI has continuously stimulated the development of GPUs, whose computing power has increased 1,000-fold in the past decade. With the development of GPUs, it improves the feasibility of using more complex Neural Network (NN) models in high-energy physics experiments with lower latency to achieve higher physics accuracy.

Compared to the industrial field, high-energy physics experiments have lower training requirements for ultra-large scale models and lower communication performance requirements for GPUs. GPU acceleration in high-energy physics focuses more on the computational power of a single GPU and the size of the GPU memory, so the cost of the required GPU clusters is relatively controllable. Taking LHCb as an example, LHCb performs pure software triggering at 30 MHz with a nominal data throughput of 40 Tb/s. The first stage of HLT is implemented on regular on-shelf GPUs where the full LHCb physics program can be achieved by using around 300 GPUs, while about 500 GPUs are used in production allowing to extend LHCb physics reach above original proposal. This solution to HLT was found to be very high cost efficient and scalable to the experiment.

In collider experiments, some typical advanced algorithms such as track reconstruction and particle identification algorithms are well suited for acceleration using GPUs. By using GPU acceleration in HLT, the system is able to parallelize the processing of multiple starting points, different candidate tracks, or particle feature extraction by different threads. This parallel processing not only improves the overall computational efficiency, but also provides an effective solution for processing larger and more complex experimental data. Therefore, the application of GPU acceleration in HLT for high-energy physics collider experiments is a key way to optimize the performance of the experiments and obtain the analysis results quickly.

### 12.3.4 FPGA acceleration

There are specific advantages that FPGA can maximize the performance of HLT system, especially, the bandwidth of raw data-flow is very high even after the L1 trigger, FPGA acceleration of real-time data processing is capable of handling the data-flow. In addition, FPGA technique has been widely developed with the features utilized for computation acceleration, while keeping a significantly low latency. The following specific rationales for the FPGA usage are showing the HLT system can maximize its performance by employing FPGA technique.

- Low Latency: FPGAs execute tasks in hardware with deterministic timing, ensuring ultra-low latency.
- Parallelism: Their architecture supports simultaneous processing of multiple data streams.
- Customization: Pipelines can be tailored to specific algorithms, maximizing resource efficiency.
- Real-Time Processing: Ideal for preprocessing and intermediate calculations in streaming data workflows.

As shown in Figure 12.5, the main algorithms running on HLT are track finding and fitting, calorimeter clustering. Algorithmic design for FPGA tasks could both refer to the L1 trigger system and the offline reconstructions. FPGAs could be used in the stage of data preprocessing, for instance, noise filtering, feature extraction, clustering, track seed finding. As the first step of tracking reconstruction, noise filtering needs large amount of computing power, and it is very sensitive to the beam background conditions. It has been known that a NN based algorithm is sufficient to reduce the beam background for the Belle II [15]. The NN model is able to be implemented to FPGA (Virtex UltraScale, XCVU160) with the high level synthesis. Implementation of a simplified Hough transform for quick identification of straight-line patterns on FPGA has already been used in the L1 trigger system of the Belle II for the tracking reconstruction. The neural clustering finding algorithm designed with light-weighted ML model, is also a good example that can be deployed on FPGA for acceleration.

Development and deployment of FPGA algorithm on HLT is still difficult compared to CPU and GPU based algorithm. It is necessary to keep the development of both the FPGA based algorithm (e.g. tracking and neural clustering reconstruction), and the software frameworks, such as the interface with software frameworks for HLT pipeline integration.

## 12.4 Trigger simulation and algorithms

### 12.4.1 Physics signatures and trigger primitives

The detector response and the physics signatures of each process are studied using the MC simulations. In the Higgs mode, the simulated physical processes include $e^+e^- \to f\bar{f}H$, where the $f$ can be electron, muon, tau, neutrino, or quark [16, 17]. Besides the Higgs boson production processes, other processes in the Higgs mode include two-fermion final states , four-fermion final states , and the hadronic di-photon events. In the $Z$ mode, the simulated physical processes focus on two-fermion final state processes and the hadronic di-photon events. The di-photon events are generated by BesTwogam [1], while all other simulated physical samples are generated by Whizard generator [2, 3, 4]. The detector simulation and response to the final states are done by a Geant4-based software called CEPCSW, as mentioned in Chapter 13.

The main sources of background in both Higgs and $Z$ modes are beam backgrounds and detector noise. For both the Higgs and the $Z$ mode, an 92 ns (4 clocks) time window is chosen as the trigger window, and each beam background event includes 2 bunch crossings, corresponding to a safety factor of 2 mentioned in Table 3.6. The detector noise will be studied in future studies.

Charged particles, such as electrons and muons, interact with the tracker, leaving detectable hits. With this information, trigger tracks could be reconstructed quickly, determining whether the track originates from the interaction point and the track's momentum. For background processes, the tracks typically originate from random points along the beamline and generally exhibit low energy. Therefore, by analyzing these track properties, it is possible to distinguish whether the event corresponds to a process of interest.

For most of the physical events, the final states, i.e., electrons, photons, and jets, deposit significant amounts of energy in the calorimeter. In contrast, background processes typically result in a more uniform energy distribution with lower energy deposition in both the barrel and endcap regions. The ECAL units are approximately 1.5 cm $\times$ 1.5 cm $\times$ 40 cm, with layers stacked perpendicularly. The ECAL supercells, mentioned in Section 7.2, are used for the preliminary trigger studies. The barrel region of the ECAL is divided into 15 $\times$ 32 supercells along the $z$-axis and the polar angle $\phi$, and each supercell contains 18 longitudinal layers. For the endcap region, the ECAL consists of 112 supercells with 6 different types, and each supercell contains 18 $Z$ layers. Similarly, the barrel region of the HCAL is divided into 20 $\times$ 32 supercells along the $z$-axis and the polar angle $\phi$, and each supercell contains 48 longitudinal layers. For the endcap region, the HCAL is located in the $xy$-plane within the coordinate range of [-3500 mm, 3500 mm] for both the $x$-axis and $y$-axis. This 7000 $\times$ 7000 mm$^2$ area is divided into a 20 $\times$ 20 grid of uniform square blocks, with each block measuring 350 $\times$ 350 mm$^2$. The current segmentation scheme of the HCAL endcap has not yet incorporated electronics constraints, though this aspect will be addressed in the future study.

Muons deposit little energy in the calorimeter, usually less than 1 GeV in the supercell, but they penetrate to the muon detector and generate hits.

## 12.4.2 Sub-detectors trigger algorithms

The development of trigger algorithms critically depends on the design and implementation of the accelerator, detector, and associated systems. It requires obtaining comprehensive technical specifications and parameters, followed by detailed simulations of the detector response to complex physical processes. The goal is to differentiate between signal and background processes, and to explore various criteria and methods to develop a robust suite of trigger algorithms. This section will describe the preliminary trigger algorithm using the Calorimeter and Muon detectors for the L1 trigger.

### 12.4.2.1 Calorimeter

The ECAL comprises 18 layers, while the HCAL comprises 48 layers. The energy sum of all the ECAL and HCAL cells inside each supercell is calculated separately as the trigger feature for that supercell. For each event, the supercells with the largest total energy in the ECAL barrel and endcap are selected as representatives respectively, resulting in two parameters for triggering decisions in the ECAL. Similarly, the supercells with the largest total energy are selected to represent the barrel and endcap for the HCAL, adding another two parameters for triggering decisions. In total, four parameters are used for the calorimeter trigger.

Figure 12.6 shows the distribution of the maximum energy deposition in the supercell for the $e^+e^- \to Z(\nu\bar{\nu})H(b\bar{b})$, $e^+e^- \to Z(\nu\bar{\nu})H(\gamma\gamma)$ processes and the beam background. Large energy deposition, for example, more than 10 GeV in the ECAL barrel region, is observed for these two physical processes. For the beam background, in most cases, the maximum energy value does not exceed 0.5 GeV, except for the ECAL endcap region, where the energy deposition can be larger than 2 GeV. Therefore, to distinguish between the signal and the beam background, a set of energy threshold requirements is applied to these four parameters.

**Table 12.6:** Supercell energy baseline and optimal thresholds for calorimeter subdetectors at the Higgs mode for 50 MW and the $Z$ mode for 12.1 MW. An event is classified as the signal if the energy measured in any calorimeter supercell is higher than the corresponding thresholds. The threshold for endcap is higher than barrel, since the beam background has higher energy deposition in the endcap region.

| **Subdetector** | **Baseline threshold** | | **Subdetector** | **Baseline threshold** | |
|---|---|---|---|---|---|
| | **Higgs mode** | ***Z* mode** | | **Higgs mode** | ***Z* mode** |
| ECAL Barrel | > 0.20 GeV | >0.113 GeV | HCAL Barrel | > 0.023 GeV | >0.017 GeV |
| ECAL Endcap | > 2.57 GeV | >0.96 GeV | HCAL Endcap | > 0.16 GeV | >0.059 GeV |

Two sets of energy thresholds, a baseline and an optimal, are shown in Table 12.6 for Higgs mode and $Z$ mode. An event is classified as the signal if the energy measured in any calorimeter supercell is higher than the corresponding thresholds. The baseline threshold is chosen so that the background efficiency is less than 0.2% for the Higgs mode and 0.2% for the $Z$ mode when any single threshold is used alone. The threshold for the endcap region is higher than the barrel region, because of the higher energy deposition in the endcap region from the beam background.

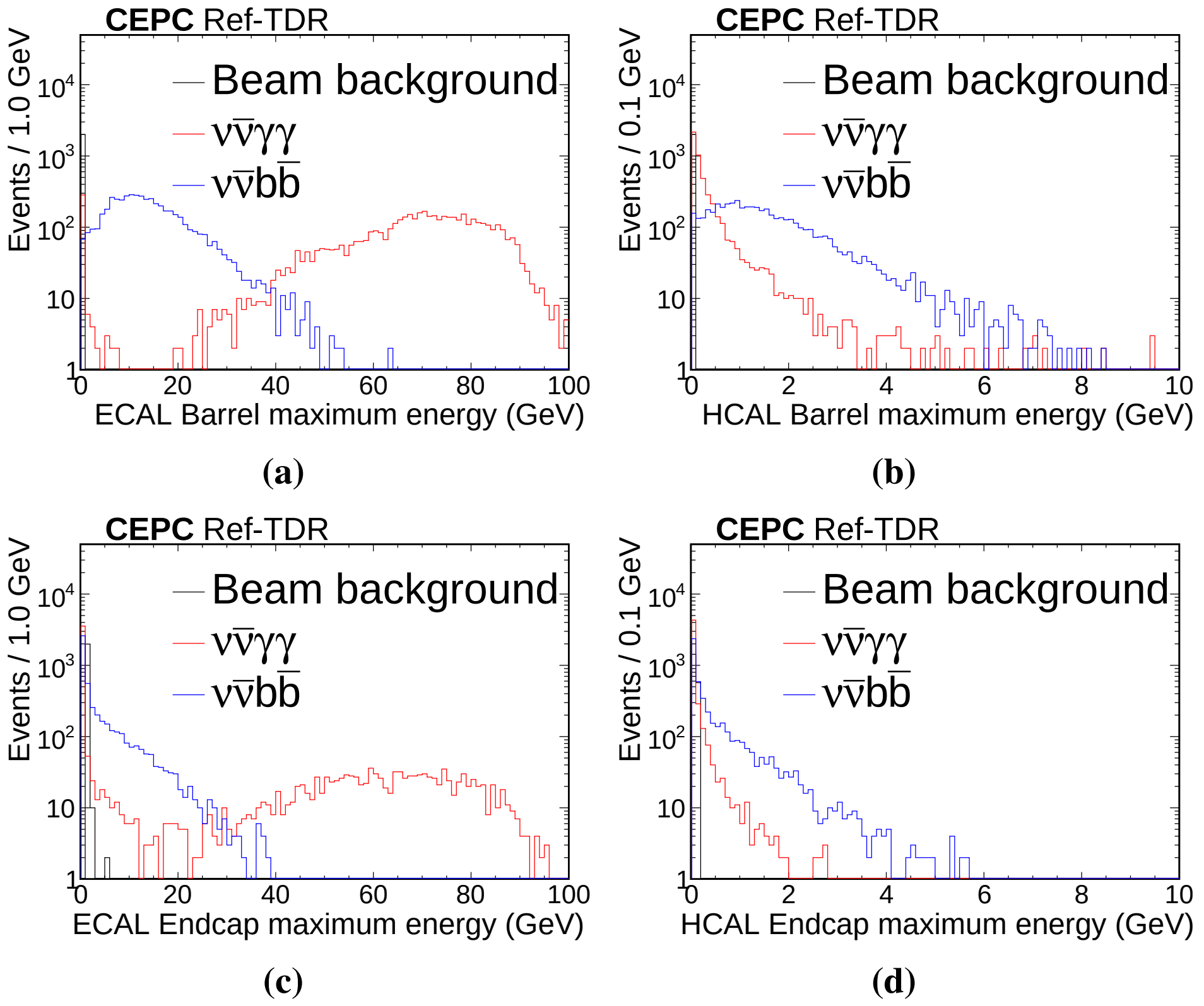

**Figure 12.6:** Maximum energy distribution of the supercell of beam background, $e^+e^- \to Z(\nu\bar{\nu})H(\gamma\gamma)$ and $e^+e^- \to Z(\nu\bar{\nu})H(b\bar{b})$. (a) ECAL Barrel; (b) HCAL Barrel; (c) ECAL Endcap; (d) HCAL Endcap.

**Table 12.7:** Baseline calorimeter energy threshold efficiency at the Higgs mode for 50 MW and the $Z$ mode for 12.1 MW.

| Process | Efficiency(%) | | Process | Efficiency(%) | | Process | Efficiency(%) | |
|---|---|---|---|---|---|---|---|---|
| **Higgs production** | | | | | | | | |
| $Z(\nu\bar{\nu})H(\gamma\gamma)$ | >99.9 | | $Z(\nu\bar{\nu})H(\gamma Z)$ | >99.9 | | $Z(\nu\bar{\nu})H(b\bar{b})$ | >99.9 | |
| $Z(\nu\bar{\nu})H(\mu^+\mu^-)$ | >99.9 | | $Z(\nu\bar{\nu})H(\tau^+\tau^-)$ | >99.9 | | $Z(\nu\bar{\nu})H(W^+W^-)$ | >99.9 | |
| $Z(\nu\bar{\nu})Z(W^+W^-)$lep | >99.9 | | $Z(\nu\bar{\nu})H(ZZ)$ | >99.9 | | $Z(\nu\bar{\nu})H(ZZ)$lep | 99.9 | |
| **Two Fermions** | **Higgs mode** | **Z mode** | | **Higgs mode** | **Z mode** | | **Higgs mode** | **Z mode** |
| $q\bar{q}$ | >99.9 | >99.9 | $\mu^+\mu^-$ | 99.6 | >99.9 | $\tau^+\tau^-$ | 99.6 | >99.9 |
| Bhabha | 99.9 | >99.9 | | | | | | |
| **Di-photon process** | **Higgs mode** | **Z mode** | | **Higgs mode** | **Z mode** | | **Higgs mode** | **Z mode** |
| $\gamma\gamma \to b\bar{b}$ | 99.6 | >99.9 | $\gamma\gamma \to c\bar{c}$ | 98.4 | 99.7 | $\gamma\gamma \to q\bar{q}$ | 81.6 | 90.7 |
| $\gamma\gamma \to \mu^+\mu^-$ | 26.9 | 29.6 | $\gamma\gamma \to \tau^+\tau^-$ | 78.0 | 92.1 | | | |
| **Background** | **Veto efficiency** | | | | | | | |
| | **Higgs mode** | **Z mode** | | | | | | |
| Beam Background | 99.3 | 99.3 | | | | | | |

Table 12.7 shows the efficiency for the baseline set of energy thresholds for the Higgs and $Z$ modes. For most of the physical processes, the efficiency is higher than 99%, except for the processes where the final state particles contain only neutrino or muon. For the di-photon processes, since they are low energy events, this calorimeter threshold algorithm will drop about 50% of the $\gamma\gamma \to q\bar{q}$.

To investigate the correlation between the efficiency and threshold, a threshold factor is defined. Different sets of energy thresholds are obtained by multiplying the threshold factor by the baseline thresholds. Figure 12.7 shows the efficiency as a function of threshold factor for Higgs mode, the range for the threshold factor is from 0.8 to 1.2, and "1" represents the baseline set of energy thresholds. In Figure 12.7, only the Higgs production with an efficiency

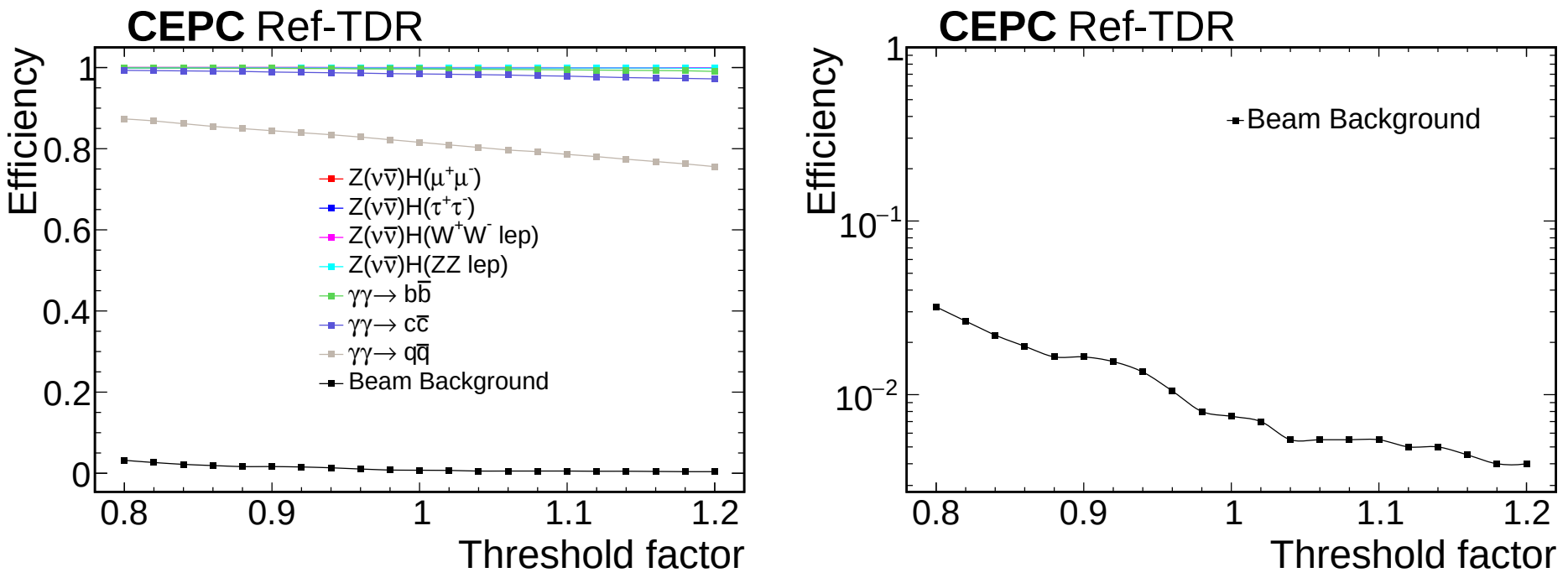

**(a)** **(b)**

**Figure 12.7:** Only those Higgs production with an efficiency below 99%, the di-photon processes and beam background are shown. Different sets of thresholds are obtained by multiplying the threshold factor, which is from 0.8 to 1.2, by the baseline set of energy thresholds. This figure shows the efficiency as a function of threshold.

below 99%, the di-photon processes and the beam background are shown in the Figure. So the variation of the threshold doesn't affect the efficiency for most of the physical events, since they leave significant energy deposition in the calorimeter, except for the low energy di-photon events and the processes contain only neutrino or muon at the final state.

#### 12.4.2.2 Muon detector

The signal processes that contain only $\mu$ as the visible final state particle, such as $e^+e^- \to Z(\nu\bar{\nu})H(\mu^+\mu^-)$ and $e^+e^- \to \mu^+\mu^-$, deposit only a negligible amount of energy deposition in the calorimeter. Consequently, such processes may fail the calorimeter trigger selection, but can be detected by the Muon detector. For the beam background, on average, it will generate less than 1 hit in the barrel region, but 400 hits in the endcap region. And in the endcap region, more than 99% of the beam background hits are within a radius smaller 1 meter in the $xy$-plane. To better separate the signal and background, the Muon endcap region is divided into two sub-regions: Endcap with radius larger than 1 meter and Endcap with radius smaller than 1 meter.

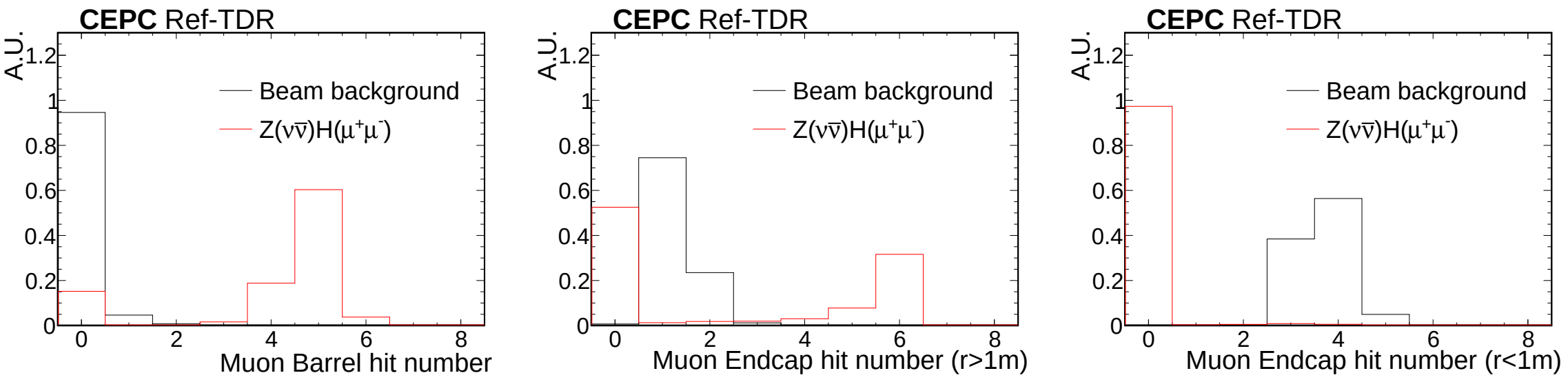

**(a)** **(b)** **(c)**

**Figure 12.8:** Number of hits for the $e^+e^- \to Z(\nu\bar{\nu})H(\mu^+\mu^-)$ and the beam background on the (a): Muon Barrel (ΔR<0.05), (b): Endcap with radius larger than 1 meter (ΔR<0.02) and (c): Endcap with radius smaller than 1 meter (ΔR<0.01). All the histograms are normalized to 1.

Define the angular separation $\Delta R = \sqrt{\theta^2 + \phi^2}$, where $\theta$ is the polar angle, and $\phi$ is the azimuth angle. By counting the maximum number of hits within ΔR<0.05 in the barrel region, and ΔR<0.02 in the Endcap with radius larger than 1 meter, it is able to tag the real $\mu$. Figure 12.8 shows the number of hits for the $e^+e^- \to Z(\nu\bar{\nu})H(\mu^+\mu^-)$ process and beam background on the Muon Barrel (a), Endcap with radius larger than 1 meter (b), and Endcap with radius smaller than 1 meter (c). For $e^+e^- \to Z(\nu\bar{\nu})H(\mu^+\mu^-)$, more than 99% of the events contain more than two hits in either Barrel or Endcap with radius larger than 1 meter. In contrast, for the beam background, nearly all the events have zero hits in the Barrel region, and less than three hits in the Endcap with a radius larger than 1 meter.

So for the baseline threshold, the Muon track is tagged if there is more than two hits in either the Barrel region or more than three hits in the Endcap region with a radius larger than 1 meter for both Higgs and $Z$ modes. The Endcap with a radius smaller than 1 meter of the Muon detector is not used at this moment, further study will be done in the future to use the information in this region. An event passes the Muon trigger if it contains at least one tagged Muon track. These thresholds are chosen as the baseline thresholds to suppress about 99% of the beam background. Table 12.8 shows the efficiency for the Higgs and $Z$ modes with Muon trigger. The efficiency for the $\mu^+\mu^-$ is only about 98%, since $\mu^+\mu^-$ is more forward than $Z(\nu\bar{\nu})H(\mu^+\mu^-)$ process, and it will fail the selection if both of the two muon are inside the Endcap with radius smaller than 1 meter region.

**Table 12.8:** Muon trigger efficiency for baseline threshold for the Higgs mode for 50 MW and the $Z$ mode for 12.1 MW.

| **Process** | **Efficiency(%)** | **Process** | **Efficiency(%)** | | **Background** | **Veto efficiency(%)** | |
|---|---|---|---|---|---|---|---|
| | | | **Higgs mode** | **Z mode** | | **Higgs mode** | **Low Lumi. Z mode** |
| $Z(\nu\bar{\nu})H(\mu^+\mu^-)$ | 99.8 | $\mu^+\mu^-$ | 97.6 | 98.5 | Beam background | 99.9 | >99.9 |
| | | $\gamma\gamma \to \mu^+\mu^-$ | 7.8 | 2.9 | | | |

### 12.4.3 Global trigger algorithms

The Global Trigger integrates trigger conditions from one or more sub-detectors to make a final trigger decision. By combining information from multiple trigger conditions, the system can more effectively distinguish between signal and background events, thereby further suppressing the beam background trigger rate. Furthermore, the injection background may be also needed to be considered as the newly injected beam is usually not very stable. A time window could be implemented in the trigger decision to suppress effect from the injection background.

Characters of the physics process would be utilized as the key point to separate the signal and beam background. In the case of beam background event, the number of trigger tracks or energetic supercells in the calorimeters is normally in the level of one or two based on the experience of the electron-positron collider (BESIII, Belle II, etc). In the contrary, for a real physics event, the number of tracks or the energy of the supercells is normally higher, depending on different process.

Table 12.9 summarizes the baseline L1 trigger algorithm, which uses the information from the Calorimeter and Muon detectors. Further studies for the track trigger for L1 need to be done.

For the calorimeter, a set of energy thresholds, as shown in Table 12.6, is applied to each subdetector. An event is classified as the signal if it satisfies any one of the energy threshold requirements. For the Muon detector, Muon track is tagged as mentioned in Section 12.4.2.2, and the event is classified as the signal if it contains at least one tagged Muon track. And the event is classified as the signal if it satisfies either the tracker, calorimeter or muon detector requirement.

**Table 12.9:** Baseline L1 global trigger algorithm. The Calorimeter and Muon detectors are used for the baseline L1 global trigger. Further study for the track trigger for L1 will be done in the future.

| **Detector** | **Algorithm** | **Threshold** |
|---|---|---|
| Calorimeter | Energy of supercell | See Table 12.6 |
| Muon Detector | Muon track tag | Number of Muon track > 0 |

Table 12.10 shows the expected L1 trigger rate for the baseline threshold for the Higgs and $Z$ mode. With the combination of the Calorimeter and Muon trigger, the L1 trigger efficiency for all the signal processes for the Higgs modes exceeds higher than 99%. Since the event production rate is relatively low compared with the designed trigger rate, which is less than 500 Hz for Higgs mode and 10 kHz for $Z$ mode, it is possible to collect most of the physical processes, including the Higgs resonance, two fermions for both Higgs and $Z$ mode, and four fermions for Higgs mode. For the beam background, the trigger veto efficiency is expected to be more than 99% at the L1 and more than 90% at the HLT.

Although the current trigger algorithm can already capture most of the standard model processes, specialized algorithms need to be developed to trigger new physical processes, such as the long-lived particles. In the future, further optimization of trigger algorithms, including the use of machine learning and neural network-based approaches, can be considered to enhance trigger efficiency while further suppressing the background rate.

**Table 12.10:** Expected L1 trigger rate for the baseline threshold for the Higgs mode for 50 MW and at the $Z$ mode for 12.1 MW. Samples are simulated with CEPCSW. Electronic noise is not added. Pile up event is not added.

| **Higgs mode** | **Efficiency(%)** | **$Z$ mode** | **Efficiency(%)** |
|---|---|---|---|
| Higgs production | >99.9 | $q\bar{q}$ | >99.9 |
| $q\bar{q}$ | >99.9 | $\mu^+\mu^-$ | >99.9 |
| $\mu^+\mu^-$ | 99.8 | $\tau^+\tau^-$ | >99.9 |
| $\tau^+\tau^-$ | 99.6 | Bhabha | >99.9 |
| Bhabha | 99.9 | | |
| **Di-photon processes** | | **Di-photon processes** | |
| **Di-photon event rate** | **Efficiency(%)** | **Di-photon event rate** | **Efficiency(%)** |
| 9.1 kHz | 40.3 | Low Lumi: 18.5 kHz | 42.9 |
| **Beam Background** | | **Beam Background** | |
| **Background event rate** | **Veto efficiency(%)** | **Background event rate** | **Veto efficiency(%)** |
| 11.4 kHz | 99.2 | Low Lumi: 90.0 kHz | 99.3 |
| **Total** | | **Total** | |
| 20.6 kHz | | Low Lumi: 118.9 kHz | |

### 12.4.4 Preliminary study for tracker

Besides the Calorimeter and Muon detector trigger, a preliminary track trigger is studied. The tracker comprises VTX, ITK, TPC and OTK. Charged particles produce hits across these sub-detectors. The track trigger study described in this section focuses on ITK, OTK and the outermost four layers of VTX. The innermost two VTX layers exhibit excessive background hits, while the TPC's 34 µs time window necessitates specialized matching algorithms; these components need to be addressed. The charged particles follow helical trajectories due to the magnetic field, which can be described by the equation 12.2 in the cylindrical coordinate system ($r\phi z$), with the assumption that the IP is at the origin:

$$\frac{r\kappa}{2} = \sin(\phi - \phi_0), z = \tan\lambda \cdot s \tag{12.2}$$

where $\kappa$ is the helix curvature, $\phi_0$ is the azimuth angle of the track at the IP, $\lambda$ is the helix angle, and $s$ is the helix arc length from the IP to the hit. For high transverse momentum tracks ($p_T$>2 GeV), $\phi - \phi_0$ becomes small, so $\frac{r\kappa}{2} \approx \phi - \phi_0$. And the helix arc length is equal to the radius of curvature times the central angle: $s = 2\frac{\phi-\phi_0}{\kappa}$. Combining these equations, one obtains the equation 12.3:

$$\frac{r\kappa}{2} \approx \phi - \phi_0, z \approx \tan\lambda \cdot r \tag{12.3}$$

The preliminary track trigger employs a 3D Hough transform [18, 19], mapping the hits position from the cylindrical coordinate space to the parameter space ($\kappa\phi_0\lambda$) within predefined ranges: $0 < \kappa < 0.0025$, $-\pi < \phi_0 < \pi$, $-\pi < \lambda < \pi$. The parameter space is divided into 50 bins per parameter, creating a 125,000-cell grid. A track is identified if a cell accumulates more than two hits from the VTX, more than two hits from the ITK, and at least one hit from the OTK. An event passes the track trigger selection if at least one track is found by the Hough transform. Table 12.11 summarizes the efficiency for the Higgs and $Z$ modes with the track trigger for the $e^+e^- \to Z(\nu\bar{\nu})H(\mu^+\mu^-)$, $e^+e^- \to \mu^+\mu^-$ and beam background processes.

**Table 12.11:** Track trigger efficiency at the Higgs mode for 50 MW and the $Z$ mode for 12.1 MW. Each background event includes one bunch crossing only.

| **Process** | **Efficiency(%)** | **Process** | **Efficiency(%)** | | **Background** | **Veto efficiency(%)** | |
|---|---|---|---|---|---|---|---|
| | | | **Higgs mode** | **Z mode** | | **Higgs mode** | **Z mode** |
| $Z(\nu\bar{\nu})H(b\bar{b})$ | 99.5 | $q\bar{q}$ | 98.0 | 99.5 | Beam background | 98.8 | 99.7 |
| $Z(\nu\bar{\nu})H(\mu^+\mu^-)$ | 93.5 | $\mu^+\mu^-$ | 77.0 | 85.0 | | | |

## 12.5 Data acquisition

### 12.5.1 Architecture design

The data acquisition system for the Reference Detector is specifically engineered to tackle the demanding requirements of high throughput and bandwidth characteristic of such large-scale experiments. At the heart of this system is an advanced online data processing framework, which is depicted in the accompanying Figure 12.9. This framework is organized into a two-level processing hierarchy, effectively separated by an essential buffering mechanism.

At the first level of this architecture, processing nodes are charged with reading raw data at phenomenal rates, reaching terabytes per second (TB/s). To alleviate the strain on network resources, these nodes conduct crucial initial data reconstruction and triggering. This processing phase handles both hardware-triggered data and more complex trigger-less data streams. Trigger-less data streams demand substantial processing capabilities, which are essential not only for the first-level but also for second-level nodes. To enhance processing efficiency across both levels, heterogeneous computing models are employed, providing significant acceleration in data handling and processing speeds.

The design of the core buffer in our two-level processing system serves two primary functions. Firstly, it facilitates system decoupling by segregating the processing of the two-level nodes, simplifying system development and maintenance. It also provides cache space, data access interfaces, and an asynchronous data processing mechanism, which allows algorithms developed offline to be seamlessly migrated to an online environment. This flexibility enhances the real-time data processing capability of the system. Secondly, the buffer plays a critical role in fine-tuning data through a calibration process that takes advantage of the buffer's ability to retain data over time. This facilitates complex reconstruction operations at the secondary nodes. The buffer, which can be implemented using either disk or memory in a distributed fashion, forms the core of our system's caching architecture. Its presence allows for online data analysis and the subsequent integration of processed results into the cached data, enabling more precise calibration and correction of the data in real time. This cache-based online computing approach exemplifies the progressive trend toward more adaptable and efficient system architectures in high-energy physics.

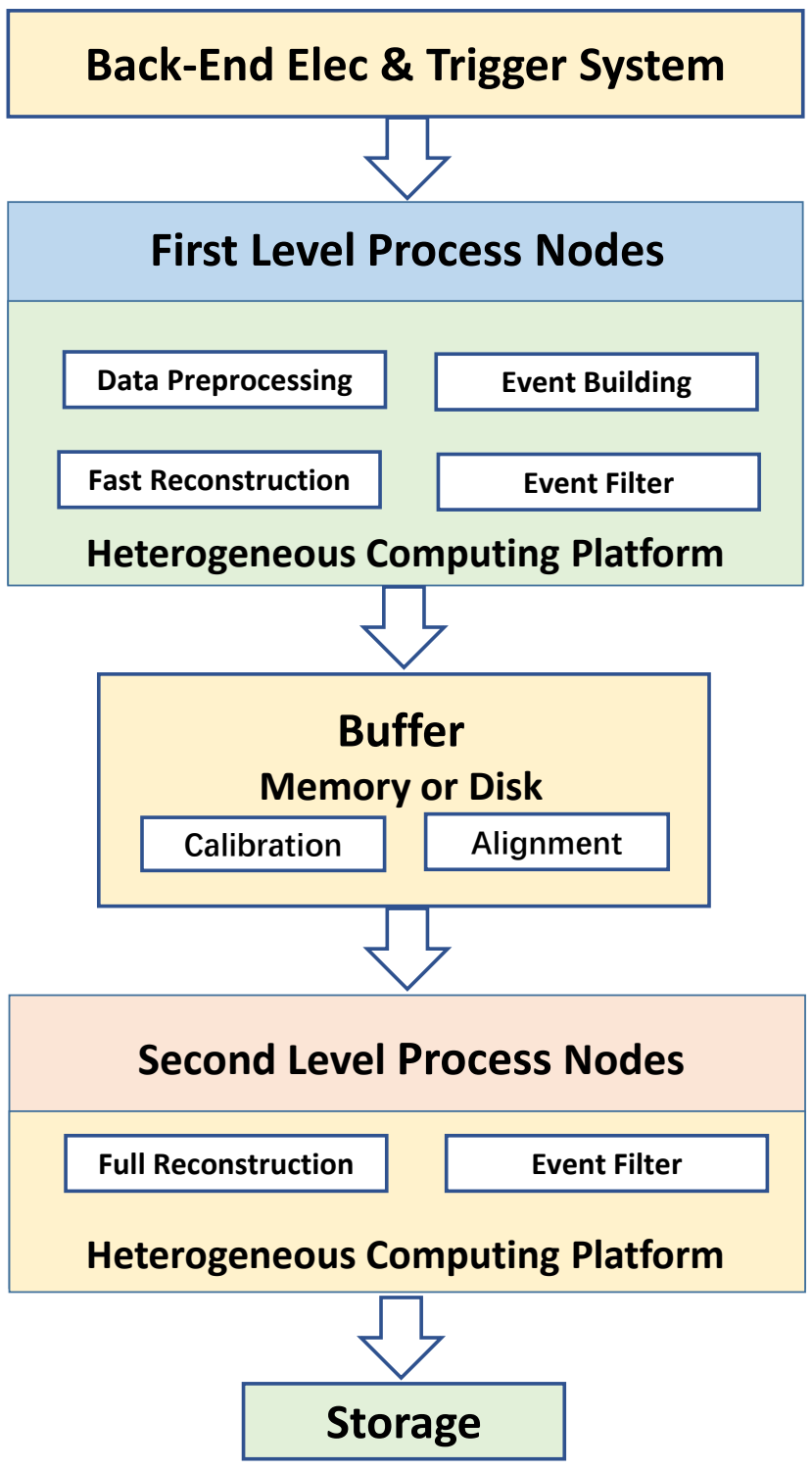

**Figure 12.9:** DAQ Architecture Design. This architecture includes a two-level processing framework with an intermediate buffer for decoupling stages, enhancing throughput and enabling real-time capabilities.

This buffer, which can be implemented in a distributed manner using disk-based or memory-based storage, forms the cornerstone of our system's caching architecture. Disk storage offers advantages in terms of larger capacity and cost-effectiveness, while memory storage provides faster access speeds and reduced latency. Ongoing investigations aim to determine the optimal strategy, balancing these factors with experimental application requirements, budgetary constraints, data access frequency, performance benchmarks, and system scalability.

Data entering the second-level nodes comes from this highly efficient buffer. Having been pre-filtered and processed at the first level, data volumes are substantially reduced. This reduction enables second-level nodes to deploy more complex algorithms and engage in thorough refined reconstruction and filtering activities. The capability to handle data in the form of files also allows second-level processing to adapt seamlessly to file-based operational algorithms, matching offline algorithm requirements effectively. This setup reduces additional development work and streamlines ongoing maintenance of processing algorithms.

The clear decoupling between first- and second-level processing provided by this buffer allows each to operate with entirely separate frameworks. The first set of frameworks can focus on optimal performance execution, while the second prioritizes precision in results. This configuration is crucial for balancing performance efficiency and accuracy throughout the data processing lifecycle.

From a software architecture perspective, data acquisition system is composed of two primary components: data flow software and online software. The data flow software is critical for managing the intricate processes of data collection, assembly, processing, and storage—essential tasks given the numerous computing nodes involved. At the core of this component is the detector readout, a vital system element intimately connected with the FEE, warranting its own comprehensive focus. Meanwhile, the online software operates independently to govern distributed node management, offering essential services such as configuration, control, information exchange, and high availability. These services include monitoring functions that ensure the DAQ system maintains high operational efficiency and stability.

Subsequent sections will provide a detailed examination of these components—detector readout, data flow software, and online software—further elucidating the system's comprehensive functionality design.

### 12.5.2 Detector data readout

The data collected by BEE and hardware trigger systems of each detector will be transmitted to a computer cluster for subsequent data processing. Table 12.12 lists the average data volume after the L1 trigger, the number of BEE boards for each detector and their respective data readout requirements. The data readout requirements listed in the Table 12.12 are estimated based on the average data volume of Low Luminosity $Z$ due to the simultaneous operation of Higgs and Low Luminosity $Z$ for data acquisition.

The data transmission scheme for each BEE board and hardware triggering systems to the data acquisition computer cluster primarily consists of two options: PCIe or other bus-based solutions, and network-based solutions. Since the BEE and hardware trigger systems of the Reference Detector adopt an xTCA architecture design, data readout from the detectors will utilize a network-based approach.

As shown in Table 12.12, under the current Higgs mode and low-luminosity $Z$ mode, conventional network readout schemes based on TCP/UDP or proprietary protocols are sufficient to meet the experiment's data throughput requirements. Therefore, the baseline design of this project adopts a TCP based network readout architecture.

Considering the anticipated increase in bandwidth demand and data processing complexity in future experiments (e.g., high-luminosity $Z$ mode), the project team has initiated a parallel pre-study on RDMA technology, based on the following considerations:

- RDMA enables direct memory access, significantly reducing CPU usage during data transfers. This helps to overcome the performance bottlenecks of traditional protocols and meets the demands of future high-throughput data transmission [20, 21, 22, 23].
- Validating RDMA on the existing readout electronics platform in advance can avoid large-scale hardware upgrades triggered by data volume growth, thereby lowering the cost and technical risks of future system upgrades.
- Applying RDMA in the HLT system reduces CPU load and interrupt overhead during data transmission, improves the efficiency of trigger algorithms, and lays the technical foundation for future one-to-many data distribution and real-time event assembly.

**Table 12.12:** Requirements for Detector Readout after L1. L1 trigger rate is 20 kHz for Higgs mode and 120 kHz for low luminosity $Z$ mode, while the proportions of the beam background passed L1 are about 1%.

| Detector | Readout Data Rate after L1-Trigger (Gbps) at Higgs | Readout Data Rate after L1-Trigger (Gbps) at Low Lumi $Z$ | BEE Board Number | Data rate per BEE board (Gbps) at Low Lumi $Z$ |
|---|---|---|---|---|
| VTX | 1.94 | 6.14 | 6 | 1.02 |
| TPC | 26.4 | 57.1 | 32 | 1.78 |
| ITK | 0.317 | 0.756 | 30 | 0.0252 |
| OTK_B | 0.690 | 1.62 | 55 | 0.0295 |
| OTK_E | 0.544 | 1.15 | 34 | 0.0338 |
| ECAL_B | 4.13 | 8.68 | 60 | 0.145 |
| ECAL_E | 7.10 | 15.4 | 28 | 0.55 |
| HCAL_B | 0.0448 | 0.204 | 348 | <0.01 |
| HCAL_E | 1.54 | 3.88 | 192 | 0.0202 |
| Muon | <0.1 | <0.1 | 3 | <0.01 |
| Trigger | - | - | 103 | - |
| Sum | 42.7 (5.34 GB/s) | 95.1 (11.9 GB/s) | 891 | |

In conclusion, although the current TCP based solution is adequate for present needs, early RDMA research is a forward-looking strategy to address the high-performance data transmission requirements of future experiments, aligning with the development trends of leading international particle physics projects.

### 12.5.3 Dataflow software

Data flow software is a critical component of a data acquisition system, managing the collection, assembly, processing and storage of data. For a data acquisition system to perform optimally, stable and reliable data flow software is essential.

The data flow software for the DAQ system is built using a proprietary framework called Heterogeneous architecture of data acquisition and processing (Radar). Radar's data flow infras-tructure, shown in the Figure 12.10, is a framework for reading and processing streaming data flows. The streaming readout architecture is highly efficient in high energy physics experiments, allowing data to be collected continuously from the detector without relying on specific trigger signals. This approach ensures the acquisition of comprehensive data during complex physical processes. Fast online processing algorithms then analyse this data in real time, filtering out invalid events and increasing the flexibility and efficiency of the system. Similar to traditional large-scale experimental data flow software, the data flow in Radar consists of four basic components: Read Out, Event Building, Online Processing, and Data Storage. In addition, there is a data flow manager component that coordinates and manages the operation of these data flow components. While the current default approach for DAQ is not streaming readout, there is potential to upgrade to this method in the future. Therefore, the adoption of a streaming readout framework such as Radar meets current requirements while providing flexibility for future upgrades.

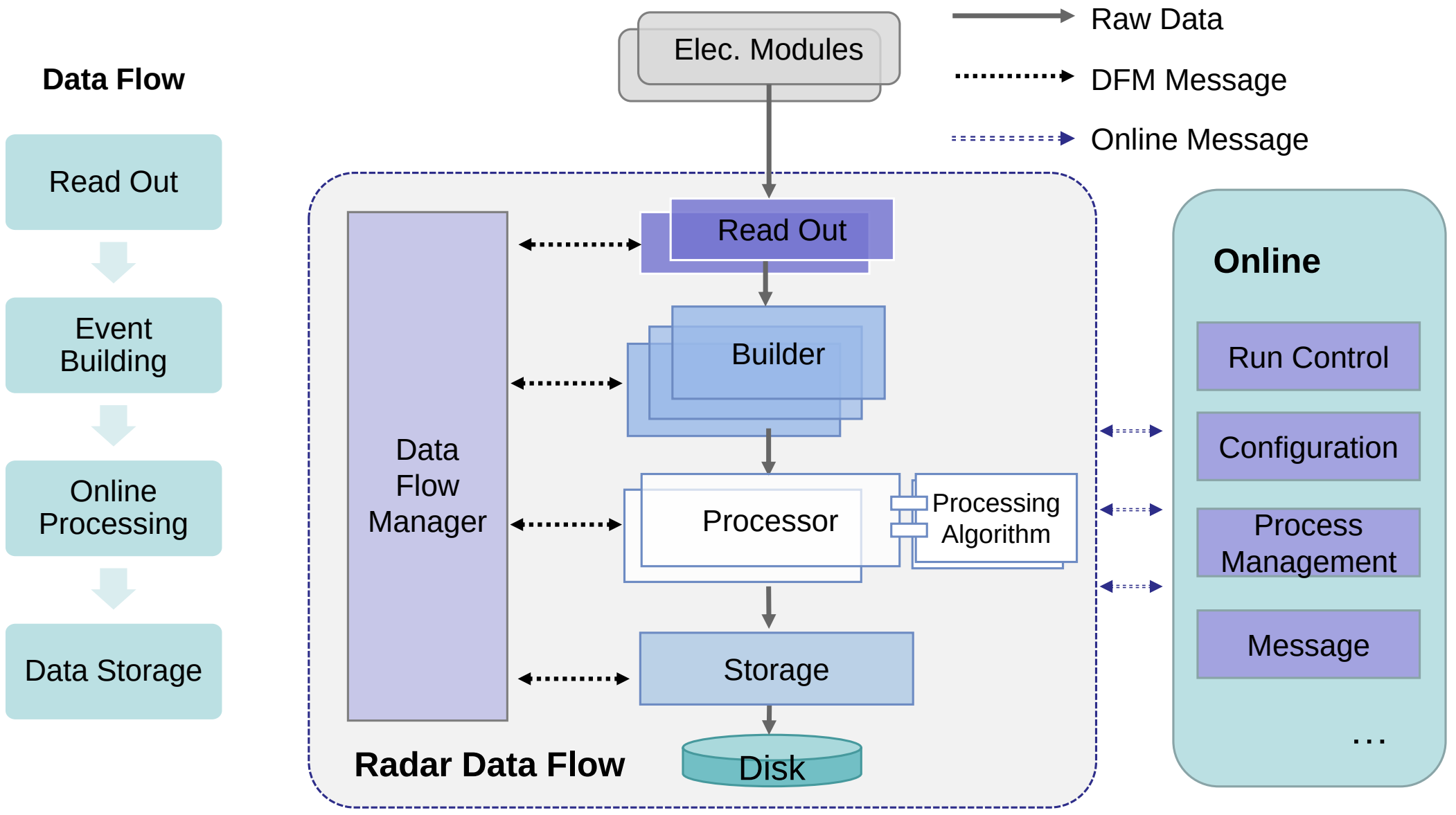

**Figure 12.10:** Dataflow Software Framework Design. The data flow in Radar is composed of four primary components: Read Out, Event Building, Online Processing, and Data Storage, all coordinated and managed by a data flow manager.

However, streaming acquisition generates numerous data streams, placing significant demands on the real-time transmission and processing capabilities of DAQ systems. Efficient and fast online data processing techniques and algorithms become essential to effectively handle incoming data streams.

The first version of Radar successfully managed the DAQ system for LHAASO for over five years without hardware triggering by analyzing all data in real time. The second version of Radar has been developed and will soon be operational for JUNO, using a hybrid model that integrates traditional hardware triggering with trigger-less streaming readout to increase flexibility and performance[24].

Efforts are currently focused on enhancing the Radar framework to meet the specific needs of the experiment. This includes extending the system to handle high-bandwidth requirements and improving real-time processing capabilities using CPU/GPU-based heterogeneous computing platforms for acceleration. Key research areas include high-bandwidth online software triggering, data compression and analysis of CPU/GPU cooperation mechanisms.

Effective scheduling strategies are essential for optimising the performance of heterogeneous computing frameworks. These strategies must take into account computation times, inter-processor data transfer times and network latency, especially for parallel tasks. In addition, adapting to changing hardware prices and performance requires support for different heterogeneous devices, including different brands of GPUs and FPGAs, to ensure long-term adaptability.

Memory management is critical to heterogeneous computing. When using multi-processor architectures, it is essential to minimise memory latency to avoid bottlenecks. Designing memory structures for parallel computing maximises performance by allowing parallel tasks to efficiently access shared resources.

There are three primary sources of algorithms implemented on heterogeneous frameworks at first-level processing nodes: optimising and accelerating offline algorithms for online use, porting hardware-based algorithms (e.g. from FPGAs) to a software environment, and developing new online algorithms directly. Key considerations include the design of interfaces for online data interaction, efficient memory management, and thread safety in parallel computations.

The primary objective is to increase the efficiency and speed of algorithms through heterogeneous computing, ensuring effective data acquisition and processing. This approach supports the development of a flexible data acquisition system that meets current requirements and can adapt to future challenges.

### 12.5.4 Online software

The data flow software for large scale high energy physics experiments runs on hundreds of computing nodes, requiring effective management and coordination of these distributed nodes as a central aspect of the DAQ system design. Addressing this challenge requires a separate layer of software that is independent of the data flow, known as online software. This online software plays a critical role in providing both internal and external services to the DAQ system and in monitoring the operation of the data flow software to ensure its smooth functioning.

A comprehensive online software solution has been developed and implemented as part of the Radar software, as shown in the Figure 12.11. Current online software is designed using a layered architecture. The interface layer provides access interfaces to the dataflow software and external systems. The distributed middleware consists of Kafka[25] and ZooKeeper[26], two third-party applications that transfer data between the dataflow and online software, decoupling their communication. The core online service consists of four modules that call each other via the HTTP[27] protocol without interfering with each other. The Supervisor, acting as the brain of the system, is responsible for the overall control and coordination of the system. This online software provides essential services such as configuration, control, information exchange, and high availability to the DAQ system. It uses a microservices architecture, which effectively minimizes module coupling and increases system flexibility and scalability. Additionally, container management mechanisms based on Kubernetes optimize software deployment and failover processes, thereby extending the reliability and lifecycle of the data acquisition software[28].

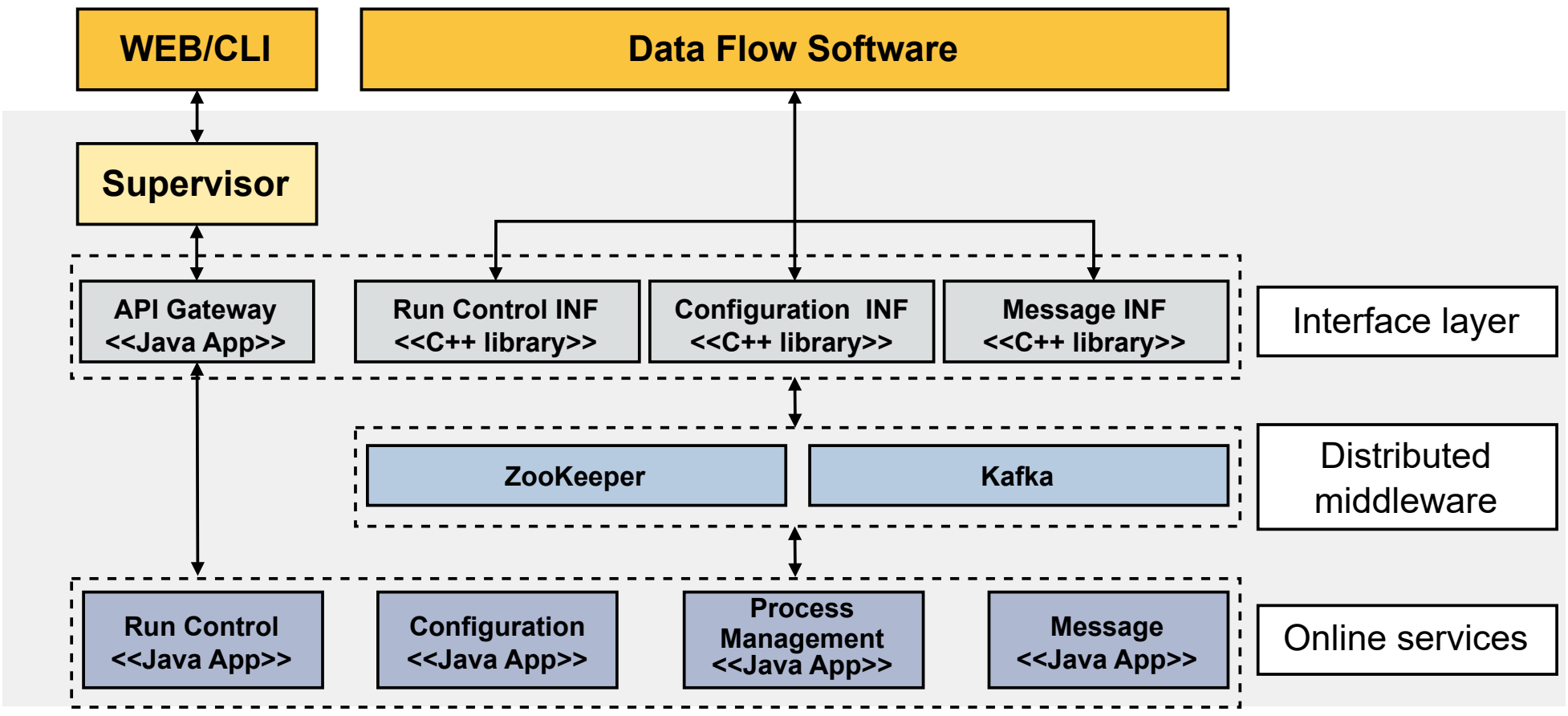

**Figure 12.11:** Architecture Design of the Online Software[28]. The online software system is structured into four layers: the Supervisor for control and coordination, the Interface Layer for accessing dataflow and external systems, distributed middleware with Kafka and ZooKeeper for communication decoupling, and online services utilizing a microservices architecture for enhanced flexibility and scalability.

This online software solution provides a high-availability service framework that is ideal for large-scale, high-energy physics experiments. It is also well suited to the DAQ system, providing a robust foundation upon which various services can be tailored and optimized to meet specific needs. As requirements evolve, this software can be further refined and enhanced to meet the particular functional and operational needs of the project, ensuring that the system remains adaptable and efficient for future challenges.

In conclusion, the data acquisition system's sophisticated integration of detector readout, data flow software, and online software ensures each component functions cohesively to meet the demanding requirements of high-energy physics experiments. Meanwhile, the system's design provides the necessary flexibility to accommodate potential upgrades and changes in experimental setups, ensuring continued effectiveness and adaptability.

## 12.6 Detector control

The detector control system aims to provide a stable, reliable, and secure operational environment for experimental operations, maximizing experimental efficiency and data quality. To achieve this, the system undertakes several key tasks.

Firstly, it continuously monitors a large number of critical parameters and immediately issues alerts upon detecting any anomalies, ensuring timely corrective actions to maintain system stability and safety. Secondly, the system efficiently executes process-oriented control based on the specific control requirements of the detector to ensure that all components of the detector operate safely and in coordination. Additionally, the system offers a unified, concise, and user-friendly interface, making monitoring and control more intuitive and efficient for operators. To enhance flexibility, the system also supports remote control and monitoring via terminal servers or web servers, allowing experts and operators of sub-detectors to conveniently manage and operate the system from remote locations. These features collectively ensure that the detector maintains optimal performance under all conditions.

### 12.6.1 Monitoring range

The detector mainly consists of VTX, Silicon Tracker, TPC, ECAL, HCAL and Muon detectors. Its control system is also divided into six sub-detector systems. Figure 12.12 shows the control requirements for each sub-detector system.

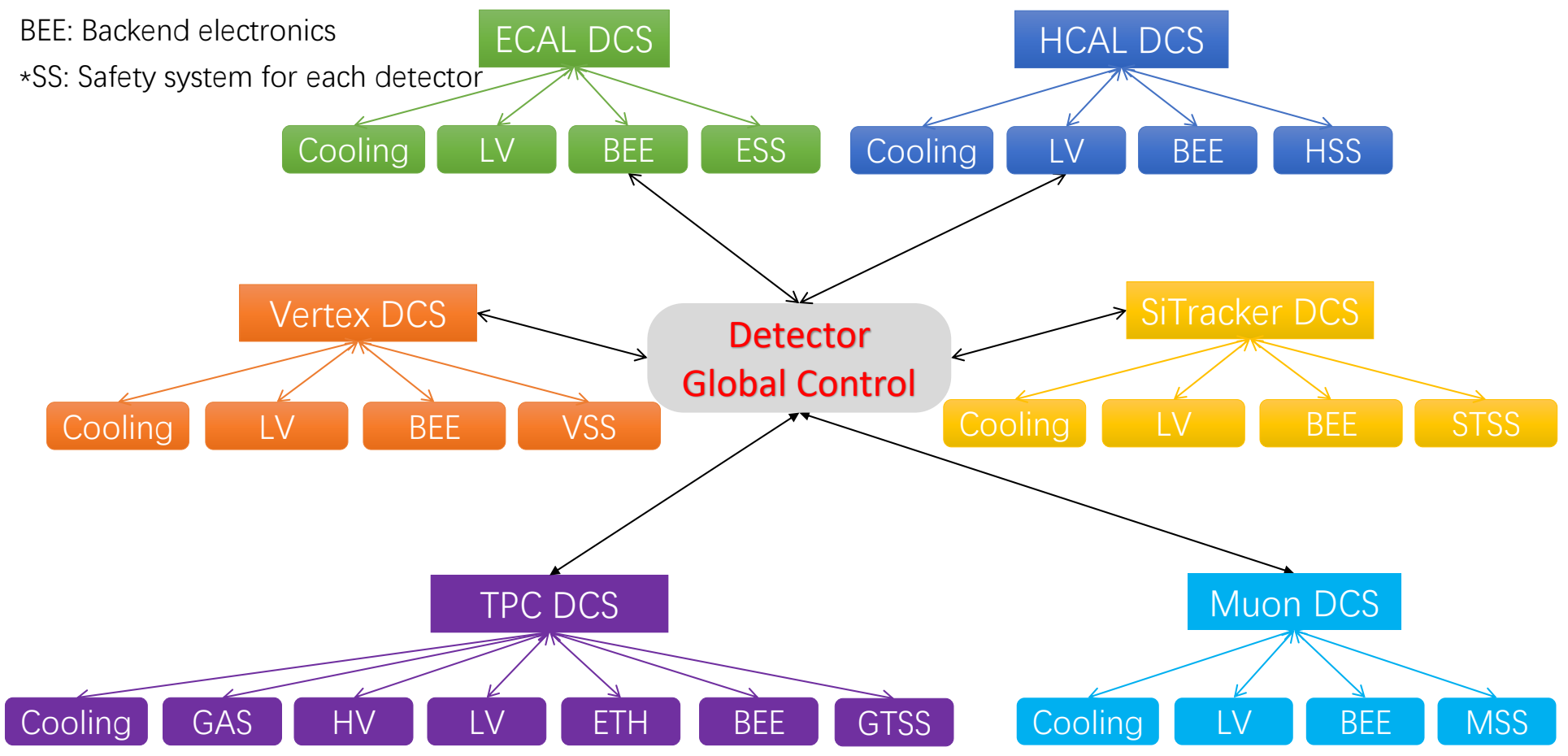

**Figure 12.12:** Schematic of sub-detector monitoring requirements. Standard systems (high-voltage, low-voltage, environmental monitoring, safety, and electronics control) with specialized modules for TPC (gas flow).

In this overview diagram, it can clearly identify the current common monitoring requirements for each sub-detector. These requirements primarily include low voltage system, cooling system, the electronics system and safety requirements. To better understand and manage these requirements, an initial compilation of the control and monitoring demands for each detector device has been prepared, along with an estimation of the number of parameters needed for monitoring, as shown in the Table 12.13. BEE mainly pertains to the electronics crates and boards, serving a critical role in controlling and monitoring the FEE boards. Through the BEE boards, the control system can precisely monitor each connection node, ensuring the stable operation of the front-end equipments. Furthermore, in addition to these general needs, the TPC detector has specific monitoring requirements. It requires additional attention to HV supply, gas flow, and environmental conditions such as temperature and humidity, to ensure precise and safe detection.

**Table 12.13:** Summary of specific devices and number of channels for detector requirements.

| Equipment | VTX | ITK | OTK | TPC | ECAL | HCAL | Muon | Sum |
|---|---|---|---|---|---|---|---|---|
| Power Crate | 2 | 10 | 56 | 6 | 9 | 90 | 2 | 175 |
| Power Channel | 96 | 202 | 1168 | 496 | 704 | 8640 | 40 | 11346 |
| BEE Crate | 1 | 3 | 8 | 4 | 10 | 56 | 1 | 83 |
| BEE Board | 6 | 30 | 89 | 32 | 88 | 540 | 3 | 788 |
| Trigger Crate | - | - | - | - | - | - | - | 17 |
| Trigger Board | - | - | - | - | - | - | - | 103 |
| Environmental | 8 | 984 | 800 | 496 | 768 | 512 | | 3568 |
| HV Crate | 1 | 4 | 6 | 4 | 6 | 40 | 1 | 62 |
| HV Channel | 66 | 3900 | 4500 | 496 | 740 | 8608 | 384 | 18694 |
| Gas | | | | ~30 | | | | ~30 |

This compilation not only helps to clarify the needs of each subsystem but also provides a reference for future system design and resource allocation. Continuous refinement and optimization of this process will ultimately lead to the efficient, reliable, and safe operation of the entire detector control system.

### 12.6.2 Common control

Before detailing the architecture, preliminary estimation of detector control parameters is essential, as outlined in the Table 12.14, with the DCS requiring real-time monitoring of 520,000 parameters – a current estimate excluding redundancy that may expand during operations. This scale necessitates robust real-time monitoring capabilities and efficient key data storage to support scientific analysis, demanding an optimized data architecture to maintain analytical accuracy and system performance under complex tasks. Consequently, the DCS employs a large-scale distributed framework with layered control units, which is illustrated in Figure 12.13, integrating EPICS and Tango Controls frameworks to enhance scalability, interoperability, and reliability while ensuring strict functional compliance at each module level.

The control unit employs a three-tiered architecture. The foundational controller module interfaces directly with hardware devices through flexible implementation options: utilizing digital communication modules (e.g., EPICS IOC services or SNMP protocols in high-energy physics equipment), adopting industrial digital controllers, or developing custom solutions. The intermediate data structure manager ensures reliable controller communication through robust drivers, enforces unified data management with established messaging frameworks, maintains database persistence, and exposes access interfaces to upper layers. The top user service layer provides intuitive interfaces that abstract underlying complexity through data encapsulation, enabling user-friendly data querying, manipulation, and monitoring without exposing low-level logic. Compatible with both EPICS and Tango Controls frameworks, this structure combines flexibility with lightweight design to optimize system performance and usability for complex detector operations.

**Table 12.14:** Summary of detector control and monitoring quantities. Multi-tiered Control Architecture Spanning Power Distribution (AC/DC, electronics), Trigger System, High-Voltage Infrastructure, and Gas/Environmental Sensors with 520,000 Parametric Monitoring Points.

| Equipment | Channel Number | Control and Monitoring Items | Total Count |
|---|---|---|---|
| AC-DC Crate | 55 | Switch, voltage, current, etc., about 6 items | 330 |
| AC-DC Channel | 550 | Switch, voltage, current, etc., about 6 items | 3,300 |
| Power Crate | 390 | Switch, voltage, current, etc., about 6 items | 2,340 |
| Power Channel | 17,696 | Switch, voltage, current, etc., about 6 items | 106,176 |
| FEE Board | 27,984 | Voltage, current, temperature, etc., about 10 items | 279,840 |
| BEE Crate | 110 | Switch, fan, status, etc., about 10 items | 1,100 |
| BEE Board | 1,018 | Switch, voltage, current, etc., about 6 items | 6,108 |
| Trigger Crate | 20 | Switch, fan, status, etc., about 10 items | 200 |
| Trigger Board | 150 | Switch, fan, status, etc., about 10 items | 1,500 |
| HV Crate | 91 | Switch, fan, status, etc., about 10 items | 910 |
| HV Channel | 18,694 | Voltage, current, status, etc., about 6 items | 112,164 |
| Gas | 1 | Flow control, humidity, etc., about 30 items | 30 |
| Environmental | 3,568 | Temperature, humidity | 5,448 |
| Total | | | ~520,000 |

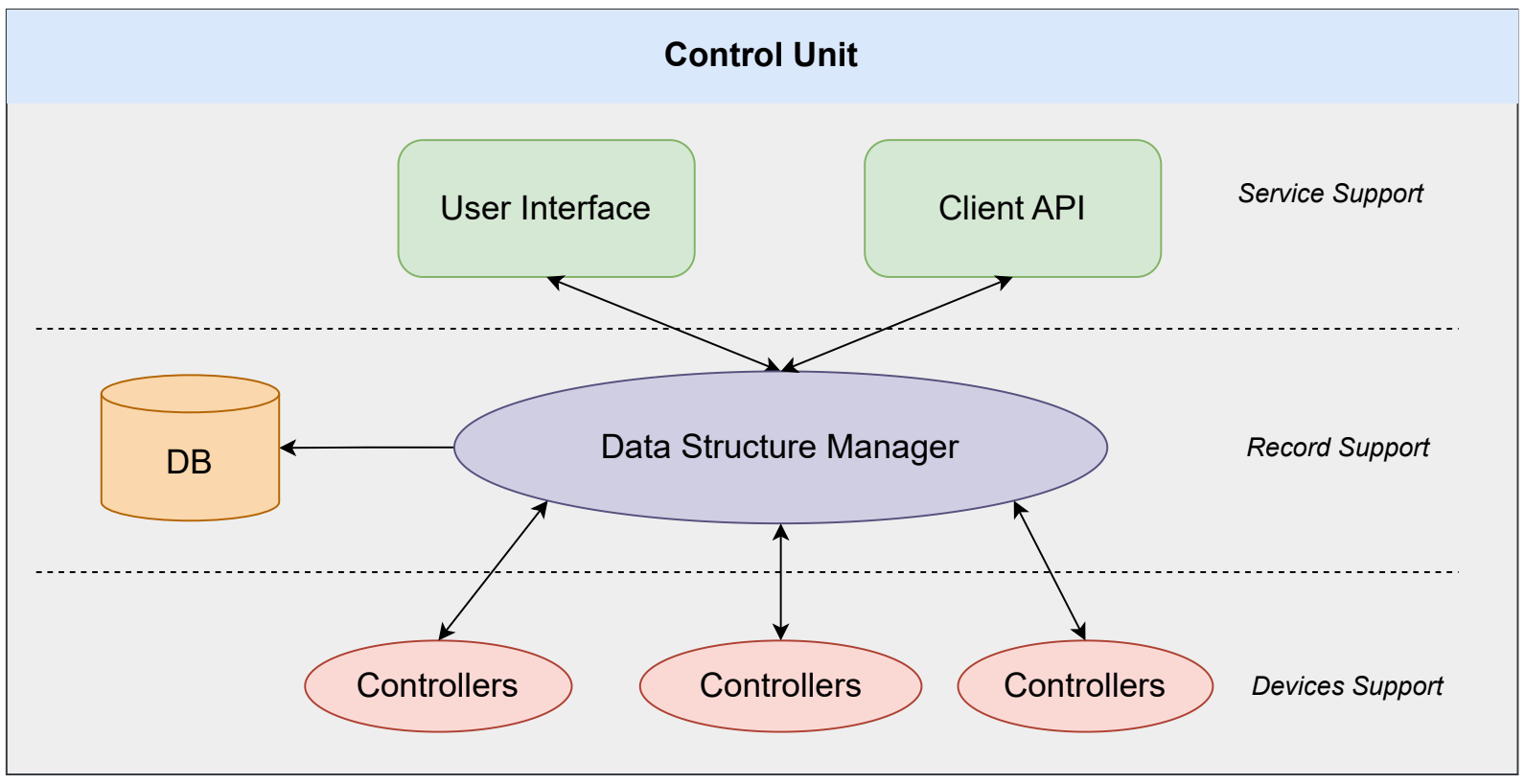

**Figure 12.13:** Software architecture of the control unit. Layered design of control units implementing functional requirement validation, ensuring modular integrity across a distributed framework..

### 12.6.3 Detector safety interlock

Detector storage and operational efficiency critically depend on environmental conditions, necessitating a highly available safety control system for protection. Silicon sensors are vulnerable to temperature extremes and humidity from excessive cooling, while gas detectors require precise gas state monitoring; all detectors must track beam conditions. This system must operate independently of the main control system to avoid software-induced safety failures, functioning continuously through power outages and maintenance periods. To enhance reliability, it implemented a two-level control scheme, as shown in Figure 12.14: the Detector Safety Module provides hardware-level emergency decision/execution, while the Detector Safety System (DSS) Control Unit software generates control strategies by integrating hardware signals and real-time monitoring data, issuing commands to coordinate auxiliary systems (e.g., HV, cooling, gas plant) to maintain safe operations.

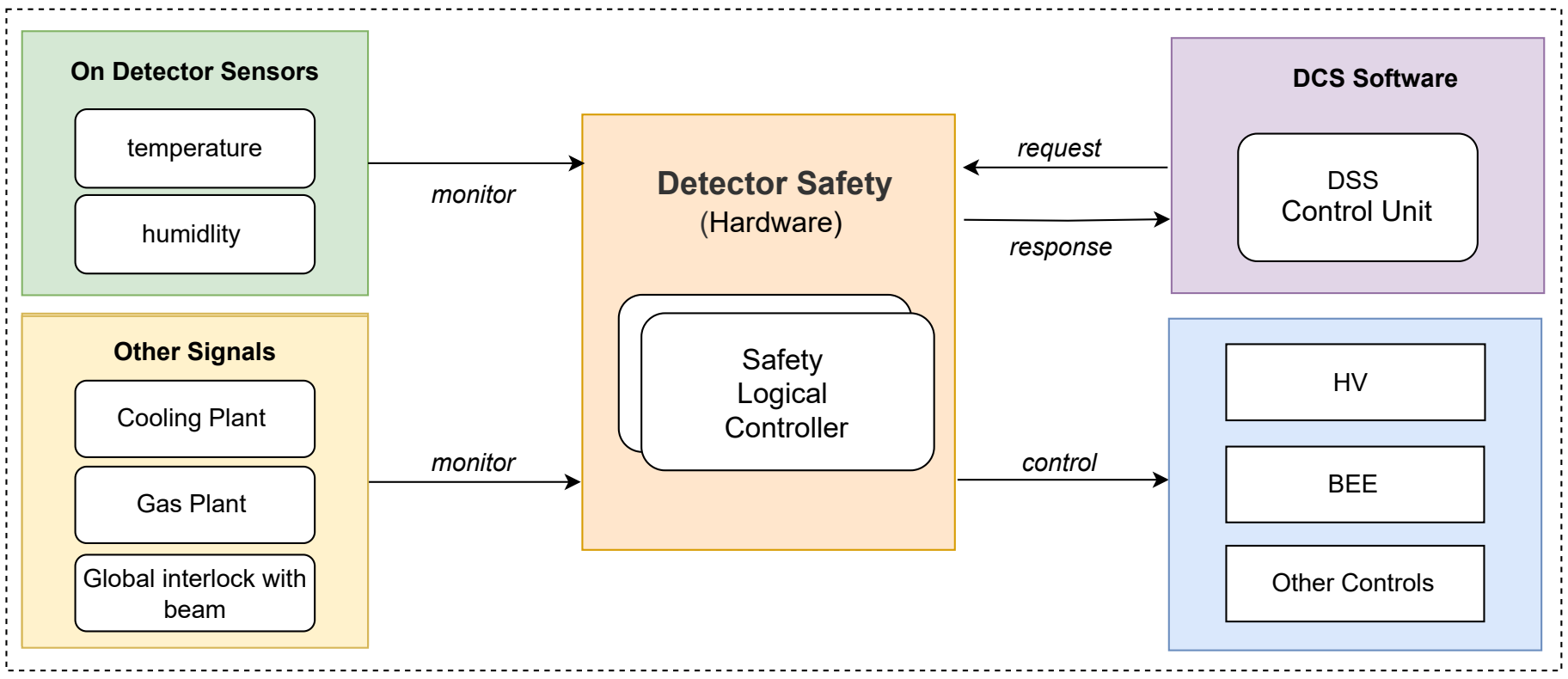

**Figure 12.14:** Detector safety control structure diagram. Hardware-driven emergency execution layer integrated with software-defined decision logic, synchronizing electronics, thermal management, and gas systems through real-time monitoring data fusion..

- Detector Safety: A hardware layer, independently handle emergency events, with fast hardware signals and trigger logic. Cut off HV at millisecond level through hardwired circuit to avoid risk diffusion caused by software delay.
- DSS: A set of control units, perform complex state machine control and the two interact and verify through request commands and response values.

The two-level security system has been achieved reliability, hardware handles millisecond emergencies to avoid the risk of software delays or crashes, software manages complex fine-state machines and provides human-machine interaction interfaces. At the same time, it also increases flexibility and expansion. The hardware layer solidifies the core security logic, and the software layer supports dynamic policy loading.

### 12.6.4 Environmental monitoring

This solution implements a hierarchical monitoring framework with differentiated technical strategies:

- Core Monitoring Unit employs industrial-grade wired transmission (anti-interference twisted-pair cabling, Modbus TCP/IP) for millisecond-level acquisition of temperature/humidity in detector environments, featuring redundant ring topology for high availability and three-tiered interlock protection triggered by self-diagnosing sensors during threshold breaches.
- Extended Monitoring Unit utilizes LoRaWAN-based[29] low-power wireless networks with self-organizing dynamic relays for wide-area coverage in peripheral spaces, enabling real-time environmental monitoring while reducing wiring complexity, as shown in Figure 12.15.

By combining industrial-grade assurance in the core unit with flexible deployment in peripheral zones, this architecture meets stringent reliability requirements for detector operations while enabling cost-effective large-scale monitoring, establishing a complete closed-loop monitoring ecosystem.

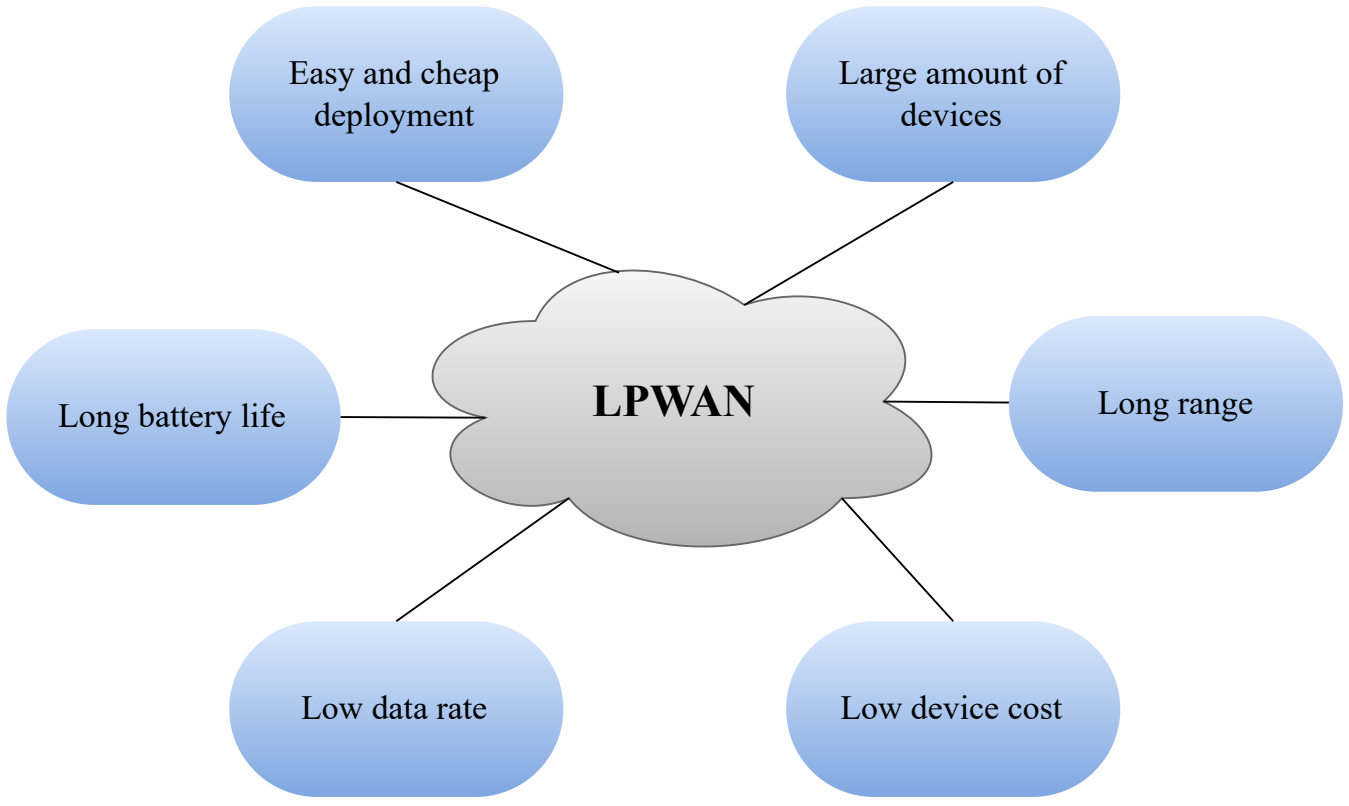

**Figure 12.15:** LPWAN technical advantages for six core competencies: long battery life, long range, low device cost, easy and cheap deployment.

## 12.7 Experiment control

The ECS is the central management platform in complex experimental environments, designed to achieve comprehensive and coordinated management of multiple scientific instruments and experimental processes. By providing a unified control platform, it ensures the real-time monitoring and management of various experimental data and equipment, thereby ensuring the safety, efficiency, and quality of the experiment.

It is responsible for global monitoring and management, integrating subsystems like TDAQ, DCS, environmental monitoring, and safety interlocks to ensure smooth and risk-free operations.

Unlike DCS, which manages detector hardware, ECS operates at a higher level, coordinating multiple DCS nodes, overseeing scheduling, and facilitating subsystem communication. It enhances system integration and response efficiency through AI technologies, such as large language models for human-machine interaction and ML for anomaly diagnosis, providing a user-friendly experience and intelligent analysis. Additionally, ECS consolidates data from various sources to offer comprehensive operational insights, enabling detailed assessments and early anomaly detection, thereby ensuring efficient and safe experiment operations, even in unmanned settings.

### 12.7.1 Architecture design

The ECS delivers efficient, safe experimental task execution through flexible and scalable management. As the central platform, it enables seamless integration of TDAQ, DCS, and environmental monitoring subsystems, reducing communication barriers via standardized protocols. ECS provides real-time monitoring of critical nodes with dynamic resource adjustment, coupled with robust fault detection, automated recovery mechanisms, and comprehensive safety interlocks. It features an intuitive graphical interface supporting advanced human-machine interactions to optimize user experience, while ensuring data integrity, traceability, and intelligent analysis for large-volume experimental data to drive decision-making and process optimization.

By addressing these requirements, ECS constructs a secure, efficient, and integrated experimental operational environment. Figure 12.16 illustrates the structure of the ECS system integrated with various subsystems.

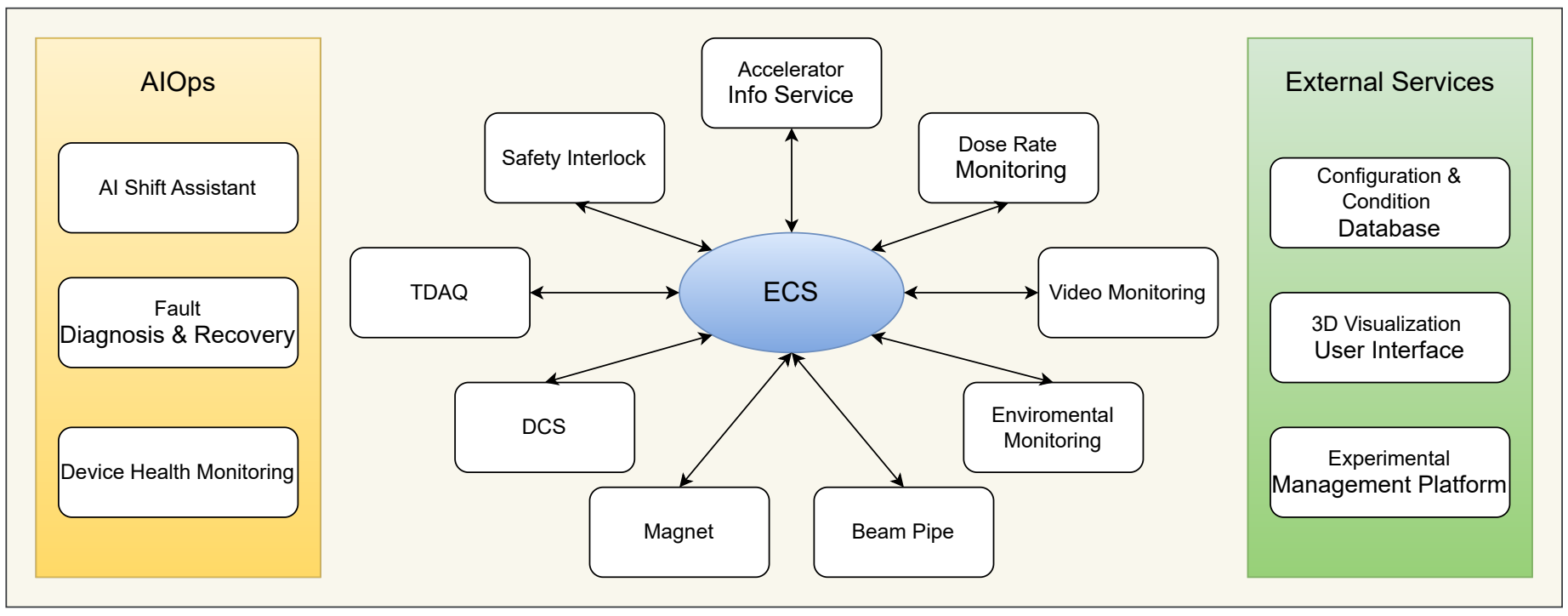

**Figure 12.16:** ECS structural overview. Integrate all systems, such as DCS, TDAQ, video, global safety interlock and many other services for infrastructure management and AI platform..

The ECS architecture delivers a scalable, reliable framework addressing complex experimental demands through modular design that enables independent functional unit development and adaptive subsystem integration via standardized protocols. Incorporating AI technologies enhances intelligent data management and process optimization, while fault tolerance mechanisms ensure operational safety. The system features: a Central Control Module orchestrating workflows and resource allocation; a Data Acquisition and Management Module preserving data integrity for real-time analysis; an intuitive User Interface Module with visualization tools; an AI Integration Layer implementing Large Language Models (LLMs) and ML for automation; and Safety Mechanisms with automated anomaly detection/interlocks. This unified structure supports current requirements and future technological evolution, with detailed module functionality explored in subsequent sections.

### 12.7.2 Module design

The ECS architecture integrates five key modules to ensure coordinated functionality:

- The Central Control Module, which is shown in Figure 12.17 orchestrates subsystem operations through its Process Manager handling production processes and Event Scheduler enabling automated control via Kafka/RabbitMQ[30] message brokerage.

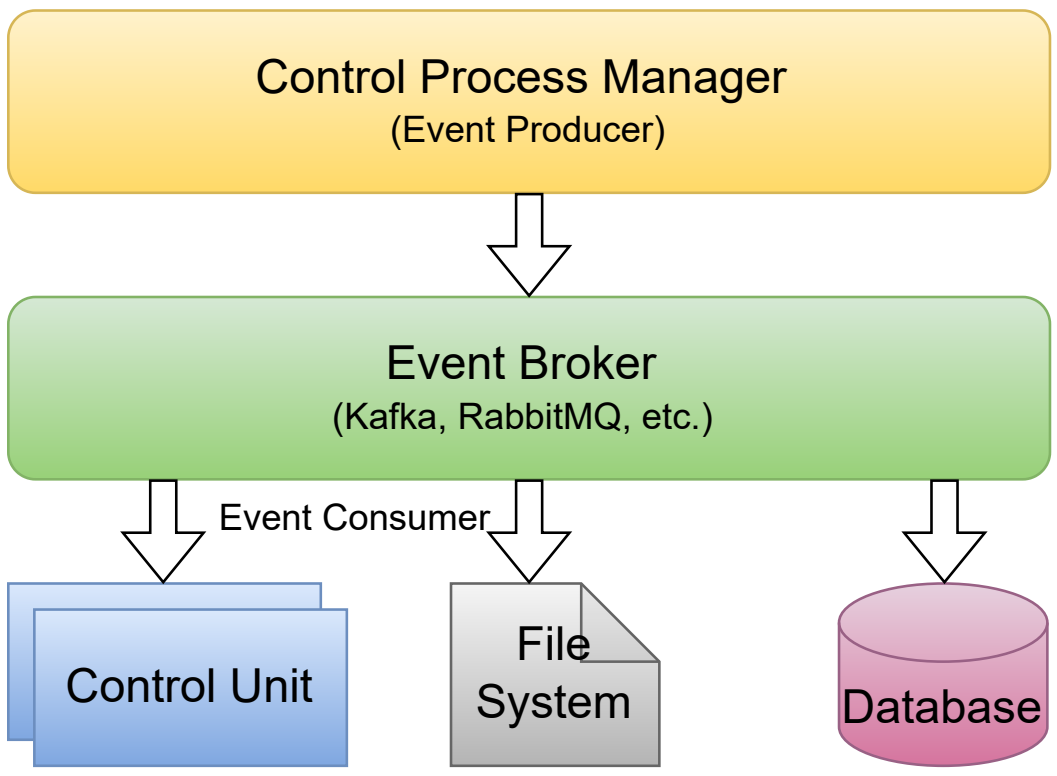

**Figure 12.17:** Central control software architecture. Decoupled component coordination through message brokers, enabling flexible integration of control processes, data services, and hardware units via producer-consumer patterns..

- The Data Acquisition Module centralizes operations by collecting subsystem data, interfacing with DCS/TDAQ systems while ensuring data integrity for analysis.
- The User Interface Module, which is shown as Figure 12.18 features React[31]/ Grafana[32] frontends and Flask[33]/ROOT[34] backends for intuitive control and ML-enhanced operations.
- The AI Integration Layer embeds ML/LLMs for anomaly prediction and Natural Language Processing (NLP)-driven task automation.
- Safety Mechanisms implement automated interlocks and redundancy designs for equipment protection.

Collectively, these components deliver secure, efficient experimental control in demanding scientific environments.

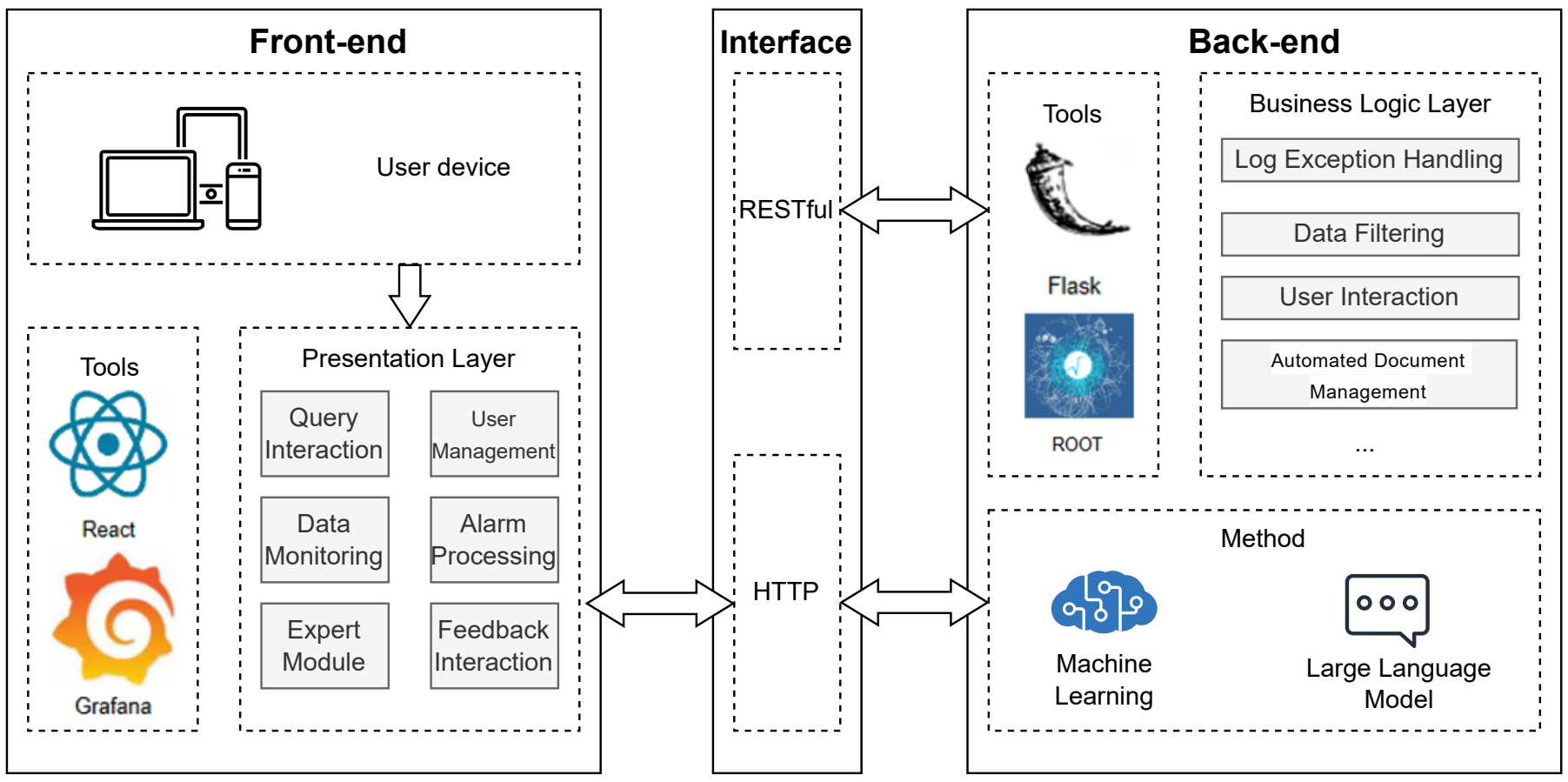

**Figure 12.18:** ECS user interface framework design. The user interface module integrates React, Grafana, RESTful APIs, Flask, and ROOT to ensure intuitive interactions, real-time data updates, and enhanced system intelligence, providing an efficient platform for experimental operation management.

## 12.8 Summary

This chapter outlines the requirements and overall architectural design of TDAQ and online system for the Reference Detector. Subsequently, it details the design specifications for each component and presents an implementation solution. The system must accommodate multiple operating modes and data rate variations from 1.34 MHz to 40 MHz collision frequencies across varying luminosity levels. Background data throughput reaches approximately 100 GB/s at Higgs mode and escalates to 1 TB/s at $Z$ pole.

The current simulation results of background and trigger algorithm shows the data rate could be reduced to the range of few to tens GB/s while maintaining high trigger efficiency via the trigger system. The L1 trigger system hardware utilizes a general-purpose processing board to build a three-stage trigger system. The first stage receives trigger data from BEE, performing module-level trigger condition generation. Subsequent stages integrate these conditions across sub-detector channels and execute global trigger selection for the entire detector. The data acquisition system handles post-trigger electronics data at terabyte/s rates over the network. Through multi-level parallel processing, it accomplishes HLT processing steps such as online event assembly, classification, and analysis, further reducing event rates and sizes before storing the filtered event data to disk.

The ECS provides unified monitoring and coordinated control of all experimental facilities, incorporating artificial intelligence methods to streamline system operation. Preliminary technical solution exploration has been conducted for these tasks. The detector control system, tasked with collecting data from control interfaces, directly interfaces with hardware systems for basic monitoring and transmitting information to the ECS. Additionally, the ECS supplies centralized IT infrastructure management, databases, and monitoring services to offer platform support for both data acquisition and detector control systems.

Following the recent update results of sub-detector design and simulation, significant progress in the technical development of the TDAQ and online systems has been achieved. No critical technological risks have been identified in the current design framework. The existing technical design demonstrates the capability to achieve trigger, control, and data acquisition functions, satisfying the operational requirements of the Reference Detector configuration. However, challenges persist in developing efficient trigger algorithms and managing data throughput within manageable hardware costs.

To propel the system's development, targeted R&D initiatives will be pursued to address critical experimental requirements. This initiative will concentrate on prototyping a high-speed universal trigger processor and establishing a compact experimental testbed. The platform will implement a TDAQ system based on modern ATCA standards, integrating backend readout subsystems, high-efficiency data transmission, and distributed parallel computing infrastructure. Key research priorities include:

- Simulation and Optimization of Typical Trigger Algorithms
  - Conduct in-depth simulations of detector-specific trigger algorithms, assess system throughput and real-time processing performance, and verify the integrated trigger-acquisition architecture.
- High-Speed Data Transmission Technology for Trigger Electronics
  - Develop high-speed data transmission interfaces operating at line rate and deliver parallel computing resources optimized for trigger processing.
- Distributed High-Bandwidth Parallel Computing Solutions
  - Design and implement distributed computing frameworks with terabit-per-second data transmission capabilities and parallel processing architectures.

These objectives need establish a systematic R&D strategy to achieve full TDAQ system functionality.

# Chapter 13 Offline Software and Computing

The CEPCSW [1] is designed for processing and analyzing data from the detector at the CEPC collider. The development of CEPCSW is based on several foundational High Energy Physics (HEP) software packages, including the Gaudi [2] software framework for event processing, EDM4hep [3] for the event data model, DD4hep [4] for detector description, Geant4 [5] for detector simulation, and ROOT [6] for data analysis. A key aspect of this endeavor is the development of CEPC-specific software, designed to meet the experiment's unique requirements. The core software will be introduced in Section 13.1, while the applications for simulation, reconstruction, analysis, and visualization will be described in Sections 13.2, 13.3, and 13.4, respectively.

To tackle the growing complexity of data processing tasks, emerging technologies are being actively explored. Research and development efforts focus on areas such as concurrent computing, machine learning, and quantum computing, which will be discussed in Section 13.5.

Grid computing technology, successfully implemented in the LHC experiments, connects computing and storage resources distributed across laboratories worldwide, facilitating the processing and analysis of EB-level data. Leveraging this technology, a Distributed Computing Infrastructure (DCI) have been established to support detector R&D activities. On this prototype, the latest Grid computing technologies are being investigated and evaluated to ensure robust support for the experiment's future operations. These aspects will be discussed in Sections 13.7 and 13.10.

In the current project, the software and computing team is responsible for providing the core software, including the framework, various services, data management, and computing systems. The development of detector-specific algorithmic code for simulation, calibration, and reconstruction is a shared responsibility between the software and computing team and the sub-detector teams. Close collaboration with the physics group ensures the smooth development of global event-reconstruction code and software tools for physics analysis.

## 13.1 Core software

### 13.1.1 Software architecture

The experiment initially began its detector design using the iLCSoft [7] software, which laid the foundation for the Conceptual Design Report [8]. To support ongoing R&D efforts, the experiment later transitioned to Gaudi as its underlying software framework.

The Gaudi framework, originally developed by the LHCb experiment [9], has now evolved into a common project used by many other experiments, including ATLAS [10], FCC-ee [11], BESIII [12], Dayabay Reactor Neutrino Experiment (DYB) [13], and others. Gaudi features a component-based architecture designed to support a wide range of physics data-processing applications, including event generation, simulation, reconstruction, and analysis. Its component-based nature provides flexibility, enabling the development of both shared components and, where necessary, experiment-specific components tailored to meet the unique requirements of each experiment.

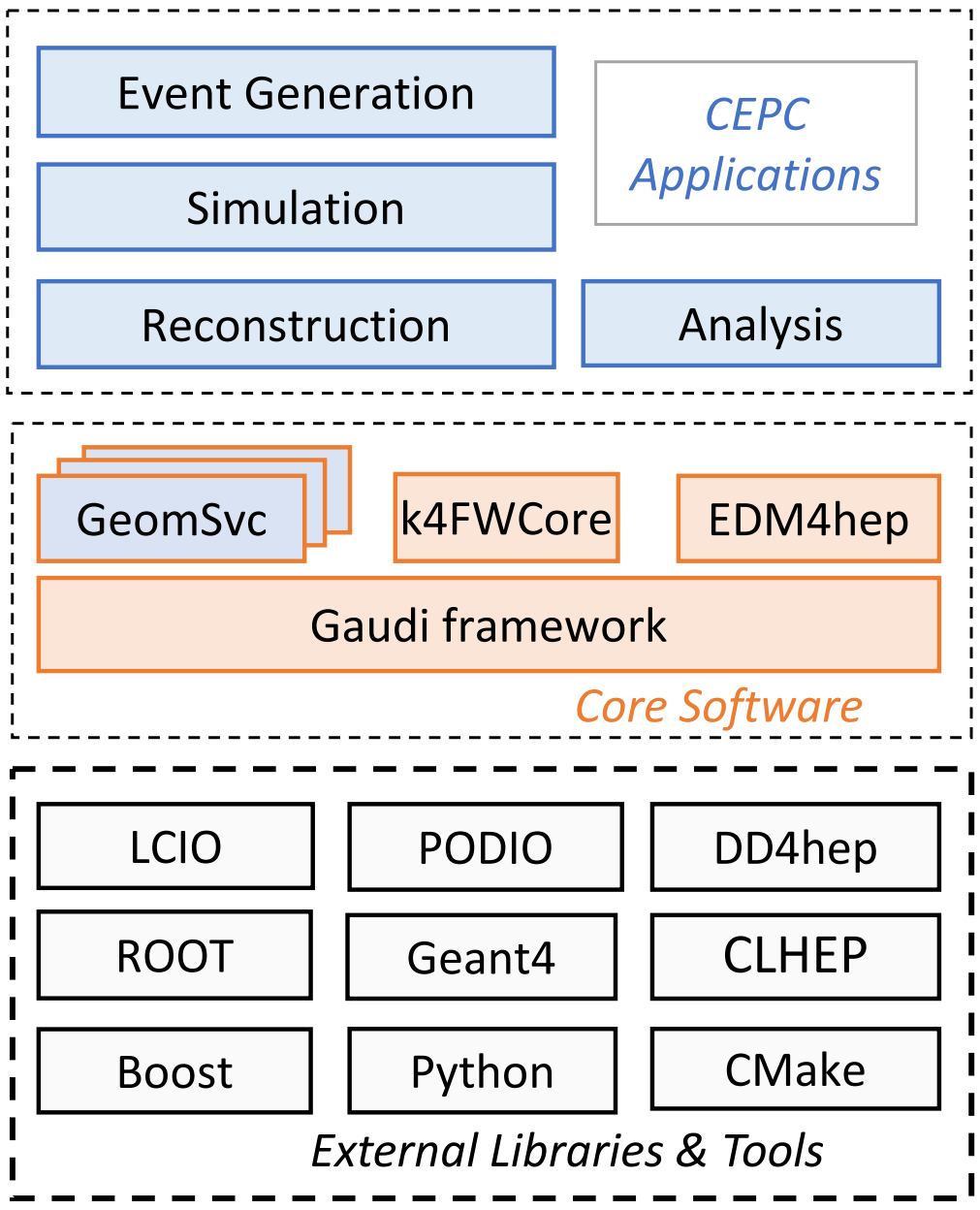

**Figure 13.1:** Schematic view of the CEPCSW software structure: The top layer contains applications for physics generation, simulation, reconstruction, and analysis. The middle layer provides the essential infrastructure and tools for building these applications. The bottom layer comprises third-party packages and libraries.

During the migration from iLCSoft to Gaudi, the Processors from the previous framework were re-implemented as Gaudi Algorithms. The entire data processing chain—including the physics event generator, detector simulation, digitization, and reconstruction—has been successfully migrated to the Gaudi framework. The migration was validated by comparing results from identical inputs with the same reconstruction algorithm implemented within the two frameworks.

The CEPC experiment is one of the pioneers in integrating with Key4hep [14], a unified software stack developed for future collider experiments such as CEPC, Compact Linear Collider (CLIC) [15], FCC [16], and ILC [17]. Through the adoption of the common event data model, EDM4hep, and standardized interfaces for accessing event data, the sharing of tools, algorithms, and software components across various experiments has been significantly streamlined.

Building on these advancements, CEPCSW has developed into a comprehensive software suite based on the Gaudi framework, the EDM4hep event data model, and a diverse array of its own dedicated services. These components underpin the simulation and reconstruction modules, which are responsible for generating and processing event data. This enables physicists to carry out in-depth analyses with precision and efficiency.

The overall structure of the CEPCSW is shown in Figure 13.1. It consists of three layers, defined as follows:

- Application Layer (Top Layer): This layer contains the core business logic and application-specific functionalities. It interacts with event data, enabling physicists to perform tasks such as physics generation, simulation, event reconstruction, and data analysis using the framework and external libraries.
- Core Software Layer (Middle Layer): This layer provides the essential infrastructure and tools for building applications. The Gaudi framework offers interfaces for much of the boilerplate code, allowing developers to focus on implementing application-specific logic. Additionally, `k4FWCore` supports smooth use of EDM4hep collections in Gaudi algorithms. Many other experiment-specific services are also provided, such as the `GeomSvc` which facilitates access to detector geometry information.
- External Library Layer (Bottom Layer): This layer consists of third-party packages and libraries integrated into CEPCSW to ensure full functionality. It includes common HEP packages such as `ROOT`, `Geant4`, `DD4hep`, `CLHEP` [18], `Linear Collider I/O (LCIO)` [19], and `PODIO` [20], along with non-HEP libraries like `CMake` [21], `Python` [22], and `Boost` [23].

CEPCSW provides two modes for software distribution: LHC Computing Grid (LCG)-based and Spack-based deployment. In the LCG-based deployment, most packages are sourced from CERN's LCG releases, while the remaining packages are built and deployed at IHEP. This layered architecture enables the deployment of multiple CEPCSW versions while ensuring consistency across external library versions. In the Spack-based deployment, CEPCSW is built using Spack as part of the Key4hep stack, which is hosted at CERN. With every new CEPCSW version release, the Spack recipe for CEPCSW must be updated to include the latest version in the Key4hep stack.

### 13.1.2 Detector description

Detector geometry is a crucial aspect of offline software, required by simulation, reconstruction, and event display. Maintaining consistency in geometry, materials, and magnetic fields across these applications can be challenging. For instance, the current Reference Detector comprises several sub-detectors. As the sub-detectors are being optimized, various designs need to be evaluated. Due to the numerous configurations, maintaining consistency across different applications is essential.

The DD4hep toolkit addresses this issue by offering detector information to various applications from a unified source. Instead of building everything from the ground up, DD4hep utilizes the ROOT geometry package as its core component. Based on the ROOT TGeo module, it introduces a Generic Detector Description Model, which extends ROOT's existing function-ality. Each detector element can be created using Extensible Markup Language (XML)-based compact detector descriptions and C++-based detector constructors. Additionally, any detector element can be augmented with extensions, such as providing extra information for reconstruction algorithms.

DD4hep has been integrated into the Key4hep project, and consequently, it is utilized in CEPCSW. A geometry service has been developed to manage the DD4hep instance. By specifying the path to the compact detector description file in the geometry service, the Generic Detector Description Model is created during initialization. The simulation and reconstruction applications can then access this information through the geometry service.DD4hep provides a simulation extension called DDG4, which translates DD4hep geometry into Geant4 geometry. The reconstruction software also leverages DD4hep for geometry information. The geometry service facilitates access to the detector description, enabling reconstruction algorithms to interact with the geometry model. This approach ensures consistency between simulation and reconstruction.

DD4hep can also be used to represent more realistic geometry, rather than idealized models, by integrating alignment and calibration data. Misalignment corrections can be applied to reflect actual detector shifts, and calibration constants are included to ensure accurate detector responses. By incorporating these elements, DD4hep offers a comprehensive and precise detector description for the experiment.

Magnetic field configurations in DD4hep are defined either through compact XML files or via C++ plugins. DD4hep supports several types of magnetic field setups. Currently, a uniform magnetic field of 3 Tesla is used for simulation studies.

### 13.1.3 Event data model

For CEPCSW, EDM4hep has been officially adopted as the event data model. EDM4hep is specifically designed to provide a unified framework for managing event data across various experiments. This standardized approach reduces fragmentation and ensures that data can be effectively shared, compared, and reused across different experiments and projects.

EDM4hep supports various backend data formats, including the ROOT format, LCIO format, and other custom data formats. These formats can be selected and configured according to the user's needs and the characteristics of the experimental data, ensuring efficient processing and analysis of high-energy physics experimental data within EDM4hep. By accommodating multiple backend data formats, EDM4hep delivers flexibility and scalability, empowering users to select the most appropriate data format for data management and processing based on their specific requirements. This feature also simplifies data exchange and enhances collaboration across different research projects. In CEPCSW, the ROOT format is designated as the default backend for data storage.

EDM4hep defines models for tracker and calorimeter data and supports all aspects of event reconstruction, ranging from raw data to fully reconstructed particles. Moreover, EDM4hep is designed to be extendable and customizable to meet the specific needs of individual experiments and several CEPC-specific classes have been added to EDM4hep.

### 13.1.4 Software development and validation

Modern software development practices are applied in CEPCSW, following High Energy Physics Software Foundation (HSF) [24] best practices. The focus is on open-source tools rather than HEP-specific solutions, with LLM [25] assisting developers using GitHub [26] and Stack Overflow [27] resources.

Git manages version control, with source code hosted on GitHub and IHEP GitLab. Contributions are managed through a Pull Request (PR) and Merge Request (MR) model, while issues are used to track bugs and enhancements. Automated building and testing are performed using GitHub Actions and GitLab Runners.

CEPCSW uses CMake for configuration and supports both Make and Ninja for building. The framework is organized into packages that produce applications, libraries, and Python modules. Installation is streamlined through Key4hep-Spack, precompiled CernVM File System (CVMFS) binaries, and a supplied Dockerfile.

To ensure software quality, a comprehensive validation system is implemented, operating at three distinct levels. Unit testing examines the smallest testable software components, serving as the foundation of the system. Integrated testing assesses overall CEPCSW functionality across platforms, as well as complete offline workflows, including simulation, calibration, reconstruction, and analysis. Finally, physics performance validation verifies characteristic physical quantity distributions to evaluate software health and measure the impact of code changes.

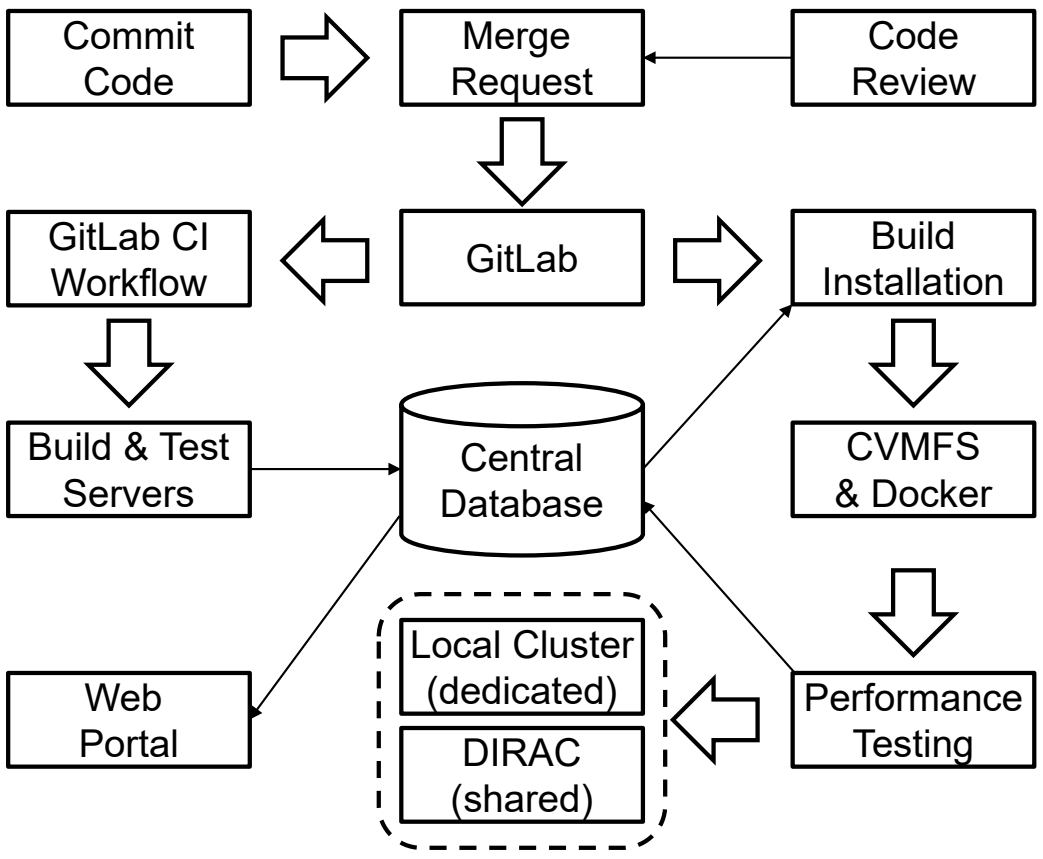

**Figure 13.2:** Automated validation infrastructure: designed to support unit testing, integrated testing, and physics performance validation, ensuring robust and reliable functionality across all aspects.

The automated validation infrastructure integrates with GitHub/GitLab Continuous Inte-gration (CI)/Continuous Deployment (CD) systems, as illustrated in Figure 13.2, and provides the following capabilities:

- Development testing: Offers configurable frequency, with lightweight unit and integration tests triggered on each commit or merge request to ensure prompt issue detection. Physics validation runs are scheduled daily or weekly. Test results are stored in a centralized repository and visualized through a web interface for real-time monitoring.
- Release validation: Large-scale data production is manually initiated using predefined workflows submitted to local or distributed computing systems. Key physical quantities are compared with standard reference distributions to assess and validate detector performance.
- Automated releases: The latest CEPCSW builds, documentation, and container images are automatically published daily to CVMFS using GitHub/GitLab CI/CD pipelines.
- Central database: Records all test results with automatic test matrix supporting multi-platform testing and real-time web status display.

The CEPCSW documentation is written in Markdown and built using Sphinx with the Read the Docs theme, generating outputs in both HTML and PDF formats. Markdown files are maintained within the CEPCSW repositories, enabling documentation to be updated alongside code changes.

## 13.2 Simulation

### 13.2.1 Simulation workflow

The CEPCSW simulation workflow comprises three stages (Figure 13.3): physics event generation, detector simulation, and digitization. All stages use Gaudi algorithms and tools connected through the EDM4hep data model. Physics Event Generation generates initial particles using tools like WHIZARD [28], Pythia6 [29], and Babayaga [30]. GenAlg reads generator files and converts them to MCParticle objects via GenTools. Particle gun generation is also supported. Detector simulation processes initial particles to generate detector hits. Full simulation uses Geant4 with DDG4 extension of DD4hep for geometry conversion and sensitive detector setup, recording Monte Carlo truth information for hit-particle relationships. Fast simulation employs parameterization or machine learning methods. Digitization converts simulation hits into data objects matching real detector data. Sub-detector-specific algorithms handle silicon trackers, TPC, calorimeters, and muon detectors. Beam-induced background is incorporated through three mixing modes: pre-simulation particle mixing, post-simulation hit mixing before digitization, or post-digitization overlay enabling data-simulation mixing.

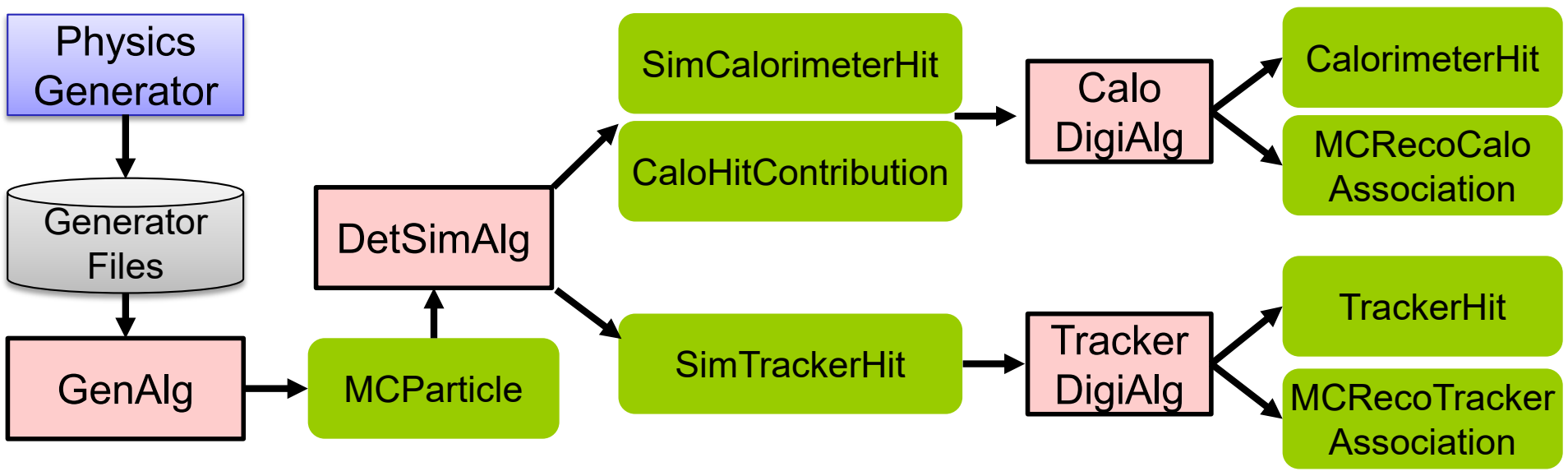

**Figure 13.3:** Simulation data flow. The simulation begins with the physics generator, which produces files containing the initial particles information. At the detector simulation level, GenAlg and DetSimAlg generate MC truth data, including MCParticle, SimTrackerHit, SimCalorimeterHit and CaloHitContribution. Finally, digitization algorithms, such as TrackerDigiAlg and CaloDigiAlg, simulate TrackerHit and CalorimeterHit data, which are similar to real experimental data.

## 13.2.2 Physics generators

During physics and detector R&D, a variety of event generators are utilized to generate physics events and save them for further processing. Among them, WHIZARD plays a significant role in the study of physics performance and optimization of detectors. WHIZARD is a versatile MC event generator capable of simulating particle collision processes in high-energy physics experiments, including particle production, decay, and interactions. WHIZARD is employed in several aspects:

- Physical Performance Research: WHIZARD can simulate various physical processes generated by electron-positron collisions at CEPC, such as $H$ boson production and the production of $W$ and $Z$ bosons. This aids researchers in understanding the physics characteristics and potential signals of these processes.
- Detector Optimization: By simulating different detector designs and configurations, WHIZARD assists researchers in evaluating the performance of detectors, including momentum resolution of charged particles, energy resolution of photons, jet energy resolution, particle identification capabilities, etc., thereby optimizing the detector design.
- Event Simulation: WHIZARD is capable of generating a large number of physics events, which can be used to test and verify the detector system, ensuring its ability to accurately measure and reconstruct physical events.
- New Physics Exploration: WHIZARD can also be used to explore new physics phenomena beyond the Standard Model, such as dark matter, electroweak phase transition, and supersymmetry particles. By simulating these new physics processes, researchers can design experiments to search for evidence of new physics.

In addition to WHIZARD, Pythia6 [29] is extensively used for simulating hadronization processes, where the hadronization parameters of Pythia6 have been optimized and tuned using the Omni-Purpose Apparatus (OPAL) data [31]. Babayaga is used for generating Bhabha pro-cesses, di-gamma processes, and di-muon processes, and so on. TwoGam [32] and DIAG36 [33] are employed to simulate two-photon collision processes. All these generators, based on their characteristics, are used in different aspects of CEPC R&D, yielding reliable cross-sections and a large number of physical samples for study.

Furthermore, an interface has been developed within CEPCSW that can directly call Pythia8 [34], making it very convenient to generate many physical processes without the need to save intermediate results, thus saving storage space.

The utilization of these event generators is crucial for the R&D, as they provide the necessary tools to simulate and understand the complex physics that will be encountered in the experiments. This comprehensive approach ensures that the detectors are designed and optimized to meet the challenges of high-energy physics research at the CEPC.

### 13.2.3 Detector simulation

The development of detector simulation software based on Geant4 aims to support simulations for various detector designs. Geant4 is a versatile toolkit used for simulating the interaction of particles with matter, widely applied in particle physics, nuclear physics, astrophysics, and medical physics. Since Geant4 itself is a toolkit and does not provide a complete simulation program, software development needs to be tailored based on experimental requirements. An independent Geant4 program requires a main function, detector construction, physics list, and physics generation action. Additionally, to extract information from Geant4 during the simulation process, User Actions at the Run, Event, Tracking, and Stepping levels must be provided, as well as a Sensitive Detector module for simulating detector responses. In the Sensitive Detector module, corresponding hit types need to be implemented and placed into a hit collection. Geant4 simulations also offer fast simulation interfaces, allowing users to customize the simulation process when particles enter specific regions. As introduced earlier, in CEPCSW, the detector geometry is described using the DD4hep package, which provides a unified and modular way to define geometry and materials. These descriptions are then converted into Geant4-compatible representations for simulation. Merely completing the aforementioned Geant4 components is insufficient to run programs in CEPCSW. The main issue is that Geant4 has its own Run Manager responsible for the simulation process, controlled by Geant4 itself, whereas in CEPCSW, the underlying framework Gaudi is responsible. This necessitates the development of additional software modules to integrate Geant4 and Gaudi for smooth operation. Figure 13.4 illustrates the design of the current simulation framework software.

The algorithm `DetSimAlg` is the main entry point for detector simulation. It is responsible to initialize multiple components, which will be used by Geant4. Since Gaudi and Geant4 have their own base classes, the so called Geant4 wrapper layer is developed to integrate both parts. For example, the Geant4 run manager needs a class derived from `G4VUserDetectorConstruction` to initialize a detector geometry, so a derived class is implemented. Within the implementation of this class, an instance of Gaudi component `DetSimTool` is invoked. In the algorithm, the Gaudi component and the Geant4 wrapper class are created and associated. After all the Gaudi components and Geant4 wrapper classes are created, a service named `DetSimSvc` is then invoked by the algorithm to initialize the Geant4 run manager. As mentioned before, the Geant4 run manager will be initialized with these wrapper classes. After the initialization, the Geant4 run manager is ready for simulation. During the event loop, the algorithm invokes the service, which invokes the run manager to simulate one event only.

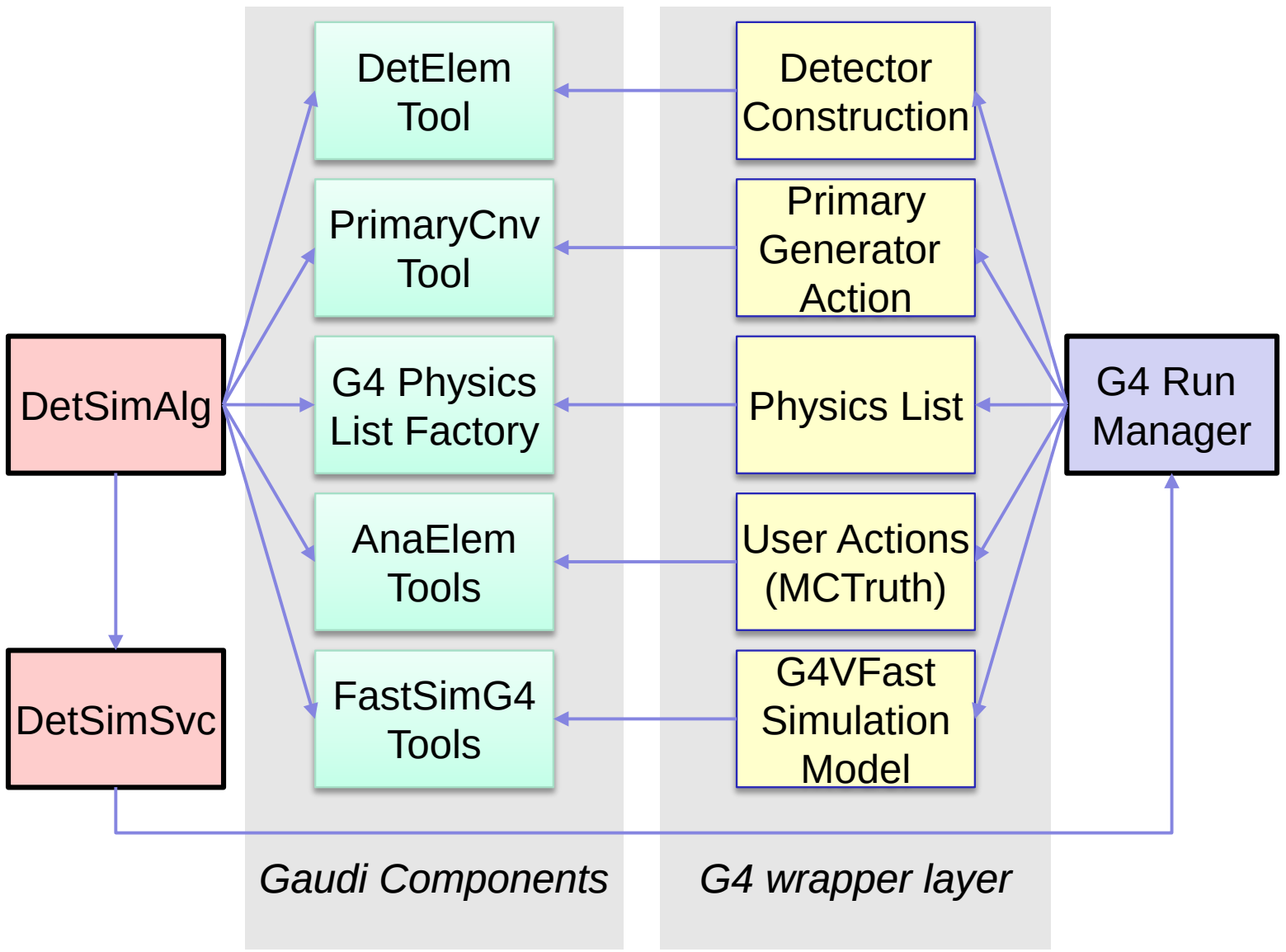

**Figure 13.4:** Simulation framework. At the initialization stage, DetSimAlg creates instances of Gaudi tools and Geant4 objects, registering the pointers of Gaudi tools to the corresponding Geant4 objects. Once all tools are initialized, the Geant4 objects are assigned to the Geant4 Run Manager, which is managed by DetSimSvc. During event simulation, DetSimAlg invokes the Geant4 Run Manager, which handles the simulation process and triggers the corresponding User Actions.

### 13.2.4 Beam-induced background simulation

Investigating the impact of background is essential for optimizing detector design, improving reconstruction algorithms and enhancing physics studies. A background simulation software offers capability to studies them. The CEPC background simulation considered several possibilities. Firstly, there are many collisions in the same bunch crossing as the signal event. Secondly, many of the subsystems have readout windows longer than the bunch spacing, which is the interval between two bunch crossings. Thirdly, some types of background are not related to the bunch. There are several modes to simulate the event mixing, including the track level mixing at generation stage, the hit level mixing at detector simulation stage and the digit level mixing after digitization. These modes can be combined according to the requirements.

For the track level mixing at generation stage, several `GenTools` have been developed.

The MDI group provides the different types of background using different generators. As the generators yield different file formats, several loaders have been implemented to parse these files and load the data into memory. The dedicated `GenTools` are responsible to sample the track time information and then load tracks from the loaders. After the generation stage, all the necessary tracks are then passed to the Geant4 simulation. After the simulation, all the hits will be recorded in files. This mode is useful to study the detector performance and estimate the data rate.

For the hit level mixing at detector simulation stage, all the components need to be simulated individually, then an event mixing algorithm is used to combine the inputs to yield the hits. Due to the large total size of all the input events the first implementation of this approach had excessive memory usage. Therefore, instead of loading all the input events, the event mixing algorithm generates metadata of these input events. The timestamps are assigned to them at the beginning. Then, each input event is loaded into memory and only selected hits are copied, for example this can avoid memory consumption by hits which are outside the readout time window. Some types of background are related to the bunch crossing, therefore "batch" is defined in the software to organize the batch of bunch crossings. The number of bunch crossings in a batch can be tuned according to the readout windows and the bunch spacing. This mode is useful to the physics studies, as the background hits can be reused and avoid simulation from track level.

Digit level mixing is not yet implemented. The plan is to prepare background data without physics signals and then overlay them with the physics signals. One possibility is that using the random trigger to record the real data, which is a data driven method. According to the experiences from other experiments, such data should not include zero compression, so that all the channels are saved into the random trigger file.

### 13.2.5 Digitization

The digitization step, which follows the hit creation step, mimics the behavior of readout electronics. It begins with the simulated hit position, time and deposit energy in the sensitive detectors, producing in the same format that the real front-end electronics and DAQ system would produce. Information from the generation stage, such as particle type and momentum, is retained during the digitization process.

The silicon tracker digitization simulation aims to emulate particle detection responses with a balance between accuracy and computational efficiency. It is based on particle trajectories through sensitive materials during simulation. Currently, a simplified method is used, relying on spatial resolution from silicon tracker design parameters or beam test data. This models position detection and includes measurement uncertainties. The framework allows flexible parameterization to suit different detector modules. However, it does not fully capture ionization processes during particle interaction. To improve this, a fast digitization approach will be developed that incorporates the pixel and strip segmentation of silicon detectors. This method will standardize ionization modeling along particle paths, independent of chip design, enhancing realism. The goal is to build a comprehensive digital model that accurately reflects ionization and charge collection. By combining both approaches, the system will deliver efficient and precise digitization to support detector evaluation, data analysis, and physics research.

Two primary methods are available for digitizing the TPC, each offering a different balance between realism and computational efficiency. The current approach divides the TPC volume into layers, using spatial resolution data from full simulations with `Garfield` (see Section 6.4.1). As particles pass through the TPC, energy deposits are recorded as hit points, which are then smeared using Gaussian functions based on resolution parameters. This method is simple, efficient, and well-suited for large-scale simulations. The future approach employs the `Garfield::Heed` module to simulate detailed primary ionization events along particle paths. It accurately models particle-gas interactions and calculates ionization positions. The full `Garfield` simulation then applies smearing to both spatial coordinates and drift times, producing realistic hit data that reflect the complete detection process, including ionization, charge drift, and readout effects. This method is ideal for detailed performance studies and complex event reconstruction. Each method has its strengths: the current approach favors speed for large datasets, while the detailed method offers high fidelity for in-depth analyses. Depending on simulation needs, either method can be used independently or in combination.

The ECAL digitization simulates light scintillation in BGO crystals, SiPM responses, and electronics behavior to produce timing and charge readouts. Geant4 records energy deposits and arrival times in the BGO. To reduce CPU load, scintillation is modeled with a simplified method instead of full optical simulations. Readout timing includes light transport effects and is smeared with a Gaussian to match time resolution. Charge is estimated from deposited energy, factoring in scintillation yield, light transport, SiPM efficiency, and electronics noise—critical for high-precision ECAL performance. Temperature and radiation effects are also included (see Chapter 7).

HCAL digitization follows a similar structure but omits timing. Due to short attenuation in glass scintillators and the 4-SiPM readout scheme, ECAL's simplified transport model is not applicable. Instead, a dedicated optical simulation generates a response map for HCAL digitization. Additional effects like quenching, light yield, and tile non-uniformity are discussed in Chapter 8.

Muon digitization simulates the response of plastic scintillators and SiPMs to muon energy deposits. As muons pass through scintillators, they generate photons transported to SiPMs via optical fibers. To avoid the high cost of simulating individual photons with `OpticalPhysics` in Geant4, the number of detected photons is estimated from the energy deposited ($E_{\mathrm{dep}}$). Each valid interaction is recorded as a hit with position, time, momentum, and energy. Hits with out-of-range deposits or low strip energy are excluded. Photon counts are modeled to decay exponentially with distance from the SiPM and scale with total energy deposited. The resulting signal is converted to ADC units, with Landau fluctuations to reflect statistical variation. This method balances realism and efficiency, enabling accurate muon digitization in CEPCSW.

## 13.3 Reconstruction

Reconstruction is the process of deriving physical quantities from the raw data collected during an experiment. This includes steps such as track reconstruction, particle identification, reconstruction in the calorimeter system, and muon reconstruction, among others.

### 13.3.1 Track reconstruction

Reconstruction of charged particles is a crucial and complex component of event reconstruction. The performance of track reconstruction directly affects the efficiency and resolution of subsequent reconstruction algorithms, such as particle identification algorithms and particle flow algorithms. To meet the goals of the experiment, it's important to create fast, reliable, and easy-to-maintain tracking software.

Currently, track reconstruction algorithms ported from ILCSoft [35] have been integrated into CEPCSW. This includes standalone reconstruction algorithms for silicon detectors and time projection chambers, as well as their combined reconstruction methodologies. Specific reconstruction geometry modules have been established for the various tracking sub-detectors, enabling the use of geometric information during the reconstruction process.

Building on this foundation, a track reconstruction chain for the baseline tracking detectors has been developed, as illustrated in Figure 13.5. The default comprehensive track reconstruction consists of three algorithms: `SiliconTracking`, `Clupatra`, and `FullLDCTracking`, all optimized for the tracking detectors. The SiliconTracking algorithm operates on digitized tracker hits from the VTX and the ITK to generate track candidates stored in the collection `SiTracks` through pattern recognition and track fitting. Subsequently, `Clupatra` employs a nearest neighbor search to identify track candidates within the digitized TPC hits in `TPCTrackerHits`, resulting in the collection `ClupatraTracks`. The `FullLDCTracking` algorithm, also developed by ILD [35], combines matching silicon and TPC candidates from the same tracks while retaining any independent track candidates in the final track collection named as `CompleteTracks`. The joint reconstruction of candidate tracks is finalized using the Kalman filter, which also incorporates hits from the outer tracking detectors during the final track fitting stage. It also supports changing the options to reconstruct different tracker sets such as VTX+ITK, VTX+ITK+OTK, VTX+ITK+TPC, and so on. This capability enables performance evaluation of different tracking detectors.

In the reconstruction workflow, the processes of track finding and track fitting are decoupled in Gaudi algorithm via the Gaudi tool interface. This separation permits modifications to the tracking algorithms during different phases of reconstruction, which also supports refitting any reconstructed tracks—an essential feature for enhancing precision measurements post-particle identification. Specifically, conducting Kalman filter fitting while assuming a particle type can lead to more accurate momentum measurements; By default, alternative particles are assumed to be pions for the Kalman filter fitting prior to any particle type assumptions. This approach facilitates easy updates to both track finding algorithms and track fitters, such as ACTS Common Tracking Software (ACTS) [36] and traccc [37], and it will be developed continuously. In the presence of a non-uniform magnetic field, track fitting using the Kalman filter method can effectively account for field variations, provided a precisely measured field map is available. Simulation studies with the drift chamber [38] indicate no significant degradation in performance.

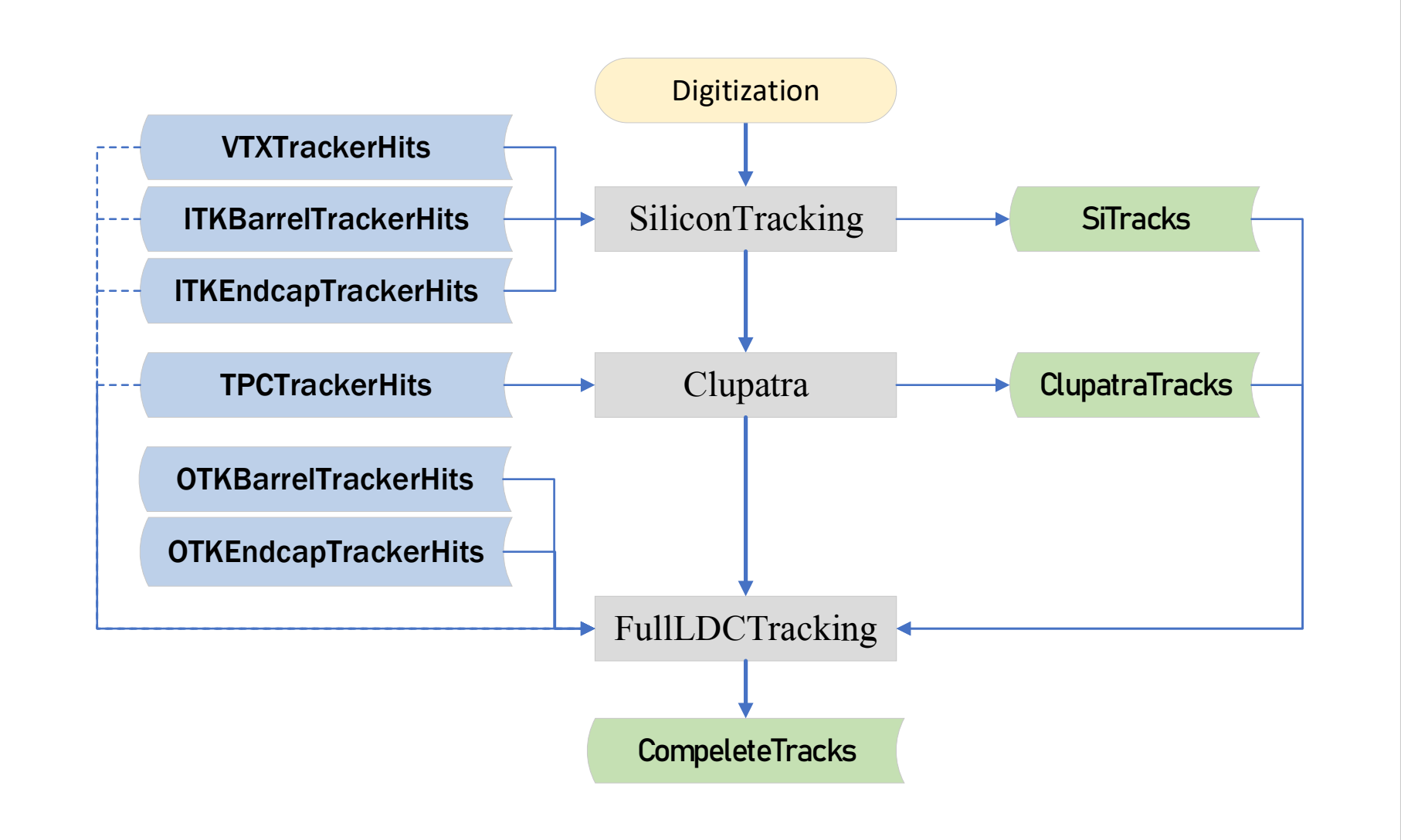

**Figure 13.5:** Flow of tracking reconstruction from digitized `TrackerHit` objects: the input hits are processed through track finding to create track candidates and fit track objects and the non-assigned hits undergo a secondary tracking test during the track combining phase to ascertain their potential association with specific tracks.

For the diverse schemes of track detectors, spanning from planar vertex detectors to curved-plate composite vertex detectors, tailored reconstruction processes have been established. The algorithms have been adapted for various readout schemes, such as charge centroid and pixel readout, to enhance track performance accuracy. Furthermore, a machine learning-based clustering algorithm specifically designed for track reconstruction in high-granularity-readout TPC is under development, with the implementation of machine learning-based track finding algorithms planned for the future.

## 13.3.2 Particle identification

Particle identification is one of the core requirements of an experiment, especially in cutting-edge research areas such as jet flavor tagging and flavor physics. It imposes stringent demands on the detector's capability to distinguish charged hadrons across a wide momentum range of tens of GeV, a challenge that current technologies struggle to meet. The high-granularity readout TPC, with a pad size of $0.5 \times 0.5$ mm$^2$, provides ionization measurements via both $\mathrm{d}N_p/\mathrm{d}x$ and $\mathrm{d}E/\mathrm{d}x$ methods that outperform traditional large-pad $\mathrm{d}E/\mathrm{d}x$ techniques.

However, due to the intrinsic ionization behavior among different particle species, there are some "blind spots" in low-momentum region for the $\mathrm{d}N_p/\mathrm{d}x$ methods. It is necessary to have a ToF detector to cover the PID in that region. The AC-LGAD detector [39] in the outer tracker could provide timing information with a resolution of 50 ps.

As a result, the hadron identification system is based on an ionization measurement in the gaseous detector and a ToF measurement in the outer tracker. The reconstruction process not only extracts the ionization and timing information from the detector responses, but also provides the $\chi^2$ values for the physics analysis. Details of the reconstruction approach will be presented in the following sections.

**Ionization measurement in the TPC** Ionization of matter by charged particles is the principle mechanism behind the TPC. As a charged particle traverses the chamber, it produces cascades of ionizations along its track, which are recorded by the high-granularity readout. The number of activated pads reflects the ionization density (used for $\mathrm{d}N_p/\mathrm{d}x$ reconstruction), while the total energy deposited is used for $\mathrm{d}E/\mathrm{d}x$ reconstruction.

As detailed in section 13.2.5, the activated pads create TPC hit objects during digitization, which encapsulate 3D positional and energy loss data. Following TPC tracking, a set of TPC hits can be linked to a charged particle track, facilitating ionization reconstruction. The hit-level $\mathrm{d}N_p/\mathrm{d}x$ can be determined by counting the number of associated TPC hits that exceeds a specific energy threshold and dividing by the track length. Conversely, the hit-level $\mathrm{d}E/\mathrm{d}x$ is calculated by aggregating the energies of the associated TPC hits and dividing by the track length. However, due to the presence of secondary ionizations, the measured hit-level $\mathrm{d}N_p/\mathrm{d}x$ or $\mathrm{d}E/\mathrm{d}x$ follows a Landau distribution, which diminishes resolution. A prevalent approach to mitigate this issue is the truncated mean method, which excludes 30-40% hits with highest $\mathrm{d}N_p/\mathrm{d}x$ or $\mathrm{d}E/\mathrm{d}x$ values. The detailed reconstruction algorithm is described in section 6.4.4.1.

With the reconstructed $\mathrm{d}N_p/\mathrm{d}x$ or $\mathrm{d}E/\mathrm{d}x$ values, the $\chi^2$ is calculated using the following equation:

$$\chi_R = \frac{\langle R\rangle_{\mathrm{meas}} - \langle R\rangle^i_{\mathrm{exp}}}{\sigma^i_{\mathrm{exp}}}, \tag{13.1}$$

where $R$ stands for $\mathrm{d}N_p/\mathrm{d}x$ or $\mathrm{d}E/\mathrm{d}x$ $\langle R\rangle^i_{\mathrm{exp}}$ and $\sigma^i_{\mathrm{exp}}$ are the expected measured value and the expected resolution for hypotheses involving $e^\pm$, $\mu^\pm$, $\pi^\pm$, $K^\pm$ and $p^\pm$.

**Time-of-Flight in the Outer Tracker** The outer tracker provides not only precise position measurements but also high-precision timing using the AC-LGAD technique. The ToF system measures the time for a charged particle to travel from its point of origin to the timing detector. Combined with momentum and trajectory information from tracking detectors, this timing information enables the calculation of the particle's mass. The relationship is given by the

formula:

$$t_{ToF} = \frac{s}{c}\sqrt{\frac{m^2}{p^2} - 1}, \tag{13.2}$$

where $t_{ToF}$ denotes the time of flight, $s$ is the trajectory length, $c$ is the speed of light, and $m$ and $p$ represent the particle's mass and momentum, respectively. By comparing the measured and expected $t_{ToF}$, one can compute the $\chi^2$ values according to Equation 13.1.

In the high-luminosity Z phase, the bunch spacing in time is 23 ns. The ToF system can help determine which bunch crossing a certain event belongs to, as well as the event start time, $t_0$. Multi-jet events are the most dominant type of events at the CEPC, with charged pions making up the majority of the particle population. These pions are well-detected by the detector, and their momenta and time-of-flight are accurately measured. The event starting time, $t_0$, can be estimated using these tracks, and its uncertainty is proportional to the factor of $\frac{1}{\sqrt{N}}$, where $N$ is the number of tracks, meaning that $t_0$ can be measured with high precision $\mathcal{O}$ (10) ps. This excellent accuracy can contribute to the study of the time uncertainty of the bunch crossings themselves. Additionally, the time offset of the event starting time within a bunch crossing can be further studied and calibrated.

### 13.3.3 Particle flow algorithm

The unique 4D crystal bar ECAL design necessitates the development of a new particle flow algorithm, **CyberPFA**. The existing particle flow algorithms, Pandora PFA (PandoraPFA) [40] and Arbor [41], are developed for the conventional high granularity calorimeter and unable to reconstruct the objects through the orthogonal bar structure. CyberPFA adheres to the fundamental particle flow principle—utilizing the optimal detector subsystem to determine the energy/momentum of each particle within a jet, meanwhile addresses critical challenges specific to the ECAL design.

In homogeneous calorimeters, crystal's Molière radius is larger than that in sampling calorimeters, increasing shower overlap probability. Additionally, orthogonal crystal bars' intersecting geometry causes the ghost-hit issue in cluster reconstruction, a challenge unaddressed by prior particle flow algorithms. CyberPFA is newly developed following the particle flow idea and addresses these through two key innovations: 1) Shower-core-initiated cluster formation via controlled energy propagation, enhancing pattern recognition efficiency through dedicated core identification algorithms, and 2) Multivariate ghost-hit suppression combining longitudinal shower profiles, module geometry constraints, and SiPM timing information. The experience of clustering in high granularity calorimeter and integrating track and calorimeter information from PandoraPFA and Arbor are also adopted in CyberPFA' concept. It maintains complete independence from existing code-bases.

The current CyberPFA implementation consists of five primary processing stages as shown in Figure 13.6, with additional specialized sub-algorithms:

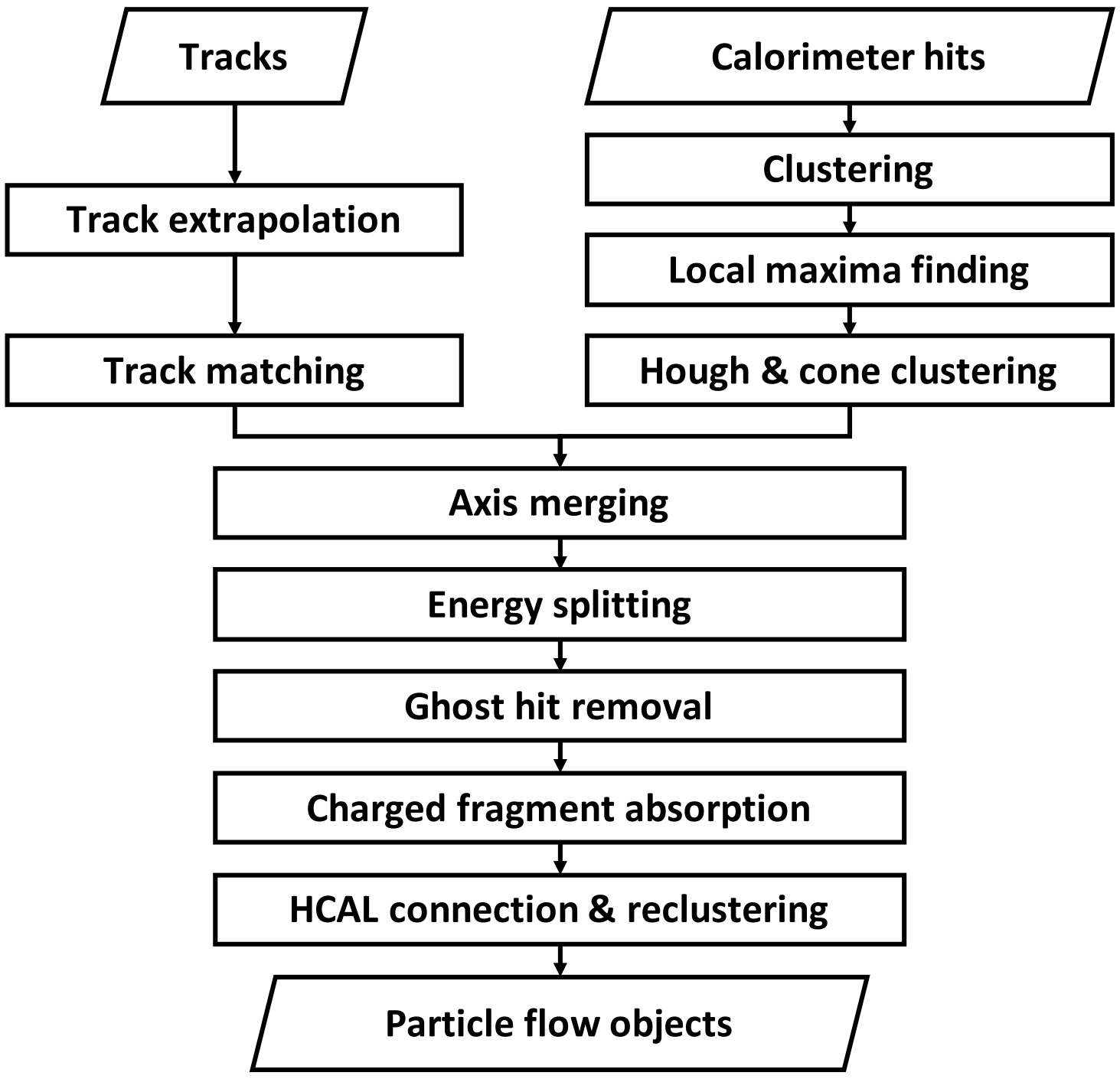

**Figure 13.6:** Work flow of CyberPFA, including the processing of input tracks and calorimeter hits, the advanced cluster core recognition, the splitting of overlapped showers, the ghost hit removal as well as the final reclustering combined with track, ECAL and HCAL. The particle flow objects are output into Podio.

**Event pre-processing** Following the conventional particle flow approach, the algorithm processes reconstructed tracks and calorimeter hits. Low-quality tracks are filtered at this stage using a trained Boosted Decision Trees (BDT) to prevent mismatches in subsequent particle flow analysis. Neighboring crystals are clustered, with local maxima identified as potential shower cores.

**Pattern recognition with local maxima** Three sub-algorithms have been developed to identify showers from different particle types. For charged particles, showers are reconstructed following the extrapolated tracks. Meanwhile the track-cluster associations are established. Neutral EM showers are identified via Hough transforms [42]. Remaining hits are clustered using an adaptive cone algorithm with layer-dependent parameters. Reconstructed structures serve as shower cores, with topological merging applied to enhance hadronic shower identification.

**Overlapped energy splitting** The expected energy deposition is predicted using EM/hadronic shower profiles with recognized shower cores:

$$E_{i\mu}^{\text{exp}} = E_{\mu}^{\text{core}} \times f(|x_i - x_c|)$$

where $E_{\mu}^{\text{core}}$ represents the energy of shower $\mu$ at its core, $E_{i\mu}^{\text{exp}}$ denotes the expected energy deposition from shower $\mu$ in crystal $i$, and $f(|x_i - x_c|)$ is the shower profile function obtained from shower development studies [43]. This information determines the weighting factor of shower $\mu$ in crystal $i$: $w_{i\mu} = \frac{E_{i\mu}^{\text{exp}}}{\sum_\mu E_{i\mu}^{\text{exp}}}$.

**Ghost hit removal** Multiple approaches are developed to utilize the 5D information from the calorimeter to veto the ghosts:

- **Longitudinal shower development**: Since shower starting depths follow stochastic distributions, ghost hits can be identified by matching energy deposits in orthogonal crystals within the same layer. This matching quality is quantified by a $\chi^2$ statistic computed across all 9 double-layers.
- **Module structure**: Ghost hits only occur when multiple particles enter a single module. When showers extend beyond module boundaries, adjacent modules provide additional spatial information about the shower position.
- **Timing information**: SiPMs coupled at both ends of each crystal measure the arrival time difference of scintillation light, enabling reconstruction of the shower position along the bar. This provides an additional dimension for $\chi^2$ calculations and ghost hit rejection.

**Reclustering** The high granularity of the HCAL enables conventional cone-based clustering for particle flow reconstruction. Local clustering with a cone approach is performed using HCAL hits, which are then associated with ECAL clusters using topological information. The reclustering process, inherited from Arbor, is conducted to compare track momenta with cluster energies and refine clusters using track information.

Following these steps, the PFOs are created, containing information such as four-momentum, and stored in PODIO data format. Several functional plugins have been developed as additional Gaudi processors, including particle identification and fine energy corrections. The performance is initially validated using the di-jet invariant mass in the $e^+e^- \to ZH \to \nu\nu gg$ process, where $\nu$ denotes a neutrino and $g$ a gluon, and is further evaluated through various benchmark studies.

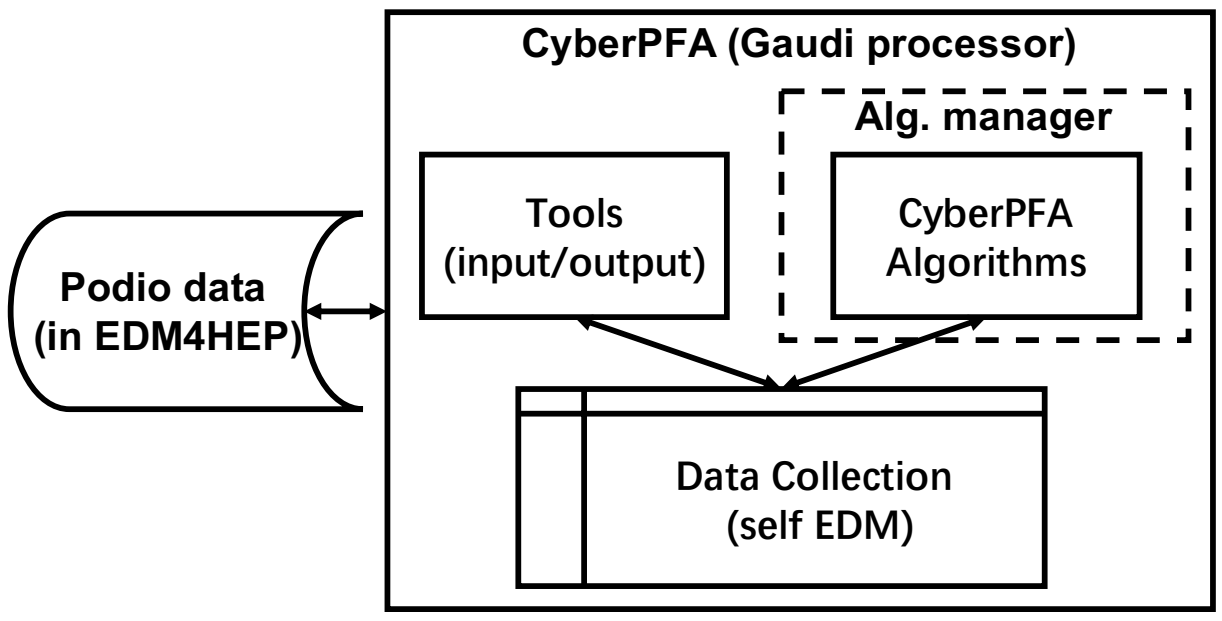

**Figure 13.7:** Algorithm framework of CyberPFA, including centralized control, interface with PODIO data, core algorithms and an independent data collection.

CyberPFA follows Key4hep's interoperability principles through its Gaudi processor within the CEPCSW framework, maintaining dependencies on core HEP components (Gaudi, EDM4hep, ROOT). The architecture inherits the advantages of the Pandora Software Development Kit (PandoraSDK) [44], including its building-block algorithm arrangement, data management capabil-ities, and potential for future extensions. Given that Key4hep already ensures code robustness and reusability across the community, and considering the numerous extensions required for the crystal bar ECAL, PandoraSDK itself is not reused.

The CyberPFA system features a centralized control mechanism where a main program coordinates execution through a management module (*Manager*). This module handles the registration and dispatch of specialized sub-algorithms, as well as the pattern recognition logic.

Additionally, several *Tools* are constructed as interfaces between CyberPFA and EDM4hep data. Figure 13.7 provides a detailed visualization of this software structure, illustrating the workflow and component interactions.

### 13.3.4 Muon reconstruction

For the muon detector, the reconstruction algorithms mainly involves track reconstruction and muon identification. Track reconstruction analyzes the muon's hit information across multiple scintillator strips and applies geometric matching algorithms to reconstruct individual track segments. These segments are then matched with corresponding tracks identified in the silicon trackers and the TPC. Given the high precision of the silicon trackers and the TPC, the muon's momentum is determined based on their measurements.

Muon identification integrates information from multiple sub-detectors, including the tracker, calorimeters (ECAL and HCAL), and the dedicated muon detector. Particle Flow algorithms are employed to analyze the Energy/Momentum ratio and the energy deposition patterns in the ECAL and HCAL. Additionally, the muon detector plays a critical role by providing confirmation of muon signatures. A muon is typically identified when it traverses the tracker, deposits energy in the ECAL and HCAL, and produces at least three hits in the muon detector along the extrapolated track. This multi-detector approach enables the reconstruction software to achieve precise muon identification with high efficiency and minimal background contamination.

## 13.4 Analysis tools and visualization

### 13.4.1 Analysis tool

To effectively utilize available resources on local user machines and shared computing clusters for efficient analysis of the CEPC scientific data, a high-performance data analysis framework, RDataFrame Analysis (RDFA), is developed based on ROOT's RDataFrame to speed up physics analysis. This framework will leverage parallel computing and heterogeneous computing technologies to fully utilize the computational power of multi-core CPUs, while also simplifying the physics analysis process through declarative programming techniques.

RDFA will be developed based on the RDataFrame provided by ROOT, with the design outlined in Figure 13.8. RDFA primarily consists of the RDataFrame library, external libraries such as EDM4hep, ROOT, FastJet etc., low-level C++ code that implements main functionalities, as well as Python analysis scripts for user interaction and configuration. The RDataFrame mainly provides basic functionalities, such as Task scheduler with Intel's Threading Building Blocks (TBB) [45], event filtering, event objects I/O, and histogram plotting etc. To cater to the specific needs of physics analysis, RDFA will integrate additional high-level toolkits, such as jet reconstruction, particle identification, jet tagging, kinematics fitting, vertex fitting etc. These toolkits primarily developed in C++, will be wrapped using Python Binding technology, making them accessible within the Python environment. This integration streamlines the use of these tools in RDFA and simplifies their integration with other algorithms or tools through standardized interfaces.

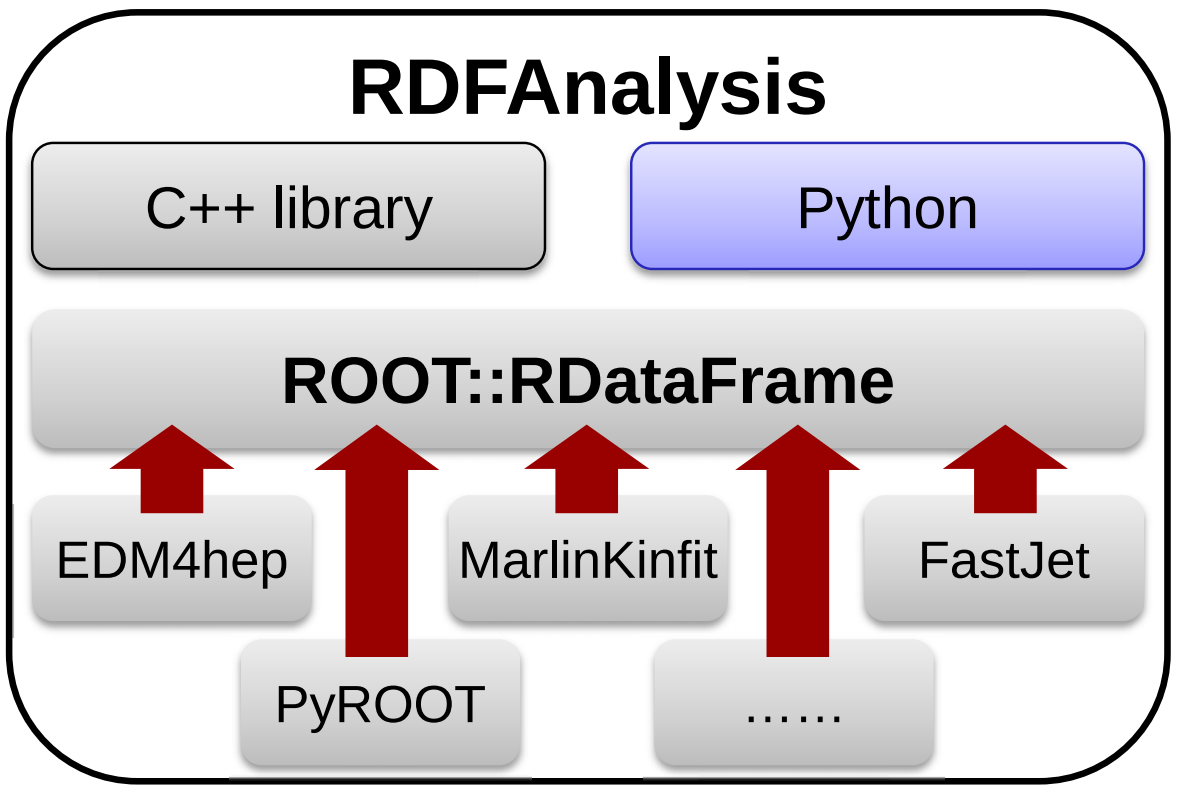

**Figure 13.8:** Design scheme of RDFA Framework: RDFA uses the RDataFrame library for its main functions, along with external libraries like EDM4hep, ROOT, and FastJet. It also includes C++ code for core tasks and Python scripts for user interaction and job configuration.

Moreover, RDFA is designed to be adaptable and versatile. Given that the event data model is based on the EDM4hep design as well as some extensions, while a significant volume of MC data has been generated using the older LCIO-based event data model, RDFA must be capable of analyzing data in both formats. This flexibility ensures that the framework can handle a diverse range of data types, enhancing its utility and versatility. To further maximize the computational efficiency, RDFA will incorporate Message Passing Interface (MPI) [46] technology for parallelizing RDataFrame, enabling parallel execution across multiple computational nodes, significantly boosting the framework's performance and throughput.

In summary, by integrating high-level physics analysis tools, supporting multiple data formats, and ensuring cross-platform compatibility, RDFA aims to significantly enhance the efficiency and productivity of physicists in their data analysis endeavors. Furthermore, DeepSeek will be leveraged with RDFA and declarative programming to boost productivity and streamline the learning curve for ROOT and declarative workflows.

### 13.4.2 Visualization software

Detector and event visualization software is a crucial component of HEP experiment [47]. It plays an important role throughout the life cycle of the experiment, including detector design, validating physical metrics, detector simulation, debugging reconstruction algorithms, data quality monitoring, data physics analysis, as well as in scientific outreach and education [48, 49, 50].

Currently, there are three major directions for the development of visualization software in high-energy physics experiments, namely, Phoenix [51], ROOT EVE [52], and industrial visualization platforms such as Unity [53] and UnReal engine [54]. Each of the three methods has their own advantages. To meet the visualization requirements, the Phoenix based event display and Unity based event display have been selected for development because of their specific and complementary advantages.

The first approach is based on the general-purpose event display platform Phoenix, where visualization modules are developed using JavaScript. Phoenix focuses on being experiment agnostic by design, with common tools (such as custom menus, controls, propagators) and the possibility to add experiment specific extensions.

This platform supports detector modeling inputs in formats such as ROOT, as well as event data inputs in EDM4hep format. Its web-based display and simulation allow for real-time visualization of experimental data. The Phoenix platform has been applied to large-scale experiments like ATLAS, CMS, and LHCb. It is also recommended by the HSF [55]. One of the great advantages of using Phoenix to develop visualization software is its general-purpose detector description with ROOT and event data model with EDM4hep, making it convenient to be implemented to different HEP experiments with universal data interfaces. Phoenix also provides the common interactive functions for Graphic User Interface (GUI) control, such as zoom in, zoom out, rotation, color theme, view options and animation. These features significantly reduce the effort required in the development of visualization software.

The second approach uses multimedia engines commonly employed in the industrial sector, such as Unity [53] and Unreal [54], which are popular in gaming, animation, and film production. Since the industrial community is much larger than the HEP community and has more resources in financial and technical supports, these platforms offer superior display quality and dynamic features, making them suitable for cross-platform deployment and web integration. Some HEP projects have already successfully implemented visualizations using Unity, demonstrating its feasibility for visualizing HEP experiment detectors, including ATLAS [56, 57], BESIII [58], Belle II [59] and JUNO [60]. The detector description can be converted into the 3D modeling format with automatic conversion interfaces[61]. They also provide strong support for main-stream industrial display devices, making them ideal for applications in Virtual Reality (VR)[62] and Augmented Reality (AR). For non-profit purpose, such as scientific research or education, Unity and Unreal are free to use for software development.

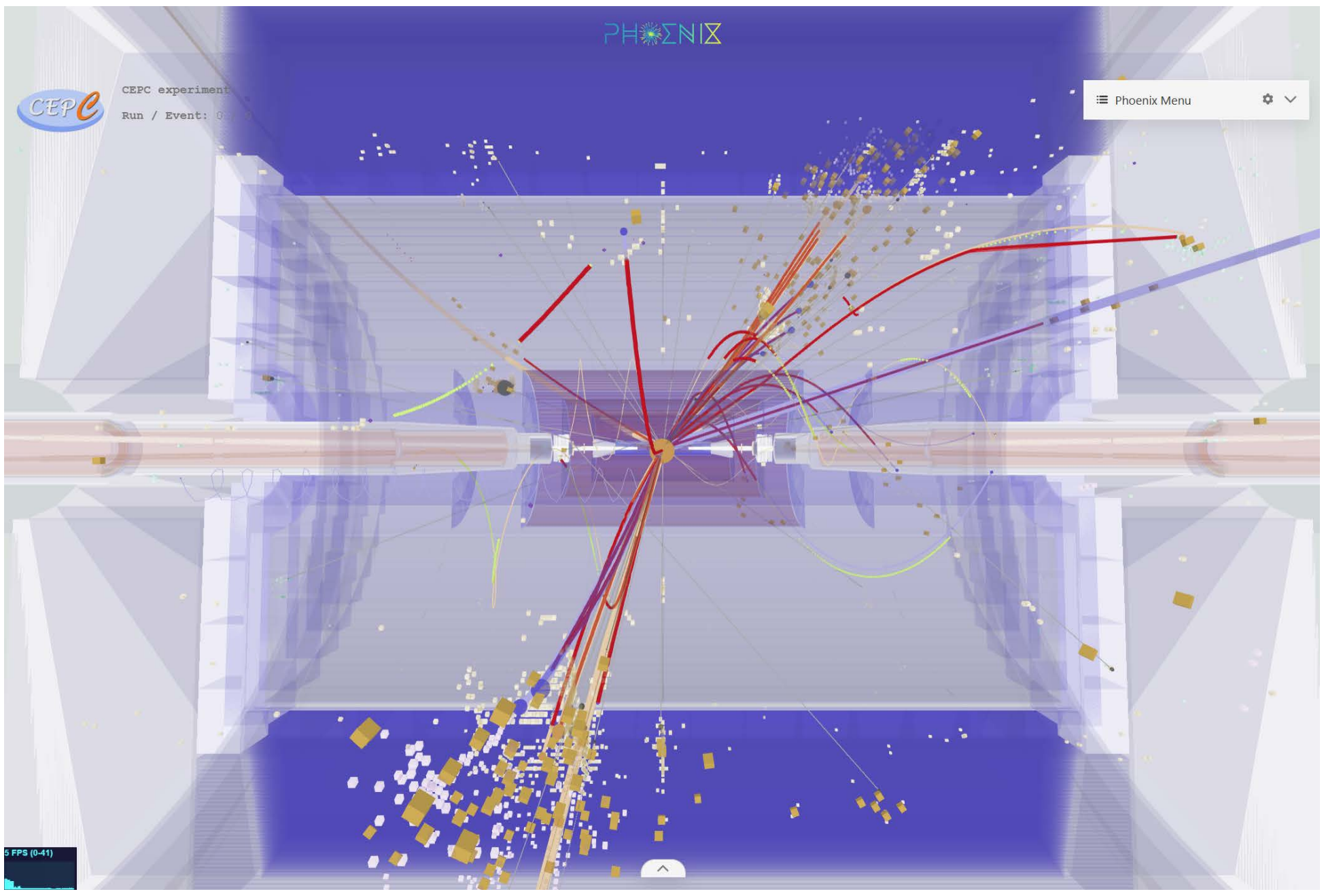

**Figure 13.9:** Visualization of the CEPC detector and display of an $e^+e^- \to ZH, Z \to \mu^+\mu^-, H \to b\bar{b}$ event in the Phoenix platform. The detector geometry derives from the DD4hep detector description in the offline software and is converted to ROOT format automatically as input of geometry for Phoenix. The red curves are reconstructed tracks in EDM4hep format generated with offline software, the small and white cubes are calorimeter cells, and the yellow pointing cuboids are clusters.

For the development of visualization software, two parallel approaches are planned. On one hand, a lightweight event display tool will be built based on the Phoenix platform. This tool will import detector geometry from the existing offline software, as shown in Figure 13.9, and integrate it with the event data model to enable web-based visualization of the detector and event data. On the other hand, research will focus on Unity-based visualization software, developing an automatic conversion interface to transform offline software geometry into industrial standard formats such as FBX supported by Unity. This will enable the automatic conversion of detector geometry for visualization in Unity, as well as the display of event information (such as tracks, jets, hits, and clusters) within the Unity environment.

Building on this event display software, further development will focus on creating immersive experiences, such as VR, to provide a new level of interaction for detector design, data collection, physics research, and scientific outreach education.

## 13.5 Software R&D activities

The trend in computing hardware is shifting toward increased CPU cores and more diverse architectures, including GPU, FPGAs, and other accelerators. Machine learning plays a crucial role in HEP experiments. Consequently, the development of software capable of efficiently handling concurrent computations and leveraging machine learning techniques to enhance data processing and analysis has become an increasingly important topic.

During $Z$-pole operation at CEPC, computing demands are expected to reach the exabyte scale, posing technical challenges comparable to those encountered at the High-Luminosity Large Hadron Collider. Advances in artificial intelligence, along with the integration of concurrent and heterogeneous computing, will form the operational backbone of future collider experiments. Meanwhile, quantum computing and its associated algorithmic frameworks present a potential paradigm shift in addressing these challenges, offering a promising—though ambitious—solution for the CEPC initiative.

This section will present relevant research efforts exploring these advancements.

### 13.5.1 R&D on concurrent computing

**Multi-threaded simulation** Precise detector simulation can both CPU and memory intensive due to large detector geometries and complex physics models. A detector geometry can consume significant memory because of the precise description required for each detector volume and the large number of volumes invoked. Particle propagation through such geometries can also be time-consuming due to navigation through numerous volumes. At each step, the corresponding physics models are invoked. These physics models can be memory-intensive due to large associated data tables. Multi-threaded simulation could significantly reduce memory usage by sharing geometries and physics models among threads.

Recent Geant4 releases support both inter- and intra-event parallelism. The underlying principle is that events or a sub-events are independent of one another. Dedicated Geant4 run managers have been developed to handle the simulation of individual events or sub-events within separate threads.

To leverage parallelism effectively, the existing simulation framework requires significant evolution to handle Geant4's run manager alongside Gaudi's Functional and Hive components. Gaussino [63], derived from LHCb's Gauss framework, represents a promising solution to this challenge. The Key4hep project already considers Gaussino as the next-generation simulation framework. For CEPC, one key activity focuses on demonstrating the integration of Gaussino within Key4hep ecosystem. A CEPC-on-Gaussino prototype has been developed by adapting part of external libraries and data models from LHCb.

In the future, adopting the intra-event parallelism into Gaussino will be valuable for accelerating the simulation. The primary challenge in intra-event parallelism is that secondary particles depend on parent particles. Effective load balancing across sub-events is crucial for optimizing computational performance.

**Tracking with ACTS** ACTS is designed to support concurrent computing by leveraging modern computing architectures with multiple cores and accelerators. It provides thread-safe algorithms and optimized data structures to enable efficient parallel execution of track reconstruction tasks.

The first research topic related to ACTS focuses on leveraging the ACTS library to implement track reconstruction for the Reference Detector and to demonstrate track reconstruction in a multi-threaded environment. As illustrated in Figure 13.10, a complete track reconstruction chain is realized using the Gaudi Algorithm. This process includes Event Data Preparation, which retrieves Track Hits from the Gaudi Event Store; Seeding; Track Parameter Estimation for the identified seeds; track finding using the Combinatorial Kalman Filter (CKF); and track fitting based on the Kalman Filter (KF). Finally, the tracking results are registered with the Gaudi Event Store. The next step involves enabling parallel execution of multiple algorithms using Gaudi's multi-threading scheduler, optimizing performance and computational efficiency.

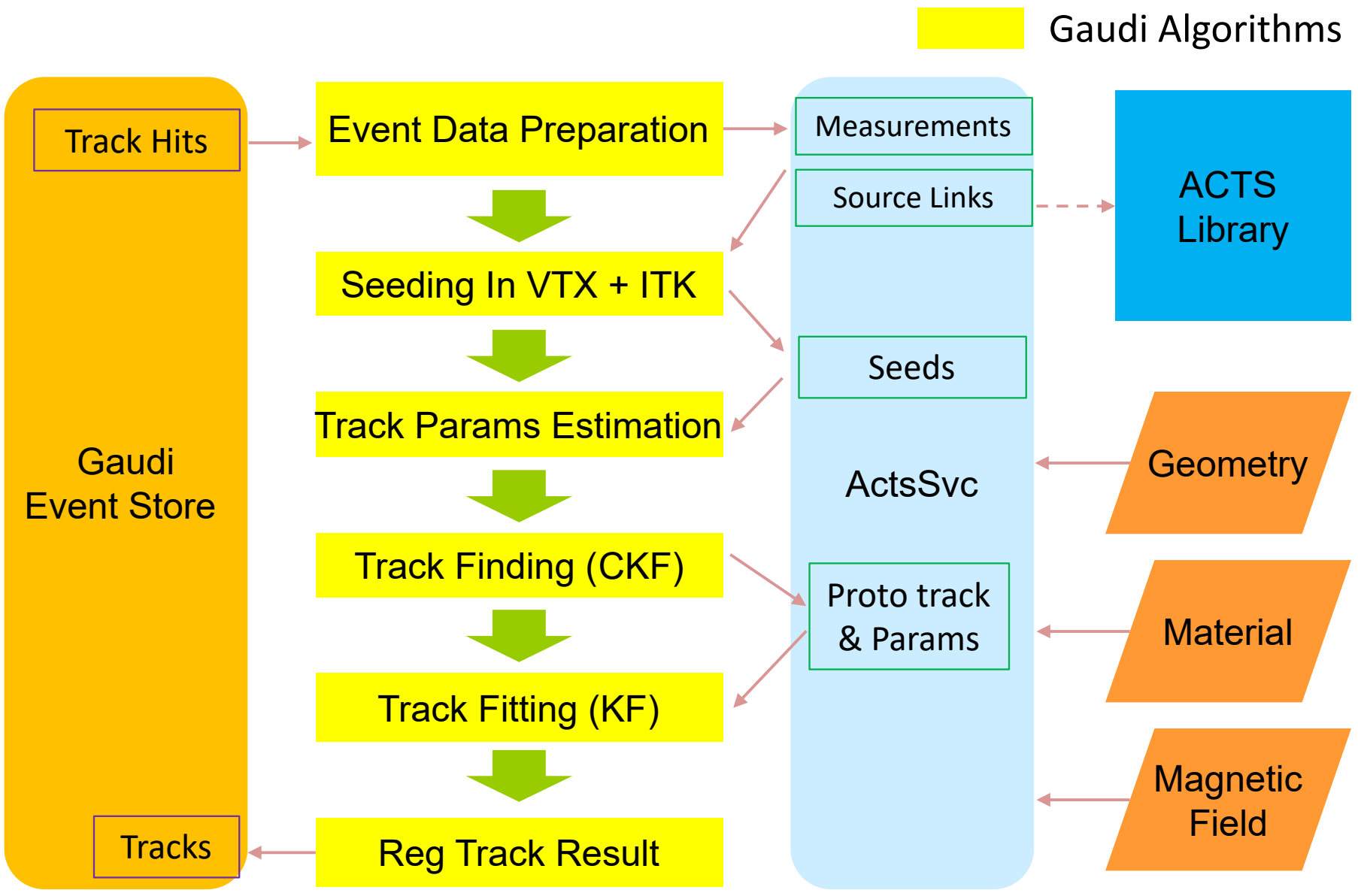

**Figure 13.10:** Diagram illustrating the implementation of track reconstruction using ACTS within CEPCSW. The core tracking process comprises three key stages: Seeding, Seed Track Parameter Estimation, and Track Finding, which leverages the Combinatorial Kalman Filter.

The ACTS project includes traccc, a subproject dedicated to optimizing track reconstruction algorithms for highly parallel hardware, with a particular focus on GPU. The second research topic focuses on integrating the traccc seeding algorithm into the CEPCSW framework and evaluating its performance on GPU platforms. This integration has led to several algorithmic enhancements, with a key achievement being the optimization of data communication between CPUs and GPU. Modifications to EDM4hep 's data structure were made to enable direct memory sharing and ensure smooth compatibility with ACTS' VecMem data model [64]. VecMem provides a comprehensive library for vectorized data structures, efficiently utilized across various computing devices to help traccc manage vectorized data and enhance computational performance. This optimization eliminates conversion overhead between EDM4hep and traccc formats. By enabling traccc to directly access EDM4hep data via shared memory, the system significantly reduces data transfer latency and format conversion overhead

### 13.5.2 R&D on machine learning applications

The widespread use of neural network based machine learning techniques together with GPU acceleration can be regarded as the emergence of a new computational paradigm. Neural networks have demonstrated superior performance across a wide range of tasks, leveraging the parallel processing capabilities of GPU to achieve significant advancements in computational efficiency and effectiveness.

**Fast calorimeter simulation** High-energy electrons or photons entering an ECAL induce showers via bremsstrahlung and electron–positron pair production. Secondary particles ionize the detector medium and deposit energy, which is converted into an electrical signal through scintillation. Geant4's MC algorithm traces each primary and secondary interaction step by step, accurately reproducing full shower development; however, this detailed simulation is computationally expensive, limiting sample-generation efficiency and slowing physics analyses.

Advanced neural network architectures such as CaloFlow [65], which employs normalizing flows, offer several advantages over Generative Adversarial Network (GAN) and Variational Autoencoder (VAE) [66], including a tractable likelihood, stable training, and principled model selection [67]. By establishing a bijective mapping between data and latent spaces, normalizing flows are also promising for applications like detector unfolding. Bounded Information Bottleneck Autoencoder (BIB-AE) [68] further advances latent interpretability by correlating its variables with physical observables of calorimeter showers (e.g., the centroid position along the $z$-axis). In their seminal work, Paganini et al. [69] introduced a multilayer-perceptron–based generator–discriminator pair to model 3D energy deposition in electromagnetic calorimeters, achieving speedups of two to three orders of magnitude relative to conventional MC methods, and up to five orders of magnitude when deployed on GPU, while maintaining acceptable accuracy. The subsequent 3D Generative-Adversarial Network (3DGAN) variant demonstrated that generated showers reproduce a wide range of physical features to within 10 % of the precision afforded by standard MC simulations [70]. Other studies have also shown that incorporating a pretrained regressor network can further improve training stability. More recently, there some point cloud-based diffusion models [71, 72, 73] have been explored in HEP. For example, the CaloClouds [74] have been shown to provide improved performance, particularly in highly granular calorimeters.

GAN, composed of generator and discriminator networks, learn the underlying data distribution by adversarial training and can rapidly generate realistic simulation data from random noise and conditional inputs. To investigate the application of GAN to fast simulation of the ECAL with silicon tungsten, single photons were simulated with various energies and incident angles. Some preliminary results have been checked, such as the deposited energy distributions along different axes. These results indicate that the GAN is capable of simulating the calorimeter and producing high-fidelity simulation results.

We will further apply this approach to simulate the crystal-bar ECAL, characterized by its crisscrossing geometry that introduces substantial complexity in modeling. A rectangular region of interest, RoI, centered on the incident point and spanning the entire calorimeter thickness, will be voxelized into three-dimensional cubes. The energy deposited in each voxel will serve as input to the GAN model, enabling a structured, high-resolution representation of the detector response to support accurate simulation.

To further enhance the accuracy and speed of simulations, the transport of scintillation photons can be traced using NVIDIA OptiX [75] during the digitization process, resulting in improved energy deposition feedback. Additionally, new temporal data prediction architectures, such as Transformers [76], can be explored.

**Gaseous detector simulation** When a charged particle passes through the gaseous detector, it ionizes the gas, producing primary and secondary ionization electrons. These electrons then drift to the signal amplification region under the influence of electric and magnetic fields, where they undergo an avalanche multiplication process, ultimately generating an electrical signal.

To achieve a realistic simulation of the gaseous detector, a combination of Geant4 and Garfield++ [77] is required. Geant4 simulates particle transport and interactions with detector materials, while Garfield++ handles ionization and detector response simulation. Simulating the avalanche multiplication process with Garfield++ is highly computationally intensive, making it impractical for the detector response simulation of the gaseous detector. To overcome this limitation, machine learning is employed to accelerate the avalanche process simulation, effectively addressing the computational bottlenecks inherent in Garfield++ simulations.

The proposed simulation chain for the drift chamber is shown in Figure 13.11. Firstly, the Geant4 step information is passed to Garfield++ simulation tool. This tool will use the TrackHeed to simulate the ionization process for this Geant4 particle, including primary and secondary ionization. Next, a neural network is applied to simulate the pulse, estimating its timing and amplitude based on the local position of the ionized electron. The simulated pulse is then convolved with a predefined pulse shape template. Finally, all the resulting convoluted pulses are combined to produce the final simulated waveform.

The neural network model is developed based on Normalizing Flow (NF) [67] to do the simulation. The structure of our NF network is similar to that of CaloFlow [65], which uses Rational Quadratic Spline (RQS) [78] for transformation and uses a Masked Autoencoder for Density Estimation (MADE) [79] block for learning the parameters of the RQS transformation.

The Garfield++ simulation data was used for training the NF model, and the Adam optimizer [80] was used for the optimization.

During training, input data from Garfield++ simulations—including amplitude and drift

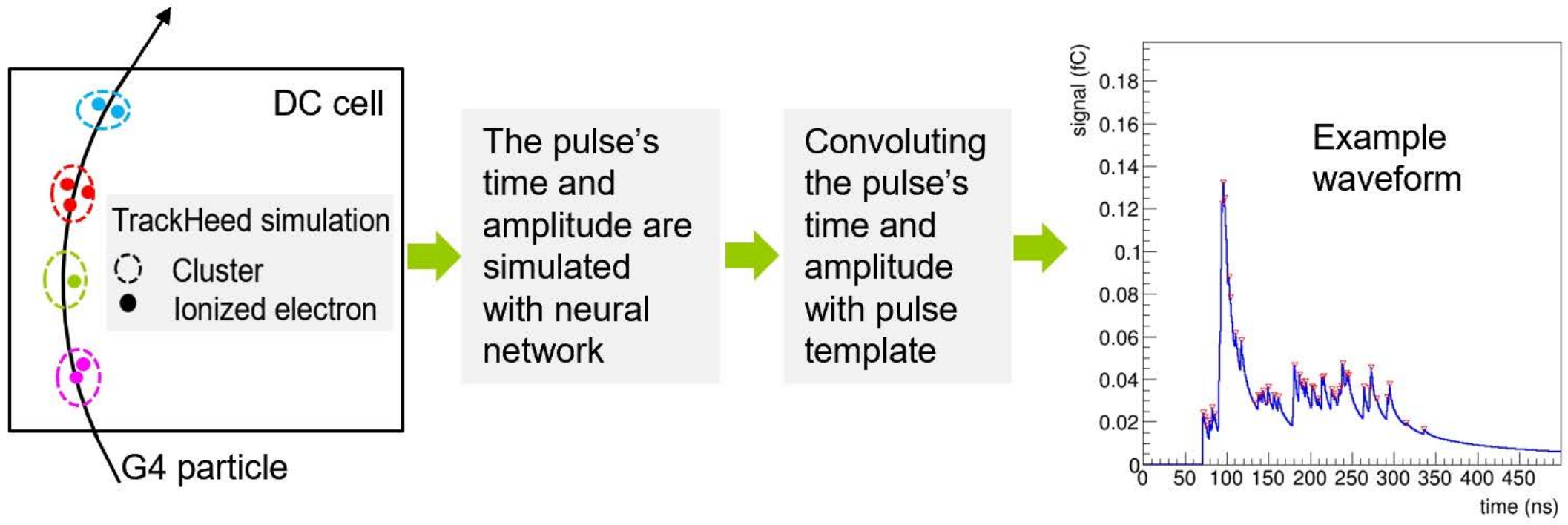

**Figure 13.11:** Flow of the drift chamber simulation: Geant4 step information is passed to the Garfield++ tool, which uses TrackHeed to simulate primary and secondary ionization for the particle. A neural network then simulates pulse time and amplitude based on the ionized electron's position. Afterward, a pulse shape template is used for convolution, and all pulses are combined to create the final waveform.

time—are fed into the NF network, alongside local x and y coordinates as conditional inputs for the MADE block. The MADE block learns the parameters of the RQS transformation, which is subsequently applied to reshape the distributions of amplitude and drift time. By stacking this process iteratively, the transformed distributions gradually converge to the base distributions—specifically, two independent Normal distributions in our case.

During inference, simulation is achieved by sampling from these two Normal distributions and applying the inverse RQS transformations to recover the simulated amplitude and drift time.The NF model was trained using Garfield++ data, with optimization performed via the Adam optimizer [80].

### 13.5.3 Application of quantum algorithms

Quantum computational approaches can be categorized into three primary classes: (1) gate-based quantum computing, (2) quantum annealing, and (3) quantum-inspired algorithms. Gate-based quantum computers utilize quantum logic gates to realize universal computation. In contrast, quantum annealing leverages the adiabatic theorem to seek the ground state of a specified Hamiltonian, rendering it suitable only for combinatorial optimization problems rather than universal computation. Meanwhile, quantum-inspired algorithms—executed on classical architectures—emulate aspects of quantum mechanics to solve optimization tasks; prominent examples include simulated annealing, simulated coherent Ising machines, and simulated bifurcation techniques. When specifically tailored for optimization, both quantum annealing and these quantum-inspired methods are collectively referred to as Ising machines.

The anticipated exponential speed-up of quantum computers arises from uniquely quantum mechanical phenomena such as superposition, entanglement, and interference. For instance, in quantum annealing, quantum tunneling facilitates the escape from local minima during the search for the Hamiltonian's ground state. Furthermore, the integration of quantum comput-ing with machine learning—often termed quantum machine learning—leverages the expansive Hilbert space to potentially enhance data analysis and pattern recognition. Another promising application is quantum simulation, whereby quantum computers are employed to model the time evolution of complex Hamiltonians. Recent reviews, such as Ref. [81], have provided comprehensive overviews of quantum computing applications in high-energy physics.

**Multijet reconstruction with quantum-annealing-inspired algorithms** Various computationally challenging tasks in high energy physics can be formulated as Quadratic Unconstrained Binary Optimization (QUBO) or Ising problems. The problem is designed so that the ground state of the QUBO/Ising model provides the correct answer. Jet reconstruction is a complicated clustering procedure and can be formulated as an Ising problem. Applying quantum computing and algorithms to track [82, 83, 84, 85, 86, 87, 88, 89, 90] and jet reconstruction [91, 92, 93, 94, 95, 96, 97, 98] has been considered actively in the past several years.

A recent study used a novel quantum-annealing-inspired approach, Simulated Bifurcation (SB), to reconstruct multijet events formulated as an Ising problem [98]. SB is a novel quantum-annealing-inspired approach and predicts the Hamiltonian ground state by classically emulating the quantum adiabatic evolution of Kerr-nonlinear parametric oscillators, exhibiting bifurcation phenomena to represent the two Ising spin states. SB can update all the Ising problem spins in parallel, allowing us to achieve computational acceleration. The original version of SB, Adiabatic Simulated Bifurcation (aSB), is prone to errors originating from the continuous treatment of the spins in the classical differential equations. Two variants of SB are introduced to suppress such analog errors: ballistic Simulated Bifurcation (bSB) introduces inelastic potential walls to constrain the spin within ±1. Discrete SB (dSB) further discretizes spins in the mean-field term to suppress the error.

Jet reconstruction is formulated as the QUBO below [98]:

$$O^{\text{multijet}}_{\text{QUBO}}(s_i^{(n)}) = \sum_{n=1}^{n_{\text{jet}}} \sum_{i,j=1}^{N_{\text{input}}} Q_{ij} s_i^{(n)} s_j^{(n)} + \lambda \sum_{i=1}^{N_{\text{input}}} \left(1 - \sum_{n=1}^{n_{\text{jet}}} s_i^{(n)}\right)^2 , \tag{13.3}$$

where $n$ considers the jet multiplicity and the binary $s_i^{(n)}$ is defined for each $i$-th jet constituent and $n$-th jet. $Q_{ij}$ is the distance between the $i$-th and $j$-th jet constituents. In the study, the $ee$-$k_t$ distance is adopted and compared to a simple angled-based method as a benchmark. The second term is introduced as the constraint to ensure that each jet constituent is assigned to a jet only once. This study uses Delphes fast simulated datasets using the 4th detector concept card for three physics processes: $e^+e^- \to Z \to q\bar{q}$, $e^+e^- \to ZH \to q\bar{q}b\bar{b}$ and $e^+e^- \to t\bar{t} \to bq\bar{q}\bar{b}\bar{q}q$. As is the case at the electron-positron colliders, the jet multiplicity is fixed to a certain value for reconstruction based on the physics process under consideration.

By using bSB and the $ee$-$k_t$ distance, we, for the first time, succeed in multijet reconstruction with a QUBO model, a global jet reconstruction in contrast to the traditional iterative approach. In addition, bSB also provides slightly better invariant mass resolution in $t\bar{t}$ events than FastJet.

The results indicate that the global QUBO jet reconstruction using bSB may provide more precise jet clustering.

SB is a promising option for high energy physics due to its capability to handle large dataset at high speed and to predict ground state with outstanding performance even for fully connected QUBO problems.

**Fast calorimeter simulation with quantum generator** High-energy physics relies on large and accurate samples of simulated events, but generating these samples with Geant4 is CPU intensive. Various classical machine learning algorithms have been utilized for fast calorimeter simulation, which is an important approach to solving the problem. Quantum algorithms, leveraging the advantages of quantum computing, have the potential to outperform the classical counterpart.

Various quantum algorithms, such as quantum Generative Adversarial Network (qGAN) and Quantum Ansatz Generator (QAG), have been used for fast calorimeter simulation. Quantum angle generator achieves the most excellent results due to the simplicity of training. However, the original version of QAG [99] has a few limitations. On one hand, the QAG model has limited expressibility, resulting in relatively poor agreement in the pixel-wise energy distributions. On the other hand, the QAG model has a poor scaling, requiring n qubits for the generation of n pixels.

We have introduced several modifications to the QAG architecture. In the original version, the latent vector $z$ is uploaded to the quantum generator as the rotation angles of the RY gates only in the first layer. It has been verified that such uploading scheme has limited expressibility. To increase the expressibility, the latent vector $z$ is uploaded multiple times. For each qubit, the original version only measures the expectation values of Pauli Z operator, requiring n qubits for the generation of n pixels. In the modified version, the expected values of Pauli Z and Pauli X operators are measured, requiring n/2 qubits for the generation of n pixels. The modified version achieves better results in the pixel-wise energy distributions.

By using data re-uploading and multiple measurements, the performance of the QAG model is improved and the required number of qubits is reduced. These architectural improvements represent a meaningful step toward the practical deployment of quantum generative networks.

## 13.6 Computing requirements

### 13.6.1 Event data

During data acquisition, raw data from particle collisions is recorded by the detector and processed through the Trigger/DAQ system. The filtered data is then transferred to storage systems for offline processing. Once stored in distributed storage systems, detector calibration and alignment are performed, followed by raw data processing to reconstruct particle tracks, energy measurements, and other relevant properties. Subsequently, physics information is extracted from the reconstructed data, leading to the generation of event tag data for analysis. Ultimately, data analysis provides valuable physical insights by comparing experimental results with theoretical predictions.

For the simulation, a physics generator produces a list of particles originating from electron-positron interactions. These generated events are then fed into the detector simulation, where each particle is propagated through the detector using Geant4. During this process, the simulated interactions between particles and the detector are recorded. In the digitization step, the responses of individual detector modules are modeled. This step incorporates MC hits from signal events, along with hits from beam-induced background events, as input.

The types of data corresponding to real data processing and simulation are defined as follows:

- GenData (physics generator information): a collection of four-vectors for a physics event provided by a physics event generator
- RawData (raw data): the real detector data of the event data from the high level trigger
- SimData (simulated data): output of the digitization step in the detector simulation
- RecData (reconstructed data): which contains the detailed output of the detector reconstruction and is produced from RawData or SimData
- AnaData (analysis data): which is a summary of the reconstructed event, and contains sufficient information for common analyses
- TagData (event tag): summary of some general features of the event, allowing rapid access to the events and performing fast selection

The target size for RecData has been determined based on insights gained from current reconstructed data. However, the size and content of RecData are still evolving rapidly due to ongoing improvements in our understanding of what is needed for the experiment. AnaData and TagData are expected to be relatively smaller in scale compared to RecData. Given their importance for data analysis, they will be defined as soon as event-level reconstruction, following data reconstruction in sub-detectors, becomes available. Events generated from simulated datasets also include an additional component, the MC truth, which enables a detailed comparison between the reconstructed data and the original event. This extra information significantly increases the event size, but is essential for gaining a thorough understanding of the performance of the reconstruction software.

### 13.6.2 Resource estimation

The computing requirements outlined are derived from the projected data volume during the first five years of data collection, encompassing 3 600 operational hours per year in the Higgs boson phase. During this period, the majority of data will be collected in the ZH mode, with a smaller fraction obtained in the Low-luminosity Z mode. The exceptionally large cross-section at the Z pole enables rapid accumulation of substantial lepton and jet data, which are essential for calibrating both the detector and software systems. Each year, it is estimated that 2–3 months of data collection will take place at the Z pole, operating at an accelerator output power of 12.1 MW. As a result, the total data volume for the first five years is expected to comprise roughly four years of Higgs boson data and one year of Low-luminosity Z data collection.

Each year, the number of MC events is assumed to closely approximate the number of collected data events. Both MC generation and the reconstruction of raw data are conducted twice each year. This approach ensures the utilization of the most finely tuned simulations and optimized reconstructions, thereby enhancing result accuracy. Notably, the volume of reconstructed data accounts for 10% of the raw data and that of analysis data for 15% of raw data.

In the Higgs boson mode, the estimated size per RawData event is 405 kB, corresponding to a DAQ storage output rate of 0.405 GB/s. Consequently, the annual raw data volume during the Higgs factory stage is estimated to be 5.25 PB, accompanied by reconstructed and analysis data volume of 2.63 PB. Additionally, the annual MC simulation data volume for the Higgs boson is projected to be 2 PB, culminating in a total data volume of 48 PB over the four-year Higgs factory stage.

In the Low-luminosity Z phase, the estimated raw data volume is 88 PB, while the MC simulation data volume is projected to be 140 PB. The total data volume for one year of Low-luminosity Z data collection is 412 PB.

With a disk-to-tape storage ratio of 1:1 and a CPU utilization rate of 80%, along with an additional 10% allocated for physics analysis requirements, the estimated total resource demand over the five-year period is summarized in Table 13.1.

**Table 13.1:** Estimation of computational and storage resources

| Mode | Disk (PB) | Tape (PB) | CPU (kHS23) |
|---|---|---|---|
| Higgs boson | 48 | 48 | 220 |
| Low-lumi Z | 412 | 412 | 1,760 |
| Total | 460 | 460 | 1,980 |

Detailed estimates of computing resource requirements will be obtained through a series of Data Challenges—large-scale Monte Carlo productions designed to validate computational needs before building the final system. These challenges will be conducted in multiple rounds, each increasing in complexity and data volume to more accurately simulate future CEPC operational conditions. Early stages will focus on prototype tests of offline software, profiling CPU, memory and storage demands for core algorithms, followed by full-chain processing from simulation to reconstruction, with particular emphasis on bottleneck and performance of key workflows. As the Data Challenges progress, later rounds will emphasize the generation of approximately 1 million Higgs events, complemented by inclusive Monte Carlo datasets that simulate one month of low-luminosity Z data collection. Those rounds will incorporate Grid computing integration, significantly enhancing the realism of both data processing workflows and analysis pipelines.

The final phase of the program will feature collaborative efforts with the CEPC international sites, coordinated through a Common Computing Readiness Test. This structured approach ensures comprehensive validation of the computing model, overall operational readiness, paving the way for the full-scale deployment of the CEPC.

## 13.7 Distributed computing infrastructure

### 13.7.1 Computing model

During the Higgs factory and low-luminosity Z phases, data volumes could reach up to 420 petabytes. In contrast, high-luminosity Z data taking is expected to generate data on the exabyte scale. The computing model for the Higgs boson phase must be both flexible and scalable, with resources and system architecture designed to enable faster data processing at lower cost—while carefully balancing technological capabilities and financial constraints.

As illustrated in Figure 13.12, the computing model will adopt distributed computing paradigm and organizes resources in a hierarchy model, which allows us to extend resources easily. The hierarchy model allows flexible connections among different tiers based on network performance. In the high-luminosity Z phase, the width and depth of the tiers must be extended to meet larger amount of data. To address short-term resource gaps, facilities such as supercomputing centers, commercial cloud platforms, and other opportunistic sources will be leveraged to provide burst capacity in an elastic manner. These resources can serve as a supplementary tier within the computing model, supporting large-scale Monte Carlo simulations and interactive bulk data analysis during peak demand periods.

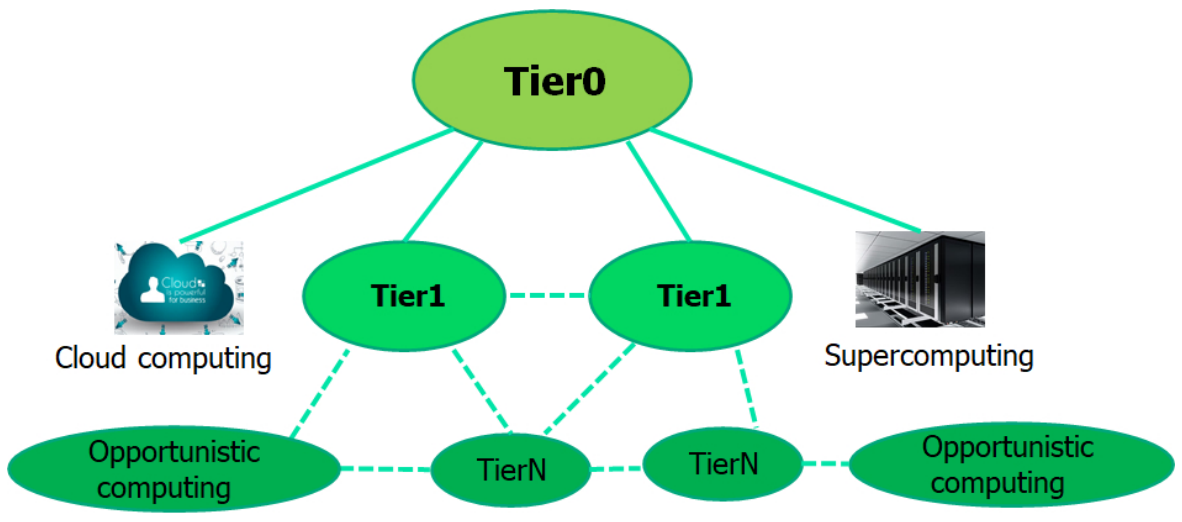

**Figure 13.12:** The computing model adopts a hierarchical multi-tier structure, enabling flexible extensions and dynamic resource connections tailored to network performance, including cloud computing, supercomputing and opportunistic computing.

On the other hand, machine learning and deep learning techniques are emerging as key methods alongside traditional approaches, offering significant acceleration in data processing and analysis. When deployed in production environments, these applications introduce workflows that differ substantially from conventional methods and often require alternative hardware such as GPUs, FPGAs, and other specialized accelerators. Integrating these new workflows and dataflows with traditional ones adds complexity and presents fresh challenges for the distributed computing model.

A key requirement of the computing model is to ensure rapid access to all relevant data for reconstruction and analysis by all members of the experiment, along with appropriate access to conditions data. To meet this need, a central data storage architecture with a multi-tier caching system is planned. Given the massive volume of data involved, the model must incorporate cost-effective and resilient storage solutions, as illustrated in Figure 13.13. The central storage service will be geographically distributed across large data centers, interconnected by fast networks to optimize both hardware utilization and operational costs. Thanks to high-speed connectivity, processing resources—including heterogeneous systems and smaller sites—need not be co-located with the storage infrastructure. Data will be delivered through a multi-tier caching system, enabling processing centers to access content via a distributed delivery and caching mechanism. This approach effectively mitigates latency issues that arise when compute and storage are physically separated.

An efficient data distribution strategy within the data model is essential for optimizing data storage, access, and processing. The RawData will reside in central storage, while the RecData, SimData, and AnaData will be distributed across Tier-1 storage. These datasets will be accessible on demand via a tiered storage and caching system, ensuring efficient availability across diverse sites. A progressive reduction in data format for user analysis, combined with dedicated analysis facilities on both Grid and non-Grid resources, establishes a resilient framework that supports fast and reliable user analyses.

### 13.7.2 Workload management system

In the distributed computing infrastructure, the Workload Management Systems (WMS) distributes and manages computational tasks across a Grid computing environment. This includes resource allocation, dynamic scheduling, job monitoring, and fault tolerance. Both DIRAC [100] and Production ANd Distributed Analysis (PanDA) [101] have demonstrated the capability to efficiently handle tens of thousands of jobs in parallel. However, managing a wider variety of hardware, resources, and more complex workflows may pose significant challenges.

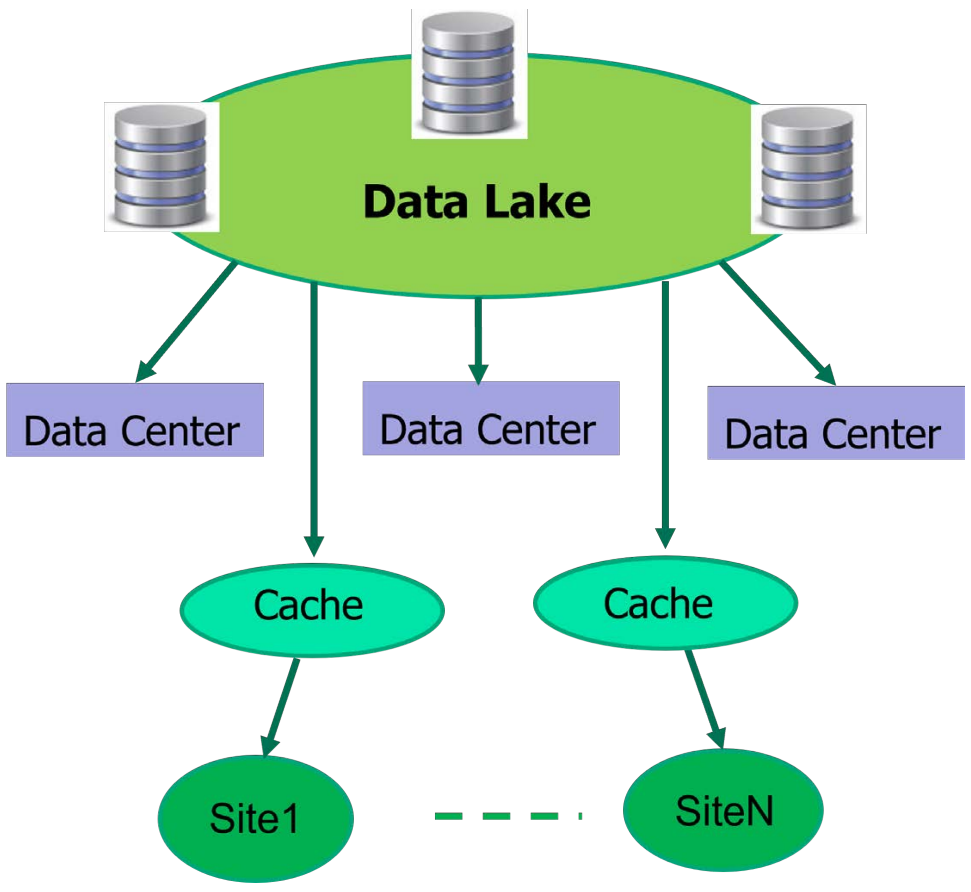

**Figure 13.13:** To ensure robustness, accessibility, and low operational costs, the data model adopts a centralized data lake with hierarchical cache layers, enabling fast data access across all tiers.

**DIRAC** DIRAC provides a complete distributed computing framework and solutions for organizing various resources to fulfill experimental data processing tasks. A prototype based on DIRAC for the experiment has been established for evaluation and CEPC-specific applications is developed in IHEPDIRAC which is an extension of DIRAC. Leveraging the flexibility and extensibility of the DIRAC WMS, the distributed heterogeneous resources have been successfully integrated and tested. Currently, this prototype has been used for Research and Development, which employs a centralized one-tier distributed computing model. IHEP serves as the main center responsible for storing the complete dataset and conducting MC production, while other sites contribute to this production. All output MC data is copied back to IHEP. To facilitate the use of Grid resources, the frontend tool JSUB [102] and the production system ProdSys [103] have been developed to accommodate a large volume of user job submissions and management.

**PanDA** PanDA is another option to act as a WMS system for the experiment. PanDA originates from ATLAS distributed computing, and is engineered to operate at the LHC data processing scale. The prototype based on PanDA was developed and managed by the JINR group, and it is deployed alongside Rucio Data Management System (Rucio) [104] as its data management system. This prototype has demonstrated effective integration with cloud resources based on OpenNebula. A similar setup was utilized during the R&D phase of the JINR Nica/SPD [105] experiment, where it efficiently handled the complex workflows and data flows required for various data processing activities. The comparison between PanDA and DIRAC will continue, focusing on functionality and performance, while considering the requirements of the computing model as well as factors such as complexity and manpower.

### 13.7.3 Data management system

A Data Management System (DMS) in a Grid computing environment is designed to efficiently handle, store, retrieve, and manage large volumes of data across distributed resources.

**DIRAC DMS** The DIRAC DMS [106] is a crucial component of the DIRAC framework, designed to handle the storage, transfer, DIRAC DMS provides a comprehensive data catalog that allows users to register, search, and manage datasets efficiently. The DIRAC DMS has been successfully evaluated for the experiment. As part of the distributed computing system based on DIRAC, the DMS effectively managed over 2PB of simulated data during the R&D period, enabling smooth transfers between sites. It also worked in close coordination with the WMS to support MC production. The DIRAC File Catalogue, part of the DIRAC DMS, is employed to register and manage the dataset, while FTS3 manages file-level replication.

**Rucio** Rucio is a distributed data management system designed to handle exabyte-scale data volumes. A functional Rucio prototype has been deployed for the experiment, incorporating critical infrastructure services (FTS3, Identity and Access Management (IAM), SE) via Docker. Tests involving data injection and multi-site distribution confirmed the system's usability and maintainability, requiring minimal operational staff. Upcoming work aims to integrate Rucio into the DIRAC system and collaborate with both communities to support smooth adoption for the experiment.

### 13.7.4 Opportunistic computing

Resource usage during high-energy experiments is not uniform across different periods. At times, urgent and centralized data processing demands a substantial amount of CPU resources within a short time frame, necessitating greater flexibility and scalability in our infrastructure. Opportunistic computing offers a dynamic and adaptable approach to resource utilization in distributed systems, helping to address the increasing demand for low-latency access to computing resources. Supercomputing centers and commercial clouds are two of the most commonly utilized opportunistic resources.

Both DIRAC and PanDA are able to support connection to these two resources in an elastic way. In a cloud computing environment, virtual computing nodes are dynamically provisioned as needed. The cloud resource management system handles the lifecycle management and configuration of virtual machines. One feasible model is to design and develop a cloud resource scheduling system that works in conjunction with the cloud resource management system to create virtual resources in real-time. The model enables resources to scale instantly and as needed.

### 13.7.5 Computing operations

Data processing activities must be conducted in cooperation with WMS and DMS, together with offline database, in accordance with the tiered computing and data model.

Tier-0 is responsible for the initial data reconstruction, archiving, and distribution of primary RawData and RecData, along with calibration and alignment datasets. It includes a dedicated computing farm that processes RawData streams into RecData and manages conditions records within the central database.

Tier-1 acts as full-service computing centers for storing RawData and SimData, providing resources for reprocessing and analysis. Each Tier-1 must store a significant portion of the RawData and RecData datasets and ensure access for the entire collaboration.

Tier-2 plays various roles in simulation and analysis, with production managers overseeing simulation production and working groups managing analysis activities.

To handle concurrent requests in distributed data model, the database infrastructure aims to utilize a single database technology with multi-cache layers to ensure efficient and reliable data access.

## 13.8 Open data

The experiment needs to share specific data with both the global collaboration network and the public. The goal of the open data research project is to enable multi-tiered data interoperability and transparent knowledge sharing, in alignment with global open science frameworks. Based on the DPHEP Study Group's taxonomy for data granularity, the open data will be organized into four hierarchical levels: Level 1 includes datasets linked to published results (e.g., extended figures and tables for validation); Level 2 offers simplified formats for outreach and training (such as basic four-vector event-level data); Level 3 provides reconstructed collision and simulated data with analysis-level, experiment-specific software for full scientific analyses; and Level 4 comprises raw data, if not already covered in Level 3, along with reconstruction and simulation software to enable new signal generation or data reprocessing. Data dissemination will prioritize curated releases of Levels 1–3 to ensure accessibility and reproducibility, while benchmark subsets of Level 4 may be selectively shared to support cross-verification and collaborative innovation. As a member of the DPHEP community, IHEP will contribute to building a comprehensive open data policy and infrastructure.

## 13.9 Long term data preservation

The CEPC experiment faces significant demands and challenges in long-term data preservation. According to the four-level data preservation model of Data Preservation in HEP (DPHEP), it is essential to preserve not only experimental and reconstructed data but also reconstruction and simulation software. To ensure flexibility for future research, all data must be saved. Storing reconstruction software and analysis processes is crucial for maintaining the ability to derive new correlations, efficiencies, corrections, and to conduct comprehensive systematic error analysis.

In terms of the storage architecture, a multi-layer storage system will be adopted, integrating disk storage for medium to long-term data preservation with tape libraries for permanent backup. This ensures quick access and high reliability. For the data management, all experimental data, reconstructed data, reconstruction software, and simulation software will be systematically stored and managed. This includes implementing effective data cataloging and retrieval mechanisms to handle the vast amounts of data generated by the experiment. In addition to the requirement of saving all the data, it is also necessary to preserve the software and the data processing procedures. The reconstruction software and analysis workflows will be maintained to support future execution and modification. This may involve containerization technologies to ensure software environments remain consistent and executable over time.

This data preservation strategy will achieve reliable long-term storage of all experimental and reconstructed data, along with the necessary software and analysis tools. It ensures data accessibility and usability for future research. By combining disk storage with tape libraries backup, the solution offers robust protection against data loss and supports the evolving needs of the experiment. It enables researchers to revisit and reanalyze data as needed. The scheme provides the performance and advanced features necessary to meet the future requirements of the experiment. It ensures comprehensive data storage, supports in-depth analysis and reanalysis, and is adaptable to technological advancements and increasing data volumes.

## 13.10 Resource provision

### 13.10.1 Computing

The architecture of computing resources is becoming increasingly heterogeneous, incorporating CPUs, GPU, Neural Processing Unit (NPU) [107], and DCU [108] to support diverse data processing requirements of the experiment. The computing system must scale to manage million-core heterogeneous resources, dynamically matching millions of concurrent computing tasks. Additionally, the system requires dual-mode computing support: High-Throughput Computing for large-scale data processing and High-Performance Computing for latency-sensitive parallel workloads. Proven technologies such as HTCondor and Slurm have been deployed in the IHEP computing cluster to meet these demands.

The experiment will integrate AI-driven applications including fast calorimeter simulation, track reconstruction, and particle identification. An intelligent scientist workstation provides fundamental components: pre-trained algorithms, curated datasets, and scalable inference services. Its modular interface unifies workflow design, parameter optimization, and collaborative analysis, while standardized tools streamline data lifecycle management, execution scheduling, and result dissemination. An embedded LLM-based scientific assistant automates documentation generation, optimizes experimental parameters, and delivers domain-specific guidance, accelerating AI adoption and international collaboration in HEP research.

### 13.10.2 Storage

The experiment requires exabyte-scale storage capacity and large-scale of parallel data access. The data storage system should support large-scale scientific computing, high-speed access, and long-term archiving. The system has four key components: user home directories, software repositories, disk storage, and tape archives. User home storage handles high-concurrency access using NVMe SSD [109] and a distributed architecture, offering low latency and scalability. Future integration of all-flash file systems will enhance data management capabilities. Software repository storage utilizes the CVMFS [110] for efficient software distribution, cache optimization, and version consistency, widely adopted in Grid computing environments. Disk storage will be built on EOS [111, 112], providing high throughput, availability, and fault tolerance. It serves as the data backbone for distributed computing with local and remote access capabilities. Tape storage leverages the CERN Tape Archive (CTA) [113, 114] system for high-capacity, cost-effective archival, fully integrated with Grid environments to ensure long-term preservation and collaborative data access. Currently, EOS and CVMFS have been deployed as core components, with plans to expand toward intelligent tiered storage and exabyte-scale capacity to meet evolving performance requirements and support global collaboration.

### 13.10.3 Network

To address complex network requirements of the experiment, spanning data production, transfer, storage, analysis, and sharing, the network system targets a unified environment delivering super high-bandwidth, lossless, and ultra-low latency capabilities. A dedicated high-bandwidth infrastructure will be implemented within and between data centers to support batch processing systems essential for cloud computing. Concurrently, a lossless network will be deployed both intra-data center and across geographically dispersed facilities to enable the large-scale data processing and AI services. Critically, an ultra-low latency network will support the Intelligent Scientist Workstation for interactive computing and real-time Internet Of Things (IoT) operations – including accelerator tunnel monitoring. Currently, IHEP maintains a globally interconnected network topology, with novel technologies under evaluation to enhance these capabilities.

As a world-renowned large-scale scientific facility, the experiment's data processing systems, networks, and information infrastructure face diverse cyber threats necessitating a comprehensive cybersecurity framework. This integrated structure comprises three pillars: 1) A cybersecurity management system establishing defined policies, governance protocols, and procedural standards; 2) A technical security architecture implementing logically segmented security domains with layered defenses—including boundary protection, IDS/IPS, host hardening, vulnerability management, and log auditing; and 3) A security operations system deploying a big data analytics-driven security platform alongside a cross-facility cybersecurity alliance with joint defense coordination. This robust framework ensures resilient operation of all information systems, with validated implementation across multiple IHEP large-scale scientific facilities demonstrating efficacy in protecting operational stability.

### 13.10.4 Smart data center infrastructure

A dedicated data center will be established to support the experiment's local storage and processing of massive datasets. To improve power usage effectiveness, the facility will incorporate innovative energy-conservation and emission-reduction technologies, as described in [115].

The data center integrates cold plate and immersion high-efficiency liquid cooling systems, leveraging liquid's superior thermal conductivity to resolve heat dissipation issues at the heat source, addressing ultra-high rack power densities (exceeding 100 kW).By utilizing free natural cold sources (e.g., outdoor air, lake water) for cooling, cooling-related energy consumption (40–50% of total usage) is reduced, achieving Power Usage Effectiveness (PUE) below 1.1. The above solutions achieve highly efficient and green energy efficiency and thermal management.

Green data centers prioritize the adoption of renewable energy sources, such as photovoltaic solar and wind power, to replace fossil fuel-based electricity. Utilizing heat pump waste heat recovery technology, the data center's waste heat provides district heating and domestic hot water, significantly reducing dependence on fossil fuels and lowering operational expenses, transforming data centers from high-carbon units into hubs of energy efficiency and low environmental impact.

AI powers intelligent data center operations. Machine learning models predict cooling requirements and optimize energy distribution, minimizing manual intervention and reducing operational risks. By analyzing historical data and simulating various scenarios, these systems enhance resource utilization, enabling adaptive, energy-efficient management.

## 13.11 Summary

The Reference Detector project represents a significant advancement in high-energy physics research, with its software and computing infrastructure serving as a cornerstone for scientific exploration. This chapter has detailed the comprehensive approach taken to develop the offline software and computing systems necessary to support the CEPC's ambitious scientific goals.

The core software architecture, built upon the Gaudi framework and Key4hep common software stack, demonstrates the project's commitment to leveraging established HEP software packages while developing innovative solutions tailored to the experiment's requirements. The simulation, reconstruction, and analysis tools represent state-of-the-art approaches to handling the complex data processing challenges inherent in modern particle physics experiments.

The distributed computing infrastructure offers a robust solution for managing the massive data volumes anticipated from the detector, with meticulous planning for resource allocation across multiple computing centers. By incorporating cutting-edge Grid computing technologies from the LHC and embracing advancements in artificial intelligence, the efficient management and utilization of computational resources for the future experiment is entirely achievable.

Looking ahead, the software and computing team faces several critical challenges and opportunities. As the project transitions toward the engineering phase, continuous refinement of the software stack will be crucial to ensure optimal performance and reliability. Adopting new computing paradigms, including expanded use of heterogeneous computing resources and quantum computing techniques, will demand both technical innovation and interdisciplinary collaboration. Additionally, the technologies and methodologies developed for the experiment are poised to make a profound and enduring impact, influencing future experimental designs and advancing computing strategies in high-energy physics and beyond.

# Chapter 14 Mechanics and Integration

This chapter introduces the mechanical design and integration of the CEPC Reference Detector. Section 14.1 gives the overall mechanical layout. The boundaries and connections between sub-detectors are given in Section 14.2. To facilitate maintenance and upgrades, the installation scheme of each sub-detector and the standardized installation tooling can be found in Section 14.3. The layout of both the underground experimental hall for the detector and the auxiliary hall for the electronics and supporting facilities is described in Section 14.4.

## 14.1 Overall mechanical design of the detector

The CEPC Reference Detector incorporates various sub-detectors with distinct functions and superior particle-resolving capabilities. These sub-detectors exhibit distinct mechanical properties and requirements, as well as significant differences in mass and size. This section describes the integration of these sub-detectors into a state-of-the-art device with a stable and reliable overall structure, which meets the requirements of installation and alignment, and achieves the physics goals.

### 14.1.1 Overall layout

The CEPC Reference Detector comprises two parts: the barrel and the endcaps, as shown in Figure 14.1. The barrel region is constructed from the outside inward, with the following sequence of components: the barrel yoke (which incorporates the muon detector), the detector superconducting solenoid magnet (Magnet), the HCAL, the ECAL, the TPC (including the barrel OTK), the ITK, and the beam pipe assembly (including the VTX and the LumiCal). Similarly, the endcap part consists of the endcap yoke (with the muon detector), the endcap HCAL, the endcap ECAL (including the endcap OTK) and the accelerator magnet (Acc magnet).

The total weight of the detector is about 5,205 tons. The yoke is 3,110 tons, consisting of 3,010 tons of 10# carbon steel and 100 tons of the muon detector. The superconducting solenoid magnet accounts for 250 tons, the HCAL for 1655 tonnes, and the ECAL for 180 tonnes. The remaining sub-detectors, water pipes and cables add up to about 10 tons.

### 14.1.2 Design requirements

The CEPC Reference Detector mechanical design is based on the design parameters and technical requirements of the detector described in previous chapters and reference to the CDR[1]. The mechanical overall layout and the sub-detector connections was developed, along with the clarification of requirements for boundary dimensional tolerance, installation clearance, coaxiality alignment, connecting and the detector-accelerator interface.

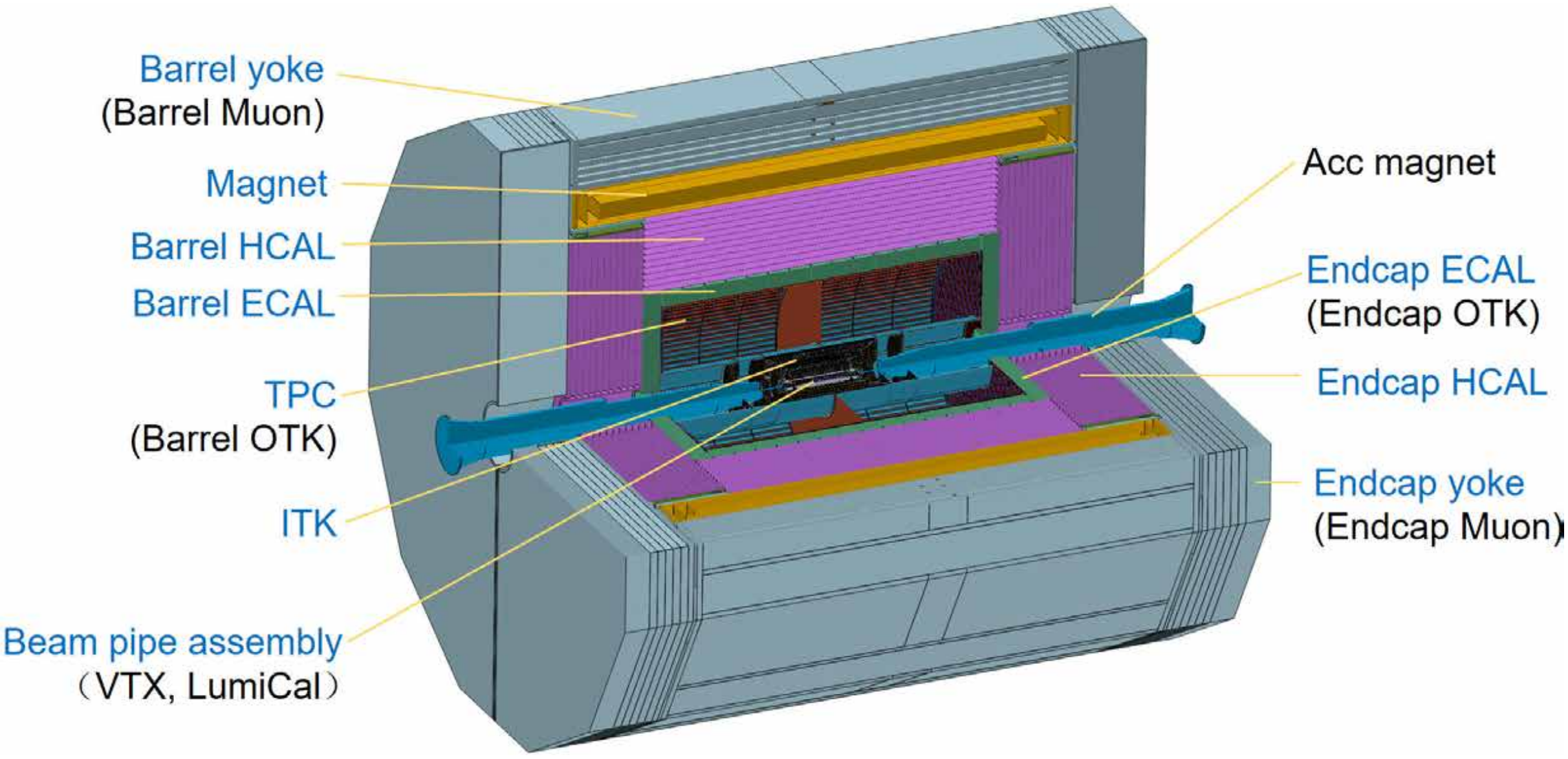

**Figure 14.1:** CEPC Reference Detector highlighting the sub-detector assemblies for installation. The yoke with the muon detectors, the TPC with the barrel OTK, the beam pipe assembly containing the VTX and LumiCal, the endcap ECAL with the endcap OTK, will first be individually assembled before proceeding to installation.

1. Sub-detectors boundary dimensions

   The overall layout of the CEPC Reference Detector and dimensions of each sub-detector are shown in Figure 14.2. The detector is a duodecagon prism, 11,970 mm in length, 10,370 mm in height. The design requirements for each sub-detector are as follows:

   (1) Each sub-detector shall undergo detailed design based on the boundary dimensions specified in the overall layout, and demonstrate the rationality and feasibility of these dimensions and parameter.

   (2) The design of each sub-detector shall incorporate the cooling structures, cabling and piping, etc.

   The beam pipe assembly has a length of 1,400 mm with a maximum outer diameter of 153 mm. It is part of the detector and accelerator interface specifications. More dimensional requirements for this interface are introduced below.

2. Boundary tolerance, installation clearance and coaxiality of sub-detectors

   To reduce the detection loss, the clearance between the sub-detectors is minimized. As shown in Table 14.1, the nominal clearance between sub-detectors in the overall mechanical design is set to 10 mm, with the exception of the 2 mm clearance between the beam pipe assembly and the ITK. These clearance parameters are based on the successful experience at BESIII. They will be optimized and adjusted as the design is refined.

   To ensure smooth installation, adjustment and alignment of each sub-detector, while accounting for errors in the process and assembling, the contour tolerance of each sub-detector is defined in Table 14.1. The tolerance of the outer diameter (outer boundary) dimension of each sub-detector is negative, and that of the inner diameter (inner boundary) dimension is positive, so that the installation clearance of each sub-detector can be guaranteed to be larger than the nominal dimension.

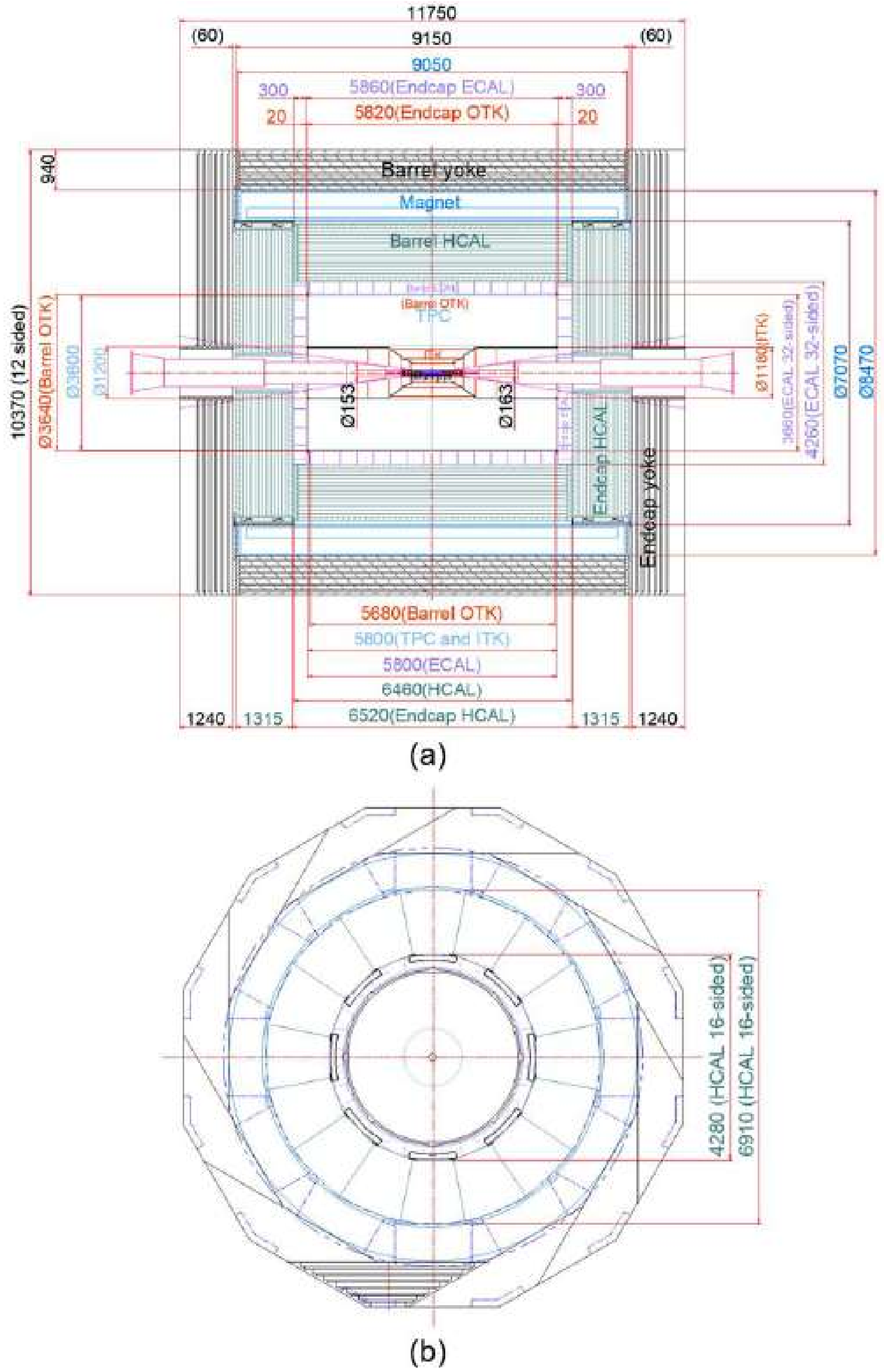

**Figure 14.2:** The overall layout of the CEPC Reference Detector and the dimensions of each sub-detectors. (a) Front view; (b) Side view. Units are in millimeters. Paraffin layers at the two ends of the detector increase its total length to 11,970 mm.

**Table 14.1:** Boundary tolerance, installation clearance and coaxiality of sub-detectors.

| Detector | Boundary dimensional tolerance (mm) | | Installation clearance (mm) | Coaxiality (mm) |
|---|---|---|---|---|
| | Outer | Inner | | |
| Yoke | – | ±2 | – | ±0.5 |
| Magnet | 0 to -5 | 0 to +5 | 10 | ±1 |
| HCAL | 0 to -5 | 0 to +5 | 10 | ±0.5 |
| ECAL | 0 to -5 | 0 to +5 | 10 | ±0.5 |
| TPC | 0 to -2 | 0 to +2 | 10 | ±0.1 |
| ITK | 0 to -2 | 0 to +2 | 10 | ±0.1 |
| Beam pipe assembly | 0 to -0.3 | 0 to +0.3 | 2 | ±0.1 |

3. Design requirements for the connection of each sub-detector
   The main structural body is the barrel yoke, with a direct connection to the magnet and barrel HCAL. Each sub-detector connects to neighboring sub-detectors with a clearance of 10 mm. The connection relationships between sub-detectors are listed in Table 14.2. The connection configuration shall be readily adjustable with adequate clearance.

**Table 14.2:** Sub-detector connection relation.

| Detector | Fixture point |
|---|---|
| Barrel yoke | Bottom base |
| Magnet | Barrel yoke |
| Barrel HCAL | Barrel yoke |
| Barrel ECAL | Barrel HCAL |
| TPC (Barrel OTK) | Barrel ECAL |
| ITK | TPC |
| Beam pipe assembly (VTX, LumiCal) | ITK |
| Endcap ECAL (Endcap OTK) | Barrel HCAL |
| Endcap HCAL | Barrel HCAL |
| Endcap yoke | Base |

4. Interface between detector and accelerator
   The interface between the detector and the accelerator is shown in Figure 14.3. The design boundary mainly consists of a section of tapered hole and three sections of step holes:
   (1) The tapered hole has an opening angle of about 8.1°, and a length of 2,900 mm.
   (2) The endcap ECAL has a square hole with dimensions of 700 mm × 700 mm.
   (3) The endcap HCAL has a round hole with a diameter of 800 mm.
   (4) The endcap yoke has a round hole with a diameter of 1,100 mm.

This space is reserved for the accelerator's superconducting focusing magnet and associated pipelines. All components need to facilitate installation, alignment and maintenance during the design in the area of the MDI.

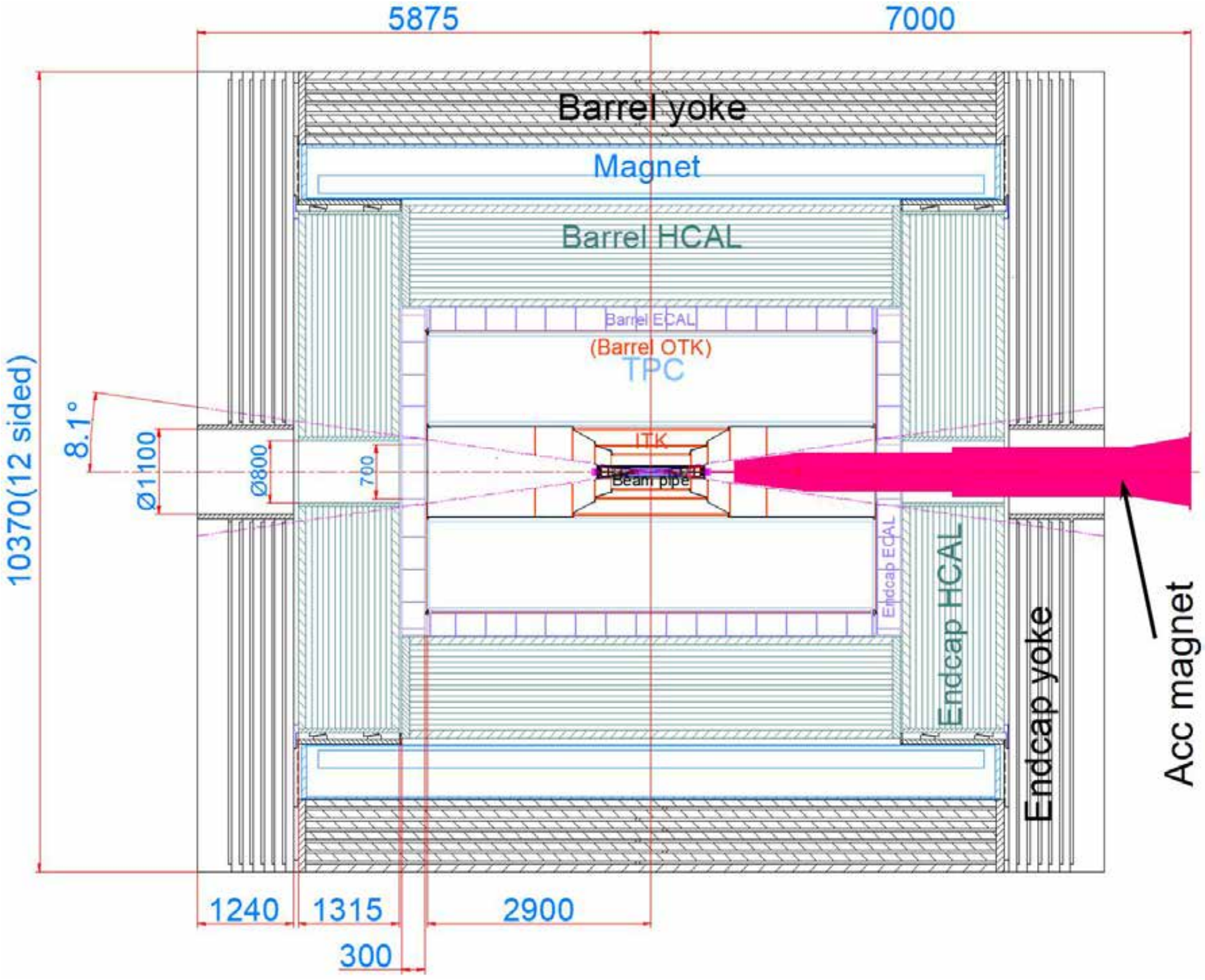

**Figure 14.3:** Overall detector layout emphasizing the interface between detector and accelerator. The detector and accelerator envelope boundary is defined by one tapered hole within the ITK and three step holes within endcap calorimeters and the yoke. Units are in millimeters.

## 14.2 Sub-detector connections

For the convenience of future maintenance and upgrades, the connection structure between sub-detectors must be designed to take into account the positioning and adjustment, installation and removal, with repeatable and reversible installation accuracy.

The detector installation sequence generally proceeds from the outside inward. This process begins with the yoke to guarantee the safe and reliable positioning and operation of all sub-detectors. The following subsections will introduce the connection structure of each sub-detector.

### 14.2.1 Yoke

The yoke consists of the barrel and the endcap parts. The yoke material uses 10# carbon steel with excellent magnetic properties (high permeability and low coercivity), good machinability, stable material properties and low cost.

Each barrel yoke module and endcap yoke module has seven layers, six of which are equipped with muon detectors. Details of the muon detector are provided in Chapter 9. The yoke modules are assembled in the ground hall and then integrated with the muon detector modules into the sectors. These sectors should be tested and secured before being transported underground for final installation. At the endcap, an additional outer layer holds paraffin for shielding neutron backgrounds from accelerator tunnel. All other sub-detector components are directly or indirectly anchored and supported by the barrel yoke. The endcap yokes are placed at both ends of the barrel yoke, one on each side.

1. Barrel yoke

   The barrel yoke with a total mass of 1,610 tons is composed of 12 modules assembled in a dodecagonal structure. As shown in Figure 14.4, it has an axial length of 9,150 mm and a thickness of 940 mm. The total height is 10,370 mm and the bore vertical dimension is 8,490 mm. The optimized yoke design incorporates two additional flanges at both ends, which serve as the support structure for barrel yoke modules to effectively reduce the deformation of the barrel yoke.

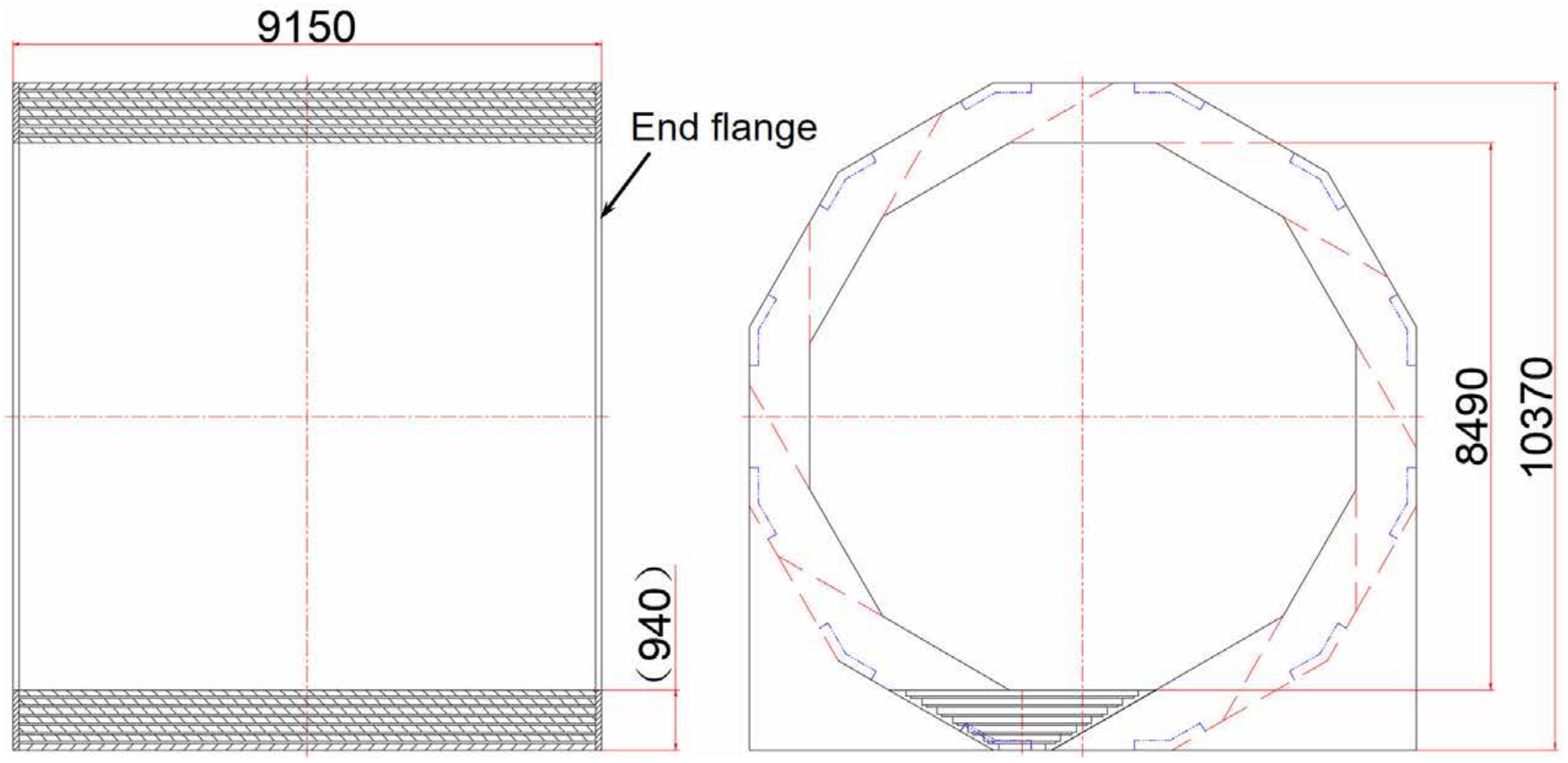

**Figure 14.4:** Axial and transversal view of the barrel yoke structure. In the transverse plane, the yoke has a regular 12-sided shape. Two additional end flanges serve as the support to the barrel yoke modules. Six layers of muon detectors are embedded in the yoke, and are seen through a cut-out on the end flange at the bottom. Units are in millimeters.

   The key point of this design is the inter-supporting between the polygonal structure, as shown in Figure 14.5. The two end flanges serve not only to reduce the deformation, but also to act as tooling to facilitate the installation of barrel yoke modules. For more details, refer to the yoke installation in Section 14.3.3. The side panel of each barrel module is made of removable metal slabs to provide access to service the muon detectors after installation.

   Figure 14.6 compares the overall deformation frin FEA for the barrel yoke without and with the end flanges. In this calculation, the barrel yoke is supported on the base and subjected to its self-weight. Results show that the maximum deformation without the end flanges is about 2.5 mm, while that with the end flanges is only 0.9 mm. Adding the end flanges reduces the self-weight deformation significantly, which will improve the installation accuracy of each sub-detector. This design shows advantages in both overall safety and the structural stability (refer to the Section 14.2.8 of reliability and safety assessment for details).

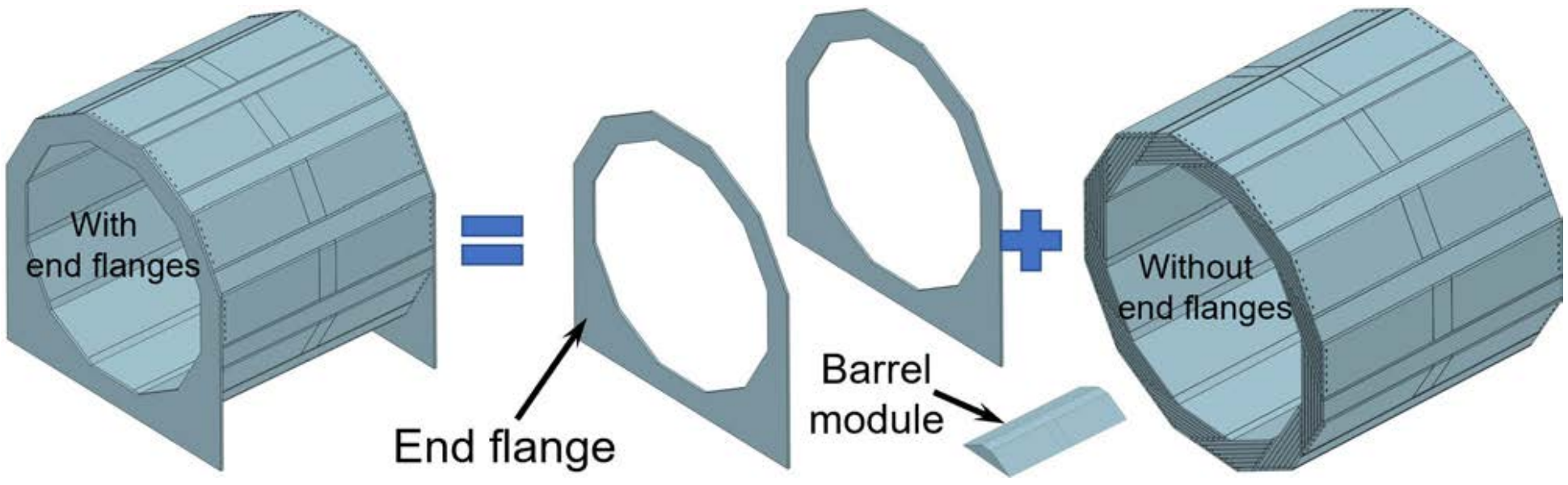

**Figure 14.5:** The idea of the inter-supporting connection structure. The two end flanges provide structural reinforcement and facilitate the yoke installation.

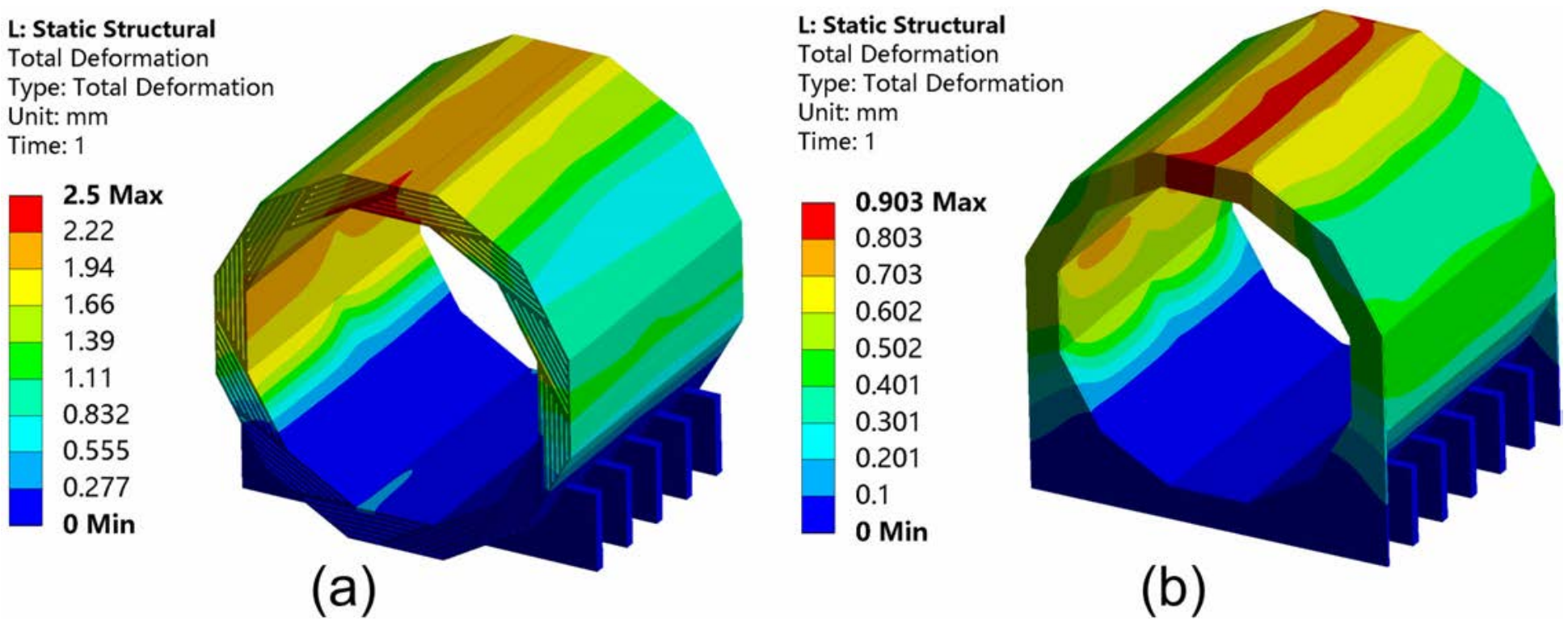

**Figure 14.6:** Overall deformation comparison of FEA results for the barrel yoke without and with the end flanges in gravity load. (a) Maximum deformation is 2.5 mm without the end flanges; (b) Maximum deformation is 0.9 mm with the end flanges.

2. Endcap yoke

   Each endcap yoke is a 12-sided polygon assembled structure, which is divided into two halves along the vertical center line as shown in Figure 14.7. Each half weighs 350 tons and can be moved along the guide rail to open or close, enabling maintenance of the sub-detectors.

   Each half has a thickness of 1,240 mm, a width of 5,185 mm, and a height of 10,370 mm. The clearance between the endcap yoke and the barrel yoke is 60 mm for the internal sub-detectors' piping and cabling. The central hole of 1,100 mm in diameter of the endcap yoke accommodates to install the accelerator superconducting magnet. A 100-mm paraffin shield is packed in a steel casing, and then secured to the outside of each endcap yoke with screws, to mitigate beam background from neutrons originating from the accelerator

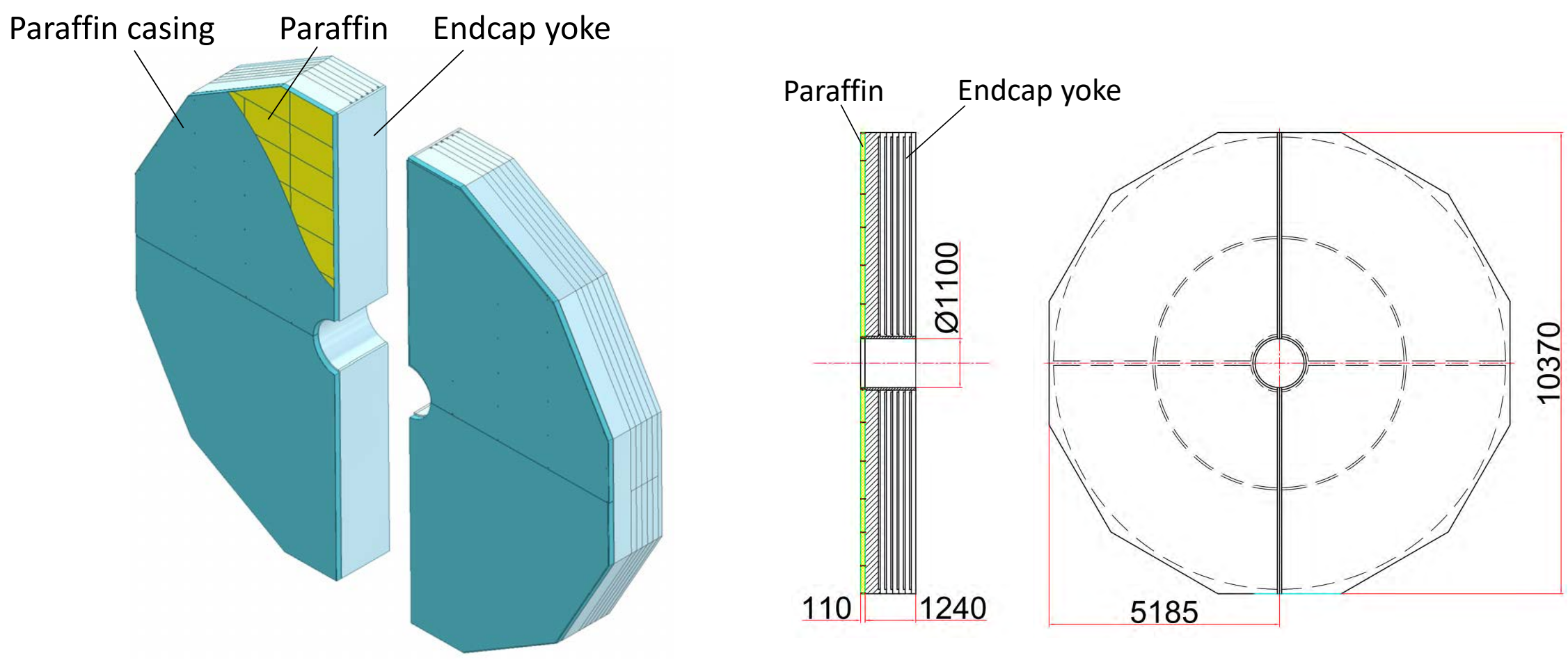

(a) (b)

**Figure 14.7:** Endcap yoke. (a) 3D view of endcap yoke in open position; (b) 2D drawing in closed position. The dashed lines show the location of the muon detectors inside the yoke. The total thickness of the paraffin-filled casing is 110 mm. Units are in millimeters.

tunnel.

### 14.2.2 Superconducting solenoid magnet

Figure 14.8 shows the boundary dimensions of the magnet. It weighs 250 tons and is mounted inside the barrel yoke, with a minimum clearance of 10 mm. The dewar is made from low-permeability stainless steel with overall dimensions of 9,050 mm in length, 8,470 mm in outer diameter. Its structure features a 25-mm-thick outer wall, a 15-mm-thick inner wall, and 70-mm-thick end rings.

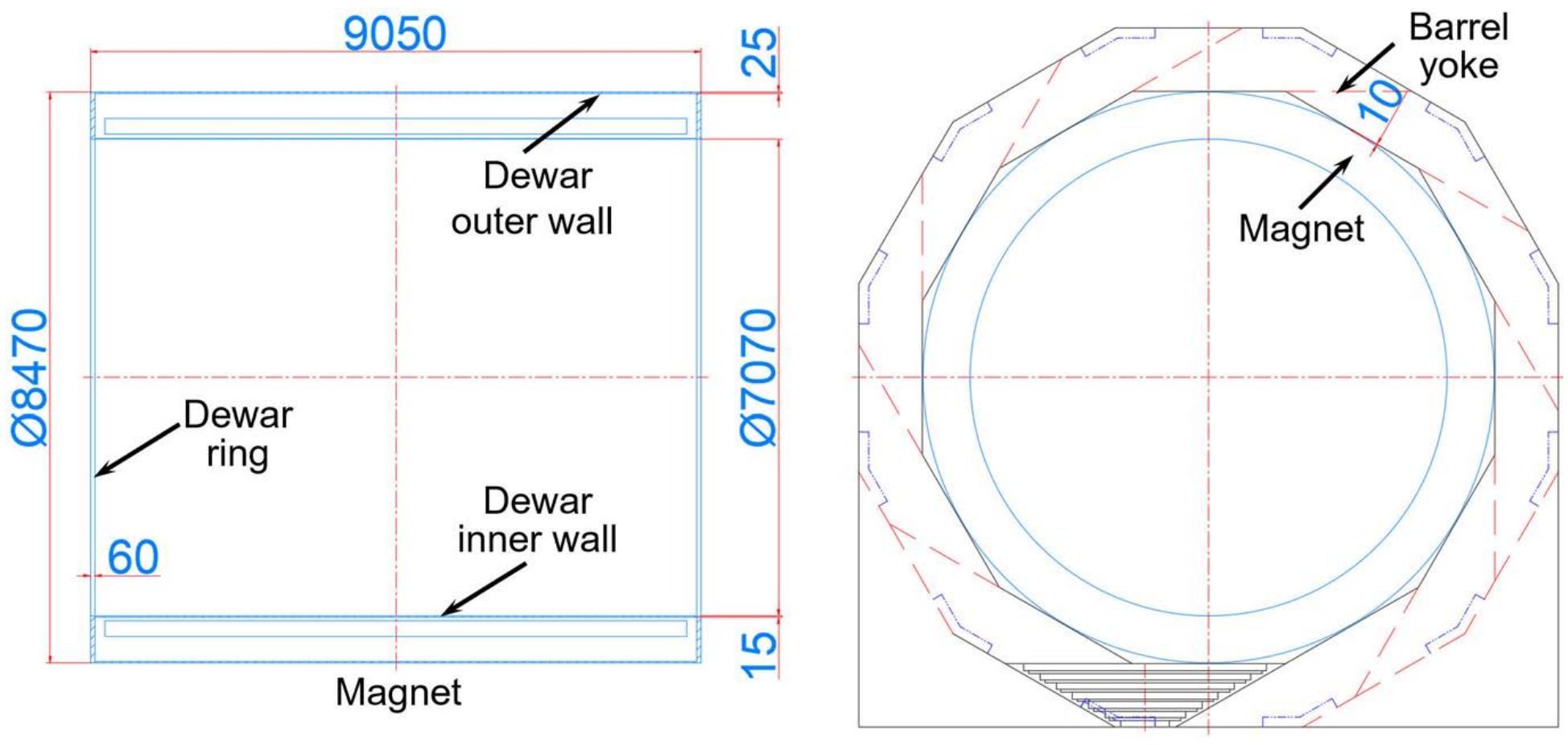

(a) (b)

**Figure 14.8:** Superconducting solenoid magnet inside the dewar. (a) Dewar dimensions, including the 60-mm thick rings used for connection to the barrel yoke; (b) The minimum clearance with the barrel yoke is 10 mm. Units are in millimeters.

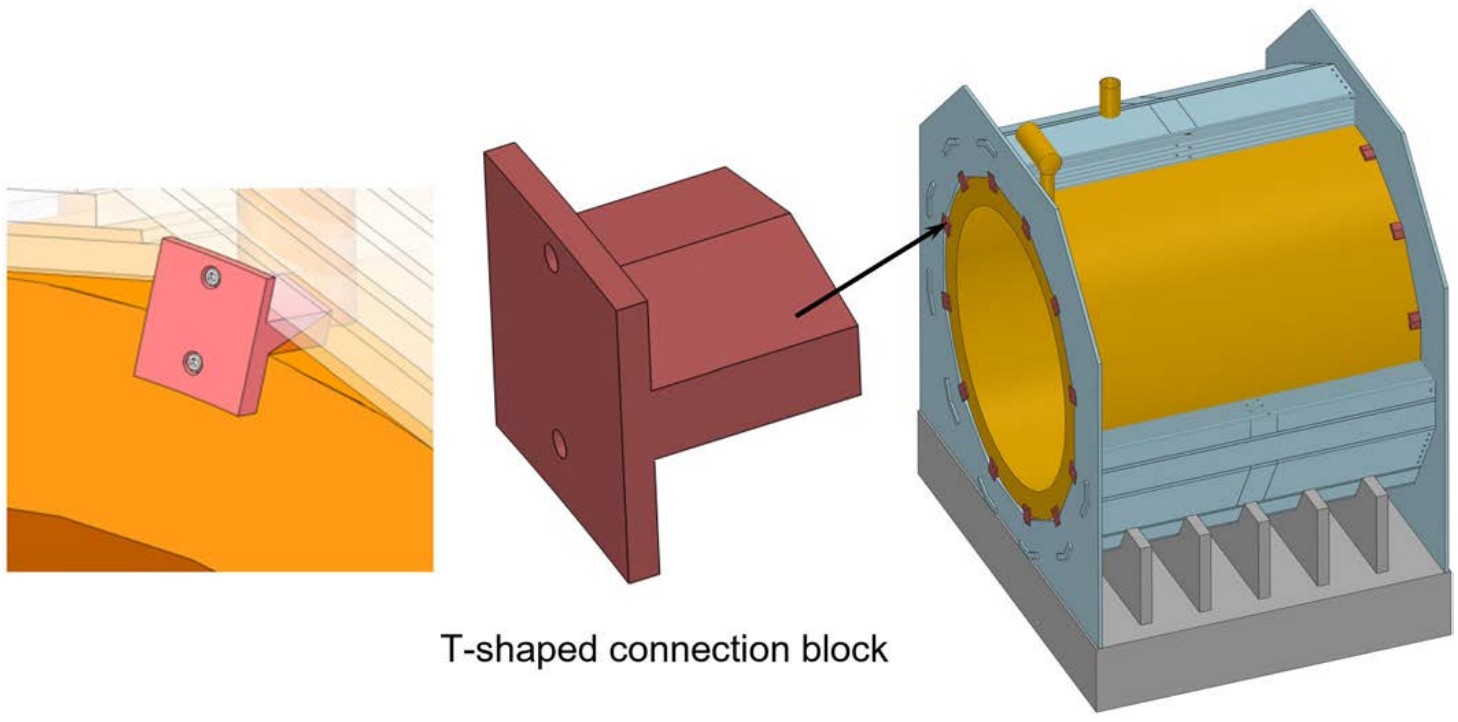

**Figure 14.9:** Connection structure of the superconducting solenoid magnet. The solenoid is connected to the yoke end flange by twelve T-blocks placed in the triangular spaces between yoke polygons and the magnet ring.

The details of the magnet are described in Chapter 10. The connection between the magnet and the barrel yoke was designed to facilitate adjustment and installation, while securely constraining the magnet position within tolerance during vacuum, cryogenic, and electrical loading. Figure 14.9 shows the twelve equally spaced T-blocks, positioned in the triangular spaces between yoke polygons and the magnet ring, to connect the magnet to yoke end flanges by M36 bolts. The design utilizes end flanges to enforce the axial and circumferential positioning and fixation of the magnet. The FEA results of the magnet connection structure in gravity load are shown in Figure 14.10. The maximum deformation is 0.06 mm and the maximum stress is 390 MPa. Table 14.3 lists the deformation and stress of the T-block under different coil offset conditions. T-blocks are made of titanium alloy with an allowable strength of approximately 550 MPa. This ensures compliance with the requirements during both axial and radial offsets. For angular tilt, the allowable limit is 0.14 degrees.

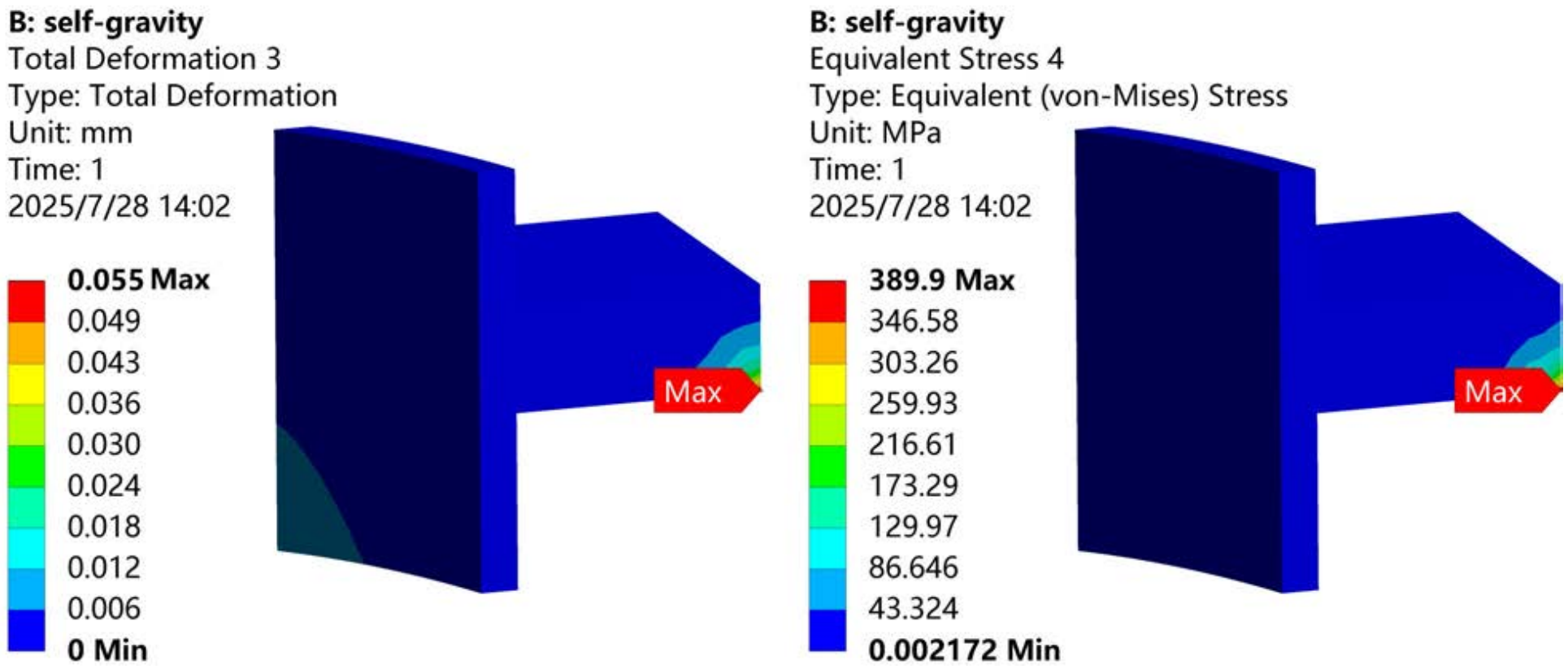

**Figure 14.10:** FEA of magnet connection structure in gravity load. The maximum deformation is 0.06 mm and the maximum stress is 390 MPa.

### 14.2.3 HCAL

1. Barrel HCAL

**Table 14.3:** Deformation and stress of connection structure under different coil offset.

| Parameter | 1 cm | 3 cm | 10 cm |
|---|---|---|---|
| **Axial displacement** | | | |
| Deformation ( mm) | 0.0556 | 0.0559 | 0.0568 |
| Stress (MPa) | 391 | 394 | 402 |
| **Radial displacement** | | | |
| Deformation ( mm) | 0.056 | 0.057 | 0.059 |
| Stress (MPa) | 392 | 398 | 416 |
| **Angular tilt** | | | |
| Angular tilt (degrees) | 0.14 | 0.42 | 1.41 |
| Deformation ( mm) | 0.064 | 0.25 | 0.62 |
| Stress (MPa) | 423 | 1308 | 3201 |

The barrel HCAL weighs 955 tons. It is installed inside the magnet, as shown in Figure 14.11. Given its thin and structurally weak design, the magnet cannot support the barrel HCAL. Consequently it must be protected from potential damage during installation and other unexpected cases. Therefore, the barrel HCAL connection is designed to bypass the magnet and is fixed directly to the barrel yoke using 60 mm thick rings with an outer diameter of 8690 mm and an inner diameter of 6930 mm.

As shown in the Figure 14.11, the barrel HCAL has a 16 sided polygonal structure. It has a total length is 6,460 mm, with external and internal opposite-side dimensions of 6,910 mm and 4,280 mm, respectively. The L-shaped auxiliary hollow cylinders at both ends are made of Q345 steel, each with a length of 1345 mm, an outer diameter of 7,050 mm and an inner diameter of 6930 mm. The cylinders serve as both assembly fixture and the key connection interface, with 10 mm clearance from the magnet.

The 3D model of the connection structure is shown in the Figure 14.12. The barrel HCAL matches the barrel yoke in length. At each end, there is a steel circular ring for connecting the auxiliary hollow cylinder of the HCAL and the end flange of the barrel yoke. For easy installation, the ring is designed as two detachable halves, which is described in HCAL installation Section 14.3.3. The ring features some slots for cable routing. The FEA result of the barrel HCAL connecting structure is shown in Figure 14.13. The maximum deformation and stress are 1.9 mm and 201 MPa, respectively. The caculated stress is below the allowable limit of 230 Mpa for the 18Mn18Cr material, which is derived using a safety factor of 1.5.

2. Endcap HCAL

   The endcap HCAL is located at the ends of the barrel HCAL, as shown in Figure 14.14. Each endcap HCAL weighs about 350 tons and is installed in the auxiliary hollow cylinder of the barrel HCAL. The endcap HCAL is constructed as a 16-sided polygon, with a thickness of 1315 mm, a width of 6,770 mm across opposite sides, and a central hole diameter of 800 mm. To accommodate the piping and cabling of internal sub-detectors, a clearance ranging from 14 mm to 80 mm is provided between the endcap HCAL and the auxiliary ring, as illustrated in Figure 14.14.

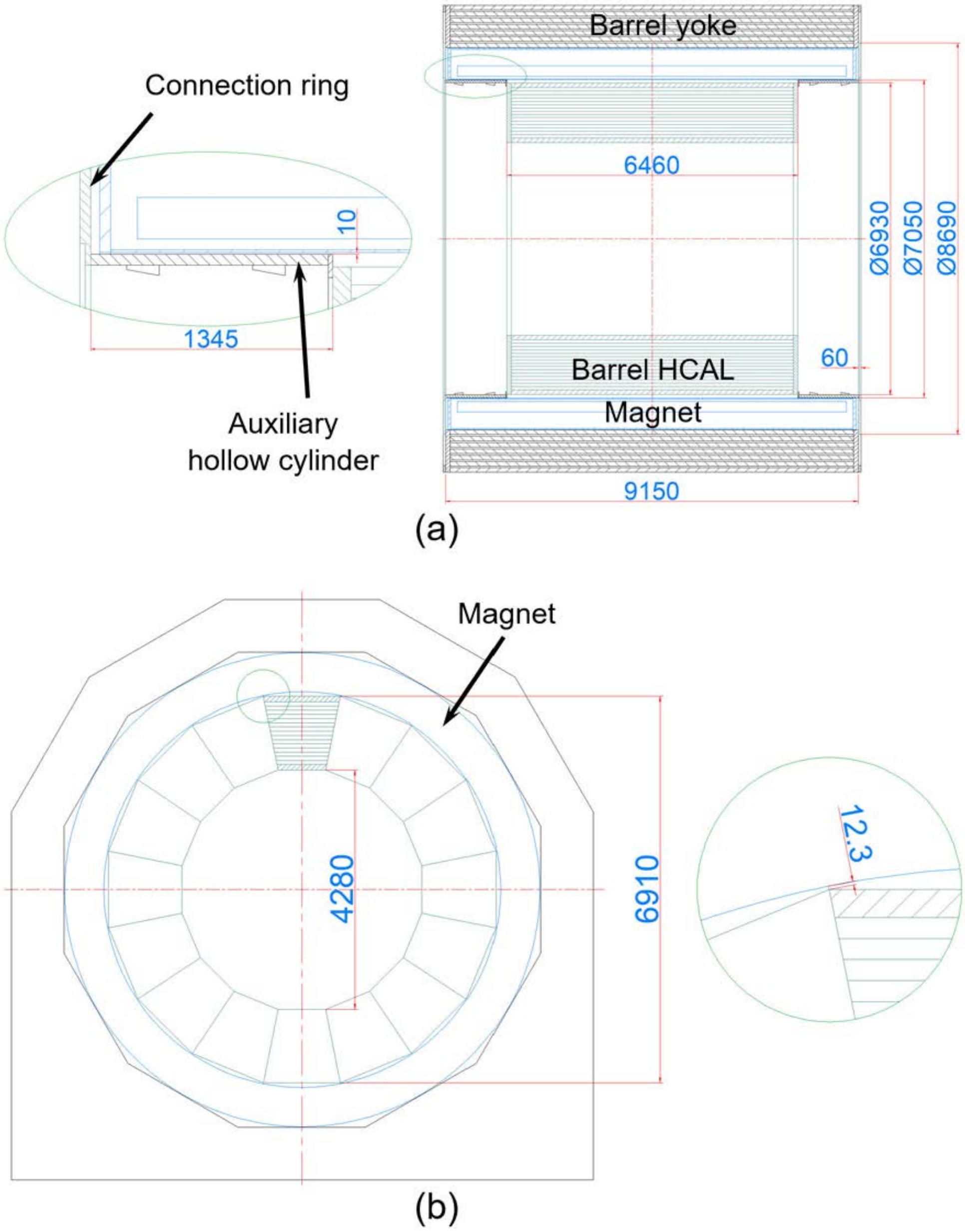

**Figure 14.11:** Barrel HCAL inside the magnet. Units are in millimeters. (a) The L-shaped auxiliary hollow cylinders at both ends of the barrel HCAL are directly connected to the barrel yoke through 60-mm-thick connecting rings, with 10 mm clearance from the magnet. (b) The barrel HCAL inside the magnet as a 16-sided polygonal structure.

A reserved circular hole in the endcap HCAL is designed for the installation and support of the accelerator superconducting magnet, which requires nanoscale vibration stability and a stable and reliable connection system. To achieve this, 16 pairs of adjustable steel pads are designed evenly distributed along the circumference of the endcap HCAL for concentric adjustment and fixation. Two pairs of adjustable pads are arranged axially to ensure the reliability of the connection between the endcap HCAL and the steel auxiliary hollow cylinder. At the rear end of the endcap HCAL, a twelve-tooth flange is designed for connection to the end flange of the barrel yoke. The endcap HCAL, when fixed in the circumferential, axial and rear directions, forms a stable and rigid connection with the barrel yoke. This connection provides stable support for the accelerator superconducting magnet. Figure 14.16 shows the FEA result of the endcap HCAL connection structure. The maximum deformation is 2.8 mm, and the maximum stress is 94 MPa, and thus below the allowable limit.

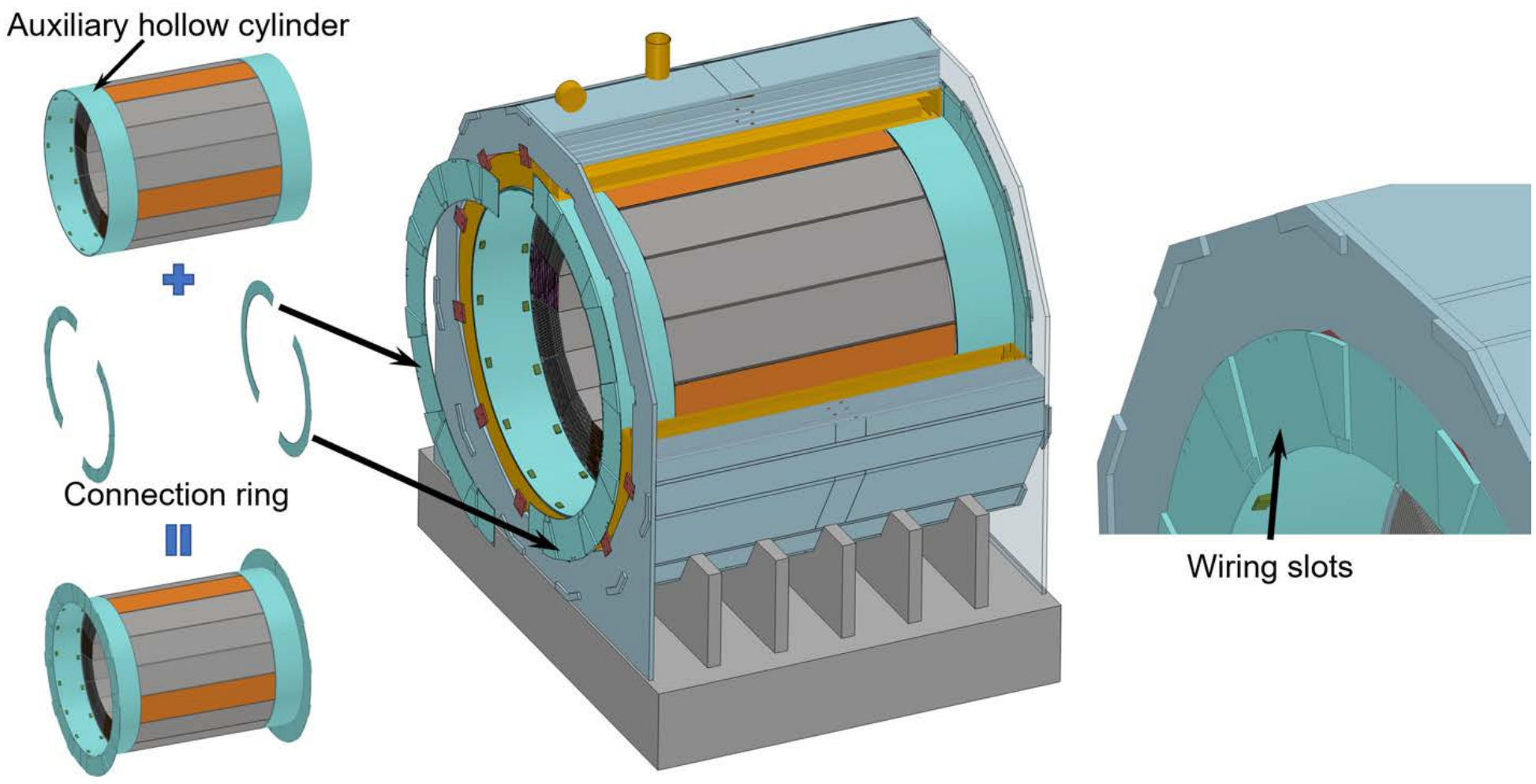

**Figure 14.12:** Connection structure of the barrel HCAL. There is a steel circular ring as two detachable halves at each end for connecting the auxiliary hollow cylinder of the HCAL's and the end flange of the barrel yoke's. The ring features some slots for cable routing.

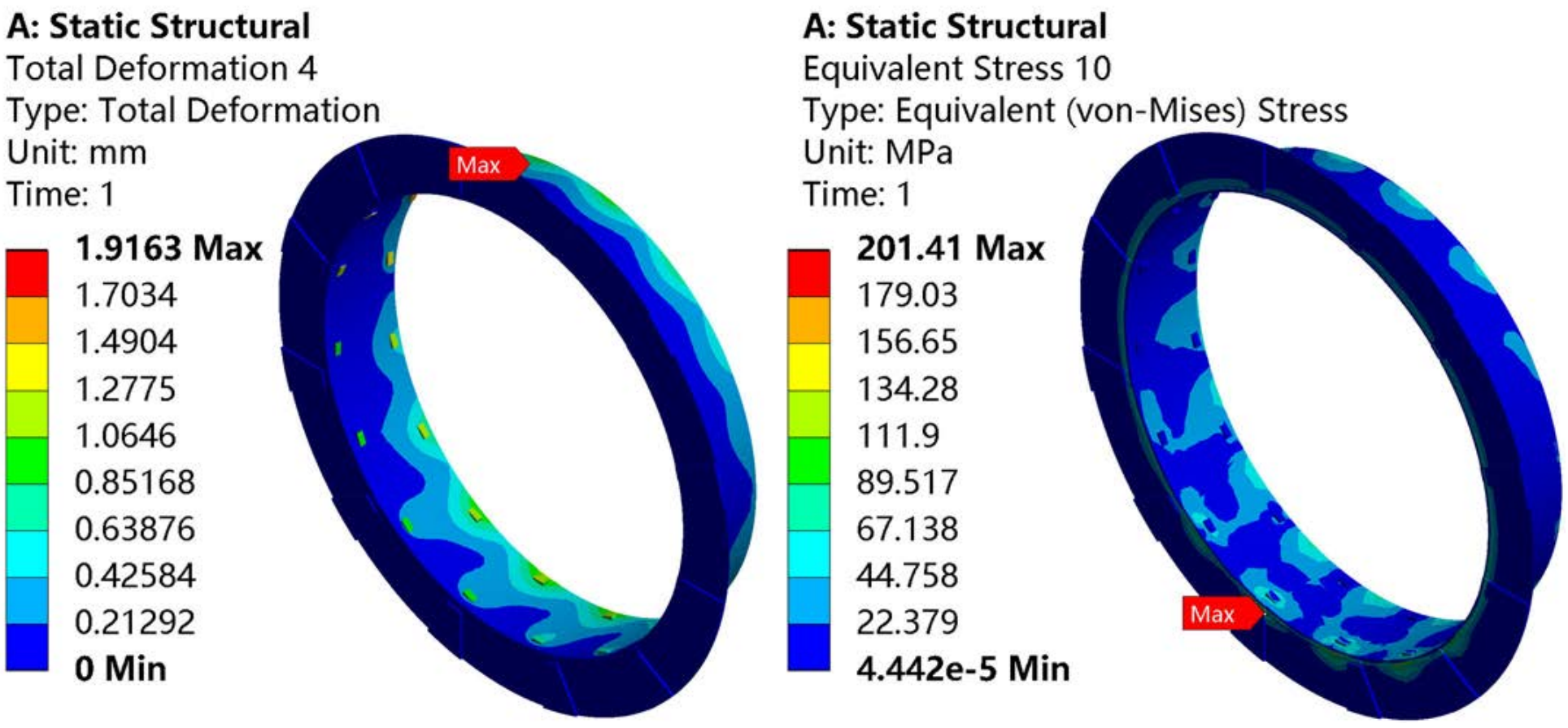

**Figure 14.13:** FEA result of the barrel HCAL connection structure. The maximum deformation is 1.9 mm, and the maximum stress is 201 MPa

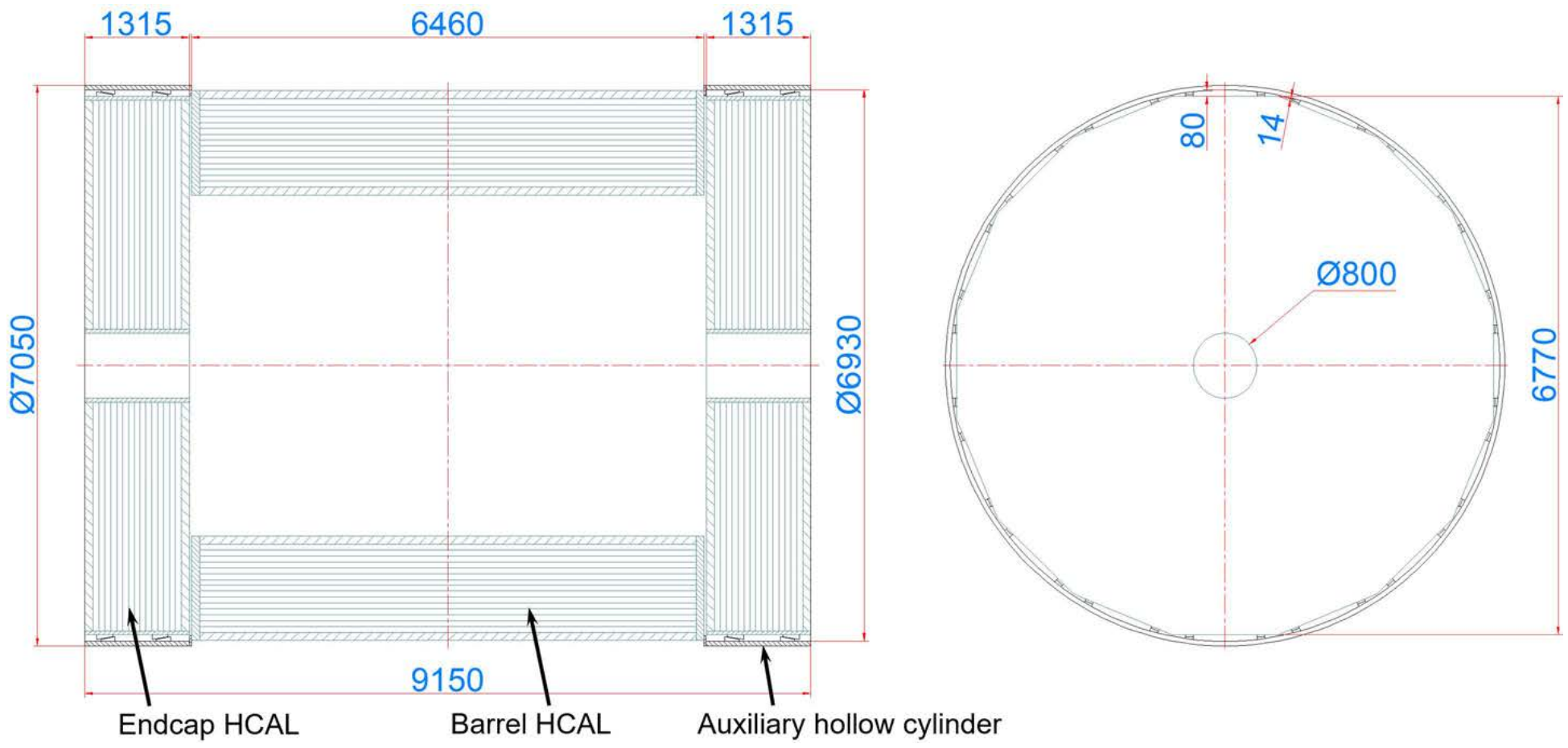

**Figure 14.14:** Endcap HCAL and barrel HCAL. Endcap HCAL is installed in the auxiliary hollow cylinder of the barrel HCAL. The clearance between them ranges from 14 mm to 80 mm for routing space. Units are in millimeters.

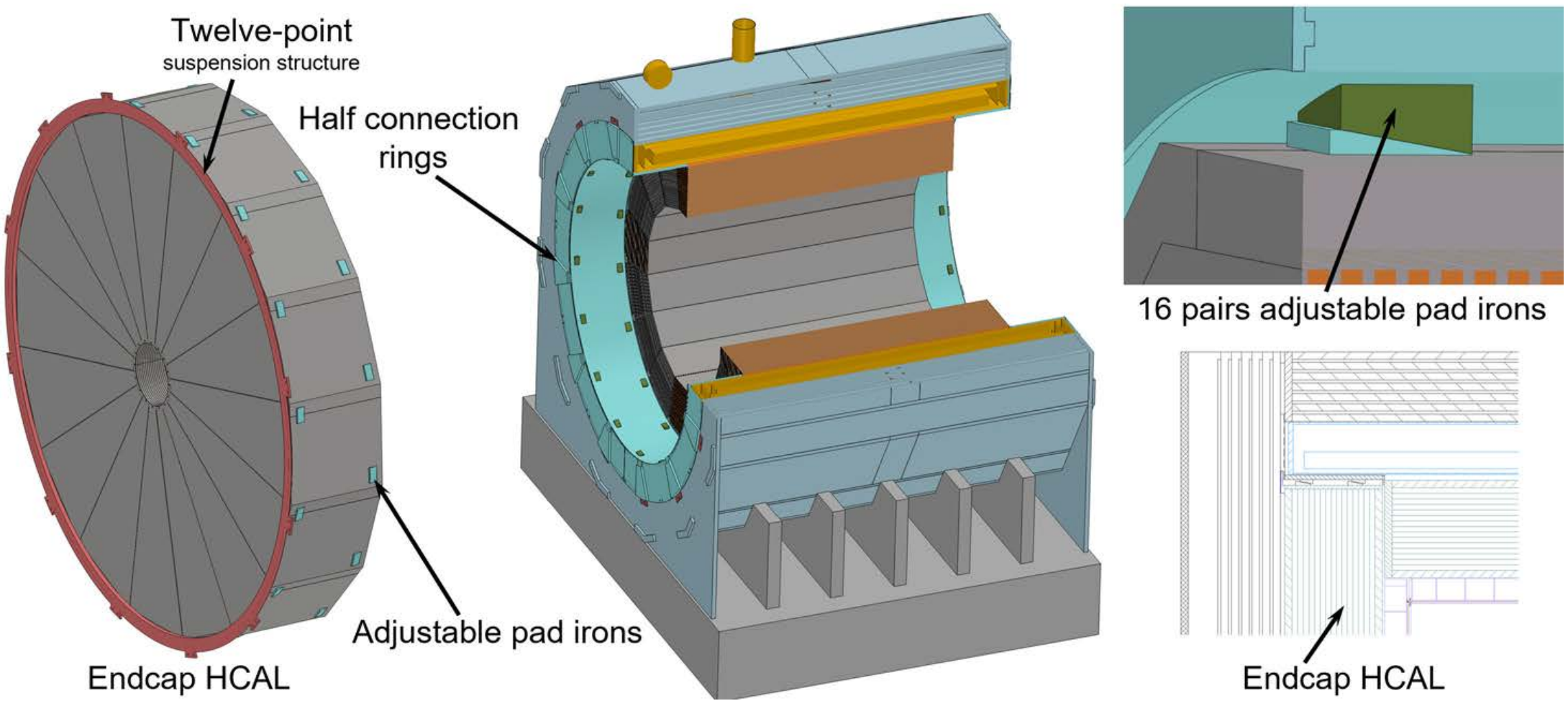

**Figure 14.15:** Connection structure of the endcap HCAL. The endcap HCAL is rigidly connected to the barrel yoke through 16 pairs of adjustable pad irons, two pairs of axial pads, and a twelve-tooth flange, ensuring stable support for the superconducting magnet.

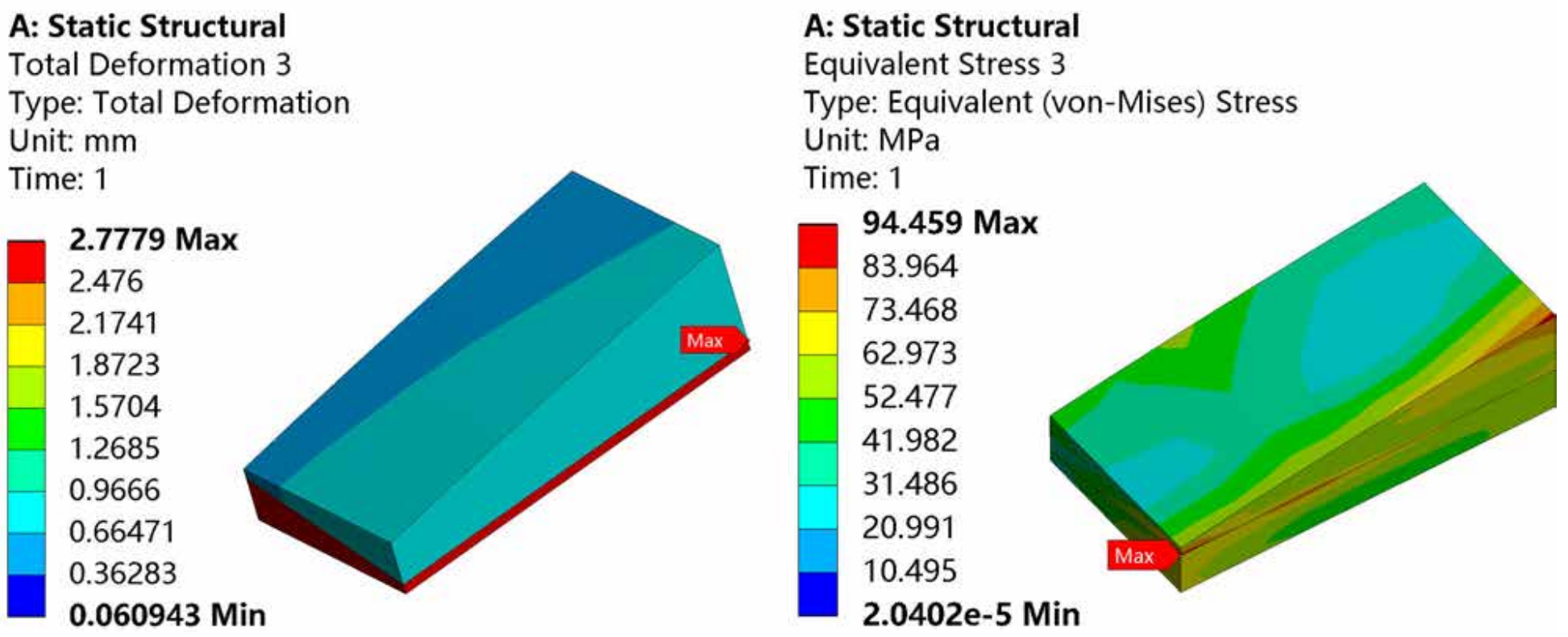

**Figure 14.16:** The FEA result of the endcap HCAL connection structure.

### 14.2.4 ECAL

1. Barrel ECAL

The barrel ECAL, described in Section 7.2.5, weighs 130 tons. As shown in Figure 14.17, the barrel ECAL is installed inside the HCAL, with a minimum clearance of 10 mm. The barrel ECAL is a regular 32-sided polygon structure, with a total height of 4,260 mm, a bore dimension of 3,660 mm, and a total length of 5,800 mm.

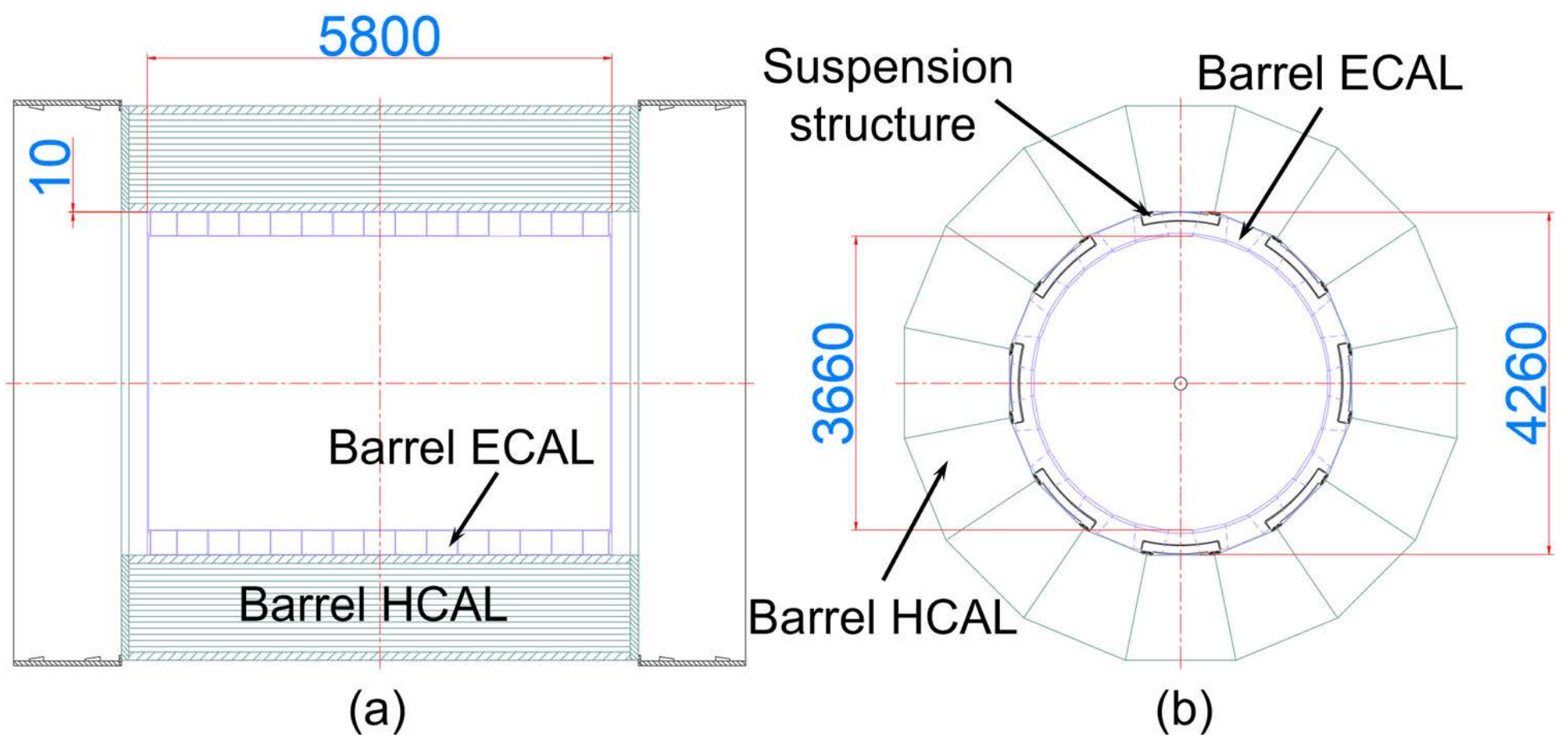

**Figure 14.17:** Barrel ECAL inside the barrel HCAL with a minimum clearance of 10 mm. (a) Front view. (b) Side view. Units are in millimeters.

As shown in Figure 14.18, the barrel ECAL features eight suspension structures, each connecting to two Z-shaped connection blocks on the barrel HCAL to distribute weight. This design reduces the deformation caused by the stress concentration and improves the reliability of the connection. The shape of each suspension structure matches the slots in the flanges at both ends of the barrel ECAL, to minimize the gap between the sub-detectors. The FEA result of the barrel ECAL connection structure is shown in Figure 14.19. The maximum deformation is 3 mm and the maximum stress is 101 MPa, below the allowable limit.

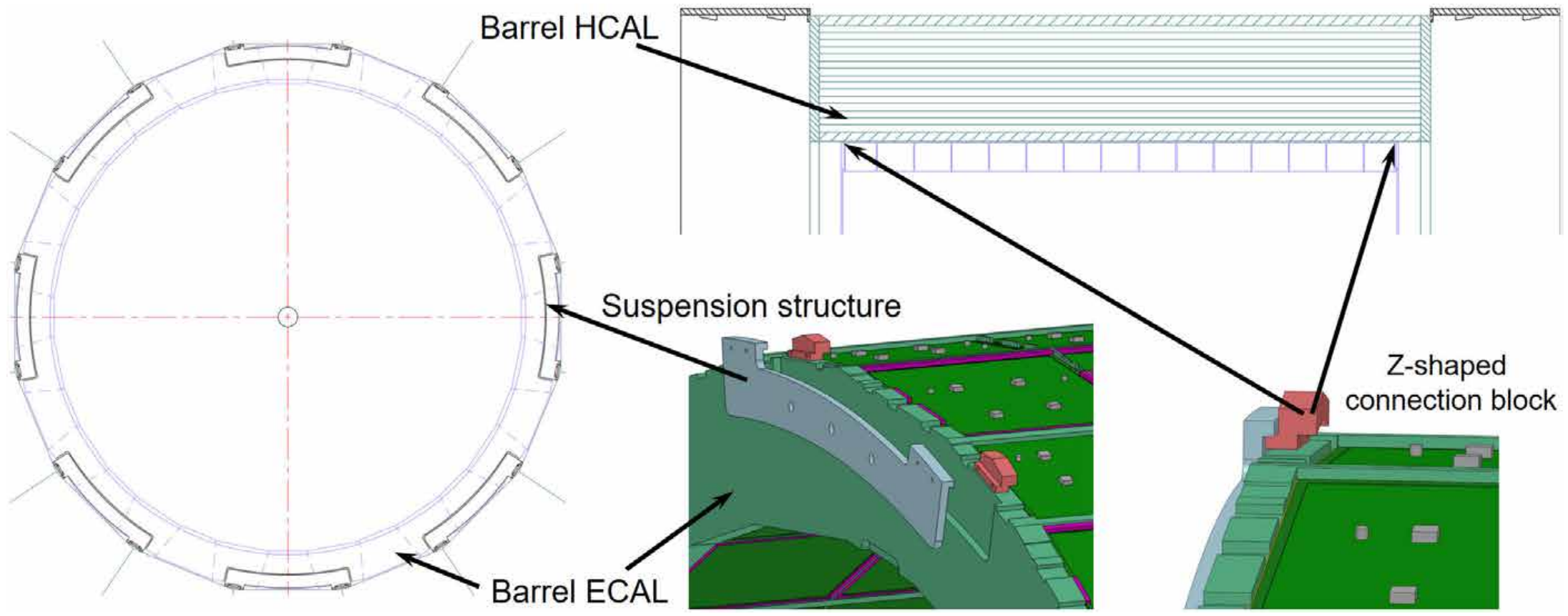

**Figure 14.18:** Connection structure of the barrel ECAL. Eight-point circular suspension structure are adopted to connect the barrel ECAL and the barrel HCAL.

2. Endcap ECAL (endcap OTK)

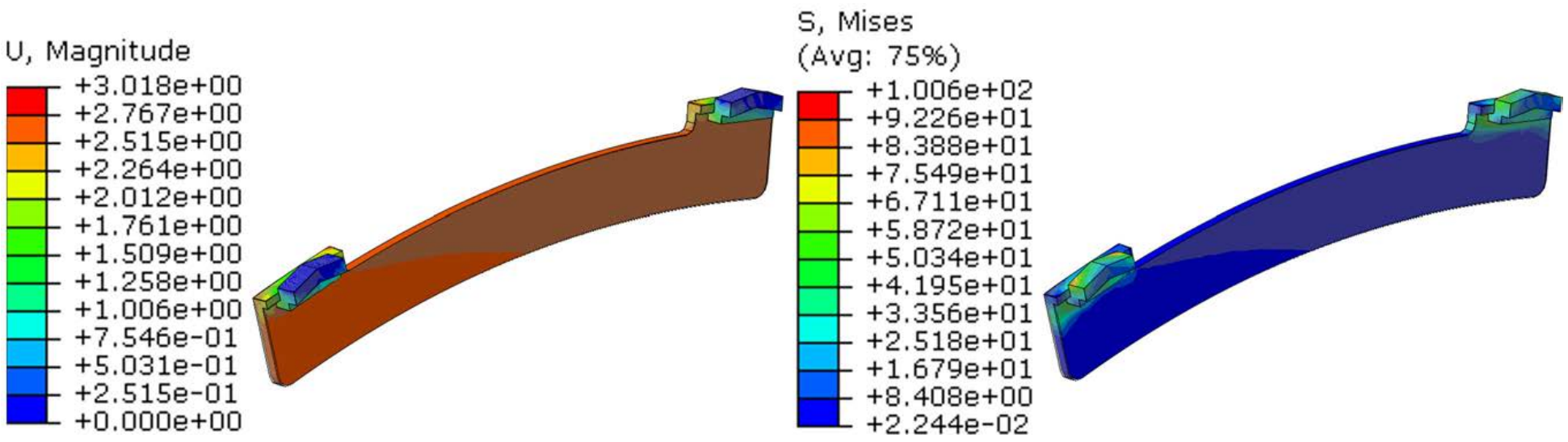

**Figure 14.19:** Connection structure FEA results of the barrel ECAL. The maximum deformation is 3 mm and the maximum stress is 100 Mpa.

The endcap ECAL is located at both ends of the barrel ECAL as shown in Figure 14.20. Each ECAL weighs approximately 25 tons and is mounted inside the barrel HCAL. The endcap ECAL features a 16-sided polygonal geometry, a thickness of 300 mm, a width of 4220 mm across opposite sides, and a central square hole measuring 700 mm × 700 mm. The minimum clearance between the endcap ECAL and the barrel HCAL is 30 mm, with redundant space for the routing of pipes and cables.

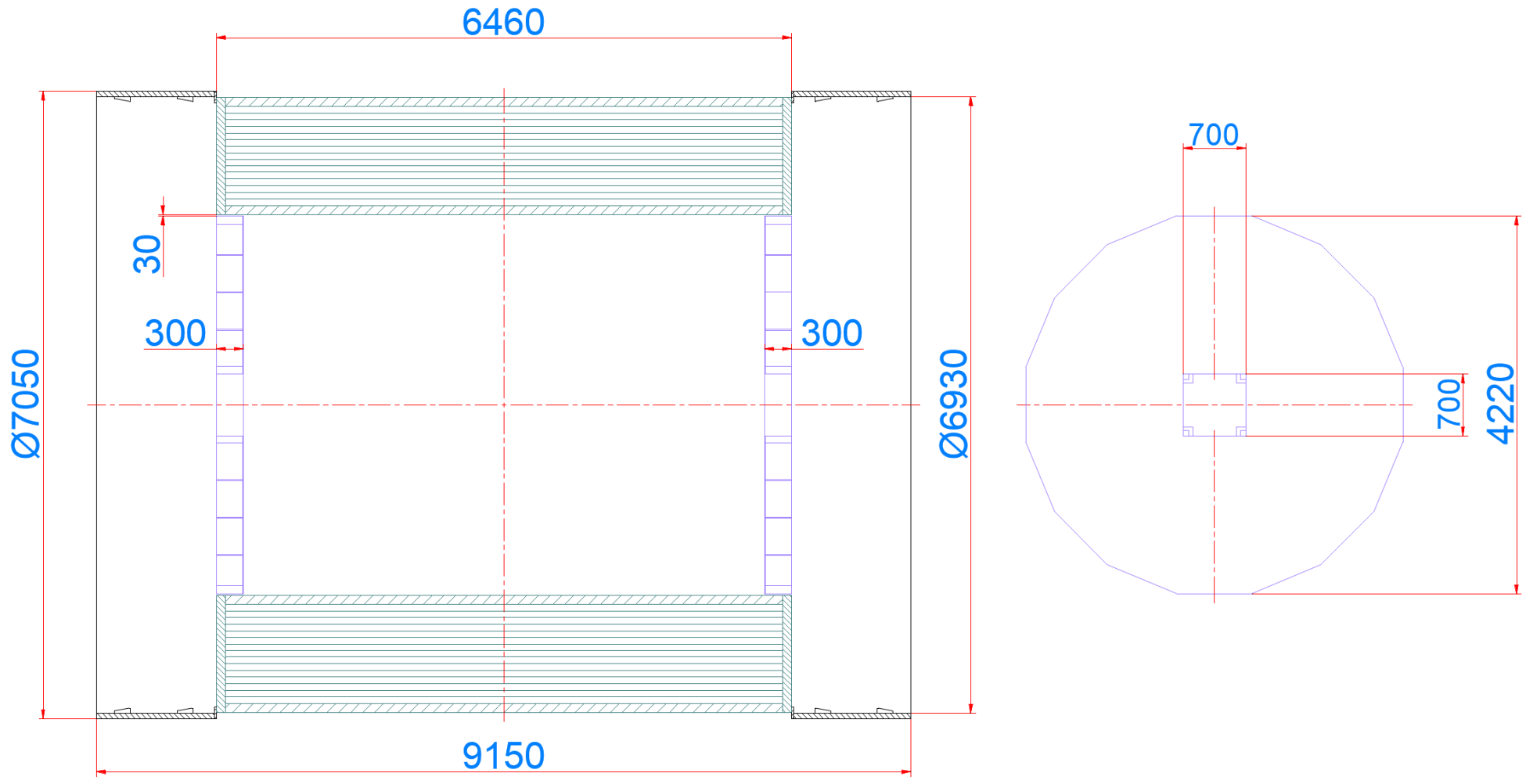

**Figure 14.20:** Endcap ECAL is located at both ends of the barrel ECAL, with the minimum clearance of 30 mm for the routing of cables and pipes. Units are in millimeters.

The endcap ECAL is connected to the barrel HCAL and positioned by the four pairs of adjustable steel pads uniformly distributed around the circumference. Due to its 25-ton weight, the adjustment operation must rely on a hydraulic system or external tooling assistance to support the endcap ECAL. The FEA results of the connection structure are shown in Figure 14.21. The maximum deformation is 0.01 mm and the maximum stress is 63 MPa, below the allowable limit.

The endcap OTK is a thin-walled circular structure, as shown in Figure 14.22, and is connected to the endcap ECAL by 16 L-shaped connection blocks to form a unified assembly. The endcap OTK can be installed and maintained concurrently with the endcap ECAL.

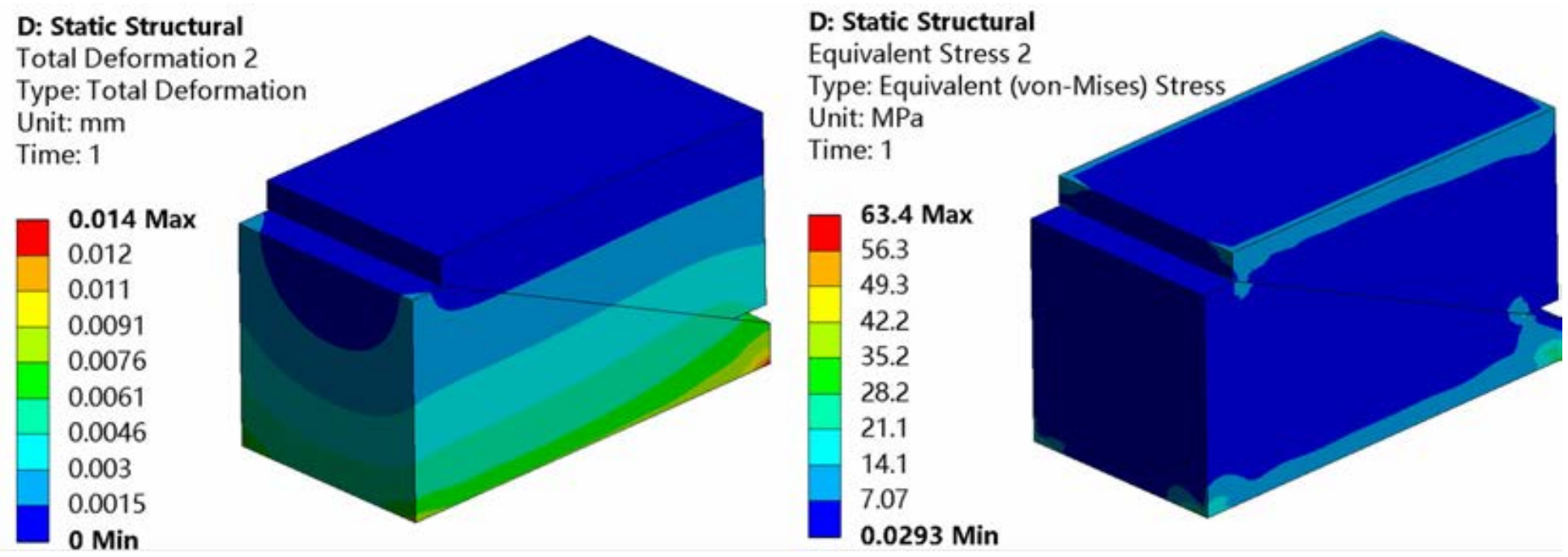

**Figure 14.21:** FEA results of the endcap ECAL connection structure.

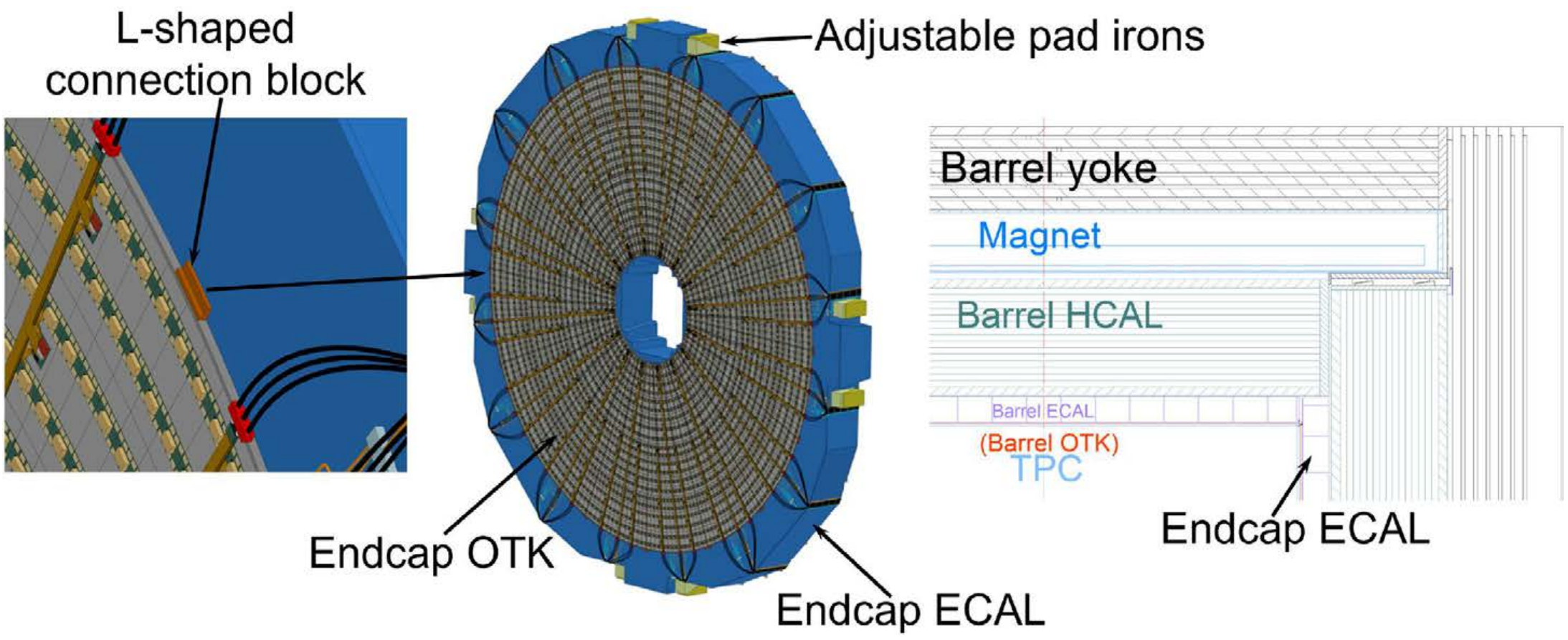

**Figure 14.22:** Connection structure between the endcap ECAL and the endcap OTK. They are connected and positioned by four pairs of adjustable pad irons uniformly distributed around the circumference.

### 14.2.5 TPC (barrel OTK)

The TPC weighs about 1.5 tons. As shown in Figure 14.23, two aluminium alloy end flanges densely packed with electronics are the main structural components. Two carbon fibre cylinders connect the flanges: on with an outer diameter of 3,600 mm and a length of 5800 mm, and the other with an outer diameter of 1,200 mm and a length of 5,800 mm. The TPC is mounted inside the barrel ECAL, with a minimum clearance of 10 mm.

As shown in Figure 14.24, The TPC is connected to the barrel ECAL by an aluminium alloy circular ring through eight-point radial suspension structures, with slots for cabling and piping. The FEA result of this connection ring is shown in Figure 14.25. Both stress and deformation are significantly below the allowable limits.

The detailed design of the TPC is described in Chapter 6. Since the TPC features a lightweight design, the two thin-walled carbon fibre cylinders are insufficient to support the end flanges (each weighing approximately one ton) and are unstable during installation. Therefore, during the TPC assembly, a support fixture with sufficient strength and stiffness to support both end flanges and ensure the coaxiality is required. This fixture must remain engaged throughout the installation and cannot be removed until the TPC is positioned and connected to the barrel

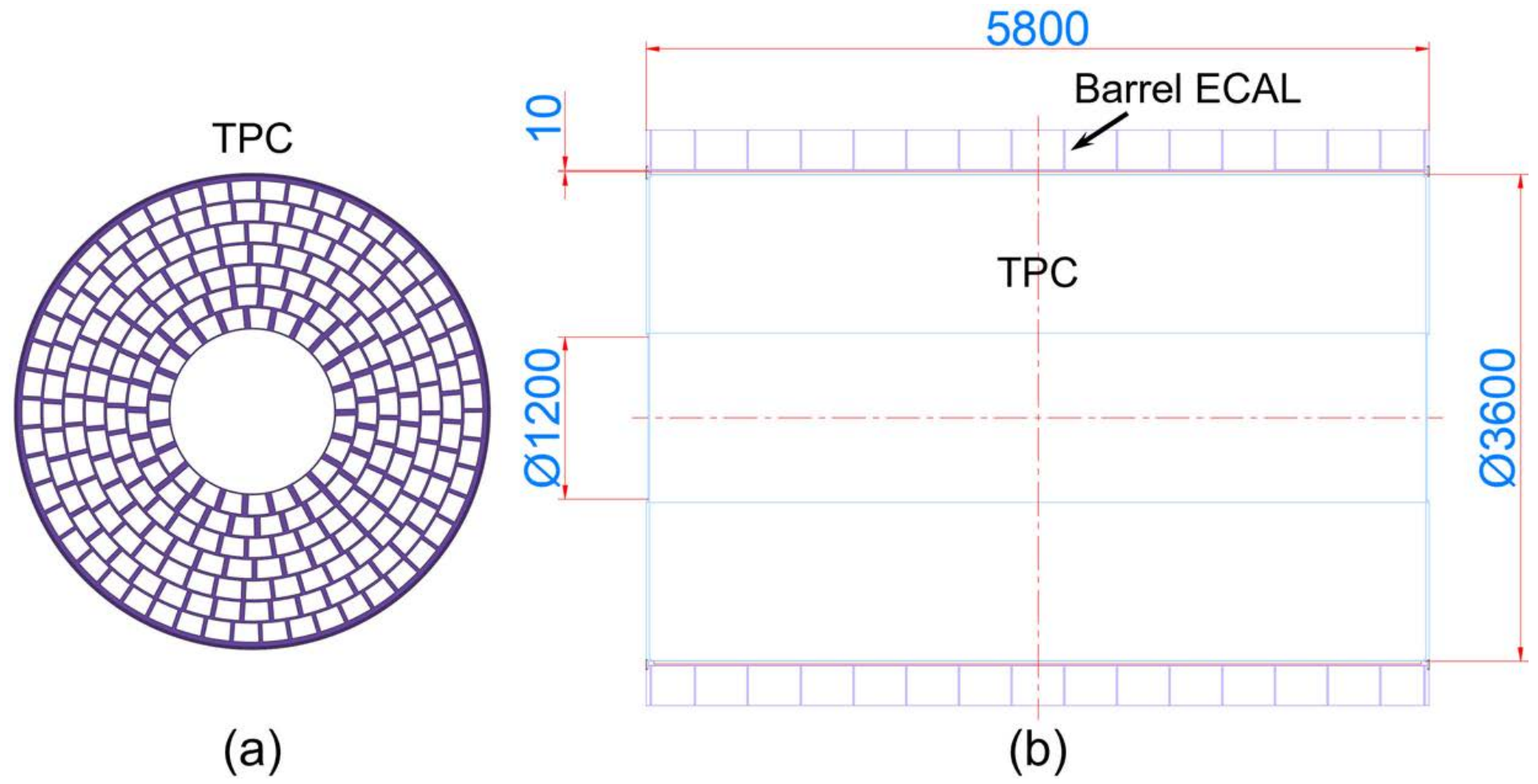

**Figure 14.23:** (a) The TPC sideview showing the aluminium alloy end flange. (b) The TPC mounted inside the barrel ECAL with a minimum clearance of 10 mm. Units are in millimeters.

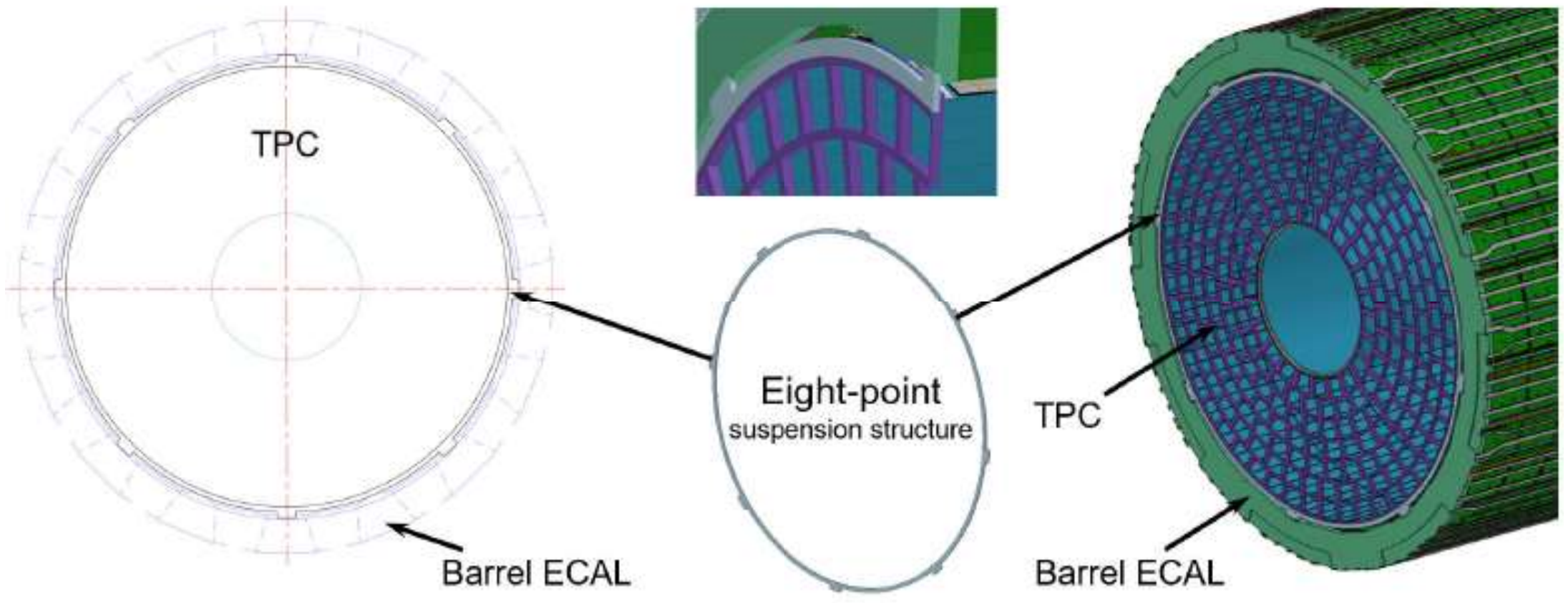

**Figure 14.24:** Connection structure of the TPC. The TPC is connected to the barrel ECAL by an eight-point suspension structure with the slots for cabling and piping.

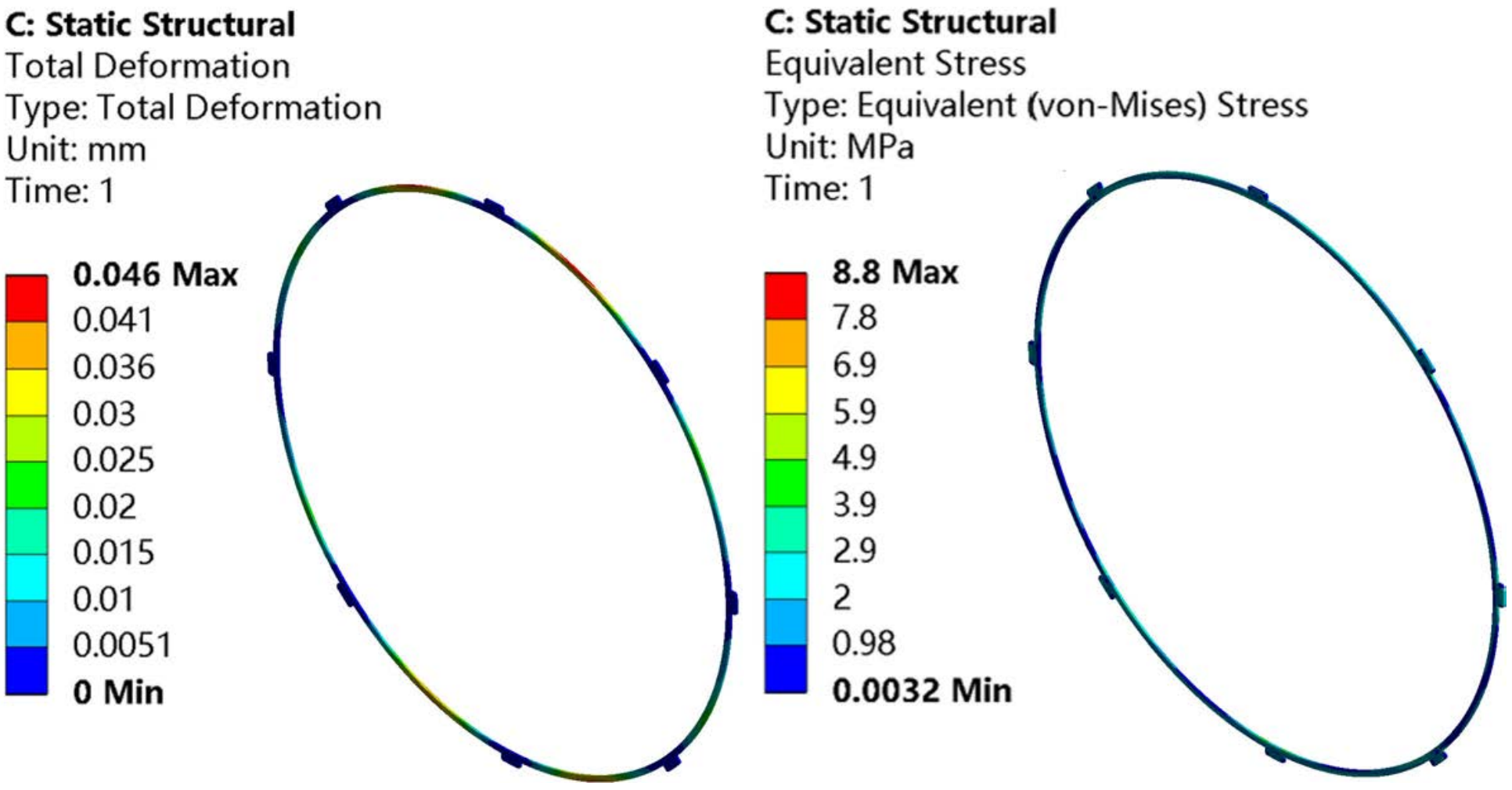

**Figure 14.25:** FEA of the TPC connection structure.

ECAL.

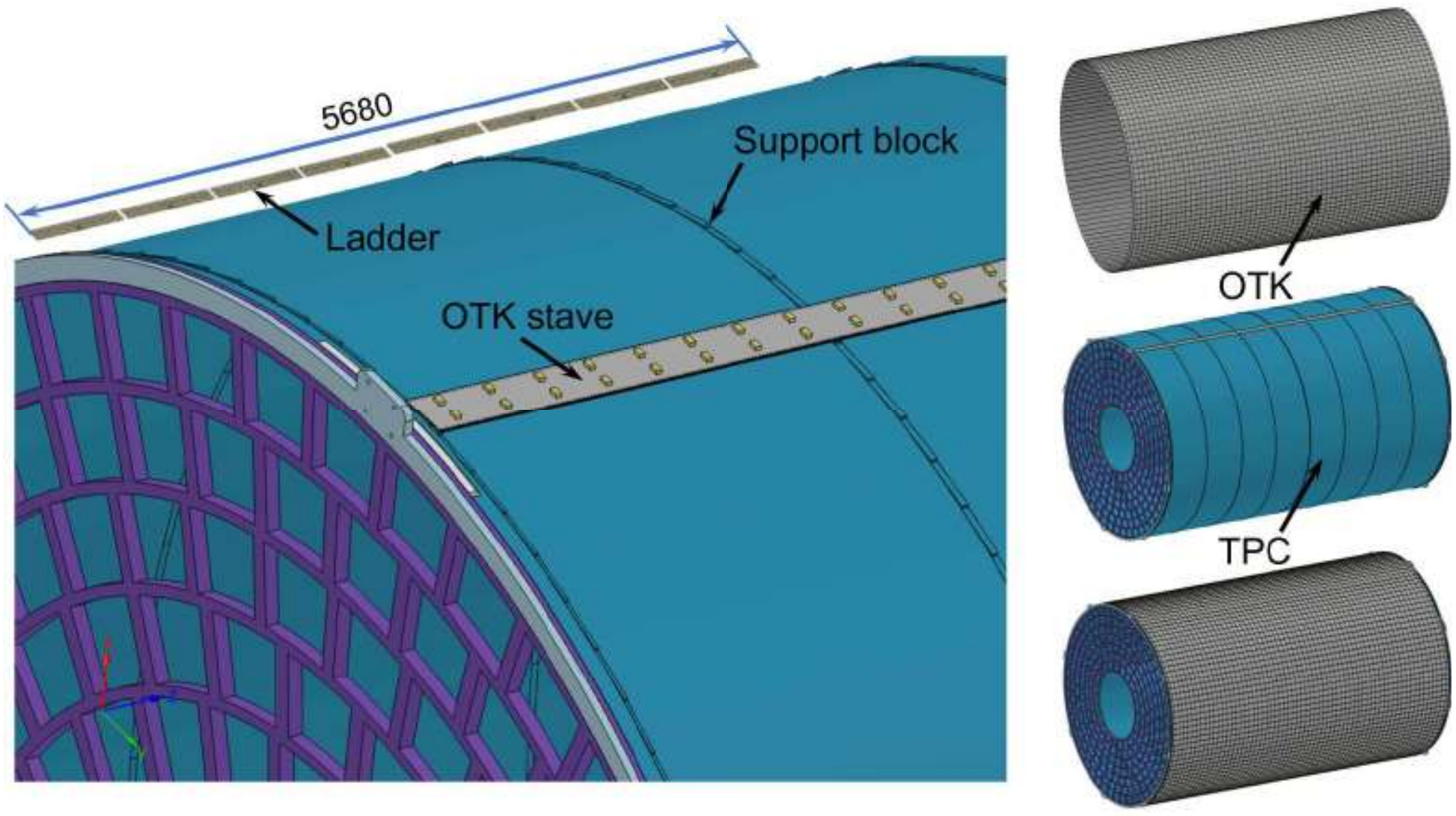

**Figure 14.26:** Connection structure of the barrel OTK. The barrel OTK contains 110 staves,each 5680 mm in length and divided into eight ladders, which are fixed to the TPC's outer cylinder through support blocks. The barrel OTK is assembled with TPC for integrated installation. Units are in millimeters.

The barrel OTK weighs approximately 300 kg and is mounted on the exterior of the TPC carbon fiber outer cylinder. Due to its large diameter, single-layer and thin-walled cylinder structure, the barrel OTK cannot be installed independently and must be integrated with the TPC during installation. As shown in Figure 14.26, the barrel OTK contains 110 staves with a total length of 5680 mm, evenly spaced on the TPC outer cylinder. Each stave is divided axially into eight long ladders. Both ends of each ladder are fixed to the corresponding inclined support blocks pre-mounted on the TPC, which integrates the OTK with the TPC and constrains the relative position between them within tolerance. however, the final positioning accuracy depends on the geometric precision of the TPC cylinder and the machining accuracy of the pre-mounted support blocks.

### 14.2.6 ITK

The ITK weighs 80 kg, as shown in Figure 14.27, with a total length of 5,770 mm and an outer diameter of 1,180 mm, maintaining a 10 mm clearance with the TPC inner cylinder. The ITK consists of three barrel layers and four endcap layers, featuring a lightweight structural design with thin, precisely manufactured components. The assembly process includes multiple stages with precision requirements, maintaining critical dimensional tolerances within 100 μm. The connection structures utilize carbon fiber characterized by low density and high strength to minimize the material mass.

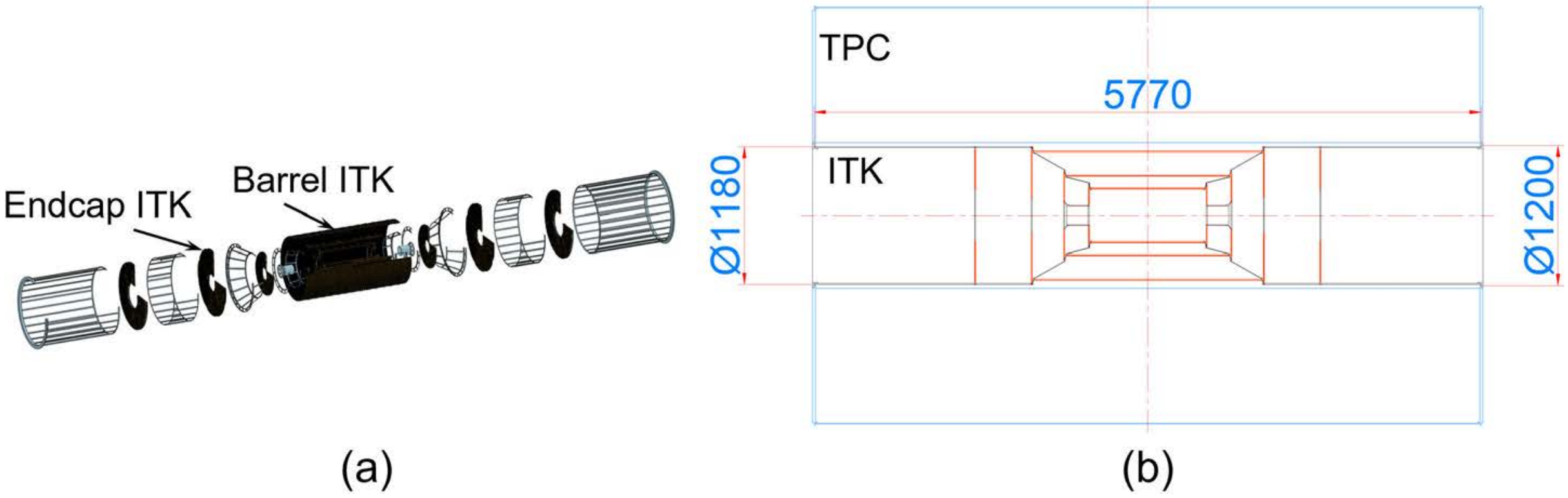

**Figure 14.27:** (a) Exploded view for showing the ITK components and their assembly relationship. (b) The ITK inside the TPC with a 10 mm clearance, a total length of 5770 mm and an outer diameter of 1180 mm. Units are in millimeters.

As shown in Figure 14.28, the ITK is installed inside the TPC. The L-shaped connection rings at both ends are axially positioned with the inner ring of the TPC flange by the mounting shoulder, while maintaining radial adjustment clearance. The assembly is secured with screws.

### 14.2.7 Beam pipe assembly (VTX and LumiCal)

The beam pipe assembly weighs about 10 kg and is installed inside the ITK. As shown in Figure 14.29, the beam pipe assembly has a total length of 1,400 mm and an outer diameter of 153 mm, while maintaining a 2 mm clearance with the ITK's connection hollow cylinder.

The detector beam pipe is a vacuum assembly comprising a beryllium pipe and an aluminum extension pipe. Both feature a double-walled structure with circulating purified water for cooling. Both ends are equipped with compound vacuum flanges, connecting with accelerator vacuum flanges for connection. The VTX and the LumiCal are installed surrounding the detector beam pipe. The five layers of VTX are arranged near the central beryllium pipe, while two LumiCals are placed near the two end flanges. Two electrodes Beam Position Monitors (BPMs) are placed inside end compound flange to monitor the beam position. The outermost VTX is a carbon fiber protective cylinder, which not only secures the assembly during installation, commissioning, and transportation, but also provides an outer cylinder to enclose the duct for air cooling of the VTX.

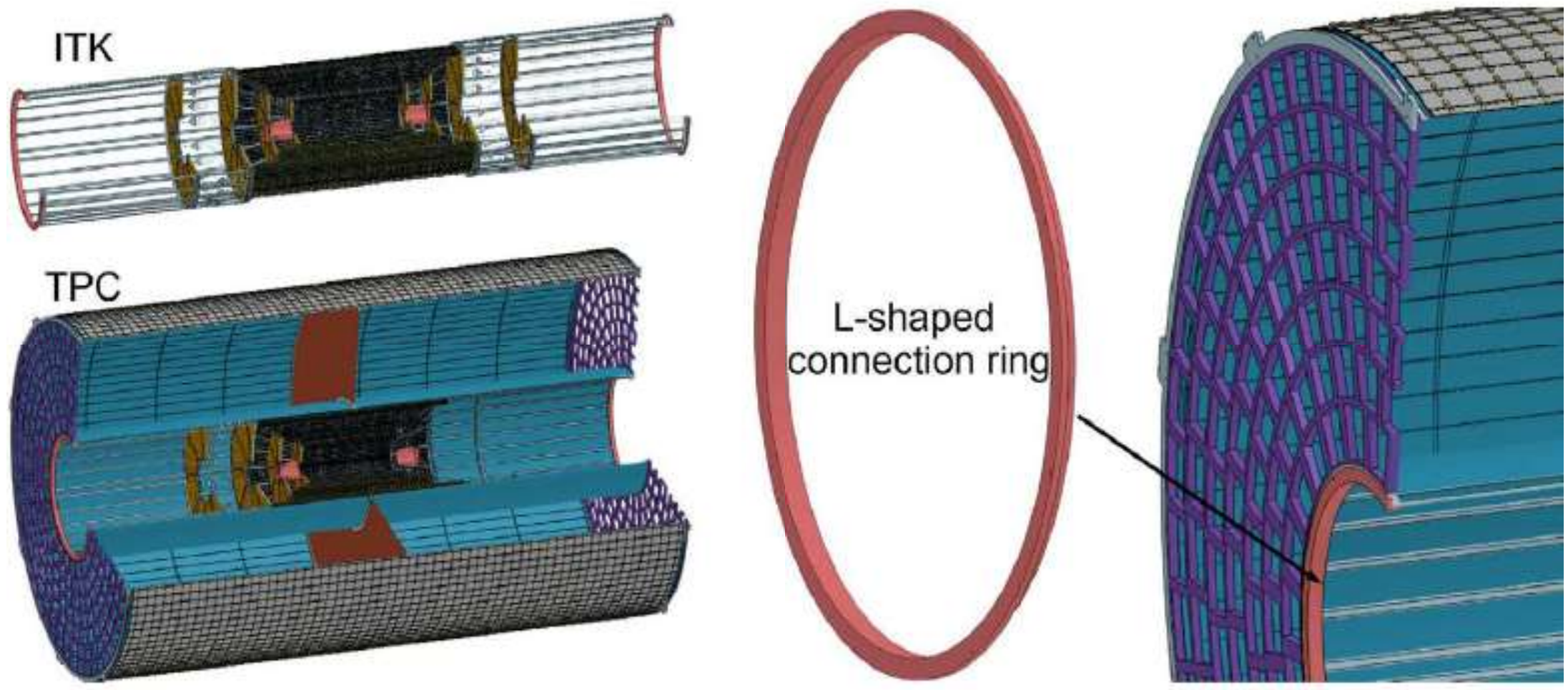

**Figure 14.28:** Connection structure of the ITK. The L-shaped connection rings at both ends connect the ITK to the TPC.

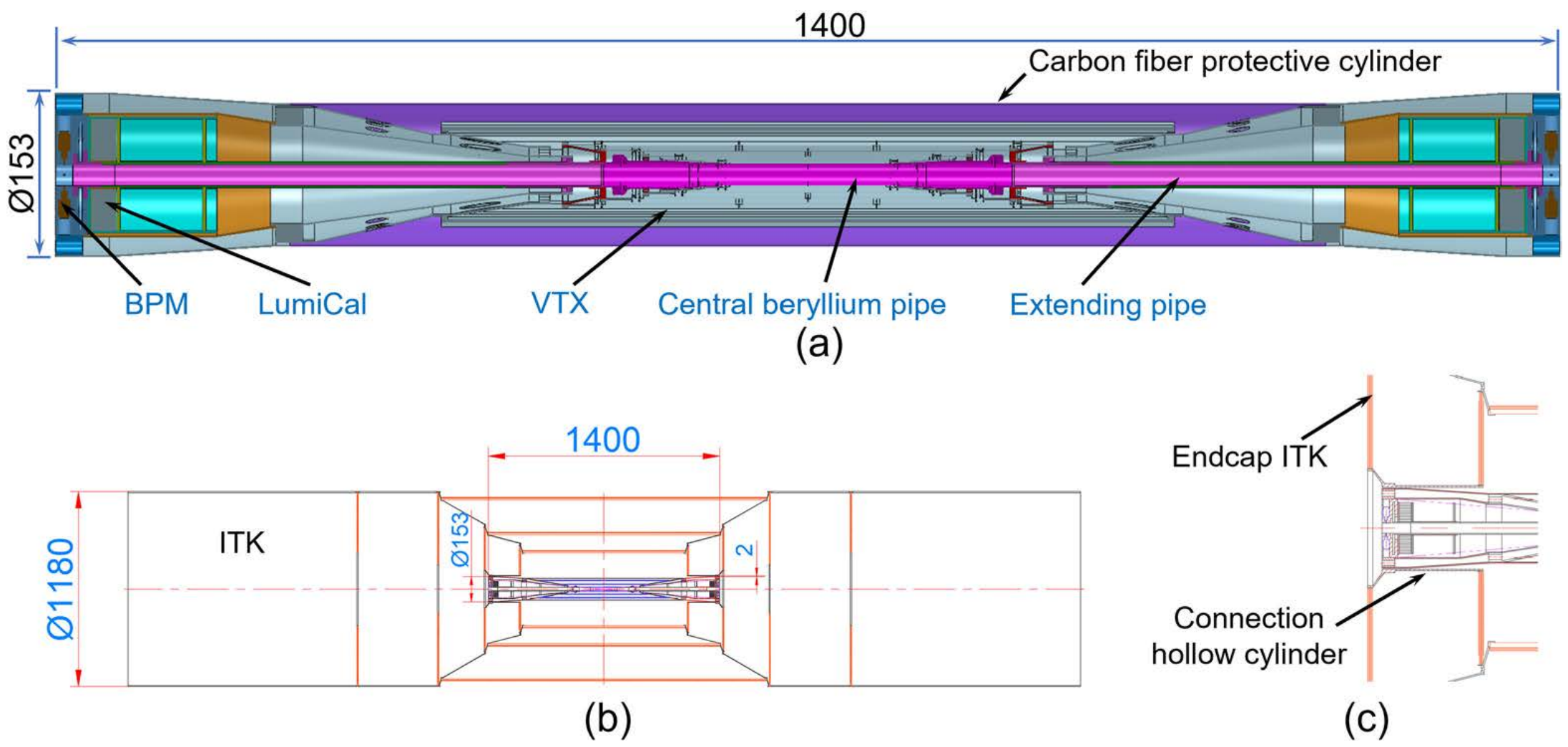

**Figure 14.29:** Beam pipe assembly. (a) The 3D model shows the components of beam pipe assembly. (b) The beam pipe assembly is located inside the ITK with 2 mm clearance, iwith a length of 1400 mm and an outer diameter of 153 mm. (c) connection between the detector beam pipe and the ITK. Units are in millimeters.

The beam pipe assembly is connected and aligned to the ITK using a four-point radial adjustment system, as shown in Figure 14.30. Both the beam pipe assembly flanges and the ITK connection hollow cylinders feature four countersunk grooves and four threaded holes at orthogonal positions. The alignment procedure is as follows: First, screws are installed in the ITK's threaded holes. Then, these four screws are used to adjust the beam pipe assembly until its axis aligns with the beam center within tolerance. Finally, the screws are locked to secure the alignment.

The detector beam pipe and the accelerator vacuum pipe are connected via a pillow seal. This specialized vaccuum coupling, shown in Figure 14.30, provides an automatic compensation mechanism. The end of the detector beam pipe has a stainless steel flat flange with high surface smoothness, while the end of the accelerator vacuum pipe has a stainless steel membrane with a spring structure, also requiring high surface smoothness. During the connection, the pillow seal is pressurized via an inflatable tube, causing its stainless steel membrane to expand until it contacts the detector's flat flange, creating a vacuum seal. This method enables remote automatic sealing and is currently the simplest design. However, further research is needed to verify its ultra-high vacuum.

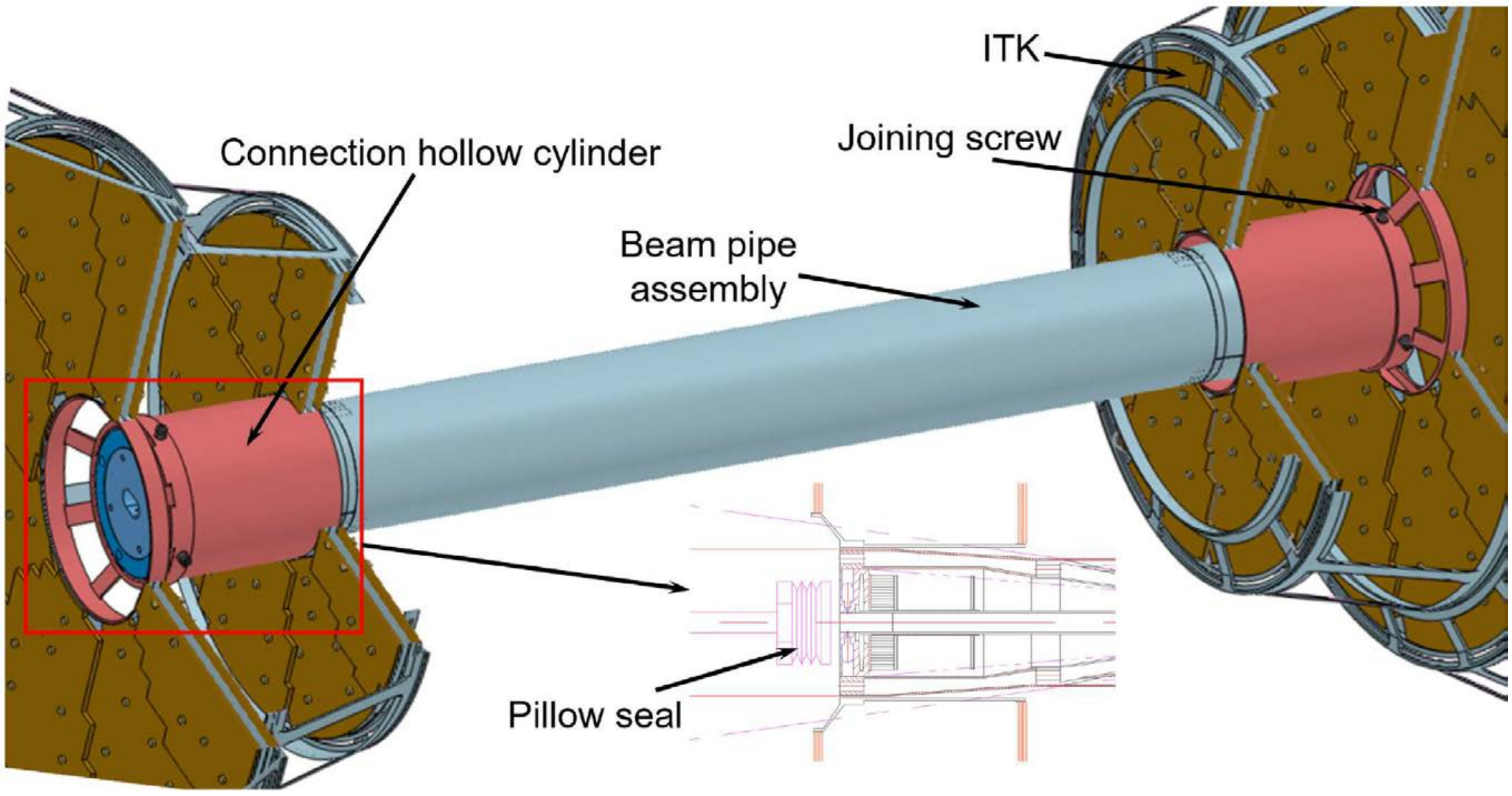

**Figure 14.30:** The detector beam pipe and the accelerator vacuum pipe use an automatic compensation connection of a pillow seal.

### 14.2.8 Reliability and safety assessment

The overall reliability and safety assessment of the yoke assembly primarily evaluates whether the deformation and stress values of the barrel yoke remain within safety limits. The yield strength of the 10# carbon steel is 205 MPa, which, after applying a safety factor of 1.5, results in an allowable stress of 136.7 MPa [2].

After the whole detector is installed, the barrel yoke will undergo these two loading conditions: without and with electromagnetic force. Figure 14.31 gives the FEA results of the deformation and stress of the barrel yoke under two conditions.

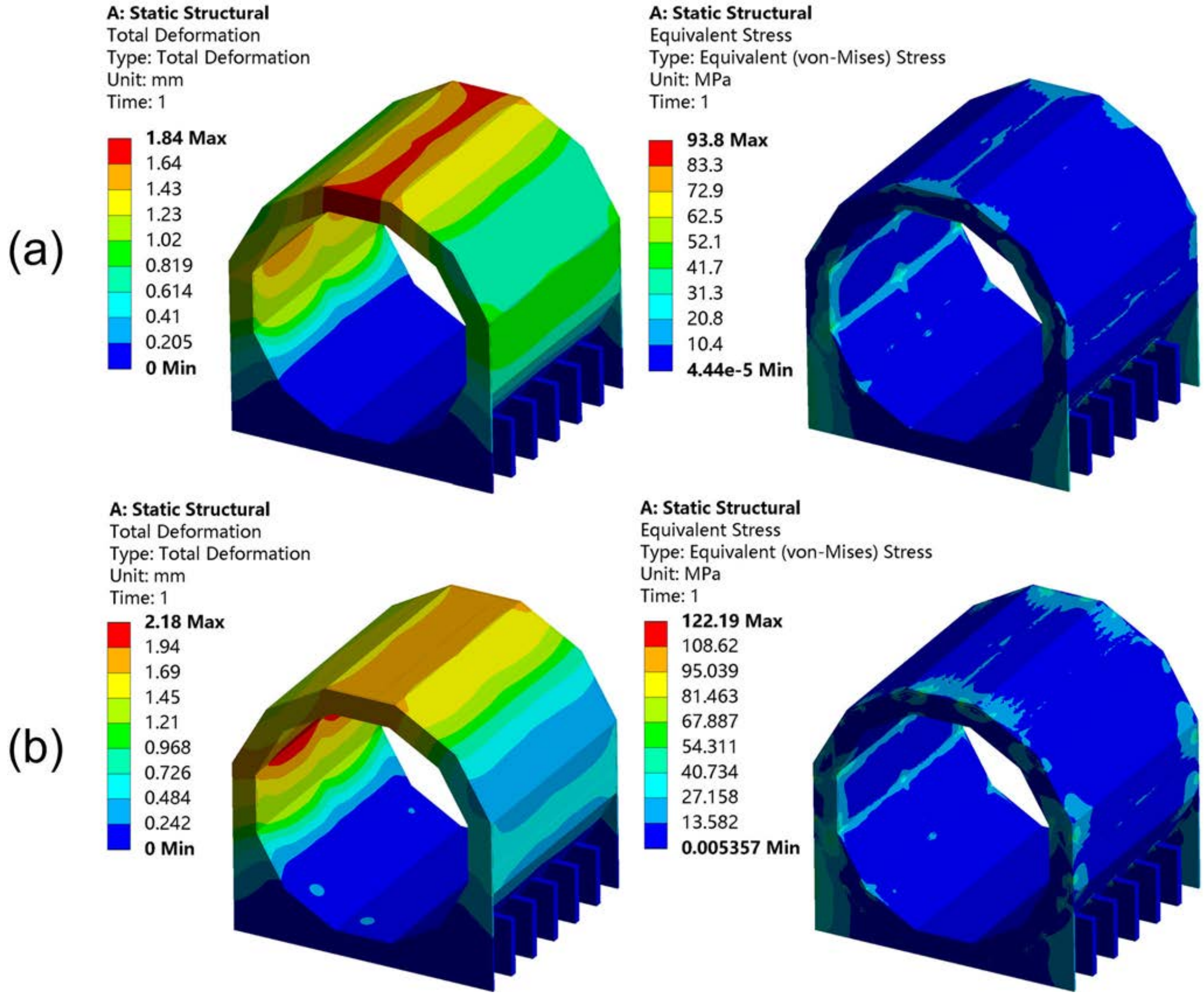

**Figure 14.31:** FEA results of the barrel yoke deformation and stress when 2145 tons of sub-detectors installed (including the superconducting solenoid magnet, the barrel HCAL, the barrel ECAL, the endcap HCAL, the endcap ECAL and the barrel muon detectors. The weight of the lightweight tracker detector is ignored.) (a) without electromagnet force. (b) with electromagnet force.

Based on the FEA results, there are following conclusions:

(1) When 2,145-ton of sub-detectors load is applied to the yoke , the maximum deformation is 1.8 mm. The maximum stress is approximately 94 MPa, which is below the allowable stress of 136.7 MPa, thereby meeting the design and safety requirement.

(2) When all the sub-detectors are installed and subjected to electromagnetic force from the

superconducting solenoid magnet, the maximum deformation is 2.2 mm. The maximum stress is about 122 MPa, less than the the allowable stress 136.7 MPa, remaining within safe operational limits. .

## 14.3 Detector installation

The overall installation methodology and technical scheme of the detector are essential to the final installation quality and precision. The installation process should ensure safety, reliability and repeatability in order to meet requirements of future operation, maintenance, and upgrades.

### 14.3.1 Technical scheme

The detector is finally installed in the underground experimental hall. A sound installation scheme should combine conditions of the ground hall and underground hall to effectively coordinate both sub-detector pre-assembly, final installation, and integration.

The technical scheme of detector's installation involves: overall installation sequence, sub-detector installation order, and pre-assembly point deployment plan.

1. Overall installation sequence

   The overall installation follows a surface-to-underground sequence:

   - Step 1: Pre-assembly in the ground hall

     Pre-assembly enables efficient underground installation by first integrating and assembling mechanical supports, detector components, and electronics for each sub-detector. After assembly verification and alignment measurements, the pre-assembled sub-detectors are prepared for vertical transfer to the underground experimental hall.
   - Step 2: Lifting through the vertical shaft

     Safe and reliable equipment lifting from the ground to underground is essential for a smooth installation. Early mechanical design must systematically plan and standardize lifting requirement for all sub-detectors.
   - Step 3: Installation in the underground experimental hall

     Installation of each sub-detector in the underground experimental hall is the most critical phase, where confined space constraints the operation of lifting, transportation, installation, and connection, thereby posing significant design challenges for safe and efficient implementation.
2. Sub-detector installation sequence

   The detector installation starts from the yokes (muon detector). Figure 14.32 shows the explosive decomposition of sub-detectors inside the barrel yoke. The installation proceeds from the outside inward. Installtion begins with the barrel sub-detectors, in the order of:

superconducting solenoid magnet, barrel HCAL, barrel ECAL, TPC (barrel OTK), and finally the ITK and beam pipe assembly (VTX, LumiCal). Next, installation of the endcap sub-detectors proceeds with the endcap ECAL (and its OTK) followed by the endcap HCAL). The deformation of the two end flanges of the barrel yoke is monitored by the laser tracker throughout the installation process to confirm that the deformation of the barrel yoke is within design tolerances after each installation stage.

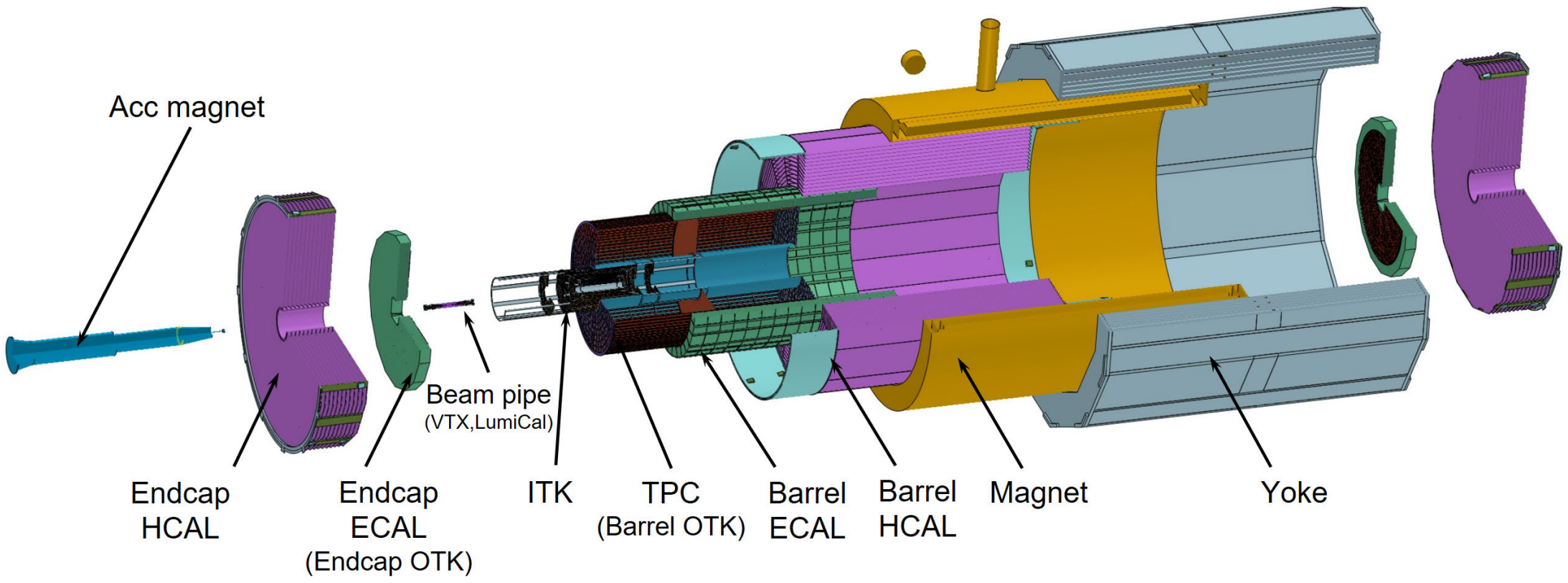

**Figure 14.32:** The exploded view of sub-detectors installation.

3. Pre-assembly point installation scheme

   Figure 14.33 shows the layout of the pre-assembly point and the installation scheme. This scheme enables concurrent sub-detector installation and accelerator beam commissioning. Temporary shielding wall is placed between the pre-assembly point and the collision point to ensure the safety of personnel and the detector during the beam commissioning.

   Once all sub-detectors are installed at the pre-assembly point, the complete detector assembly is moved to the collision point on a dual guide rail system. Secondary alignment and adjustment are then performed to facilitate future maintenance and replacement of the sub-detectors at the collision point.

### 14.3.2 Services

This section describes the external service routing of sub-detectors, while internal cabling and piping of sub-detectors are explained in their respective chapters. These services are still being finalized and optimized. Envelopes was assigned, while detailed routing to assure that all services can reach various routing requirements is still ongoing. A list of all external cables and cooling pipes required to service the detector is shown in Table 14.4. The number in the table represents only one end, therefore the values should be doubled for the whole detector.

The detector cables and pipes will be routed when they are located and positioned. Figure 14.34 shows the routing paths of them. Routing arrangement follows the detector installation sequence: Barrel HCAL, barrel ECAL, TPC + barrel OTK, ITK + VTX + LumiCal + beam pipe, endcap ECAL + endcap OTK, endcap HCAL. Cables and pipes are routed from the in-nermost sub-detectors outward, progressively coverging with those from outer systems before finally exiting the detector through the gap between the barrel and endcap yokes. In contrast, the muon detector cables are routed directly outside the yoke, without occupying space inside the envelope. To prevent cable damage and accommodate the minimum bending radii required for cables and water pipes, all turns in the cable routing channels are designed with chamfered edges.

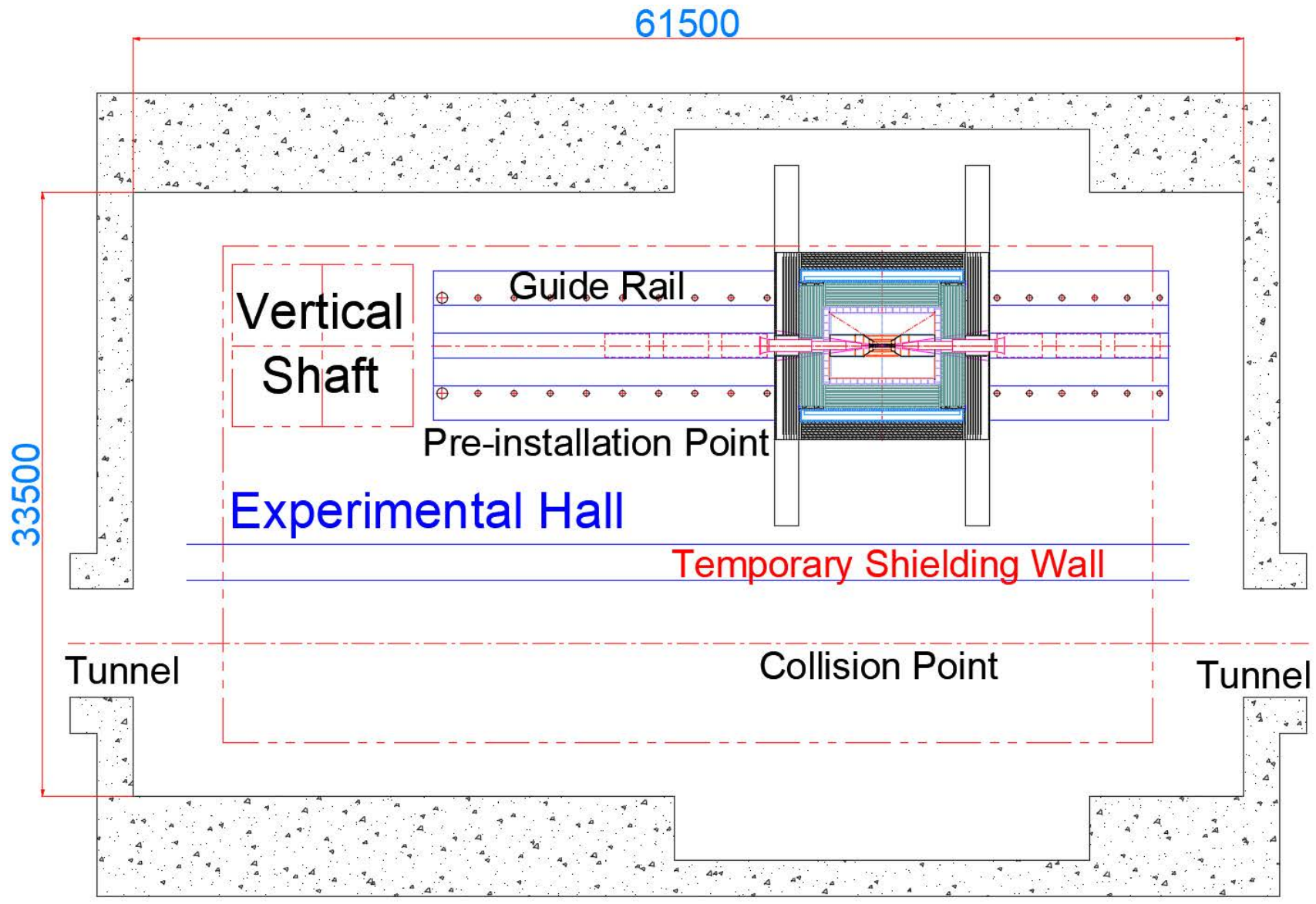

**Figure 14.33:** The detector pre-assembly point installation scheme. All sub-detectors are installed at the pre-assembly point to become a complete detector assembly and then moved to the collision point on rails. Units are in millimeters.

**Table 14.4:** Inventory of cables and pipes. The quantity in the table represents only one end.

| | **Cooling pipe** | | **Cable** | |
|---|---|---|---|---|
| | Size (mm) | Number | Size (mm) | Number |
| Barrel muon detector | / | / | $\phi 5$ | 12 |
| Barrel HCAL | $\phi 12$ | 96 | $\phi 5$ | 2784 |
| Barrel ECAL | $\phi 10$ | 80 | $\phi 5$ | 240 |
| Barrel OTK | $\phi 5.5$ | 220 | $\phi 5$ | 440 |
| TPC | $\phi 6$ | 12 | $\phi 5$ | 248 |
| ITK | $\phi 5$ | 69 | $\phi 5$ | 101 |
| VTX | $\phi 12$ | 16 | $\phi 5$ | 48 |
| LumiCal | / | / | 30×6 | 4 |
| Detector beam pipe | $\phi 8$ | 8 | / | / |
| Endcap OTK | $\phi 6$ | 32 | $\phi 5$ | 144 |
| Endcap ECAL | $\phi 10$ | 16 | $\phi 5$ | 112 |
| Endcap HCAL | $\phi 12$ | 32 | $\phi 5$ | 1536 |
| Endcap muon detector | / | / | $\phi 5$ | 8 |
| Total | / | 581 | / | 5677 |

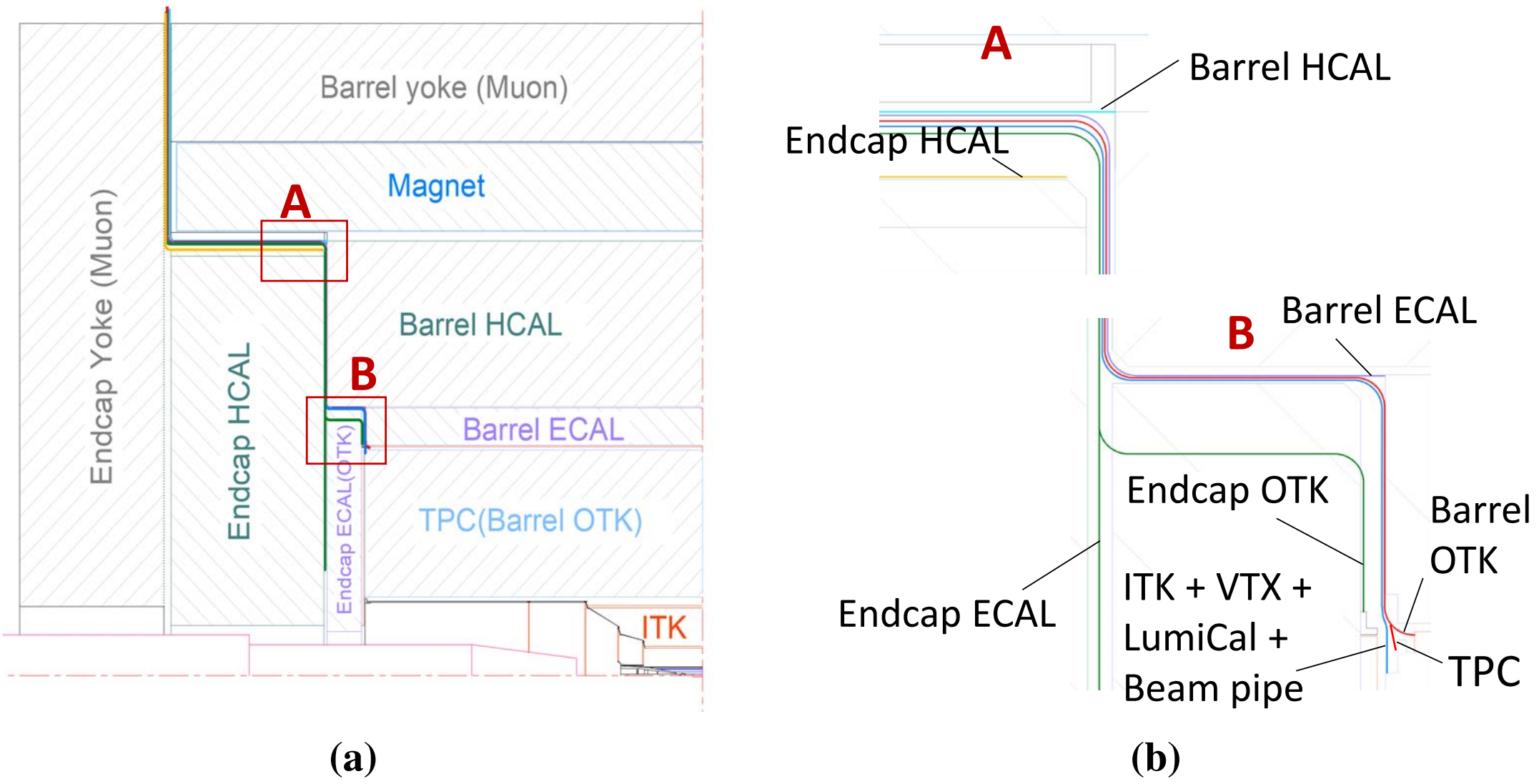

**Figure 14.34:** External routing paths of sub-detector. (a) The cables and pipes are routed from their sub-detectors along a designated shared path to the yoke before exiting the detector. The muon detector cables are routed directly outside the yoke, without occupying space inside the envelope. (b) The cables and pipes of sub-detector merge at separate locations. Endcap OTK has a shoulder structure for cable routing, which provides a dedicated cable routing channel before converging with other cables.

The routing path contains five channels, as shown in Figure 14.35. The gap width of each channel and the quantity of cables and pipes passing through their sections are listed in Table 14.5. All cables are ultimately divided into bundles and routed radially outward. In channels 1, 3, and 5, cables and pipes are arranged radially along the end face with enough space. However, in channels 2 and 4, cables run axially and encounter some connection structures between sub-detectors that impact the routing path, a clearance check is therefore required for these two routing paths. After evaluating the cross-sectional areas of the most constrained zones (A3 and B1), it is confirmed that the net service cross-section is sufficient for the routing.

Cables emerging from the yoke undergo interconnection before reaching the electronics room. To allow for future endcap yoke access, all patch panel boxes are placed on the barrel yoke, as shown in Figure 14.36. To address the unsuitability of the bottom position for cable routing, 44 patch panel boxes are installed on the barrel yoke exterior. They are arranged in groups or four, with each group offering 128 channels. Here, the outgoing cables are grouped and interconnected before entering the electronics room.

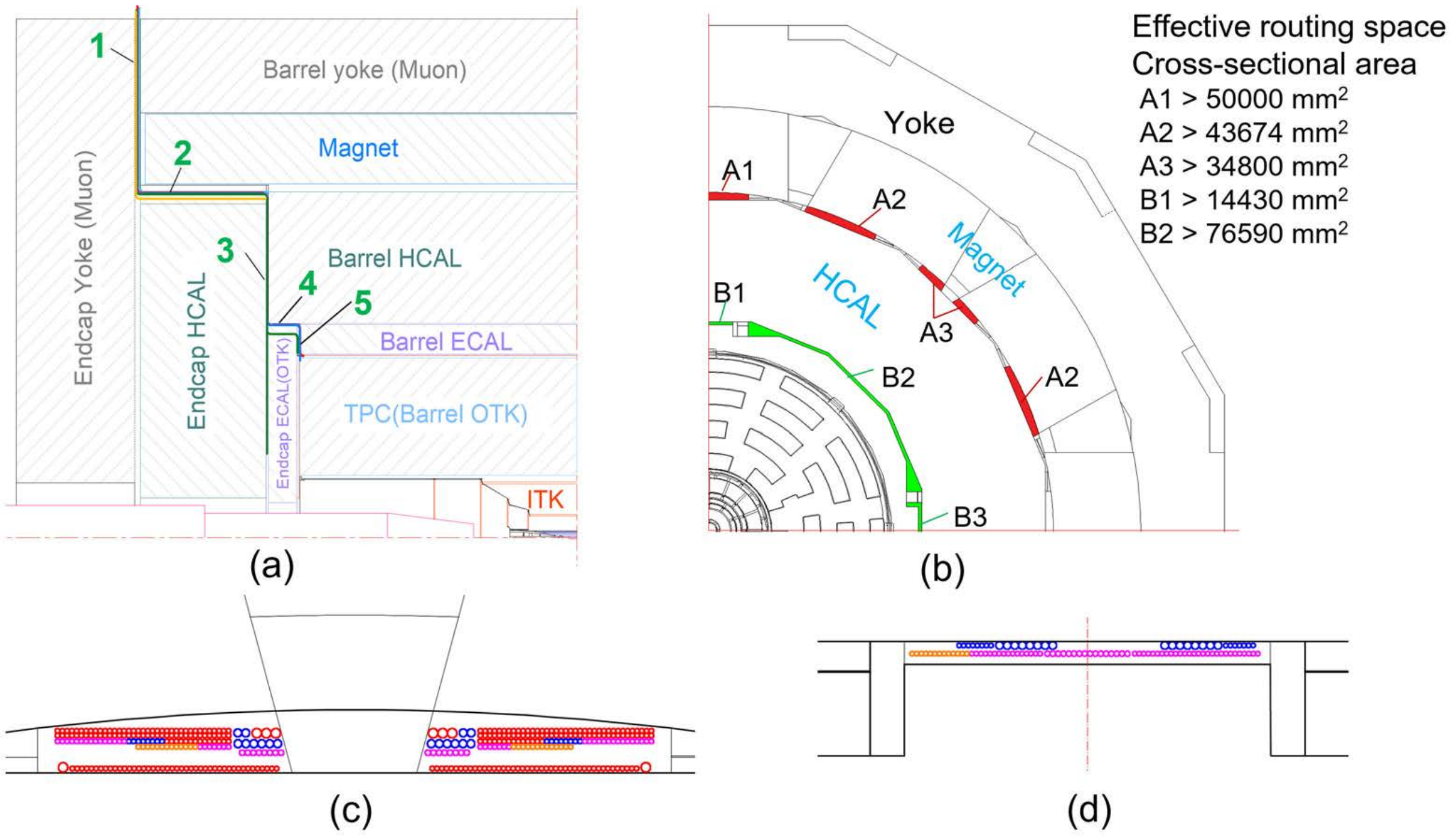

**Figure 14.35:** Routing space checking. (a) Five channels in routing path. Radial routing along the end face in the channels 1,3,5 has more spaces than axial routing in channels 2,4. (b) The red and green highlighted regions represent the permissible routing areas for channels 2 and 4 with the cross-sectional areas of the most constrained zones, A3 and B1, being of primary concern. (c) Arrangement of cables and pipes in A3 region, with a total net occupied sectional area of 10,000 square millimeters. (d) Arrangement of cables and pipes in B1 region, with a total net occupied sectional area of 4,000 square millimeters.

**Table 14.5:** Routing channel gap, cables and pipes in channel sections.

| Channel | Gap (mm) | In channel | | Cross section ($mm^2$) | |
|---|---|---|---|---|---|
| | | Cables | Pipes | Channel | Cables and Pipes |
| 1 | 60 | 5677 | 581 | $1.3 \times 10^6$ | $1.4 \times 10^5$ (11%) |
| 2 | 14–80 | 5677 | 581 | $5.1 \times 10^5$ | $1.4 \times 10^5$ (27%) |
| 3 | 30 | 1336 | 453 | $4.0 \times 10^5$ | $4.4 \times 10^4$ (11%) |
| 4 | 30 (Min) | 1080 | 405 | $3.6 \times 10^5$ | $3.7 \times 10^4$ (10%) |
| 5 | 30 | 984 | 357 | $3.4 \times 10^5$ | $2.9 \times 10^4$ (8.5%) |

The detector cooling pipes are connected to flexible hoses, which are then supported by brackets and insulated. Shielded penetrations provide radiation protection and airtight sealing when passing through walls. The pipes are routed along seismic-resistant cable trays in corridors, maintaining separation from electrical cables. Isolation valves, filters, and vibration-damping connectors are installed before entering the chilled water plant, with leak detection sensors and labeling systems implemented throughout the line.

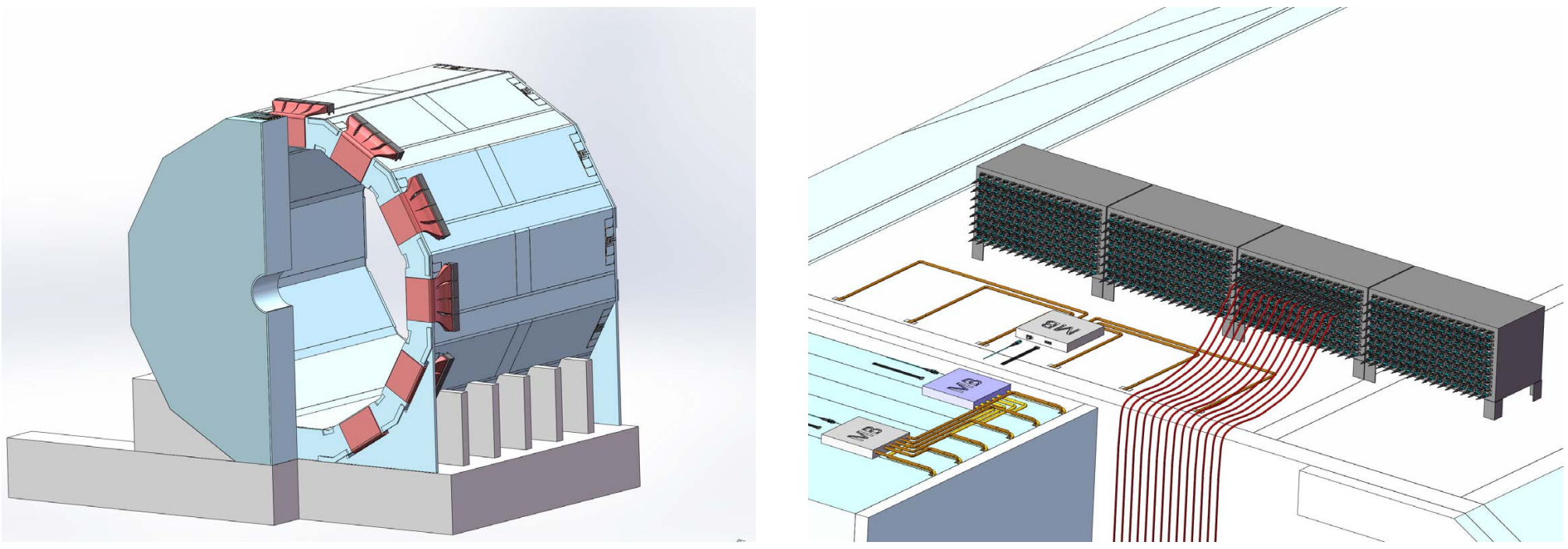

(a) (b)

**Figure 14.36:** Cable interconnection outside the barrel yoke. (a) 44 patch panel boxes are installed on the exterior of the 11 barrel yoke modules. Cables will undergo interconnection before reaching the electronics room. (b) Four boxes form a group, with 128 channels per group.

### 14.3.3 Installation of sub-detectors

This section details the installation process for each sub-detector in the underground experimental hall. As shown in Figure 14.37, the assembled sub-detectors are pushed into the barrel yoke one by one along the guide rails.

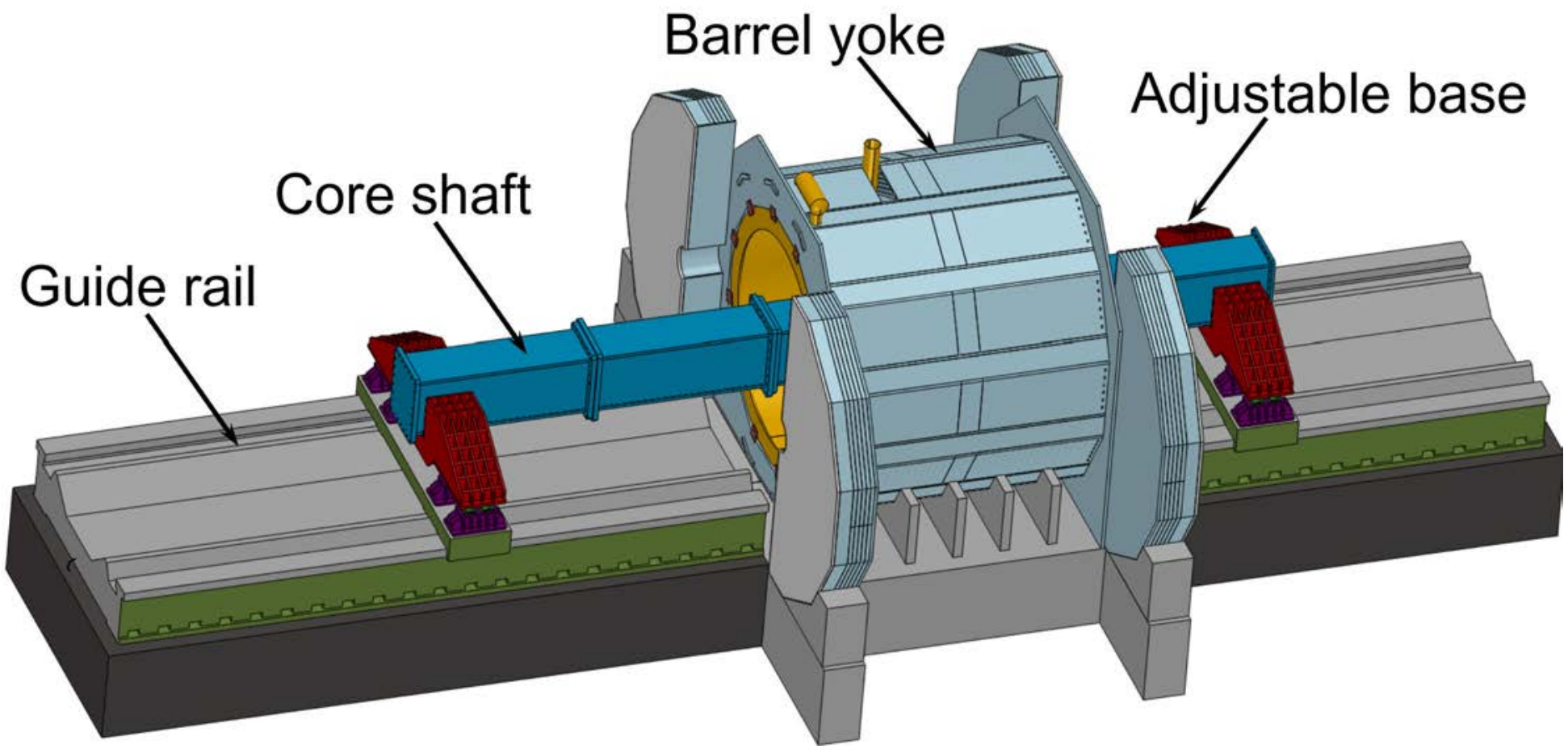

**Figure 14.37:** Detector installation in the underground experimental hall. The assembled sub-detectors are pushed into the barrel yoke one by one along the guide rails.

To achieve efficient installation at the pre-assembly point and easy future maintenance and replacement, the installation shall follow the four design principles:

- Minimization of installation steps;
- Simplification of installation operations;
- Reduction of tooling;
- Quick connection of electronics cables, etc.

Based on the overall installation sequence and stage-specific tooling requirements, the sub-detector installation is divided into five phases: alternating installation, core shaft installation, cantilever installation, positioning and survey during detector installation, and final translation and alignment of the entire detector.

1. Alternating installation

   The alternating installation method enables simultaneous installation of two components using identical installation process, where one component's installation is completed during the other's installation. The installations of the barrel yoke and superconducting solenoid magnet apply a typical "no-tooling" installation idea.

   Installing barrel yoke, superconducting solenoid magnet and endcap yoke using the alternating installation are accomplished in 6 steps as shown in Figure 14.38:

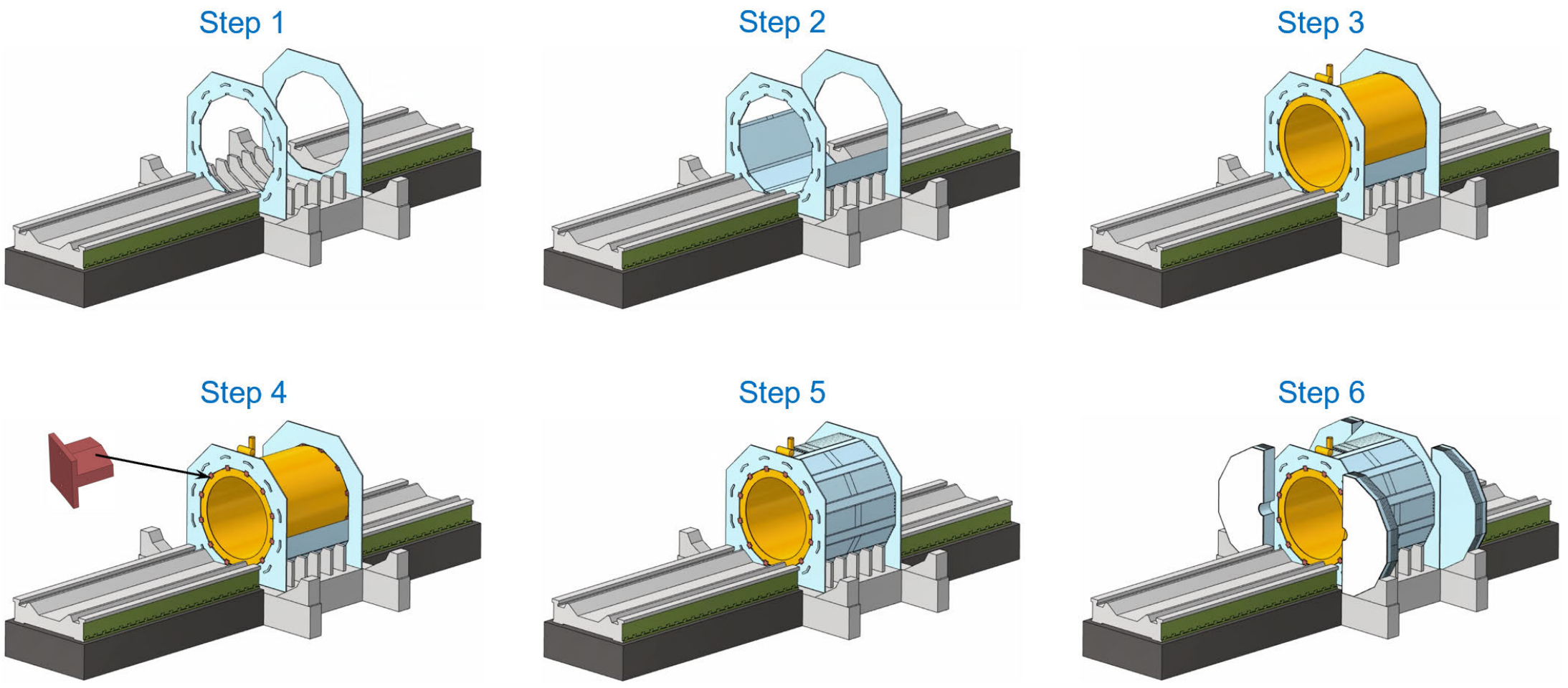

**Figure 14.38:** Alternating installation. The red part in Step 4 is the T-shaped connection block.

   - Step 1: Install both end flanges.
   - Step 2: Install the barrel yoke sectors located in the lower position and secure them to the end flanges, and then test the muon detectors to make sure there is no any damage caused during installation.
   - Step 3: Lift the superconducting solenoid magnet between the two flanges of the barrel yoke.
   - Step 4: Mount the magnet to the end flanges by T blocks.
   - Step 5: Install the barrel yoke sectors located in the upper position and secure them to the end flanges, and then test the muon detector.
   - Step 6: Install the endcap yokes and muon detectors.

   The alternating installation allows for the simultaneous installation of the yoke, muon detectors, and solenoid magnet, as well as the wiring and commissioning of their electronics, which greatly improves overall efficiency.

2. Core shaft installation

   The core shaft installation mounts the parts onto the core shaft, as shown in Figure 14.37, which is then precisely positioned on the external guide rails to move the parts into the place. This method is employed for installing the barrel HCAL and ECAL.

The key challenge of core shaft installation lies in assembling or disassembling components from the core shaft when both ends are constrained with limited space. To address this, the shaft is designed as a three-section modular structure. This design allows for pre-assembly of components with one shaft section before integrating with other sections for moving and positioning. During installation, each section of the core shaft and installed parts are first installed independently, they are then combined, moved into their final position, and integrated. Finally, the temporary core shaft sections are separately removed.

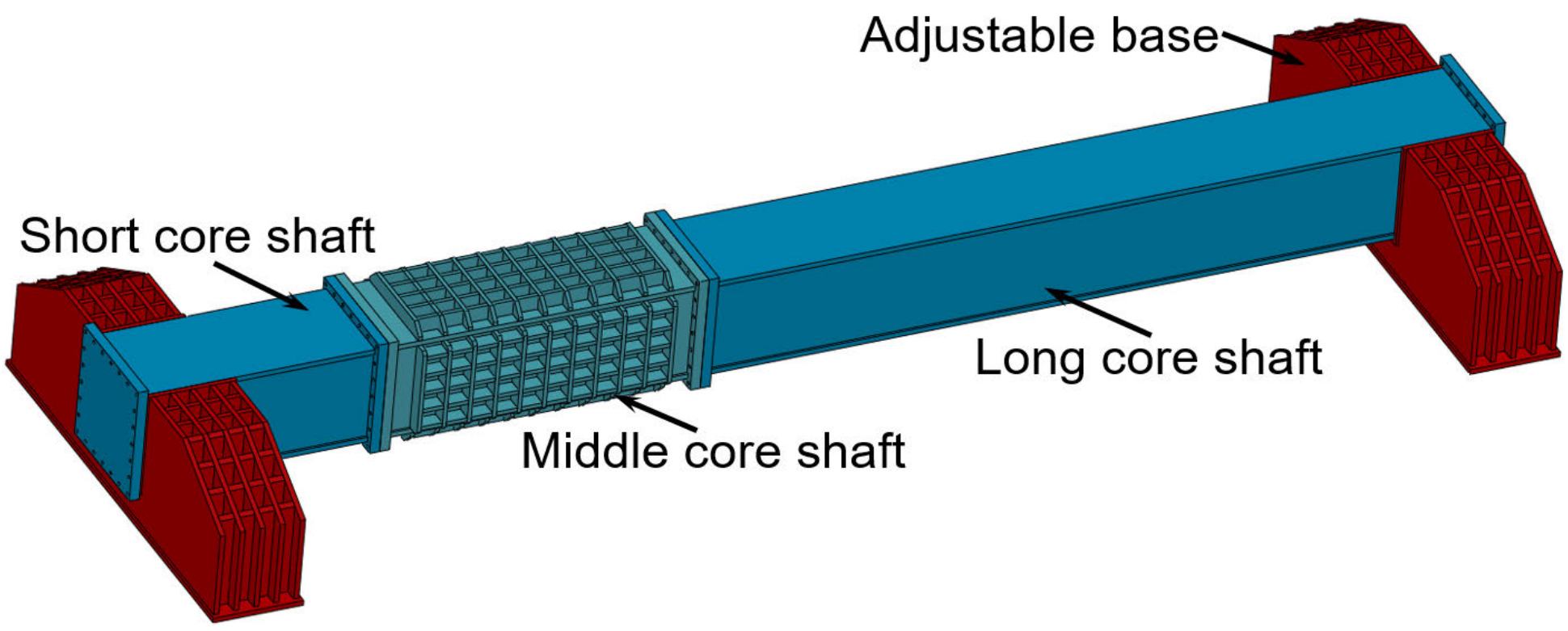

**Figure 14.39:** Three-section modular structure of the core shaft. Each section is bolted with the other one for easy assembly and disassembly.

(1) Barrel HCAL installation

The steps to install the barrel HCAL are shown in Figure 14.40:

- Step 1: Install the auxiliary hollow cylinders on both ends.
- Step 2: Install the lower half of the barrel module.
- Step 3: Mount the long and short sections of the core shaft on the supporting plate of the auxiliary hollow cylinder from each end. Then, lift and connect the middle section.
- Step 4: Install the upper half of the barrel module.
- Step 5: Drive the core shaft and push the barrel HCAL into the barrel yoke.
- Step 6: Once in place, connect the HCAL to the barrel yoke's end flanges using the half connection rings.
- Step 7: Remove the three yellow "$\pi$" brackets at the nearside, and then remove the core shaft, followed by removal of the three yellow "$\pi$" brackets at the other side.
- Step 8: After the installation of the barrel HCAL, withdraw the middle shaft and transfer the long shaft to the guide rail at the other end, leaving the designated space for the barrel ECAL installation.

(2) Barrel ECAL installation

For the barrel ECAL installation, the combination of the core shaft follows the same procedure as that for the barrel HCAL, as shown in Figure 14.41:

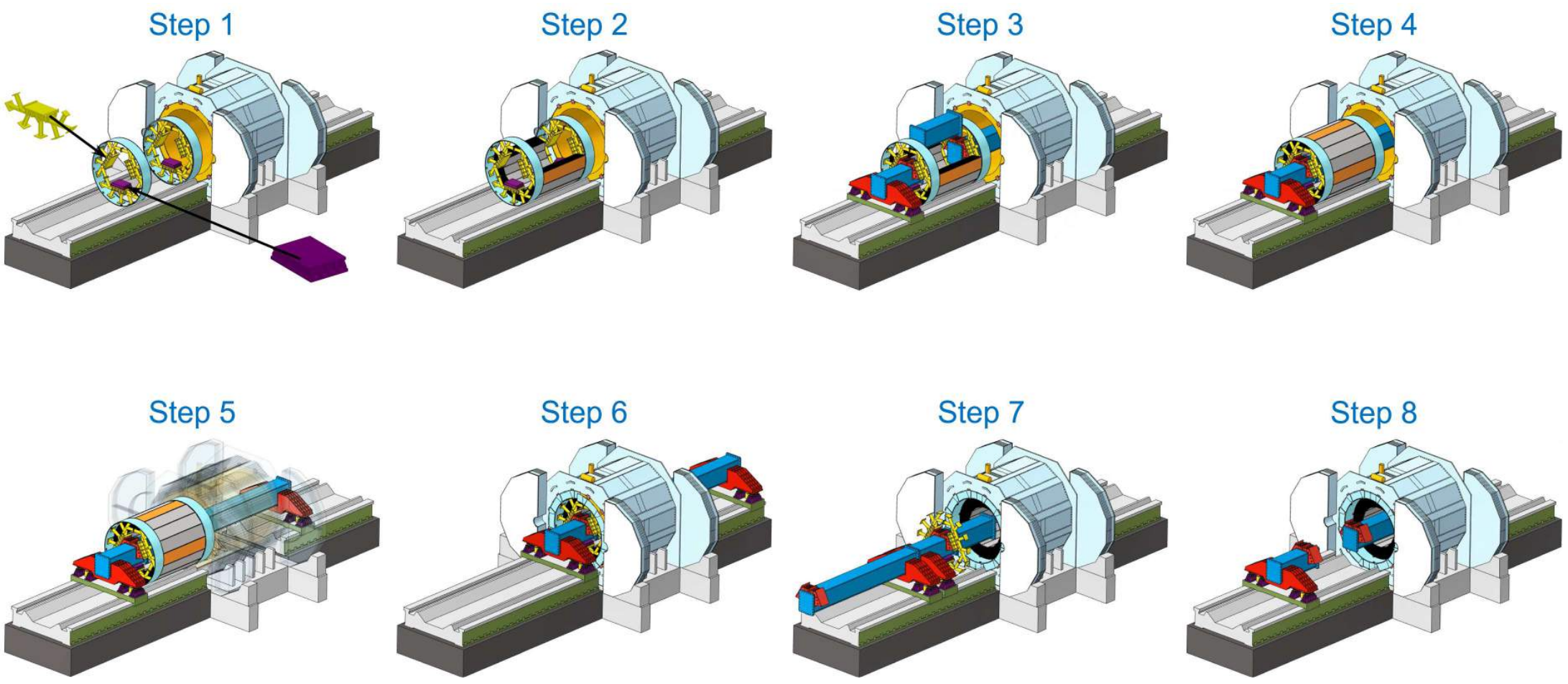

**Figure 14.40:** Installation process for the barrel HCAL. The yellow component in Step 1 is "$\pi$" bracket. The purple component in Step 1 is support plate.

- Step 1: After assembling the middle-section shaft with the barrel ECAL, lift it into position between the long short sections.
- Step 2: After assembling and connecting with the long shaft and short shaft sections, push the barrel ECAL into the barrel yoke.
- Step 3: Once in place, connect the barrel ECAL to the barrel HCAL with the suspension connection blocks.
- Step 4: After connection, withdraw the core shaft to finalize the barrel ECAL installation.

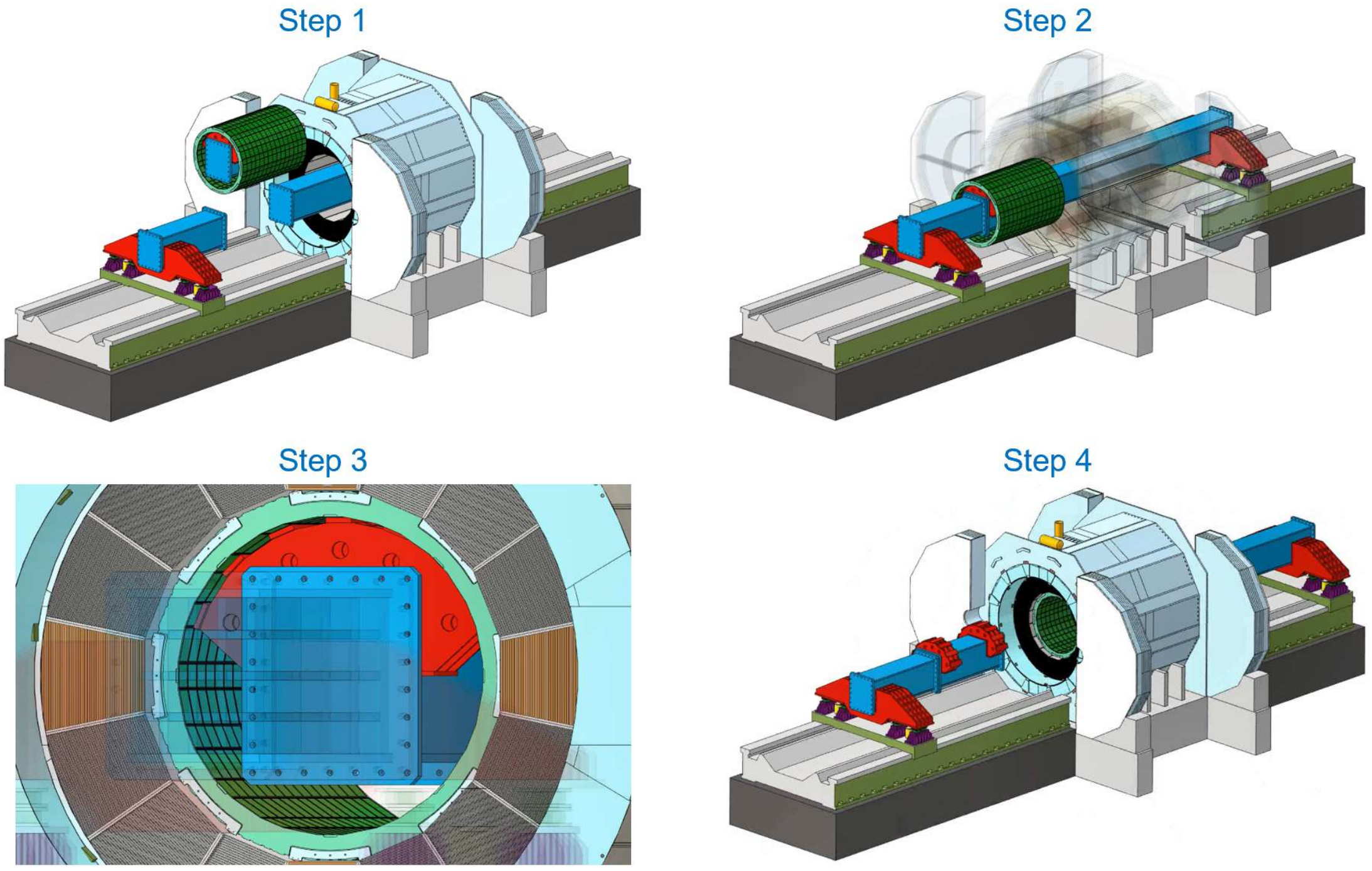

**Figure 14.41:** Installation process for the barrel ECAL. The barrel ECAL are assembled with the middle-section shaft and then lifted to connect with other shaft sections.

(3) Verification calculation of the core shaft

For the core shaft installation, the worst load case occurs during the installation of barrel HCAL, where a 1000-ton weight is directly applied to the shaft. The verification calculation for this case is conducted. The material of the shaft is Q355B structural steel. Its allowable stress is 237 MPa. In the FEA simulation, the connections between the shaft adjustable base, and between adjacent shaft sections, are modeled as bonded contacts. The bottom of the adjustable bases at both ends is defined as fixed support constraint. An equivalent force of $1\times10^7$ N weight of the barrel HCAL is applied to the top of the middle shaft, while a gravity load is applied to the entire structure.

The simulation results are shown in Figure 14.42. The maximum deformation is approximately 3.9 mm, which is within 10 mm installation clearance. The maximum stress is about 115 MPa, which is below the allowable stress. This design meets the design requirement.

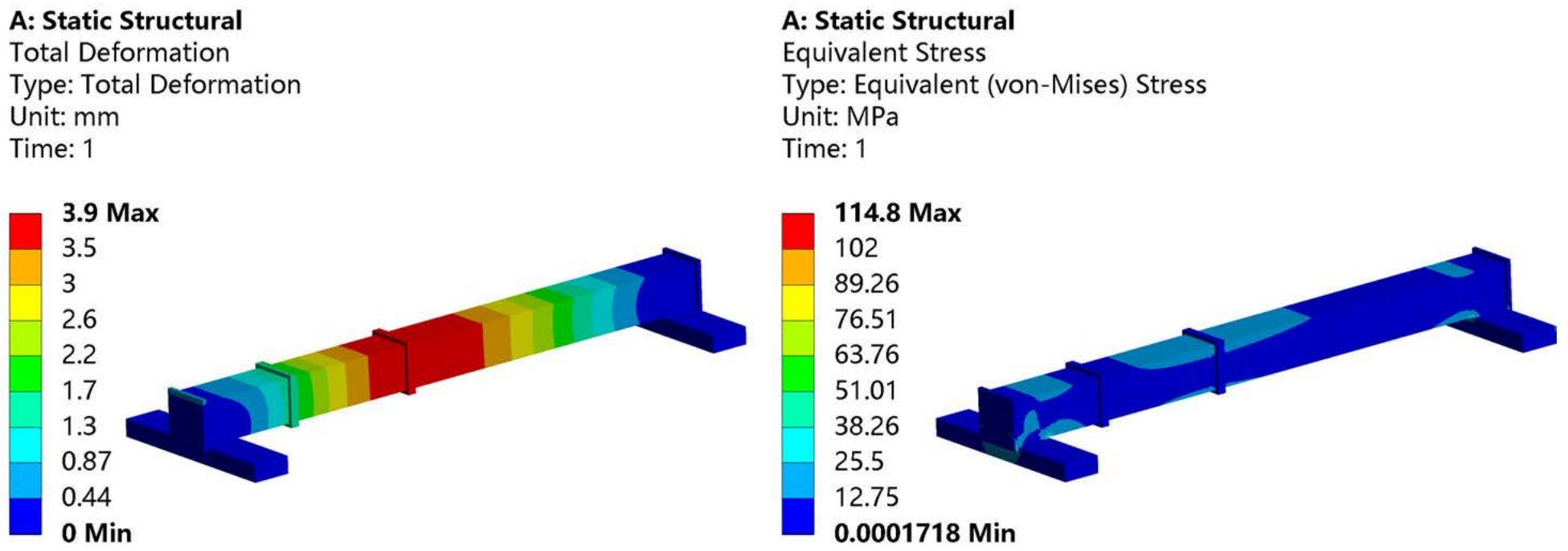

**Figure 14.42:** The simulation result of the core shaft in the worst case of installing the barrel HCAL where a 1000 tons weight is applied to the shaft.

3. Cantilever installation

Cantilever installation is a method in which a component is fixed at one end to the core shaft while the other end remains free, as illustrated in Figure 14.43. Components installed utilizing the cantilever installation method include TPC, ITK, endcap ECAL, endcap HCAL, and beam pipe assembly.

(1) TPC installation

The installation of the TPC, shown in Figure 14.44, is accomplished in three steps:

- Step 1: Lift the TPC assembly and secure it to one end of the long shaft.
- Step 2: Push the TPC into the barrel yoke. Then connect and secure it to the barrel ECAL using the connecting ring.
- Step 3: Once installation is complete, withdraw the core shaft.

During the deformation simulation for the TPC cantilever installation, the materials of both the long shaft and the TPC auxiliary tooling are structural steel. The contact settings are all set as bonded contacts. The area of contact between the long shaft and the adjustable base is set as fixed support, and the gravity field is applied to the whole model. The above material selection, boundary and constraint settings, and the gravity load application also apply to the simulation of the ITK, endcap ECAL, and endcap HCAL installation.

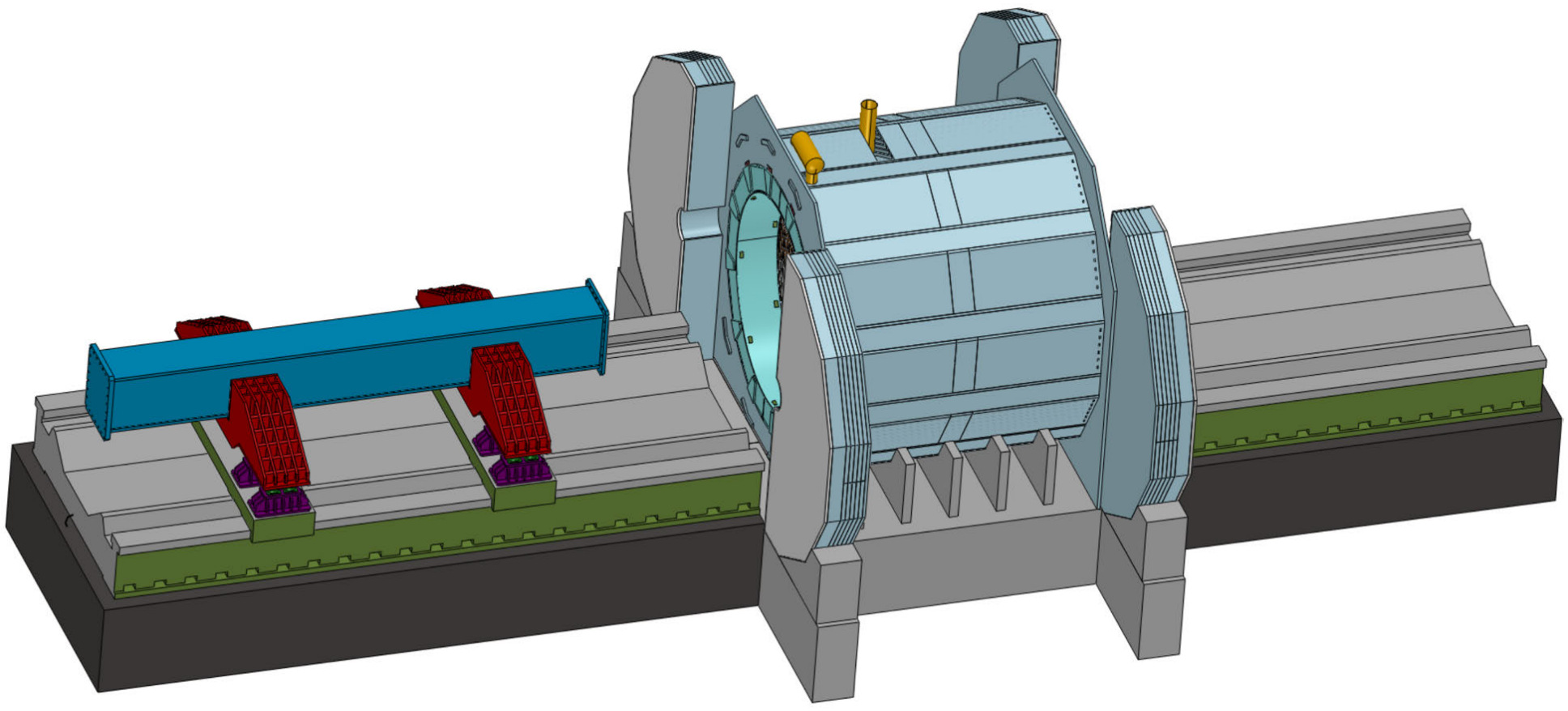

**Figure 14.43:** Cantilever installation. This method means when the sub-detector component is installed, only one end is fixed to the core shaft while the other end keeps free.

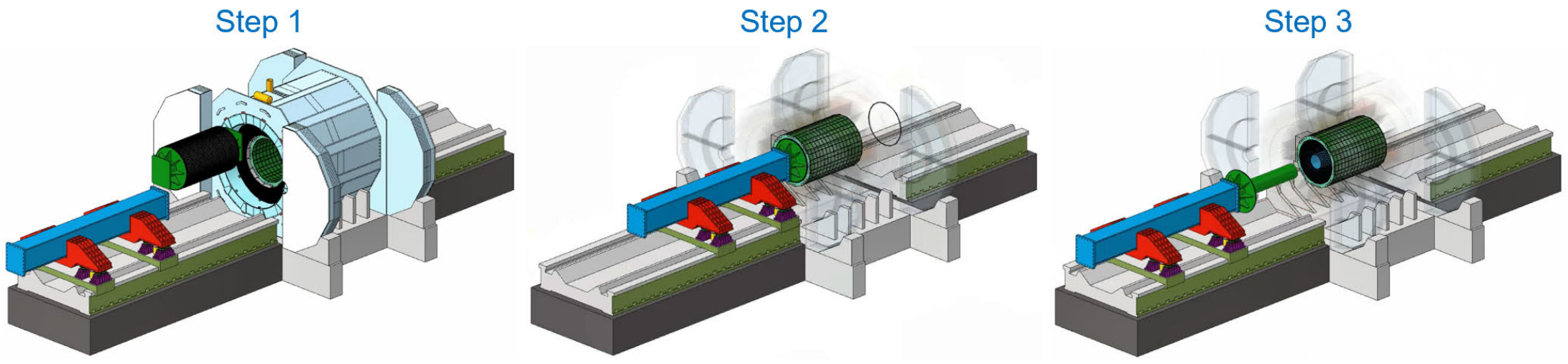

**Figure 14.44:** The installation process for the TPC. It will be secured to the side of the long shaft and then pushed into designed place.

The simulation result of the TPC cantilever installation is shown in Figure 14.45. The maximum deformation is 1.9 mm, less than the 10 mm installation clearance.

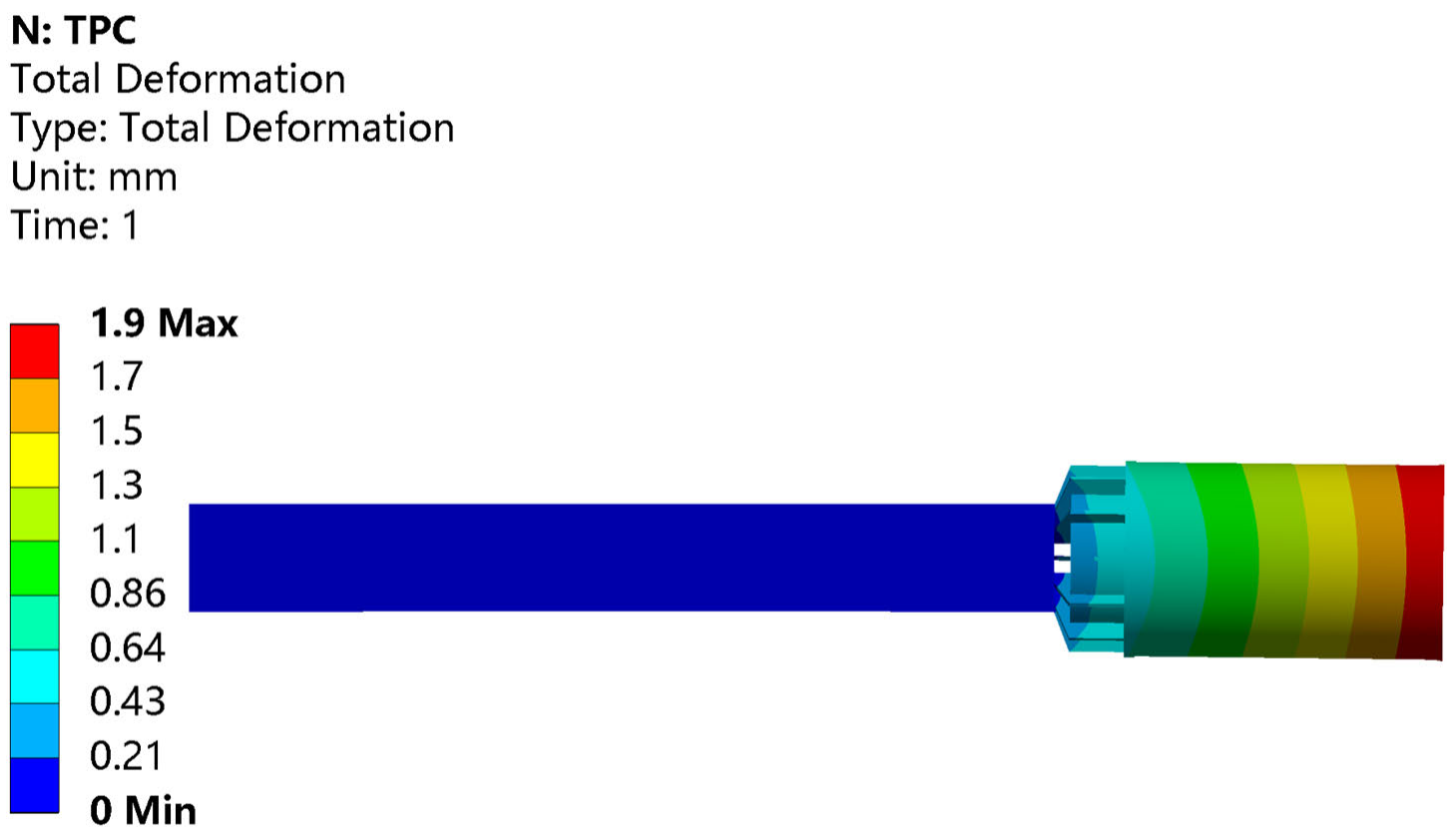

**Figure 14.45:** The simulation result of TPC cantilever installation.

(2) ITK installation

The installation of the ITK, shown in Figure 14.46, has five steps. The first two steps are conducted in the ground hall. These pre-assembled sections are then transported underground, where the remaining three steps of final installation are completed:

- Step 1: In the ground hall, assemble the beam pipe assembly with the ITK.
- Step 2: Adjust the beam pipe assembly into position with the screws and lock it in place.
- Step 3: Lift the entire ITK assembled with the beam pipe assembly into its installation position.
- Step 4: Push the ITK into the barrel yoke, and then connect and secure the ITK to the TPC with the L shape connection ring.
- Step 5: Once installation is complete, withdraw the core shaft.

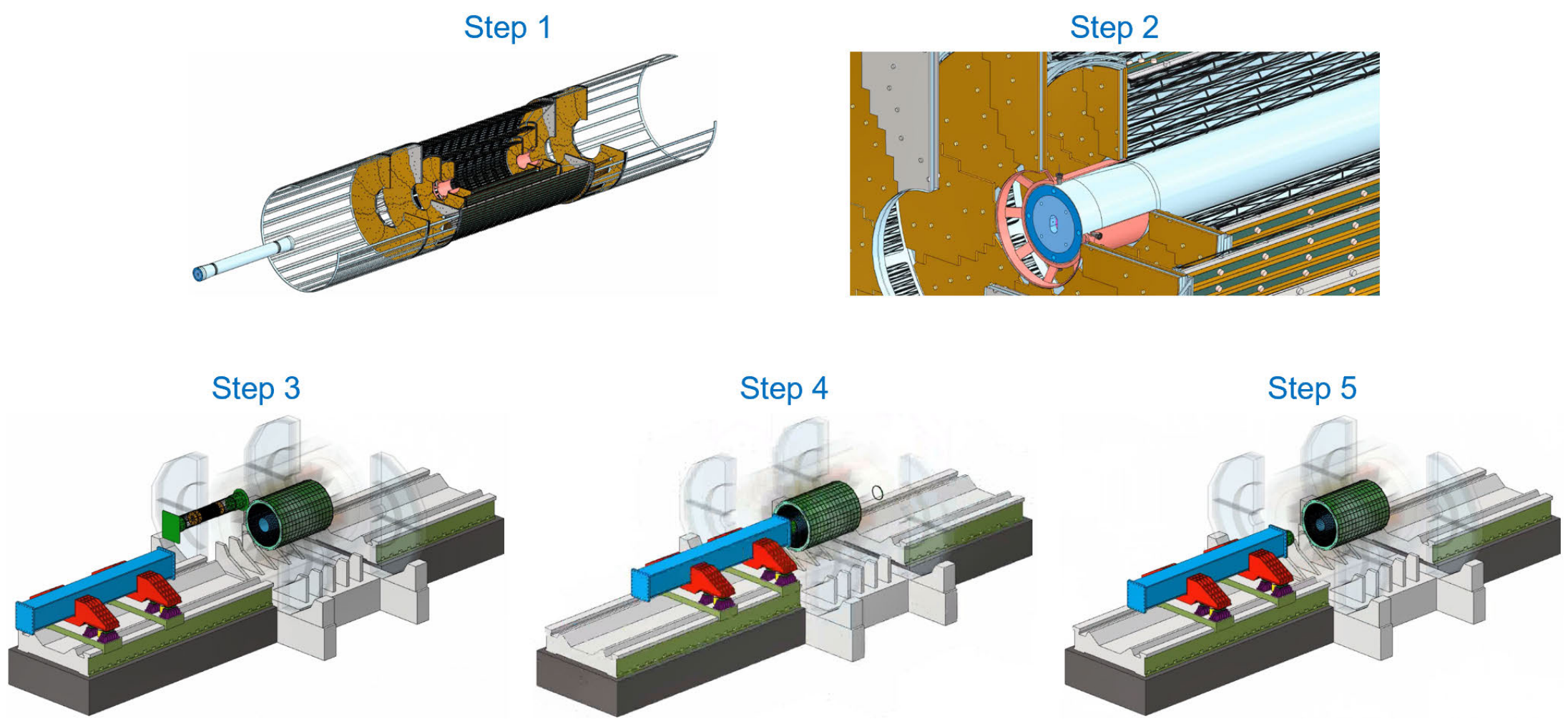

**Figure 14.46:** Installation process for the ITK. ITK will be assembled with the beam pipe assembly in the ground hall and then transported to the underground for the final installation. Cables and pipes not shown in figure.

The simulation result of ITK cantilever installation is shown in Figure 14.47. The maximum deformation is 3.3 mm, less than the 10 mm installation clearance.

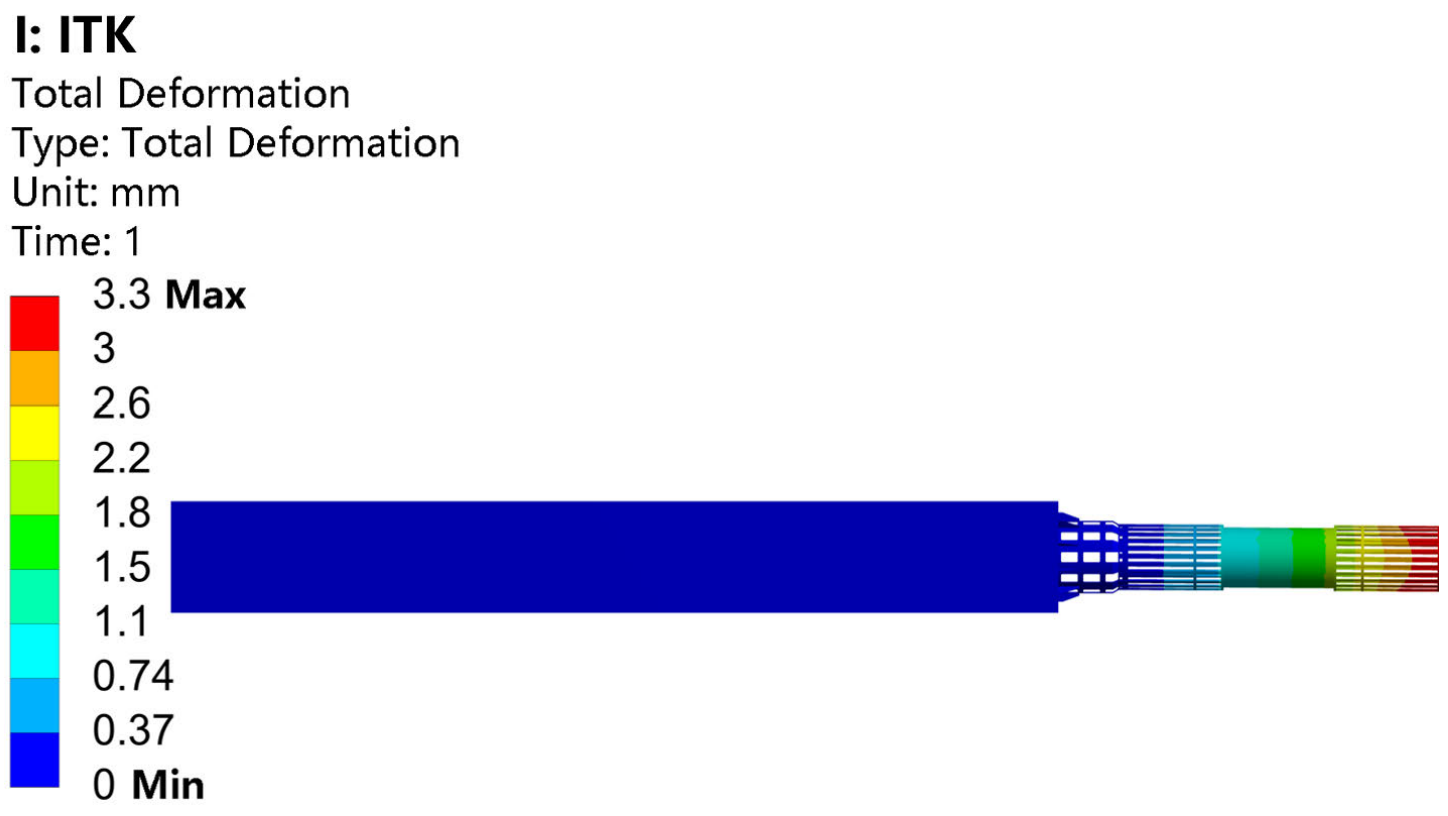

**Figure 14.47:** The simulation result of ITK cantilever installation.

(3) Endcap ECAL installation

The installation of the endcap ECAL, shown in Figure 14.48, is accomplished in four steps:

- Step 1: Lift the assembled endcap ECAL into the installation position.
- Step 2: Push the endcap ECAL into the barrel yoke.
- Step 3: Install the adjusting pad irons. Then, connect, adjust and secure them to the barrel HCAL.
- Step 4: Withdraw the core shaft and then install the endcap ECAL on the other end.

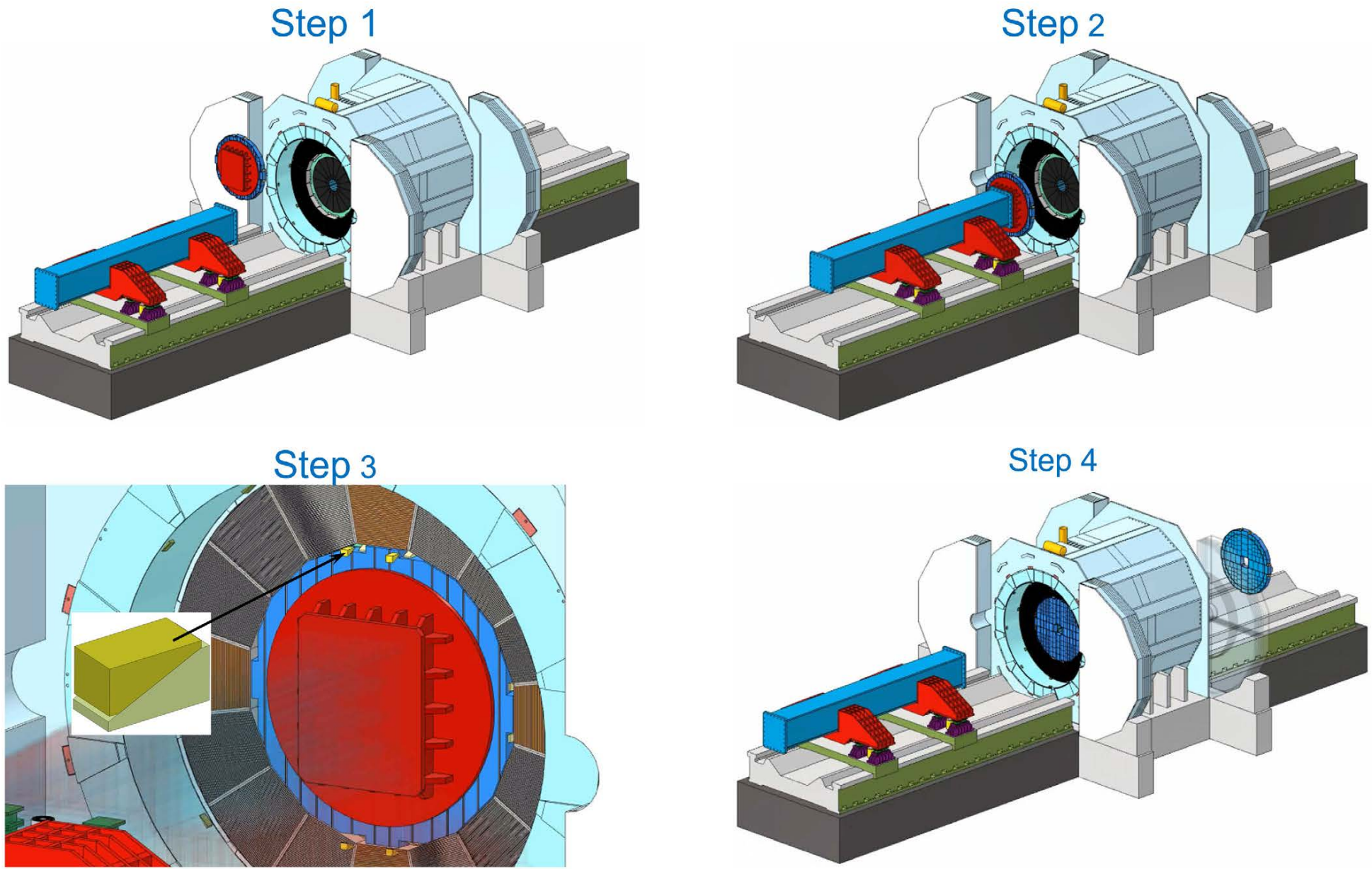

**Figure 14.48:** Installation process for the endcap ECAL installation. When one side installation finished, the core shaft will be moved to the other side to install the endcap ECAL on the other end.

The simulation result of the endcap ECAL cantilever installation is shown in Figure 14.49. The maximum deformation is 0.2 mm, which meets the 10 mm installation clearance requirement.

(4) Endcap HCAL installation

The installation of the endcap HCAL, shown in Figure 14.50, is accomplished in four steps:

- Step 1: Lift the assembled endcap HCAL into the intended installation position.
- Step 2: Push the endcap HCAL into the barrel yoke.
- Step 3: Connect, adjust, and lock the endcap HCAL to the auxiliary hollow cylinder of the barrel HCAL. Then, connect and lock its rear end flange to the half-connection-ring structure of the barrel HCAL.
- Step 4: Withdraw the core shaft and then install the endcap HCAL on the other end.

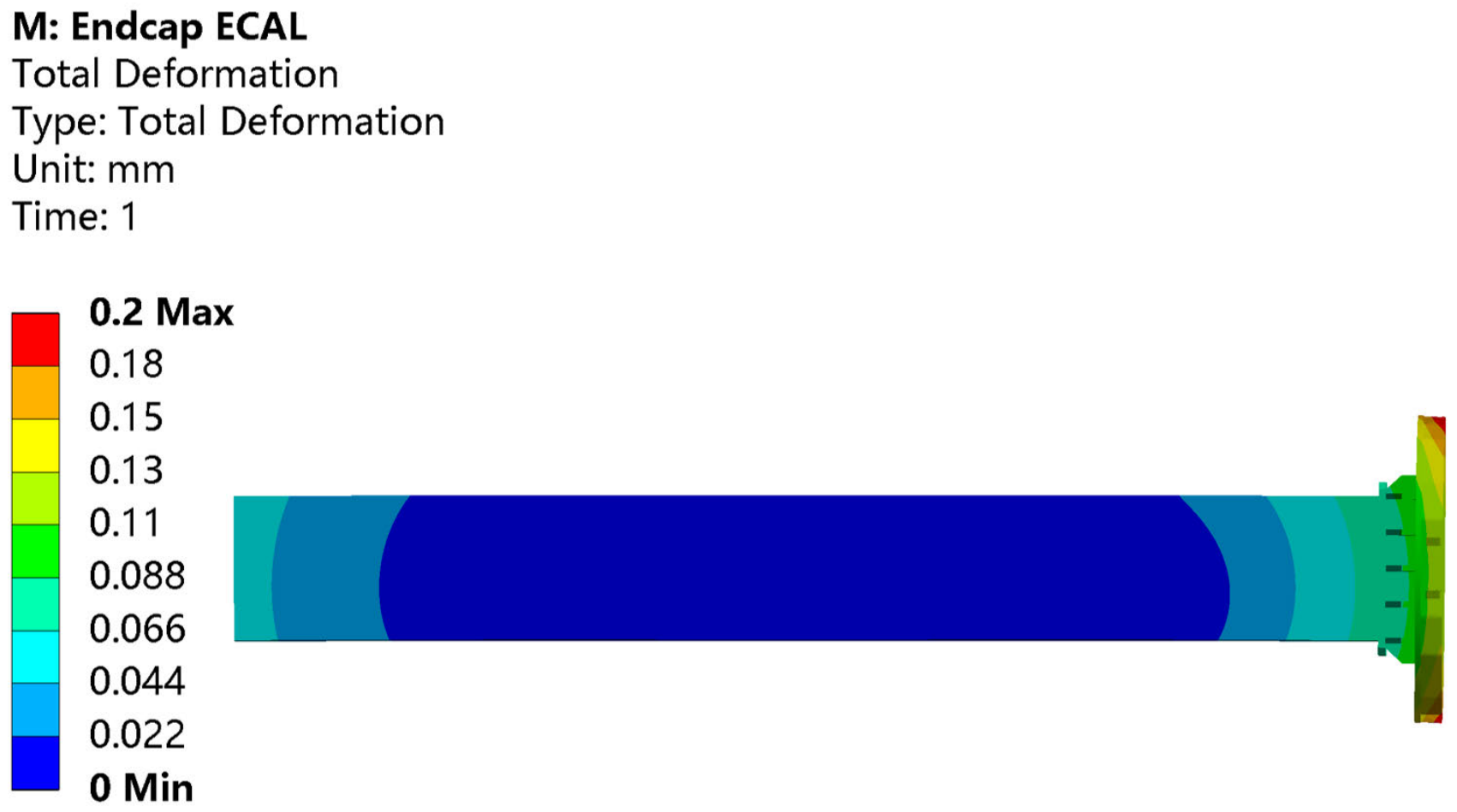

**Figure 14.49:** The simulation result of endcap ECAL cantilever installation.

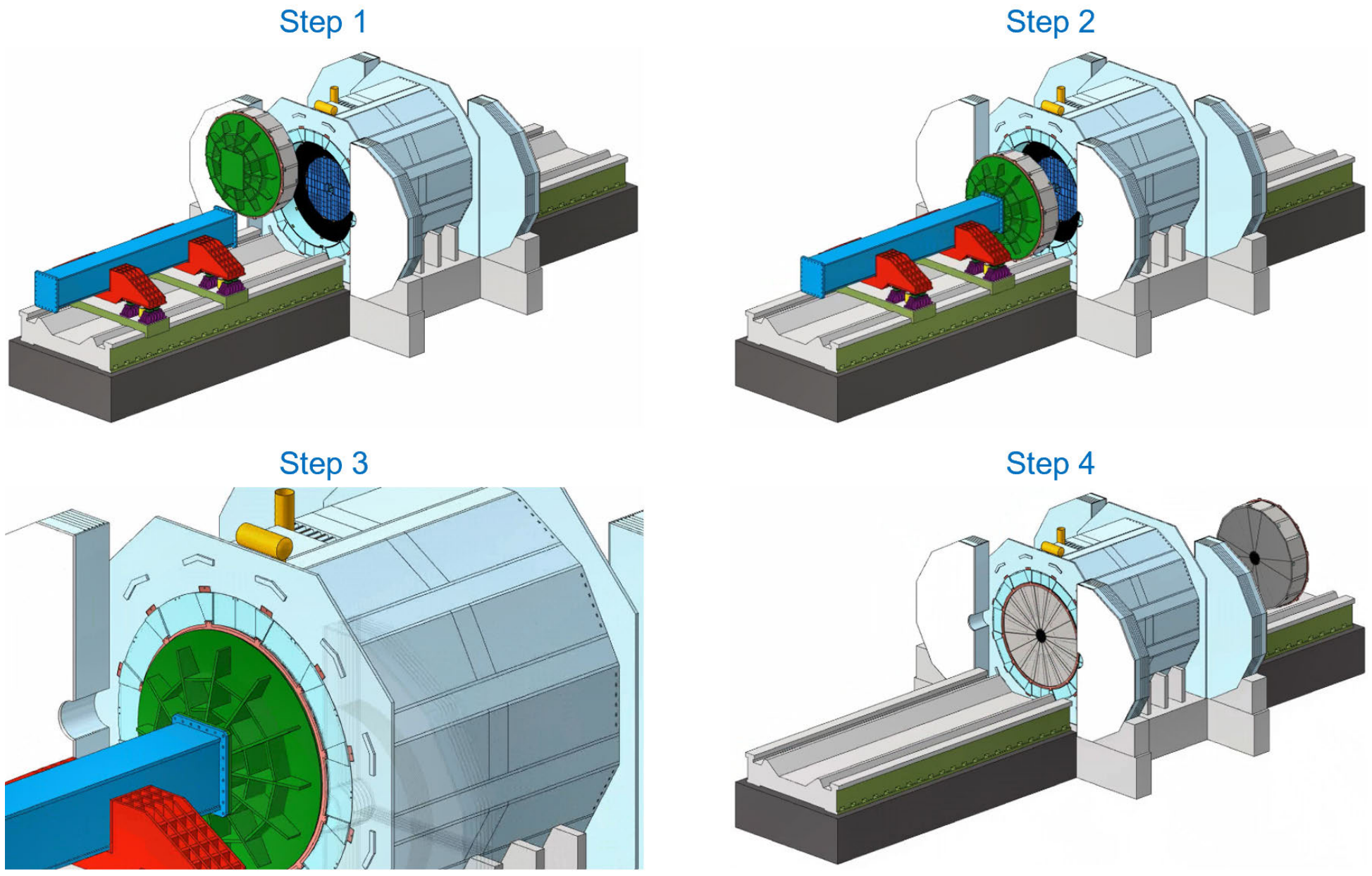

**Figure 14.50:** Installation process for the endcap HCAL installation. When one side installation finished, the core shaft will be moved to the other side to install the endcap HCAL on the other end.

The simulation result of the endcap HCAL cantilever installation is shown in Figure 14.51. The maximum deformation is 2.7 mm, which meets the 10 mm installation clearance requirement.

4. Positioning and survey during detector installation
   The detector installation and survey start from the barrel yoke. Some benchmark points will be established on the base to set up a local installation survey reference. For each sub-detector, some alignment targets are designed, such as the aligning holes or the marks, as shown in Figure 14.52 . The locating and positioning of the sub-detector will be implemented with a laser tracker, locating pins and some adjusting fixtures.

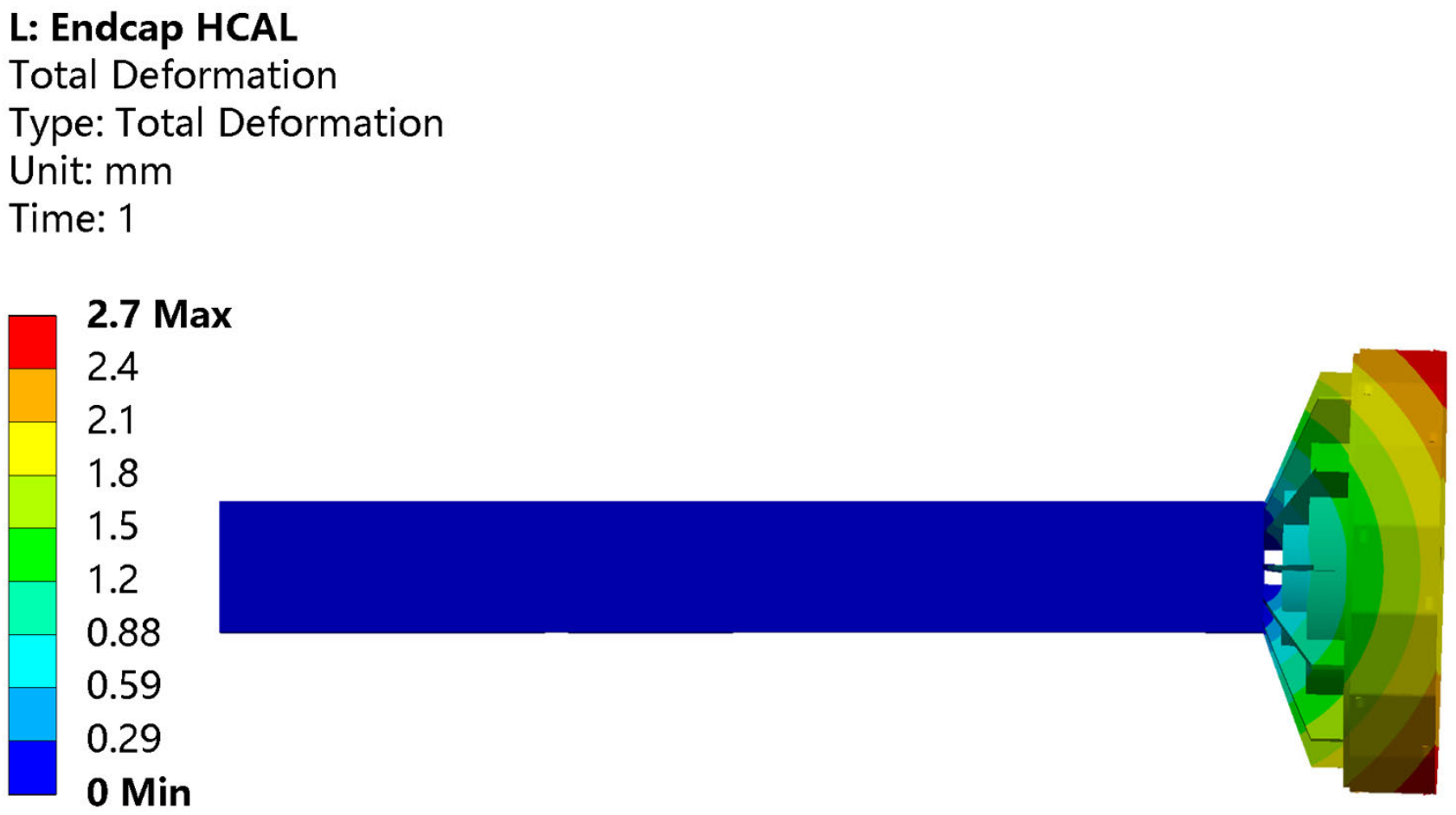

**Figure 14.51:** The simulation result of endcap HCAL cantilever installation.

Additionally, the position of barrel yoke end-flanges will be monitored throughout the detector installation process to confirm its deformation is within design tolerances after each installation stage. For some critical sub-detectors, such as the VTX and silicon tracker, the precise knowledge of the sensor positions is essential for achieving micron-level position resolution. So the precise alignment during their assembly, installation and running is necessary. In the assembly and installation stage, optical metrology techniques, including reference targets, laser tracking, and other precise referencing method, are employed for initial positioning. During operation, track-based alignment is implemented to maintain and verify the required positioning accuracy. For more details, refer to the alignment in Section 4.4 and 5.4.

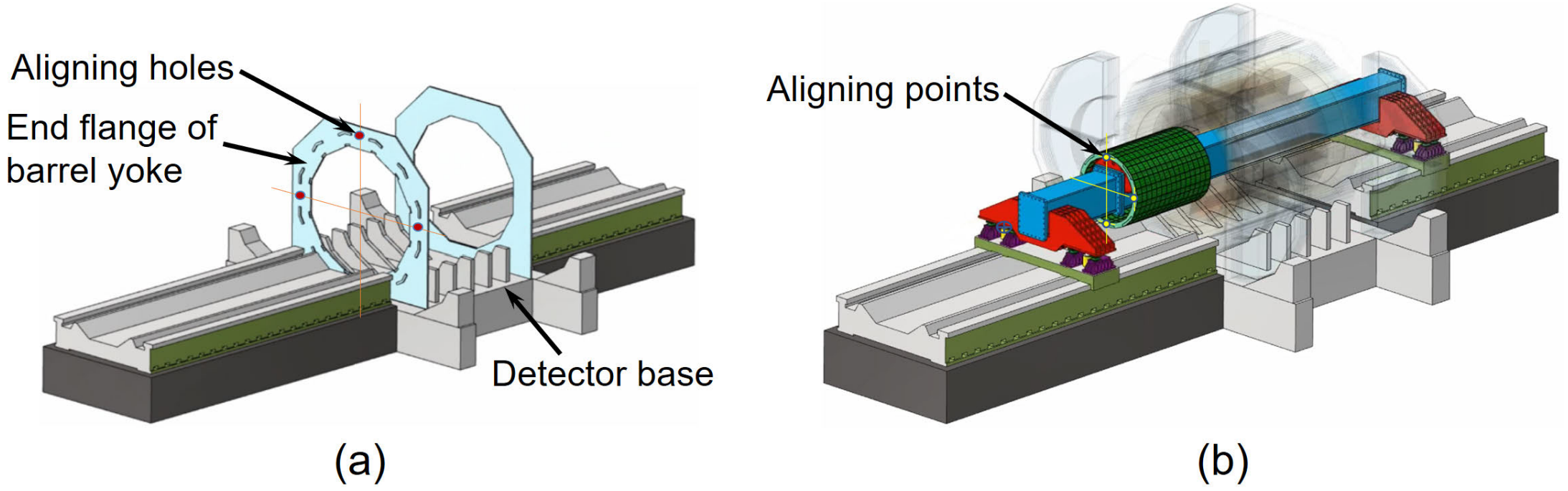

**Figure 14.52:** a) Alignment targets in the end flanges of barrel yoke. b) Aligning targets during sub-detector installation

5. Final translation and alignment of the entire detector

   The whole process of moving the entire detector from the pre-assembly point to the collision point, as shown in Figure 14.53, is operated in three steps:
   - Step 1: Before moving the whole detector, close the endcap yoke.
   - Step 2: Transport the detector to the collision point and perform alignment and adjustment. Then, position the rails and execute final alignment to set a unified reference for future detector maintenance.

- Step 3: Open the endcap yoke for interface installation of the accelerator vacuum components.

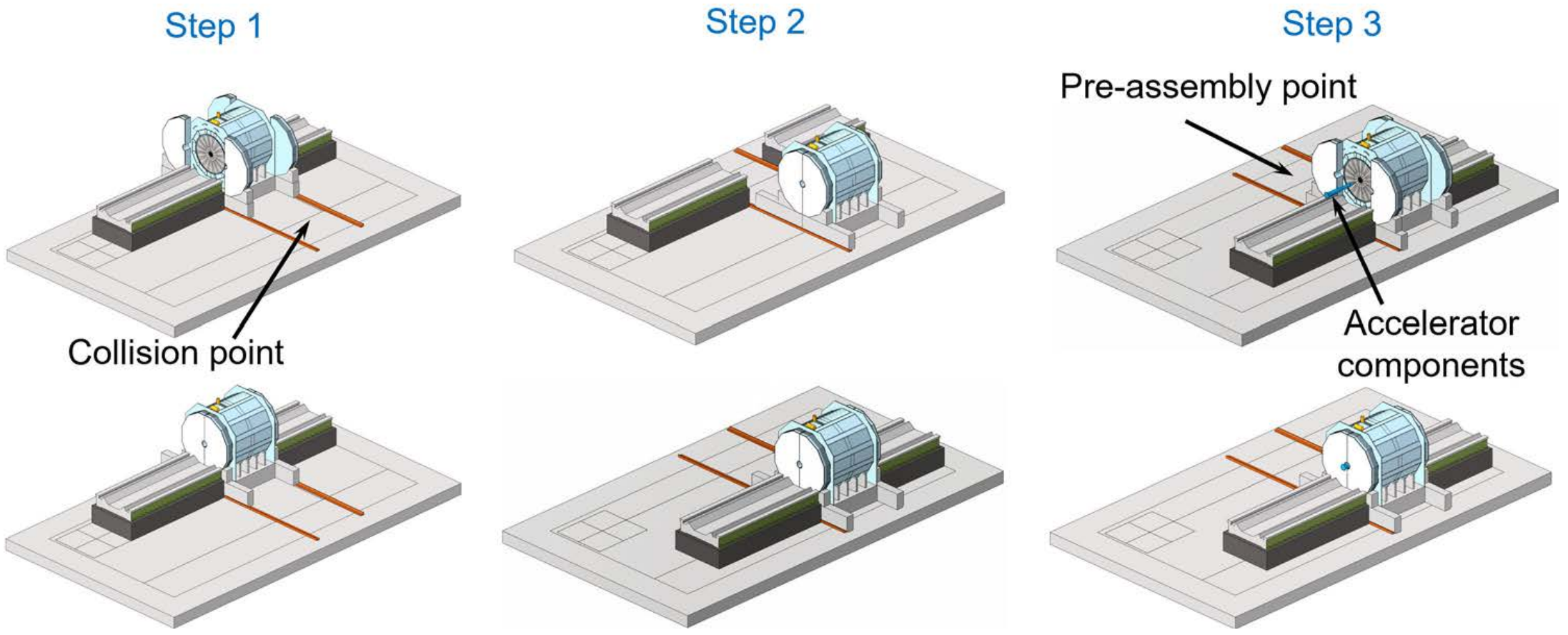

**Figure 14.53:** Moving and alignment of the detector.

## 14.4 Layout of the underground hall

The underground experimental hall, including the underground experimental hall and the underground auxiliary hall is located about 100 m underground, as shown in Figure 14.54. The underground auxiliary hall is divided into Zone 1 and Zone 2. For cost consideration, one vertical shaft is tentatively planned for each underground hall, which serves as a transportation passage for equipment and personnel. The quantity and sizing of the shafts will undergo iterative refinement during detailed design in the future.

The underground experimental hall measures 61.5 m (length) × 33.5 m (width), with an overhead crane coverage of 51.5 m × 27.5 m , taking into account lifting dead zone.

The auxiliary hall (Zone 1 and Zone 2) each measures 50 m × 10 m.

### 14.4.1 Experimental hall

As shown in Figure 14.55, the underground experimental hall is configured with two overhead cranes, two multi-degree robotic arms, two levels of surrounding corridor, and one vertical shaft. During overall installation, operators remotely control the overhead cranes and robotic arms from the corridor to perform alignment and adjustment, and precision installation of sub-detectors.

The configuration of the alignment control network in the underground experimental hall is shown in Figure 14.56. A number of control points are established on the walls and the floor in the experimental hall. The laser tracker is applied to survey these control points by using multi-station lap survey, interconnecting all control points into a unified coordinate system.

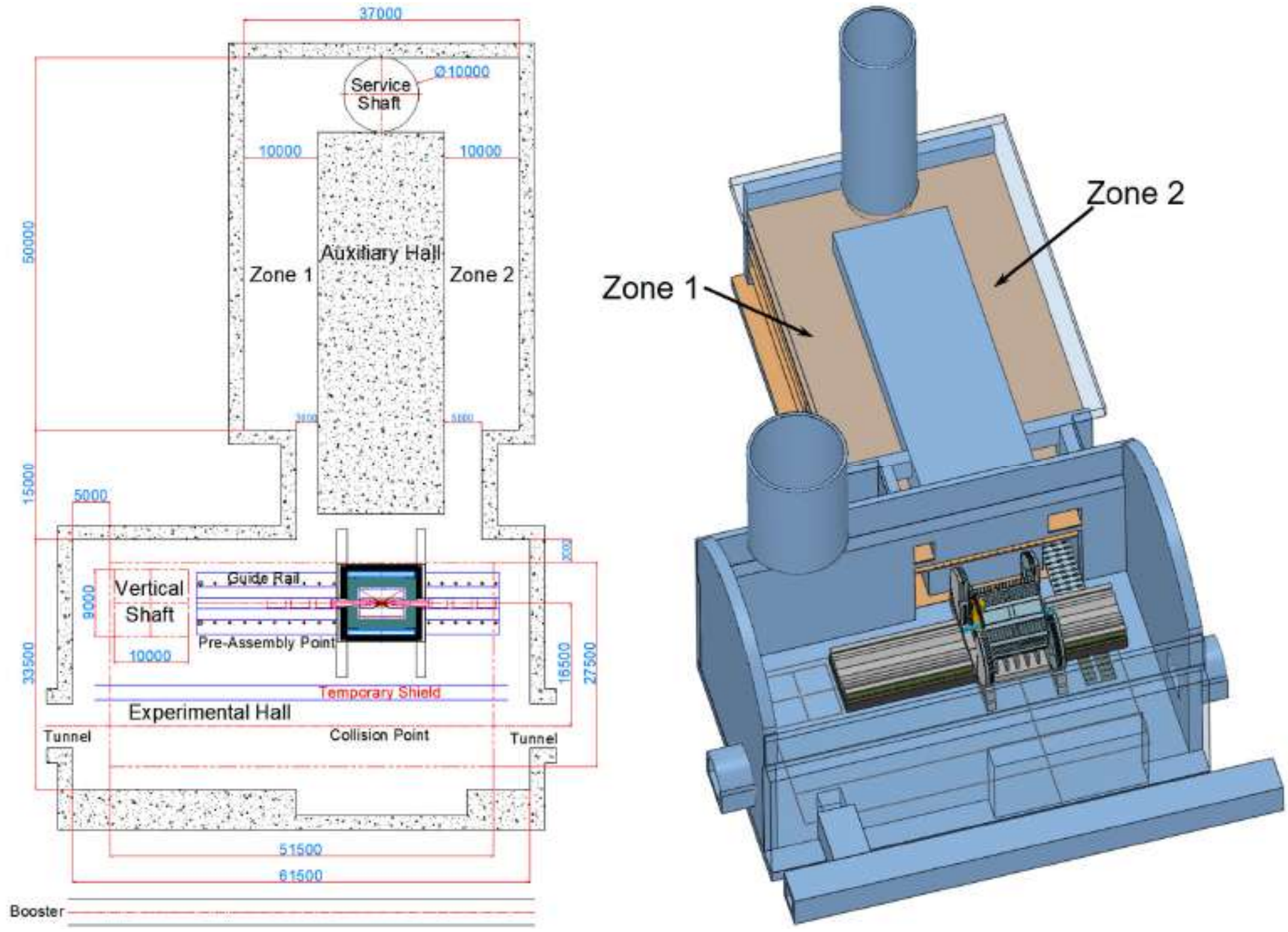

**Figure 14.54:** Layout of the underground hall including experimental hall and auxiliary hall. Units are in millimeters.

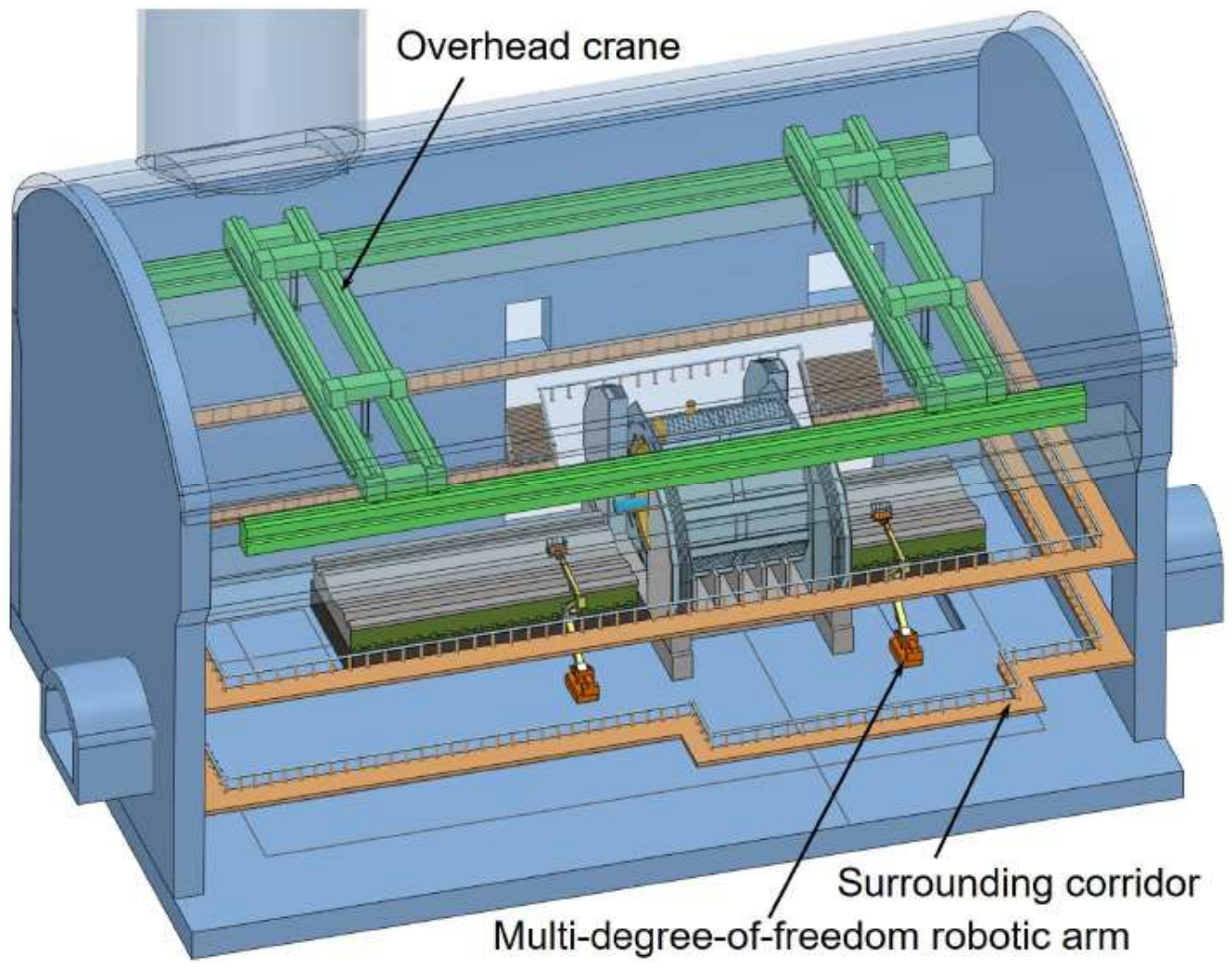

**Figure 14.55:** The layout of the underground experimental hall.

By surveying the control points of the tunnel at both ends of the experimental hall, the alignment control network between the experimental hall and the accelerator tunnel establishes the geometric link. Through inter-survey connection, the coordinates of these control points in the underground experimental hall are established in the global coordinate system, providing a positional references for detector alignment.

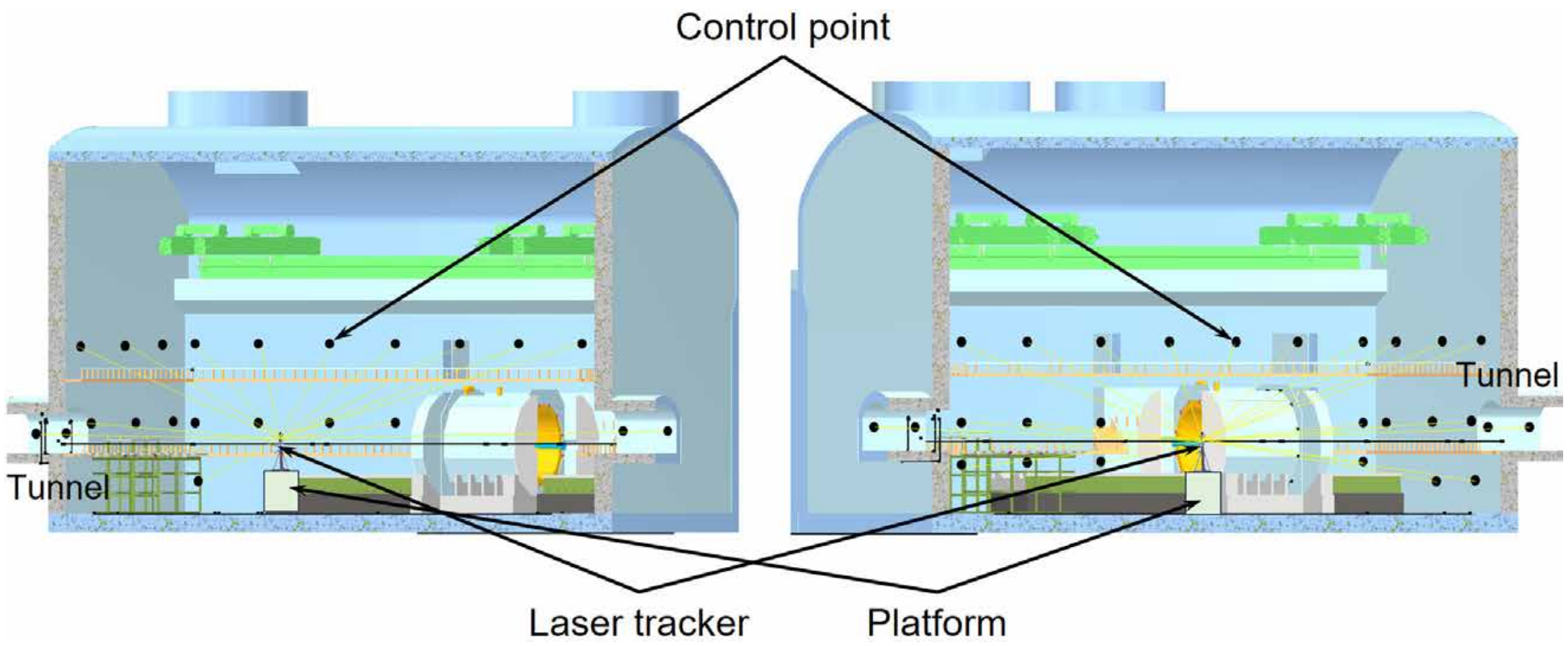

**Figure 14.56:** Alignment control network in the underground experimental hall. Control points on the experimental hall walls and floor are surveyed with a laser tracker via multi-station measurements, integrating them into a unified coordinate system.

### 14.4.2 Auxiliary hall

The underground auxiliary hall accommodates the auxiliary facilities as support systems for installation and operation of sub-detectors and accelerator, while maintaining design coordination with the ground hall.

**1. Auxiliary facility**

The auxiliary facilities that support the installation and normal operation of each sub-detector mainly include hydraulic pump station, air conditioning system and cooling system, electronics system, TPC gas system, cryogenic system, and power distribution system. Key functions of these systems are as follows:

(1) Hydraulic pump station

The hydraulic pump station facilitates detector moving and positioning from pre-assembly point to collision point, sub-detector alignment on guide rails, and final positioning and locking of sub-detectors. The multi-function integrated design of the hydraulic pump system will optimize the maintenance and upgrades access and life-cycle costs.

The integrated pump station system is equipped with three sets of high-pressure pumps and three sets of low-pressure pumps, with a total power of approximately 235 kW.

(2) Air conditioning system and cooling system

The air conditioning system is designed to regulate the temperature and humidity in the underground experimental hall and auxiliary hall, preventing equipment performance degradation and failures caused by excessive temperature and humidity. The system also provides dust filtration, fresh air supply, and hazardous gas exhausts to ensure equipment reliability and personnel safety. It maintains 17 ℃ / ≤70% relative humidity in the experimental hall and 20 ℃ / ≤60% relative humidity in the auxiliary hall [3].

The cooling system is designed for heat dissipation of front-end electronic power-consuming components in each subsystem to ensure normal operation of electronic devices. It includes the water-cooling system and the air-cooling system. The water-cooling system provides pure water with an inlet temperature of 5 ℃ for ITK and OTK, and 15 ℃ pure water for HCAL, ECAL, TPC, and beam pipe assembly. The air-cooling system supplies 5 ℃ dry air to VTX. Air ducts and water pipes operating at 5°C are insulated to prevent condensation.

The air conditioning system and cooling system adopt an integrated, multi-stage design to ensure operational stability, reliability, and energy-efficiency. The systems are equipped with standby parallel equipment. As shown in Figure 14.57, the standby parallel equipment automatically activates when primary system fails, maintaining stable temperature / humidity control and detector cooling in the underground hall.

The air conditioning system and cooling system have a rough estimated total power consumption of 2,050 kW.

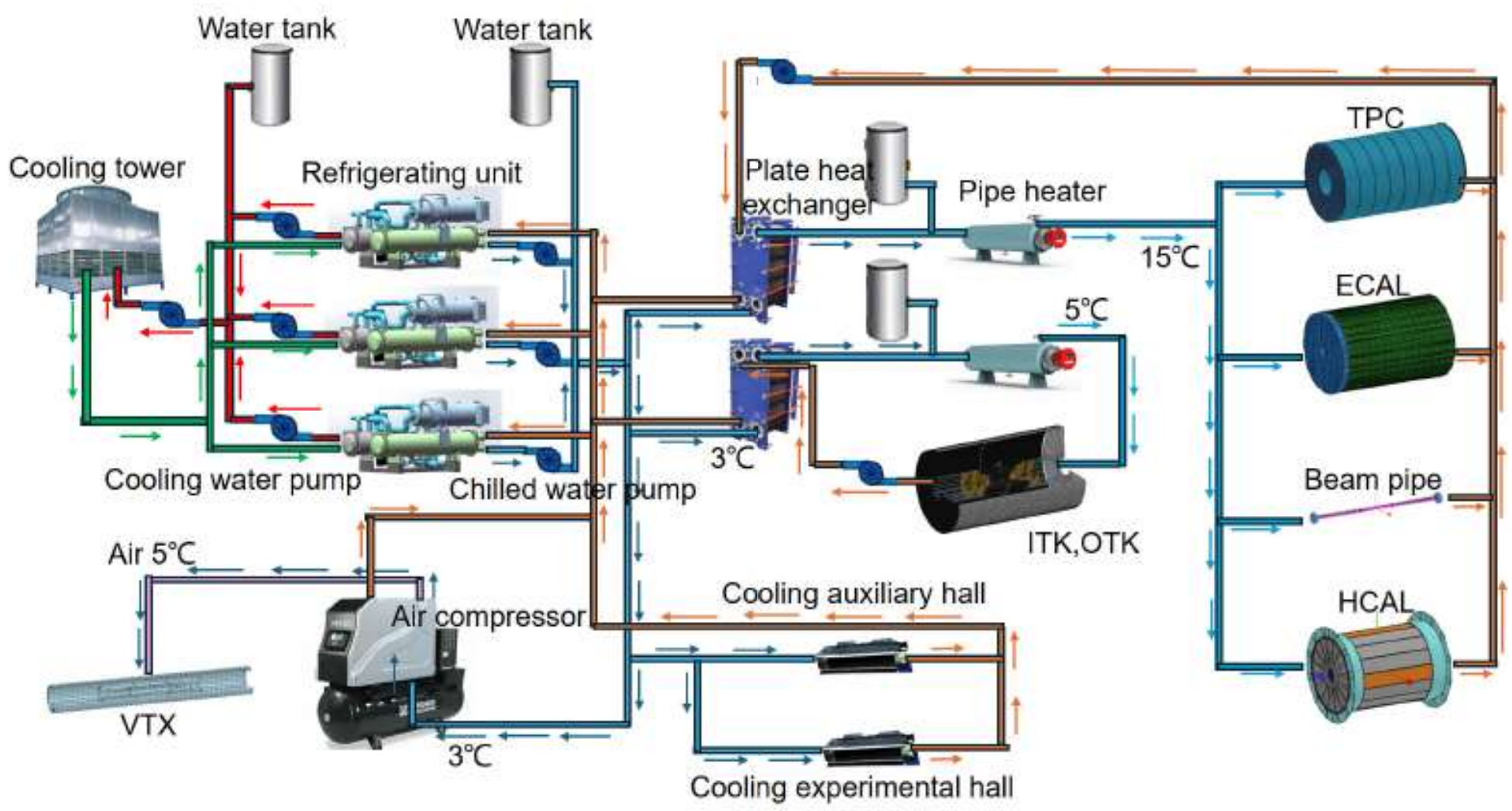

**Figure 14.57:** Flow chart of air conditioning system and cooling system.

(3) Electronics system

The electronics system acquires and processes detector signals for transmission to the DAQ system. Its key components include: readout ASICs for various detector types, signal processing systems; low-voltage power supplies for the readout circuits; high-voltage bias systems for the detectors, and assorted cables for signal and power, etc.

The total power of the electronics system equipment is approximately 830 kW.

(4) TPC gas system

The TPC gas system supplies the working gas for TPC detector with a stable argon-based gas mixture of 0.5 L/min. This platform is equipped with a cascade gas cylinder group, a gas buffer storage tank, and a control and monitoring system (gas composition, flow, temperature, humidity). A connection channel to the shaft in the auxiliary hall is reserved for maintenance access.

The total power of the TPC gas system equipment is approximately 5 kW.

(5) Cryogenic system

The core function of the cryogenic system is to maintain a stable and efficient low-temperature environment for the superconducting magnet. The cryostat ensures thermal stability, keeping the superconducting magnet within its operational temperature range. A high-performance cryogenic transfer and circulation system is required to ensure effective coolant flow thermal equilibrium.

The cryogenic system comprises two subsystems: one for the detector's superconducting magnet and another for the accelerator's magnet, with a total power of approximately 30 kW.

(6) Power distribution system

The power distribution system delivers electricity to all equipment according to their specific requirements. It also incorporates protection devices that immediately isolate faulty circuits in the event of overloads or short-circuits, thereby ensuring equipment safety. Integrated energy monitoring equipment is installed to provide real-time consumption data, enabling energy consumption analysis and management for optimized efficiency and cost reduction.

Currently, the total demand for detector auxiliary facilities is 3,150 kW, of which 235 kW is for the hydraulic pumping station system, 2,050 kW for the air conditioning system and cooling system, 830 kW for the electronics system, 5 kW for the TPC gas system, and 30 kW for the cryogenic system.

**2. Layout of the auxiliary hall**

The underground auxiliary hall is divided into Zone 1 and Zone 2, each comprising 5 floors with a floor area of 500 $m^2$ per level, totaling 5,000 $m^2$ of floor area. As shown in Figure 14.58, the service vertical shaft is located between Zone 1 and Zone 2, to enable personnel access and equipment transportation.

Base on multiple engineering requirements including wiring specifications, installation and transportation constraints, floor load-bearing capacities, and interface coordination with ground facilities, the auxiliary facilities are strategically arranged across different zones and floors, as shown in Figure 14.58:

(1) The electronic system is placed on the first, second and third floor of Zone 2.

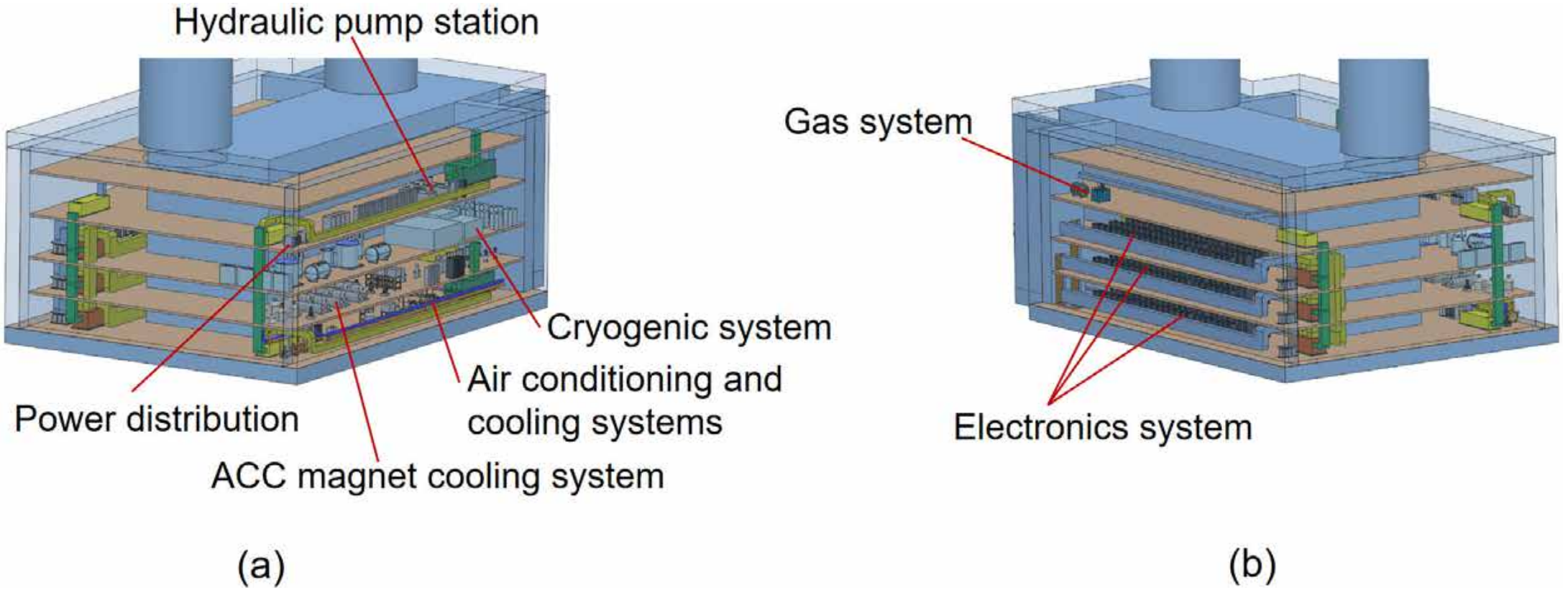

**Figure 14.58:** Layout of the auxiliary hall and arrangement of the auxiliary facilities. The hall has two zones (Zone 1 and zone 2), each with 5 floors. A service vertical shaft between zones provides access and equipment transport. (a) Equipment arrangement in Zone 1. (b) Equipment arrangement in Zone 2.

(2) The TPC gas system is placed on the fourth floor of Zone 2.
(3) The air conditioning system and the detector cooling system are placed on the first floor of Zone 1.
(4) The accelerator magnet water cooling system is placed on the second floor of Zone 1.
(5) The cryogenic system is placed on the third floor of Zone 1.
(6) The hydraulic pump station system and power distribution system are placed on the fourth floor of Zone 1.
(7) The fifth floor of Zone 1 and Zone 2 is a common-usage area for the detector and the accelerator during installation and operation, used for storing tools, spare parts, etc.

### 14.4.3 Passageway for the underground hall

The passageway between the ground hall, underground experimental hall and auxiliary hall shall meet the requirements for personnel access and equipment hoisting and transporting, while also considering radiation safety as well as fire prevention and emergency escape routes.

As shown in Figure 14.59, the underground experimental hall is equipped with a vertical shaft and two surrounding corridors, while the auxiliary hall is equipped with a service shaft. The two halls are interconnected via a passageway system that links the ground level, two surrounding corridors of the underground experimental hall and the first, third and fourth levels of the underground auxiliary hall. This integrated connection system facilitates the personnel access, equipment transfer as well as cable and pipeline routing between two halls.

After the entire detector and installation guide rails are in place at the collision point, as shown in Figure 14.60, the temporary shielding wall is moved to the entrance of the passageway between the experimental hall and the auxiliary hall for permanent use. This solution provides both cost and time savings, while facilitating personnel access, equipment transfer, and cable and pipe installation. The design of the temporary shielding wall should therefore take into account its permanent function and incorporate a safety interlocking gate to prevent accident entry from the auxiliary hall during detector operation.

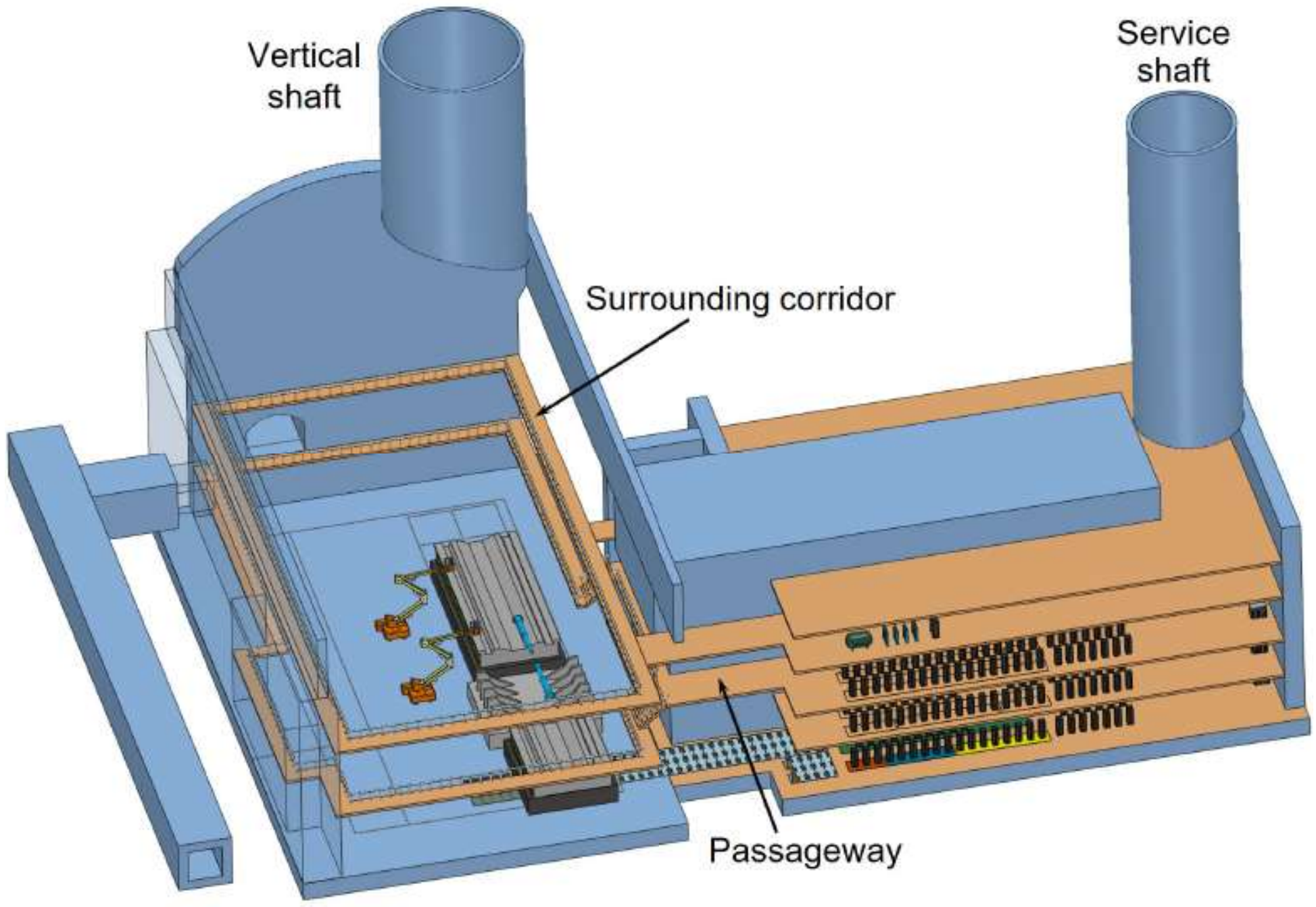

**Figure 14.59:** The design of passageway to the experimental hall and the auxiliary hall.

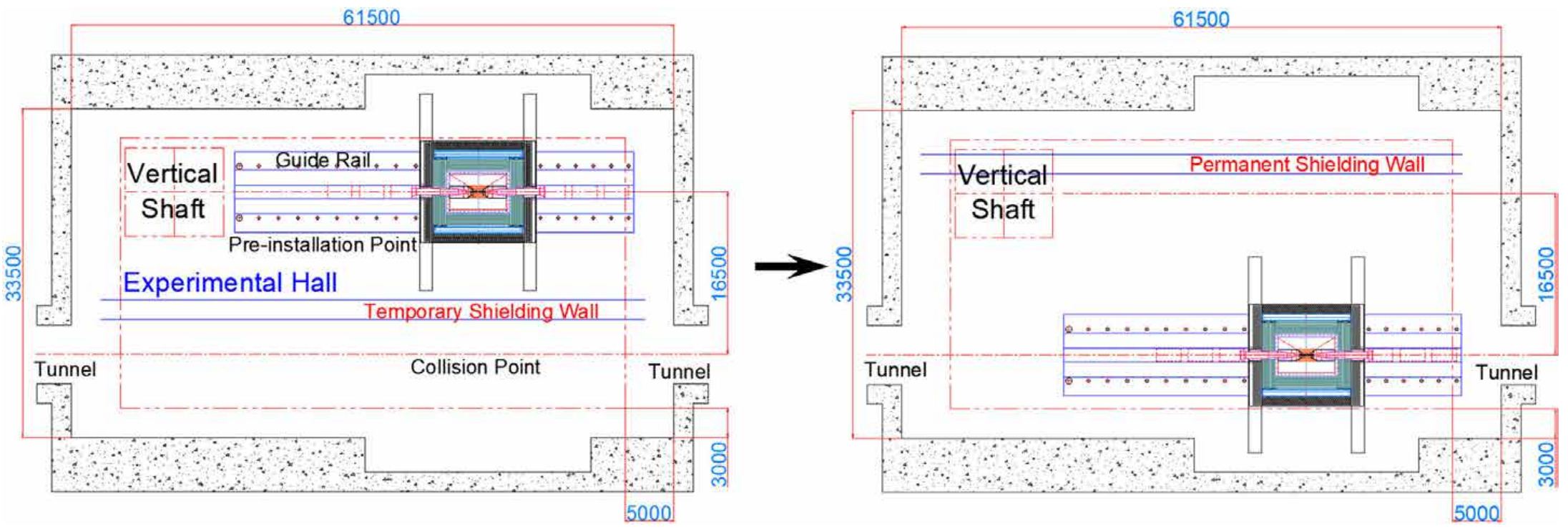

**Figure 14.60:** Permanent use design of the temporary shielding wall. After positioning at the collision point, the temporary shielding wall will be relocated to the entrance of the passageway between the experimental hall and the auxiliary hall for permanent use. Units are in millimeters.

## 14.5 Summary

The mechanical system design for the CEPC Reference Detector has established three key outcomes: the overall detector layout, the integration scheme for sub-detector connections, and the pre-assembly installation plan. This foundational work guides all subsequent detailed mechanical design of the sub-detectors. Preliminary designs for the underground experimental hall and auxiliary hall now accommodate detector installation, operation, and maintenance, while the design of the ground hall remains pending.

With ongoing advancements in detector technology and electronics, the mechanical design will be further optimized, particular in the alignment accuracy measurement, structural optimization of carbon fiber, cooling system development, and enhancement of mechanical system stability and performance.

International collaboration will address critical technical challenges. Participation in international workshops and technical exchanges will foster knowledge sharing, diversify funding, and mitigate project risks while promoting scientific achievements.

# Chapter 15 Detector and physics performance

The overall performance of the Reference Detector at the CEPC is evaluated using a detailed Geant4 model [1] and complete reconstruction of the simulated events. Most of the sub-detectors are designed with detailed engineering considerations, including mechanical support structures, electronics, cabling, dead material and cracks. The material budget associated with the support structures and services is based on the best current estimates from the detector R&D groups. Using full simulation and thoroughly developed reconstruction algorithms helps ensure that the performance is as realistic as possible. Events are reconstructed using a sophisticated algorithmic chain, which includes Kalman-filter-based track reconstruction and the CyberPFA particle flow algorithm, as described in Section 13.3. A description of the detector parameters and reconstruction software can be found in Chapter 13.

The performance of physics objects from the Reference Detector[1] is discussed in Section 15.1.8.2. Then, a series of physics studies, pursued using fully simulated Monte Carlo samples at various center-of-mass energies ranging from 91 to 360 GeV, are presented in Section 15.2.11. These analyses are selected for their complex and complementary final states, thereby demonstrating the Reference Detector capability to fulfill the CEPC physics program. Section 15.3 outlines strategies for measuring the absolute luminosity, usage of resonant depolarization to measure the $Z$ and $W$ boson masses with high precision, and methods for the calibration and alignment for the sub-detectors. Section 15.4 provides a summary of the key performance metrics achieved with the Reference Detector and the primary areas where detector configuration optimizations and technology decisions could be further explored.

## 15.1 Detector performance

To assess the efficacy of the Reference Detector and associated reconstruction software, an in-depth examination of the performance metrics for tracking, PID, vertexing, and particle flow algorithms is conducted.

### 15.1.1 Tracking

The tracking system of the Reference Detector consists of three subsystems capable of standalone reconstruction: the VTX, the ITK, and the TPC. These are augmented by an auxiliary tracking system, the OTK, which uses the AC-LGAD and provides both additional high-resolution measurement points and time-of-flight information. The TPC provides full coverage down to $\theta \approx 32°$, beyond which the number of measurement points decreases. The forefront

[1] The reference cartesian system is chosen to be aligned with the $z$-axis along the incoming $e^-$ beam and the $y$-axis perpendicular to the accelerator plane. The forward (backward) direction indicates particles at the polar angles $\theta$ close to 0° (180°).

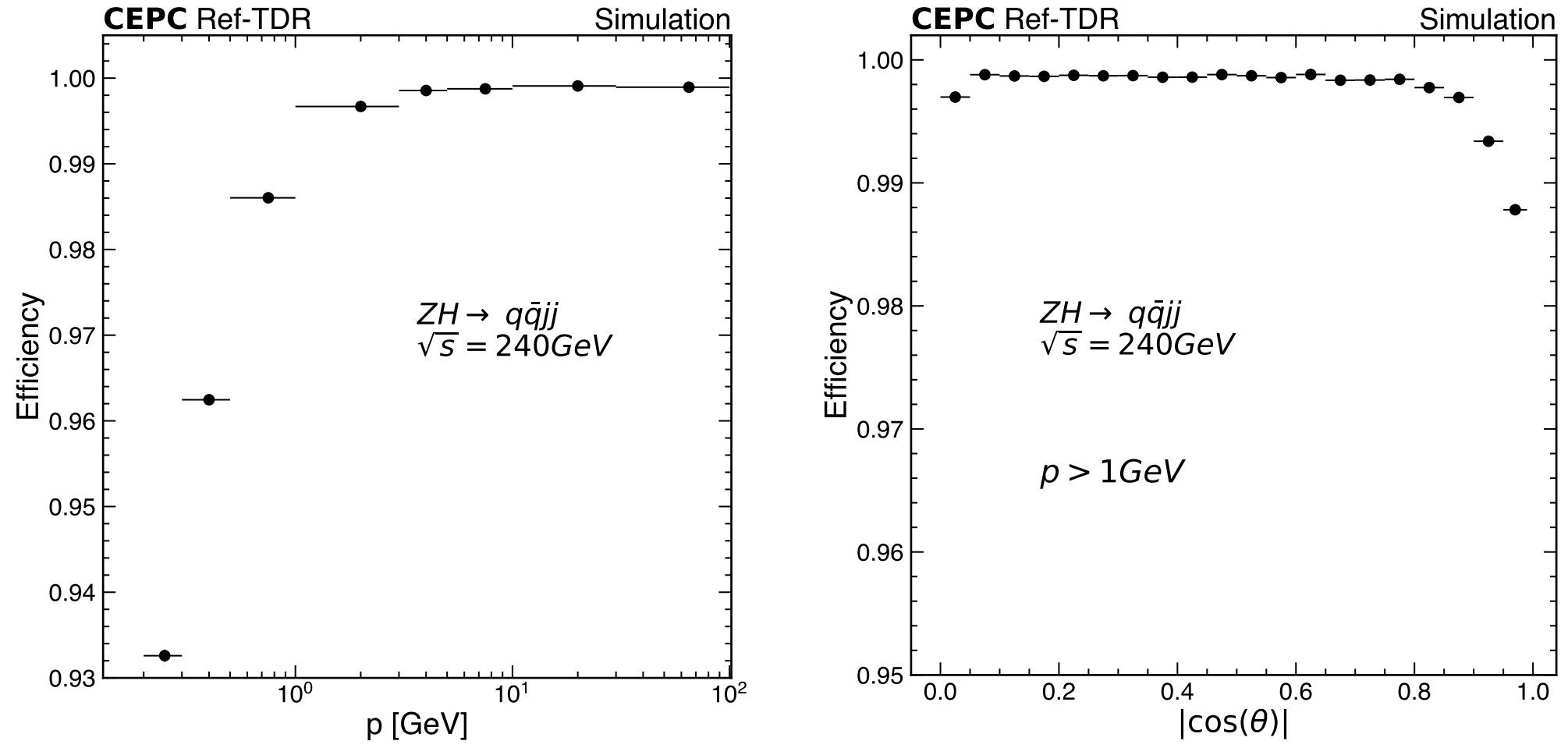

**(a)** **(b)**

**Figure 15.1:** Track reconstruction efficiency as a function of the (a) track momentum and (b) track $\cos(\theta)$ in $ZH \longrightarrow q\bar{q}jj$ events at $\sqrt{s}$ = 240 GeV.

measurement point provided by the TPC is at $\theta \approx 11.7°$. The barrel inner tracking system, comprising the six-layer VTX and the three-layer ITK, offers nine precise measurements down to $\theta \approx 32°$. The ITK endcap supplies up to a maximum of four measurement points for tracks at small polar angles. The OTK provides a single high-precision measurement point with a large lever arm outside the TPC volume down to $\theta \approx 8.1°$.

#### 15.1.1.1 Tracking efficiency

With numerous continuous readout layers, track reconstruction in TPCs is generally efficient, even in environments with significant background noise. Moreover, the standalone tracking capabilities provided by the VTX and ITK allow for reconstruction of low-transverse-momentum ($p_T$) tracks that do not reach the TPC. The endcap coverage of the ITK facilitates track reconstruction at polar angles as low as approximately $\theta \approx 8.1°$.

Figure 15.1 shows the track reconstruction efficiency as a function of the momentum and polar angle in $ZH \longrightarrow q\bar{q}jj$ events at the center-of-mass energy of $\sqrt{s}$ = 240 GeV. Efficiencies are calculated relative to the stable charged particles at the generator level, originating within 10 cm around the IP, with $p_T$ > 100 MeV and $|\cos(\theta)| < 0.99$. This selection excludes decays-in-flight and requires hit purity of at least 90%. The hit purity is defined as the fraction of a track hits that are associated with its corresponding generator-level particle. On average, the combined tracking system achieves the track reconstruction efficiency of 99.7% for track momenta greater than 1 GeV across the whole polar angle. When $|\cos(\theta)| < 0.05$, the efficiency decreases because of the membrane cathode situated between the two rings at the TPC center. Figure 15.2 shows that the efficiency drops at low momentum and at large $\cos(\theta)$.

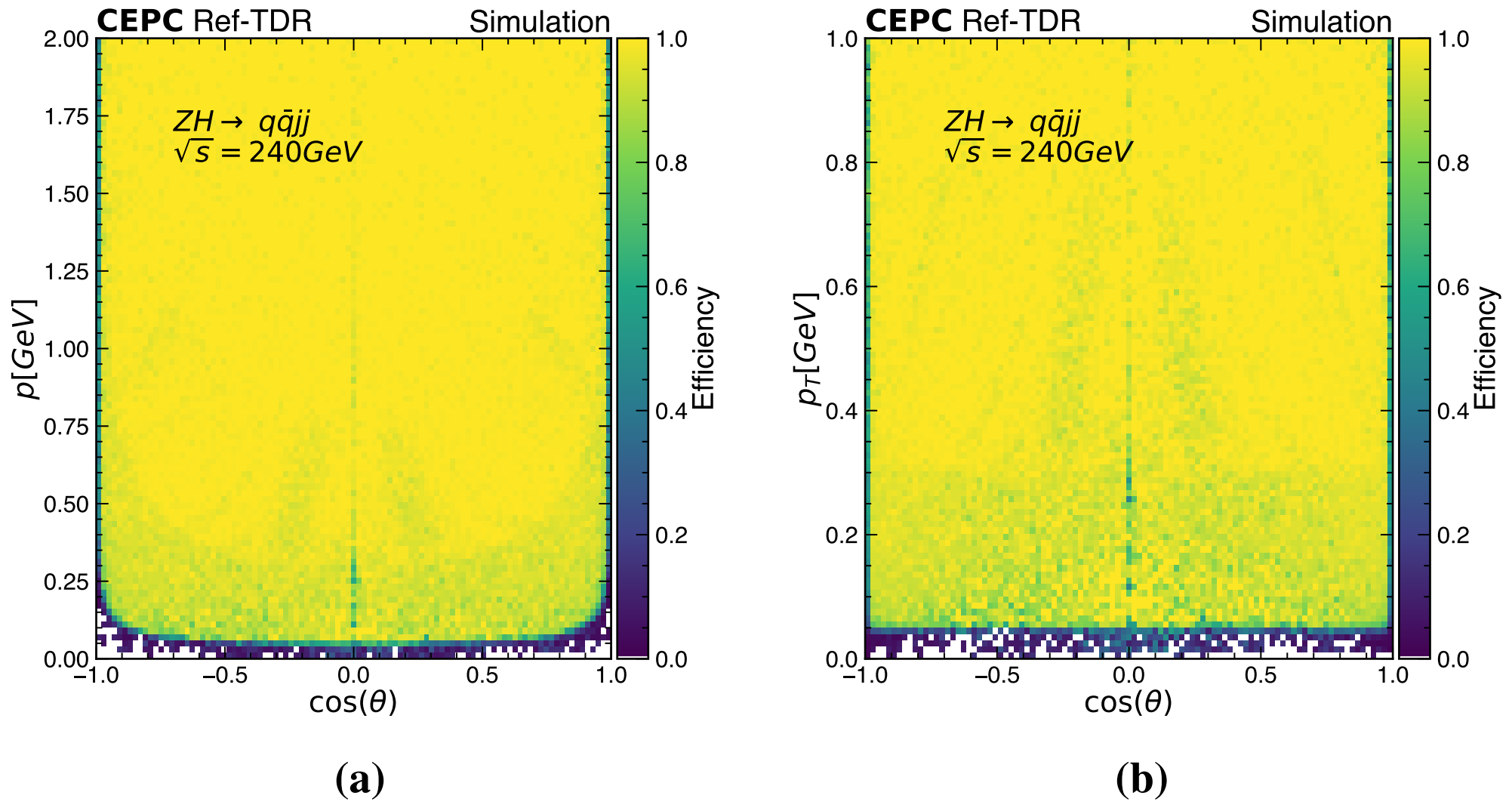

**Figure 15.2:** Track reconstruction efficiency in the low momentum region for $ZH \to q\bar{q}jj$ at $\sqrt{s}$ = 240 GeV presented against the 2D plane of (a) momentum $p$ or (b) transverse momentum $p_{\mathrm{T}}$ versus $\cos(\theta)$.

#### 15.1.1.2 Momentum resolution

Figure 15.3a shows track momentum resolution across various configurations of the tracking system. The TPC (OTK) significantly improves resolution for track $p_{\mathrm{T}} \gtrsim 0.3$ (0.8) GeV, where tracks start to reach the detector. In particular, OTK provides the longest possible radial lever arm for the track fit.

The momentum resolution with the entire tracking system at different polar angles is shown in Figure 15.3b. The study is conducted with muons generated at fixed polar angles of $\theta = 20°$, $40°$ and $85°$, with momenta varying from 1 to 100 GeV. At the polar angle of 85°, its resolution is compared with the parametric form of $\sigma_{1/p_{\mathrm{T}}} = a \oplus b/(p \cdot \sin^{3/2}\theta)$, with $a = 2.9 \times 10^{-5}$ GeV$^{-1}$ and $b = 1.2 \times 10^{-3}$. For high momentum tracks, the asymptotic value of the momentum resolution is $\sigma_{1/p_{\mathrm{T}}} = 3 \times 10^{-5}$ GeV$^{-1}$. In the forward region, the momentum resolution is inevitably degraded because of the relatively small angles between the magnetic field and the track momenta.

#### 15.1.1.3 Impact parameter resolution

The performance of track impact parameters is detailed in Section 4.7. For $r$-$\phi$ impact parameter resolution, the required performance is achieved for tracks with momentum down to 0.5 GeV, while it exceeds expectations for high-momentum tracks where the asymptotic resolution is close to 2 μm. The $z$ impact parameter resolution is similar to the $r$-$\phi$ resolution and reaches an asymptotic value of ≤ 3 μm for the whole barrel region.

The effects of beam-induced background and electronics noise are evaluated and found

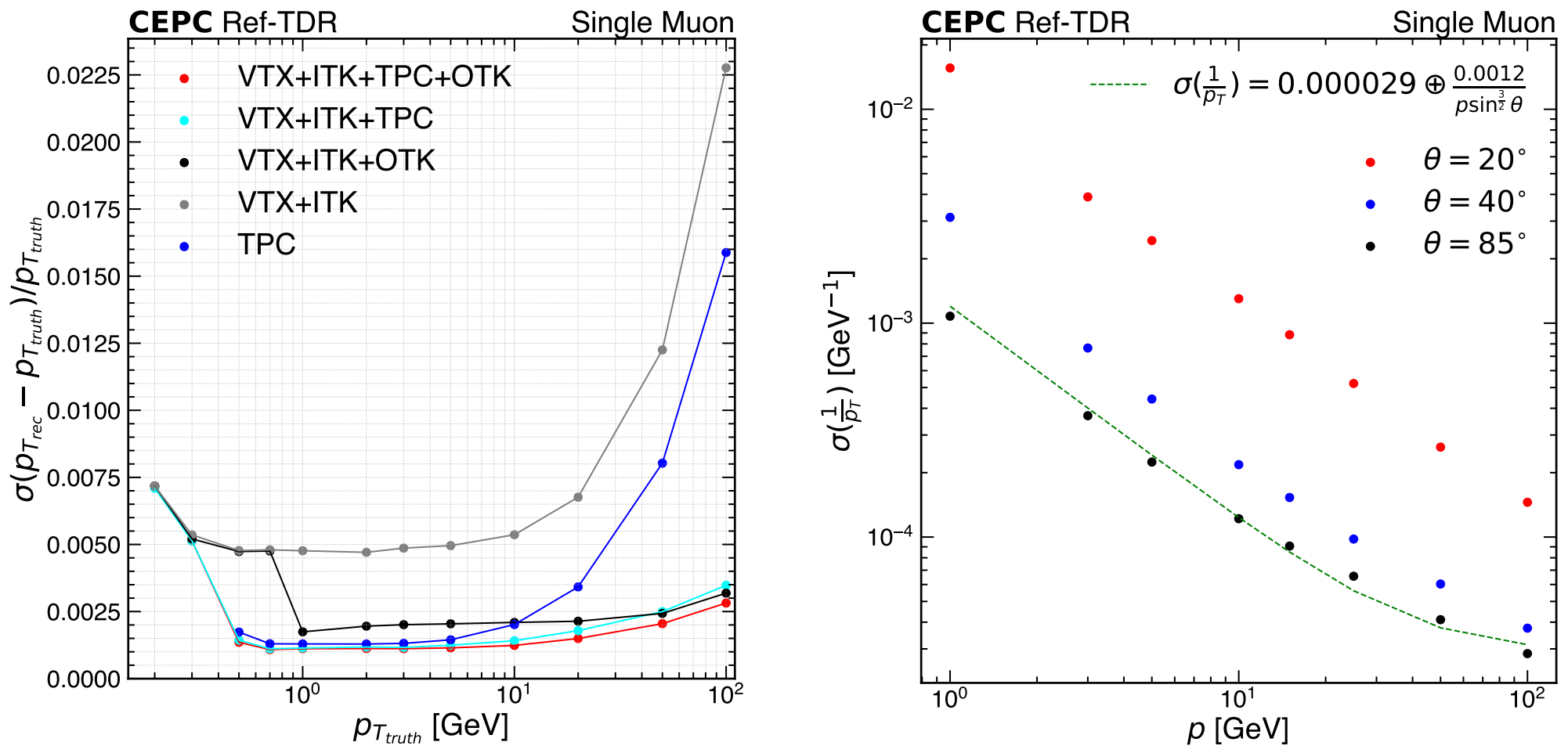

**(a)** **(b)**

**Figure 15.3:** Track transverse momentum resolution for (a) different configurations of the tracking system at the polar angle of 85° and (b) single muon events as a function of the momentum for different polar angles. The line shows $\sigma_{1/p_{\mathrm{T}}} = 2.9{\times}10^{-5} \oplus 1.2{\times}10^{-3}/(p{\cdot}\sin^{3/2}\theta)$.

to have only very few percent increase on the resolutions and less than 0.5% reduction on the efficiency for tracks with $p_{\mathrm{T}} \leq 1$ GeV. The impact from a non-uniform magnetic field on the tracking is evaluated using ACTS framework and found to be negligible on both the efficiency (reduced by $\leq 0.1\%$) and resolution (a maximum degradation of 5%).

## 15.1.2 Leptons

The Reference Detector particularly with its highly segmented calorimetry, provides a wealth of data crucial for identifying leptons. Electrons can be identified by their characteristics: the narrow electromagnetic shower patterns in the ECAL, which correlate with tracks detected in the tracking system. Muons are recognized as particles with minimal ionization in the calorimeter, confirmed by matching tracks in both the tracker and the muon detector.

An eXtreme Gradient Boosting (XGBoost) machine learning library [2] is employed to distinguish muons and electrons from charged hadrons, leveraging information from multiple sub-detectors. Energy deposit and cluster shape in the ECAL and HCAL, various PFO information, track consistency with the muon detector hits, $\mathrm{d}N_p/\mathrm{d}x$ in TPC, and ToF measurement in OTK are considered as inputs for the classifier.

In particular, the $\mathrm{d}N_p/\mathrm{d}x$ and time-of-flight variables, described in Section 15.1.4, can offer additional means to discriminate electrons, muons, and hadrons by identifying their masses.

The XGBoost models are trained and tested with the $ee \to ZH$ simulated samples, in bins of the PFO momentum and $\theta$. The model outputs the probabilities for each PFO to be a muon, an electron, or a hadron. Different Working Points (WPs) are provided: either with a fixed efficiency, such as 50%, 70%, 90% and 98%, or with BEST WP, the optimal working point, which identifies the PFO according to its highest probability among the three categories. In all

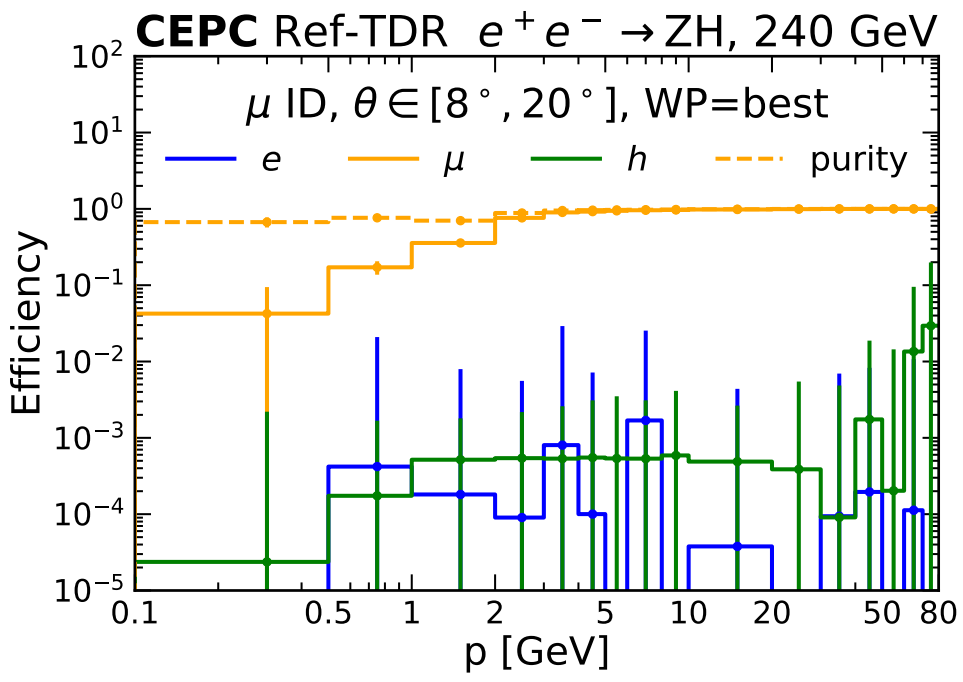

**(a)**

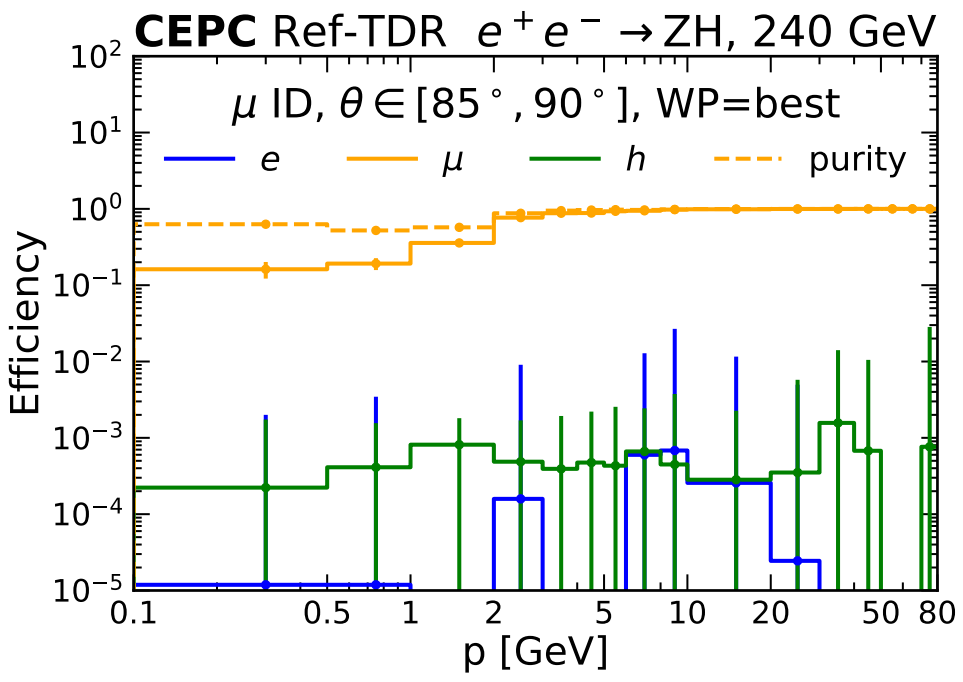

**(b)**

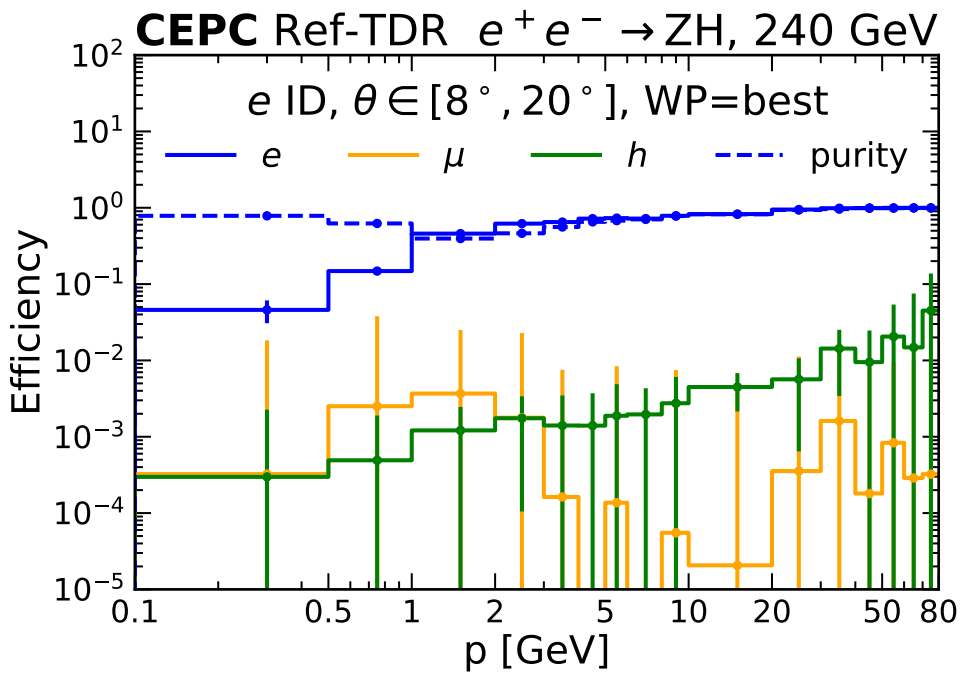

**(c)**

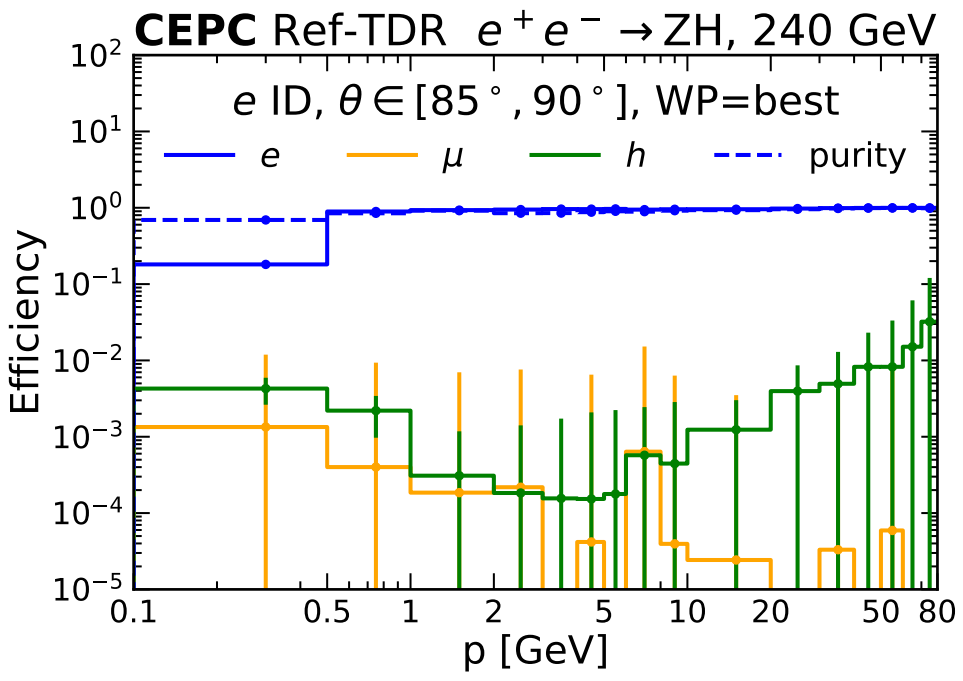

**(d)**

**Figure 15.4:** The ID efficiency, purity, and misidentification rates with the BEST WP as a function of particle momentum, obtained with the XGBoost models for (a, b) muons and (c, d) electrons with (a, c) $\theta \in [8, 20]^\circ$ and (b, d) $\theta \in [85, 90]^\circ$. "h" represents the charged hadrons.

cases, each PFO is identified in the order of a muon, an electron, and a charged hadron.

The performance metrics for the BEST WP are presented in Figure 15.4. The muon identification efficiency falls below 50% for momenta less than 2 GeV; notably, the efficiency in the endcap region is lower than that in the barrel region. However, as the momentum surpasses 2 GeV, the efficiency rises significantly, reaching levels above 90% and approaching 100%. The misidentification rate for electrons and charged hadrons is predominantly below 0.1%, demonstrating the excellent discriminative capabilities of the algorithm. Similar behaviors can be observed in electron identification, although the misidentification rate from the charged hadrons is higher, approximately 1% at high momenta. The purity of both electron and muon identification is predominantly above 90%, indicating a high level of accuracy. For isolated leptons with momenta greater than 5 GeV, the misidentification rate from charged hadrons is smaller than 2% for a 99% efficiency of lepton identification.

Beyond this initial lepton ID development, further optimizations are left for future studies. They will occur alongside the improvement of other event reconstruction algorithms, which include the electron tracking algorithm using Gaussian Sum Filter, enhanced clustering algorithms of energy deposits in the HCAL, and standalone track reconstruction in the muon detector.

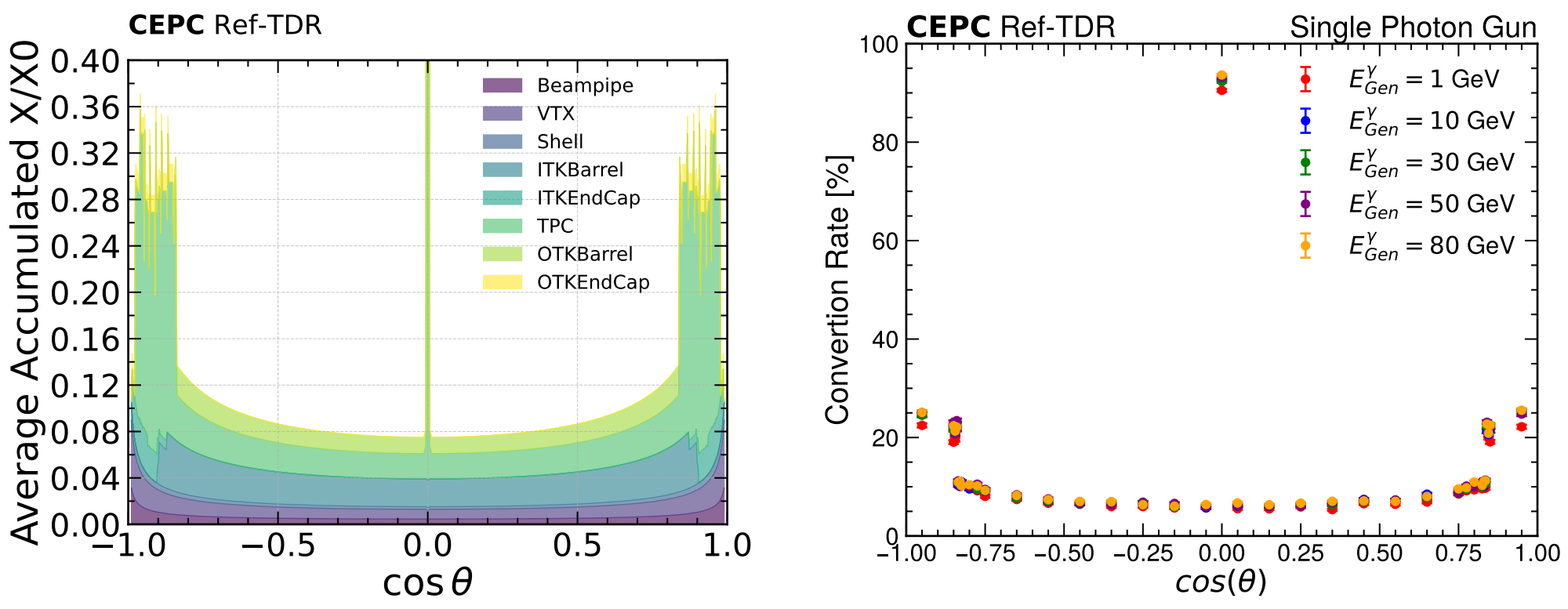

**(a)** **(b)**

**Figure 15.5:** (a) The amount of material in the unit of radiation length inside the tracker and (b) the conversion rate of photons with different energies as a function of polar angles, with bins representing fixed values of the cosine of the polar angle.

### 15.1.3 Photons

Photons have similar signatures to electrons in the calorimeter, but generally do not have tracks that match the EM clusters. However, 6–10% of photons in the central region and about 25% of photons in the forward region convert to $e^+e^-$ pairs through their interaction with the materials in front of the calorimeter. Some of these converted photons may have tracks matching the EM clusters. Figure 15.5 shows the amount of material in units of radiation length and the photon conversion rates at different polar angles.

For unconverted photons, their reconstruction efficiency and energy resolution are described in Section 7.3.4. For photons with energy above 3 GeV, the reconstruction efficiency reaches above 99% within statistical uncertainty. Due to the minimal stochastic term inherent in homogeneous calorimeters, the photon energy resolution reaches a level of well below 1% for photons with energy above 10 GeV.

Unconverted photons, reconstructed as neutral PFOs, need to be distinguished from other neutral PFOs, which are predominantly $K^0_{\mathrm{L}}$ in the $e^+e^-$ collision environment. Similar to lepton identification, an XGBoost-based algorithm is employed, utilizing similar input features as those described in Section 15.1.2, except for those originating from the tracker and muon detector. The XGBoost models are trained in bins of particle momentum and $\theta$ using single-particle simulation samples of photons and $K^0_{\mathrm{L}}$. Particles with a probability higher than 0.5 of being a photon are regarded as photons.

Examples of photon ID efficiency and $K^0_{\mathrm{L}}$ misidentification rate are shown in Figure 15.6. For the photons with energy above 3 GeV, the ID efficiency remains stable above 95%. The $K^0_{\mathrm{L}}$ misidentification rate is around 2% at $p < 10$ GeV, and around 1% at $p > 10$ GeV. No significant difference is observed between the barrel and endcap regions. Photon ID against merged photons from high-$p_{\mathrm{T}}$ neutral pion decays is under investigation.

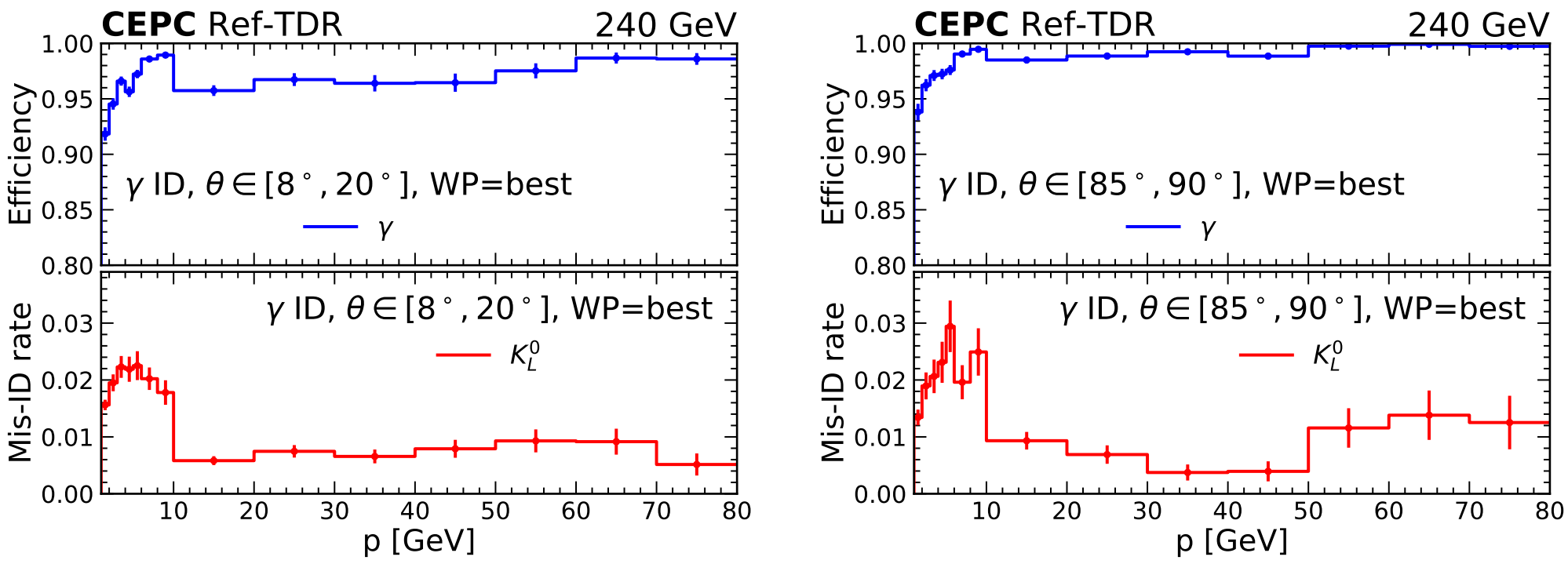

**(a)** **(b)**

**Figure 15.6:** The photon ID efficiency and $K_{\mathrm{L}}^0$ misidentification rate as a function of particle momentum with (a) $\theta \in [8, 20]^\circ$ in the endcap region and (b) $\theta \in [85, 90]^\circ$ in the barrel region.

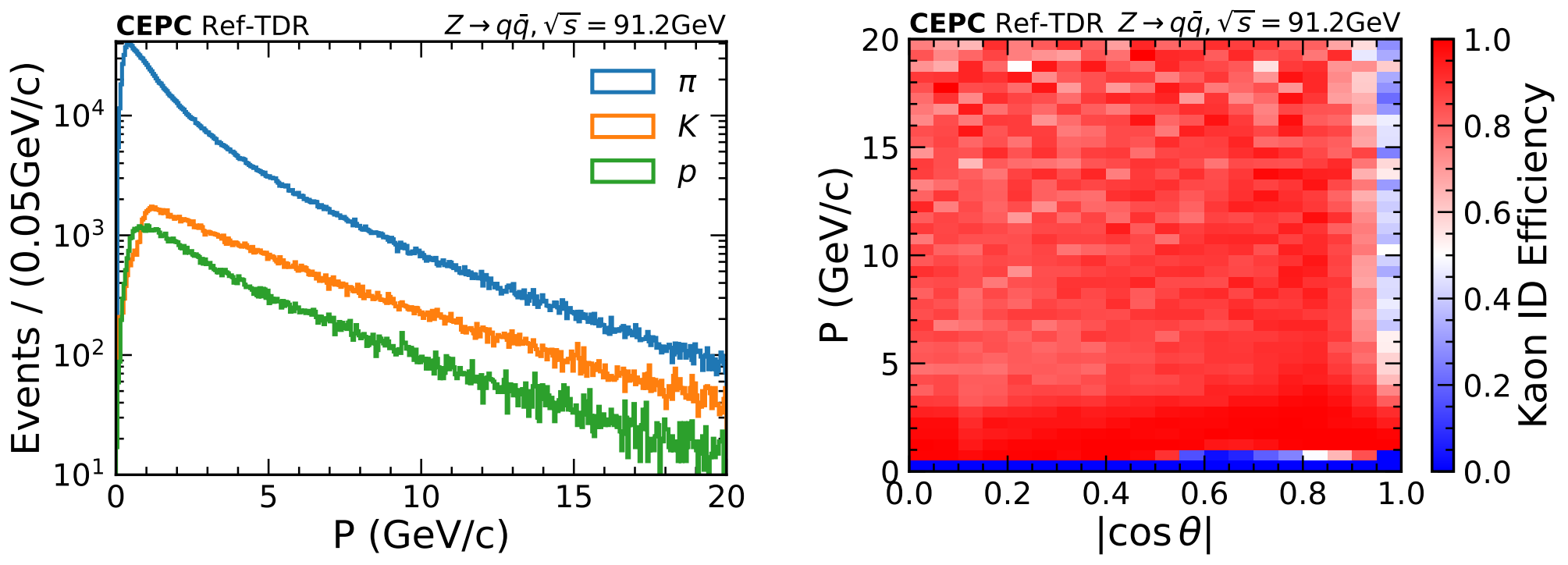

**(a)** **(b)**

**Figure 15.7:** (a) Generator-level momentum distributions of pions, kaons and protons after requiring no decay in the $Z \to q\bar{q}$ sample. (b) Kaon efficiency in $Z \to q\bar{q}$ events is shown as a function of momentum and $|\cos(\theta)|$.

### 15.1.4 Charged hadrons

Measuring the energy loss of charged particles in the TPC is a powerful tool for identifying particle species. Compared with the $\mathrm{d}E/\mathrm{d}x$, the $\mathrm{d}N_p/\mathrm{d}x$ improves particle identification by reducing statistical fluctuations in the secondary ionization charges. The $\mathrm{d}N_p/\mathrm{d}x$ reconstruction algorithm and its separation powers for pions, kaons and protons is described in Section 6.4.4.2. The ToF measurement provides additional separation power, particularly in the low-momentum region below 3 GeV as shown in Section 5.5. The performance achieved by combining $\mathrm{d}N_p/\mathrm{d}x$ with the ToF measurement is presented in Section 6.4.4.2.

The performance of kaon identification is evaluated with $Z \to q\bar{q}$ events. Figure 15.7a presents the momentum distributions of pions, kaons and protons. Figure 15.7b illustrates the identification efficiency of kaons as a function of momentum and $|\cos(\theta)|$ based on the combined $\chi^2$ of $\mathrm{d}N_p/\mathrm{d}x$ and ToF. The kaon identification efficiency and purity reach 91% and 86.7%, respectively, in the $Z \to q\bar{q}$ events.

Alternatively, similar to lepton identification, XGBoost models are trained using only the ToF and $\mathrm{d}N_p/\mathrm{d}x$ features and are categorized in regions of particle momentum and $\theta$. The overall kaon efficiency is improved to around 92%, and the purity is increased to approximately 91%.

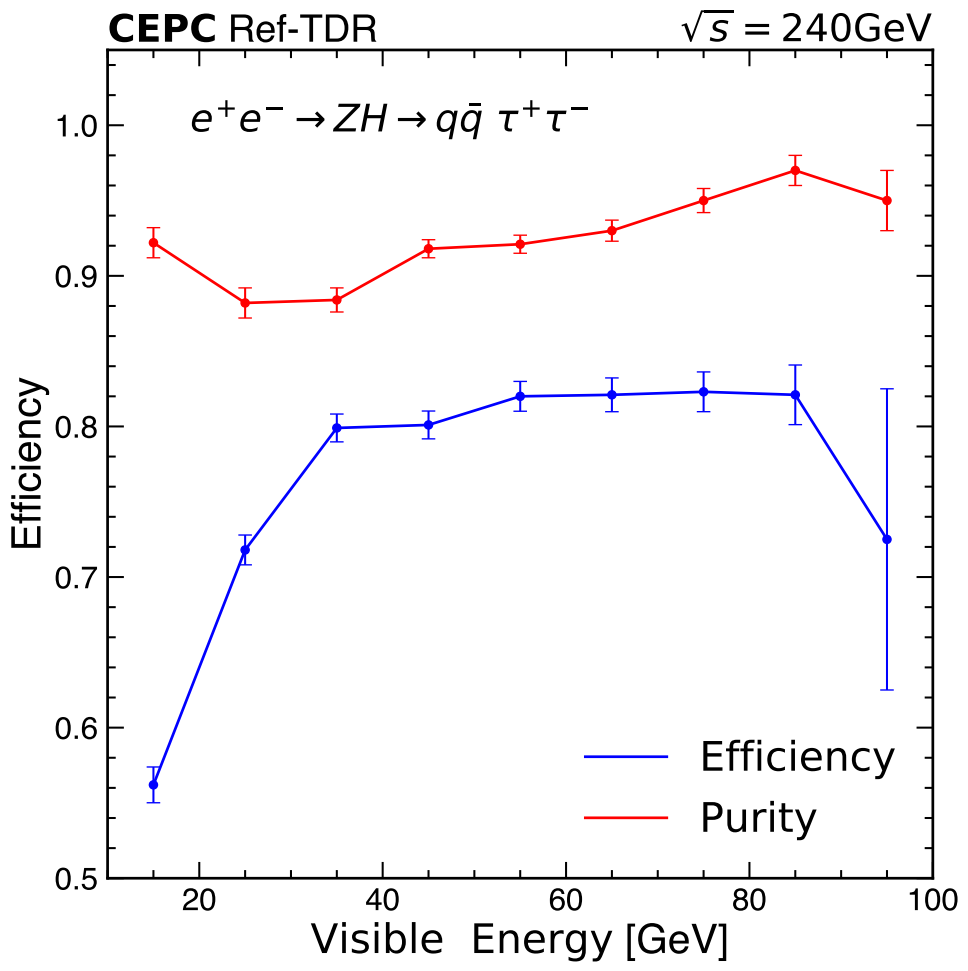

**Figure 15.8:** The efficiency and purity as functions of the visible energy of the $\tau$-lepton candidates. Both the efficiency and purity are determined from $e^+e^- \to ZH \to q\bar{q}\tau^+\tau^-$ events at $\sqrt{s}$ = 240 GeV.

### 15.1.5 Tau leptons

Tau leptons, being the heaviest leptons, play a distinctive role in Higgs boson physics research. Their leptonic decays, namely $\tau \to e\nu_e\nu_\tau$ and $\tau \to \mu\nu_\mu\nu_\tau$, are not distinguishable from prompt electrons or muons. The hadronic decays of tau leptons appear in the detector as narrow jets with a low particle multiplicity. A preliminary identification algorithm is developed for those hadronic decays. This algorithm starts from a seed track whose energy exceeds 1.5 GeV, and gathers charged and neutral PFOs within a small cone of 0.12 radians to form a $\tau$-lepton candidate. The invariant mass of the PFOs within this cone must fall within the range of 0.01–2 GeV, aligning with the $\tau$-lepton mass. Lastly, the $\tau$-lepton candidate must be isolated, with the total energy within a surrounding cone of 0.12–0.31 radians to be less than 8% of the $\tau$-lepton candidate energy. The primary background consists of jets originating from partons, which highly resemble the signal. The efficiency and purity as functions of the visible energy of the $\tau$-lepton candidates are presented in Figure 15.8, measured from $e^+e^- \to ZH \to q\bar{q}\tau^+\tau^-$ events. For visible energy between 10–100 GeV, the efficiency approaches 80%, and the purity surpasses 90%. The efficiency loss is primarily attributed to the large cone size required for isolation. Further optimizations are left for future studies.

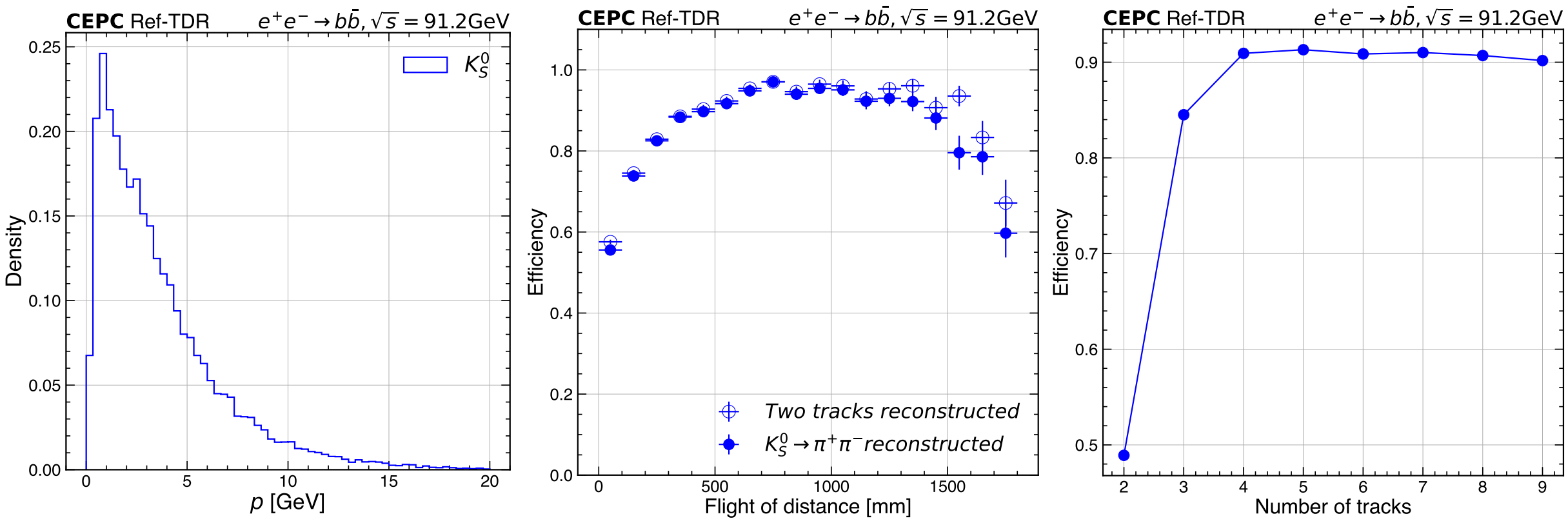

**(a)** **(b)** **(c)**

**Figure 15.9:** (a) Momentum spectrum of $K_S^0$, (b) reconstruction efficiency of $K_S^0 \longrightarrow \pi^+\pi^-$ as a function of the $K_S^0$ flight distance and (c) inclusive reconstruction efficiency of the secondary vertices as a function of the number of tracks associated with the vertex in the $e^+e^- \longrightarrow b\bar{b}$ events.

## 15.1.6 Vertexing

Vertexing plays a crucial role in event reconstruction by identifying the points where particles originate or decay. Following LCFIPlus [3], a similar algorithm is developed.

### 15.1.6.1 Vertex efficiency

In flavor physics, identifying mesons such as $B$ or $D$ from a dense track environment near the interaction point is challenging. Accurately reconstructing secondary vertices helps suppress combinatorial background and enhance signal purity. The performance benchmark for flavor physics (Section 15.2.10) demonstrates how effectively the vertexing algorithm can distinguish signal decays from random track combinations.

For particles with a relatively longer lifetime, they can travel a macroscopic distance before decaying, often producing displaced vertices that serve as key signatures. Efficiency in this context is evaluated using $K_S^0 \longrightarrow \pi^+\pi^-$ events. For the momentum spectrum shown in Figure 15.9a, the efficiency is about 70% and is presented as a function of the particle flight distance in Figure 15.9b.

Contrary to exclusive reconstruction mentioned above, inclusive reconstruction is performed without any pre-defined characteristics of the target events. Similar to Ref. [3], the algorithm examines all track pairs, discards false candidates based on vertex fitting, and then attempts to attach previously discarded tracks to remaining vertex candidates. This procedure iterates until no more tracks can be connected to the vertex candidates while meeting specific selection criteria as follows. These criteria include the collinearity between the candidate position vector and the total momentum of all associated tracks, and constraints on the energy and invariant mass of those tracks. Due to the complexity of vertices and tracks in jets, defining

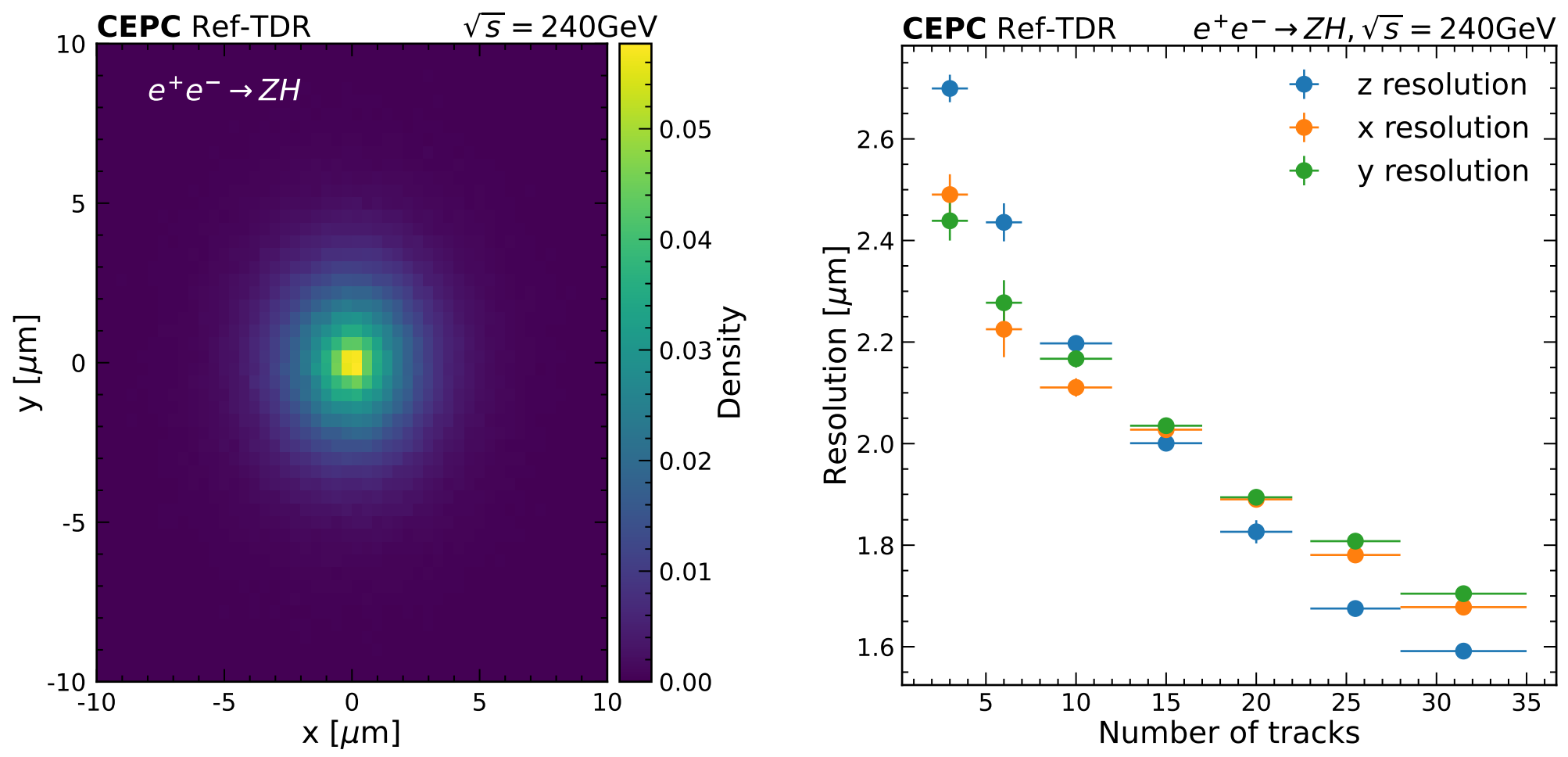

**(a)** **(b)**

**Figure 15.10:** (a) Position of the reconstructed primary vertex and (b) resolution of the primary vertex position as a function of the number of tracks originating from that vertex.

the efficiency of inclusive vertex reconstruction is a challenging task. For simplicity, a true secondary vertex is considered reconstructed if it is found within 200 μm of a reconstructed secondary vertex. Additionally, if a true vertex has more than two tracks, at least two corresponding reconstructed tracks must be used to form this reconstructed secondary vertex. The inclusive efficiency for $e^+e^- \longrightarrow b\bar{b}$ events is 75% and varies with the number of tracks associated with the vertices, as shown in Figure 15.9c.

#### 15.1.6.2 Vertex resolution

The vertex resolution is assessed using the $ZH$ events. Figure 15.10a shows the position of the reconstructed primary vertices in the events containing two isolated leptons and two $b$-quarks. The electron-positron interaction is simulated at the position (0, 0, 0). Figure 15.10b shows the resolution of the primary vertex position versus the number of tracks originating from the primary interaction. The resolution is below 3 μm for low-multiplicity events and approaches 2 μm for high-multiplicity events.

The precision of the secondary vertices is studied using $e^+e^- \longrightarrow b\bar{b}$ events. It is examined along its orientation to analyze the impact of resolution effects caused by boosted events, as represented in the Cartesian coordinate system. The orientation is derived from the following equation:

$$\vec{e_L} = \frac{\vec{r}}{|\vec{r}|}, \ \vec{e_{T1}} = \vec{r} \times \vec{z}, \ \vec{e_{T2}} = \vec{e_L} \times \vec{e_{T1}}, \tag{15.1}$$

where $\vec{e_{T1}}$ and $\vec{e_{T2}}$ represent the transverse directions and $\vec{e_L}$ is the longitudinal direction. $\vec{r}$ is the location vector of the corresponding generator-level vertex and $\vec{z}$ is the beam direction. The precision improves as more tracks are associated, as shown in Figure 15.11. The overall performance of the Reference Detector in vertexing is excellent, as expected from its single-track

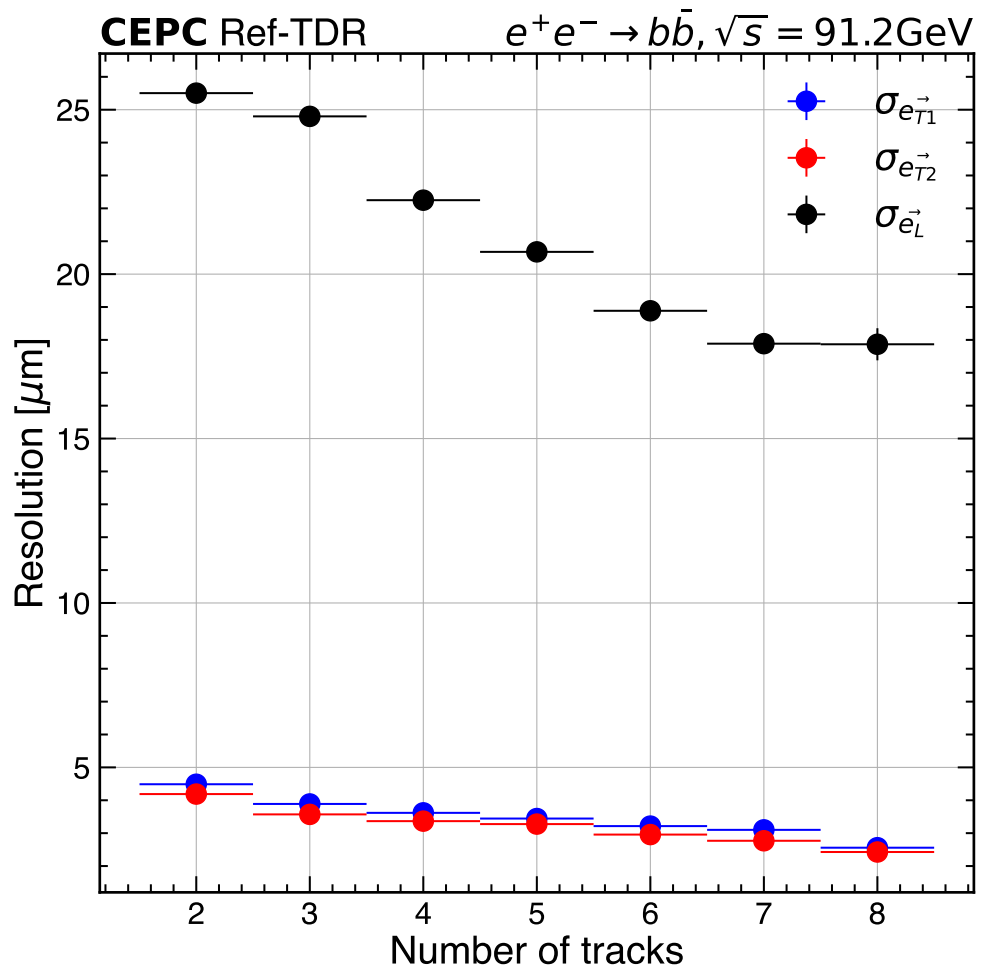

**Figure 15.11:** Resolution of the transverse and longitudinal components of the secondary vertices as a function of the number of associated tracks.

impact parameter resolution.

## 15.1.7 Jets

The design of the Reference Detector is optimized for jet energy resolution using the particle flow approach, which requires a strong interplay among various sub-detectors. This optimization has led to the choice of calorimeters with a high degree of longitudinal and transverse segmentation. To achieve a jet energy resolution that allows for the separation of $W$ and $Z$ boson decays, sophisticated reconstruction algorithms are necessary.

The $ee$-$k_t$ algorithm, also known as the Durham algorithm [4], is employed as the baseline jet clustering algorithm using the FastJet library [5]. "GenJets" are the clustered generator-level MC particles produced from the hadronization of partons simulated by Pythia 6.3 [6], including subsequent decay products such as photons, leptons, or light hadrons. Neutrinos are excluded from the clustering due to their non-interacting nature, which does not contribute to the energy measurements. "RecoJets" are clustered from the reconstructed particle flow candidates using the $ee$-$k_t$ algorithm in the same manner as "GenJets", allowing for a comparison of the clustering process at different levels.

### 15.1.7.1 Jet energy and angular resolution

The jet reconstruction performance of $ZH$ events at $\sqrt{s}$ = 240 GeV is analyzed. The GenJet and RecoJet with the closest angular distance are matched. For a given pair, the relative difference is expressed in terms of the Jet Energy Resolution (JER) and the Jet Angular Resolution (JAR). The relative energy difference is modelled with the Double-Sided Crystal Ball (DSCB) function, while the angular (polar and azimuth angle) differences of the GenJet-RecoJet pairs are modelled with the Gaussian function. The JER is extracted from the standard deviation ($\sigma$) of the DSCB

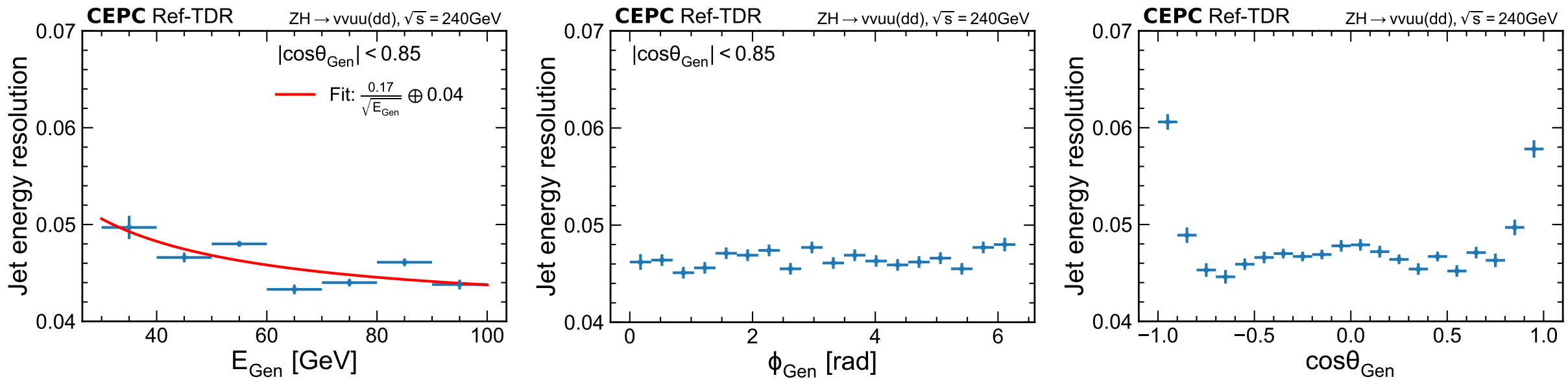

**(a)** **(b)** **(c)**

**Figure 15.12:** Jet energy resolution as a function of the (a) GenJet energy, (b) azimuth angle, and (c) cos$\theta_{\text{Gen}}$ for the $ZH \to \nu\bar{\nu}u\bar{u}$ and $\nu\bar{\nu}d\bar{d}$ processes. Only the statistical uncertainty is presented.

fit to the relative energy difference between RecoJet and GenJet, and the JAR is the $\sigma$ of the fitted Gaussian distribution to the angular difference.

Figure 15.12 shows the jet energy resolution as a function of the GenJet energy, azimuth angle, and cos$\theta_{\text{Gen}}$ in the $ZH \to \nu\bar{\nu}u\bar{u}$ and $\nu\bar{\nu}d\bar{d}$ events. The JER ranges from 4.5% to 5.0% in the barrel region, and improves as the energy increases, reaching 4.5% for jets with energy greater than 70 GeV, as shown in Figure 15.12 (a). The JER is stable against $\phi$ and slightly worse in the endcap region, as presented in Figures 15.12 (b) and (c). The jet angular resolution JAR is generally around 0.01 radian for $\theta$, and 0.012 radian for $\phi$.

#### 15.1.7.2 Boson mass resolution

Figure 15.13 shows the invariant mass distribution of the Higgs bosons in the $ZH \to \nu\bar{\nu}gg$ events. When both jets are selected in the barrel region ($|\cos\theta_{\text{jet}}| < 0.85$), the BMR reaches 3.87%, which surpasses the design goal of 4%. In the endcap region, the BMR is approximately 6%. Figure 15.14 shows a clear separation among the $W$, $Z$, and Higgs bosons with hadronic final states in their reconstructed invariant mass spectra.

With further development of the particle flow algorithm, including the integration of PID information, reducing the confusion of photons through $\pi^0$ identification, and correcting the neutrino component in the semileptonic decays of the heavy-flavor jets, the BMR performance is expected to exceed the design goal significantly.

### 15.1.8 Jet flavor tagging

Identifying bottom, charm, and other quark decay processes is crucial for evaluating detector performance. This capability plays a crucial role in precisely determining the Higgs boson couplings and electroweak observables at CEPC. Two specific approaches are developed.

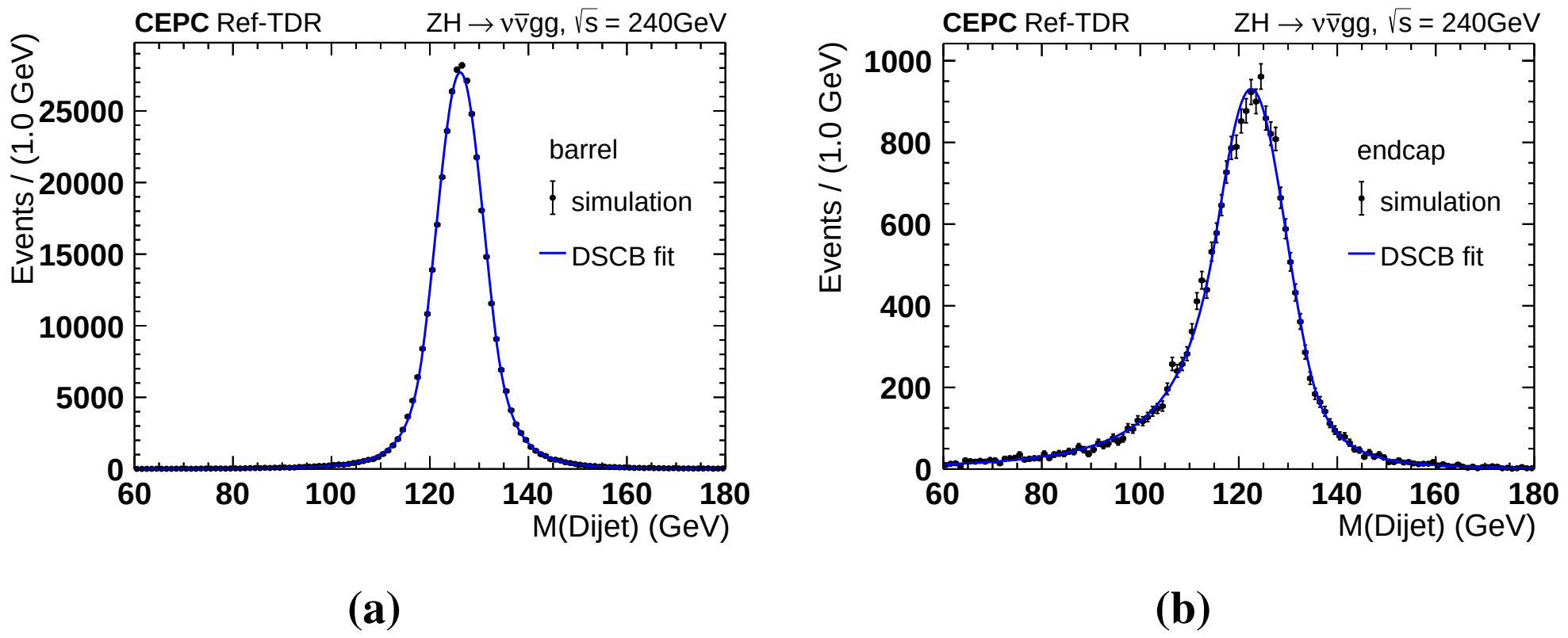

**Figure 15.13:** Reconstructed dijet invariant mass distributions in the $ZH \rightarrow \nu\bar{\nu}gg$ process at $\sqrt{s}$ = 240 GeV for jets selected in (a) the barrel and (b) the endcap.

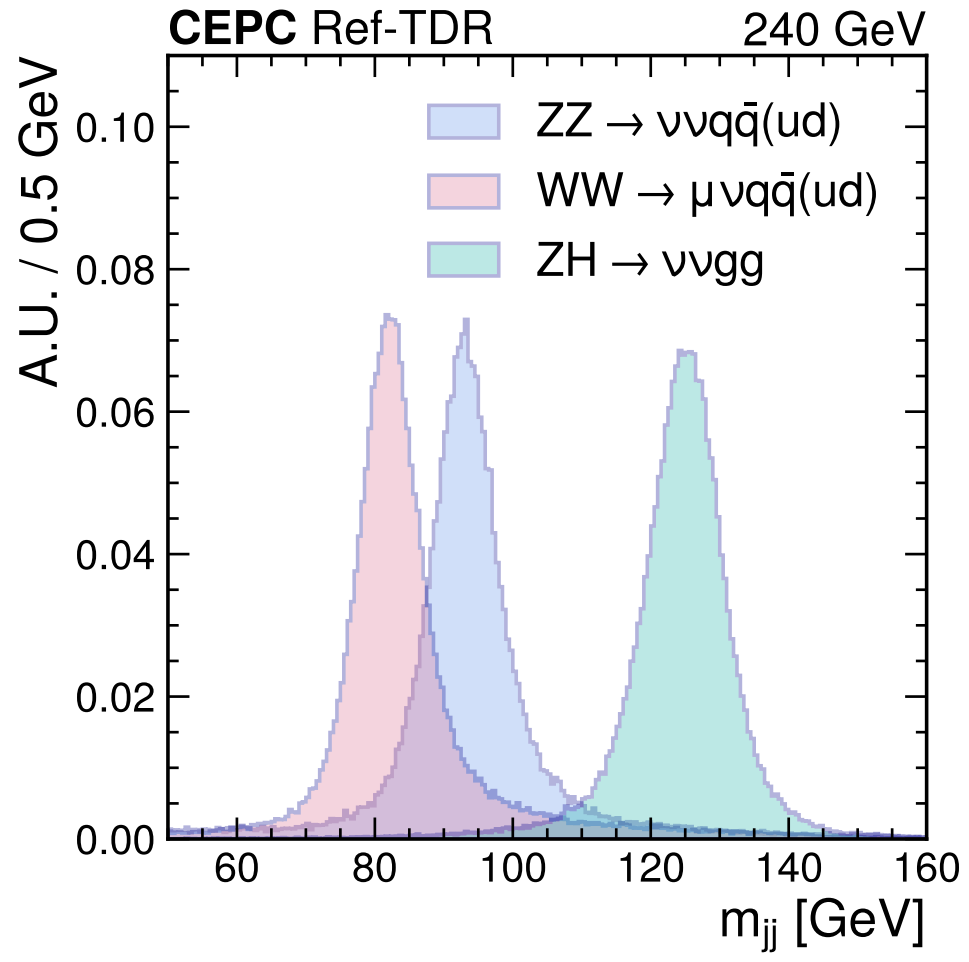

**Figure 15.14:** Reconstructed dijet mass distribution from $WW$, $ZZ$ and $ZH$ processes at $\sqrt{s}$ = 240 GeV.

#### 15.1.8.1 Jet flavor tagging with XGBoost

XGBoost is well-suited for jet flavor tagging, offering both strong classification performance and greater interpretability than deep learning models, which are essential for understanding the underlying detector performance.

In this study, the XGBoost classifier is trained for three jet categories: $b$-jet, $c$-jet and $uds$-jet. It employs a total of 22 input variables related to track and vertex information, similar to those adopted in Ref. [3]. The flavor tagging performance in $e^+e^- \rightarrow q\bar{q}$ events at $\sqrt{s}$ = 91.2 GeV is summarized in the next section along with the other algorithm, with separate evaluations for $b$-tagging and $c$-tagging. For $b$-tagging, the misidentification rates at $\epsilon_{b\text{-jet}}$ = 80% (50%) are 2.2% (0.11)% for $c$-jets and 0.24% (0.03%) for $uds$-jets. For $c$-tagging, the misidentification rates at $\epsilon_{c\text{-jet}}$ = 80% (50%) are 13.6% (2.9%) for $b$-jets and 13.9% (0.78%) for $uds$-jets.

**Table 15.1:** Input variables used in the ParT algorithm for jet flavor tagging, including kinematic and particle-specific features. The particle 4-momenta are only used to compute particle-pairwise features [11].

| Variable | Definition |
|---|---|
| $(p_x, p_y, p_z, E)$ | particle 4-momentum, with mass derived from PID considered for energy $E$. |
| $\Delta\eta, \Delta\phi, \Delta R$ | angle differences and angular distance between the particle and the jet axis |
| $\log p_{\mathrm{T}}, \log \frac{p_{\mathrm{T}}}{p_{\mathrm{T}}(\mathrm{jet})}$ | logarithms of the particle $p_{\mathrm{T}}$ and of its ratio to the jet $p_{\mathrm{T}}$ |
| $\log E, \log \frac{E}{E(\mathrm{jet})}$ | logarithms of the particle energy and of its ratio to the jet energy |
| $d_0, \sigma_{d_0}, z_0, \sigma_{z_0}$ | impact parameters and their uncertainties of the track |
| charge | electric charge of the particle |
| PID | Reconstructed particle type of $e, \mu, \pi, k, p, \gamma$ and neutral hadron |

#### 15.1.8.2 Jet Origin Identification

Jet Origin Identification (JOI) [7] is a novel concept for distinguishing jets originating from different quarks and gluons. It could use any advanced artificial intelligence algorithm such as ParticleNet [8, 9] and LLM [10] designed for jet flavor tagging and jet charge measurement. In this study, the ParticleTransformer (ParT) algorithm [11] is adopted and its implementation is optimized for the Reference Detector. JOI enables the simultaneous identification of 11 distinct jet species: five quarks, five antiquarks, and gluons.

To evaluate JOI performance, $\nu\bar{\nu}H$ production with $H \to u\bar{u}$, $d\bar{d}$, $s\bar{s}$, $c\bar{c}$, $b\bar{b}$, and $gg$ is fully simulated at the center-of-mass energy of 240 GeV. The $u\bar{u}$ and $d\bar{d}$ processes are also considered despite their extremely small branching ratios, as it allows for a more comprehensive understanding of the detector's response to different light-quark jets. For each process, one million physics events are generated, with 800k used for training, 100k for validation, and 100k for testing. The reconstructed final-state particles are clustered into two jets using the $ee\text{-}k_t$ algorithm. For each jet, kinematic and particle species information—including track impact parameters for charged particles, and particle identification described in Section 15.1.2 are fed into the ParT algorithm. The comprehensive set of input variables is listed in Table 15.1.

The algorithm computes likelihoods for classifying each jet into 11 categories and assigns each jet to the type with the maximum likelihood value. The model is trained for 30 epochs, and the epoch yielding the highest accuracy on the validation sample is applied to the test data sample to extract the final numerical results.

Figure 15.15 shows the JOI performance via an 11-dimensional "confusion matrix" $M_{11}$, where jets are categorized based on their highest likelihood. Overall, The matrix primarily exhibits an approximate quark-antiquark symmetry and can be viewed as block-diagonalized into $2 \times 2$ sub-matrices corresponding to specific quark species. This comprehensive assessment is based on a test sample where each of the 11 categories contains 100k events, and the statistical uncertainty on the probabilities is at the level of $O(10^{-3})$.

To further quantify JOI performance and assess its robustness, three distinct scenarios are evaluated as shown in Figure 15.16: (1) perfect PID, (2) realistic PID reconstruction, and (3) tracks only, excluding neutral components. While the realistic PID scenario shows slightly degraded performance compared to the ideal case, primarily due to lost PID information for low-energy tracks, the b- and c-jet tagging efficiencies remain stable as their vertex information ($d_0$ and $z_0$) provides strong discrimination power. The track-only analysis demonstrates that while the exclusion of neutral particles reduces the performance, which reflects the impact of the calorimeters, the JOI remains functional, confirming the robustness of the algorithm even with partial detector information.

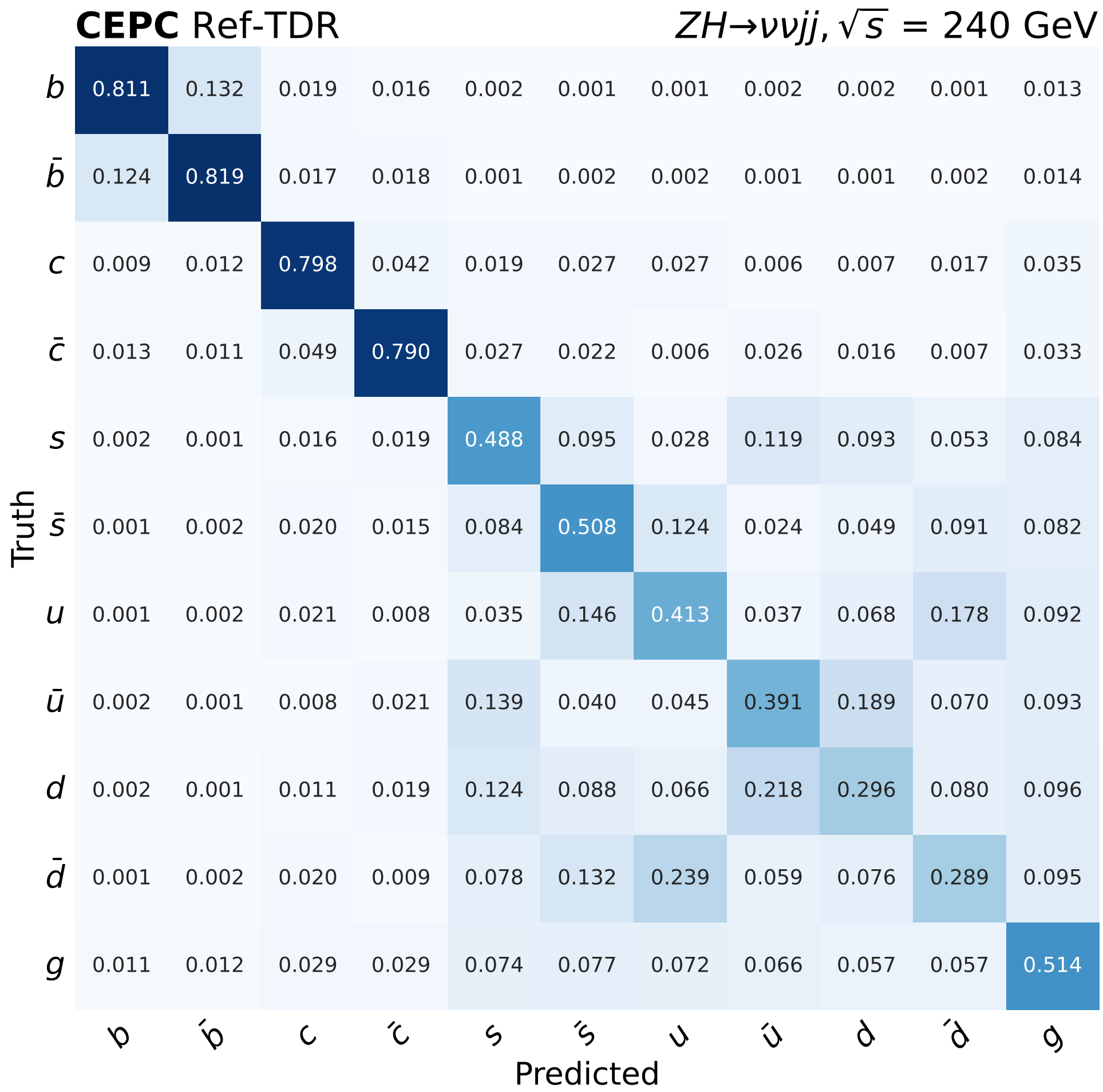

**Figure 15.15:** The confusion matrix $M_{11}$ of JOI using realistic PID of leptons and charged hadrons for $\nu\bar{\nu}H$, $H \to qq$ events at $\sqrt{s}$ = 240 GeV, with the Reference Detector. The matrix is normalized to unity for each parton label.

Figure 15.17 compares the JOI performance with the XGBoost method. Generally, JOI performance has lower mis-identification rates than the XGBoost method by one to two orders of magnitude for the fixed efficiencies. Remarkably, JOI achieves a $b$-jet tagging efficiency of 94% with a mis-identification rate of only 0.2% for light-quark jets.

This JOI model, optimized within a Higgs-boson production environment, demonstrates sufficient universality and capability across diverse kinematic energy scales. Its application to

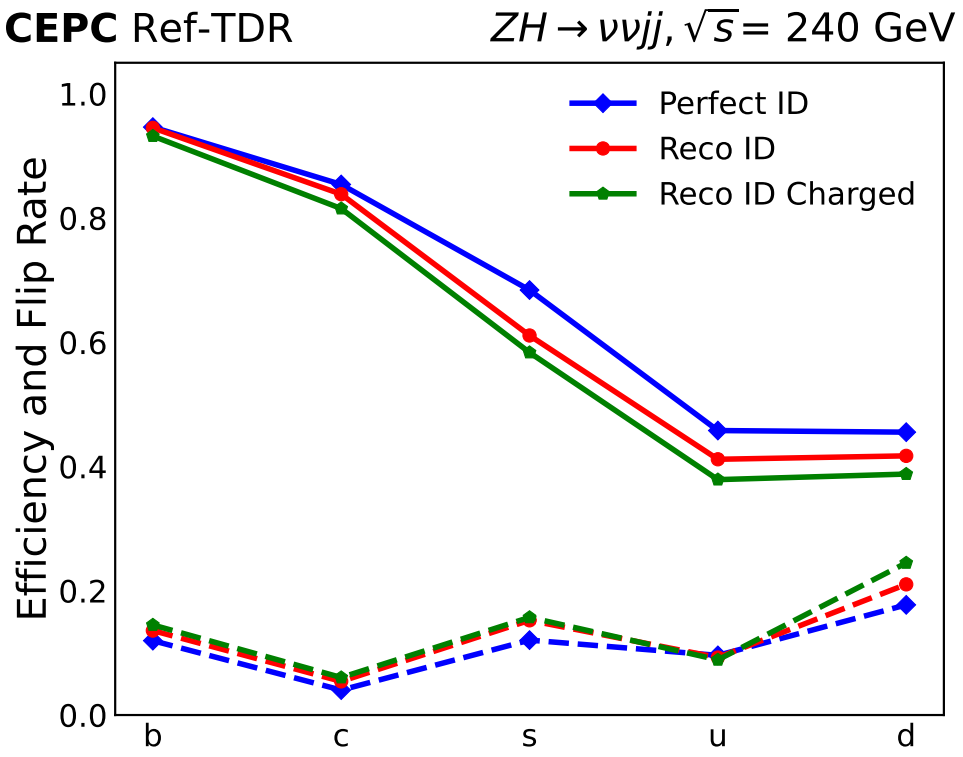

**Figure 15.16:** Comparison of JOI performance on the efficiency (solid lines) and charge flip rate (dashed lines) under three conditions: perfect PID (blue), realistic PID reconstruction (red), and realistic PID reconstruction using only charged tracks (green).

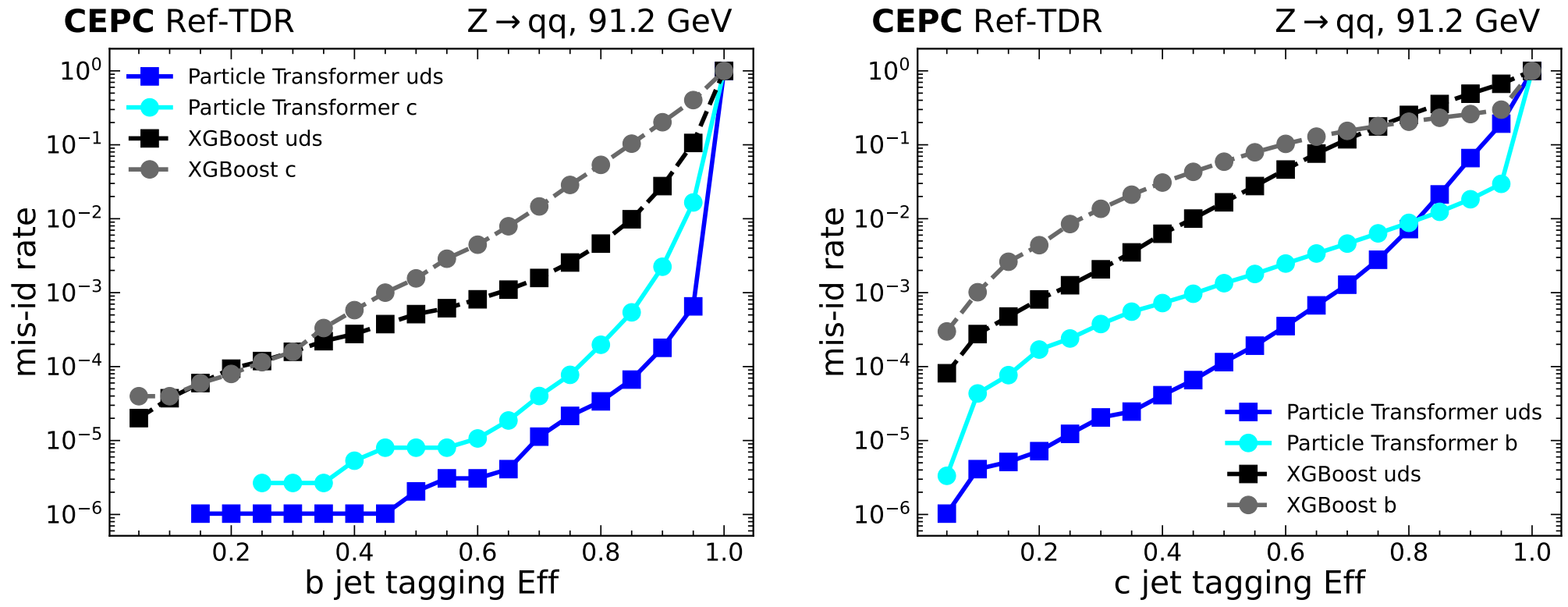

**(a)** **(b)**

**Figure 15.17:** Jet tagging efficiency versus mis-identification (mis-id) rate for (a) $b$-tagging and (b) $c$-tagging in the $Z$ pole. Solid blue/cyan for JOI with ParticleTransformer, dashed black/gray for XGBoost, solid circle point for mis-id rate from heavy quarks ($c$- or $b$-jets), solid square for mis-id rate from light quarks ($uds$).

$Z \to q\bar{q}$ final states reveals no significant performance degradation when compared against the JOI model tailored explicitly for $Z \to q\bar{q}$ data. Furthermore, consistent performance of JOI is validated for $e^+e^- \to q\bar{q}$ jet samples at $\sqrt{s}$ = 240 and 360 GeV, highlighting its energy-independent robustness beyond the training environment.

### 15.1.9 Missing energy, momentum, mass

Neutrinos interact with detectors through weak interactions and evade detection, leaving no discernible trace. This characteristic is also attributed to hypothetical dark matter particles. However, the presence of these elusive particles can be deduced from observable particles and plays a crucial role in the CEPC physics program. Approximately 20% of $Z$ bosons and 30% of $W$ bosons decay into final states with neutrinos. The search for Higgs boson decays into dark matter particles represents a pivotal objective of the Higgs factory. Reconstruction of the missing energy, momentum, or mass can also show the detector coverage capacity and the overall performance of the particle flow algorithm.

With the Reference Detector thanks to its excellent energy and momentum resolution, extensive coverage of the solid angle, and complete knowledge of the initial state, the missing energy and momentum can be determined with high precision. It is computed as the following four-momentum difference:

$$p_{\text{missing}} = p_{\text{total}} - p_{\text{visible}}, \tag{15.2}$$

where the total 4-momentum in the lab frame is:

$$p_{\text{total}} = (0, 0, 0, \sqrt{s})\text{GeV}, \tag{15.3}$$

and the total visible 4-momentum is computed by summing up all PFOs:

$$p_{\text{visible}} = \sum_{i}^{\text{PFOs}} p_i. \tag{15.4}$$

The $z$ component is called the Undetected Momentum In Beam Direction (MEZ) and the missing mass is computed as the mass of $p_{\text{missing}}$.

The Higgs boson invisible decay at $\sqrt{s} = 240$ GeV is used to illustrate the performance. The Higgs-strahlung production is considered, with $Z \to \mu^+\mu^-, e^+e^-$ or $q\bar{q}$, and $H \to ZZ^* \to 4\nu$. Additionally, the Higgs-strahlung production with $Z \to \nu\bar{\nu}$ and $H \to q\bar{q}$ is also considered. The distributions of the reconstructed missing mass are shown in Figure 15.18. For $H \to 4\nu$, the missing mass distributions are always around 125 GeV, while for $Z \to \nu\bar{\nu}$, they are around 91 GeV. The missing mass is fitted with a DSCB function in each channel. The $Z \to \mu^+\mu^-$ channel has the best missing mass resolution of 0.29 GeV, while the $e^+e^-$ channel gives a slightly worse resolution of 0.40 GeV because the tracking for electrons is more complicated due to their higher energy loss in the tracker and the higher rate of final state radiations. Among the hadronic channels, those associated with light-flavor quarks and gluons provide the best resolution, which is around 6.4 GeV for $Z \to$ light quarks, $H \to 4\nu$ and 9.2 GeV for $Z \to \nu\bar{\nu}, H \to gg$.

The missing energy and mass are sensitive to BIBs. The BIBs induce additional neutral objects in the calorimeters, and lead to an increase of around 7 GeV to the visible energy. With $Z \to q\bar{q}$, $H \to 4\nu$, the mean of the missing mass distribution decreases to around 118 GeV, whereas the resolution degrades to 7.2 GeV. Most neutral objects from BIBs lie in very forward and backward regions. By discarding neutral objects with $|\cos\theta| > 0.98$, the effect can be suppressed, with the missing mass mean value back to around 125 GeV, and the resolution restored to 6.4 GeV.

### 15.1.10 Summary of physics object performance

The Reference Detector provides excellent performance and achieves all the requirements for the physics objects, and enables a rich physics program to be conducted at CEPC. Table 15.2 summarizes the performance.

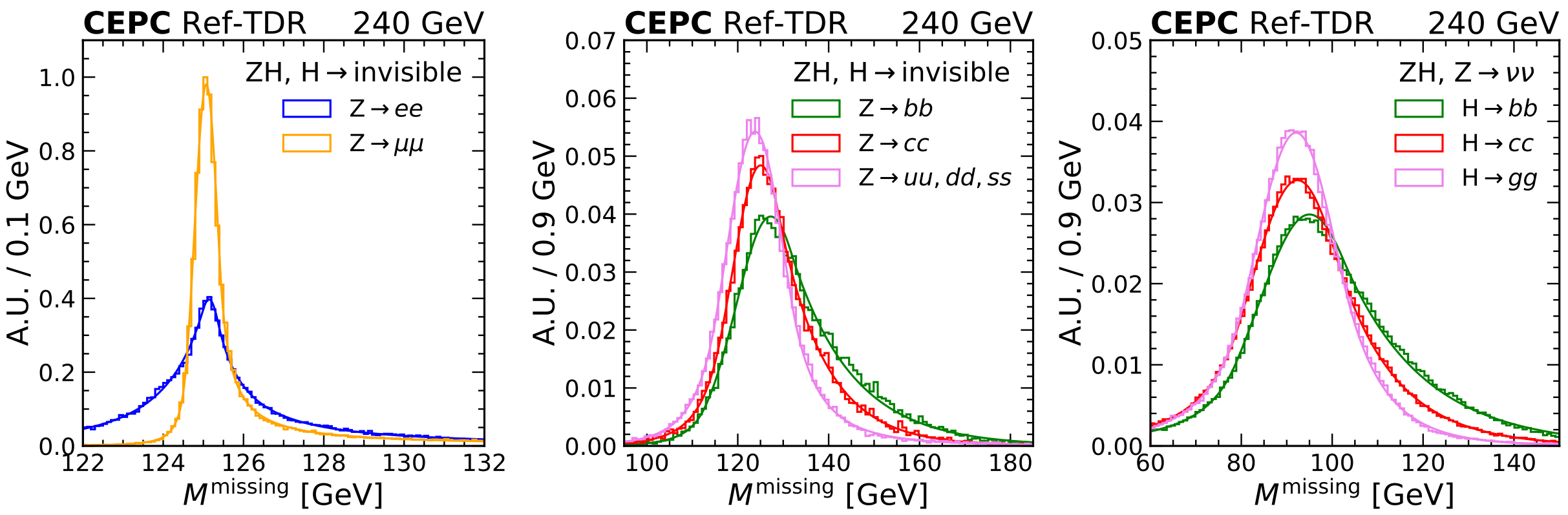

**(a)** **(b)** **(c)**

**Figure 15.18:** The distributions of reconstructed missing mass in the $ZH$ production, with (a) $Z \to e^+e^-/\mu^+\mu^-, H \to 4\nu$, (b) $Z \to q\bar{q}, H \to 4\nu$ or (c) $Z \to \nu\bar{\nu}, H \to q\bar{q}$. The solid lines show the fitted DSCB functions. In the (b) and (c) plots, different quark flavors are drawn and fitted separately.

**Table 15.2:** Performance of physics object reconstruction and identification

| **Physics Object** | **Processes considered** | **$E_{\text{c.m.}}$(GeV)** | **Performance** |
|---|---|---|---|
| Track efficiency | $ZH \to q\bar{q}jj$ | 240 | 99.7% for track $p$ >1 GeV |
| Track $p$ resolution | Single muon gun | – | $\sigma_{1/p_T} = 2.9 \times 10^{-5} \oplus 1.2 \times 10^{-3}/(p \cdot \sin^{3/2}\theta)$ |
| Electron ID | $ZH$ inclusive | 240 | eff.>90% for $p > 2$ GeV, misID <1% (BEST WP) |
| Muon ID | $ZH$ inclusive | 240 | eff.>90% for $p > 2$ GeV, misID <0.1% (BEST WP) |
| Photon ID | Single photon gun | – | eff.>95% for $E > 3$ GeV, misID $\lesssim$2% ($\gamma$ vs. $K_L^0$) |
| Charged hadron ID | $Z \to q\bar{q}$ | 91.2 | overall kaon eff.=92% and purity=90.7% |
| Tau lepton ID | $ZH \to q\bar{q}\tau^+\tau^-$ | 240 | eff.~80% and purity$\gtrsim$90 GeV (visible $E$=30−90 GeV) |
| Vertex reconstruction | $b^+b^-$ | 91.2 | eff.>90% for number of tracks $\geq 4$ |
| Primary vertices | $ZH \to \ell^+\ell^- b\bar{b}$ | 240 | resolution < 3 μm |
| Secondary vertices | $b^+b^-$ | 91.2 | longitudinal (transverse) resolution $\lesssim$ 25(5) μm |
| JER | $ZH \to \nu\bar{\nu}u\bar{u}$ or $\nu\bar{\nu}d\bar{d}$ | 240 | $\sim 0.2/\sqrt{E_{\text{Gen}}} \oplus 0.04$ GeV ($\lvert\cos(\theta)_{\text{Gen}}\rvert < 0.85$) |
| JAR | $ZH \to \nu\bar{\nu}u\bar{u}$ or $\nu\bar{\nu}d\bar{d}$ | 240 | 0.01 radian ($\theta$) and 0.012 radian ($\phi$) |
| BMR | $ZH \to \nu\bar{\nu}gg$ | 240 | 3.87% (barrel), approximately 6% (endcap) |
| Jet flavor tagging | $ZH \to \nu\bar{\nu}jj$, $Z \to q\bar{q}$ | 240, 91.2 | $b$-tag eff.=94% with mis-ID rate=0.2% for $uds$ |
| Missing mass resolution | $ZH$ | 240 | 0.29 GeV ($e^+e^-4\nu$), 0.40 GeV ($\mu^+\mu^-4\nu$) 6.4 GeV ($q\bar{q}4\nu$), 9.2 GeV ($\nu\bar{\nu}q\bar{q}$) |

## 15.2 Physics benchmarks

Detailed simulation studies with the Reference Detector are presented in this section. Benchmark studies are conducted at various center-of-mass energies of 91 GeV, 240 GeV, and 345 GeV, as shown in Table 15.3. Table 15.4 summarizes the integrated luminosity assumed for each analysis. These studies cover essential physics areas such as Higgs, electroweak, flavor, top, and new physics. They were selected to stress different detector capabilities, the most relevant sub-detectors are highlighted for each benchmark in the Table 15.3.

**Table 15.3:** Physics benchmarks and relevant detector performances

| Physics Benchmarks | Process | $E_{\rm c.m.}$(GeV) | Domain | Relevant Det. Performance |
|---|---|---|---|---|
| Recoil $H$ mass | $\mu\mu H$ | 240 | Higgs | Tracking |
| $H \to$ hadronic decays | $\mu\mu H$ | 240 | Higgs | PID, Vertexing, PFA |
| $H \to \gamma\gamma$ | $ZH$ | 240 | Higgs | photon ID, EM resolution |
| $H \to$ invisible | $ZH$ | 240 | Higgs/BSM | PFA, missing energy, BMR |
| $H \to$ LLP | $ZH$ | 240 | BSM | Vertexing, TOF, muon detectors |
| Smuon pair | $\tilde{\mu}^+\tilde{\mu}^-$ | 240 | BSM | muon ID, missing energy |
| $A^{\mu}_{FB}$ | $\mu^+\mu^-$ | 91.2 | EW | Tracking, muon ID |
| $R_b$ | $Z \to q\bar{q}$ | 91.2 | EW | PFA, jet flavor tagging |
| CPV in $D^0 \to h^+h^-\pi^0$ | $Z \to q\bar{q}$ | 91.2 | Flavor | PID, vertex, $\pi^0$, EM resolution |
| Top quark mass | $t\bar{t}$ | $\sim 345$ | Top | Beam energy |

**Table 15.4:** CEPC luminosity assumptions for the physics benchmarks.

| $E_{\rm c.m.}$(GeV) | Total $\int \mathcal{L}$(ab$^{-1}$) |
|---|---|
| 240 | 21.6 |
| 91 | 1 or 100 |
| Top quark mass threshold | 0.1 |

### 15.2.1 Event generation

The production of simulated events for the benchmarking analyses includes the MC generation of a comprehensive set of the SM processes. The mass of the SM Higgs boson is set to 125 GeV.

The benchmarking studies primarily use the WHIZARD 1.9.5 [12] event generator for cross-section calculations and MC sample generation. Fragmentation and hadronization are then simulated using Pythia 6.3 [6]. Alternatively, the Bhabha cross section is calculated using the BABAYAGA event generator [13], requiring the final-state particles to have $|\cos\theta| < 0.99$.

At the $Z$ pole ($\sqrt{s}$ = 91.2 GeV), the total hadronic cross section ($\sigma_{had}$) is approximately 29.6 nb. For $t\bar{t}$ threshold scan, the cross section at $\sqrt{s}$ = 345 GeV is 534.8 fb, calculated by QQbar_threshold [14], which assumes a top quark mass of 171.5 GeV, and incorporates corrections for the ISR and light spectrum effects.

For $\sqrt{s}$ = 240 GeV, Higgs signal production mechanisms manifest through three principal channels: Higgsstrahlung ($e^+e^- \to ZH$), $W$-boson fusion ($e^+e^- \to \nu\bar{\nu}H$), and $Z$-boson fusion ($e^+e^- \to e^+e^-H$). The background processes, excluding Higgs signals, can be broadly classified into two major categories based on the number of final-state particles: (1) $e^+e^- \to$ 2-fermion processes, encompassing leptonic pair and quark-antiquark production, and (2) $e^+e^- \to$ 4-fermion processes, which show significant complexity, necessitating a detailed treatment of electroweak interference effects. These 4-fermion backgrounds are systematically categorized by their intermediate bosonic contributions: $WW$, $ZZ$, single $W$ and $Z$ productions.

**Table 15.5:** Cross sections for the Higgs production and background processes at $\sqrt{s}$ = 240 GeV.

| **Process** | $\sqrt{s}$ **= 240 GeV** |
|---|---|
| **Higgs boson production, cross section in fb** | |
| $e^+e^- \to ZH$ | 196.9 |
| $e^+e^- \to \nu_e\bar{\nu}_e H$ | 6.2 |
| $e^+e^- \to e^+e^-H$ | 0.5 |
| **Total Higgs** | **203.6** |
| **background processes, cross section in pb** | |
| $e^+e^- \to e^+e^-(\gamma)$ (Bhabha) | 930 |
| $e^+e^- \to q\bar{q}(\gamma)$ | 54.1 |
| $e^+e^- \to \mu^+\mu^-(\gamma)$ | 5.30 |
| $e^+e^- \to \tau^+\tau^-(\gamma)$ | 4.75 |
| $e^+e^- \to WW$ | 16.7 |
| $e^+e^- \to ZZ$ | 1.1 |
| $e^+e^- \to e^+e^-Z$ | 4.54 |
| $e^+e^- \to e^+\nu W^-$ + c.c. | 5.09 |

The cross sections at $\sqrt{s}$ = 240 GeVis presented in Table 15.5. Photons, if any, are required to have $E_\gamma > 0.1$ GeV and $|\cos\theta_{e^\pm\gamma}| < 0.99$. The ISR and FSR effects are included in all the final states. All the cross sections are calculated using WHIZARD. Notably, WHIZARD demonstrates particular efficacy in calculating gauge boson production processes, interference effects, and initial-state radiation correction, which are critical for precision measurements at CEPC.

For photon-related processes, WHIZARD automatically manages the initial and final state photons by default. For example, final states such as $e^+e^- \to q\bar{q}\gamma\gamma$ are included in $e^+e^- \to q\bar{q}$ process, with diphoton invariant mass and individual photon energy required to be above 10 GeV. Similarly, the process $e^+e^- \to e^+e^-e^+e^-$ considers the effect from the diphoton pair production with the same selection.

The 4-fermion samples are categorized into 40 individual types, ensuring no overlap or omission during the event generation. Without an overlap and disregarding the interference, the overall 4-fermion cross section is 19.3 pb at $\sqrt{s}$ = 240 GeV.

Finally the generated events go through the full detector simulation and reconstruction chains in CEPCSW. The detailed information of the MC samples produced, including their statistics and data paths, is maintained in a centralized database using the GitLab service provided by IHEP.

### 15.2.2 Higgs boson mass measurement through recoil mass

Since the discovery of the Higgs boson at the LHC in 2012 [15, 16], extensive efforts have been made to measure its mass ($m_H$) accurately. The most precise measurement to date is $m_H$ = 125.11 ± 0.11 GeV [17], obtained by the ATLAS experiment. The projected precision at the HL-LHC is 21 MeV [18], combining the ATLAS and CMS results. These measurements rely on specific Higgs decays, where the signals are either limited by low event yields or affected by significant hadronic backgrounds and pileup in the hadron collider environment.

This study aims to report the expected precision of the $m_H$ measurement using the $e^+e^- \rightarrow ZH \rightarrow \mu^+\mu^- X$ events, where $X$ represents inclusive final states. Because of the well-known initial-state particles and beam energy, this method only requires reconstruction and identification of the two muons from the $Z$-boson decay, which recoil against the Higgs boson, making the result independent of the Higgs decay modes. This study is crucial for validating the detector design and performance, particularly in terms of tracking efficiency, momentum resolution, and muon identification.

Signal events are characterized by two oppositely charged, high-momentum muons whose invariant mass is around $m_Z$ = 91.19 GeV. As discussed in Section 15.1.1, within the detector acceptance, these muons are expected to be efficiently reconstructed with a transverse momentum precision better than 0.3%. The recoil mass of the muon pair $M_{\text{rec}}$ can be calculated according to Equation 15.5, requiring no explicit knowledge of the Higgs boson decay components:

$$M_{\text{rec}}^2 = (\sqrt{s} - E_{\mu^-} - E_{\mu^+})^2 - |\vec{p}_{\mu^-} + \vec{p}_{\mu^+}|^2. \tag{15.5}$$

To achieve high purity and efficiency of the signal events, muons are selected with momenta between 20 and 80 GeV. The lower threshold suppresses soft muons from jets, while the upper limit enhances purity by excluding higher-momentum backgrounds, such as $e^+e^- \rightarrow \mu^+\mu^-$ events. The invariant mass of the muon pair is required to be in the range of 50–120 GeV to be consistent with the $Z$ pole. The momentum of the di-muon system is limited to the range of 20–60 GeV since the momentum of $Z$ or $H$ tends to be around 50 GeV for the signal events.

Given the significantly large cross section of the $e^+e^- \rightarrow \mu^+\mu^-$ background process, further suppression of its contribution is necessary. Most muons from this background have nearly a half of the center-of-mass energy; thus, an upper limit of 110 GeV on the di-muon energy is imposed. Furthermore, muons from this background with ISR would have lower energy and contribute to the $Z$ boson peak. As those events tend to have large MEZ, MEZ < 50 GeV is applied. By applying the selections mentioned above, this background can be effectively suppressed across nearly the whole energy range. The event selection criteria are summarized in Table 15.6, where the four-fermion backgrounds are categorized by their final states.

After applying the event selections, the recoil mass spectrum of the signal is modeled using the DSCB function. Ideally, the peak position of this function should correspond to $m_H$. The background is modeled using a Chebyshev polynomial, which is obtained by fitting the background-only $M_{\text{rec}}$ distribution. The spectra within the signal region 120–140 GeV and the signal-plus-background shape are shown in Figure 15.19.

**Table 15.6:** Summary of event selections and cutflow. Four-fermion backgrounds are categorized by their final states

| **Final States** | $2\nu 2\mu$ | $4\mu$ | $2\mu 2e$ | $2\mu 2\tau$ | $2\mu 2q$ | $2\mu$ | $\mu\mu H$ |
|---|---|---|---|---|---|---|---|
| Events number | 120000 | 40000 | 40000 | 40000 | 80000 | 100000 | 40000 |
| Muon pair | 31.4% | 41.7% | 6.7% | 25.8% | 29.5% | 88.2% | 95.6% |
| MEZ $\in [0, 50]$ GeV | 26.5% | 29.9% | 4.7% | 17.1% | 25.7% | 54.2% | 94.4% |
| $E_{\mu\mu} \in [0, 110]$ GeV | 12.7% | 16.5% | 4.1% | 10.0% | 15.0% | 53.0% | 93.8% |
| $p_{\mu\mu} \in [20, 60]$ GeV | 7.5% | 9.9% | 2.1% | 5.6% | 8.0% | 9.8% | 83.7% |
| $m_{\mu\mu} \in [50, 120]$ GeV | 5.8% | 6.7% | 0.4% | 3.1% | 3.1% | 9.8% | 83.6% |
| $M_{\text{rec}} \in [120, 140]$ GeV | 2.6% | 3.2% | 0.1% | 1.6% | 1.7% | 6.5% | 78.7% |

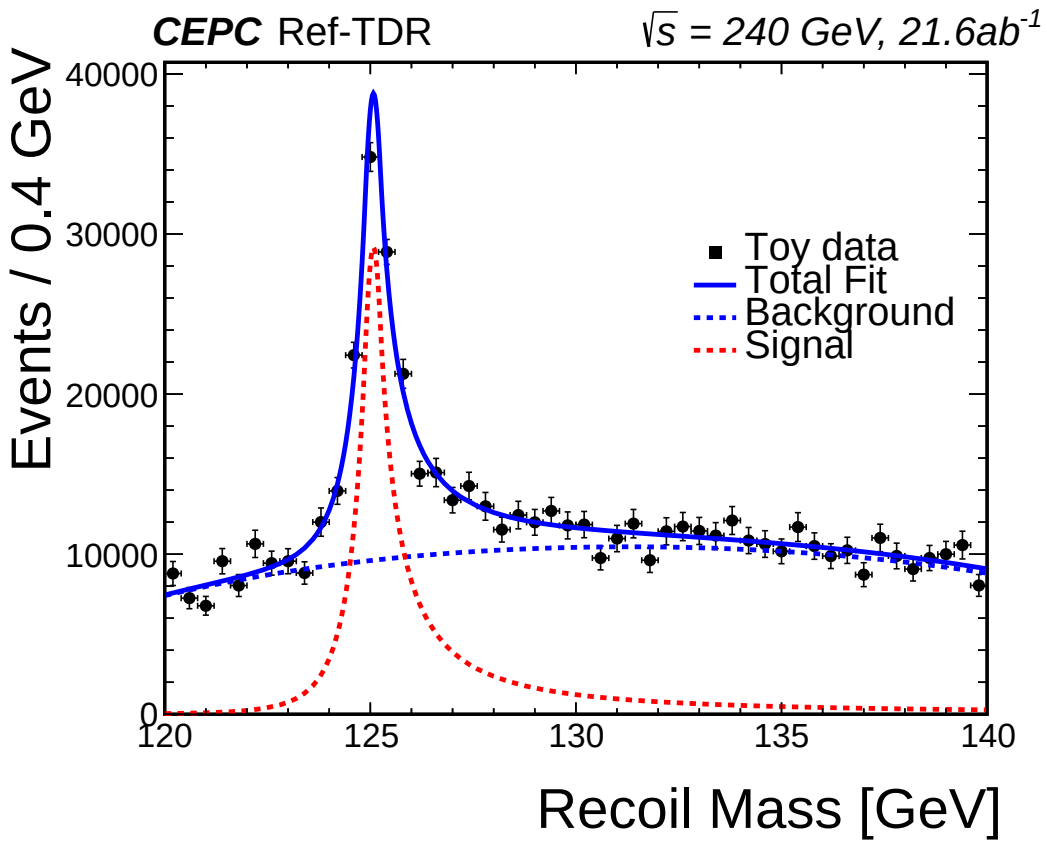

**Figure 15.19:** $M_{\text{rec}}$ distribution and the models for signal and background. The beam energy spread is taken into account.

The precision of $m_H$ measurement is estimated based on the projected statistics at an integrated luminosity of 21.6 ab$^{-1}$. The statistical uncertainty is $\Delta m_H = \pm 3.1$ MeV.

Major foreseeable systematic uncertainties are considered below:

- Momentum scale: During the whole $ZH$ data taking of 21.6 ab$^{-1}$, $3 \times 10^7$ radiative return events ($\mu\mu\gamma$ final state) will be produced. This high-statistics process acts as a standard candle for precision measurements. Given a 0.2% tracking resolution, the $Z$ boson peak shift can be monitored with $10^{-6}$ relative precision. The momentum scale uncertainty is estimated to be 2 MeV, accounting for potential limitations from the magnetic field and tracker alignment.
- Center-of-mass energy: The uncertainty in the center-of-mass energy directly propagates into the Higgs recoil mass spectrum, as shown in Equation 15.5. It can be monitored using $e^+e^- \to Z\gamma, Z \to f\bar{f}$ events, with precise knowledge of the $Z$ boson mass. It is assumed to be 2 MeV.
- Beam energy spread: Beam energy spread is expected to be 0.17% at $\sqrt{s} = 240$ GeV and

is included in the signal modeling. Its uncertainty can be precisely measured using the radiative return events. A perturbed sample with 0.18% spread shows a negligible impact on $m_H$.

- ISR: ISR can modify the Higgs recoil mass spectrum by reducing the effective collision energy. As a theoretical uncertainty in QED, its impact on $m_H$ is conservatively assumed to be 1 MeV.
- BIB: After mixing BIB with signal events, the recoil mass spectrum of $e^+e^- \to \mu^+\mu^- HX$ is found to be 5 MeV wider than before. To account for the beam-induced background effects, a nuisance parameter of 5 MeV is tentatively added to the width parameter in the signal model. Its contribution to $\Delta m_H$ is negligible.

These systematic uncertainties contribute to an additional 2.5 MeV to $\Delta m_H$, resulting in the final precision of $\Delta m_H = \pm$ 4.0 MeV, which is a significant improvement from the projected precision of 21 MeV at the HL-LHC [18].

### 15.2.3 Branching ratios of the Higgs boson in hadronic final states

According to the theoretical predictions, the branching fractions of the 125 GeV Higgs boson into $b\bar{b}$, $c\bar{c}$, $s\bar{s}$, $gg$, $WW^*$, $ZZ^*$ are 57.7%, 2.91%, 0.024%, 8.57%, 21.5% and 2.64%, respectively. The measurements can be conducted simultaneously for all hadronic decay channels, including the full hadronic decay modes of $WW^*/ZZ^*$ [19]. The main backgrounds are from processes with two-fermion and four-fermion final states. The Particle Flow Network (PFN) was previously [19] employed for the direct separation among different decay modes with the whole event reconstruction information, in contrast with traditional jet tagging methods. In this study, a more complete simultaneous measurement is performed with the inclusion of $H \to s\bar{s}$ decays and an advanced event-level machine learning technique of ParticleTransformer [11].

In this study, only the $ZH$ production with $Z$ decaying into a pair of muons is considered. At least one muon pair with the opposite charge is required. To further reduce the four-fermion and two-fermion background, the polar angle of the muon pair system is required to be in the range of $|\cos\theta_{\mu\mu}| < 0.996$, and the number of tracks is required to be at least seven. The muon pair with the invariant mass closest to the $Z$ boson mass and in the mass window of 75 to 105 GeV is chosen as the $Z$ candidate. The final Higgs boson signal is selected by requiring the $Z$ recoiling mass in the window of 120 to 140 GeV.

Table 15.7 presents the event selection efficiencies for various signal and remaining background processes, detailing the efficiency at each selection step. For signal processes, an efficiency of over 82% is observed, while the efficiency of four-fermion backgrounds $(ZZ)_{sl}$ is suppressed to 0.7% and the other backgrounds are found to be negligible.

After these selection, the training process involves a seven-classification task. 300,000 events for each of the seven processes are provided to the model whose weights are all equal to 1, with data split into training, validation and test sets in an 8:1:1 ratio. The training variables include the reconstructed energy, momentum, azimuth angle, polar angle, and PID of each PFO, the charge and impact parameter of each track, as well as whether the muons originated from the $Z$ boson decays. The migration matrix $M_{\text{mig}}$ is defined as the probability $\epsilon_{i,j}$ of genuine signal with class $i$ reconstructed as category $j$ as shown in Figure 15.20. The migration matrix reflects the overall high accuracy of the model. By considering all signal and background processes, the expected number of events for each process can be calculated as follows:

$$\mathbf{N}^T = (M_{\text{mig}}^T M_s)^{-1}\mathbf{n}^T, \tag{15.6}$$

**Table 15.7:** The cutflow selection efficiency for signal processes and remaining background process $(ZZ)_{sl}$. Other backgrounds were significantly suppressed.

| Process | $b\bar{b}$ | $c\bar{c}$ | $gg$ | $WW^*$ | $ZZ^*$ | $s\bar{s}$ | $(ZZ)_{sl}$ |
|---|---|---|---|---|---|---|---|
| Muon pair | 96.9% | 96.7% | 96.7% | 96.7% | 96.7% | 96.6% | 21.1% |
| Isolation | 90.3% | 90.3% | 90.5% | 90.4% | 90.7% | 90.5% | 19.7% |
| $\lvert\cos\theta_{\mu\mu}\rvert$ <0.996 | 90.0% | 90.0% | 90.2% | 90.1% | 90.4% | 90.1% | 3.0% |
| $N_{\text{tracks}}$ >7 | 90.0% | 90.0% | 90.2% | 90.1% | 90.4% | 90.1% | 3.0% |
| $Z$ mass window | 86.4% | 86.4% | 86.5% | 86.4% | 86.7% | 86.5% | 1.4% |
| $H$ mass window | 82.4% | 82.3% | 82.5% | 82.4% | 82.8% | 82.4% | 0.7% |

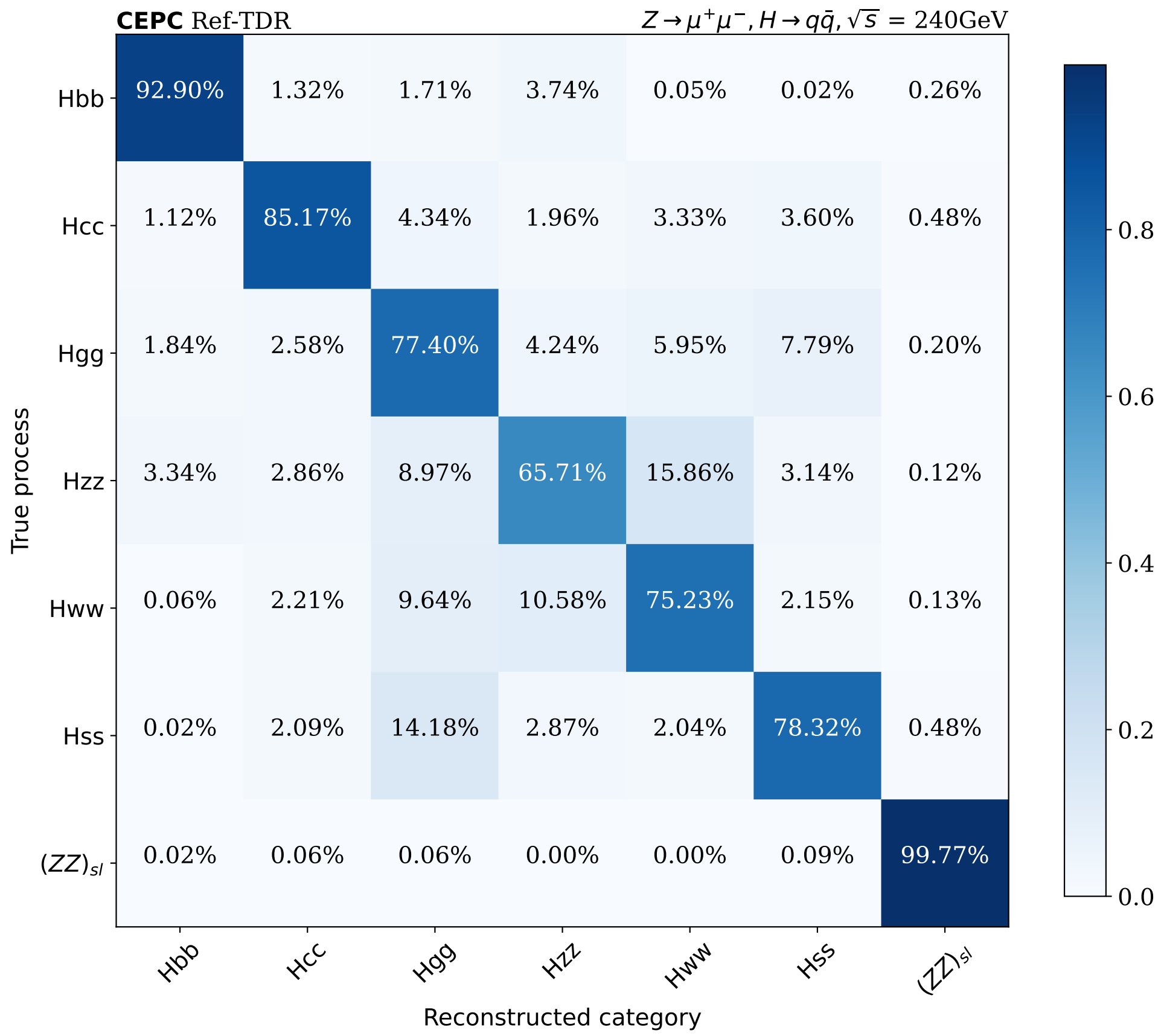

**Figure 15.20:** The migration matrix for the seven classes is shown. The horizontal axis represents the prediction of the model for each event in the test set, while the vertical axis indicates the true labels. The sum of values in each row equals 1.

**Table 15.8:** The measured branching fractions for the Higgs boson decays, along with their relative statistical and systematic uncertainties, are shown.

| **Decay channels** | $b\bar{b}$ | $c\bar{c}$ | $gg$ | $WW^*$ | $ZZ^*$ |
|---|---|---|---|---|---|
| BR | 57.7% | 2.9% | 8.6% | 21.5% | 2.6% |
| Rel. Stat. Un. | 0.3% | 2.2% | 1.3% | 1.1% | 7.6% |
| Rel. Syst. Un. | 0.1% | 3.6% | 1.8% | 0.4% | 4.3% |
| Rel. Total Un. | 0.3% | 4.2% | 2.2% | 1.2% | 8.7% |

where $n_i$ and $N_i$ are the numbers of reconstructed and expected events in the category $i$, respectively. The $M_s$ is a diagonal matrix containing the selection efficiencies, while $M^T_{\text{mig}}$ denotes the transposed migration matrix.

The branching fractions of $H \to b\bar{b}/c\bar{c}/gg/WW^*/ZZ^*$ at the CEPC, with the center-of-mass energy of 240 GeV and luminosity of 21.6 ab$^{-1}$, are measured to be 57.7%, 2.9%, 8.6%, 21.5%, and 2.6%, with the relative statistical uncertainty of 0.3%, 2.2%, 1.3%, 1.1%, and 7.6% respectively. To account for detector-related effects, particularly those arising from vertex reconstruction and tracking performance, the spatial resolution of each track is conservatively smeared by 20%. The resulting relative systematic uncertainties in the branching fraction measurements are estimated to be 0.1%, 3.6%, 1.8%, 0.4%, and 4.3%, depending on the decay channel. An upper limit of 0.2% at 95% Confidence Level (CL) is found for the $H \to s\bar{s}$ channel. CEPC provides unprecedented precisions to measure all the main decay channels and sensitivity to channels that are not accessible at the LHC [20, 21].

## 15.2.4 $H \to \gamma\gamma$

The diphoton decay channel of the Higgs boson is an important benchmark to evaluate the electromagnetic calorimeter performance. The branching ratio is small due to the top-quark and massive-boson loop diagram, but it has a very clean final state with two energetic photons.

Following the strategy in CDR [22], this analysis focuses on the $ZH$ production at $\sqrt{s} = 240$ GeV, with Higgs boson decaying into two photons. Three sub-channels are considered based on the $Z$ boson decays: $Z \to q\bar{q}$, $\mu^+\mu^-$, and $\nu\bar{\nu}$. The $Z \to e^+e^-$ channel is excluded due to the overwhelming Bhabha background, and $Z \to \tau^+\tau^-$ is omitted due to the complexity of $\tau$ identification. The dominant background ($e^+e^- \to f\bar{f}$) with two ISR/FSR photons is considered, while the others, including the Higgs boson resonant contributions, four fermion processes and reducible backgrounds from photon mis-identification are expected to be negligible, as discussed in Ref. [22]. The events are generated as described in Section 15.2.1. The signal samples are generated through full simulation, while the background samples are processed using the Delphes fast simulation for large statistics.

Regarding the event topology of the photons from the Higgs boson and fermions from the $Z$ boson decay, the events are categorized into three channels: $q\bar{q}\gamma\gamma$, $\mu^+\mu^-\gamma\gamma$ and $\nu\bar{\nu}\gamma\gamma$. Event selection is applied to reject the backgrounds and mis-tagged events. The leading (subleading) photon energy is required to be greater than 30 (20) GeV, and the invariant mass of the diphoton must be within [110, 140] GeV in all three channels. Additional requirements on the angles and energies of photons, fermions, or missing mass are applied individually, depending on the final state. The cut-based selection efficiency and expected yields are listed in Table 15.9. The contamination among the three sub-channels is found to be minor (<0.3%) after the event selection. A Boosted Decision Trees with Gradient boosting (BDTG) is trained to improve the analysis sensitivity. The input variables are the same as in Ref. [22]. The output of training is shown in Figure 15.21.

**Table 15.9:** The cut-based selection efficiency and expected event yields in three sub-channels.

| | **Selection efficiency** | **Expected yield at 21.6 ab$^{-1}$** |
|---|---|---|
| $q\bar{q}\gamma\gamma$ signal | 64.6% | 4340 |
| $q\bar{q}\gamma\gamma$ background | 0.06% | 710200 |
| $\mu^+\mu^-\gamma\gamma$ signal | 50.4% | 167 |
| $\mu^+\mu^-\gamma\gamma$ background | 0.01% | 12600 |
| $\nu\bar{\nu}\gamma\gamma$ signal | 59.0% | 1360 |
| $\nu\bar{\nu}\gamma\gamma$ background | 0.002% | 21200 |

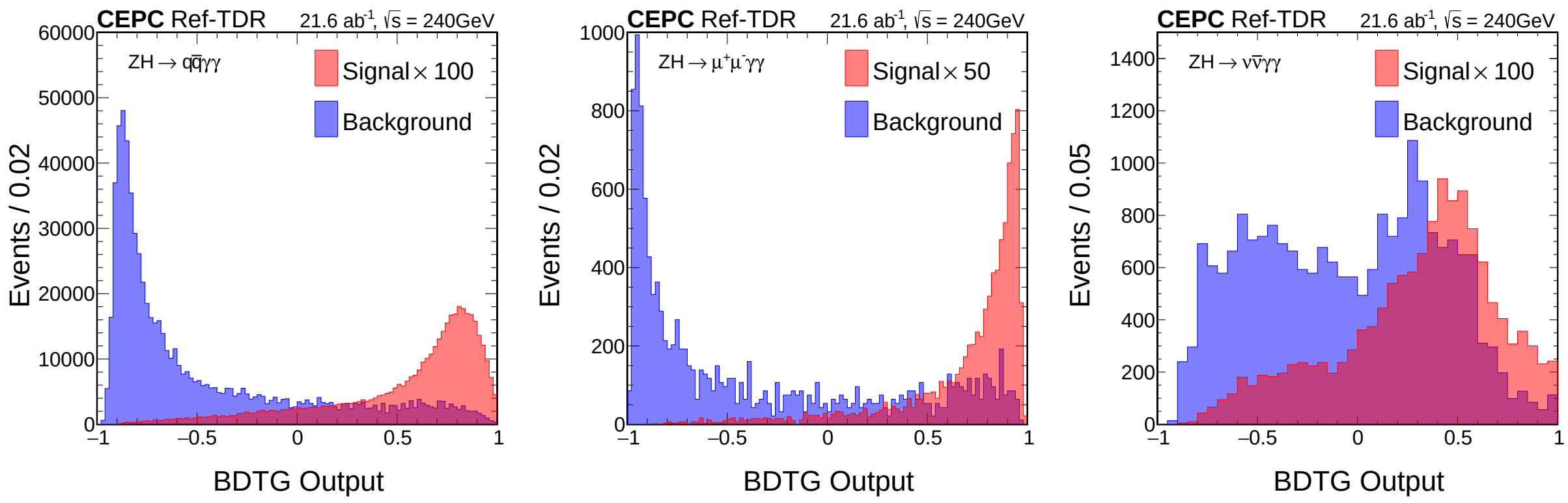

**(a)** **(b)** **(c)**

**Figure 15.21:** Output of BDTG training for signal and background in three channels: (a) $q\bar{q}\gamma\gamma$, (b) $\mu^+\mu^-\gamma\gamma$, (c) $\nu\bar{\nu}\gamma\gamma$. For visibility, the signal histograms are scaled to 100, 50, and 10, respectively.

The signal is extracted by fitting the diphoton invariant mass $m_{\gamma\gamma}$ and the BDTG score. The DSCB function is chosen to describe the signal $m_{\gamma\gamma}$ shape. The fitted $m_{\gamma\gamma}$ resolution is 0.34 GeV in the $q\bar{q}\gamma\gamma$ channel, benefiting from the excellent EM energy resolution in the Reference Detector. Several continuum functions are tested for the background modeling, and the one with minimum $\chi^2/N_{dof}$ is chosen. A two-dimensional model is constructed by multiplying the $m_{\gamma\gamma}$ and BDTG models by their binned Probability Density Functions (PDFs). This two-dimensional

**Table 15.10:** Expected statistic precision on $\sigma(ZH) \times \text{Br}(H \to \gamma\gamma)$ from Asimov data fitting in the three channels and their combination with 21.6 ab$^{-1}$ data.

| | $\Delta(\sigma \times \text{Br})/(\sigma \times \text{Br})_{SM}$ |
|---|---|
| $q\bar{q}\gamma\gamma$ | 0.038 |
| $\mu^+\mu^-\gamma\gamma$ | 0.153 |
| $\nu\bar{\nu}\gamma\gamma$ | 0.048 |
| Combined | 0.031 |

model is used to fit diphoton invariant mass $m_{\gamma\gamma}$ as shown in Figure 15.22.

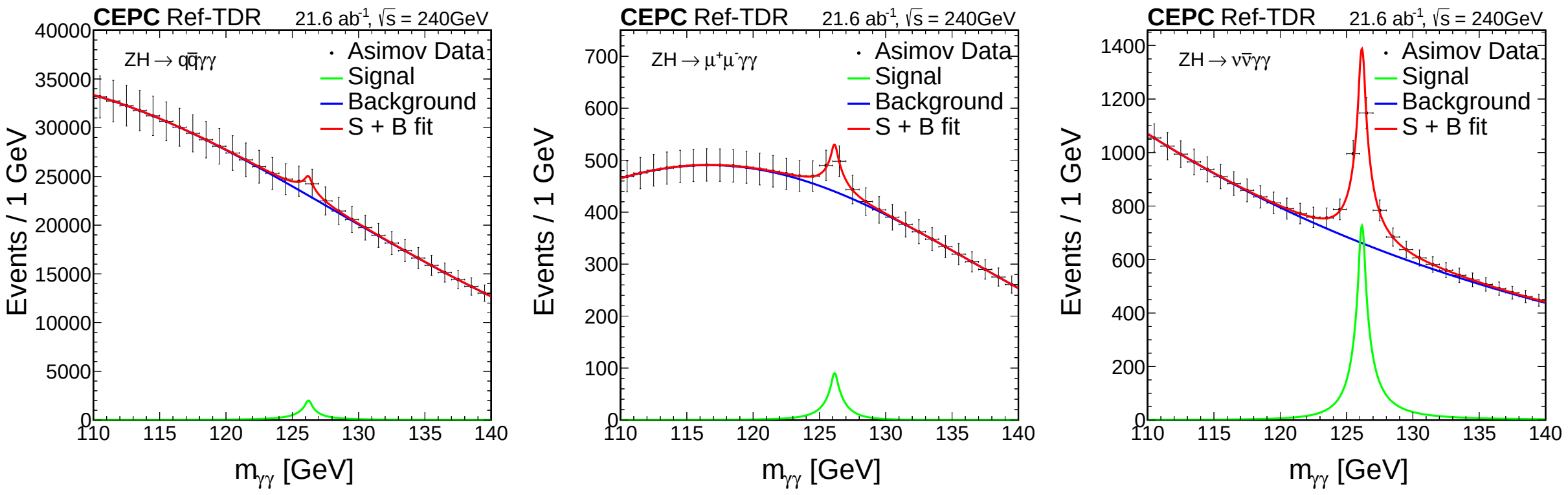

**(a)** **(b)** **(c)**

**Figure 15.22:** Combined fitting of diphoton invariant mass $m_{\gamma\gamma}$ to the Asimov data in the three channels: (a) $q\bar{q}\gamma\gamma$, (b) $\mu^+\mu^-\gamma\gamma$, and (c) $\nu\bar{\nu}\gamma\gamma$. The error bars indicate the statistical uncertainty from the Asimov data.

The relative precision of $\sigma \times Br(H \to \gamma\gamma)$ obtained by fitting the Asimov data is summarized in Table 15.10, with only statistical uncertainties considered. Dominant systematic uncertainties are from the photon reconstruction efficiency, energy scale and resolution, and the mis-modeling introduced by the analysis strategy. Their impact was discussed in Ref. [22] and a 15% degradation on the precision is expected for 21.6 ab$^{-1}$ data, conservatively. As expected from the homogeneous ECAL, slight improvement from the CDR considering the sampling ECAL is obtained.

An equivalent comparison to the current LHC results [23, 21] is not possible due to the different production kinematics between the lepton and hadron colliders, but overall, significantly improved precision is expected, attributed to a better energy resolution and lower background.

### 15.2.5 $H \to$ invisible

In the SM, the Higgs boson can decay into two $Z$ bosons, each decaying into two neutrinos, with a Branching Ratio (BR) of around 0.1%. On the other hand, BSM models predict more scenarios of the Higgs boson's invisible decays, including those into dark matter or supersymmetric particles, among others. Compared with the LHC and HL-LHC, future electron-positron colliders would provide significantly improved sensitivities [24, 25, 26] due to their low background and complete reconstruction of the missing information.

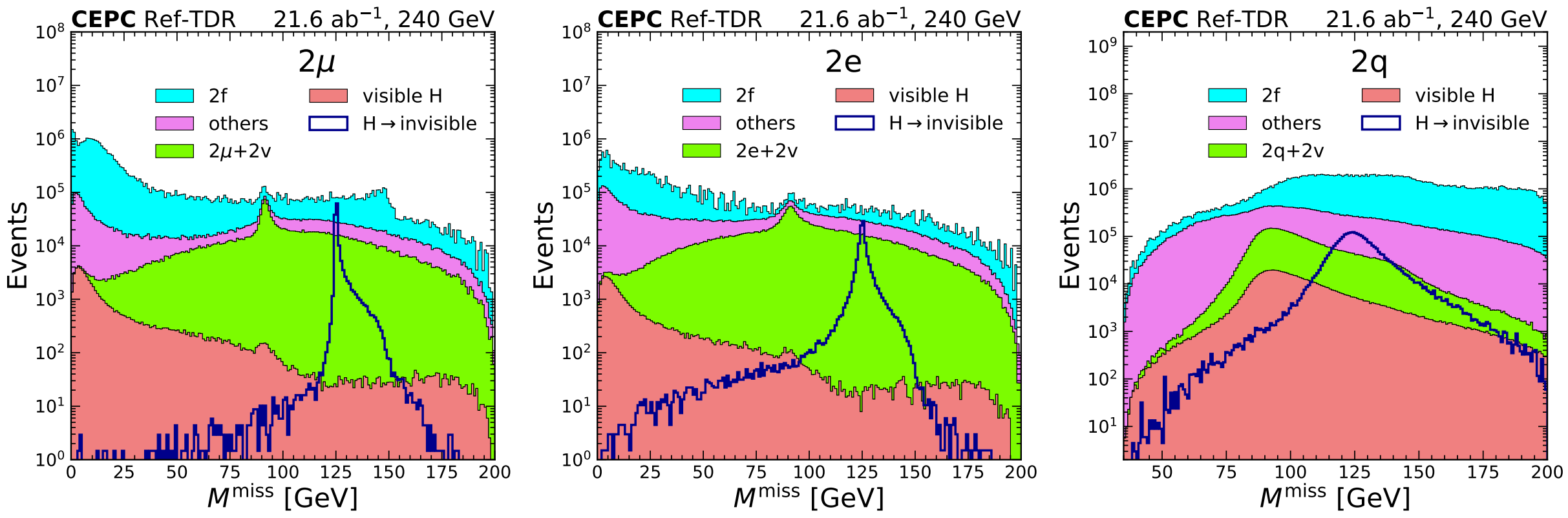

(a) (b) (c)

**Figure 15.23:** The $M^{\rm miss}$ distributions of signal and background processes in the (a) $2\mu$, (b) $2e$, and (c) $2q$ channels after baseline selection, estimated with simulation samples. To visualize the signal distribution shape, its invisible decay branching ratio is set to be 1.

Searches for the Higgs boson invisible decay are performed with full simulation samples at $\sqrt{s}$ = 240 GeV, described in Section 15.2.1. For the signal, the Higgs-strahlung process is considered, where the $Z$ boson decays into two muons, two electrons or two quarks, and the Higgs boson decays into four neutrinos. All background processes are considered, including the other Higgs boson production and decay channels, 4-fermion final state, and 2-fermion final state processes.

Events first satisfy the baseline selection and are categorized into three channels, selected successively:

- The $2\mu$ channel: events should contain exactly two PFOs satisfying the BEST muon ID WP and with $|\cos\theta| < 0.99$. The two muons should have the opposite charge with their invariant mass between 40 and 120 GeV.
- The $2e$ channel: events should satisfy the same criteria as the $2\mu$ channel, except for the lepton ID being replaced to that for the electrons.
- The $2q$ channel: events should not belong to the channels above, but should have visible mass between 30 and 130 GeV, and visible momentum between 10 and 80 GeV.

The missing mass $M^{\rm miss}$, introduced in Section 15.1.9 (Equation 15.2), has the strongest sensitivity to the signal among the kinematic variables considered in the analysis. Its distributions are presented in Figure 15.23 for the three channels. The signal processes are distributed around 125 GeV, whereas backgrounds are broadly distributed with different features depending on their physics processes. Especially, the irreducible backgrounds that have the same final states as the signal, shown as the green histograms in each plot, are mainly composed of the $ZZ$ production, and therefore distributed around 91 GeV.

**Table 15.11:** Total generated yields, baseline selection efficiency, kinematic selection efficiency (yields after both selections divided by yields after baseline selection), and yields after both selections. For the signal process, only the final state corresponding to the channel is considered.

| | **process** | **signal** | **$2(\mu/e/q)$+$2\nu$** | **2-fermion** | **visible $H$** | **others** |
|---|---|---|---|---|---|---|
| $2\mu$ | Total yield | $1.56\times10^{2}$ | $6.13\times10^{6}$ | $1.92\times10^{9}$ | $4.40\times10^{6}$ | $4.09\times10^{8}$ |
| | Baseline sel | 96.1% | 32.0% | 2.35% | 2.55% | 0.88% |
| | Kinematic sel | 98.0% | 19.8% | 3.40% | 0.44% | 5.31% |
| | Selected yield | $1.46\times10^{2}$ | $3.88\times10^{5}$ | $1.53\times10^{6}$ | $4.91\times10^{2}$ | $1.92\times10^{5}$ |
| $2e$ | Total yield | $1.61\times10^{2}$ | $6.02\times10^{6}$ | $1.92\times10^{9}$ | $4.40\times10^{6}$ | $4.09\times10^{8}$ |
| | Baseline sel | 83.8% | 41.7% | 1.03% | 1.96% | 1.60% |
| | Kinematic sel | 95.3% | 23.0% | 3.35% | 2.19% | 5.77% |
| | Selected yield | $1.29\times10^{2}$ | $5.78\times10^{5}$ | $6.62\times10^{5}$ | $1.89\times10^{3}$ | $3.77\times10^{5}$ |
| $2q$ | Total yield | $3.13\times10^{3}$ | $7.98\times10^{6}$ | $1.92\times10^{9}$ | $4.40\times10^{6}$ | $4.07\times10^{8}$ |
| | Baseline sel | 99.0% | 66.1% | 9.24% | 19.8% | 8.35% |
| | Kinematic sel | 95.4% | 38.1% | 37.3% | 37.8% | 12.9% |
| | Selected yield | $2.96\times10^{3}$ | $2.01\times10^{6}$ | $6.62\times10^{7}$ | $3.28\times10^{5}$ | $4.37\times10^{6}$ |

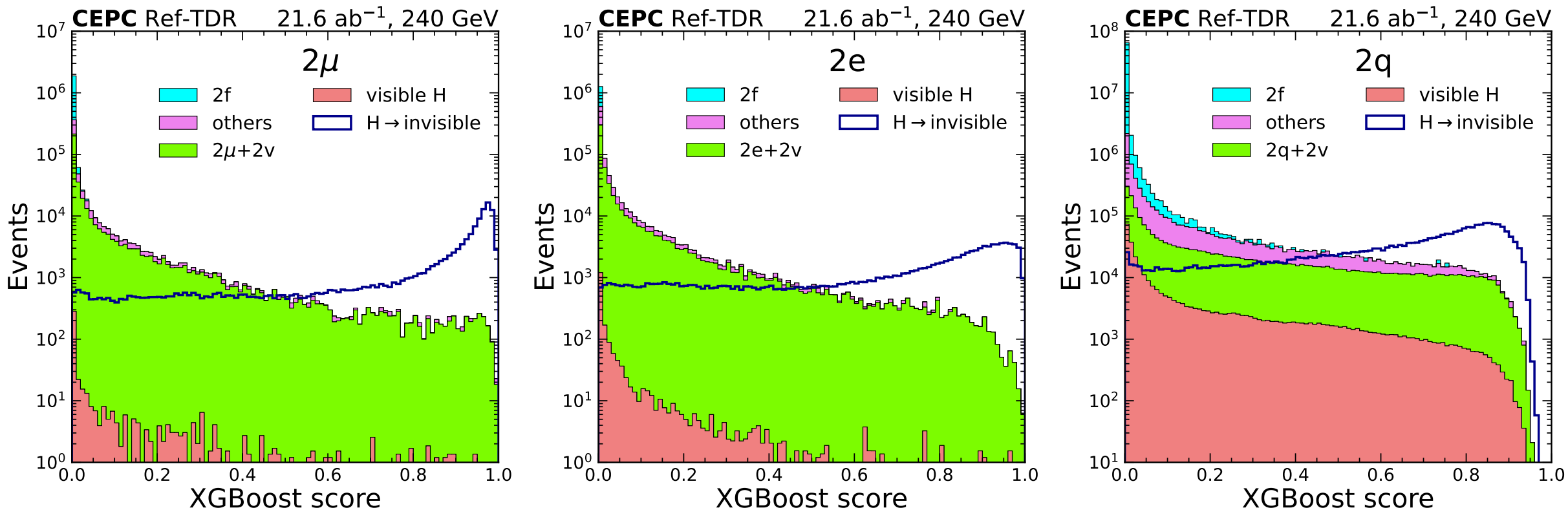

(a) (b) (c)

**Figure 15.24:** The XGBoost score distributions of signal and background processes in the (a) $2\mu$, (b) $2e$ and (c) $2q$ channels. To visualize the signal distribution shape, its invisible decay branching ratio is set to be 1.

Events are further selected based on the kinematics including $M^{\text{miss}}$, visible energy and momentum, missing energy and momentum, as well as the number and total energy of charged or neutral particles. The selection efficiency for the signal and background processes is summarized in Table 15.11.

After event selection, an XGBoost model is trained in each channel to discriminate signal and background, exploiting all variables mentioned above, as well as lepton impact parameters and jet substructure variables. The distributions of the output scores are shown in Figure 15.24. Most backgrounds are concentrated around zero. The major backgrounds that have large contamination in the high score region are the irreducible backgrounds. The distributions of the XGBoost score are directly fitted to do statistical analyses. Systematic uncertainties, including luminosity, beam energy measurements, efficiencies, and resolutions, are estimated to be less than 1%, which is negligible compared with the statistical uncertainty. For the SM invisible decay, expected uncertainties in the decay branching ratio and statistical significance are computed; for BSM scenarios, the SM signal is added as an additional background, and expected Upper Limits (ULs) at 95% CL on the decay branching ratio are computed.

**Table 15.12:** Expected relative uncertainties, statistical significance of the SM Higgs boson invisible decay, and upper limits on the BSM invisible decay branching ratio for 20 ab$^{-1}$ in each channel and all channels combined. Only the statistical uncertainty is included.

| **channel** | **uncertainties** | **significance of SM obs.** | **UL on BSM BR** |
|---|---|---|---|
| $2\mu$ | −42%/+43% | $2.5\sigma$ | 0.089% |
| $2e$ | −60%/+62% | $1.7\sigma$ | 0.13% |
| $2q$ | −29%/+29% | $3.4\sigma$ | 0.061% |
| combine | −22%/+22% | $4.5\sigma$ | 0.047% |

Results with only the statistical uncertainty considered are shown in Table 15.12. Overall, the $2q$ channel has the highest sensitivity. The combined significance of observing the SM invisible decay is expected to reach 4.5 $\sigma$ with 21.6 ab$^{-1}$, significantly improved compared with Ref. [26], because of the multivariate analysis approach. The expected UL at 95% CL on the BSM invisible decay branching ratio is 0.049%, which is about two orders of magnitude more stringent than the reach by the HL-LHC experiments [18].

Considering beam induced backgrounds, as mentioned in Section 15.1.9, their effects on the missing mass are eliminated by discarding neutral PFOs with $|\cos\theta| > 0.98$. Therefore, the effects on sensitivities to the Higgs boson invisible decay should be under control.

### 15.2.6 Long-lived particles searches

The hypothesis that BSM particles could have prolonged lifetimes, escaping immediate detection, has evolved from a speculative concept to a major, actively pursued direction in particle physics. These LLPs are now widely recognized as sensitive probes to BSM phenomena.

At lepton colliders, the Higgs bosons can be produced via the process $e^+e^- \rightarrow ZH$, offering a clean and well-defined environment to search for rare LLP decays. Thanks to the precise initial-state conditions and significantly reduced background compared with hadron colliders, this process provides an ideal basis for exploring exotic decay signatures of the Higgs boson [27, 28, 29]. The Higgs boson can decay into a pair of LLPs (denoted by $\chi$), each of which later decays into final state objects such as jets, charged leptons, or neutrinos. For the multijet final state, the search for LLPs at lepton colliders was studied previously using machine learning techniques [27]. The leptonic final state proceeds via the cascade $H \rightarrow \chi\chi \rightarrow \ell^+\ell^- + \ldots$, where each long-lived particle $\chi$ can decay into either two charged leptons ($\ell^+\ell^-$) or neutrinos. With

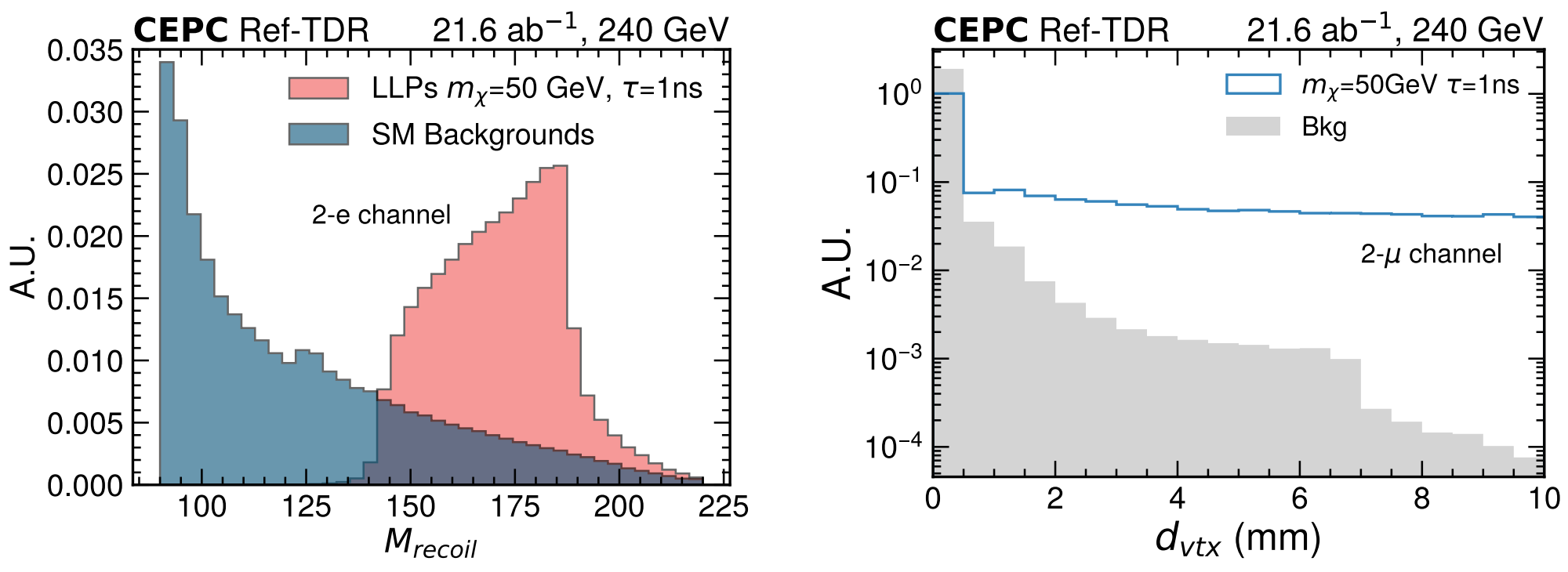

**(a)** **(b)**

**Figure 15.25:** The signal and SM background distributions of (a) $M_{\text{recoil}}$ for $2e$-channel and (b) $d_{vtx}$ for 2-$\mu$ channel. The LLP signal is at $m_\chi = 50$ GeV and 1 ns lifetime.

$Z$ bosons decaying into neutrinos, the entire process may yield either a 2-lepton or 4-lepton final state, depending on the specific $\chi$ decay model. This work focuses on the 2-lepton channels, including both dielectron (2-$e$) and dimuon (2-$\mu$) final states. This choice allows clean reconstruction and efficient background suppression.

The experimental signature for the leptonic decay model of the LLPs is a displaced vertex accompanied by two leptons. Fifteen signal datasets are generated to cover the LLP parameter space with LLP masses of 1, 10, and 50 GeV and lifetimes of 0.001, 0.1, 1, 10, and 100 ns. The dominant background processes include $e^+e^- \to ZH \to$ inclusive process, $e^+e^- \to \ell^+\ell^-$ process, diboson production processes such as $e^+e^- \to ZZ \to \ell^+\ell^-\nu\bar{\nu}$ and $e^+e^- \to W^+W^- \to \ell^+\nu\ell^-\bar{\nu}$, single boson production processes such as $e^+e^- \to \nu\bar{\nu}Z$ and $e^+e^- \to e^\pm\nu W^\mp$.

Events are required to contain an oppositely charged lepton pair, with each lepton having momentum ($p_\ell$) greater than 3 GeV and satisfying the BEST WP criteria. Additional selections are applied with the following variables to suppress SM backgrounds:

- nPFOs: the total number of reconstructed particle-flow objects,
- $d_{vtx}$: the distance between the lepton-pair vertex and the IP,
- $\Delta\theta_{\ell\ell}$: the polar angle difference between the two leptons,
- $M_{\ell\ell}$: invariant mass of the dilepton system,
- $M_{\text{recoil}}$: inferred mass for the system recoiling against the identified leptons,
- $\Delta T = \sum_i \left(t_{\text{hit},i} - r_{\text{hit},i}/c\right)$: the time-of-flight delay, where $t_{\text{hit},i}$ is the hit time of the $i$-th hit in the ToF detector, $r_{\text{hit},i}$ is the distance of that hit to the IP, and $c$ is the speed of light in the vacuum.

The nPFO $< 20$ requirement suppresses SM processes involving jets, while the $Z$ mass window selection ($Z$ veto) with $|M_{\ell\ell} - 90| > 5$ GeV reduces $Z$ related backgrounds. The $M_{\text{recoil}}$ selection further suppresses $Z/W$ related backgrounds. The timing resolution of the ToF detector enables background rejection through the time-of-flight delay ($\Delta T$), defined as the difference between the measured hit time in the timing layers and the expected arrival time for

**Table 15.13:** Event selection criteria for LLP signals.

| Selection | Category | Criteria |
|---|---|---|
| Jet veto | 2-$e$ and 2-$\mu$ channels | nPFO < 20 |
| Lepton ID | 2-$e$ and 2-$\mu$ channels | two oppositely charged leptons satisfying **BEST WP** |
| Lepton momentum | 2-$e$ and 2-$\mu$ channels | $p_\ell > 3$ GeV |
| Polar angle difference | 2-$e$ and 2-$\mu$ channels | $1^\circ < \Delta\theta_{\ell\ell} < 60^\circ$ |
| $Z$ veto | 2-$e$ and 2-$\mu$ channels | $\lvert M_{\ell\ell} - 90\rvert > 5$ GeV |
| Recoil mass | 2-$e$ channel<br>2-$\mu$ channel | $M_{\text{recoil}} > 130$ GeV<br>$M_{\text{recoil}} > 140$ GeV |
| Vertex displacement | $m_\chi = 1$ GeV<br>$m_\chi = 10, 50$ GeV | $d_{vtx} > 3.5$ mm<br>$d_{vtx} > 1$ mm |
| Invariant mass window | $m_\chi = 1$ GeV<br>$m_\chi = 10, 50$ GeV | $\lvert M_{\ell\ell} - m_\chi\rvert < 0.6$ GeV<br>$\left\lvert \frac{M_{\ell\ell}-m_\chi}{m_\chi} \right\rvert < 10\%$ |
| Time-of-flight delay | $m_\chi = 1$ GeV and 2-$\mu$ channel | $\Delta T > 0.05$ ns |

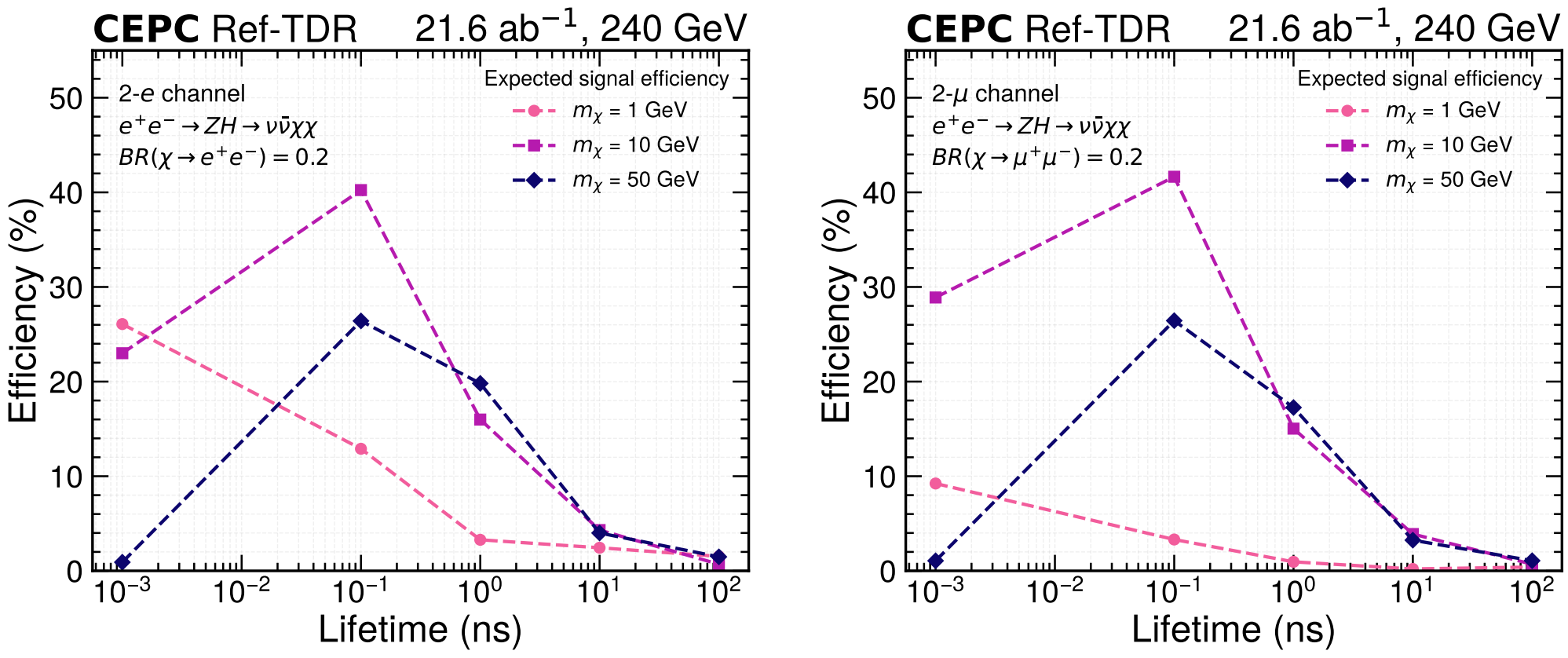

**(a)** **(b)**

**Figure 15.26:** LLPs signal efficiencies for the (a) 2-$e$ channel and (b) 2-$\mu$ channel.

prompt particles traveling at the speed of light from the interaction point. Figure 15.25 shows the distributions of $M_{\text{recoil}}$ and $d_{vtx}$ for the LLP signal and SM backgrounds.

The event selection in the 2-lepton channels is summarized in Table 15.13. Figure 15.26 shows the signal efficiencies after all selections. The signal efficiency reaches up to 40% in both channels at $m_\chi = 10$ GeV under background-free conditions.

With an integrated luminosity of 21.6 ab$^{-1}$ at $\sqrt{s} = 240$ GeV for CEPC, 95% CL upper limits on the BR for the Higgs boson decay into pairs of LLPs are shown in Figure 15.27, assuming BR($\chi \to \ell\ell$) = 20%. The most stringent limits reach a BR of $4 \times 10^{-5}$ for $m_\chi = 10$ GeV with a lifetime of 0.1 ns, where the LLP signal achieves optimal detector acceptance and selection efficiency. The search sensitivity to lighter LLPs such as those with $m_\chi = 1$ GeV degrades because their decays produce softer, more collimated, and more prompt signatures, which reduces both reconstruction efficiency and the effectiveness of background suppression

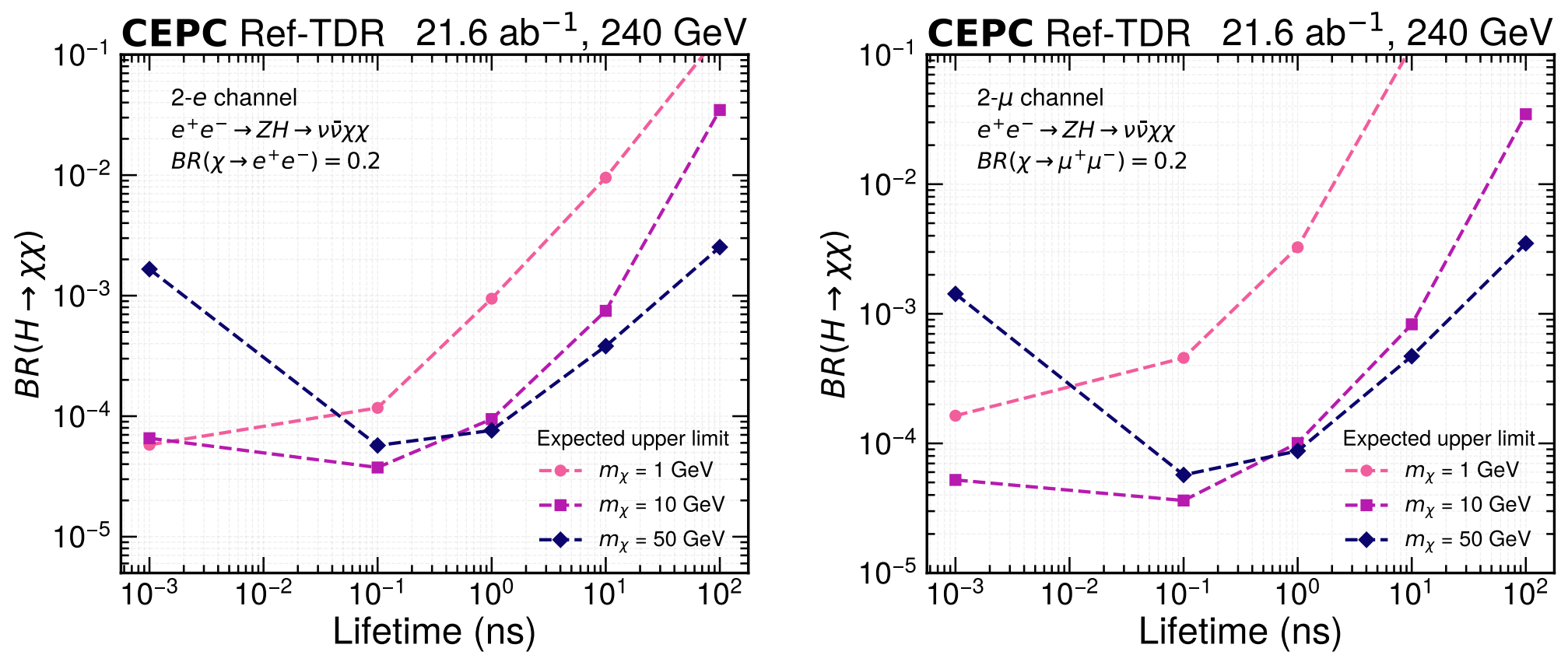

**(a)** **(b)**

**Figure 15.27:** 95% CL upper limits on BR$(H \to \chi\chi)$ for the (a) 2-$e$ and (b) 2-$\mu$ channels.

techniques such as vertex displacement and timing discrimination. For LLPs with lifetimes longer than 100 ns, their search sensitivity can benefit from an external detector located far away from the detector [27].

As shown in Figure 15.27, the sensitivity in the 2-$e$ and 2-$\mu$ channels reaches branching ratio limits on the order of $10^{-4}$ to $10^{-5}$ for $m_\chi = 10$ GeV and lifetimes near 0.1–1 ns, comparable to the latest LHC constraints [30, 31, 32]. For the optimal mass–lifetime region, the projected exclusion limit at the CEPC can be improved significantly, by up to a factor of 5, when additional final states from the $Z$ decay are included and combined.

### 15.2.7 Supersymmetric muon

Supersymmetry (SUSY), a symmetry relating bosons and fermions, predicts a superpartner particle for every SM particle. As one of the most compelling frameworks beyond the SM, it offers potential solutions to critical challenges including the hierarchy problem, dark matter, and the unification of the fundamental forces.

Designed for the electroweak energy scale, the CEPC operating at $\sqrt{s}$ = 240 GeV can probe crucial low-mass parameter spaces—particularly compressed mass scenarios challenging to explore at hadron colliders. Light smuons, favored by dark matter relic density constraints, represent promising targets for searches at the CEPC. This analysis [33] focuses on the pair production of charged smuons, which decay into a final state containing two muons and missing energy from two Lightest Supersymmetric Particles (LSPs), specifically $\tilde{\chi}_1^0$. It can also demonstrate the muon and missing energy performance from the corresponding sub-detectors, such as the muon detector and calorimeter.

MadGraph is used for the generation of the signal process, and the theoretical cross sections at the leading order is 2.17 fb for smuon with mass of 119 GeV [34]. Events are selected with exactly two Opposite Sign (OS) muons, each with the momentum exceeding 1.0 GeV. The recoil system comprises all particles excluding these two OS charged leptons, including invisible particles such as neutrinos and neutralinos. The following variables are employed to distinguish signal events from SM backgrounds:

- $\Delta R(\mu, \text{recoil})$: the angular distance between the leading muon and the recoil system.
- $E_\mu$: muon energy.
- $M_{\mu\mu}$: the invariant mass of two muons.
- $M_{\text{recoil}}$: the invariant mass of the recoil system.

**Table 15.14:** Summary of selection requirements for the signal regions to search for the direct smuon production.

| **SR-$\Delta M^h$** | **SR-$\Delta M^m$** | **SR-$\Delta M^l$** |
|---|---|---|
| $E_{\mu 1,2} > 40$ GeV | $9 < E_{\mu 1,2} < 48$ GeV | $E_{\mu 1,2} > 1$ GeV |
| $E_{\mu 1,2} \in (40 - -50, > 50)$ GeV | $E_{\mu 1,2} \in (9 - -25, 25 - -48)$ GeV | – |
| $\Delta R(\mu, \text{recoil}) < 2.9$ | $1.5 < \Delta R(\mu, \text{recoil}) < 2.8$ | $1.5 < \Delta R(\mu, \text{recoil}) < 2.8$ |
| $M_{\mu\mu} < 60$ GeV | $M_{\mu\mu} < 80$ GeV | – |
| $M_{\text{recoil}} > 40$ GeV | – | $M_{\text{recoil}} > 220$ GeV |

Signal regions are optimized through multi-dimensional scans of the kinematic variables defined above with varied thresholds to minimize variable correlations. The most sensitive selection criterion is adopted, validated by re-evaluating the kinematic distributions, and is used to define the signal regions. To estimate the sensitivity to the signals, the median significance denoted by $Z_n$ [35] is used, which is a reliable approximation when the number of signal events is sizable. In the calculation of $Z_n$ for defining and optimizing signal regions, both the statistical uncertainty and a 5% flat systematic uncertainty are taken into account.

Table 15.14 summarizes three Signal Regions (SRs) developed to cover different mass splittings between $\tilde{\mu}$ and $\tilde{\chi}_1^0$ ($\Delta M$). The SR-$\Delta M^{h(m,\,l)}$ covers regions with high (medium, low) $\Delta M$. To enhance signal sensitivity, the SR-$\Delta M^{h(m)}$ regions are further subdivided into intervals based on the energies of the leading and sub-leading muons ($E_{\mu 1}$ and $E_{\mu 2}$, collectively denoted by $E_{\mu 1,2}$), where $\mu 1$ ($\mu 2$) is the muon with the highest (second-highest) energy. The $E_{\mu 1,2}$ selections are implemented to reject $\tau^+\tau^-$ and $\nu\bar{\nu}Z$ backgrounds. Requirements on $\Delta R(\mu, \text{recoil})$ are employed to suppress $\tau^+\tau^-$, $\mu^+\mu^-$ and $ZZ$ processes, while $M_{\mu\mu}$ selections serve to suppress $W^+W^-$, $\mu^+\mu-$ processes and additional backgrounds including $Z$ bosons. Based on the signal event topology, most signal events exhibit a large recoil mass. Therefore, events with low $M_{\text{recoil}}$ are excluded to suppress $\mu^+\mu^-$, $ZZ$, $W^+W^-$ processes, and other SM processes within this region. The main background contributions to the SRs originate from the processes of $ZZ \to \mu^+\mu^-\nu\bar{\nu}$, $W^+W^- \to \mu^+\mu^-\nu\bar{\nu}$, $\mu^+\mu^-$, $\tau^+\tau^-$ and $\nu\bar{\nu}Z, Z \to \mu^+\mu^-$.

Figure 15.28 presents the projected exclusion limits ($2\sigma$) and discovery potential ($5\sigma$) contours for the direct production of smuons, with different assumptions of the systematic uncertainty 0%, 3% and 5%. For each signal point, the signal region yielding the optimal $Z_n$ value is selected. To evaluate the impact of detector-related systematic uncertainties, a conservative 5% systematic uncertainty—consistent with the LEP experiment—is employed. The discovery sensitivity reaches up to 119 GeV, varying with the masses of the LSP and smuon. Notably, this sensitivity is not significantly impacted by the detector-related systematic uncertainties. Since the most sensitive signal region is selected for each individual signal point, adjacent signal points in overlapping regions may be associated with different signal regions, potentially leading to non-smooth exclusion contours.

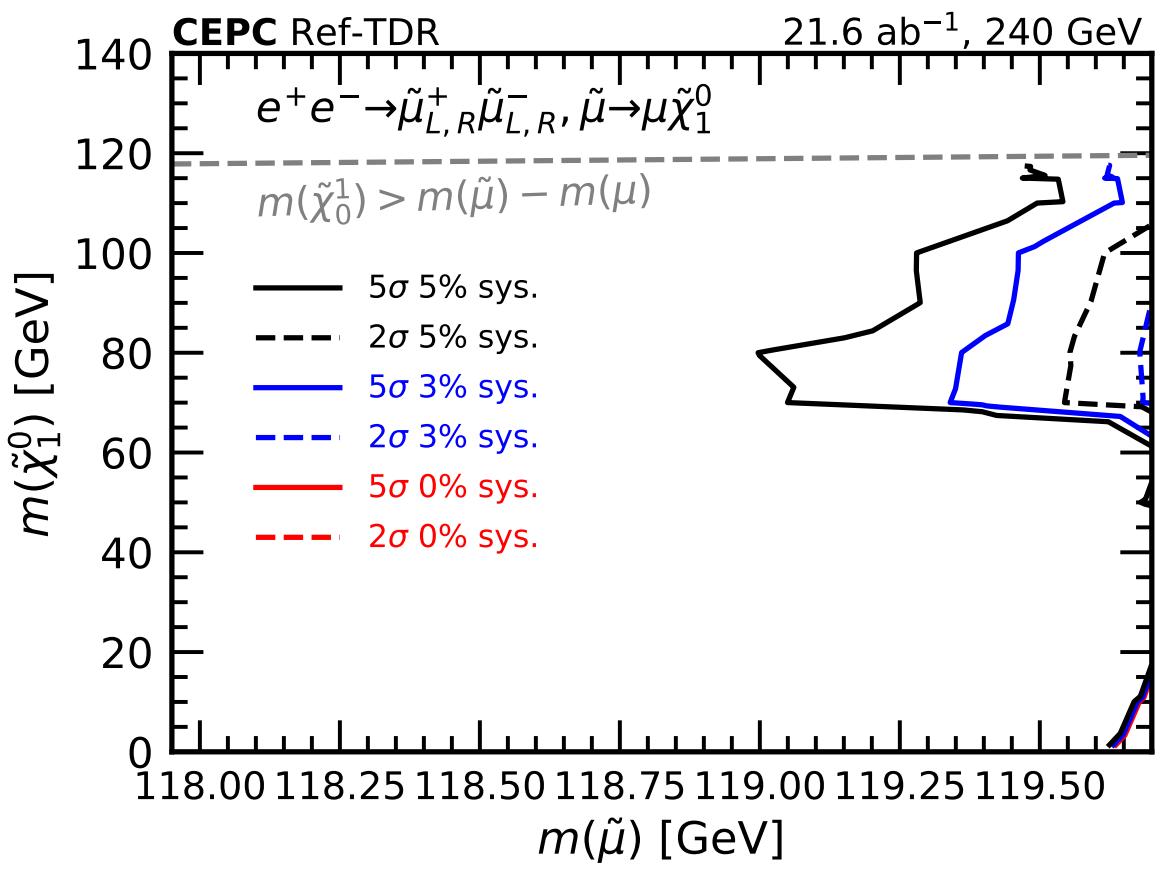

**Figure 15.28:** The projected exclusion ($2\sigma$) and discovery ($5\sigma$) contours for the direct $\tilde{\mu}$ production at the CEPC, considering fixed systematic uncertainties ranging from 0% to 5%. The red contours correspond to the statistics-only scenario. The gap in the exclusion limits arises because signal points are capped at a maximum mass of 119.7 GeV; beyond this value, the sensitivity exceeds the $5\sigma$ discovery threshold.

The $\tilde{\mu}$ mass limit extends approximately 20 GeV beyond the previous constraints set by the LEP in the high $\tilde{\mu}$ mass region. Additionally, this analysis can probe the compressed parameter space where the mass difference between the smuon and the lightest supersymmetric particle (LSP) is small—a region that is challenging to explore by ATLAS and CMS at the LHC.

### 15.2.8 $A^{\mu}_{FB}$ ($e^+e^- \to \mu^+\mu^-$) at $Z$ pole

The CEPC data at the $Z$-pole energy allow for high precision electroweak measurements of the $Z$ boson properties, such as the forward–backward charge asymmetry ($A_{FB}$) as a function of the effective weak mixing angle. The $\mu^+\mu^-$ channel is one of the cleanest final states at the $Z$ pole.

The forward–backward asymmetry is defined in terms of the angle $\theta_{CM}$ between the negatively charged final-state muon and the initial-state electron in the dilepton center-of-mass frame:

$$A_{FB} = \frac{\sigma_F - \sigma_B}{\sigma_F + \sigma_B}, \tag{15.7}$$

where $\sigma_F$ ($\sigma_B$) is the total cross section for the forward (backward) events, defined by $\cos\theta_{CM} > 0$ ($< 0$).

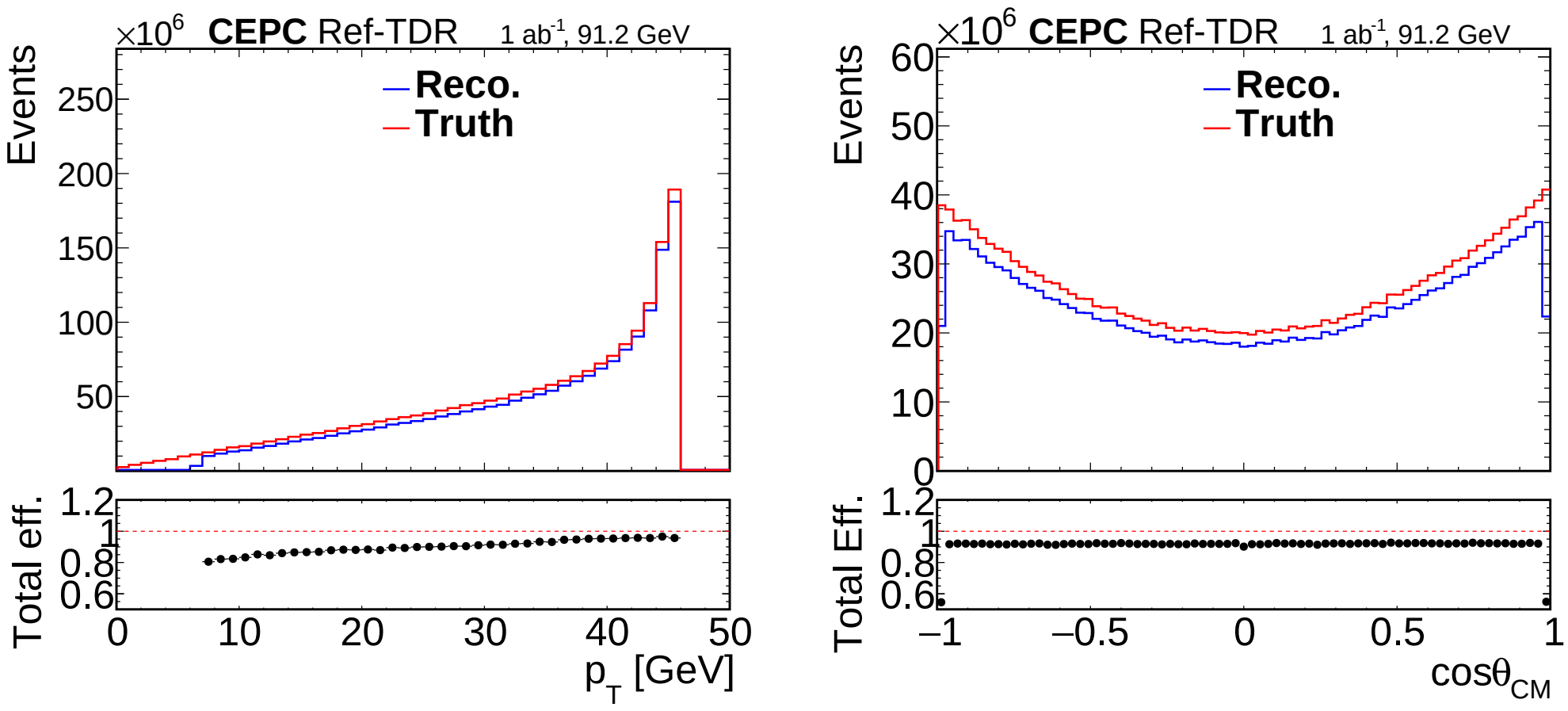

**(a)** **(b)**

**Figure 15.29:** Distributions of (a) $\mu^-$ transverse momentum at the lab frame and (b) $\mu^-$ $\cos(\theta_{CM})$ at the center-of-mass frame. The bottom panels reflects the total event selection efficiency.

When the collision energy is close to the $Z$ boson mass, $A_{FB}$ reflects a pure $Z$ exchange and is close to zero because of the small value of the charged-lepton vector coupling to the $Z$ bosons. The combination of LEP measurements yields a value of $A^{\mu}_{FB} = 0.0169 \pm 0.0013$ in the $\mu^+\mu^-$ channel [36]. At the CEPC, the statistical uncertainty of $A^{\mu}_{FB}$ can be significantly reduced, and the total uncertainty is expected to be dominated by systematic uncertainties, such as the beam energy spread and the energy and angular resolution of the reconstructed PFO [37].

The signal and background events are simulated as described in Section 15.2.1. The interference between $Z$ and $\gamma^*$ is included, and the ISR and FSR are enabled in the simulation. The predicted $A^{\mu}_{FB}$ is $0.0161 \pm 0.0010$ by simulating one million events, which is consistent with the LEP result.

The $e^+e^- \to \mu^+\mu^-$ events are selected by identifying the opposite-charge muon pairs. The muons are required to satisfy transverse momentum $p_T > 1$ GeV, $\cos\theta < 0.99$, and the muon identification requirements (the "BEST" working point). A $\pm 10$ GeV $Z$ boson mass window is required to reject the background with genuine muons, which mainly originate from $e^+e^- \to \tau^+\tau^-$ and $e^+e^- \to b\bar{b}$ events. Finally, a $|\cos(\theta_{\mu^-})| > 0.05$ selection is applied to remove migrations between forward and backward regions.

Figure 15.29 presents the transverse momentum and $\cos(\theta_{CM})$ distributions of the genuine and reconstructed $\mu^-$, and the bottom panels reflects the total event selection efficiency. Table 15.15 shows the event yields and selection efficiencies of signal and background. The signal efficiency is approximately 90%, while the background efficiencies are lower than 0.005%. The main background is $e^+e^- \to \tau^+\tau^-$ and the others are negligible.

Since the fraction of signal events is close to 100% after the event selection, the $A^{\mu}_{FB}$ is directly calculated by counting the forward (backward) events, by assuming $\cos(\theta_{CM}) > 0$ ($<$ 0). The asymmetry estimated with this method $A^{obs}_{FB}$ corresponds to a restricted phase space, mainly due to the $Z$ boson mass window constraint, and has to be corrected for this effect. A multiplicative phase space correction factor is calculated using MC events, as the ratio of the obtained $A_{FB}$ with no event selection, versus the value obtained by applying the event selection (Table 15.16). This correction factor deviates from unity by around $9\times10^{-6}$, which is considered as a systematic uncertainty.

**Table 15.15:** The cross section, number of simulated events, and selection efficiencies of signal, $e^+e^- \rightarrow \mu^+\mu^-$, and background, $e^+e^- \rightarrow \tau^+\tau^-$ and $e^+e^- \rightarrow b\bar{b}$.

| | $e^+e^- \rightarrow \mu^+\mu^-$ | $e^+e^- \rightarrow \tau^+\tau^-$ | $e^+e^- \rightarrow b\bar{b}$ | $e^+e^- \rightarrow e^+e^-$ |
|---|---|---|---|---|
| Cross section | 1.5 nb | 1.5 nb | 6.7 nb | 6.6 nb |
| Simulated events | 982476 | 185855 | 44550 | 32397 |
| A pair of muons | 967262 | 5135 | 1035 | 0 |
| $Z$ mass window | 903640 | 5 | 0 | 0 |
| Muon $\|\cos(\theta)\| > 0.05$ | 869450 (88.5%) | 5 (0.003%) | 0 (<0.002%) | 0 (<0.003%) |

The statistical uncertainty of $A_{FB}^{\mu}$ is extrapolated from $10^6$ muon pairs in the simulated sample for the following data-taking scenario: one-month low-luminosity $Z$ pole data taking in the first year of $ZH$ operation, with $4 \times 10^{10}$ $Z$ bosons expected during this period. Several systematic uncertainties are considered and estimated as follows:

- Muon mis-identification: the probability and impact of selecting incorrect pairs of PFO from the signal events is found to be negligible.
- Background: by comparing $A_{FB}^{obs}$ with and without background events, the uncertainty is measured to be $1 \times 10^{-6}$.
- Detector acceptance and resolution of $|\cos(\theta_{\mu^-})|$ and $p_{\mathrm{T}}^{\mu^-}$: by comparing $A_{FB}^{obs}$ with event selections at the generator-level or PFO, the uncertainty is estimated to be $9 \times 10^{-6}$.
- Beam energy calibration and spread: The uncertainty from the beam energy spread is estimated by comparing $A_{FB}^{\mu}$ with a set of simulated samples generated at different energy values, considering a spread of 0.13% and 0.14% [38]. The beam energy spread is assumed to follow a Gaussian distribution. The uncertainty is measured to be $1 \times 10^{-6}$. A 300 keV beam energy calibration uncertainty is also assigned, as a conservative assumption based on a 100 keV goal of the FCC-ee experiment. The impact on $A_{FB}^{\mu}$ from energy calibration is measured to be $2.7 \times 10^{-5}$.

An alternative method fitting the $\cos(\theta_{CM}^{\mu^-})$ distribution, which is totally independent on acceptance, is also investigated and found to be consistent with the nominal counting method within the statistical uncertainty.

In summary, the uncertainty of $A_{FB}^{\mu}$ measurement is ±0.000031 (stat.) ±0.000028 (syst.), based on the data from the one-month low-luminosity $Z$-pole data taking during the first year of $ZH$ operation. The CEPC result improves the precision of the LEP result (±0.0013) [36] by two orders of magnitude.

**Table 15.16:** The number of forward (backward) events and the corresponding $A^{\mu}_{FB}$, from simulation, after selection, and after reconstruction. A correction of phase space was developed based on MC particles, and applied to the PFO, the remained difference is treated as a systematic uncertainty.

| | No selection | After selections | Using PFO |
|---|---|---|---|
| Forward ($\cos(\theta_{CM}) > 0$) | 499136 | 442727 | 442723 |
| Backward ($\cos(\theta_{CM}) < 0$) | 483340 | 426723 | 426727 |
| $A^{\mu}_{FB}$ or $A^{obs}_{FB}$ | 0.016078 | 0.018407 | 0.018398 |
| $A^{\mu}_{FB}$ after a phase space correction | N/A | 0.016078 | 0.016070 |

### 15.2.9 $R_b$ at $Z$ pole

The measurement of $R_{b\ (c,\ q)}$, which represents the relative partial decay width of the $Z$ boson into the $b\bar{b}$ ($c\bar{c}$, $q\bar{q}$) final states to the total hadronic decay, holds paramount physics significance. It serves as a fundamental cornerstone for testing the SM and a highly sensitive probe for uncovering new physics phenomena [39, 40, 41, 42].

The current experimental landscape for $R_b$ measurements is shaped by data from experiments conducted at the LEP and the Stanford Linear Collider (SLC) [43, 44, 45, 46, 47, 36]. At the CEPC, jets and their flavor tagging play a key role in the measurement of $R_b$ [48], which can serve as a key benchmark reflecting the vertex detector performance. In this study, jet flavor tagging is performed using the JOI tagger described in Section 15.1.8.2. However, instead of using the full 11 categories as in Section 15.1.8.2, several categories are merged with the following scheme: quark and antiquark categories for the same flavor are merged; categories besides $b$- and $c$-quarks are merged into the light-quark ($q$) category. With such a merged scheme, the 11 categories are reduced to three. Thus, the jet origin is determined with three probabilities: $p_b$, $p_c$, and $p_q$.

The criteria used to tag $b$, $c$, and $q$ are the same as Section 15.1.8.2. A jet is tagged as a $b$-jet if the $p_b$ is the highest among the three probabilities. The same strategy is applied to select $c$- and $q$-tagged jets. The events are split into two hemispheres by the plane perpendicular to the thrust axis and containing the interaction point. They are further required to have two jets both tagged as $b$, $c$, or $q$, located in the opposite hemispheres. Namely, events are grouped into three categories: $bb$, $cc$, and $qq$ (light-quarks), and the event categorization distributions are used as the fit templates. A binned likelihood fit is performed to extract $R_b$, $R_c$, and $R_q$ from the event categorization distributions:

$$\begin{aligned} \mathcal{L}(N_{\text{obs}}|R_b, R_c, \boldsymbol{\theta}) &= \prod_{i=b,c,q} \text{Pois}(N^{ii}_{\text{obs}}|\nu^{ii}) \times \prod_{\theta_j \in \boldsymbol{\theta}} \text{Gaus}(0|\theta_j), \\ \nu^{ii} &= R_b N^{ii}_{bb}(\boldsymbol{\theta}) + R_c N^{ii}_{cc}(\boldsymbol{\theta}) + R_q N^{ii}_{qq}(\boldsymbol{\theta}), \end{aligned} \quad (15.8)$$

where $\nu^{ii}$ is the expected number of events in category $ii$, and $N^{ii}_{\text{obs}}$ is the observed number of events in the category. The $N^{ii}_{bb\ (cc,qq)}$ represents the expected number of events selected

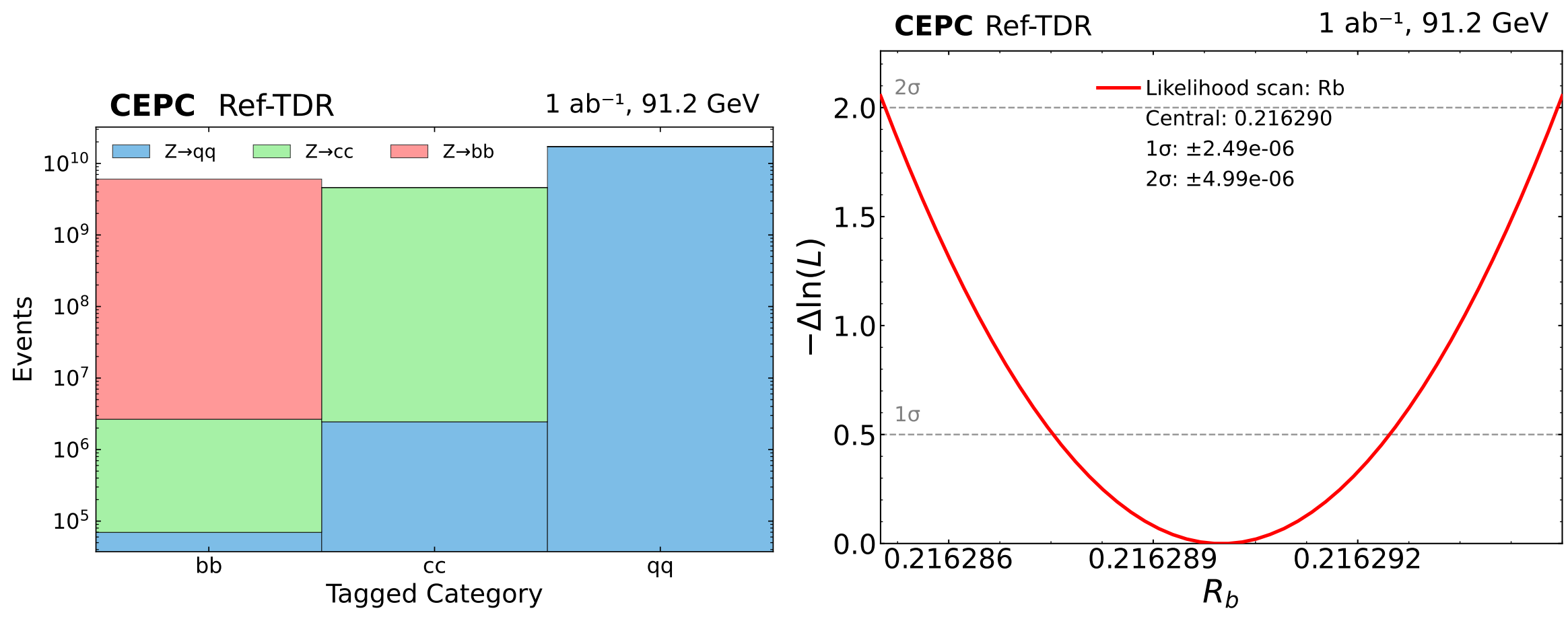

**(a)** **(b)**

**Figure 15.30:** (a) The comparison of tagged category distributions from simulated samples. Events are normalized to an integrated luminosity of 1 ab$^{-1}$. The $R_b$, $R_c$, and $R_q$ values from Particle Data Group (PDG) are used to scale the corresponding simulated samples. (b) The likelihood scan of $R_b$ with statistical-only uncertainty using 1 ab$^{-1}$ $Z$ pole data.

**Table 15.17:** Summary of $R_b$, $R_c$ measurement precision from CEPC data. Results are obtained by only including statistical uncertainties. $n_Z$ is the number of $Z$ boson events.

| **Mode** | **Luminosity** | $n_Z$ | $\sigma(R_b)$ | $\sigma(R_c)$ |
|---|---|---|---|---|
| Low luminosity $Z$ pole | 1 ab$^{-1}$ | $4 \times 10^{10}$ | $2.5 \times 10^{-6}$ | $2.4 \times 10^{-6}$ |

in category $ii$ from the $Z \to b\bar{b}$ $(cc, qq)$ events, which is determined from the MC simulated sample. The Parameters of Interest (POI) are $R_b$ and $R_c$, whereas $R_q$ is given as $1 - R_b - R_c$ in the fit. The systematic uncertainties are modeled with nuisance parameters $\boldsymbol{\theta} = \{\theta_j\}$ according to the Gaussian distribution.

Results with the low-luminosity $Z$ pole run corresponding to an integrated luminosity of 1 ab$^{-1}$ are evaluated. Figure 15.30a shows the compositions of the tagged categories. Most events are correctly tagged, and the mis-tagging rate for the $Z \to b\bar{b}$ events is at a level of $\mathcal{O}(10^{-3})$. Figure 15.30b shows the likelihood scan of $R_b$ with only the statistical uncertainty considered for 1 ab$^{-1}$. Table 15.17 summarizes the measurement precision of $R_b$ and $R_c$ at the CEPC considering only the statistical uncertainty.

A detailed evaluation of systematic uncertainty is left for future studies. The dominant source of experimental systematic uncertainty originates from JOI, arising mainly from the track impact parameter and the correlation between the two hemispheres in the events. The impact of JOI uncertainty is estimated to be at a level of $\mathcal{O}(10^{-5})$ by assuming a variation of 0.1‰ on $p_b$, $p_c$ and $p_q$ respectively, and could be reduced by a detailed study of the vertex detector performance. The modeling of gluon radiation from quarks, the splitting of gluons into $b\bar{b}$, and parton shower would be major theoretical systematic uncertainties. It is foreseeable that these uncertainties would be reduced with an improved theoretical calculation of the QCD process and constraints with data from the LHC and CEPC experiments in the future. Even in the conservative scenario, the $R_b$ ($R_c$) measurement at CEPC is anticipated to demonstrate a one (two) order-of-magnitude improvement compared with the previous experiments at LEP and SLC [36].

### 15.2.10 CP violation searches in $D^0 \to h^- h^+ \pi^0$

The yields of heavy-flavor hadrons from the CEPC $Z$-pole operation mode are larger than those from the existing electron-positron colliders by one to several orders of magnitude [49], and CEPC can access heavy hadrons that those colliders cannot produce. Due to the large production cross section at hadron colliders, LHCb can measure more heavy-flavor hadrons than the CEPC. However, the complex collision environment at LHCb leads to significantly lower reconstruction and selection efficiencies compared to CEPC. Consequently, CEPC still holds advantages in many scenarios, particularly for heavy-flavor decays involving missing energy, tau leptons, or neutral final-state particles. Given the high luminosity, low background, and excellent detector performance, CEPC may significantly enhance the precision of heavy-flavor physics.

To reconstruct heavy-flavor decay events, track and vertex reconstruction, particle identification, and neutral particle reconstruction are all crucial. Here, $D^0 \to h_1^- h_2^+ \pi^0$ (where "$h_{1/2}$" represents either a kaon or a pion) process is considered as a benchmark, to show the impact of detector performance on flavor physics studies.

Charge Conjugation And Parity (CP) violation in charm meson two-body decays was recently discovered by the LHCb experiment [50]. However, multi-body decays provide rich resonance structures, which could be helpful in understanding the source of CP violation. The sensitivity of CP violation searches largely depends on the sample statistics. Due to the large branching fraction of the singly-Cabibbo-suppressed decay $D^0 \to \pi^- \pi^+ \pi^0$, it could potentially be a sensitive channel to CP violation in multi-body decays. The Cabibbo-favored decay channel $D^0 \to K^- \pi^+ \pi^0$ will serve as a reference channel.

To obtain a reasonable estimate of the yields, a quantitative study of the efficiency for reconstructing and selecting $D^0$ decays must be pursued. An inclusive $Z \to q\bar{q}$ full simulation sample using CEPCSW is produced. In this sample, two charged-particle tracks that originate from the same displaced vertex can be reconstructed. In $D^0 \to h^- h^+ \pi^0$ decays, an additional neutral pion is reconstructed using two photon clusters recorded by ECAL, and an invariant mass constraint is applied to the two charged-particle tracks and the neutral pion. The reconstruction efficiency of those photon clusters is generally low because of their low momenta. To suppress background and maintain efficiency, the following strategy is used:

- Using the generator-level information, the opening angle between the two photons is

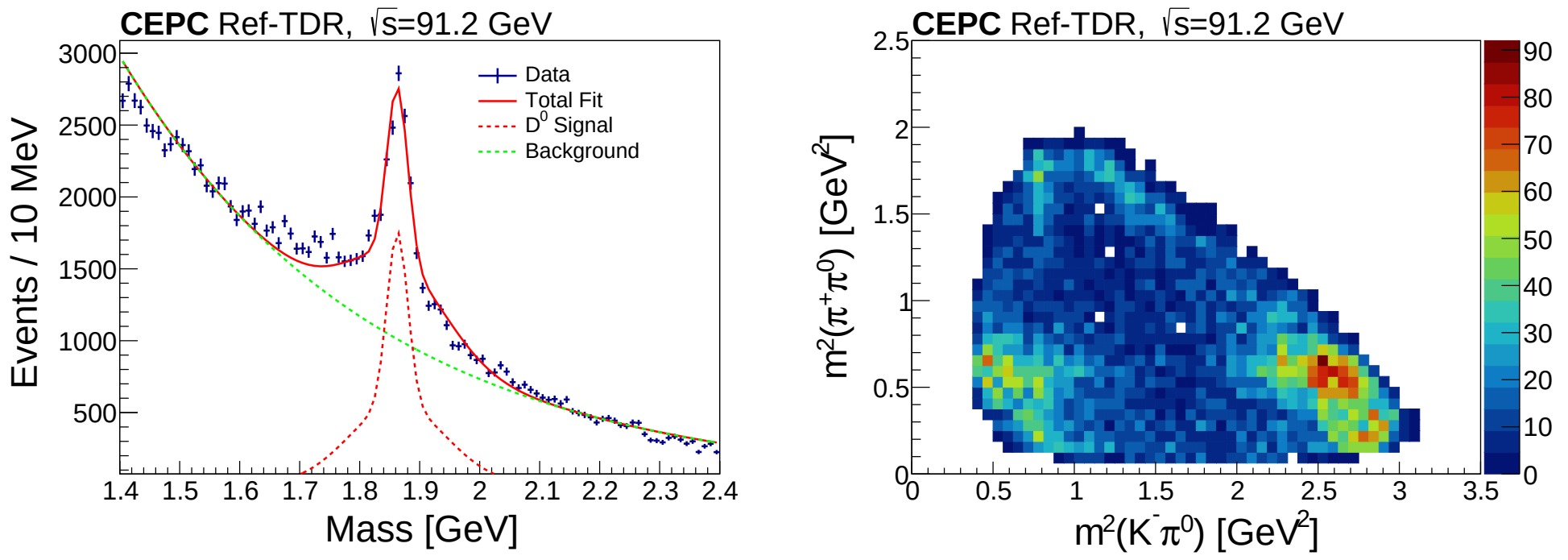

**(a)** **(b)**

**Figure 15.31:** (a) Invariant mass from selected $D^0 \to K^-\pi^+\pi^0$ decay candidates reconstructed from simulated PFO objects and (b) a Dalitz plot of the $D^0 \to K^-\pi^+\pi^0$ decay sample.

required to be below 10° as the criterion to suppress random combinations.

- The highest energy photon is required to have its energy above 0.5 GeV and is considered as the leading photon.
- The photon with $E > 0.1$ GeV closest to the leading photon is chosen as the sub-leading.

A dedicated study is then performed using $\mathrm{d}N_p/\mathrm{d}x$-only and $\mathrm{d}N_p/\mathrm{d}x$+ToF PID information, to estimate the impact of different detector designs on the final efficiency. With these selections, most background can be suppressed, while the signal efficiency can be $\mathcal{O}(0.1)$.

The invariant mass of $D^0 \to K^-\pi^+\pi^0$ decay candidates reconstructed from PFOs is shown in Figure 15.31a, and the Dalitz plot for this decay channel is shown in Figure 15.31b. A preliminary qualitative estimate of several $D^0$-decay yields is shown in Table 15.18. These decays have fully hadronic final states. Data collected by the LHCb experiment during its Run 2 period (approximately 6 fb$^{-1}$) and the expected data to be collected over the entire lifetime of the LHC and LHCb (approximately 300 fb$^{-1}$), as well as the number of corresponding decay modes expected to be collected at the CEPC $Z$-pole operation mode are shown. Furthermore, a comparison of the number of relevant decay modes reconstructed in specific physics analyses is conducted. The total yields at LHCb are estimated by using the cross section measured by Ref. [51]. The reconstructed and selected events from LHCb are obtained from Refs. [52, 53], while the reconstruction and selection efficiency at CEPC is assumed to be 10%.

Despite the lower reconstruction efficiency, LHCb has a significant statistical advantage over CEPC for $D^0$ decays into fully charged hadronic final states. However, as a hadron collider experiment, LHCb has particularly low efficiency for reconstructing $\pi^0$ particles; thus, it does not have a statistical advantage over CEPC in terms of measuring decay modes with $\pi^0$ final states. Current LHCb sensitivity to CP violation in the $D \to \pi\pi\pi^0$ decay is 0.5° asymmetry in phase or 0.5% asymmetry in magnitude in the dominating resonance. With statistics more than 100 times larger than the current LHCb statistics, the sensitivity to CP violation in the same channel at CEPC is expected to be about 0.05° asymmetry in phase or 0.05% asymmetry in magnitude in the dominating resonance. Charm CP violation is expected to be at the per mille

**Table 15.18:** The number of $D^0$ and related fully hadronic final states, and their reconstructed and selected yields at the LHCb experiment in Run 2, the projected yields in the entire lifetime of the LHC and LHCb, and CEPC $Z$-pole operation.

| **Decays** | **LHCb (6 fb$^{-1}$)** | **LHCb (300 fb$^{-1}$)** | **CEPC (4 Tera $Z$)** |
|---|---|---|---|
| $D^{*+}$ | $4.7\times10^{12}$ | $2.4\times10^{14}$ | $4.6\times10^{11}$ |
| $D^0$ from $D^{*+}$ | $3.2\times10^{12}$ | $1.6\times10^{14}$ | $3.1\times10^{11}$ |
| $D^{*+}\to(D^0\to K^-K^+)\pi^+$ | $1.6\times10^{10}$ | $6.5\times10^{11}$ | $1.3\times10^{9}$ |
| $D^{*+}\to(D^0\to \pi^-\pi^+)\pi^+$ | $4.6\times10^{9}$ | $2.3\times10^{11}$ | $4.5\times10^{8}$ |
| $D^{*+}\to(D^0\to K^-\pi^+)\pi^+$ | $1.6\times10^{11}$ | $6.3\times10^{12}$ | $1.2\times10^{10}$ |
| $D^{*+}\to(D^0\to \pi^-\pi^+\pi^0)\pi^+$ | $4.8\times10^{10}$ | $2.4\times10^{12}$ | $4.6\times10^{9}$ |
| $D^{*+}\to(D^0\to K^-\pi^+\pi^0)\pi^+$ | $4.6\times10^{11}$ | $2.3\times10^{13}$ | $4.4\times10^{10}$ |
| Reco. & Sel. $D^0\to K^-K^+$ | $5.8\times10^{7}$ [52] | $2.9\times10^{9}$ | $1.3\times10^{8}$ |
| Reco. & Sel. $D^0\to \pi^-\pi^+$ | $1.8\times10^{7}$ [52] | $9\times10^{8}$ | $4.5\times10^{7}$ |
| Reco. & Sel. $D^0\to K^-\pi^+$ | $5.2\times10^{8}$ [52] | $2.6\times10^{10}$ | $1.2\times10^{9}$ |
| Reco. & Sel. $D^0\to \pi^-\pi^+\pi^0$ | $2.5\times10^{6}$ [53] | $1.2\times10^{8}$ | $4.6\times10^{8}$ |
| Reco. & Sel. $D^0\to K^-\pi^+\pi^0$ | $1.9\times10^{7}$ [53] | $9.6\times10^{8}$ | $4.4\times10^{9}$ |

level; therefore, CEPC can observe CP violation in multi-body charm decays.

### 15.2.11 Top quark mass

The top quark, the most massive elementary particle in the Standard Model, has the strongest coupling to the SM Higgs boson, providing an excellent probe for precision measurements and new physics beyond the SM. To date, the top quark mass has been measured at hadron collider experiments, such as those conducted at the Tevatron and the LHC, through direct reconstruction of the invariant mass of decay products. The latest top quark mass precision is down to less than half a GeV [54, 55, 56, 57], and it is mainly limited by the systematic uncertainties, such as jet energy scale, which are very challenging to reduce. The expected uncertainty at the HL-LHC is projected to be 200 MeV [18] for the direct mass measurement. In addition, the direct mass measurement predicts the top-quark mass in the MC generator and thus suffers from a fundamental mass-scheme ambuiguity of around 0.5–1.0 GeV, largely induced by non-perturbative and other QCD effects. This irreducible ambiguity cannot be eliminated at the hadron colliders, whereas scanning the $t\bar{t}$ production threshold at the electron-positron colliders would provide not only a high experimental precision but also a rigorous interpretation of the mass renormalization schemes.

Electron-positron colliders will enable not only direct reconstruction measurements but also an alternative approach using a threshold scan of the center-of-mass energy near the $t\bar{t}$ production threshold. At the energy threshold of $t\bar{t}$, the $t\bar{t}$ production cross section increases sharply as shown in Figure 15.32 and is highly sensitive to the top quark mass, its decay width, and the strong coupling constant $\alpha_S$. The threshold scan method has been extensively discussed in the literature as a precise method for determining the top quark mass at ILC, CLIC, FCC-ee and CEPC [58, 59, 60, 61, 62, 63, 64].

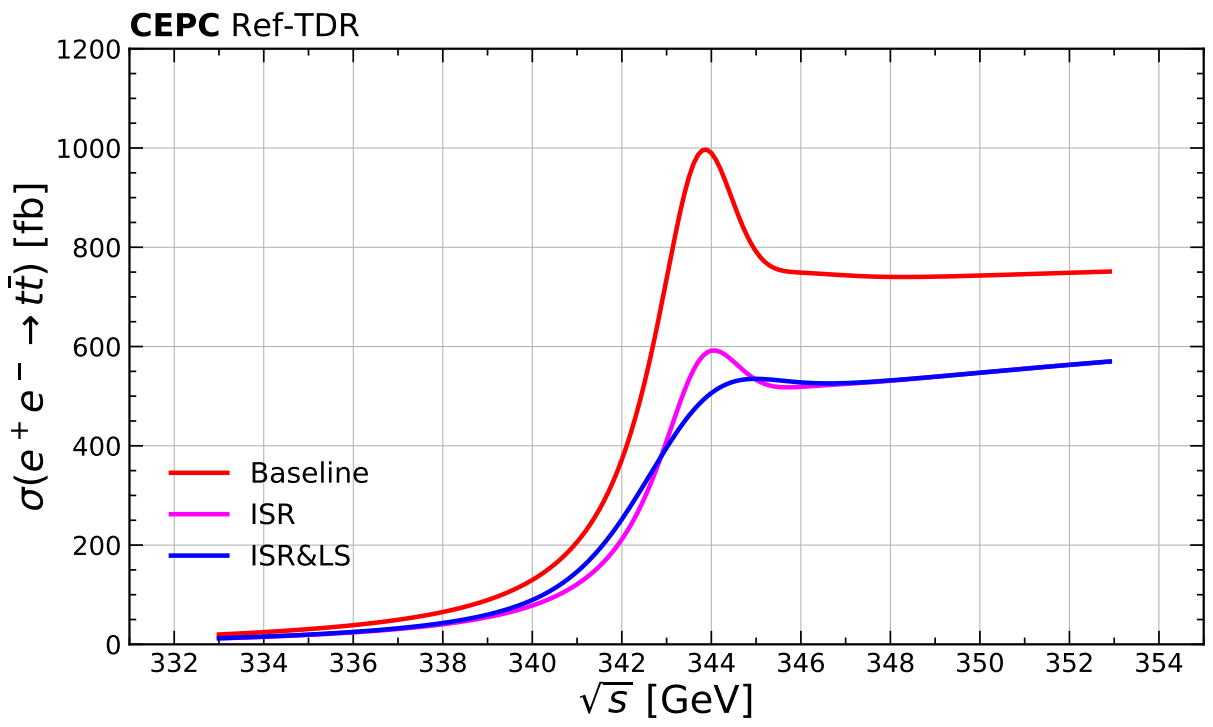

**Figure 15.32:** Cross section of $t\bar{t}$ production as a function of center-of-mass energy at CEPC [64], including the cross-section values without ISR or Luminosity Spectrum (LS) (baseline), the ones with ISR only and the ones with both ISR and LS.

**Table 15.19:** The expected statistical and systematic uncertainties of the top quark mass measurement in optimistic and conservative scenarios at CEPC.

| Source | $m_{\text{top}}$ precision (MeV) | |
|---|---|---|
| | Optimistic | Conservative |
| Statistical | 7 | 7 |
| Theory | 8 | 24 |
| Quick scan | 2 | 2 |
| $\alpha_S$ | 3 | 16 |
| Top width | 5 | 5 |
| Experimental efficiency | 4 | 9 |
| Background | 1 | 3 |
| Beam energy | 2 | 2 |
| Luminosity spectrum | 3 | 6 |
| Total (without theory) | 11 | 21 |
| Total | 14 | 32 |

In this study, the semileptonic and all-hadronic channels are considered. At CEPC, realistic scan strategies at the threshold are discussed to maximize the sensitivity in the one-dimensional measurement of top quark mass and in the two-dimensional measurements of top quark mass with other parameters such as the top quark width or $\alpha_S$, assuming a total luminosity limited to 100 fb$^{-1}$ [64]. Given a fixed total amount of luminosity, scanning with only one energy point, which is most sensitive to the top quark mass, outperforms scanning with more energy points as discussed in Ref. [64].

Table 15.19 summarizes the two scenarios of systematic uncertainties considered: the optimistic and conservative ones. The experimental efficiency of future detectors is not yet precisely known. Given the large number of data events taken for vector bosons, the dominant uncertainties in the efficiencies, such as jet energy scale and $b$-jet tagging, should be constrained strictly down to a level of 1% or below. Accordingly, the uncertainty of the experimental efficiency is assumed to be 0.5% and 1% for the two scenarios, separately. The theoretical uncertainty in the cross-section calculation is assumed to be 3%, based on conservative estimates from Refs. [65, 66], and 1%, expected to be achievable by the time of the experiments. This assumption aligns with that of Ref. [59]. In the conservative scenario, the uncertainty of $\alpha_S$ is taken as 0.0007 from the world summary in 2012 [67]. In the optimistic scenario, a projection using the observables from $Z$ bosons [68] and the energy correlators of final state particles [69] is considered, and the precision of $\alpha_S$ measurements at $e^+e^-$-colliders could possibly reach 0.0001. The corresponding impact on the top quark mass will decrease to 3 MeV. In this study, one-dimensional fit is applied to evaluate the uncertainty in the top quark mass in Table 15.19, while two-dimensional fits on the top quark mass versus the decay width or $\alpha_S$ were tested in Ref. [64] and higher-dimensional fits including the top Yukawa coupling have also been studied recently in Ref. [63]. Further details of other uncertainties considered in Table 15.19 can be found in Ref. [64].

All these estimations are based on the Reference Detector design, using a $b$-jet tagging efficiency of about 90% and a lepton identification efficiency of about 98% (see Section 15.1.8), which are the leading factors in determining the overall acceptance. The acceptance times efficiency is 44% and 62% for the semileptonic and all-hadronic channels, respectively. In terms of the beam parameters, the beam energy could vary by 2.6 MeV as estimated from the accelerator team, which is comparable to LEP [70, 71] and ILC [72]. Its impact on the top quark mass measurement is below the statistical uncertainty, consistent with the CLIC study [59]. The variations in the spread of the luminosity spectrum are assumed to be 10% and 20%, respectively. This assumption of 20% represents a conservative choice, using the same input uncertainty as that adopted in the CLIC scenario [59], while 10% is assumed in the optimistic scenario given a better control of the luminosity spectrum with the bending magnets in circular colliders.

In summary, the top quark mass precision is expected to be 7 MeV, considering only the statistical uncertainty. Taking into account the systematic uncertainties, the top quark mass can be measured at the precision of 14 MeV optimistically and 32 MeV conservatively at CEPC, which is significantly improved from the HL-LHC projection of 200 MeV [18].

## 15.3 Further performance aspects

This section outlines strategies for measuring the absolute luminosity, the usage of resonant depolarization to measure the $Z$ and $W$ boson masses with high precision, methods for the calibration and alignment of the sub-detectors, and the primary areas where detector configuration optimizations and technology decisions could be further explored.

### 15.3.1 Strategy for measuring absolute luminosity

Precision measurement of the integrated luminosity $\mathcal{L}_{\text{int}}$ is crucial for the realization of the physics program at CEPC. This is particularly true for the $Z$-pole cross-section measurement, determination of the $Z$-boson's width from the line shape of $e^+e^- \to 2f$ production and the measurements of the $W$ boson mass and width from the line shape of the $W$-pair production near the threshold. At the $Z$-pole, the relative uncertainty of $\mathcal{L}_{\text{int}}$ is required to be of the order of $10^{-4}$, while $10^{-3}$ should suffice at higher center-of-mass energies.

The strategy for achieving the required precision in the absolute luminosity measurement involves an iterative process, including the identification of SM processes, monitoring and calibration of the detector response, and high-precision calculations.

A compact EM calorimeter is designed to function as a luminometer, as discussed in Section 3.4. The experimental precision of the integrated luminosity depends on several sources of uncertainty, including the luminometer resolution in terms of position and energy measurement of Bhabha showers. The position resolution of Bhabha hits in the luminometer front plane can be improved to a micron level by placing a Si-tracking plane in front of the luminometer, which provides enhanced electron-photon separation.

Uncertainties in the $\mathcal{L}_{\text{int}}$ measurement may arise from various misalignments of the detector arms relative to each other and the IP, as well as from the uncertainties related to finite beam sizes and beam delivery to the IP. These uncertainties, collectively referred to as metrological uncertainties, are discussed in Section 3.4. Detailed studies [73] have shown that achieving the targeted precision is feasible at CEPC, both at $\sqrt{s}$ = 240 GeV and during the $Z$-pole operation. The major challenge at the $Z$-pole arises from the need to control the inner aperture of the luminometer at the micron level. A dedicated laser-based system for luminometer position monitoring must be developed, with technologically feasible precision margins set in Ref. [73]. Systematic uncertainties in the integrated luminosity measurement originating from beam-beam interactions have also been studied [74], showing a comparable impact as those at the FCC-ee [75], and being less pronounced than those at the linear $e^+e^-$ colliders due to less compact bunches. The impact of the beam energy spread on the $\mathcal{L}_{\text{int}}$ measurement is discussed in Ref. [76], as well as the precision required to determine it from the luminosity spectrum obtained from di-muon production at CEPC. The possibility of using other SM Electroweak (EW) processes, such as $e^+e^- \to \gamma\gamma$, to determine the luminosity is also under investigation.

For the $e^+e^- \to e^+e^-$ process, theoretical uncertainty is limited by the hadronic vacuum polarization at the level of $10^{-4}$, while for the $e^+e^- \to \gamma\gamma$ process, the contribution of hadronic loops is less than $10^{-5}$ [77]. The current uncertainty of MC generators for the $e^+e^- \to \gamma\gamma$ process is $10^{-3}$ at $M_Z$, for instance, at the BABAYaga NLO [78]. When Next-To-Next-To-Leading-Order (NNLO) corrections from Ref. [77] are applied, reaching an accuracy of around $10^{-4}$ could be possible. To achieve an accuracy of $10^{-4}$ to $10^{-5}$, a comprehensive calculation of NNLO QED corrections and eventually two-loop weak contributions is required.

The $e^+e^- \rightarrow \gamma\gamma$ process at energies above the $Z$ boson mass was studied at the LEP collider [79]. Each of the four detectors observed approximately 5,000 events, indicating that the process was primarily observed, with statistical uncertainty being the dominant factor. The most accurate result was obtained with the OPAL detector [80], where the major component of the systematic uncertainty was related to the photon conversion probability, limited by the available statistics of photon events. At CEPC, the conversion probability can be expected to be well studied due to the large statistics of photon events.

At CEPC, the process $e^+e^- \rightarrow \gamma\gamma$ can be used to determine luminosity offline, provided there is a sufficiently well large data set; at least 3.3 ab$^{-1}$ is needed to achieve a $10^{-4}$ accuracy. This process can help reduce systematic uncertainties in the relative luminosity measurements, which can be confirmed by comparing the ratio of the number of $\gamma\gamma$ events to $e^+e^-$ events under consistent selection conditions with theoretical predictions.

### 15.3.2 Application of the resonant depolarization method for the $W/Z$ boson mass determination

At the CEPC, measurements of the $W$ and $Z$ boson masses via the beam energy scan method are expected to achieve very small statistical uncertainties due to its high luminosity and low background. The dominant source of systematic uncertainty will stem from the precise monitoring and measurement of the beam energy. To address this demand, the CEPC plans to use a resonant depolarization technique for the beam energy measurement, with the expected precision reaching 0.3 MeV at the beam energy of 80 GeV and 0.1 MeV at the beam energy of 45 GeV respectively. Consequently, the measurements of the $W$ boson mass and $Z$ boson mass at the CEPC are expected to achieve accuracies better than 0.5 MeV [81, 82] and 0.2 MeV [82], respectively.

The resonant depolarization technique employs a pulsed ”depolarizer” with a frequency-scan capability to induce a narrow artificial spin resonance for beam depolarization. It measures the location of depolarization with the precisely known depolarizer frequency, which now provides the most precise measurements of the beam energy. Its implementation at the CEPC would require achieving transverse polarization levels of at least 5% to 10% for both beams. This also requires the implementation of laser-Compton polarimeters capable of measuring the vertical beam polarization with a resolution of 1% every few seconds. Additionally, monitoring the evolution of beam energies throughout the physics runs is essential for the precision measurements of $W/Z$ masses. This requires conducting resonant depolarization measurements frequently, approximately every 10–15 minutes, where each measurement depolarizes one or two bunches for each particle species and determines the instantaneous beam energy. Interpolation can then be used to model the evolution of the beam energies more precisely. The systematic uncertainties in the center-of-mass energy, which are propagated from the beam energy measurements, are estimated to be around 100 keV at the beam energy of 45 GeV and 300 keV at the beam energy of 80 GeV respectively, according to the studies [83, 84] by the FCC-ee EPOL Collaboration.

Besides the scheme[2] of using self-polarization in the collider rings [83], it is also viable to prepare polarized lepton beams from the source, as detailed in the CEPC Accelerator TDR [38] and references therein. These beams can be transported throughout the injector chain and injected into the collider rings to meet the requirements of the beam energy calibration. The strategy involves using one or two bunches per species that can be efficiently depolarized. Once depolarized, they can be readily removed and replaced with new polarized bunches. This process aligns with the capabilities of the injector chain, ensuring a streamlined calibration procedure, and can deliver electron bunches with a polarization of above 50% and positron bunches with a polarization of above 20%. Availability studies [85], including failures of systems such as the RF systems or power converters, suggest that the latter approach could guarantee much less physics dead time and a substantial increase in the integrated luminosity.

This approach requires some modifications to the entire injector chain as outlined in the CEPC Accelerator TDR and should be implemented in an updated design. Additionally, preparing electron beams with over 85% polarization from the source is achievable with state-of-the-art technology. This approach can achieve over 50% longitudinal polarization for colliding beam experiments.

To prepare for the high performance of polarized electron sources, Compton polarimetry, and resonant depolarization techniques to be used at the CEPC, dedicated R&D efforts are underway in the Engineering Design Report (EDR) phase of the CEPC. There is a 400 kV photocathode DC gun at the Platform of Advanced Photon Source Technology R&D (PAPS) managed by IHEP. There are ongoing efforts to convert it to a polarized electron source and build an associated beamline to measure the beam polarization, in addition to domestic fabrication of the superlattice GaAs/GaAsP photocathodes. The plan is to conduct the first experiment on generating polarized electron beams in 2027.

A Compton polarimeter is constructed at BEPCII, reusing the hutch and beamline of a dismantled wiggler, to measure the vertical polarization of the electron beam in the storage ring due to the self-polarization buildup. Although this Compton polarimeter is based on detecting backscattered $\gamma$ photons, rather than the preferred solution of detecting backscattered electrons as planned for CEPC, it will establish the know-how to operate such a delicate instrument, and advance laser polarization control and pixel detector technologies for Compton polarimetry. In the previous round of beam commissioning, the signal of backscattered $\gamma$ from laser-electron collisions was detected, the first polarization measurement is foreseen in the upcoming BEPCII run.

Once reliable beam polarization measurements are achieved, demonstration of the beam energy calibration using the resonant depolarization technique is foreseen at BEPCII over the next 2–3 years. Being a double-ring collider with many bunches similar to CEPC, BEPCII will serve as an ideal testbed for the operational concepts of resonant depolarization for CEPC: using dedicated pilot bunches for resonant depolarization and continuous monitoring of the beam energy throughout physics runs, for example.

[2] It can generate above 10% polarization in about two hours, particularly with the help of asymmetric wigglers at the $Z$-pole.

The knowledge gained through these R&D activities will help establish a solid foundation for the practical design of Compton polarimetry and preparation for resonant depolarization applications at CEPC.

### 15.3.3 Methods and considerations for calibration, alignment

This subsection summarizes the calibration and alignment strategies for the subdetectors, combining physics-driven methods and technical monitoring to meet the required performance for precision measurements.

**Vertex Detector Calibration and Alignment** Vertex alignment is based on $Z \to \mu^+\mu^-$ events, minimizing track-hit residuals through iterative geometry updates. A sub-10 μm accuracy is achieved, with the impact parameter resolution validated using displaced $K^0_S \to \pi^+\pi^-$ vertices. Thermal and mechanical shifts are monitored over time, with correction maps derived from residual drifts. Temperature variations of 1 °C can induce micron-level misalignments. Temporal stability is cross-checked using photon conversions and $K^0_S$ events.

Charge collection efficiency is monitored with $e^+e^-$ pairs from conversions, and pixel gain equalization uses MIP tracks to reduce response variation. Inter-layer timing alignment is calibrated using relativistic muons, reaching synchronization within 0.5 ns. Radiation effects are tracked via MIP signal trends, with all sensors pre-tested for long-term stability.

**Tracker Calibration and Alignment** The tracking system, comprising silicon trackers and a TPC, is calibrated for geometry, momentum scale, and timing. Initial alignment uses cosmic ray and beam halo muons to reduce residuals below 10 μm (silicon) and 100 μm (TPC). Momentum calibration employs $Z$, $J/\psi$, and $\Upsilon$ resonances. Deviations in invariant mass are used to extract per-module corrections, combined via momentum-weighted averages. OTK and TPC timing is calibrated using $\tau$-lepton decay chains and cosmic ray muons. In the TPC, UV-laser tracks calibrate drift velocity and correct space charge effects. Timing resolution reaches 50 ps for the OTK and 100 μm spatial precision for the TPC. A Kalman filter refines alignment using $Z \to \mu^+\mu^-$ events, reducing residuals to design-level precision. Real-time monitoring involves tracking channel-level hit efficiency and making environmental corrections based on both temperature and radiation dose, ensuring optimal performance. Laser interferometry tracks mechanical shifts at micron scale. Structural expansion is modeled and combined with track-based alignment to maintain long-term stability.

**ECAL Calibration and Alignment** ECAL calibration is performed using physics-driven and monitoring-based methods to ensure an electromagnetic energy resolution better than $3\%/\sqrt{E}$.

The $E/p$ technique, utilizing electrons from $e^+e^-$ and $W/Z$ boson decays, compares ECAL energy to tracker momentum to monitor gain drifts. Corrections are updated regularly based on time-dependent $E/p$ trends.

Intercalibration employs $\pi^0 \to \gamma\gamma$ decays, reconstructing invariant masses and weighting energy contributions by crystal to derive per-channel constants. Iterative averaging and $\eta$-ring smoothing reduce local statistical fluctuations. The absolute scale is determined from $Z \to e^+e^-$ events by comparing reconstructed masses to the nominal $Z$ boson mass. $J/\psi \to e^+e^-$ decays provide cross-checks in the low-electron-energy region.

Cosmic ray muons supply continuous monitoring of SiPM gain and crystal transparency changes. Timing synchronization between channels is refined using relativistic muons, reaching alignment within 0.5 ns.

Track-based ECAL alignment extrapolates high-$p_T$ tracker tracks to the calorimeter surface, minimizing residuals to refine module positions. Mechanical stability is monitored with strain sensors and laser lines, ensuring long-term alignment consistency.

**HCAL Calibration and Alignment** The HCAL is calibrated to achieve 5% jet energy resolution at 100 GeV, beginning with radioactive source scans and test beam calibration of modules during construction. These define initial channel gains and validate response uniformity across $\eta$ and $\phi$.

During operation, gain stability is maintained via LED/laser monitoring. Energy reconstruction applies corrections for light-yield variation, intercalibration factors, and $\eta$-dependent scaling. Pileup suppression is performed with pulse-shape fits.

The $\phi$-symmetry is exploited to equalize response across the $\eta$ rings, using statistical moment methods. The absolute scale is tuned using isolated charged hadrons, comparing their HCAL energies to tracker momenta after the ECAL subtraction. Dijet and multijet events provide additional cross-checks via transverse balance and reconstructed $W/Z$ boson masses.

SiPM gain and radiation damage are tracked with the MIP response and temperature modeling. Alignment is derived from cosmic-ray muons traversing multiple layers, with geometry updated to minimize hit residuals. Structural sensors ensure sub-mm mechanical stability.

**Muon Detector Calibration and Alignment** For the muon system, channel efficiency and energy response are calibrated with $Z \to \mu^+\mu^-$ tracks, extrapolated from the tracker. MIP peaks and detection efficiency are extracted layer-by-layer. Cosmic-ray muons complement this with absolute intercalibration, and responses are smoothed over $\eta$ regions.

Muon ID performance is maintained using likelihood templates based on hit multiplicity and penetration depth, trained on muon and pion control samples. SiPM gain is monitored via cosmic MIPs and LED pulses, with corrections applied for temperature-induced variation and optical degradation.

**Table 15.20:** Physics benchmarks, relevant detector performances and expected precision.

| **Physics Benchmarks** | **Relevant Det. Performance** | **Expected Precision** |
|---|---|---|
| Recoil $H$ mass | Tracking | $\Delta m_H$=±4.0 MeV |
| $H \to$ hadronic decays | PID, Vertexing, PFA | relative unc. of BR = 0.3% $(b\bar{b})$ − 8.7% $(ZZ^*)$ |
| $H \to \gamma\gamma$ | photon ID, EM resolution | $\Delta(\sigma \times \mathrm{BR})/(\sigma \times \mathrm{BR})_{\mathrm{SM}}$=3.1% |
| $H \to$ invisible | PFA, missing energy, BMR | 4.5$\sigma$ to SM, BR<0.047% to BSM at 95% CL |
| $H \to$ LLP | Vertexing, TOF, muon detectors | BR($H \to \chi\chi$) < $4 \times 10^{-5}$ for $m_\chi$=10 GeVand lifetime=0.1 ns |
| Smuon pair | muon ID, missing energy | Discovery reach up to $m(\tilde{\mu})$=119 GeV |
| $A^{\mu}_{FB}$ | Tracking, muon ID | ±0.000031 (stat.) ± 0.000028 (syst.) |
| $R_b$ | PFA, jet flavor tagging | $2.5 \times 10^{-6}$ (stat. unc. only) |
| CPV in $D^0 \to h^+h^-\pi^0$ | PID, vertex, $\pi^0$, EM resolution | Sensitivity to 0.05° (0.05%) asymmetry in phase (magnitude) |
| Top quark mass | Beam energy | $\Delta m_{\mathrm{top}}$=14 MeV (optimistic syst.), 32 MeV (conservative syst.) |

Timing calibration uses dual-ended strip readout and waveform fitting, achieving sub-ns resolution after correcting propagation delays. Geometry alignment is refined by minimizing residuals between extrapolated tracks and measured hits using $Z$ boson events and cosmic-ray muons. Embedded mechanical sensors help monitor structural deformation, ensuring alignment stability over time.

## 15.4 Summary and plans

With the Reference Detector design and full simulation, detailed performance of the physics objects is evaluated, and a set of benchmark studies are conducted. They illustrate the detector performance for center-of-mass energies in the range from 91 GeV up to 360 GeV. The results of physics benchmark studies are summarized in Table 15.20. In addition, areas of challenges and plans beyond the current TDR are also identified.

The period from the present TDR and the final production design will be used for further studies and optimizations. The following areas will be considered:

- ECAL transverse granularity for boosted $\pi^0/\gamma$ separation: Evaluate the transverse granularity of the ECAL long crystal bars to improve discrimination between boosted $\pi^0$ decays and single photons. This study will measure the impact of finer segmentation on reconstruction algorithms and photon purity, particularly in high-occupancy regions.
- HCAL thickness versus the polar angle for jet energy resolution: A parameterized analysis of HCAL thickness as a function of the polar angle will be conducted to balance jet energy resolution against cost constraints. This study includes optimizing the radial and longitudinal segmentation to mitigate energy leakage while minimizing cost.
- Muon system optimization for long-lived particle searches: The muon identification performance will be studied as a function of the number of muon detector layers. The trade-offs between the layer count, spatial coverage, and sensitivity to long-lived particles (e.g., displaced vertices) will be assessed to ensure compatibility with background rejection requirements.

- Low-momentum charged hadron PID with AC-LGAD timing layers: To enhance the identification of low-momentum charged hadrons, which could not reach the OTK, a proposal is made to replace the outermost layer of the ITK with an AC-LGAD-based timing layer. The combined spatial and timing resolution of 10 µm and 50 ps respectively from this technology could significantly improve the $\pi/K/p$ separation without compromising tracking performance.

In terms of technology choices, several key performance-driven evaluations needed for detector subsystems include:

- Pixel versus strip ITK for PID requirements: A comparative study will assess the impact of using strip-based ITK instead of pixel sensors on the PID capabilities. This study will include evaluating resolution trade-offs, occupancy limits, and compatibility with the timing layer integration proposed above.
- Beam-induced background and TPC versus Drift CHamber (DCH) at the high-luminosity $Z$ pole run: Beam-induced background will be simulated to measure their impact on tracking subsystem choices, namely TPC or DCH, during the high-luminosity $Z$-pole running. Critical metrics include occupancy, hit reconstruction efficiency, and robustness against pileup.
- HTS ultra-thin magnet feasibility: The potential use of HTS magnets will be explored to reduce the solenoid's radial thickness, while maintaining the field strength. This study will address the mechanical stability, quench protection, and integration with the overall material budget in the detector.

# Chapter 16 Cost of the Reference Detector

The construction cost analysis for the CEPC Reference Detector employs an internationally standardized Work Breakdown Structure (WBS). The overall cost is estimated as 333.3 MCHF, with a breakdown of each detector system provided in Table 16.1. The analysis includes a 3% installation cost, while contingency is not included. This calculation derives from a comprehensive current market survey and incorporates projections spanning the next five to ten years. All prices are converted from CNY to CHF using an exchange rate 8:1 from February, 2025.

**Table 16.1:** The CEPC Reference Detector cost breakdown.

| **System** | **Cost (MCHF)** |
|---|---:|
| The Reference Detector | 333.3 |
| Machine detector interface and luminosity measurement | 1.8 |
| Vertex detector | 4.5 |
| Silicon tracker | 29.7 |
| Time projection chamber | 6.2 |
| Electronmagnetic calorimeter | 115.0 |
| Hadron calorimeter | 68.3 |
| Muon detector | 2.5 |
| Detector magnet system | 22.0 |
| Readout electronics | 19.3 |
| Trigger and data acquisition | 12.2 |
| Offline software and computing | 23.1 |
| Mechanics and integration | 28.9 |

# Chapter 17 Timeline and plans

A proposal for the CEPC project will be formally submitted to the Chinese central government for review and approval. As part of this critical application, the TDR for the Reference Detector serves as one of the supporting documents to substantiate the project's viability. This report aims to demonstrate the technical readiness and feasibility of the proposed detector technologies, also providing a more realistic evaluation of the required personnel resources, timeline, and budgetary commitments to construct the detector.

This chapter outlines the strategic roadmap for advancing the project. Section 17.1 presents a timeline of the project under optimal conditions. Following the completion of the TDR, ongoing R&D efforts will focus on refining detector technologies as reported in Section 17.2. Finally, Section 17.3 details the prototyping of individual detector modules.

## 17.1 The CEPC timeline

The CEPC project, first proposed in 2012, has advanced significantly through years of dedicated effort by its international working group and supports from the global high energy physics community. While this progress makes an important milestone, substantial work remains to realize the ambitious scientific objectives. Continued collaboration and sustained investment will be essential in the coming years.

A feasible timeline of the project is shown in Figure 17.1. The CDRs for the accelerator, physics and detector were released in 2018. The TDR for the accelerator was released in 2023. The TDR for the Reference Detector is expected to be published in 2025. Both TDRs will be included as supporting documents for the CEPC proposals to be submitted to the Chinese central government for approval. Once approved, one detector will be constructed as the baseline scenario presented in Chapter 1. In the upgraded scenario, two international collaborations will be established, corresponding to the two experiments to be operated at the interaction points. The design of the two detectors will be proposed, reviewed, and further developed through an international collaborative effort. The planned schedule aims to start construction during the "15th five-year plan (2026-2030)" (e.g., around 2027) and to complete around 2035. Physics operation is expected to start in 2037 after two years of commissioning.

## 17.2 Further R&D and performance study

**MDI and luminosity measurement** The current design encounters certain challenges. For instance, after implementing shielding measures, the overall cryogenic components have become relatively heavy, and some mitigation techniques, such as collimators, could be further optimized. Additionally, critical technologies associated with the luminosity measurement detectors and central beam pipe require further validation. A prototype could be developed following additional design optimization to verify the feasibility of these key technologies, and thereby establishing a more robust foundation for future construction.

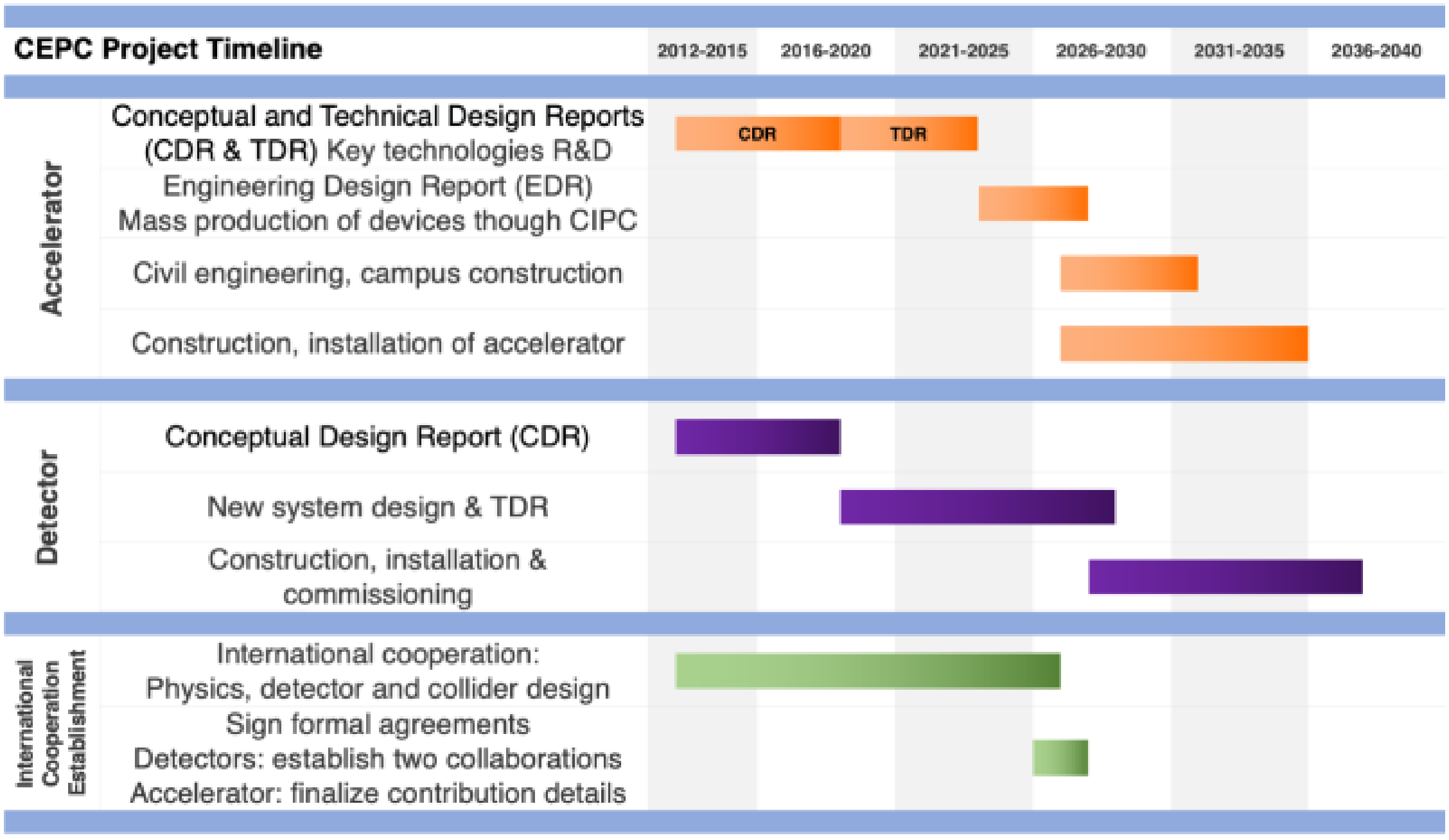

**Figure 17.1:** The timeline of the CEPC project since it was first proposed in 2012, including the accelerator, the detector, and international cooperation. In an optimal scenario, the detector construction will start in 2027 and complete around 2035. Physics operation is expected to start in 2037 after commissioning.

**Vertex detector** The upcoming R&D efforts will focus on the development of wafer-scale stitched monolithic active pixel sensors, ultra-thin mechanical supports, and low-mass integration techniques. A full-scale vertex detector prototype will be constructed to address challenges related to mechanical precision, cooling performance, and system-level integration.

**Silicon tracker** The ongoing progress ranges from the construction of prototype detectors to the refinement of the final design. Continuous innovations in sensor technology, readout electronics, mechanical design, and the development of the cooling system will ensure that the silicon tracker meet the requirements of CEPC physics.

**TPC** A prototype with finalized high granularity readout modules will be built, and dedicated experiments are planned focusing on resolution, ion backflow, and electron transmission. De-tailed simulation are intended to investigate distortion and non-uniform magnetic field effects and their correction. The ongoing work involves development of low-noise and low-power consumption ASIC based on a 65 nm process. Subsequently, beam tests are scheduled using 5 GeV electrons and 1.0 T magnetic field to access the detector's performances.

**ECAL** Future ECAL R&D activities will focus primarily on key performance studies of the high-granularity crystal calorimeter. The R&D plan will be prioritised development of a full-scale technological prototype, beam tests, performance validation, and particle flow algorithm optimization for the full demonstration. Essential R&D topics include mass production and quality control of crystals and SiPMs, development of BSO crystals, and performance study of readout chips.

**HCAL** To meet detector design requirements, several key steps are essential. For the GS-HCAL, the energy resolution constant term aims to be reduced below 3% through improved calibration, energy weighting algorithms, position-dependent corrections, and machine learning techniques. Research on GS components should focus on enhancing light yield and attenuation length, while advancements in SiPM technology can optimize performance and reduce costs. Customized front-end electronics must address charge range, timing resolution, and power consumption, integrating with the electronics system. Extensive testing of mini and full-scale prototypes will validate designs and ensure HCAL meets stringent physics requirements, enabling high-energy physics discoveries.

**Muon detector** Our R&D plan includes the following items: enhance the light yield and photon collection efficiency of the 5 m PS strips; construct multiple detector modules for comprehensive system testing, with a focus on stable long-term operation; conduct radiation hardness assessments on the PS strips, WLS fibers, and SiPMs; refine track reconstruction algorithms; improve the measurement of low-momentum muons, and expand physics potentials, specifically in the search for LLPs.

**Detector magnet system** The production of the detector superconducting magnet requires foundational systems such as aluminum-stabilized cables, coil winding platforms, cryogenic systems, and mechanical components. We will develop the full-size aluminum-stabilized NbTi Rutherford cable samples by 2025. R&D focuses on enhancing mechanical strength, achieving self-supporting structures, and ensuring bonding strength between Rutherford cables and aluminum bodies. For the cryogenic system, R&D is needed on the welding of multi-channel tubes with gas collection tanks and liquid accumulation tanks to determine the processing technology. The mechanical system faces the challenge of suspending the approximately 145-ton coil concentrically; therefore, R&D will explore different types of titanium-alloy or carbon-fiber suspension rods, which will be designed, processed, and tested for mechanical strength.

**Readout electronics** The current electronics system aims to establish a preliminary baseline scheme. The development will proceed in stages, with the TDR milestone providing the final design scheme based on detector requirements and background simulations. Short-term (3-year) and long-term (5-year) goals will focus on prototype verification, front-end chip development, and the electronics common interface.

**TDAQ** Key R&D areas include the simulation and development of advanced trigger algorithms, high-speed transmission technology for trigger electronics, and high-bandwidth distributed parallel computing. A small-scale experimental verification system will be constructed, implementing a trigger and data acquisition framework based on the new ATCA standard architecture.

**Offline software and computing** The software and computing team will face critical challenges and opportunities. As the project moves into the engineering phase, it will be essential to continuously refine the software stack to maintain optimal performance and reliability. Embracing new computing paradigms, such as the expanded use of heterogeneous computing resources and quantum computing techniques, will require both technical innovation and interdisciplinary collaboration. Furthermore, the technologies and methodologies developed for the experiment are expected to have a significant and lasting impact, shaping future experimental designs and advancing computing strategies in high-energy physics and beyond.

**Mechanics and integration** With ongoing advancements in detector technology and electronics, the mechanical design will undergo further optimization. Key research priorities include alignment accuracy measurement, structural optimization of carbon fiber materials, cooling system development, and enhancement of mechanical system stability and performance. Throughout the mechanical system design process, international collaboration will be strengthened through joint research initiatives addressing critical technical challenges such as carbon dioxide cooling system integration, remote automated control systems, and robotic remote operations.

## 17.3 Detector prototyping

### Integration strategy and system architecture

A comprehensive mock detector system will be constructed to validate the integration process of all subsystems, including mechanical interfaces, assembly, and cabling. The prototype will incorporate all detector elements, arranged as in the final detector according to their distance from the interaction point:

- MDI and luminosity monitors
- Vertex detector

- ITK
- TPC
- OTK
- ECAL
- HCAL
- Magnet system
- Muon detector

## Prototyping

Detector modules will be constructed to verify the technologies outlined in the TDR and to demonstrate, through beam tests, that the detectors meet the required performance parameters.

**Inner detector systems (MDI to TPC)** The innermost region integrates the beryllium beampipe (0.8 mm wall thickness) within the MDI shielding assembly. Main goals:

- **Vertex detector**: A full-stack prototype with stitched MAPS sensors (180 nm technology) on carbon foam supports will be constructed to achieves $3-5$ μm spatial resolution.
- **ITK/OTK**: Monolithic silicon modules (ITK) with a strip-to-pixel transition at $R$ = 350 mm will be developed. The design will target a material budget of $< 0.15\ X_0$ per layer with ±5 μm mechanical alignment. AC-LGAD strip modules (OTK) will be implemented to achieve good spatial and time resolution.
- **TPC**: A seven-module prototype with pixelated readout in a 1 T field will be used to validated that ion-backflow suppression can reach $< 0.1\%$ during DESY beam tests.

**Calorimetry and muon systems (ECAL to Muon)**

- **ECAL**: A full-scale crystal calorimeter module will be constructed to achieve electron energy resolution $\sigma_E/E \leq 3\%/\sqrt{E(\mathrm{GeV})} \oplus 1\%$.
- **HCAL**: A full-scale GS module will be developed to demonstrate hadron energy resolution $\sigma_E/E \approx 30\%/\sqrt{E(\mathrm{GeV})}$ with an additional constant term of 3%-5%.
- **Muon system**: A PS module will be integrated within magnet flux return to achieve the designed spatial resolution.

## Conclusion and outlook

The detector prototyping program will represent a crucial step toward validating both the integration feasibility and the performance targets of the detector concept. Outstanding issues that will require further R&D include validating the radiation hardness of vertex sensors, ensuring uniformity in large-scale crystal and GS production, and optimizing both online and offline algorithms.

# Part II

# Other CEPC Detector Concepts

# Chapter 18 ILD Detector for CEPC

The International Large Detector, ILD, is a detector concept that has been developed for more than two decades by a broad and international community of scientists, to equip future electron-positron colliders both linear such as ILC [1], LCF [2] and circular such FCC-ee [3] and is proposed here to equip one of the two interaction points (IP) of the CEPC.

In this chapter, the considerations which have guided the ILD concept group in the design of the detector are summarised. The main challenges for the realisation of the concept proposed here for CEPC are described, and possible technological solutions are sketched. The presentation of ILD in this chapter closely follows the document presented by the ILD collaboration to the 2026 update of the European Strategy for Particle Physics [4].

Historically, ILD was originally conceived as part of the ILC project. However, an ILD-like concept was already adopted by CEPC as one of its baseline options in its CDR [5]. Detailed simulation studies have shown the excellent performances that such a detector can provide in the framework of a circular collider like CEPC. Yet, R&D on some of the ILD detector components and their integration is still needed to reach the full potential that ILD can offer for CEPC.

The ambitious requirements of the ILC detectors sparked a world-wide R&D program to develop and demonstrate the different technologies needed [6] to fulfil these requirements. The scope and the needs of the R&D required for detectors at future colliders including including linear and circular ones have been more recently summarised in a report by ECFA [7] following the 2019 update of the European Strategy recommendation. Consequently, a new generation of R&D groups have been initiated, the so-called DRD collaborations. The former R&D collaborations for the most part have been integrated into these international DRD colaborations.

The ILD concept group from its beginning has collaborated very closely with R&D groups, and has organised the needed R&D work through and with the relevant R&D collaborations. This close and extremely fruitful cooperation will also continue with the new mode of operation and will be used to reach the performance goals within CEPC.

## 18.1 The ILD Detector design: requirements

ILD has been conceived and designed to study Higgs physics in great detail, and to improve our general understanding of electroweak and top-quark physics at high energies. The state-of-the-art regarding physics at such a high energy lepton collider has been summarized in the recent ECFA study report

[8]. The science case is strongly dominated by the quest for high precision in measurements of the properties of

the Higgs boson, the weak gauge bosons, and the top quark (see for example [8],[9],[10], [11] ).

The design of the ILD is driven by this quest for precision measurements and its realisation through the concept of "particle flow". In this approach, all particles in an event, both charged and neutral, are individually reconstructed as much as possible. This requires a detector which can distinguish single particles even within dense jets, and which separates and efficiently measures the energy of neutral and charged particles. For the detector, this implies that particle separation power is of the utmost importance. The calorimeters, in particular, need to be designed with very high granularity, both in the transverse and longitudinal directions. In addition, the linking between the tracking and the calorimeter systems should be highly efficient. Particle flow is particularly important at higher center-of-mass energies, where events are generally less spherical and more jet-like. The design of the ILD detector fulfils these requirements and is described in detail in [12]

The design drivers of the ILD detector for CEPC can be summarized by general requirements that apply for the different kinds of colliders, followed by a few specific ones to the circular nature of CEPC collider:

### 18.1.1 Requirements at future Collider

- **Impact parameter resolution:** An impact parameter resolution of $5\ \mu\text{m} \oplus 10\ \mu\text{m}/[p(\text{ GeV}/c)\sin^{3/2}\theta]$ has been defined as a goal, where $\theta$ is the angle between the particle and the beamline. This ensures excellent identification of displaced vertices for the identification of e.g. b- and c-quarks.
- **Momentum resolution:** An inverse momentum resolution of $\Delta(1/p) = 2\times10^{-5}(\text{ GeV/c})^{-1}$ asymptotically at high momenta should be reached with the combined silicon - TPC tracker. Maintaining excellent tracking efficiency and very good momentum resolution at lower momenta will be achieved by an aggressive design to minimise the detector's material budget. This requirement is driven by the Higgs recoil mass measurement, ensuring that its resolution is not dominated by the tracking performance.
- **Jet energy resolution:** Using the paradigm of particle flow, a jet energy resolution $\Delta E/E = 3\%$ or better for light-flavour jets should be reached. The resolution is defined in reference to light-quark jets, as the Root Mean Square (RMS) of the inner 90% of the energy distribution. This resolution allows hadronic decays of the $W$, $Z$ and Higgs bosons to be distinguished on a statistical basis.
- **Hermiticity:** The detector should cover as much as possible of the solid angle around the collision point, including in the very forward region. Ideally, the only uncovered regions should be the in- and out-going beampipes. This allows efficient identification of very forward-going particles, which, if missed, would restrict the ability to probe signals with missing energy.
- **Readout:** The detector readout should avoid as much as possible using a hardware trigger, ensuring maximal efficiency for all possible event topologies.

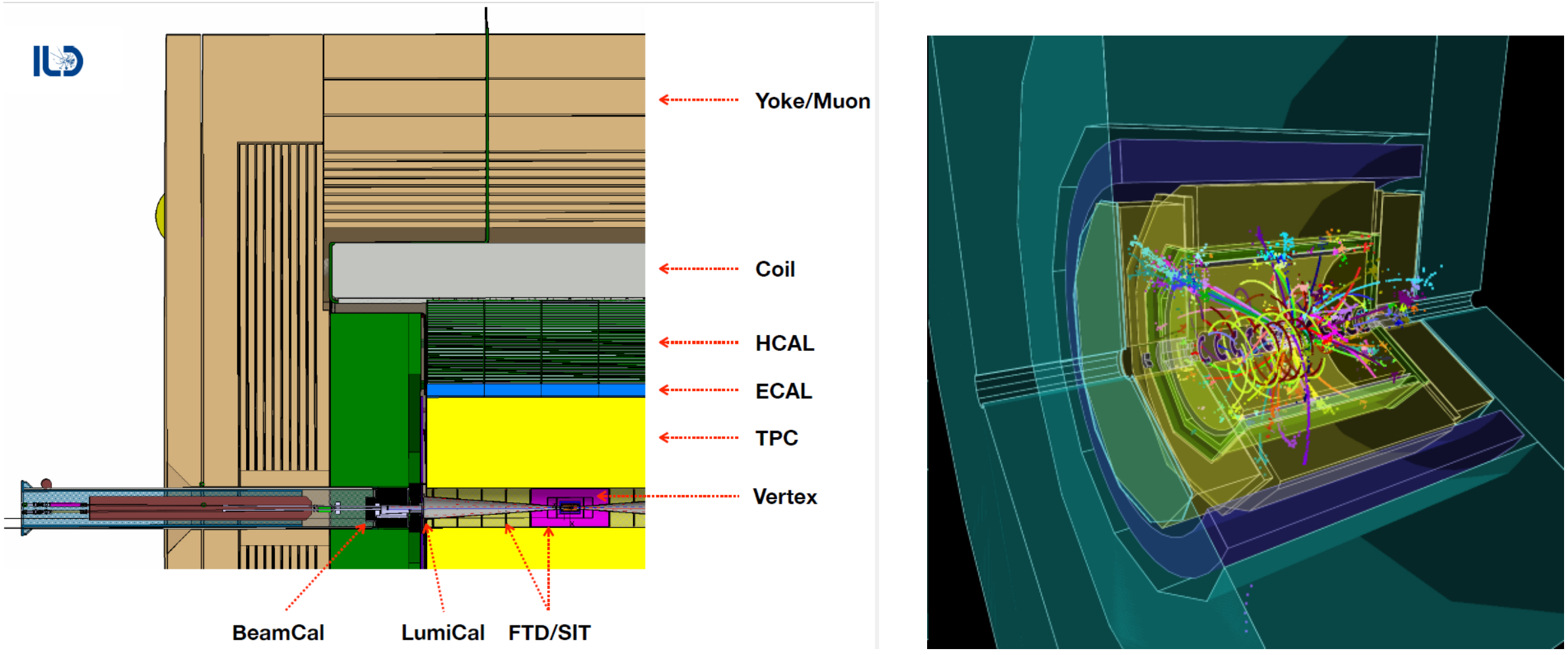

(a) (b)

**Figure 18.1:** (a) Single quadrant view of the ILD detector. (b) Event display of a simulated hadronic decay of a $t\bar{t}$ event in ILD. The colouring of the tracks show the results of the reconstruction, each colour corresponding to a reconstructed particle.

## 18.1.2 Requirements specific to Circular Collider

ILD proposes to use a derivative of its concept elaborated for ILC to suit the circular nature of CEPC. Indeed, the bunch train mode adopted for ILC is to be replaced by a continuous mode. The time interval between collisions is as short as as 577 ns when running in Higgs factory mode and goes to 23 ns when running at the $Z$-pole, significantly less than the one foreseen for ILC ( $\approx$ 199 ms). Therefore, the ILD for CEPC should include the following important points:

- The collision repetition rate at the CEPC forbids the use of power pulsing, significantly increasing (by about a factor of 100) the power dissipation in detectors, compared to ILC.
- The continuous operation of the collider makes a triggerless operation significantly more challenging.
- The lower top-energy of the colliders allows for a design of a more compact detector.
- The final beamline elements are significantly closer to the IP, and will intrude into the active detector region. A very careful design of the innermost part of the detector is needed to minimize the backgrounds in the detector, and to maximise the detector acceptance.
- As the beams are less tightly focused, the beamstrahlung per collision is less intense. Due to the larger number of collisions per time however, the integral background levels are comparable or larger that those at a linear collider.
- The CEPC proposal foresees running at the $Z$-pole energies at rates 25 times higher than that of the Higgs mode. $Z$-pole running mode therefore puts additional requirements on the detector in particular in the innermost region. Operation of ILD at the rate, expected at the $Z$-pole, will need more study. $Z$ operation might also imply a weaker magnetic field.

## 18.2 The ILD Detector system

The ILD detector is a multi-purpose detector in which the different requirements are addressed by a combination of different sub-detector systems. The optimization for ultimate precision in the reconstruction of charged and neutral particles requires that

- All major systems are contained within a strong solenoidal magnetic field, of 3.5 T. However, a 3 T field as proposed in CEPC CDR still provides excellent performances);
- To separate the impact into the calorimeter of nearby charged and neutral particles;
- To remove low-energy background from the main part of the detector.

Ultimate precision also requires that as little material as possible is introduced before the calorimeters, which implies that the tracking and calorimeter systems should be within the coil.

Over the past few years, an intense effort has been undertaken by the ILD concept group, to optimize the size of the ILD detector (see [12] for a review of the optimization).

A quadrant view of the detector model is shown in Figure 18.1 (a), together with an event display in this detector of a $t\bar{t}$ event (b).

The ILD concept from its inception has been open to new technologies. No final decision on subdetector technologies has been taken at this time, and in many cases several options are currently under consideration. For any technology to be accepted by ILD however its capabilities have to have been demonstrated experimentally, including demonstartion of its performance with prototypes and under as realistic conditions as possible.

The main parameters of the ILD detector are summarised in Table 18.1, together with the different technological options currently under consideration.

### 18.2.1 Vertexing system

The system closest to the interaction region is a pixel detector designed to reconstruct decay vertices of short lived particles with great precision. ILD has chosen a system consisting of three double layers of pixel detectors. The innermost layer is only half as long as the others to reduce the exposure to background hits. Each double layer will provide a spatial resolution around 3 µm at a pitch of about 22 µm, and a timing resolution per double layer of around 2–4 µs (see e.g. [14, 15]). To this end, the two layers in one double layer are optimized individually, one towards best spatial resolution, the other towards excellent timing resolution. R&D is directed towards improving this even further, to a point which would allow hits from individual bunch crossings to be resolved.

Over the last 10 years the MAPS technology has matured close to a point where all the requirements (material budget, readout speed, granularity) needed for an ILC detector can be met and efforts are ongoing to adapt this technology to circular colliders such CEPC, namley by reducing the pixel size and power consumption at the same time. The technology has seen a first large scale use in the STAR vertex detector [16], and more recently in the upgrade of the ALICE vertex detector. MAPS technology in general is undergoing very rapid progress and development, with many promising avenues being explored. To minimize the material in the system, sensors are routinely thinned to 50 μm.

**Table 18.1:** Key parameters of the ILD detector. All numbers from [13]. “Start” and “Stop” refer to the minimum and maximum extent of subdetectors in radius and/or $z$-value.

| **Technology** | **Detector** | **Start (mm)** | **Stop (mm)** | **Comment** |
|---|---|---|---|---|
| Pixel detectors | Vertex | $r_{in}$ = 16 | $r_{out}$ = 58 | 3 double layers of silicon pixels |
| | Forward tracking | $z_{in}$ = 220 | $z_{out}$ = 371 | 2 Pixel disks |
| | SIT | $r_{in}$ = 153 | $r_{out}$ = 303 | 2 double layers of Si pixels |
| Silicon strip | Forward tracking | $z_{in}$ = 645 | $z_{out}$ = 2212 | 5 layers of Si strips |
| | SET | $r_{in}$ = 1773 | $r_{out}$ = 1776 | 1 double layer of Si strips |
| Gaseous tracking | TPC | $r_{in}$ = 329 | $r_{out}$ = 1770 | MPGD readout, 220 points along the track, Alternative: pixel readout |
| Silicon tungsten calorimeter | ECAL option | $r_{in}$ = 1805 | $r_{out}$ = 2028 | 30 layers of $5 \times 5$ mm$^2$ pixels |
| | ECAL EC option | $z_{in}$ = 2411 | $z_{out}$ = 2635 | 30 layers of $5 \times 5$ mm$^2$ pixels |
| | Luminosity calorimeter | $r_{in}$ = 83<br>$z_{in}$ = 2412 | $r_{out}$ = 194<br>$z_{out}$ = 2541 | 30 layers |
| Diamond tungsten or GaAs calorimeter | Beam calorimeter | $r_{in}$ = 18<br>$z_{in}$ = 3115 | $r_{out}$ = 140<br>$z_{out}$ = 3315 | 30 layers |
| SiPM-on-Tile | ECAL alternative | $r_{in}$ = 1805 | $r_{out}$ = 2028 | 30 layers, 5 mm strips, crossed |
| | ECAL EC alternative | $z_{in}$ = 2411 | $z_{out}$ = 2635 | 30 layers, 5 mm strips, crossed |
| | HCAL option | $r_{in}$ = 2058 | $r_{out}$ = 3345 | 48 layers, $3 \times 3$ cm$^2$ pixels |
| | HCAL EC option | $z_{in}$ = 2650 | $z_{out}$ = 3937 | 48 layers, $3 \times 3$ cm$^2$ pixels |
| (M)RPC | (T)HCAL option | $r_{in}$ = 2058 | $r_{out}$ = 3234 | 48 layers, $1 \times 1$ cm$^2$ pixels |
| | (T)HCAL EC option | $z_{in}$ = 2650 | $z_{out}$ = 3937 | 48 layers, $1 \times 1$ cm$^2$ pixels |
| SiPM on scintillator bar | Muon | $r_{in}$ = 4450 | $r_{out}$ = 7755 | 14 layers |
| | Muon EC | $z_{in}$ = 4072 | $z_{out}$ = 6712 | up to 12 layers |

Advancements in technology make it possible to revisit the baseline geometry (three double-sided layers). In particular, stitched sensors, explored within the ALICE-ITS3 upgrade, offer the potential to incorporate bent sensors to optimize the material budget. However, significant challenges remain in demonstrating the feasibility of this approach in the context of a lepton collider, including azimuthal acceptance and bending at a radius close to 12 mm.

Other technologies under consideration for ILD are DEPFET, which is also currently being deployed in the Belle II vertex detector [17], fine pitch CCDs [18], and also less mature technologies such as Silicon On Insulator (SOI) and Chronopix [6]. Very light weight support

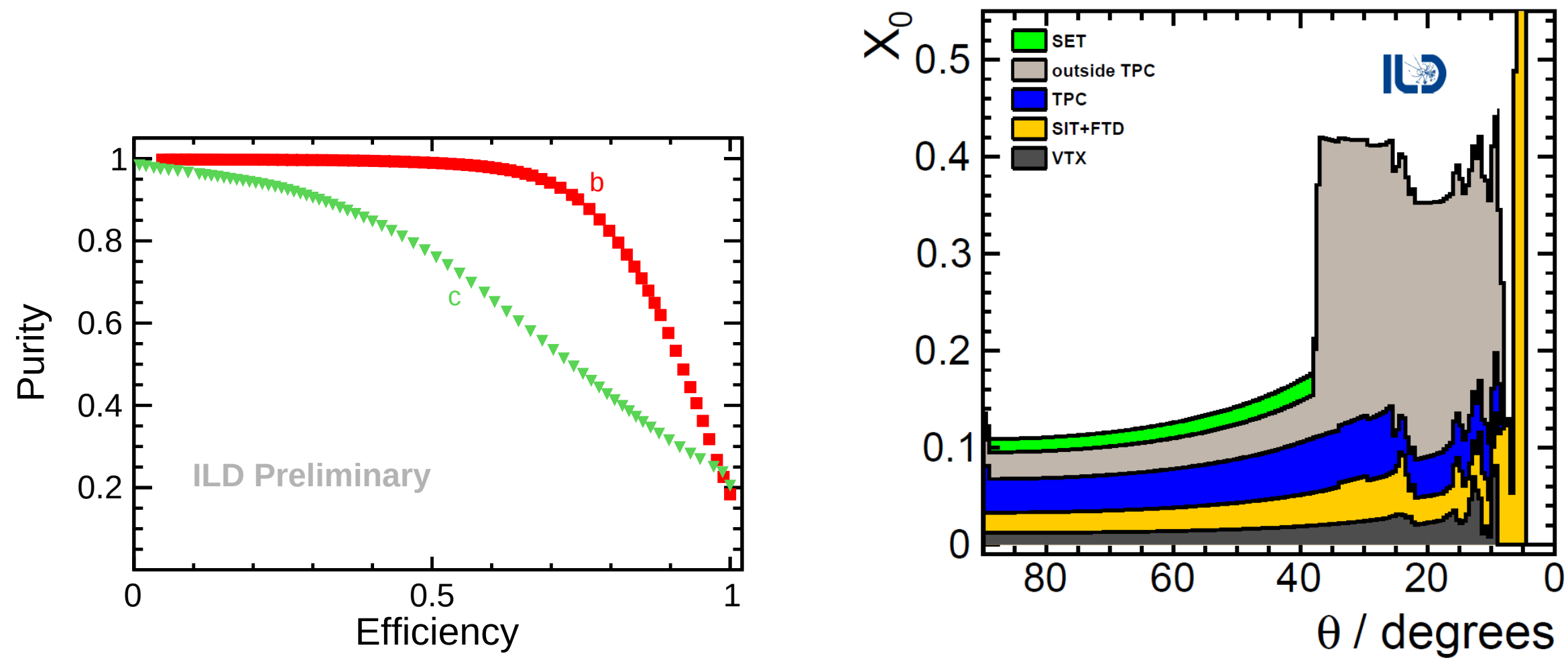

**(a)** **(b)**

**Figure 18.2:** (a) Purity of the flavour tag as a function of the efficiency, for different flavours tagged. (b) Cumulative material budget in ILD up to the calorimeter, in fraction of a radiation length. Figures are taken from [12].

structures have been developed, which bring the goal of 0.15% of a radiation length per layer within reach [19]. Such structures are now used in the Belle II vertex detector.

In Figure 18.2 (a) the purity of the flavour identification in ILD is shown as a function of its efficiency. The performance for b-jet identification is excellent, and charm-jet identification is also good, providing a purity of about 70% at an efficiency of 60%. The system also allows the accurate determination of the charge of displaced vertices, and contributes strongly to the low-momentum tracking capabilities of the overall system, down to a few 10s of MeV. An important aspect of the system leading to superb flavour tagging is the small amount of material in the tracker. This is shown in Figure 18.2 (b).

### 18.2.2 Tracking system

ILD has decided to approach the problem of charged particle tracking with a hybrid solution, which combines a high resolution TPC with a few layers of strategically placed strip or pixel detectors before and after the TPC (for a recent review, see [20]). The TPC will fill a large volume about 4.6 m in length, spanning radii from 33 to 180 cm. In this volume the TPC provides up to 220 three dimensional points for continuous tracking with a single-hit resolution of better than 100 µm in $r\phi$, and about 1 mm in $z$. This high number of points allows a reconstruction of the charged particle component of the event with high accuracy, including the reconstruction of secondaries, long lived particles, kinks, etc. For momenta above 100 MeV, and within the acceptance of the TPC, greater than 99.9% tracking efficiency has been found in events simulated realistically with full ILC backgrounds and similar performances are expected at CEPC running in Higgs mode. At the same time the complete TPC system will introduce only about 10% of a radiation length into the detector [21].

Inside and outside of the TPC volume a few layers of silicon detectors provide additional high resolution points, at a point resolution of around 10 μm. Combined with the TPC track, this will result in an asymptotic momentum resolution of $\delta p_t/p_t^2 = 2 \times 10^{-5}$ (GeV/c)$^{-1}$ for the complete system. Since the material in the system is very low, a significantly better resolution at low momenta can be achieved than is possible with a silicon-only tracker. The achievable resolution is illustrated in Figure 18.3 (a), where the $1/p_t$-resolution is shown as a function of the momentum of the charged particle. In the forward direction, extending the coverage down to the beampipe, a system of two inner pixel disks (point resolution 3 μm), followed by five strip disks (resolution 7 μm) provide tracking coverage down to the beam-pipe.

The silicon layer outside the TPC could be instrumented by sensors with excellent timing precision to measure particles' time-of-flight. Several sensor technologies which would be applicable are currently under active R&D within the DRD3 collaboration, including Trench-Isolated Low Gain Avalanche Detectors (TI-LGADs), AC-coupled LGADs (AC-LGADs), Resistive DC-coupled Silicon Detectors (RSD-LGADs), and Inverse Low Gain Avalanche Detectors (iLGADs).

In parallel, the development of low-power front-end readout electronics is essential to ensure power-efficient operation while preserving the timing resolution, a critical factor for large-scale detectors. Sensor concepts relying on signal-sharing mechanisms (such as AC-LGADs and RSD-LGADs) which enable a reduced number of readout channels may offer advantages in optimizing the balance between spatial resolution, power dissipation, and timing performance.

The TPC also enables the identification of particle type by the measurement of the specific energy loss, $\mathrm{d}E/\mathrm{d}x$ for tracks at intermediate momenta [22]. The achievable performance is shown in Figure 18.3 (b). The relatively new approach of cluster counting in the TPC promises a significantly improved resolution. If the outer silicon layers can provide timing with about 100 ps resolution, time of flight measurements can provide additional information, which is particularly effective in the momentum regime which is problematic for $\mathrm{d}E/\mathrm{d}x$ as shown in Figure 18.3 (b).

The design and performance of the TPC has been the subject of intense R&D over the last 15 years. Several MPGD

technologies for the readout of the TPC have been successfully developed, and have demonstrated the required performance in test beam experiments. The readout of the charge signals is realised either through traditional pad-based systems, or by directly attaching pixel readout-ASICs to the endplate. A large volume field cage has been built to demonstrate the low mass technology needed to meet the 10% $X_0$ goal discussed above. Most recently the performance of the specific energy loss, $\mathrm{d}E/\mathrm{d}x$ has been validated in test beam data. Based on these results, the TPC technology is sufficiently mature for use in the ILD detector, and can deliver the required performance when operating at ILC (see e.g. [23, 24]). For CEPC, running at $Z$-pole, the

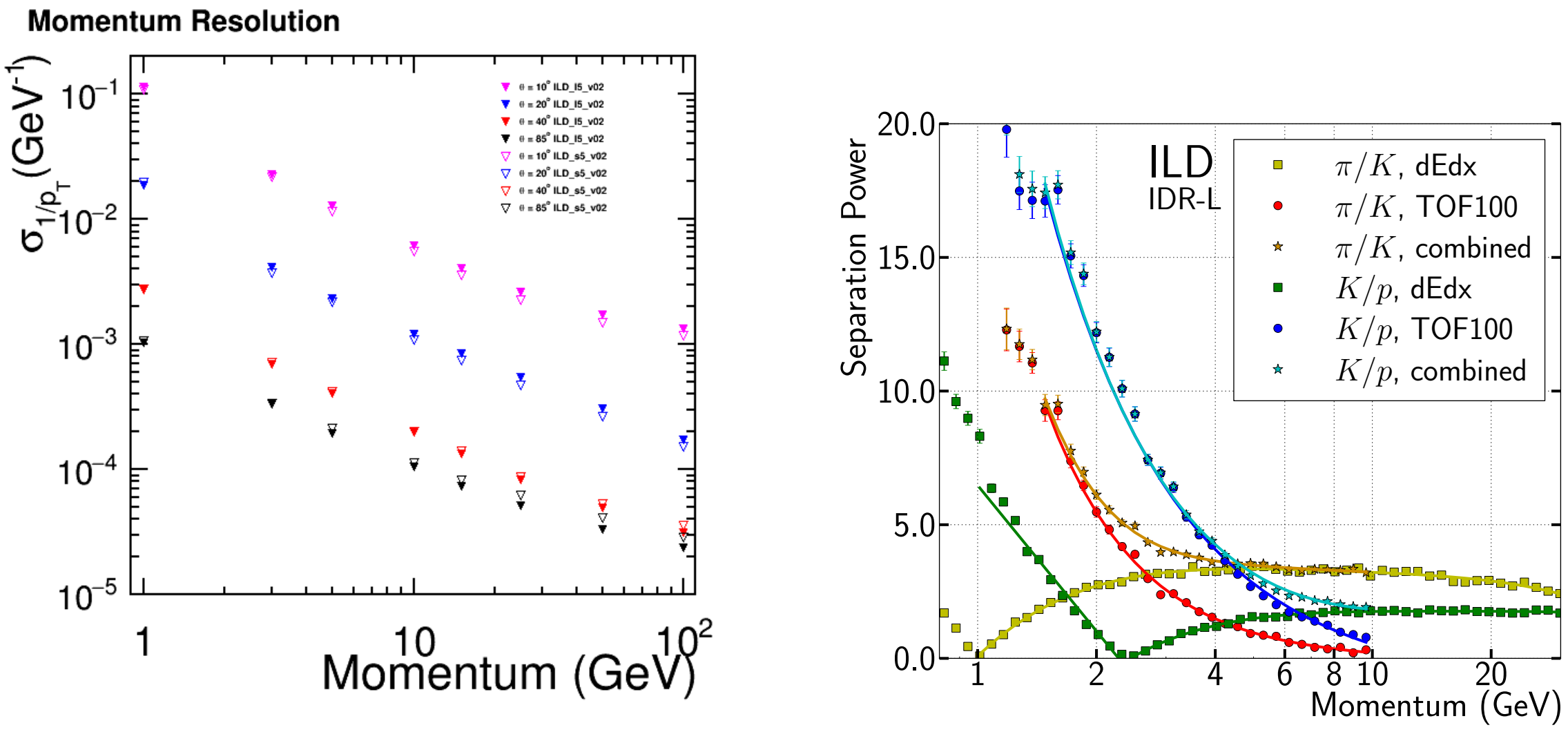

**(a)** **(b)**

**Figure 18.3:** (a) Simulated resolution in $1/p_t$ as a function of the momentum for single muons. The different curves correspond to different polar angles - the open symbols are for a smaller version of ILD studied in [12]. (b) Simulated separation power (probability for a pion(kaon) to be reconstructed as a kaon(proton)) between pions and kaons, from d$E$/d$x$ and from timing, assuming a 100 ps timing resolution of the first ECAL layer. Figures are taken from [12].

increase of the rate may have may lead to a distortion of the electric field due to increased number of ions issued from the ionisation as well as from the amplification process, the so-called Ion Back Flow (IBF). Studies are ongoing to quantitatively estimate how large an impact these distortions will have on the final performance of the system.

### 18.2.3 Calorimeter system

A very powerful calorimeter system is essential to reach the needed performance of the detector. Particle flow, which is driving the design of ILD to a significant extent, relies on the ability to separate individual particles in a jet, both charged and neutral. This puts the imaging capabilities of the system at a premium, and pushes the calorimeter development in the direction of a system with very high granularity as the solution to this challenge [25]. The conceptual and technological developments of the particle flow calorimeter have been largely done by the CALICE collaboration (for a review of recent CALICE results see e.g. [26]).

ILD has chosen a sampling calorimeter readout with silicon diodes as one option for the electromagnetic calorimeter. Diodes with pads of about (5×5) mm$^2$ are used, to sample a shower up to 30 times in the electromagnetic section. In 2018, a test beam experiment demonstrated the large scale feasibility of this technology, by showing not only that the anticipated resolution can be reached, but also by demonstrating that a sizeable system can be built and operated. The test has provided confidence that scaling this to an ILD-sized system will be possible. An alternative using (50 × 50) μm$^2$ MAPS-based active layer is also considered [27].

As an alternative to the silicon-based system, sensitive layers made from thin scintillator strips are also investigated (ScW-ECAL). Orienting the strips perpendicular to each other has the potential to realize an effective cell size of $5 \times 5$ mm$^2$, with the number of readout channels reduced by an order of magnitude.A full 32-layer prototype was recently constructed. It was tested in beams at CERN-PS/SPS in 2022 and 2023 and its performance is being evaluated [28].

For the hadronic part of the calorimeter of the ILD detector, two technologies are studied, based on either silicon SiPM

on scintillator tile technology [29] or resistive plate chambers [30]. The SiPM-on-tile option has a moderate granularity, with $3 \times 3$ cm$^2$ tiles, and provides an analogue readout of the signal in each tile (AHCAL). The RPC technology has a better granularity, of $1 \times 1$ cm$^2$, but provides only 2-bit amplitude information (SDHCAL). For both technologies, significant prototypes have been built and operated. Both follow the engineering design anticipated for the final detector, and demonstrate thus not only the performance, but also the scalability of the technology to a large detector.

While the original design of the system was done for the ILC, its adaptation to running at the CEPC and other circular colliders is under study. The main challenge in the adaptation lies in coping with the continuous readout and with the much higher rate, bandwidth, power, and cooling requirements, without compromising granularity and compactness. For a circular collider like CEPC, the readout rates and, more critically, heat dissipation have to be reassessed, a work on-going; the preliminary conclusions hint strongly at the need for an active cooling, especially for the ECAL. Practical cooling aspects were reviewed for the HCAL [31]. More recently, studies toward a thin and uniform active cooling for the ECAL have started.

The potential of adding precise timing to the calorimeters, or at least to a few layers, is actively being investigated – from the sensor to physics performances – for most of the options, T-SDHCAL, AHCAL, SiW-ECAL, and ScW-ECAL. For the T-SDHCAL, this is accompanied by new developments in sensors, moving from RPCs to multigap RPCs, which offer better time resolution of only tens of picoseconds and enhanced rate capability. Other alternatives under R&D, such as GEM or Micromegas, could also be promising options. For the AHCAL, the ongoing construction of a significant calorimeter in this technology for the CMS upgrade is providing important information on system design and performance, which will go a long way towards an optimized design for a future lepton collider detector.

A new generation of ASICs is also under development at the Omega laboratory, focusing on reducing power consumption and data volume, including auto-trigger and data-driven readout, as well as improving time resolution down to tens of picoseconds. Transitioning from the AMS130 CMOS technology used in the first ASICs to more modern technologies, such as TSMC 130 or 65, could meet the sensor requirements for all types of future lepton colliders.

It has been a major success in the past years that the technologies needed for a true particle flow calorimeter have been successfully demonstrated in a design which is suitable for the ILD detector. With this demonstration, a major hurdle towards the realization of ILD has been overcome [32]. The simulated particle flow performance is shown in Figure 18.4.

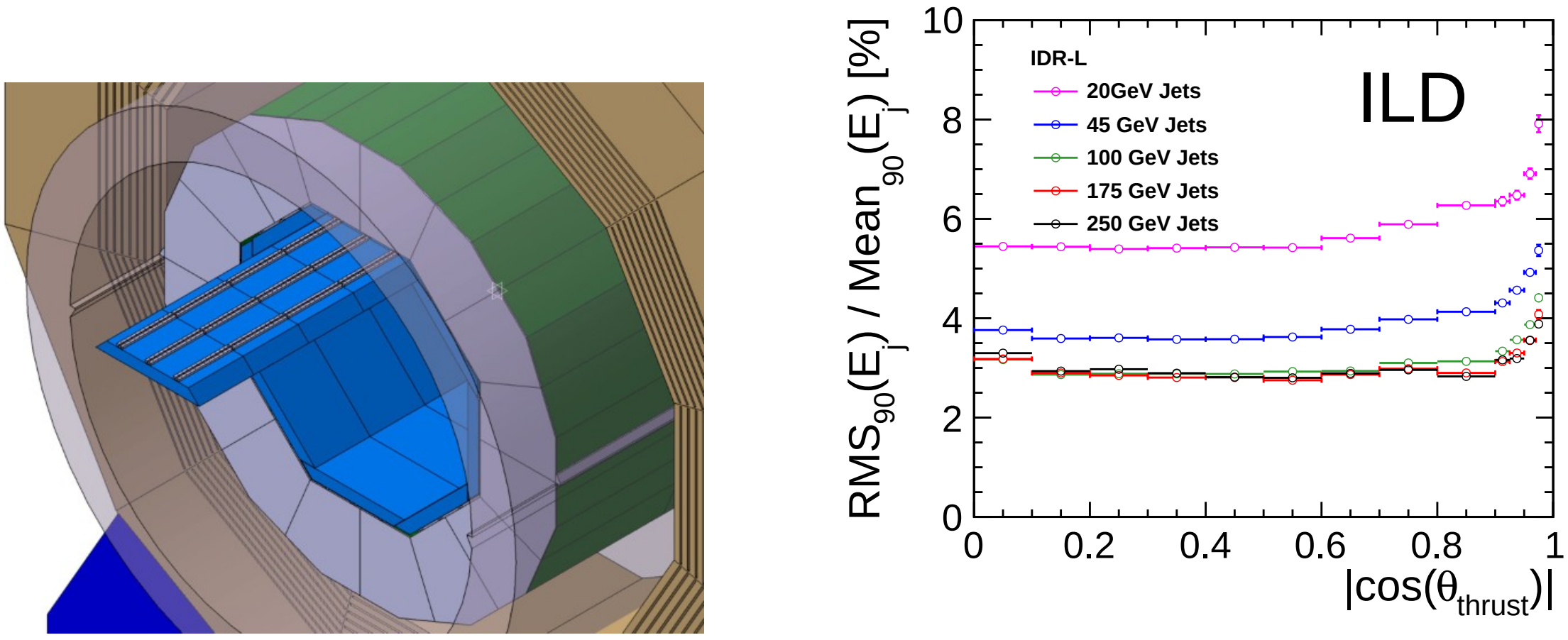

**(a)** **(b)**

**Figure 18.4:** (a) Three-dimensional rendering of the barrel calorimeter system, with one ECAL module partially extracted. (b) Particle flow performance, measured as the energy resolution in two-jet light flavour events, for different jet energies as a function of $\cos\theta$). The resolution is defined as the rms of the distribution truncated so that 90% of the total jet energy is contained inside the distribution. Figures are taken from [12].

### 18.2.4 The Forward system

Three rather specific calorimeter systems are foreseen for the very forward region of the ILD detector [33]. LumiCal is a high precision fine sampling silicon tungsten calorimeter primarily designed to measure electrons from Bhabha scattering, and to precisely determine the integrated luminosity [34]. The Luminosity Hadronic Calorimeter) (LHCAL) just outside the LumiCal extends the reach of the endcap calorimeter system to smaller angles relative to the beam, and closes the gap between the inner edge of the ECAL endcap and the luminosity calorimeter, LumiCal. Below the LumiCal acceptance, where background from beamstrahlung rises sharply, BeamCal, placed further downstream from the interaction point, provides added coverage and is used to provide a fast feedback on the beam position at the interaction region. As the systems move close to the beampipe, the requirements on radiation hardness and on speed become more and more challenging. Indeed this very forward region in ILD is the only region where radiation hardness of the systems is a key requirement. In placing the different detector components, particular care has been taken to allow the bulk of the beamstrahlung photons and pairs to leave the detector through the outgoing beampipe.

In particular the BeamCal has been positioned in such a way that backscattering of particles from the BeamCal face into the active part of the detector is minimised. As the inner region of the detector at a circular collider like CEPC is different, the layout of the forward system will need to be re-optimized. Work on this has started.

### 18.2.5 The Muon system

The iron return yoke of the detector, located outside of the coil, is instrumented to act as a calorimeter tail catcher and as a muon identification system. Several technologies are possible for the instrumented layers. Both RPC chambers and scintillator strips readout with SiPMs have been investigated. Up to 14 active layers, located mostly in the inner half of the iron yoke (see Table 18.1 and Figure 18.1 for more details) could be instrumented.

## 18.3 The software environment for ILD

ILD has from the start actively contributed to the community driven software project iLCSoft [35] and more recently to the larger Key4hep [36] software ecosystem. A strong focus has always been on defining realistic simulation models of the ILD detector that are defined in DD4hep [37] used to produce event samples in full simulation with Geant4. These event samples are used in physics analyses to obtain sensitivities to various aspects of particle physics, to develop data reconstruction algorithms, to optimise the layout of the detector or to investigate the potential advantages of new or improved technologies.

The WHIZARD [38] event generator is typically used to generate the hard scattering event, interfaced with CIRCE2 to describe the beam energy spectrum at the future colliders such as CEPC, and including the effects of initial state radiation. Pythia is then used to describe hadronisation and final state radiation, and Geant4 to simulate the detailed detector. The Marlin software package is then used to process the simulated events. A number of ILD models have been developed in DD4hep as part of the detector optimisation activities using the chain described above.

An alternative approach based on fast simulation using the SGV [39] is used in the case of very high statistics samples for which the CPU cost or time of full Geant4 simulation is prohibitive.

A suite of digitisation and reconstruction algorithms has been developed within the Marlin framework. These convert the energy deposits in sensitive detectors simulated by Geant4 into realistic detector signals that are validated in comparisons to the performance of real sub-detector prototype test beam results. These signals are then reconstructed into particle tracks, calorimeter clusters and then to particle flow objects corresponding to single particles, followed by clustering into multiple-particle systems such as hadronic jets, tau leptons or converted photons. Superimposing the effects of background processes such as beamstrahlung and photo-production of hadrons is the job of dedicated overlay processors. Algorithms for particle identification (leptons, different hadron flavours) are run, as well as jet flavour identification.

Over the past years, large samples of events from Standard Model processes have been produced at various centre-of-mass energies, corresponding to proposed ILC energy points 250, 350, 500, 550, and 1000 GeV. These large-scale productions have used grid computing resources, organised via the LCDirac infrastructure. These samples have been instrumental in ensuring that the majority of ILD analyses are done with fully simulated event samples.

All tools used by ILD are incorporated in Key4hep, where work is currently ongoing to transfer some of the older tools to more modern replacements, including incorporating the seamless use of modern AI tools, further strengthening the use of Key4hep in the workflow.

Computational resource estimates based on the software chain above including background simulations suggest that the total raw data rate of the ILC will be $\approx$ 1.5 GB/s and the total estimated storage needs will be a few tens of PB/y [10], at least an order of magnitude smaller than for the LHC.

## 18.4 Science with ILD

ILD has been designed to operate with electron-positron collisions between 90 GeVand 1 TeV. The science goals of the future lepton collider have been recently reviewed in detail in [8], and will not be repeated here. The ILD concept group contributed many results to this report and it should be pointed out that most of them were based on fully simulated events, using a realistic detector model and advanced reconstruction software, and in many cases included estimates of key systematic effects.

The detector model used for the ILD studies was tested against performance of the prototype detectors. The key performance numbers for the vertexing, tracking and calorimeter systems are all based on results from test beam experiments. Key aspects of the particle flow performance, includes the single particle resolution for neutral and charged particles, the particle separation in jets, the matching between tracking and calorimetry, were also verified in dedicated measurements.

A significant number of benchmark studies were undertaken to fully understand the performance of the ILD detector, to determine in particular the correlations between detector response, reconstruction results and science objectives, and to optimise the detector design parameters. The center-of-mass energy for the first stage of the lepton collider is about 250 GeV, nevertheless, the ILD detector is designed to meet the more challenging requirements of higher center-of-mass energies, since major parts of the detector, e.g. the coil, the yoke and the main calorimeters will not be replaced when upgrading the accelerator. Therefore, benchmark analyses were also performed at a center-of-mass energy of 500 GeVand even 1 TeV. The potential of new features incorporated in the detector design, e.g. time-of-flight measurement, was also considered. Results of all these studies were published in the ILD Interim Design Report [12].

Developments in the detector design and in analysis methods are also very important for precision studies and BSM searches.

Considered in the recent study [40] was the impact of the TPC design on the efficiency of the charged hadron identification in the quark-antiquark forward-backward asymmetry measurement in heavy quark pair-production.

A novel, optimised TPC design, with higher readout granularity, allows for the use of the cluster counting reconstruction ($\mathrm{d}N_p/\mathrm{d}x$) method for charged-hadron identification, resulting in better resolution than the mean energy loss ($\mathrm{d}E/\mathrm{d}x$) measurement used in the earlier studies.

The impact of the charged hadron particle identification (PID) on the size of the statistical discrimination power between different gauge-Higgs unification (GHU) scenarios [41, 42] and the Standard Model is shown in Figure 18.5. In particular in the studies of reactions involving charm quarks PID has the largest impact, and thus improvements by using $\mathrm{d}N_p/\mathrm{d}x$ are seen most strongly.

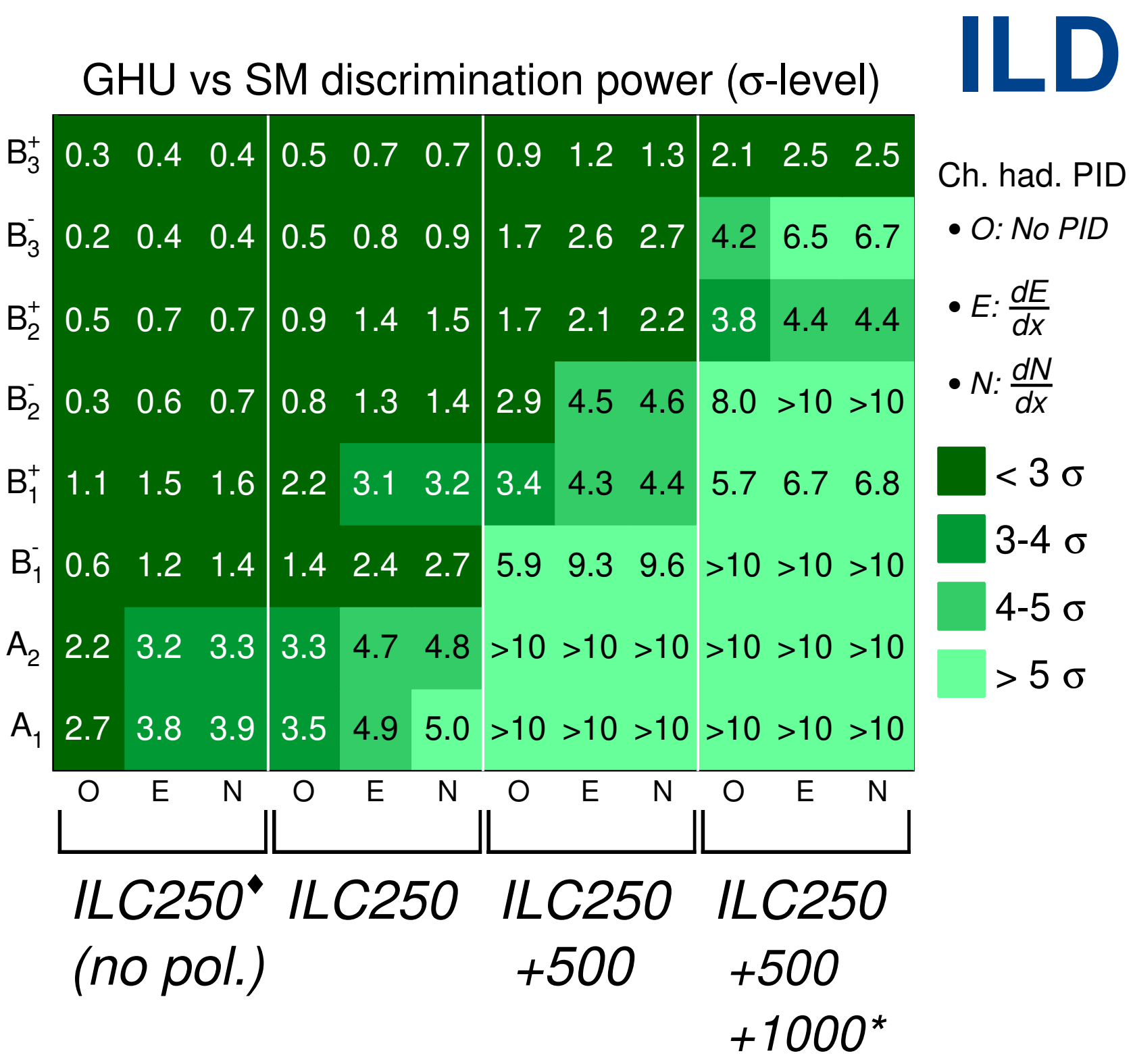

**Figure 18.5:** Statistical discrimination power between different gauge-Higgs unification (GHU) scenarios and the SM, for different ILC running scenarios and three different charged hadron particle identification capabilities (PID) considered: O for no PID used, $E$ for PID based on the $\mathrm{d}E/\mathrm{d}x$ reconstruction, and $N$ for an optimized TPC design with cluster counting reconstruction $\mathrm{d}N_p/\mathrm{d}x$.

First physics studies has also been completed comparing the ILD performance at a linear and at a circular collider. The most relevant difference here is the design of the forward region of the detector. For a circular collider, elements from the final focus system reach further into the active volume of the detector, and limit to total hermeticity of the detector. Studies which strongly rely on the forward region of the detector are thus most strongly affected. In addition polarisation of the lepton beams is currently not anticipated for a circular collider.

## 18.5 Integration of ILD

A detailed concept has been worked out for the ILD integration. This includes a mechanical model of the detector, and of the infrastructure needed to operate the detector, and a model of how the detector would be assembled and installed.

The current model assumes an initial assembly of the detector on the surface, similar to the construction of CMS at the LHC. A vertical shaft from the surface into the underground experimental cavern allows ILD to be lowered in five large segments, corresponding to the five yoke rings.

ILD is self-shielding with respect to radiation and magnetic fields to enable the operation and maintenance of equipment surrounding the detector, e.g. cryogenics. The current design of ILD was strongly influenced by the proposal to operate two detectors at the ILD in a push-pull mode. In case this requirement is released, the integration concept of the ILD detector could be re-optimised.

The cost of the ILD detector has been estimated in a dedicated and detailed costing study in 2012. The cost has been last updated in 2019 for the ILD IDR [12]. The total detector cost is about US$390 million in 2012 costs. The cost of the detector is strongly dominated by the cost of the calorimeter system and the yoke, which together account for about 60% of the total cost.

## 18.6 The ILD concept group

The ILD concept group comprises around 275 people from 59 member institutes, and a few individual guest members, from around the world. ILD has given itself a structure, with groups who want to join do sign a memorandum of participation. Scientists who want to participate in ILD but whose institutes cannot join have the option to ask for guest membership in ILD.

A map indicating the location of the ILD member institutes is shown in Figure 18.6.

## 18.7 Conclusion and outlook

The ILD concept is a well developed integrated detector optimised for use at a future electron-positron collider. It is based on advanced detector technology, and driven by the science requirements at future lepton collider. Most of its major components have been fully demonstrated through prototyping and test beam experiments. The physics performance of ILD has been validated using detailed simulation systems. It is the expressed intention of the ILD concept group to contribute a detector to a future lepton collider, regardless of how and where it is going to be implemented.

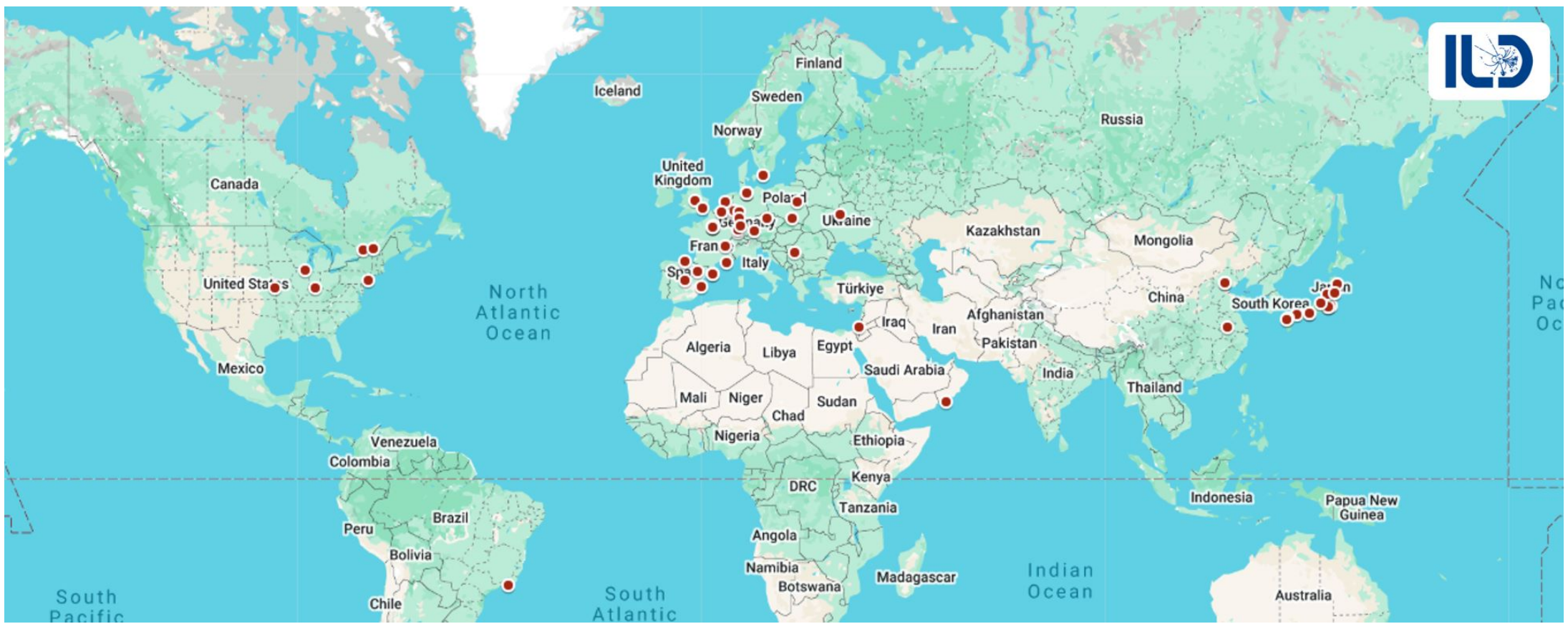

**Figure 18.6:** Locations of the ILD member institutes, as of March 2025.

# Chapter 19 IDEA Detector for CEPC

An alternative detector concept, named IDEA [1], optimized for the physics and running conditions of a large circular $e^+e^-$ collider is presented. This detector design is based on demonstrated technologies and is already supported by a large international collaboration. In the following we recall the detector requirements and, after an overview of the full detector, give a brief description of all the individual detector elements.

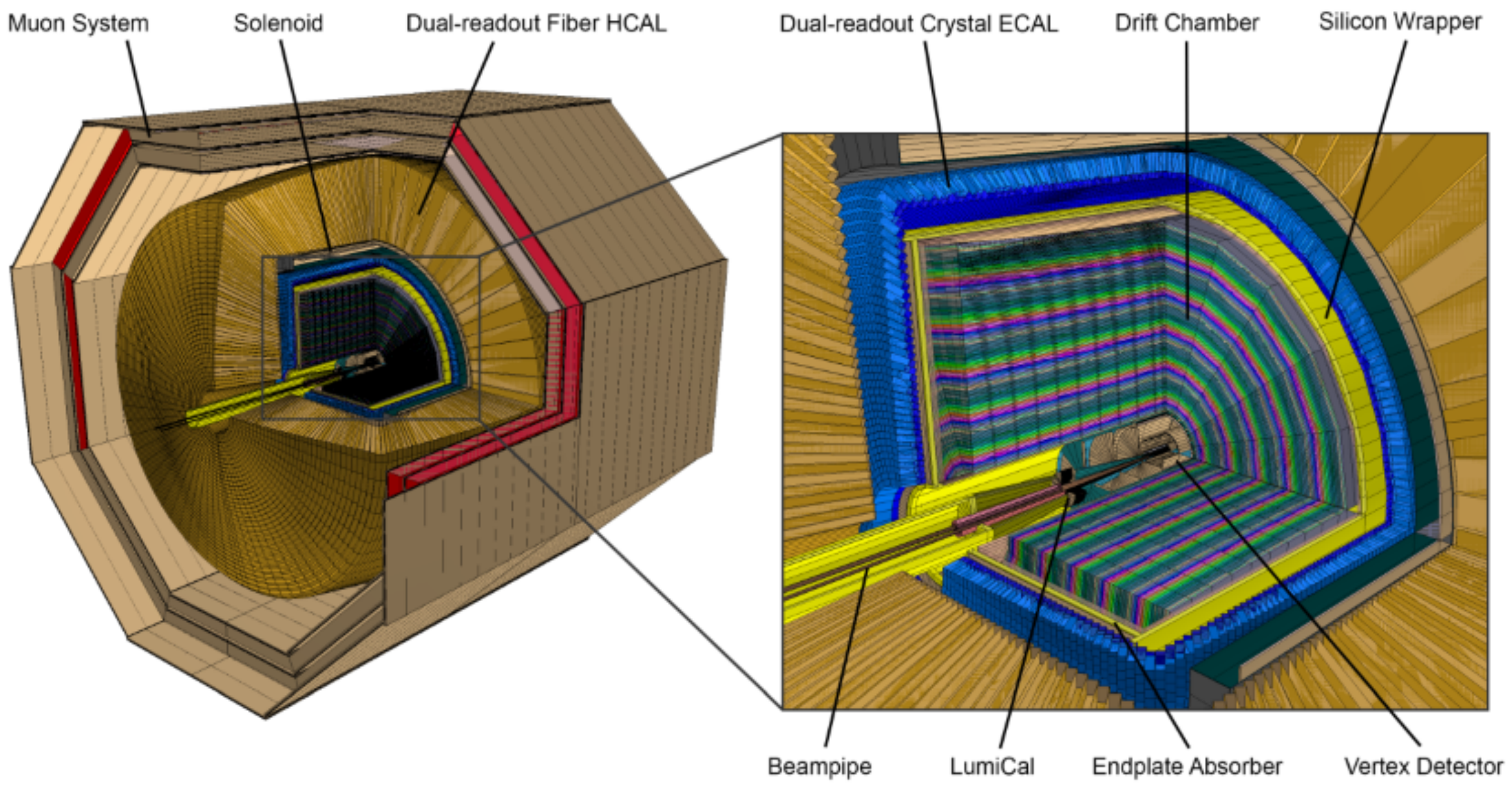

**Figure 19.1:** 3D cutout view of the IDEA baseline detector design with subdetector labels.

## 19.1 Requirements

### 19.1.1 Running conditions

The CEPC is planned to operate at or around four center of mass energies: the $Z$ boson mass, the $WW$ production threshold, the $Z$-Higgs associated production and the top quark pair production threshold. The luminosity, in excess of $10^{36}$ cm$^{-2}s^{-1}$, is highest at the $Z$ resonance with a short bunch spacing in the tens of ns range. It then drops rapidly while the energy increases, as the beam currents are reduced to keep the maximum energy radiated by the beams below 30 MW/beam. The bunch spacing grows as the energy increases, reaching about 1 µs at the ZH associated production energy.

These running conditions are particularly challenging for the detectors at the $Z$ energy, due to the fast beam-related backgrounds and the rather high rate of interesting physics, which is expected to be in the order of 100 kHz. Limitations also exist in the current final focus configuration due to the magnetic field of the detector solenoid, which contributes to increase the beam emittances. A field of 2 T is currently considered to be the maximum acceptable to avoid degrading the luminosity when running at the $Z$ pole. This restriction becomes much weaker at higher energies, so magnets that can provide up to 3 T when running above the $Z$ resonance are being considered.

The beam structure at CEPC does not have significant gaps in time with no beam as foreseen at ILC or CLIC. This has a large impact on the detector design, because powering on and off the detectors to reduce power consumption is not an option, thus increasing cooling requirements.

Relative to linear colliders the beamstrahlung contribution [2, 3] to the center-of-mass energy uncertainty is much smaller, being in the order of 1‰ around the $ZH$ energy scale, thus allowing tighter constraints from the beam energy. A notable example where this helps significantly is the determination of the Higgs recoil mass.

### 19.1.2 Physics drivers

The physics accessible at FCC-$ee$ falls into a few major categories: detailed measurement of the Higgs boson properties, high precision electro-weak and heavy flavor studies at the $Z$ boson center of mass energy, precision measurements of the $W$ boson and top quark parameters, and searches for new physics.

The main objective for the tracking systems is to provide good momentum, angle and impact parameter resolution for tracks with $p_{\mathrm{T}}$ below 100 GeV/c. At higher momenta the requirement on the momentum resolution is driven mostly by the measurement of the Higgs recoil mass in $HZ$ events with Z→ $\mu^+\mu^-$. The precision of this measurement depends on the center of mass energy spread due to beamstrahlung, that is $\sim 0.1\%$ at CEPC, and the experimental error on the momentum measurement of the two muons from the decay of the $Z$ boson. A track momentum resolution, $\sigma_{p_{\mathrm{T}}}/p_{\mathrm{T}}$, of a few times $10^{-5} \times p_{\mathrm{T}}(\text{ GeV})$ is required to preserve the precision driven by the beam energy spread. Given the constraints on the magnetic field of the detector solenoid this leads to a rather large tracking volume.

Heavy flavor physics and flavor tagging, all profit from high accuracy in secondary vertex measurements implying the need for powerful vertex detectors with hit resolutions at the few micron level. Typical momenta are generally lower for these types of physics, necessitating low mass tracking detectors. Particle identification, in particular separation of pions from kaons up to a few tens of GeV, is also important to enhance the signal over background for many specific rare processes and to allow some level of strange flavor tagging.

Jet energy and angular resolution are of key importance to exploit final states involving decays to high energy quarks. Typical examples are $Z$, $W$ or Higgs boson decays to two jets or hadronic decays of top quarks. A useful figure of merit is the separation of $Z$ from $W$ decays using the di-jet invariant mass, leading to a required jet energy resolution $\sim 30 - 40\%/\sqrt{E}$. A new generation of calorimeters based on dual readout is used to achieve such precisions.

Extreme electromagnetic resolutions, attainable only with crystals, can provide high quality neutral pion identification that is important for the measurement of many heavy flavor quark and tau final states. Resolved neutral pions also play a role in improving the jet resolution. A good photon energy resolution is also relevant in improving the signal over noise of the process $H \to \gamma\gamma$ and in correcting the electron track energy for bremsstrahlung radiation.

## 19.2 Detector overview

The IDEA is a new, innovative detector concept, specifically designed for CEPC and FCC-$ee$. It is based on a sophisticated tracker and a crystal Dual Readout (DR) electromagnetic calorimeter within a superconducting solenoid magnet, followed by a dual readout fiber calorimeter and then completed by a muon detection system placed within the iron yoke that closes the magnetic field.

The tracker is composed of a high resolution silicon pixel vertex detector, a large-volume and extremely low-mass drift chamber for central tracking and particle identification, and a surrounding wrapper made from silicon sensors to ensure the best overall momentum resolution. The wire chamber provides up to 112 space-point measurements along a charged particle trajectory with excellent particle identification capabilities provided by the cluster counting technique. An outstanding energy resolution for electrons and photons is provided by a finely segmented crystal electromagnetic calorimeter. The particle identification capabilities of the drift chamber are complemented by a time-of-flight measurement provided either by LGAD technologies in the Si wrapper or by a first layer of LYSO crystals in the ECAL. Outside an ultra-thin, low-mass superconducting solenoid the calorimeter system is completed by the dual-readout fiber calorimeter. The hybrid dual-readout approach promises an outstanding energy resolution for hadronic showers without loss of EM performance.

The muon detection system is based on Micro-Resistive WELL ($\mu$-RWELL) detectors, a new micro pattern gas detector that will provide a space resolution of a few hundred of microns, giving the possibility of selecting muons with high precision and also reconstructing secondary vertices at a large distance from the primary interaction point.

A 3-D view of the IDEA detector is shown in Figure 19.1 and its cross-section is shown in Figure 19.2.

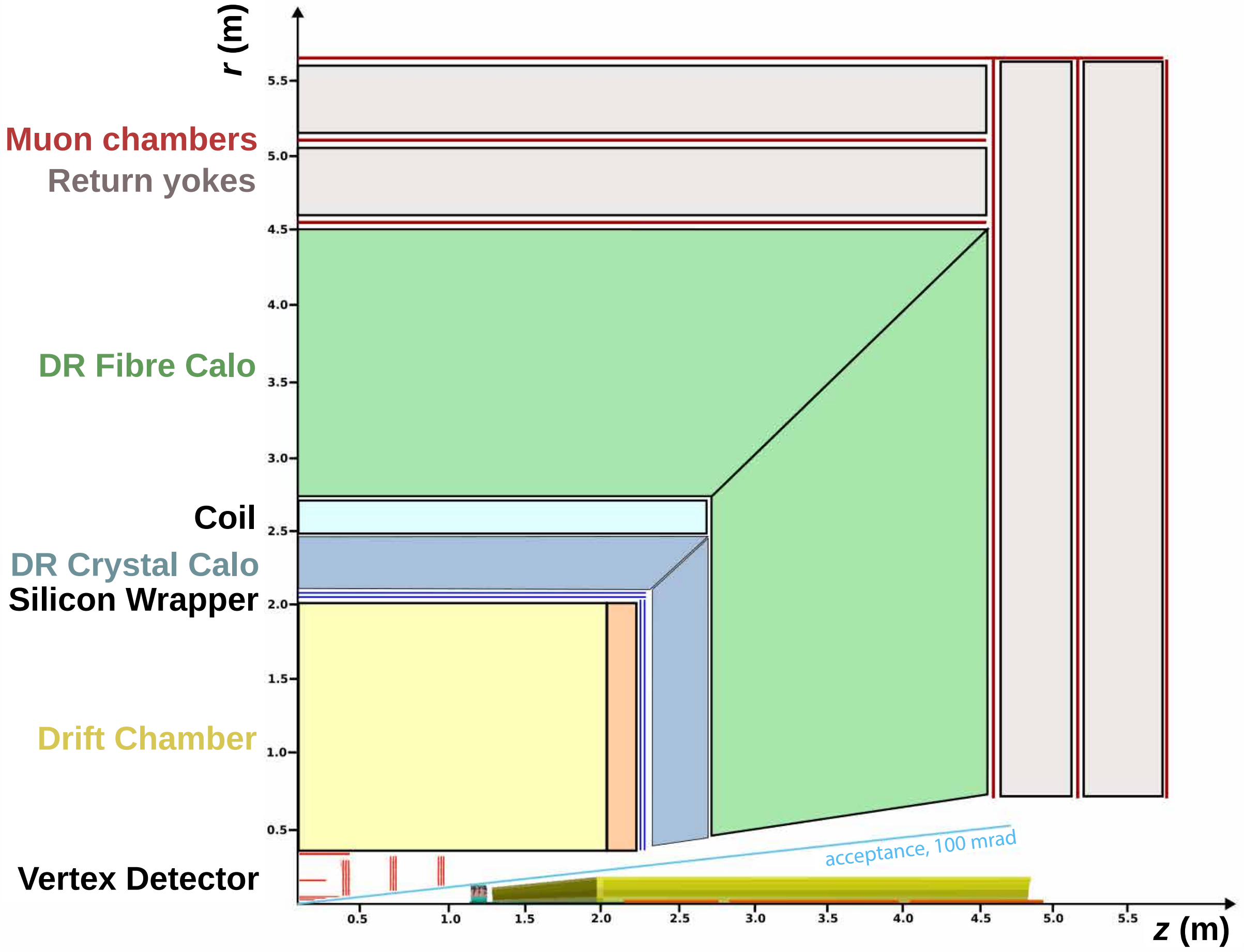

**Figure 19.2:** Overview of the IDEA detector layout.

## 19.3 The sub-detectors

More details on the detector elements of IDEA are given in the following.

### 19.3.1 Vertex Detector

The IDEA silicon vertex detector features two main subsystems, whose active elements are based on 50 µm MAPS:

- an inner vertex detector, located close to the beam pipe, at radii between 13.7 and 35.6 mm, covering an angular acceptance of about $|\cos(\theta)| < 0.99$;
- an outer vertex detector, located at larger radii between 130 and 315 mm, composed of a barrel section and forward disks.

Two alternative layouts are being explored for the inner vertex detector. The first layout uses a traditional approach, in which modules are mounted on a carbon fiber support [4], with overlapping structures to allow full coverage and alignment. The second layout, more advanced but still under development, is based on curved detectors, using a concept similar to the ALICE ITS3 [5].

The baseline vertex detector layout is illustrated in Figure 19.3.

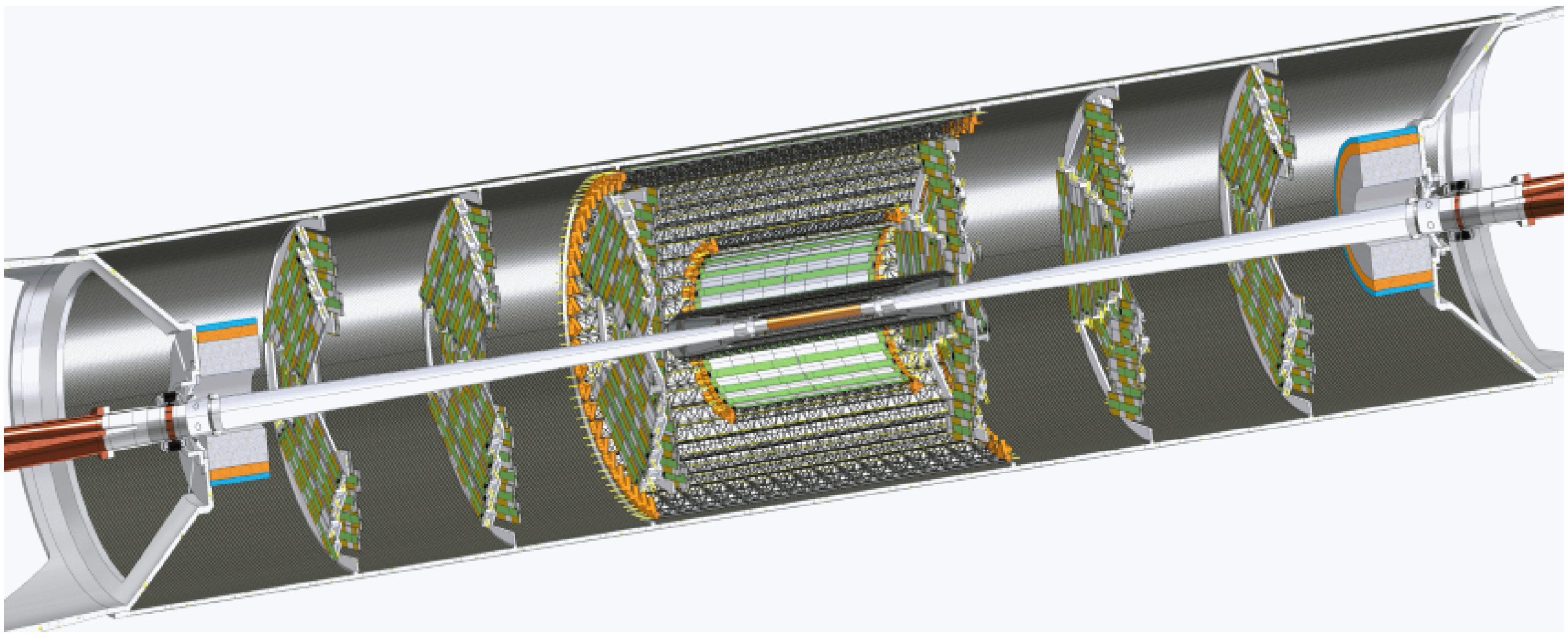

**Figure 19.3:** Vertex detector layout

The inner vertex detector is composed of three concentric barrel layers mounted on carbon fiber support structures. The elementary unit is a module of dimensions 32 $(z) \times 8.4$ $(r\text{–}\phi)$ mm$^2$. Each module has two chips abutted in $z$ inspired by the ARCADIA INFN R&D program [6]. The active area is made of 640 $(z) \times 256$ $(r\text{–}\phi)$ pixels of 25 μm × 25 μm size. The power consumption of the current ARCADIA prototype, read out at 100 MHz/cm$^2$, is measured to be about 30 mW/cm$^2$, thus allowing air cooling.

The outer vertex detector is composed of two barrel layers and three disks on either side of the IP. The elementary unit is a module of dimensions $40.6(z) \times 42.2(r - \phi)$ mm$^2$. Each module has four HV-CMOS monolithic pixel sensors inspired by the ATLASpix3 design [7]. The active area of the chip consists of 132 columns of 372 pixels each, with square pixels of 150 μm × 50 μm. The sensor thickness of the modules is 50 μm and their power consumption is assumed to be 100 mW/cm$^2$, half of the level observed at the current stage of development, but still too high (by at least a factor of two) to be handled by air cooling.

The same module type is used for the barrel layers and the disks. One barrel layer is placed at 13 cm radius (middle barrel) and is made of 22 ladders, each with 8 modules, The outermost layer (outer barrel) is placed at 31.5 cm and is composed of 51 ladders of 16 modules each. The outer layer is supported by a flange that is attached to the external support tube. The flange also supports the middle barrel as well as the first disks. Three disks per side, located at z=±29.25, ±62 and ± 93 cm, complete the outer vertex detector. The inner disk is located inside the barrel.

### 19.3.2 Drift Chamber

The drift chamber for the IDEA detector concept is designed to provide precise tracking, high-precision momentum measurement and excellent particle identification by adopting the "cluster counting" technique [8]. The main peculiarity of this drift chamber is its high transparency, which is a crucial feature for the charged particle momentum range, from several tens of GeV/c to a few hundred of MeV/c, where the Multiple Scattering (MS) contribution is far from being negligible, but also to limit $\gamma$ conversions and hadronic interactions. This high transparency is obtained mainly thanks to a novel approach adopted for the wiring and assembly procedures [9]. The total amount of material in terms of radiation lengths is about 1.6% $X_0$ in the radial direction towards the barrel calorimeter, mostly due to the outer wall. In the forward direction it is about 5.0% $X_0$, with 75% of this located in the end plates, which are instrumented with front-end electronics.

The chamber design inherits some aspects of drift chambers previously built and operated, like the ones of the KLOE experiment [10] and the MEG2 experiment [11]. However it also presents fully innovative features, aimed to combine granularity and transparency needs.

The IDEA drift chamber is a single-volume, cylindrical wire chamber, co-axial with the 2 T solenoidal field and operated with a very light gas mixture, 90% He–10% $C_4H_{10}$. It extends from an inner radius $R_{in}$ = 350 mm to an outer radius $R_{out}$ = 2000 mm, for a length of $L$ = 4000 mm. Coverage in polar angle extends down to ~13°. The chamber is a full-stereo wire device, and it consists of 14 co-axial super-layers with 8 layers each, for a total of 112 layers, strung at alternating sign stereo angles ranging from 50 to 250 mrad. Wires are arranged azimuthally in 24 identical angular sectors, 15° wide. The cells are approximately square, since this geometry ensures a more uniform cell distribution inside the chamber volume for a full stereo configuration. The cell size ranges between 12.0 and 14.5 mm, thus implying a maximum drift time of about 400 ns, assuming a drift velocity of 2.2 cm/μm in a typical 90% He–10% $C_4H_{10}$ gas mixture [10]. The total number of drift cells is 56,448, and the field to sense wires ratio is 5:1.

The mean number of ionization clusters per cm generated by a minimum ionizing particle in the drift chamber gas mixture is about 12.5 at atmospheric pressure. By counting the number of ionization events per unit length ($\mathrm{d}N_p/\mathrm{d}x$) [8] particles can be separated and identified with superior resolution compared to the conventional method based on ionization loss per unit length ($\mathrm{d}E/\mathrm{d}x$). Since in Helium-based gas mixtures the signals generated by ionization events and collected at the anode are separated in time from a few nanoseconds to a few tens of nanoseconds (depending on the drift distance from the anode), fast read-out electronics (~ GHz sampling) and sophisticated peak-finding algorithms are required to efficiently identify the ionization clusters by acquiring and analyzing the chamber signal waveforms, an example of which is shown in Figure 19.4.

Assuming an empirical parametrization of the $\mathrm{d}E/\mathrm{d}x$ resolution obtained with the truncated mean technique [12] and a Poisson distribution for the number of the ionization clusters in a 90% He – 10% $iC_4H_{10}$ gas mixture at atmospheric pressure, the resolution expected with the cluster counting method is about 2 times better than the usual $\mathrm{d}E/\mathrm{d}x$ method. Analytical calculations assuming 2 m long tracks predict excellent $K/\pi$ separation for momenta up to ~30 GeV/c except in the $0.85 < p < 1.05$ GeV/c range, where separation could be recovered by a timing layer with a modest 100 ps resolution as shown in Figure 19.5.

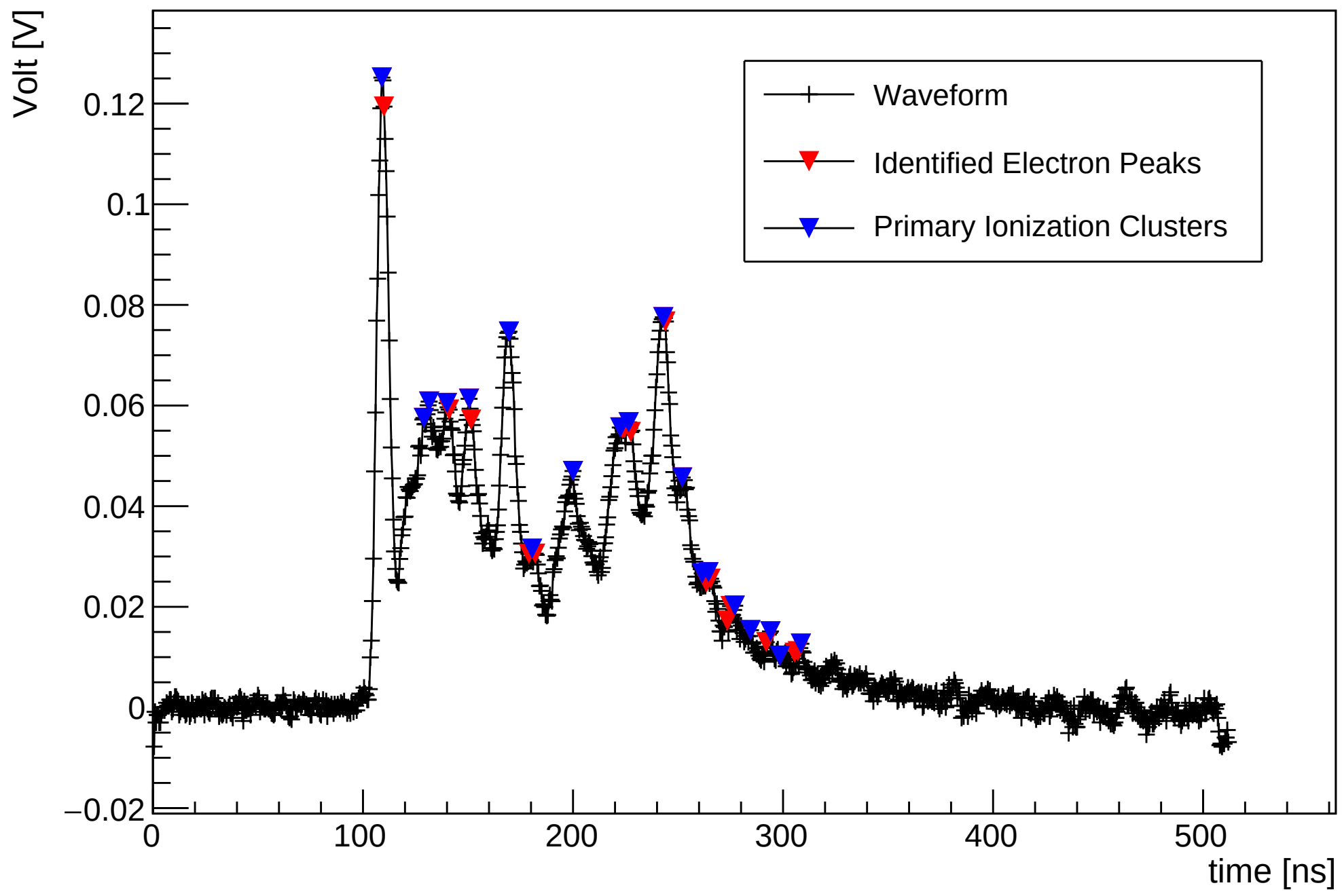

**Figure 19.4:** Typical signal waveform collected in Helium-Isobutane gas mixtures. Ionization clusters (marked in blue) can be counted after a clustering algorithm identifies individual ionization electrons (marked in red) belonging to the same cluster. The peaks of individual electrons are recognized by peak-finding algorithms.

### 19.3.3 Silicon Wrapper

The Silicon Wrapper is the last piece of the IDEA tracking system. It surrounds the drift chamber ($|z| \leq 2.25$ m and $r < 2$ m), covering an area of more than 100 $m^2$.

**Table 19.1:** Main parameters of the current IDEA Silicon Wrapper layout, assuming ATLASpix3-sized sensors ($42.2 \times 40.2$ $cm^2$). The numbers for the disks are per side.

| Layer | Radius [m] | $z$ [m] | Number of sensors | Silicon area [$m^2$] | Coverage |
|---|---|---|---|---|---|
| Barrel 1 | 2.04 | $\|z\| < 2.4$ | $O(20,000)$ | $\sim 32$ | $\|\cos(\theta)\| < 0.762$ |
| Barrel 2 | 2.08 | $\|z\| < 2.4$ | $O(20,000)$ | $\sim 32$ | $\|\cos(\theta)\| < 0.756$ |
| Disk 1 | 0.35 – 2.04 | ±2.30 | $O(7,500)$ | $\sim 12 \times 2$ | $0.748 < \|\cos(\theta)\| < 0.989$ |
| Disk 2 | 0.35 – 2.04 | ±2.32 | $O(7,500)$ | $\sim 12 \times 2$ | $0.751 < \|\cos(\theta)\| < 0.989$ |
| **Total** | | | $O(70,000)$ | $\sim 112$ | |

The sensor technology for the Silicon Wrapper has not been fixed yet. Candidates are, for example, two layers of silicon microstrip detectors or one layer of MAPS or LGADs. In case the LGAD option is chosen, the goal would be to reach a time resolution per track of $\sim 100$ ps for time-of-flight particle identification which, together with the cluster counting from the drift chamber, ensures a $K$-$\pi$ separation power of more than three sigma up to momenta of ~30 GeV/c. The time-of-flight measurement will also be crucial for long-lived particle searches at the FCC-ee [13]. The other options do not provide timing within the tracking system; in this case timing would be supplied by a LYSO layer in front of the EM calorimeter.

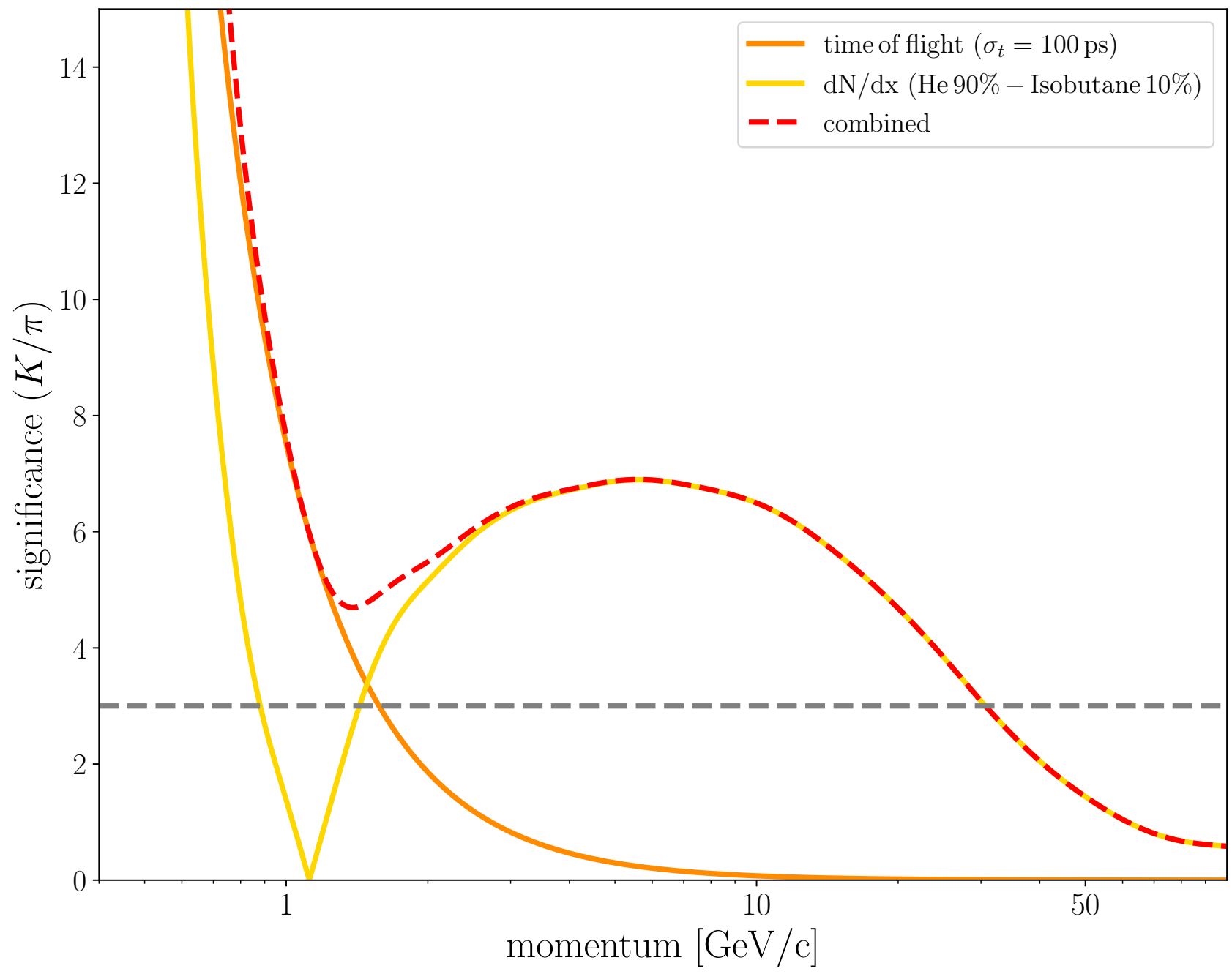

**Figure 19.5:** $K/\pi$ separation significance (in standard deviations) of the IDEA drift chamber, from a fast DELPHES simulation study, with cluster counting and using a drift chamber performance parametrised from GARFIELD++ results. A better than 3 $\sigma$ separation is obtained up to momenta ~ 30 GeV. The $p < 1.6$ GeV region is covered by a time-of-flight system, extending over a distance of 2 m and having a timing resolution assumed to be 100 ps.

The requirement on the spatial resolution of the Silicon Wrapper is $\mathcal{O}(10\,\mu\text{m})$. This enables enables a precise momentum measurement, especially for high-momenta particles and in the forward region where the drift chamber only provides a limited number of hits. The reasoning behind this small spatial resolution in $\theta$ is to allow the Silicon Wrapper to act as a precise and stable ruler for the detector acceptance definition ($< \mathcal{O}(10\,\mu\text{rad})$, which is critical for cross-section measurements. The transverse momentum resolution with and without the silicon wrapper in the central region is shown in Figure19.6.

The current implementation of the Silicon Wrapper is based on MAPS and consists of two barrel layers and two disks per side that overlap at their interface. Table 19.1 lists the properties of the current layout. The barrel layers (disks) are made up of staves (modules) with sensors on both sides, flex circuits on top, and a support and cooling structure in the middle. The sensors have a thickness of 50 µm. Each stave/module has a 1.4 mm carbon fiber support, the same cooling pipes, and four readout flex circuits as for the outer vertex detector described in Section 19.3.1. The stave/module is inspired by the ATLASpix3 tile proposed for the CEPC tracker [14, 15], and the global layout of the disks follows the one of the CMS endcap timing layer [16]. The current implementation of the Silicon Wrapper features a material budget of $\approx 0.7\%$ of $X_0$ per barrel layer/disk.

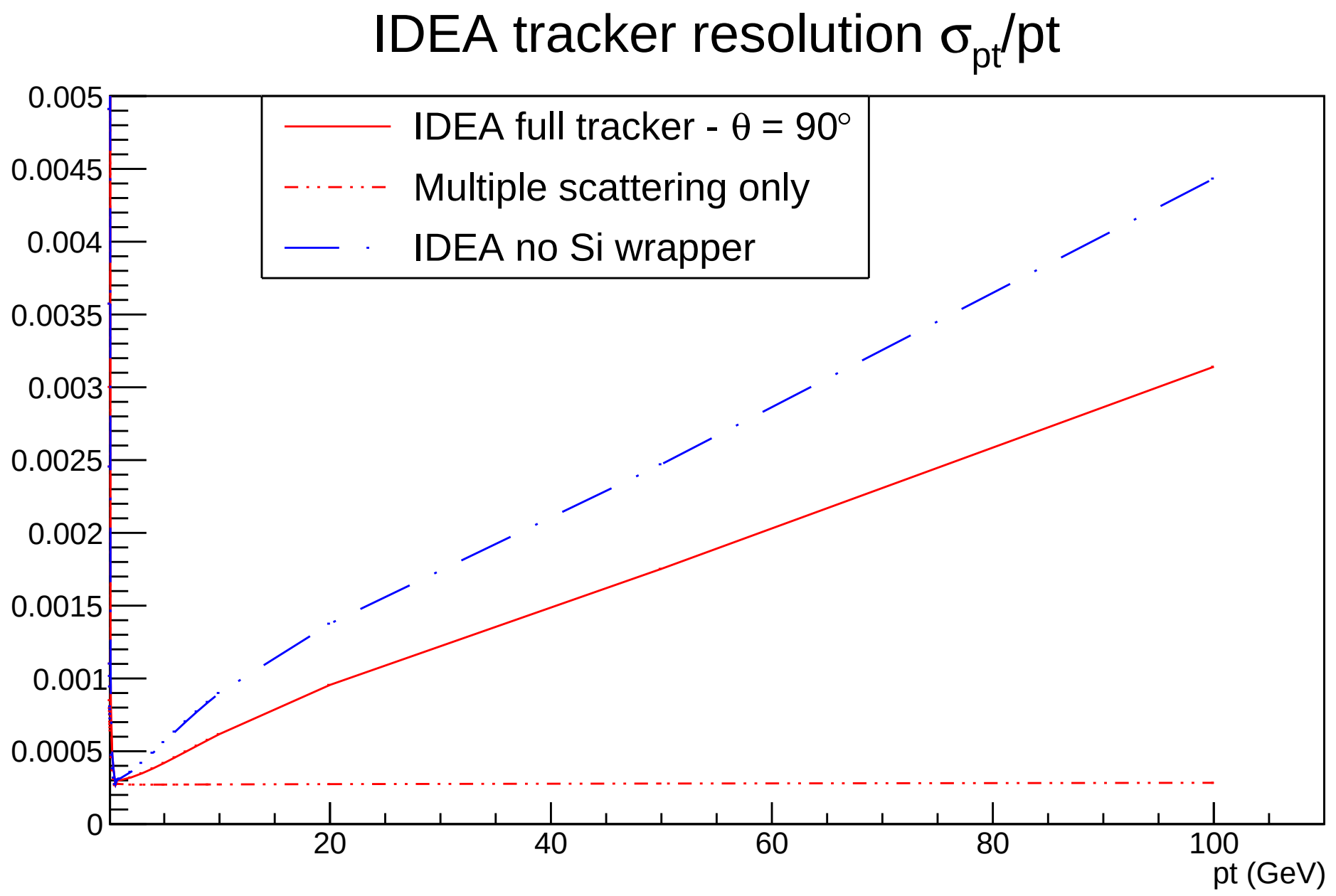

**Figure 19.6:** Transverse momentum resolution as a function of transverse momentum for the IDEA tracking system (Delphes). The (red) continuous line is the resolution of the baseline detector, while the (blue) dashed line shows the resolution without the silicon wrapper. The (red) dotted horizontal line shows the multiple scattering contribution.

### 19.3.4 EM Calorimeter

The IDEA Dual-Readout Crystal calorimeter section is designed to achieve an electromagnetic energy resolution better than $3\%/\sqrt{E} \oplus 1\%$ exploiting the use of high density inorganic scintillators such as PWO or BG(S)O. The crystals are read out using Silicon Photomultipliers combined with optical filters to enable the simultaneous readout of scintillation and Cherenkov photons which, integrated with the DR fiber hadronic calorimeter section, is a feature required to achieve a hadronic energy resolution of about $30\%/\sqrt{E} \oplus 3\%$ [17, 18].

Another key feature compared to state-of-the-art homogeneous electromagnetic calorimeters is the increased transverse granularity and longitudinal segmentation which constitute powerful handles for particle identification and global event reconstruction based on the particle flow approach. The calorimeter can provide a time resolution at the level of 30 ps for electromagnetic showers with energy above 30 GeV. An additional layer consisting of a 6 mm thick LYSO crystal grid can be added in front of the calorimeter to provide cost-effective time resolution at the level of 20 ps for all charged particles, as described in [17], and exploiting a technology similar to that of the CMS MTD barrel [19, 20, 21].

The crystal calorimeter is located between the tracking system and the solenoid with a radial envelope of 2150-2500 mm, and is divided into a central barrel section and two endcaps. The $\phi$ profile of the barrel is segmented as a regular polyhedron by a fixed number of global $\phi$ segments to allow for flat surfaces upon which a precision timing layer may later be instrumented. All crystal towers are initially oriented as projective trapezoids with respect to the interaction point.

Each crystal tower has nominal front face dimensions of $10 \times 10$ mm$^2$ and is composed of two longitudinal segments of 6 $X_0$ (front, E1) and 16 $X_0$ (rear, E2) for a total depth of 22 $X_0$.

The barrel has a $z$-length of 4.8 m and the center point of the crystal front faces are located at a radius of 2250 mm. The endcaps connect continuously with the barrel and are therefore also segmented in $\phi$ with an outermost rear crystal radius of $\approx$ 2386 mm and a polar angle of 8.5 degrees designed to coincide with the Silicon Wrapper endcap disks, corresponding to an inner radius of $\approx$ 360 mm.

### 19.3.5 Solenoid

The original IDEA particle detector concept magnet is evolving from a 2 T solenoidal superconducting magnet, based on the reliable and well-established technology of aluminum stabilized NbTi Rutherford cables, to an HTS solenoidal magnet capable to reach up to 3 T.

The recent conceptual design, implementing two separate dual readout calorimeters (a crystal calorimeter for the electromagnetic shower and a fiber calorimeter for the hadronic shower) increases the internal radius of the superconducting coil, now placed between the two calorimeters, see Figure 19.2, and reduces the constraint on the solenoid transparency. In this context, a new preliminary design for a HTS solenoid operating at T > 20 K is being developed by the INFN LASA [22] superconducting magnet group.

The electromagnetic design of the HTS solenoid is optimized to produce a 3 T central field leading to improved particle tracking resolution of the detector at particle collision energies above the $Z$ resonance while being able to provide a 2 T solenoidal magnetic field within the requirements on field quality and superconducting coil stability. Using the equation for the magnetic field produced by a finite length solenoid with L = 5.3 m, $R_{in}$ = 2.5 m, to generate a magnetic field value of $B_z$ = 3 T, a total of NI = 17.4 MAturns or, equivalently, an average engineering current density of $J_{eng}$ = 37 A/mm$^2$ ($J_{eng}$ = 25 A/mm$^2$ in case of $B_z$ = 2 T) is required.

A classical approach for the cable design is to use a "U-shaped" channel made of aluminum alloy Al-2.0% Ni for the cable stabilizer offering high mechanical strength (E = 70 [GPa], yield strength = 170 [MPa] ) compared to other aluminum alloys and high thermal stability during a quench event with a still acceptable value of RRR = 170. The number of superconducting 12 mm tapes within the cable has been optimized to maximize the operating margin at the peak field on the coil volume. A maximum of $I/I_c$ = 51.5% at the edge of the solenoid has been obtained from the electromagnetic optimization of the solenoid dimension with most of the coil volume operating below 25% of the critical current ($I_c$) of the superconducting tape,

The total thickness of the cold mass equals 116.7 mm and the normalized radiation lengths for each material are reported in Table 19.2. The total thickness of the cold mass, equivalent to $X/X_0$= 1.492, is within the superconducting coil design constraints and does not affect the detector energy resolution. Preliminary mechanical structure and cryostat dimensioning evaluations are under development at LASA where research and development activities will be carried out to enhance the technological readiness level (TRL) of HTS superconducting magnets supporting sustainable solutions for future collider projects.

**Table 19.2:** Coil Transparency

| Material | Thickness (mm) | $X/X_0$ |
|---|---|---|
| HTS tape | 4.0 | 0.296 |
| Al2.0%wtNi | 72 | 0.862 |
| Insulation | 16 | 0.056 |
| Shell | 25 | 0.278 |
| TOTAL | 117.3 | 1.492 |

### 19.3.6 Hadronic Calorimeter

The proposed hadronic section for the IDEA calorimeter is based on a fiber-sampling dual-readout calorimeter [23], consisting of alternate rows of scintillating and clear (Cherenkov) fibers inserted in an iron capillary tube, that serves both as an absorber and to capture part of the magnetic field flux.

The calorimeter is divided in a barrel and two endcaps. In the baseline design, the transition from one to the other is assumed to occur at a 45° angle with respect to the beam axis. The inner radius (lowest z coordinate) of the barrel (endcap) is 2.8 m, and the tower length is 1.8 m. The tube outer radius is 1 mm and the fiber radius is 0.5 mm. In total, around 80 million tubes are placed in this geometry, roughly divided into 60 million for the barrel and 20 million for the two endcaps.

The hadronic energy resolution was estimated, following a full simulation of the IDEA (standalone) fiber-sampling calorimeter [24], at about $30\%/\sqrt{E}$, with a small constant term. The expected achievable separation for the $W/Z/H \to jj$ peaks is shown in Figure 19.7 for an ideal fiber-sampling calorimeter. Simulation studies show that this detector can also provide excellent standalone particle-identification capabilities [25].

Additional studies integrating the crystal EM calorimeter and using a dedicated particle flow method described in [18] confirm this expected performance on hadronic jets as shown in Figure 19.8.

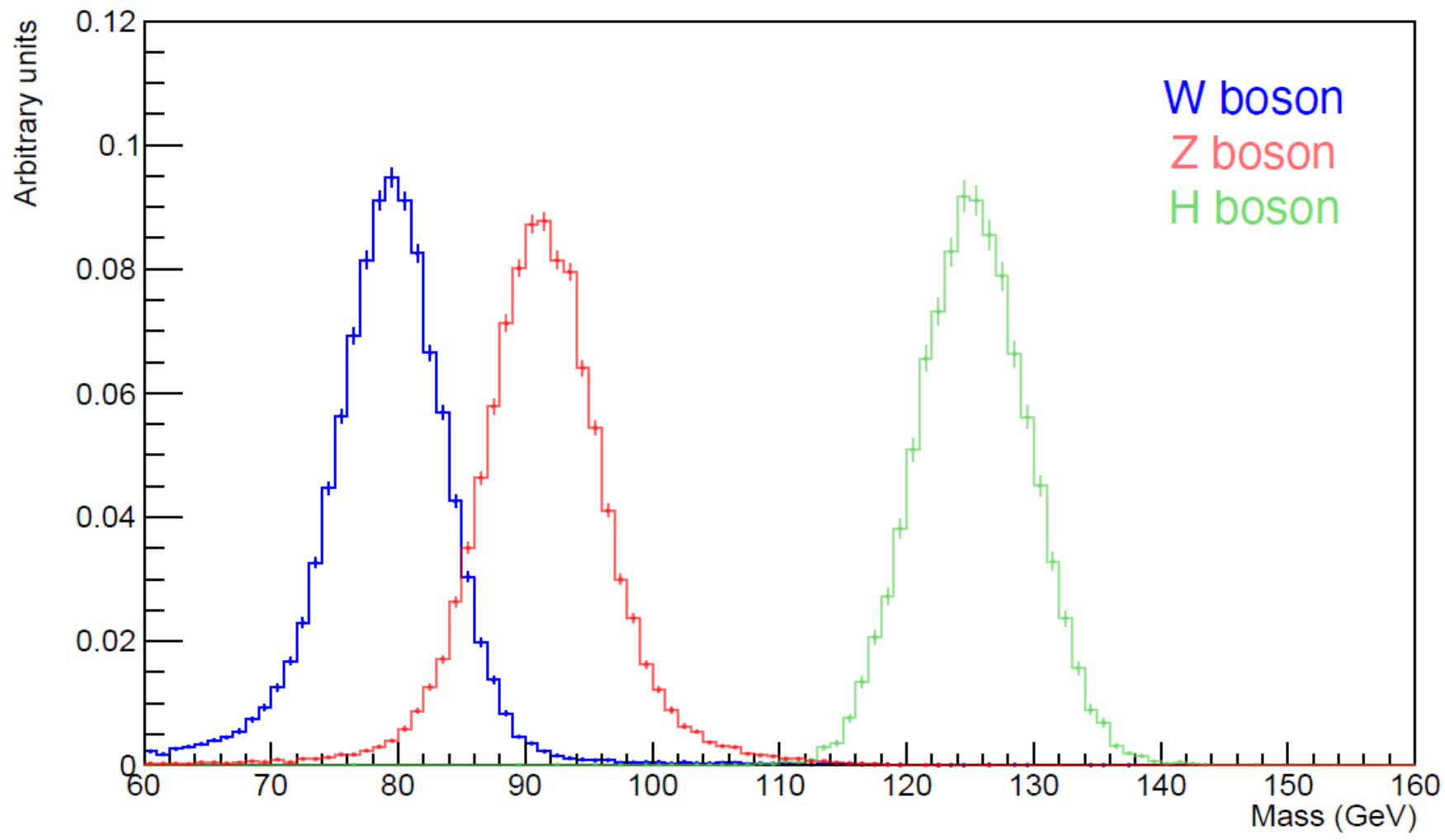

**Figure 19.7:** Reconstructed invariant mass distributions for $W/Z/H \to jj$ fully-simulated events in a highly granular DR fiber-sampling calorimeter [24].

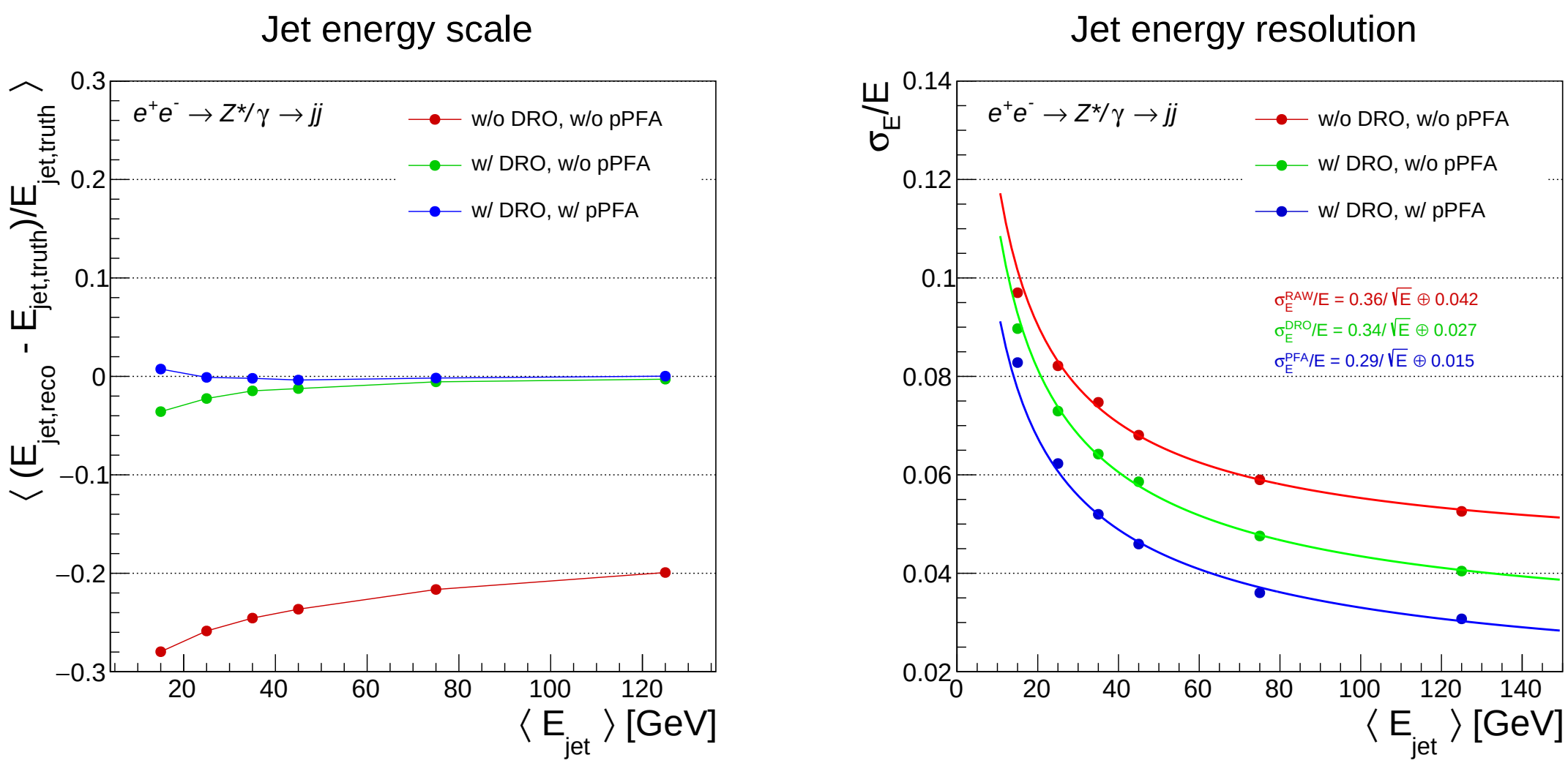

**(a)** **(b)**

**Figure 19.8:** Hadronic jet energy (a) linearity and (b) resolution as measured with the scintillation signal only (red), the dual-readout corrected calorimeter signal (green), and the dual-readout plus particle-flow corrected estimation of the jet energy (blue). Figures from Ref [18].

## 19.3.7 Muon System

The Muon detection system consists of three layers of detectors covering the barrel and endcap regions, housed within the iron yoke that encloses the solenoidal magnetic field. This detector is thought to allow detection of long-lived particles decaying beyond the tracker, besides tagging muons originating from the tracking region. Simulation studies indicate that a spacial resolution of a few hundred microns per point is required after accounting for multiple scattering effects. This resolution can be achieved with $\mu$-RWELL detectors [26], innovative, single-stage and compact gaseous detectors belonging to the MPGD family. To take advantage of the industrial production capabilities of this technology, a modular design has been adopted for the muon detection layers. Each basic $\mu$-RWELL tile features an active area of 50×50 cm$^2$ with a two dimensional strip readout. A strip pitch of approximately 0.4÷1.5 mm provides a typical spatial resolution in the range of 100÷500 µm, which corresponds to 2,500÷640 readout channels per tile.

The choice of detector tile size, strip pitch, and strip width represents a compromise among several factors: the largest $\mu$-RWELL detector that can be industrially mass-produced, the maximum input detector capacitance tolerable by the FEE to maintain an adequate signal-to-noise ratio, the spatial resolution required by the IDEA experiment, and the costs of electronics for the readout channels, which need to remain within reasonable budget constraints.

# Glossary

**$\mu$-RWELL** Micro-Resistive WELL 645, 655

**$\mu$ID** muon identification 313, 314, 333–337, 343

**3DGAN** 3D Generative-Adversarial Network 492

**AC** alternating current 131, 132, 143, 144, 152–154, 227

**AC-LGAD** AC-coupled Low Gain Avalanche Detector 16, 26, 29, 32, 61, 111–113, 135, 136, 152–158, 163–165, 482, 559, 609, 621, 631

**ACTS** ACTS Common Tracking Software 481, 490, 491, 562

**ADC** Analog-to-Digital Converter 186, 187, 193, 227, 288, 289, 293, 294, 305, 306, 327, 333, 479

**AERD** Address-Encoder and Reset-Decoder 96, 97

**AFE** Analog Front-End 186, 192, 193

**AHCAL** Analog HCAL 302, 304–307

**AI** Artificial Intelligence 441, 462, 463

**ALDD** Array Laser Diode Driver 388, 395–397

**ALPIDE** ALICE Pixel Detector 80, 83, 96–98

**APEP** Associated Proton Beam Experiment Platform 283

**AR** Augmented Reality 488

**aSB** Adiabatic Simulated Bifurcation 495

**ASIC** Application Specific Integrated Circuit 113, 119, 120, 134, 136–139, 141–146, 149, 158, 159, 162–164, 177, 182, 183, 186–193, 203, 218, 221, 225–227, 231, 235, 246, 252, 263, 270–273, 296, 303, 304, 319, 320, 385–388, 392, 395, 397, 410, 418, 420, 421

**ASICs** Application-Specific Integrated Circuits 188, 190, 191, 193, 432, 553

**ASTC** Aluminum stabilized Stacked REBCO Tape Conductor 377–379

**ASU** Active Sensor Unit 303

**ATCA** Advanced Telecommunications Computing Architecture 415, 417, 620

**ATIA** Array Transimpedance Amplifier 388, 395–397

**ATLAS** A Toroidal LHC ApparatuS 342, 343, 385, 419, 440, 469, 488

**AWG** American Wire Gauge 416

**BDC** Bidirectional Control 391

**BDT** Boosted Decision Trees 484

**BDTG** Boosted Decision Trees with Gradient boosting 584

**BEE** Back-End Electronics 87, 116, 119, 120, 143, 186, 383–388, 405, 412–414, 418, 421, 428–430, 433–438, 452, 453, 458, 459, 464

**Belle II** Belle II experiment 385, 435, 438, 442, 448, 488

**BEPCII** Beijing Electron-Positron Collider 46, 47, 605, 606

**BER** Bit Error Rate 415

**BESIII** Beijing Spectrometer 37, 38, 46, 47, 50, 385, 418, 440, 448, 469, 488, 514

**BGB** Beam-Gas Bremsstrahlung 43, 49

**BGC** Beam-Gas Coulomb scattering 43, 49

**BGO** Bismuth Germanate 16, 30, 32, 217, 218, 221, 222, 224, 225, 228, 236, 242, 244, 245, 247–252, 256, 271, 280–282, 479

**BGS** Beam-Gas Scattering 43, 45–47

**BIB** Beam Induced Background 35, 36, 42–48, 50–52, 236, 253–255, 257–259, 575, 581

**BIB-AE** Bounded Information Bottleneck Autoencoder 492

**BMR** Boson Mass Resolution 23, 24, 30, 241, 296, 301, 307, 570

**BNL** Brookhaven National Laboratory 152

**BPM** Beam Position Monitor 36, 37, 49, 50, 53, 57, 63, 64

**BPMs** Beam Position Monitors 533

**BR** Branching Ratio 585, 590

**bSB** ballistic Simulated Bifurcation 495, 496

**BSC** Beam Stay Clear 36

**BSM** Beyond the Standard Model 3, 5, 7, 8, 10, 21, 25, 32, 585, 588

**BSO** Bismuth Silicate 218, 224, 244, 619

**BTH** Beam Thermal photon scattering 43, 45, 49

**BX** Bunch Crossing 80, 197

**CALICE** Calorimeter for the Linear Collider Experiment 279, 285, 302, 304, 306

**CaloFlow** A fast detector simulation framework based on normalizing flows 492, 493

**CC** Concentrator Card 145, 146

**CD** Continuous Deployment 474

**CDR** Clock Data Recovery 389, 392, 393

**CDR** Conceptual Design Report 4, 14, 304, 316, 513, 583, 585, 617

**CDS** Correlated Double Sampling 186

**CEPC** Circular Electron Positron Collider xxv, xxxvii, 3, 6–8, 10–12, 14, 16, 21, 22, 24–26, 28, 32, 35, 37, 38, 42, 46, 69, 70, 96, 98, 111, 112, 129–131, 158, 164, 168, 169, 171, 175, 177, 186–189, 195–198, 200–203, 208, 212, 213, 222, 253, 267, 269, 277, 278, 297, 302, 304, 307, 308, 313, 327, 345, 362, 377, 426, 469, 470, 473, 475–477, 483, 486, 489, 490, 498, 499, 501, 503, 506, 513–515, 556, 559, 570, 574, 575, 577, 578, 583, 590, 591, 593–606, 615, 617, 618

**CEPCSW** CEPC Software 46, 50, 95, 104, 105, 196, 202, 208, 214, 219, 242–244, 252, 260, 264, 296, 301, 330, 332, 337, 443, 469–474, 476, 480, 485, 491, 578, 598

**CERN** European Organization for Nuclear Research 3, 213, 247, 251, 279, 295, 302–306, 308, 388, 419, 440

**CFRP** Carbon Fiber Reinforced Polymer 74, 88, 89, 228–233, 263

**CI** Continuous Integration 473, 474

**CIS** CMOS Image Sensor 129

**CKF** Combinatorial Kalman Filter 491

**CL** Confidence Level 583, 588, 590, 591

**CLIC** Compact Linear Collider 470, 601, 602

**CML** Current Mode Logic 391

**CMM** Coordinate Measuring Machine 165

**CMOS** Complementary Metal Oxide Semiconductor 79, 96, 100, 101, 112, 114, 129, 130, 132, 134, 135, 188, 192, 394

**CMS** Compact Muon Solenoid 284, 285, 341, 385, 419, 440, 651

**COB** Chip-On-Board 397

**COG** Center Of Gravity 206, 207

**COTS** Commercial Off-The-Shelf 407, 408, 415

**CP** Charge Conjugation And Parity 598–600

**CPC** Clock Phase Control 389, 393

**CPU** Central Processing Unit 438–440, 442, 453, 455

**CRX** Clock Receiver Circuit 394

**CSA** Charge-Sensitive Preamplifier 132, 186

**CSNS** China Spallation Neutron Source 295, 308

**CTA** CERN Tape Archive 505

**CVD** Chemical Vapor Deposition 55

**CVMFS** CernVM File System 473, 474, 505

**CVTX** Curved Vertex Layer 70–74, 77, 88, 90, 92, 95, 104

**CyberPFA** Crystal bar ECAL reconstruction PFA 223, 224, 483–486, 559

**DAC** Digital-to-Analog Converter 100, 159, 160, 162, 227

**DAQ** Data Acquisition 186, 197, 292, 293, 412, 414, 416, 425, 428, 430–432, 435, 439, 451–456, 478, 496, 553

**DC** Direct Current 156, 605

**DC** Drift Chamber 211, 212

**DC-DC** Direct Current-Direct Current 114–116, 118–122, 136–138, 141–146, 149, 164

**DCH** Drift CHamber 609

**DCI** Distributed Computing Infrastructure 469

**DCOC** DC Offset Cancellation 395

**Dcols** Double Columns 83–85, 98

**DCR** Dark Count Rate 225, 226, 245, 248, 258, 260, 285, 286, 320, 327–329

**DCS** Detector Control System 412, 414, 430, 458, 461–463

**DCTD** Data Concentrate and Timing Distribution 436, 438

**DCU** Deep Computing Unit, a high-performance computing accelerator card 504

**DD4hep** Detector Description for high energy physics, a software tool that provides a complete solution for detector description 243, 332, 469, 471, 472, 474, 476, 489

**DDG4** A tool for simulating particle interactions in detectors, built within the DD4hep framework and working alongside Geant4 472, 474

**DDMTD** Digital Dual Mixer Time Difference 406, 407

**DDR** Dynamic Random Access 413

**DEMUX** Demultiplexer 394

**DESY** Deutsches Elektronen-Synchrotron 98, 99, 107, 213, 252

**DHCAL** Digital HCAL 302, 307

**DIRAC** Distributed Infrastructure with Remote Agent Control, interware for distributed computing platform 500–502

**DLL** Delay-Locked Loop 160, 394, 395

**DMS** Data Management System 502, 503

**DPC** Data Phase Control 394, 395

**DPHEP** Data Preservation in HEP 503

**DQM** Data Quality Monitor 320

**DR** Dual Readout 645

**DRD** CERN Detector R&D Collaboration 280

**DRD1** R&D Collaboration on gaseous detector 213

**DRD7** R&D Collaboration on electronics and on-detector processing 419

**DRX** Data Receiver Circuit 394

**DSB** Barium Di-Silicate BaO-2Si$O_2$ 281

**DSCB** Double-Sided Crystal Ball 569, 575, 576, 579, 584

**DSP** Digital Signal Processors 186

**DSS** Detector Safety System 460

**DUT** Device Under Test 99, 155

**DWDM** Dense Wavelength Division Multiplexing 401, 404, 405

**DYB** Dayabay Reactor Neutrino Experiment 469

**ECAL** Electromagnetic Calorimeter 4, 16, 24–26, 30–32, 166, 181, 184, 217–234, 236–249, 252–254, 256–258, 260–263, 271, 274–276, 279, 285, 290, 291, 298, 300, 384, 386, 418–420, 427–430, 433, 436–438, 443–445, 453, 457, 458, 479, 483–486, 492, 493, 513, 514, 516, 525–530, 534, 536, 537, 541–545, 547, 548, 553, 562, 585, 598, 607, 608, 619, 621

**ECS** Experiment Control System 425, 430, 432, 461–464

**EDM4hep** Event Data Model for high energy physics 469–474, 485–489, 491, 492

**EDR** Engineering Design Report 605

**EIC** Electron Ion Collider 175

**EM** Electromagnetic 217, 221, 222, 239, 243, 246, 252, 253, 257, 262, 263, 564, 584, 603

**EMI** Electro-Magnetic Interference 411

**ENC** Equivalent Noise Charge 187, 193

**EOS** EOS Open Storage provides a service for storing large amounts of physics data and user files, with a focus on interactive and batch analysis 505

**EQR** Epitaxial Quenching Resistor 327, 328

**ESR** Enhanced Specular Reflector 247, 252, 288, 305, 306

**EW** Electroweak 603

**FastLumi** Fast Luminosity Monitor 36, 53–55, 87

**FBK** Fondazione Bruno Kessler 152

**FCC** Future Circular Collider 14, 469, 470, 595, 601, 603

**FD** Frequency Detector 393

**FDR** Final Design Review 421

**FE** Front-End 143–146

**FEA** Finite Element Analysis 180, 184, 185, 190, 191, 228–231, 233, 359, 367, 518, 519, 521, 522, 524–528, 530, 534, 544

**FEB** Front-End Electronics Board 319, 320, 322–325, 339, 340

**FEC** Forward Error Correction 391, 392

**FEDA** Front-End Data Aggregator 120, 121, 143–145, 164, 387–390, 420

**FEDI** Front-End Data Interface 120, 121, 143–146, 162, 164, 387–394, 398, 405, 406, 420

**FEE** Front-End Electronics 86, 87, 182–187, 194, 293, 341, 342, 383–388, 390, 405, 406, 408, 409, 412, 413, 416, 428, 432, 433, 452, 458, 459, 655

**FEM** Finite Element Method 355

**FIFO** First In First Out 84

**FPC** Flexible Printed Circuit 28, 74, 75, 86–89, 92, 93, 99, 114–116, 118, 119, 121–124, 127, 137, 138, 141–143, 145, 146, 165

**FPGA** Field-Programmable Gate Array 133, 193, 386, 401, 405, 407, 412–415, 431, 435, 439, 440, 442, 455

**FSO** Free-Space Optical Communication 404

**FSR** Final State Radiation 59, 61–63, 578, 583, 594

**FT** Fast Trigger 430, 433–436

**GAN** Generative Adversarial Network 492, 493

**Gaudi** A software framework designed for high-energy physics experiments, particularly for event data processing 469–471, 474, 476, 477, 480, 485, 490, 506

**Geant4** GEometry ANd Tracking, toolkit for the simulation of the passage of particles through matter 46, 61, 220, 221, 243, 252, 262, 263, 469, 471, 472, 474, 476–479, 490, 492–494, 496, 497, 559

**GEM** Gas Electron Multiplier 191

**GFO** Gadolinium Fluoro-Oxide 280–283, 285, 307

**GNN** Graph Neural Network 440

**GPU** Graphics Processing Unit 438–442, 455, 491, 492, 504

**Grid** A distributed computing model that connects multiple computers across different locations to work together on complex tasks 469, 500–502, 505, 507

**GRPC** Glass Resistive Plate Chambers 303

**GS** Glass Scintillator 268–275, 277–294, 296, 297, 299, 302, 307, 308, 619, 621

**GS-HCAL** Glass Scintillator HCAL 267–280, 282, 284–286, 290–298, 300–302, 307, 308, 619

**GTX** Gigabit transceiver 404, 406, 415

**HBU** HCAL Base Unit 306

**HCAL** Hadronic Calorimeter 4, 16, 24, 25, 30–32, 227–229, 244, 267, 268, 271, 274, 279, 281, 286, 290, 291, 293, 294, 297, 298, 303, 304, 307, 308, 335, 377, 379, 384, 416, 418–420, 427–430, 433, 436–438, 443–445, 453, 457, 458, 479, 484–486, 513, 516, 521–527, 534, 536, 537, 541–545, 547–549, 553, 562, 563, 607, 608, 619, 621

**HEP** High Energy Physics 469, 471, 485, 488, 505, 506

**HFG:Ce** Cerium-Doped Heavy Metal Fluoride Glass 279

**HGCAL** High Granularity Calorimeter 284, 285

**HGTD** High Granularity Timing Detector 152

**HL-LHC** High-Luminosity LHC 440, 579, 581, 586, 588, 600, 602

**HLMC** Huali Microelectronics Corporation 71

**HLT** High Level Trigger 425, 430–432, 438–442, 448, 453, 464

**HOM** High Order Mode 40

**HPDS** Hybrid Photodiodes 285

**HPK** Hamamatsu 152, 285, 286, 288, 289, 305, 327–329

**HSF** High Energy Physics Software Foundation 473, 488

**HTCondor** a software system that creates a High-Throughput Computing (HTC) environment 504

**HTS** High Temperature Superconductor 377–379, 609, 652

**HV** High Voltage 116, 119, 137, 141, 144–146, 179, 183, 184, 319, 320, 341, 343, 458–460

**HV-CMOS** High Voltage Complementary Metal-Oxide-Semiconductor 111, 112, 114, 115, 117, 119, 129, 130, 134, 135

**I/O** Input/Output 121, 145, 189

**I2C** Inter-Integrated Circuit 100, 121, 144, 145, 159, 162

**IAM** Identity and Access Management 502

**IBF** Ion Backflow 176, 188, 194, 195, 200–202, 213

**Ic** Critical Current 349, 350, 379, 380

**ID** Identification 144, 161, 563–565, 577, 586, 607, 608

**IDEA** Innovative Detector for Electron-positron Accelerator xxv, 14, 262, 645–647, 650, 651,

655

**IEEE** Institute of Electrical and Electronics Engineers 401, 406, 407

**IHEP** Institute of High Energy Physics 152, 153, 156, 279, 280, 295, 305, 377, 471, 473, 501, 504–506, 578, 605

**ILC** International Linear Collider 14, 470, 601, 602, 625

**ILD** International Large Detector xxv, 14, 175, 262

**iLGADs** Inverse Low Gain Avalanche Detectors 631

**INFN** National Institute for Nuclear Physics 152

**INL** Integral Non-linearity 193

**IoT** Internet Of Things 505

**IP** Interaction Point 12, 26, 27, 35, 41–43, 45, 49, 50, 53–57, 60, 64, 69, 218, 223, 229, 232, 239, 316, 321, 323, 355, 356, 560, 589, 603

**IR** Interaction Region 35, 36, 41–43, 46, 48–50, 52, 60

**ISR** Initial State Radiation 59, 61–63, 578, 579, 581, 583, 594, 601

**ITK** Inner Silicon Tracker 26–29, 32, 50, 111–129, 135, 146, 148, 151, 165–167, 175, 201, 207, 208, 357, 358, 418, 428–430, 433, 436–438, 449, 450, 453, 458, 513, 514, 516, 517, 531–533, 536, 537, 544–546, 553, 559, 560, 609, 621

**ITS** Inner Tracker System 77, 78, 80, 83, 98

**JAR** Jet Angular Resolution 569, 570

**JER** Jet Energy Resolution 569, 570

**JOI** Jet Origin Identification 572–574, 596, 597

**JTE** Junction Termination Extension 152, 153

**JUNO** Jiangmen Underground Neutrino Observatory 385, 419, 454, 488

**JUNO-TAO** Jiangmen Underground Neutrino Observatory-Taishan Antineutrino Observatory 325

**Key4hep** Turnkey Software for Future Colliders 470–473, 485, 486, 490, 506

**KF** Kalman Filter 491

**L/S** Line/Space 189

**L1** First Level Hardware Trigger 425, 430–439, 442, 444, 448, 449, 452, 453, 464

**LA** Limiting Amplifier 395, 396

**LATRIC** LGAD Timing and Readout Integrated Chip 143, 144, 158–161, 163, 164

**LCG** LHC Computing Grid 471

**LCIO** Linear Collider I/O 471, 472, 487

**LDO** Low Dropout Regulator 100, 101, 409

**LED** Light Emitting Diode 269, 291, 292, 296, 305, 306, 607, 608

**LEP** Large Electron-Positron Collider 6, 24, 263, 593–596, 598, 602, 604

**LGAD** Low Gain Avalanche Detector 143, 144, 152, 153, 155–160, 162–164, 211, 649

**LHAASO** Large High Altitude Air Shower Observatory 325, 385, 418, 454
**LHC** Large Hadron Collider 3, 5, 6, 24, 42, 263, 284, 341, 342, 469, 501, 507, 579, 583, 585, 586, 591, 598–600
**LHCAL** Luminosity Hadronic Calorimeter) 634
**LHCb** Large Hadron Collider beauty 385, 438, 440, 441, 469, 488, 490, 598–600
**LLM** Large Language Model 463, 473, 505, 572
**LLP** Long-Lived Particle 313, 314, 343, 344, 588–591, 619
**LRB** Left-end Readout Block 73, 81, 84–87, 91
**LS** Luminosity Spectrum 601
**LSP** Lightest Supersymmetric Particle 591, 593
**LTS** Low Temperature Superconductor 345, 379
**LumiCal** Luminosity Calorimeter 36–38, 40, 49, 50, 53, 54, 56–58, 60, 61, 63–65, 513, 514, 516, 533, 536, 537
**LV** Low Voltage 116, 119, 121, 137, 141, 144, 145, 183, 184, 319, 322, 325, 339
**LVDS** Low-Voltage Differential Signaling 100, 394
**LYSO** Lutetium Yttrium Oxyorthosilicate 26, 36, 53, 54, 59–63, 87, 218, 651

**MADE** Masked Autoencoder for Density Estimation 493, 494
**MAPS** Monolithic Active Pixel Sensor 70–75, 77, 82, 83, 86, 88–104, 107, 129, 621, 646
**MB** Management Board 319, 320, 324–326, 339, 340
**MC** Monte Carlo 45–48, 203, 204, 251, 253, 475, 487, 492, 497, 498, 501, 502, 569, 577, 578, 596, 597, 600, 603
**MDC** Main Drift Chamber 46, 47
**MDI** Machine-Detector Interface 4, 26, 35, 49, 86, 92, 194, 196, 218, 330, 383, 384, 386, 478, 517, 617, 620, 621
**MEZ** Undetected Momentum In Beam Direction 575, 579
**Micromegas** Micro-Mesh Gaseous Structure 176, 177, 182, 183, 185, 186, 188, 194–196, 205
**MIITs** Multi-Integral of Inductance and Time 362
**MIMOSA** Minimum Ionising MOS Active pixel sensor 97
**MIP** Minimum Ionizing Particle 155, 220, 221, 237, 246–249, 252, 268, 269, 277, 286–293, 296, 305, 332, 606–608
**ML** Machine Learning 212, 433, 439–442, 462, 463
**MLI** Multilayer Insulation 366
**MOSS** MOnolithic Stitched Sensor 80
**MOST** MOnolithic Stitched sensor with Timing 80
**MPGD** Micro Pattern Gaseous Detector 176, 186, 194, 302, 631, 655
**MPI** Message Passing Interface 487
**MPPC** Multi-Pixel Photon Counter 327–330
**MPV** Most Probable Value 158, 243, 287

**MPW** Multi-Project Wafer 129, 130, 134, 135
**MQE** Minimum Quench Energy 364
**MR** Merge Request 473
**MT** Mechanical Transfer 397
**MUX** Multiplexer 394

**NDL** Novel Device Laboratory 285, 286, 288, 289, 294, 327–329
**nEXO** next-generation Enriched Xenon Observatory 419
**NIEL** Non-Ionizing Energy Loss 46, 51, 52, 235, 253, 397
**NLO** Next-To-Leading Order 52, 603
**NLP** Natural Language Processing 463
**NMOS** Negative channel Metal-Oxide-Semiconductor 100, 132, 134, 135
**NN** Neural Network 441, 442
**NNLO** Next-To-Next-To-Leading-Order 603
**nPE** number of photoelectrons 327–331, 333
**NPU** Neural Processing Unit 504
**NVMe SSD** A high-performance solid-state storage device that uses the NVMe protocol to communicate directly via a computer's PCIe bus 505

**OAT** Optical Array Tranceiver 86, 87, 120–122, 143, 145, 146, 164, 387–390, 397, 398, 416, 420
**OPAL** Omni-Purpose Apparatus 475, 604
**OS** Opposite Sign 591
**OTK** Outer Silicon Tracker 26–29, 50, 111–113, 135–153, 158, 163–168, 170, 171, 175, 184, 185, 201, 207–209, 211, 228, 405, 416, 418–420, 428–430, 433, 436–438, 449, 450, 453, 458, 513, 514, 516, 526–528, 530, 531, 536–538, 553, 559–562, 606, 609, 621

**p.e.** Photoelectron 268, 277, 286–290, 293, 296, 305
**PAL** Power-at-Load 87, 120–122, 143–146, 164, 387, 408, 410, 411, 416, 420
**PanDA** Production ANd Distributed Analysis 500–502
**PandoraPFA** Pandora PFA 483
**PAPS** Platform of Advanced Photon Source Technology R&D 605
**ParT** ParticleTransformer 572
**$PbWO_4$** Lead Tungstate 217
**PCB** Printed Circuit Board 118, 136–138, 141, 142, 146, 148, 149, 151, 179, 180, 186, 192, 193, 232, 244, 270–274, 287, 293, 294, 305, 306, 320, 321, 387, 404
**PD** Phase Detector 393
**PD** Photodiode 395, 397
**PDE** Photon Detection Efficiency 225, 226, 242, 249, 278, 285, 288, 305, 328
**PDF** Probability Density Function 584

**PDG** Particle Data Group 597
**PEEK** Polyether Ether Ketone 126, 150
**PFA** Particle Flow Algorithm xxvii, 4, 16, 21, 25, 26, 30–32, 217, 218, 220, 222, 223, 240, 262, 267, 271, 279, 285, 294, 297, 300–304, 307
**PFO** Particle Flow Object 223, 224, 485, 562–564, 566, 575, 582, 586, 588, 594–596, 599
**Phoenix** A web-based event display framework for high-energy physics 488, 489
**PID** Particle Identification 26, 29, 32, 165, 168, 170, 175, 177, 185, 186, 202, 203, 209, 211–213, 439, 440, 482, 559, 570, 572–574, 582, 599, 609
**PLL** Phase-Locked Loop 100, 145, 159, 161, 162, 389, 393
**PMOS** Positive channel Metal-Oxide-Semiconductor 132, 134
**PMT** Photomultiplier Tube 281
**PODIO** Plain Old Data Input Output 471, 485
**POI** Parameters of Interest 597
**PR** Pull Request 473
**PRBS** Pseudo-Random Binary Sequence 161, 394, 415
**PRR** Production Readiness Review 421
**PS** Plastic Scintillator 313–316, 318–321, 324, 325, 327, 328, 330–334, 338, 339, 343, 344, 619, 621
**PS-HCAL** Plastic Scintillator Hadronic Calorimeter 271, 279, 298, 302
**PSU** Plastic Scintillator Unit 314–316, 318, 320–325, 328, 330, 331, 333, 338–340
**PTP** Precision Time Protocol 406, 407
**PUE** Power Usage Effectiveness 506
**PVTX** Planar Vertex Layer 70, 73, 74, 77, 88, 90–92, 103, 104
**PWM** Pulse Width Modulation 410, 411

**QA/QC** Quality Assurance, Quality Control 235, 340
**QAG** Quantum Ansatz Generator 496
**QCD** Quantum ChromoDynamics 5, 9, 10, 23, 598, 600
**QED** Quantum ElectroDynamics 10, 52, 581, 603
**qGAN** quantum Generative Adversarial Network 496
**QSFP+** Quad Small Form-factor Pluggable Plus 414
**QUBO** Quadratic Unconstrained Binary Optimization 495, 496

**R&D** Research and Development xxxvii, 4, 37, 129, 134, 135, 153, 154, 163, 164, 175, 191, 192, 213, 345, 350, 375, 380, 469, 475, 476, 501, 502, 559, 605, 606, 617–621
**Radar** Heterogeneous architecture of data acquisition and processing 453–455
**RBB** Radiative Bhabha Scattering 43, 44, 56
**RDFA** RDataFrame Analysis 486, 487
**RDMA** Remote Direct Memory Access 435, 453
**RF** Radio Frequency 402, 403, 605

**RIA** Radiation Induced Attenuation 397–399
**RMS** Root Mean Square 626
**RoI** Regions Of Interest 431, 432, 438, 440, 493
**ROOT** An software framework developed for high-energy physics data analysis 469, 471, 472, 485–489
**RPB** Right-end Power Block 72, 73, 81
**RPC** Resistive Plate Chamber 302, 303, 316, 341–343
**RPWELL** Resistive Plate WELL 302
**RQS** Rational Quadratic Spline 493, 494
**RRR** Residual Resistivity Ratio 347, 349, 351, 362, 363, 375, 376, 378, 380, 652
**RSD-LGADs** Resistive DC-coupled Silicon Detectors 631
**RSU** Repeated Sensor Unit 70–73, 75, 78, 81, 82, 84–87, 91
**Rucio** Rucio Data Management System 501, 502
**RXs** Receivers 394

**S2D** Single-Ended-To-Differential 394
**SAR** Successive Approximation Register 193
**SB** Simulated Bifurcation 495
**ScW** Scintillator-Tungsten 262, 263
**SDHCAL** Semi-Digital Hadronic CALorimeter 302–304
**SEL** Single Event Latch-Up 188
**SerDes** Serializer/Deserializer 389, 390
**SEU** Single Event Upset 187, 392
**SFP** Small Form-factor Pluggable 404, 414
**Si-W ECAL** Silicon-Tungsten Electromagnetic Calorimeter 306
**SiD** Silicon Detector 14
**SIPAC** SiPM ASIC for Calorimeter 227
**SiPM** Silicon Photomultiplier 30–32, 217–219, 221, 225–227, 234, 235, 238, 239, 242, 245–250, 252, 258, 260, 262, 263, 268–271, 277–280, 284–294, 296, 302, 305–308, 313, 314, 316, 319–324, 327–330, 332, 333, 337, 339, 343, 344, 479, 483, 607, 619, 633
**SiW** Silicon Tungsten 262
**SLC** Stanford Linear Collider 596, 598
**Slurm** A highly scalable cluster management and job scheduling system for large and small Linux clusters 504
**SM** Standard Model 3, 5, 7, 12, 14, 52, 313, 577, 585, 588–592, 596, 600, 603
**SoC** System-On-Chip 401
**SOI** Silicon On Insulator 629
**Spack** A flexible package manager designed for supercomputers, Linux, and macOS 471, 473
**SPI** Serial Peripheral Interface 162, 401

**SR** Signal Region 592

**SR** Synchrotron Radiation 11, 37, 43, 45, 47–51

**SUSY** Supersymmetry 591

**TCAD** Technology Computer Aided Design 132, 153, 163

**TCDS** Trigger Control and Distribution System 433–438

**TCT** Transient Current Technique 155, 157

**TDAQ** Trigger and Data Acquisition 120, 384–386, 405, 412, 417, 425, 426, 429, 430, 435, 461–465, 620

**TDC** Time-To-Digital Converter 134, 158–163, 227

**TDR** Technical Design Report 3, 253, 283, 420–422, 605, 608, 617, 620, 621

**TI-LGADs** Trench-Isolated Low Gain Avalanche Detectors 631

**TIA** Transimpedance Amplifier 395–397

**TID** Total Ionizing Dose 46, 51, 52, 188, 193, 235, 236, 253, 254, 256, 397, 398

**TIE** Time Interval Error 407

**TOA** Time-of-arrival 158–162, 164, 227

**ToF** Time-of-Flight 23, 26, 27, 29, 32, 111, 135, 153, 165, 168, 170, 171, 175, 209, 211, 482, 483, 562, 565, 566, 589, 599

**TOT** Time-over-threshold 158–162, 164

**TPC** Time Projection Chamber 4, 14, 16, 23–30, 32, 50, 111–113, 122, 137, 138, 146, 147, 165, 166, 168, 169, 175–179, 181–189, 191, 192, 194–209, 211, 213, 345, 357, 358, 361, 362, 416, 418–420, 428–431, 433, 436–439, 449, 453, 457, 458, 474, 479–482, 486, 513, 514, 516, 528–532, 536, 537, 544–546, 552–555, 559–562, 565, 606, 609, 618, 621, 630

**TPSCo** Tower Partners Semiconductor Co. 71, 78, 108

**traccc** A GPU-based track reconstruction library developed as part of the ACTS project 481, 491, 492

**TRL** technological readiness level 653

**TSK** Touscheck Scattering 43–47, 49

**TTC** Timing, Trigger, and Clock 405, 406, 415

**TTCD** TTC Distributor 405, 406

**TX** Transmitter 394

**UDC** Unidirectional Control 391

**UL** Upper Limit 588

**UV** Ultraviolet 191, 192, 606

**VAE** Variational Autoencoder 492

**VCDL** Voltage-Controlled Delay Line 394, 395

**VCO** Voltage-Controlled Oscillator 393

**VCSEL** Vertical-Cavity Surface-Emitting Laser 122, 388, 395, 397

**VecMem** A heterogeneous memory management library eveloped as part of the ACTS project 491

**VPI** Vacuum Pressure Impregnation 352–354

**VR** Virtual Reality 488, 489

**VTX** Vertex Detector 25–28, 37, 40, 48, 50, 69–73, 75–80, 84, 86–88, 90–96, 98–100, 102–108, 111, 113, 129, 168, 169, 384, 386, 387, 408, 410, 418–420, 428–430, 433, 436–438, 449, 450, 453, 457, 458, 513, 514, 516, 533, 536, 537, 549, 553, 559, 560

**WBS** Work Breakdown Structure 615

**WLS** Wavelength-Shifting 313, 314, 316, 320–322, 324, 327, 328, 330, 333, 338, 339, 343, 344, 619

**WMS** Workload Management Systems 500–503

**WP** Working Point 562, 563, 586, 589, 590

**XEL** X-Ray Excited Luminescence 280, 282

**XGBoost** eXtreme Gradient Boosting 562–565, 571, 573, 574, 587, 588

**XML** Extensible Markup Language 472

**YFS** Yennie-Frautschi-Suura 59